%% file: main.tex
\documentclass{ucetd}

\usepackage[T1]{fontenc}
\usepackage{subcaption,graphicx}
\usepackage[square, numbers, comma, sort&compress]{natbib} 

\usepackage{mathtools}  
\usepackage{amssymb}    
\usepackage{amsthm}

\usepackage{tikz}
\usepackage[compat=1.1.0]{tikz-feynman}
\tikzfeynmanset{warn luatex=false}

\usepackage{silence}
\usepackage{booktabs}
\usepackage{multirow}

\usepackage{listings}
\usepackage{color}

\definecolor{dkgreen}{rgb}{0,0.6,0}
\definecolor{gray}{rgb}{0.5,0.5,0.5}
\definecolor{mauve}{rgb}{0.58,0,0.82}

\usepackage{lineno}

\usepackage[percent]{overpic}

\usepackage{comment}
\usepackage{hyperref}
\usepackage{float}
\usepackage{xcolor}
\usepackage{makecell}

\newcommand{\thesistitle}{Search for an Anomalous Excess of Single Photons in the MicroBooNE Neutrino Experiment}
\newcommand{\thesisauthor}{Lee Hagaman}
\department{Physics}
\division{Physical Sciences}
\degree{Doctor Of Philosophy}
\date{AUGUST 2025}

\title{\thesistitle}
\author{\thesisauthor}

\dedication{}
\epigraph{}

\usepackage{doi}
\usepackage{xurl}
\hypersetup{bookmarksnumbered,
            unicode,
            linktoc=all,
            pdftitle={\thesistitle},
            pdfauthor={\thesisauthor},
            pdfsubject={},                                
            pdfborder={0 0 0}}
\makeatletter
\let\ORG@hyper@linkstart\hyper@linkstart
\protected\def\hyper@linkstart#1#2{%
  \lowercase{\ORG@hyper@linkstart{#1}{#2}}}
\makeatother

\begin{document}

\maketitle
\makecopyright

\tableofcontents
\listoffigures
\listoftables

\acknowledgments

First of all, I would like to thank my advisor, Bonnie Fleming, for a lot of support, guidance, and advice, both at Yale and at the University of Chicago. Moving between institutions in the middle of a PhD could have been stressful, but Bonnie gave me a lot of freedom throughout the process, and I think it could not have gone more smoothly. 

I would like to thank all of the other students and postdocs at these instutitions who have worked with me on MicroBooNE, Giacomo Scanavini, London Cooper-Troendle, Kaicheng Li, Jay Hyun Jo, and Avinay Bhat, for a lot of help and support over the years. I would also like to thank everyone I have worked with at Yale and the University of Chicago who have worked on other LArTPC experiments, Matthew King, Angela White, Mun Jung Jung, Gray Putnam, Lynn Tung, Elise Hinkle, Nathaniel Rowe, Yinrui Liu, Thomas Wester, David Schmitz, and Edward Blucher, all of whom have taught me a lot in our many interactions.

I am especially thankful to Xin Qian, who inspired and guided much of the research in this thesis. He introduced me to much of the work going on in MicroBooNE, and welcomed me at Brookhaven National Lab when I visited for a year. I gained a lot of invaluable advice, and appreciated many interesting conversations about lots of science.

I owe a great deal of appreciation to those who worked on MicroBooNE single photon searches before me, including Mark Ross-Lonergan and Kathryn Sutton who laid a lot of groundwork for this thesis. I would like to thank Benjamin Bogart and Giacomo Scanavini for a lot of work related to NC $\pi^0$ events using Wire-Cell reconstruction, which was critically important for this single photon search. I would like to thank Mark Ross-Lonergan, Guanqun Ge, Pawel Guzowski, and Erin Yandel for countless valuable interactions as we developed MicroBooNE photon-like analyses at the same time.

I would also like to thank other members of the MicroBooNE Wire-Cell team, including Hanyu Wei, Wenqiang Gu, Chao Zhang, Haiwang Yu, Sergey Martynenko, Xiangpan Ji, Nitish Nayak, Mohamed Ismail, and Jesse Mendez who created a lot of the infrastructure used in this thesis, and helped me get past roadblocks countless times. I also thank Afroditi Papadopoulou for helpful answers to many questions about Pandora reconstruction and our neutrino interaction modeling.

I also very much appreciate the MicroBooNE leadership, including the analysis coordinators David Caratelli, Kirsty Duffy, and Steven Gardiner, and the spokespeople Matthew Toups and Justin Evans, for helping to create such a fun, helpful and productive work environment, which I do not take for granted.

More broadly, I would like to thank everyone who worked on the MicroBooNE experiment in any way. I joined the experiment after it was designed and built and operating and calibrated, and after most of the analysis toolchain was already working well; this required a huge amount of work that was vitally important for my research.

Even more broadly, I thank everyone who contributes to society in any way they can, for allowing the funding of science and giving me the privilege to work on this basic research.

\abstract
Neutrinos are some of the most elusive particles in the standard model, being incredibly common throughout the universe, but interacting with detectors incredibly rarely. Certain properties of neutrinos remain difficult to measure, including their masses, their CP violation properties, and whether or not they are their own antiparticles. Additionally, there have been several anomalous results in neutrino experiments which remain unexplained. MicroBooNE was built in order to study these anomalous results using a more capable detector technology, the Liquid Argon Time Projection Chamber. Specifically, MicroBooNE is able to search for an anomalous excess of low energy electromagnetic showers, which was previously observed by the MiniBooNE experiment. In particular, MicroBooNE is able to study whether the excess could consist of electron showers or photon showers. In this thesis, I describe a search for this anomalous excess by targeting neutral current Delta radiative decays, the largest expected source of single photons in MicroBooNE. We observe data consistent with our nominal expectation, but cannot rule out all potential sources of additional single photon events, particularly those with no visible proton activity. There remains significant potential to probe this channel in even more detail using MicroBooNE and other experiments in the near future.

\mainmatter

\include{chapters/01_intro.tex}
\include{chapters/02_microboone.tex}
\include{chapters/03_eLEE.tex}
\include{chapters/04_nc_delta.tex}

\include{chapters/05_more_photon.tex}

\include{chapters/06_conclusions.tex}
\appendix
\include{chapters/07_cross_sections.tex}

\include{chapters/08_sbnd_dune.tex}
\include{chapters/09_WC_BDT_vars.tex}

\bibliographystyle{apsrev4-1-with-titles}

\addcontentsline{toc}{chapter}{REFERENCES}
\bibliography{references}

\end{document}

%% file: chapters/01_intro.tex
\chapter{Introduction}

\section{Neutrino Discoveries}

\begin{figure}[H]
    \centering
    \includegraphics[width=0.49\textwidth]{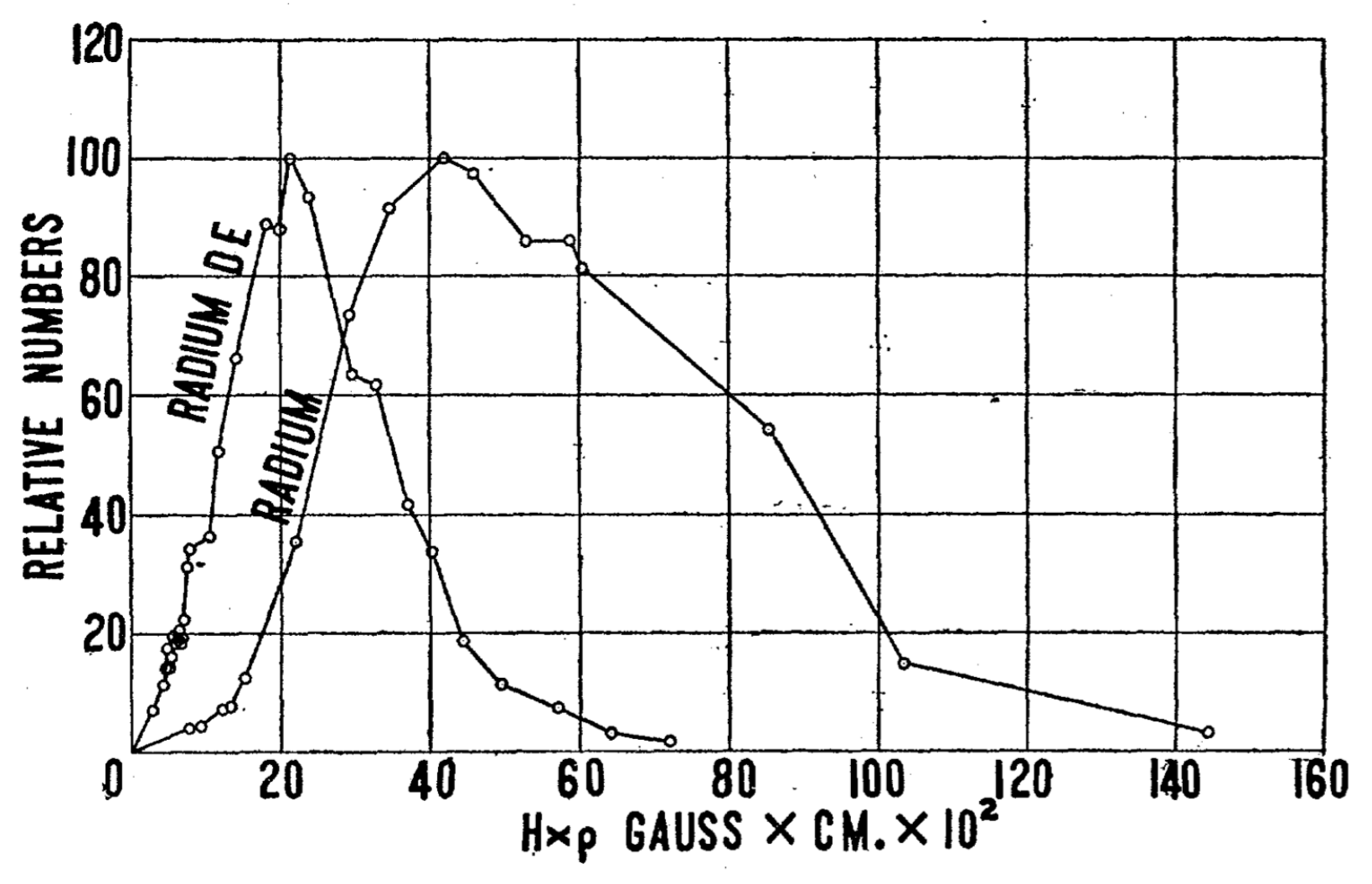}
    \caption[Early beta decay spectrum]{Beta decay spectrum of radium and radium daughters, showing a continuous spectrum, published in 1916 \cite{continuous_beta_decay_english}. Essentially, this is some of the earliest evidence for the existence of neutrinos, but this was not understood until much later.}
    \label{fig:beta_spectrum}
\end{figure}

The first hints of the neutrino's existence was the observation of a continuous beta decay spectrum as observed by Lise Meitner and Otto Hahn in 1911 \cite{continuous_beta_decay_german}, and later by others as shown in Fig. \ref{fig:beta_spectrum}. A beta decay is when a nucleus spontaneously emits an electron as a neutron turns into a proton. With energy conservation, it was hard to explain how a single ejected final state electron could have a continuous distribution of energies in different decays, with a discrete set of possible nuclear final states.

\begin{figure}[H]
    \centering
    \includegraphics[width=0.49\textwidth]{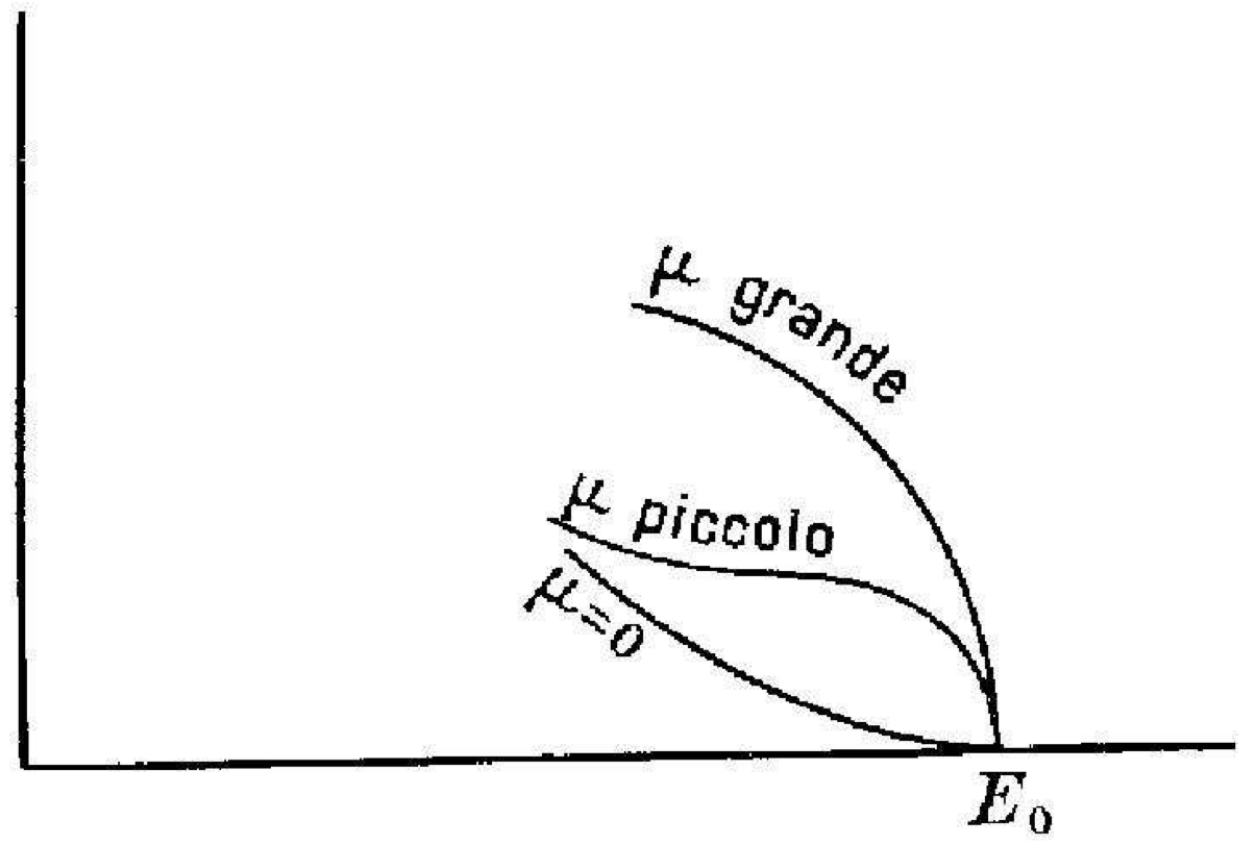}
    \caption[Early beta decay spectrum prediction]{Beta decay spectrum for different neutrino masses, predicted by Fermi in 1933 \cite{fermi_prediction}.}
    \label{fig:fermi_spectrum}
\end{figure}

In 1930, Wolfgang Pauli came up with a ``desperate remedy to save [...] the law of
conservation of energy'', hypothesizing the existence of neutrinos (which he initially called neutrons) \cite{Pauli1930}. The idea was that if there are additional particles emitted in beta decays which invisibly carry away extra energy, this would allow energy to be conserved. Pauli's theory was further developed by Enrico Fermi in 1933 \cite{fermi_prediction}, including quantitative predictions for the shape of beta decay spectra, and considering the potential effects of a nonzero neutrino mass, as shown in Fig. \ref{fig:fermi_spectrum}. On the neutrino mass, Fermi concluded ``Therefore we reach the conclusion the the neutrino mass is zero or, in any case, is small compared to the electron mass''.

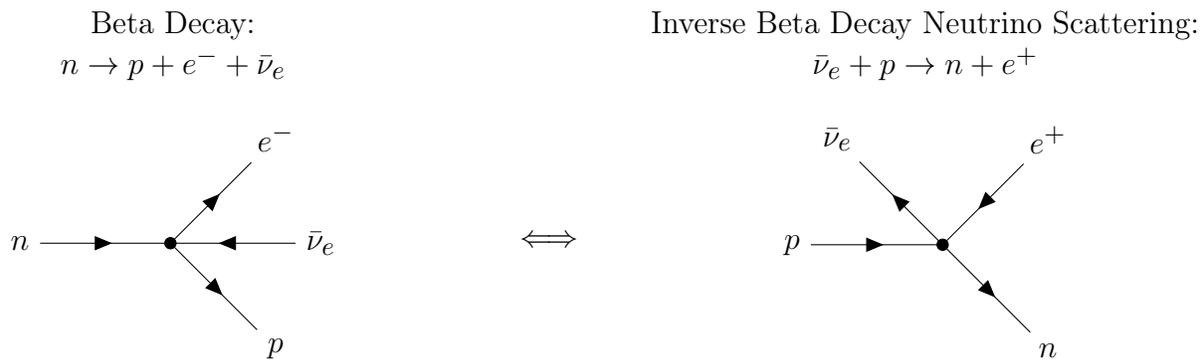
\begin{figure}[H]
    \centering
    \begin{tikzpicture}
        \node (beta) at (0,0) {
        \begin{tikzpicture}[baseline=(v)]
            \begin{feynman}
            \vertex [dot] (v) at (0,0) {};
            \vertex (n) at (-2,0) {\(n\)};
            \vertex (p) at (1.4,-1.4) {\(p\)};
            \vertex (e) at (1.4,1.4) {\(e^-\)};
            \vertex (nubar) at (2,0) {\(\bar{\nu}_e\)};
            \diagram* {
                (n) -- [fermion] (v),
                (v) -- [fermion] (p),
                (v) -- [fermion] (e),
                (v) -- [anti fermion] (nubar)
            };
            \end{feynman}
        \end{tikzpicture}
        };
    
        \node (scatter) at (10,0) {
        \begin{tikzpicture}[baseline=(v)]
            \begin{feynman}
            \vertex [dot] (v) at (0,0) {};
            \vertex (n) at (1.4,-1.4) {\(n\)};
            \vertex (nu) at (-1.4,1.4) {\(\bar{\nu}_e\)};
            \vertex (p) at (-2,0) {\(p\)};
            \vertex (e) at (1.4,1.4) {\(e^+\)};
            \diagram* {
                (p) -- [fermion] (v),
                (v) -- [fermion] (nu),
                (v) -- [fermion] (n),
                (e) -- [fermion] (v)
            };
            \end{feynman}
        \end{tikzpicture}
        };
    
        \node at (5,0) {$\Longleftrightarrow$};
    
        \node[above] at (0,2) {\shortstack{Beta Decay:\\ \(n\to p+e^-+\bar{\nu}_e\)}};
        \node[above] at (10,2) {\shortstack{Inverse Beta Decay Neutrino Scattering:\\ \(\bar{\nu}_e+p\to n+e^+\)}};
    \end{tikzpicture}
    \caption[Beta decay crossing symmetry]{Crossing symmetry between beta decay and inverse beta decay neutrino scattering. Note that this early understanding of neutrino interactions did not involve any $W$ or $Z$ bosons.}
    \label{fig:crossing_symmetry}
\end{figure}

Despite the fact that the weak interaction was not at all understood, and nobody had even proposed the existence of $W$ or $Z$ bosons, it was understood that there should exist a crossing symmetry, implying that the same calculations will affect interactions even if you rearrange the particles in a Feynman diagram, as shown in Fig. \ref{fig:crossing_symmetry}. This allowed Hans Bethe and Rudolf Peierls to predict an upper limit on the inverse beta decay interaction cross section in 1934, $\sigma < 10^{-44}\ \mathrm{cm}^2$ \cite{bethe_neutrino_xs_prediction}. In their paper, they state ``It is therefore absolutely impossible to observe processes of this kind with the neutrinos created in nuclear transformations.''

In 1956, Frederick Reines and Clyde Cowan proved Bethe and Peierls wrong, observing neutrino interactions from nuclear decays. They had two large advantages that Bethe and Peierls did not predict: the existence of nuclear fission chain reactions, allowing very large rates of nuclear decays in concentrated volumes, and the technique of observing a double coincidence between the positron and the neutron each being detected from a single inverse beta decay interaction, allowing a large reduction in background rates. This method of inverse beta decay detection is still widely used in reactor neutrino experiments today.

Reines and Cowan first planned to observe neutrinos emitted in a nuclear bomb detonation, but after devising their double coincidence technique, decided to look for neutrinos at a nuclear reactor instead \cite{first_neutrino_detection}. Their experiment was successful, winning the Nobel Prize in 1995 alongside Martin Perl for the discovery of the tau lepton. 

This represented the first observation of neutrinos, using the inverse beta decay process. We now understand this interaction via the exchange of a charged $W^-$ boson, and we call this a charged-current (CC) interaction. In 1973, neutral-current (NC) interactions operating by the exchange of a neutral $Z$ boson were discovered by GARGAMELLE \cite{GARGAMELLE_NC_discovery,GARGAMELLE_review}. Since then, many CC and NC interactions have been measured at many energy scales.

\section{Neutrino Oscillations}

Neutrino oscillations were first postulated by Bruno Pontecorvo in 1957 \cite{pontecorvo_oscillation, pontecorvo_review}. The first hint of the existence of neutrino oscillations came in 1967 with the Homestake experiment \cite{homestake}, which observed a lower rate than expected of solar neutrino interactions on $^{37}\mathrm{Cl}$, winning Raymond Davis Jr. the Nobel Prize in 2002, alongside Masatoshi Koshiba for the discovery of neutrinos in cosmic radiation and Riccardo Giacconi for pioneering x-ray astrophysics. The first evidence that truly convinced people of neutrino oscillations came from the Sudbury Neutrino Observatory (SNO) and Super-Kamiokande. In 2001, SNO directly observed a non-$\nu_e$ flux of solar neutrinos by comparing charged current interactions (measuring $\nu_e$ specifically) and neutrino-electron elastic scattering (measuring a combination of $\nu_e$, $\nu_\mu$, and $\nu_\tau$) \cite{sno}. This almost coincided with the first observation of atmospheric neutrino oscillations from Super-Kamiokande, which compared the rates of muon neutrinos coming from above or below the detector, corresponding to short distances through the atmosphere, or long distances through the entire earth \cite{super_k_first_oscillation}. These SNO and Super-Kamiokande results won the Nobel Prize in 2015.

Neutrino oscillations occur because neutrinos have mass, and the mass eigenstates do not coincide with flavor eigenstates. Therefore, when a neutrino is produced as a flavor eigenstate, it consists of a superposition of mass eigenstates, and these mass eigenstate wavefunctions accumulate phase changes at different rates according to Schrodinger's equation, so when the neutrino interacts it can collapse to a different flavor eigenstate.

Today, the study of neutrino oscillations is entering the precision era. Most experiments fit into a consistent picture of neutrino oscillations with three flavors, described by the Pontecorvo-Maki-Nakagawa-Sakata (PMNS) matrix \cite{pmns},

\begin{equation}
    \begin{pmatrix}
        \nu_e \\
        \nu_\mu \\
        \nu_\tau
    \end{pmatrix} 
    = 
    \begin{pmatrix}
        U_{e 1} & U_{e 2} & U_{e 3} \\ 
        U_{\mu 1} & U_{\mu 2} & U_{\mu 3} \\ 
        U_{\tau 1} & U_{\tau 2} & U_{\tau 3}
    \end{pmatrix}
    \begin{pmatrix} 
        \nu_1 \\ 
        \nu_2 \\ 
        \nu_3 
    \end{pmatrix}.
\end{equation}

This matrix is commonly parametrized as

\begin{equation}
\begin{split}
 & \begin{pmatrix} 1 & 0 & 0 \\ 0 & c_{23} & s_{23} \\ 0 & -s_{23} & c_{23} \end{pmatrix}
 \begin{pmatrix} c_{13} & 0 & s_{13}e^{-i\delta_\text{CP}} \\ 0 & 1 & 0 \\ -s_{13}e^{i\delta_\text{CP}} & 0 & c_{13} \end{pmatrix}
 \begin{pmatrix} c_{12} & s_{12} & 0 \\ -s_{12} & c_{12} & 0 \\ 0 & 0 & 1 \end{pmatrix} \\
 & = \begin{pmatrix} c_{12}c_{13} & s_{12} c_{13} & s_{13}e^{-i\delta_\text{CP}} \\
 -s_{12}c_{23} - c_{12}s_{23}s_{13}e^{i\delta_\text{CP}} & c_{12}c_{23} - s_{12}s_{23}s_{13}e^{i\delta_\text{CP}} & s_{23}c_{13}\\
 s_{12}s_{23} - c_{12}c_{23}s_{13}e^{i\delta_\text{CP}} & -c_{12}s_{23} - s_{12}c_{23}s_{13}e^{i\delta_\text{CP}} & c_{23}c_{13} \end{pmatrix},
\end{split}
\end{equation}
where $s_{ij}$ and $c_{ij}$ mean $\sin\theta_{ij}$ and $\cos\theta_{ij}$. If neutrinos are Majorana particles, then we multiply by an additional matrix containing two majorana phases, $\mathrm{diag}(1, e^{i\alpha}, e^{i\beta})$, but these phases would not affect neutrino oscillations.

The PMNS matrix (denoted by $U$), along with the squared differences in masses between neutrino mass eigenstates, tells us how neutrinos should change flavor (oscillate) when traveling over a distance $L$ with an energy $E$ via the following equation (with $c=\hbar=1$) \cite{pmns},
\begin{equation}
\begin{split}
  P_{\alpha\rightarrow\beta} = 
  \delta_{\alpha\beta}
    & - 4\sum_{i<j} {\rm Re}\left(U_{\alpha i} U_{\beta i}^* U_{\alpha j}^* U_{\beta j}\right) \sin^2\left(\frac{\Delta m_{ij}^2 L}{4E}\right) \\
    & + 2\sum_{i<j}{\rm Im}\left(U_{\alpha i} U_{\beta i}^* U_{\alpha j}^* U_{\beta j}\right) \sin\left(\frac{\Delta m_{ij}^2 L}{2E}\right).
\end{split}
\end{equation}
However, we note that the Mikheyev-Smirnov-Wolfenstein effect (MSW effect, or matter effect) alters predictions for neutrino oscillations in the presence of matter (for example, neutrinos fired in a beam traveling through the earth's crust) \cite{MSW_effect}. This is because neutrinos undergo coherent forward scattering with the electrons in matter, and only electron-flavor neutrinos interact via charged-current scattering with these electrons.

One particularly interesting and difficult quantity to measure for neutrino oscillations is $\delta_\text{CP}$, the $CP$ violating phase. This phase describes how antineutrinos might oscillate differently from neutrinos. Note that $CPT$ symmetry, which is required to be preserved in any quantum field theory, requires that $P(\nu_\alpha \rightarrow \nu_\beta) = P(\overline{\nu_\beta} \rightarrow \overline{\nu_\alpha})$, since we have applied $CP$ by swapping to antineutrinos and we have applied $T$ by reversing time. With $\alpha=\beta$, this implies that there can be no difference between neutrino and antineutrino oscillations when only one neutrino flavor is used; this means that we need high energy experiments involving both muon and electron flavors in order to study this parameter. Note that this $CP$ violation is considering how flavor eigenstates might oscillate differently from their corresponding antiparticles; this is an entirely independent question from whether neutrino mass eigenstates might be indistinguishable from their antiparticles making them Majorana rather than Dirac particles, a question which neutrinoless double beta decay experiments are studying. If neutrinoless double beta decay experiments discover that neutrinos are Majorana particles, and oscillation experiments discover a nonzero $\delta_\text{CP}$, then that would provide circumstantial evidence for the process of ``leptogenesis'', a proposed mechanism that might have contributed to the matter-antimatter asymmetry in the early universe \cite{leptogenesis, dune_fd_cdr_intro}.

Measurements of neutrino oscillations can also let us search for nonstandard interactions that would alter oscillation probabilities \cite{nsi_status_report}. There are also a variety of theoretical predictions of neutrino mixing and mass parameters, so more precise measurements of these parameters will directly test many of these hypotheses, for example those in Refs. \cite{delta_cp_prediction_axion, delta_cp_prediction_su6, delta_cp_prediction_a4, delta_cp_prediction_a4_2, analytic_PMNS_calculation, analytic_PMNS_calculation_2, lepton_mixing_predictions_s4, mixing_sum_rules, pmns_from_entanglement, predicting_cp_discrete_flavor, neutrino_parameters_minimal_model_lin_comb}. A simple example of this type of theoretically motivated prediction is that of an exact muon-tau symmetry in the PMNS matrix, which would require $\theta_{23}$ to be exactly $45^\circ$ and is possible according to our current experimental observations  \cite{mu_tau_symmetry}.

\begin{figure}[H]
    \centering
    \begin{subfigure}[b]{0.49\textwidth}
        \centering
        \includegraphics[width=\textwidth]{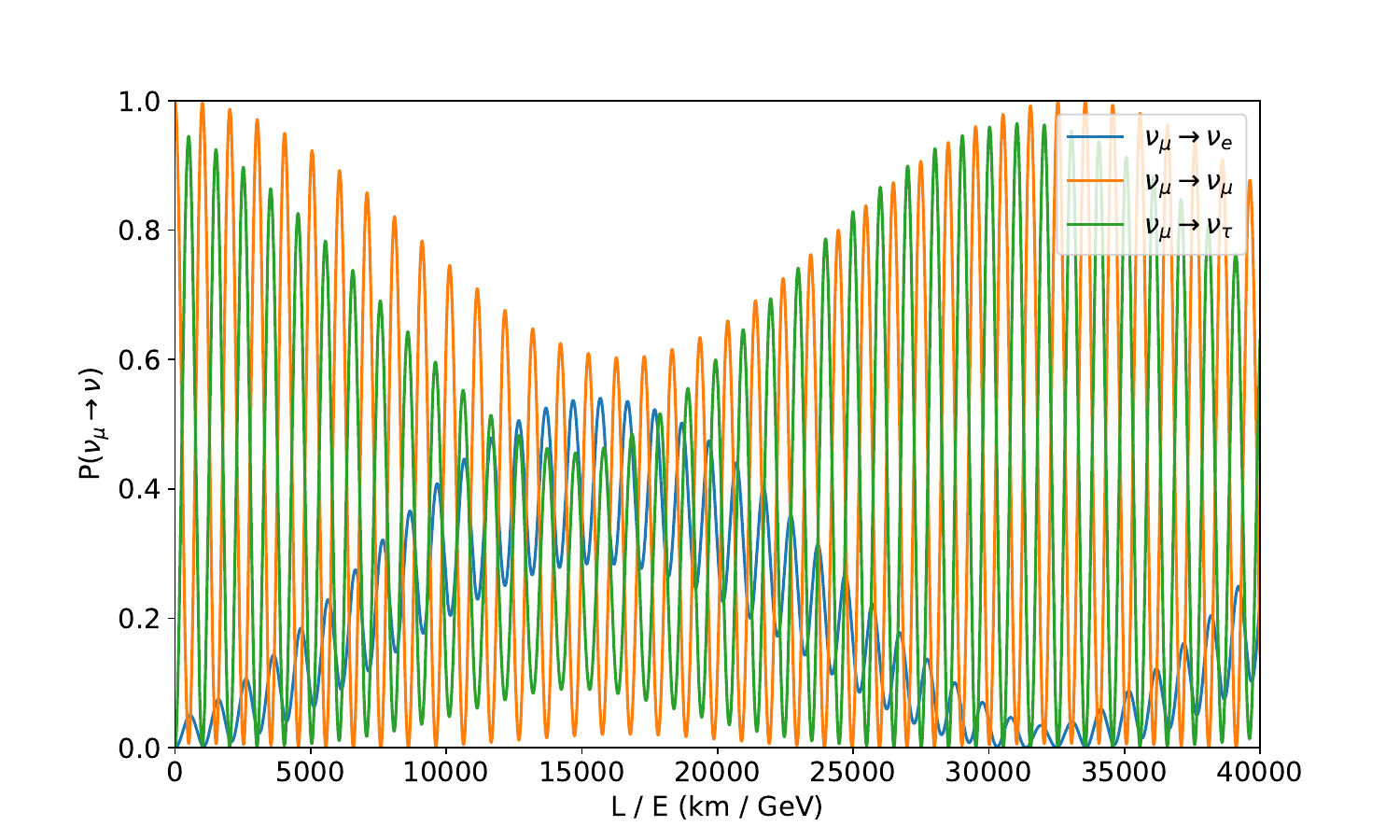}
        \caption{}
        \label{fig:oscillations_a}
    \end{subfigure}
    \hfill
    \begin{subfigure}[b]{0.49\textwidth}
        \centering
        \includegraphics[width=\textwidth]{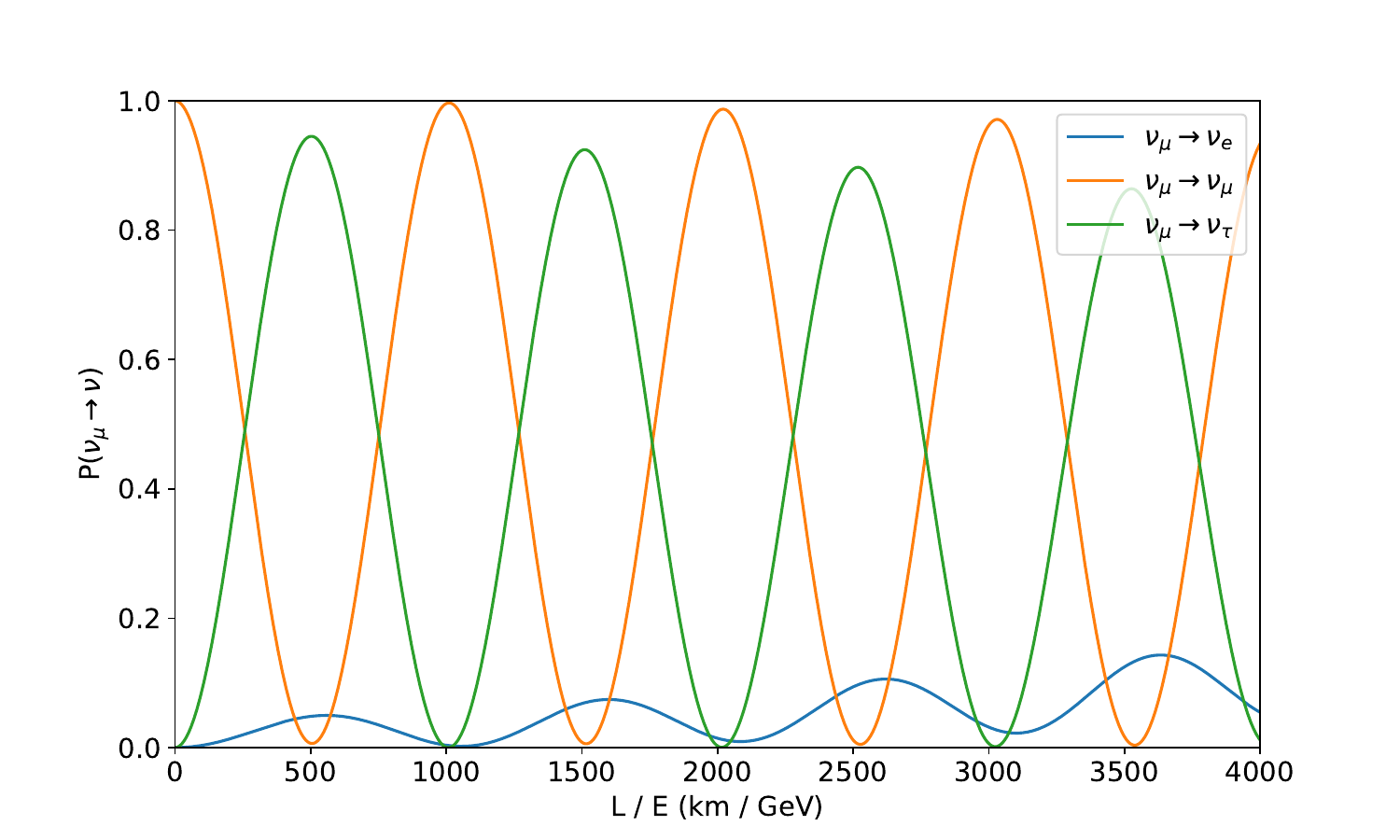}
        \caption{}
        \label{fig:oscillations_b}
    \end{subfigure}
    \caption[Neutrino oscillation probabilities]{Neutrino oscillation probabilities starting from the $\nu_\mu$ state, according to the best fit to world data in the normal hierarchy from NuFit 6.0 \cite{nufit, nufit_website}. Panel (a) shows a larger $L/E$ range, while panel (b) zooms into lower $L/E$ values. Code for these plots is available at \cite{lhagaman_neutrino_mixing_plots}.}
    \label{fig:oscillations}
\end{figure}

\begin{figure}[H]
    \centering
    \includegraphics[width=0.5\textwidth]{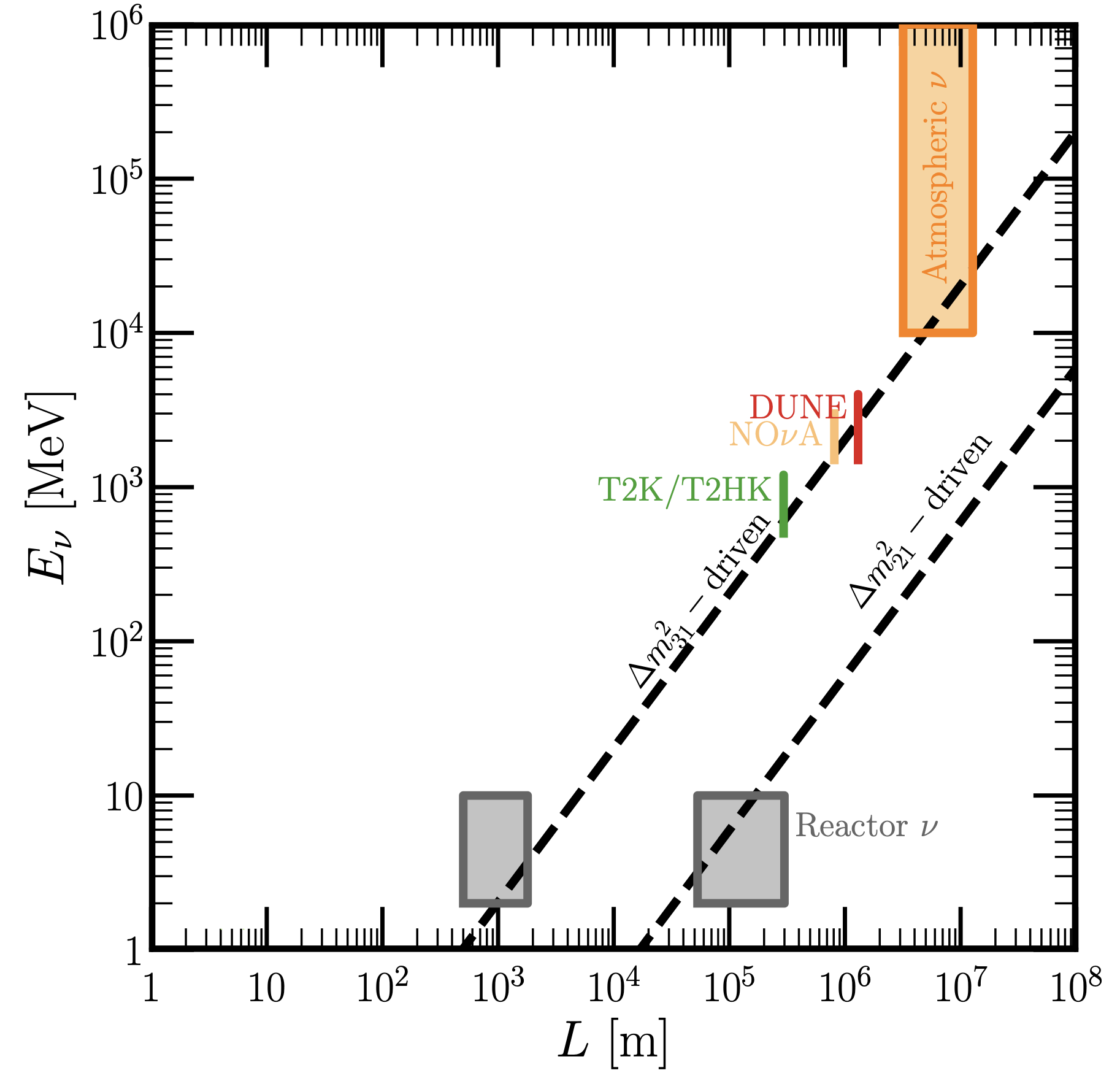}
    \caption[Neutrino oscillation experiment baselines and energies]{Visualization of some neutrino oscillation experiment baselines and energies. Figure taken from Ref. \cite{Kelly2021MiniSBNTH}.}
    \label{fig:L_vs_E}
\end{figure}

\begin{figure}[H]
    \centering
    \includegraphics[width=0.6\textwidth]{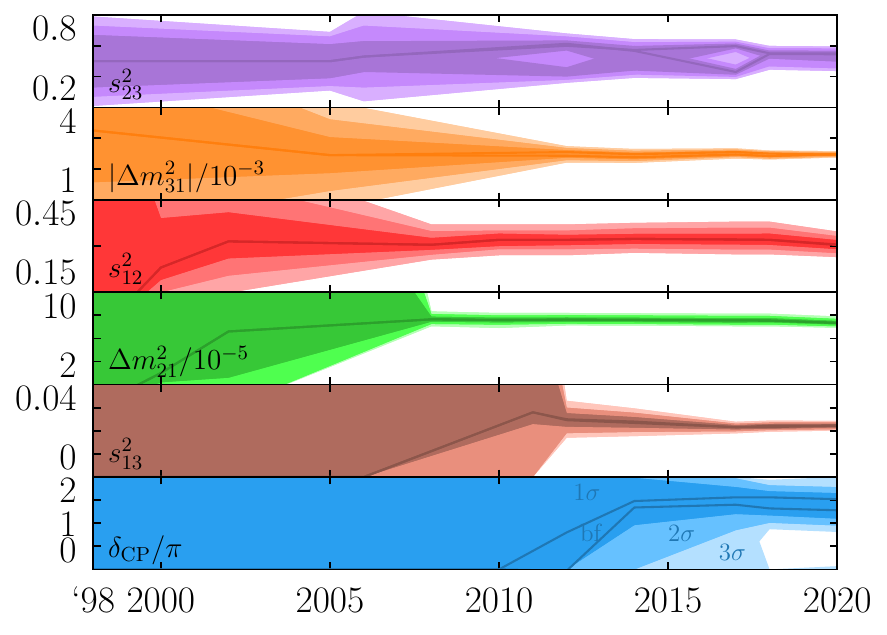}
    \caption[Neutrino oscillation parameter uncertainties over time]{Neutrino oscillation parameter uncertainties over time. Figure taken from Ref. \cite{NF01_snowmass}.}
    \label{fig:osc_params_over_time}
\end{figure}

The current best-fit understanding of neutrino oscillations is illustrated in Fig. \ref{fig:oscillations} for muon neutrinos as a function of $L/E$. In Fig. \ref{fig:oscillations_a}, we can see the longer-baseline oscillation, observed by solar neutrino experiments and the reactor experiment KamLAND \cite{sno_solar_oscillation, kamland}. This oscillation is governed by $\Delta m^2_{21}$, measured to be $(7.49\pm0.19) \cdot 10^{-5}\ \mathrm{eV}^2$ \cite{nufit, nufit_website}. In Fig. \ref{fig:oscillations_b}, we zoom in to see the shorter-baseline oscillation, observed by reactor experiments like Daya Bay \cite{daya_bay}, accelerator experiments like T2K \cite{T2K} and NOvA \cite{NOvA}, and atmospheric experiments like Super-Kamiokande \cite{super_k} and IceCube \cite{icecube}. This oscillation is governed by $\Delta m^2_{31}$, measured to be $(2.534\pm0.03) \cdot 10^{-3}\ \mathrm{eV}^2$ \cite{nufit, nufit_website}. The uncertainties in these parameters have rapidly decreased over time as shown in Fig. \ref{fig:osc_params_over_time}, and is expected to continue to decrease in the future, particularly with observations from the Deep Underground Neutrino Experiment (DUNE), Hyper-Kamiokande, and Jiangmen Underground Neutrino Observatory (JUNO) experiments in the future. A visualization of the baseline and energy ranges for some experiments and how they compare to the first oscillation maximum for these mass splittings is shown in Fig. \ref{fig:L_vs_E}.

One more interesting phenomenon to consider is neutrino decoherence. We normally assume that experiments measure the flavor of neutrinos rather than the mass, but this is not always entirely the case. If the energy and time of flight of a neutrino is measured very precisely (not necessarily measured in a detector, but measured in the quantum mechanical sense, i.e. an interaction with the larger environment), then effectively the mass of the neutrino is measured. Since mass and flavor are conjugate variables for neutrinos, this causes the Heisenberg uncertainty principle to affect the resulting flavor. Another way to think about this process: we can imagine a neutrino as three overlapping wave packets, one for each mass eigenstate. When these wave packets are overlapping, we get normal neutrino oscillations, but they will travel at different speeds and can therefore separate from each other at longer distances, altering the oscillation behavior. How much these wavepackets separate will depend on the neutrino production mechanism which affects the initial size of the wavepacket; for example, the BeEST experiment recently set a 6.2 picometer lower limit on the spatial extent of the neutrino wavepacket in beryllium electron capture decays \cite{beest_wavepacket_size}. For pion decay in flight neutrino beams, which are used for the main topics in this thesis, this decoherence is expected to have a negligible effect, but it could be an important consideration for some nuclear reactor and radiactive source neutrino experiments \cite{neutrino_decoherence}.

It is informative to ask the question, why do neutrinos oscillate while charged leptons do not? This asymmetry between neutrinos and charged leptons arises due to the fact that flavor eigenstates are defined to be the weak eigenstates exactly coinciding with the mass eigenstates of charged leptons. This definition is useful, since electromagnetic interactions and time-of-flight wave packet separation cause rapid collapse of charged lepton mass superpositions, while neutrinos easily stay in mass superpositions for extended distances. But ultimately this definiton for flavor is a choice, and one could use a different definition, for example defining ``odor'' as an analogous quantity to flavor, but now defined such that odor eigenstates are the weak eigenstates exactly coinciding with neutrino mass eigenstates \cite{charged_lepton_oscillation}. In this formulation, odor is conserved in weak interactions; neutrinos do not oscillate, maintaining the same odor throughout their trajectories; and charged leptons experience odor oscillations as they travel. Of course, neutrino masses are much harder to measure experimentally than charged lepton masses, and therefore odor is much harder to measure experimentally than flavor, so this redefinition is not be particularly useful. This choice of what particles should oscillate is analogous to the quark sector and the Cabibbo-Kobayashi-Maskawa (CKM) matrix. In this case, we define $u$, $c$, and $t$ to be both mass and weak eigenstates, and let $d$, $s$, and $b$ be mass eigenstates differing from $d'$, $s'$, and $b'$ weak eigenstates; therefore $d$, $s$, and $b$ oscillate while $u$, $c$, and $t$ do not, but this definition can be easily flipped, causing oscillations in the opposite set of particles.

\section{Neutrino Mass}

The discovery of neutrino oscillations marked the discovery of a nonzero neutrino mass, the first concrete discovery of physics Beyond the Standard Model (BSM) (although nowadays many people have updated their working definition of ``Standard Model'' to include nonzero neutrino masses).

\begin{figure}[H]
    \centering
    \centering
    \includegraphics[width=0.7\textwidth]{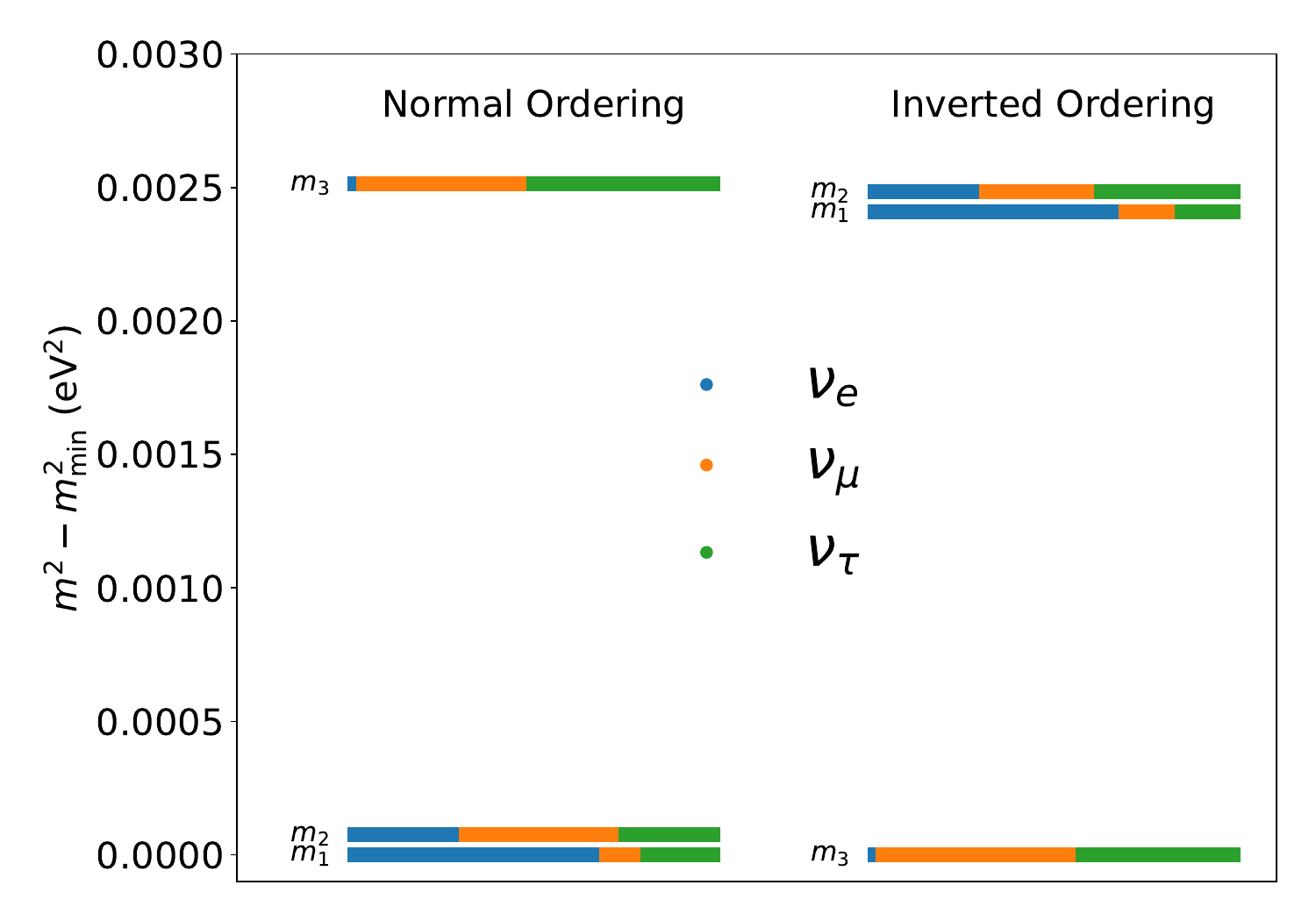}
    \caption[Neutrino mass ordering]{Neutrino mass ordering possibilities, showing flavor fractions for each mass eigenstate according to the central value from NuFit 6.0 \cite{nufit, nufit_website}. The minimum neutrino mass is unknown, so this figure shows mass squared differences with respect to the minimum neutrino mass. Code for this plot is available at \cite{lhagaman_neutrino_mixing_plots}.}
    \label{fig:neutrino_mass_ordering}
\end{figure}

The neutrino mass ordering is not known; the minimum-electron-flavor neutrino mass eigenstate could be either the heaviest or the lightest, as shown in Fig. \ref{fig:neutrino_mass_ordering}. This mass ordering can in principle be measured by oscillation experiments like NOvA, T2K, DUNE, Hyper-K, and JUNO, and we do not expect this question to remain a mystery for much longer.

Neutrino oscillations tell us the values of the difference in masses squared between pairs of neutrinos, $\Delta m^2$, but do not tell us the absolute values of the masses. So the lightest neutrino mass eigenstate, $m_\mathrm{min}$, is unknown. $m_\mathrm{min}$ could even be identically zero, as unnatural as that might sound with the existence of two other massive neutrino states.

The sum of neutrino masses $\Sigma m_i$ can be measured by cosmology experiments, since this quantity will affect how neutrinos transitioned from non-relativistic to relativistic and impacted structure formation in the early universe. Recently, the Dark Energy Spectroscopic Instrument (DESI) alongside other cosmology experiments set an upper limit $\Sigma_i m_i < 0.064\ \mathrm{eV}/c^2$ at 95\% CL in a $\Lambda$CDM model, placing tensions on the inverted ordering possibility \cite{DESI_new}.

There are also experiments trying to measure the effective electron neutrino mass from beta decays, $m_\beta = \sqrt{\Sigma_i |U_{ei}^2 m_i^2|}$. This mass is relevant for the kinematics of beta decays, where there is a minimum amount of energy that the neutrino can take away from the nuclear decay due to the neutrino's rest mass, and this influences the maximum energy of the emitted electron. This $m_\beta$ value is the mass that Fermi determined must be small when he considered beta decay spectra, as shown in Fig. \ref{fig:fermi_spectrum} \cite{fermi_prediction}. Experiments like KATRIN are now measuring these beta decay spectrum endpoints much more precisely, with the most recent constraint of $m_\beta < 0.45\ \mathrm{eV}/c^2$ at 90\% CL \cite{KATRIN_neutrino_mass}. These beta decay spectra can also be sensitive to the existence of heavier neutrino states that mix with electron neutrinos, for example sterile neutrinos \cite{KATRIN_sterile}.

Further, there are experiments trying to measure the effective Majorana mass for double beta decays $m_{\beta\beta} = |\Sigma_i U_{ei}^2 m_i|$. If neutrinos are Majorana fermions, they are their own antiparticles, and therefore two neutrinos produced in a double beta decay can annihilate and take no energy away from the electrons. In this case, the effective Majorana mass determines the rate of neutrinoless double beta decays. The current best limits are from KamLAND-ZEN, which looks for a peak at the highest energies of the double beta decay energy spectrum for the $^{136}\mathrm{Xe}$ nucleus. Their limit is $m_{\beta\beta}<28-122\ \mathrm{eV}/c^2$ at 90\% CL, with the range coming from large uncertainties in xenon nuclear matrix element calculations \cite{kamland_zen_neutrinoless_double_beta_decay}.

\begin{figure}[H]
    \centering
    \centering
    \includegraphics[width=\textwidth]{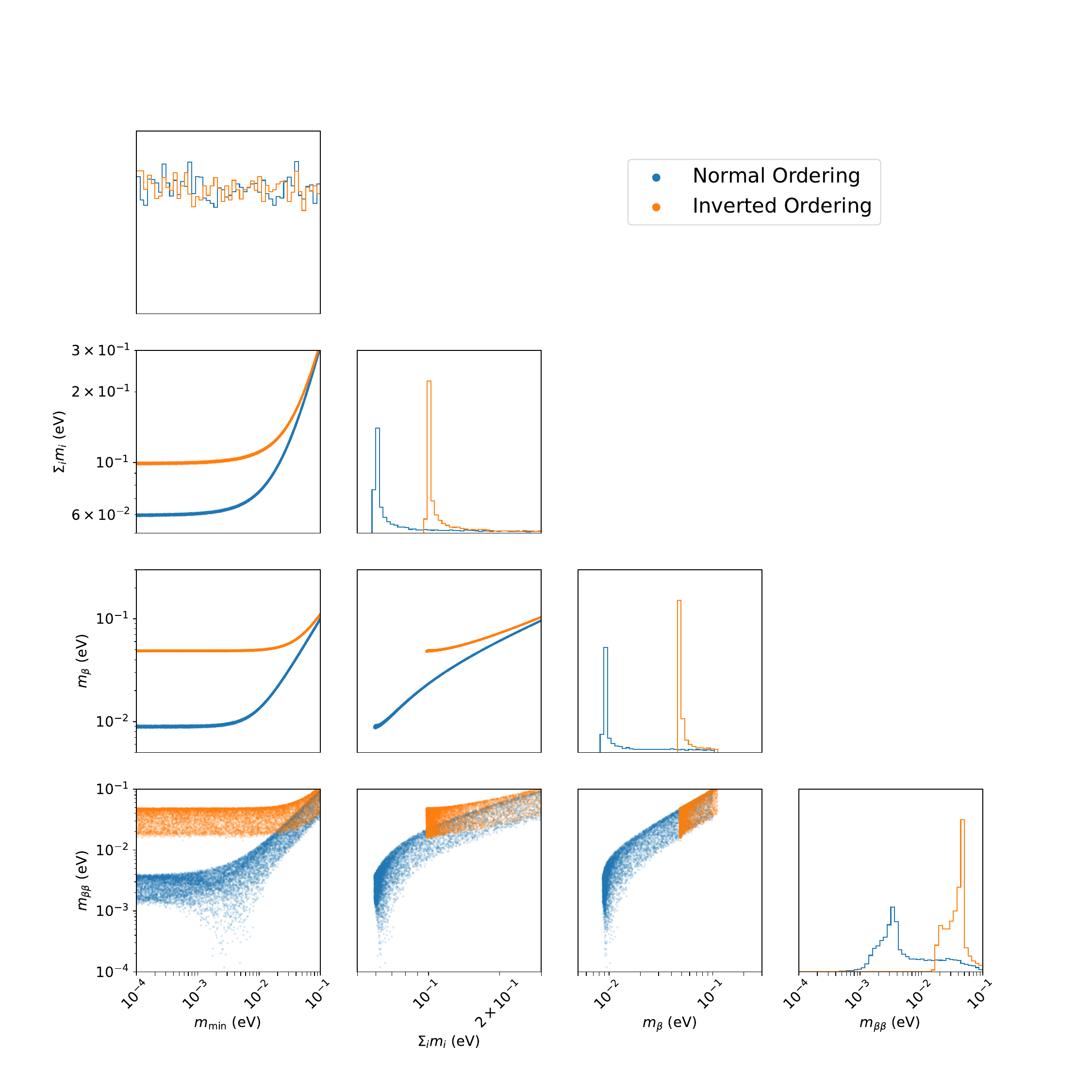}
    \caption[Neutrino mass correlations]{Neutrino mass correlations. Code for this plot is available at \cite{lhagaman_neutrino_mixing_plots}.}
    \label{fig:neutrino_mass_correlations}
\end{figure}

All of these definitions of neutrino masses have certain correlations with each other, which I show in Fig. \ref{fig:neutrino_mass_correlations} for both the inverted hierarchy and the normal hierarchy. I generate 10,000 samples, varying $m_\mathrm{min}$ uniformly in logarithmic space from $10^{-4}\ \mathrm{eV}/c^2$ to $10^{-1}\ \mathrm{eV}/c^2$, varying the standard PMNS matrix parameters and mass splittings from the central values with gaussian uncertainties associated with the NuFit 6.0 fit to world data \cite{nufit, nufit_website}, and varying the unmeasured $\alpha$ and $\beta$ phases (present only in the case of Majorana neutrinos) uniformly from $0$ to $2\pi$. 

This lets us visualize a variety of interesting conclusions about neutrino masses. Looking at the top rows of Fig. \ref{fig:neutrino_mass_correlations}, can see that $m_\mathrm{min}$, $\Sigma_i m_i$, and $m_\beta$ are all strongly correlated with each other, with only small uncertainties from the PMNS matrix parameters and mass splittings. We also see that $\Sigma_i m_i$ and $m_\beta$ have nonzero minimum possible values.

In contrast, looking at the bottom row of Fig. \ref{fig:neutrino_mass_correlations}, $m_{\beta\beta}$ has weaker correlations with more uncertainties due to the Majorana phases $\alpha$ and $\beta$. Looking at the correlation with $m_\mathrm{min}$ in the bottom left, we see the famous ``Lobster Plot'', named because the normal ordering in blue vaguely resembles a lobster claw reaching toward the bottom left. This plot illustrates the fact that if neutrinos are Majorana particles, and follow the normal hierarchy, and if we get very unlucky in the values for $m_\mathrm{min}$ and the $\alpha$ and $\beta$ phases, it could be that $m_{\beta\beta}$ is very small, meaning the corresponding rate of neutrinoless double beta decay could be very hard to observe. Theoretically, $m_{\beta\beta}$ could even be zero, and neutrinoless double beta decay could be impossible to observe despite the existence of Majorana neutrinos, although that would require an unexpected fine tuning of the mass eigenstates and PMNS matrix parameters. This region of small $m_{\beta\beta}$ masses is sometimes called the ``funnel of death'' \cite{neutrinoless_double_beta_decay_funnel}. 

These are the commonly discussed quantities related to neutrino mass, but other quantities are also possible to consider. For example, in analogy to the effective Majorana mass for double beta decays $m_{\beta\beta}$, we could consider the effective Majorana mass for double muon decays $m_{\mu\mu} = |\Sigma_i U_{\mu i}^2 m_i|$. This mass would determine the rate of $K^+ \rightarrow \pi^- \mu^+ \mu^+$, which has two muon neutrinos annihilating in the final state. This is potentially another way to search for the Majorana nature of neutrinos, although it is much more difficult than normal neutrinoless double beta decay searches because large numbers of atomic nuclei appear in the universe naturally, unlike kaons. The current best limit on this type of process is from the NA62 experiment, with $\mathcal{B}(K^+ \rightarrow \pi^- \mu^+ \mu^+) < 4.2 \times 10^{-11}$ \cite{NA62_LNV}, which would translate into a very weak upper limit on $m_{\mu\mu}$. Given our current understanding of the PMNS matrix, it is impossible for both $m_{\beta\beta}$ and $m_{\mu\mu}$ to both be zero, so if we fall into the ``funnel of death'', the Majorana nature of neutrinos will still be in principle possible to detect with very futuristic kaon decay experiments or other types of lepton number violation searches. There is no possible ``double funnel of death'' where both muon and electron lepton number violation searches would hide the Majorana nature of neutrinos.

\section{Sterile Neutrinos}\label{sec:sterile}

Some experimental observations have not easily fit into this three neutrino picture. Several notable anomalies appear to show signs of neutrino oscillations, despite the fact that the length and energy values are such that negligible neutrino oscillations are expected. This has been interpreted as evidence for additional neutrino mass states causing additional oscillation. From electron-positron colliders analyzing $Z$ boson decays, we can precisely measure the number of light neutrino species which interact with the weak force, $2.9840 \pm 0.0082$, consistent with the three expected in the standard model \cite{Z_resonance}. This tells us that if there is an additional light neutrino participating in neutrino oscillations, it must be ``sterile'', not interacting with the weak force. This is often considered in a ``3+1'' framework, where we have three active neutrinos mixing with just one sterile neutrino. This would expand the PMNS matrix and allow for more complex neutrino oscillation behaviors:

\begin{equation}
    \begin{pmatrix}
        \nu_e \\
        \nu_\mu \\
        \nu_\tau \\
        \nu_s
    \end{pmatrix} 
    = 
    \begin{pmatrix}
        U_{e 1} & U_{e 2} & U_{e 3} & U_{e 4}\\ 
        U_{\mu 1} & U_{\mu 2} & U_{\mu 3} & U_{\mu 4} \\ 
        U_{\tau 1} & U_{\tau 2} & U_{\tau 3} & U_{\tau 4} \\
        U_{s 1} & U_{s 2} & U_{s 3} & U_{s 4} \\
    \end{pmatrix}
    \begin{pmatrix} 
        \nu_1 \\ 
        \nu_2 \\ 
        \nu_3 \\
        \nu_4 
    \end{pmatrix}.
\end{equation}

This is parameterized in a similar way as the three-neutrino PMNS matrix, but now with three additional rotations, which are described by three new mixing angles $\theta_{14}$, $\theta_{24}$, and $\theta_{34}$, as well as two additional phases $\delta_{24}$ and $\delta_{34}$:

\begin{equation}
U = R_{34}(\theta_{34}, \delta_{34}) R_{24}(\theta_{24}, \delta_{24}) R_{14}(\theta_{14}, 0)
R_{23}(\theta_{23}, 0) R_{13}(\theta_{13}, \delta_{CP}) R_{12}(\theta_{12}, 0)
\end{equation}

These sterile neutrino mixings are often assembled into $\sin^2 2\theta_{ee}$, $\sin^2 2\theta_{\mu\mu}$, and $\sin^2 2\theta_{\mu e}$, which approximately represent the amplitudes of the probability oscillations at short baselines for $\nu_e\rightarrow \nu_e$ disappearance, $\nu_\mu\rightarrow \nu_\mu$ disappearance, and $\nu_\mu\rightarrow \nu_e$ appearance. These are calculated from the mixing angles and PMNS matrix elements:

\begin{alignat}{3}
    \label{eqn:theta_ee}
    \sin^2 2\theta_{ee} \quad{} & = {} & \quad \sin^2 2\theta_{14} \quad{} & = {} & \quad 4(1 - |U_{e4}|^2)|U_{e4}|^2 \\
    \label{eqn:theta_mumu}
    \sin^2 2\theta_{\mu\mu} \quad{} & = {} & \quad 4\cos^2 \theta_{14}\sin^2 \theta_{24}(1-\cos^2 \theta_{14}\sin^2 \theta_{24}) \quad{} & = {} & \quad 4(1-|U_{\mu 4}|^2)|U_{\mu 4}|^2 \\
    \label{eqn:theta_mue}
    \sin^2 2\theta_{\mu e} \quad{} & = {} & \quad \sin^2 2\theta_{14}\sin^2 \theta_{24} \quad{} & = {} & \quad 4|U_{\mu 4}|^2|U_{e4}|^2
\end{alignat}

\begin{figure}[H]
    \centering
    \begin{subfigure}[b]{0.4\textwidth}
        \centering
        \includegraphics[width=\textwidth]{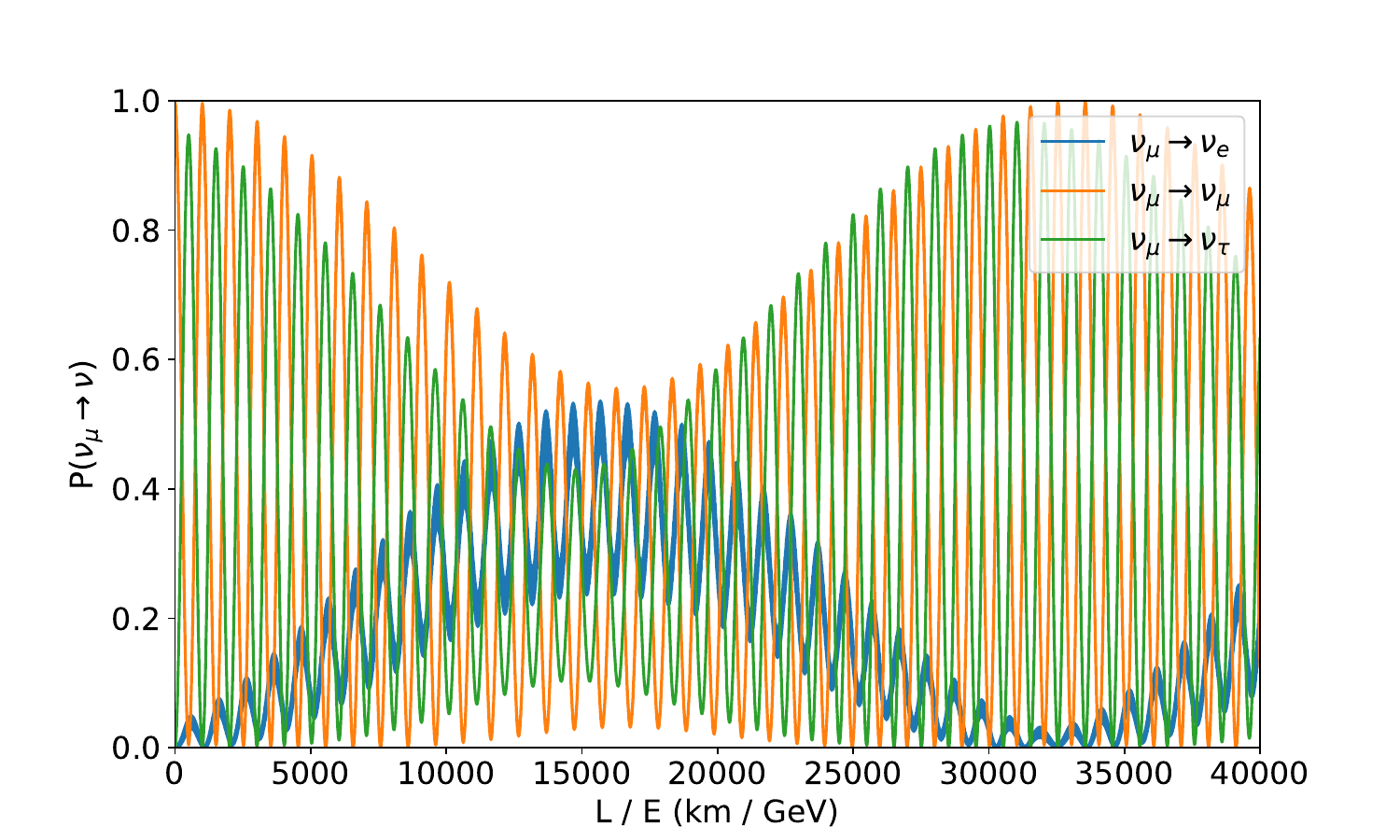}
        \caption{}
        \label{fig:sterile_oscillations_a}
    \end{subfigure}
    \begin{subfigure}[b]{0.4\textwidth}
        \centering
        \includegraphics[width=\textwidth]{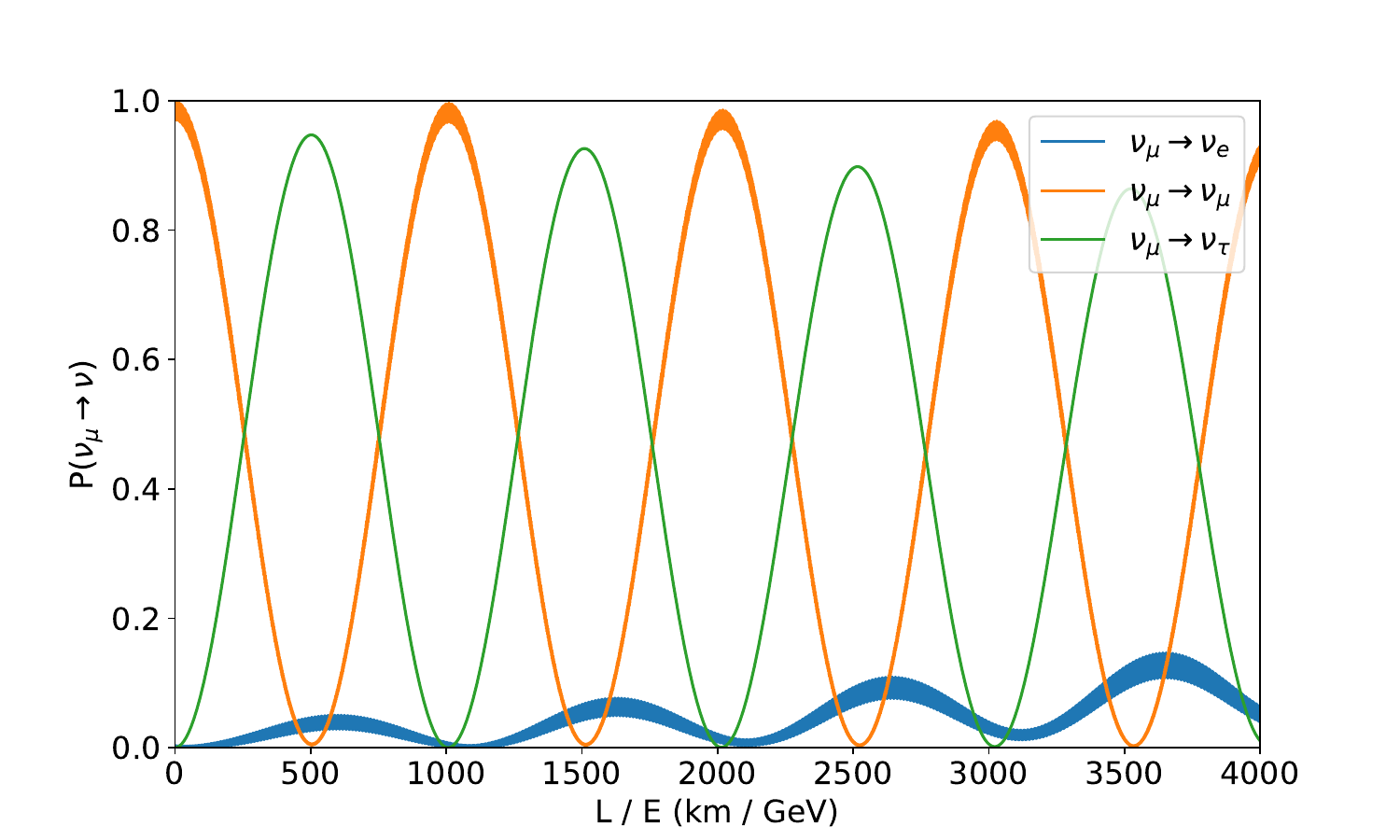}
        \caption{}
        \label{fig:sterile_oscillations_b}
    \end{subfigure}
    \begin{subfigure}[b]{0.4\textwidth}
        \centering
        \includegraphics[width=\textwidth]{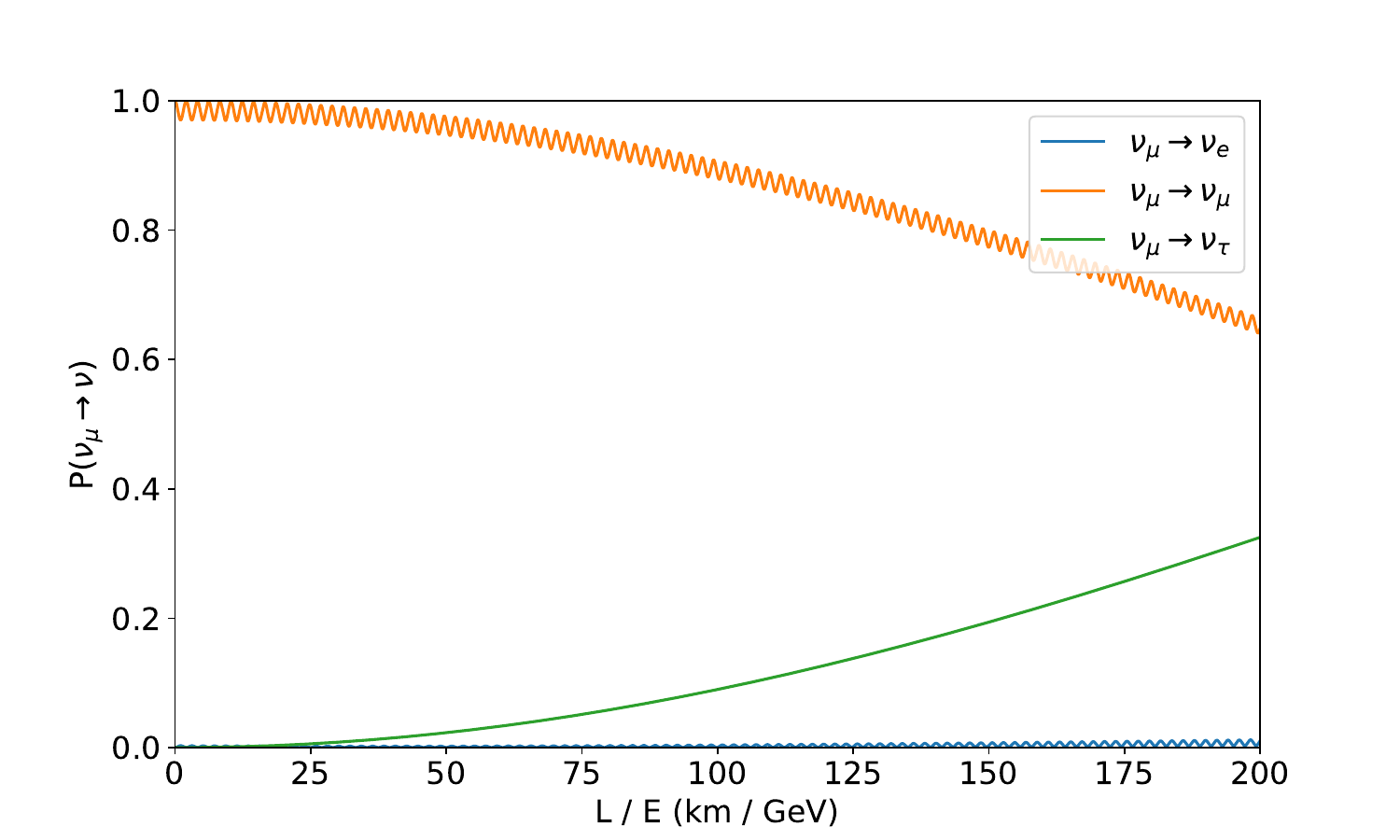}
        \caption{}
        \label{fig:sterile_oscillations_c}
    \end{subfigure}
    \begin{subfigure}[b]{0.4\textwidth}
        \centering
        \includegraphics[width=\textwidth]{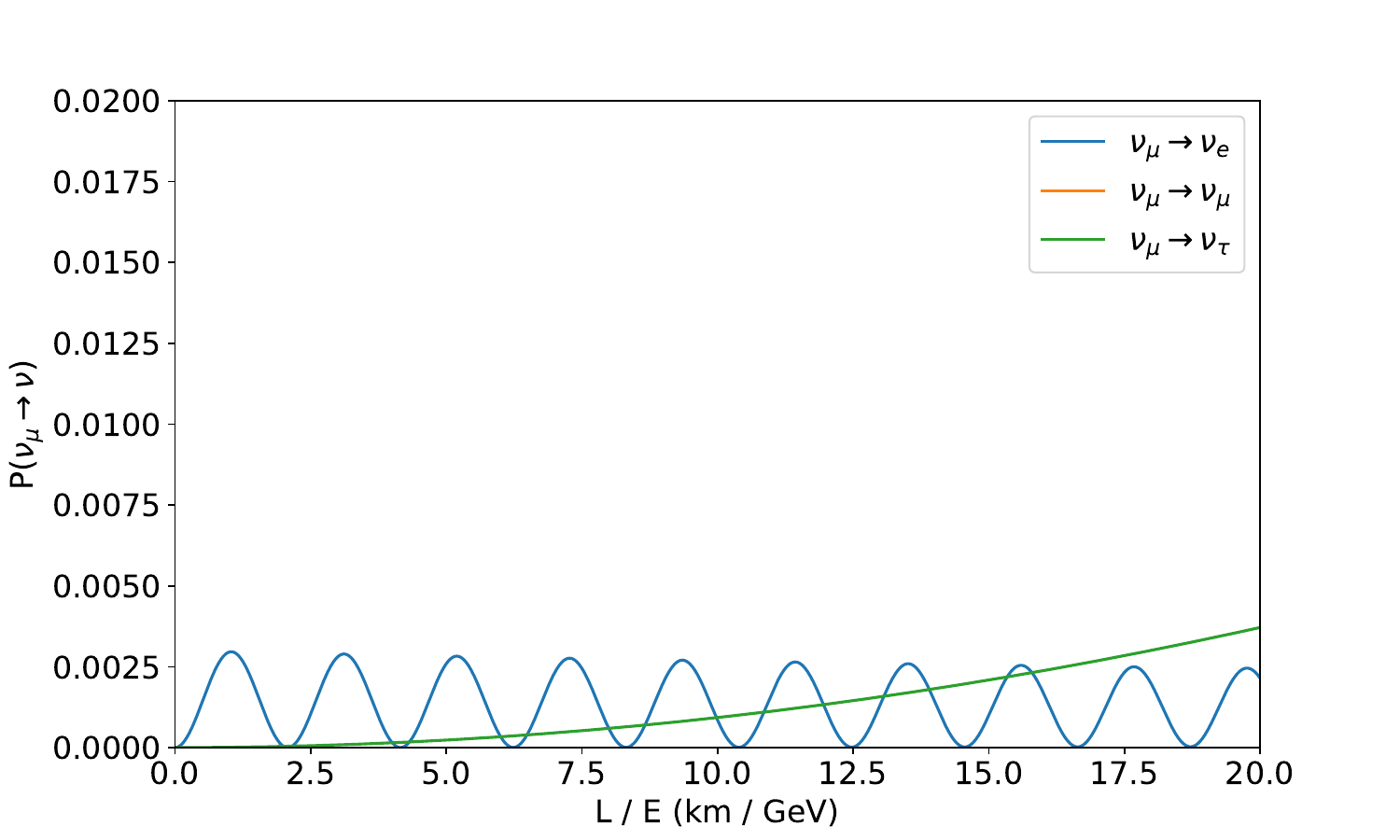}
        \caption{}
        \label{fig:sterile_oscillations_d}
    \end{subfigure}
    \begin{subfigure}[b]{0.4\textwidth}
        \centering
        \includegraphics[width=\textwidth]{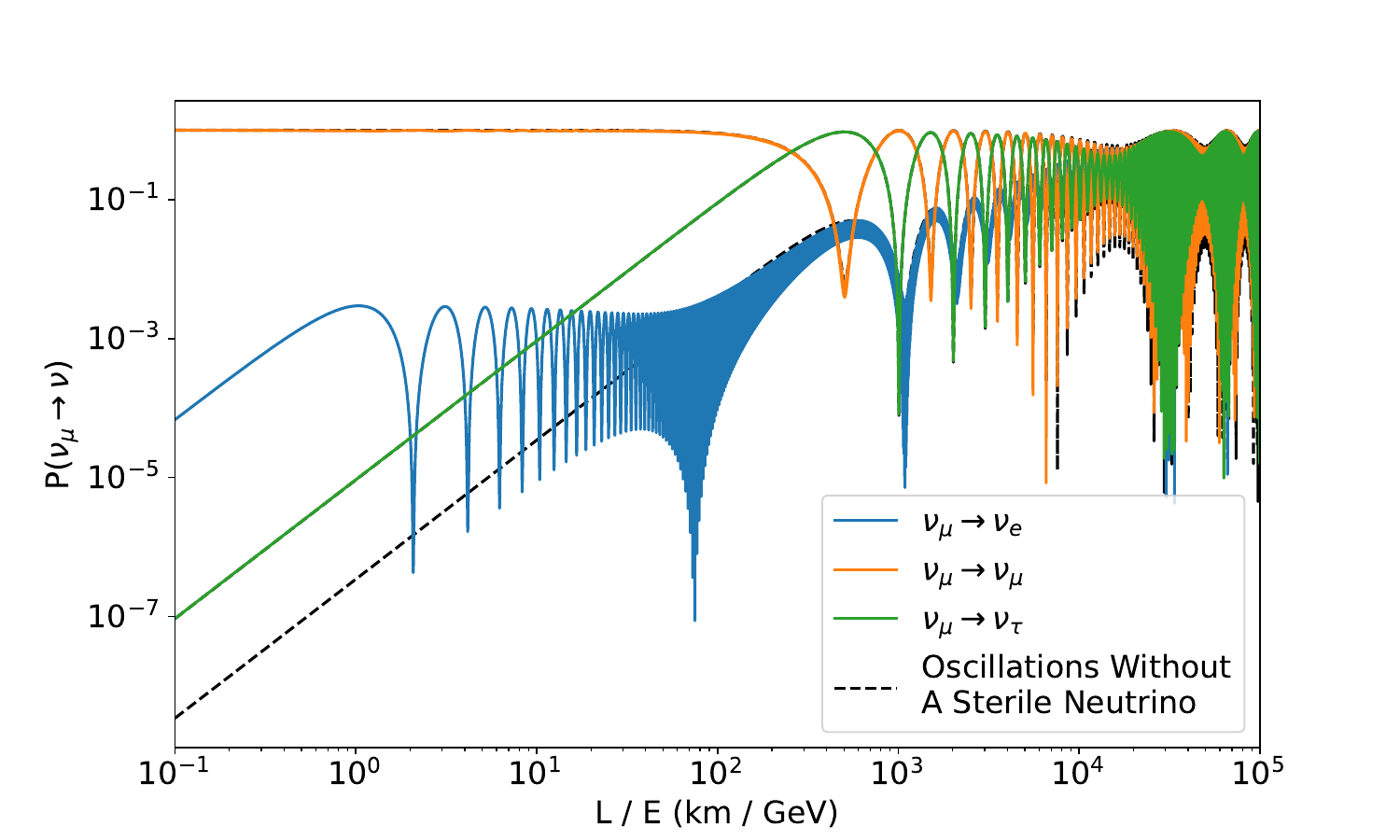}
        \caption{}
        \label{fig:sterile_oscillations_e}
    \end{subfigure}
    \begin{subfigure}[b]{0.4\textwidth}
        \centering
        \includegraphics[width=\textwidth]{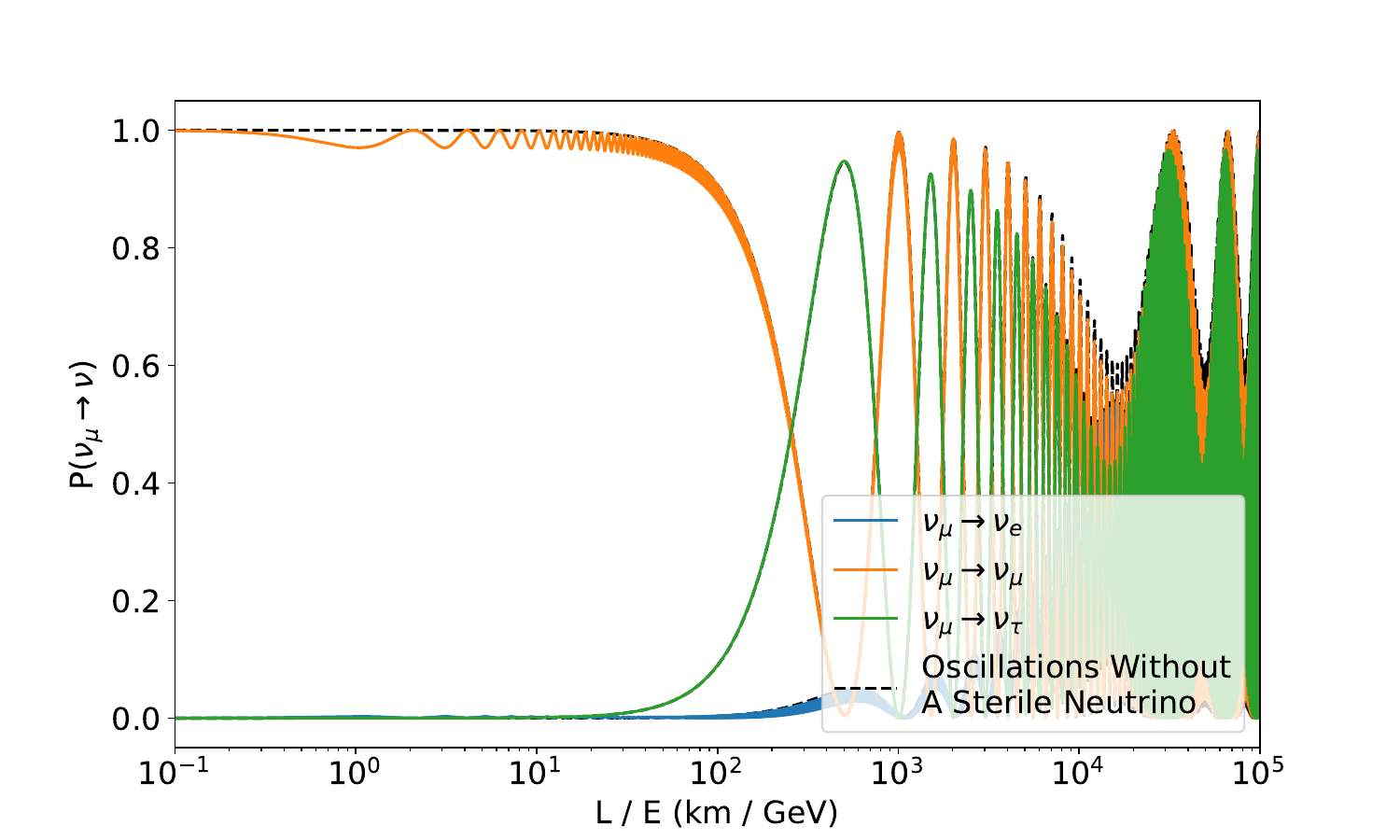}
        \caption{}
        \label{fig:sterile_oscillations_f}
    \end{subfigure}
    \caption[Sterile neutrino oscillation probabilities]{Neutrino oscillation probabilities starting from the $\nu_\mu$ state, according to the normal ordering best fit to world data from NuFit 6.0 \cite{nufit, nufit_website}, with an additional sterile state consistent with the LSND best-fit point \cite{LSND}, 
    $\Delta m^2_{14} = 1.2\ \mathrm{eV}^2/c^4$, $\sin^2 2\theta_{\mu e} = 0.003$. Specifically, the full 3+1 oscillation paramaters we show are
    $\Delta m^2_{14} = 1.2\ \mathrm{eV}^2/c^4$,
    $\theta_{14} = 18.4^\circ$,
    $\theta_{24} = 5.2^\circ$, 
    $\theta_{34}=0$, 
    $\delta_{24}=0$,
    $\delta_{34}=0$. This corresponds to 
    $\sin^2(2 \theta_{\mu e}) = 0.003$,
    $\sin^2(2 \theta_{e e}) = 0.36$, and
    $\sin^2(2 \theta_{\mu \mu}) = 0.0298$.
    Panels (a)-(d) successively zoom into smaller $L/E$ values. Panels (e) and (f) show a broad range of $L/E$ values on a logarithmic scale, with and without a logarithmic y-axis. Code for these plots is available at \cite{lhagaman_neutrino_mixing_plots}.}
    \label{fig:sterile_oscillations}
\end{figure}

\begin{figure}[H]
    \centering
    \includegraphics[width=0.5\textwidth]{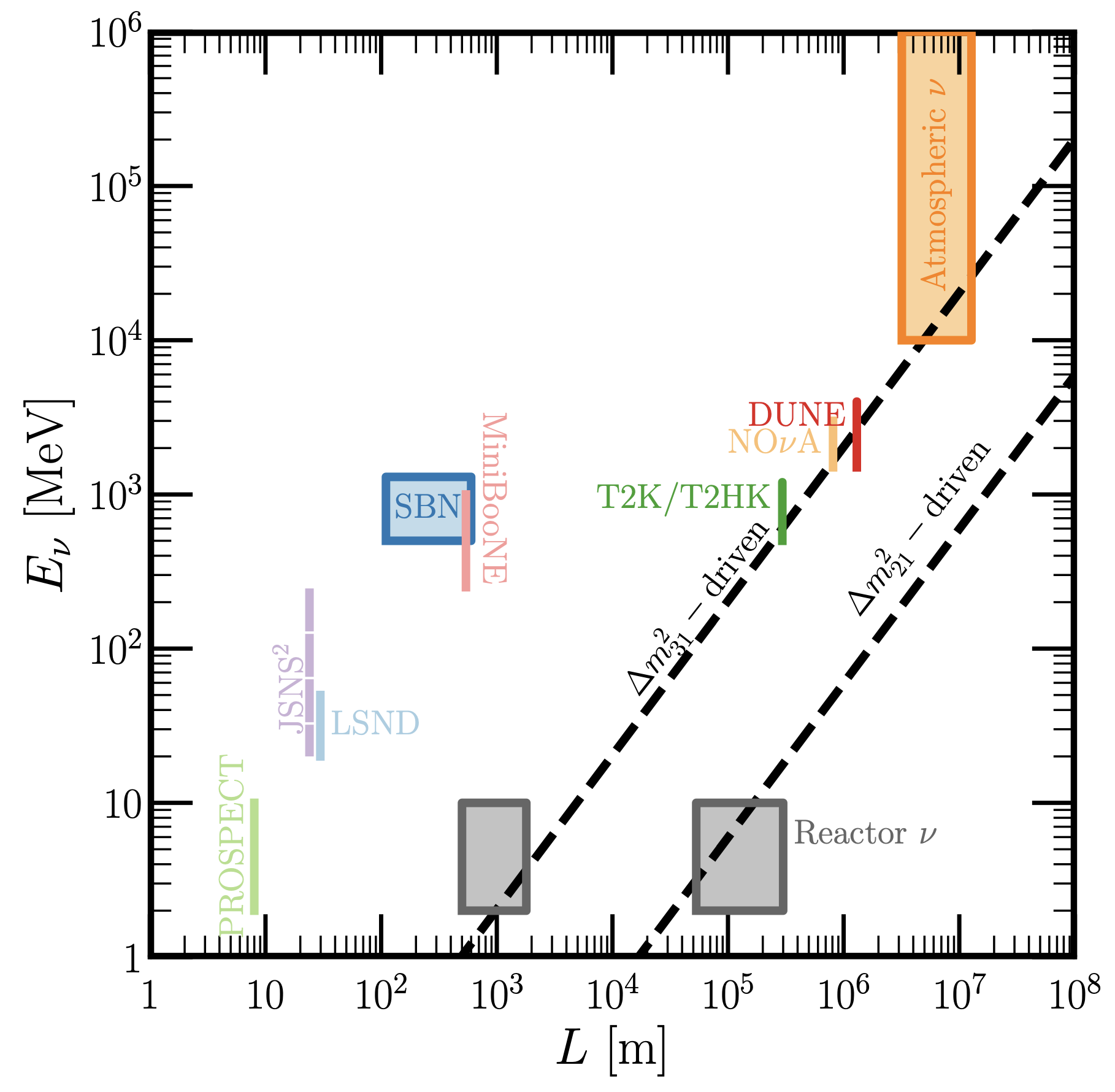}
    \caption[Sterile neutrino oscillation experiment baselines and energies]{Visualization of some neutrino oscillation experiment baselines and energies, including sterile neutrino searches. Figure taken from Ref. \cite{Kelly2021MiniSBNTH}.}
    \label{fig:L_vs_E_w_sterile}
\end{figure}

As an example, Fig. \ref{fig:sterile_oscillations} shows how an $\mathcal{O}(\mathrm{eV}^2)$ scale sterile neutrino could affect oscillation probabilities. We primarily expect this effect to appear as small changes in the oscillation probability at short baselines corresponding to small $L/E$ values. Figure \ref{fig:L_vs_E_w_sterile} illustrates how some sterile neutrino experiments probe smaller $L/E$ regions compared to traditional oscillation experiments.

We will now outline several experiments whose results could potentially hint at the existence of a sterile neutrino.

\subsection{Reactor Antineutrino Anomaly}\label{sec:RAA}

\begin{figure}[H]
    \centering
    \includegraphics[width=0.7\textwidth]{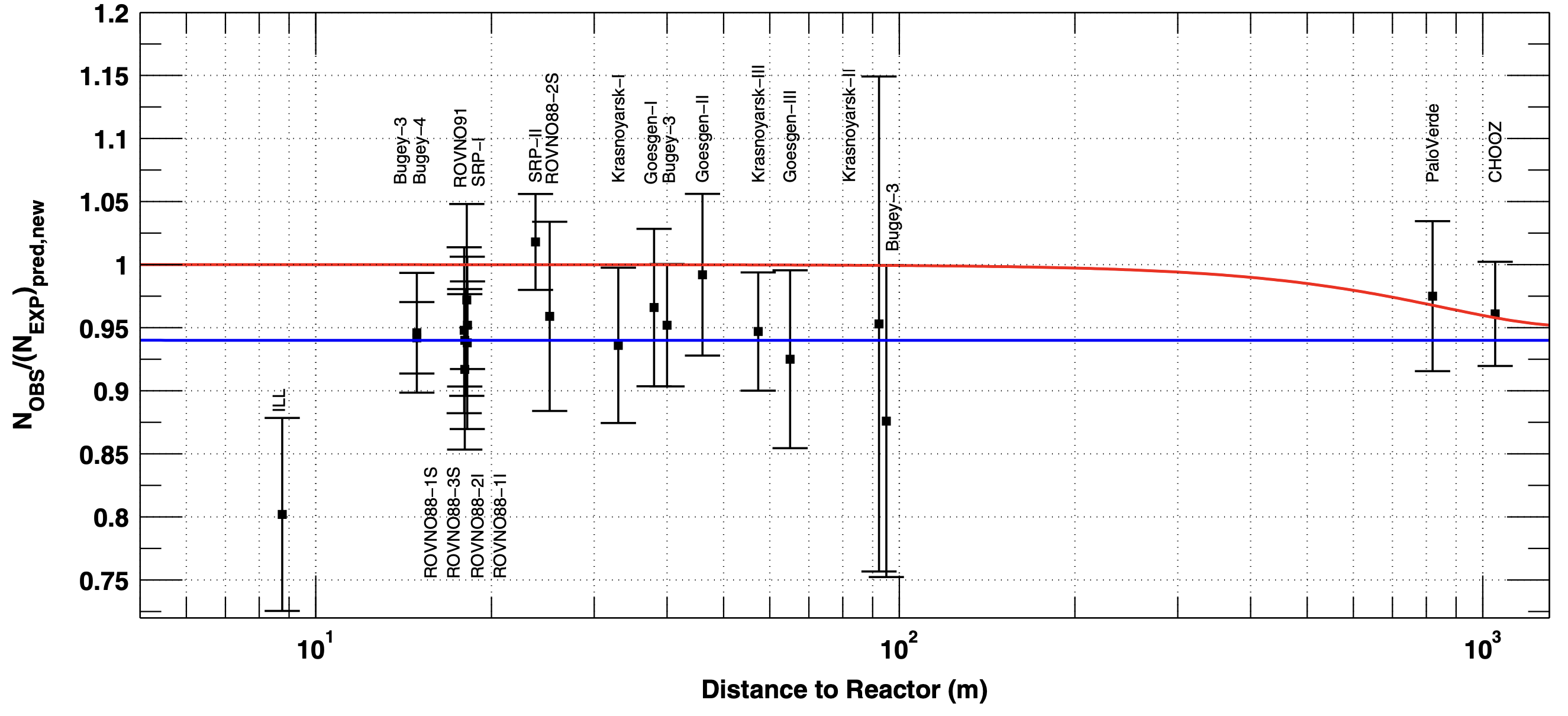}
    \caption[Reactor Antineutrino Anomaly as a function of distance]{Reactor Neutrino Anamaly as a function of distance, taken from Ref. \cite{reactor_anomaly}.}
    \label{fig:RAA_distances}
\end{figure}

\begin{figure}[H]
    \centering
    \includegraphics[width=0.6\textwidth]{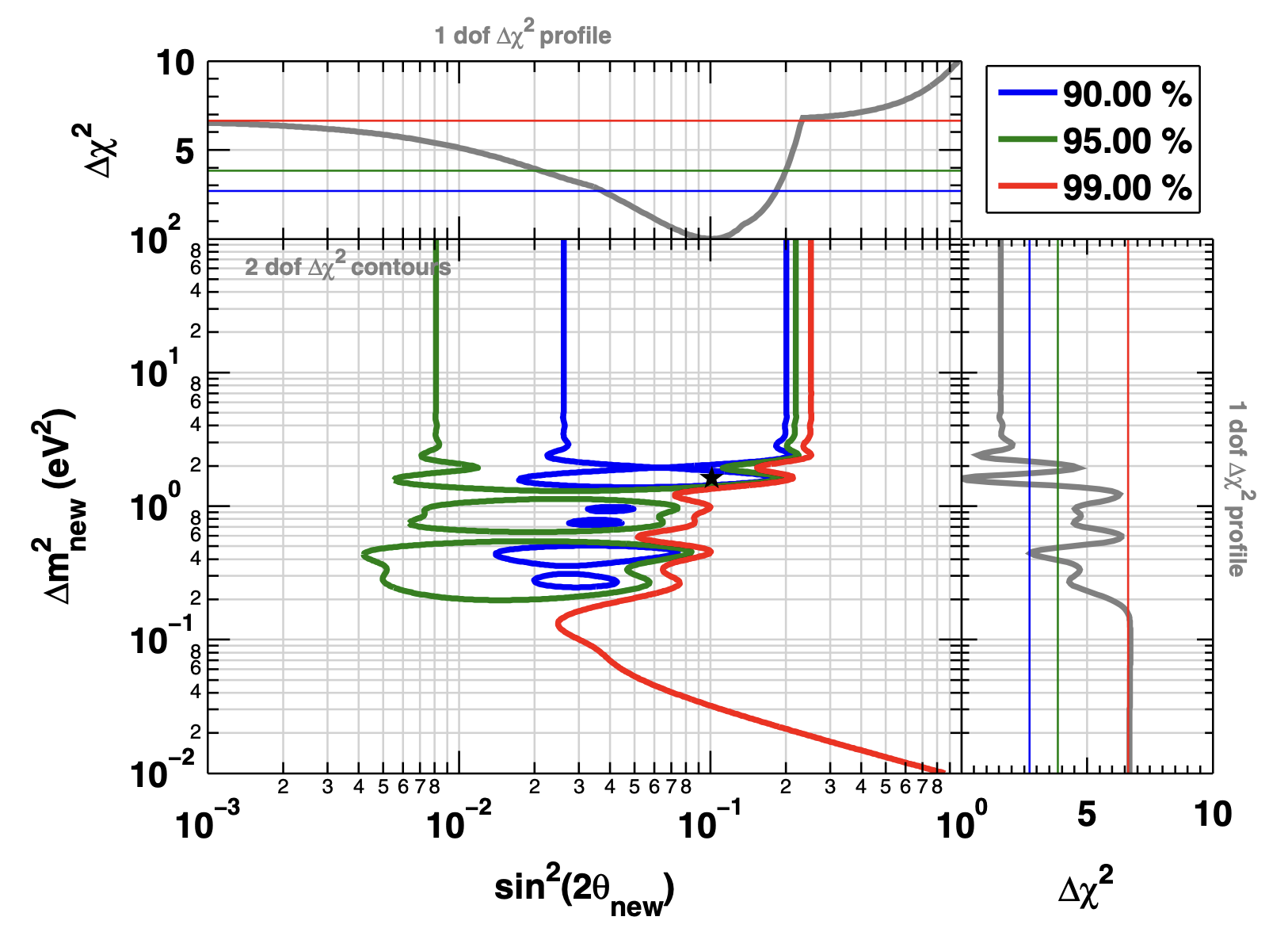}
    \caption[Reactor Antineutrino Anomaly allowed 3+1 phase space]{Reactor Antineutrino Anamaly allowed 3+1 sterile neutrino phase space, taken from Ref. \cite{reactor_anomaly}.}
    \label{fig:RAA_allowed}
\end{figure}

The Reactor Antineutrino Anomaly (RAA) describes the fact that neutrino experiments at nuclear reactors have seen fewer $\overline{\nu}_e$ events than expected \cite{reactor_anomaly}, as shown in Fig. \ref{fig:RAA_distances}. There has been speculation that this could be due to short-baseline $\overline{\nu_e}\rightarrow \overline{\nu_e}$ disappearance due to a sterile neutrino, with an allowed space shown in Fig. \ref{fig:RAA_allowed}.

Nuclear reactors produce antineutrinos from beta decays of many isotopes along the $^{235}\mathrm{U}$ and $^{239}\mathrm{Pu}$ fission decay chains. This involves many rare short lived isotopes which have not had their beta decay energy spectrums measured independently, so the calculation of the precise neutrino flux from all of these beta decays is very challenging. A common calculation is the Huber-Mueller model \cite{huber_reactor_flux, mueller_reactor_flux}, which relies on measurements of the total beta decay spectrum from all fission-induced nuclei at experiments like the Institut Laue-Langevin (ILL) reactor \cite{ILL_reactor_beta_spectrum}.

In addition to this deficit, many reactor neutrino experiments have seen a ``5 MeV bump'', an excess of data over prediction between 4-6 MeV. This observation has been consistent at many experiments and many baselines, and it cannot be explained by a sterile neutrino hypothesis. It has therefore been understood as evidence of a potential miscalculation of the Huber-Mueller model and its uncertainties.

Recently, new neutrino flux calculations have been performed by directly summing predicted beta spectra from different nuclei in the fission decay chains, known as a ``summation method'' \cite{raa_summation}. These new flux calculations significantly reduce the tension between reactor neutrino data and experiment.

Reactor experiments which can simultaneously probe multiple baselines, and therefore ignore the details of the flux modeling, have fairly consistently seen no evidence for sterile neutrinos, and have significantly constrained the available phase space of the RAA. This includes BUGEY \cite{BUGEY_sterile}, NEOS \cite{NEOS_sterile}, DANSS \cite{DANSS_sterile}, RENO \cite{RENO_sterile}, STEREO \cite{STEREO_sterile}, PROSPECT \cite{PROSPECT_sterile}, and Double-Chooz and Daya Bay \cite{double_chooz_daya_bay_reno_sterile}.

Today, the community no longer sees the RAA as compelling evidence for the existence of sterile neutrinos. A comprehensive review is available in Ref. \cite{RAA_review}.

\subsection{Neutrino-4 Anomaly}

\begin{figure}[H]
    \centering
    \begin{subfigure}[b]{0.49\textwidth}
        \includegraphics[width=\textwidth]{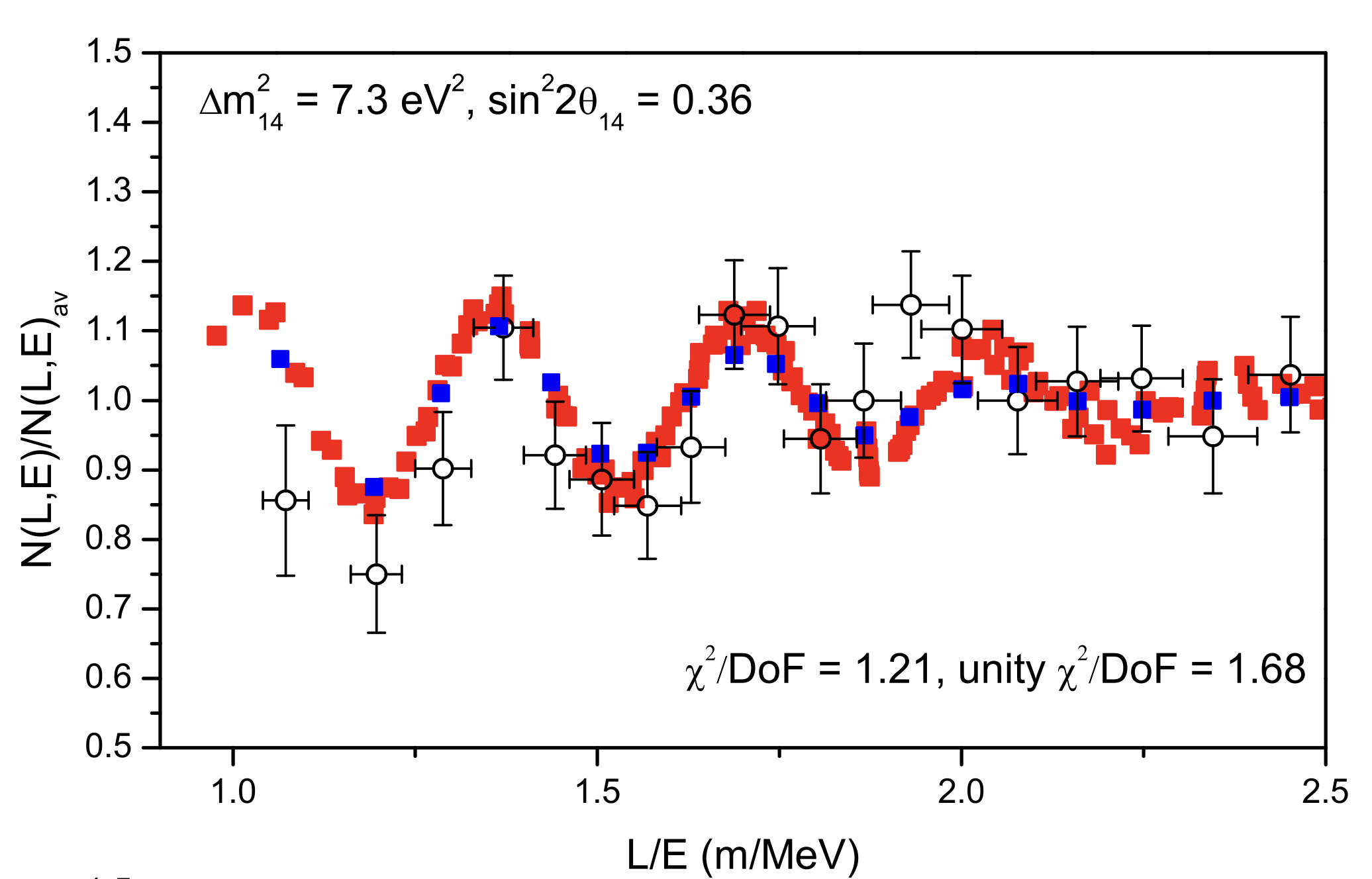}
        \caption{}
        \label{fig:neutrino_4_oscillation}
    \end{subfigure}
    \hfill
    \begin{subfigure}[b]{0.49\textwidth}
        \includegraphics[width=\textwidth]{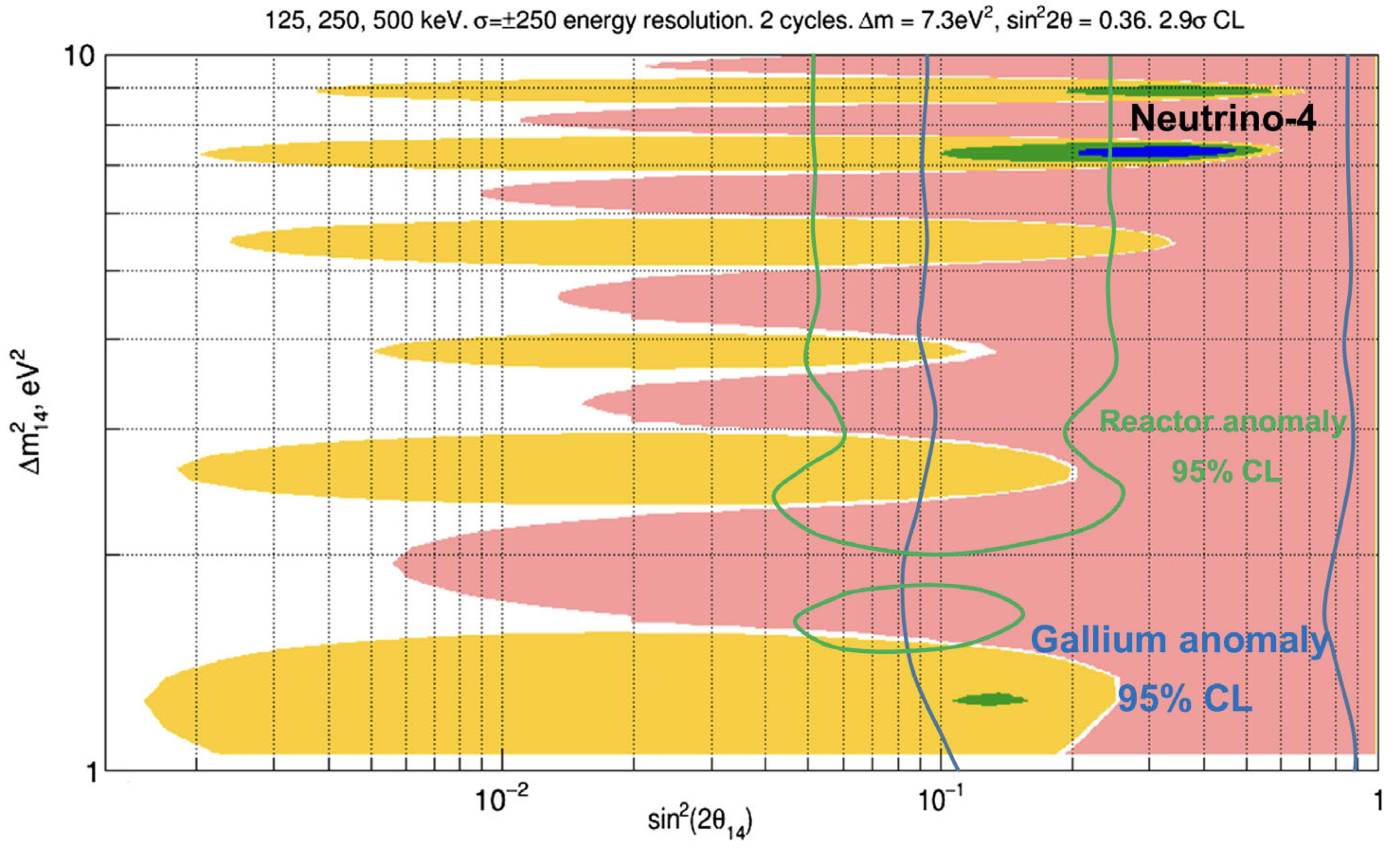}
        \caption{}
        \label{fig:neutrino_4_phase_space}
    \end{subfigure}
    \caption[Neutrino-4 anomaly]{Evidence for sterile neutrinos from Neutrino-4 \cite{neutrino_4_sterile}. Panel (a) shows the spectrum of $L/E$ showing oscillatory behavior. Panel (b) shows the preferred sterile neutrino phase space of $\Delta m^2$ and mixing angle.}
    \label{fig:neutrino_4}
\end{figure}

As described in Section \ref{sec:RAA}, most short baseline reactor experiments do not see evidence of sterile neutrinos, but there is one exception. The Neutrino-4 experiment does claim to see evidence of sterile oscillations \cite{neutrino_4_sterile} with $2.9 \sigma$ confidence, as shown in Fig. \ref{fig:neutrino_4}. A clear oscillatory pattern in $L/E$ is present, corresponding to a fairly narrow allowed set of $\Delta m^2$ frequencies near $7.3 \mathrm{eV}^2$, but a fairly large allowed range of mixing angles.

This result has received scrutiny from physicists outside of the experiment, who implement a simulation that considers detector energy resolution and estimate a significance of only $2.2\sigma$, and find that the Neutrino-4 results are in strong tension with KATRIN, PROSPECT, STEREO, and solar $\nu_e$ bounds \cite{neutrino_4_criticism}.

\subsection{Gallium Anomaly}

The ``gallium anomaly'' describes the fact that GALLEX \cite{gallex}, SAGE \cite{sage}, and most recently BEST \cite{BEST_results} observe lower rates of $\nu_e$\ than expected from calibrated $^{51}$Cr and $^{37}$Ar radioactive electron capture decay sources. These experiments observe about 80\% as many results as expected, an overall significance of about $4 \sigma$ \cite{gallium_review}.

\begin{figure}[H]
    \centering
    \begin{subfigure}[b]{0.49\textwidth}
        \includegraphics[width=\textwidth]{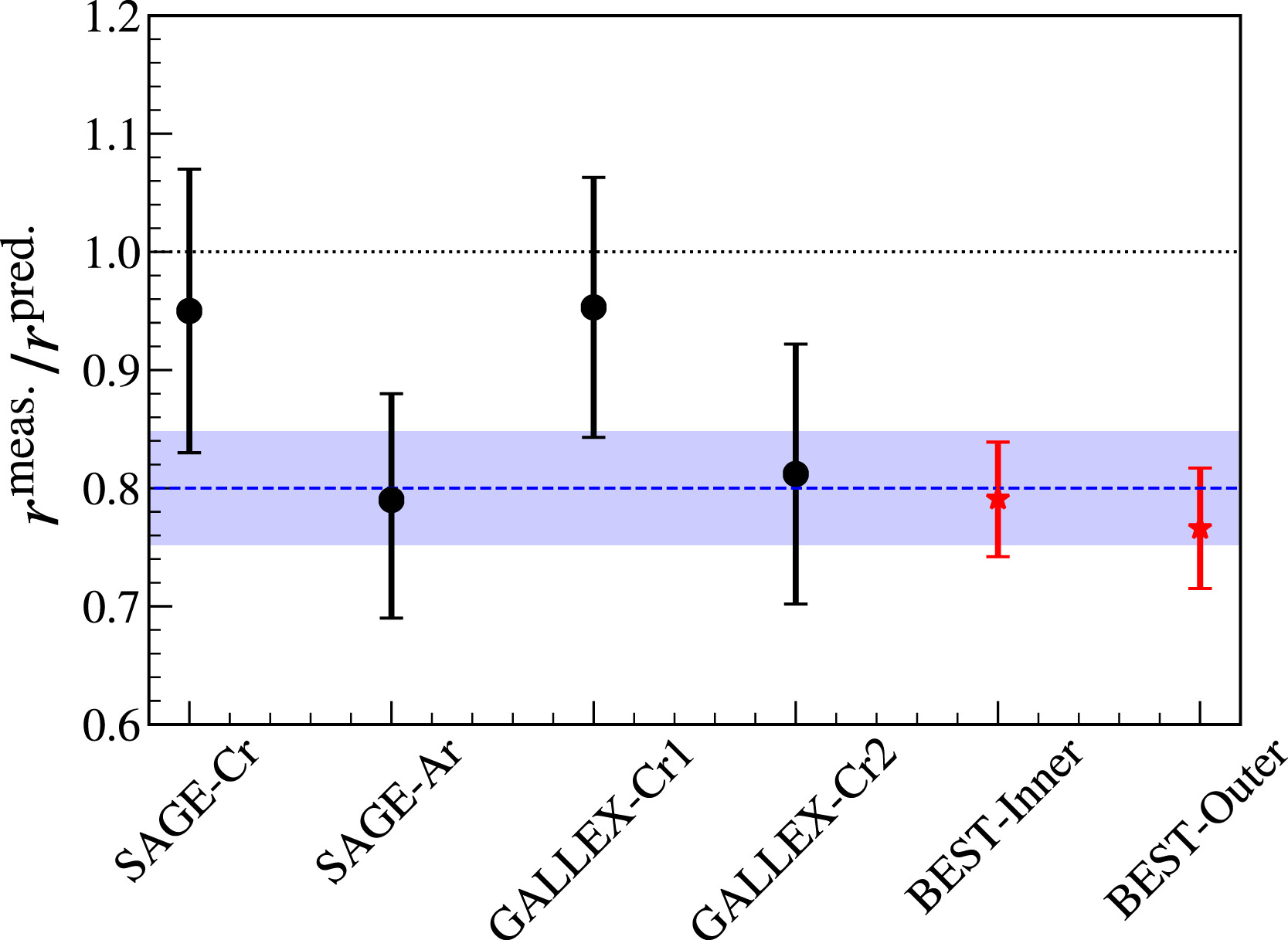}
        \caption{}
        \label{fig:gallium_norms}
    \end{subfigure}
    \hfill
    \begin{subfigure}[b]{0.49\textwidth}
        \includegraphics[width=\textwidth]{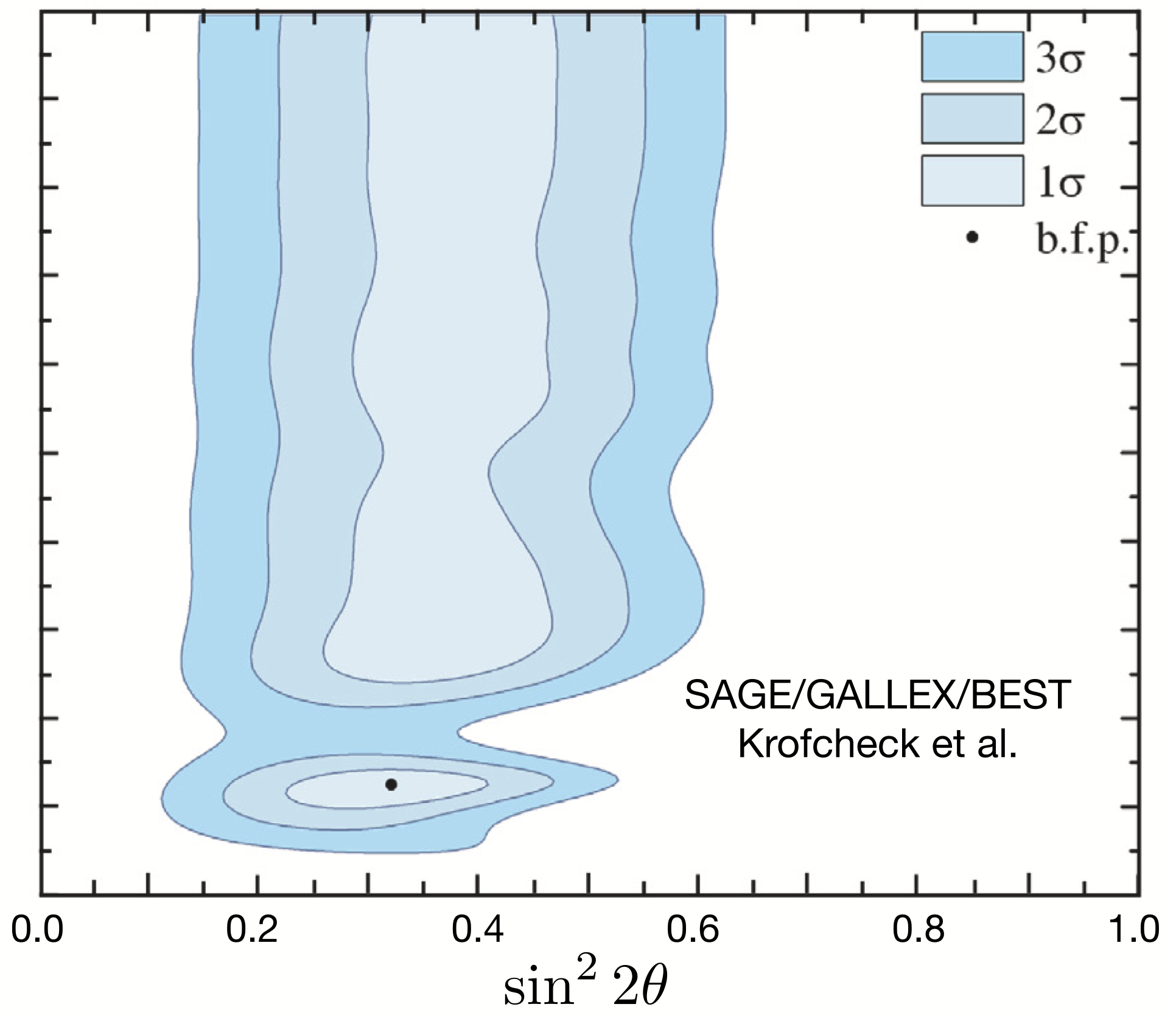}
        \caption{}
        \label{fig:gallium_allowed}
    \end{subfigure}
    \caption[Gallium anomaly]{Evidence for sterile neutrinos from GALLEX, SAGE, and BEST \cite{gallium_review}. Panel (a) shows the normalization of events observed by each experiment. Panel (b) shows the preferred sterile neutrino phase space of $\Delta m^2$ and mixing angle.}
    \label{fig:gallium_anomaly}
\end{figure}

These results are compelling, but could potentially suffer from miscalibrations in the neutrino flux or the neutrino cross section, and no oscillatory $L/E$ behavior has been observed. Like the other anomalies we have discussed, these results are in tension with KATRIN, reactor neutrino experiments, and solar neutrino experiments.

\subsection{LSND Anomaly}

\begin{figure}[H]
    \centering
    \begin{subfigure}[b]{0.49\textwidth}
        \includegraphics[width=\textwidth]{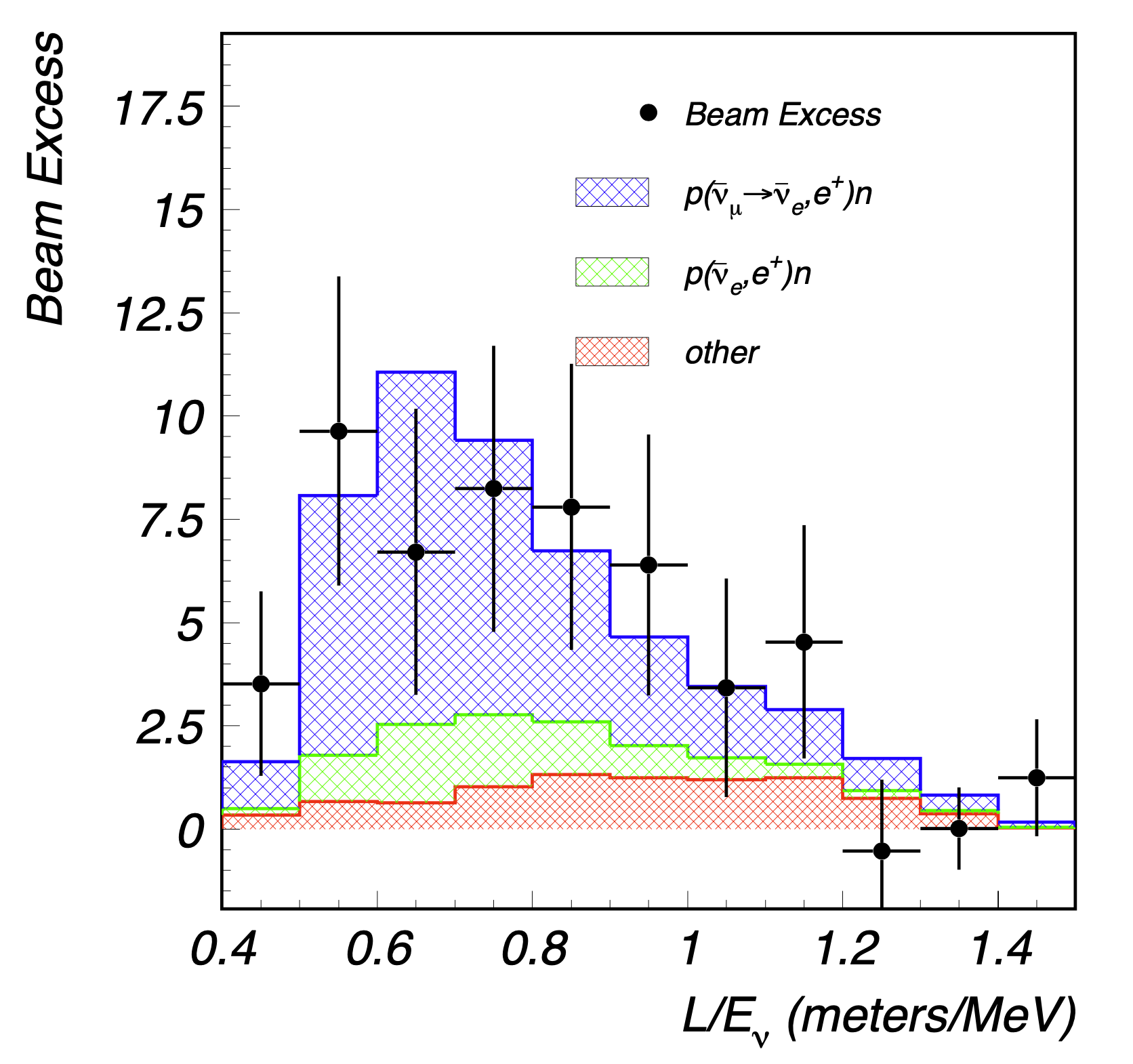}
        \caption{}
        \label{fig:LSND_spectrum}
    \end{subfigure}
    \hfill
    \begin{subfigure}[b]{0.49\textwidth}
        \includegraphics[width=\textwidth]{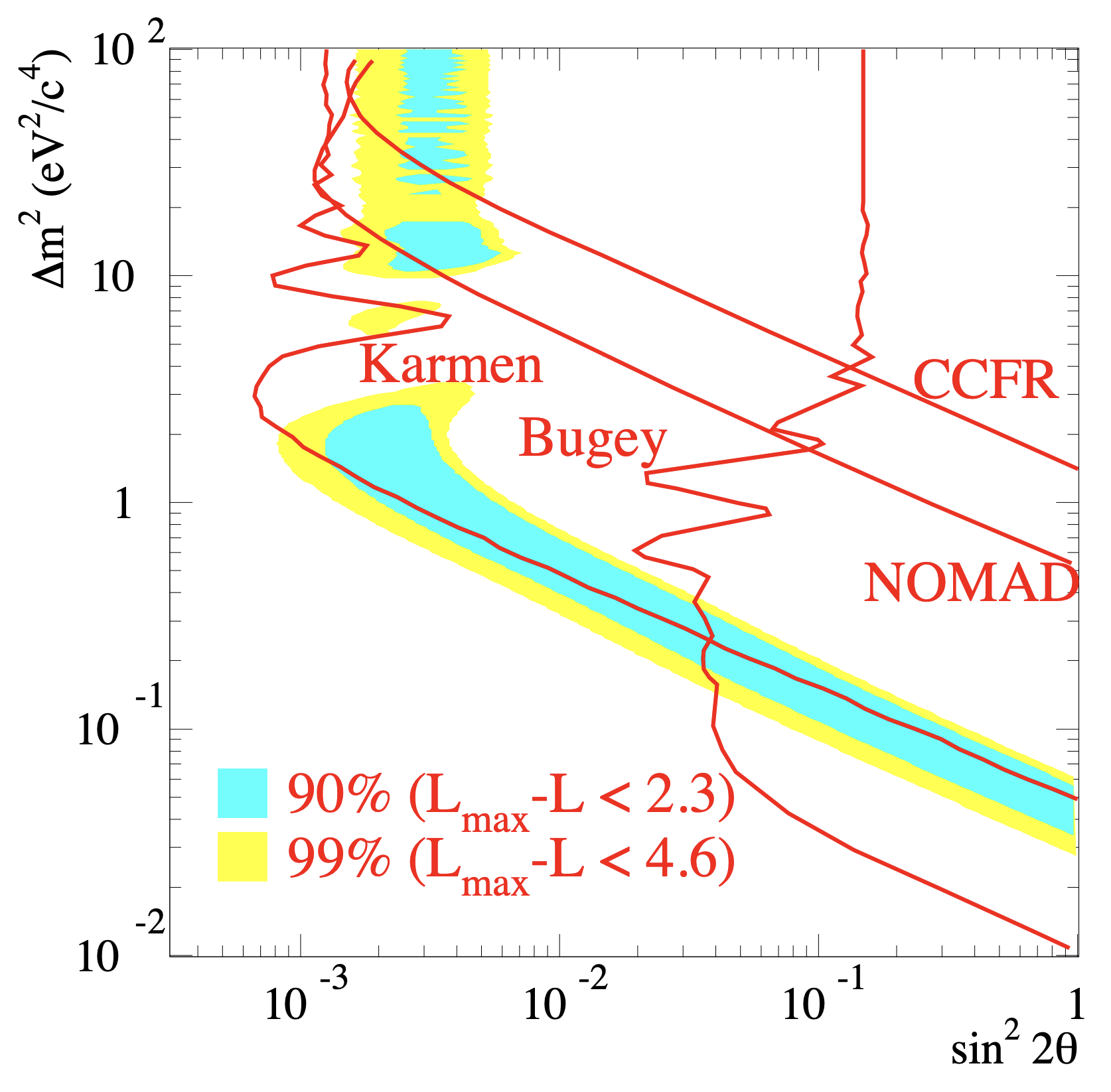}
        \caption{}
        \label{fig:LSND_allowed}
    \end{subfigure}
    \caption[LSND anomaly]{Evidence for sterile neutrinos from the LSND experiment \cite{LSND}. Panel (a) shows the $L/E$ spectrum, with the prediction in green and red, and the extra prediction from the best fit model with a sterile neutrino in blue. Panel (b) shows the preferred sterile neutrino phase space of $\Delta m^2$ and mixing angle.}
    \label{fig:LSND}
\end{figure}

The Liquid Scintillator Neutrino Detector (LSND) is an experiment observing neutrinos from a stopped $\pi^+$ beam and secondary stopped $\mu^+$ beam. Like reactor experiments, the LSND detector uses the inverse beta decay process, $\overline{\nu_e} + p \rightarrow e^+ + n$, and is therefore only sensitive to $\overline{\nu_e}$. The stopped $\mu^+$ beam produces neutrinos via $\mu^+ \rightarrow e^+ + \overline{\nu_\mu} + \nu_e$, creating no neutrinos that can interact via the inverse beta decay process. However, in a $\overline{\nu_e}$ search, LSND observed a 3.8$\sigma$ excess, as shown in Fig. \ref{fig:LSND}. This can be interpreted as evidence for the existence of short baseline $\overline{\nu_\mu} \rightarrow \overline{\nu_e}$ oscillations due to a sterile neutrino with $\Delta m^2$ in the 0.2-10 $\text{eV}^2/c^4$ range. The best fit oscillation parameters are $\text{sin}^2\ 2\theta = 0.003$, $\Delta m^2 = 1.2 \mathrm{eV}^2/c^4$.

The LSND results have still not been conclusively explained. The J-PARC Sterile Neutrino Search at J-PARC Spallation Neutron Source ($\mathrm{JSNS}^2$) experiment is currently operating using the same beam type and detection principle, and could shed light on these results in the near future.

\subsection{MiniBooNE Low Energy Excess}\label{sec:MB_LEE}

\begin{figure}[H]
    \centering
    \includegraphics[width=0.7\textwidth]{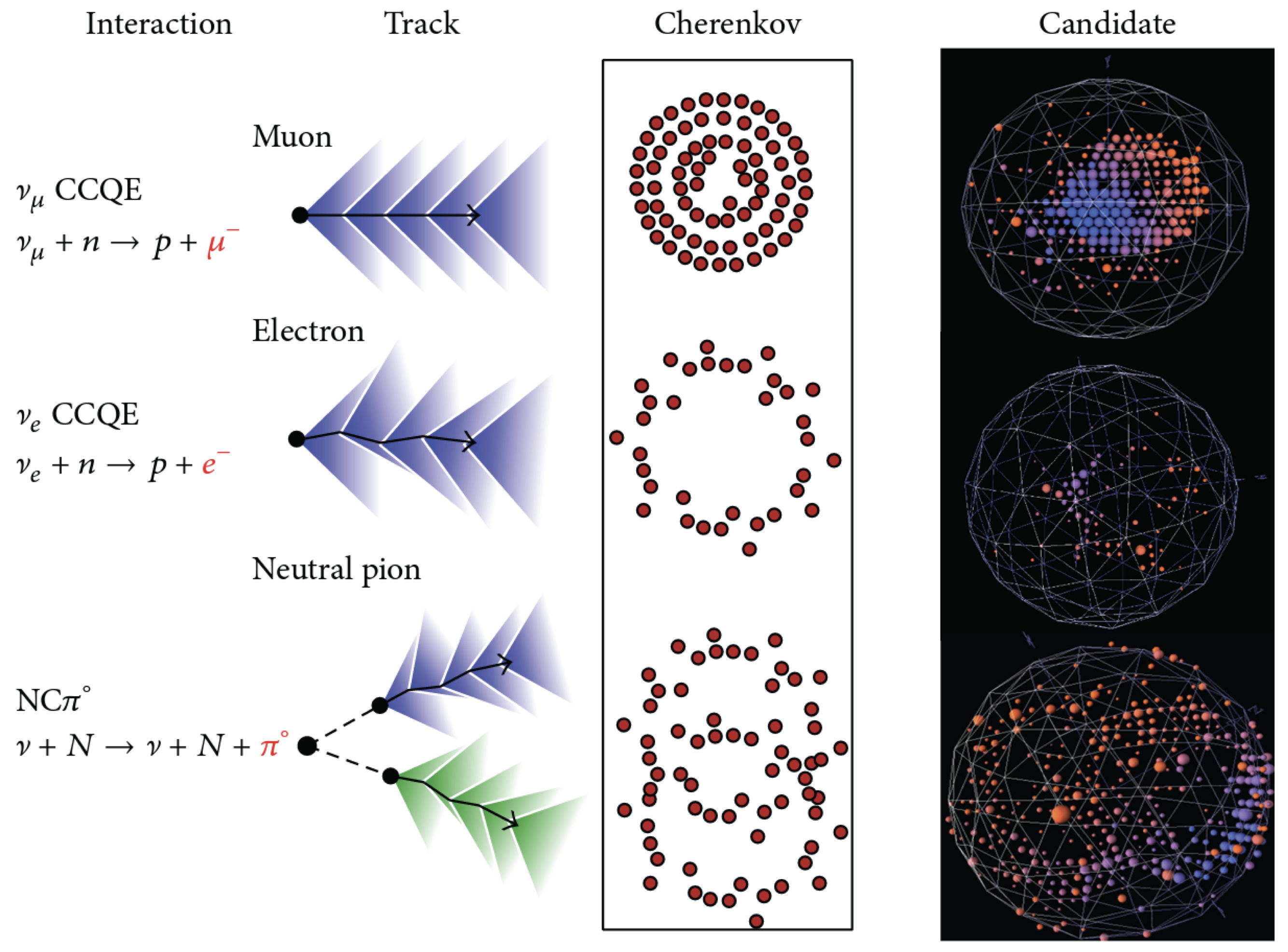}
    \caption[MiniBooNE Cherenkov reconstruction]{Illustration of MiniBooNE Cherenkov cone event reconstruction. Figure from Ref. \cite{foppiani_thesis}.}
    \label{fig:miniboone_cherenkov_mechanism}
\end{figure}

The MiniBooNE experiment at Fermi National Accelerator Laboratory (Fermilab) was designed to examine the LSND anomaly by increasing both the baseline and energy by about a factor of ten, therefore probing the same $L/E$ range while significantly changing the types of systematic uncertainties involved. It primarily used Cherenkov light to identify particles produced in neutrino interactions from the Booster Neutrino Beam (BNB), a beam of primarily $\nu_\mu$ or $\overline{\nu}_\mu$ neutrinos, in order to search for $\nu_\mu \rightarrow \nu_e$\ and $\overline{\nu}_\mu \rightarrow \overline{\nu}_e$ oscillations. In MiniBooNE, smoother Cherenkov rings represent tracks from heavy particles like muons, while fuzzier rings represent electromagnetic showers, from electrons or photons, as shown in Fig. \ref{fig:miniboone_cherenkov_mechanism}.

\begin{figure}[H]
    \centering
    \begin{subfigure}[b]{0.49\textwidth}
        \includegraphics[width=\textwidth]{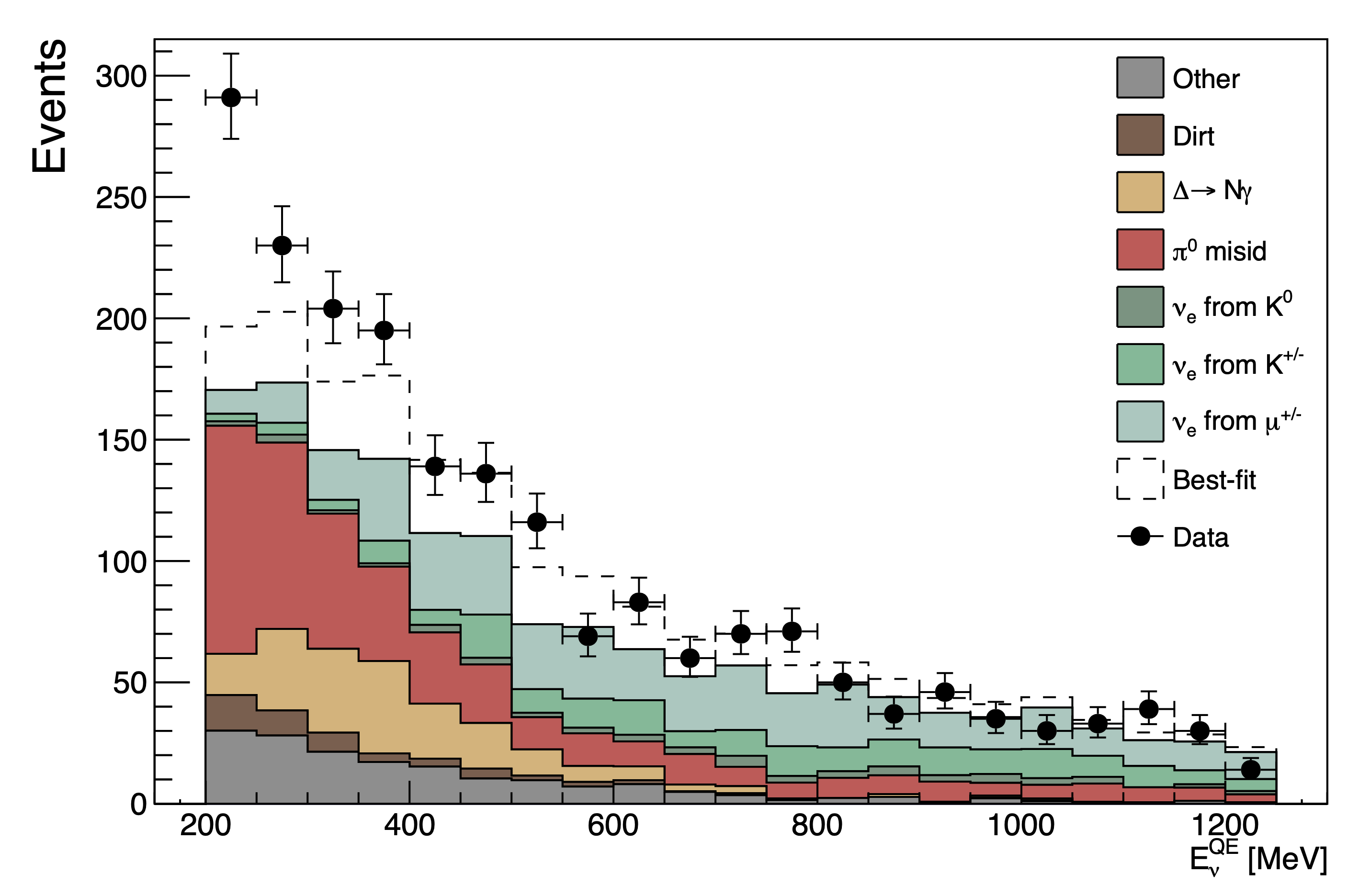}
        \caption{}
        \label{fig:miniboone_E_nu_QE}
    \end{subfigure}
    \hfill
    \begin{subfigure}[b]{0.49\textwidth}
        \includegraphics[width=\textwidth]{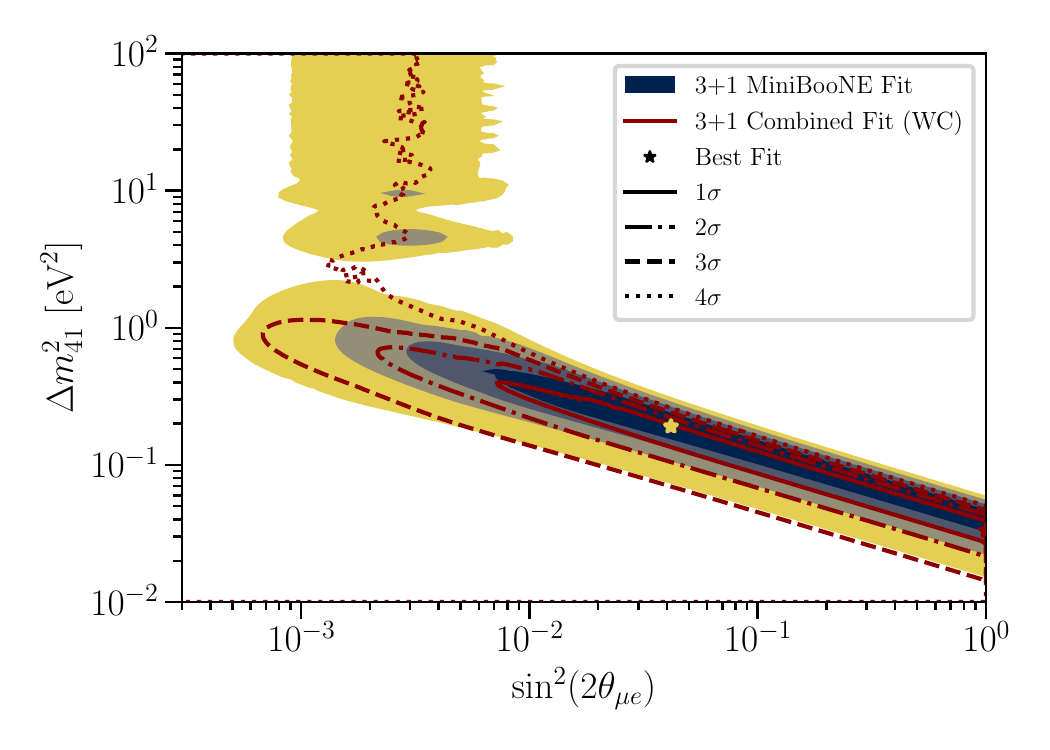}
        \caption{}
        \label{fig:miniboone_allowed}
    \end{subfigure}
    \caption[MiniBooNE Low Energy Excess under a 3+1 hypothesis]{Panel (a) shows the MiniBooNE Low Energy Excess as a function of $E_\nu^{QE}$, an estimate of neutrino energy using just the lepton energy and angle, from Ref. \cite{miniboone_lee}. The best-fit sterile oscillation parameters from a simple two-neutrino oscillation $\text{sin}^2 2\theta = 0.807$ and $\Delta m^2 = 0.043\ \mathrm{eV}^2/c^4$ are indicated with a dashed line. Panel (b) shows the shaded allowed regions in a more complete 3+1 sterile neutrino analysis from Ref. \cite{miniboone_sterile} (ignore the dashed lines which involve adding the MicroBooNE experiment to the fit).}
    \label{fig:miniboone_sterile}
\end{figure}

MiniBooNE saw a 4.8$\sigma$ excess of low energy $\nu_e$-like events \cite{miniboone_lee}, which is now commonly referred to as the Low Energy Excess (LEE), shown in Fig. \ref{fig:miniboone_E_nu_QE}. This can be interpreted as additional evidence supporting the LSND observation for short baseline oscillation via the existence of a sterile neutrino. The best fit oscillation parameters are $\text{sin}^2 2\theta = 0.807$, $\Delta m^2 = 0.043\ \mathrm{eV}^2/c^4$. The allowed region in a full 3+1 sterile oscillation framework is shown in Fig. \ref{fig:miniboone_allowed}.

We will have a lot more to say about the MiniBooNE LEE in Sec. \ref{sec:miniboone_photonlike} and the rest of this thesis.

\subsection{IceCube Hint}

Typically, atmospheric neutrinos have very large $L$ values, on the scale of the entire Earth, so atmospheric neutrino experiments can only observe the collective effect of many sterile oscillations in $L/E$. This is what IceCube DeepCore and Super-Kamiokande have reported, exclusions in the mixing of short baseline $\nu_\mu$ oscillations with no sensitivity to the $\Delta m^2_{41}$ mass splitting \cite{deepcore_sterile, super_k_sterile}.

\begin{figure}[H]
    \centering
    \begin{subfigure}[b]{0.49\textwidth}
        \includegraphics[width=\textwidth]{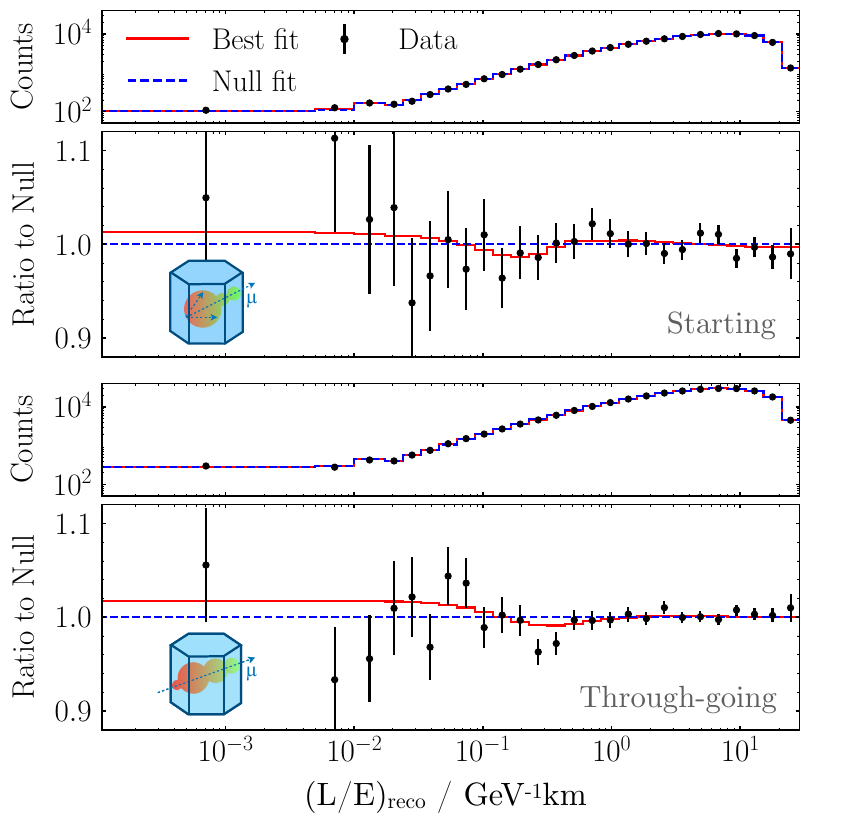}
        \caption{}
        \label{fig:icecube_spectrum}
    \end{subfigure}
    \hfill
    \begin{subfigure}[b]{0.49\textwidth}
        \includegraphics[width=\textwidth]{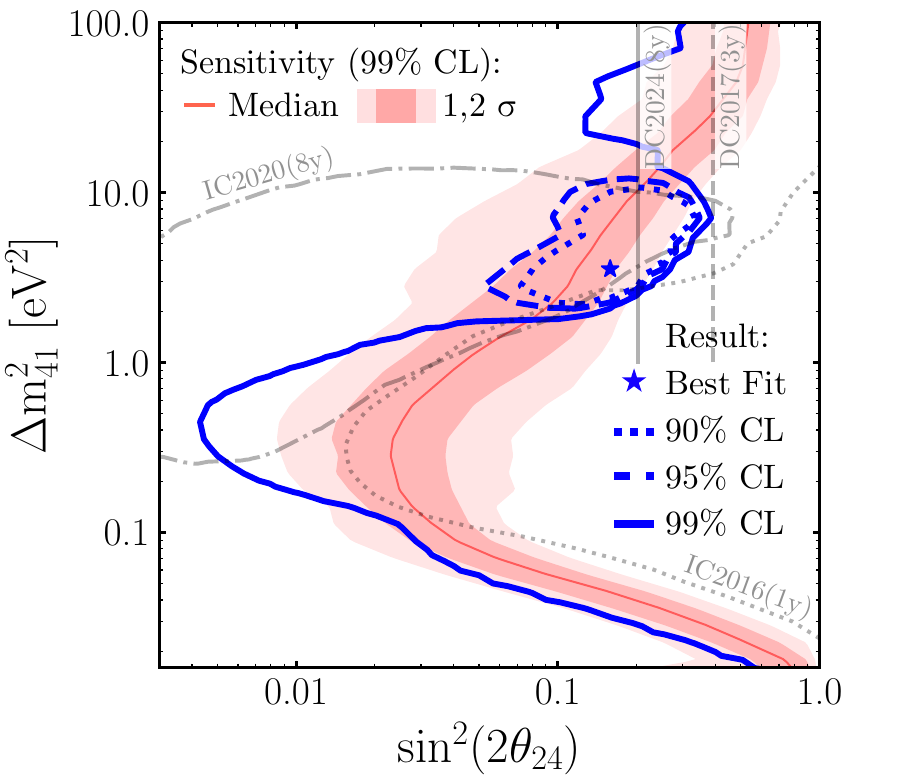}
        \caption{}
        \label{fig:icecube_allowed}
    \end{subfigure}
    \caption[IceCube hint]{Panel (a) shows the IceCube $L/E$ spectrum. Panel (b) shows the IceCube 3+1 allowed region. From Ref. \cite{icecube_sterile}.}
    \label{fig:icecube}
\end{figure}

However, IceCube is a very large detector, giving it the unique capability to measure very high neutrino energies, and therefore probe small $L/E$ values for sterile oscillation searches. They performed this search and saw short baseline $\nu_\mu\rightarrow \nu_\mu$ dissapearance oscillations at $2.2\sigma$ significance \cite{icecube_sterile, icecube_sterile_methods}. The $L/E$ spectra and allowed sterile parameters are shown in Fig. \ref{fig:icecube}.

At less than $3\sigma$ significance, this result is less of an anomaly and more of a hint relative to the other experiments discussed in this section. This hint does remain present after many ways of slicing the IceCube data, into samples from different years, into separate starting muon and throughgoing muon samples, into different energy regions, and into different azimuth regions. This hint is in tension with other $\nu_\mu\rightarrow \nu_\mu$ disappearance searches, particularly MINOS+.

\subsection{Summary Of 3+1 Sterile Neutrino Hints}

\begin{figure}[H]
    \centering
    \includegraphics[width=\textwidth]{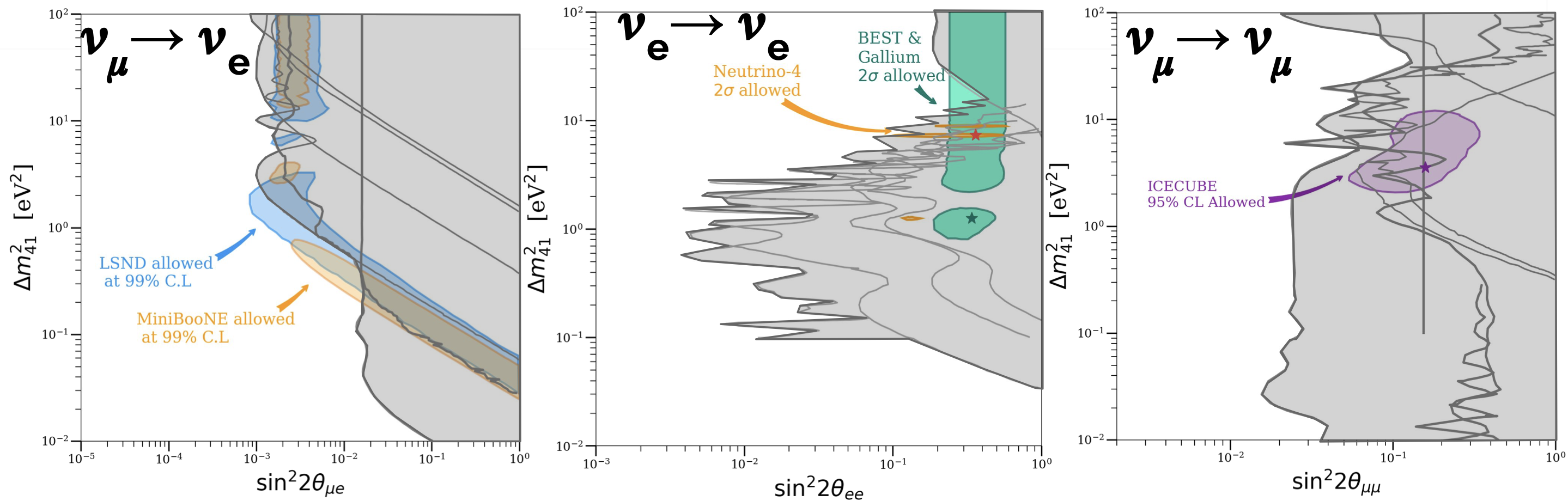}
    \caption[3+1 sterile neutrino summary]{3+1 sterile neutrino summary from Ref. \cite{mark_nnn_sterile}.}
    \label{fig:sterile_global_picture}
\end{figure}

In all three of these commonly explored channels, $\nu_\mu\rightarrow \nu_e$ appearance, $\nu_e\rightarrow \nu_e$ disappearance, and $\nu_\mu\rightarrow \nu_\mu$ disappearance, the number of null results outweighs the number of anomalies. As shown in Fig. \ref{fig:sterile_global_picture}, these null results disfavor much of the allowed phase space from these anomalies.

Additionally, there are some potential exclusions not shown on these plots. If we consider an updated reactor flux normalization model (Estienne-Fallot), what previously was an anomaly hinting towards sterile neutrinos, the RAA, now becomes a strong exclusion disfavoring the entire $2\sigma$ allowed BEST $\nu_e\rightarrow\nu_e$ disappearance region \cite{RAA_update_sterile_exclusion}; however, the best model for reactor flux normalizations is not a completely settled issue. Also, if we consider cosmological data which is sensitive to neutrino masses via their impact on structure formation in the early universe, there is a very significant tension with these preferred regions \cite{cosmology_sterile}; however, this depends on the $\Lambda$CDM model, and other new physics could potentially change these cosmological bounds.

\begin{figure}[H]
    \centering
    \includegraphics[width=0.49\textwidth]{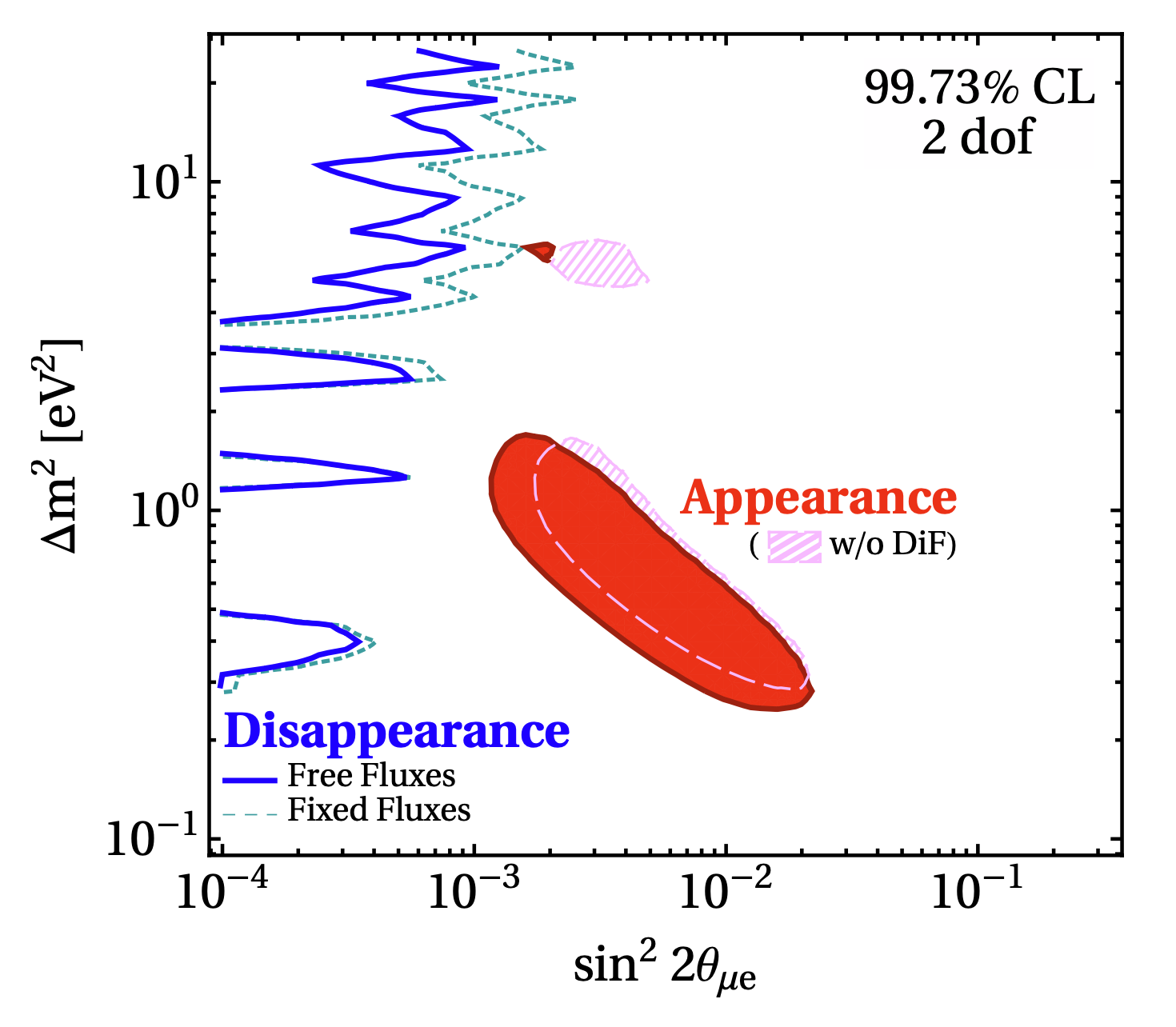}
    \caption[Tensions in the 3+1 sterile neutrino model]{Tension between $\nu_e$ and $\nu_\mu$ disappearance searches and $\nu_e$ appearance searches. The red shows the region allowed by $\nu_e$ appearance searches, and the area to the left of the blue lines is the region allowed by $\nu_\mu$ disappearance searches. From Ref. \cite{sterile_global_fit}}
    \label{fig:sterile_global_tension}
\end{figure}

Normally, each of these sterile oscillation channels is considered independently, but there have also been efforts to combine these hints into a consistent global picture \cite{sterile_global_fit}. These fits have found significant tension between $\nu_e$ and $\nu_\mu$ disappearance searches, as shown in Fig. \ref{fig:sterile_global_tension}.

To summarize, there are several hints pointing toward $\mathcal{O}(\mathrm{eV}^2)$ 3+1 sterile neutrino oscillations, but currently no consistent global understanding of these anomalies. There are also many results that are more consistent with the standard model. In Chapter \ref{sec:nueCC}, I will describe how we used MicroBooNE data to expand our understanding of these anomalies in the context of 3+1 sterile neutrino oscillations.

\section{Photon-like Models of the MiniBooNE LEE}\label{sec:miniboone_photonlike}

The MiniBooNE anomaly is especially compelling because of its high significance. For the main topic of this thesis in Chapter \ref{sec:nc_delta}, I will use data from the MicroBooNE experiment to shed light on the MiniBooNE anomaly from a perspective other than sterile neutrino oscillations.

In the MiniBooNE selection, there is a lot of predicted background from photon showers, which look identical to electron showers in a Cherenkov detector. There exist a lot of these photon backgrounds due to neutrino-induced neutral pion decay $\pi^0\rightarrow \gamma + \gamma$ and rare Delta baryon decay to a nucleon and a photon, $\Delta\rightarrow N \gamma$. This leaves the door open for the excess to be caused by photon showers rather than electron showers, and therefore caused by processes other than sterile neutrino oscillations.

\begin{figure}[H]
    \centering
    \begin{subfigure}[b]{0.49\textwidth}
        \includegraphics[width=\textwidth]{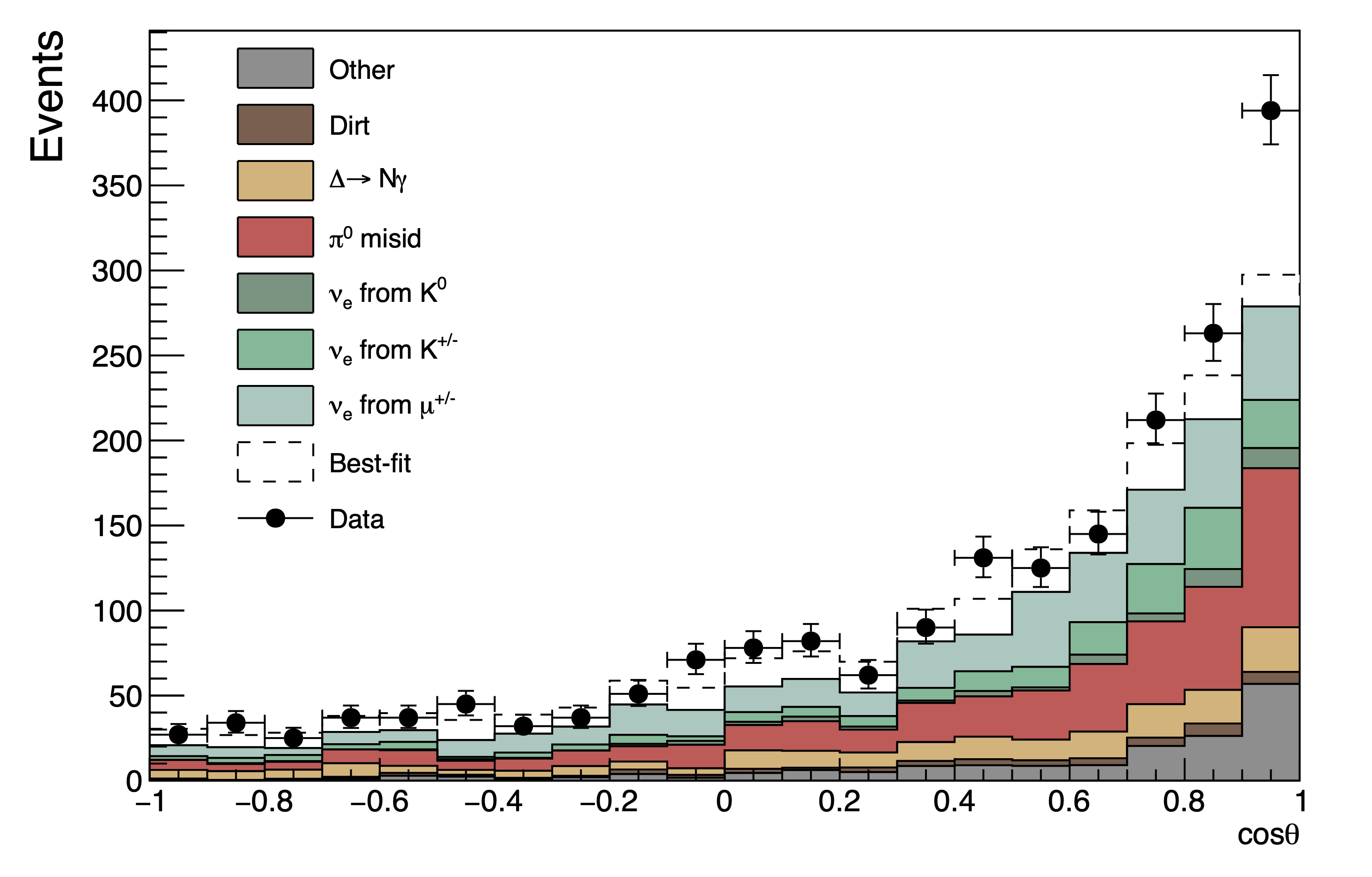}
        \caption{}
        \label{fig:miniboone_angle_costheta}
    \end{subfigure}
    \hfill
    \begin{subfigure}[b]{0.49\textwidth}
        \includegraphics[width=\textwidth]{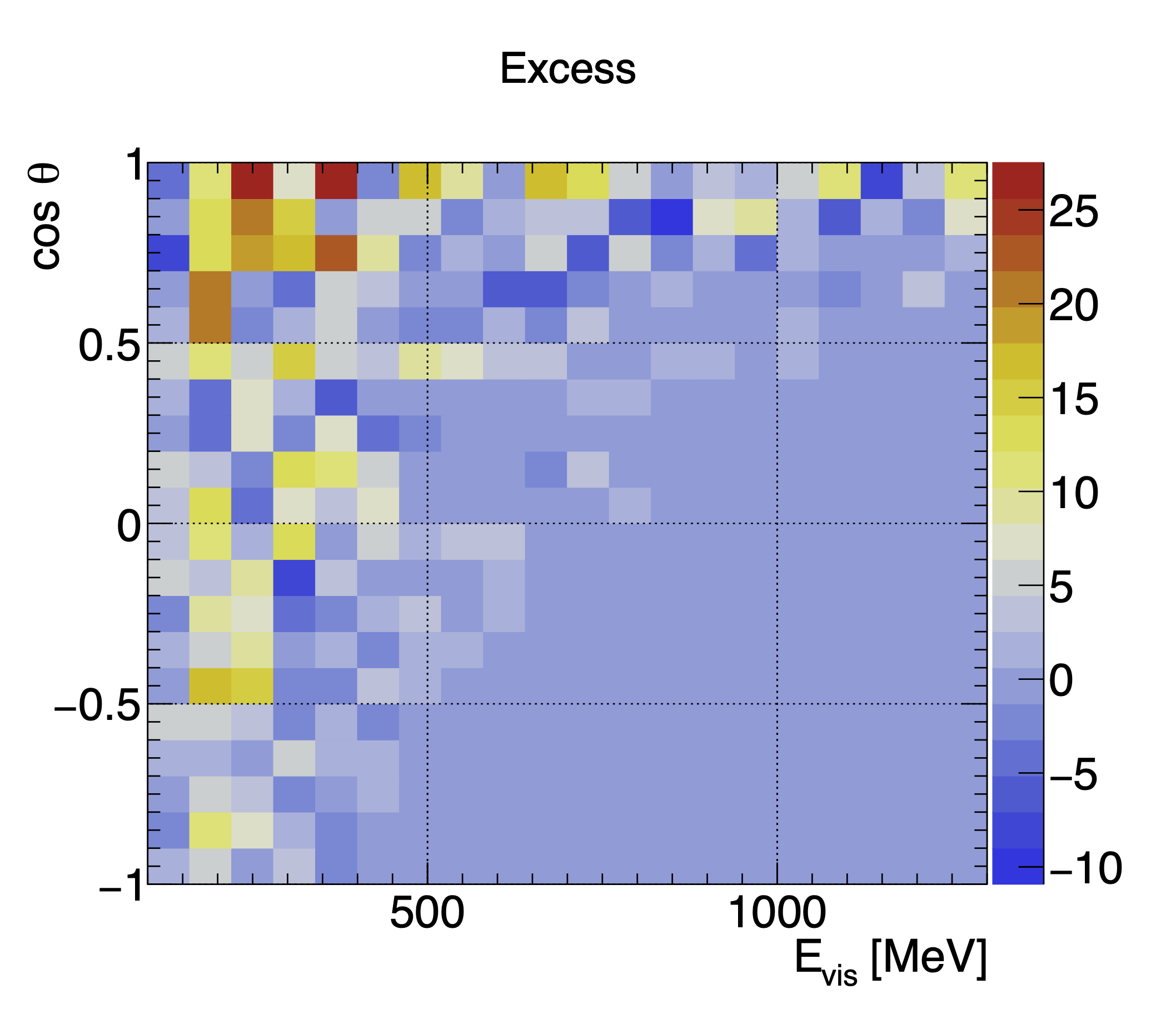}
        \caption{}
        \label{fig:miniboone_2d}
    \end{subfigure}
    \caption[MiniBooNE LEE angular distributions]{Panel (a) shows the MiniBooNE $\nu_e$-like $\cos \theta$ distribution. The best-fit sterile oscillation parameters $\text{sin}^2 2\theta = 0.807$ and $\Delta m^2 = 0.043\ \mathrm{eV}^2/c^4$ are indicated with a dashed line. Panel (b) shows the MiniBooNE excess visible energy and $\cos \theta$ distribution. From Ref. \cite{miniboone_lee}.}
    \label{fig:miniboone_angle}
\end{figure}

We can see in Fig. \ref{fig:miniboone_angle} that the excess is concentrated along the beam direction to a greater extent than would be predicted by sterile neutrino oscillations, again indicating that other processes could be more likely explanations of the LEE.

Besides sterile neutrinos, there are a variety of other ideas for the source of the MiniBooNE LEE. If the excess is not coming from a true $\nu_e$ excess, then most likely it is coming from a different source of electromagnetic showers, either photons or exotic $e^+ e^-$ production. If an $e^+ e^-$ pair has a sufficiently small opening angle, it will appear as one merged shower rather than two independent showers in MiniBooNE, and therefore will be indistinguishable from a photon. It is also possible for a highly asymmetric $e^+ e^-$ decay, where one electron or positron has very low energy, to resemble a single shower. In either case, these events could be identified as $\nu_e$-like by the MiniBooNE experiment. 

\begin{figure}[H]
    \centering
    \includegraphics[width=0.6\textwidth]{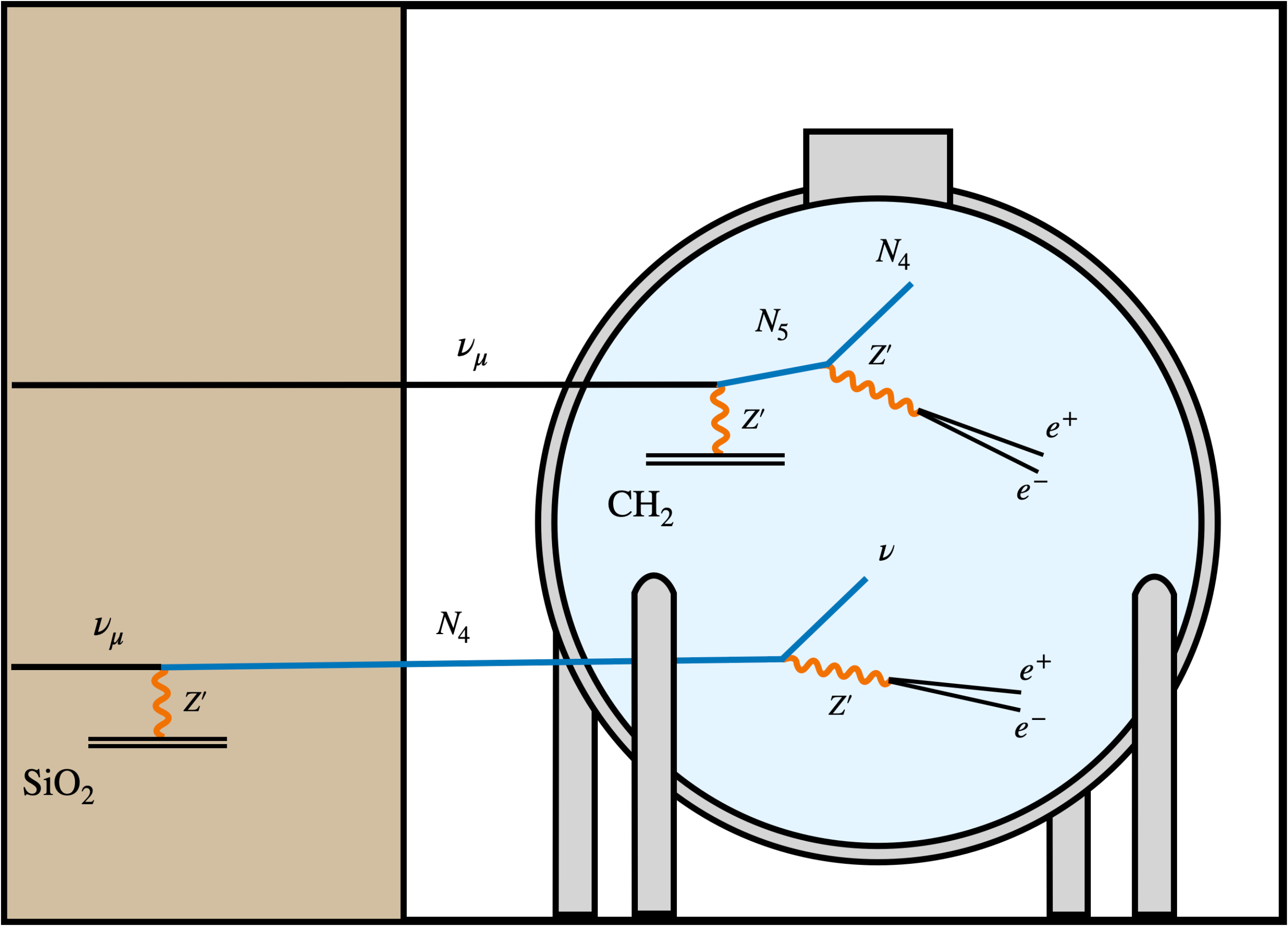}
    \caption[MiniBooNE dark neutrino diagram]{Illustration of the dark neutrino model from Ref. \cite{miniboone_new_physics_panorama}. The top interaction illustrates the case of a short lived heavier dark neutrino being produced in the MiniBooNE detector before decaying into a lighter sterile neutrino, and the bottom interaction illustrates the case of a longer lived dark neutrino being produced outside of the MiniBooNE detector before entering and decaying to a standard neutrino.}
    \label{fig:miniboone_dark_neutrino_diagram}
\end{figure}

\begin{figure}[H]
    \centering
    \begin{tikzpicture}[line width=1.0 pt, scale=1]

        \coordinate (nucleus_start) at (0, 0);
        \coordinate (nucleus_int) at (2.5, 0);
        \coordinate (nucleus_end) at (5, 0);

        \coordinate (neutrino_start) at (0, 2);
        \coordinate (upscatter) at (2.5, 2);
        \coordinate (hnl_decay) at (4, 2.5);
        \coordinate (neutrino_end) at (5, 3.5);

        \coordinate (epem_start) at (5.2, 1.7);
        \coordinate (ep_end) at (7, 1.7);
        \coordinate (em_end) at (7, 1.2);

        \draw[ultra thick, double distance=2pt] (nucleus_start) -- (nucleus_int);
        \draw[ultra thick, double distance=2pt] (nucleus_int) -- (nucleus_end);

        \draw[ultra thick] (neutrino_start) -- (upscatter) node[at start, above] {$\nu$};
        \draw[ultra thick] (upscatter) -- (hnl_decay) node[midway, above] {$N$};
        \draw[ultra thick] (hnl_decay) -- (neutrino_end) node[at end, above] {$\nu$};

        \draw[ultra thick, decorate, decoration={snake, amplitude=2pt, segment length=6pt}] 
        (upscatter) -- (nucleus_int) node[midway, left] {$I$};
        \fill (nucleus_int) circle (3pt); 

        \draw[ultra thick, decorate, decoration={snake, amplitude=2pt, segment length=6pt}] 
        (hnl_decay) -- (epem_start) node[midway, above, xshift=3pt] {$I$};

        \draw[ultra thick] (epem_start) -- (ep_end) node[at end, above] {$e^+$};
        \draw[ultra thick] (epem_start) -- (em_end) node[at end, below] {$e^-$};

    \end{tikzpicture}
    \caption[HNL photon-like Feynman diagram]{HNL $e^+e^-$ production Feynman diagram, covering a wide range of models. The upscattering of a neutrino $\nu$ interacting with a nucleus to an HNL $N$ can happen either before or after entering the MiniBooNE detector. The final neutrino $\nu$ could be an additional sterile state $N'$ in some models. The intermediate state $I$ could be a $Z'$ boson \cite{miniboone_new_physics_panorama, dark_neutrino_portal, heavy_sterile, dark_seesaw, dark_neutrino_three_portal, dark_Z_prime_sterile}, a new Higgs state \cite{two_higgs_doublet, extended_higgs}, a light scalar \cite{dark_sector_s_ee_gammagamma}, or a light pseudo-scalar \cite{17_mev_pseudoscalar}.}
    \label{fig:hnl_photon_like_feynman_diagram}
\end{figure}
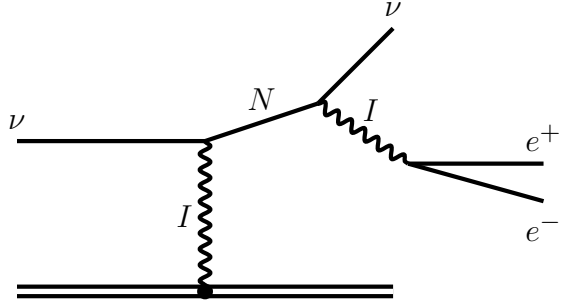

Several models introduce heavy neutral leptons (HNLs), also called ``dark neutrinos'' or ``heavy sterile neutrinos'' or ``neutrissimos'', which are produced via upscattering in neutrino-nucleus interactions, and then decay to produce a neutrino (active or sterile) and an intermediate particle which then decays to an $e^+e^-$ pair. This intermediate particle can be a dark $Z'$ boson (also called a ``dark photon'') \cite{miniboone_new_physics_panorama, dark_neutrino_portal, heavy_sterile, dark_seesaw, dark_neutrino_three_portal, dark_Z_prime_sterile}, a new Higgs state \cite{two_higgs_doublet, extended_higgs}, a light scalar \cite{dark_sector_s_ee_gammagamma}, or a light pseudo-scalar \cite{17_mev_pseudoscalar}. Depending on the lifetime of the intermediate particle, these models can involve upscattering in the soil upstream of MiniBooNE, or upscattering in the MiniBooNE detector itself, as illustrated in Fig. \ref{fig:miniboone_dark_neutrino_diagram}. The Feynman diagram for these models is illustrated in Fig. \ref{fig:hnl_photon_like_feynman_diagram}. 

There are also similar theories involving an HNL, but with a slightly different Feynman diagram. Some models have the HNL decay directly into a neutrino and a photon, possibly via a transition magnetic moment \cite{heavy_sterile_radiative_PRD, heavy_sterile_radiative_PRL, heavy_neutrino_radiative, heavy_neutrino_decay_MB, constraints_EM_neutrino_MB, dipole_portal_HNL, neutrissimo}, and one model considers this simultaneously with 3+1 sterile neutrino oscillations \cite{mixed_oscillation_decay}. Ref. \cite{dark_sector_s_ee_gammagamma} also considers the possibility of the intermediate light scalar particle decaying to two photons rather than $e^+e^-$. One model involves neutrino upscattering to an HNL with a dark $Z'$, but instead of HNL decay, the dark $Z'$ interacts with a new Higgs state that decays to an $e^+e^-$ \cite{higgs_doublet_through_Z}.

Some models produce BSM particles directly from meson decays in the beam rather than via upscattering in neutrino-nucleus interactions. In one model, a $K^+$ decays to an HNL which decays to a leptophilic axion-like particle which produces an $e^+ e^-$ pair \cite{axion_like}. In another model, a kaon decays to an HNL which decays directly into a neutrino and a photon with no mediator \cite{250_MeV_sterile_decay}. In one model, charged mesons decay to vector-portal dark-matter which produces an $e^+e^-$ after upscattering, or charged mesons decay to a long-lived scalar or pseudoscalar which produces a photon via ``dark Primakoff'' scattering \cite{MB_charged_meson_decays}. In a similar model, charged or neutral meson decays produce vector, scalar, or pseudoscalars which scatter off a nucleus to produce a photon \cite{meson_portal_MB_CCM}.

In Chapter \ref{sec:nc_delta}, I will describe a search for potential photon-like excesses in the MicroBooNE experiment, which could have sensitivity to many of these specific types of theoretical models of the LEE.

%% file: chapters/02_microboone.tex
\chapter{MicroBooNE}

\section{Neutrino Detectors}

Neutrino detectors represent an interesting challenge in high energy particle physics. In collider experiments, it is known beforehand almost exactly where particles will be produced, so we can place high resolution trackers, electromagnetic calorimeters, hadronic calorimeters, and muon spectrometers sequentially in the path of those particles. In neutrino detectors, the interactions are very rare, and therefore we have no idea where the neutrino will interact, and therefore no knowledge of where the resulting high energy particles will be produced \footnote{Some experiments like MINERvA and DUNE-ND SAND do utilize separate targets outside of their detector materials, but this is only feasible in very high-neutrino-flux environments, and not for any long-baseline oscillation experiments.}. This is particularly apparent for neutrino detectors, but also applies to other types of fixed-target experiments with wide beams and low interaction rates. This means that we cannot have these separate components, and to get detailed information about all the particles we must have one type of detection mechanism playing many roles, acting as the tracker, electromagnetic calorimeter, hadronic calorimeter, and muon spectrometer.

To measure enough interactions, neutrino detectors also have to be very large, particularly for long-baseline oscillation experiment far detectors. To scale a technology up to a very large detector, one has to be mindful of how the number of electronic readout channels scales. Imagine a detector of volume $L^3$, and a number of readout channels $N$. A three dimensional array of pixels doing imaging is not feasible; with $N\propto L^3$, the number of readout channels, and the associated power usage and cost, will simply be too large for any reasonable spatial resolution. So we need to use some way of reducing the dimensionality of our detection.

An experiment like T2K reduces this dimensionality by only putting their photomultiplier tube (PMT) detection channels at the surface of their very cyclindrical detector (Super-Kamiokande), achieving $N\propto L^2$. In this case, the number of photodetectors is mostly driven by light collection area needs rather than by spatial resolution goals. This mechanism relies on Cherenkov imaging, so a lot of information about lower energy particles is lost; this makes detection of the neutrino energy particularly difficult for example. There is eventually a theoretical limit to this $N\propto L^2$ scaling of course, because at some point the detector will become large enough that Cherenkov light scatters or is absorbed too often, but this is not a practical concern for detectors today.

An experiment like NOvA reduces this dimensionality by imaging scintillation light only in projected two dimensional views, again achieving $N\propto L^2$. This can potentially create some ambiguity in the reconstructed three dimensional image of the interaction products, but this is a challenge that we can overcome. There is again a theoretical limit to this $N\propto L^2$ scaling, because at some point the detector will become large enough that the scintillation light will become too attenuated to reach the photodetectors at the boundaries, but this is not a practical concern for detectors today.

With this scaling for imaging detectors like NOvA, an increase in detector mass would likely mean a decrease in detector resolution; a painful tradeoff to consider. However, the Time Projection Chamber (TPC) changes this picture.

The key advantage of a TPC is that it can essentially reduce the dimensionality once more, to $N\propto L$, while maintaining the capability for imaging with high-resolution local calorimetry, which is often critical for particle identification. It achieves this by reconstructing a two-dimensional projected view of an event, now using just one spatial dimension (the position of a wire connected to sensitive electronics) and the time dimension (how long it takes ionization charge to drift to the wire). This does sacrifice the very high timing resolution of other detector technologies, so it often needs to be paired with light detectors as well. This $N\propto L$ scaling has limits, due to the increasing electronic noise associated with very long wires, and due to the increasing high-voltage and electron attenuation associated with very long drift times; this does start to limit the scaling for current experiments, with DUNE for example placing several TPCs next to each other rather than scaling one up to be even larger. However, even with the limits to this scaling, the TPC does allow for a much higher spatial resolution compared to other similarly large-scale technologies. 

It is possible that detector technology will advance to the point that very large numbers of readout channels become feasible, and this $N\propto L$ scaling is no longer important. For example, TPCs that use charge detecting pixels rather than wires are planned for the DUNE near detector, and have been proposed for the DUNE far detector. In that case, the TPC can have even higher resolution and fewer ambiguities.

\section{LArTPCs}

For neutrino experiments, the Liquid Argon Time Projection Chamber (LArTPC) is a particularly compelling detection technology. The LArTPC was first proposed by Carlo Rubbia in 1977 \cite{rubbia_lartpc}. For a liquid TPC, a noble gas is desirable due to its minimal interactions with drifting electrons. Among noble gases, argon is a particularly good choice, primarily due to its relatively low price per kilogram, due to its high natural abundance at around 0.93\% of the Earth's atmosphere. Argon's relatively high density, high boiling point, and high scintillation wavelength also make it preferable to lighter noble gases, as shown in Table \ref{tab:noble_gases}. Some experiments do use other noble gases than argon for different physics goals. In particular, xenon is often for dark matter searches, due to its high density and lack of strongly radioactive isotopes similar to $^{39}\mathrm{Ar}$ beta decays. Some dark matter experiments will use helium in order to try to detect lower energy recoils from lower mass dark matter pairticles \cite{helium_dm}.

\begin{table}[H]
    \centering
    \small
    \begin{tabular}{lcccccc}
        \toprule
        Noble Gas & \makecell{Atomic\\Number} & \makecell{Price\\(/kg)} & \makecell{Density\\($\mathrm{kg}/\mathrm{m}^3$)} & \makecell{Melting\\Point (K)} & \makecell{Boiling\\Point (K)} & \makecell{Scintillation\\ Wavelength (nm)} \\
        \midrule
        Helium (He) & 2& \$85  & 141.1 & - & 4.22 & 80 \\
        Neon (Ne)   & 10&  \$600 & 1224 & 24.56 &  27.07 & 77 \\
        Argon (Ar)  &18 & \$0.48 & 1401 & 83.9 & 87.4 & 128 \\
        Krypton (Kr) & 36& \$220  & 2423 & 115.79 & 119.93  & 150 \\
        Xenon (Xe)  & 54 & \$8,300 & 2948 & 161.4 & 165 & 175 \\
        \bottomrule
    \end{tabular}
    \caption[Noble gas comparison]{Comparison of noble gases based on price and physical properties. The helium price estimate is from Ref. \cite{USGS_Helium}, the argon price estimate is from Ref. \cite{BusinessAnalytiq_Argon2025}, and the neon, krypton, and xenon price estimates are from Ref. \cite{economist_rare_gas}. These prices are relatively volatile over time, and can also vary with purity requirements. Radon is excluded due to its rarity and radioactivity (and oganesson has been excluded for similar reasons).}
    \label{tab:noble_gases}
\end{table}

\begin{figure}[h]
    \centering
    \includegraphics[width=0.7\textwidth]{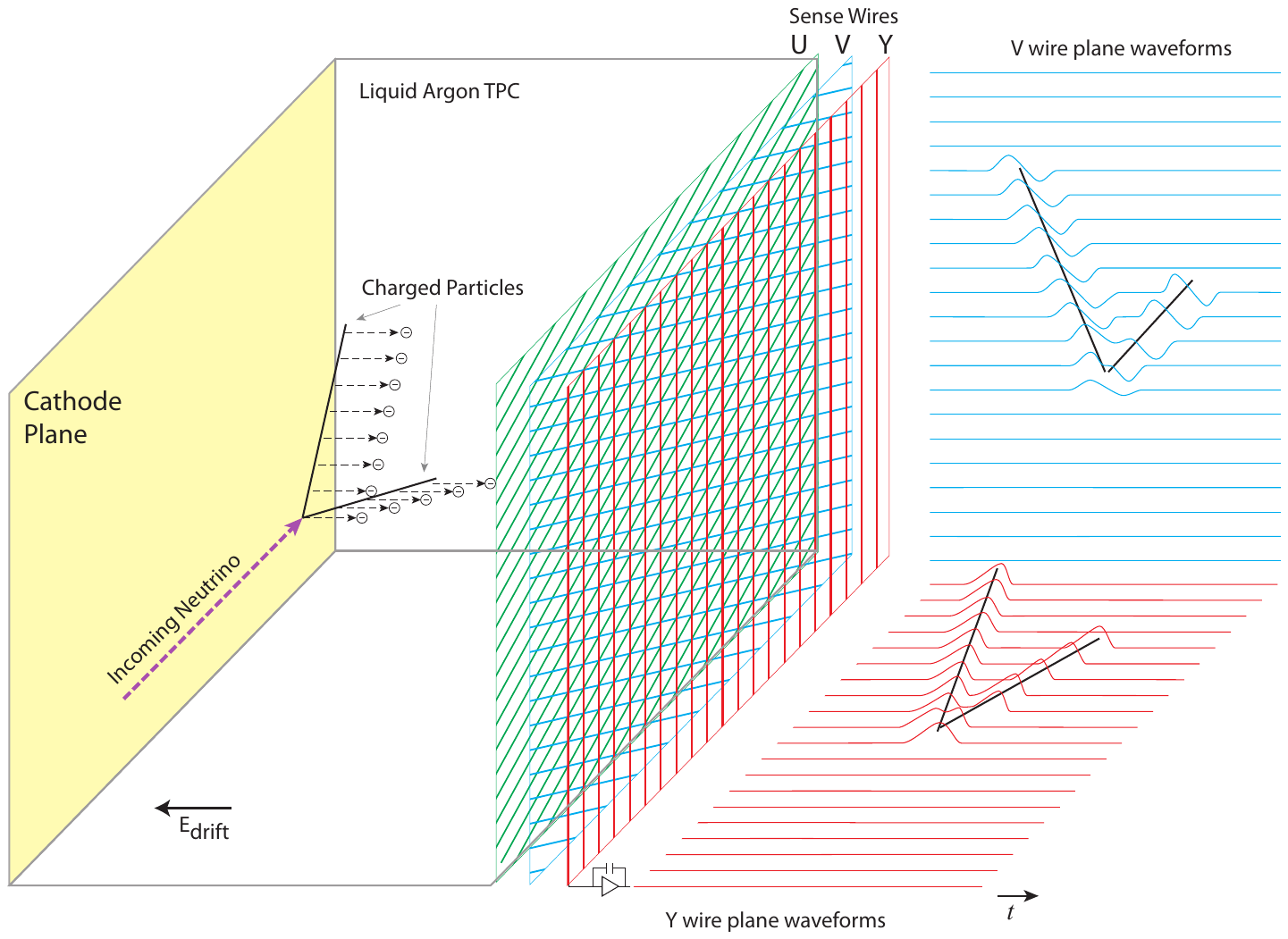}
    \caption[LArTPC Concept]{LArTPC operation principle illustration. Figure from Ref. \cite{microboone_design_constuction}.}
    \label{fig:lartpc_concept}
\end{figure}

In a LArTPC, charged particles are produced and scatter inside of the liquid argon, ionizing argon atoms as they pass by. Some of the freed ionization charge immediately recombines with positively charged ions, producing scintillation light. Some of the charge does not immediately recombine, and instead drifts toward wire planes due to an externally applied electric field. These wire planes have bias voltages applied that cause a bipolar induction signal as the electrons pass the first two wire planes, and a unipolar signal as the electrons are collected on the third wire plane. There are variations on this design, with some LArTPCs using only one or two wire planes, some using printed circuit board (PCB) plates with holes rather than wires, and some using charge collecting pixels rather than wires. Since a wire cannot tell where along its path charge arrived, each wire plane gives a 2D projected view of the 3D ionization charge activity in the detector; you can imagine looking down the direction of the wires in order to visualize the projection. This detection principle is illustrated in Fig. \ref{fig:lartpc_concept}.

The first large-scale LArTPC was ICARUS-T600, commissioned in 2001 and running in Laboratori Nazionali del Gran Sasso (LNGS), studying astroparticle and neutrino physics and searching for nucleon decays \cite{icarus_t600}. The first LArTPC in the US was constructed and operated at Yale in 2007 \cite{yale_lartpc}. That same year, the LArTPCs ArgoNeuT \cite{argoneut_mou} and MicroBooNE \cite{microboone_proposal} were proposed to operate in neutrino beams at Fermilab.

\section{MicroBooNE Detector}

MicroBooNE was designed to investigate the MiniBooNE LEE, as was described in Sec. \ref{sec:MB_LEE}. The LArTPC technology allows much more information to be collected for each neutrino event when compared to the simpler Cherenkov technology in MiniBooNE. LArTPCs allow us to do much more detailed particle ID, by imaging all charged particles and observing the deposited energy per unit length (dE/dx) of each track, including for particles well below MiniBooNE's Cherenkov energy threshold. In particular, this allows us to distinguish electron showers from photon showers, allowing us to carefully study the source of the MiniBooNE LEE in new ways. These capabilities allow for a wide range of physics analyses, including many oscillation searches, cross-section studies, and other beyond-the-standard-model investigations. 

MicroBooNE is located just 72.5 m upstream of the MiniBooNE detector in the same accelerator beam at Fermilab. Its active volume contains 85 metric tons of liquid argon, and a 273 V/cm electric field drifts ionization charge to a set of three wire planes containing 8,192 wires. The wires in these planes are offset from each other by 60 degrees, allowing us to reconstruct 3D views from the 2D projections. Each wire plane has a pitch (distance between adjacent parallel wires) of 3 mm. MicroBooNE also contains 32 photomultiplier tubes (PMTs) which help to establish precise timing information for measured charge activity. MicroBooNE is the first LArTPC to collect high statistics data of neutrino-argon interactions, and it has pioneered many technologies and analysis techniques that are being used in newer LArTPCs like ICARUS and SBND, and that will be used in future LArTPCs like DUNE.

\begin{figure}[H]
    \centering
    \includegraphics[width=0.5\textwidth]{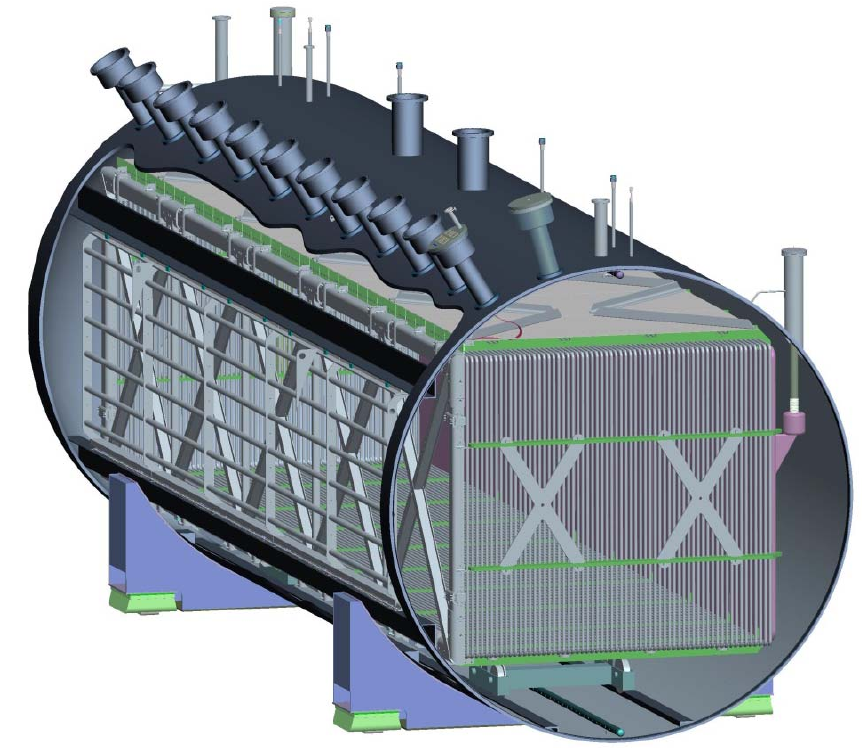}
    \caption[MicroBooNE render]{MicroBooNE render, showing the field cage inside the cryostat. From Ref. \cite{microboone_design_constuction}.}
    \label{fig:microboone_render}
\end{figure}

Figure \ref{fig:microboone_render} shows a rendering of the MicroBooNE detector showing the field cage inside of the cryostat.

\begin{figure}[H]
    \centering
    \begin{subfigure}[b]{0.49\textwidth}
        \includegraphics[width=\textwidth]{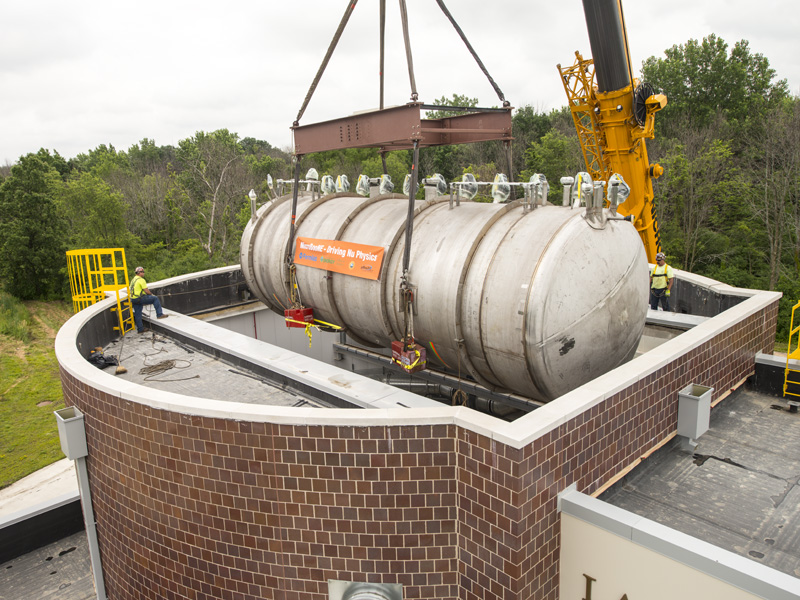}
        \caption{}
        \label{fig:cryostat_crane}
    \end{subfigure}
    \hfill
    \begin{subfigure}[b]{0.49\textwidth}
        \includegraphics[width=\textwidth]{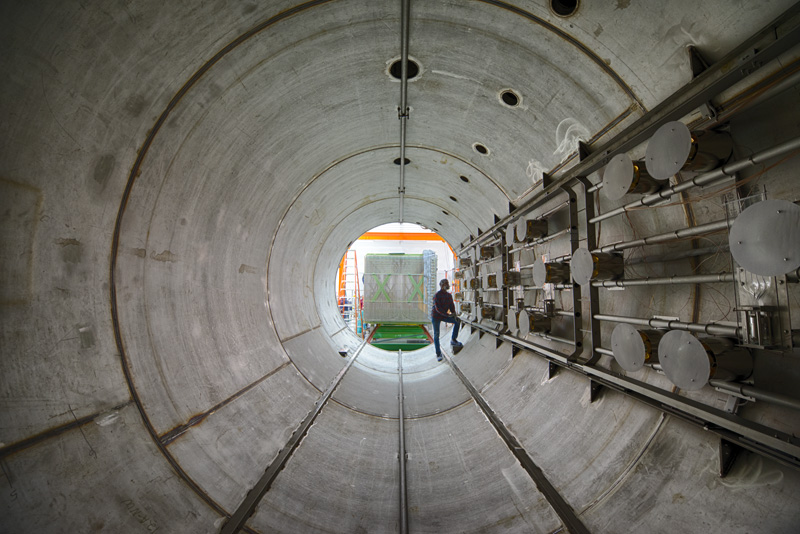}
        \caption{}
        \label{fig:pmts_in_cryostat}
    \end{subfigure}
    \begin{subfigure}[b]{0.35\textwidth}
        \includegraphics[width=\textwidth]{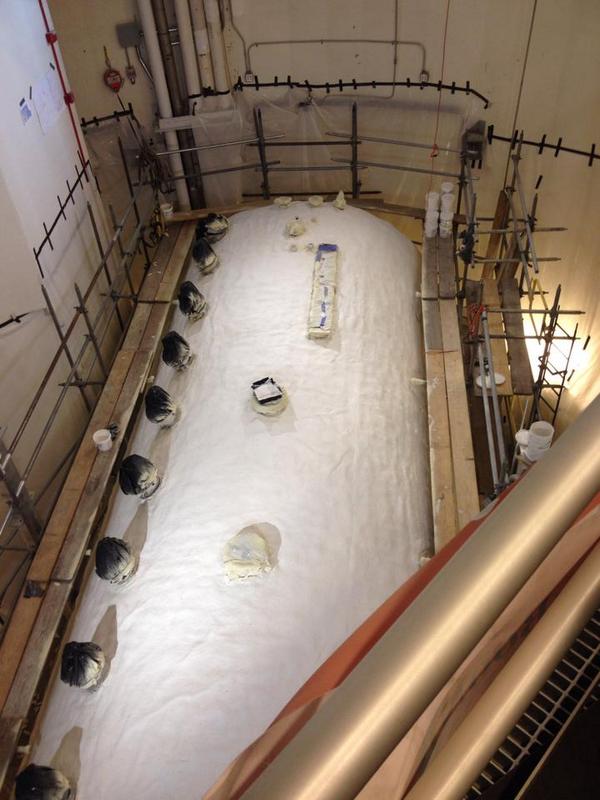}
        \caption{}
        \label{fig:foam_cryostat_above}
    \end{subfigure}
    \hfill
    \begin{subfigure}[b]{0.6\textwidth}
        \includegraphics[width=\textwidth]{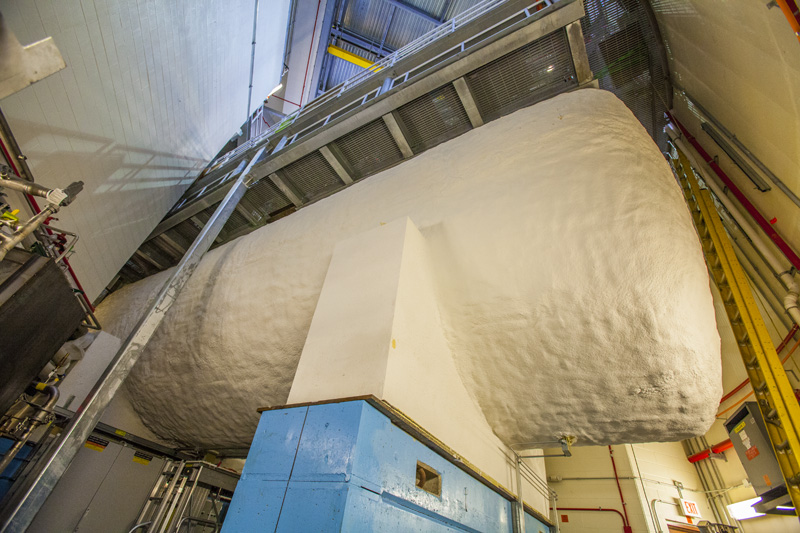}
        \caption{}
        \label{fig:foam_cryostat_below}
    \end{subfigure}
    \caption[MicroBooNE cryostat]{MicroBooNE's steel cryostat. Panel (a) shows the cryostat being lowered into the LArTF building at Fermilab. Panel (b) shows the cryostat from the inside, with PMTs mounted on one side. Panel (c) shows the cryostat from above, installed with foam insulation. Panel (d) shows the cryostat from below, installed with foam insulation. Panels (a)-(b) images are from Ref. \cite{microboone_first_results_press_release}, and panels (c)-(d) images are from Ref. \cite{microboone_design_constuction}.}
    \label{fig:cryostat}
\end{figure}

Figure \ref{fig:cryostat} shows images of MicroBooNE's cryostat and surrounding insulation foam, designed to minimize the heat load below 15 $\mathrm{W}/\mathrm{m}^2$ in order to minimize convection currents and prevent bubbles from liquid argon boiling. The steel cryostat is 12.2 m long, has an inner diameter of 3.81 m, and has a wall thickness of 11.1 mm.

MicroBooNE's TPC consists of a cathode plane, a field cage, and anode wires. The cathode is made out of 2.3 mm thick steel plates, and has a -70 kV potential applied to it, causing a 273 V/cm electric field throughout the TPC. The field cage consists of 64 thin walled stainless steel tubes and helps to maintain the uniformity of the electric field between the anode and cathode. The field cage is held in place by a G-10 rib support structure \footnote{In MicroBooNE ambient isolated low energy ionization deposits, we can see evidence of 2.614 MeV gamma rays originating from $^{208}$Tl decays near the surface of these G-10 fiber glass supports \cite{microboone_ambient_blips}.}. A voltage divider chain consisting of $250 \mathrm{M}\Omega$ resistors steps the voltage down in even steps between each field cage tube.

\begin{figure}[H]
    \centering
    \begin{subfigure}[b]{0.49\textwidth}
        \includegraphics[width=\textwidth]{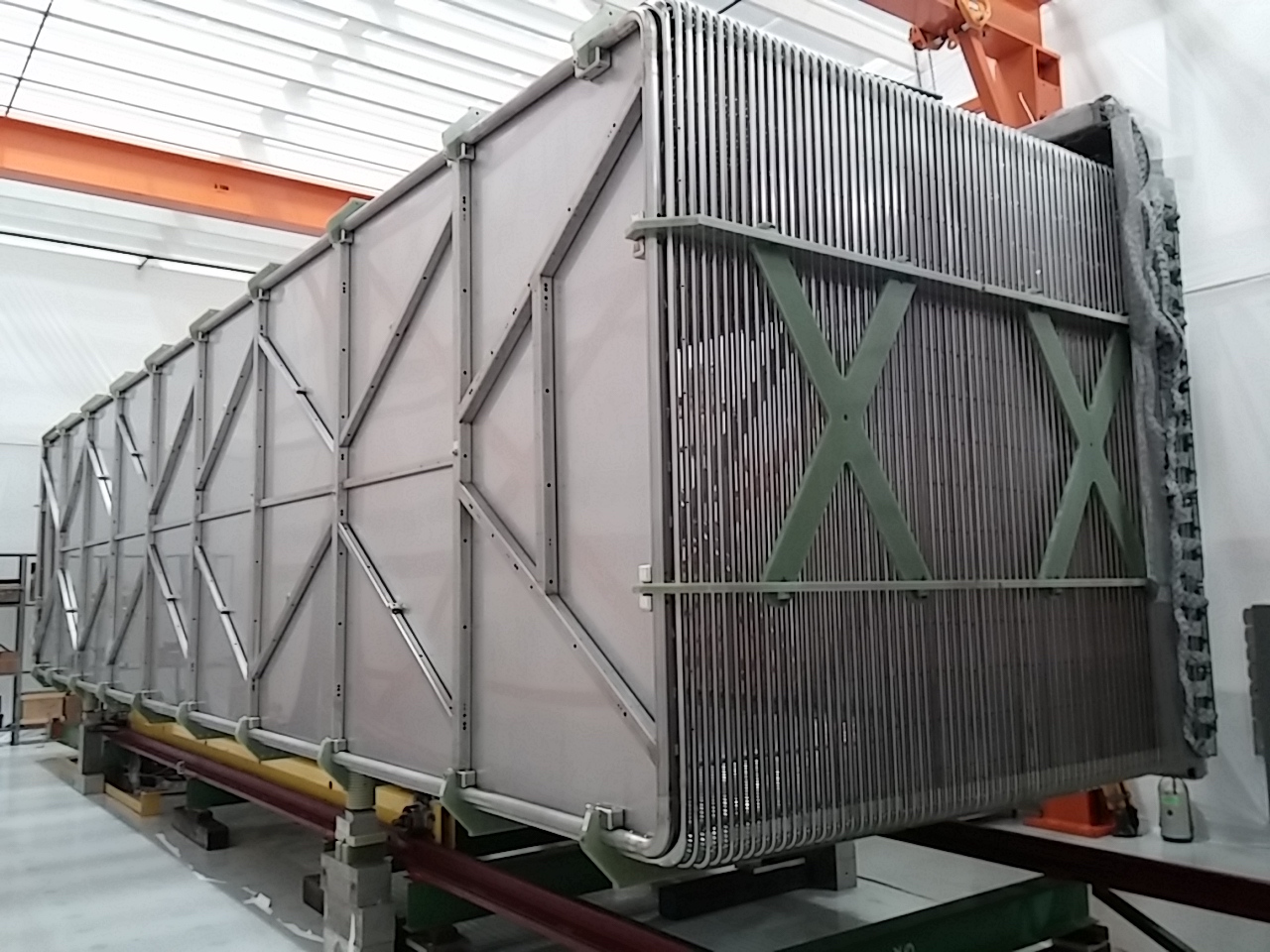}
        \caption{}
        \label{fig:tpc_left}
    \end{subfigure}
    \hfill
    \begin{subfigure}[b]{0.49\textwidth}
        \includegraphics[width=\textwidth]{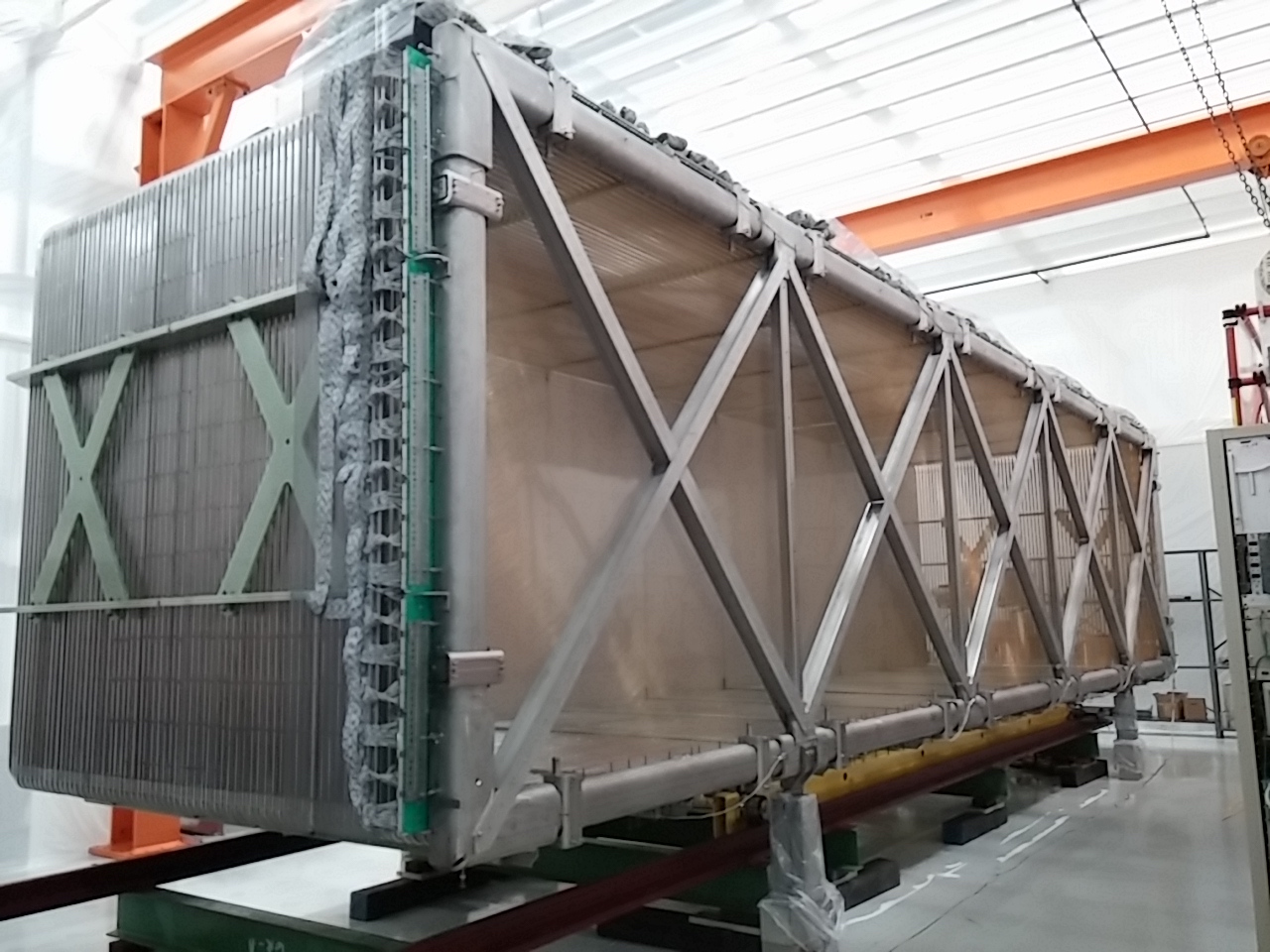}
        \caption{}
        \label{fig:tpc_right}
    \end{subfigure}
    \begin{subfigure}[b]{0.35\textwidth}
        \includegraphics[width=\textwidth]{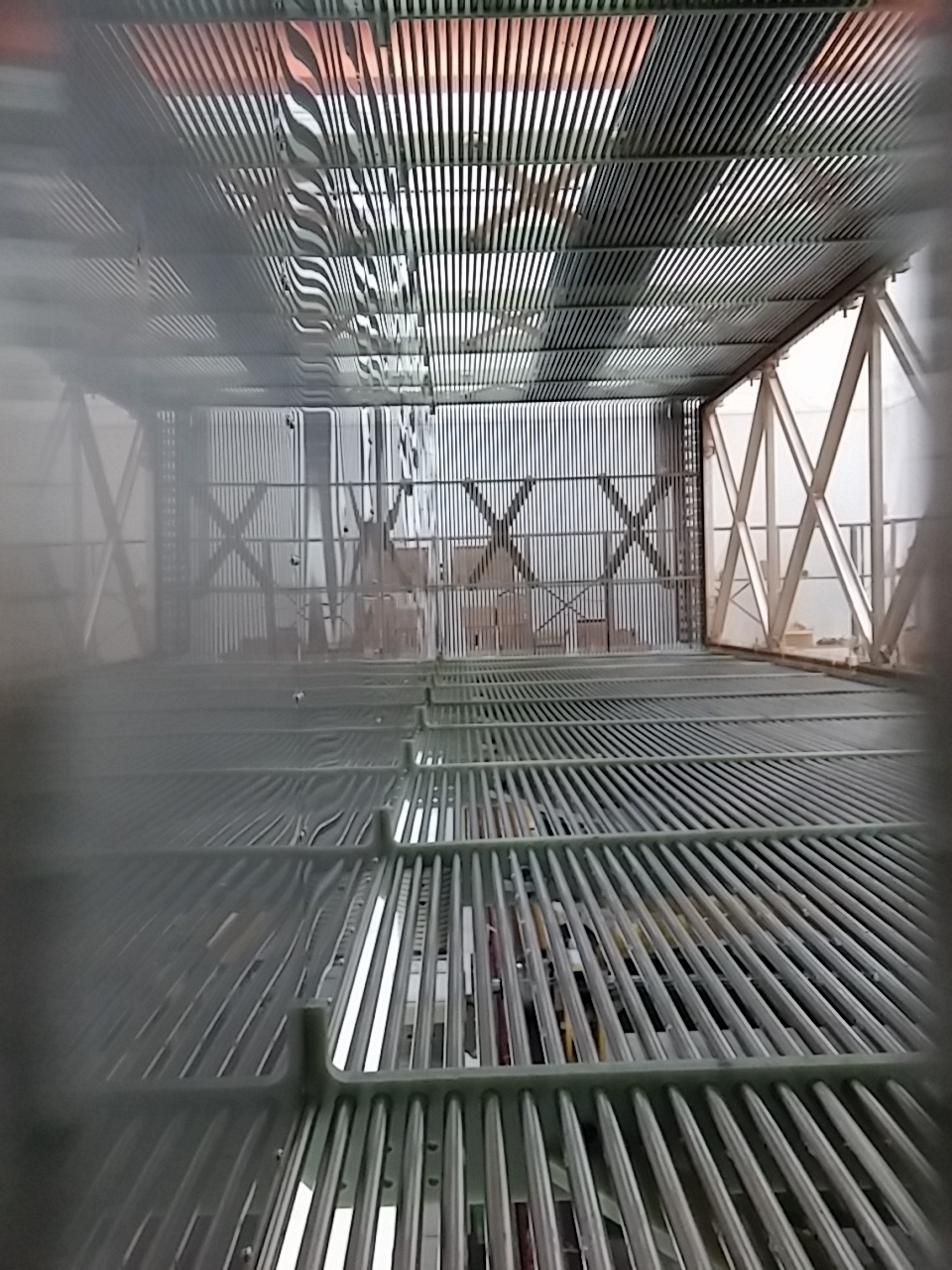}
        \caption{}
        \label{fig:tpc_inside}
    \end{subfigure}
    \hfill
    \begin{subfigure}[b]{0.6\textwidth}
        \includegraphics[width=\textwidth]{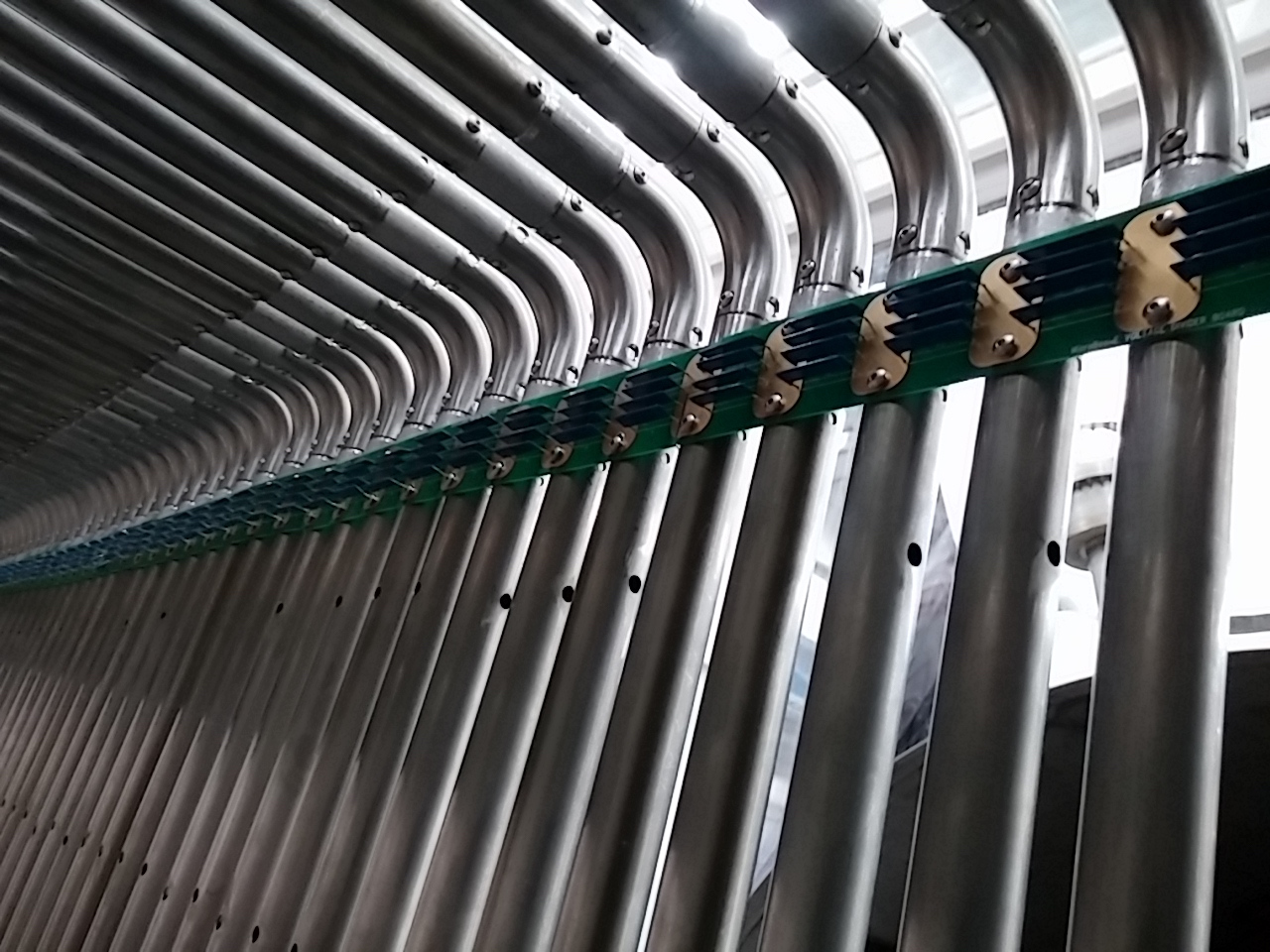}
        \caption{}
        \label{fig:tpc_voltage_dividers}
    \end{subfigure}
    \caption[MicroBooNE TPC]{MicroBooNE's TPC consisting of the cathode plane, field cage, and anode wires. Panel (a) shows the cathode and field cage. Panel (b) shows the field cage and anode wires. Panel (c) shows the field cage from an inside view. Panel (d) shows the field cage, zoomed in on the resistor divider chain. Images are from Ref. \cite{microboone_design_constuction}.}
    \label{fig:field_cage}
\end{figure}

\begin{figure}[H]
    \centering
    \begin{subfigure}[b]{0.62\textwidth}
        \includegraphics[width=\textwidth]{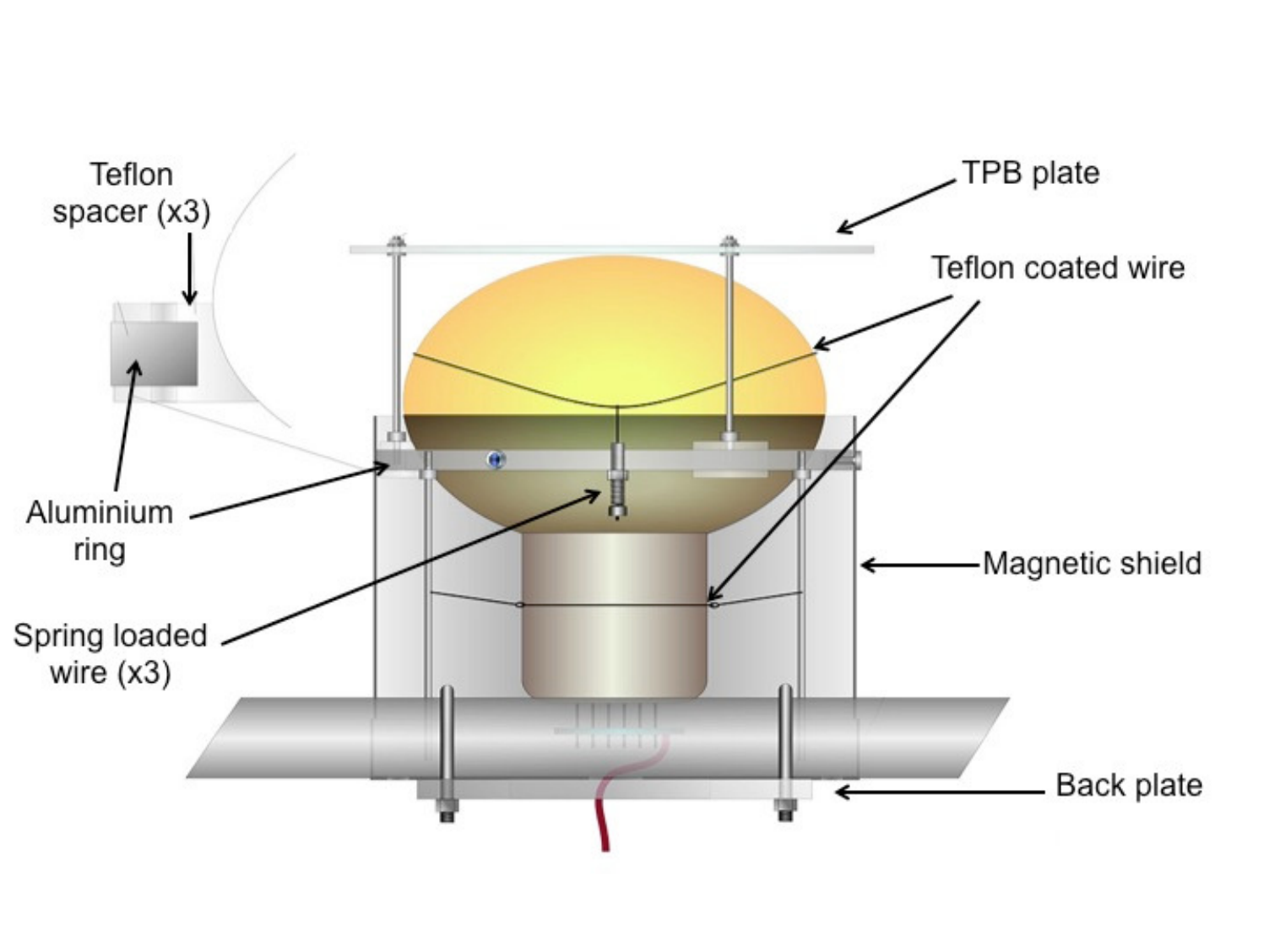}
        \caption{}
        \label{fig:pmt_diagram}
    \end{subfigure}
    \hfill
    \begin{subfigure}[b]{0.37\textwidth}
        \includegraphics[width=\textwidth]{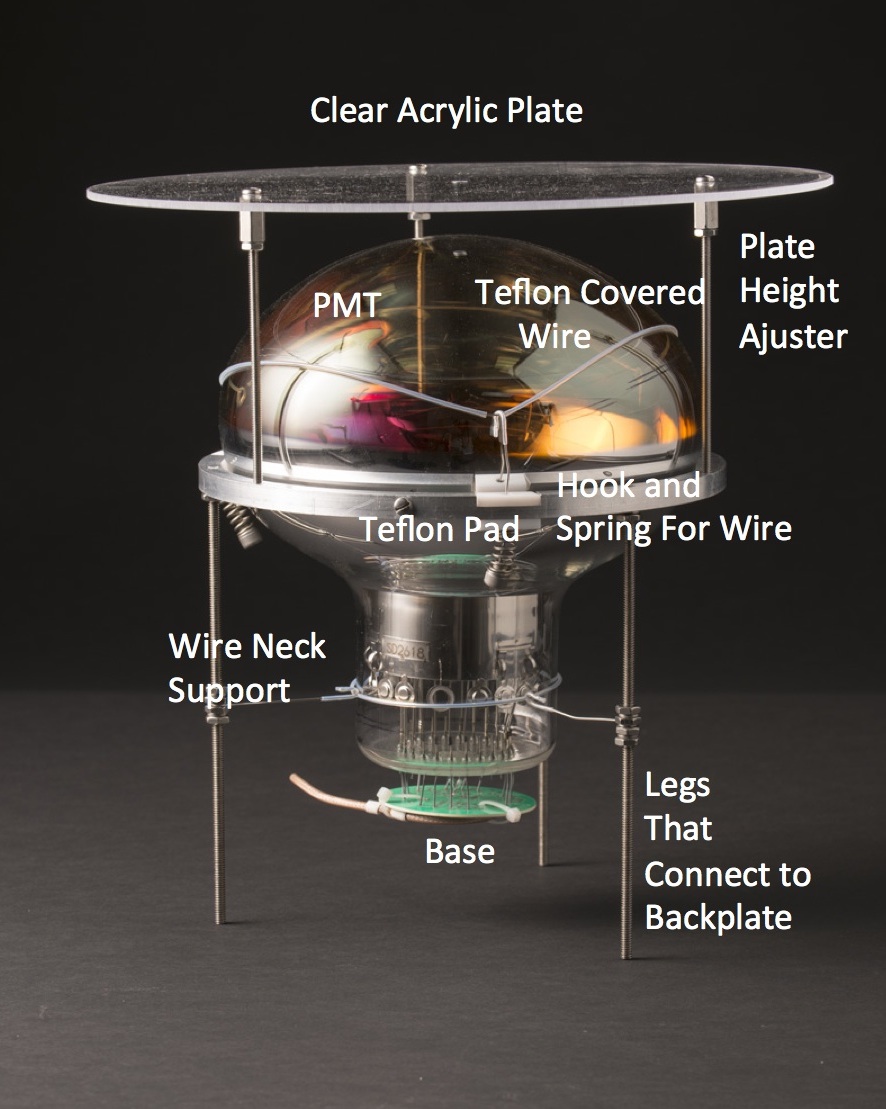}
        \caption{}
        \label{fig:pmt_image}
    \end{subfigure}
    \begin{subfigure}[b]{0.6\textwidth}
        \includegraphics[width=\textwidth]{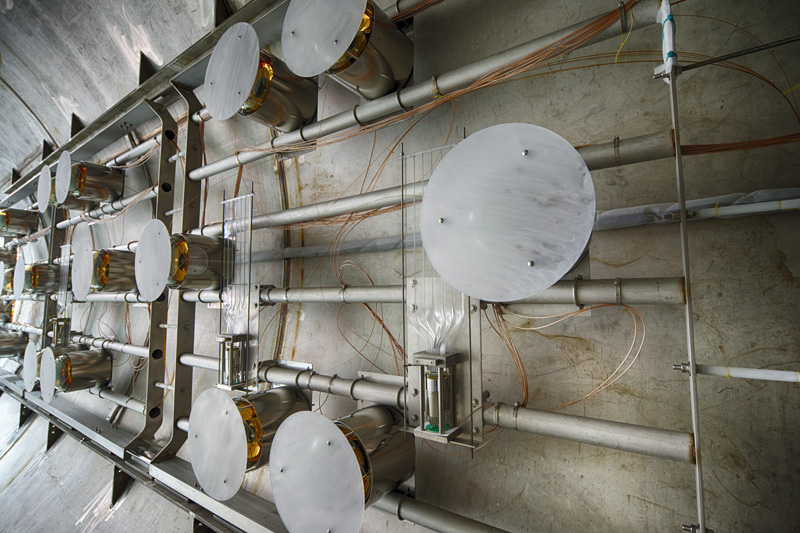}
        \caption{}
        \label{fig:mounted_pmts}
    \end{subfigure}
    \caption[MicroBooNE PMTs]{MicroBooNE's PMTs. Panel (a) shows a diagram of a PMT assembly. Panel (b) shows a labeled image of a PMT assembly, before the TPB coating has been applied to the acrylic plate. Panel (c) shows the PMT assemblies installed in the cryostat. Images are from Ref. \cite{microboone_design_constuction}.}
    \label{fig:pmts}
\end{figure}

\begin{figure}[H]
    \centering
    \includegraphics[width=0.7\textwidth]{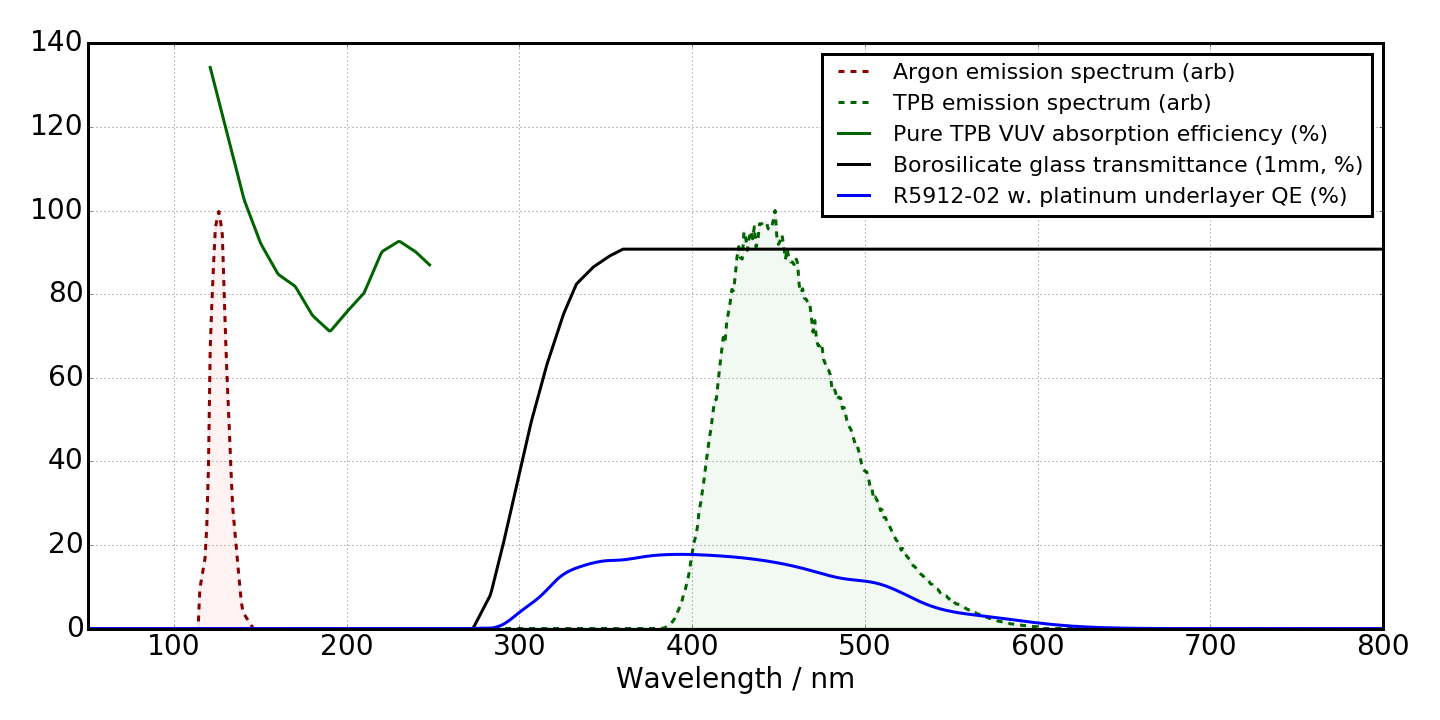}
    \caption[MicroBooNE PMT wavelength response]{MicroBooNE's PMT wavelength response. This plot illustrates how the argon emits scintillation light at VUV wavelengths, the TPB absorbs this light, and re-emits light at a much longer wavelengths, which can pass through the PMT glass and cause an electronic response. From Ref. \cite{microboone_design_constuction}.}
    \label{fig:wavelength_response}
\end{figure}

In a LArTPC, light is produced from Cherenkov light and scintillation light. Most of the light signal comes from scintillation light when the normally monoatomic argon gets excited and forms dimers, also called excimers, which emit 128 nm photons as they de-excite. Singlet dimer states emit light quickly with a decay time of 6 ns, and triplet dimer states emit light more slowly with a decay time of 1.6 $\mu$s. Liquid argon is very transparent to its own scintillation light. MicroBooNE's 32 8 inch Hamamatsu 5912-02MOD PMTs are mounted behind the anode wires on one side of the cryostat, as shown in Fig. \ref{fig:pmts}. These light detectors help us establish the precise timing of a neutrino interaction, within 2.16 ns \cite{microboone_nanosecond_timing}. To detect the 128 nm vacuum ultraviolet (VUV) light from scintillation, the PMTs are mounted behind acrylic plates coated in tetraphenyl-butadiene (TPB), which is able to absorb scintillation light and re-emit it in wavelengths that can pass through the PMT glass and produce electronic signals, as shown in Fig. \ref{fig:wavelength_response} \footnote{The angular dependence of this TPB emission light has been studied in Ref. \cite{tpb_angular} using the same optical and cryogenic apparatus that I helped construct as an undergrad at UC Berkeley, first used to study Polytetrafluoroethylene (PTFE) reflectivity to xenon scintillation light when immersed in liquid xenon, as described in Ref. \cite{xenon_ptfe_angular}.}. These PMTs need to have continuous gain calibrations, as described in Sec. \ref{sec:pmt_gains}.

\begin{figure}[H]
    \centering
    \includegraphics[width=0.7\textwidth]{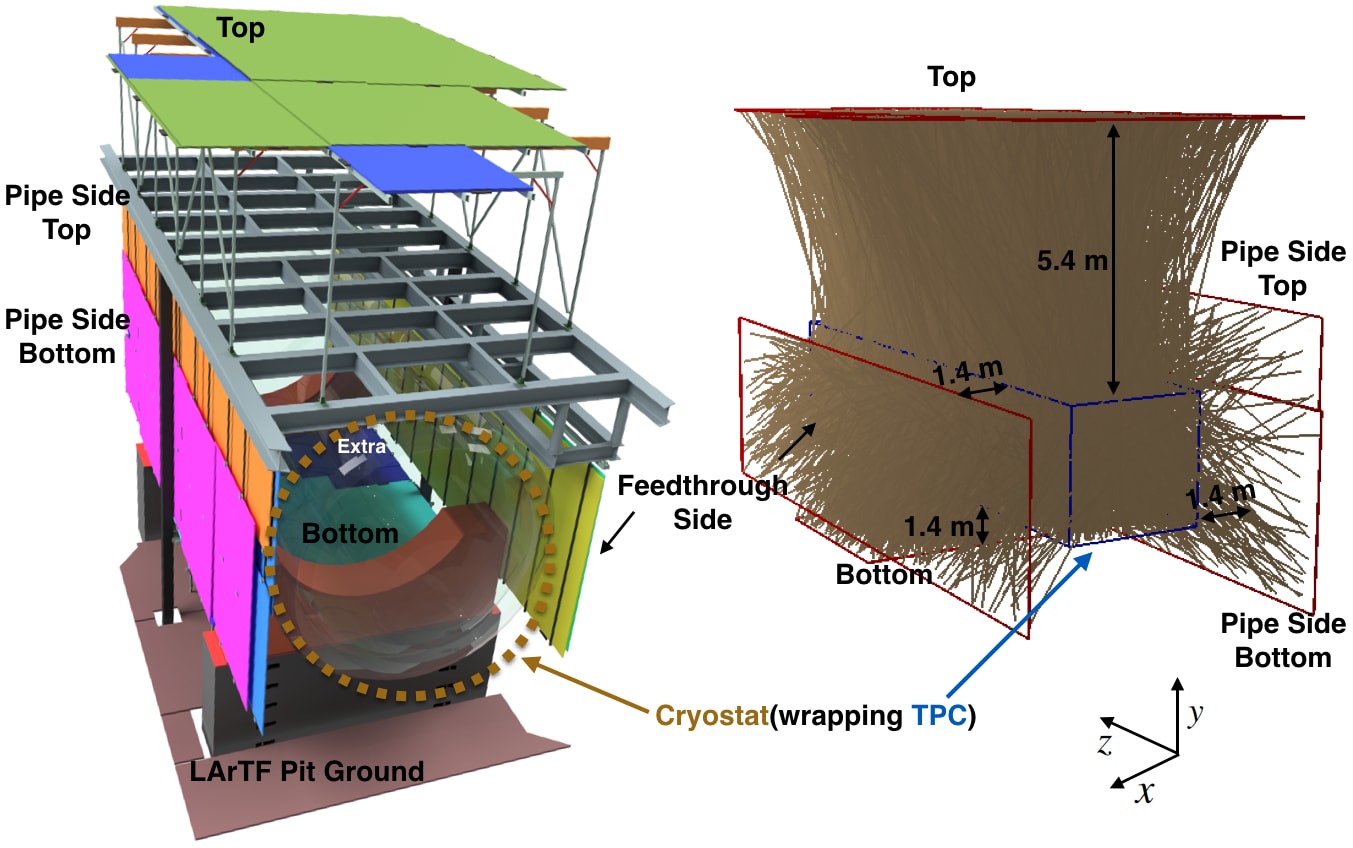}
    \caption[MicroBooNE CRT]{Diagram of MicroBooNE's cosmic ray tagger system, consisting of tob, left, right, and bottom panels. On the right, we show simulated cosmic rays passing through the CRT panels. This figure is from Ref. \cite{microboone_crt}.}
    \label{fig:microboone_crt}
\end{figure}

Part way through MicroBooNE operations, a cosmic ray tagger (CRT) system was completed. Sitting at the surface with no overburden, MicroBooNE is exposed to around 4.4 kHz of cosmic ray muons \cite{microboone_atmospheric_muon_rate}. These cosmic rays can mimic neutrino interactions if they happen to arrive in time with a beam spill window, or they can mimic additional particles present alongside neutrino-induced particles, so it is beneficial to be able to identify this activity using extra information with precise timing. This CRT system is entirely external to the cryostat, and consists of a top plane located 5.4 m above the TPC, a bottom plane located 1.4 m under the TPC, and two side planes located 1.4 m from each side of the TPC, as shown in Fig. \ref{fig:microboone_crt}. Each CRT plane consists of two layers of CRT modules. Each CRT module consists of 16 10.8 cm wide and 2 cm thick scintillating strips. Each strip is made of a USMS-03 polystyrene-based mixture, two Kuraray Y11(200)M 1mm diameter wavelength shifting fibers, and two Hamamatsu S12825-050P silicon photomultipliers (SiPMs) \cite{microboone_crt}. This lets us reconstruct cosmic ray particles that enter or exit the TPC with fairly precise position and time information. Due to the fact that this CRT system was only installed and operational after two years of MicroBooNE's beam data collection, it was not utilized for the primary analyses described in this thesis.

\begin{figure}[H]
    \centering
    \begin{subfigure}[b]{0.6\textwidth}
        \includegraphics[width=\textwidth]{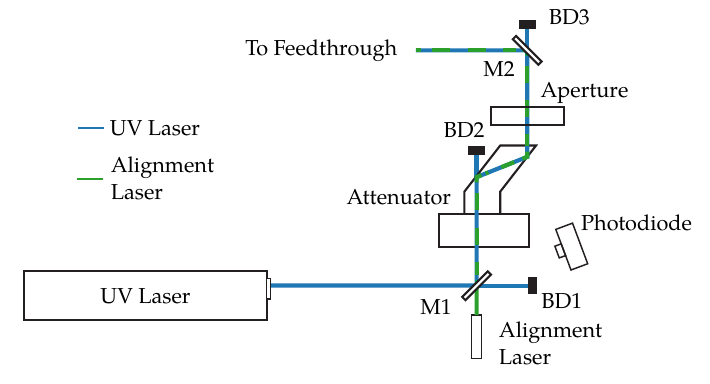}
        \caption{}
        \label{fig:laser_optics_diagram}
    \end{subfigure}
    \hfill
    \begin{subfigure}[b]{0.39\textwidth}
        \includegraphics[width=\textwidth]{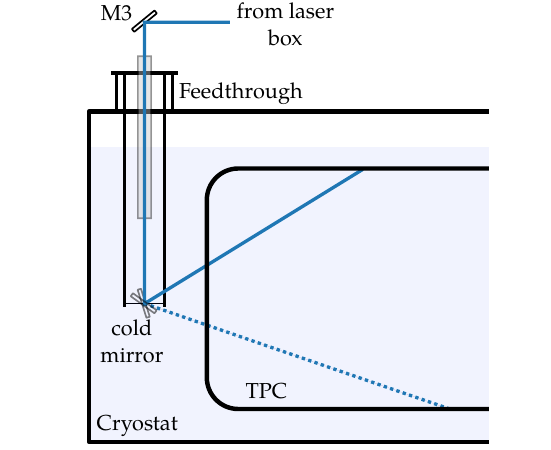}
        \caption{}
        \label{fig:laser_feedthrough_diagram}
    \end{subfigure}
    \caption[MicroBooNE laser]{MicroBooNE's laser. Panel (a) shows a diagram of the external laser optical system. Panel (b) shows a diagram of the laser's path through a feedthrough and into the MicroBooNE crystat and TPC. Images are from Ref. \cite{microboone_sce_laser}.}
    \label{fig:microboone_laser}
\end{figure}

MicroBooNE contains a system of lasers for calibration purposes. There are two identical lasers, one upstream and one downstream of the TPC. Each is a commercial Nd:YAG laser module Surelite I-10, which initially generated infrared light but passes through harmonic generators and dichroic mirrors to isolate 266 nm UV light, which is high enough energy to ionize the argon in the beam path \cite{microboone_sce_laser}. This laser light passes through a feedthrough into the cryostat, where it hits a steering mirror before crossing through the TPC as shown in Fig. \ref{fig:microboone_laser}. This steering mirror is then adjusted to aim through various parts of the TPC for electric field calibrations, as described in Sec. \ref{sec:sce}.

MicroBooNE contains a circulation system in order to purify the argon, removing contaminants that could absorb drifting electrons as described in Sec. \ref{sec:electron_lifetime} as well as contaminants that could reduce the scintillation light output as described in Sec. \ref{sec:light_detvar}. This system was designed to recirculate the entire argon volume every 2.5 days, reduce $\mathrm{O}_2$ and other electronegative impurites like $\mathrm{H}_2\mathrm{O}$ to concentrations below 100 ppt, and reduce $\mathrm{N}_2$ impurities to concentrations below 2 ppm \cite{microboone_design_constuction}. The argon is circulated through molecular sieves which primarily remove $\mathrm{H}_2\mathrm{O}$, filters consisting of pellets of copper impregnated on alumina which primarily removes oxygen. Neither of these filters are effective at removing $\mathrm{N}_2$, so extra care must be taken to avoid any significant air leaks into the argon.

\begin{figure}[H]
    \centering
    \begin{subfigure}[b]{0.7\textwidth}
        \includegraphics[width=\textwidth]{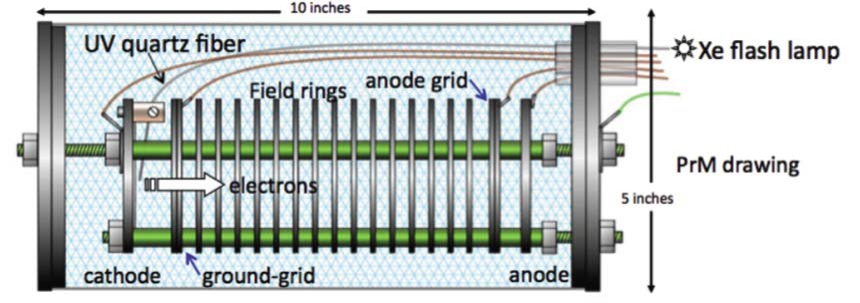}
        \caption{}
        \label{fig:purity_monitor_diagram}
    \end{subfigure}
    \hfill
    \begin{subfigure}[b]{0.29\textwidth}
        \includegraphics[width=\textwidth]{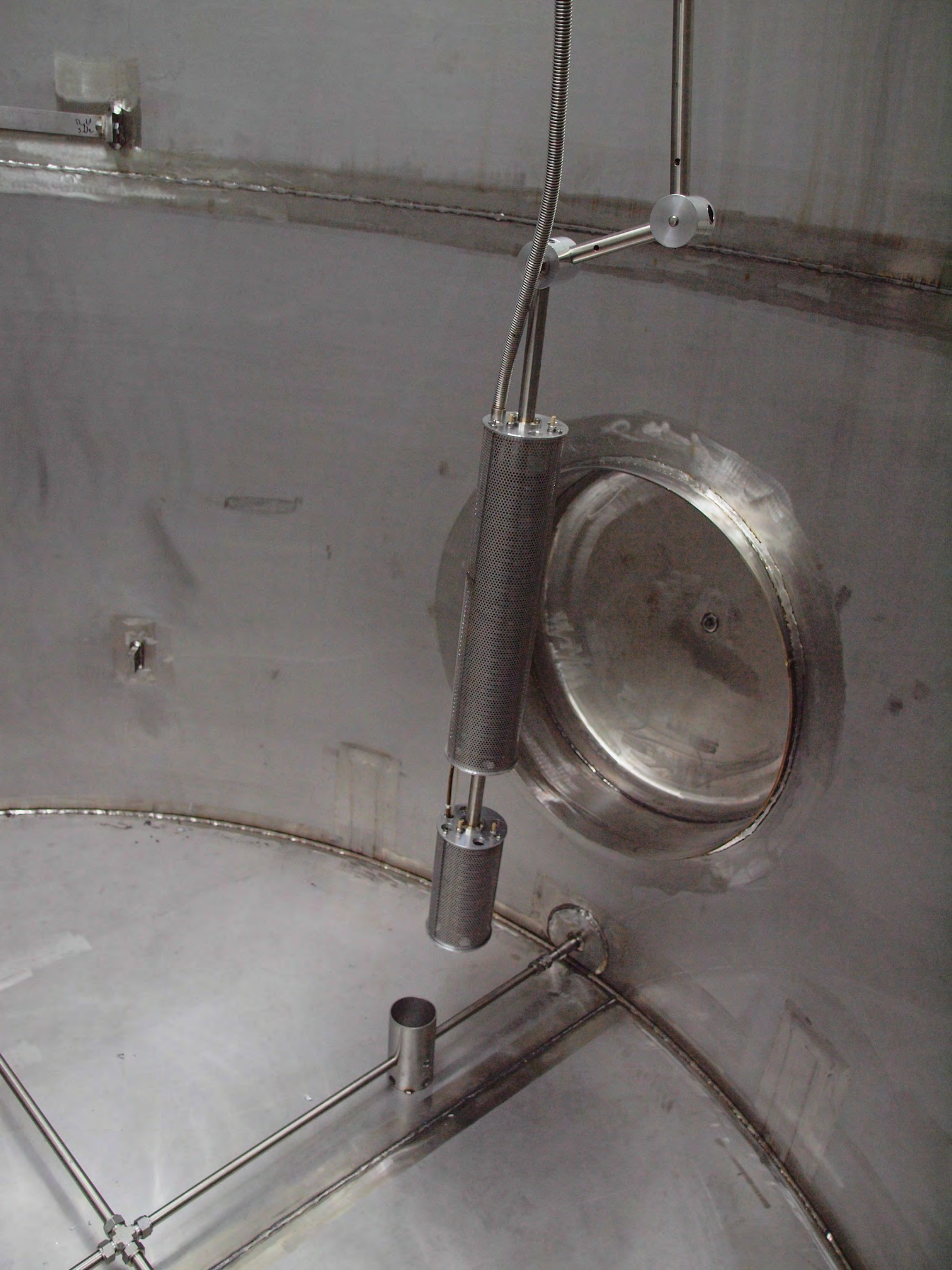}
        \caption{}
        \label{fig:purity_monitor_image}
    \end{subfigure}
    \begin{subfigure}[b]{0.8\textwidth}
        \includegraphics[width=\textwidth]{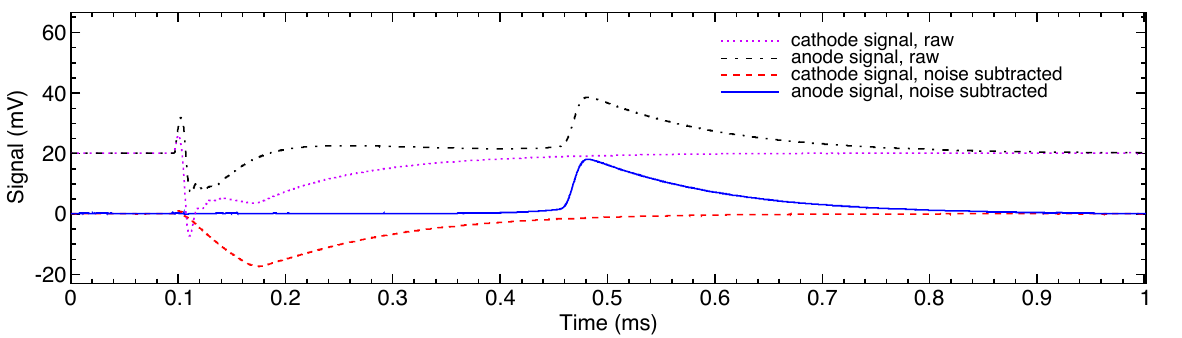}
        \caption{}
        \label{fig:purity_monitor_waveforms}
    \end{subfigure}
    \caption[Purity monitors]{Purity monitors of the same design as those in MicroBooNE. Panel (a) shows a diagram of the componentns of the purity monitor. Panel (b) shows an image containing two different lengths of purity monitor in LAPD. Panel (c) shows example waveforms from the purity monitor, from which $Q_C$, $Q_A$, and $t$ can be measured. Images are from Ref. \cite{purity_demonstrator}.}
    \label{fig:purity_monitor}
\end{figure}

The electronegative impurity was determined by dedicated purity monitors, based on the designs tested in the Liquid Argon Purity Demonstrator (LAPD) at Fermilab \cite{purity_demonstrator}, as shown in Fig. \ref{fig:purity_monitor}. MicroBooNE uses a 50 cm purity monitor just downstream of the filters, and two purity monitors with drift distances of 19 cm and 50 cm inside the cryostat at different depths. These lengths allow the measurement of purities between 50 and 300 ppt $\mathrm{O}_2$ equivalent. The purity monitors operate as double-gridded ion chambers, where charge is freed from a cathode via a xenon flash lamp, and then drifts to an anode. We measure the amount of charge freed from the cathode $Q_C$, the amount of charge collected on the anode $Q_A$, and the drift time $t$. Then, the electron lifetime $\tau$ can be calculated as 
\begin{equation}
    Q_A/Q_C = e^{-t/\tau},
\end{equation}
or
\begin{equation}
    \tau = t\ \mathrm{ln}\left(Q_C/Q_A\right).
\end{equation}
Eventually, these dedicated purity monitors were turned off, and instead cosmic muons in the full TPC were used to monitor the purity.

MicroBooNE cools its liquid argon using liquid nitrogen refrigeration via two parallel condensers, one for normal operation and one on standby. The liquid nitrogen usage is about 3,400 liters/day, corresponding to about 6 kW of heat load. The temperature of the liquid argon is important to monitor, since small temperature variations can affect the electron drift velocity. The design goal is < 0.1 K of temperature variation throughout the cryostat. The temperatures are monitored using resistive thermal devices (RTDs), with 12 along the cryostat walls, 10 attached to the TPC structure, and 18 inside the filters.

\section{Detector Uncertainties}\label{sec:DetVar}

To estimate systematic uncertainties in the detector response, we simulate the detector with a central value (CV) simulation as well as nine types of detector variation simulations. For each type of detector variation, we reconstruct and pass events through the whole analysis to a final histogram of event counts, and then estimate a $1\sigma$ uncertainty on this type of detector variation by the difference between the CV and the detector variation histogram. In the following sections, I describe the nine types of detector variations we consider.

\subsection{Space Charge Effect}\label{sec:sce}

MicroBooNE's electrons drift at 1.098 m/ms, so they quickly clear out of the TPC in between each event. However, the positive argon ions left behind by those electrons pair up via $\mathrm{Ar}^+ + \mathrm{Ar} \rightarrow \mathrm{Ar}_2^+$, and these $\mathrm{Ar}_2^+$ ions also drift, but much more slowly due to their much larger size and mass, around 4 mm/s. Therefore, there is constantly a large cloud of cosmic-ray-induced positive ions slowly drifting toward the cathode before being neutralized. This cloud is of positive ions is known as space charge, and causes electric field nonuniformity, known as the ``space charge effect'' (SCE). The velocity of these ions can be comparable to the local argon flow velocities due to circulation and convection, so the precise details of this space charge can be complex to accurately model. This nonuniformity is measured by studying the endpoints and curvature of straight tracks after they have drifted through the TPC. We study two types of straight tracks: cosmic muons \cite{microboone_sce_muons}, and ionizations from the laser calibration system \cite{microboone_sce_laser}. The difference between these two resulting electric field maps is used to generate the SCE detector variation simulation.

\subsection{Electron-Ion Recombination}

\begin{figure}[H]
    \centering
    \includegraphics[width=0.7\textwidth]{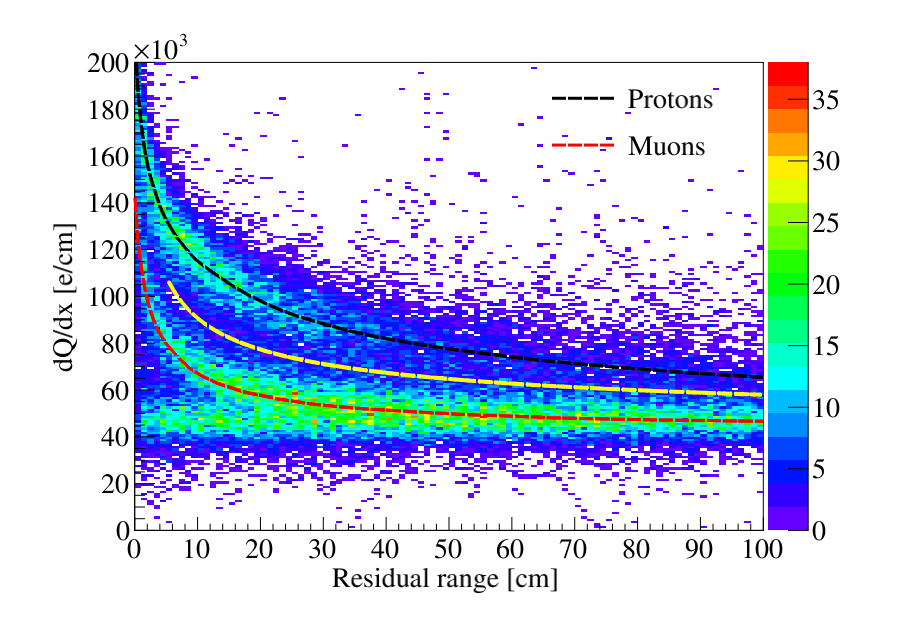}
    \caption[MicroBooNE $dQ/dx$ vs residual range]{MicroBooNE $dQ/dx$ vs residual range from data observations, with a clearly visible high $dQ/dx$ band from protons, and lower $dQ/dx$ band from muons, and a flat $dQ/dx$ band from muons exiting the TPC. The increase in $dQ/dx$ near the end of each tracks is known as the Bragg peak. From Ref. \cite{wire_cell_pattern_recognition}.}
    \label{fig:dQdx_residual_range}
\end{figure}

The amount of recombination of electrons with positive ions immediately after ionization depends on several factors. A stronger electric field makes electrons less likely to recombine. A higher local ionization charge density, affected by the ionization charge deposition per unit length $dQ/dx$, makes electrons more likely to recombine. And lastly, if a particle track is almost parallel to the drift direction, electrons are more likely to recombine \cite{icarus_angular_recombination}. All of these effects can be difficult to model precisely, so in MicroBooNE we measured this electron-ion recombination for muons (low $dQ/dx$) and protons (high $dQ/dx$) \cite{microboone_dEdx}, using the type of measurement shown in Fig. \ref{fig:dQdx_residual_range}. We then created a detector variation simulation according to an alternate recombination model that matches our data observations more closely than our CV simulation.

\subsection{Wire Modification}

The electronic signals produced on our wires by induction and collection from drifting electrons is influenced by many factors: electron-ion recombination, electron diffusion and attenuation, space charge effect, electronic response, etc. These effects can be difficult to disentangle, so we parametrize the difference between data and simulation according to different variables to arrive at a data-driven detector variation simulation \cite{microboone_wiremod}.

\begin{figure}[H]
    \centering
    \begin{subfigure}[b]{0.49\textwidth}
        \includegraphics[width=\textwidth]{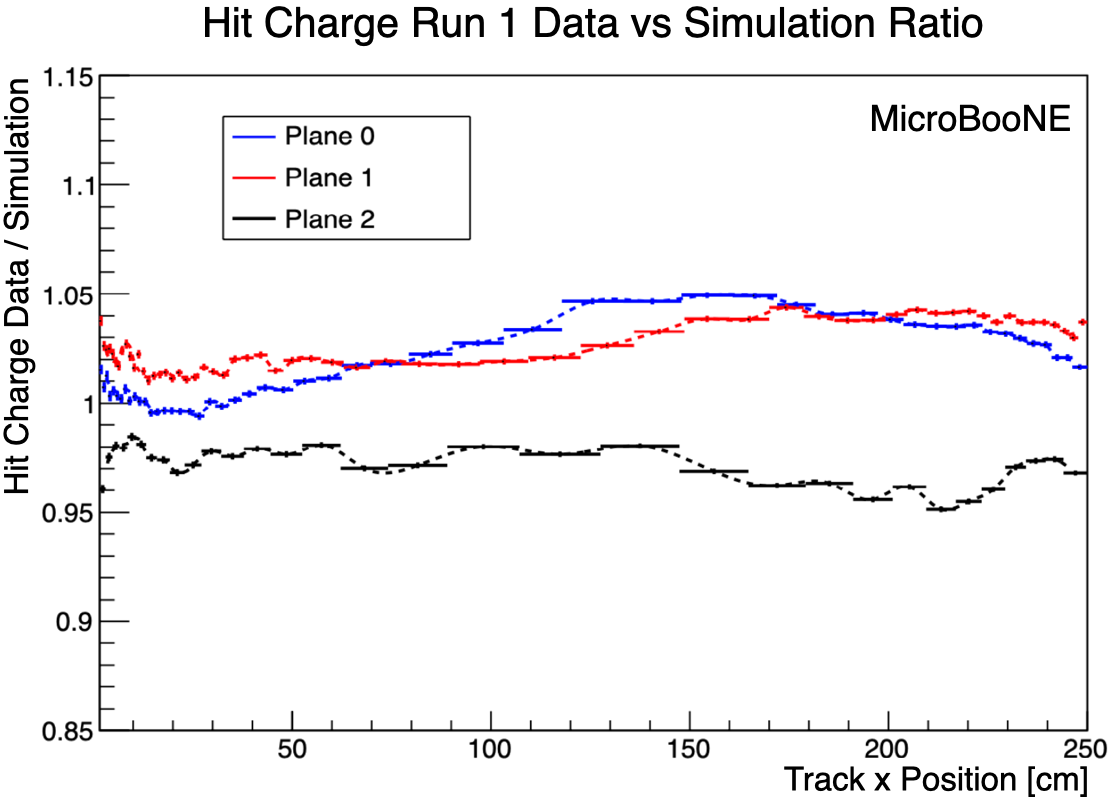}
        \caption{}
        \label{fig:wiremod_x_charge}
    \end{subfigure}
    \hfill
    \begin{subfigure}[b]{0.49\textwidth}
        \includegraphics[width=\textwidth]{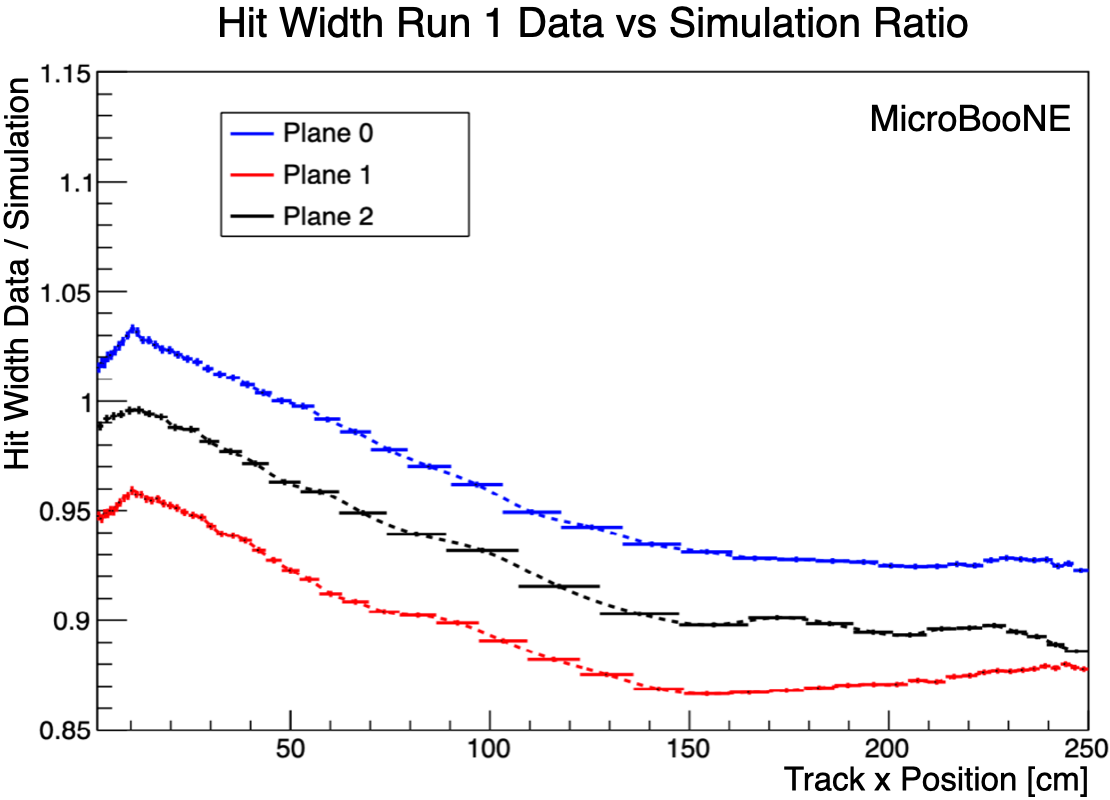}
        \caption{}
        \label{fig:wiremod_x_width}
    \end{subfigure}
    \caption[MicroBooNE wire modification detector uncertainty]{MicroBooNE wire modification hit charge and width ratios between data and simulation according to $x$ position on each of the three wire planes. Panel (a) shows the hit charge ratios, and panel (b) shows the hit width ratios. From Ref. \cite{microboone_wiremod}.}
    \label{fig:wiremodx}
\end{figure}

For example, we examine how the reconstructed integrated charge and peak width depend on $x$, the drift length coordinate, on each of our three wire planes, as shown in Fig. \ref{fig:wiremodx}. Then we modify the waveforms of our simulation (which does not require resimulating the electron drift, saving computational time) in order to get a detector variation simulation which represents waveforms that more closely match our data observations. We then repeat this process for position in the $y-z$ plane (the plane parallel to our wire planes), the angle of the track when viewed from above $\theta_{xz}$, and the angle of the track when viewed along the drift direction $\theta_{yz}$, to form four total wire modification detector variation simulations.

\subsection{Light Production and Propagation}\label{sec:light_detvar}

\begin{figure}[H]
    \centering
    \includegraphics[width=0.6\textwidth]{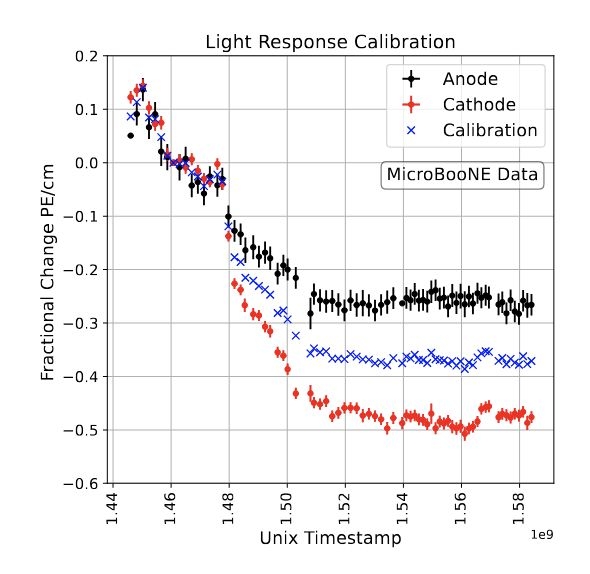}
    \caption[MicroBooNE light calibration]{MicroBooNE light calibration over time. Updated relative to Ref. \cite{microboone_light_yield_public_note} with an expanded time period.}
    \label{fig:light_calibration}
\end{figure}

Lastly, we simulate detector variations according to different scintillation light production and propagation models. MicroBooNE has seen a notable decline in measured light yield over the course of operations \cite{microboone_light_yield_public_note}. We finished finished processing the full dataset, including our earliest run periods with different trigger settings, and our latest run periods just before decomissioning, as shown in Fig. \ref{fig:light_calibration}. These new additions indicate that our measured light yield was declining steadily in the earliest runs, and that our light yield was stable at the latest runs. This light yield decline was measured by cosmic muon tracks, specifically anode piercing tracks and cathode piercing tracks. This reveals an interesting position dependence, with tracks near the cathode (further from the PMTs) experiencing significantly larger fractional light declines relative to tracks near the anode (closer to the PMTs). Interestingly, a similar study of the light yield decline using isolated point-like cosmic proton tracks showed a similar light yield decline, but much less position dependence \cite{microboone_proton_light_yield_public_note}. A variety of explanations have been hypothesized, including the introduction of dimer de-exciting or light absorbing or light scattering contaminants in the argon, degradation or dissolving of the TPB wavelength shifting coatings, or degradation of the PMTs. Further study of this light yield decline is planned, including with studies of deep-learning-based data-driven light simulation \cite{polina_dl_light_slides, siren_dl_light}, studies of our argon purity using samples provided to external laboratories \cite{ciemat_microboone_argon_purity}, and studies of the now empty MicroBooNE cryostat interior using a camera sensitive to the fluorescense and therefore possible degradation or migration of the TPB wavelength shifting material.

We account for the average light yield decline between the anode and cathode measurements when simulating events, and also simulate additional uncertainties accounting for three effects. 
We simulate an overall 25\% decrease in measured light yield, matching the type of change we have seen over time. 
We also simulate increased light attenuation during later run periods, matching the type of position dependence we have seen in our muon light yield calibration studies.
Lastly, we simulate a different Rayleigh scattering length, at 120 cm rather than the 60 cm in our nominal simulation, since Ref. \cite{LAr_rayleigh} measured longer scattering lengths around 100 cm. This generates three total light related detector variation simulations.

\section{BNB and NuMI Beams}\label{sec:beams}

\begin{figure}[H]
    \centering
    \includegraphics[width=\textwidth]{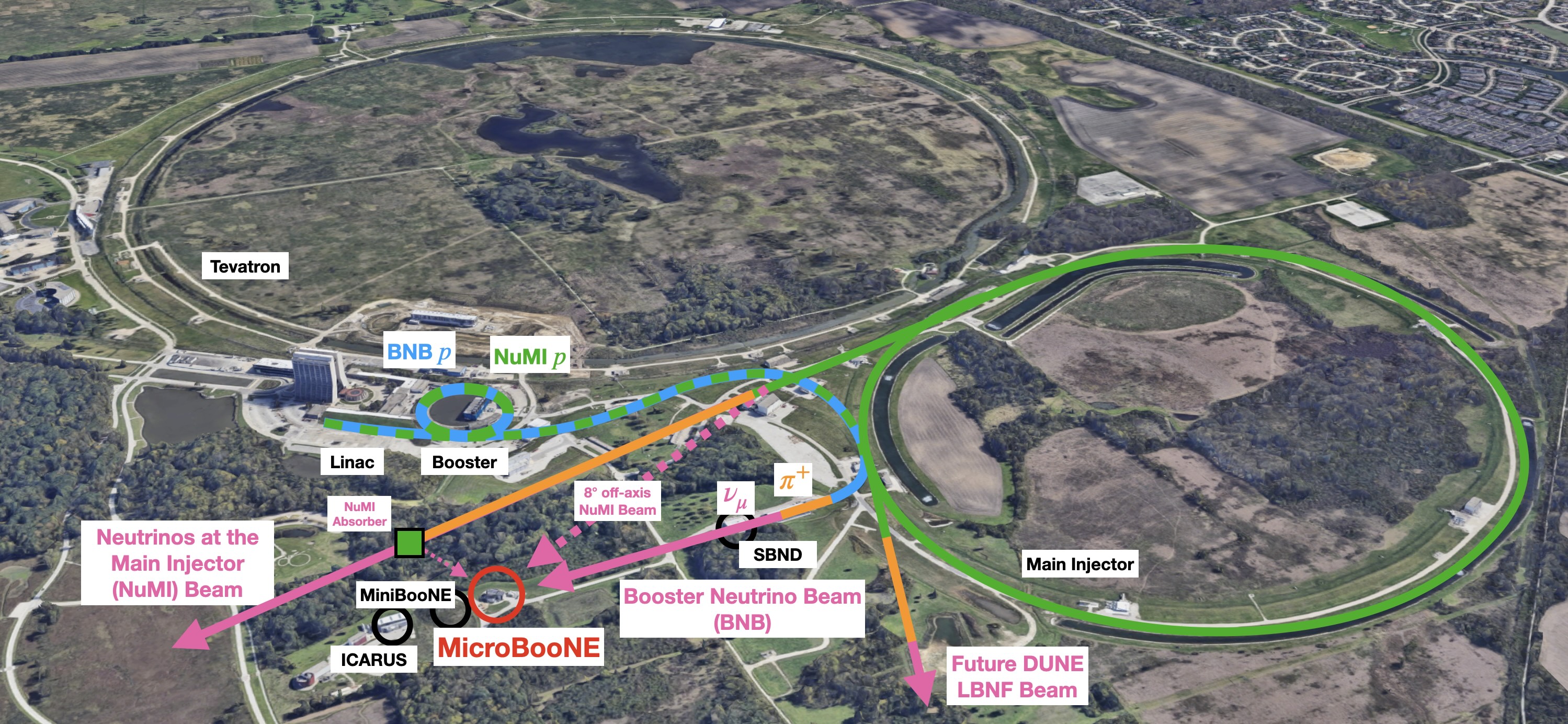}
    \caption[Fermilab neutrino beams]{Illustration of the Fermilab neutrino beams. The path of BNB protons accelerated by the Booster ring is shown in blue, while the path of the NuMI protons accelerated by the Main Injector ring is shown in green. The paths of charged pions between the target and absorber pipe is shown in orange, although protons, kaons, muons, and neutrinos also exist in these beams. The resulting neutrino beam paths are shown in pink, including the $8^\circ$ off-axis NuMI beam observed by MicroBooNE, and neutrinos from particles (for example kaons) decaying at rest in the NuMI absorber. MicroBooNE is indicated by a red circle, and the SBND, MiniBooNE, and ICARUS experiments are indicated by black circles. The future DUNE LBNF beam is also shown in the bottom portion of the image.}
    \label{fig:neutrino_beam_map}
\end{figure}

MicroBooNE sees neutrinos from two beams at Fermilab. It sits on-axis in the Booster Neutrino Beam (BNB), and sits $8^\circ$ off-axis in the Neutrinos at the Main Injector (NuMI) beam. The position of MicroBooNE relative to the Fermilab campus neutrino beams is illustrated in Fig. \ref{fig:neutrino_beam_map}.

\begin{figure}[H]
    \centering
    \begin{subfigure}[b]{0.49\textwidth}
        \includegraphics[width=\textwidth]{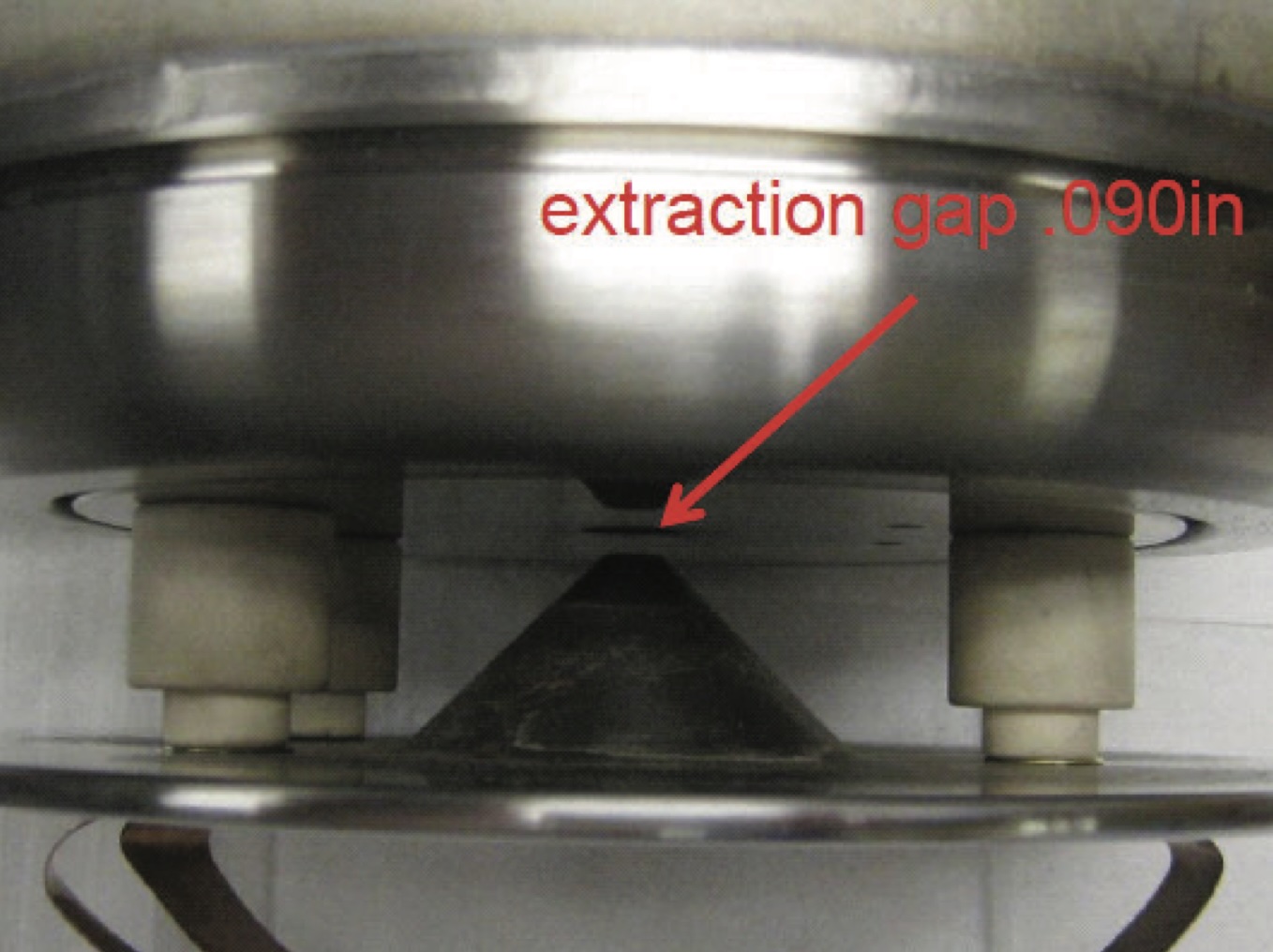}
        \caption{}
        \label{fig:extractor_cone}
    \end{subfigure}
    \hfill
    \begin{subfigure}[b]{0.49\textwidth}
        \includegraphics[width=\textwidth]{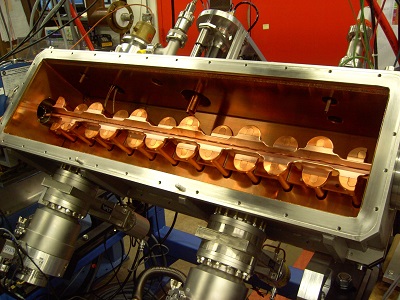}
        \caption{}
        \label{fig:fermilab_RFQ}
    \end{subfigure}
    \hfill
    \begin{subfigure}[b]{0.7\textwidth}
        \includegraphics[width=\textwidth]{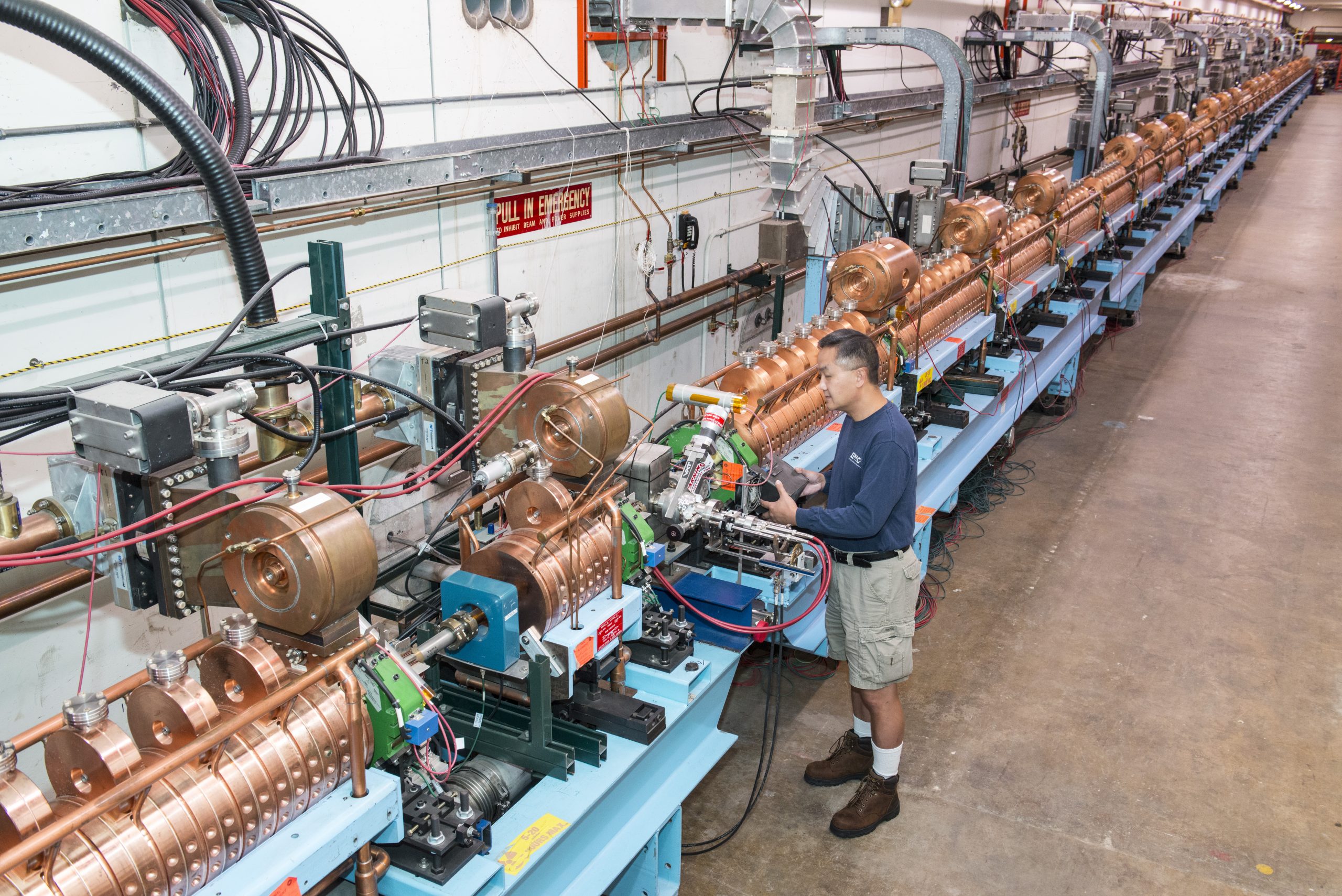}
        \caption{}
        \label{fig:fermilab_linac}
    \end{subfigure}
    \caption[Fermilab linear accelerator]{Fermilab's linear accelerators. Panel (a) shows the extraction cone electrode, which extracts and acccelerates $\mathrm{H}^-$ ions from rest, from \cite{FermilabRookieBookRIL}. Panel (b) shows the radio-frequency quadrupole (RFQ), from Ref. \cite{fermilab_RFQ}. Panel (c) shows the linear accelerator or Linac, from Ref. \cite{fermilab_linac_photo}.}
    \label{fig:fermilab_source_and_linac}
\end{figure}

These neutrino beams start as $\mathrm{H}^-$ ions, which are extracted from rest in a plasma to 35 keV using a -35 kV 250 $\mu$s high voltage pulse on a cone-shaped electrode \cite{FermilabRookieBookRIL}, then accelerated from 35 keV to 750 keV in the radio-frequency quadrupole (RFQ) \cite{fermilab_RFQ}, and then accelerated in the linear accelerator or Linac from 750 keV to 400 MeV, as shown in Fig. \ref{fig:fermilab_source_and_linac}.

\begin{figure}[H]
    \centering
    \begin{subfigure}[b]{0.49\textwidth}
        \includegraphics[width=\textwidth]{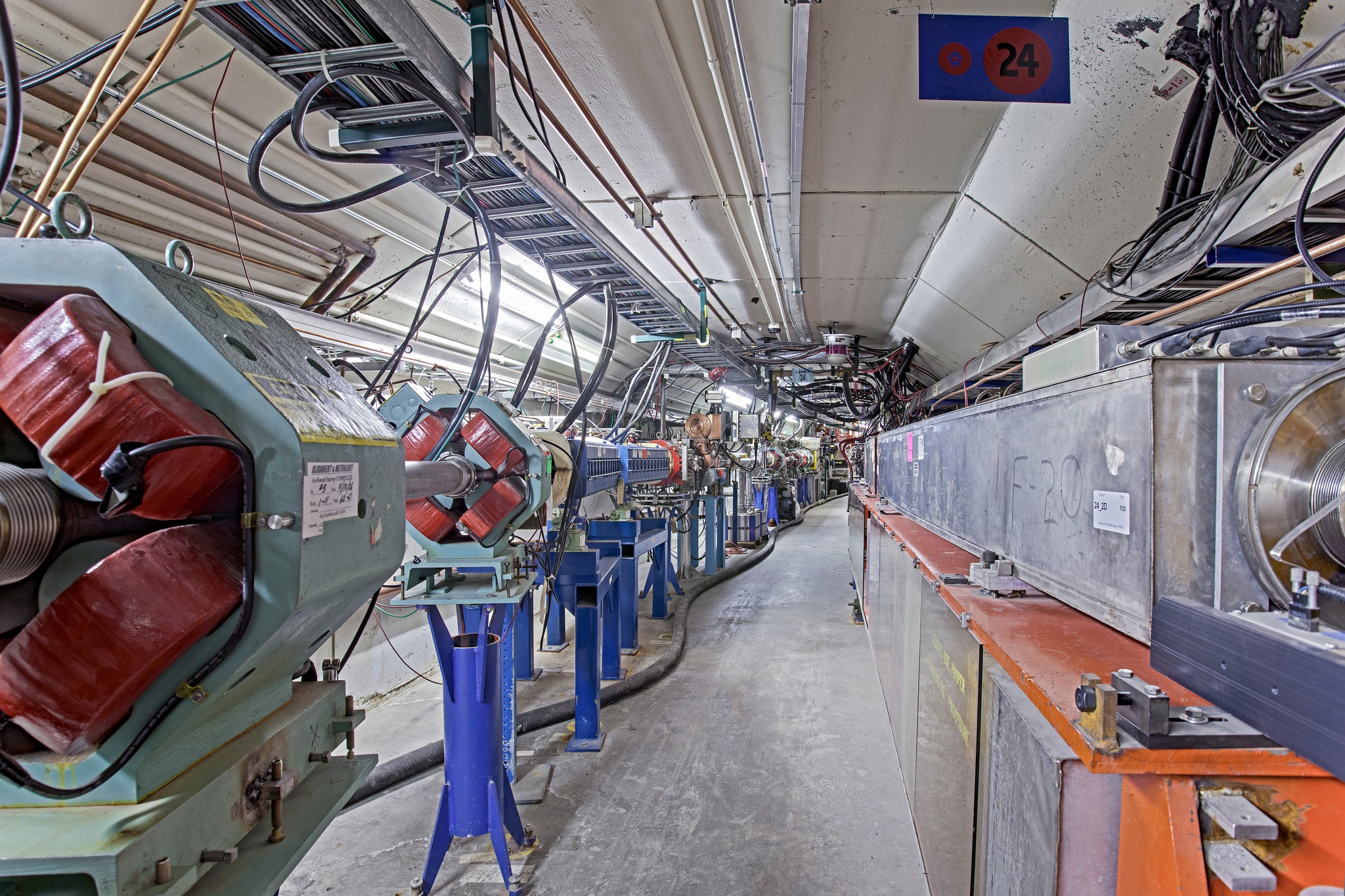}
        \caption{}
        \label{fig:booster_interior}
    \end{subfigure}
    \hfill
    \begin{subfigure}[b]{0.49\textwidth}
        \includegraphics[width=\textwidth]{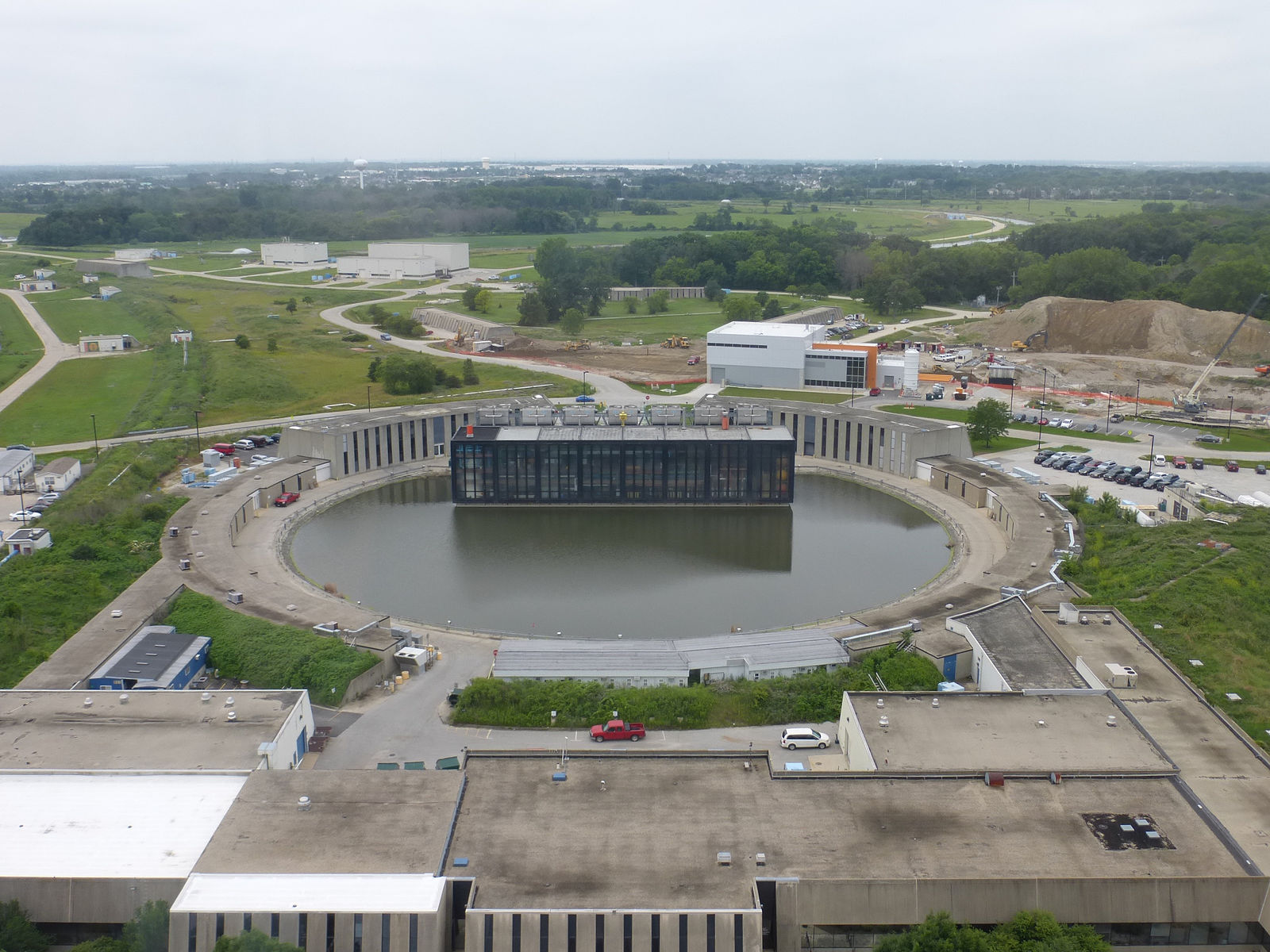}
        \caption{}
        \label{fig:booster_exterior}
    \end{subfigure}
    \caption[Fermilab Booster]{Fermilab's Booster accelerator. Panel (a) shows an interior view of the Booster ring from \cite{booster_exterior_wikimedia}, and panel (b) shows an exterior view from \cite{booster_interior_photo}.}
    \label{fig:booster}
\end{figure}

As the $\mathrm{H}^-$ ions enter the Booster ring, they are stripped of their two electrons, leaving bare protons. In the Booster, the protons are further accelerated from 400 MeV to 8 GeV in energy, as shown in Fig. \ref{fig:booster}.

\begin{figure}[H]
    \centering
    \begin{subfigure}[b]{0.37\textwidth}
        \includegraphics[width=\textwidth]{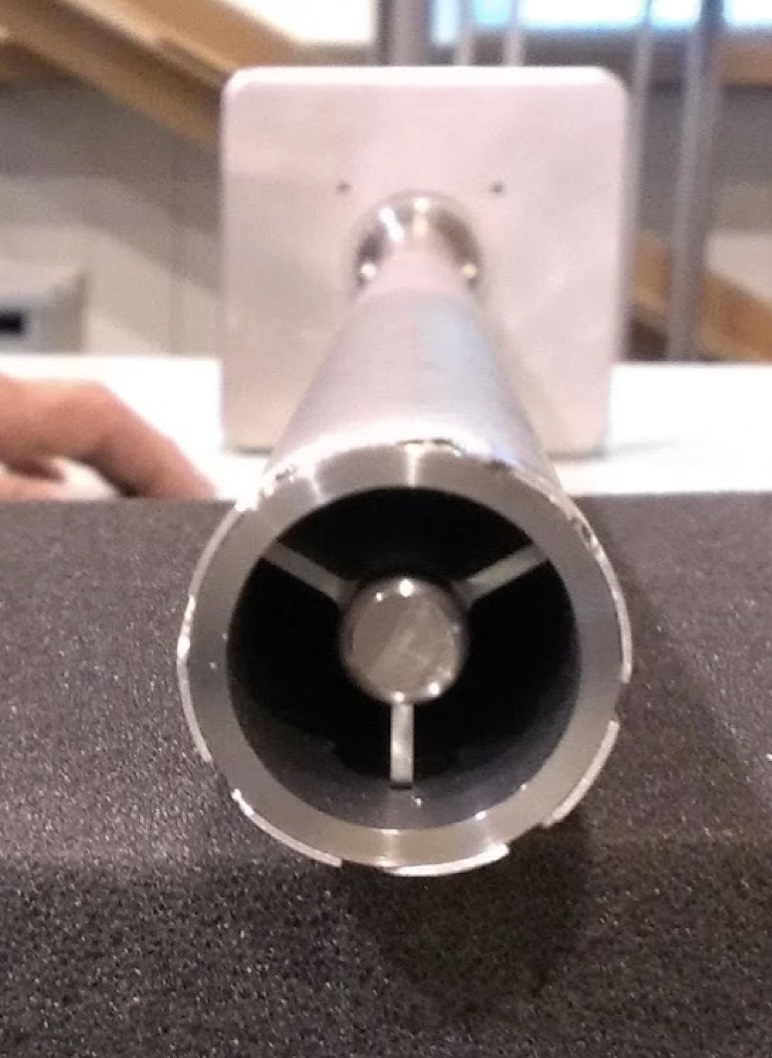}
        \caption{}
        \label{fig:bnb_target}
    \end{subfigure}
    \hfill
    \begin{subfigure}[b]{0.62\textwidth}
        \includegraphics[width=\textwidth]{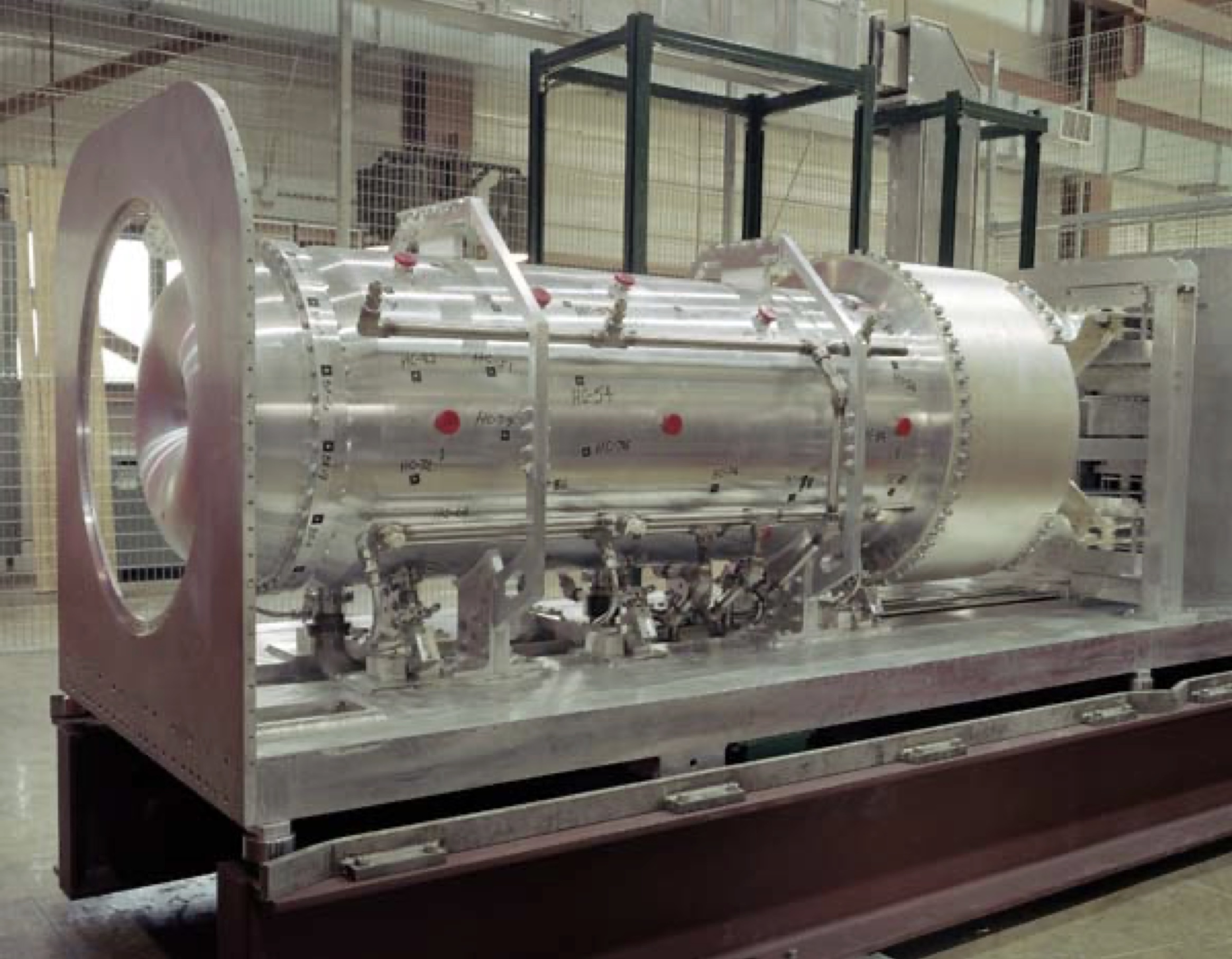}
        \caption{}
        \label{fig:bnb_horn}
    \end{subfigure}
    \caption[BNB horn and target]{Panel (a) shows the BNB beryllium air-cooled target. Panel (b) shows the BNB magnetic focusing horn. Images are from Ref. \cite{bnb_horn_target}.}
    \label{fig:bnb_horn_target}
\end{figure}

For the BNB, these protons are then directed to a target located inside a magnetic focusing horn, as shown in Fig. \ref{fig:bnb_horn_target}. The protons interact with the beryllium target, producing many hadrons at many energies and angles. The magnetic focusing horn is pulsed at 174 kA in order to produce a toroidal magnetic field which focuses positively charged particles forward along the beam path, and focuses negatively charged particles away from the beam path. This current can also flow the other way in order to focus negatively charged particles, but MicroBooNE did not take any data in this configuration of the BNB.

\begin{figure}[H]
    \centering
    \includegraphics[width=0.7\textwidth]{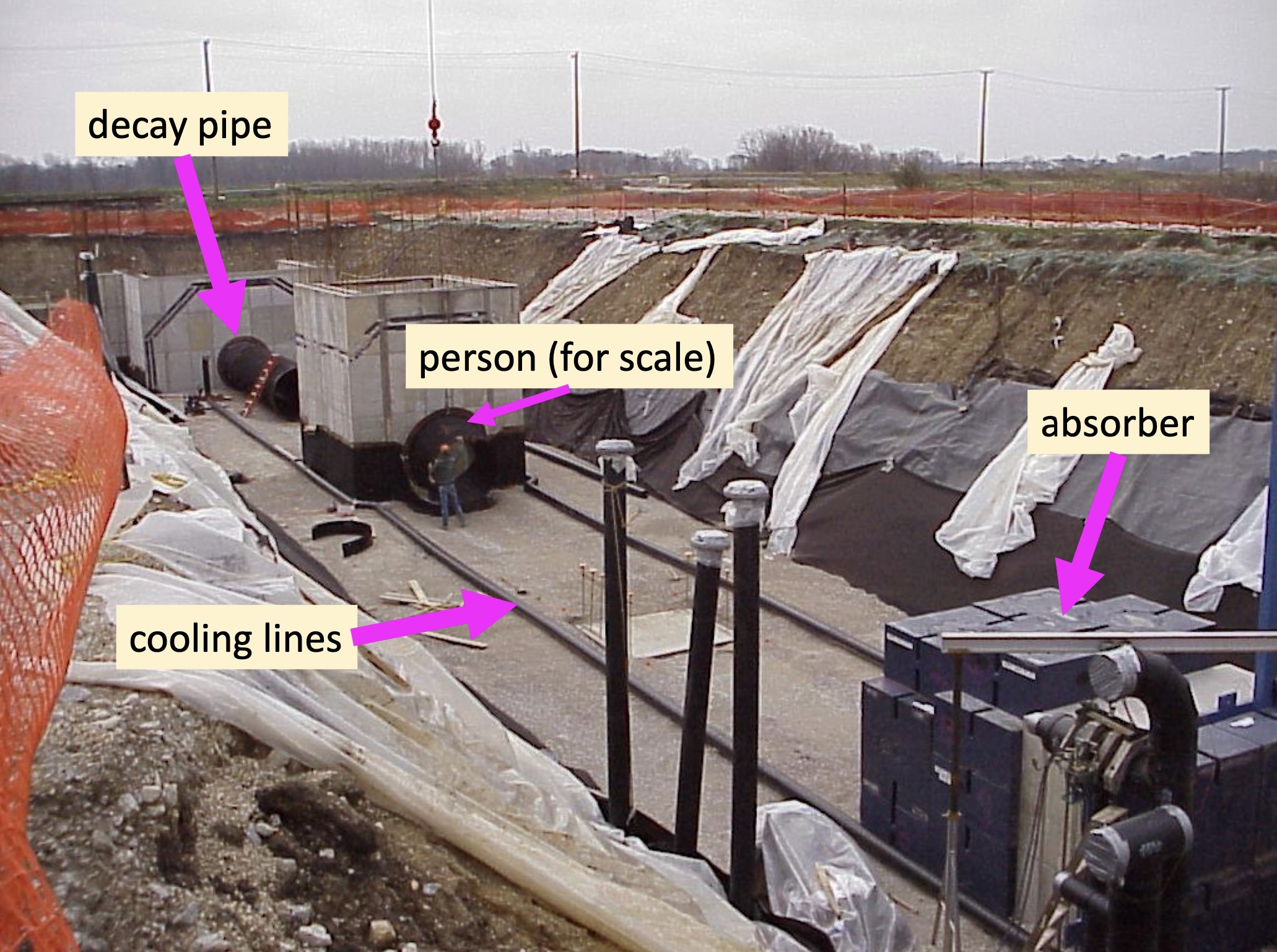}
    \caption[BNB decay pipe]{BNB decay pipe and absorber, from Ref. \cite{bnb_decay_pipe_numi_target}.}
    \label{fig:bnb_decay_pipe}
\end{figure}

This beam of positively charged hadrons then enters a 50 m decay pipe as shown in Fig. \ref{fig:bnb_decay_pipe}, where pions and kaons can decay, producing neutrinos. Specifically, the dominant neutrino production decay is $\pi^+ \rightarrow \mu^+ + \nu_\mu$. This is because the $\pi^+ \rightarrow e^+ + \nu_e$ is helicity suppressed; since the weak force only interacts with left-chiral particles and right-chiral antiparticles, this $e^+$ would have to be in a right-chiral state, while due to angular momentum conservation, it would have to be in a left-helicity state (since the neutrino will always be in a left-helicity state). This mismatch between right-chirality and left-helicity causes a suppression that depends on the charged lepton mass. It would be forbidden if the charged lepton were massless, and it suppresses the light $e^+$ much more than the heavier $\mu^+$. There are also neutrinos produced by charged kaon decays $K^+\rightarrow \mu^+ + \nu_\mu$, $K^+\rightarrow \pi^0 + e^+ + \nu_e$, neutral kaon decays $K^0_L \rightarrow \pi^\pm + e^\mp + \stackrel{\scalebox{0.6}{(}-\scalebox{0.6}{)}}{\nu_e}$, and $K^0_L \rightarrow \pi^\pm + \mu^\mp + \stackrel{\scalebox{0.6}{(}-\scalebox{0.6}{)}}{\nu_\mu}$, and muon decays $\mu^+\rightarrow e^+ + \nu_e + \overline{\nu_\mu}$, as well as other rarer decays.

\begin{figure}[H]
    \centering
    \includegraphics[width=\textwidth]{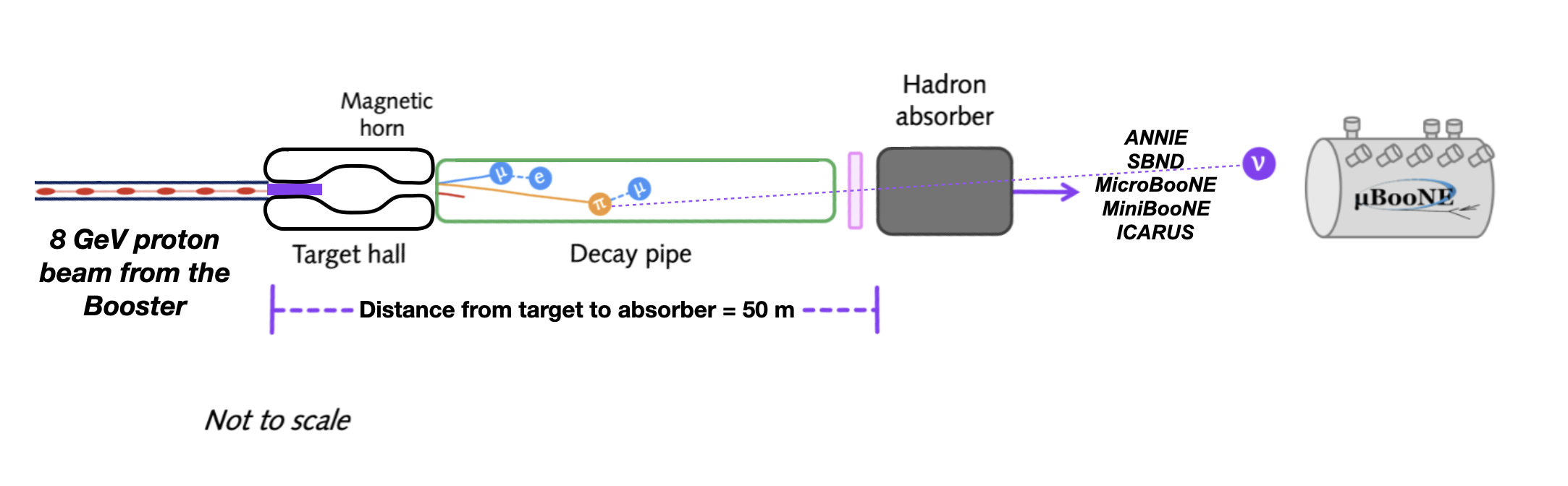}
    \caption[MicroBooNE BNB beam]{MicroBooNE BNB beam diagram. Adapted from Ref. \cite{numi_nue_Np_microboone_in_prep}.}
    \label{fig:microboone_bnb_beam}
\end{figure}

\begin{figure}[H]
    \centering
    \includegraphics[width=0.5\textwidth]{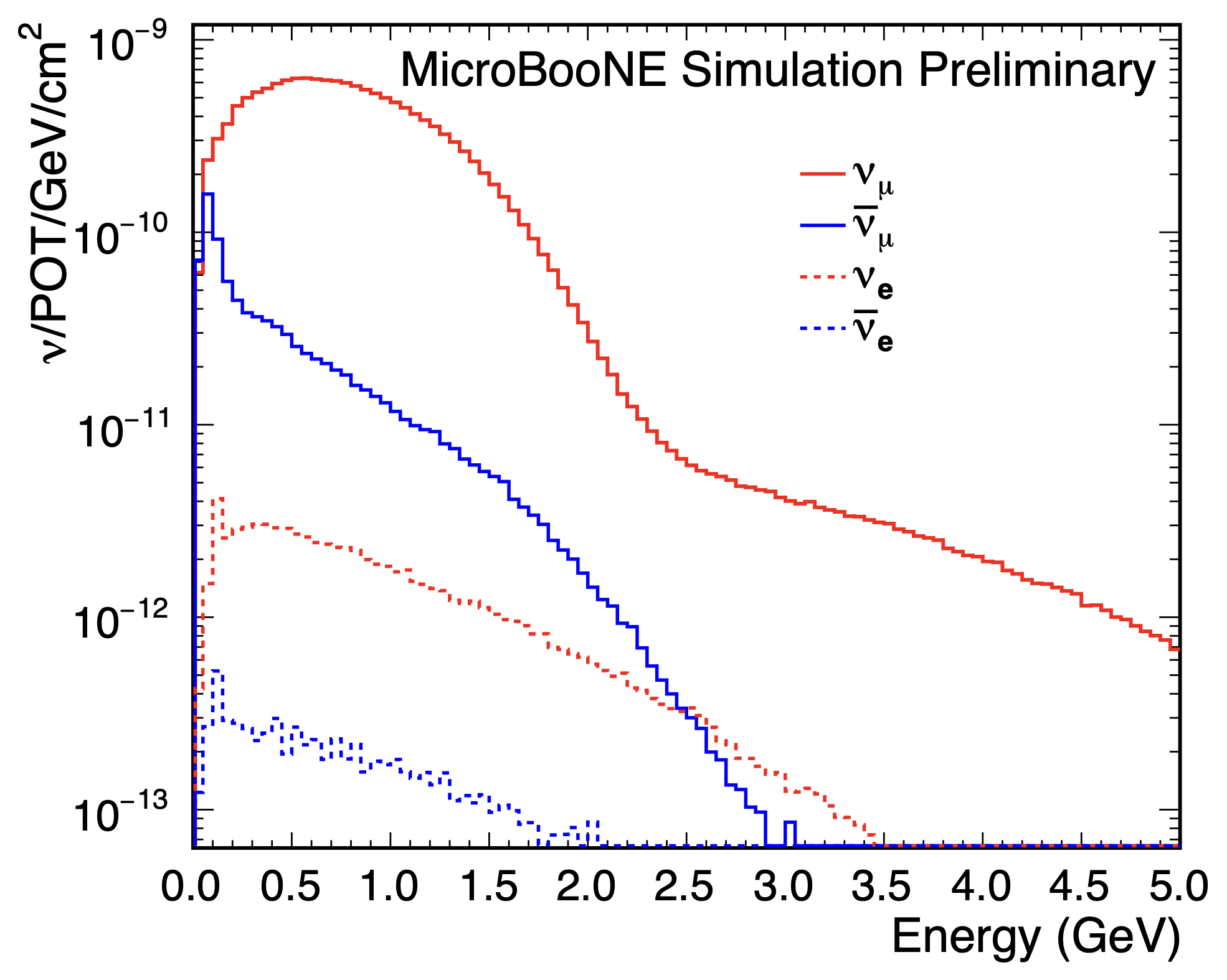}
    \caption[BNB flux spectrum]{BNB flux spectrum, from Ref. \cite{microboone_bnb_flux_public_note}.}
    \label{fig:bnb_flux_spectrum}
\end{figure}

This entire process of hadron production, focusing, and decay creating neutrinos for MicroBooNE is illustrated as a simplified diagram in Fig. \ref{fig:microboone_bnb_beam}. The resulting energy spectra of the $\nu_\mu$, $\overline{\nu_\mu}$, $\nu_e$, and $\overline{\nu_e}$ fluxes are shown in Fig. \ref{fig:bnb_flux_spectrum}.

\begin{figure}[H]
    \centering
    \includegraphics[width=\textwidth]{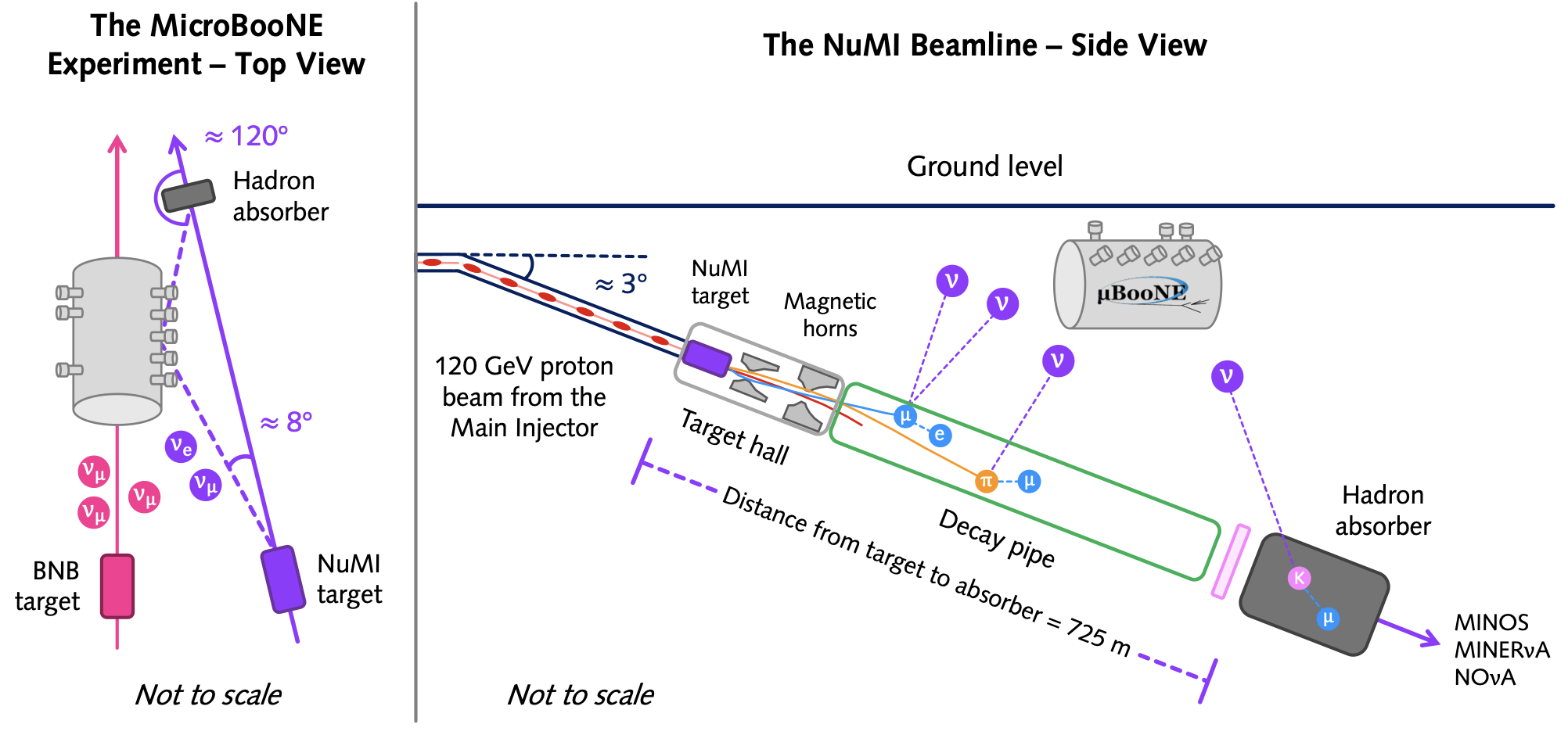}
    \caption[MicroBooNE NuMI beam]{MicroBooNE NuMI beam diagram. From Ref. \cite{numi_nue_Np_microboone_in_prep}.}
    \label{fig:microboone_numi_beam}
\end{figure}

In addition to the BNB, MicroBooNE sees $8^\circ$ off-axis flux from the NuMI beam, illustrated as a simplified diagram in Fig. \ref{fig:microboone_numi_beam}.

\begin{figure}[H]
    \centering
    \begin{subfigure}[b]{0.49\textwidth}
        \includegraphics[width=\textwidth]{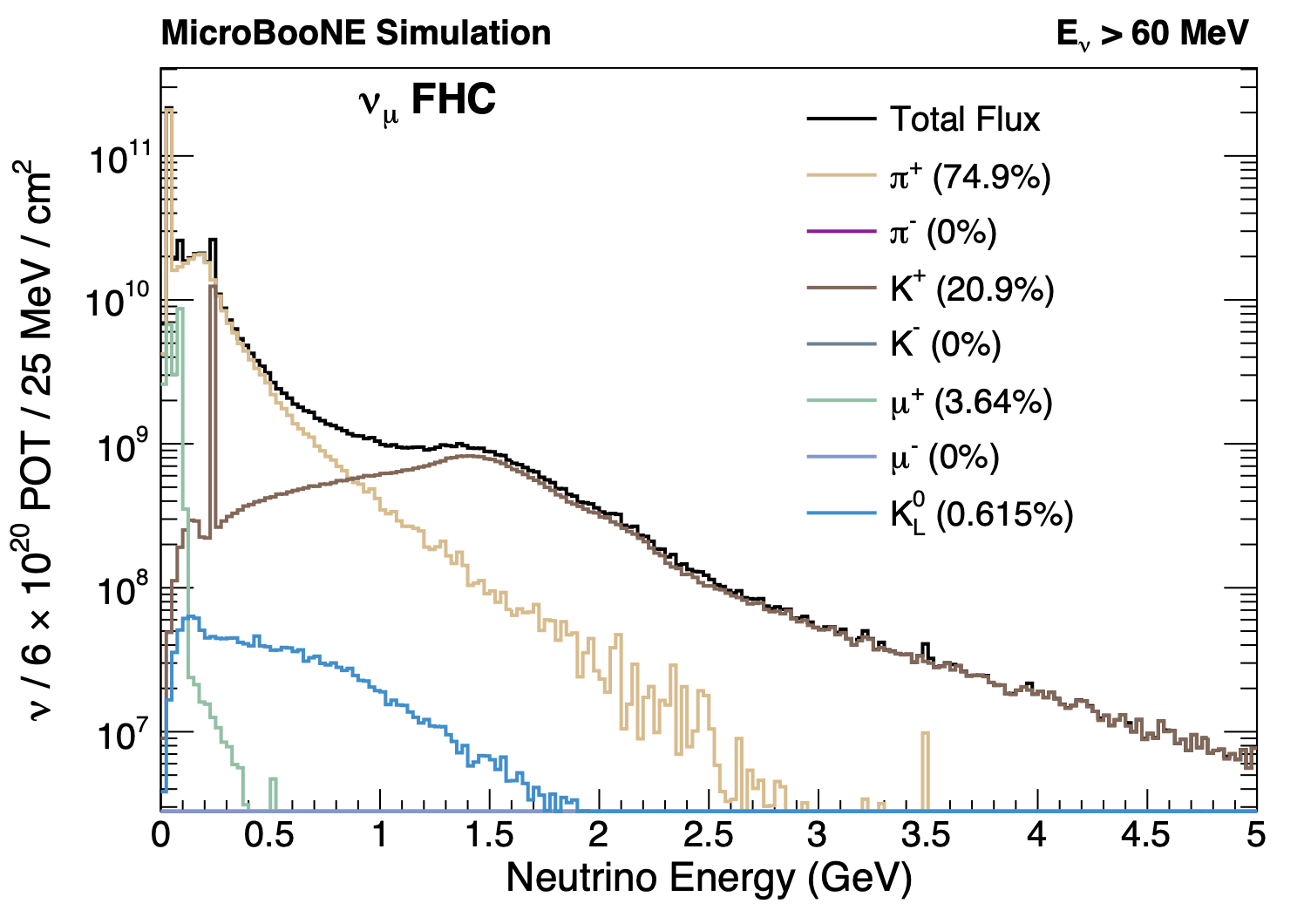}
        \caption{}
        \label{fig:numi_numu_FHC_spectrum}
    \end{subfigure}
    \hfill
    \begin{subfigure}[b]{0.49\textwidth}
        \includegraphics[width=\textwidth]{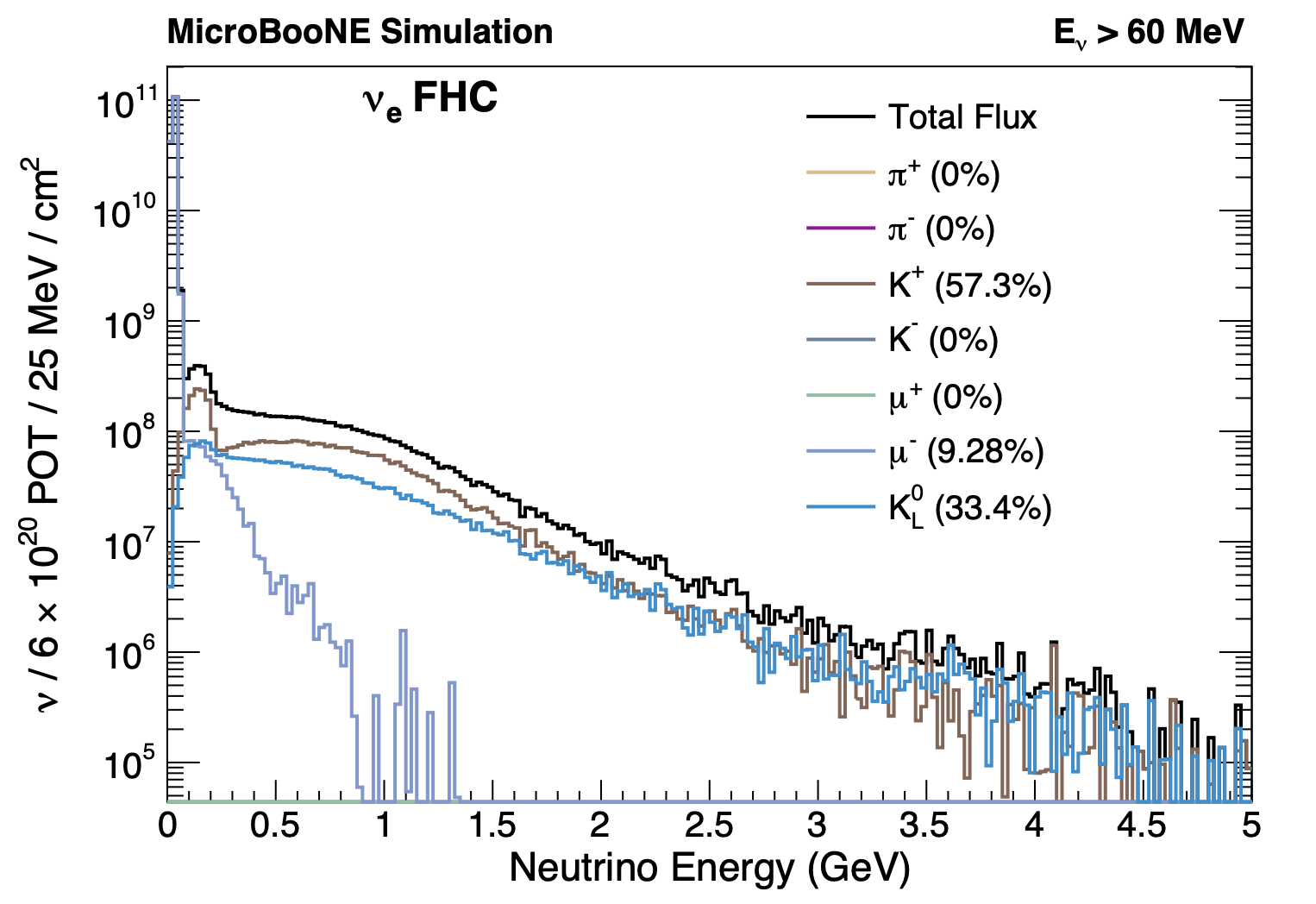}
        \caption{}
        \label{fig:numi_nue_FHC_spectrum}
    \end{subfigure}
    \caption[NuMI flux spectrum]{Panel (a) shows the NuMI flux spectrum in FHC mode for $\nu_\mu$. Panel (b) shows the NuMI flux spectrum in FHC model for $\nu_e$. From Ref. \cite{microboone_numi_flux_public_note}.}
    \label{fig:numi_flux_spectrum}
\end{figure}

The NuMI beam has significantly higher proton energies and therefore significantly higher on-axis neutrinos energies, but due to MicroBooNE's off-axis position, it sees a beam with a roughly similar energy distribution relative to the BNB, as shown in Fig. \ref{fig:numi_flux_spectrum}. Being so far off-axis, MicroBooNE sees more neutrinos from kaons, therefore having a significantly higher $\nu_e$ fraction. MicroBooNE took data in both forward horn current (FHC) and reverse horn current (RHC) modes of the NuMI beam, however the difference between these modes is actually fairly minor since we see so many neutrinos from particles which were not focused in the magnetic horns. MicroBooNE also sees neutrinos from the NuMI hadron absorber, for example kaon decay at rest (KDAR) neutrinos, which were first measured by MiniBooNE from this source \cite{miniboone_kdar}. Many features of the NuMI beam are particularly interesting for BSM particle searches, due to the fact that particles coming from the NuMI absorber will have distinct direction and timing features relative to neutrino backgrounds, as we used in Refs. \cite{HNL_scalar_microboone, HNL_microboone, scalar_microboone, scalar_microboone_2}.

\begin{figure}[H]
    \centering
    \includegraphics[width=0.7\textwidth]{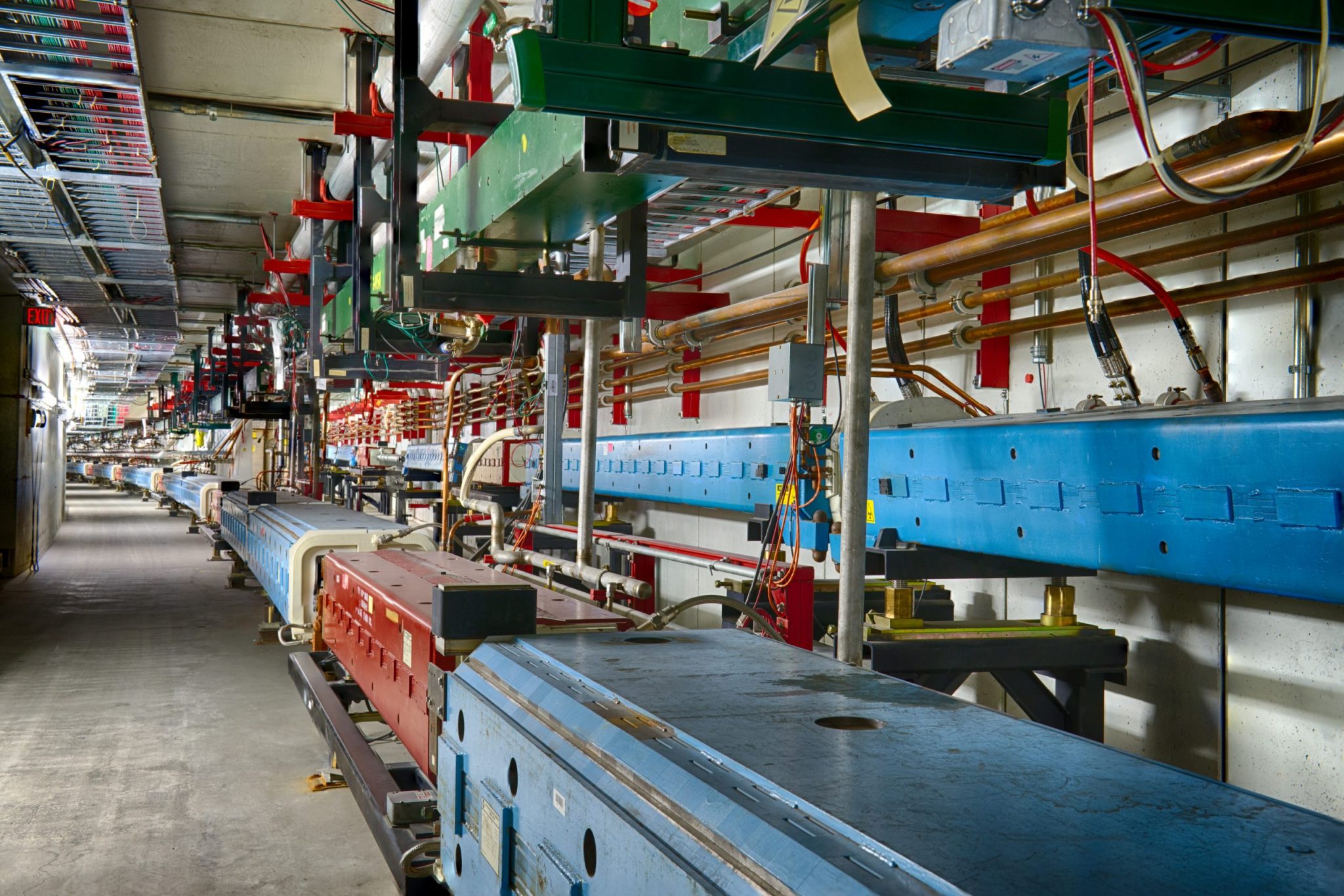}
    \caption[NuMI Main Injector]{NuMI Main Injector. From Ref. \cite{main_injector_image}.}
    \label{fig:main_injector}
\end{figure}

\begin{figure}[H]
    \centering
    \begin{subfigure}[b]{0.45\textwidth}
        \includegraphics[width=\textwidth]{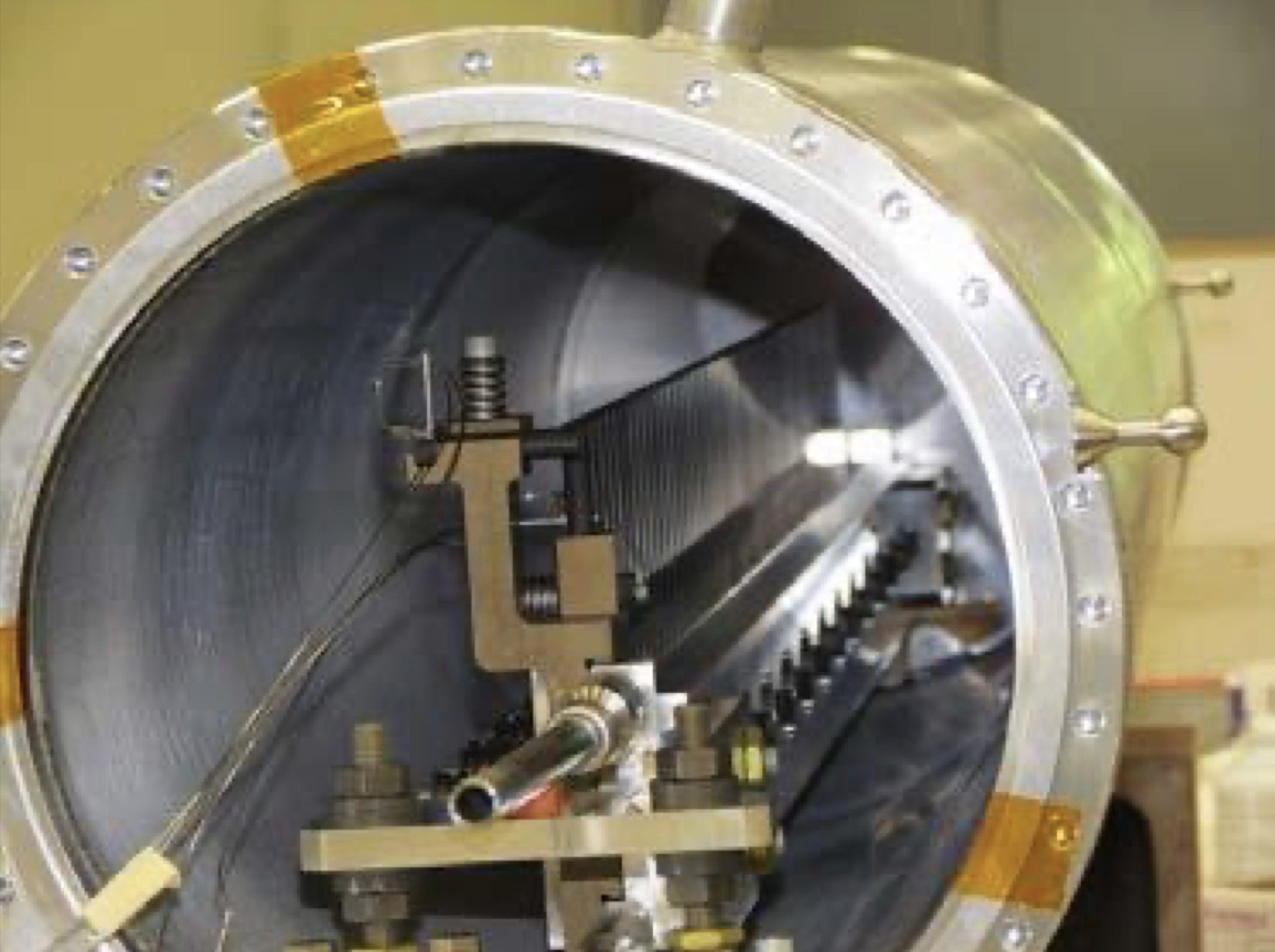}
        \caption{}
        \label{fig:numi_target}
    \end{subfigure}
    \hfill
    \begin{subfigure}[b]{0.54\textwidth}
        \includegraphics[width=\textwidth]{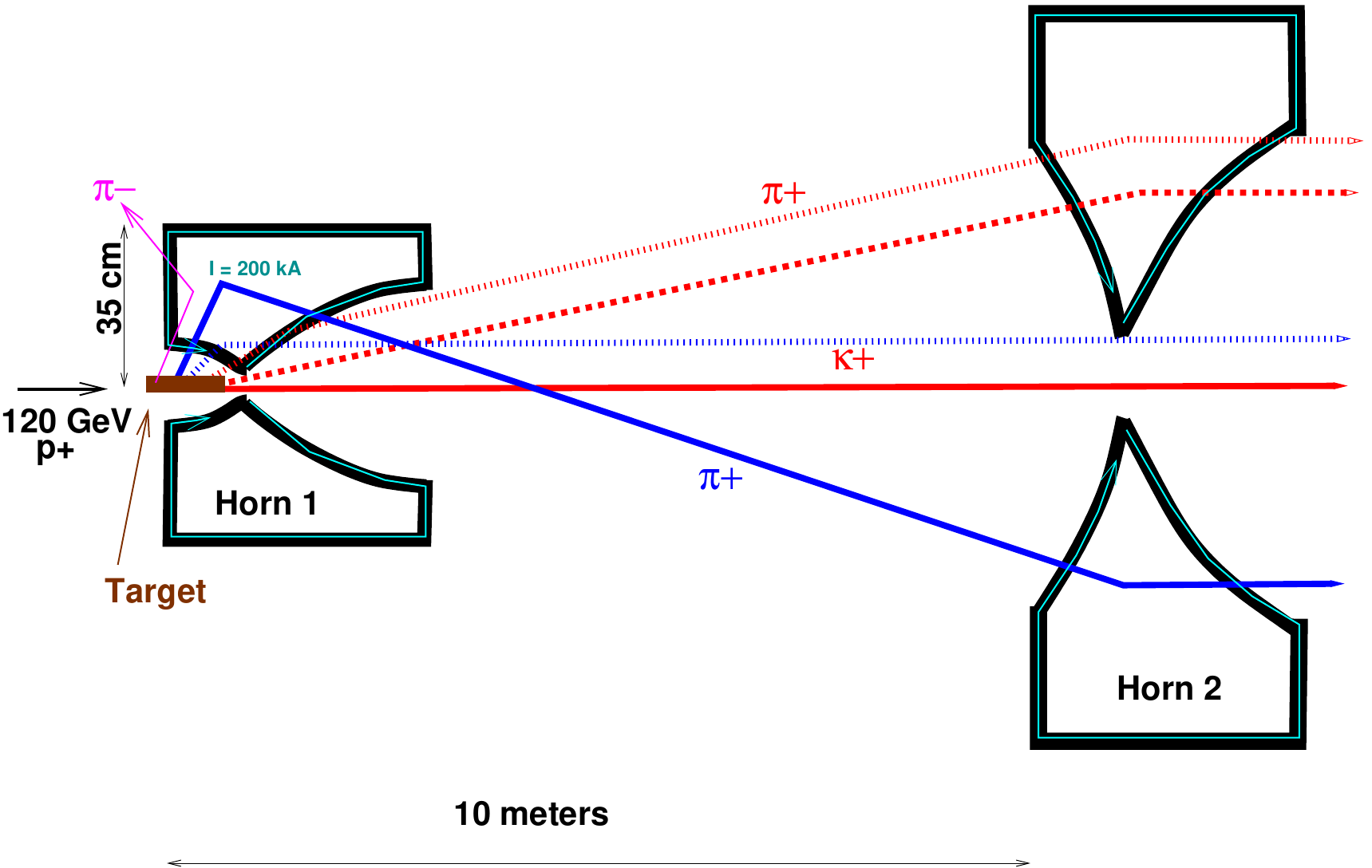}
        \caption{}
        \label{fig:numi_horns_diagram}
    \end{subfigure}
    \hfill
    \begin{subfigure}[b]{0.43\textwidth}
        \includegraphics[width=\textwidth]{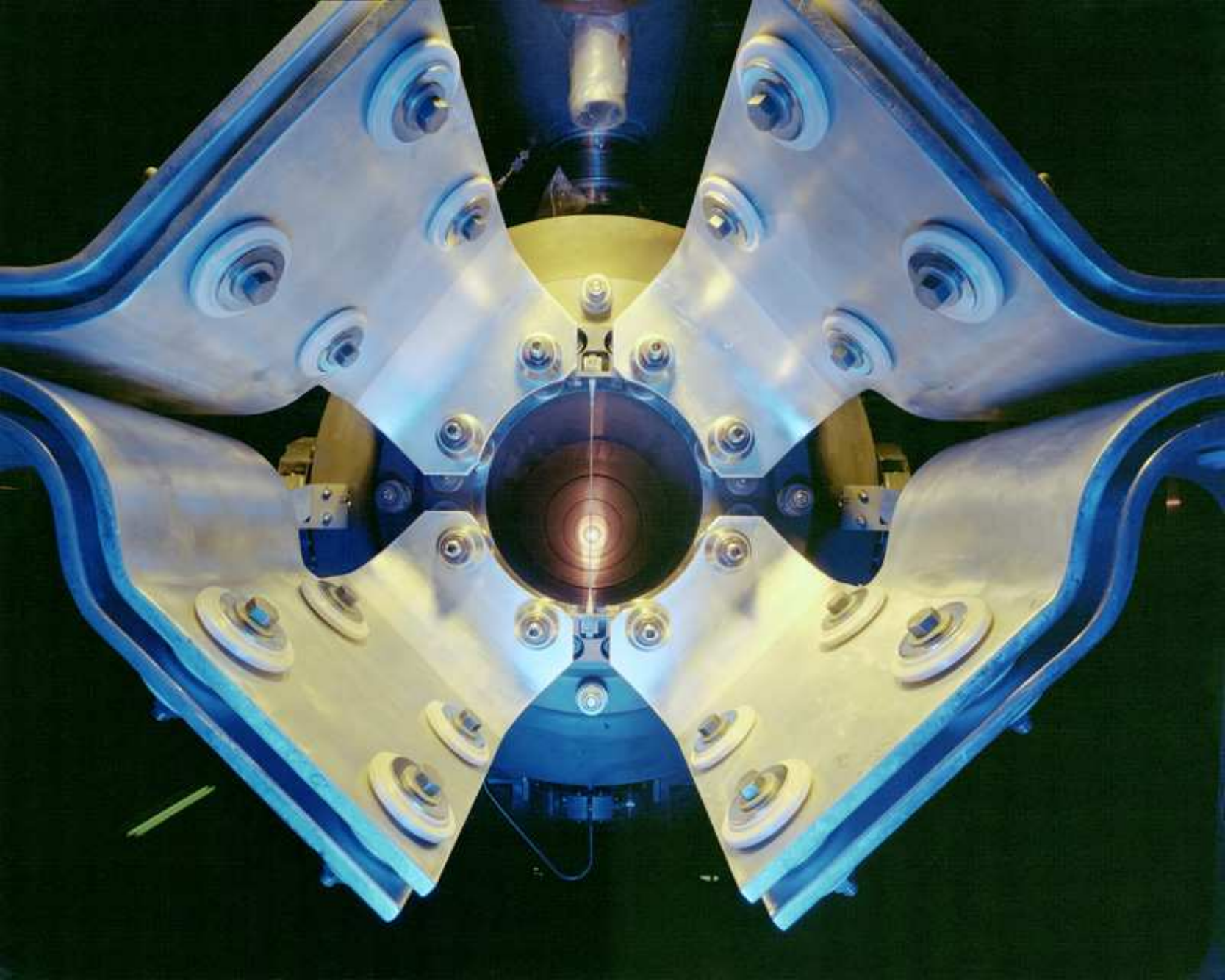}
        \caption{}
        \label{fig:numi_horn1}
    \end{subfigure}
    \begin{subfigure}[b]{0.23\textwidth}
        \includegraphics[width=\textwidth]{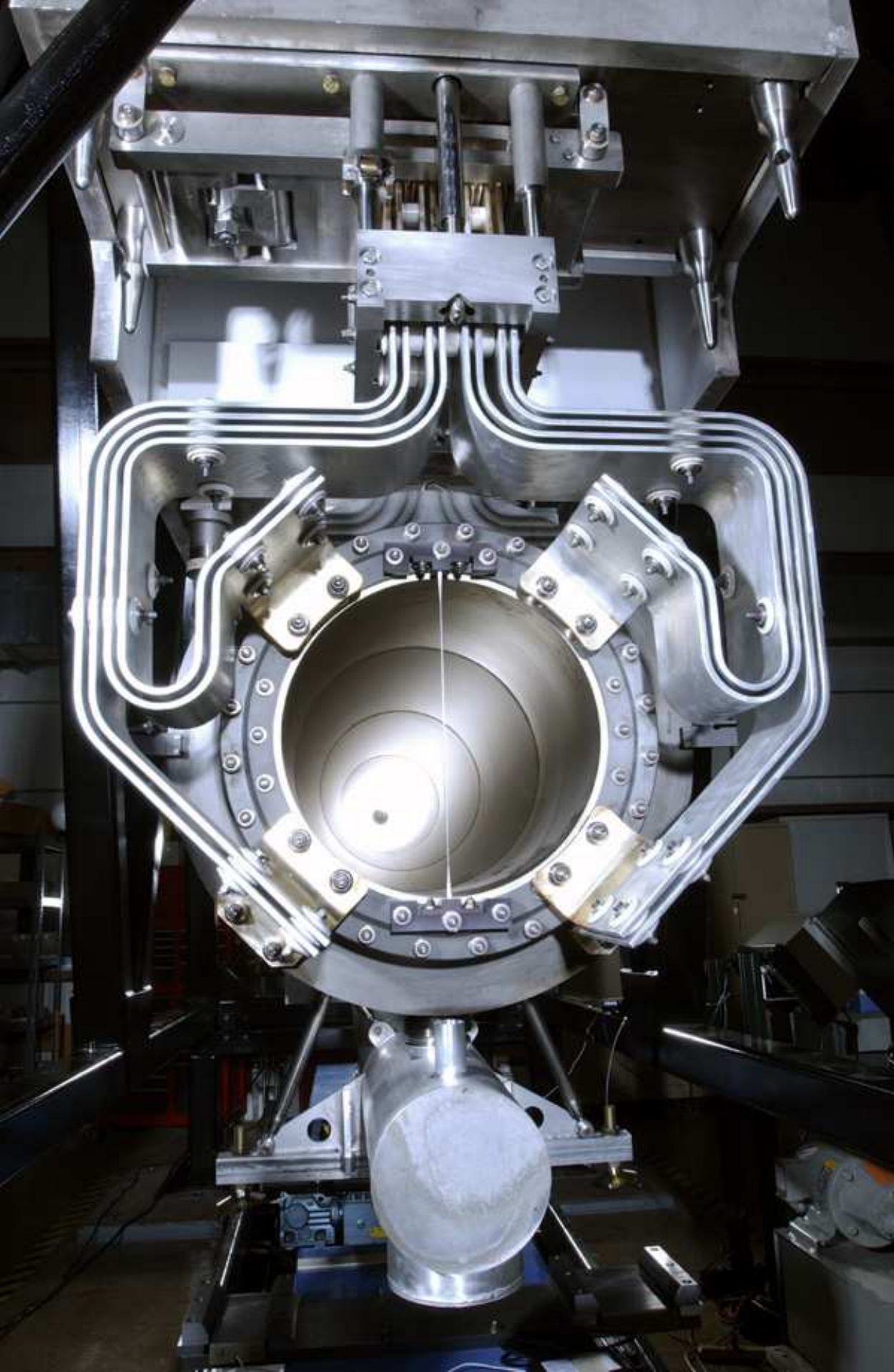}
        \caption{}
        \label{fig:numi_horn2}
    \end{subfigure}
    \caption[NuMI target and horns]{NuMI target and horns. Panel (a) shows the NuMI target, appearing as a series of graphite slabs on top of a water-cooled metal structure, from Ref. \cite{bnb_decay_pipe_numi_target}. Panel (b) shows a diagram and some charged hadron paths indicating the focusing mechanism. Panel (c) shows a photograph of the downstream side of the first horn. Panel (d) shows a photograph of the downstream side of the second horn. The panels (b-d) images are from Ref. \cite{numi_neutrino_beam}.}
    \label{fig:numi_target_horns}
\end{figure}

For the NuMI beam, after the protons emerge from the Booster, they are boosted from 8 GeV to 120 GeV in the Main Injector (named for its first use as an injector for the Tevatron) as shown in Fig. \ref{fig:main_injector}. The NuMI beam uses a water-cooled graphite target and a two-horn system as shown in Fig. \ref{fig:numi_target_horns}.

\begin{figure}[H]
    \centering
    \begin{subfigure}[b]{0.58\textwidth}
        \includegraphics[width=\textwidth]{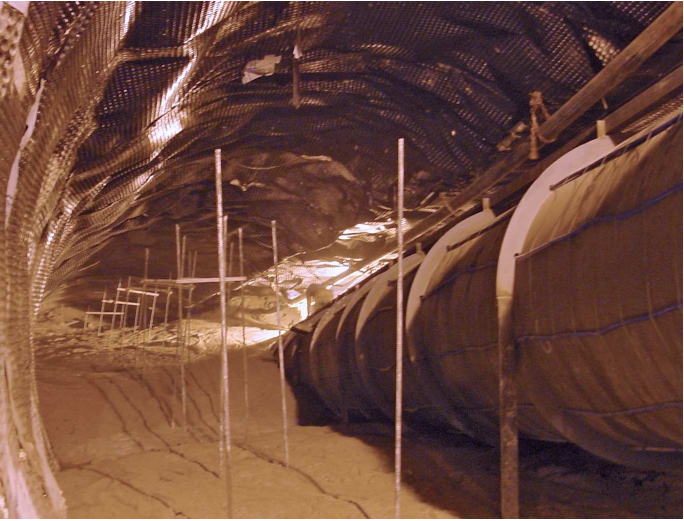}
        \caption{}
        \label{fig:numi_decay_pipe}
    \end{subfigure}
    \begin{subfigure}[b]{0.41\textwidth}
        \includegraphics[width=\textwidth]{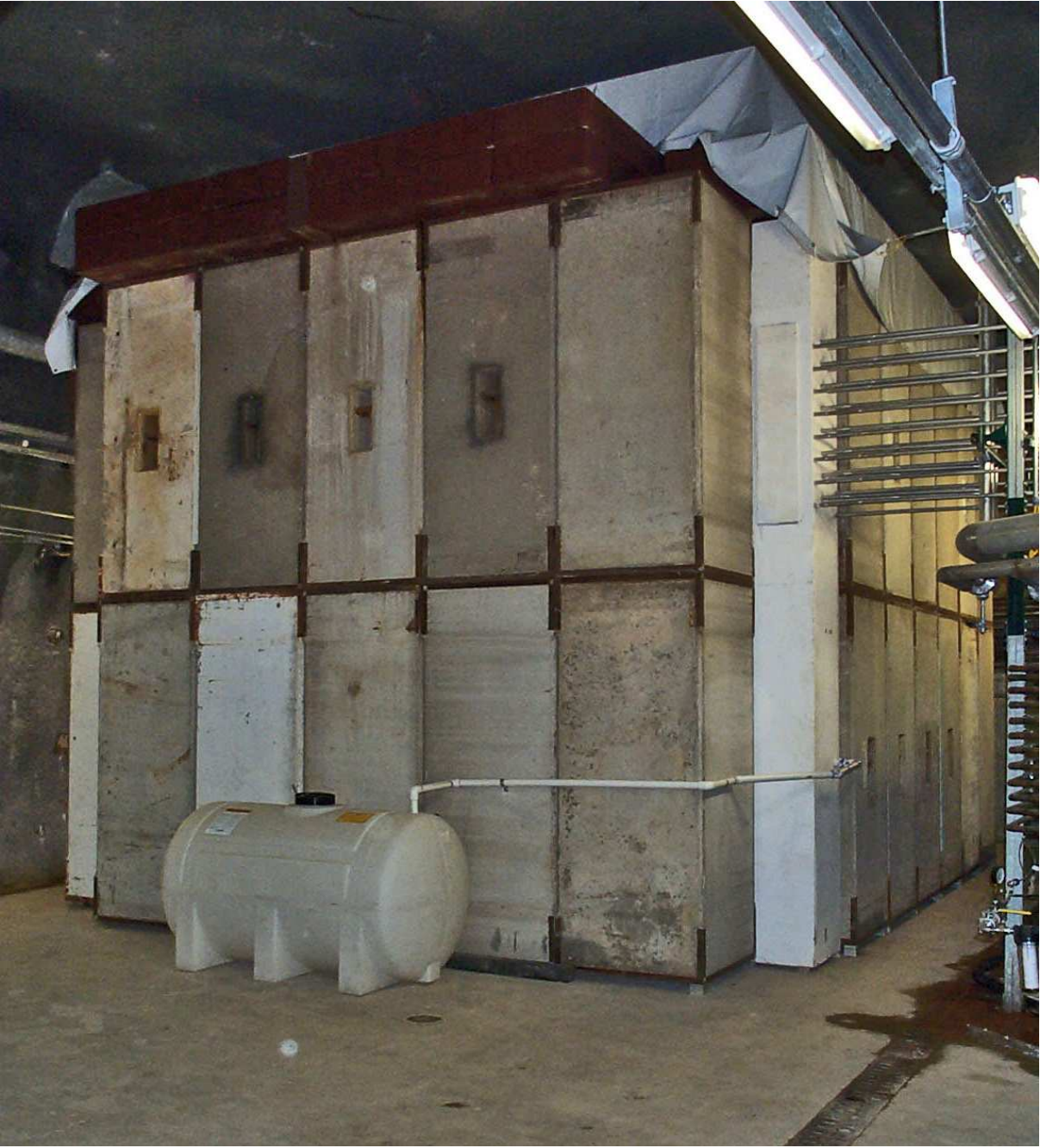}
        \caption{}
        \label{fig:numi_absorber}
    \end{subfigure}
    \caption[NuMI decay pipe and absorber]{Panel (a) shows the NuMI decay pipe, and panel (b) shows the NuMI absorber. Images are from Ref. \cite{numi_neutrino_beam}.}
    \label{fig:numi_decay_pipe_absorber}
\end{figure}

NuMI uses a helium-filled decay pipe (in contrast to the air-filled BNB decay pipe) as shown in Fig. \ref{fig:numi_decay_pipe_absorber}. The NuMI decay pipe tunnel is deep underground, from around 41 m deep at the target hall to 240 m deep at the NuMI near detector experiments like MINERvA and ArgoNeuT. The absorber is made of aluminum, steel, and concrete, as shown in Fig. \ref{fig:numi_decay_pipe_absorber}. Both the decay pipe and absorber are water cooled.

MicroBooNE's NuMI off-axis flux has recently been updated \cite{microboone_numi_flux_public_note}. In our old simulation, matching what is used by all Fermilab experiments including MINERvA, NOvA, and ICARUS, we use \textsc{Geant}4 v4.9.2.03 with FTFP-BERT hadronic processes \cite{Geant4}, and update hadron production cross sections using PPFX. In our new simulation, we added additional 40 m long steel shielding blocks which were previously missing in the beam simulation, updated \textsc{Geant}4 from v4.9.2 to v4.10.4 (which more closely matches extrapolations of kaon production measurements from the NA49 experiment), and updated PPFX to be compatible with this \textsc{Geant}4 version change.

\section{Runs 4-5 Validation}

MicroBooNE splits our data set into five main run periods, corresponding to five years of neutrino beam data, separated by regular Summer shutdowns for maintenance of the accelerator facilities. I made significant contributions to the validation of the last two years of MicroBooNE data, known as runs 4 and 5, which varied from our expectations from runs 1-3 in several important ways.

\subsection{Finite Electron Lifetime Correction}\label{sec:electron_lifetime}

\begin{figure}[H]
    \centering
    \includegraphics[width=\textwidth]{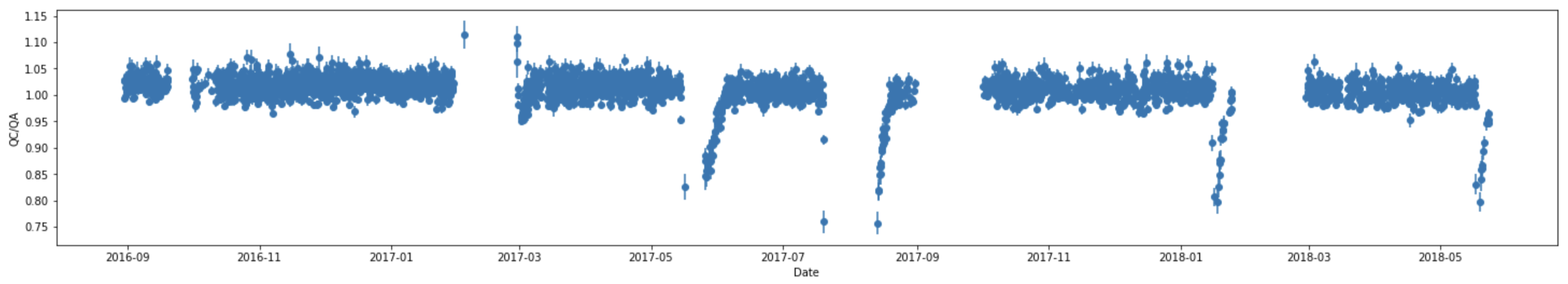}
    \caption[MicroBooNE purity over time]{MicroBooNE purity over time, as indicated by a metric from our purity monitors. Note several periods where the purity sharply decreases before the filters bring the purity back to a high level.}
    \label{fig:microboone_purity}
\end{figure}

MicroBooNE's filtration system is very good at removing electronegative impurities that reduce the electron lifetime, but this is not necessarily true in time periods where our circulation and filtration system operated in different states. In runs 1-3, the electron lifetime was very high, so any electron attenuation was negligible. However, in runs 4-5, there were time periods with significant electron attenuation, as shown in Fig. \ref{fig:microboone_purity}. Our reconstruction code did attempt to account for this, but it had not been significantly tested since it was developed purely on high-lifetime run periods.


Specifically, in the Wire-Cell reconstruction framework (described in more detail in Sec. \ref{sec:wire_cell}), we use the measured electron lifetime $\tau$ and the drift time $t$ calculated from the time between the PMT light signal and the ionization charge signal arriving on our wires, and use that to calculate the predicted electron attenuation ratio $A$ when $\tau> 35\ \mathrm{ms}$:
\begin{equation}
    A = 1 - e^{-t/\tau\ +\ t/35\ \mathrm{ms}}.
\end{equation}
The predicted fraction of charge getting absorbed by impurities is $e^{-\tau/x}$, and we added a constant offset to have a smooth function that applies no correction for lifetimes greater than 35 ms, a long lifetime which should have small attenuation effects. Once we predict this attenuation for each charge signal, we correct for this by multiplying the reconstructed charge by $1/A$, resulting in a charge measurement that should be unaffected by the finite electron lifetime. 

\begin{figure}[H]
    \centering
    \begin{subfigure}[b]{0.52\textwidth}
        \includegraphics[trim=50 50 190 75, clip, width=\textwidth]{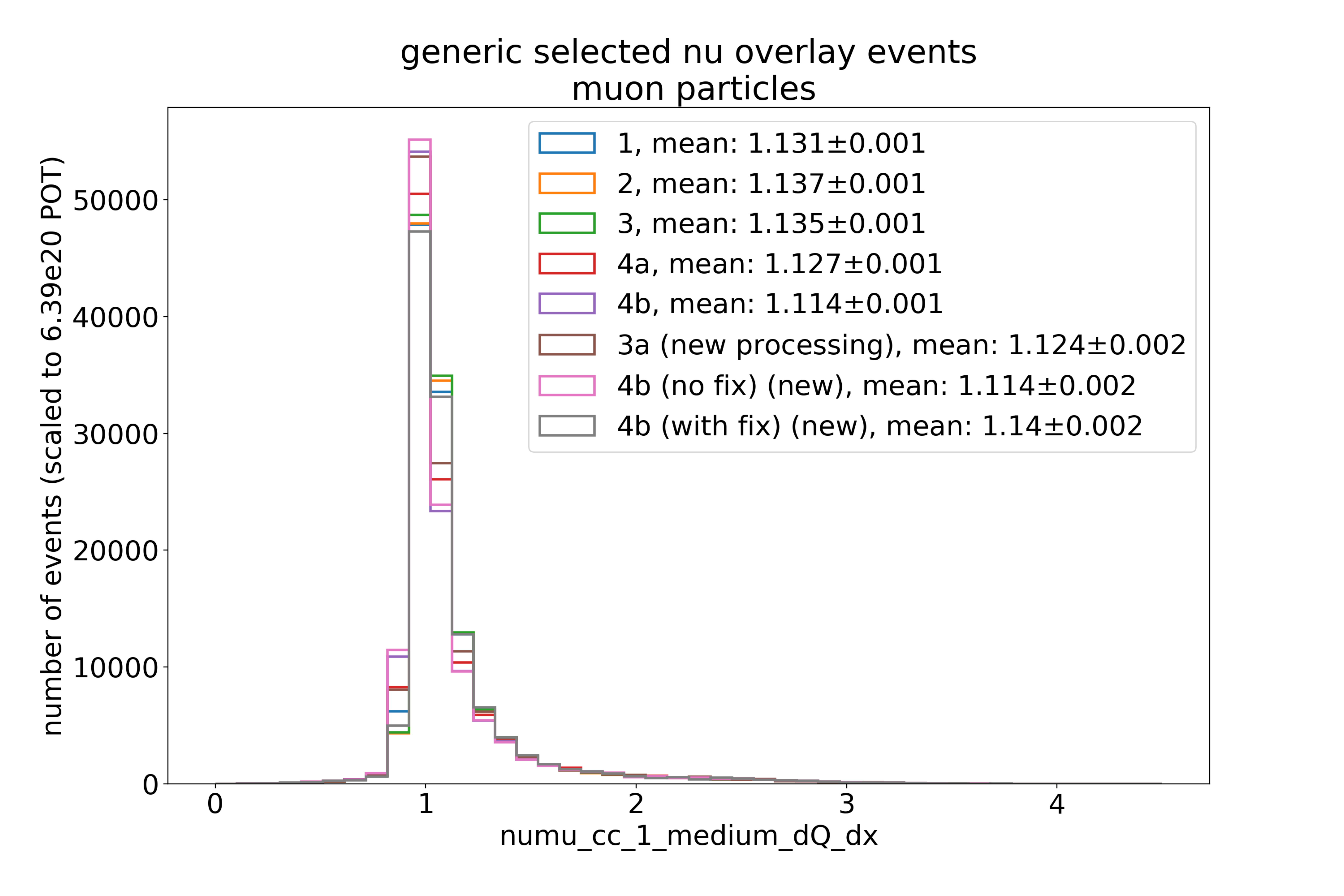}
        \caption{}
        \label{fig:runs_45_dqdx_simulation}
    \end{subfigure}
    \begin{subfigure}[b]{0.47\textwidth}
        \includegraphics[trim=50 50 190 75, clip, width=\textwidth]{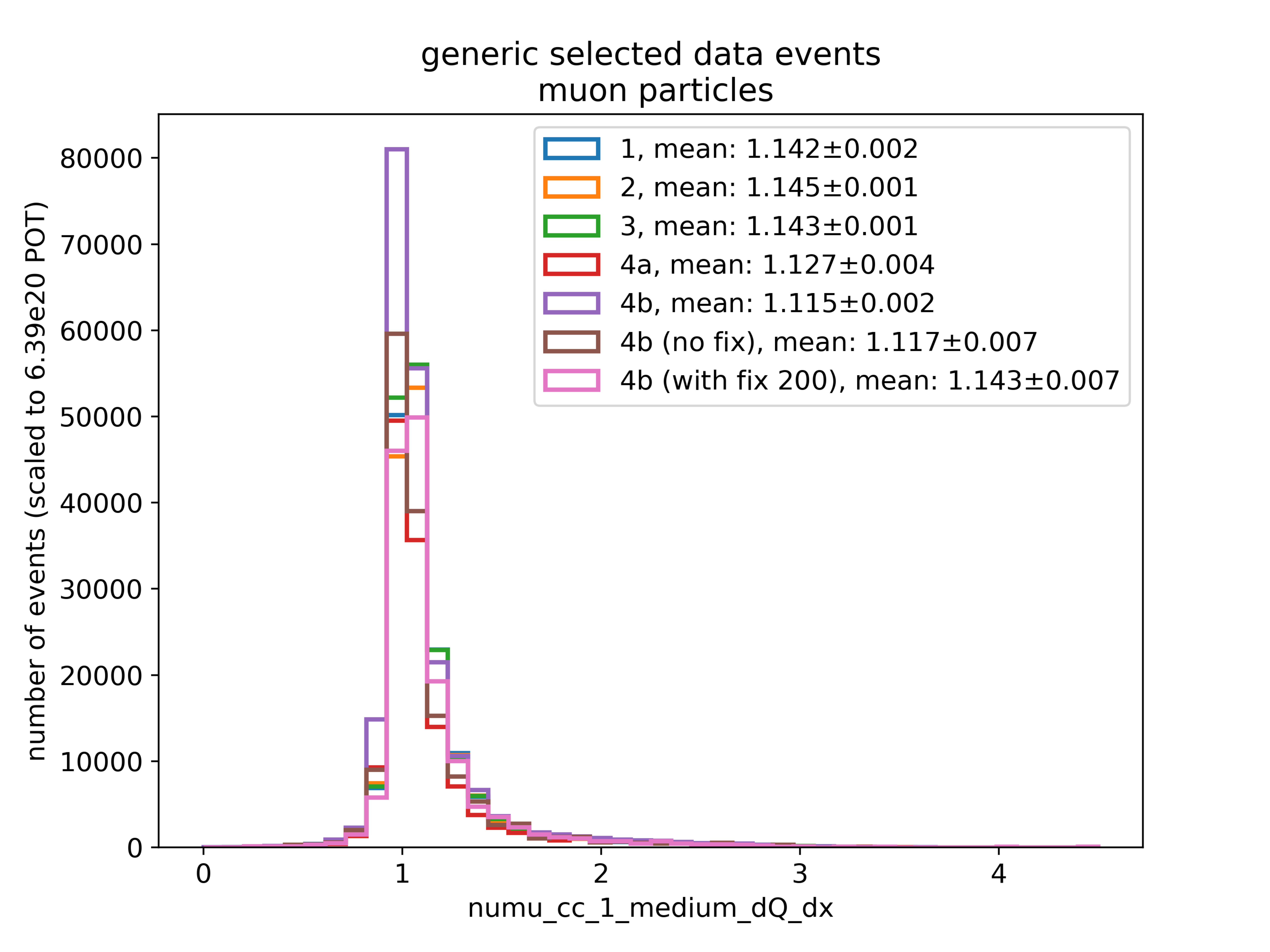}
        \caption{}
        \label{fig:runs_45_dqdx_data}
    \end{subfigure}
    \caption[Run 4 $dQ/dx$ discrepancy]{Mean $\nu_\mu$CC muon median $dQ/dx$ discrepancy between runs 1-3 and run 4b, before and after the predicted attenuation ratio fix. Panel (a) shows the discrepancy for simulation, and Panel (b) shows the descripancy for data. Runs 4a and 4b refer do different sub-periods of run 4. Run 3a is a subsample of run 3. ``New processing'' and ``new'' refer to a newer data processing campaigns. ``no fix'' refers to the old predicted attenuation ratio formula, and ``with fix'' refers to the new predicted attenuation ratio formula.}
    \label{fig:run_4_dqdx}
\end{figure}

After performing this correction, we carefully investigate the deposited charge per unit length $dQ/dx$ for simulated minimum ionizing muon tracks, specifically using a generic neutrino selection (described in more detail in Sec. \ref{sec:generic_neutrino_selection}) and looking at the reconstructed variable \texttt{numu\_cc\_1\_medium\_dQ\_dx} which identifies muon track segments and takes the median $dQ/dx$ value. The results are shown in Fig. \ref{fig:run_4_dqdx} for different run periods and different files, for both simulation and data. Runs 1-3, which had high electron lifetimes, have simulated $dQ/dx$ values of around 1.13 and data values around 1.14, while our first processing of run 4b, which had a smaller electron lifetime, has a simulated value of around 1.11 and a data value of around 1.12. This represents a relatively small but significant difference. Fortunately, this difference appears in simulation as well as data, so it should always be in principle possible to track down the source.

\begin{figure}[H]
    \centering
    \includegraphics[width=0.7\textwidth]{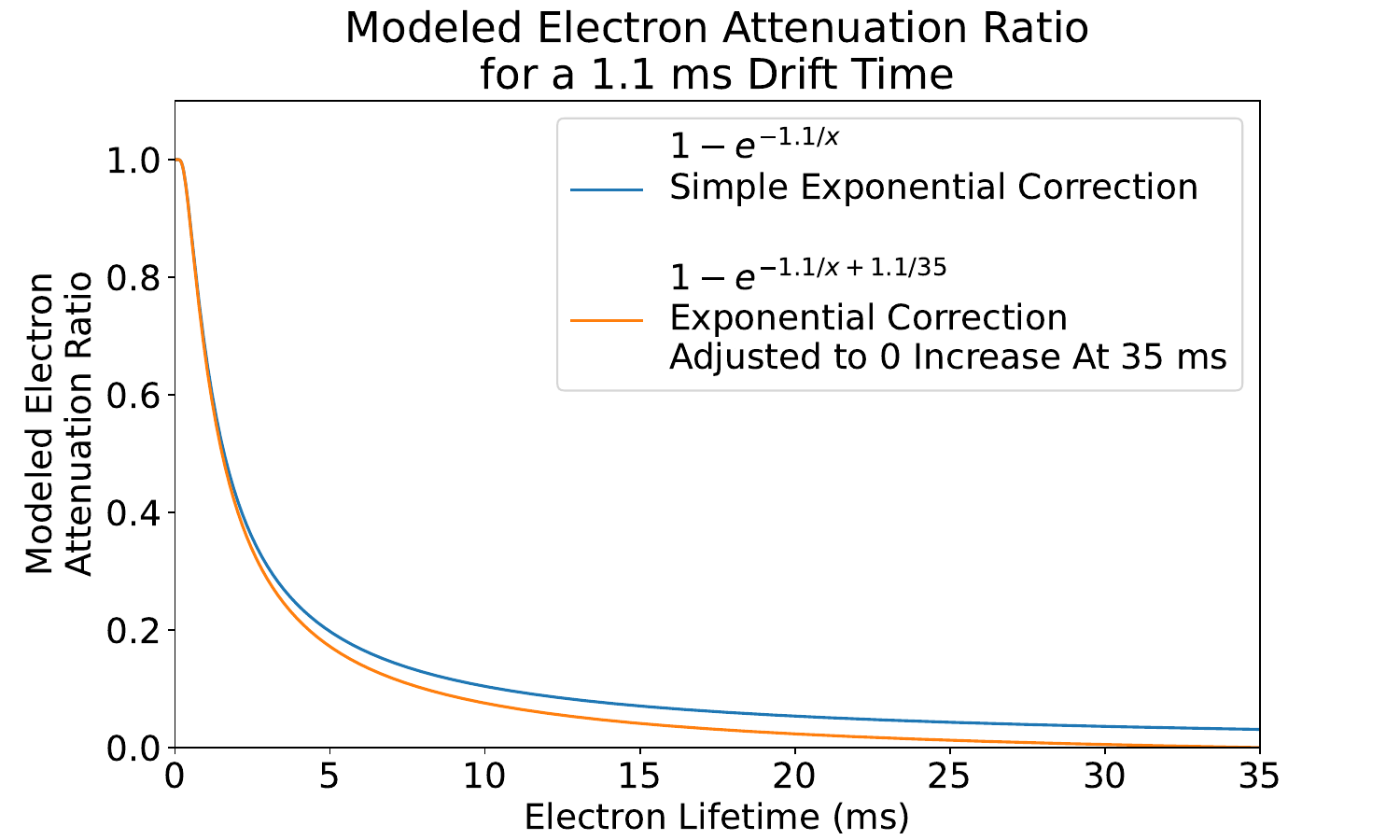}
    \caption[Electron attenuation ratio modeling]{Electron attenuation ratio modeling for a 1.1 ms drift time (the mean drift time in MicroBooNE). The older calculation is in orange, while the simpler newer calculation is in blue.}
    \label{fig:electron_attenuation_ratio}
\end{figure}

After investigations, we determined that the issue was in this 35 ms offset used in the calculation of the predicted electron attenuation ratio $A$. Even though 35 ms is a fairly long lifetime, the fact that we shifted our calculation in order to make the attenuation exactly zero at 35 ms noticeably affected the entire distribution. So, we changed to a simpler calculation with no correction,
\begin{equation}
    A = 1 - e^{-t/\tau},
\end{equation}
as illustrated in Fig. \ref{fig:electron_attenuation_ratio}. After updating the predicted attenuation ratio with the new formula, we achieve simulated and data $dQ/dx$ values of 1.14, which is much more consistent with our values in runs 1-3. So after this fix, we see very consistent $dQ/dx$ values even when we have significantly lower electron lifetimes, letting us use all of our reconstruction and particle ID tools on this more complex data set.

\subsection{PMT Gain Calibration}\label{sec:pmt_gains}

\begin{figure}[H]
    \centering
    \begin{subfigure}[b]{0.42\textwidth}
        \includegraphics[trim=50 50 190 100, clip, width=\textwidth]{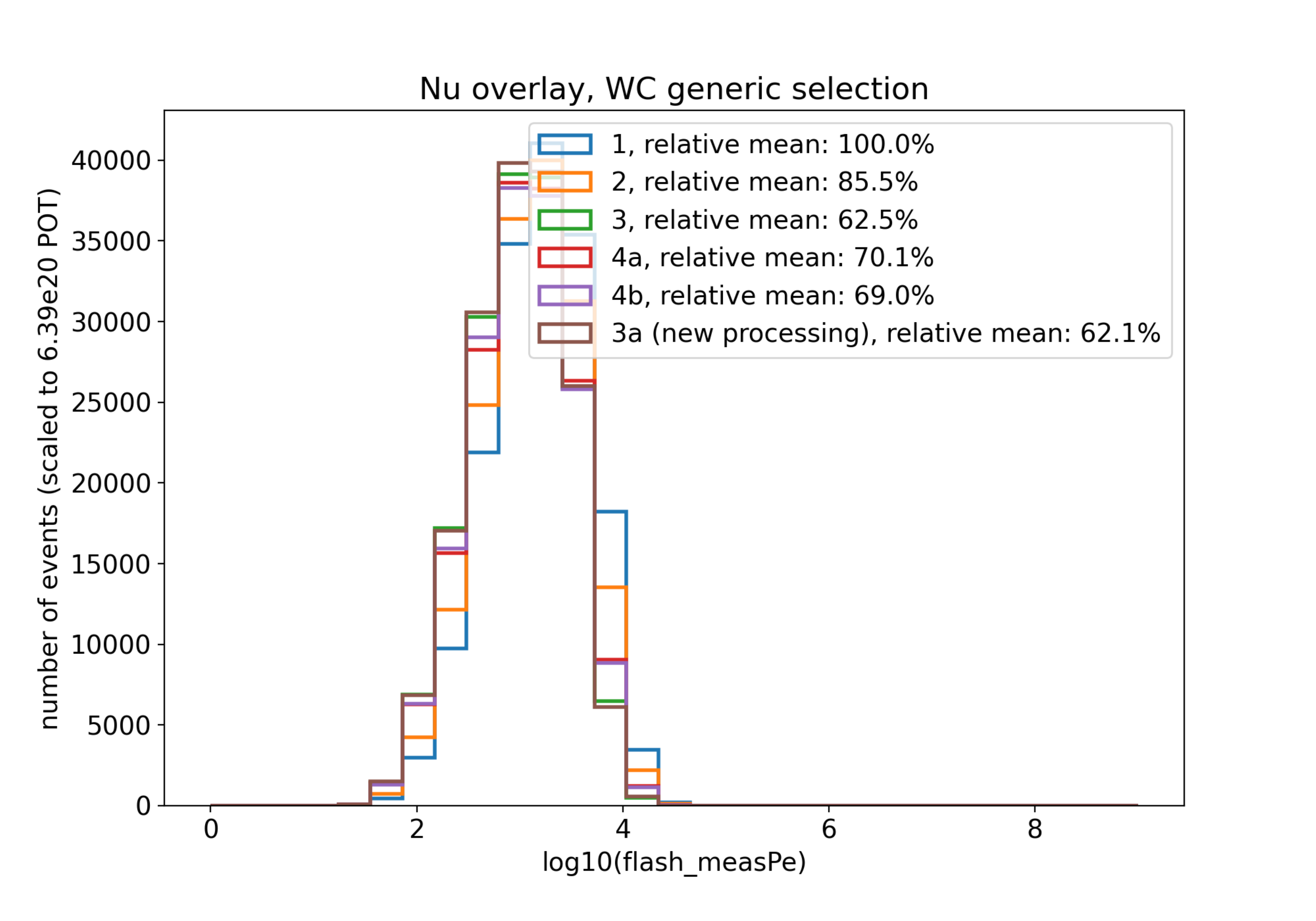}
        \caption{}
        \label{fig:overlay_meas_flashes}
    \end{subfigure}
    \begin{subfigure}[b]{0.42\textwidth}
        \includegraphics[trim=50 50 190 100, clip, width=\textwidth]{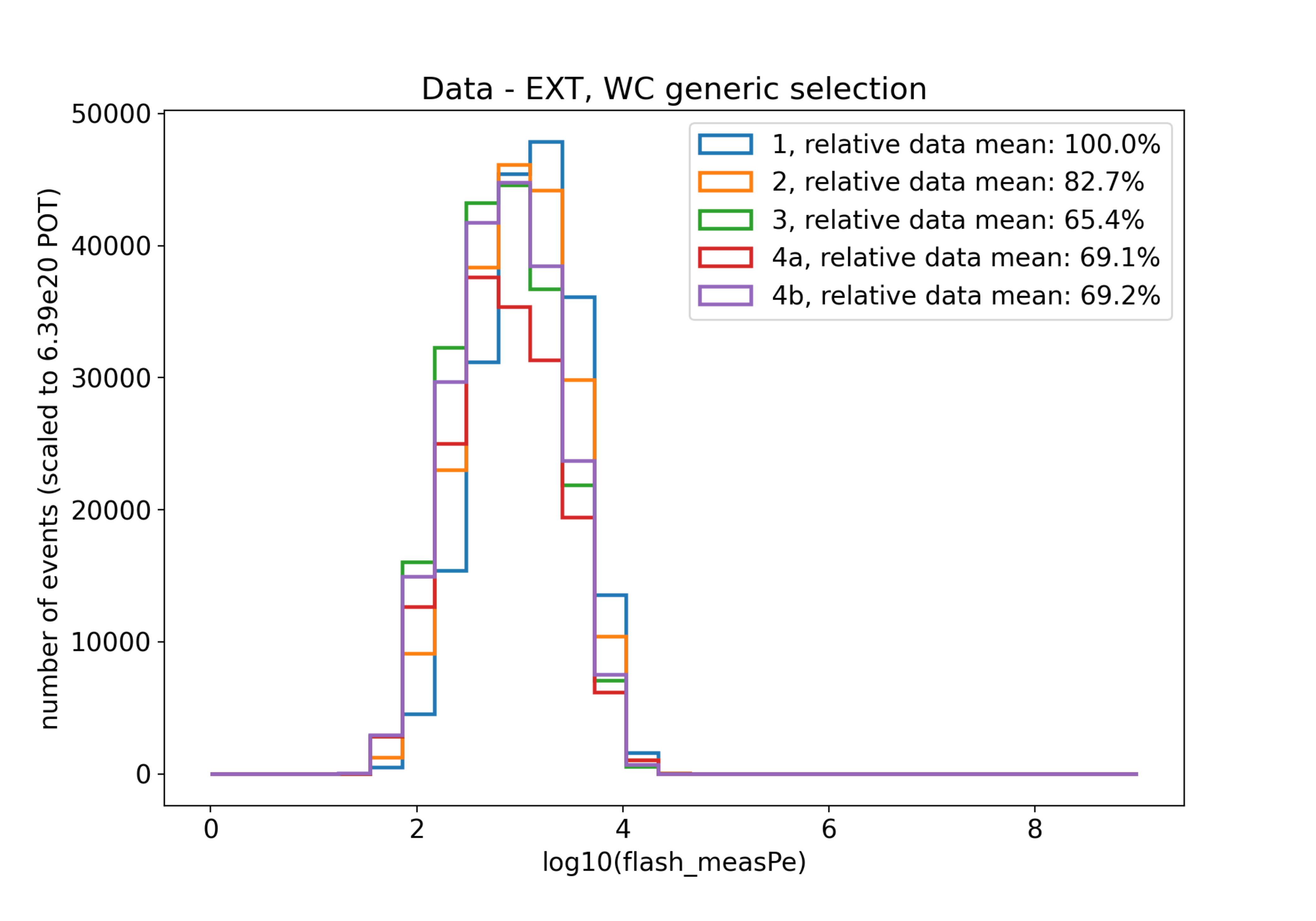}
        \caption{}
        \label{fig:data_meas_flashes}
    \end{subfigure}
    \begin{subfigure}[b]{0.42\textwidth}
        \includegraphics[trim=50 50 190 100, clip, width=\textwidth]{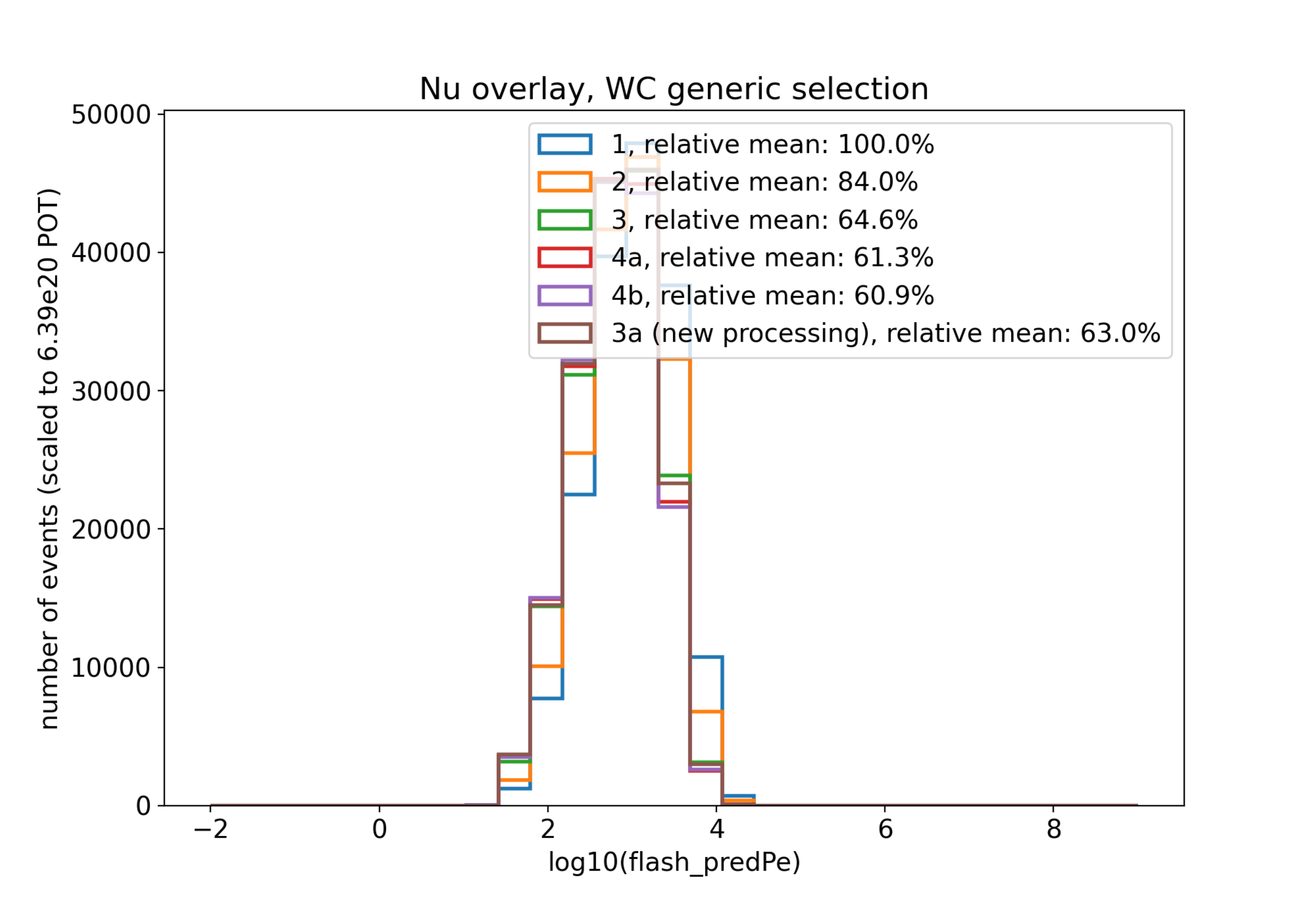}
        \caption{}
        \label{fig:overlay_pred_flashes}
    \end{subfigure}
    \begin{subfigure}[b]{0.42\textwidth}
        \includegraphics[trim=50 50 190 100, clip, width=\textwidth]{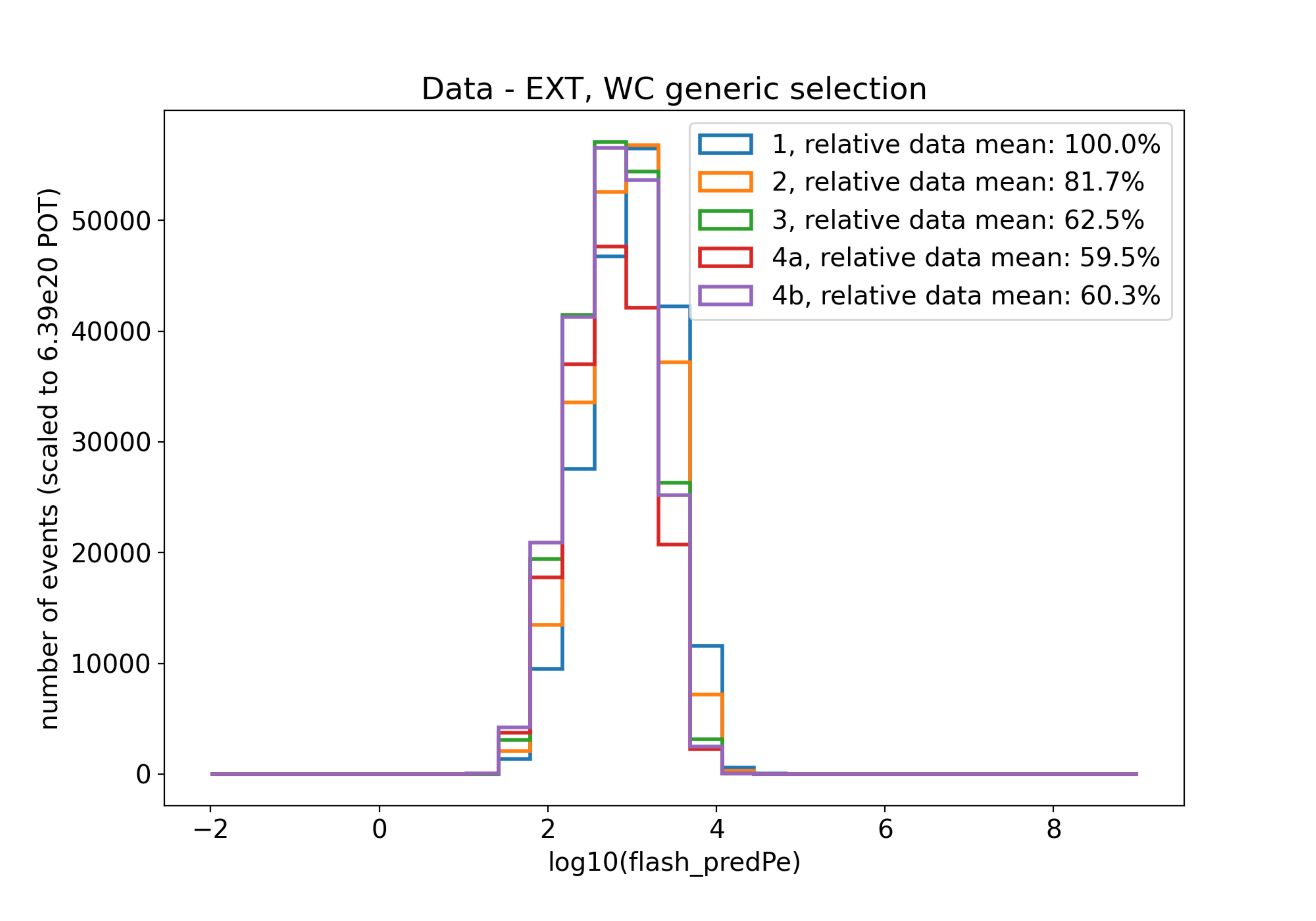}
        \caption{}
        \label{fig:data_pred_flashes}
    \end{subfigure}
    \begin{subfigure}[b]{0.42\textwidth}
        \includegraphics[trim=50 50 190 100, clip, width=\textwidth]{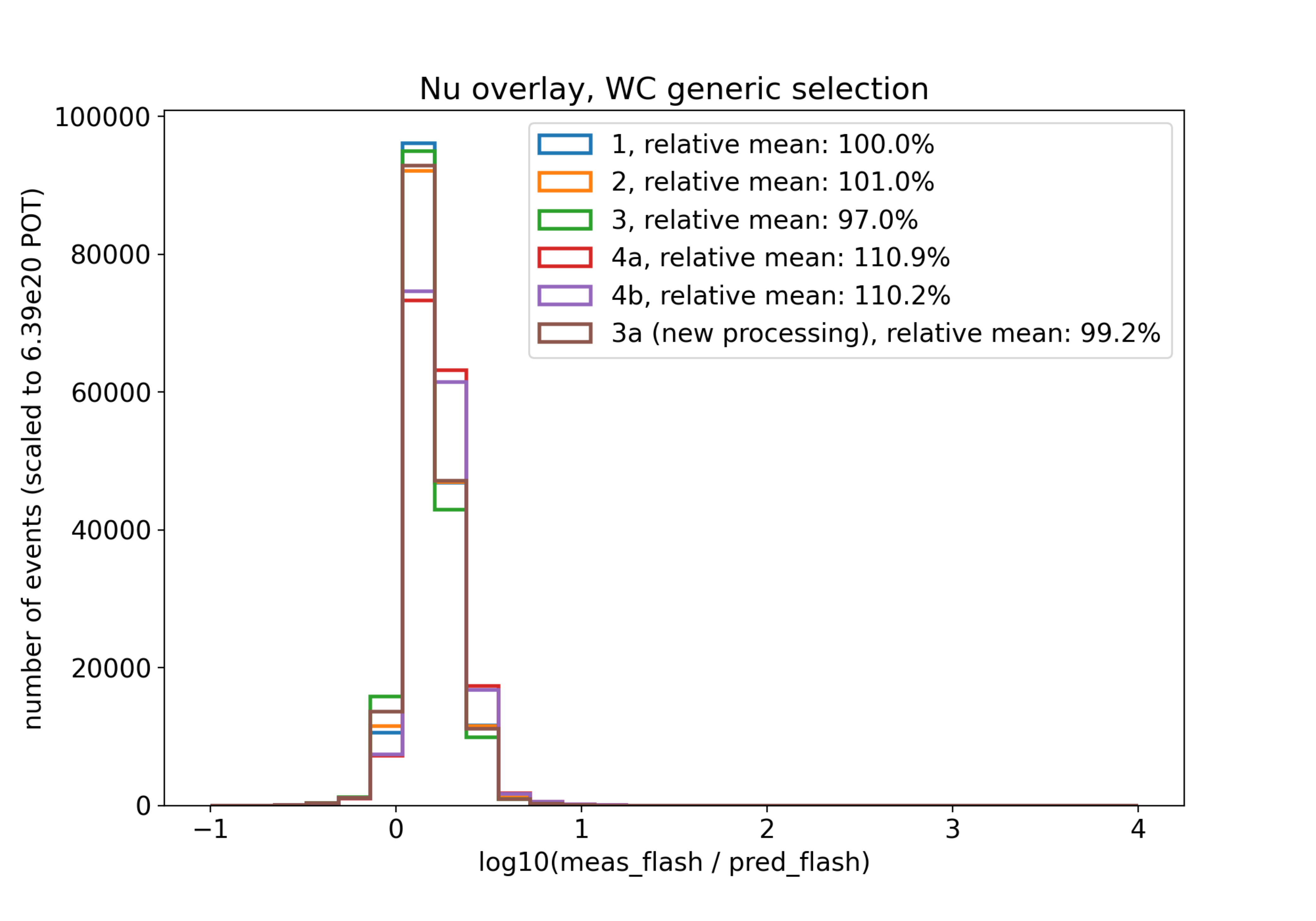}
        \caption{}
        \label{fig:overlay_ratio_flashes}
    \end{subfigure}
    \begin{subfigure}[b]{0.42\textwidth}
        \includegraphics[trim=50 50 190 100, clip, width=\textwidth]{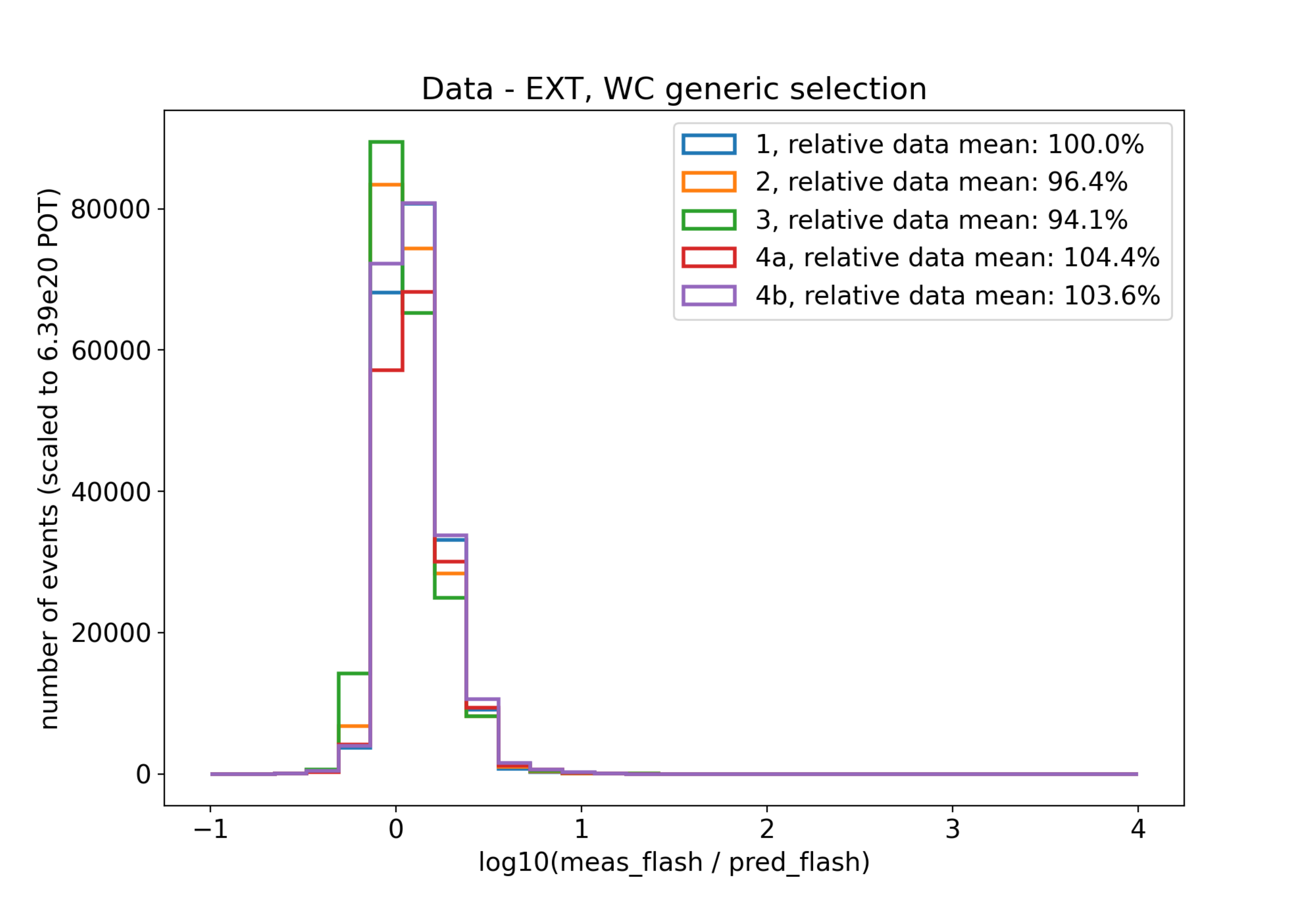}
        \caption{}
        \label{fig:data_ratio_flashes}
    \end{subfigure}
    \caption[Run dependent light yield]{Investigations of the run dependent light yield. The left panels (a), (c), and (e) show simulation, and the right panels (b), (d), and (f) show data. The top panels (a) and (b) show measured light, the middle panels (c) and (d) show predicted light, and the bottom panels (e) and (f) show the ratio of measured over predicted light. Runs 4a and 4b refer do different sub-periods of run 4. Run 3a is a subsample of run 3. ``New processing'' refers to a newer data processing campaign.}
    \label{fig:runs_45_data_and_overlay_flashes}
\end{figure}

Another validation we did on the Wire-Cell reconstruction files is looking at the run dependence of the brightness of light signals as measured by total photoelectron count on our PMTs. This light decreases over time as described in Sec. \ref{sec:light_detvar}. In each event, we identify a series of ``flashes'', each corresponding to simultaneous light activity on one or more PMTs. Wire-Cell calculates a predicted light pattern for each flash according to the ionization activity, and this accounts for the measured light yield decline over time. This predicted light pattern is used to match ionization clusters to the neutrino beam flash, with more details described in Sec. \ref{sec:wire_cell}. When we look at the distribution of these measured and predicted flash brightnesses, we see a decline over time for both the measured and predicted flashes, for both data and simulation, as shown in the top two rows of Fig. \ref{fig:runs_45_data_and_overlay_flashes}, where the mean brightness relative to run 1 declines from 100\% to around 60-70\%. This is in line with our expectations, given MicroBooNE's overall light yield decline over time. When we study the ratio of measured brightness over predicted brightness, we expect this to remain constant over time, but we see an increase as shown in the bottom row of Fig. \ref{fig:runs_45_data_and_overlay_flashes}, where the mean ratio relative to run 1 increases from 100\% to around 104-110\% in run 4.

Since this discrepancy appeared in simulation as well as data, again it should always be in principle possible to track down the source. After several hypotheses and additional tests, we were able to identify the PMT gain database as the source of the issue.

\begin{figure}[H]
    \centering
    \includegraphics[width=0.7\textwidth]{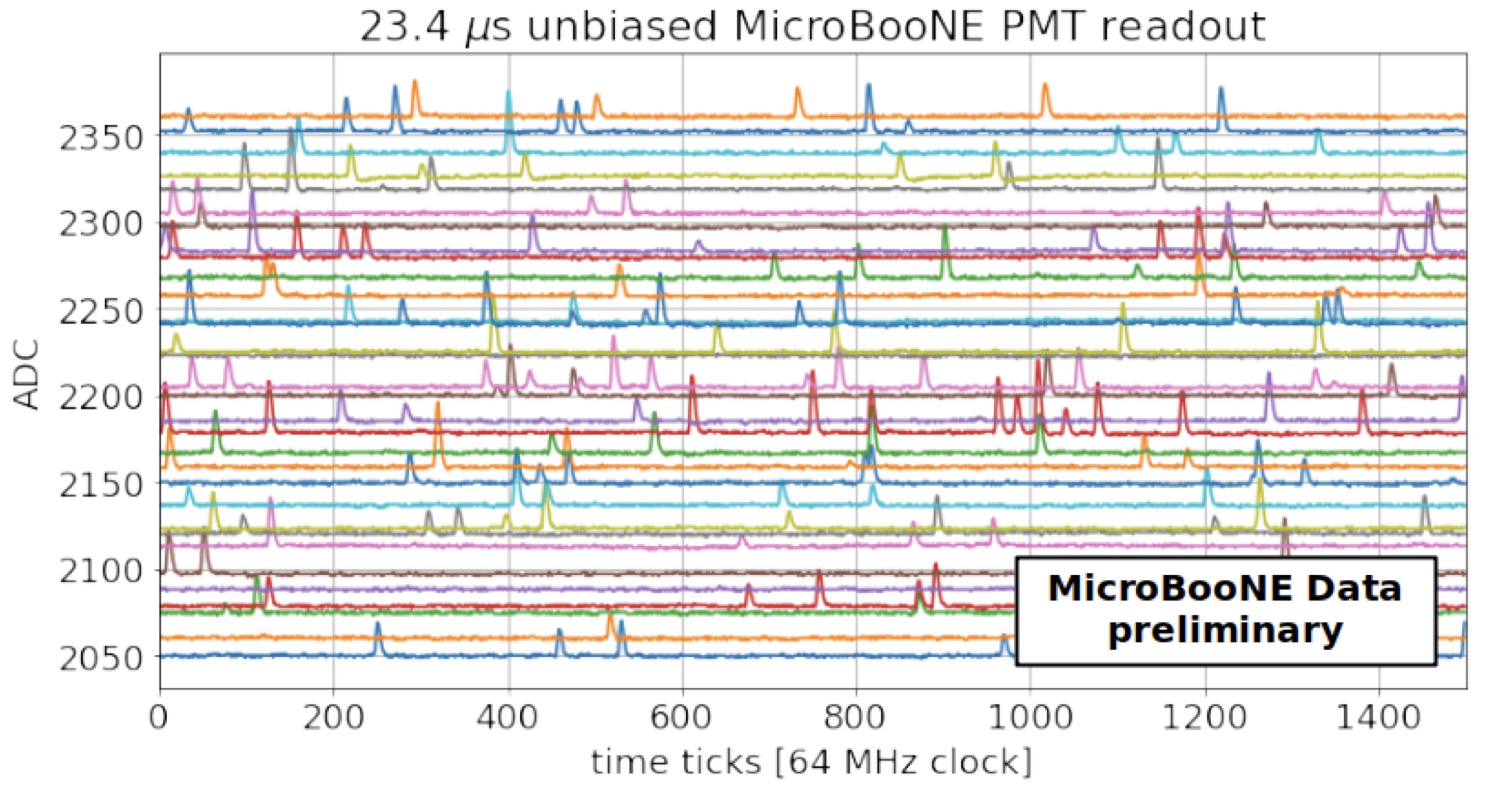}
    \caption[MicroBooNE single photoelectron waveforms]{MicroBooNE single photoelectron waveforms. Figure from Ref. \cite{microboone_pmt_gain_public_note}.}
    \label{fig:spes}
\end{figure}

PMT gains are measured using single photoelectron (SPE) pulses \cite{microboone_pmt_gain_public_note}. These pulses appear at about 200 kHz in MicroBooNE, as shown in Fig. \ref{fig:spes}. There are more SPE events than can be accounted for by the PMT dark rate, so presumably there is a true source of ambient light inside the detector. This high SPE rate has been observed in both MicroBooNE and ProtoDUNE. One potential source of these SPE events is ``volume recombination'', where a drifting electron recombines with an ambient $\mathrm{Ar}_2^+$ ion via $\mathrm{Ar}_2^+ + e^- \rightarrow \mathrm{Ar}_2^* \rightarrow 2 \mathrm{Ar} + \gamma$. Another potential source is ``mutual neutralization'', where a drifting electron interacts with an impurity like water via $\mathrm{H}_2\mathrm{O} + e^- \rightarrow \mathrm{H}_2\mathrm{O}^-$, and then this negative ion can interact with a positive ion via $\mathrm{Ar}_2^+ + \mathrm{H}_2\mathrm{O}^- \rightarrow \mathrm{Ar}_2^* + \mathrm{H}_2\mathrm{O} \rightarrow 2 \mathrm{Ar} + \mathrm{H}_2\mathrm{O} + \gamma$. In both of these cases, a single photon would be created, and the timing of this photon would be randomly distributed in time. These sources can be analyzed via an ion transport model, which can make specific predictions for how the measured SPE rate depends on different electric field strengths, different electric field directions, and different impurity concentrations \cite{ion_transport_model_SPE}. Tests of these predictions in MicroBooNE are in progress, as we took data at different electric field strengths, including a reversed electric field, just before decomissioning.

\begin{figure}[H]
    \centering
    \includegraphics[width=0.7\textwidth]{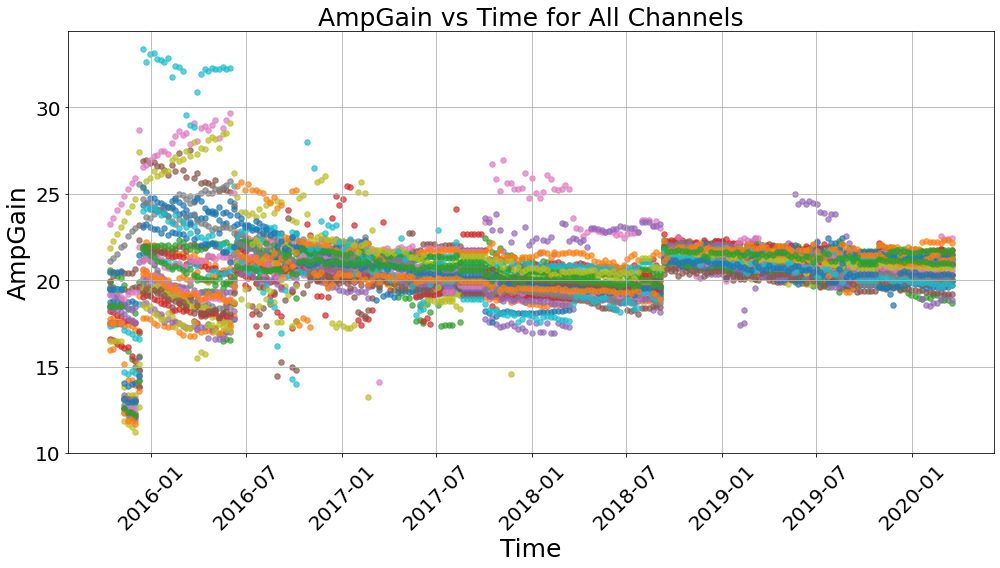}
    \caption[MicroBooNE PMT gains]{MicroBooNE PMT gains over time. Each color shows a different PMT. There are slow drifts over time, as well as sudden shifts due to changes in the hardware configuration. This plot is updated relative to \cite{microboone_pmt_gain_public_note}, including an expanded time period.}
    \label{fig:pmt_gains}
\end{figure}

Regardless of the source of these SPE events, they are very useful for calibrating the PMT gains. Each digitized waveform goes through baseline subtraction, triggering, and filtering by peak ADC and area and baseline noise. Then the distribution of amplitudes are studied in order to calibrate for any changes over time for each PMT. The gain of each PMT over the full dataset is shown in Fig. \ref{fig:pmt_gains}.

\begin{figure}[H]
    \centering
    \begin{subfigure}[b]{0.49\textwidth}
        \includegraphics[width=\textwidth]{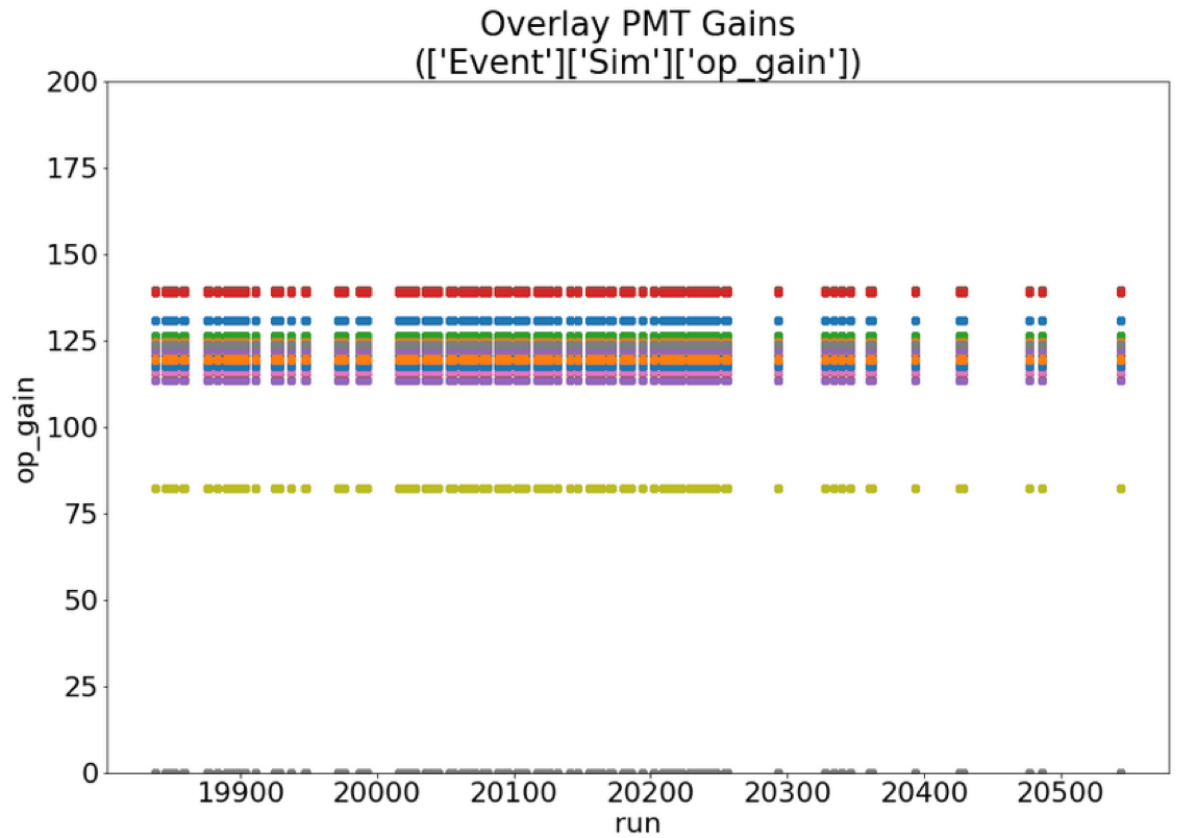}
        \caption{}
        \label{fig:old_run_4_gains}
    \end{subfigure}
    \begin{subfigure}[b]{0.49\textwidth}
        \includegraphics[width=\textwidth]{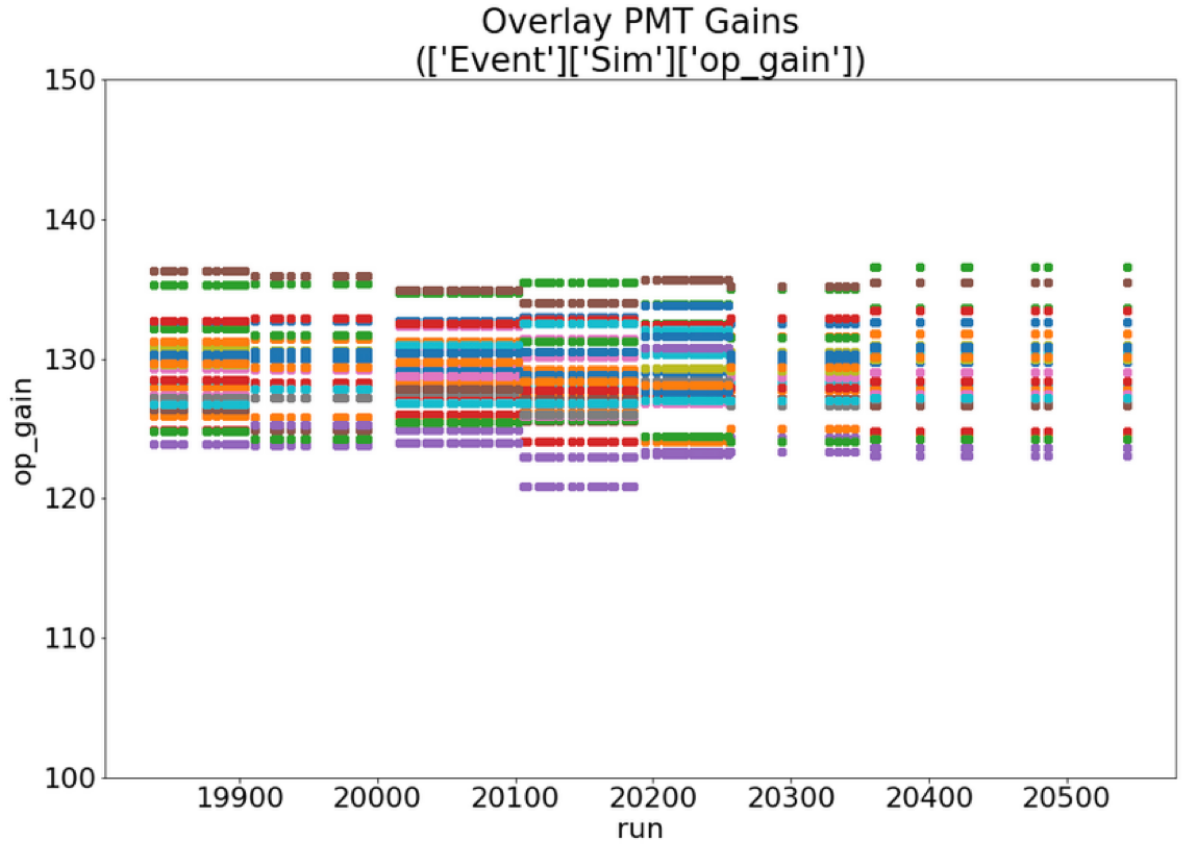}
        \caption{}
        \label{fig:new_run_4_gains}
    \end{subfigure}
    \caption[PMT gain database versions]{PMT gains over time near run 4. Panel (a) shows the older \texttt{v1r0} database tag, while panel (b) shows the newer \texttt{v1r2} database tag.}
    \label{fig:pmt_gain_database_versions}
\end{figure}

For the discrepancy in ratio of measured brightness over predicted brightness, this was tracked down to a mismatch in the PMT gain database versions. The older version of the gain database did not have updated gain values, and defaulted to a constant for newer run periods, as shown in Fig. \ref{fig:old_run_4_gains}. After updating this, we see the gains slightly changing over time as expected, as shown in Fig. \ref{fig:new_run_4_gains}. In our first processing of run 4 files, the correct newer version of the PMT gain database was used when simulating the detector, but the incorrect older version of the database was used in order to predict the brightness of the light flash from the ionization signals, and this is why there was a mismatch in both the data and simulation measured/predicted ratios.

After fixing the reconstruction code to use the correct PMT gain database version, the measured/predicted brightness ratios become stable over time, allowing us to reliably use all of our existing light reconstruction in all downstream analyses.

\subsection{Booster Neutrino Beam Mistargeting}

The last issue identified during runs 4-5 validation was the most difficult to understand, since it was only seen in the data with no issue in the simulation. 



\begin{figure}[H]
    \centering
    \includegraphics[width=0.5\textwidth]{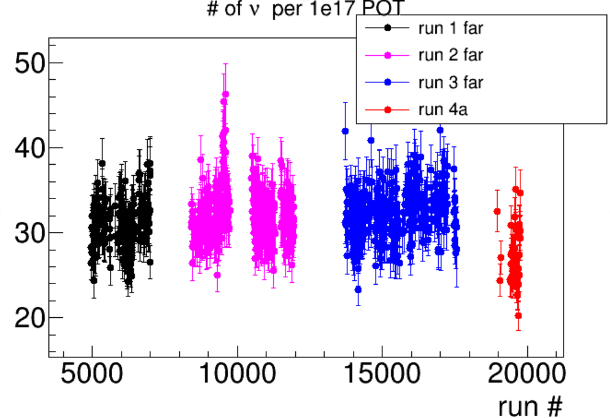}
    \caption[Neutrinos per POT in Run 4]{Neutrinos per POT over time, showing a notable decrease in run 4a.}
    \label{fig:nu_per_pot}
\end{figure}

In all neutrino distributions from all selections and reconstructions, we saw a $\sim$15\% decrease in normalization in the first period of run 4, known as run 4a. This appears after normalizing by protons on target (POT), which accounts for the fact that we expect the number of neutrinos to be proportional to the number of protons which hit our target. This deficit in neutrinos per POT in run 4a is illustrated in Fig. \ref{fig:nu_per_pot}.

After carefully ruling out many other possible explanations, we looked into logged information about the neutrino beam status more carefully. We did not identify any issues related to POT counting, so we moved on to look at more details about the proton beam.

\begin{figure}[H]
    \centering
    \begin{subfigure}[b]{0.52\textwidth}
        \includegraphics[width=\textwidth]{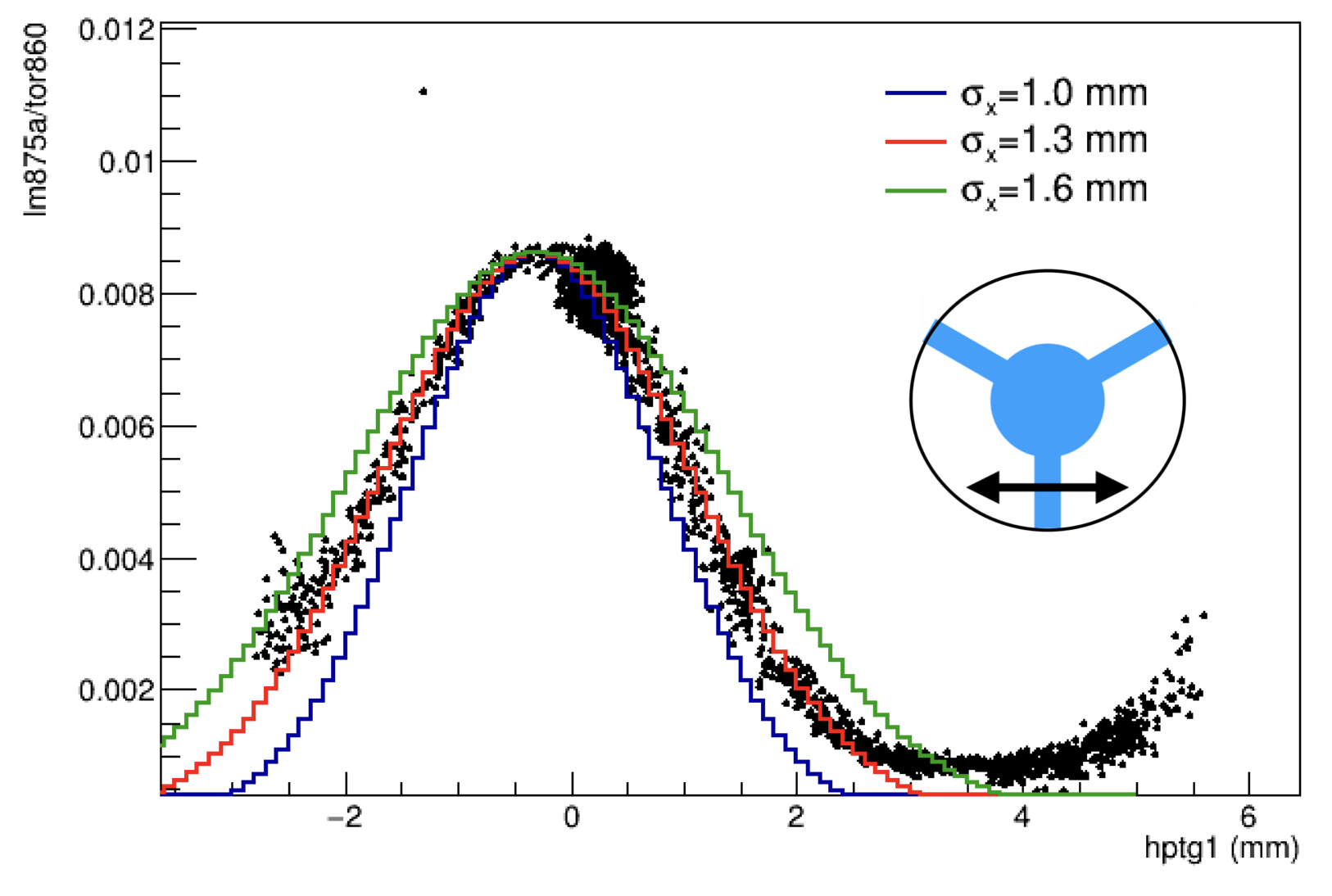}
        \caption{}
        \label{fig:horizontal_target_scan}
    \end{subfigure}
    \begin{subfigure}[b]{0.47\textwidth}
        \includegraphics[width=\textwidth]{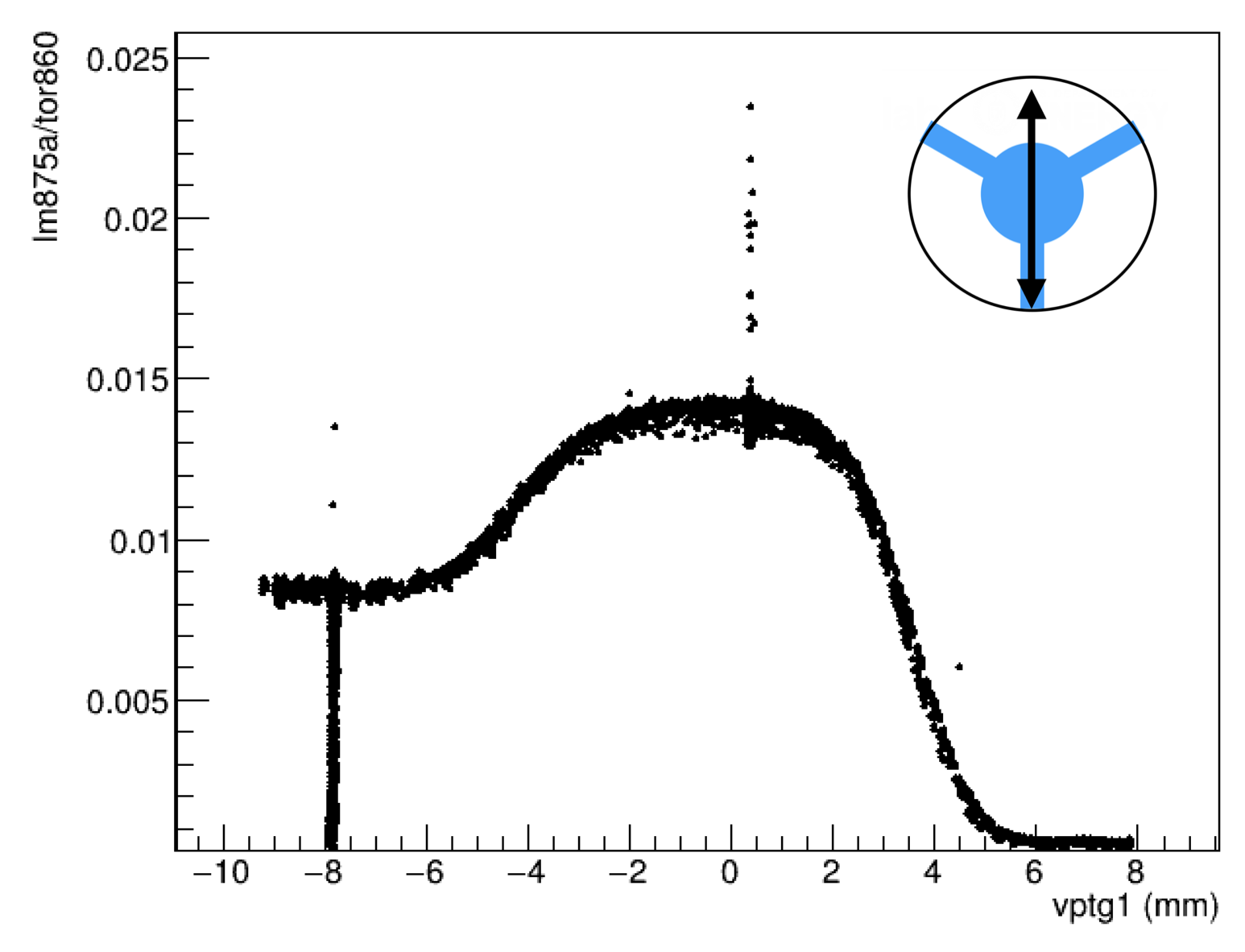}
        \caption{}
        \label{fig:vertical_target_scan}
    \end{subfigure}
    \caption[BNB target scans]{BNB target scans. Panel (a) shows the horizontal scan, centered on the vertical fin of the target, which is sensitive to the beam position and horizontal beam width, here determined to be around 1.3 mm. Panel (b) shows the vertical scan, where there is an asymmetry due to the presence of the vertical fin only on the bottom of the target.}
    \label{fig:target_scans}
\end{figure}

The BNB undergoes periodic target scans, where the proton beam is aimed at different areas on the target. When the beam overlaps the target more, there are more backscattered particles that are measured by upstream loss monitors. These scans can tell us the position and the width of the beam at the target, as shown in Fig. \ref{fig:target_scans}.



\begin{figure}[H]
    \centering
    \includegraphics[width=0.4\textwidth]{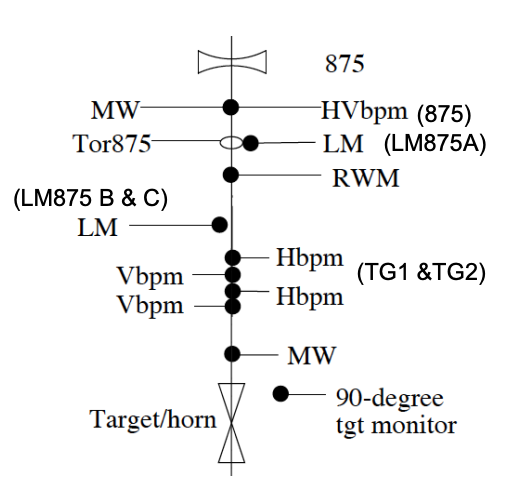}
    \caption[BNB beam monitor diagram]{BNB beam monitor diagram. The proton beam starts at the top and passes through a series of monitors before hitting the target at the bottom. The horizontal and vertical beam position monitors are labeled as Vbpm and Hbpm.}
    \label{fig:beam_monitor_diagram}
\end{figure}

However, these target scans are only done occasionally, and in between those times, the proton beam position is monitored with horizontal and vertical beam position monitors located upstream of the target, as diagrammed in Fig. \ref{fig:beam_monitor_diagram}.

\begin{figure}[H]
    \centering
    \begin{subfigure}[b]{0.49\textwidth}
        \includegraphics[trim=150 100 450 240, clip, width=\textwidth]{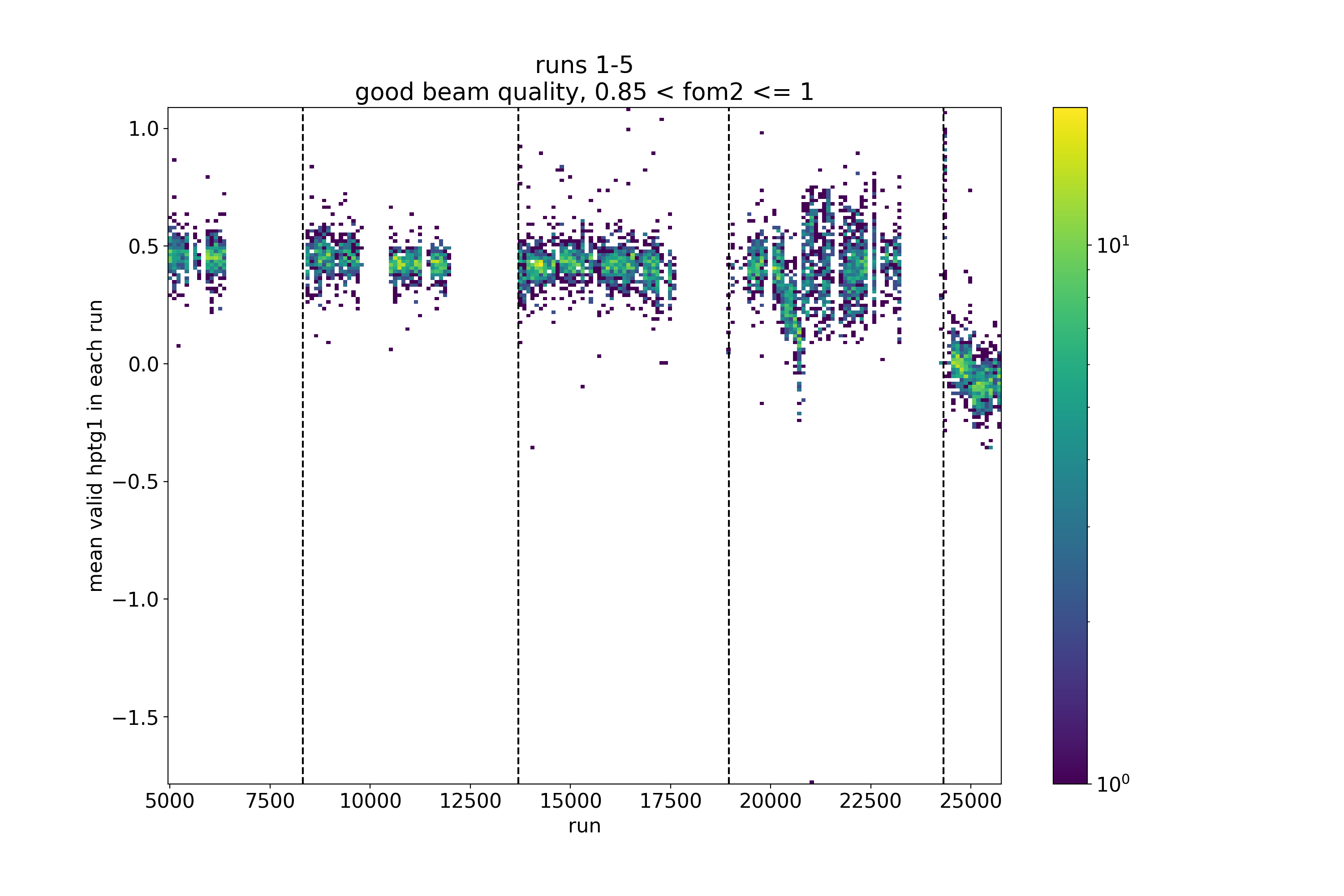}
        \caption{}
        \label{fig:hptg1}
    \end{subfigure}
    \begin{subfigure}[b]{0.49\textwidth}
        \includegraphics[trim=150 100 450 240, clip, width=\textwidth]{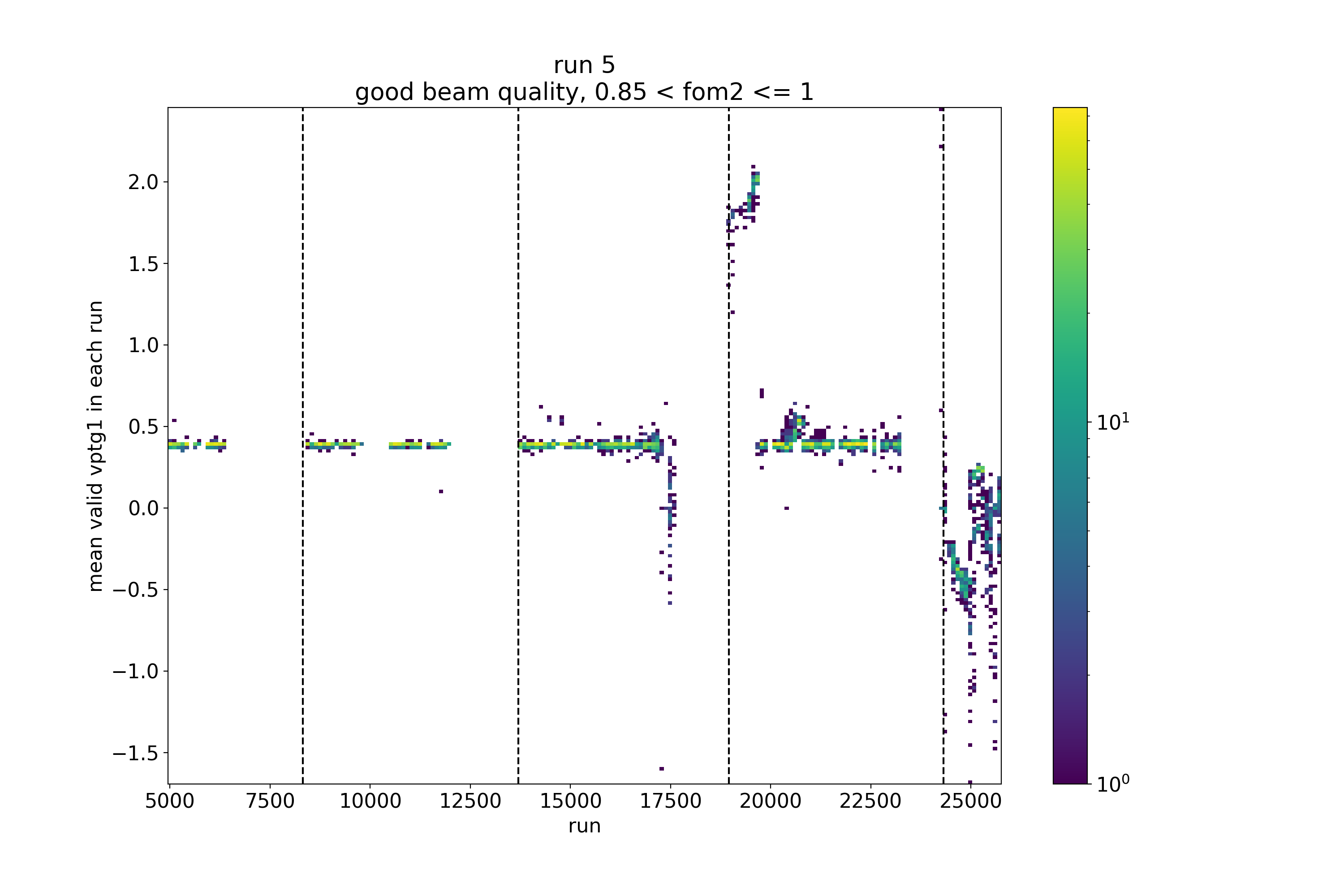}
        \caption{}
        \label{fig:vptg1}
    \end{subfigure}
    \begin{subfigure}[b]{0.49\textwidth}
        \includegraphics[trim=150 100 450 240, clip, width=\textwidth]{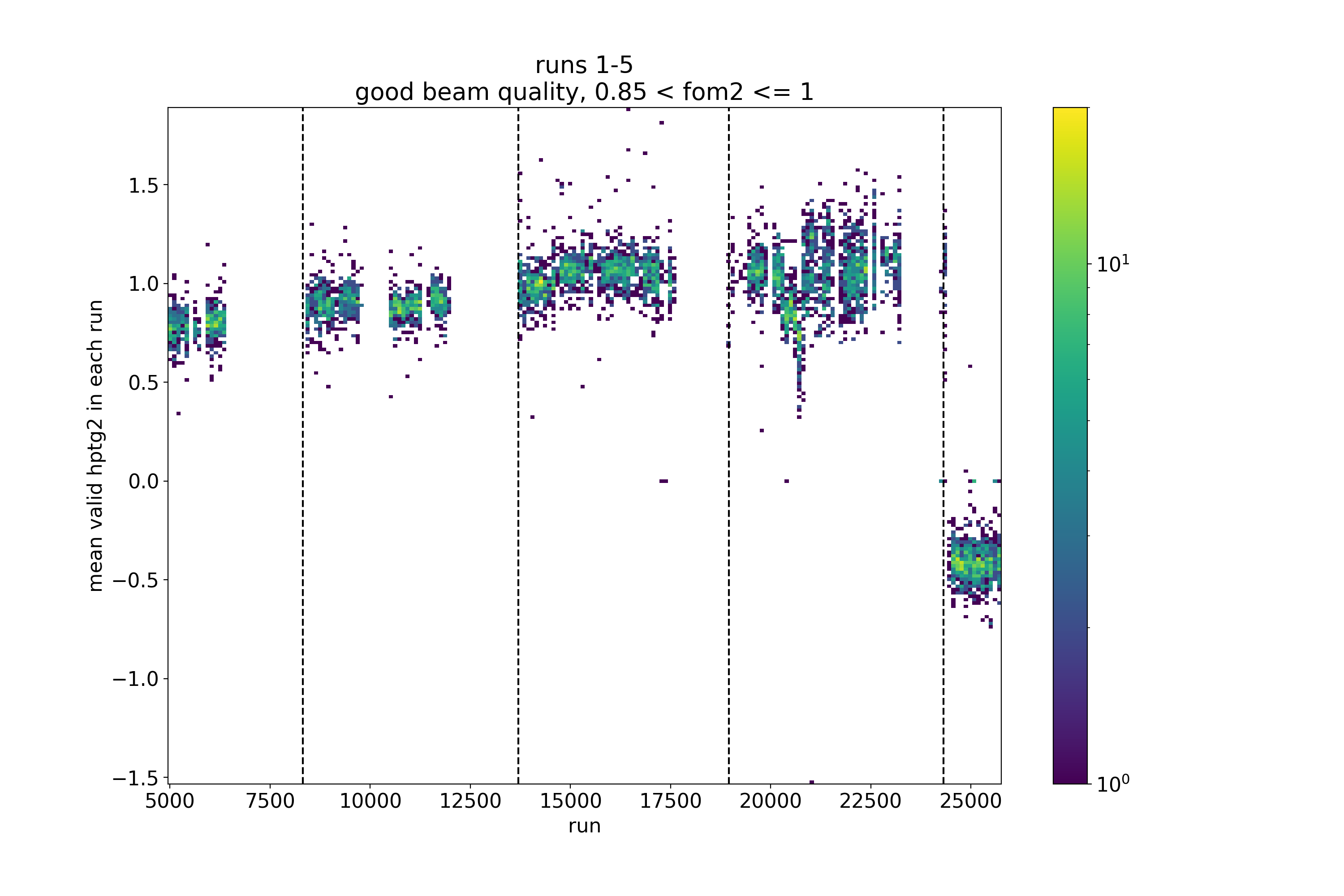}
        \caption{}
        \label{fig:hptg2}
    \end{subfigure}
    \begin{subfigure}[b]{0.49\textwidth}
        \includegraphics[trim=150 100 450 240, clip, width=\textwidth]{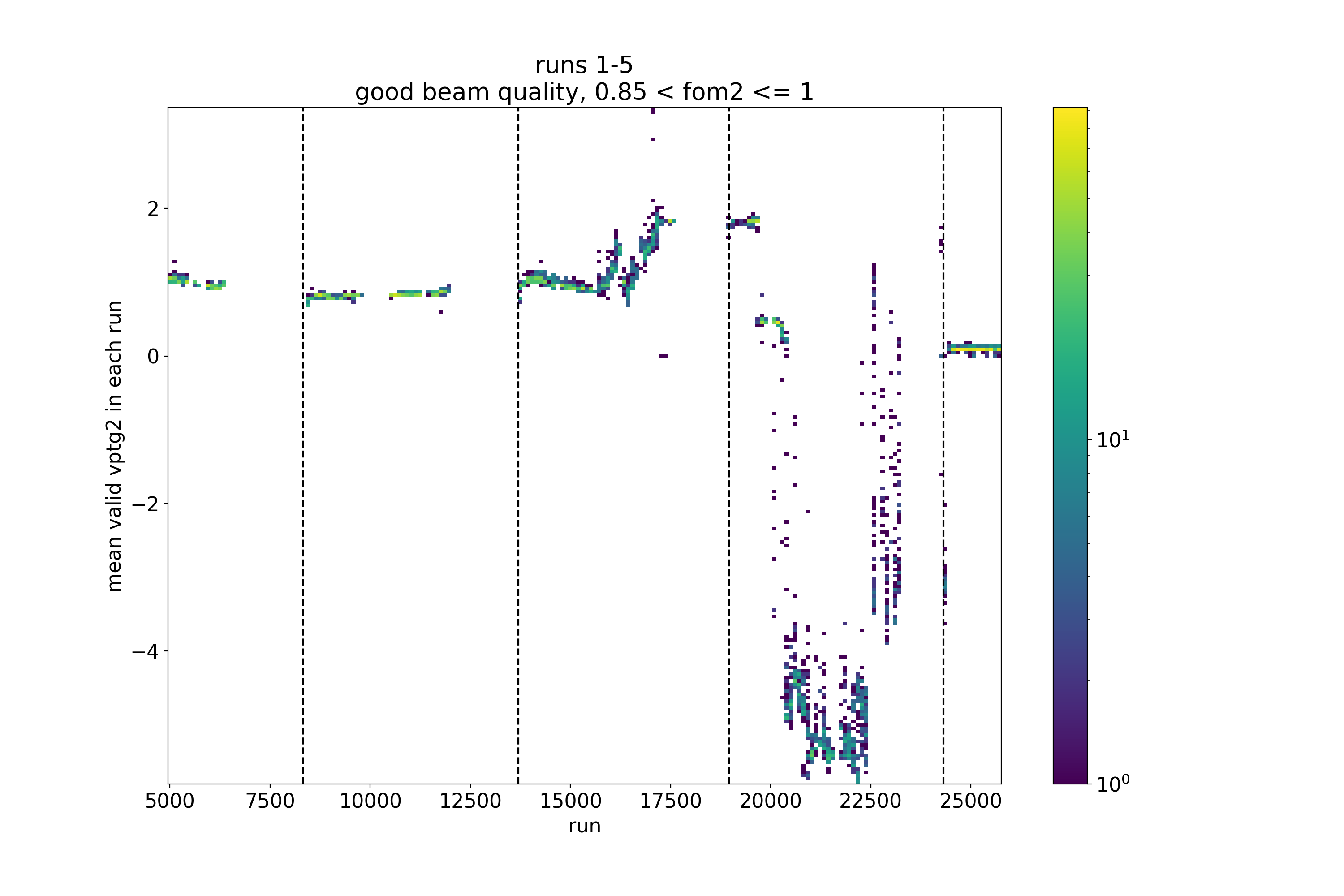}
        \caption{}
        \label{fig:vptg2}
    \end{subfigure}
    \caption[BNB position vs time]{BNB measured position over time. The vertical dashed lines indicate the boundaries separating runs 1-5 of MicroBooNE's data set. Panel (a) shows HPTG1, panel (b) shows VPTG1, panel (c) shows HPTG2, and panel (d) shows VPTG2.}
    \label{fig:bnb_position_vs_time}
\end{figure}

These beam position monitors are used to automatically maintain the beam's steering into the center of the target, known as the ``auto-tune''. This auto-tune uses only one set of beam monitors; either the first pair HPTG1 and VPTG1, or the second pair HPTG2 and VPTG2. The values from these beam position monitors over five years of MicroBooNE data are shown in Fig. \ref{fig:bnb_position_vs_time}.

\begin{figure}[H]
    \centering
    \includegraphics[width=0.5\textwidth]{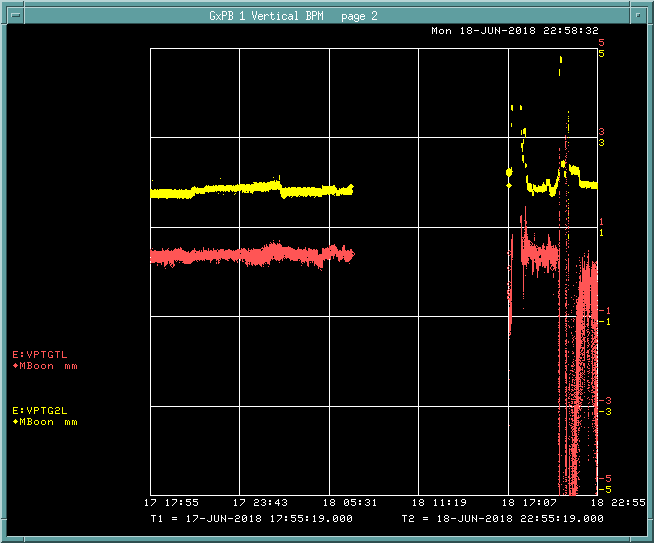}
    \caption[VPTG unreliable behavior]{VPTG unreliable behavior starting June 18, 2018.}
    \label{fig:vptg_weird}
\end{figure}

We see that the positions of the beams remained fairly stable throughout runs 1-3, and then VPTG1 sees strange behavior, with a sharp decline in vertical position measured near the end of run 3, as shown in the center of Fig. \ref{fig:vptg1}. This time period is shown in more detail from a beam monitoring plot in Fig. \ref{fig:vptg_weird}, where it becomes clear that this is just an issue with the beam position monitor, given the extreme and rapid changes in measured beam position that are not reflected in the  other beam position monitors.

Because of this strange behavior in VPTG1, the auto-tune was changed to use VPTG2 instead on June 18, 2018. When this switch happened, the nominal beam position was changed, and therefore the beam started being aimed significantly higher on the target, as shown in Figs. \ref{fig:vptg1} and \ref{fig:vptg2}.

So, the proton beam was aimed significantly off-center of the target for around four months, until the auto-tune was changed again on October 25, 2018, and there was a new target scan performed on October 31, 2018. After the new target scan, the beam's position was consistently on the center of the target. There is some other strange behavior visible in run 5 in Fig. \ref{fig:bnb_position_vs_time}, but this just reflects changes in the beam position monitors and not changes in the real beam position.

\begin{figure}[H]
    \centering
    \includegraphics[width=0.5\textwidth]{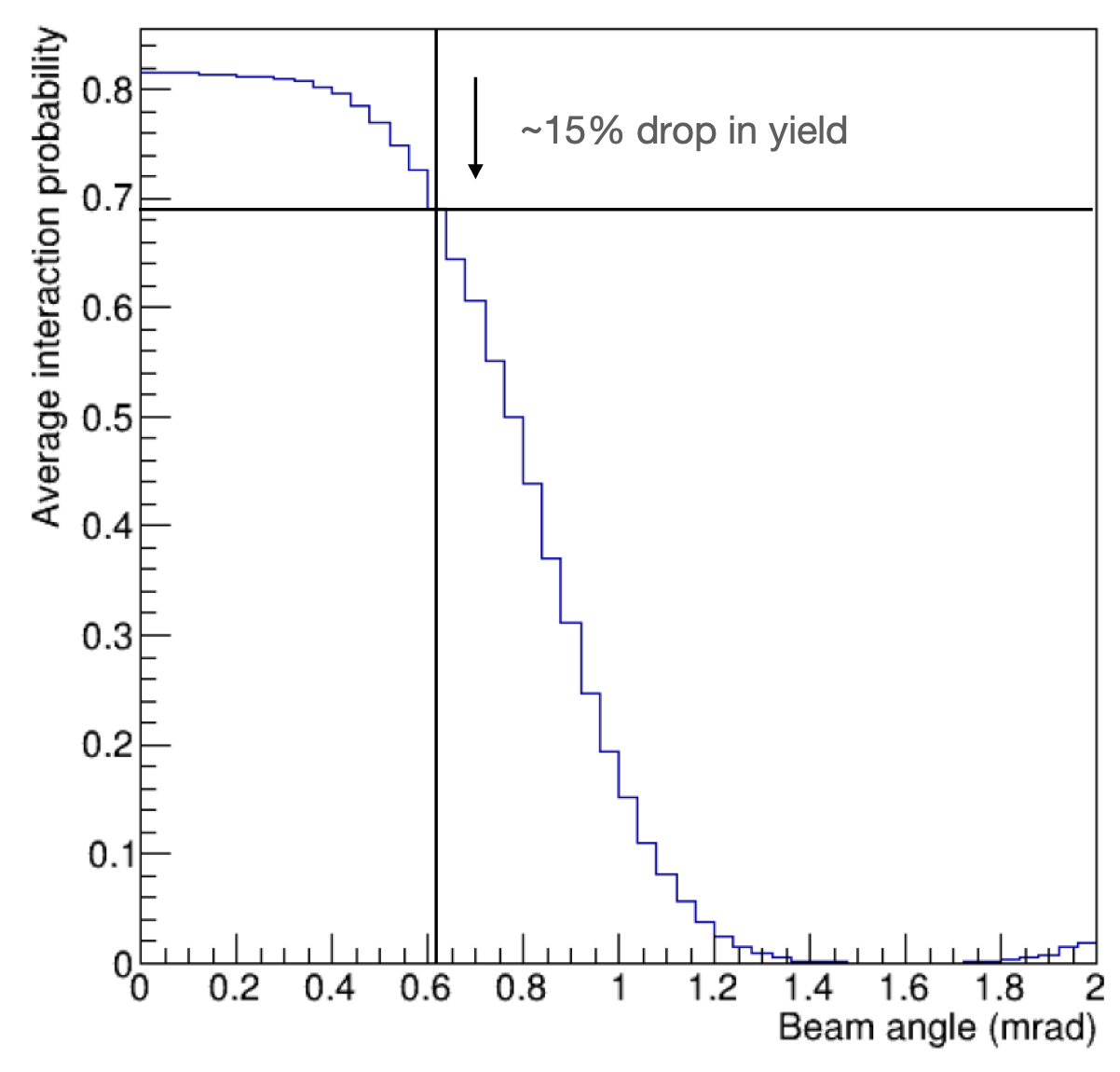}
    \caption[Mistargeted beam simulation]{Mistargeted beam simulation.}
    \label{fig:mistargeted_beam_simulation}
\end{figure}

This mis-targeting of the BNB proton beam explains the deficit we see in our run 4a neutrino dataset. In order to maintain this data for use in all of our neutrino physics analyses, we updated the neutrino beam flux simulation to account for this position offset. According to an updated beam-target interaction simulation, our observed $\sim$ 15\% drop in neutrino flux corresponds to about 0.6 mrad of beam angle offset, as shown in Fig. \ref{fig:mistargeted_beam_simulation}. This is consistent with the offset expected from our updated understanding of the data from the beam position monitors.

\begin{figure}[H]
    \centering
    \includegraphics[trim=0 0 93 0, clip, width=0.4\textwidth]{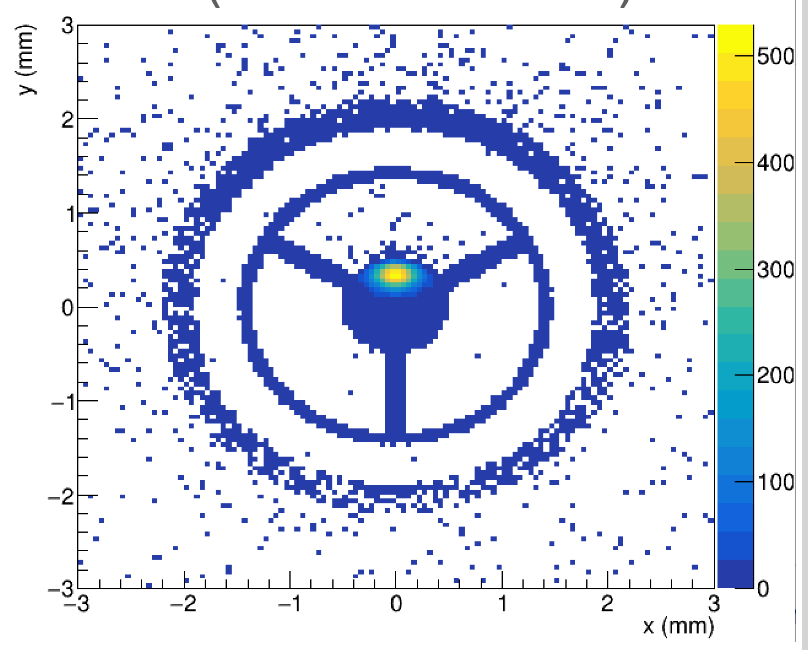}
    \caption[Run 4a simulated beam position]{Run 4a simulated beam position, shown via the density of simulated interactions. The beam is aimed near the top of the target, and there are only rare interactions in the other parts of the target such as the fins and the surrounding material.}
    \label{fig:run_4a_sim_target_position}
\end{figure}

\begin{figure}[H]
    \centering
    \begin{subfigure}[b]{0.49\textwidth}
        \includegraphics[width=\textwidth]{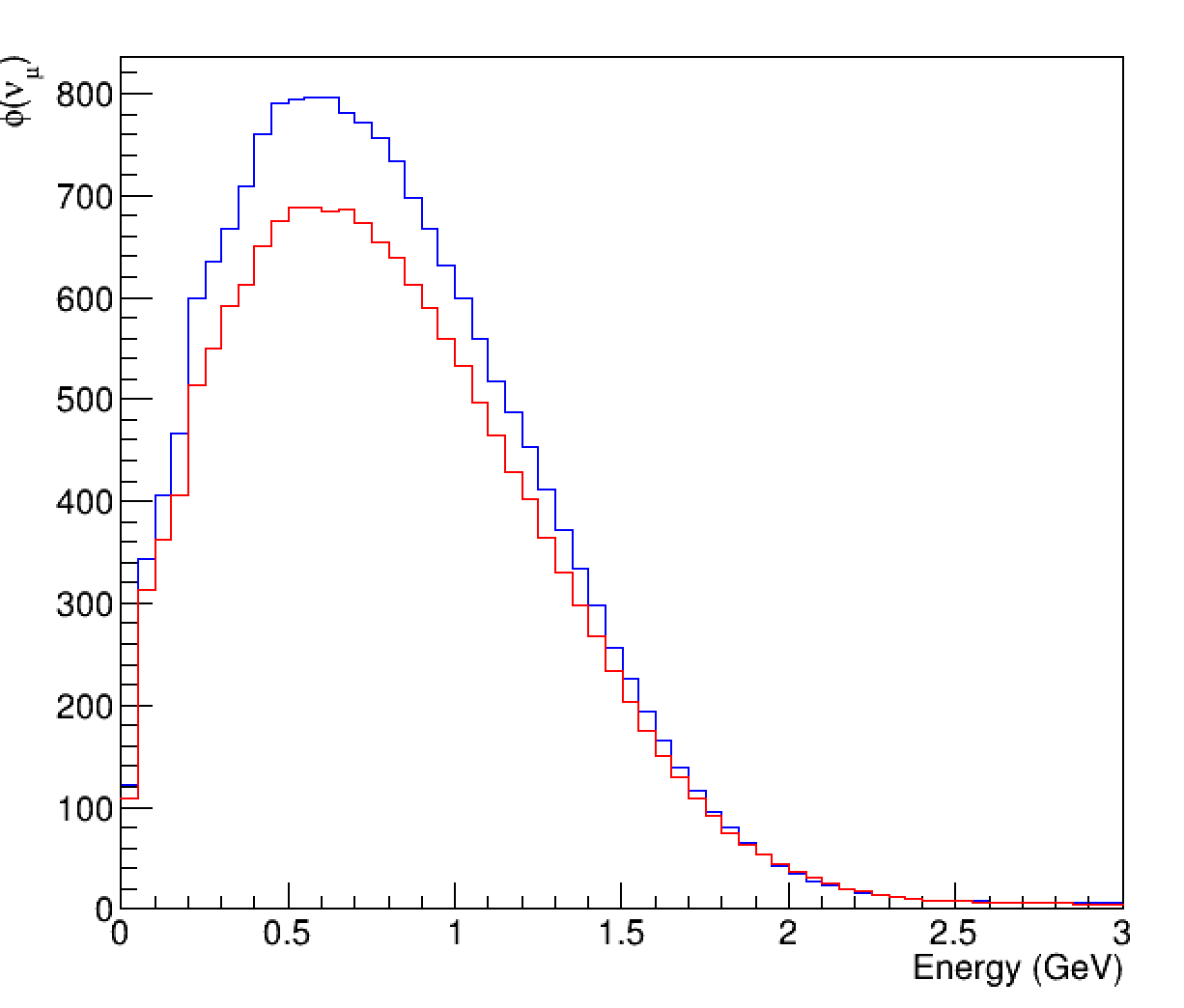}
        \caption{}
        \label{fig:new_4a_flux}
    \end{subfigure}
    \begin{subfigure}[b]{0.49\textwidth}
        \includegraphics[width=\textwidth]{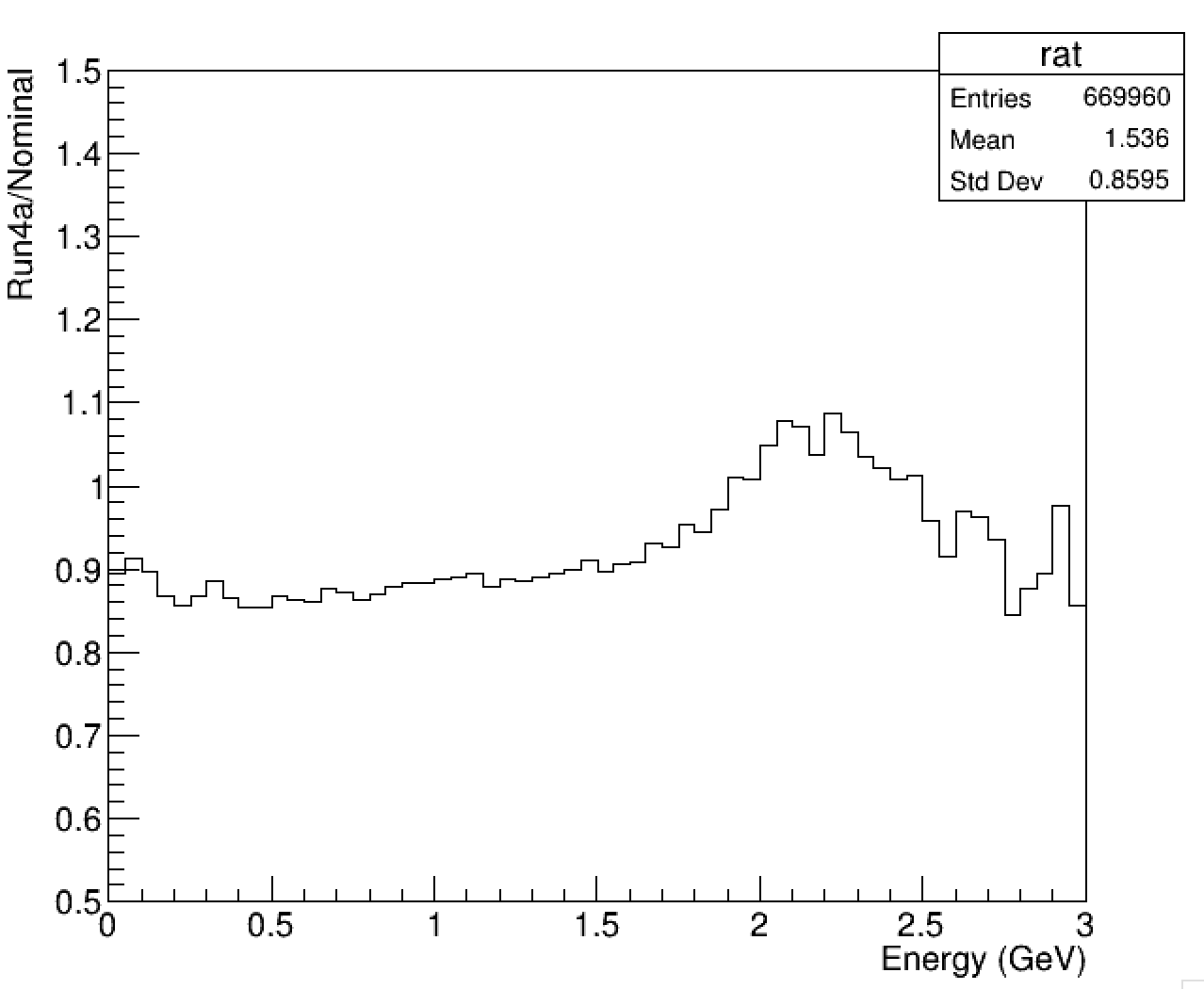}
        \caption{}
        \label{fig:new_4a_flux_ratio}
    \end{subfigure}
    \caption[Run 4a updated flux simulation]{Run 4a updated flux simulation. Panel (a) shows the on-center simulated flux in blue, and the off-center simulated flux in red. Panel (b) shows the ratio of the two, resulting in primarily a normalization shift, but some shape differences at higher neutrino energies.}
    \label{fig:new_4a_flux_and_ratio}
\end{figure}

So, now that we know that the deficit can be explained by reduced neutrino flux from this position offset, we run a full beam flux simulation, including all hadron productions and decays to investigate potential energy dependence due to the mis-targeting. The simulated beam position is shown in Fig. \ref{fig:run_4a_sim_target_position}, and the resulting $\nu_\mu$ neutrino flux simulation is shown in Fig. \ref{fig:new_4a_flux_and_ratio}.

With this new simulation, we can account for this mis-targeting which caused a change in neutrinos per POT without throwing away any of our data, letting us preserve statistical power for all of MicroBooNE's neutrino physics analyses.

%% file: chapters/03_eLEE.tex
\chapter{Anomalous \texorpdfstring{$\nu_e$}{nue}CC Search in MicroBooNE}\label{sec:nueCC}

In this chapter, I will discuss MicroBooNE's search for an anomalous $\nu_e$CC excess as an explanation of the MiniBooNE anomaly as described in Sec. \ref{sec:MB_LEE}. In particular, I will describe my contributions to the analysis that uses Wire-Cell 3D reconstruction in order to search for an anomalous excess of inclusive $\nu_e$CC events, in order to test the MiniBooNE LEE under a $\nu_e$ flux enhancement, potentially via short-baseline $\nu_\mu\rightarrow \nu_e$ sterile neutrino oscillations, as described in Sec. \ref{sec:MB_LEE}. This requires the development of high-performance event selections, which require several steps in understanding and processing data from the MicroBooNE detector.

\section{Signal Processing}

Before we can perform 3D reconstruction of an event in MicroBooNE, we need to have clean and clear 2D images from our wire planes, and this requires several steps of signal processing.

\begin{figure}[H]
    \centering
    \begin{subfigure}[b]{0.32\textwidth}
        \includegraphics[trim=0 0 50 0, clip, width=\textwidth]{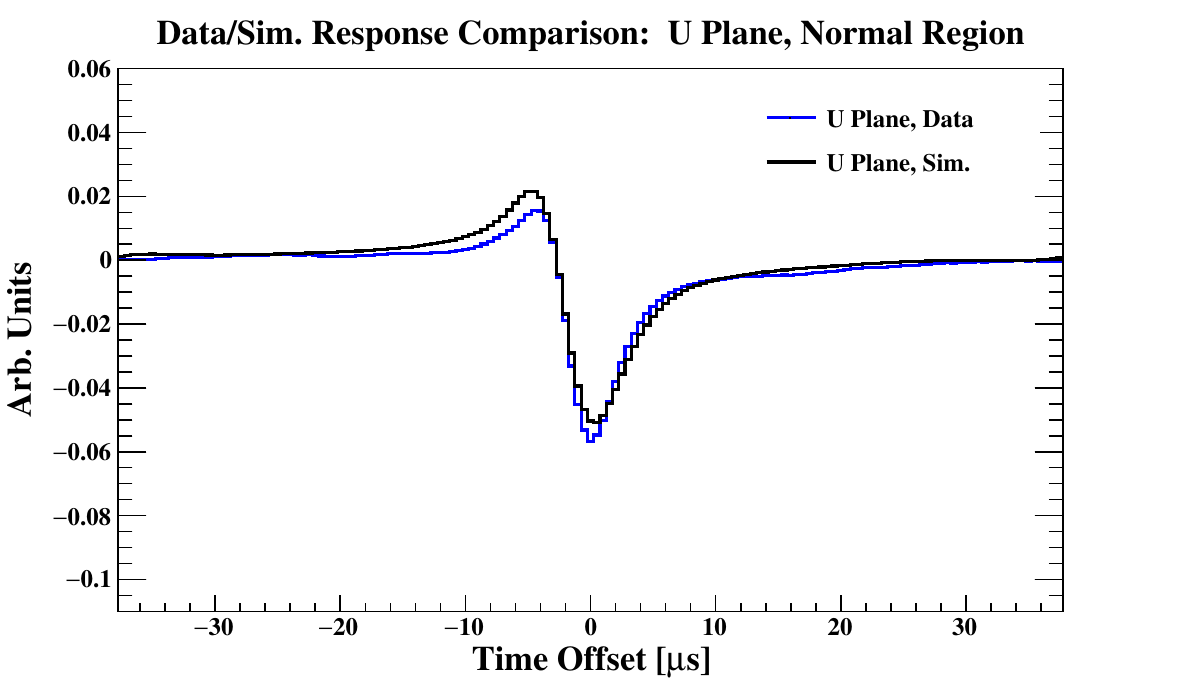}
        \caption{}
        \label{fig:wire_response_U}
    \end{subfigure}
    \begin{subfigure}[b]{0.32\textwidth}
        \includegraphics[trim=0 0 50 0, clip, width=\textwidth]{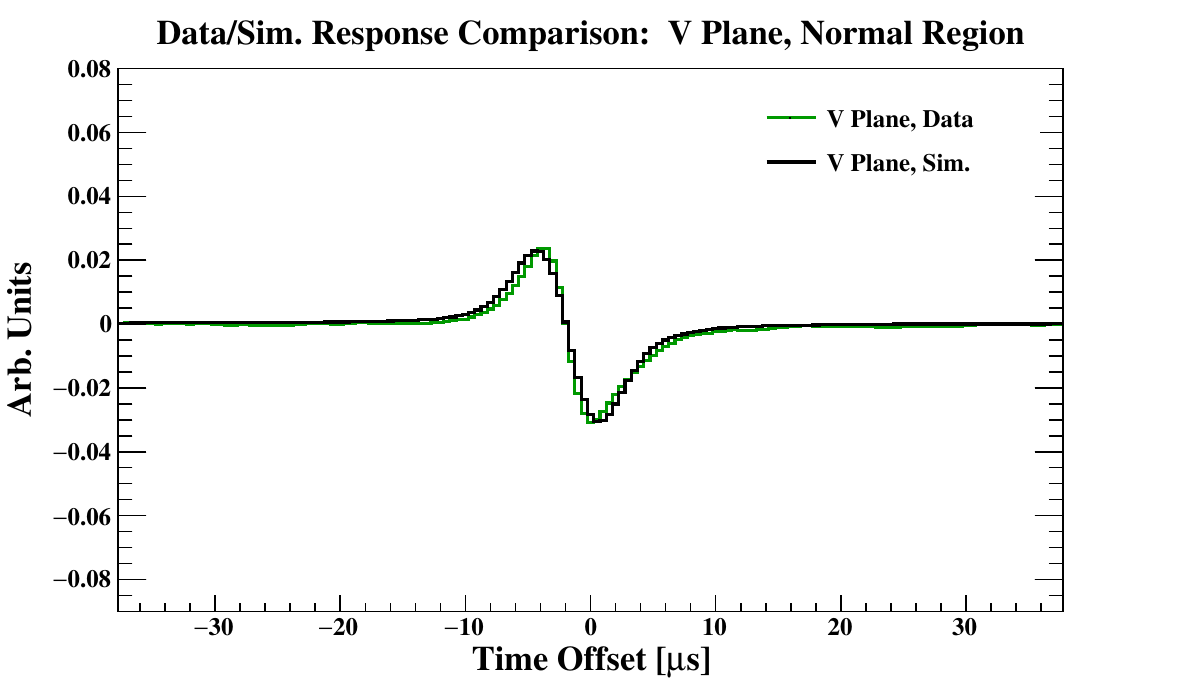}
        \caption{}
        \label{fig:wire_response_V}
    \end{subfigure}
    \begin{subfigure}[b]{0.32\textwidth}
        \includegraphics[trim=0 0 50 0, clip, width=\textwidth]{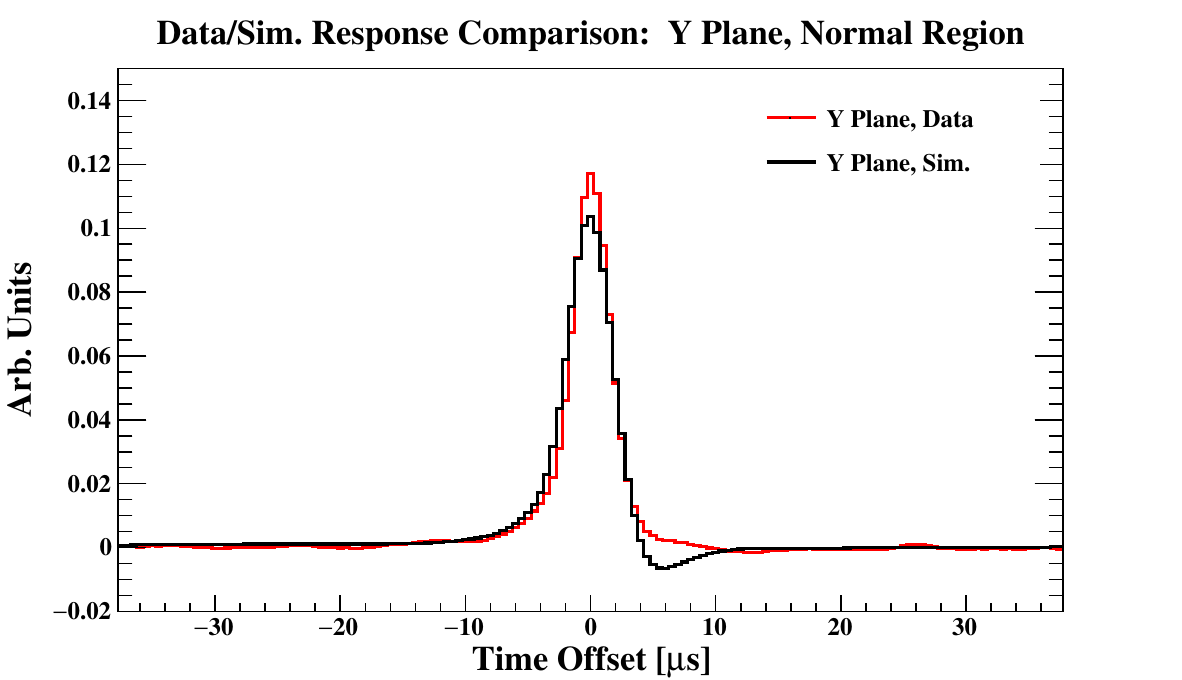}
        \caption{}
        \label{fig:wire_response_Y}
    \end{subfigure}
    \caption[MicroBooNE wire response]{MicroBooNE wire response for tracks with $40^\circ < \theta_{xz} < 50^\circ$. Some residual data-simulation differences are present, but these differences have been addressed with furher processing steps. Panel (a) shows data and simulated waveforms on the U plane, panel (b) shows data and simulated waveforms on the V plane, and panel (c) shows data and simulated waveforms on the Y plane. Figures from Ref. \cite{microboone_signal_processing_part_2}.}
    \label{fig:wire_response}
\end{figure}

The first plane that drift electrons arrive at is the U plane. The U plane first sees a positive current as the electrons attract positive charge on the wire as they approach, and then sees a negative current as the electrons drift further past the plane and the positive charge on the wire disperses. This causes a bipolar signal, where the wire current is initially positive and then negative. The V plane functions in a very similar way and also has a bipolar signal. Finally, the Y plane collects the drifting electrons, producing a purely positive current. Characteristic resulting waveforms are shown in Fig. \ref{fig:wire_response}.

Figure \ref{fig:wire_response} shows the best case scenario, and there are a variety of factors that can make the signals appear much less cleanly. 

If there are multiple groups of ionization charge arriving after each other, these bipolar signals can cancel each other out, making it particularly hard to see the induction effects on the U and V planes. For example, this happens when a track is almost perpendicular to the wire planes, which results in $\theta_{xz} \sim 90^\circ$. We refer to these as ``high-angle tracks''. In this case, there is a prolonged track which has the same amount of charge approaching and receding from the induction wires, making the signal almost invisible for much of the length of the track. 

Another complication can appear when there is a large density of charge approaching the U plane; in this case, the induction signal can start significantly earlier than the actual arrival time of the charge to the wire planes. 

\begin{figure}[H]
    \centering
    \begin{subfigure}[b]{0.9\textwidth}
        \includegraphics[width=\textwidth]{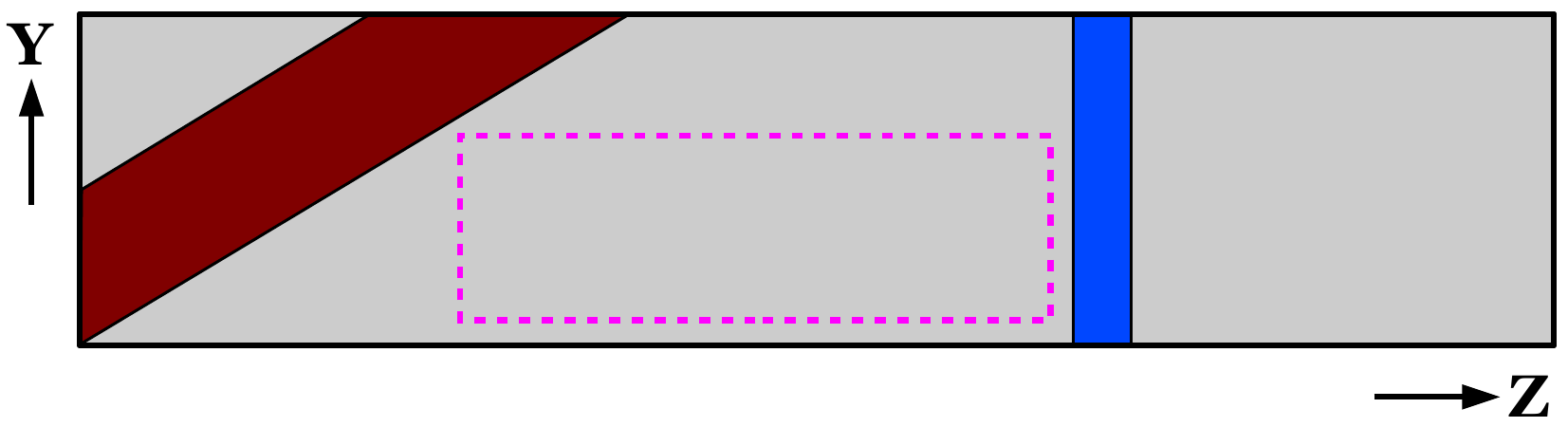}
        \caption{}
        \label{fig:shorted_regions}
    \end{subfigure}
    \begin{subfigure}[b]{0.35\textwidth}
        \includegraphics[width=\textwidth]{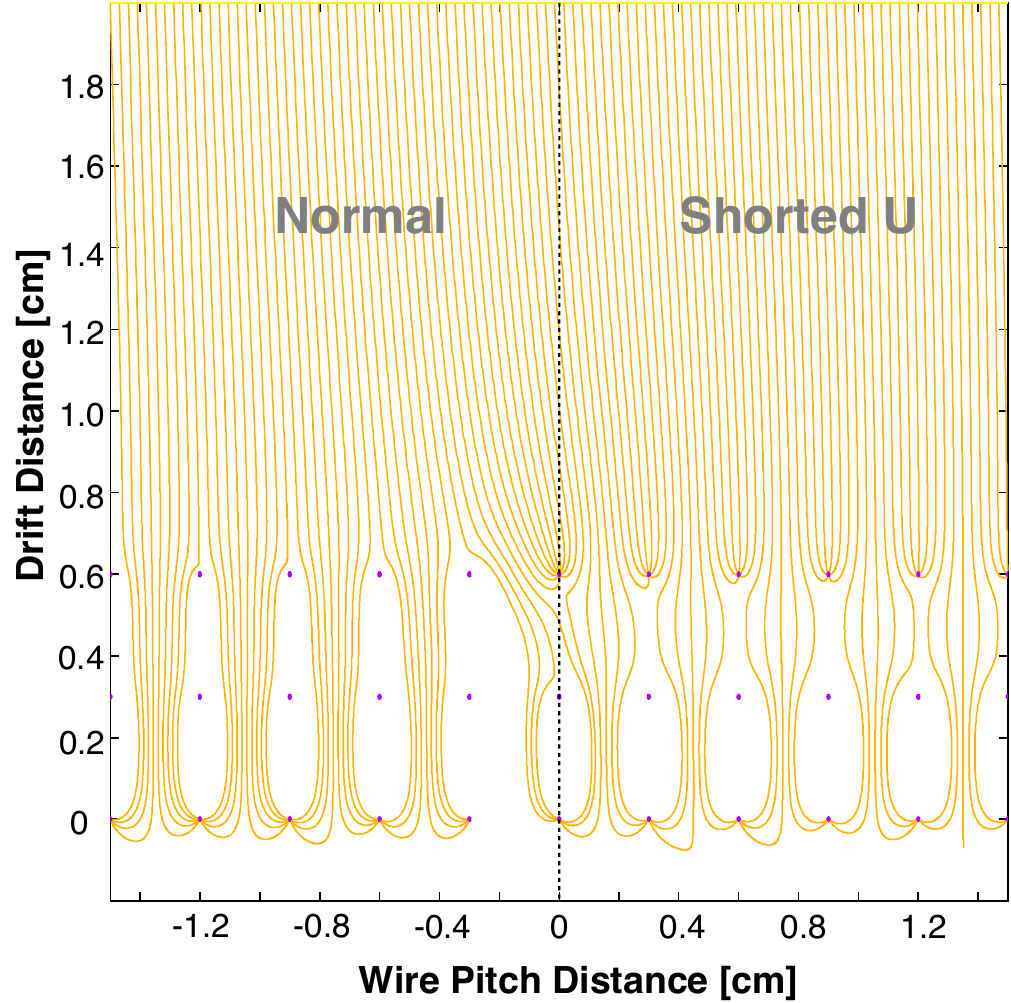}
        \caption{}
        \label{fig:shorted_U}
    \end{subfigure}
    \begin{subfigure}[b]{0.35\textwidth}
        \includegraphics[width=\textwidth]{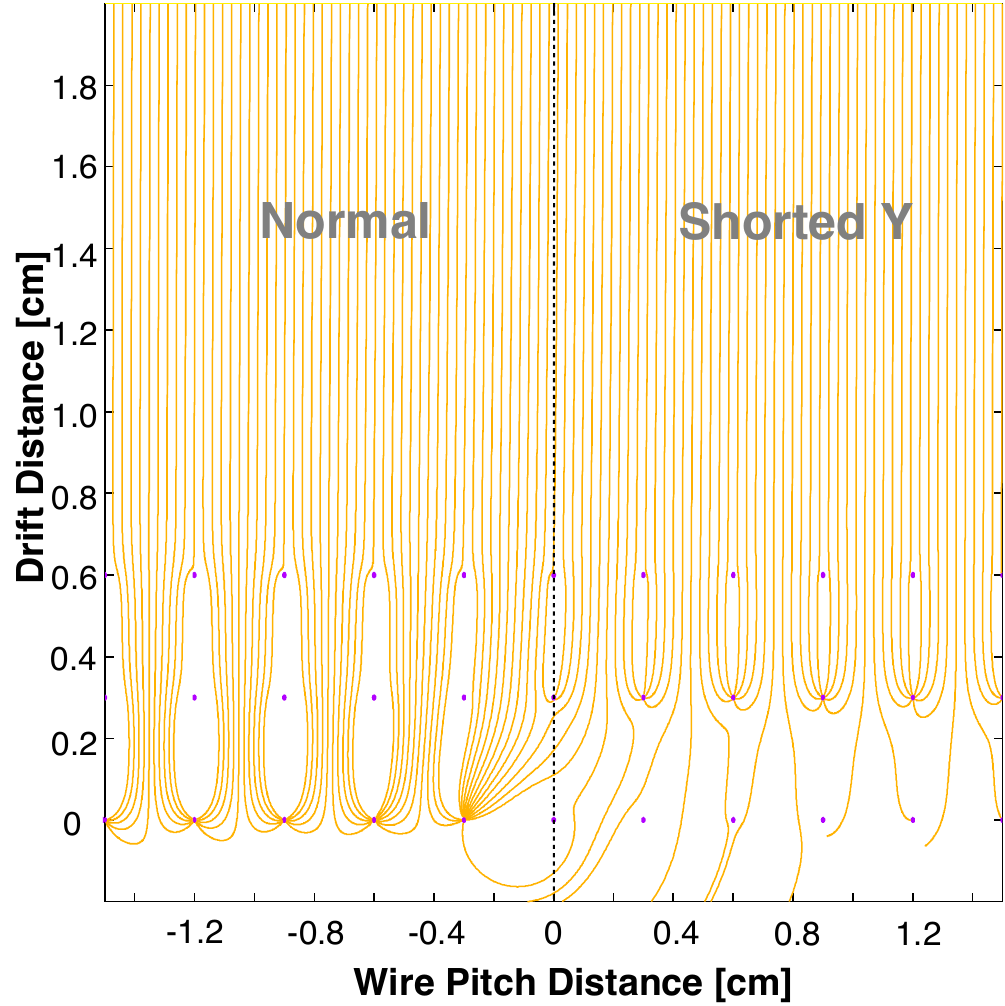}
        \caption{}
        \label{fig:shorted_Y}
    \end{subfigure}
    \caption[MicroBooNE shorted wires]{Panel (a) shows the shorted U region in red, and the shorted Y region in blue. The pink box indicates the normal region used for Fig. \ref{fig:wire_response}. Panel (b) shows simulated electron drift paths in the shorted U region, and panel (c) shows simulated electron drift paths in the shorted Y region. Figures from Ref. \cite{microboone_signal_processing_part_2}.}
    \label{fig:shorted_wires}
\end{figure}

Another complication is the presence of shorted wires. MicroBooNE places a negative bias voltage on the U plane, grounds the V plane, and places a positive bias voltage on the Y plane. There is a ``shorted U'' region, where it seems that a single V wire came into contact with a number of U plane wires, causing the U plane bias to decrease, which results in some charge collection on these non-functional U channels, and a reduced signal amplitude on the subsequent V and Y planes. This scenario is visualized in Fig. \ref{fig:shorted_U}. There is also a ``shorted Y'' region in the detector, where Y plane wires are shorted to ground, losing their bias voltage. This causes electrons in this region to be collected on the V plane rather than the Y plane, as visualized in Fig. \ref{fig:shorted_Y}. Both of these scenarios cause changes in the wire responses that need to be carefully accounted for in our simulation. More details are described in Ref. \cite{microboone_noise}.

Additionally, there are a variety of sources of noise affecting the signals on our wires. There is front-end readout electronic noise, noise from the voltage regulators of our application specific integrated circuits (ASICs), noise from the cathod high voltage power supply, noise from the PMT highvoltage power supply or the interlock system power supply, noise from capacitive coupling to scintillation-light-induced currents within the PMTs just behind the wire planes, noise from electronics associated with purity monitor operation, and noise due to voltage fluctuations on the cathode. These different noise sources have a variety of amplitudes and frequencies and correlations between channels, and in particular the removal of coherent noise can impact the reconstruction of ``isochronous'' charge activity, when a trajectory is parallel to the wire planes and lot of charge activity arrives on the wires at the same time. Overall, we have been very successful at filtering out this noise and reducing its impact on our final images. More details are described in Refs. \cite{microboone_noise, microboone_signal_processing_part_2}.

After the images have been de-noised, the signals have to be deconvolved in order to account for the complex detector response to arriving charge. A traditional 1D deconvolution uses a fourier transform of the measured signal, a fourier transform of the known response function, and a filter function to reduce high frequency noise. In MicroBooNE, we expand this procedure to account for the fact that one piece of charge can cause signals on multiple adjacent wires, leading to correlations that can be accounted for in the deconvolution process \cite{microboone_signal_processing_part_1, microboone_signal_processing_part_2}. This turns all of our images into unipolar signals that are much easier for downstream reconstruction algorithms to operate on.

\begin{figure}[H]
    \centering
    \includegraphics[trim=0 150 0 150, clip, width=0.6\textwidth]{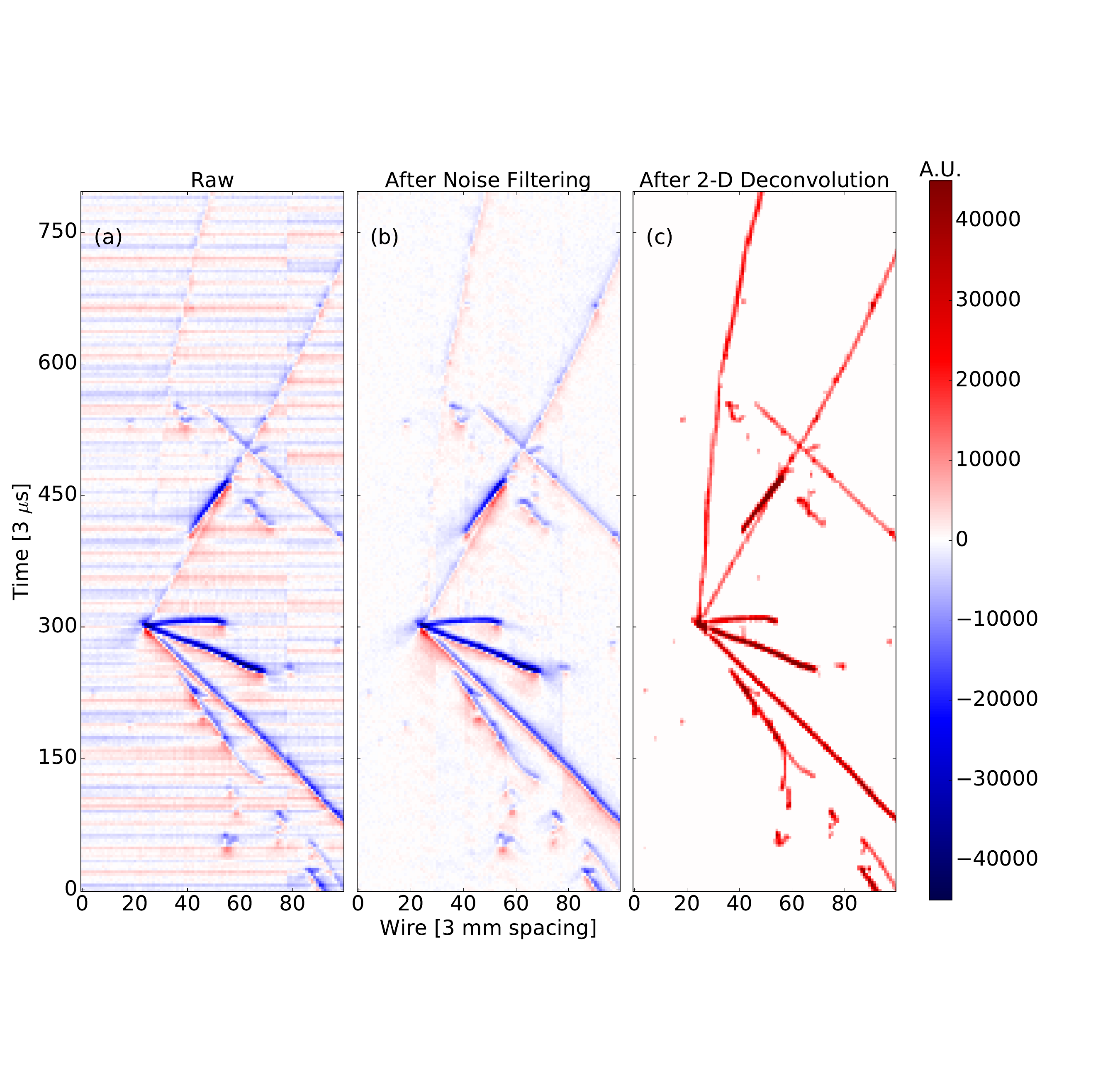}
    \caption[MicroBooNE 2D deconvolution U plane]{MicroBooNE signal processing for an example U plane image, with the left showing the raw image, the middle showing the image after noise filtering, and the right showing the image after 2D deconvolution. From Ref. \cite{microboone_signal_processing_part_2}.}
    \label{fig:U_deconvolution}
\end{figure}

\begin{figure}[H]
    \centering
    \begin{subfigure}[b]{0.47\textwidth}
        \includegraphics[trim=0 100 0 100, clip, width=\textwidth]{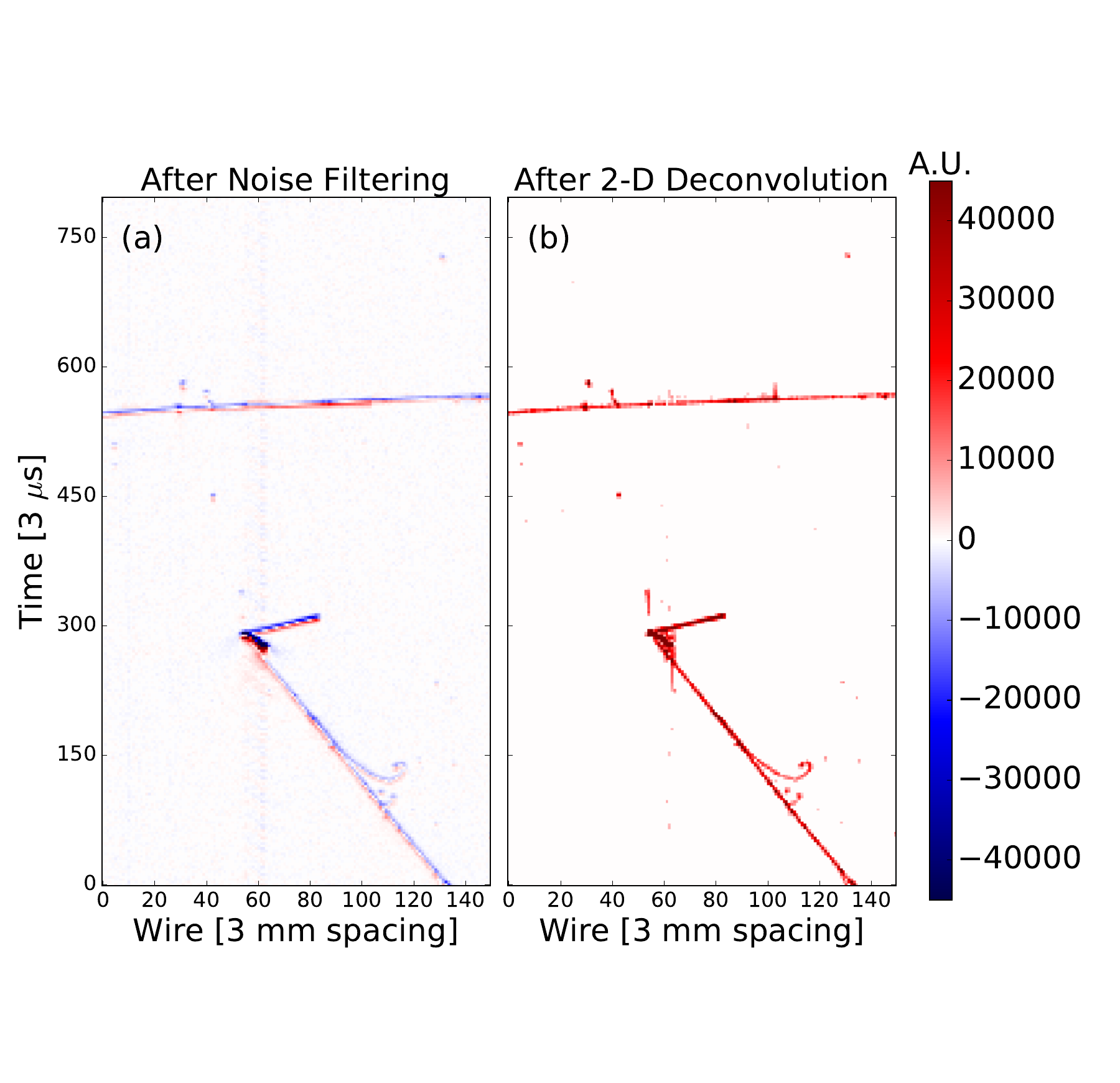}
        \caption{}
        \label{fig:V_deconvolution}
    \end{subfigure}
    \begin{subfigure}[b]{0.52\textwidth}
        \includegraphics[trim=400 50 0 100, clip, width=\textwidth]{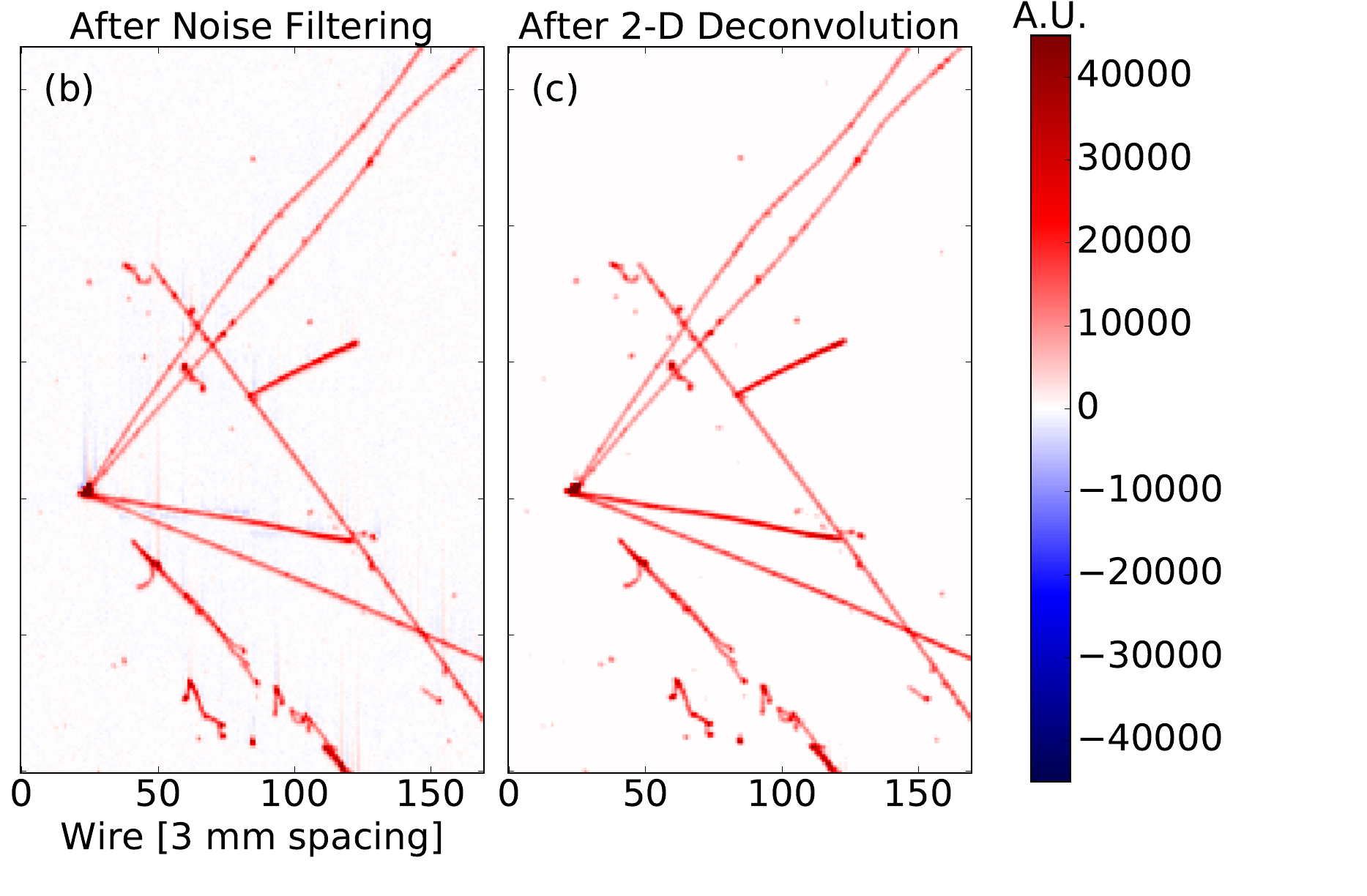}
        \caption{}
        \label{fig:Y_deconvolution}
    \end{subfigure}
    \caption[MicroBooNE 2D deconvolution VY planes]{MicroBooNE signal processing for the V and Y planes, with the left showing the image after noise filtering, and the right showing the image after 2D deconvolution. Panel (a) shows an example image from the V plane, and panel (b) shows an example image from the Y plane. Figure from Ref. \cite{microboone_signal_processing_part_2}.}
    \label{fig:UV_2D_deconvolution}
\end{figure}

Examples showing the effects of noise removal and 2D deconvolution on our images are shown for each wire plane in Figs. \ref{fig:U_deconvolution}-\ref{fig:UV_2D_deconvolution}.

\section{Wire-Cell 3D Reconstruction}\label{sec:wire_cell}

In the previous section, we discussed how we reconstruct 2D projected images of the ionization activity in MicroBooNE. In this section, we discuss the use of Wire-Cell 3D reconstruction, which takes these projected views as well as PMT information in order to reconstruct 3D particles in MicroBooNE and use these for physics analyses.

\subsection{Wire-Cell 3D Imaging}

Unlike the other popular reconstruction paradigm used by MicroBooNE, Pandora \cite{microboone_pandora}, Wire-Cell reconstruction converts all the information from the projected views to 3D as early as possible, before any clustering or other reconstruction steps. The algorithms in this section are described in more detail in Refs. \cite{wire_cell_proposal, wire_cell_imaging}.

\begin{figure}[H]
    \centering
    \begin{subfigure}[b]{0.54\textwidth}
        \includegraphics[trim=20 0 50 20, clip, width=\textwidth]{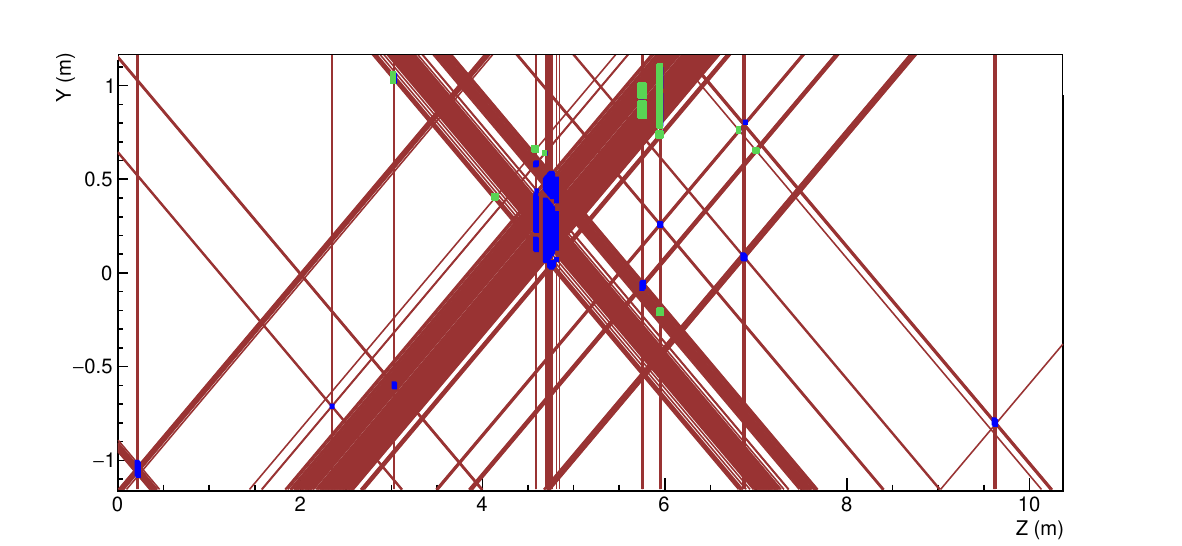}
        \caption{}
        \label{fig:tiling_zoom_out}
    \end{subfigure}
    \begin{subfigure}[b]{0.45\textwidth}
        \includegraphics[trim=100 50 100 50, clip, width=\textwidth]{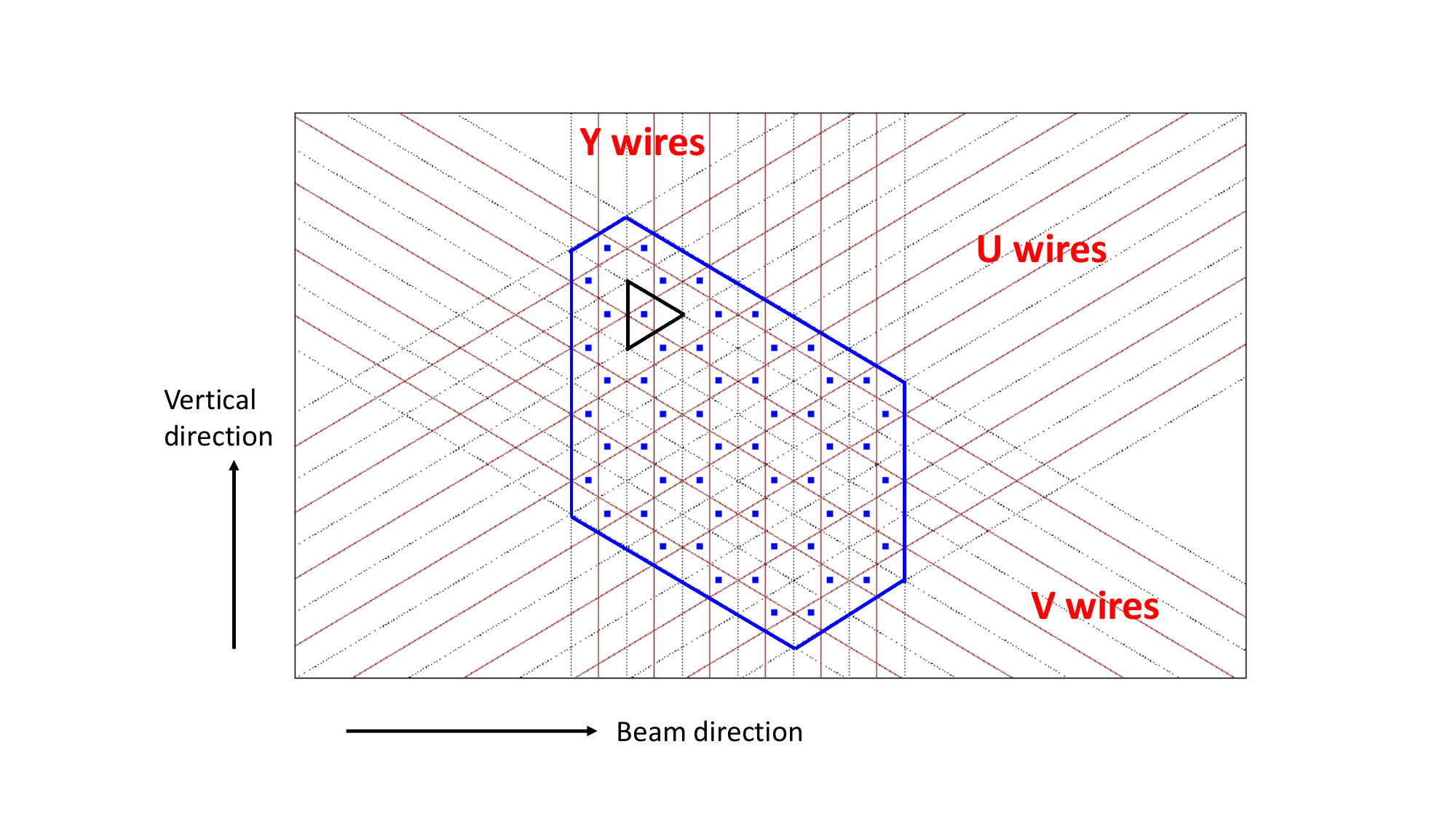}
        \caption{}
        \label{fig:tiling_zoom_in}
    \end{subfigure}
    \caption[Wire-Cell tiling]{Wire-Cell tiling. Panel (a) shows all wires which see charge at the same time in red, 3-plane tiling blobs in blue, and 2-plane tiling blobs in green. Note that some blobs shown here have been removed by a later de-ghosting step. Panel (b) shows a zoomed in example 3-plane tiling blob and the corresponding hit cells. Figures from Ref. \cite{wire_cell_imaging}.}
    \label{fig:tiling}
\end{figure}

Wire-Cell reconstruction uses a tomographic approach which operates over 2$\mu$s time slices, corresponding to about 2.2 mm of distance in the drift dimension. In each time slice, Wire-Cell operates over ``cells'', which are triangular regions on the wire planes formed by intersections of three wires. An example of a single cell can be seen as the black triangle in Fig. \ref{fig:tiling_zoom_in}. Wire-Cell performs ``tiling'', a process which groups adjacent wires which saw charge in this time slice into ``wire bundles'', and then forms ``blobs'', which are groups of cells formed by intersecting wire bundles. Note that when many wires see charge at the same time, these blobs become larger, which can make the reconstructed image harder to reconstruct; this is a reason that Wire-Cell can have a harder time properly reconstructing this isochronous charge activity. These blobs form our first preliminary 3D image, but it contains a lot of additional reconstructed 3D activity that does not match to real ionization, which we refer to as ``ghosts'', and it does not contain any information about the amount of charge in different parts of the 3D image.

\begin{figure}[H]
    \centering
    \begin{subfigure}[b]{0.6\textwidth}
        \includegraphics[trim=20 300 50 0, clip, width=\textwidth]{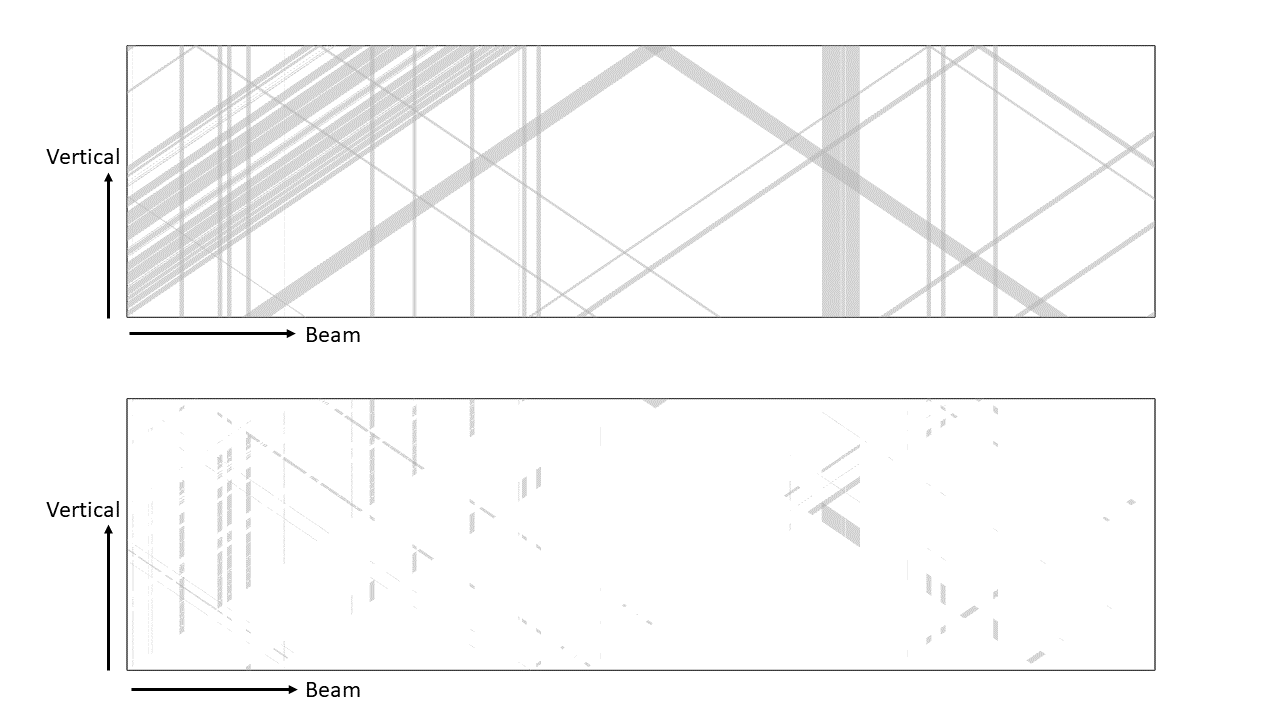}
        \caption{}
        \label{fig:1_dead}
    \end{subfigure}
    \begin{subfigure}[b]{0.6\textwidth}
        \includegraphics[trim=20 0 50 275, clip, width=\textwidth]{figs/wc_nue/tilingDeadRegion.png}
        \caption{}
        \label{fig:2_dead}
    \end{subfigure}
    \caption[MicroBooNE dead wires]{MicroBooNE dead wires. Panel (a) indicates the regions where at least one wire is dead. Panel (b) indicates the regions where at least two wires are dead. Figure from Ref. \cite{wire_cell_imaging}.}
    \label{fig:dead_wires}
\end{figure}

MicroBooNE has a significant area that is affected by dead wires, covering about 30\% of the active volume, but a much smaller area that has two overlapping planes of dead wires, which is only about 3\% of the active volume, as shown in Fig. \ref{fig:dead_wires}. Therefore, we form blobs by 2-plane in regions with dead wires, and by 3-plane tiling in regions with no dead wires. Examples of blobs are shown in Fig. \ref{fig:tiling}.

One we have blobs in each time slice, the next step is ``charge solving''. This refers to an optimization process, trying to find charges for each blob which are consistent with charges for each wire bundle. This is generally an underdetermined system, with more more blobs than wire bundles, so in this optimization process, we add a few assumptions in order to get better results. We prefer \textit{sparseness}, since we expect the vast majority of the 3D image to be free of ionization activity; \textit{positivity}, since there will be no wires that sense a negative amount of electrons; and \textit{proximity}, since we expect many blobs to be near each other for real ionization activity. All of these effects can be converted into an efficient mathematical optimization, with more details described in Ref. \cite{wire_cell_imaging}.

\begin{figure}[H]
    \centering
    \includegraphics[width=\textwidth]{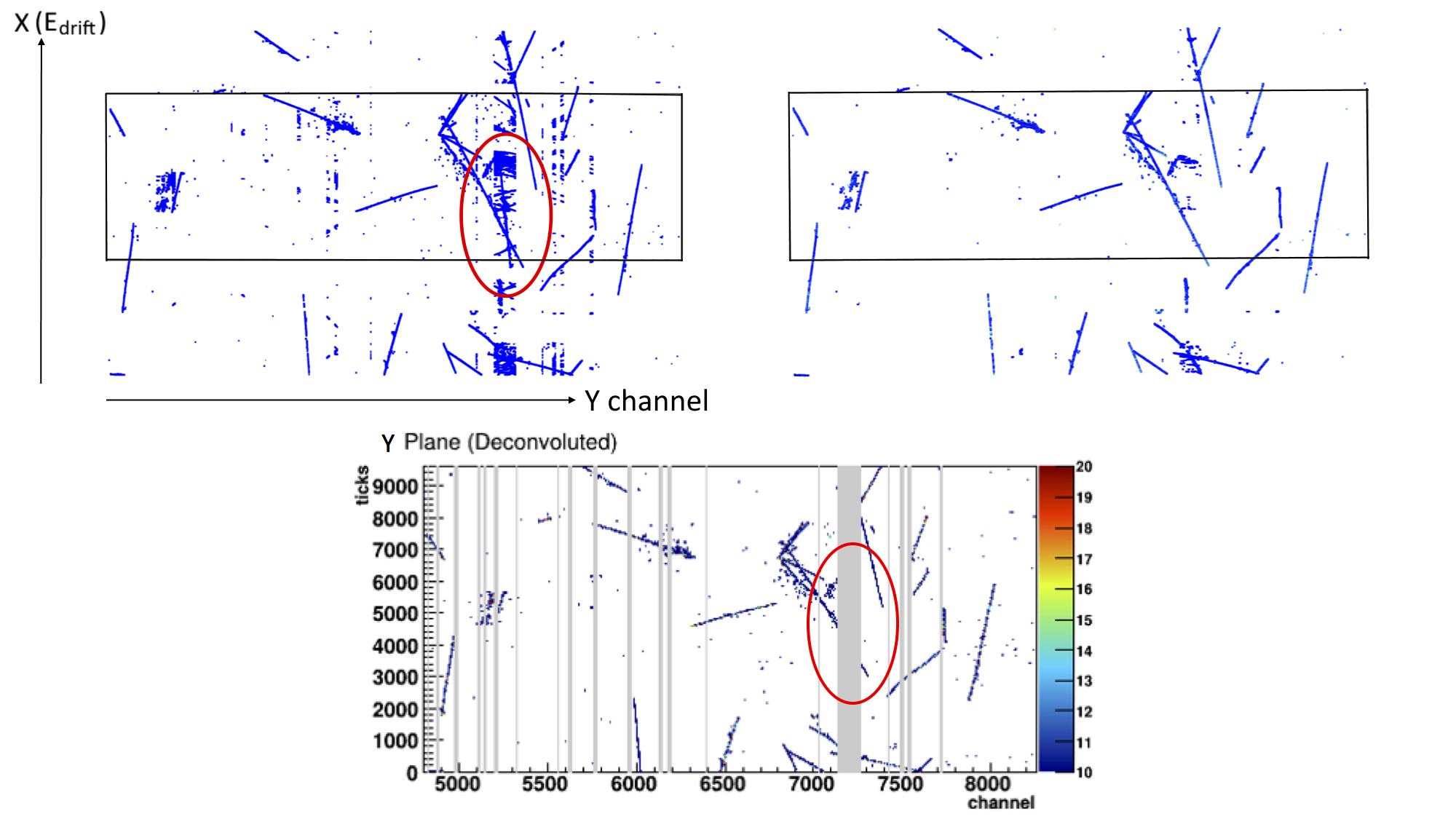}
    \caption[Wire-Cell deghosting]{Wire-Cell deghosting. Figure from Ref. \cite{wire_cell_imaging}.}
    \label{fig:deghosting}
\end{figure}

After this optimization, the ghosting will be significantly reduced, but not entirely gone. The sparsity constraint was able to remove a lot of the ghost points using local information, but at this stage there are ghosts that exist as larger ``proto-clusters'', which are connected blobs in 3D space. This is a somewhat irreducible consequence of the fact that three 2D projections are not sufficient information to uniquely determine the 3D image in all cases. However, we can further reduce this through dedicated de-ghosting algorithms which use more global information about each proto-cluster. First, a proto-cluster is removed if it exists mostly within a region of dead wires, where ghosts are much more likely to appear. Secondly, a proto-cluster is removed if they are sufficiently redundant in each projection, meaning that it has significant overlaps with other proto-clusters in all three projected views. This de-ghosting is shown in Fig. \ref{fig:deghosting}.

\begin{figure}[H]
    \centering
    \includegraphics[width=\textwidth]{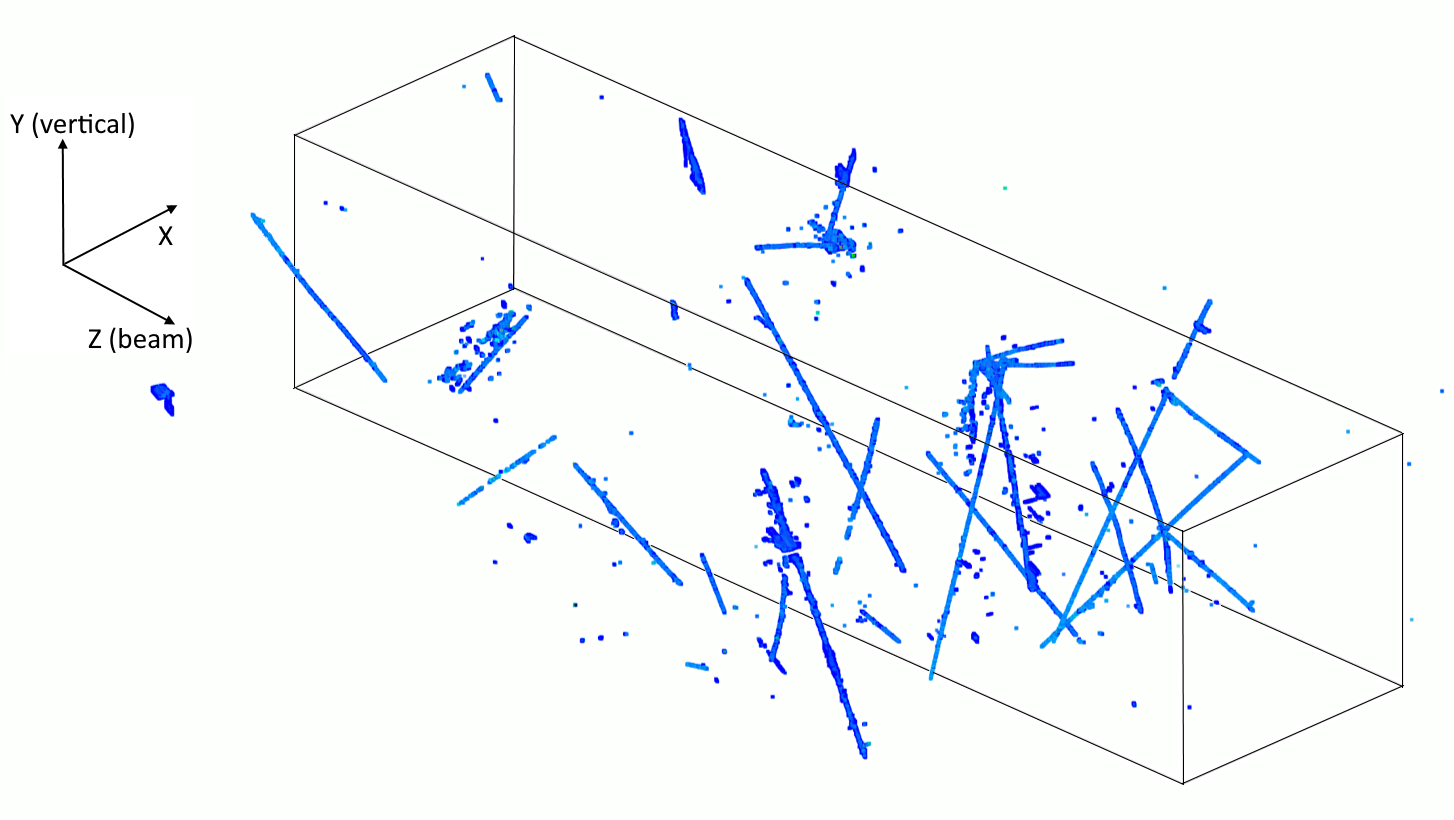}
    \caption[Wire-Cell 3D image]{Wire-Cell 3D event display after all imaging steps. This corresponds to run 3493 event 41075 from real MicroBooNE data. Figure from Ref. \cite{wire_cell_imaging}.}
    \label{fig:final_imaging}
\end{figure}

After ghosts are removed, the charge-solving step can be updated to be more accurate, so we actually perform three iterations of charge solving and deghosting before we get our final 3D image, as shown in Fig. \ref{fig:final_imaging}.

\subsection{Wire-Cell Clustering and Flash Matching}

After the imaging steps, we have a full 3D image of all the charge activity in an event. However, due to MicroBooNE's position on the surface and the relatively slow electron drift in a LArTPC, most of the activity in our image is cosmic ray induced, with only a small fraction that is possibly associated with a neutrino interaction. In order to identify this neutrino activity within the image, we first perform clustering, which has the goal of separating this image into different clusters which each have a unique true origin, either a neutrino or a cosmic ray.

From the initial 3D image, we have proto-clusters, which are just connected blobs. There are several ways in which we combine these proto-clusters in order to achieve good clustering performance. First, we identify gaps which cause a single particle to be split into multiple proto-clusters. This gap can be caused by a variety of factors, including dead regions of the detector, isochronous activity that got filtered out as coherent noise, and signal processing failures for high-angle tracks. Next, we identify X-shaped topologies where it appears that two tracks intersect. This is often a coincidental overlap, and in this case, we separate the two tracks into different clusters. Then, we run the de-ghosting step again, but now on each cluster individually in order to further remove ghosts. Finally, we cluster activity that is spatially separated but likely caused by neutrino-induced neutral particles, in particular from neutrons and photons. This procedure operates by reconstructing a preferred direction in each preliminary cluster and then looking for intersections between these directions, which could point to the neutrino vertex or a secondary interaction vertex.

\begin{figure}[H]
    \centering
    \includegraphics[width=\textwidth]{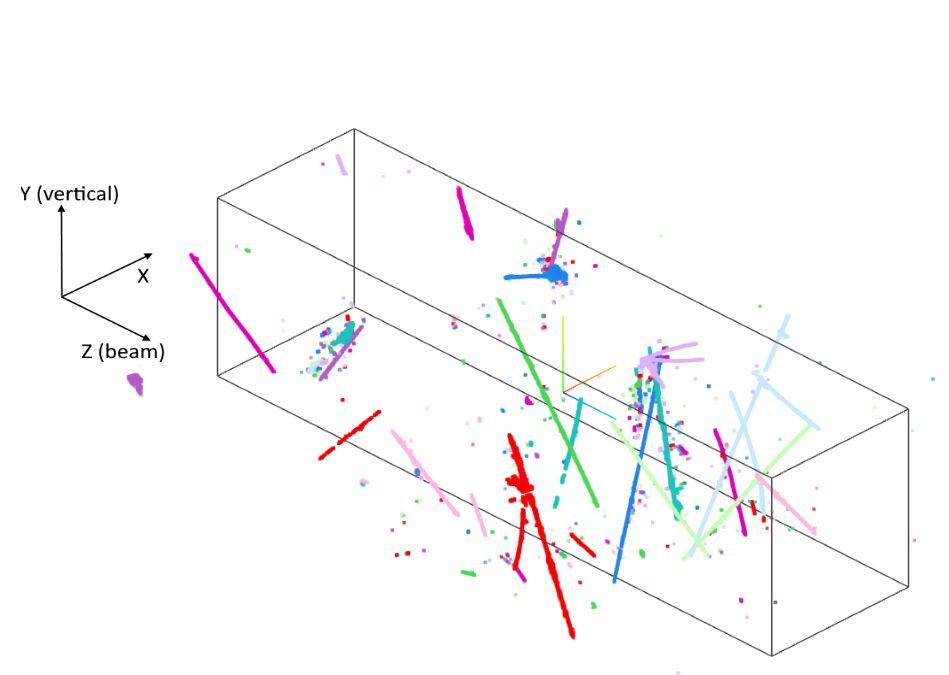}
    \caption[Wire-Cell 3D clusters]{Wire-Cell 3D event display after clustering. The TPC boundary is outlined as a black rectangular prism. Each color corresponds to a different Wire-Cell cluster. This corresponds to run 3493 event 41075 from real MicroBooNE data. Figure from Ref. \cite{wire_cell_imaging}.}
    \label{fig:final_clustering}
\end{figure}

The result of these clustering steps is shown in Fig. \ref{fig:final_clustering}, which identifies different clusters with different colors.

Our next goal is to determine which of these many clusters is neutrino-induced, with the rest being induced by cosmic rays. To do this, we use our PMTs which measure the scintillation light produced by the ionization activity. If our clustering has gone well, each cluster should have been produced at a distinct time, and therefore should be associated with a unique reconstructed pattern of PMT activity, which we call a ``flash''.

In Fig. \ref{fig:final_clustering}, you might notice that some of the charge activity is outside of the TPC boundary. This is because we do not know the true x-position of each cluster; we only know the relative time of arrival at the wires. For example, if a cosmic ray happened in the center of the detector 100 $\mu$s before the neutrino beam arrived, it would arrive at the wires relatively early, and would be drawn with a smaller x-coordinate in this figure, while if it occurred 100 $\mu$s after the beam spill, it would arrive at the wires relatively late, and would be drawn with a larger x-coordinate in this figure. This TPC boundary is really only correct for ionization activity that was produced in time with the neutrino beam pulse, and the cosmic ray activity will have shifted positions. Therefore, we will not know the true position of a cluster until we have identified the time that each cluster was produced.

In order to determine the time of each cluster, we perform many-to-many charge-light matching. At this stage, we typically have 20-30 charge clusters, and 40-50 PMT flashes, and our goal is to match them together. We allow a charge cluster to match to zero or one PMT flash, since there can be inefficiencies in the light system for low-energy activity. We allow a PMT flash to be matched to zero, one, or multiple charge clusters, since there might be ionizition happening outside of the TPC which produces light but no charge, and since there may have been underclustering in the previous steps that means multiple charge clusters actually have one true source.

\begin{figure}[H]
    \centering
    \includegraphics[width=\textwidth]{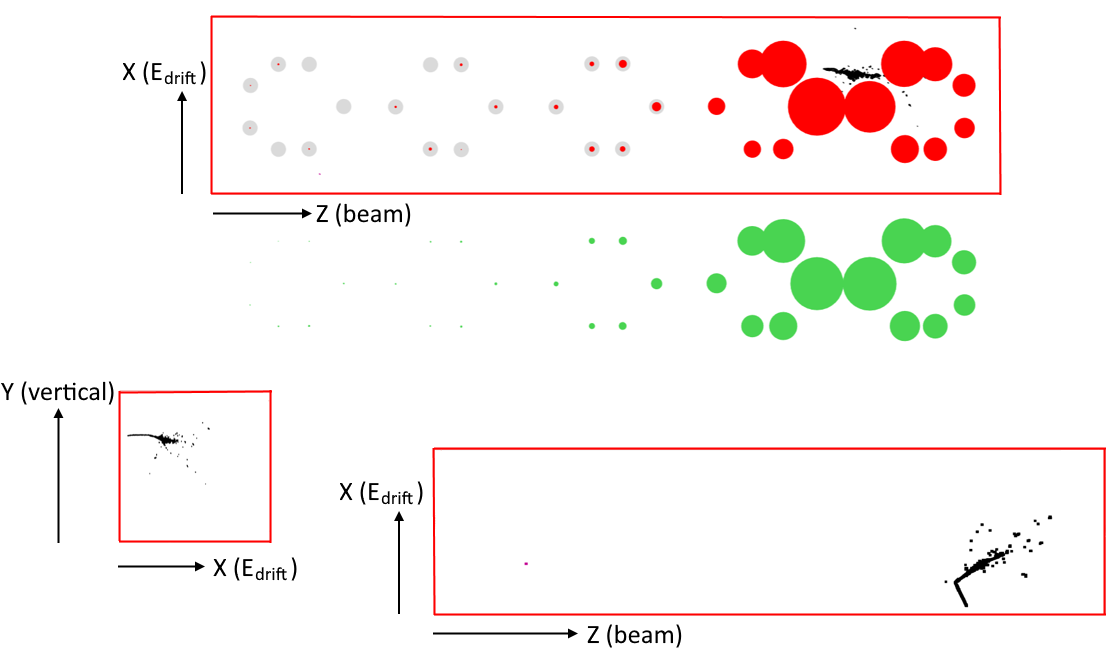}
    \caption[Wire-Cell charge-light matching]{Wire-Cell charge-light matching for a simulated electron neutrino cluster. The measured flash pattern is shown as red circles, and the predicted flash pattern is shown as green circles. The size of the circle indicates the brightness of the light signal. Only the in-beam flash is shown, and only the matching cluster is shown. Figure from Ref. \cite{wire_cell_imaging}.}
    \label{fig:charge_light_matching}
\end{figure}

Mathematically, in order to perform this matching, we use a similar algorithm as we used for charge-solving in the imaging step. We match the patterns of light between measured flashes and predicted flashes according to the observed charge activity. These measured and predicted light patterns are illustrated in Fig. \ref{fig:charge_light_matching}. This process of comparing light and charge is performed for every charge cluster and every cosmic cluster in an iterative process, which involved several rounds of matching and re-examinations between stages.

We only trigger and collect data for events with a PMT flash within the 1.6 $\mu$s neutrino beam time window, so at the end of this process, we can simply identify the cluster or clusters associated with the beam flash as a 3D image of the neutrino-induced activity.

\subsection{Pattern Recognition}\label{sec:pattern_rec}

Now that we have a 3D image of the charge activity of a neutrino interaction, we perform a variety of pattern recognition algorithms in order to reconstruct particles in the event. The algorithms in this section are described in more detail in Ref. \cite{wire_cell_pattern_recognition}.

\begin{figure}[H]
    \centering
    \includegraphics[width=\textwidth]{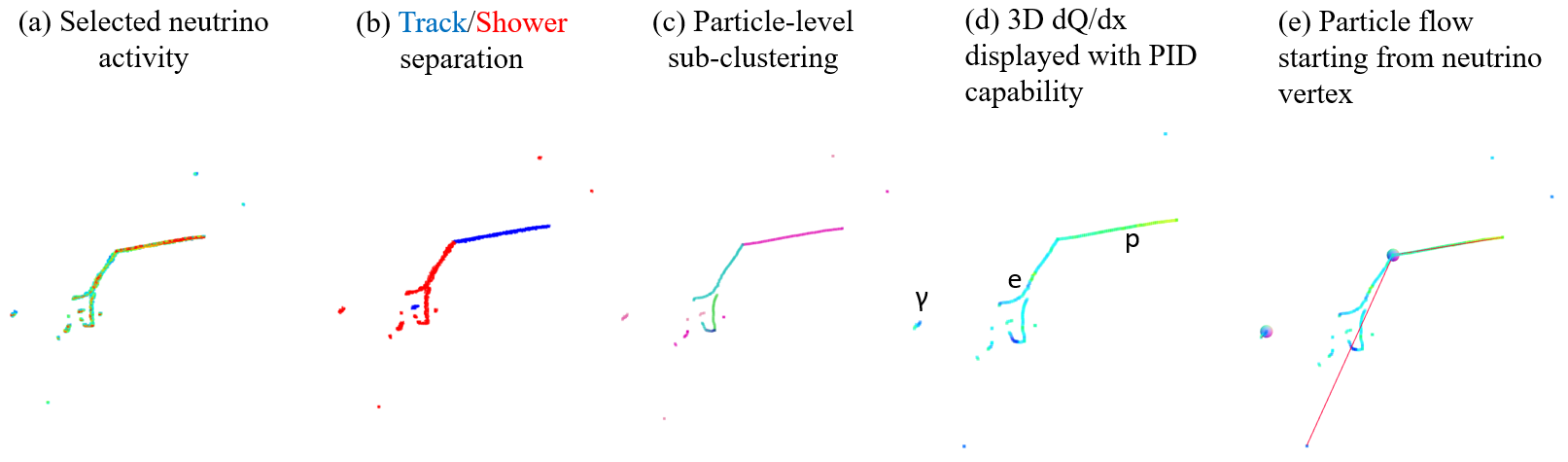}
    \caption[Wire-Cell pattern recognition stages]{Wire-Cell pattern recognition stages for an example $\nu_e$CC candidate in MicroBooNE data, run 6195 event 1020. The blue, cyan, green, yellow, and red colors correspond to roughly 1/3, 1, 2, 3, and 4 times the $dQ/dx$ of a minimum ionizing particle respectively. Figure from Ref. \cite{wc_elee_prd}.}
    \label{fig:pattern_recognition}
\end{figure}

The overall set of steps in Wire-Cell pattern recognition is illustrated in Fig. \ref{fig:pattern_recognition}. 

We start with a 3D image of the neutrino activity, with color indicating charge for each space point. We first identify vertices and track segments using an iterative approach, fitting trajectories between pairs of extreme points and identifying kinks (large angle changes) in these trajectories. The resulting trajectories follow each particle path and are made up of a single line of space points with a uniform spacing of $\sim$6 mm. This uniform spacing is especially useful in order to reconstruct the deposited charge per unit length, which is important when determining the type of particle and the direction of the particle. Next, we improve the quality of these trajectories by fitting the paths with the original 2D projected views; this is the only time after the initial 3D imaging that Wire-Cell reconstruction uses the 2D images again. 

\begin{figure}[H]
    \centering
    \includegraphics[width=0.7\textwidth]{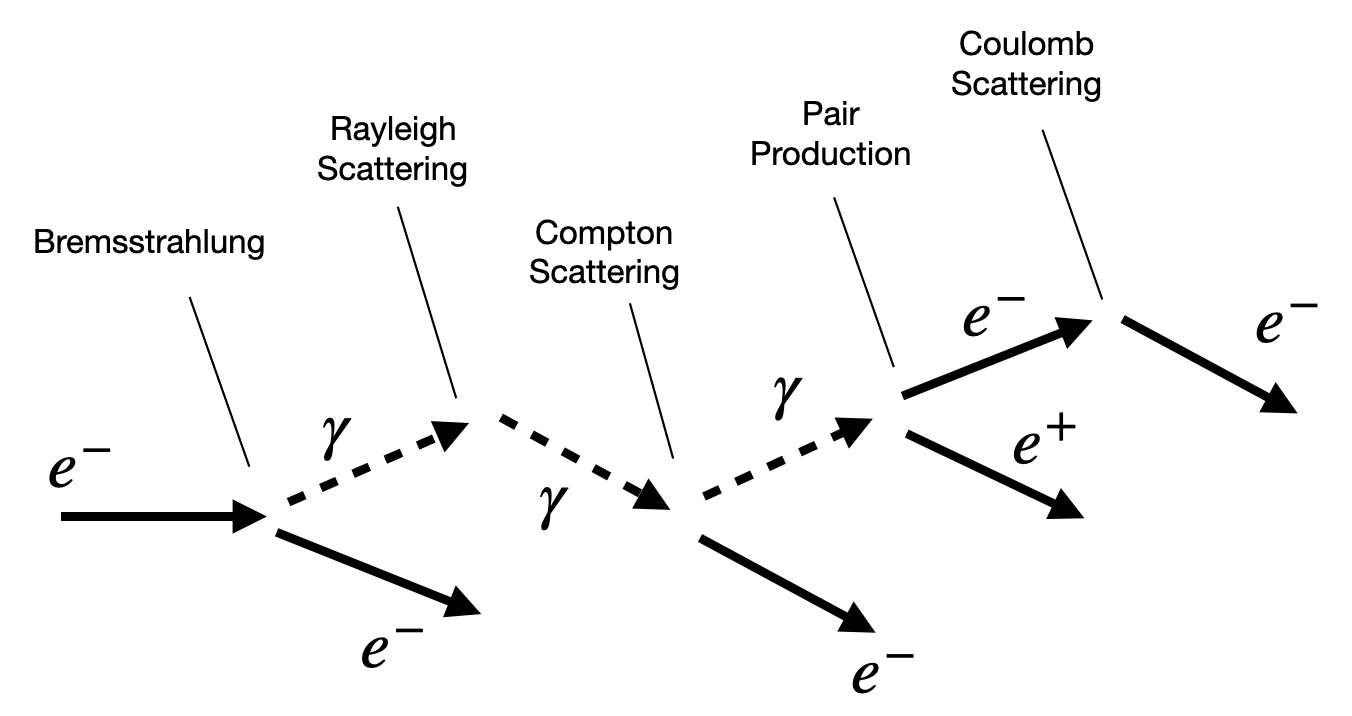}
    \caption[Electromagnetic shower diagram]{A diagram of the important processes contributing to an electromagnetic shower. Bremsstrahlung produces photons from electron scattering, and electrons can be kicked at high energies from Compton scattering of photons. An electron-positron pair can be produced by a photon via pair production. Photons and electrons/positrons can each change direction via Rayleigh scattering and Coulomb scattering respectively. Note that an electromagnetic shower can be initiated by an electron as displayed here, or by a positron, or by a photon which initiates pair production.}
    \label{fig:shower_diagram}
\end{figure}

In our reconstruction of neutrino events, an important goal is to separate ``tracks'' from ``showers''. A track refers to a single particle trajectory, appearing as a line segment with minimal scattering and no branching. A shower refers to a cascade of electromagnetically produced activity with more scattering, gaps, and branching due to the processes illustrated in Fig. \ref{fig:shower_diagram}. The goal here is essentially to distinguish electron/positron activity from other particle activity; therefore, even if an electron is low energy and makes one clear path, our goal is to classify that as a shower rather than a track. Note that in our $\mathcal{O}$(GeV) scale energies, we rarely experience significant amounts of branching due to hadronic particles, and therefore we do not use the word ``shower'' to refer to hadronic showers as you might in the context of a collider experiment.

We separate track and shower topologies using three types of inputs. First, we consider multiple coulomb scattering, which makes low energy electron trajectories appear more ``wiggly''. Second, we consider the presence of isolated charge clusters near a trajectory, which is characteristic of compton scattering or pair production from a bremsstrahlung photon in an electromagnetic shower. Lastly, we consider the width of charge along a trajectory, since an electromagnetic shower is likely to contain several electrons and positrons which will deviate from the central path of the shower.

For track topologies, we use ionization per unit length ($dQ/dx$) measurements to determine the type of particle, as shown in Fig. \ref{fig:dQdx_residual_range}. Heavier particles deposit more ionization energy per unit length than lighter particles at these energies, according to the Bethe-Bloch equation. This lets us separate tracks as muon-like or proton-like. Charged pions are identified as muon-like, since they have a very similar mass and are therefore indistinguishable by their ionization per unit length in MicroBooNE. These $dQ/dx$ measurements are also crucial to determine the direction of each particle. A particle deposits more ionization per unit length as it slows down, and results in a Bragg peak if it stops.

To determine the neutrino vertex, we first use a traditional technique that receives a variety of inputs. We consider the direction of each track using $dQ/dx$ measurements as described above. We consider the direction of each shower using the fact that the width of the charge tends to grow along the direction of a shower. We consider the position of the vertex, which is likely to be further upstream relative to the beam than most ionization activity. We consider the number of particles intersecting at each point, since primary interaction vertices tend to have higher particle multiplicities. We consider a series of connectivity rules, for example that there should only be zero or one particle entering each point, there should not be a shower entering a point and a track exiting that point, and there should rarely be very large angle changes when one particle enters and one particle leaves a point. We also exclude vertices where the local topology indicates that it is inside of an electromagnetic shower, or that it is produced by a delta ray alongside a track.

\begin{figure}[H]
    \centering
    \includegraphics[width=0.4\textwidth]{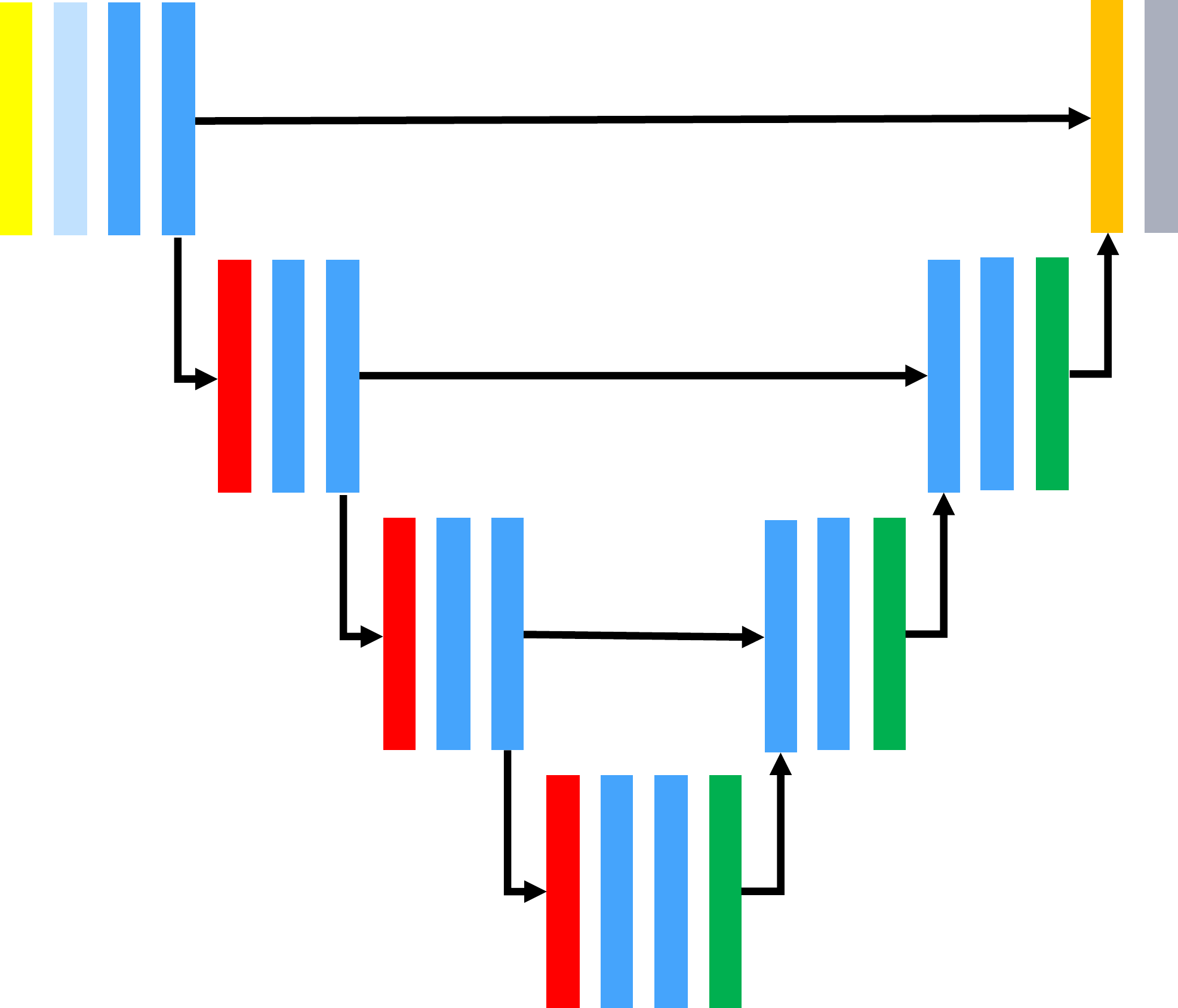}
    \hspace{0.05\textwidth}
    \includegraphics[width=0.17\textwidth]{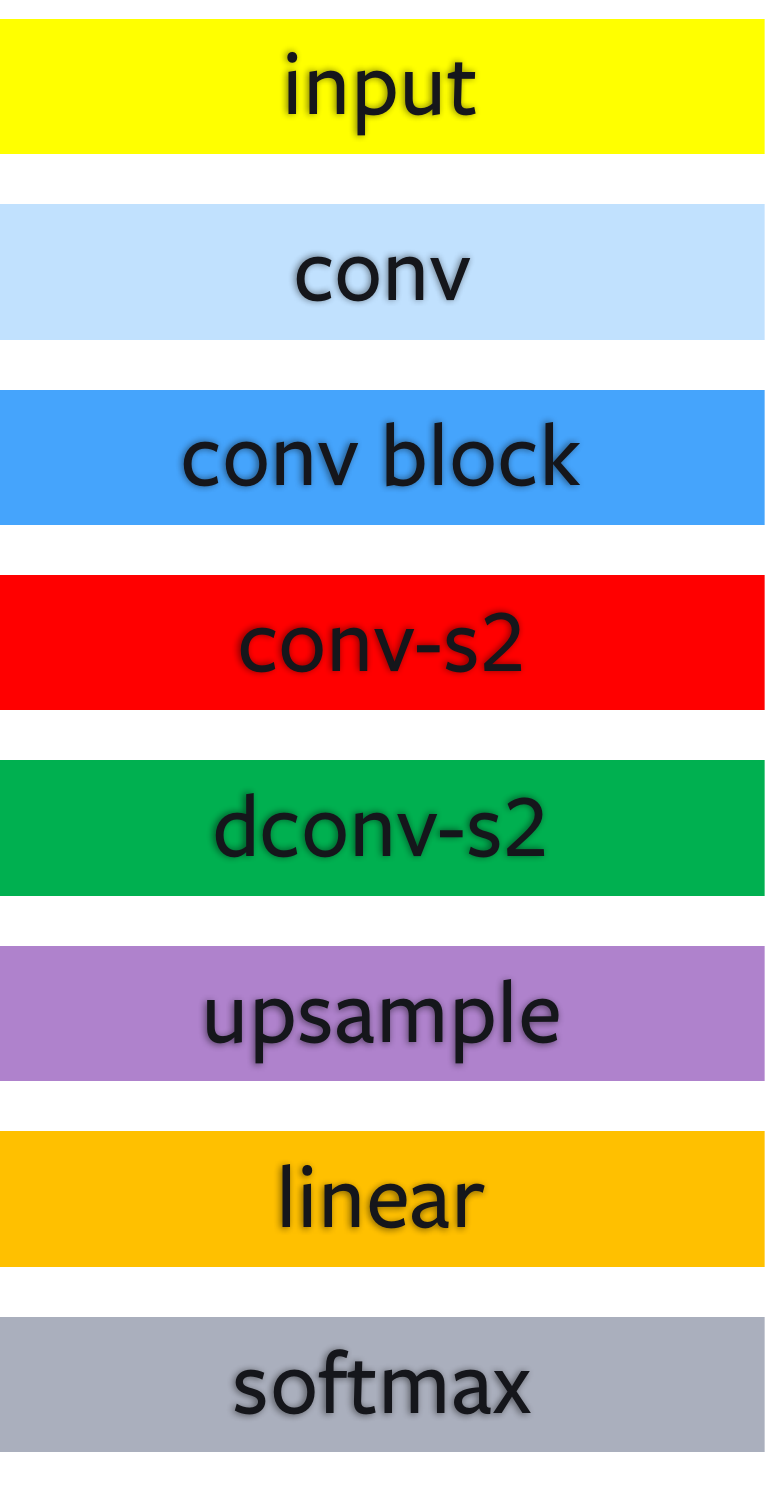}
    \caption[Sparse convolutional neural network]{Diagram of a sparse convolutional neural network of the type used for Wire-Cell DL vertexing. Figure from Ref. \cite{sparseconvnet}.}
    \label{fig:sparseconvnet}
\end{figure}

\begin{figure}[H]
    \centering
    \begin{subfigure}[b]{0.4\textwidth}
        \includegraphics[trim=0 0 450 0, clip, width=\textwidth]{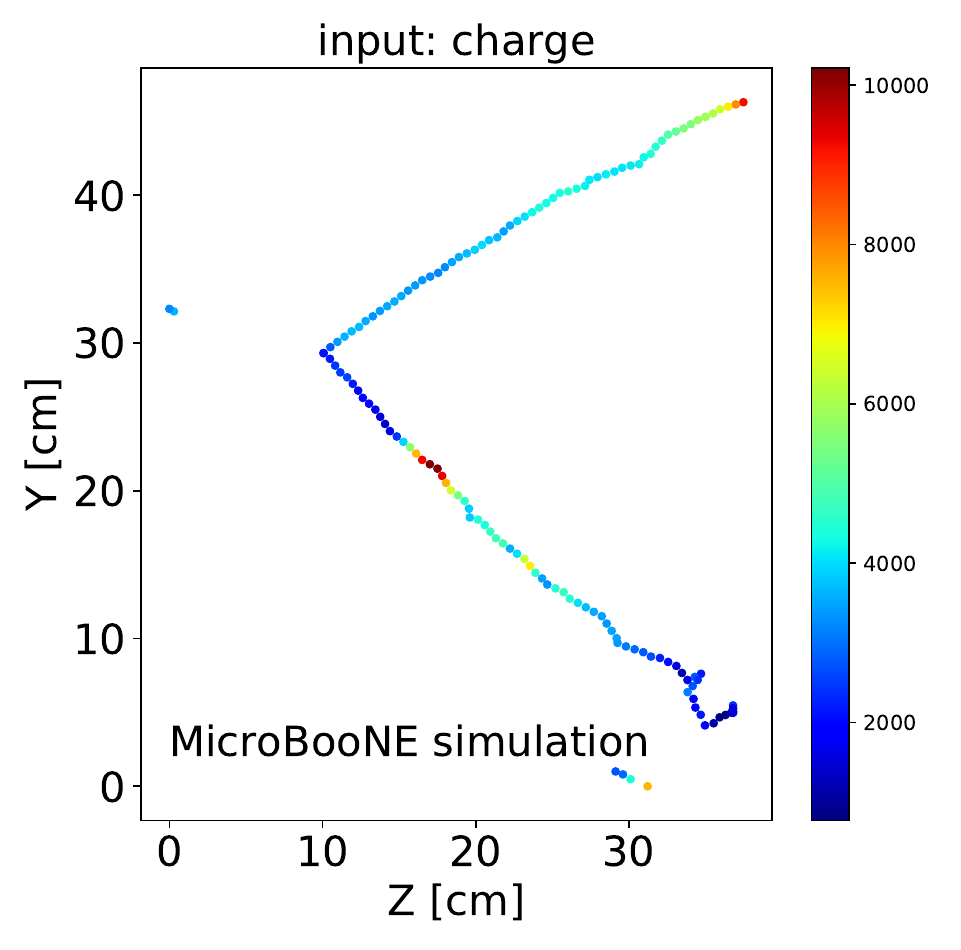}
        \caption{}
        \label{fig:dl_vertexing_charge}
    \end{subfigure}
    \begin{subfigure}[b]{0.4\textwidth}
        \includegraphics[trim=450 0 0 0, clip, width=\textwidth]{figs/wc_nue/dl_vertex_labeling.pdf}
        \caption{}
        \label{fig:dl_vertexing_label}
    \end{subfigure}
    \caption[Deep learning vertexing labels]{Panel (a) shows the charge information, and panel (b) shows the truth labeling. Each of these shows the trajectories before voxelization has occurred. Figure from Ref. \cite{wire_cell_pattern_recognition}.}
    \label{fig:dl_vertex_labeling}
\end{figure}

Next, we take neutrino vertex candidates from traditional techniques, and improve upon them using a deep learning algorithm, which we call ``DL vertexing''. We use a sparse convolutional neural network which acts on a 3D voxelization of the charge. Specifically, we use the SparseConvNet library from Facebook Research \cite{sparseconvnet}, which implements the network diagrammed in Fig. \ref{fig:sparseconvnet}. Using a sparse network increases training and inference speed significantly, since our images have a very small fraction of the 3D volume occupied by ionization activity. We create truth labels which exponentially falloff from the true vertex location as illustrated in Fig. \ref{fig:dl_vertex_labeling}. The network was trained on 48,000 $\nu_e$CC events, a choice to specifically maximize performance for our $\nu_e$CC reconstruction. This choice does have downstream consequences for other topologies, and makes Wire-Cell likely to identify the neutrino vertex as coinciding with a shower vertex; nevertheless, we still see good performance when testing the network on $\nu_\mu$CC and NC events.

For our final neutrino vertex identification, we use a hybrid approach of both the traditional and DL vertices. From a series of candidate vertices from the traditional reconstruction, we choose the one that is closest to the DL vertex as the final vertex. However, if the DL vertex is more than 2 cm from any traditional candidate, the DL vertexing is not used and the traditional vertexing takes its place. This hybrid technique was seen to maximize the accuracy of the identification, with higher performance than either the traditional or DL reconstructions alone.

Now that we have identified the neutrino vertex, shower segments, and track segments with particle identification,  We perform shower clustering, which merges shower segments into reconstructed showers. We perform dedicated $\pi^0$ reconstruction, which identifies vertices at the intersections of pairs of photon showers. We reconstruct the energy of each particle using the length for tracks over 4 cm long, using summed $dE/dx$ for shorter tracks and tracks with delta rays, and the overall amount of charge for showers. To reconstruct the neutrino energy, we sum the energy of each identified particle. For each proton, we do not include the rest mass since it previously existed inside the nucleus, and instead add a binding energy of 8.6 MeV. Note that this reconstructed neutrino energy will not include the contributions from most neutrons, from particles that exit the TPC, or from the exiting neutrino in an NC interaction, so in some ways this should be thought of as a type of visible energy variable. 

The end result of this reconstruction is a ``particle flow tree'' that labels particle types, positions, energies, momenta, and mother-daughter relationships for every particle in an event. When multiple of the same type of particle exist in this tree, we often consider the ``primary'' particle, meaning the particle of that type with the highest reconstructed energy. This tree, alongside the 3D trajectory-fitted image, is what we use to develop various selection variables for all of our Wire-Cell physics analyses.

There are also some more recent additions to Wire-Cell reconstruction which have not been used for the analyses in this thesis. The energy of exiting muons can be determined by how straight or wiggly the track is due to multiple coulomb scattering (MCS); MicroBooNE has studied this in the past \cite{MicroBooNE_MCS}, but it has not yet been widely used for physics analyses. Work is currently ongoing to implement and validate MCS muon energy reconstruction using Wire-Cell 3D tracking. We have also recently developed an improved neutrino energy construction using a recurrent neural network (RNN) which takes as input each particle in the Wire-Cell particle flow tree \cite{MicroBooNE_RNN_energy}. In particular, this reduces the bias in reconstructed energy caused by invisible particles and particles partially contained within the TPC, as the network learns how to statistically account for energy that has not alrady been reconstructed.

\section{Wire-Cell Event Selections}

Now that we have 3D images and reconstructed particles, we use these in order to select certain types of events for physics analyses.

\subsection{Generic Neutrino Selection}\label{sec:generic_neutrino_selection}

Our first step in developing physics selections is generic neutrino selection, where our primary goal is to reject comsic induced events and accept all neutrino interactions. These algorithms are described in detail in Ref. \cite{wc_generic_selection}. Note that this generic neutrino selection was developed before some of the pattern recognition tools described in Sec. \ref{sec:pattern_rec}, so at this stage we do not have access to all of the reconstructed particles, and we try to reject cosmic rays before performing all of that computation.

Our first step of rejecting cosmic activity is the light filter. Events which do not produce scintillation light within the time of our neutrino beam spill can be rejected as consisting only of cosmic rays. After this step, we will still have some cosmic rays, either due to coincidental light timing, or due to events in which there was a neutrino interaction, but it was not the cluster we identified. This can happen if the neutrino interacted outside of the TPC, producing light but no charge, or if there was a failure in charge-light matching that caused a cosmic cluster to be identified as the neutrino interaction candidate.

\begin{figure}[H]
    \centering
    \begin{subfigure}[b]{0.4\textwidth}
        \includegraphics[width=\textwidth]{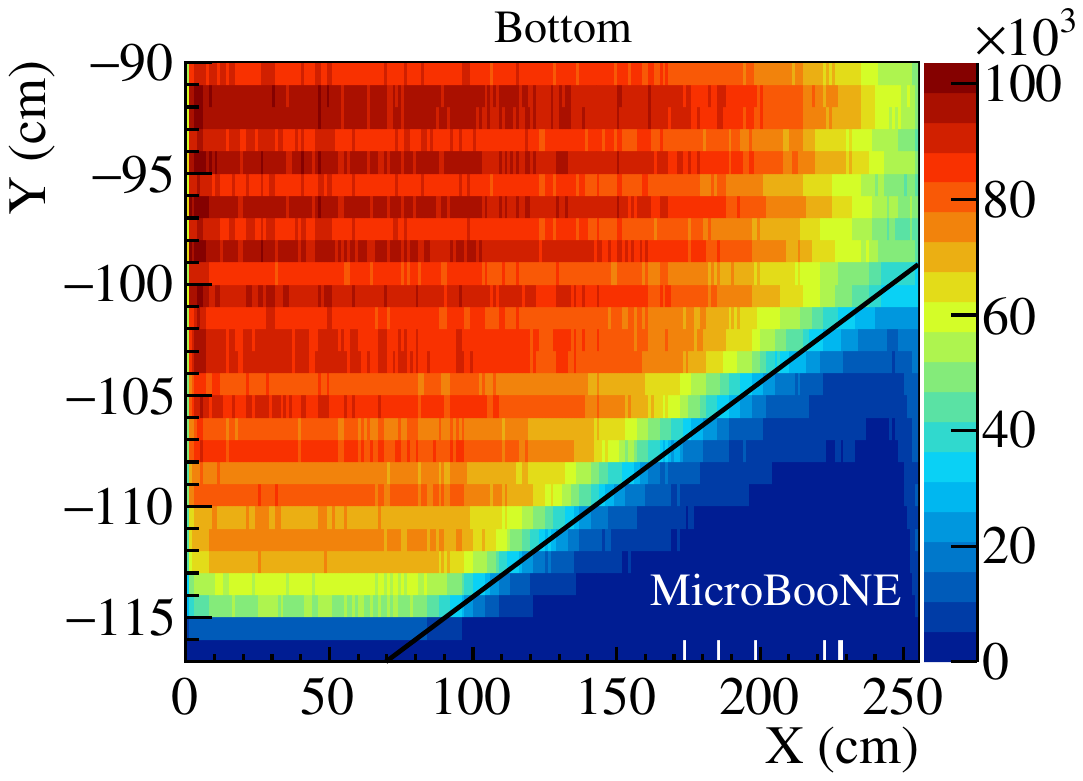}
        \caption{}
    \end{subfigure}
    \begin{subfigure}[b]{0.4\textwidth}
        \includegraphics[width=\textwidth]{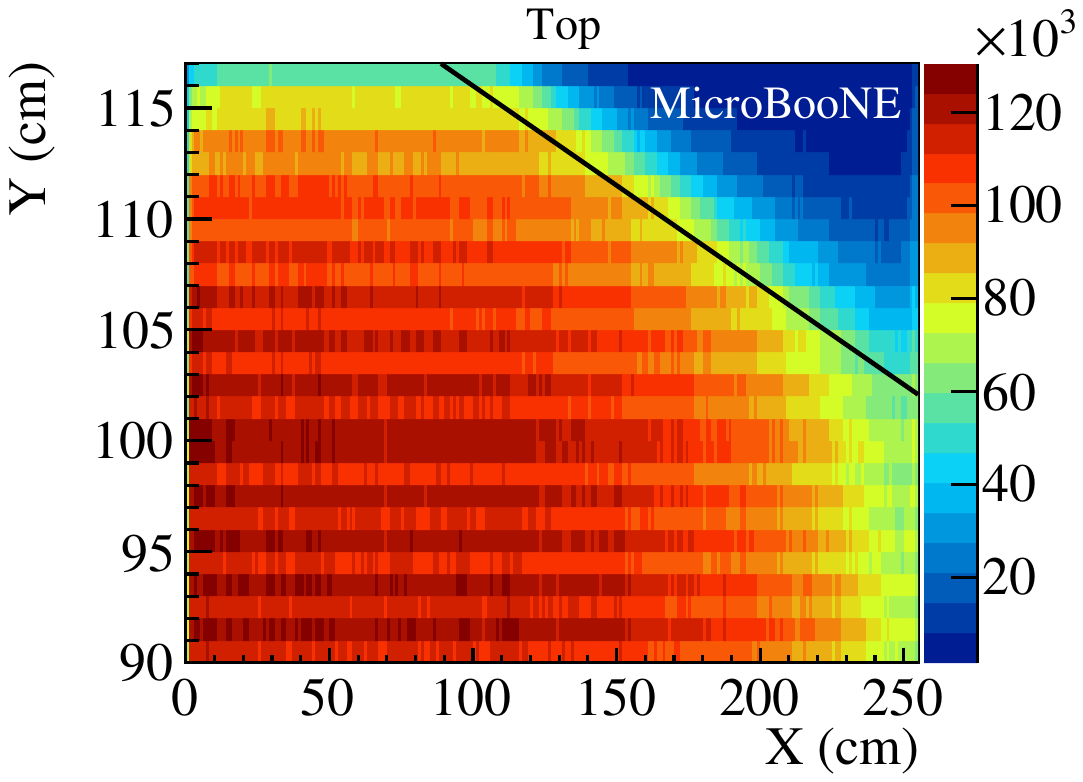}
        \caption{}
    \end{subfigure}
    \begin{subfigure}[b]{0.4\textwidth}
        \includegraphics[width=\textwidth]{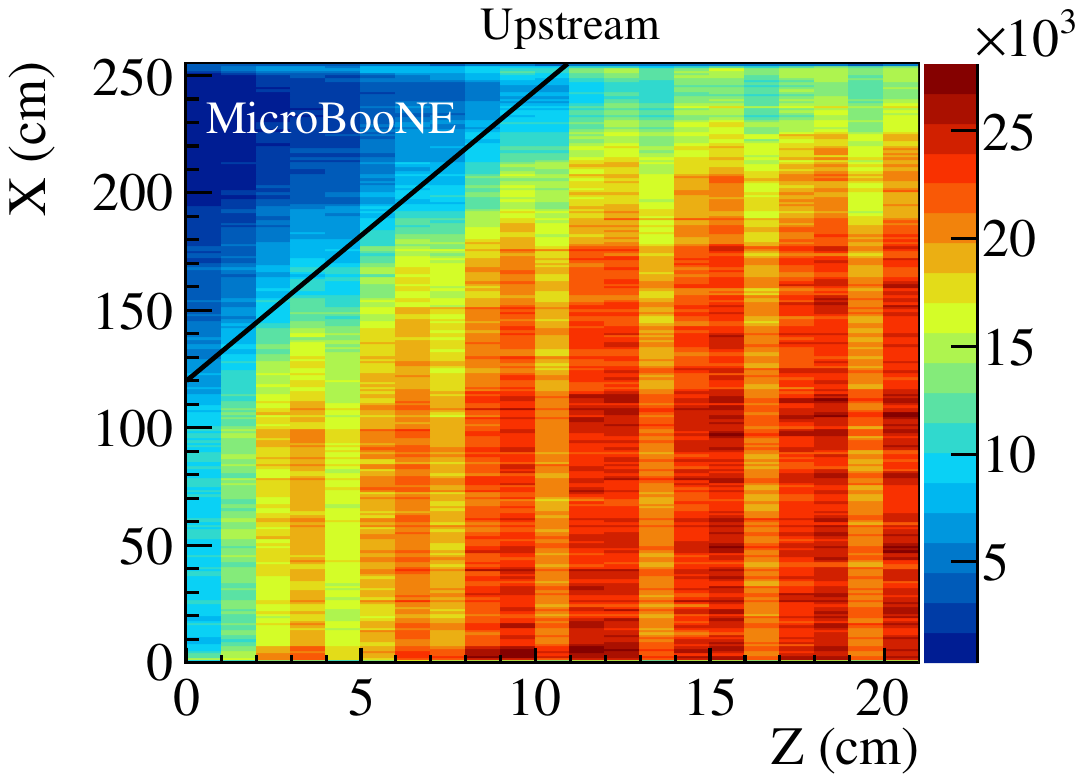}
        \caption{}
    \end{subfigure}
    \begin{subfigure}[b]{0.4\textwidth}
        \includegraphics[width=\textwidth]{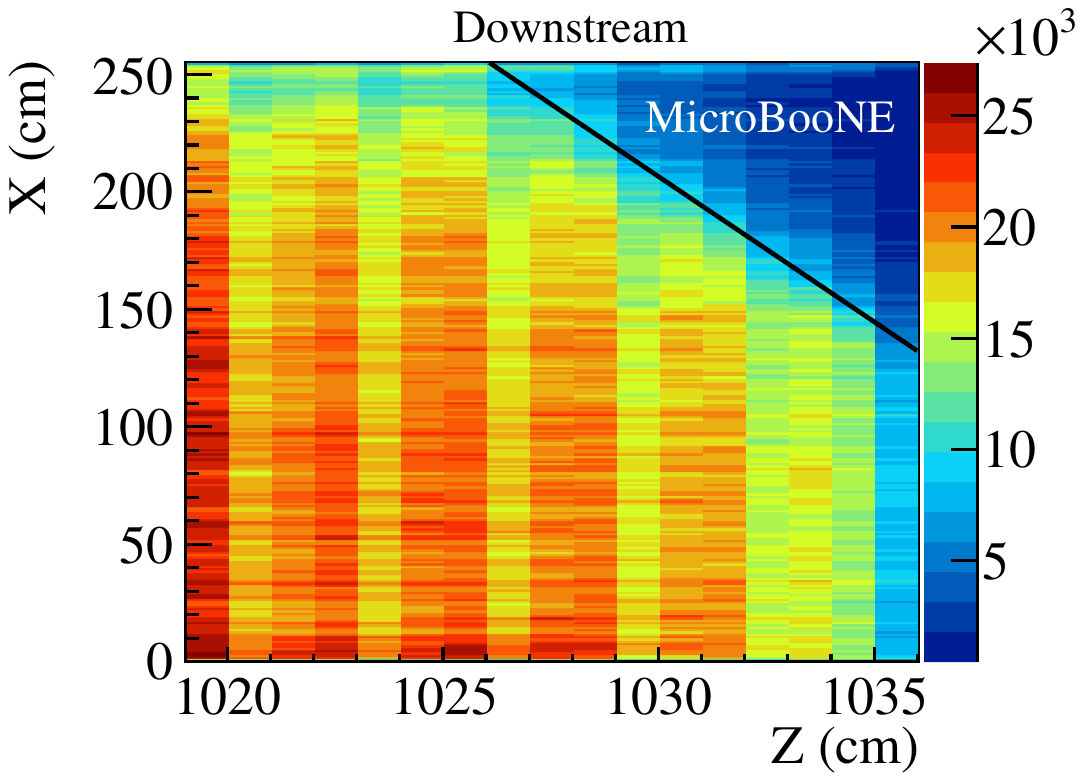}
        \caption{}
    \end{subfigure}
    \caption[Effective space charge boundary]{MicroBooNE's effective space charge boundary, determined by reconstructions of cosmic muons from about 1700 MicroBooNE events. The black line indicates the space charge boundaries used for analyses. Panels (a)-(d) show views of the space charge boundary zoomed in near the bottom, top, upstream, and downstream faces of the TPC respectively. Figure from Ref. \cite{wc_generic_selection}.}
    \label{fig:effective_space_charge_boundary}
\end{figure}

Many cosmic events will contain a cosmic muon which enters from outside the TPC. Therefore, for a generic neutrino selection, it is crucial to identify when a muon track crosses a detector boundary. In order to identify these events, we need a precise understanding of the TPC boundary, and this requires extra consideration for the effects of space charge. The cloud of positive space charge which builds up near the cathode attracts drifting electrons toward the center of the detector, curving their paths. This means that charge that starts at the edge of the TPC near the cathode will arrive at a more central location by the time it reaches the anode. This effective boundary was studied by reconstructing cosmic muons, as shown in Fig. \ref{fig:effective_space_charge_boundary}. We define a fiducial volume as 3 cm inside this effective boundary.

\begin{figure}[H]
    \centering
    \begin{subfigure}[b]{0.46\textwidth}
        \includegraphics[trim=570 20 0 30, clip, width=\textwidth]{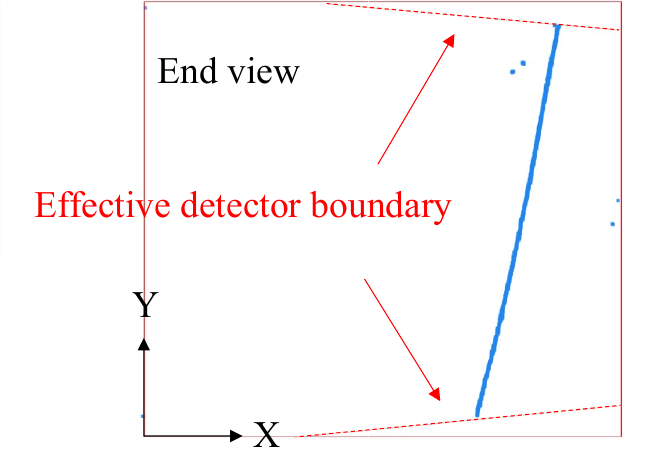}
        \caption{}
        \label{fig:TGM}
    \end{subfigure}
    \begin{subfigure}[b]{0.5\textwidth}
        \includegraphics[trim=0 10 500 37, clip, width=\textwidth]{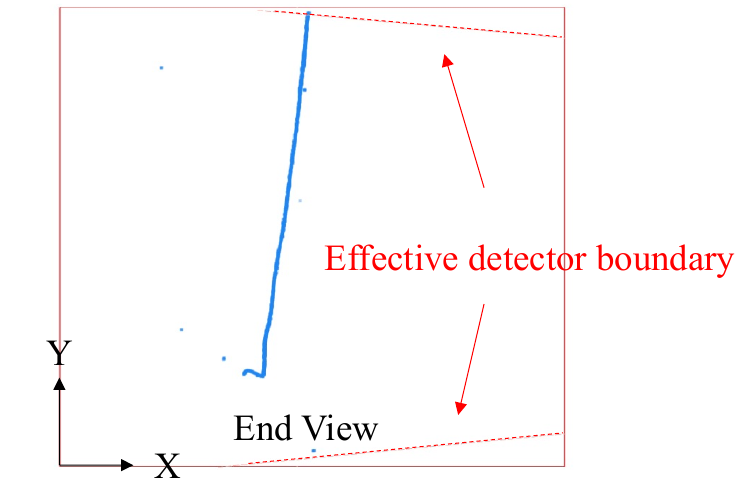}
        \caption{}
        \label{fig:STM}
    \end{subfigure}
    \caption[Cosmic muon examples with the effective space charge boundary]{Panel (a) shows an example through-going muon cosmic event. Panel (b) shows an example stopping muon cosmic event, with a Michel electron/positron present. Figure from Ref. \cite{wc_generic_selection}.}
    \label{fig:STM_TGM}
\end{figure}

We use this effective space charge boundary to determine if an event is a through-going muon, a cosmic ray muon which both enters and exits the TPC. These events are analyzed and rejected using the reconstructed positions of most extreme points along each axis, as well as the principle axis determined by principle component analysis (PCA). An example is shown in Fig. \ref{fig:TGM}. Special considerations are made for tracks that might enter or exit the TPC in a dead region and therefore appear away from the boundary, and for interactions in which two neutrino-induced particle tracks both exit the detector, but with a large angle between them.

Another large category of cosmic ray backgrounds is muons which enter and come to rest inside the TPC. In this case, we use similar techniques to identify one intersection with the effective space charge boundary. We then carefully consider the directionality; cosmic rays will often have a muon entering the detector, while neutrino interactions will often have a muon exiting the detector. This directionality is determined using the deposited charge per unit length $dQ/dx$, looking for an increase near the Bragg peak at the end of the track, as shown in Fig. \ref{fig:dQdx_residual_range}. Another consideration for stopping muons is the presence of Michel electrons or positrons from the decay of the muon, as shown in Fig. \ref{fig:STM}; this will be present at the end of all stopped $\mu^+$ tracks, and in about 25\% of stopped $\mu^-$ tracks, with the rest absorbed on argon nuclei. These are accounted for by identifying large angle changes or ``kinks'' in the track before $dQ/dx$ analysis. Special considerations are also made for delta rays which could impact $dQ/dx$ measurements, and for muons which decay in flight and therefore do not form a complete Bragg peak.

The last component of our generic neutrino selection is rejection of light-mismatched events. We reject events with very small amounts of charge, which can give unreliable light pattern predictions for our PMTs. We reject events with low consistency between the predicted and measured light patterns. If there is medium consistency, we check if the charge cluster has a better match to a different light flash; if so, we consider the possibility that this could be a mis-timed cosmic ray. In that case, we shift the charge cluster along time according to this new flash, and consider through-going and stopping muon rejections again. If there is medium predicted-measured light consistency but there is no better matching flash, we again consider the possibility of a mis-timed cosmic ray, and apply a continuous variation of the position along the drift axis and consider through-going and stopping muon rejections again at each position.

\begin{figure}[H]
    \centering
    \includegraphics[width=0.7\textwidth]{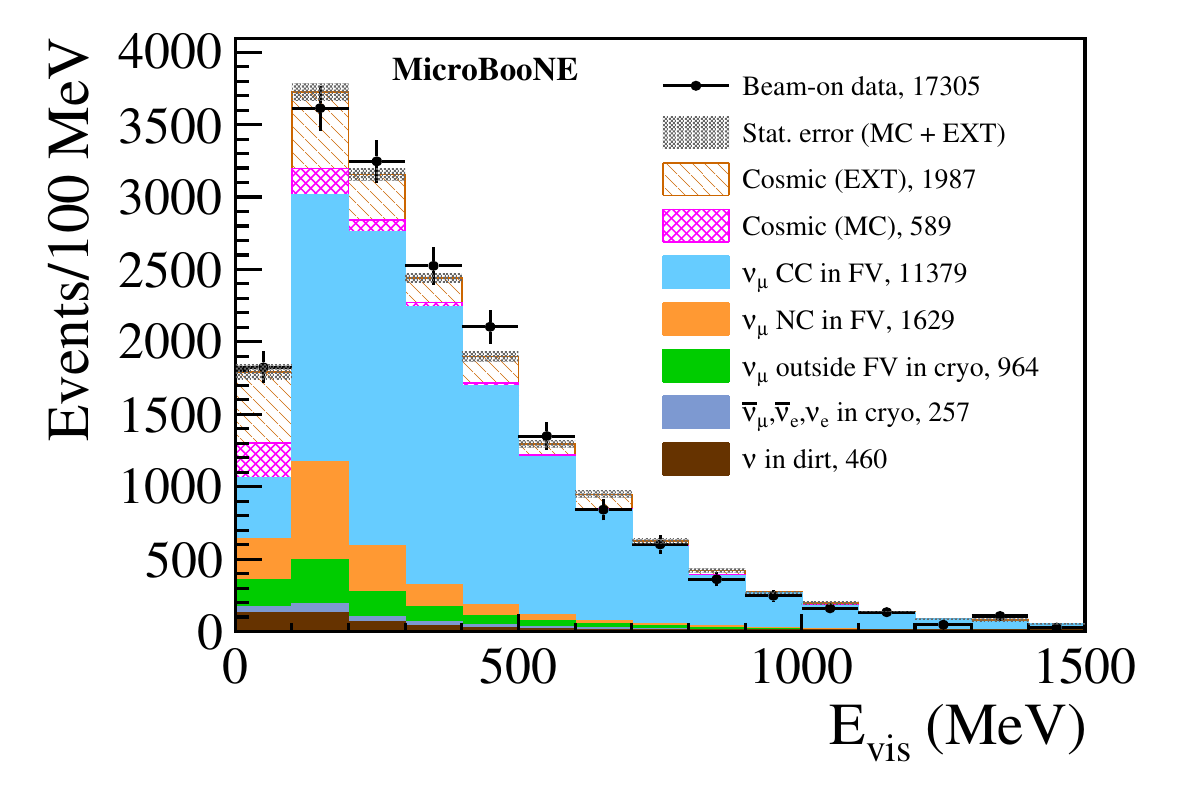}
    \caption[Generic neutrino selection]{MicroBooNE Wire-Cell generic selected neutrino events as a function of reconstructed visible energy. Figure from Ref. \cite{wc_generic_selection}.}
    \label{fig:generic_selection}
\end{figure}

The end result of this is a high performance generic neutrino selection, shown as a function of reconstructed visible energy in Fig. \ref{fig:generic_selection}. Considering all BNB beam spills in MicroBooNE, we start out with a neutrino to cosmic ray ratio of 1:20,000. After applying our light filter, charge-light matching, through-going muon rejection, stopping muon rejection, and light-mismatch rejection, we are left with a neutrino to cosmic ray ratio of 5.2:1, a factor of over 100,000 improvement. Considering events originating within our fiducial volume, which consists of 94.2\% of the active volume of the TPC, we maintain a $\nu_\mu$CC selection efficiency of 80\%, and an NC selection efficiency of 36\%. 

\subsection{Cut-based Selections}

\begin{figure}[H]
    \centering
    \includegraphics[trim=100 50 100 0, clip, width=0.7\textwidth]{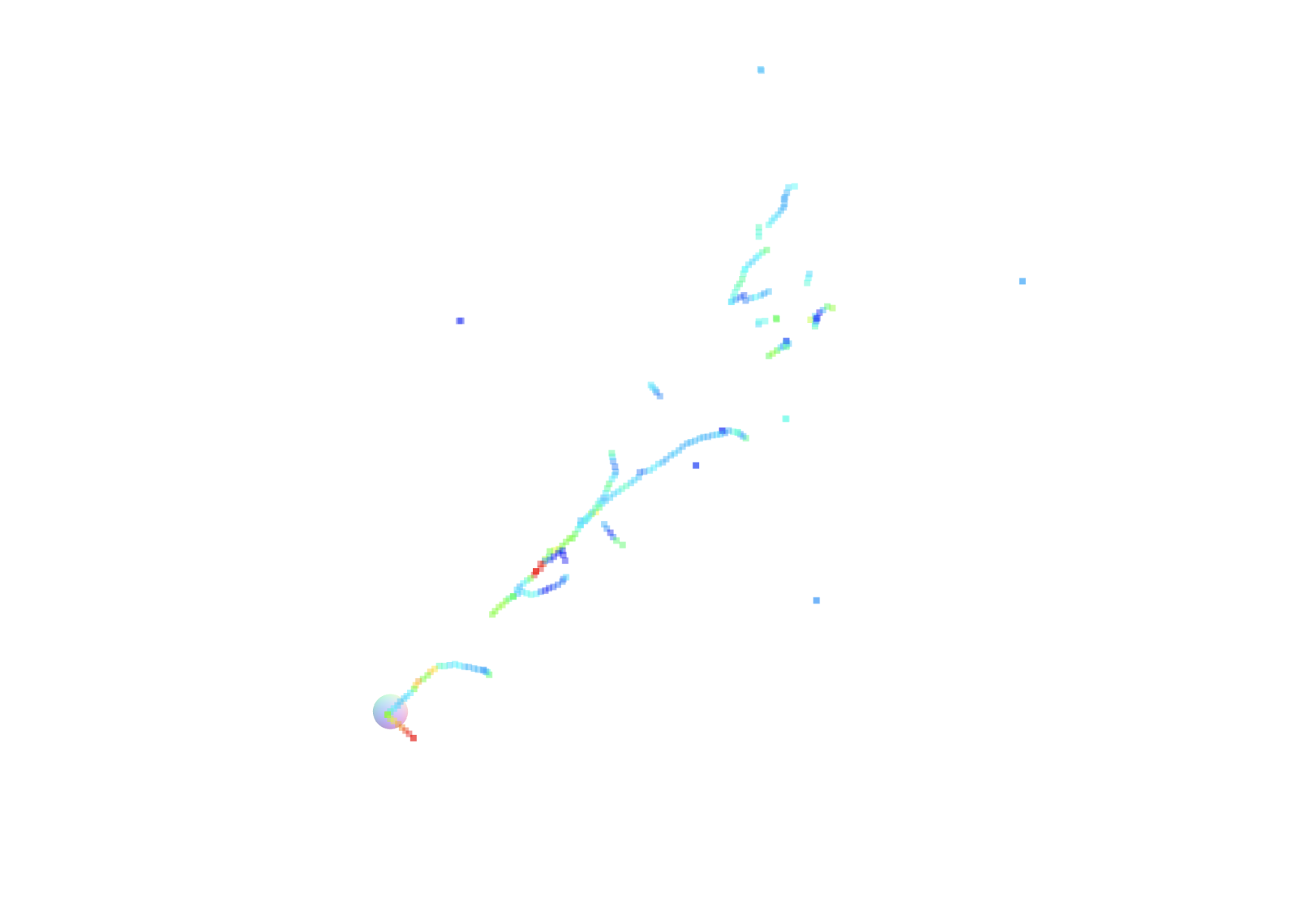}
    \caption[Example Wire-Cell $\nu_e$CC candidate]{An example $\nu_e$CC $1p$ candidate event. The candidate vertex is indicated with a sphere, with a candidate proton exiting to the bottom right and a candidate electron shower exiting to the top right. From NuMI Run 6365 subrun 0 event 21, with the 3D Wire-Cell display viewable at \url{https://www.phy.bnl.gov/twister/bee/set/uboone/reco/2021-01/numi-nue-fc-400-800-mev/event/0/}.}
    \label{fig:wc_nue_candidate}
\end{figure}

\begin{figure}[H]
    \centering
    \includegraphics[trim=0 100 0 100, clip, width=0.9\textwidth]{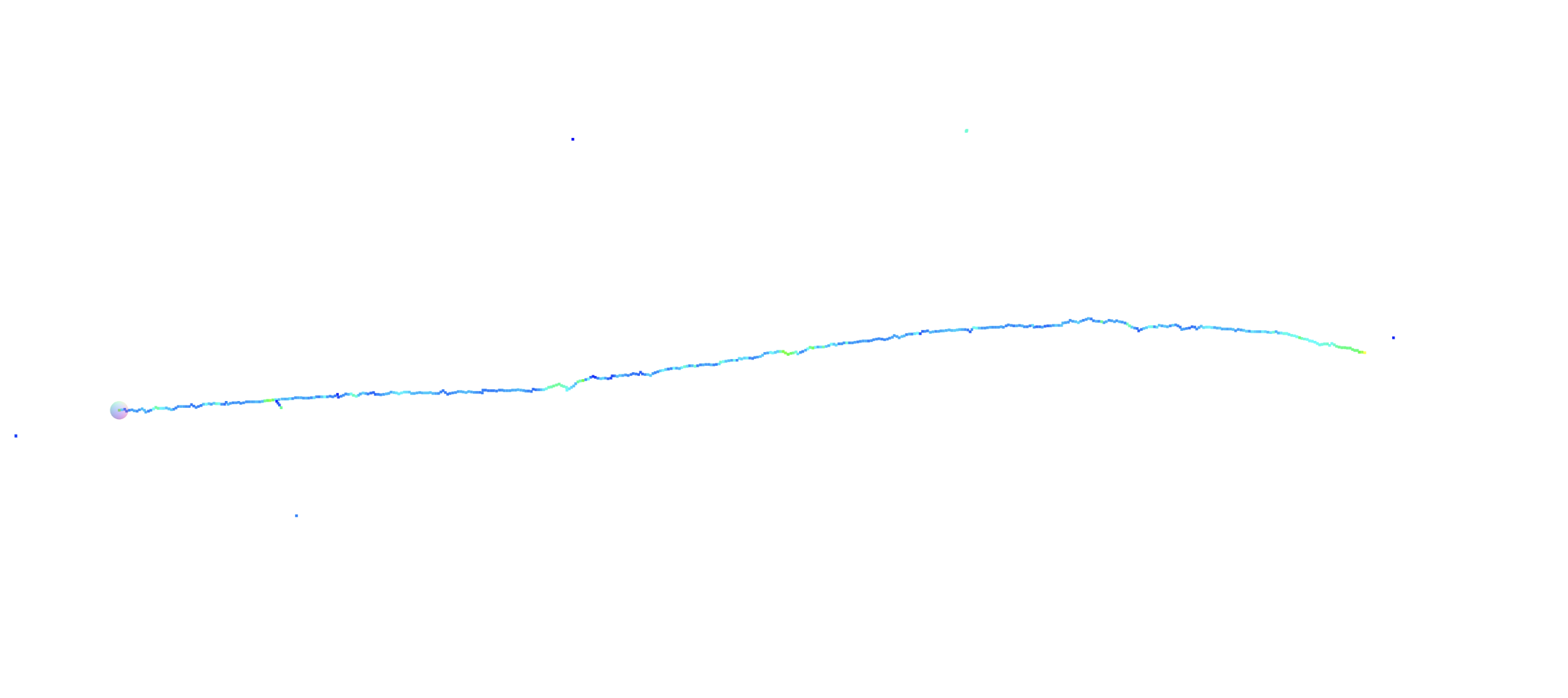}
    \caption[Example Wire-Cell $\nu_e$CC candidate]{An example $\nu_\mu$CC $0p$ candidate event. The candidate vertex is indicated with a sphere, and the candidate muon exits to the right, and experiences some distortion from multiple coulomb scattering before forming a Bragg peak and coming to rest. From BNB Run 5127 subrun 38 event 1944, with the 3D Wire-Cell display viewable at \url{https://www.phy.bnl.gov/twister/bee/set/6ca94583-0910-4a6f-90a8-ace1236edf07/event/7/}.}
    \label{fig:wc_numu_candidate}
\end{figure}

\begin{figure}[H]
    \centering
    \includegraphics[trim=0 100 0 100, clip, width=0.9\textwidth]{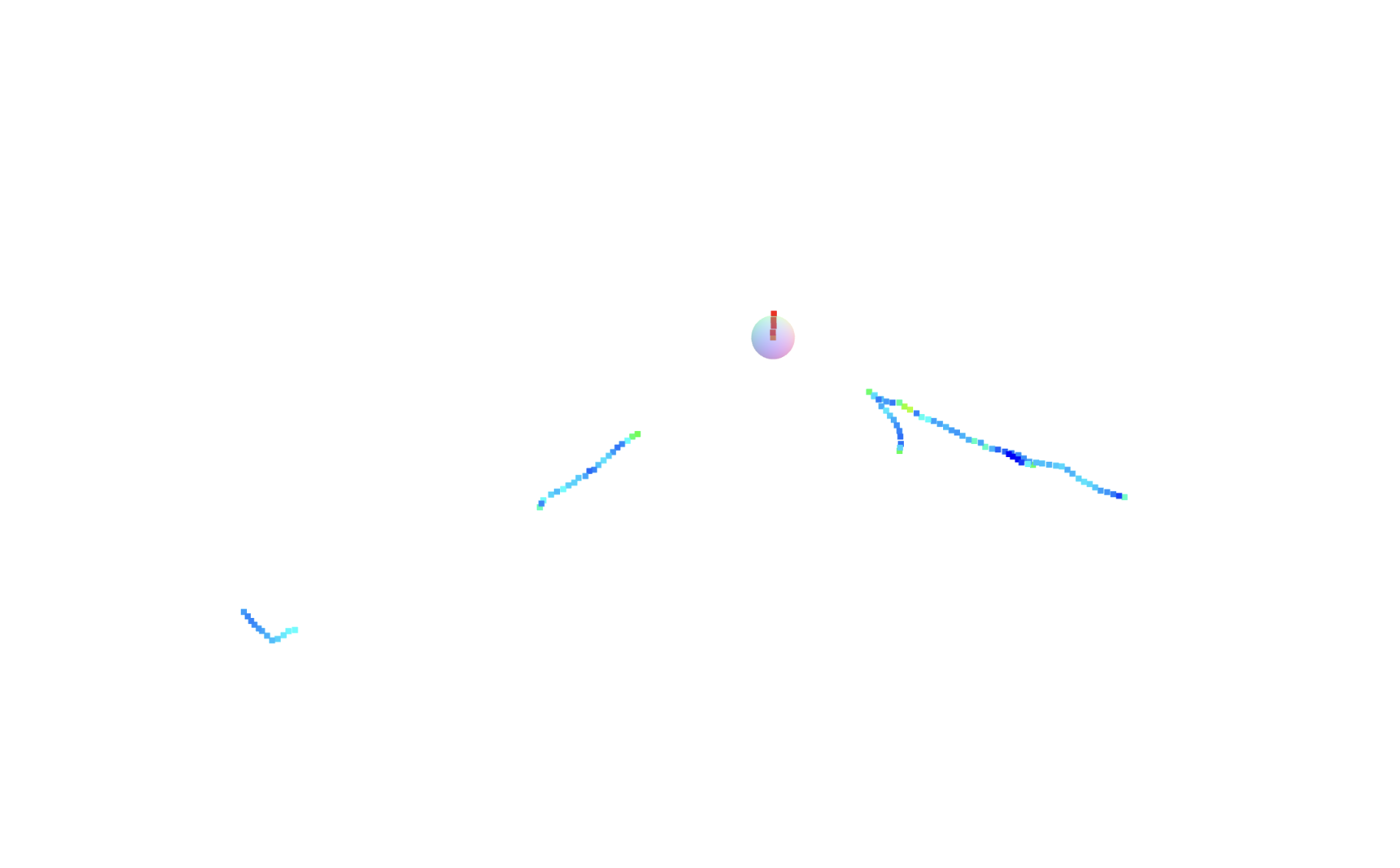}
    \caption[Example Wire-Cell $\nu_e$CC candidate]{An example NC $\pi^0$ $1p$ candidate event. The candidate vertex is indicated with a sphere, with a candidate proton exiting upward, and two photons from a candidate $\pi^0$ exiting to the bottom left and bottom right. From BNB Run 5051 Subrun 31 Event 1573, with the 3D Wire-Cell display viewable at \url{https://www.phy.bnl.gov/twister/bee/set/2bd6e1fe-b9ea-47e8-9232-358e1709d5ed/event/15/}.}
    \label{fig:wc_pi0_candidate}
\end{figure}

Now that we have our 3D image, a reconstructed particle tree, and we have rejected the vast majority of cosmic rays, we are finally ready to start reconstructing specific neutrino topologies. For this analysis, we primarily target $\nu_e$CC interactions: any event with an electron shower at the neutrino interaction vertex. There are significant flux and cross section uncertainties affecting these events, so we also measure $\nu_\mu$CC interactions, which share strong correlations with $\nu_e$CC due to the common meson parentage in the beam flux, and the common cross-section modeling enforced by lepton universality. Note that here, we use $\nu_e$CC and $\nu_\mu$CC to refer to both neutrinos and anti-neutrinos, since the different charged leptons will look almost identical in our detector. Lastly, we measure events with photon showers, since this is an important background for our search for $\nu_e$CC electron showers; specifically, we measure NC$\pi^0$ and $\nu_\mu$CC$\pi^0$ events, which each have two photon showers in the final state. Example Wire-Cell 3D images for each of these topologies we are targeting are shown in Figs. \ref{fig:wc_nue_candidate} - \ref{fig:wc_pi0_candidate}.

In this analysis, we decided to make our selections as inclusive as possible. We therefore need to consider a very broad set of possible topologies and reconstruct many types of particles. For example, we attempt to reconstruct $1\mu3p1\pi^+1\pi^0$, a possible $\nu_\mu$CC topology with a muon track, three proton tracks, a charged pion track, and two disconnected photon showers from a $\pi^0$; in contrast, a different MicroBooNE $\nu_e$CC search only considered the much simpler $1\mu1p$ and $1e1p$ topologies. The choice to target an inclusive selection has notable benefits relative to the alternatives. An inclusive selection maximizes our event statistics, important for studying very rare $\nu_e$CC interactions from the 0.5\% contribution of $\nu_e$ in our flux from the BNB. Also, an inclusive selection helps to minimize our sensitivity to our modeling of exclusive cross section predictions. Theoretical and nuclear modeling predictions of the total $\nu_e$CC cross section are much easier than exclusive cross sections. For example, it is hard to model exactly how often an interaction will produce a proton with enough kinetic energy to be visible ($\sim$35 MeV). In our inclusive selection, we select events both with and without these visible protons, so mismodeling or a large uncertainty in this migration will have minimal effects on our final selections (not zero effect, since our inclusive efficiency will not be perfectly uniform for all final states). Finally, an inclusive selection is very useful for future studies in the case that an excess is observed, allowing us to study if there could be a certain hadronic topology associated with the excess; this is the type of study that is uniquely possible in MicroBooNE, and not in MiniBooNE. We also choose to be inclusive in terms of containment in our detector, selecting events containing only particles are reconstructed as fully contained (FC) within the detector, and events containing particles which are reconstructed as exiting the detector, making the event partially contained (PC). Fully contained events will have better energy resolution, while partially contained events will help increase our statistics.

In principle, an inclusive selection is fairly simple: for our $\nu_\mu$CC and $\nu_e$CC selections, we simply identify whether or not there is an electron or muon in the final state. However, this is not as easy as it sounds, because the reconstruction can be wrong in many complicated ways. All of the pattern recognition steps described above can go wrong, and many of these cases need to be handled by dedicated algorithms.

To develop these selections, we used an iterative procedure, looking at the performance of a cut-based selection on simulated and real data events, and then adding additional variables and placing cuts on them in order to correct for poor performance observed. This is a somewhat messy process, and involves hundreds of variables, many of which correct for bad reconstruction rather than correspond to a specific physics effect. We maintained a blinded analysis, and therefore only studied data events from a small fraction of our total data sample ($\sim5\cdot 10^{19}$ POT out of $\sim6.4\cdot 10^{20}$ POT total for the final analysis). More details on this development process is described in Ref. \cite{microboone_WC_eLEE_public_note}.

\begin{figure}[H]
    \centering
    \begin{subfigure}[b]{0.52\textwidth}
        \includegraphics[width=\textwidth]{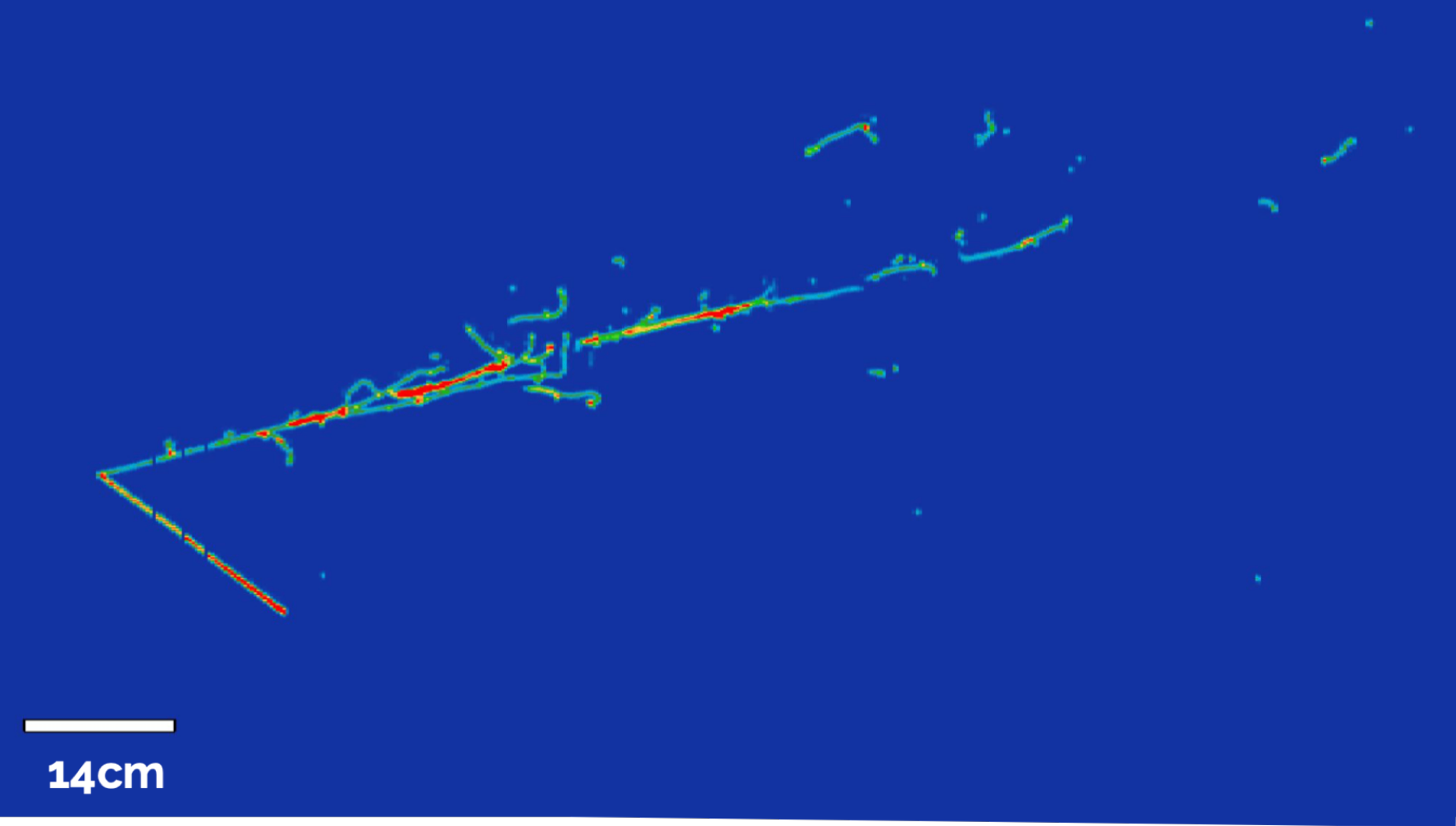}
        \caption{}
    \end{subfigure}
    \begin{subfigure}[b]{0.47\textwidth}
        \includegraphics[width=\textwidth]{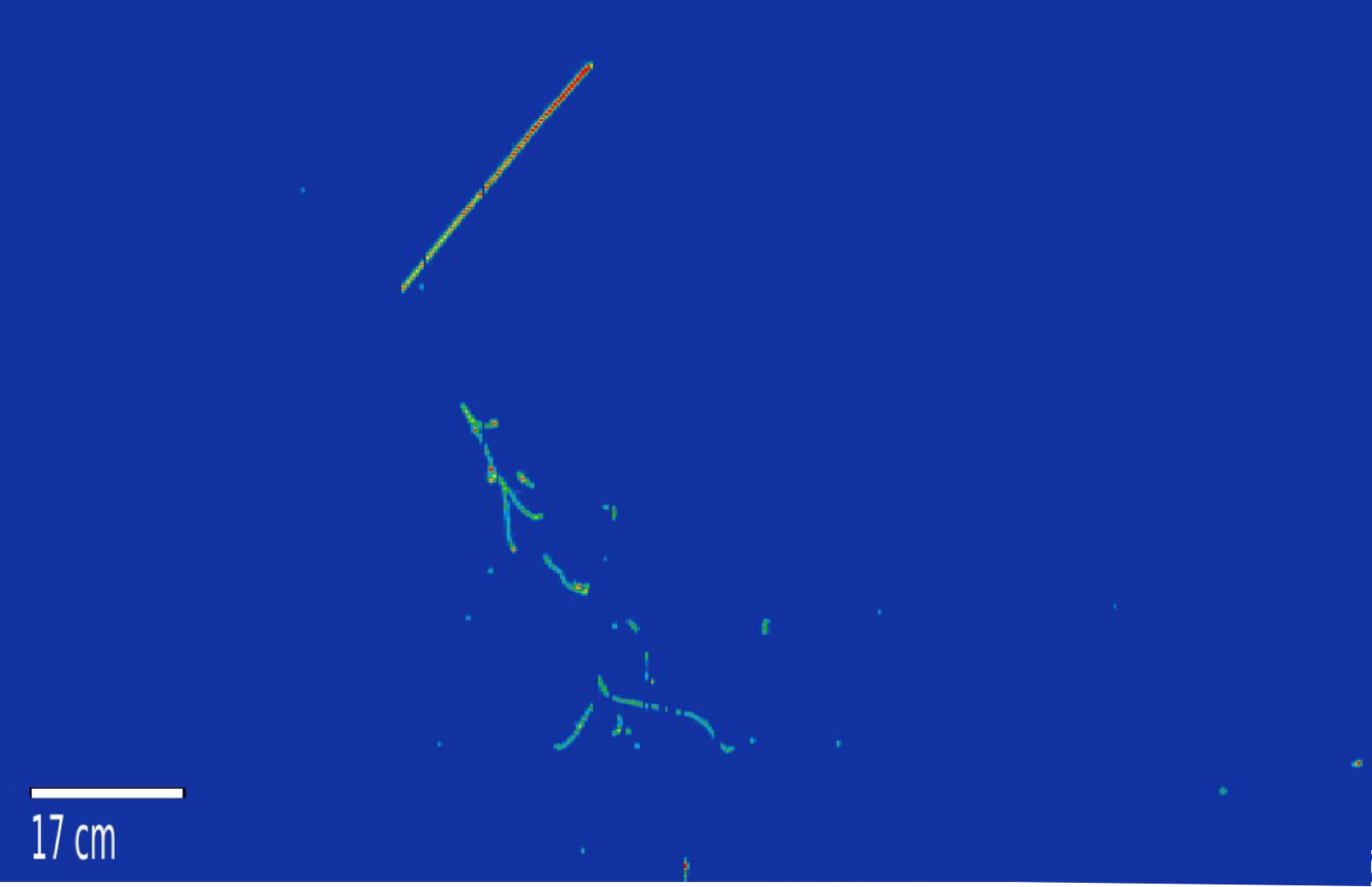}
        \caption{}
    \end{subfigure}
    \caption[MicroBooNE electron-photon separation]{MicroBooNE event displays illustrating the features which allow electron-photon separation. Panel (a) shows a candidate $\nu_e$CC event, from Run 8617 Subrun 46 Event 2329. Panel (b) shows a candidate single photon event from Run 9524 Subrun 127 Event 6375.}
    \label{fig:e_gamma_separation}
\end{figure}

First, I will describe our cut-based $\nu_e$CC selection. At a high level, the selection works by first selecting a single electromagnetic shower, similar to MiniBooNE's selection strategy. Then, we use the high resolution imaging and local calorimetry of a LArTPC to distinguish between an electron shower and a photon shower. First, we look for a gap between the neutrino vertex and the shower, which would indicate the presence of a neutral photon traveling before pair converting and forming the shower. This feature can be seen clearly in many data events, for example as shown in Fig. \ref{fig:e_gamma_separation}. Also, we look for high energy deposition per unit length at the shower stem, which would indicate the presence of an $e^+e^-$ pair produced by a photon rather than a single electron at the start of the shower. 

In reality, this process of $\nu_e$CC identification is not as simple as it sounds, given the many diverse topologies of signal and background events. For $\nu_e$CC events, our signal to background ratio is about 1:200 even after generic neutrino selection, since $\nu_\mu$ are much more prevalent than $\nu_e$ in the BNB, so rejecting this huge quantity of background is a particularly difficult task. We only attempt to reconstruct $\nu_e$CC events with an electron shower energy greater than 60 MeV, in order to avoid Michel electron backgrounds from muon decays. To select these events, we developed many variables, which can be classified into five overall types:
\begin{itemize}
    \item Primary electron identification. This includes the examination of $dQ/dx$ at the beginning or stem of the shower, which can tell us if there is one MIP (a single electron) or 2 MIP (an electron-positron pair from a photon). This also attempts to identify cases where there is a gap between the shower and the neutrino vertex, another signature of photon shower backgrounds.
    \item Multiple shower identification. This attempts to identify and reject events with two showers, typical of many $\pi^0\rightarrow \gamma \gamma$ backgrounds.
    \item Muon misidentification. This attempts to identify cases where a muon track is mis-reconstructed as a shower, potentially due to electronic noise, dead wire regions, or isochronous activity which prevents the clean reconstruction of a single track.
    \item Kinematic rejection. This uses kinematic information in order to separate signal from background. For example, certain energies and angles of showers are more likely to be associated with cosmic-induced photon showers, which primarily travel downward in the detector.
    \item Unreliable pattern recognition. This tries to identify cases where the pattern recognition failed, for example clustering a muon track as part of a reconstructed shower, which can obscure the real $dQ/dx$ and gap information for a photon shower.
\end{itemize}
The full set of $\nu_e$CC selection variables is described in App. \ref{sec:wc_bdt_vars}. 

\begin{figure}[H]
    \centering
    \includegraphics[width=0.5\textwidth]{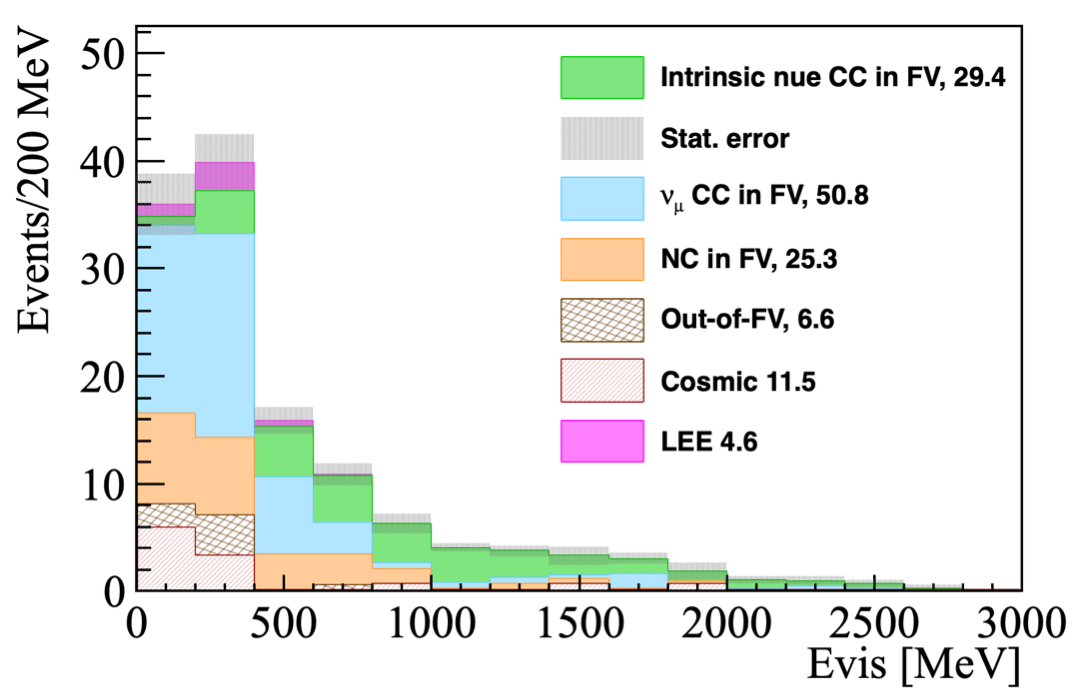}
    \caption[Cut-based $\nu_e$CC selection]{Cut-based $\nu_e$CC selection prediction for $5.3\cdot10^{19}$ POT. In pink, we show the prediction from the LEE model, described in Sec. \ref{sec:lee_model}. Figure from Ref. \cite{microboone_WC_eLEE_public_note}.}
    \label{fig:cut_based_nueCC}
\end{figure}

We show the cut-based $\nu_e$CC selection reconstructed visible energy distribution in Fig. \ref{fig:cut_based_nueCC}. We have achieved a reasonable efficiency of 32\%, and have increased the purity for $\nu_e$CC events a lot relative to the generic neutrino selection, but the purity at low energy remains poor, with many $\nu_\mu$CC backgrounds. This will be addressed and further improvements will be made in Sec. \ref{sec:bdt_selections}.

Next, I will describe our inclusive $\nu_\mu$CC selection. As shown in Fig. \ref{fig:generic_selection}, our generic neutrino selection already functions as a fairly high-performance $\nu_\mu$CC selection, with 88.4\% selection eﬃciency and 65.0\% purity. However, we can refine this selection, further reducing the small fractions of NC, cosmic ray, and $\nu_e$CC backgrounds. We develop this selection using the same methods as for the $\nu_e$CC selection. We identify events with a reconstructed muon track longer than 5 cm, and develop extra variables in order to address cases where the particles came from outside the TPC and cases where the muon was reconstructed as a charged pion or vice versa. The full set of $\nu_\mu$CC selection variables is described in App. \ref{sec:wc_bdt_vars}.

\begin{figure}[H]
    \centering
    \includegraphics[width=0.5\textwidth]{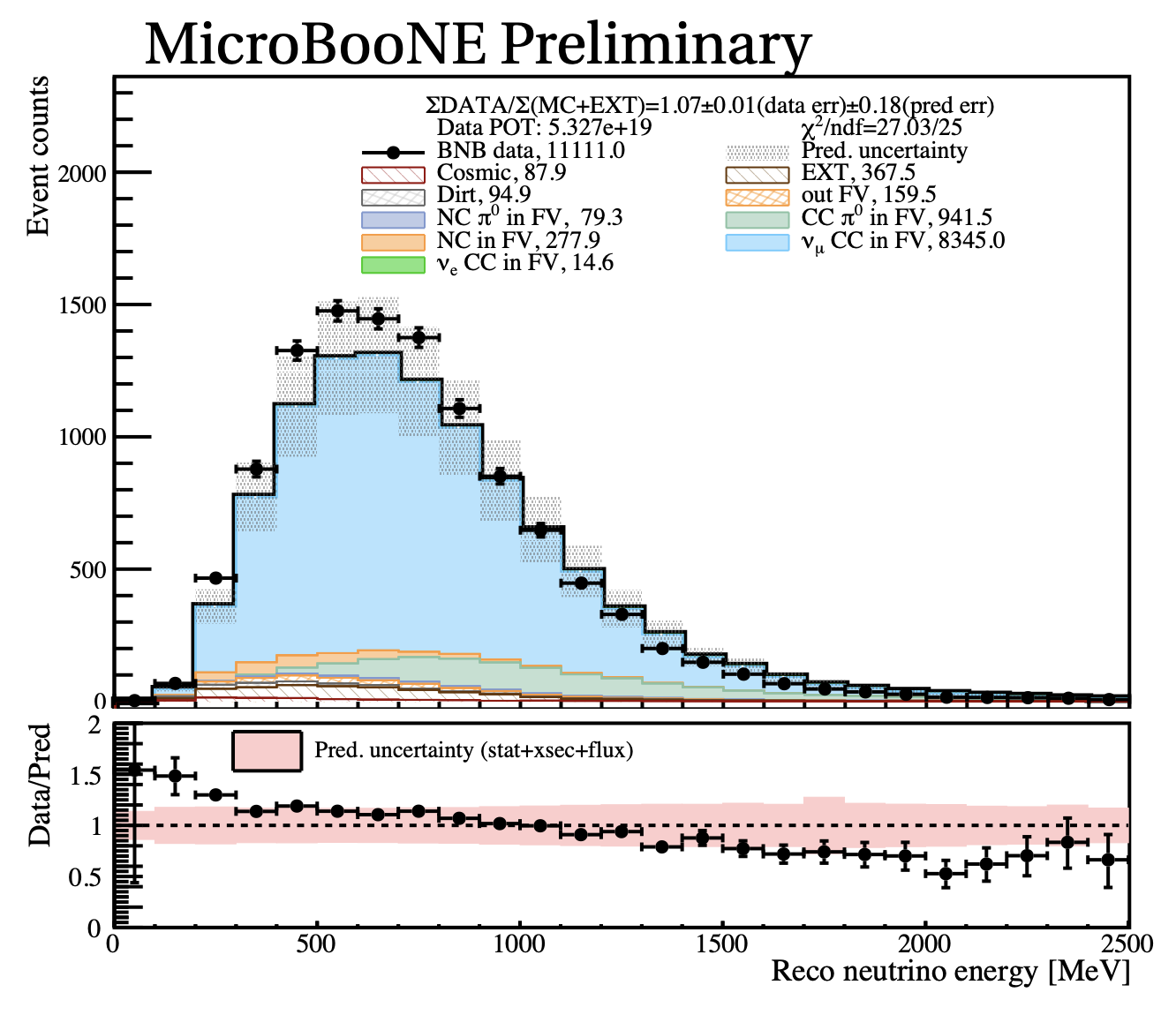}
    \caption[Cut-based $\nu_\mu$CC selection]{Cut-based $\nu_\mu$CC selection for fully contained events and $5.3\cdot10^{19}$ POT of open data. Figure from Ref. \cite{microboone_WC_eLEE_public_note}.}
    \label{fig:cut_based_numuCC}
\end{figure}

\begin{figure}[H]
    \centering
    \begin{subfigure}[b]{0.49\textwidth}
        \includegraphics[width=\textwidth]{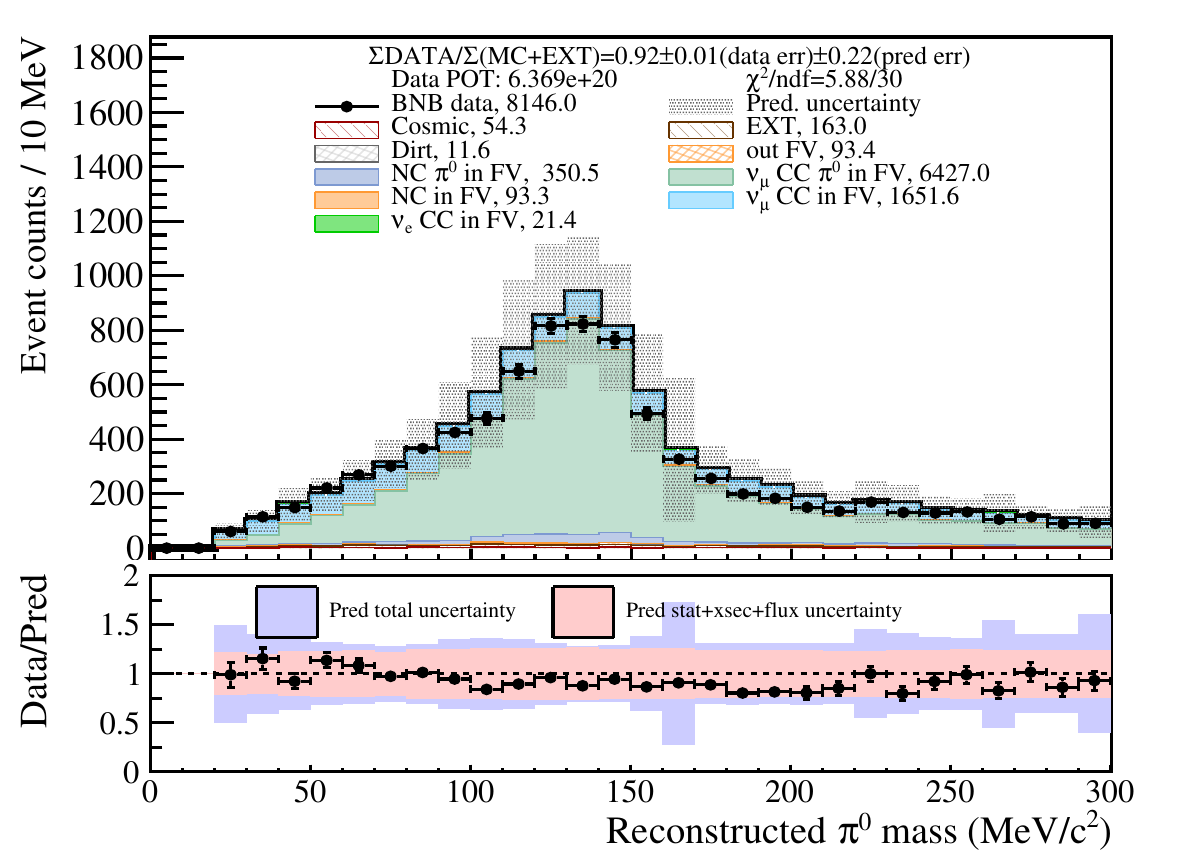}
        \caption{}
        \label{fig:ccpi0mass}
    \end{subfigure}
    \begin{subfigure}[b]{0.49\textwidth}
        \includegraphics[width=\textwidth]{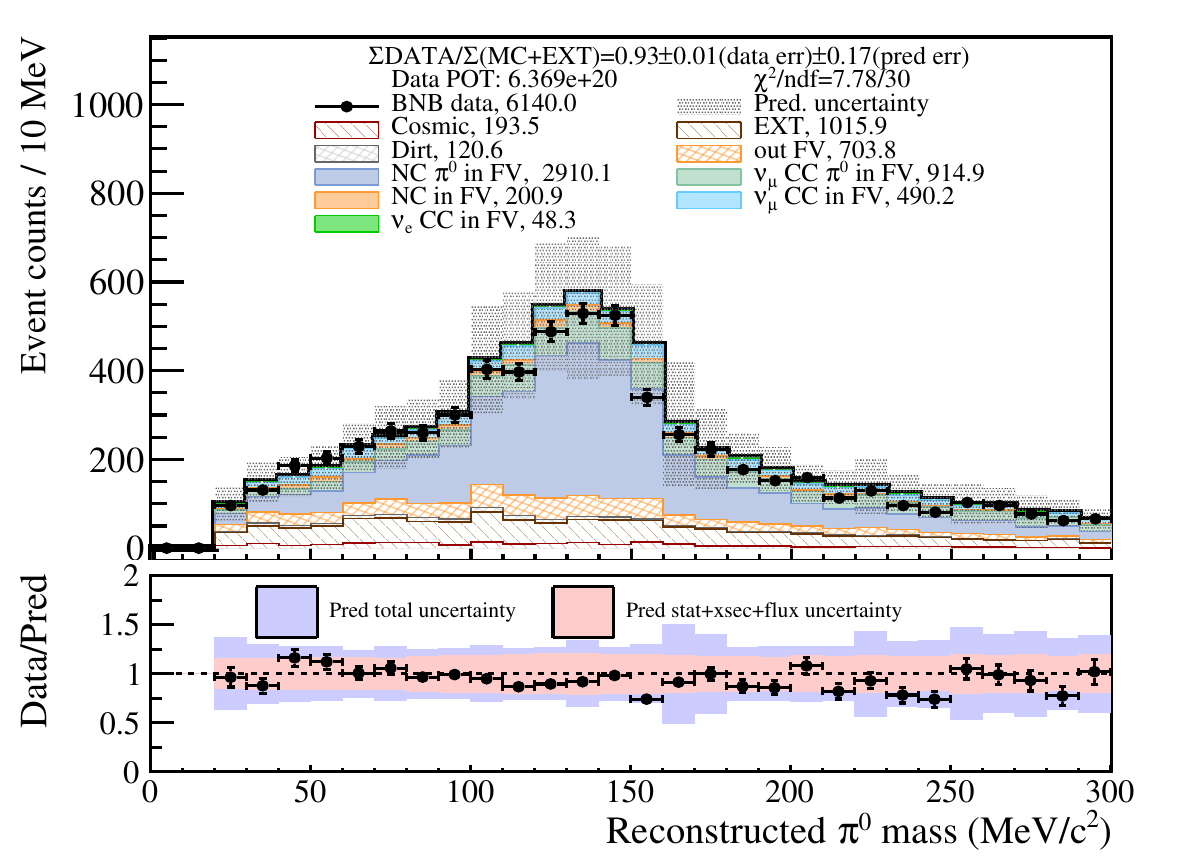}
        \caption{}
        \label{fig:ncpi0mass}
    \end{subfigure}
    \caption[$\pi^0$ mass distribution]{Figures from Ref. \cite{wc_elee_prd}.}
    \label{fig:pi0_mass_peaks}
\end{figure}

Finally, I will describe our NC$\pi^0$ and $\nu_\mu$CC$\pi^0$ selections. If an event contains at least two reconstructed showers, we create a $\pi^0$ candidate by choosing the highest energy pair of showers which approximately intersect in their upstream paths at a $\pi^0$ vertex. To select only primary $\pi^0$s, we place a cut on the distance between the $\pi^0$ vertex and the reconstructed neutrino vertex. Additional cuts are placed on the distances between the neutrino vertex and the photon shower vertices, the energies of the two showers, the opening angle between the two showers. We place a loose cut on the reconstructed invariant mass of the two showers, between 22 MeV and 300 MeV. The distribution of invariant masses is shown in Fig. \ref{fig:pi0_mass_peaks}, showing a clear peak around 135 MeV corresponding to the $\pi^0$ mass. The full set of $\pi^0$ selection variables is described in App. \ref{sec:wc_bdt_vars}.

\subsection{BDT-based Selections}\label{sec:bdt_selections}

In the process of developing the cut-based selections, we identified many key variables with the power to discriminate between signal and background events. The key advantage of this human-designed cut-based procedure is that at every step, the power of the variable to improve the performance was verified by a human, and the consistency of the behavior of each variable on data and simulation was studied by a human. The main disadvantage of this procedure is that humans can only study so many events, so the performance of cut-based selection will be limited when using such a high dimensional phase space. Specifically, in order to develop the cut-based $\nu_e$CC selection, we analyzed on the order of 1,000 events over the course of a few months, and did not achieve high purity in the selection.

We addressed this limitation by using the variables identified by the cut-based selection inside a boosted decision tree (BDT) in order to improve performance by automating the optimization of high-dimensional cuts placed on the set of variables.

In the cut-based selections, we grouped our variables into several ``taggers'', each consisting of some number of variables, and applied each tagger successively to form the selection. These taggers contain a combination of ``scalar variables'', which are floating point or integer numbers which describe features of the entire event, and ``vector variables'', which are floating point or integer numbers which describe features of certain particles within an event. For example, if there are three reconstructed showers in an event, we have certain variables which are length three vectors for the event.

\begin{figure}[H]
    \centering
    \includegraphics[width=0.9\textwidth]{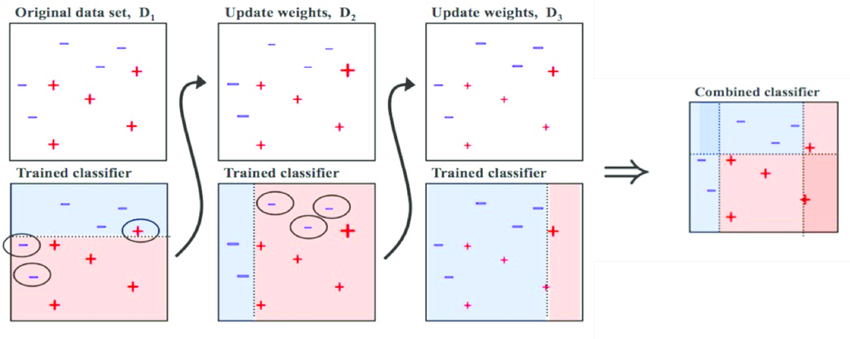}
    \caption[AdaBoost diagram]{AdaBoost diagram. Figure from Ref. \cite{adaboost_diagram}.}
    \label{fig:adaboost_diagram}
\end{figure}

We started this machine learning approach by replacing each tagger with a BDT using ROOT's TMVA toolkit using the AdaBoost algorithm \cite{TMVA}. AdaBoost uses an iterative procedure to build a forest of decision trees which each contribute weakly to the final collective decision, as illustrated in Fig. \ref{fig:adaboost_diagram}.

\begin{figure}[H]
    \centering
    \includegraphics[width=0.5\textwidth]{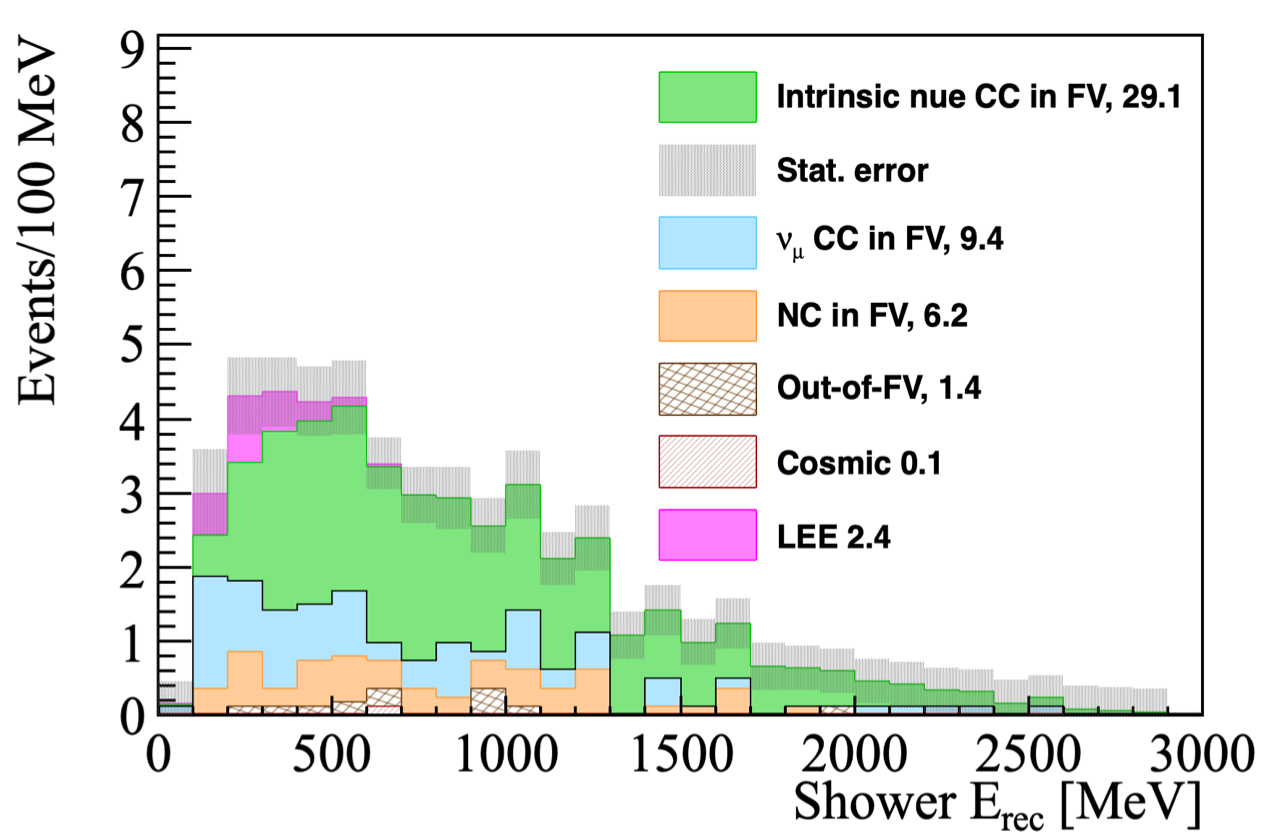}
    \caption[TMVA $\nu_e$CC selection]{$\nu_e$CC selection prediction for fully contained events using ROOT TMVA AdaBoost BDTs, for $5.3\cdot10^{19}$ POT. In pink, we show the prediction from the LEE model, described in Sec. \ref{sec:lee_model}. Figure from Ref. \cite{microboone_WC_eLEE_public_note}.}
    \label{fig:TMVA_based_nueCC}
\end{figure}

We use these TMVA BDTs in order to replace each tagger with a TMVA BDT. A BDT is designed to handle a fixed number of inputs, so for these vector variables, we evaluate the BDT on each particle successively, and take either the minimum or the maximum score as the overall result from this tagger. Using the TMVA BDT, we achieve subtantially improved performance, as shown in Fig. \ref{fig:TMVA_based_nueCC} which can be compared to the cut-based selection shown in Fig. \ref{fig:cut_based_nueCC}. For fully-contained events, we achieve a $\nu_e$CC selection efficiency of 63\% and a $\nu_e$CC purity of 32\%.

\begin{figure}[H]
    \centering
    \includegraphics[width=0.7\textwidth]{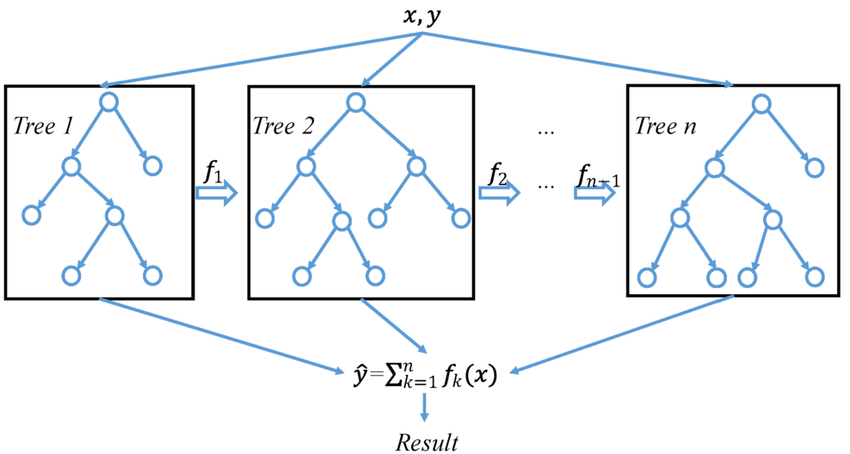}
    \caption[XGBoost diagram]{XGBoost diagram. Figure from Ref. \cite{xgboost_diagram}.}
    \label{fig:xbgoost_diagram}
\end{figure}

One of my first contributions in the Wire-Cell team was studying the performance using XGBoost \cite{xgboost}, a more modern BDT framework. XGBoost uses the machine learning method of gradient descent in order to learn more efficiently than a simple iterative procedure, and it also uses more complex multi-step decision trees as illustrated in Fig. \ref{fig:xbgoost_diagram}.

With XGBoost, we discovered that the efficiency of the algorithm was great enough that we were able to combine variables from different taggers together, handling a very large number of inputs simultaneously. So, we use one large BDT which handles all the scalar inputs, as well as ``vector BDT scores'' which come from the minimum or maximum scores from smaller BDTs acting on these vector variables for multiple particles within a single event.

To handle each set of vector varables, we start with the original cut-based tagger selection, and then train two rounds of BDTs to improve performance. Note that each tagger is not designed to reject all types of background; it is designed to reject only one specific category. To guide the vector BDT toward that purpose, we start with a ``round 1 BDT'', which is trained to distinguish $\nu_e$CC events from only the backgrounds which were rejected by the cut-based tagger. This encourages the BDT to learn to reject a specific background topology, rather than trying to do the job of a different tagger. We then do another training, developing a ``round 2 BDT'', which adds to the training additional background events which were not identified as background by the cut-based tagger, but were identified as very background-like by the round 1 BDT. This keeps the original goal of training to reject only one type of background, but expands to account for more types of events than were considered by the cut-based tagger. These round 2 BDT scores for vector variable taggers are then passed into the large BDT in order to make the final selection. Note that this procedure of iterative BDT training for vector variables means that there are still some echoes of the original human-designed cuts which contribute to our final selection.

\begin{figure}[H]
    \centering
    \includegraphics[width=0.5\textwidth]{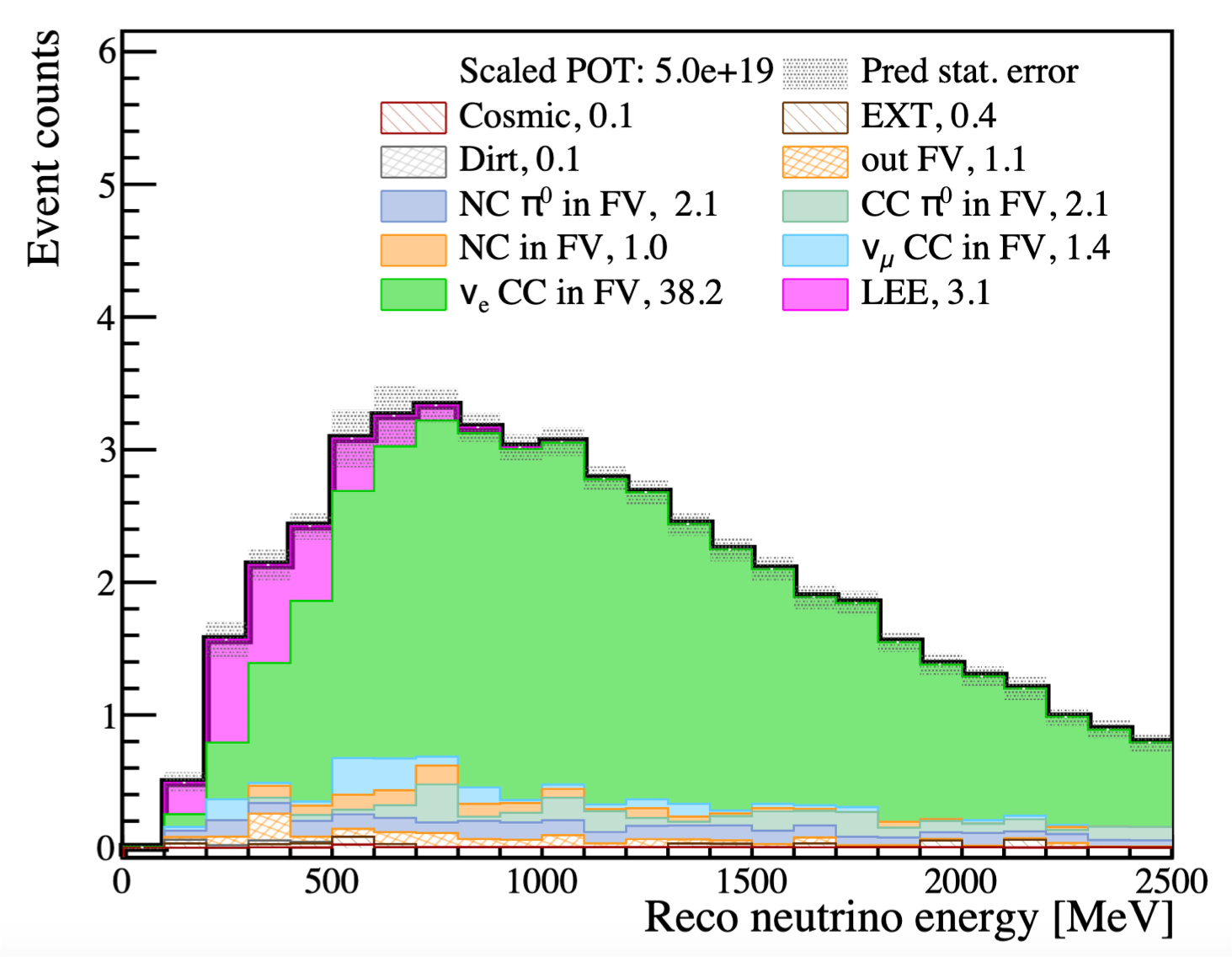}
    \caption[XGBoost $\nu_e$CC selection]{$\nu_e$CC selection prediction for fully contained events using the XGBoost BDT, for $5.3\cdot10^{19}$ POT. In pink, we show the prediction from the LEE model, described in Sec. \ref{sec:lee_model}. Figure from Ref. \cite{microboone_WC_eLEE_public_note}.}
    \label{fig:xgboost_nueCC}
\end{figure}

We use this same XGBoost BDT training strategy to improve our $\nu_e$CC and $\nu_\mu$CC selections. In our final $\nu_e$CC selection BDT, we include 243 scalar variables, and 15 vector BDT scores. In our final $\nu_\mu$CC selection BDT, we include 71 scalar variables and 3 vector BDT scores. Substantially improved performance for our $\nu_e$CC selection can be seen in Fig. \ref{fig:xgboost_nueCC} compared to Figs. \ref{fig:cut_based_numuCC} and \ref{fig:TMVA_based_nueCC}. For fully-contained events, we achieve an efficiency of 42\% and a purity of 83\%. This is our final Wire-Cell $\nu_e$CC selection which we use for the physics results in this chapter.

\begin{figure}[H]
    \centering
    \begin{subfigure}[b]{0.52\textwidth}
        \includegraphics[width=\textwidth]{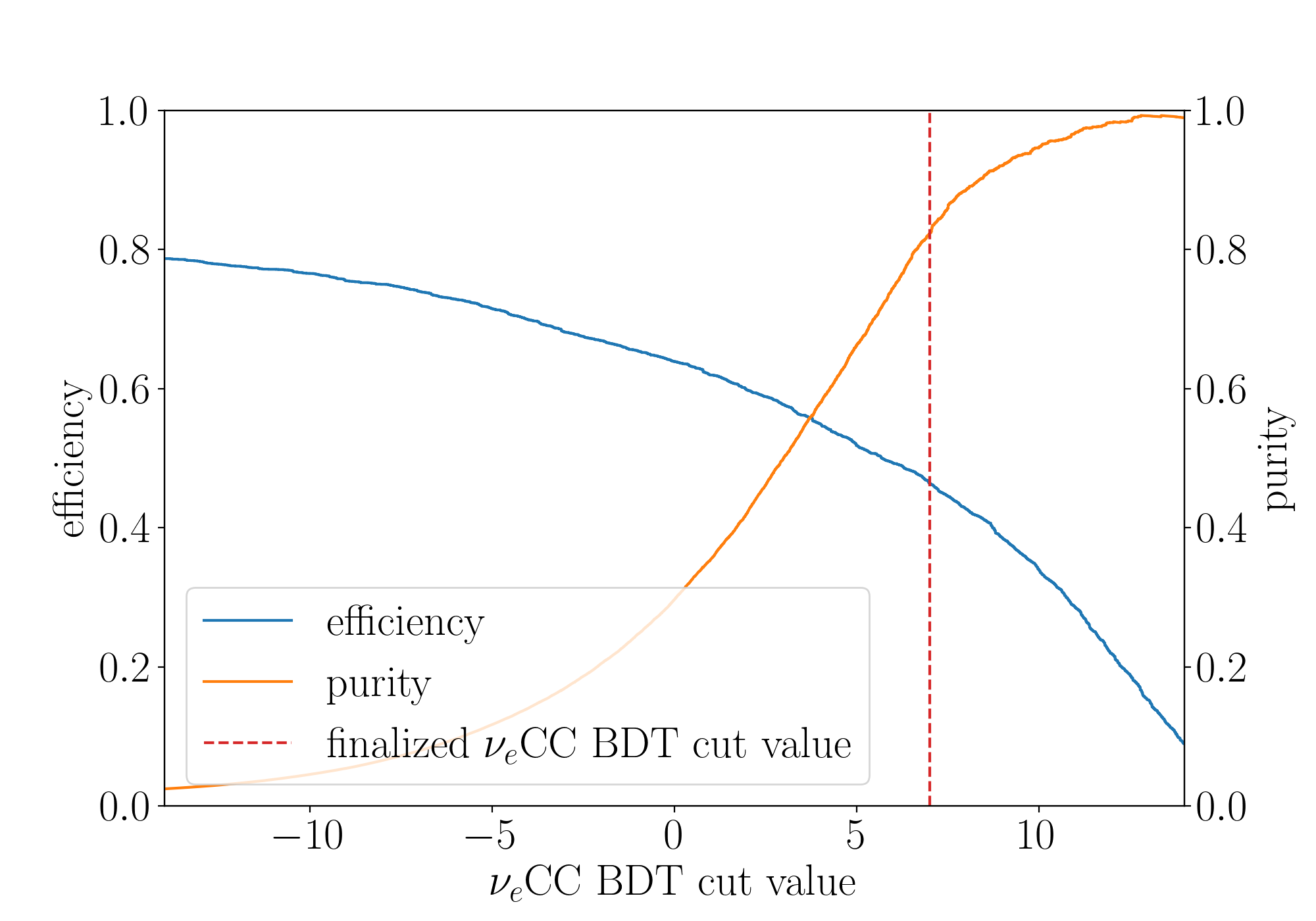}
        \caption{}
        \label{fig:nue_eff_pur}
    \end{subfigure}
    \begin{subfigure}[b]{0.47\textwidth}
        \includegraphics[width=\textwidth]{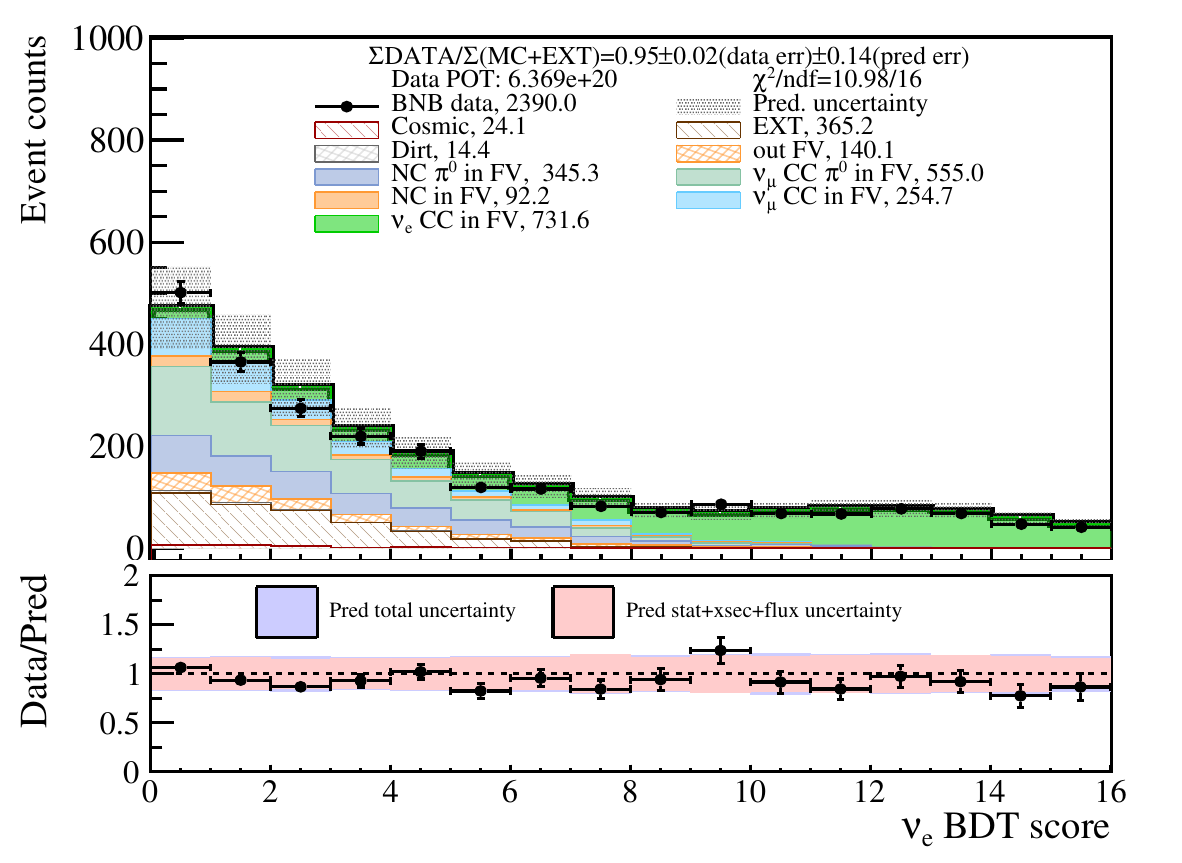}
        \caption{}
        \label{fig:nue_bdt_dist}
    \end{subfigure}
    \caption[$\nu_e$ BDT scores]{Panel (a) shows the efficiency and purity of our $\nu_e$CC BDT selection as a function of the BDT cut value, with the final cut value of 7 indicated as a red dashed line. Panel (b) shows the distribution of our $\nu_e$CC BDT score values for data and prediction. Figures from Ref. \cite{wc_elee_prd}.}
    \label{fig:nue_BDT}
\end{figure}

\begin{figure}[H]
    \centering
    \begin{subfigure}[b]{0.52\textwidth}
        \includegraphics[width=\textwidth]{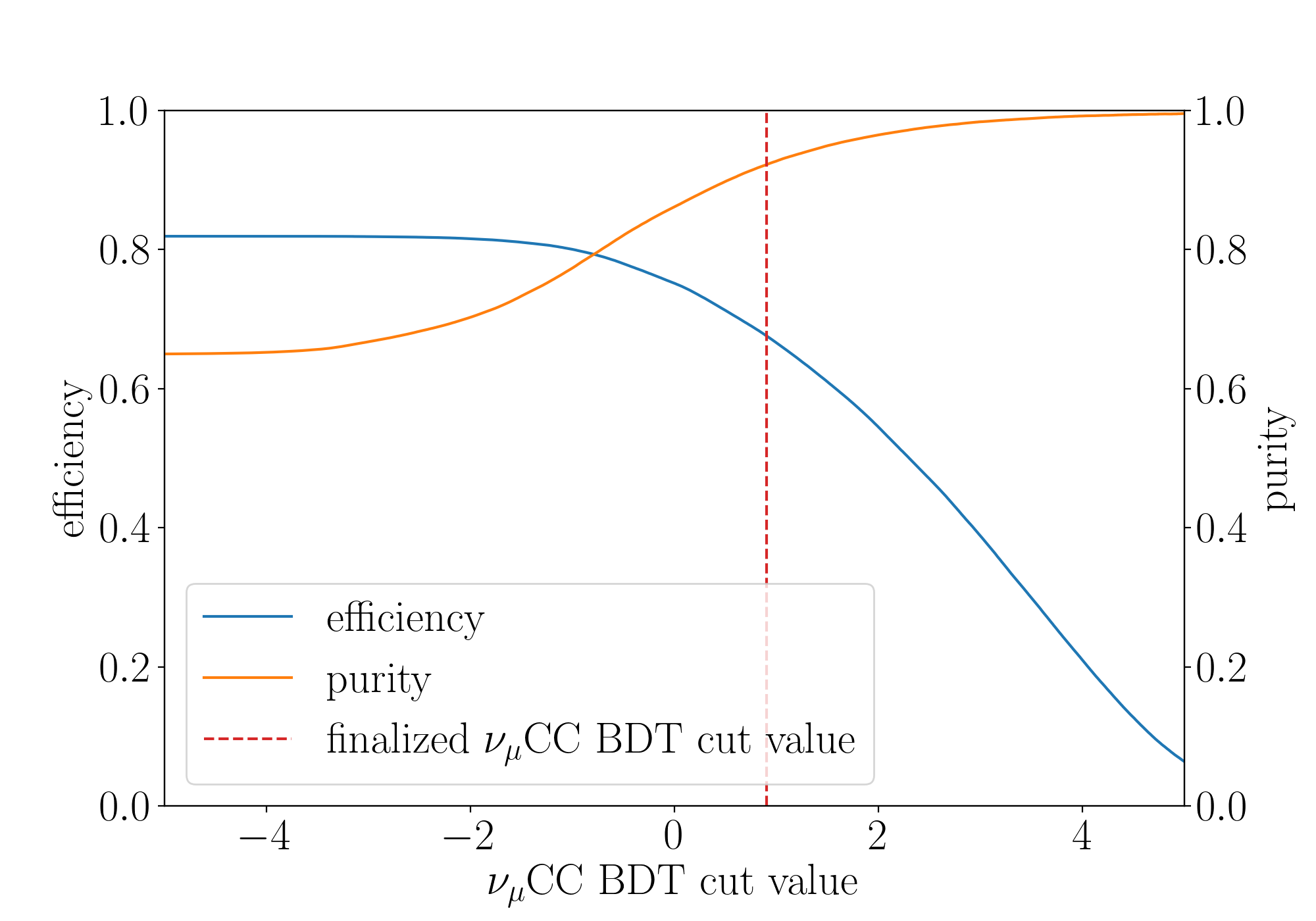}
        \caption{}
        \label{fig:numu_eff_pur}
    \end{subfigure}
    \begin{subfigure}[b]{0.47\textwidth}
        \includegraphics[width=\textwidth]{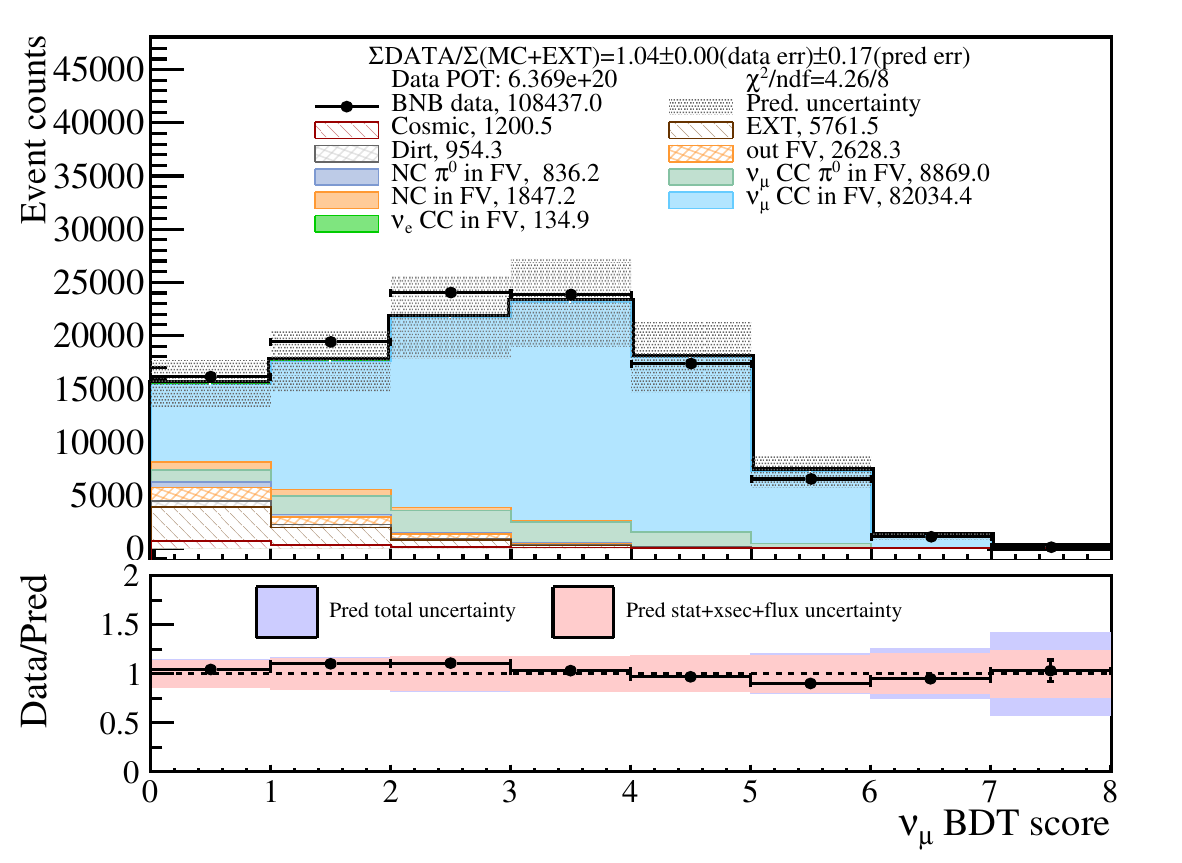}
        \caption{}
        \label{fig:numu_bdt_dist}
    \end{subfigure}
    \caption[$\nu_\mu$ BDT scores]{Panel (a) shows the efficiency and purity of our $\nu_\mu$CC BDT selection as a function of the BDT cut value, with the final cut value of 0.9 indicated as a red dashed line. Panel (b) shows the distribution of our $\nu_\mu$CC BDT score values for data and prediction. Figures from Ref. \cite{wc_elee_prd}.}
    \label{fig:numu_BDT}
\end{figure}

Figs. \ref{fig:nue_BDT} and \ref{fig:numu_BDT} show the efficiency and purity, prediction, and data as functions of the $\nu_e$CC and $\nu_\mu$CC BDT scores. We chose BDT cut values which maximize efficiency times purity. Note in Fig. \ref{fig:nue_eff_pur}, the left edge with a very loose BDT cut reflects the performance of Wire-Cell generic selection for $\nu_e$CC events, with reasonable efficiency and very low purity before applying the $\nu_e$CC BDT cut, while in Fig. \ref{fig:numu_eff_pur}, this indicates reasonable performance of Wire-Cell generic selection for $\nu_\mu$CC events, with reasonably high efficiency and purity even before applying the $\nu_\mu$CC BDT cut.

\begin{figure}[H]
    \centering
    \begin{subfigure}[b]{0.49\textwidth}
        \includegraphics[width=\textwidth]{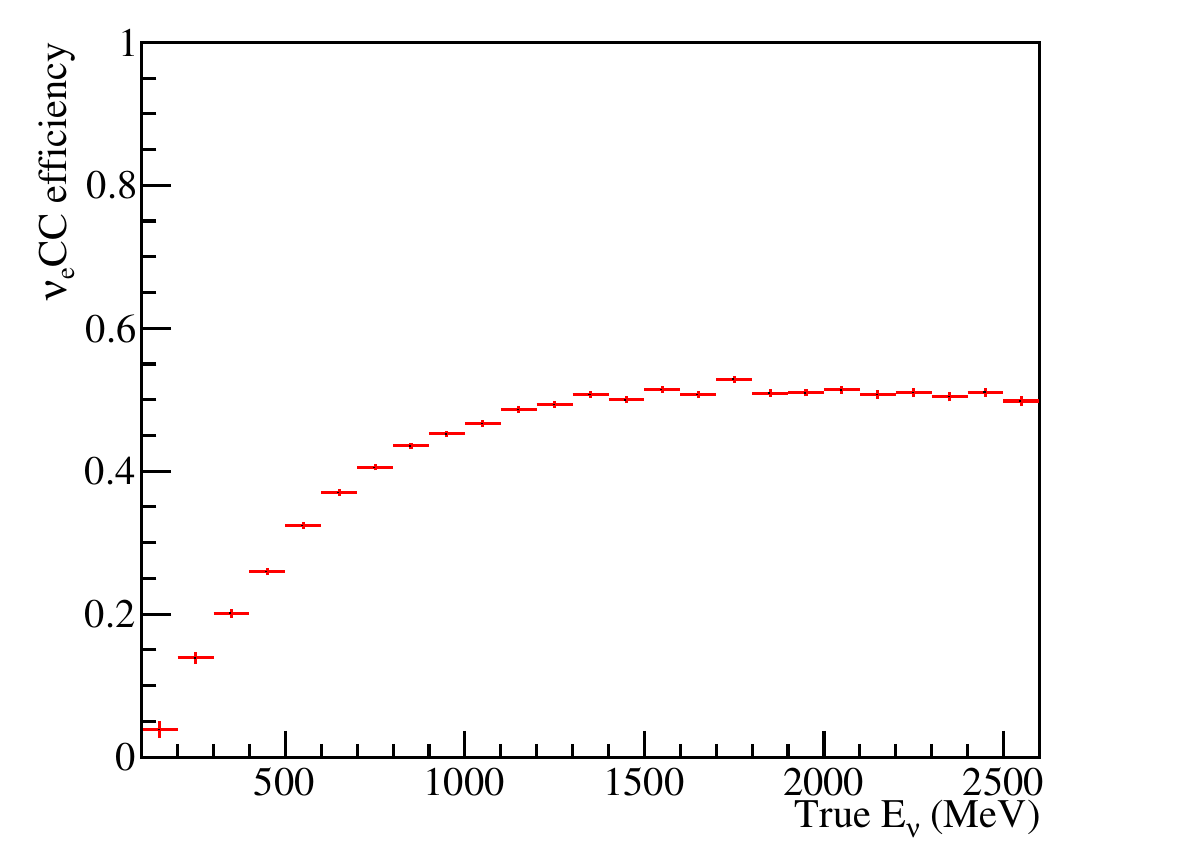}
        \caption{}
        \label{fig:nue_eff_Enu}
    \end{subfigure}
    \begin{subfigure}[b]{0.49\textwidth}
        \includegraphics[width=\textwidth]{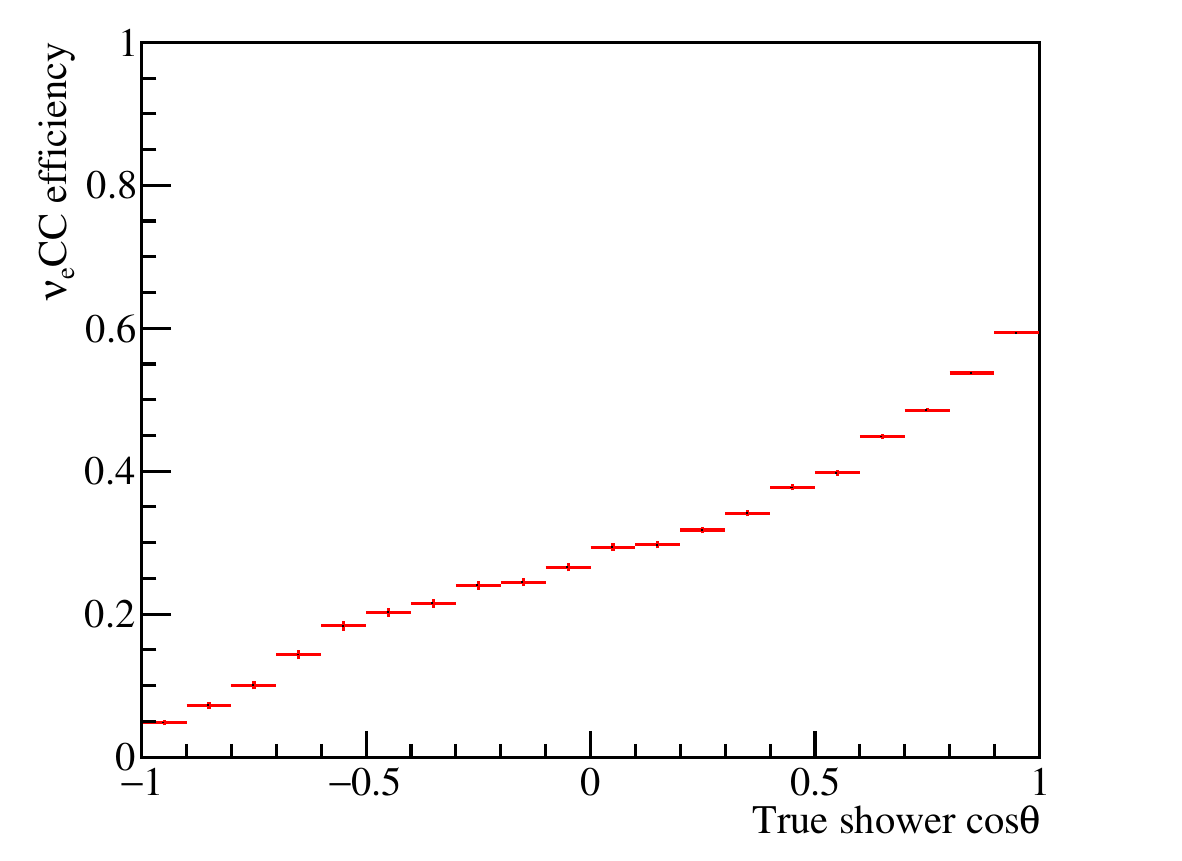}
        \caption{}
        \label{fig:nue_eff_costheta}
    \end{subfigure}
    \caption[$\nu_e$CC selection efficiencies]{Panel (a) shows the $\nu_e$CC selection efficiency as a function of the neutrino energy. Panel (b) shows the $\nu_e$CC selection efficiency as a function of the electron shower angle relative to the beam direction. Figures from Ref. \cite{wc_elee_prd}.}
    \label{fig:nue_eff_kine}
\end{figure}

\begin{figure}[H]
    \centering
    \begin{subfigure}[b]{0.49\textwidth}
        \includegraphics[width=\textwidth]{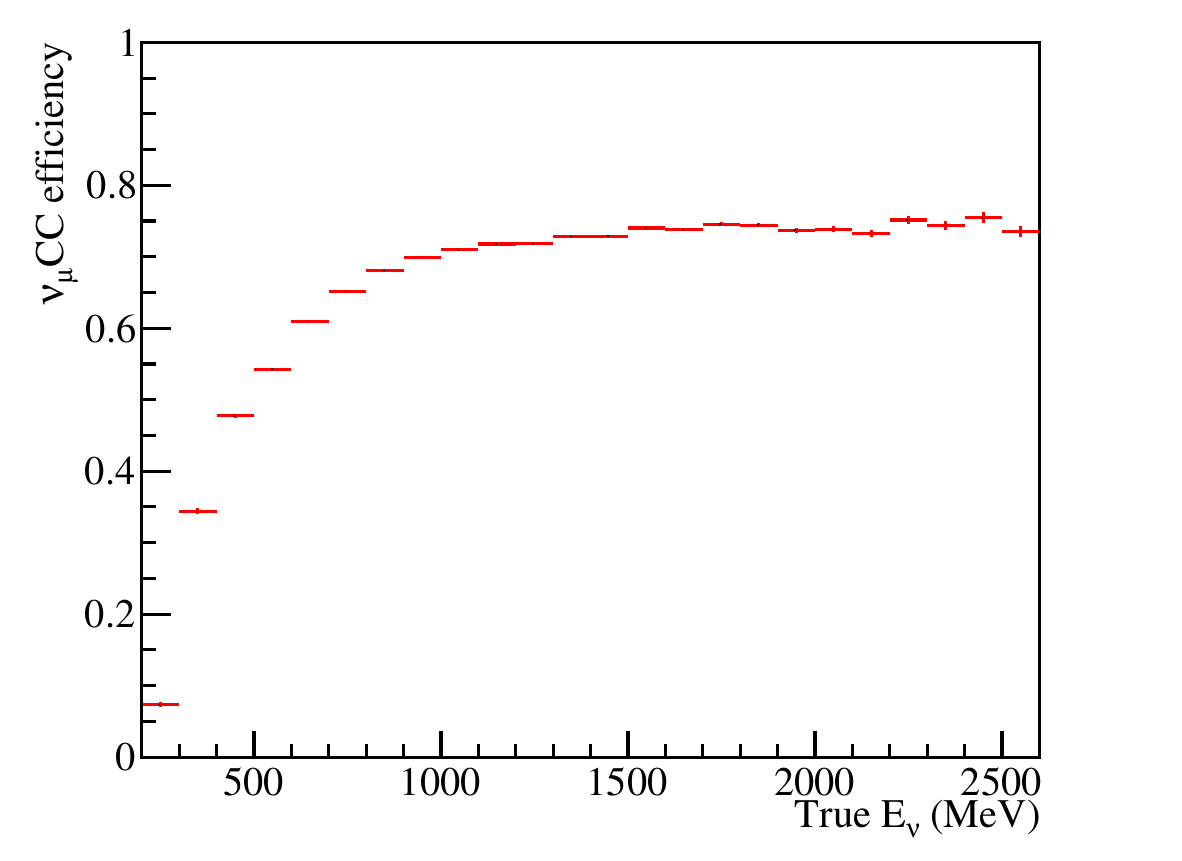}
        \caption{}
        \label{fig:numu_eff_Enu}
    \end{subfigure}
    \begin{subfigure}[b]{0.49\textwidth}
        \includegraphics[width=\textwidth]{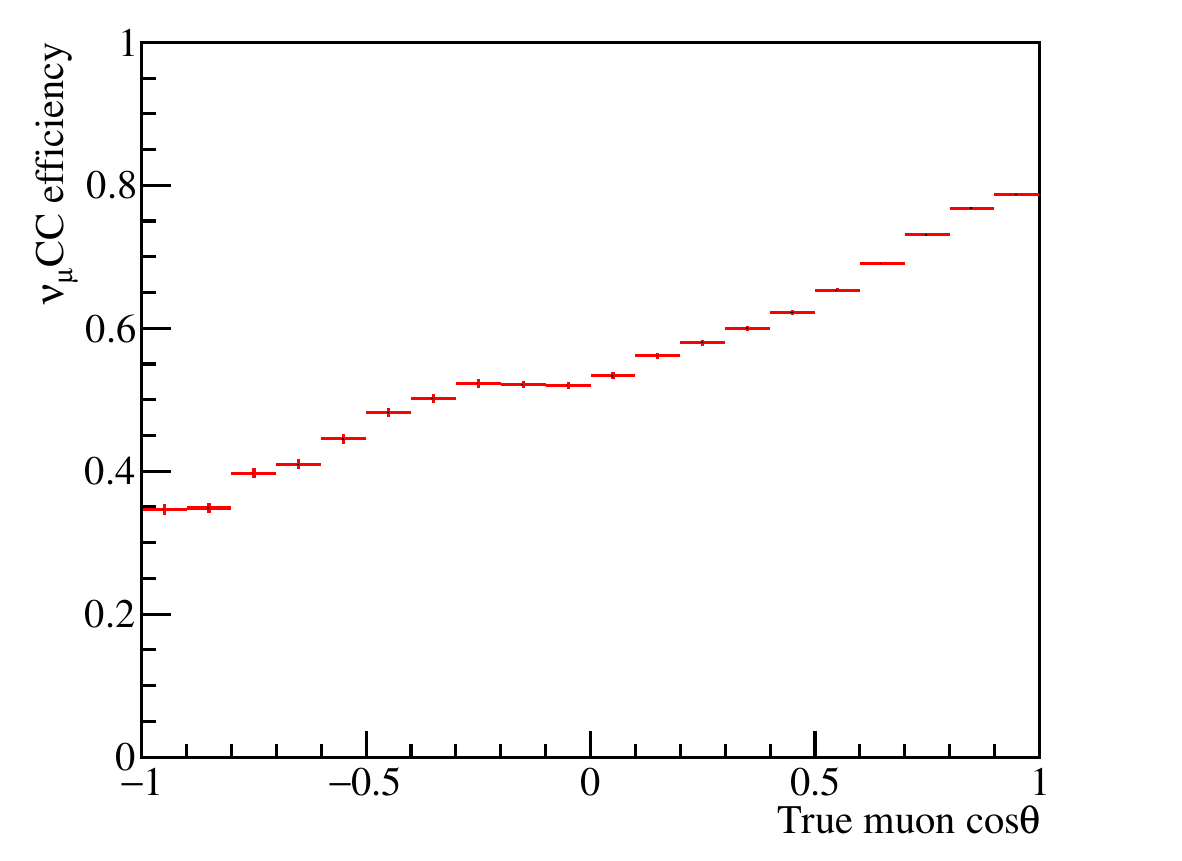}
        \caption{}
        \label{fig:numu_eff_costheta}
    \end{subfigure}
    \caption[$\nu_\mu$CC selection efficiencies]{Panel (a) shows the $\nu_\mu$CC selection efficiency as a function of the neutrino energy. Panel (b) shows the $\nu_\mu$CC selection efficiency as a function of the muon angle relative to the beam direction. Figures from Ref. \cite{wc_elee_prd}.}
    \label{fig:numu_eff_kine}
\end{figure}

Importantly, these selections result in good efficiency across a variety of lepton energies and angles for both $\nu_e$CC and $\nu_\mu$CC, as shown in Figs. \ref{fig:nue_eff_kine} and \ref{fig:numu_eff_kine}. Note that the selections struggle at the lowest energies and for backwards-going leptons, but still maintain reasonable efficiency throughout the phase space.

\section{\texorpdfstring{$\nu_e$}{nue}CC Analysis}

Now that we have high-performance $\nu_e$CC selections as well as $\nu_\mu$CC and $\pi^0$ selections in order to adress systematic uncertainties, we are ready to use these selections for a physics analysis studying the MiniBooNE LEE.

\subsection{eLEE Model}\label{sec:lee_model}

In order to understand our observations, it is helpful to have a prediction for how $\nu_e$CC events in MicroBooNE relate to the MiniBooNE excess. This can be done in many ways, including via BSM models like 3+1 sterile neutrino oscillations, or by simplified phenomenological models. For this first analysis, we used an unfolding of the MiniBooNE LEE in terms of true neutrino energy, modeling an enhancement to the $\nu_e$ flux in the BNB from some source. Note that this LEE model is just a single median unfolding; this type of model is expanded to consider MiniBooNE uncertainties in Ref. \cite{MicroBooNE_eLEE_interpretation}.

\begin{figure}[H]
    \centering
    \begin{subfigure}[b]{0.49\textwidth}
        \includegraphics[width=\textwidth]{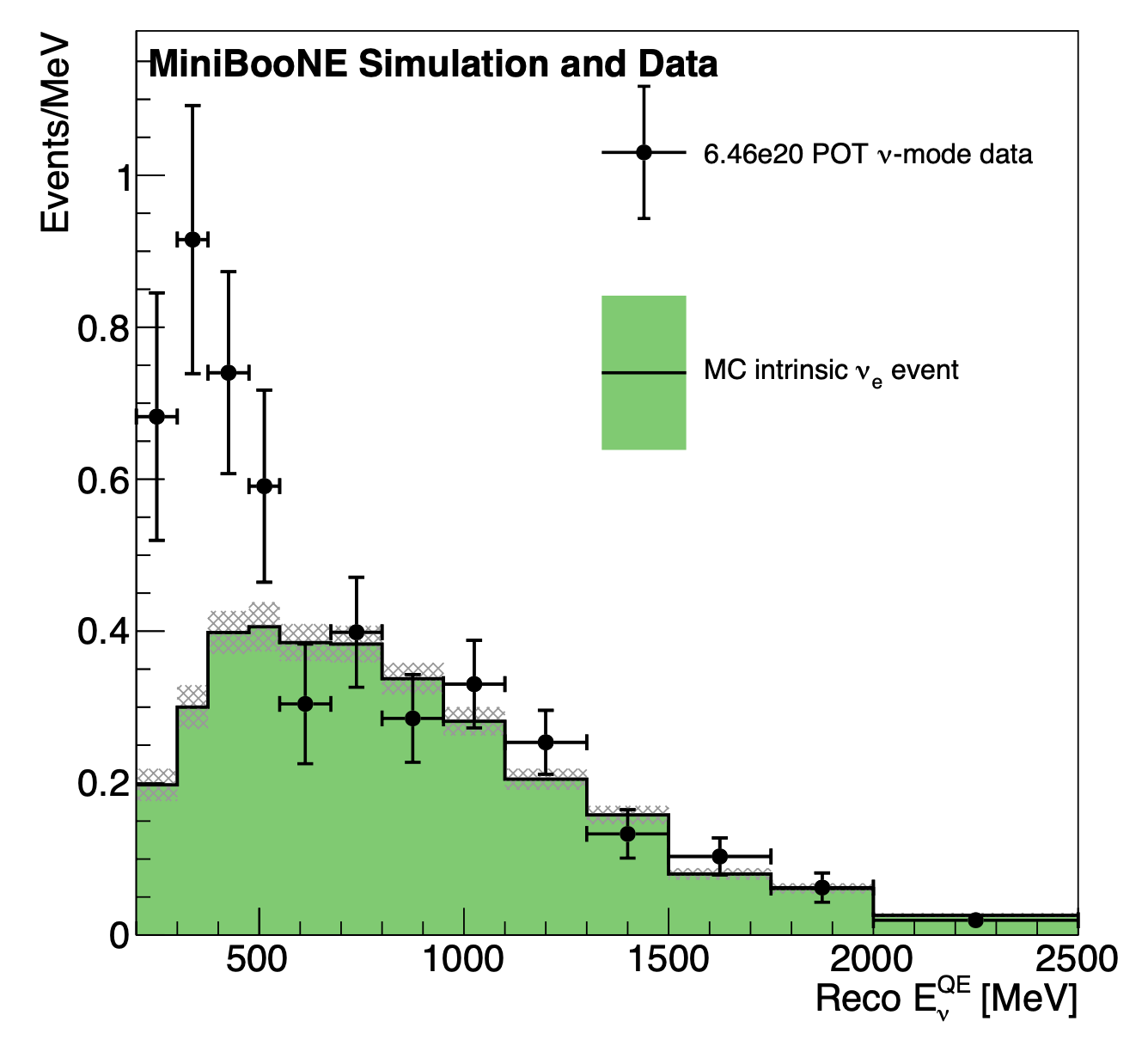}
        \caption{}
    \end{subfigure}
    \begin{subfigure}[b]{0.49\textwidth}
        \includegraphics[width=\textwidth]{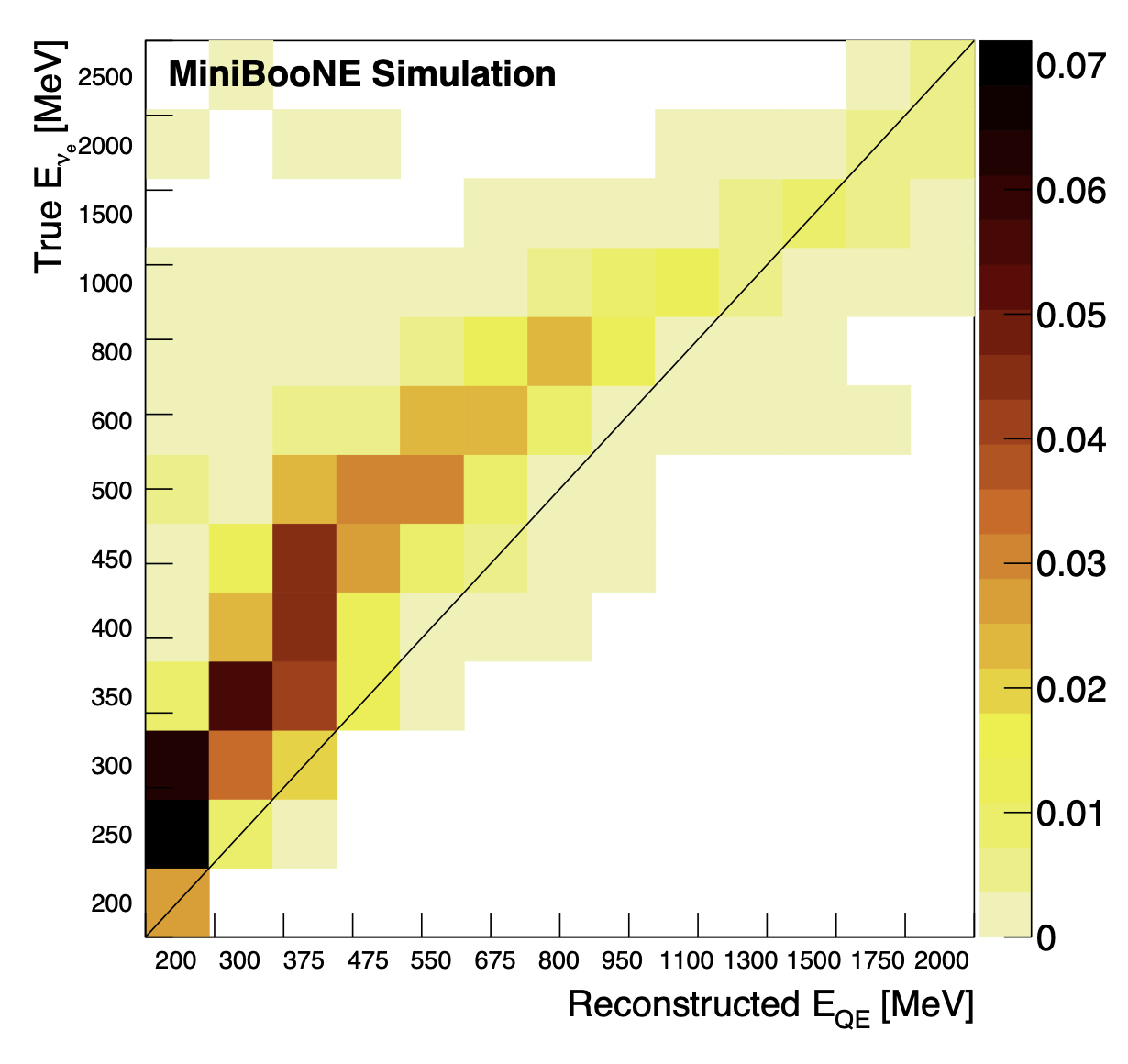}
        \caption{}
    \end{subfigure}
    \caption[MiniBooNE LEE $\nu_e$CC response]{Panel (a) shows the MiniBooNE LEE as a function of reconstructed neutrino energy.  Panel (b) shows MiniBooNE's response matrix, describing how different true $\nu_e$CC energies result in different reconstructed energy distributions. Figures from Ref. \cite{eLEE_model_public_note}.}
    \label{fig:LEE_model_response}
\end{figure}

Figure \ref{fig:LEE_model_response} shows the MiniBooNE LEE in reconstructed space, as well as the response matrix describing MiniBooNE's reconstructed energy distributions for different true $\nu_e$CC neutrino energies.

\begin{figure}[H]
    \centering
    \begin{subfigure}[b]{0.43\textwidth}
        \includegraphics[width=\textwidth]{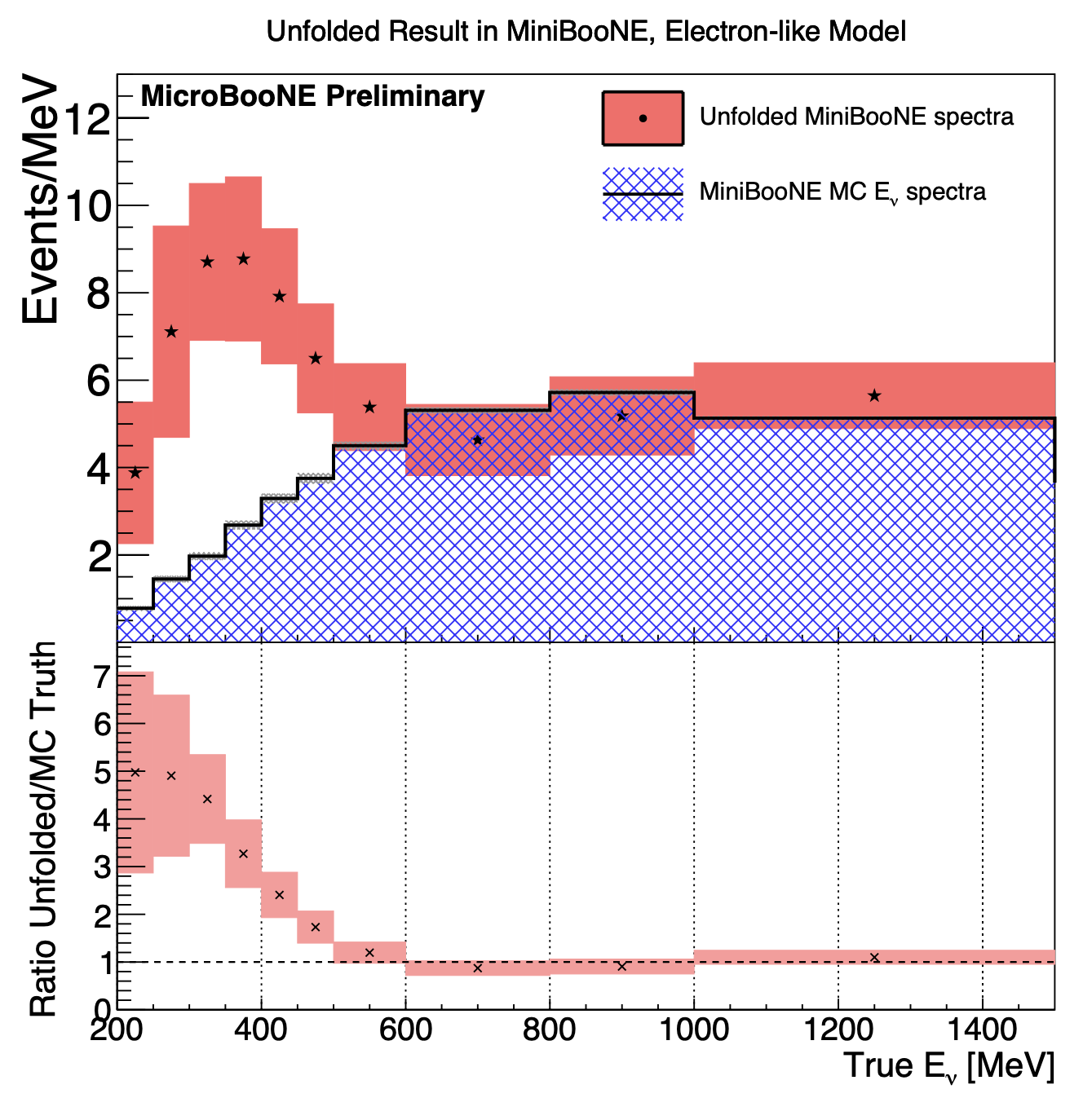}
        \caption{}
        \label{fig:MB_unfolded}
    \end{subfigure}
    \begin{subfigure}[b]{0.56\textwidth}
        \includegraphics[width=\textwidth]{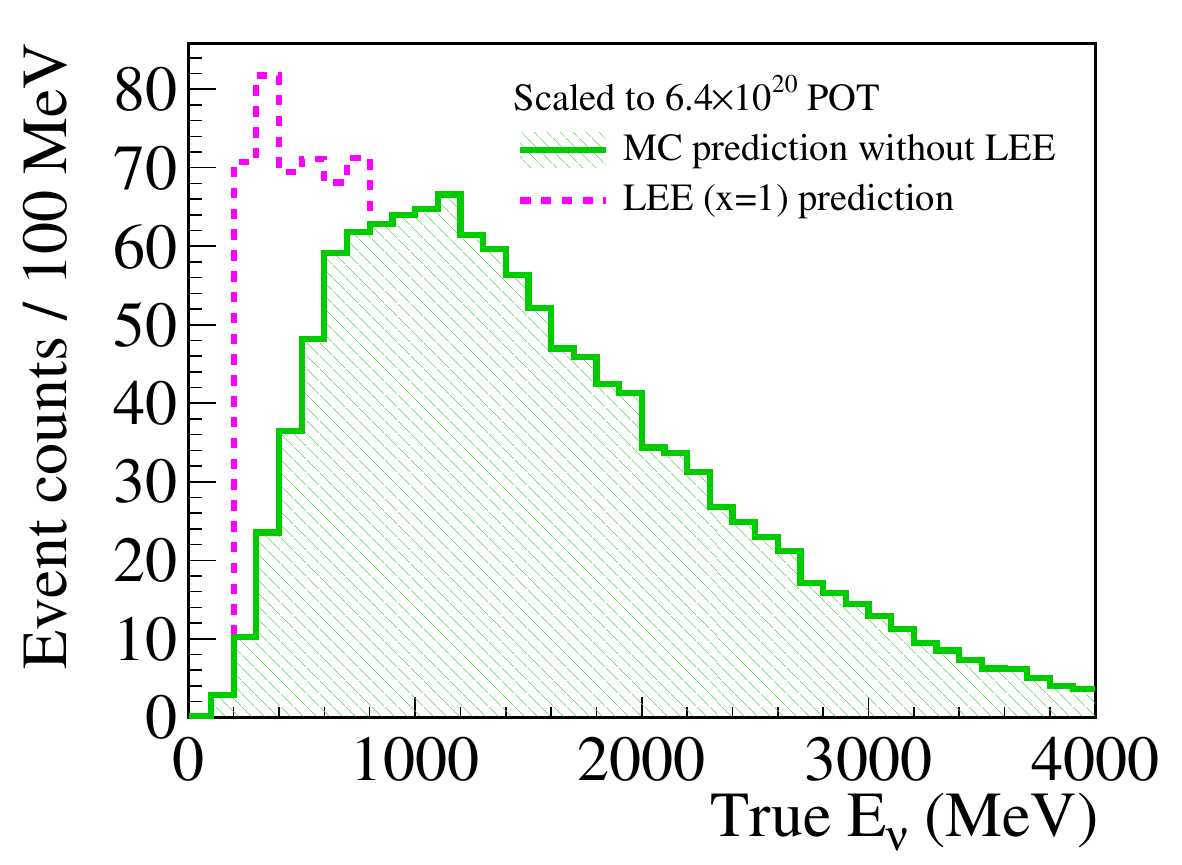}
        \caption{}
        \label{fig:uB_unfolded}
    \end{subfigure}
    \caption[Unfolded $\nu_e$CC LEE model]{Median unfolded $\nu_e$CC LEE model. Panel (a) shows the unfolded result in MiniBooNE. Panel (b) shows the median unfolded result in MicroBooNE, with no selection applied. Figures from Ref. \cite{eLEE_model_public_note} and \cite{wc_elee_prd}.}
    \label{fig:lee_model}
\end{figure}

We used this excess distribution and response matrix as shown in Fig. \ref{fig:LEE_model_response} in order to perform a median unfolding of the excess under a $\nu_e$CC hypothesis as a function of true neutrino energy, as shown in Fig. \ref{fig:MB_unfolded}. We then took the relative excess over prediction in each true neutrino energy bin and interpreted this as a relative scaling enhancement of the $\nu_e$ flux, and applied this to the MicroBooNE prediction as shown in Fig. \ref{fig:uB_unfolded}.

We label this median unfolded model as LEE (x=1), but in order to understand our sensitivity to different sizes of excesses, we also consider different fractional scalings of this electron LEE model, which we call eLEE strengths or eLEEx values. 

\subsection{Systematic Uncertainties}\label{sec:systematic_uncertainties}

This analysis includes many sources of systematic uncertainties. We treat all systematic uncertainties using covariance matrix formalism.

\begin{figure}[H]
    \centering
    \begin{subfigure}[b]{0.485\textwidth}
        \includegraphics[trim=10 0 70 0, clip, width=\textwidth]{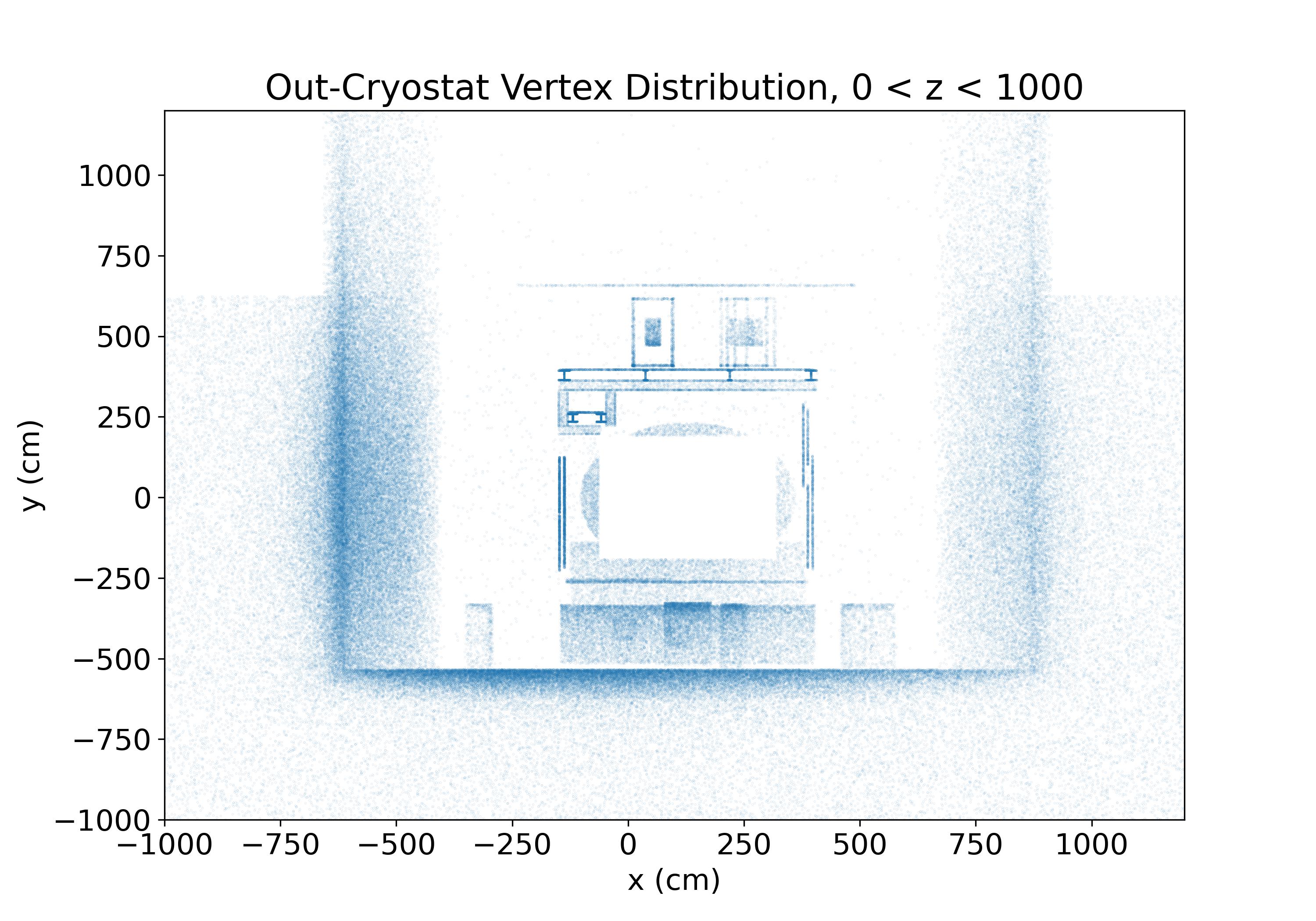}
        \caption{}
    \end{subfigure}
    \begin{subfigure}[b]{0.495\textwidth}
        \includegraphics[trim=10 0 50 0, clip, width=\textwidth]{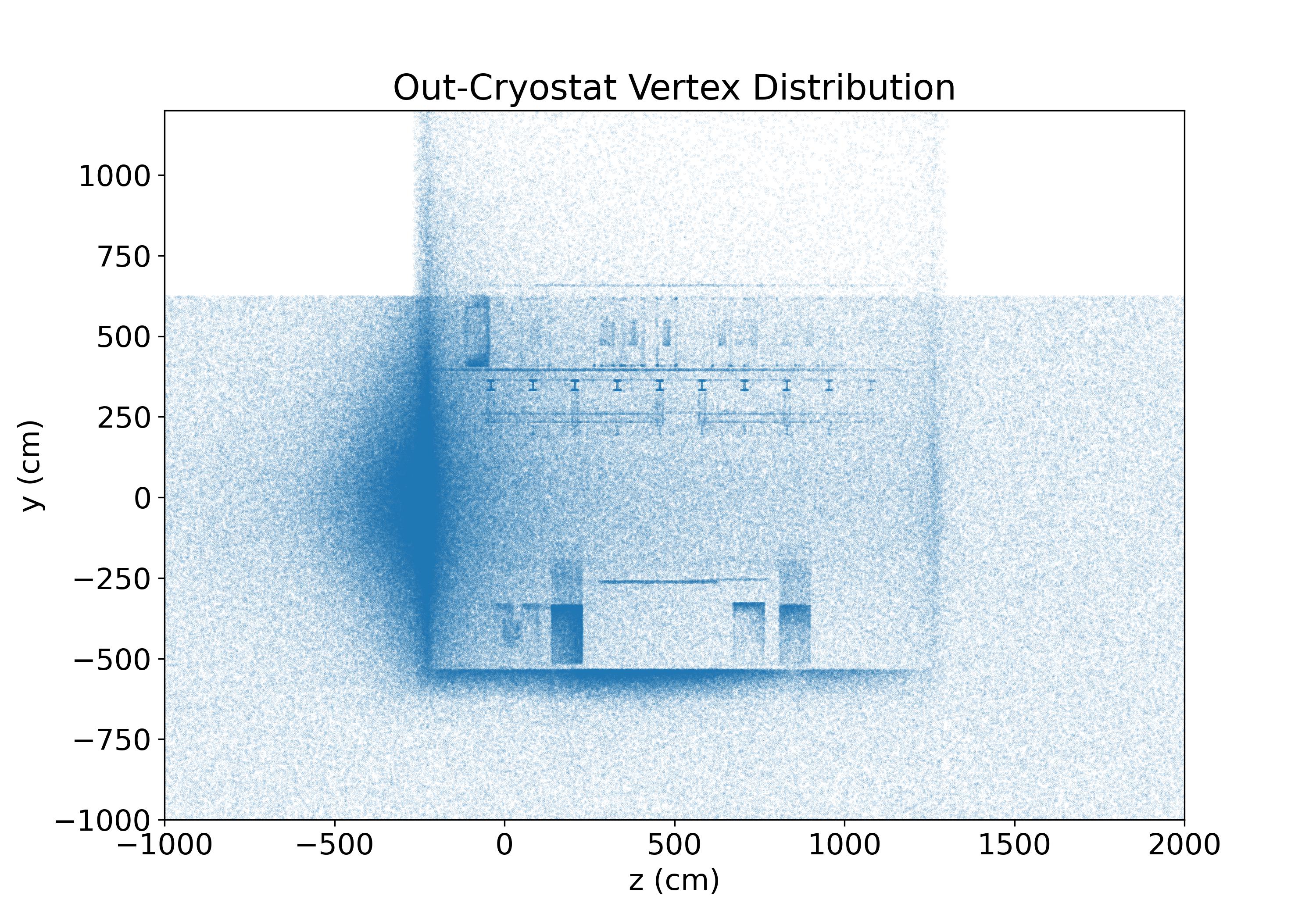}
        \caption{}
    \end{subfigure}
    \caption[Out-cryostat simulated neutrino vertex positions]{Out-cryostat simulated neutrino vertex positions, showing the effect of simulated objects surrounding the central cryostat, as well as the detector building walls. Panel (a) shows simulated x-y locations, where we are looking along the downstream neutrino beam path, with a restricted z range to highlight the cryostat surroundings. We can see an empty rectangle in the middle, which is missing to be filled in by different simulations for neutrinos interacting within the cryostat. Panel (b) shows the simulated z-y locations, where the neutrino beam goes from left to right. There is a particularly large concentration of events just upstream of the cryostat.}
    \label{fig:out_cryostat_vertices}
\end{figure}

Detector response uncertainties were discussed in Sec. \ref{sec:DetVar}. For each of the detector variations, we build a covariance matrix using the difference in the histogram prediction between the central value (CV) and the variation, and employ a repeated sampling procedure in order to account for statistical uncertainty on this difference. We also add a 50\% additional fractional uncertainty for simulated events with a neutrino interaction outside of the MicroBooNE cryostat which we refer to as ``dirt'' events, since there is more uncertainty in the modeling of the materials surrounding the detector, as illustrated in Fig. \ref{fig:out_cryostat_vertices}.

Data statistical uncertainties are considered with a Combined Neyman-Pearson (CNP) treatment \cite{CNP_statistical_uncertainty}. Monte-Carlo statistical uncertainties are calculated by $\sigma=\sqrt{\Sigma_{i=1}^N w_i^2}$ to account for unequal event weights.

For this analysis, we use neutrino flux from the BNB, described in Sec. \ref{sec:beams}. We include several systematic uncertainties on this flux prediction. We generate alternative flux variation universeses by varying the production rates of $\pi^+$, $\pi^-$, $K^+$, $K^-$, and $K_0^L$ produced by high energy proton interactions with the target. We also vary our prediction according to different possible variations of the horn current distribution and the horn current calibration, which affect the focusing of hadrons exiting the target. We vary the relative cross sections for total nucleon scattering, nucleon inelastic scattering, and nucleon quasi-elastic scattering, which affect secondary interactions of nucleons in the target and surrounding material. Finally, we vary the relative cross sections for total pion scattering, inelastic pion scattering, and quasi-elastic pion scattering, which affect secondary interactions of pions in the target and surrounding material.

We also consider variations in the neutrino-argon interaction cross section modeling. We use GENIE v3.0.6 \cite{genie_v3}, and apply a custom tune developed by adding additional parameters and fitting to T2K $\nu_\mu$CC$0\pi$ data, known as G18\_10a\_02\_11a, or the ``MicroBooNE tune'' \cite{genie-tune-paper}. With this model, we vary 46 underlying model parameters, including those related to the quasi-elastic, meson-exchange-current, resonance, deep-inelastic-scattering, coherent scattering, neutral current, and final state interaction models. 

We consider hadron re-interaction uncertainties, where particles produced by a neutrino interaction can propagate and interact with a different argon nucleus. We simulate this interaction using \textsc{Geant4} via the Bertini intranuclear cascade model \cite{geant_reinteraction}, and account for uncertainties by varying the cross sections for each of $p$, $\pi^+$, and $\pi^-$ around their central values by 20\% using Geant4Reweight \cite{geant4reweight}.

\begin{figure}[H]
    \centering
    \begin{subfigure}[b]{0.43\textwidth}
        \includegraphics[trim=30 0 310 0, clip, width=\textwidth]{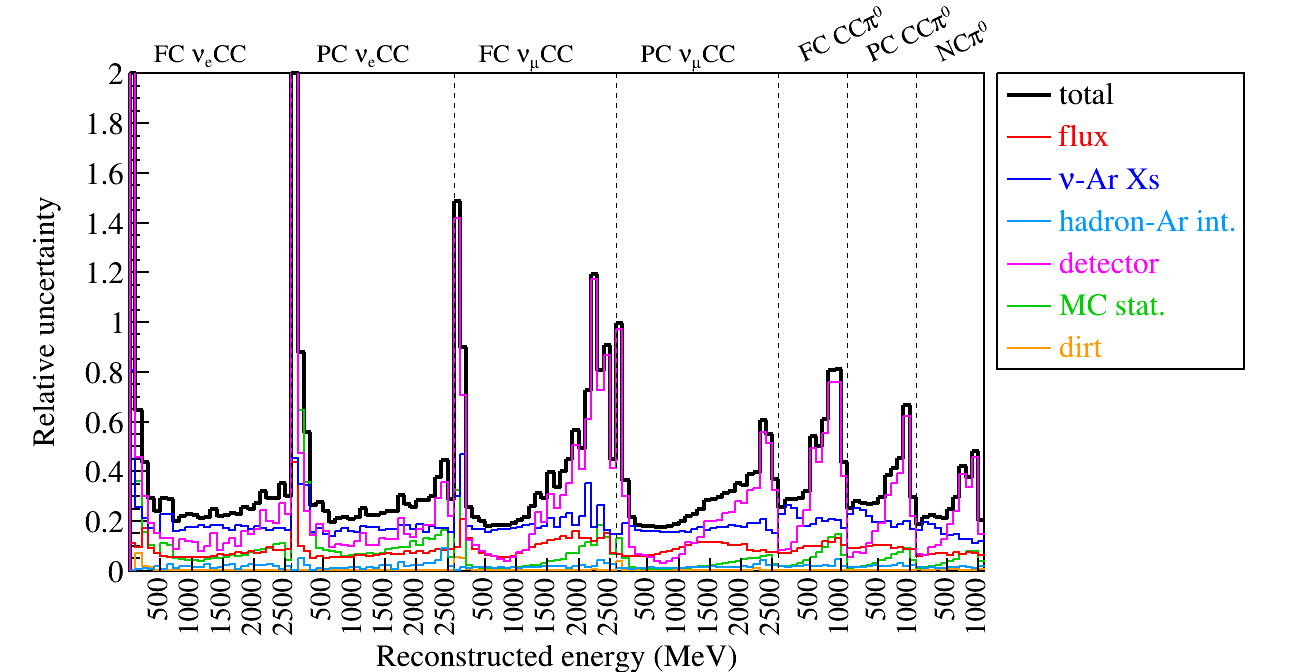}
        \caption{}
    \end{subfigure}
    \begin{subfigure}[b]{0.56\textwidth}
        \includegraphics[trim=30 0 20 0, clip, width=\textwidth]{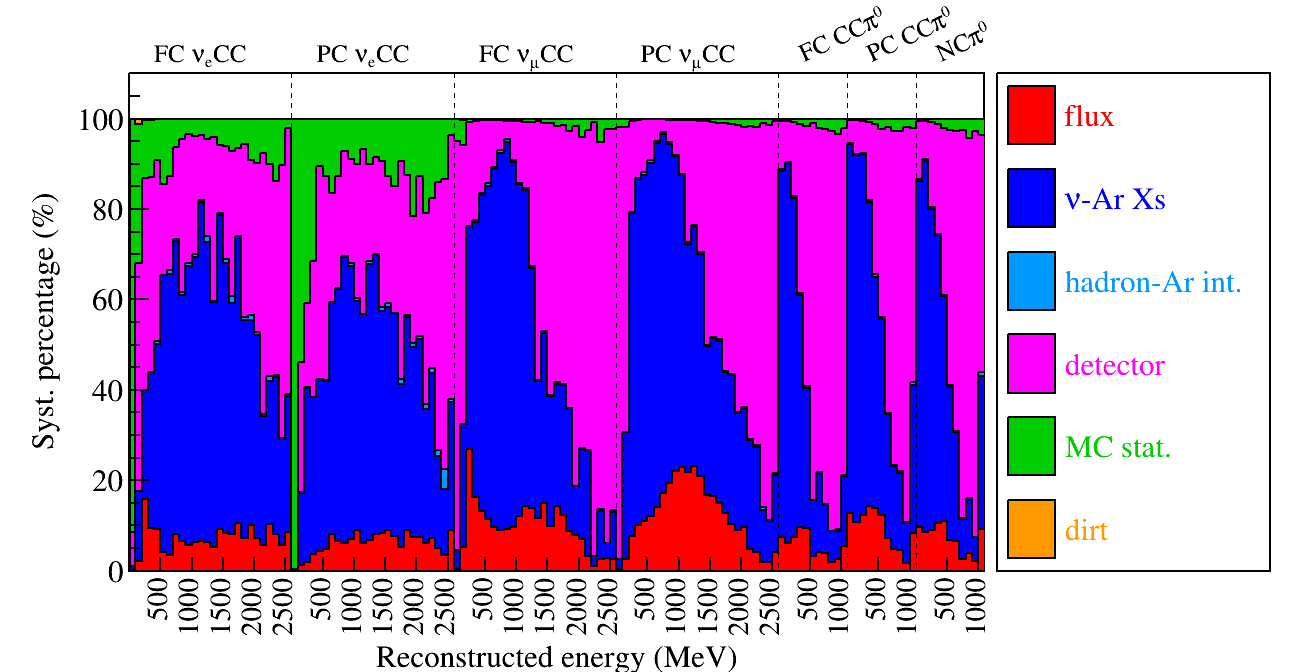}
        \caption{}
    \end{subfigure}
    \caption[$\nu_e$CC LEE systematic uncertainty breakdown]{$\nu_e$CC LEE systematic uncertainty breakdown. Panel (a) shows the fractional uncertainty from each type of systematic uncertainty, as well as the sum from all systematic uncertainties in quadrature in black. Panel (b) shows the relative contributions to each bin Figures from Ref. \cite{wc_elee_prd}.}
    \label{fig:uncertainty_breakdown}
\end{figure}

The relative sizes of these uncertainties are shown in Fig. \ref{fig:uncertainty_breakdown}. We evaluate uncertainties on all seven channels used in this analysis, which include fully contained (FC) and partially contained (PC) channels: $\nu_e$CC FC, $\nu_e$CC PC, $\nu_\mu$CC FC, $\nu_\mu$CC PC, $\nu_\mu$CC$\pi^0$ FC, $\nu_\mu$CC$\pi^0$ PC, and NC $\pi^0$ (which has FC+PC combined). The four $\nu_\mu$CC and $\nu_\mu$CC channels are binned in reconstructed neutrino energy in 25 bins from 0 to 2500 MeV plus an overflow bin, and the three $\pi^0$ channels are binned in reconstructed $\pi^0$ kinetic energy in 10 bins from 0 to 1000 MeV plus an overflow bin. We see that uncertainties are largest at low and high energies, primarily due to low Monte-Carlo statistical uncertainties in the evaluation of detector systematic variations. 

\begin{figure}[H]
    \centering
    \includegraphics[width=0.7\textwidth]{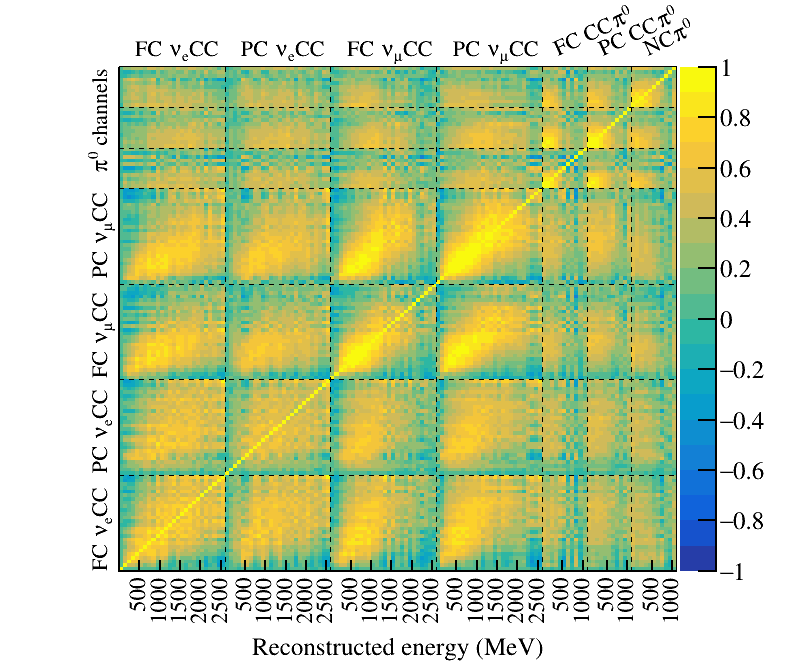}
    \caption[$\nu_e$CC LEE systematic uncertainty correlation matrix]{$\nu_e$CC LEE systematic uncertainty correlation matrix. Figure from Ref. \cite{wc_elee_prd}.}
    \label{fig:uncertainty_correlation}
\end{figure}

These systematic uncertainties are around 20\% at intermediate energies, but importantly, these uncertainties are strongly correlated between different channels as shown in Fig. \ref{fig:uncertainty_correlation}.

\subsection{Conditional Constraint}\label{sec:conditional_constraint}

We exploit these correlations between channels in order to reduce uncertainties in our signal $\nu_e$CC channels after constraining with our $\nu_\mu$CC and $\pi^0$ sideband channels. Specifically, we take our total covariance matrix for all seven channels $\Sigma$, and break it into channels containing the signal $X$ and channels containing constraining sidebands $Y$:

\begin{equation}
    \Sigma = \begin{pmatrix}
        \Sigma^{XX} & \Sigma^{XY} \\
        \Sigma^{YX} & \Sigma^{YY} 
    \end{pmatrix}
\end{equation}

We have a measurement $n$ for each bin, and a prediction $\mu$ for each bin. Using this equation, we can update our prediction and predicted uncertainty:

\begin{eqnarray}
\mu^{X,\text{const.}} &=& \mu^{X} + \Sigma^{XY} \cdot \left(\Sigma^{YY} \right)^{-1} \cdot \left( n^Y - \mu^Y \right) \\
\Sigma^{XX, \text{const.}} &=& \Sigma^{XX} - \Sigma^{XY} \cdot \left(\Sigma^{YY} \right)^{-1} \cdot \Sigma^{YX}.
\end{eqnarray}

\begin{figure}[H]
    \centering
    \includegraphics[width=0.7\textwidth]{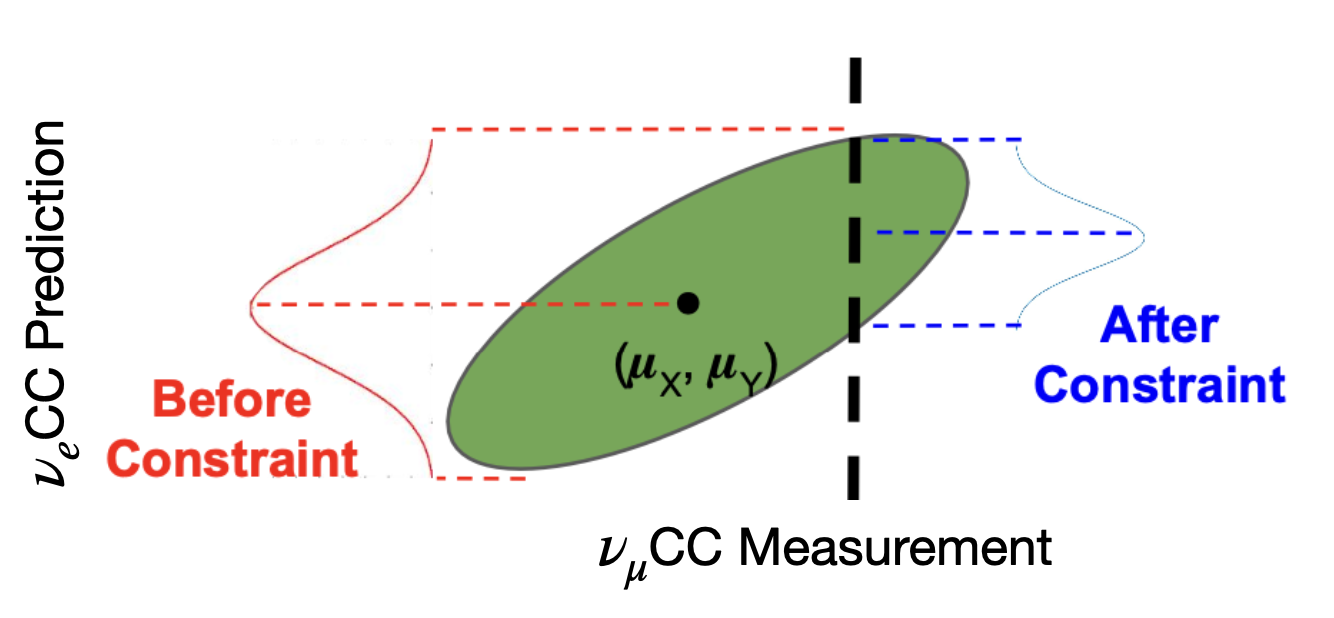}
    \caption[Conditional constraint diagram]{Conditional constraint diagram.}
    \label{fig:constraint_diagram}
\end{figure}

Conceptually, this process is illustrated in Fig. \ref{fig:constraint_diagram}. This shows a simplified example with only two bins, a measurement for $\nu_e$CC and a measurement for $\nu_\mu$CC. The correlation between these variables is illustrated as a tilted ellipse, the measurement of the $\nu_\mu$CC sideband constraining channel is illustrated as a dashed line. Given this measurement, we can see how the uncertainty in the $\nu_e$CC prediction decreases from the red distribution to the blue distribution after the constraint. The equations above let us perform a similar procedure, but now instead of two dimensions, we have 137 $(26 \cdot 4 + 3 \cdot 11)$ dimensions corresponding to every bin of our seven channels.

\begin{figure}[H]
    \centering
    \begin{subfigure}[b]{0.49\textwidth}
        \includegraphics[width=\textwidth]{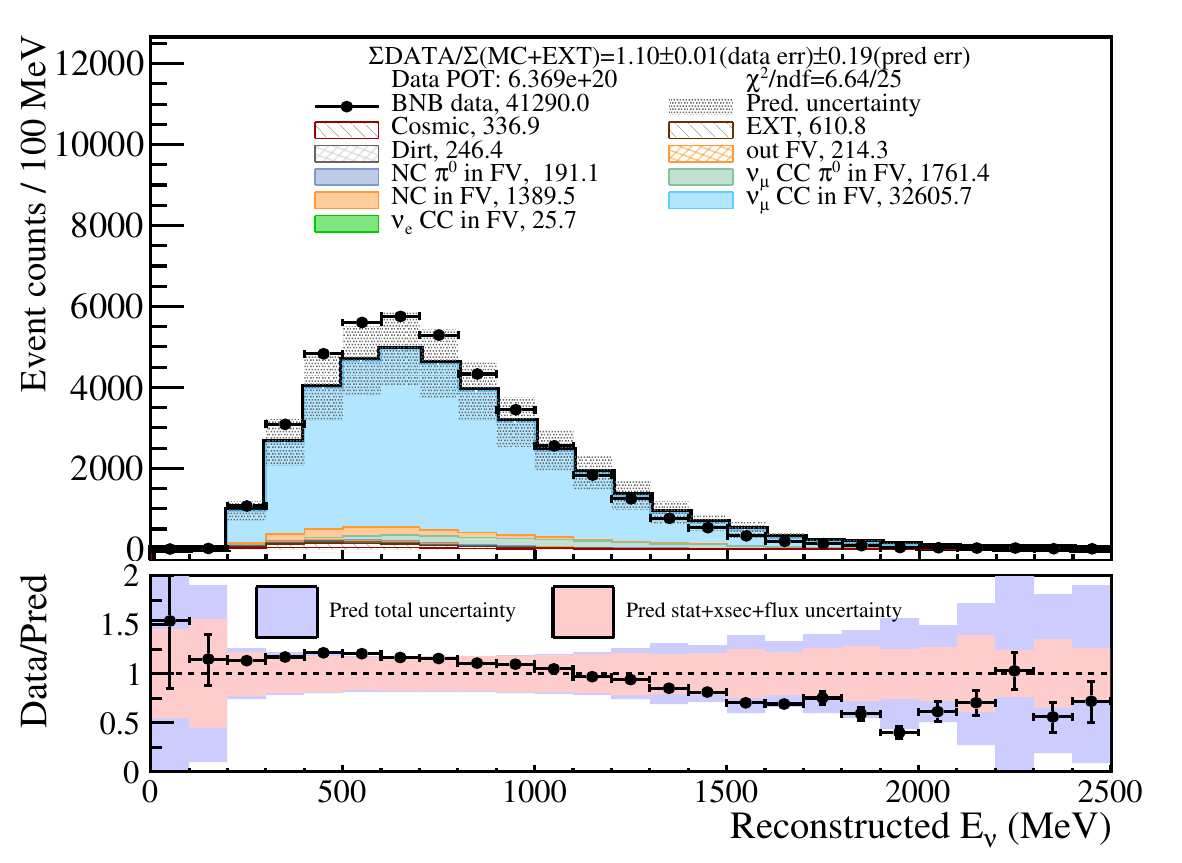}
        \caption{}
        \label{fig:numuCC_FC}
    \end{subfigure}
    \begin{subfigure}[b]{0.49\textwidth}
        \includegraphics[width=\textwidth]{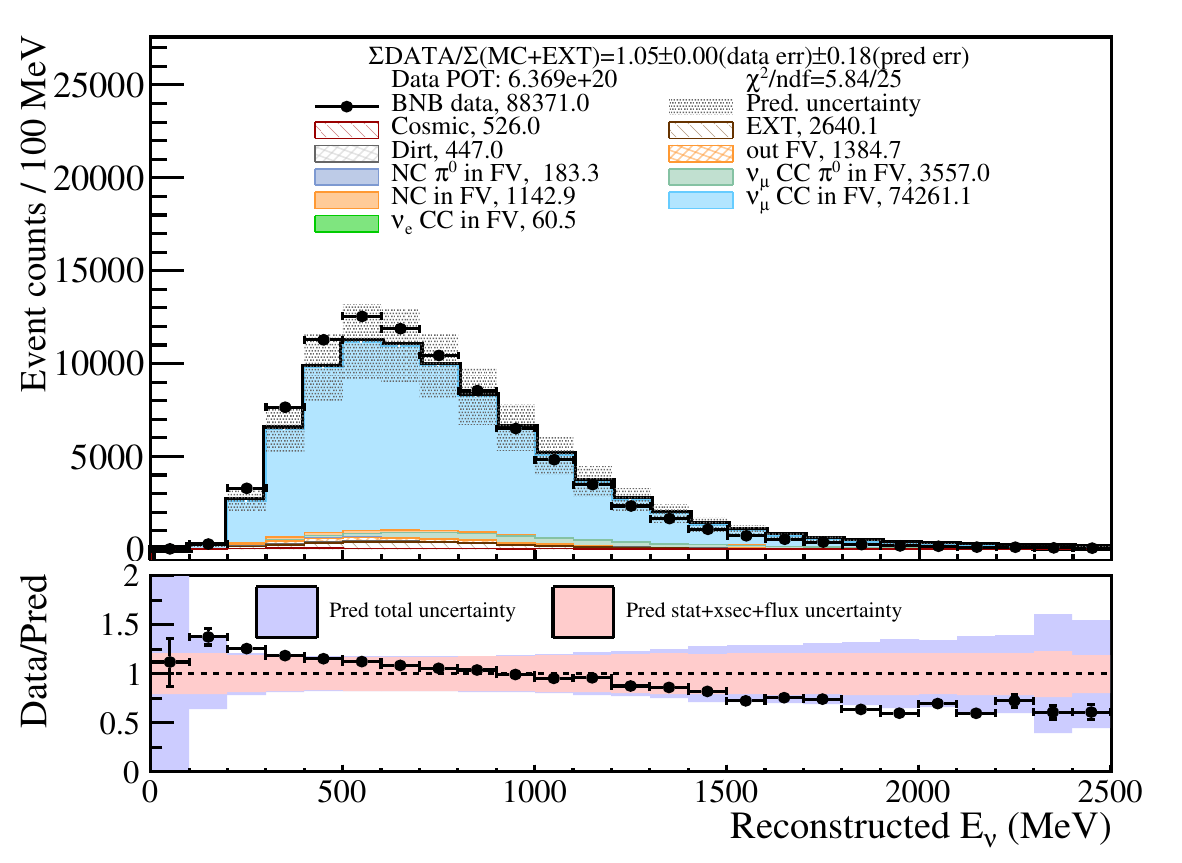}
        \caption{}
        \label{fig:numuCC_PC}
    \end{subfigure}
    \begin{subfigure}[b]{0.49\textwidth}
        \includegraphics[width=\textwidth]{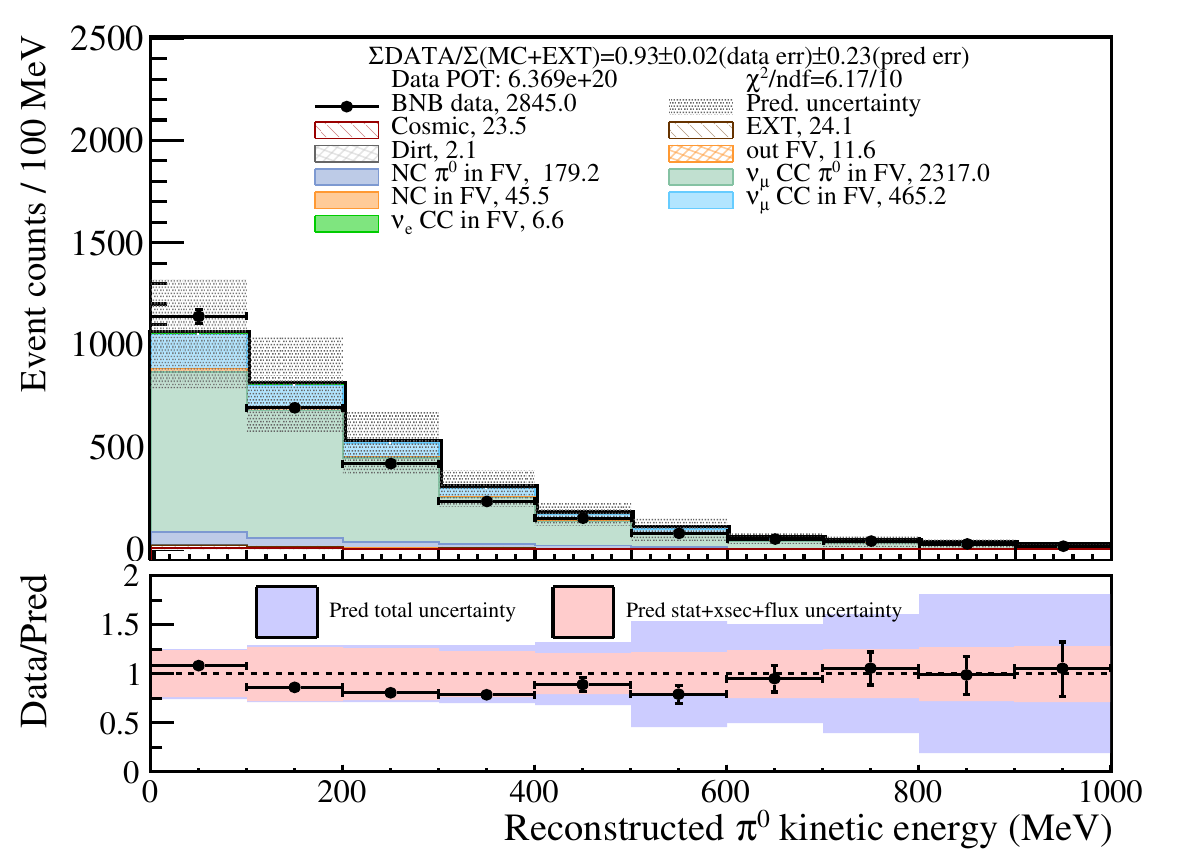}
        \caption{}
        \label{fig:ccpi0_FC_E}
    \end{subfigure}
    \begin{subfigure}[b]{0.49\textwidth}
        \includegraphics[width=\textwidth]{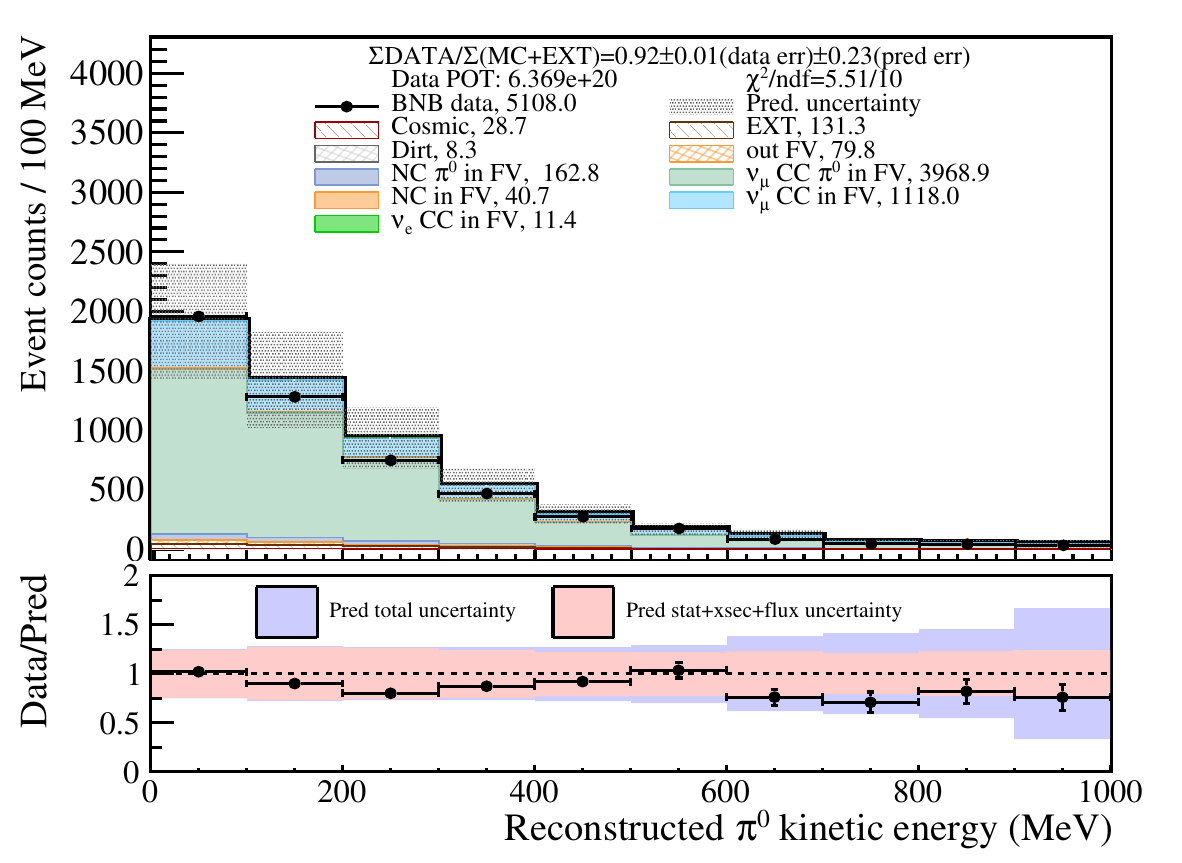}
        \caption{}
        \label{fig:ccpi0_PC_E}
    \end{subfigure}
    \begin{subfigure}[b]{0.49\textwidth}
        \includegraphics[width=\textwidth]{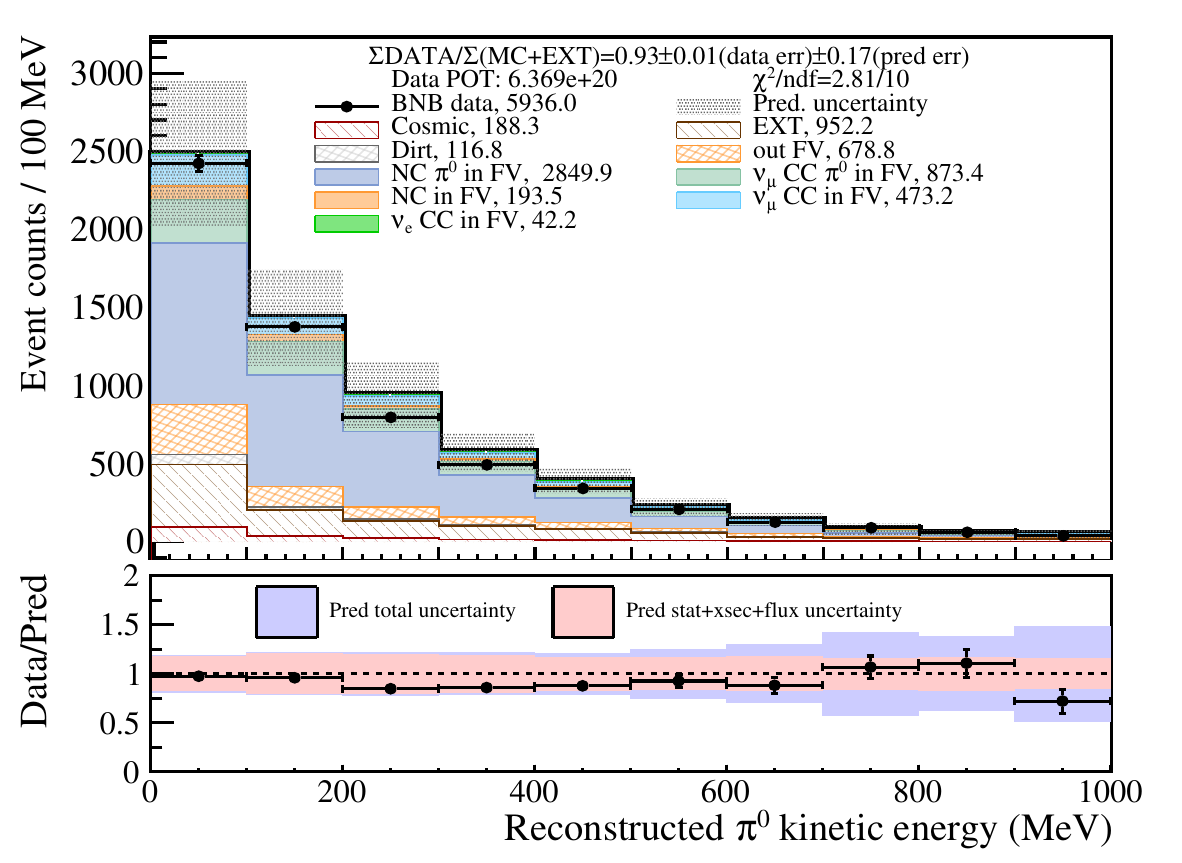}
        \caption{}
        \label{fig:ncpi0_E}
    \end{subfigure}
    \caption[$\nu_e$CC constraining sidebands]{Panel (a) shows the reconstructed neutrino energy distribution for $\nu_\mu$CC FC, and panel (b) shows the distribution for $\nu_\mu$CC PC. Panel (c) shows the reconstructed $\pi^0$ kinetic energy distribution for $\nu_\mu$CC$\pi^0$ FC, panel (d) shows the distribution for $\nu_\mu$CC$\pi^0$ PC, and panel (e) shows the distribution for $\nu_\mu$CC$\pi^0$ FC+PC. Figures from Ref. \cite{wc_elee_prd}.}
    \label{fig:constraining_distributions}
\end{figure}

Figure \ref{fig:constraining_distributions} shows all our sideband measurements used in order to constrain our $\nu_e$CC signal channels.

In order to use this constraining procedure, we need to be confident that our model accurately describes the uncertainties and correlations between channels. To validate this, we do a series of model validation tests, as shown in Table \ref{tab:eLEE_GoF}. These consist of goodness of fit (GoF) $\chi^2/ndf$ tests. We calculate a $\chi^2$ value using the covariance matrix, prediction, and data in each channel, according to this equation, where $M$ is a vector of measured data counts in each bin, $P$ is a vector of predicted data counts in each bin, and $V$ is the covariance matrix giving the correlated systematic uncertainties across all bins:
\begin{equation}
\chi^2 = (M-P)^T \times V^{-1} \times (M-P)
\end{equation}
This $\chi^2$ is an indication of how well our data match our prediction after accounting for uncertainties. When we divide by the number of degrees of freedom, equivalent to the number of bins $ndf$, a $\chi^2/ndf$ value less than one generally indicates good agreement, and can also be used to calculate more quantitative $p$ values and $\sigma$ values. We then calculate this $\chi^2$ this again after updating the prediction and reducing uncertainties from a conditional constraint. We see that in each case, the $\chi^2/ndf$ value stays below one, indicating good agreement between our data and our prediction including all types of systematic uncertainties. This gives us confidence that we can proceed when using these observations in order to constrain the $\nu_e$CC signal channels.

\renewcommand{\baselinestretch}{1.2}
\begin{table}[H]
    \centering
    \footnotesize
    \begin{tabular}{lccl}
        \toprule
        Channel & \makecell{$\chi^2/ndf$\\without constr.} & \makecell{$\chi^2/ndf$\\with constr.} & Notes \\
        \midrule
        $\nu_\mu$CC FC  & 6.64/25 & N/A & No constraint \\
        $\nu_\mu$CC PC  & 5.84/25 & 6.94/25 & Constrained by $\nu_\mu$CC FC \\
        $\nu_\mu$CC$\pi^0$ FC    & 6.17/10 & 7.39/10 & Constrained by both $\nu_\mu$CC FC and $\nu_\mu$CC PC\\
        $\nu_\mu$CC$\pi^0$ PC   & 5.51/10 & 6.80/10 &  Constrained by both $\nu_\mu$CC FC and $\nu_\mu$CC PC\\
        NC$\pi^0$  FC+PC      & 2.81/10 & 5.33/10 &  Constrained by both $\nu_\mu$CC FC and $\nu_\mu$CC PC\\
        \bottomrule
    \end{tabular}
    \caption[$\nu_e$CC constraining channel models validation tests]{$\nu_e$CC constraining channel model validation tests, showing $\chi^2/ndf$ without and with constraints, not including overflow bins. Data statistical uncertainties are calculated using the Pearson rather than CNP formulation. These numbers can be found in Ref. \cite{wc_elee_prd}.}
    \label{tab:eLEE_GoF}
\end{table}
\renewcommand{\baselinestretch}{1}

\subsection{Wire-Cell \texorpdfstring{$\nu_e$}{nue}CC Results}\label{sec:wc_nueCC_results}

Before unblinding our $\nu_e$CC results, we did many careful investigations of our selections and procedure. We investigated a small $5.3\cdot 10^{19}$ POT open data set, and confirmed that many different distributions of data were consistent with our predictions and uncertainties, including in many conditional constraint tests. We also investigated a sample of our NuMI data set. Due to our far off-axis location, we see a much higher $\nu_e/\nu_\mu$ flux ratio in NuMI, and therefore can test our $\nu_e$CC reconstruction on a larger sample of real data without unblinding any selections that are expected to be sensitive to models causing an enhanced BNB $\nu_e$ flux. We practiced our analysis procedures by investigating several fake data sets, where the simulation was altered and handed to analyzers and the nature of the alteration was not revealed until analyzers had made final plots and conclusions.

\begin{figure}[H]
    \centering
    \begin{subfigure}[b]{0.49\textwidth}
        \includegraphics[width=\textwidth]{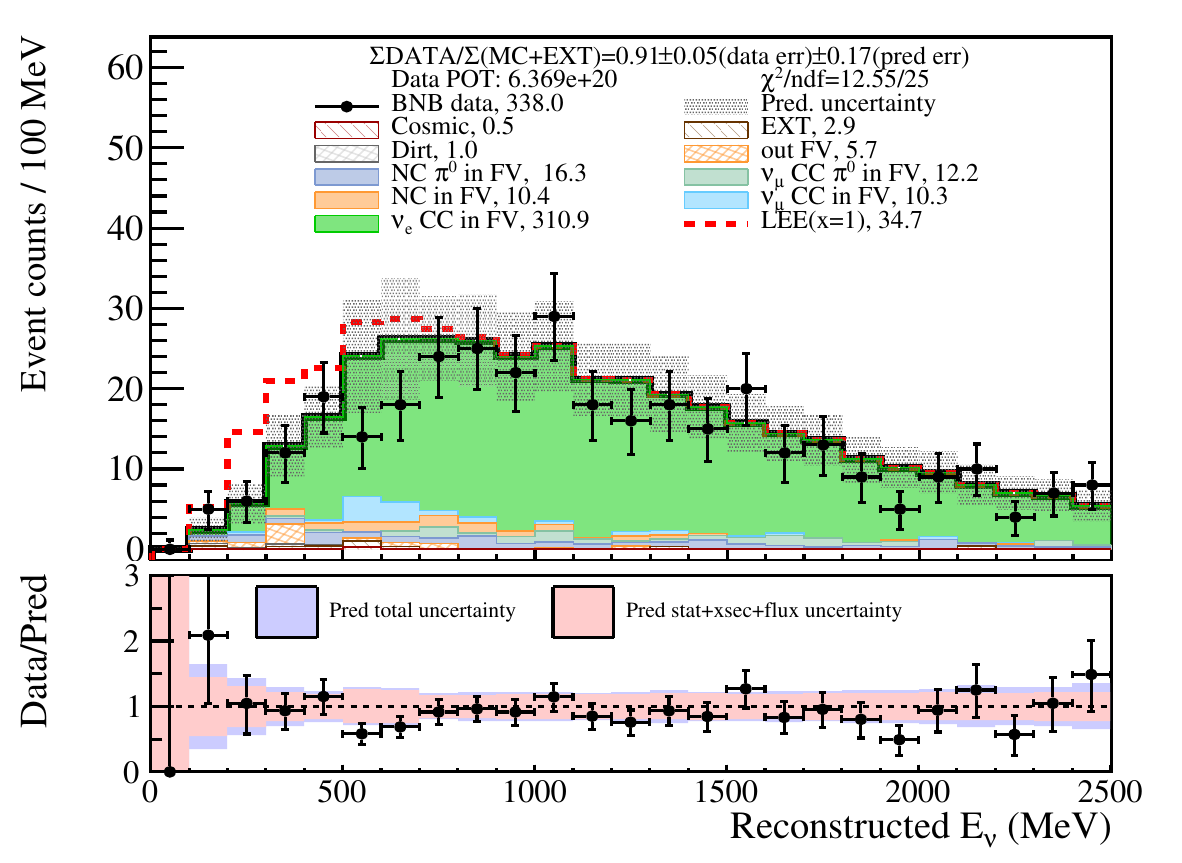}
        \caption{}
    \end{subfigure}
    \begin{subfigure}[b]{0.49\textwidth}
        \includegraphics[width=\textwidth]{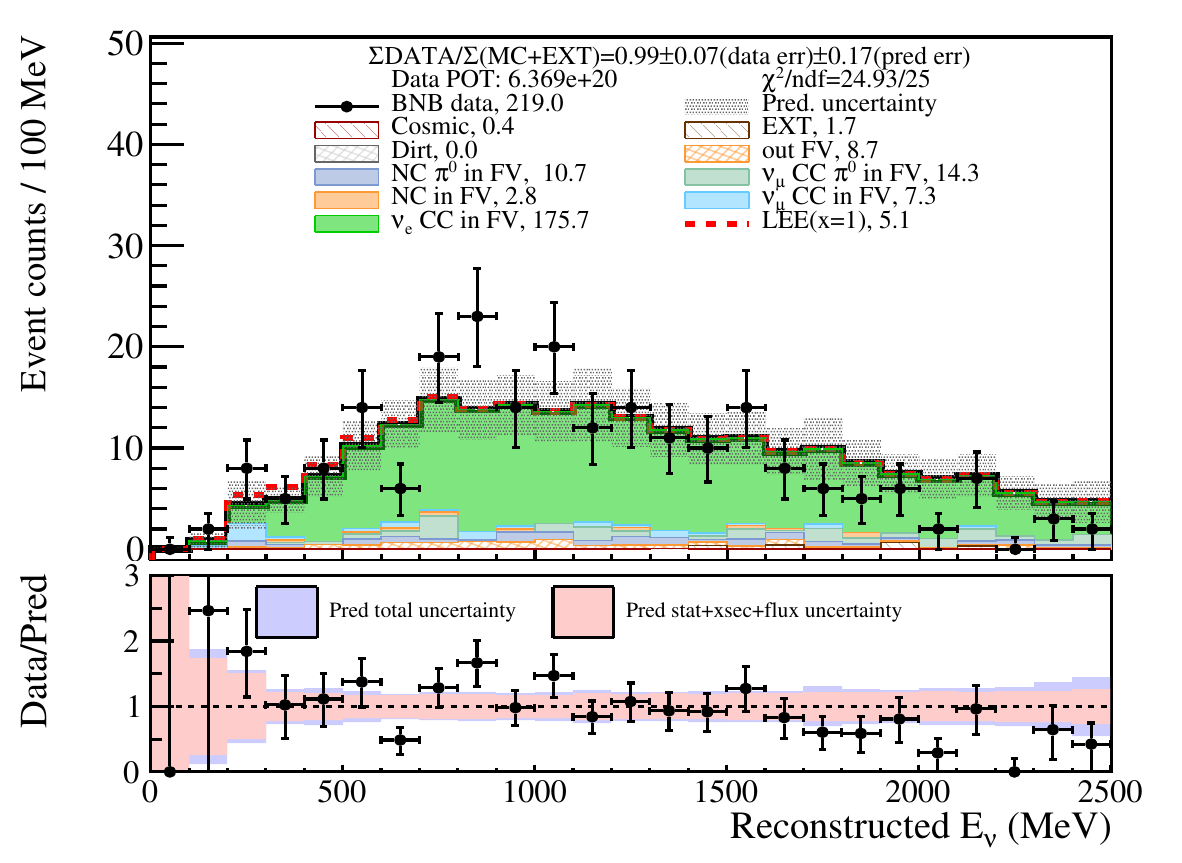}
        \caption{}
    \end{subfigure}
    \caption[$\nu_e$CC energy distributions]{Panel (a) shows the reconstructed neutrino energy distribution for $\nu_e$CC FC, and panel (b) shows the distribution for $\nu_e$CC PC. Figures from Ref. \cite{wc_elee_prd}.}
    \label{fig:nueCC_energies}
\end{figure}

Our final $\nu_e$CC selections on real data are shown in Fig. \ref{fig:nueCC_energies}. Particularly in the FC case, we see that our data match our nominal prediction at the top of the stacked histogram significantly better than the dashed red line indicating the prediction from our median unfolded MiniBooNE LEE model.

\begin{figure}[H]
    \centering
    \begin{subfigure}[b]{0.47\textwidth}
        \includegraphics[width=\textwidth]{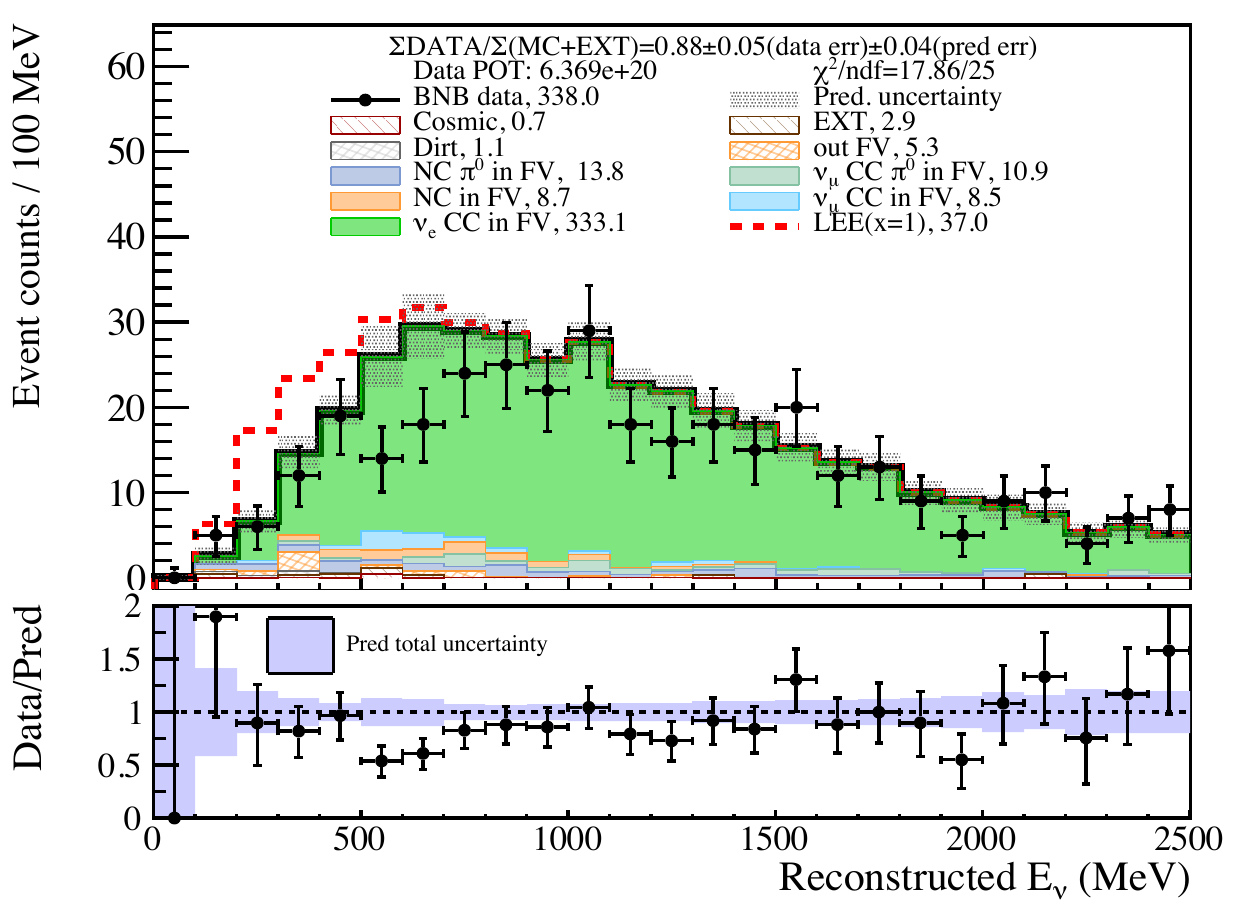}
        \caption{}
    \end{subfigure}
    \begin{subfigure}[b]{0.52\textwidth}
        \includegraphics[trim=0 30 0 0, clip, width=\textwidth]{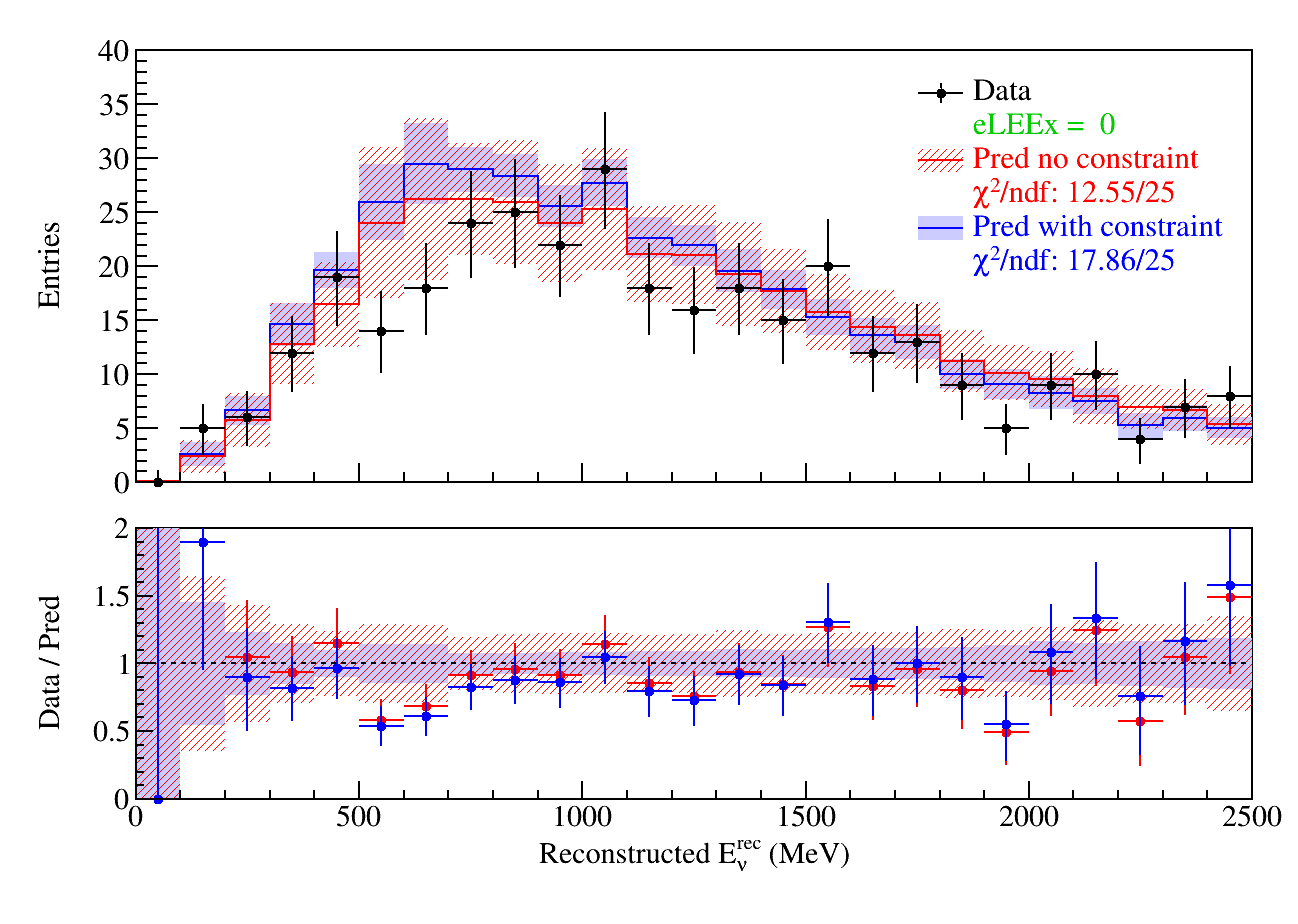}
        \caption{}
    \end{subfigure}
    \caption[Constrained $\nu_e$CC energy distributions]{Panel (a) shows the $\nu_e$CC FC reconstructed neutrino energy distribution after constraints from $\nu_\mu$CC and $\pi^0$ channels. Panel (b) compares the distribution before and after constraints. Figures from Ref. \cite{wc_elee_prd}.}
    \label{fig:nueCC_constrained}
\end{figure}

Figure \ref{fig:nueCC_constrained} shows the $\nu_e$CC FC distribution after constraints from the $\nu_\mu$CC and $\pi^0$ channels, showing a slight increase in the prediction and a significant decrease in systematic uncertainty. Our conclusion that the data matches the nominal prediction significantly better than the LEE prediction remains true.

From these spectra, we can pretty much immediately tell by eye that we are not observing a MiniBooNE-like $\nu_e$ excess. However, it can be hard to judge correlated systematic uncertainties by eye, so we perform several statistical tests to further quantify our conclusions.

\begin{figure}[H]
    \centering
    \begin{subfigure}[b]{0.49\textwidth}
        \includegraphics[trim=10 0 60 0, clip, width=\textwidth]{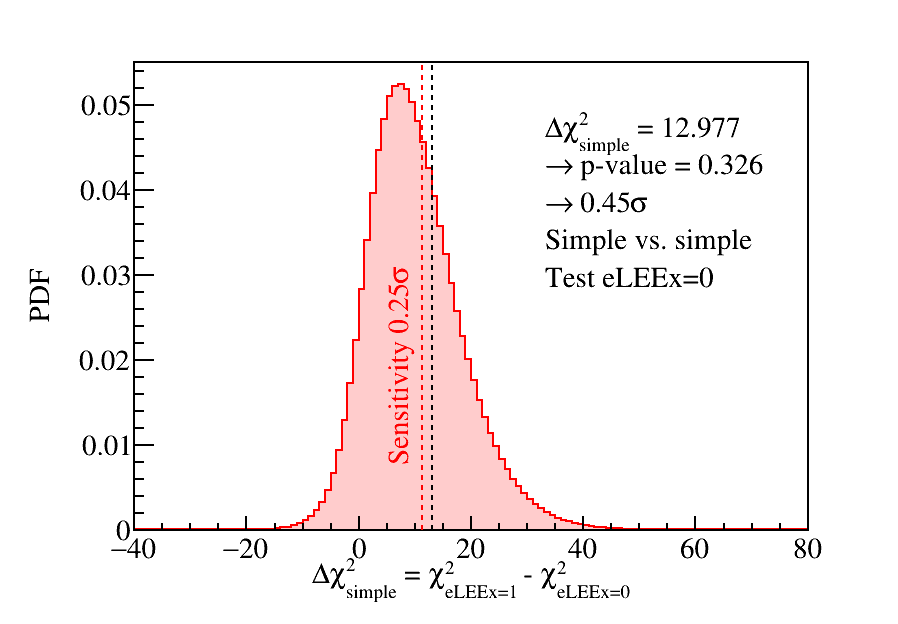}
        \caption{}
    \end{subfigure}
    \begin{subfigure}[b]{0.49\textwidth}
        \includegraphics[trim=10 0 60 0, clip, width=\textwidth]{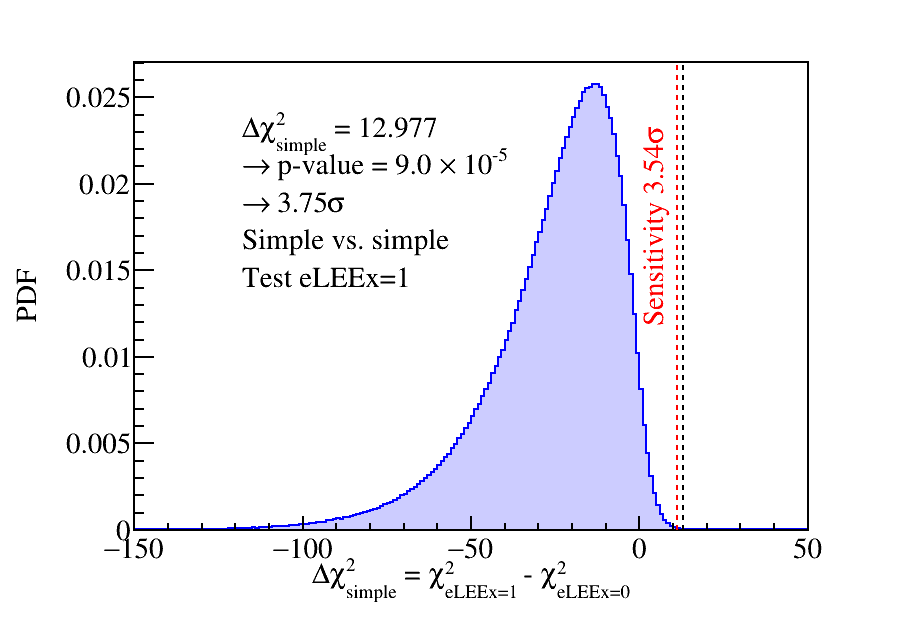}
        \caption{}
    \end{subfigure}
    \caption[$\nu_e$CC two hypothesis test]{Figures from Ref. \cite{wc_elee_prd}.}
    \label{fig:simple_simple}
\end{figure}

First, we perform a two-hypothesis test or ``simple vs simple'' test. Here, we calculate $\Delta\chi^2_\mathrm{simple} = \chi^2_\mathrm{eLEEx=1} - \chi^2_\mathrm{eLEEx=0}$, which characterizes how much better a data set matches the median unfolded LEE prediction or the nominal prediction. Our data has $\Delta\chi^2_\mathrm{simple}=12.977$, and we compare this value with pseudo-experiments with systematic and statistical fluctuations under both the nominal eLEEx=0 hypothesis and the MiniBooNE median unfolded eLEEx=1 hypothesis, as shown in Fig. \ref{fig:simple_simple}. In this statistical test, our data agrees with the nominal prediction at 0.45$\sigma$, and excludes the median unfolded MiniBooNE LEE model at 3.75$\sigma$.

\begin{figure}[H]
    \centering
    \includegraphics[width=0.7\textwidth]{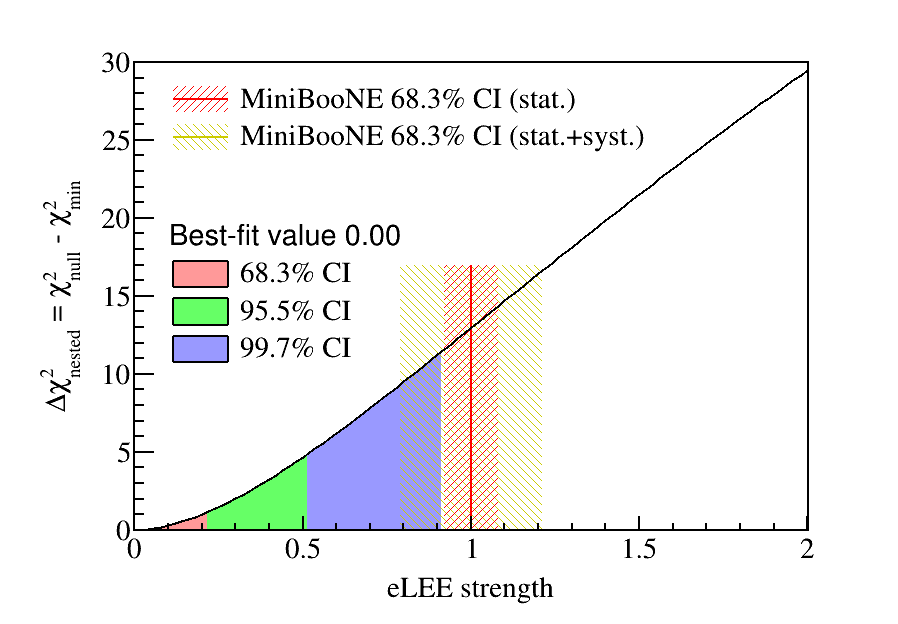}
    \caption[$\nu_e$CC LEEx exclusion]{$\nu_e$CC LEEx exclusion. Figure from Ref. \cite{wc_elee_prd}.}
    \label{fig:LEEx_exclusion}
\end{figure}

We also assess what scaling factors of the median unfolded MiniBooNE LEE model are consistent with our data. We do this by calculating $\chi^2_\mathrm{nested}=\chi^2_\mathrm{null}-\chi^2_\mathrm{min}$, the difference between the $\chi^2$ at the null hypothesis and the $\chi^2$ at the eLEE scaling value that minimizes $\chi^2$. This $\chi^2_\mathrm{nested}$ is calculated for null hypotheses at many different eLEE scaling values as shown for the black curve in Fig. \ref{fig:LEEx_exclusion}. At each of these eLEE scaling values, we also calculate the distribution of $\chi^2_\mathrm{nested}$ for many pseudo-experiments, and compare this distribution with data via the Feldman-Cousins method \cite{feldman_cousins}, giving us $1\sigma$, $2\sigma$, and $3\sigma$ confidence interval ranges as shown in red, green, and blue in Fig. \ref{fig:LEEx_exclusion}. We also estimate an uncertainty on the MiniBooNE band by using the reported significance of the excess, as shown as the yellow hashed region around the eLEE strength value of 1 in Fig. \ref{fig:LEEx_exclusion}.

\begin{figure}[H]
    \centering
    \includegraphics[width=0.7\textwidth]{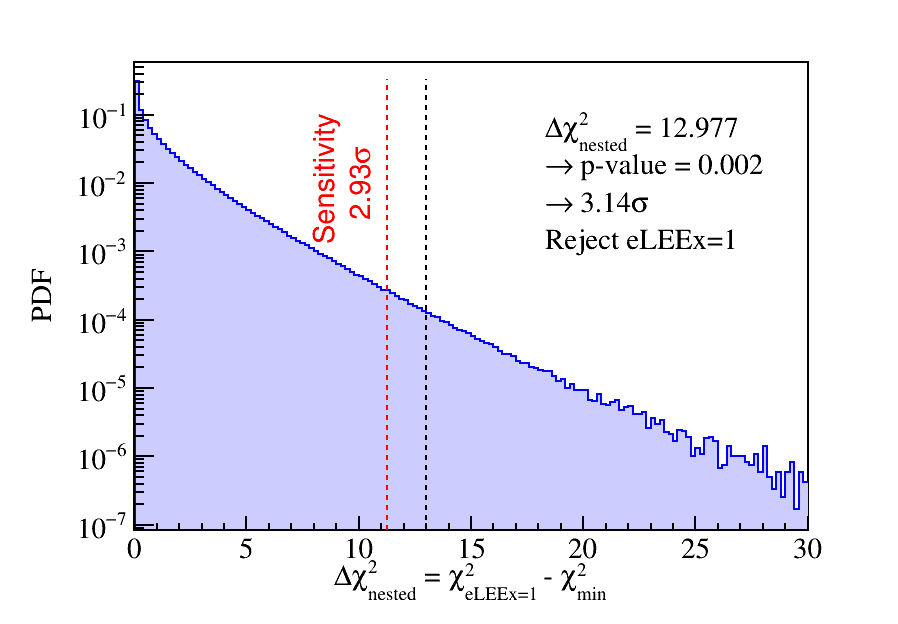}
    \caption[$\nu_e$CC nested likelihood ratio test]{$\nu_e$CC nested likelihood ratio test. Figure from Ref. \cite{wc_elee_prd}.}
    \label{fig:nested_likelihood}
\end{figure}

In Fig. \ref{fig:nested_likelihood}, we specifically examine the $\chi^2_\mathrm{nested}$ distribution for eLEEx=1, the median unfolded MiniBooNE enhanced $\nu_e$ flux. We see that this hypothesis is rejected at 3.14$\sigma$ significance.

\begin{figure}[H]
    \centering
    \begin{subfigure}[b]{0.49\textwidth}
        \includegraphics[width=\textwidth]{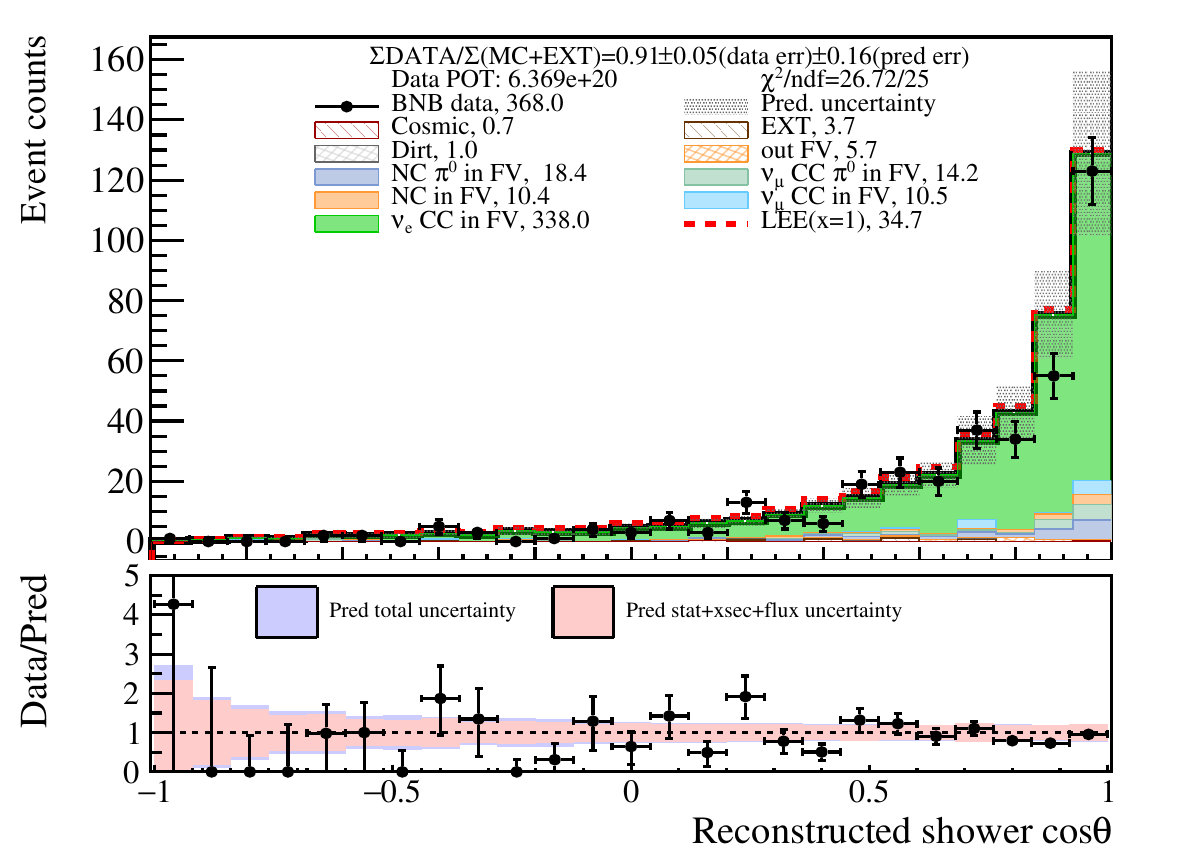}
        \caption{}
    \end{subfigure}
    \begin{subfigure}[b]{0.49\textwidth}
        \includegraphics[width=\textwidth]{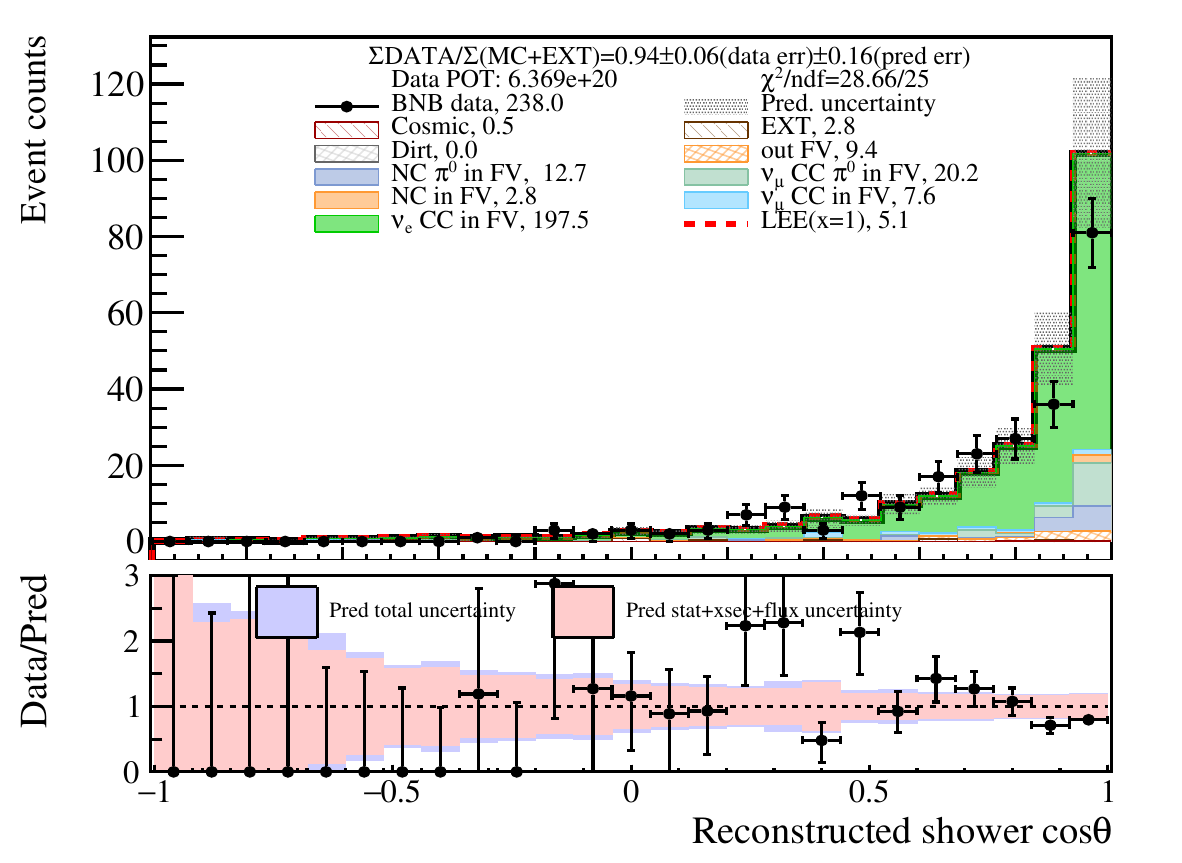}
        \caption{}
    \end{subfigure}
    \caption[$\nu_e$CC shower angle distributions]{Panel (a) shows the reconstructed shower angle distribution for $\nu_e$CC FC, and panel (b) shows the distribution for $\nu_e$CC PC. Figures from Ref. \cite{wc_elee_prd}.}
    \label{fig:nueCC_shower_angles}
\end{figure}

The primary distribution used to search for evidence of a $\nu_e$ excess is the reconstructed neutrino energy, since the excess at low energy is the primary feature of the excess identified by MiniBooNE. However, we are also able to look at the shower angle distribution, since MiniBooNE saw that the excess is somewhat concentrated at forward angles, as shown in Fig. \ref{fig:miniboone_angle}. Looking at our data in Fig. \ref{fig:nueCC_shower_angles}, we again see no sign of an excess at forward angles.

\begin{figure}[H]
    \centering
    \begin{subfigure}[b]{0.49\textwidth}
        \includegraphics[trim=10 0 18 10, clip, width=\textwidth]{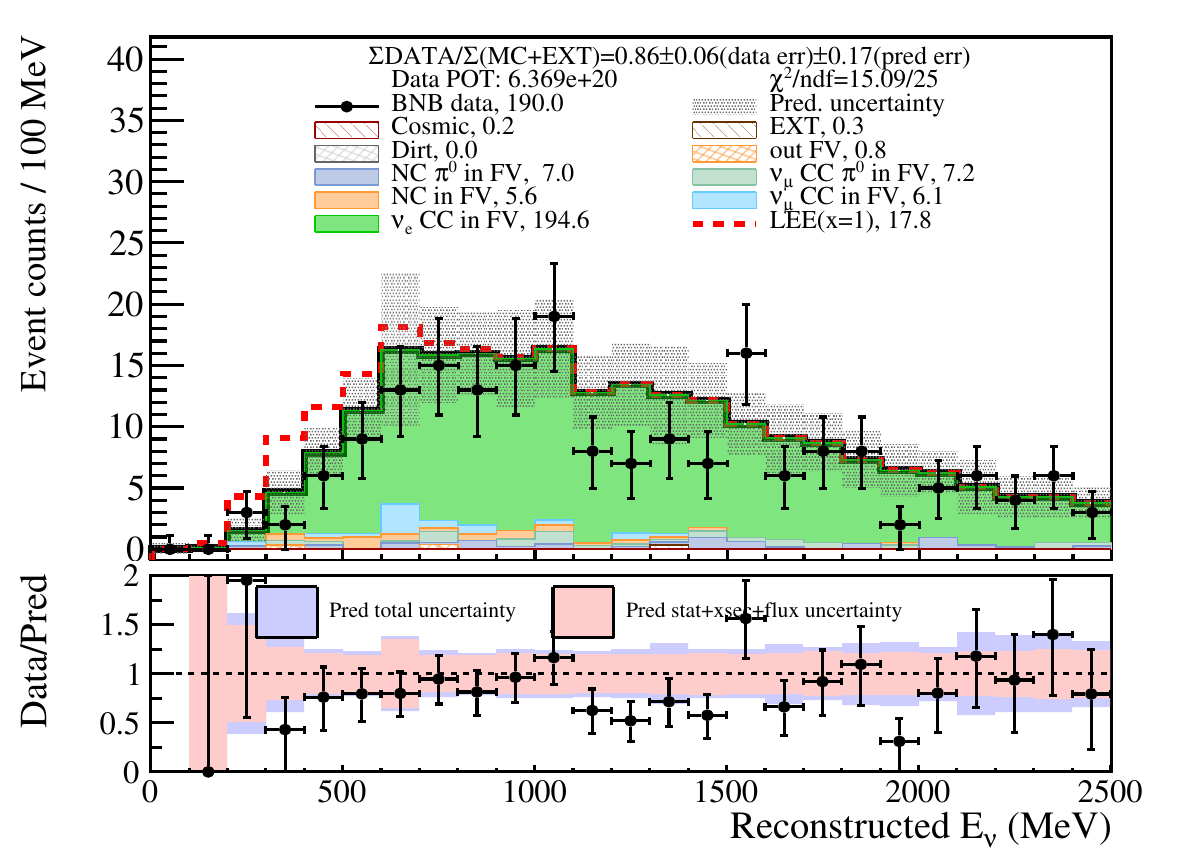}
        \caption{}
    \end{subfigure}
    \begin{subfigure}[b]{0.49\textwidth}
        \includegraphics[trim=10 0 18 10, clip, width=\textwidth]{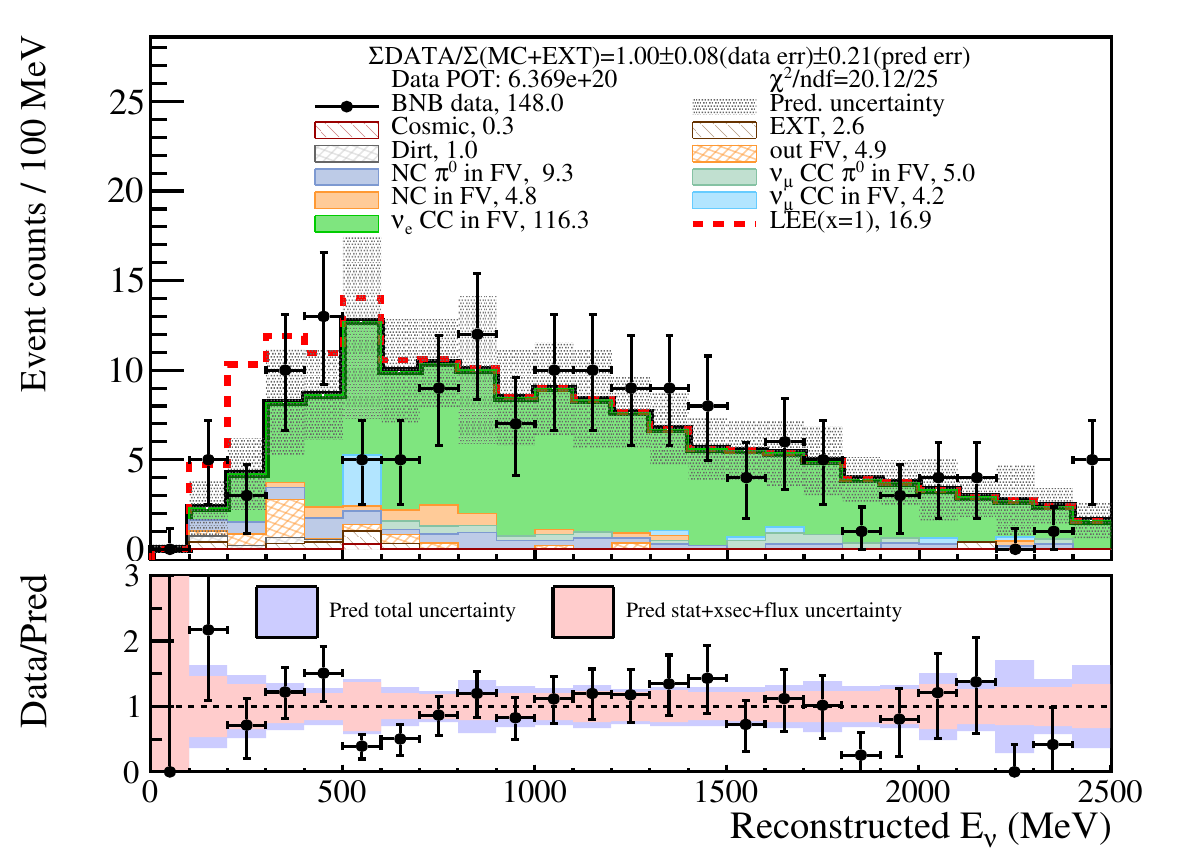}
        \caption{}
    \end{subfigure}
    \begin{subfigure}[b]{0.49\textwidth}
        \includegraphics[trim=10 0 18 10, clip, width=\textwidth]{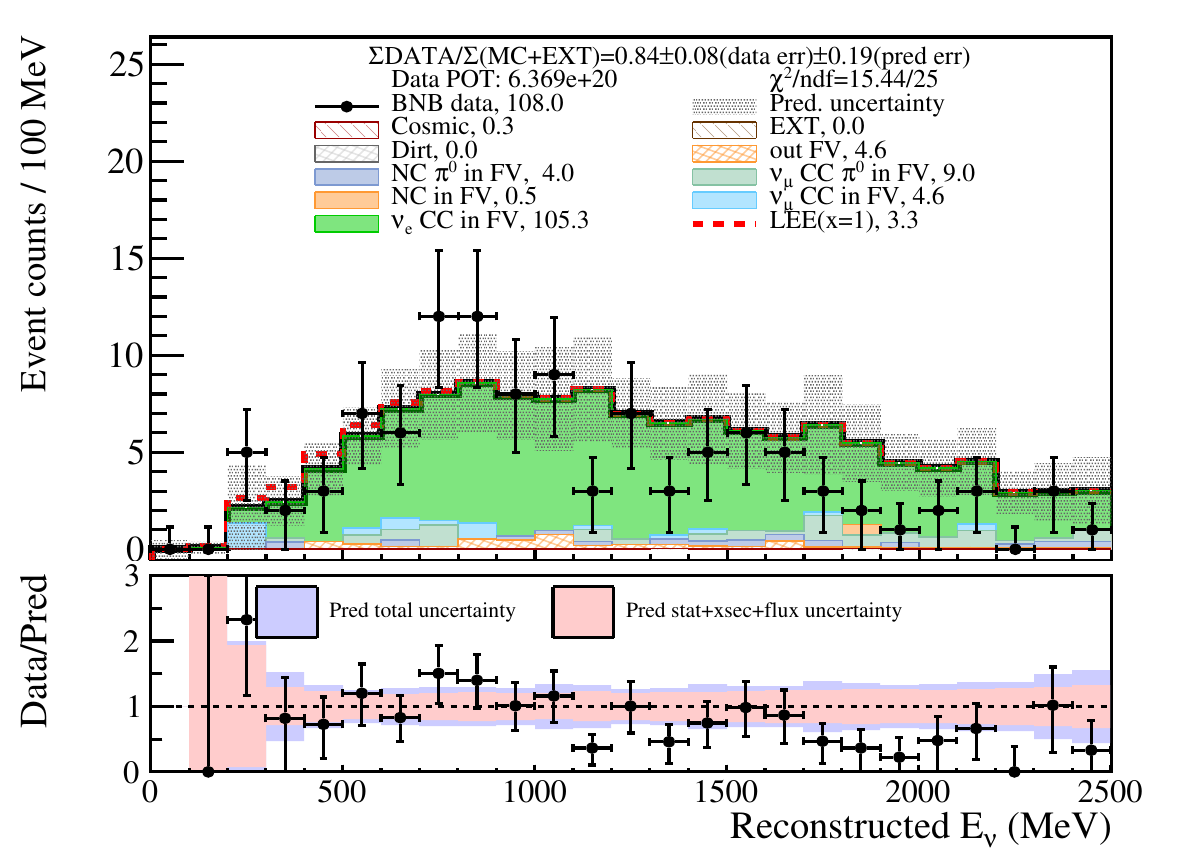}
        \caption{}
    \end{subfigure}
    \begin{subfigure}[b]{0.49\textwidth}
        \includegraphics[trim=10 0 18 10, clip, width=\textwidth]{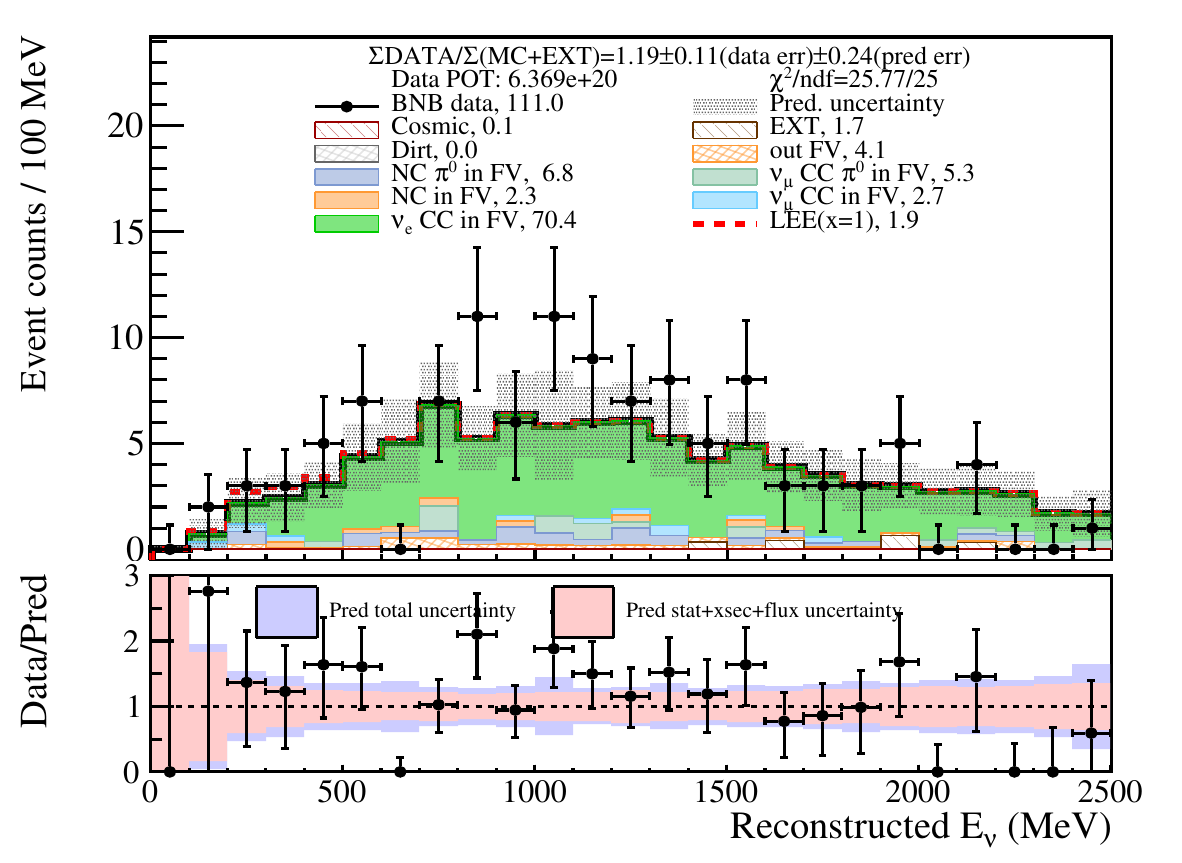}
        \caption{}
    \end{subfigure}
    \caption[$\nu_e$CC $Np$ and $0p$ energy distributions]{Reconstructed neutrino energy distributions for $\nu_e$CC selections, split by containment in the detector and by proton multiplicity.
        Panel (a) shows $\nu_e$CC FC $Np$. 
        Panel (b) shows $\nu_e$CC FC $0p$. 
        Panel (c) shows $\nu_e$CC PC $Np$. 
        Panel (d) shows $\nu_e$CC PC $0p$. 
        Figures from Ref. \cite{wc_elee_prd}.}
    \label{fig:nue_0pNp_spectra}
\end{figure}

\begin{figure}[H]
    \centering
    \begin{subfigure}[b]{0.49\textwidth}
        \includegraphics[trim=10 0 18 10, clip, width=\textwidth]{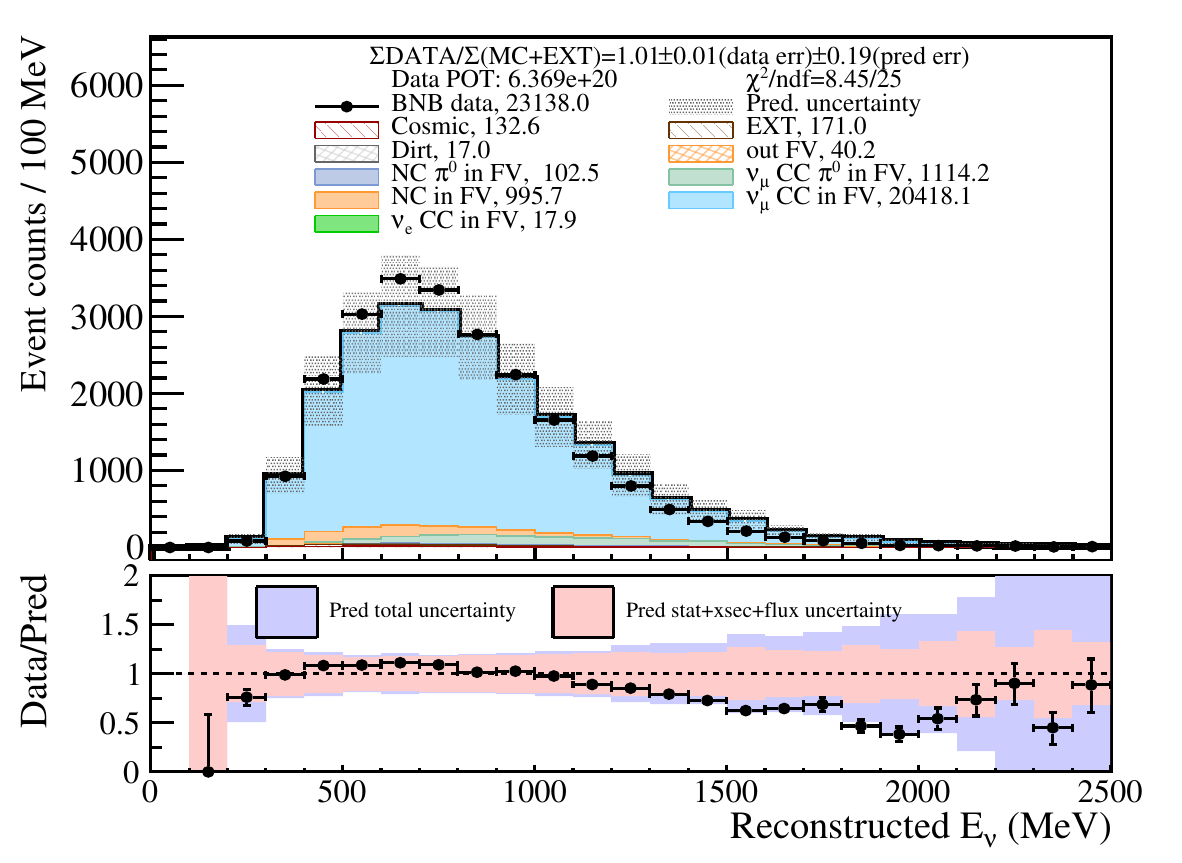}
        \caption{}
    \end{subfigure}
    \begin{subfigure}[b]{0.49\textwidth}
        \includegraphics[trim=10 0 18 10, clip, width=\textwidth]{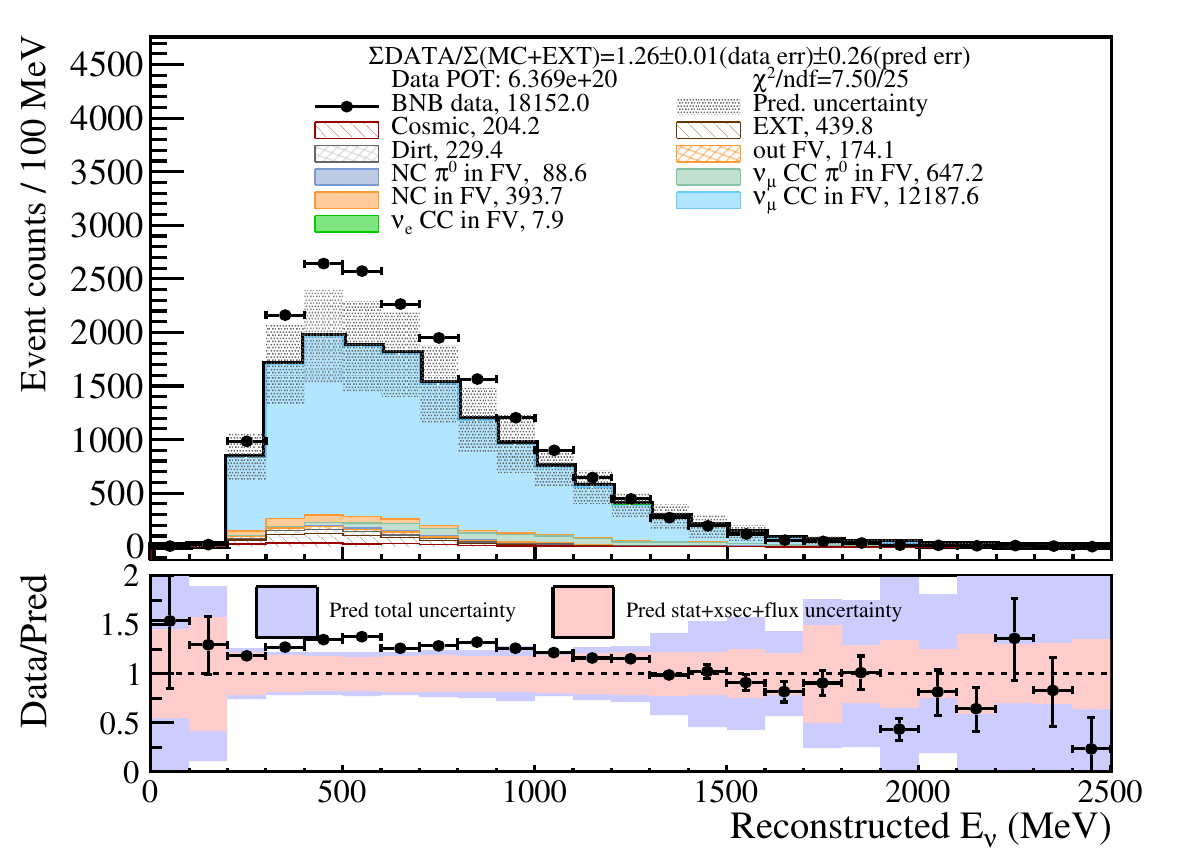}
        \caption{}
    \end{subfigure}
    \begin{subfigure}[b]{0.49\textwidth}
        \includegraphics[trim=10 0 18 10, clip, width=\textwidth]{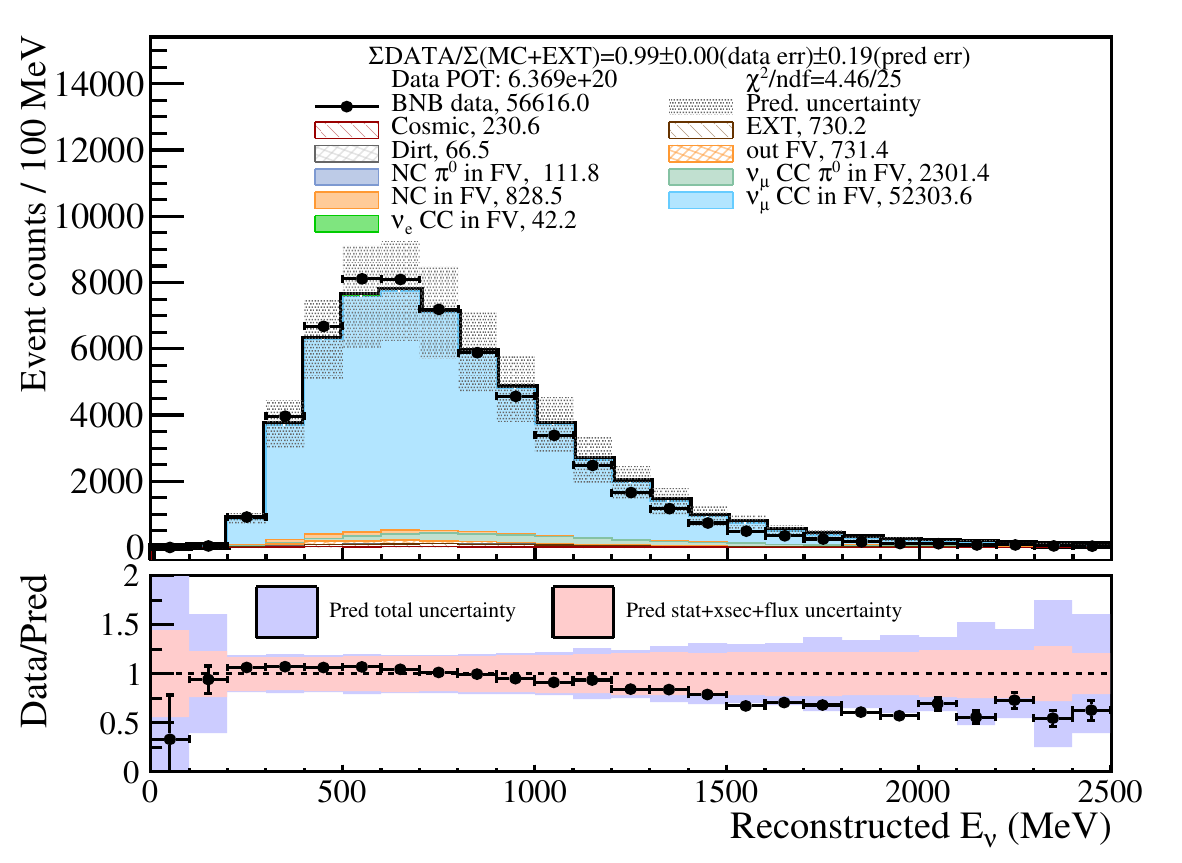}
        \caption{}
    \end{subfigure}
    \begin{subfigure}[b]{0.49\textwidth}
        \includegraphics[trim=10 0 18 10, clip, width=\textwidth]{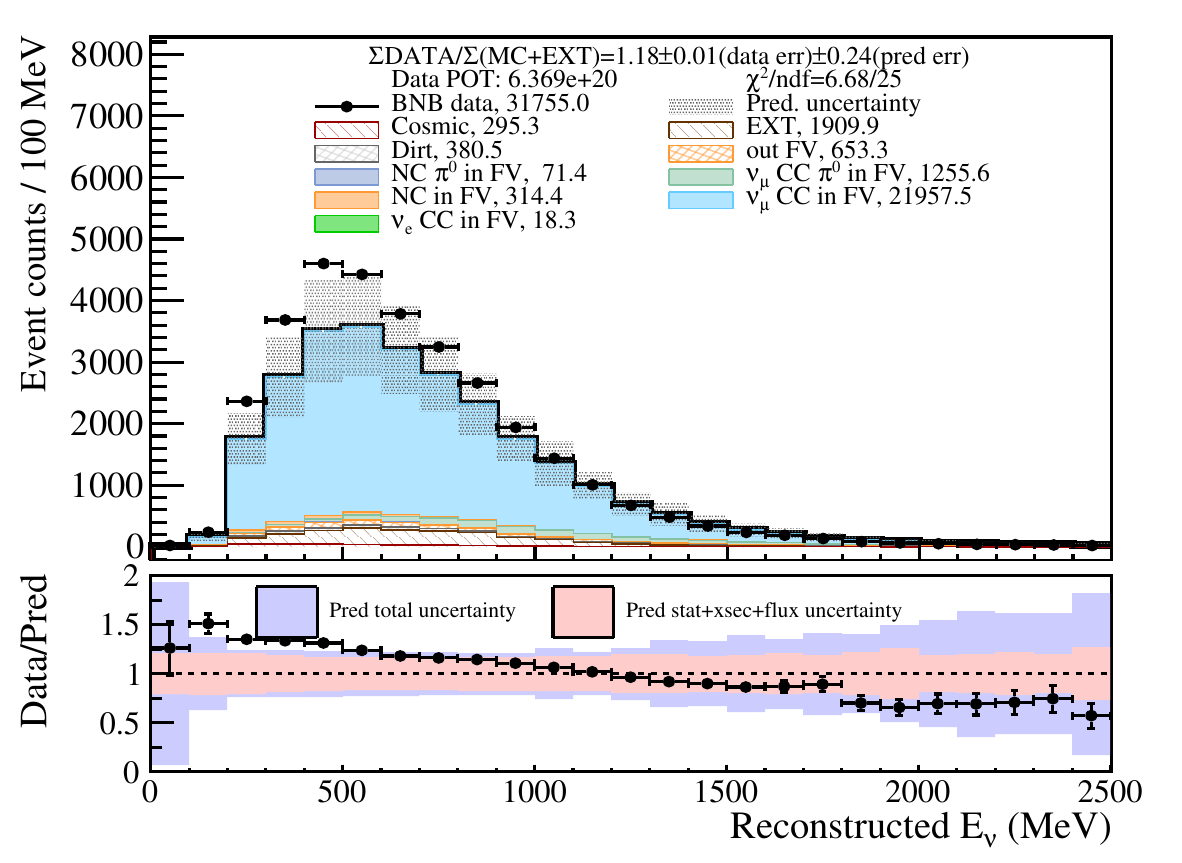}
        \caption{}
    \end{subfigure}
    \caption[$\nu_\mu$CC $Np$ and $0p$ energy distributions]{Reconstructed neutrino energy distributions for $\nu_\mu$CC selections, split by containment in the detector and by proton multiplicity.
        Panel (a) shows $\nu_\mu$CC FC $Np$. 
        Panel (b) shows $\nu_\mu$CC FC $0p$. 
        Panel (c) shows $\nu_\mu$CC PC $Np$. 
        Panel (d) shows $\nu_\mu$CC PC $0p$. 
        Figures from Ref. \cite{wc_elee_prd}.}
    \label{fig:numu_0pNp_spectra}
\end{figure}

We also investigated the proton multiplicity, a variable that MiniBooNE could not measure with their Cherenkov technology. To define $Np$ (events with reconstructed protons) and $0p$ (events without reconstructed protons), we use a 35 MeV reconstructed energy threshold, since that is the energy for a proton to travel about 1 cm and therefore cross a few of our 3 mm wire spacings, giving us a track where $dQ/dx$ can be measured and particle identification can be performed. We split our $\nu_e$CC and $\nu_\mu$CC FC and PC samples into $0p$ and $Np$, as shown in Figs. \ref{fig:nue_0pNp_spectra}-\ref{fig:numu_0pNp_spectra}. We see no sign of an excess in either $\nu_e$CC $Np$ or $\nu_e$CC $0p$. We can notice some interesting trends in proton multiplicity for $\nu_\mu$CC distributions, but these are within systematic uncertainties, and have been interpreted as a cross section effect in Refs. \cite{uboone_Np0p_PRD, uboone_Np0p_PRL}.

\begin{figure}[H]
    \centering
    \includegraphics[width=0.6\textwidth]{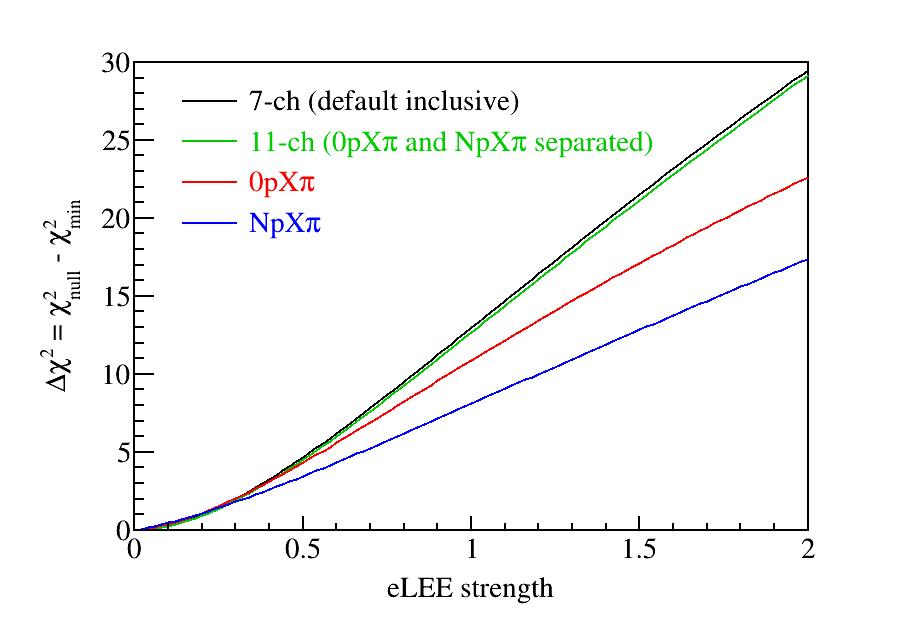}
    \caption[$\nu_e$CC 0p Np LEEx exclusion]{$\nu_e$CC nested likelihood ratio test with and without selections split by proton multiplicity. Figure from Ref. \cite{wc_elee_prd}.}
    \label{fig:0pNp_exclusion}
\end{figure}

We also use these distributions split by proton multiplicity in order to study the compatibility with different median unfolded MiniBooNE LEE model scalings. This results in four $\nu_e$CC channels, four $\nu_\mu$CC channels, and three $\pi^0$ channels, forming 11 channels total, which we compare to our original 7 channels in Fig. \ref{fig:0pNp_exclusion}. We see very consistent behavior, showing that this splitting on proton multiplicity does not significantly change any of our conclusions about this median unfolded MiniBooNE LEE model.

I synthesized the results from this Wire-Cell $\nu_e$CC search into the data release available at \url{https://www.hepdata.net/record/ins1953539}.

\subsection{Other \texorpdfstring{$\nu_e$}{nue}CC Reconstruction Results}

In addition to the Wire-Cell inclusive $\nu_e$CC search described above, MicroBooNE simultaneously had two other analyses searching for a $\nu_e$CC excess. These used the same median unfolded MiniBooNE LEE model described in Sec. \ref{sec:lee_model}, but different reconstructions and different final state topologies.

\begin{figure}[H]
    \centering
    \begin{subfigure}[b]{0.49\textwidth}
        \includegraphics[width=\textwidth]{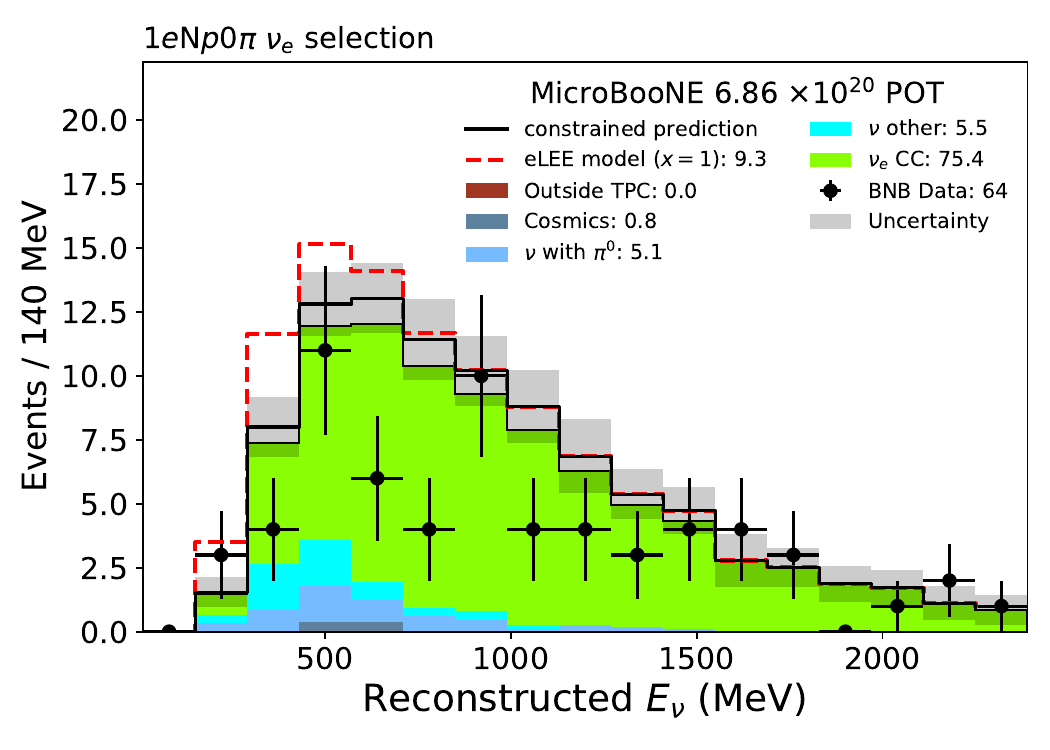}
        \caption{}
    \end{subfigure}
    \begin{subfigure}[b]{0.49\textwidth}
        \includegraphics[width=\textwidth]{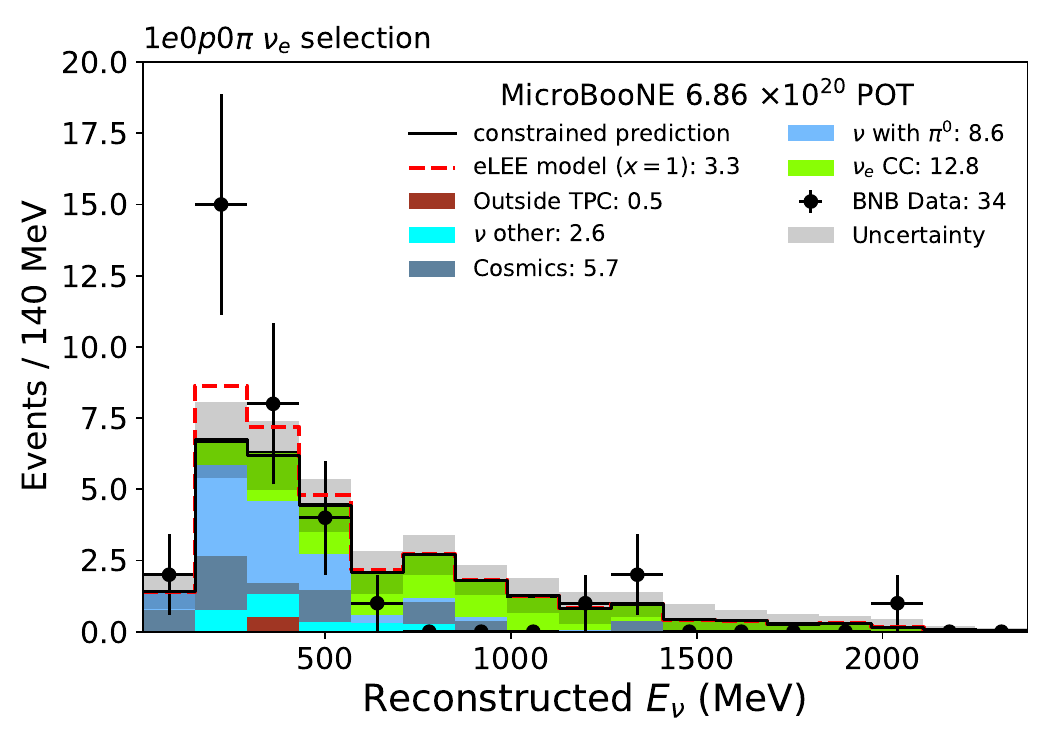}
        \caption{}
    \end{subfigure}
    \caption[$\nu_e$CC $0\pi$ results from Pandora reconstruction]{$\nu_e$CC $0\pi$ results from Pandora reconstruction. Panel (a) shows events with one or more reconstructed protons, and panel (b) shows events without reconstructed protons. Figures from Ref. \cite{eLEE_multiple_reconstructions_PRL}.}
    \label{fig:pandora_nue}
\end{figure}

There was an analysis based on Pandora reconstruction \cite{microboone_pandora} and targeting a pionless topology, split into $Np$ and $0p$. This topology most closely matches MiniBooNE's selection, since MiniBooNE removed events with charged pions which resembled muons, and removed neutral pions which create additional electromagnetic showers. The results are described in Ref. \cite{pandora_eLEE_PRD} and the resulting $\nu_e$CC reconstructed energy distributions are shown in Fig. \ref{fig:pandora_nue}. We see that in the $1eNp$ channel, this analysis saw no sign of an excess, and in the $1e0p$ channel, there is an underprediction at low energies, but it is not very significant, and it only takes place in a bin which is dominated by photon backgrounds.

\begin{figure}[H]
    \centering
    \includegraphics[width=0.49\textwidth]{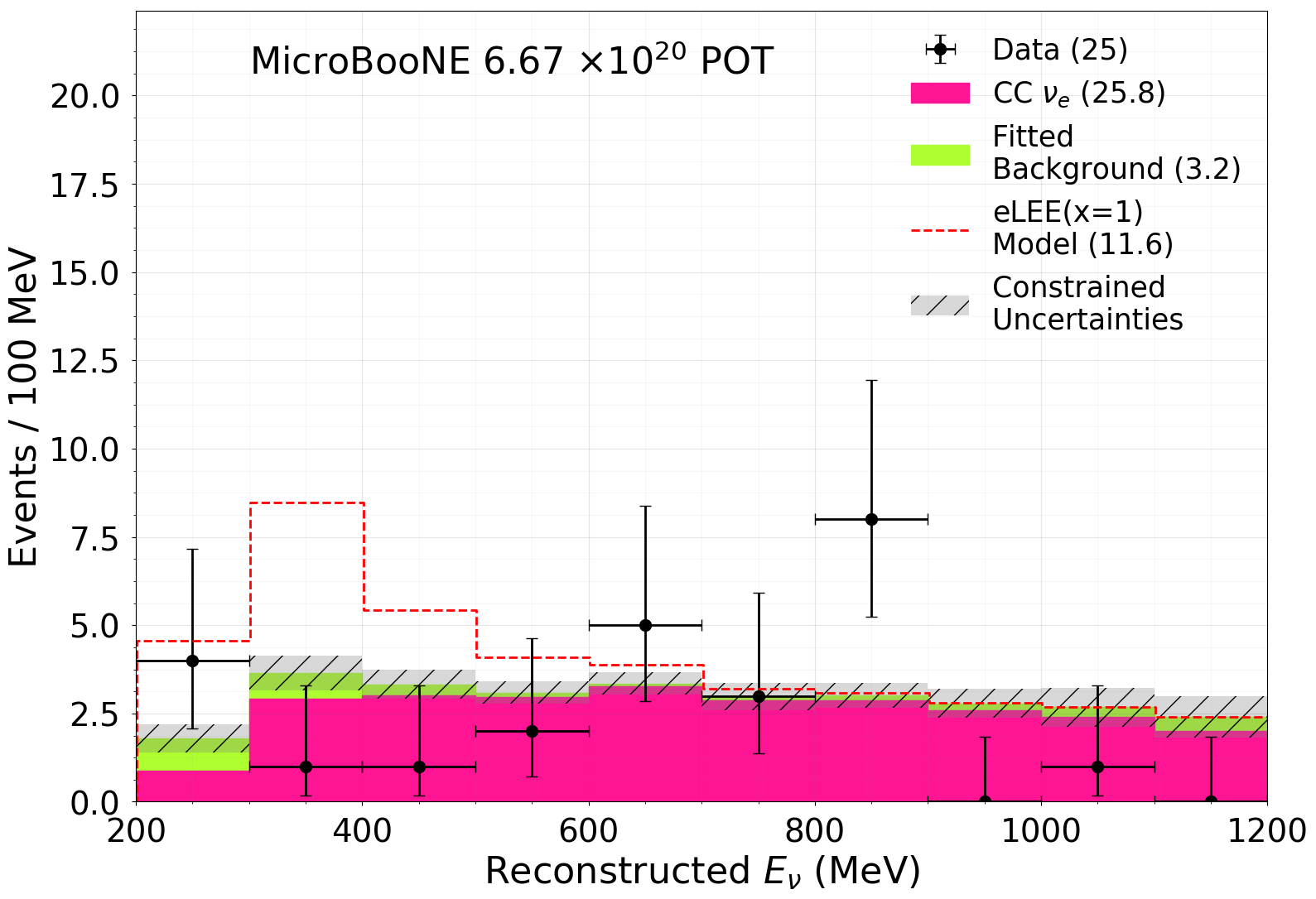}
    \caption[$\nu_e$CC $1e1p$ QE-like result from DL reconstruction]{$\nu_e$CC $1e1p$ QE-like results from DL reconstruction. Figure from Ref. \cite{eLEE_multiple_reconstructions_PRL}.}
    \label{fig:dl_nue}
\end{figure}

There was also an analysis which used deep learning based reconstruction techniques, including a sparse semantic segmentation convolutional neural network \cite{sparse_ss_net} and multiple particle identification convolutional neural network \cite{MPID}. This analysis specifically targets the $1e1p$ topology, which has a very distinctive visual signature. In addition to this topological cut, we also select events which are consistent with two-body-scattering kinematics. For simple charged current quasi-elastic events, considering a single incoming neutrino energy and azimuthal symmetry, there is only one degree of freedom describing the kinematics; the angle of the scatter in the neutrino-nucleus center of momentum frame. So, we measure four quantities, $E_p$, $E_e$, $\theta_p$, and $\theta_e$, and there are only two degrees of freedom for a quasi-elastic event, the neutrino energy and this center of momentum frame angle. We use an ensemble of BDTs to learn how correlations between these quantities can be used to categorize events as quasi-elastic or background. The results are described in Ref. \cite{DL_eLEE_PRD} and the resulting $\nu_e$CC reconstructed energy distributions are shown in Fig. \ref{fig:dl_nue}. We see that we have fairly low statistics after cutting to this specific type of event, but again we see no sign of a significant excess.

\begin{figure}[H]
    \centering
    \begin{subfigure}[b]{0.51\textwidth}
        \includegraphics[width=\textwidth]{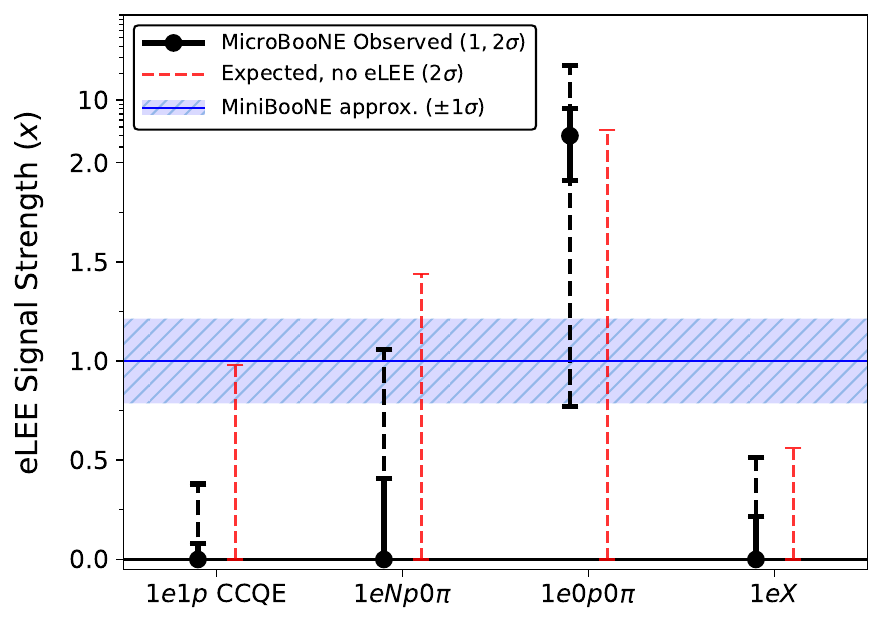}
        \caption{}
    \end{subfigure}
    \begin{subfigure}[b]{0.48\textwidth}
        \includegraphics[width=\textwidth]{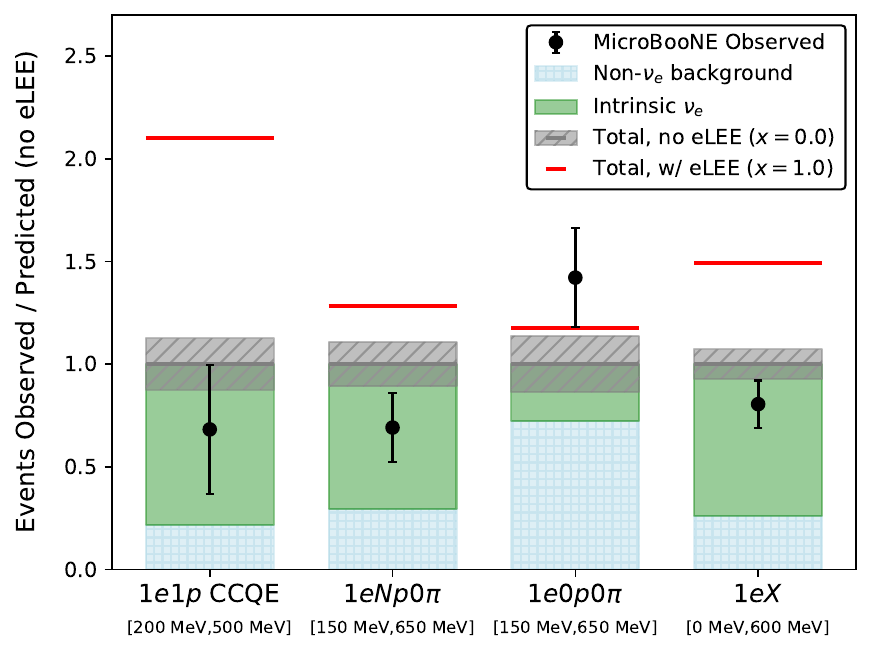}
        \caption{}
    \end{subfigure}
    \caption[$\nu_e$CC results from different reconstructions]{$\nu_e$CC results from different reconstructions. Figures from Ref. \cite{eLEE_multiple_reconstructions_PRL}.}
    \label{fig:nueCC_multiple_reconstructions}
\end{figure}

We summarize the statistical tests from all of these analyses in Fig. \ref{fig:nueCC_multiple_reconstructions}. We see that none of the selections prefer the median unfolded MiniBooNE LEE model, besides $1e0p0\pi$ which is background dominated and not statistically significant.

\begin{figure}[H]
    \centering
    \includegraphics[width=0.7\textwidth]{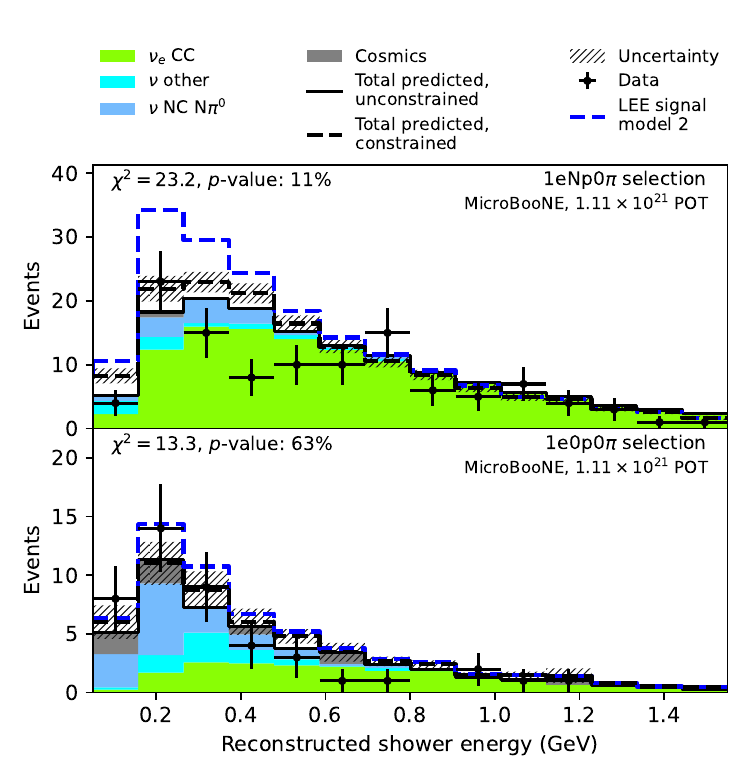}
    \caption[Updated $\nu_e$CC $0\pi$ results from Pandora reconstruction]{$\nu_e$CC $0\pi$ results from Pandora reconstruction. Figure from Ref. \cite{updated_pandora_eLEE_PRL}.}
    \label{fig:updated_pandora_nue}
\end{figure}

More recently, we have updated the Pandora pionless analysis. We expanded the median unfolding MiniBooNE LEE model described in Sec. \ref{sec:lee_model} in order to account for the electron shower energy and angle distributions observed by MiniBooNE; note that this cannot be interpreted as an enhancement of the $\nu_e$ flux in the beam, and has to be considered as part of a more complex cross section mismodeling hypothesis. This analysis also used our full runs 1-5 data set, $11.1\cdot 10^{20}$ POT rather than $6.4-6.8 \cdot 10^{20}$ for the results above. The results are described in Ref. \cite{updated_pandora_eLEE_PRL} and the resulting $\nu_e$CC reconstructed shower energy distributions are shown in Fig. \ref{fig:updated_pandora_nue}. Again, we see no significant $\nu_e$ excess.

\section{3+1 Sterile Neutrino Results}

In the previous section, we carefully searched for a $\nu_e$ flux increase that matches the MiniBooNE LEE. We used a median unfolding of the MiniBooNE excess, which does not have any precise physics interpretation. In this section, we use the 3+1 sterile neutrino model described in Sec. \ref{sec:sterile}, which allows tests of more precise models of BSM physics and allows for more robust comparisons across different experiments.

We use the same inclusive Wire-Cell $\nu_e$CC and constraining selections described above in order to search for sterile neutrino oscillations. We work in a full 3+1 framework, considering all types of oscillatory flavor changes. We compare our data with predictions in a three dimensional phase space of $\Delta m_{41}^2$, $\sin^2\theta_{14}$, and $\sin^2\theta_{24}$, which then determine $\sin^2 2\theta_{\mu e}$ and $\sin^2 2\theta_{ee}$ as described by Eqns. \ref{eqn:theta_ee}-
\ref{eqn:theta_mue}. We set $\theta_{34}=0$, since this primarily would cause small differences in neutral current events, and our selections do not have significant sensitivity to these differences. 

\begin{figure}[H]
    \centering
    \includegraphics[width=0.7\textwidth]{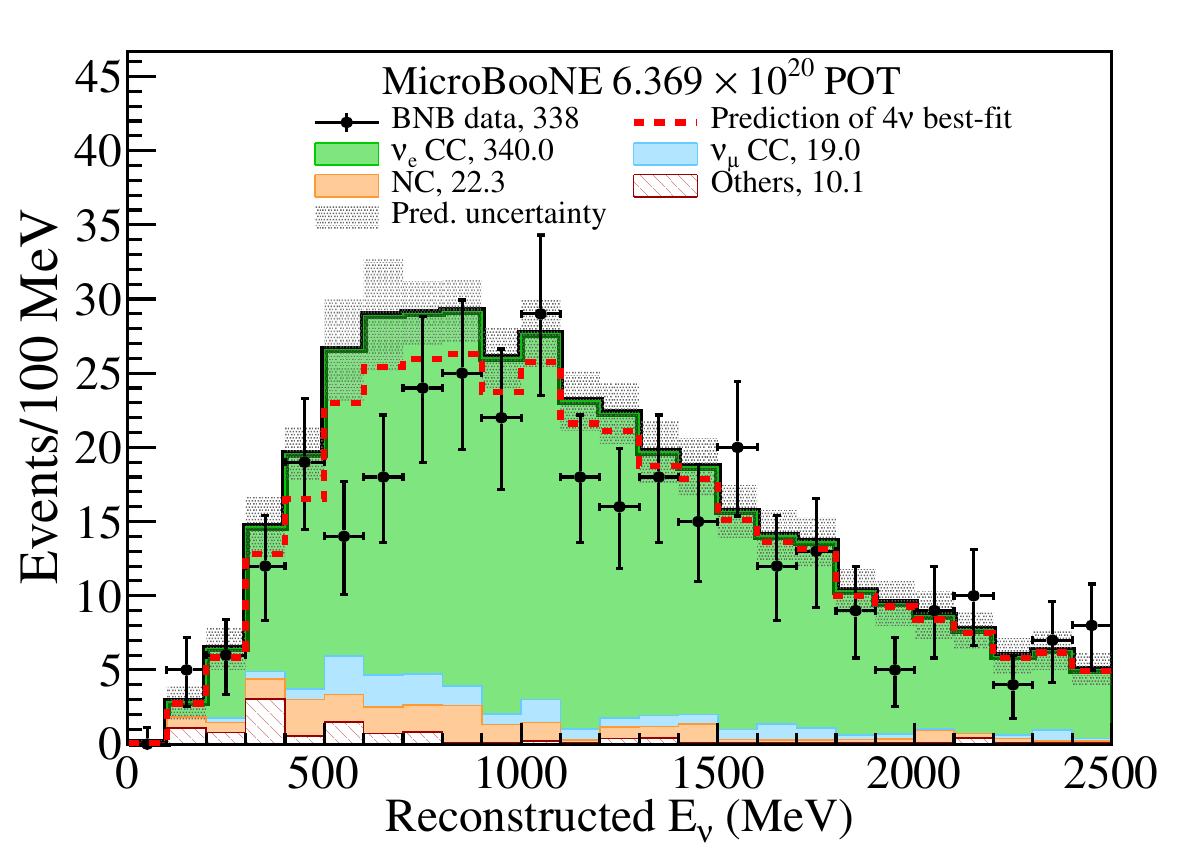}
    \caption[$\nu_e$CC 3+1 sterile neutrino best fit]{$\nu_e$CC reconstructed energy distribution compared to the 3+1 sterile neutrino best fit parameters. Figure from Ref. \cite{microboone_bnb_sterile}.}
    \label{fig:sterile_best_fit}
\end{figure}

Our best-fit value to all seven $\nu_e$CC, $\nu_\mu$CC, and $\pi^0$ channels is $\Delta m^2=1.295 \mathrm{eV}^2$, $\sin^2\theta_{14}=0.936$, $\sin^2\theta_{24}=0$, which corresponds to $\sin^2 2\theta_{ee}=0.240$ and $\sin^2 2\theta_{\mu e}=\sin^2 2\theta_{\mu \mu}=0$, as shown in Fig. \ref{fig:sterile_best_fit}. However, this best-fit 3+1 sterile neutrino oscillation hypothesis is not a significantly better fit than the no-oscillation hypothesis. We quantify this by examining $\Delta \chi^2=\chi^2_\mathrm{null}-\chi^2_\mathrm{best-fit}$. We find $\Delta \chi^2_\mathrm{data}=2.53$, which corresponds to a $p$-value of 0.426 according to the Feldman-Cousins procedure \cite{feldman_cousins}, so the no-oscillation hypothesis fits the data almost equally well.

\begin{figure}[H]
    \centering
    \includegraphics[width=0.7\textwidth]{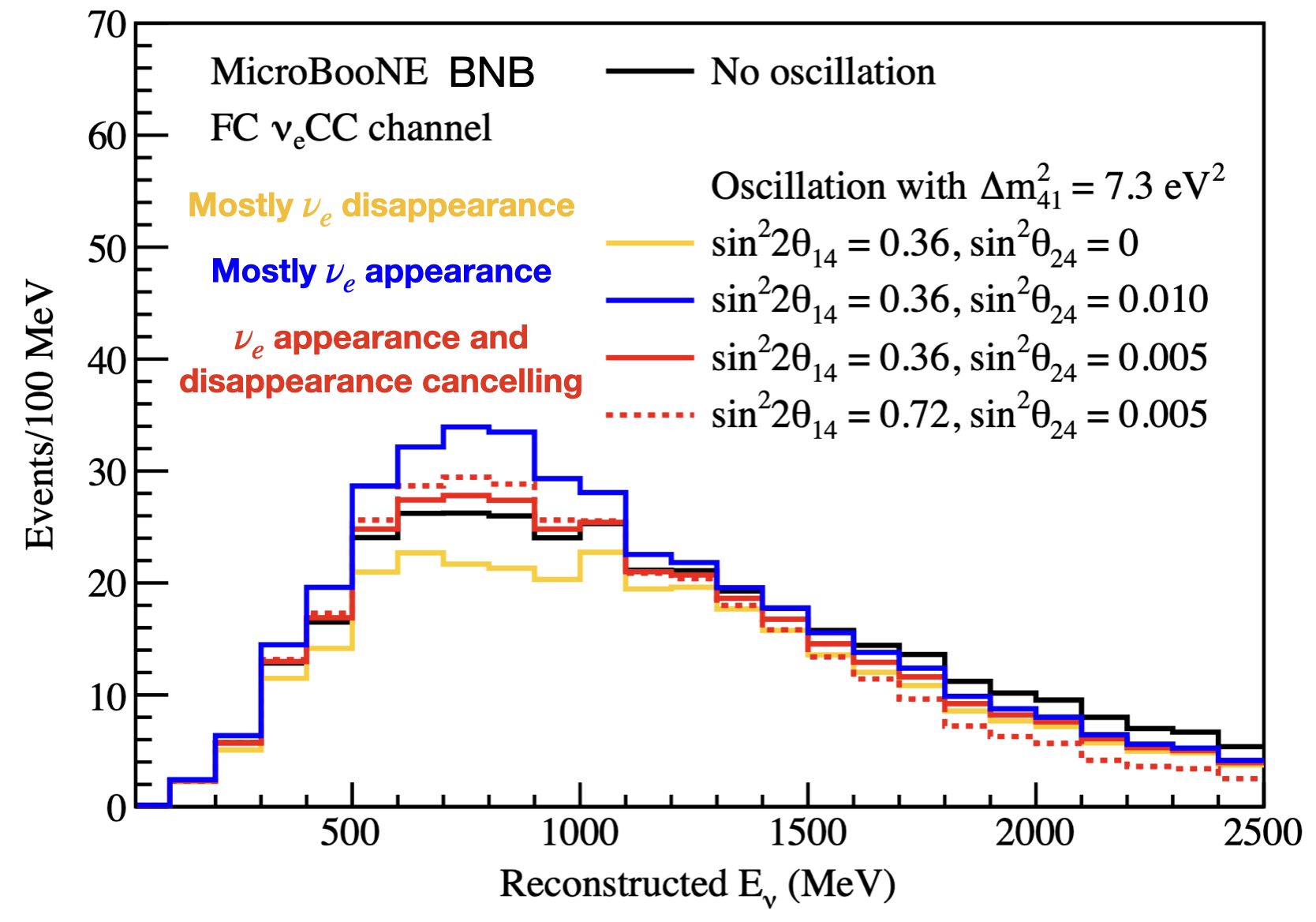}
    \caption[BNB $\nu_e$ appearance and disappearance cancellation]{BNB $\nu_e$ appearance and disappearance cancellation. Figure modified from from Ref. \cite{microboone_bnb_sterile}.}
    \label{fig:bnb_cancellation}
\end{figure}

For this analysis, there is an important degeneracy to consider: for certain parameters, $\nu_\mu\rightarrow \nu_e$ appearance and $\nu_e\rightarrow \nu_e$ disappearance can cancel each other out. This type of cancellation is important for accelerator neutrino beams which produce multiple neutrino flavors before oscillations, but not every sterile exclusion paper has considered this effect. As shown in Eqn. \ref{eqn:app_disapp_cancellation}, this cancellation happens when $\sin^2 \theta_{24}$ approaches $R_{\nu_\mu/\nu_e}$, the ratio of $\nu_e$ to $\nu_\mu$ in the initial neutrino beam. For the BNB, $R_{\nu_\mu/\nu_e}\approx 200$, but the precise ratio depends on the neutrino energy. This causes cancellation when $\sin^2 \theta_{24}\approx 1/200 = 0.005$, and the effect on the $\nu_e$CC reconstructed neutrino energy spectrum is shown in Fig. \ref{fig:bnb_cancellation}.

{\small
\begin{alignat}{2}\label{eqn:app_disapp_cancellation}
    N_{\nu_e\mathrm{\ detector}}
    &=N_{\nu_e\mathrm{\ beam}}\cdot P_{\nu_e\rightarrow\nu_e}
    +N_{\nu_\mu\mathrm{\ beam}}\cdot P_{\nu_\mu\rightarrow\nu_e}\\
    &=N_{\nu_e\mathrm{\ beam}}\left[
        1+
        \left(R_{\nu_\mu/\nu_e} \cdot \sin^2\theta_{24}-1\right)
        \cdot\sin^2 2\theta_{14}
        \cdot \sin^2\left(1.267\frac{\mathrm{eV}^2\mathrm{km}}{\mathrm{GeV}}\frac{\Delta m_{41}^2 L}{E}\right)
        \right]
\end{alignat}
}

\begin{figure}[H]
    \centering
    \begin{subfigure}[b]{0.49\textwidth}
        \includegraphics[width=\textwidth]{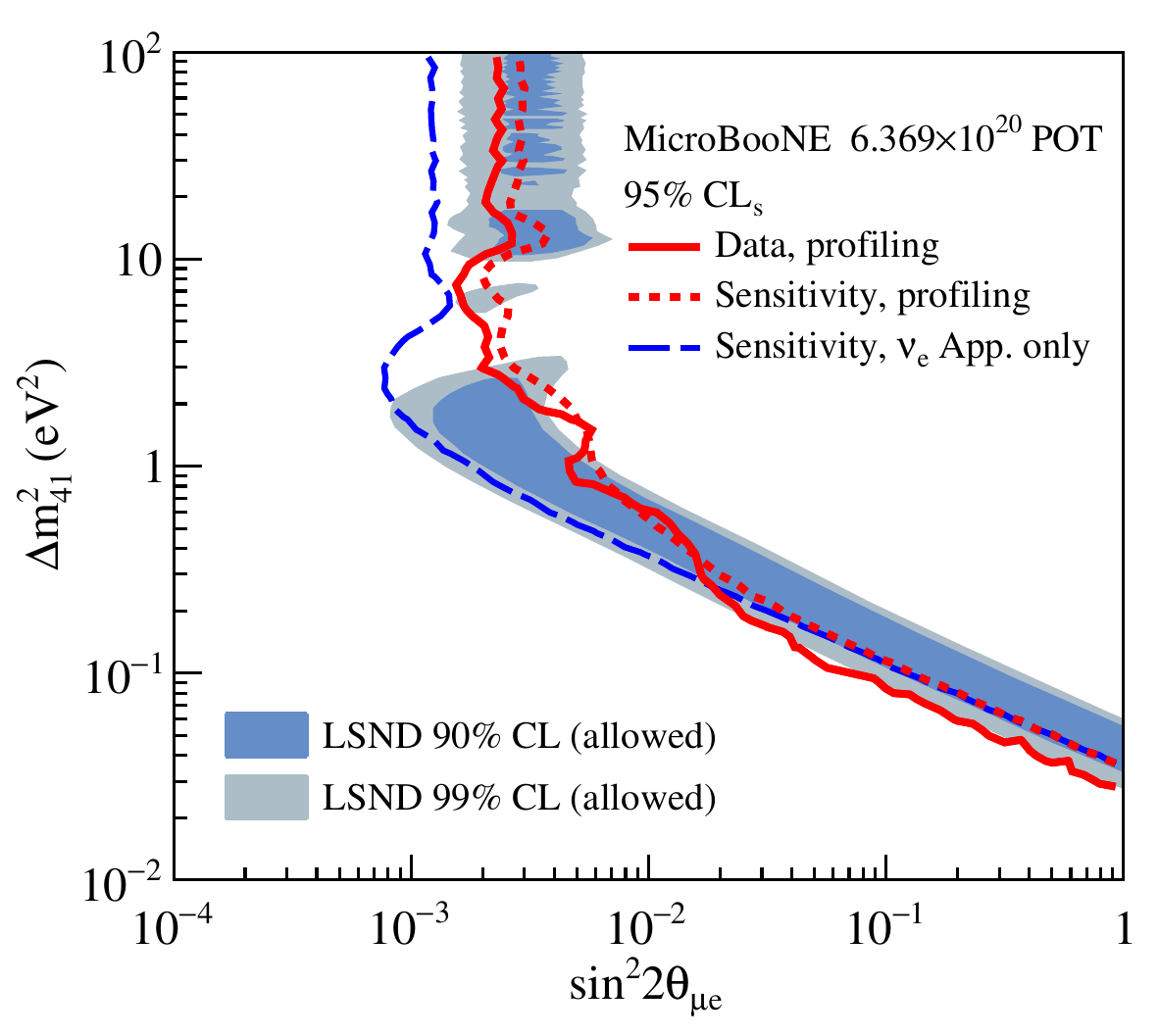}
        \caption{}
    \end{subfigure}
    \begin{subfigure}[b]{0.49\textwidth}
        \includegraphics[width=\textwidth]{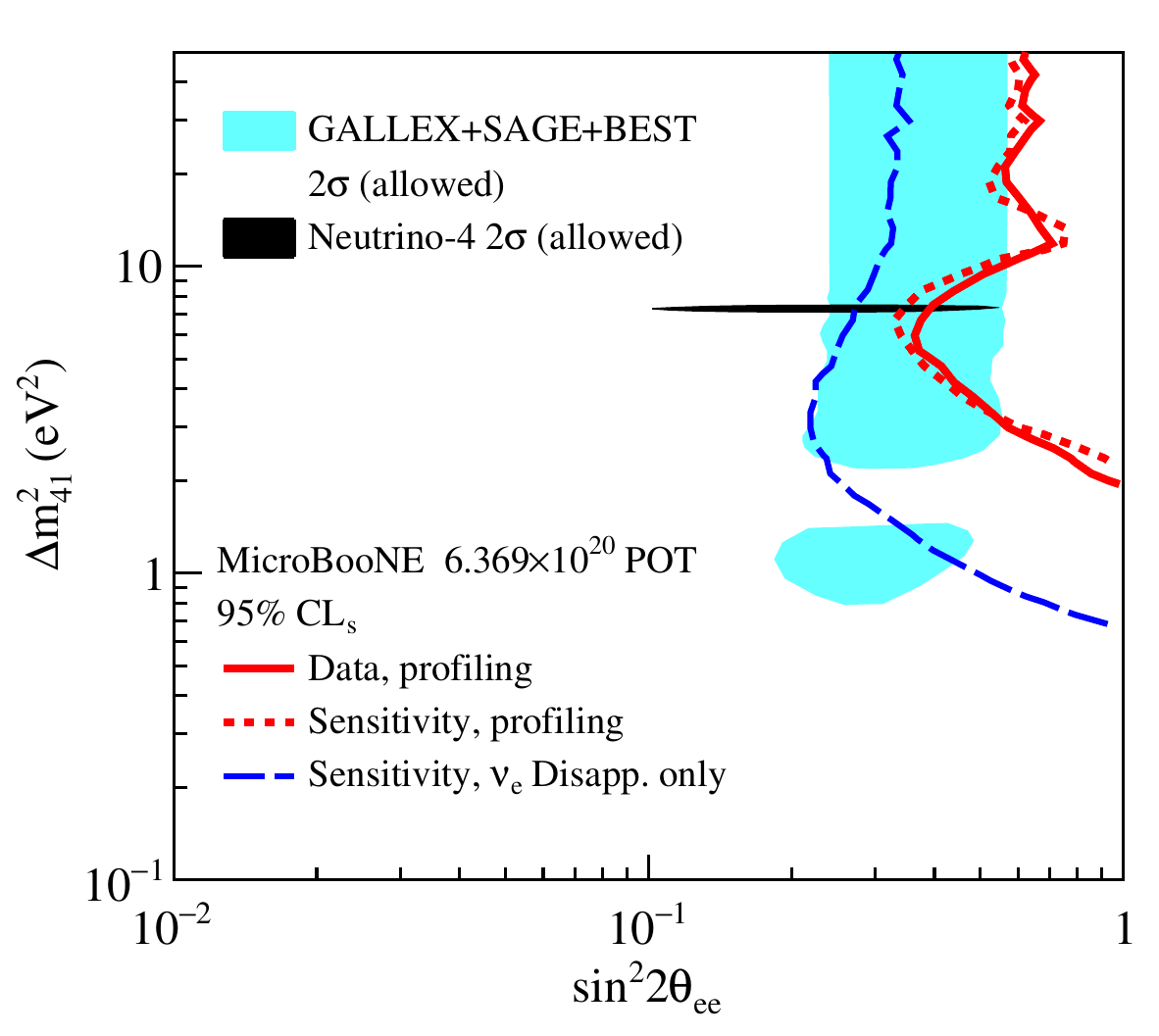}
        \caption{}
    \end{subfigure}
    \caption[MicroBooNE BNB 3+1 sterile exclusions]{MicroBooNE BNB 3+1 sterile exclusions. Panel (a) shows the exclusion for $\nu_\mu\rightarrow \nu_e$ appearance, and panel (b) shows the exclusion for $\nu_e\rightarrow \nu_e$ disappearance. Figures from Ref. \cite{microboone_bnb_sterile}.}
    \label{fig:microboone_bnb_sterile_exclusions}
\end{figure}

We form exclusion limits using the frequentist-motivated $\mathrm{CL}_\mathrm{s}$ method \cite{CL_s_method}. For each 3+1 sterile neutrino hypothesis point, we generate pseudo-experiments for both the standard three neutrino hypothesis as well as the four neutrino sterile hypothesis, and for each get a distribution of $\Delta \chi^2_{\mathrm{CL}_\mathrm{s}}=\chi^2_{4\nu}-\chi^2_{3\nu}$. Comparing each distribution to our $\Delta \chi^2$ value for data, we get $p_{4\nu}$ and $p_{3\nu}$ describing the data's compatibility with each hypothesis. We then calculate $\mathrm{CL}_\mathrm{s}$ via
\begin{equation}
\mathrm{CL}_\mathrm{s} = \frac{1-p_{4\nu}}{1-p_{3\nu}}
\end{equation}
and use this value to define our final exclusion for this point. Because we consider a 3D phase space of sterile hypotheses ($\Delta m_{41}^2$, $\sin^2\theta_{14}$, and $\sin^2\theta_{24}$) but only display 2D exclusions, we profile by choosing the minimum $\chi^2_{4\nu}$ out of all choices along the third dimension. Because of the $\nu_e$ appearance and disappearance degeneracy explained above, this profiling causes our exclusions are significantly weaker than they would be in an appearance-only framework, as shown in Fig. \ref{fig:microboone_bnb_sterile_exclusions}. We are able to exclude portions of the LSND, Neutrino-4, and gallium anomaly preferred regions.

At high $\Delta m^2$ values, the exclusions and preferred regions each become fairly constant. This is due to the very rapid oscillations in this region, meaning that these experiments do not resolve any $L/E$ oscillatory behavior and instead see a constant normalization shift for all events. At low $\Delta m^2$ values for $\nu_\mu\rightarrow \nu_e$ appearance, a similar trend appears, where the exclusion and preferred region both become fairly linear; this is the region where we do not observe a full $L/E$ oscillation, and instead see just a smooth increase in oscillation probability as $L/E$ increases. At both these high and low $\Delta m^2$ regions, we see slightly better than expected $\sin^2 2 \theta_{\mu e}$ exclusion due to our overprediction of $\nu_e$CC events. At intermediate $\Delta m^2$ values, the detailed shape of the data plays a more important role.

\begin{figure}[H]
    \centering
    \includegraphics[width=0.49\textwidth]{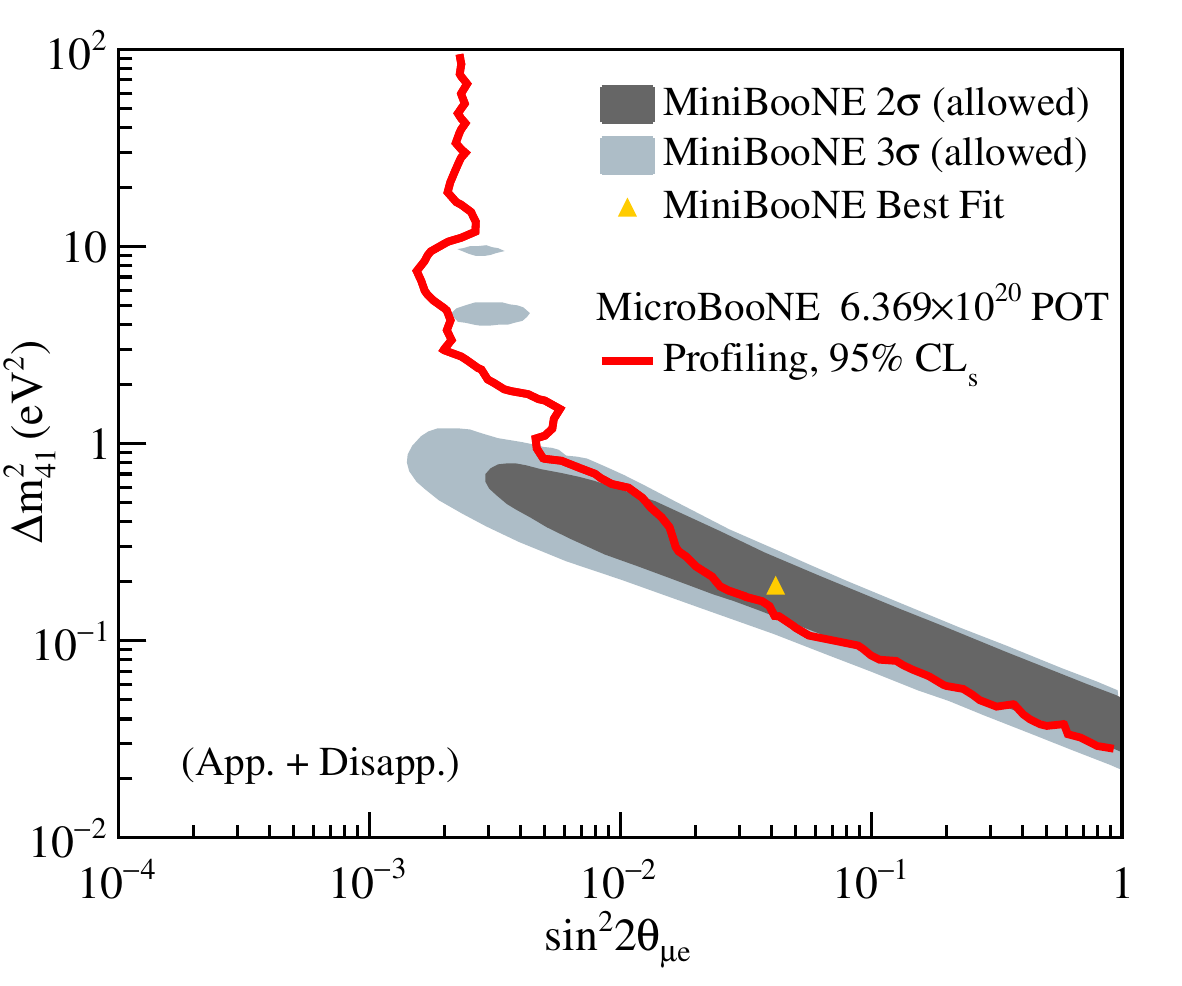}
    \caption[MicroBooNE BNB 3+1 sterile exclusion with MiniBooNE]{MicroBooNE BNB 3+1 sterile exclusions compared with the MiniBooNE LEE allowed region. Figure from Ref. \cite{microboone_bnb_sterile}.}
    \label{fig:sterile_compare_miniboone}
\end{figure}

We are also able to exclude a portion of the MiniBooNE allowed region, as shown in Fig. \ref{fig:sterile_compare_miniboone}. Note that these two results have correlated beam flux uncertainties which cannot be easily visualized on this plot.

This profiling which collapses from 3D to 2D does lose some information. We provide $\chi^2$ values for the full 3D space, which can be used for more detailed comparisons. For example, Ref. \cite{sterile_neutrino_microboone_prospect_no_profiling} considers this full 3D space and compares between MicroBooNE and MiniBooNE, concluding that MicroBooNE results exclude MiniBooNE's entire 2.3$\sigma$ allowed region at greater than or equal to 2.3$\sigma$. This reference also shows how we can combine MicroBooNE measurements with reactor neutrino information from PROSPECT in order to significantly reduce the effect of the $\nu_e$ appearance and disappearance cancellation degeneracy.

\begin{figure}[H]
    \centering
    \begin{subfigure}[b]{0.49\textwidth}
        \includegraphics[width=\textwidth]{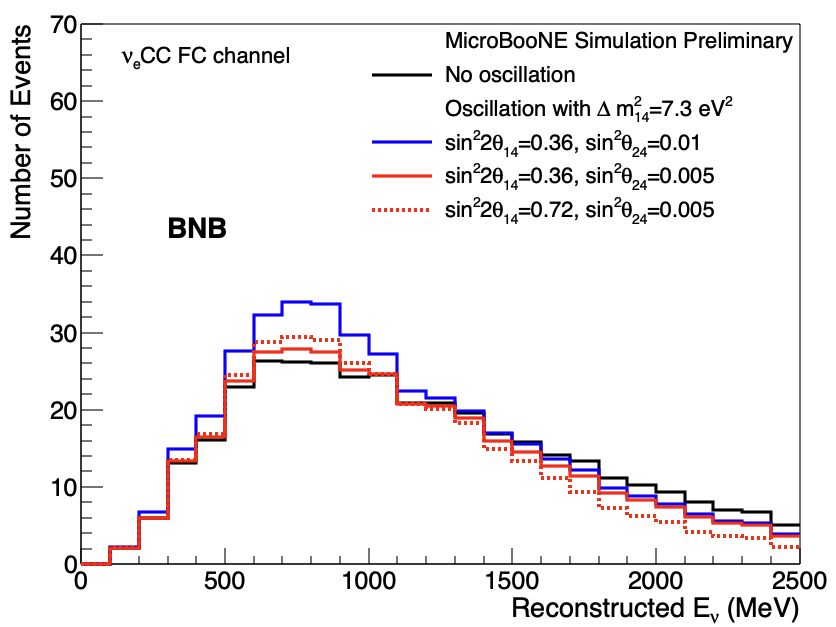}
        \caption{}
    \end{subfigure}
    \begin{subfigure}[b]{0.49\textwidth}
        \includegraphics[width=\textwidth]{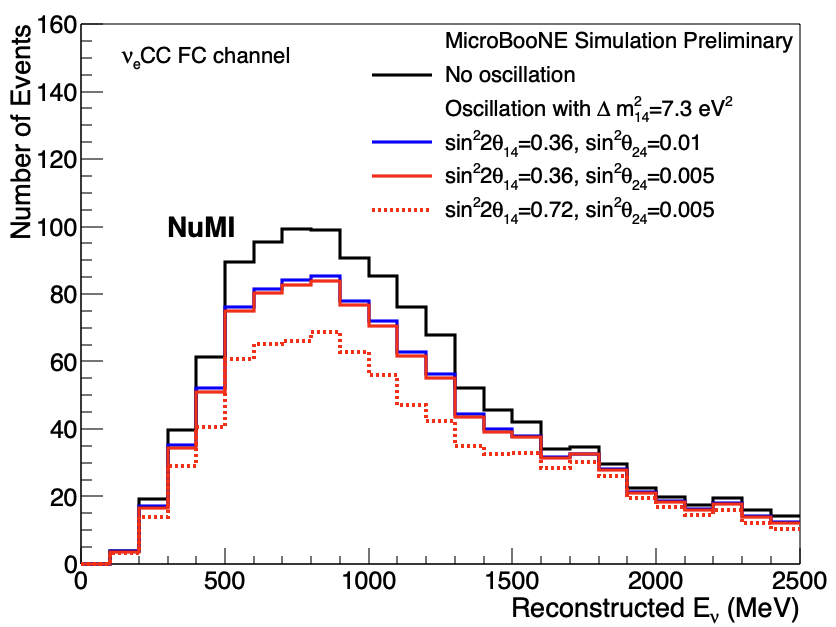}
        \caption{}
    \end{subfigure}
    \caption[$\nu_e$CC appearance and disappearance cancellation]{$\nu_e$CC appearance and disappearance cancellation. Figures from Ref. \cite{bnb_numi_sterile_public_note}.}
    \label{fig:app_disapp_cancellation}
\end{figure}

Another way to address the $\nu_e$ appearance and disappearance cancellation degeneracy is by simultaneously using the BNB and NuMI beams. For NuMI, $R_{\nu_\mu/\nu_e}\approx 25$, in contrast to $\approx 200$ for BNB. This lets us address the degeneracy, since the cancellation cannot happen in both neutrino beams at the same time, as shown in Fig. \ref{fig:app_disapp_cancellation}.

\begin{figure}[H]
    \centering
    \begin{subfigure}[b]{0.50\textwidth}
        \includegraphics[trim=0 0 50 0, clip, width=\textwidth]{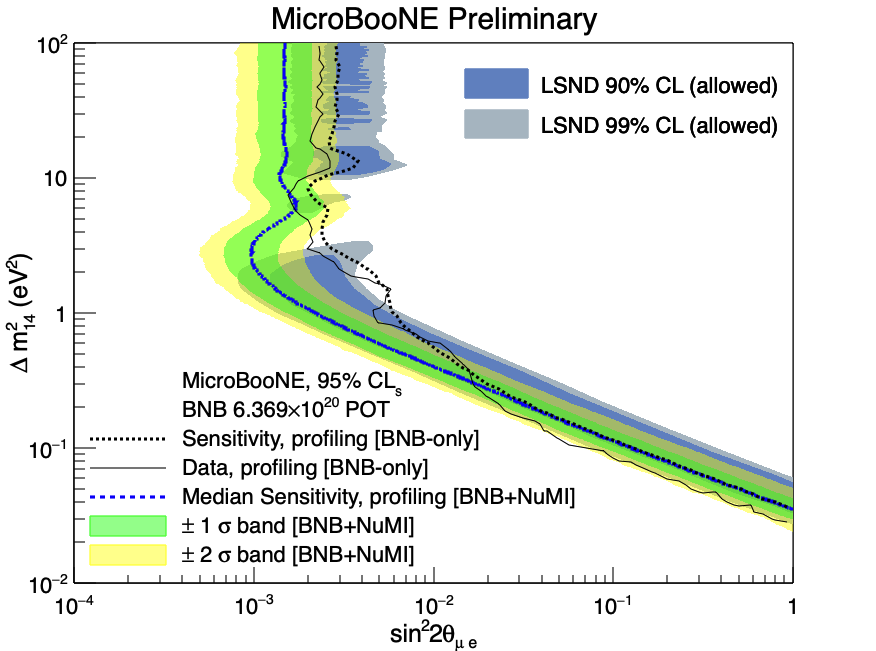}
        \caption{}
    \end{subfigure}
    \begin{subfigure}[b]{0.49\textwidth}
        \includegraphics[trim=0 0 50 0, clip, width=\textwidth]{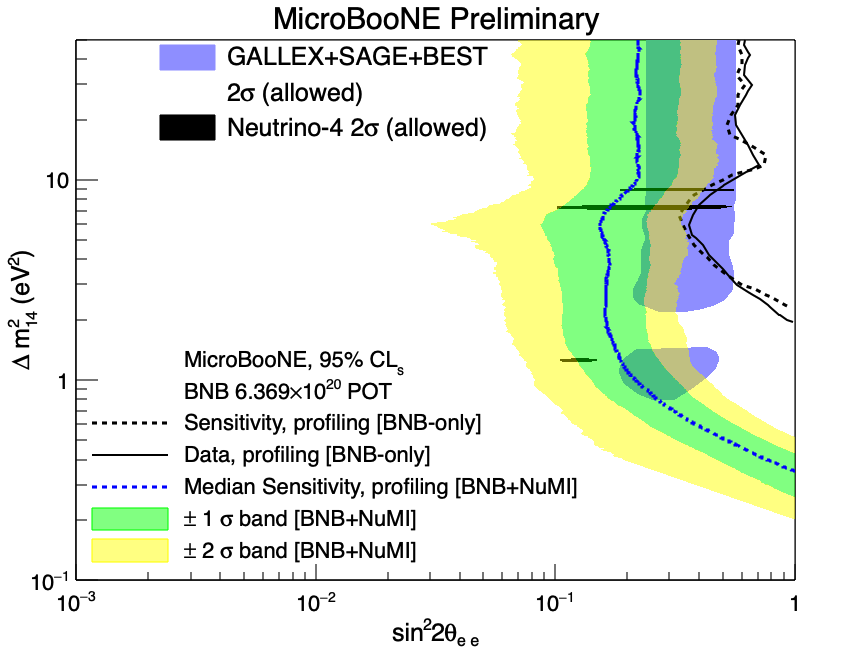}
        \caption{}
    \end{subfigure}
    \caption[BNB+NuMI 3+1 sterile neutrino exclusion sensitivities]{BNB+NuMI 3+1 sterile neutrino exclusion sensitivities. Figures from Ref. \cite{bnb_numi_sterile_public_note}.}
    \label{fig:bnb_numi_sensitivity}
\end{figure}

This leads to a significantly improved sensitivity when we include both the BNB and NuMI beams, as shown in Fig. \ref{fig:bnb_numi_sensitivity}. This analysis is currently underway, and relies on a very detailed understanding of the NuMI beam flux as described in Sec. \ref{sec:beams}.

%% file: chapters/04_nc_delta.tex
\chapter{NC \texorpdfstring{$\Delta$}{Delta} Radiative Decay Search in MicroBooNE}\label{sec:nc_delta}

In the previous chapter, I described analyses searching for an electron shower explanation of the MiniBooNE LEE. As discussed in Sec. \ref{sec:miniboone_photonlike}, there are also a variety of potential photon-like interpretations of the MiniBooNE LEE. In this chapter, I will describe my main thesis topic, a search for a single photons in MicroBooNE, in particular via the NC $\Delta$ radiative decay process.

\section{Difficulties in Single Photon Searches}

In the previous chapter, I described the process of searching for electron showers. This required rejecting a lot of background from cosmic rays, NC neutrino interactions, and $\nu_\mu$CC neutrino interactions, which required several iterations of BDTs and hundreds of selection variables. However, there was no background which we were not able to successfully reject; every low $dQ/dx$ shower connected to the neutrino vertex really is a $\nu_e$CC event, and identifying all electron showers gives you a very pure selection.

\begin{figure}[H]
    \centering
    \includegraphics[width=0.6\textwidth]{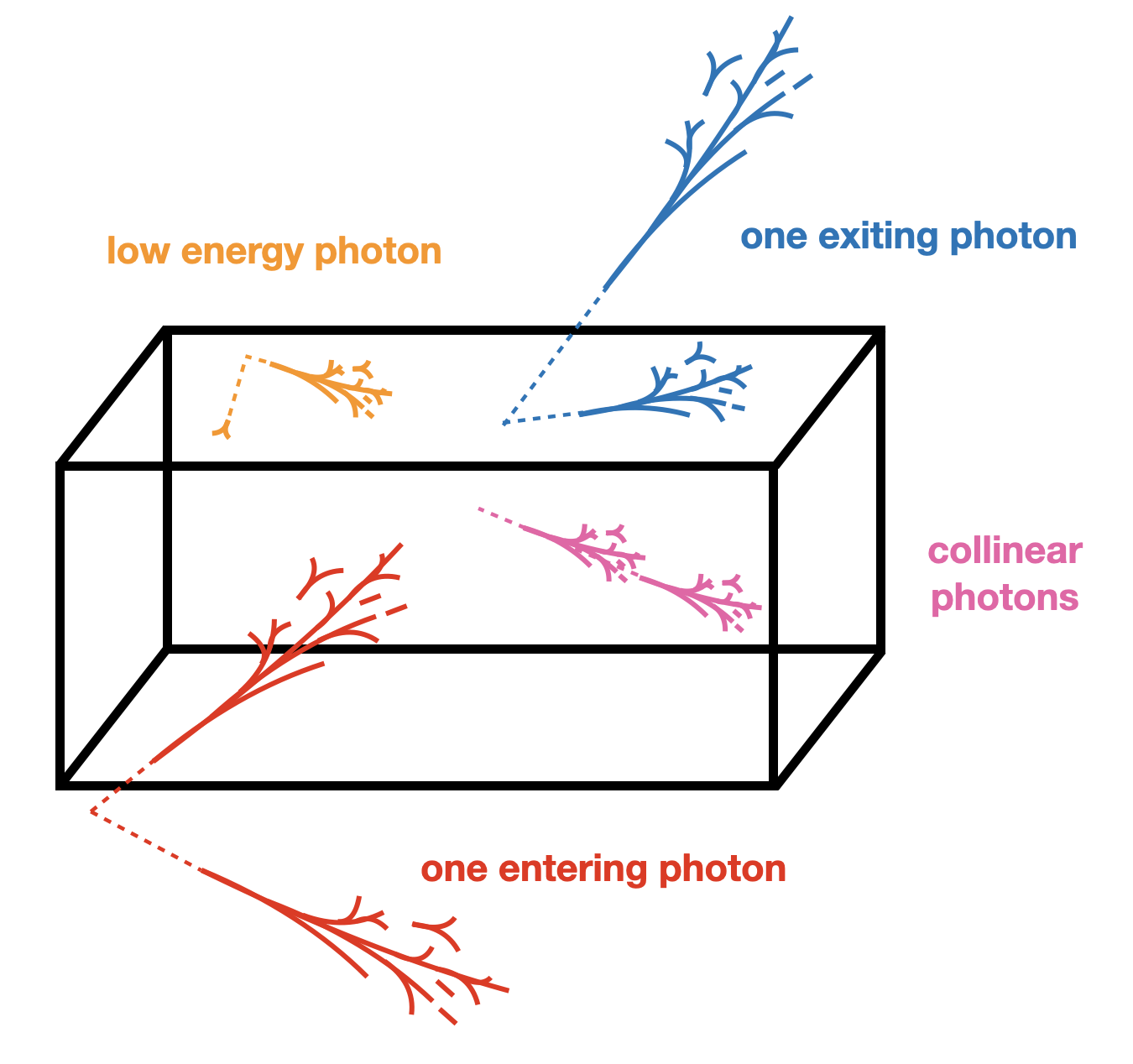}
    \caption[$\pi^0$ background diagram]{Examples of how $\pi^0\rightarrow \gamma \gamma$ events can be mistaken for single photons in neutrino detectors.}
    \label{fig:nc_pi0_diagram}
\end{figure}

Searching for single photons is more difficult; there are many backgrounds from multi-photon events, in particular the $\pi^0\rightarrow \gamma \gamma$ process which produces many two-photon pairs in our detector. This is an overwhelming background to all neutrino-induced single photon searches, since there are factors that can cause a two-photon event to appear as a one-photon event. Several of these possibilities are illustrated in Fig. \ref{fig:nc_pi0_diagram}. Photons travel invisibly before pair converting and creating an $e^+e^-$ electromagnetic shower, so one photon could enter the detector from a neutrino interaction outside, or one photon could leave the detector from a neutrino interaction inside. One photon could be too low energy to reconstruct, or two photon showers could merge together if they have similar angles. These are examples of effectively irreducible backgrounds, but there are also ways in which our existing reconstruction can fail to properly cluster and reconstruct two photons together in an event. The potential for $\pi^0$ mis-reconstruction explanations of experimental neutrino-induced single photon excesses was recognized as early as the 1980s, when $10$ single shower events were observed by Gargamelle with an expected background of $0.2\pm 0.2$ \cite{gargamelle_shower_excess} and this was interpreted as potentially due to either true single photon production or $\pi^0$ photon pairs with one getting lost \cite{gargamelle_photon_explanation_1,gargamelle_photon_explanation_2}.

\begin{figure}[H]
    \centering
    \begin{subfigure}[b]{0.48\textwidth}
        \includegraphics[width=\textwidth]{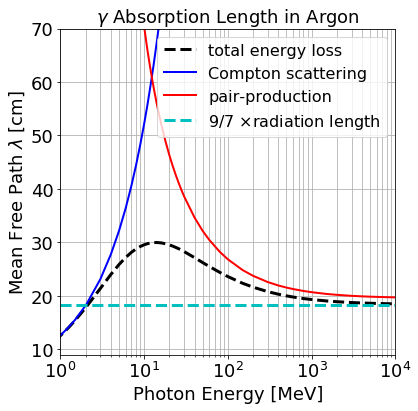}
        \caption{}
    \end{subfigure}
    \begin{subfigure}[b]{0.51\textwidth}
        \includegraphics[width=\textwidth]{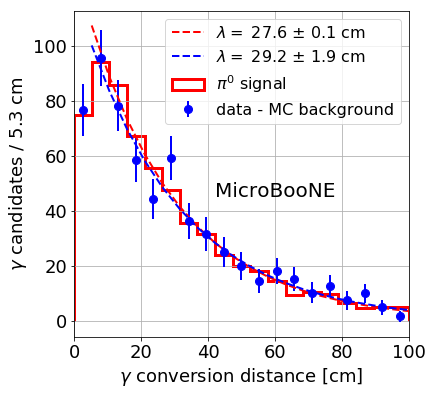}
        \caption{}
    \end{subfigure}
    \caption[Photon radiation length in argon]{Photon radiation length in argon. Panel (a) shows the predicted mean free path as a function of photon energy. Panel (b) shows a measurement of the average photon conversion distance from MicroBooNE $\pi^0$ data, 29.3 $\pm$ 1.9 cm. Figures from Ref. \cite{microboone_photon_scattering_length}.}
    \label{fig:photon_rad_length}
\end{figure}

This means that for the single photon search described in this chapter, we should not have such high expectations as we had for the $\nu_e$CC searches. In this more difficult topology, we will not succeed in attaining a similarly high efficiency and purity as we saw before. For these events, an important component of the irreducible background is entering and exiting photons. MicroBooNE is a relatively small detector, about 85 metric tons of active volume compared to about 800 metric tons for MiniBooNE; this means that these exiting photons affect our experiment to a greater degree. Also, consider MicroBooNE's TPC shape, a long rectangular prism with dimensions of approximately $2.6\times 2.3 \times 10.4$ m. This means that no location in the detector is more than 116 cm from a TPC boundary. Compare this with the average photon conversion distance of around 29.3 cm \cite{microboone_photon_scattering_length} as shown in Fig. \ref{fig:photon_rad_length}. This means that even in the best case, we are never more than about 3.9 photon conversion distances away from the detector boundary, so we can never be extremely confident that a photon could not have exited the detector before pair converting. MicroBooNE's shape means than fiducialization is very difficult relative to a more cubical or spherical detector, and we would have to lose a fairly large fraction of our volume in order to decrease the probability of entering and exiting photons.

\section{NC \texorpdfstring{$\Delta$}{Delta} Radiative Decay Single Photons}

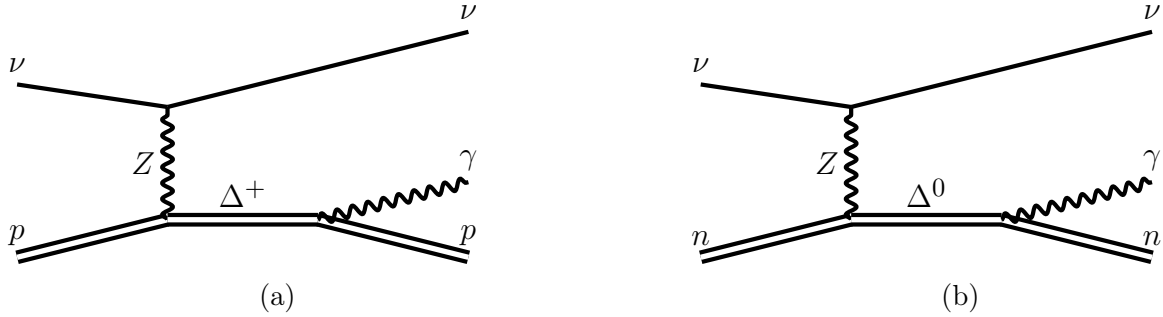
\begin{figure}[H]
    \centering
    \begin{subfigure}[b]{0.45\textwidth}
        \begin{tikzpicture}[line width=1.0 pt, scale=1]
            \coordinate (proton_start) at (0, 0);
            \coordinate (proton_to_delta) at (2, 0.5);
            \coordinate (delta_decay) at (4, 0.5);
            \coordinate (photon_end) at (6, 1);
            \coordinate (proton_end) at (6, 0);
            \coordinate (neutrino_start) at (0, 2.3);
            \coordinate (neutrino_interaction) at (2, 2);
            \coordinate (neutrino_end) at (6, 3);
            \draw[ultra thick, double distance=2pt] (proton_start) -- (proton_to_delta) node[at start, above] {$p$};
            \draw[ultra thick] (neutrino_start) -- (neutrino_interaction) node[at start, above] {$\nu$};
            \draw[ultra thick] (neutrino_interaction) -- (neutrino_end) node[at end, above] {$\nu$};
            \draw[ultra thick, decorate, decoration={snake, amplitude=2pt, segment length=6pt}] 
            (proton_to_delta) -- (neutrino_interaction) node[midway, left] {$Z$};
            \draw[ultra thick, double distance=2pt] (proton_to_delta) -- (delta_decay) node[midway, above] {$\Delta^+$};
            \draw[ultra thick, double distance=2pt] (delta_decay) -- (proton_end) node[at end, above] {$p$};
            \draw[ultra thick, decorate, decoration={snake, amplitude=2pt, segment length=6pt}] (delta_decay) -- (photon_end) node[at end, above] {$\gamma$};
        \end{tikzpicture}
        \caption{}
    \end{subfigure}
    \hfill
    \begin{subfigure}[b]{0.45\textwidth}
        \begin{tikzpicture}[line width=1.0 pt, scale=1]

            \coordinate (neutron_start) at (0, 0);
            \coordinate (neutron_to_delta) at (2, 0.5);
            \coordinate (delta_decay) at (4, 0.5);
            \coordinate (photon_end) at (6, 1);
            \coordinate (neutron_end) at (6, 0);
            \coordinate (neutrino_start) at (0, 2.3);
            \coordinate (neutrino_interaction) at (2, 2);
            \coordinate (neutrino_end) at (6, 3);
            \draw[ultra thick, double distance=2pt] (neutron_start) -- (neutron_to_delta) node[at start, above] {$n$};
            \draw[ultra thick] (neutrino_start) -- (neutrino_interaction) node[at start, above] {$\nu$};
            \draw[ultra thick] (neutrino_interaction) -- (neutrino_end) node[at end, above] {$\nu$};
            \draw[ultra thick, decorate, decoration={snake, amplitude=2pt, segment length=6pt}] 
            (neutron_to_delta) -- (neutrino_interaction) node[midway, left] {$Z$};
            \draw[ultra thick, double distance=2pt] (neutron_to_delta) -- (delta_decay) node[midway, above] {$\Delta^0$};
            \draw[ultra thick, double distance=2pt] (delta_decay) -- (neutron_end) node[at end, above] {$n$};
            \draw[ultra thick, decorate, decoration={snake, amplitude=2pt, segment length=6pt}] (delta_decay) -- (photon_end) node[at end, above] {$\gamma$};
        \end{tikzpicture}
        \caption{}
    \end{subfigure}
    \caption[NC $\Delta\rightarrow N \gamma$ Feynman diagrams]{NC $\Delta\rightarrow N \gamma$ Feynman diagrams. Panel (a) shows a neutrino exciting a proton to a $\Delta^+$ resonance which radiatively decays to a proton and photon, and panel (b) shows a neutrino exciting a neutron to a $\Delta^0$ resonance which radiatively decays to a neutron and photon.}
    \label{fig:nc_delta_feynman}
\end{figure}

Neutral current $\Delta$ radiative decay is by far the most common expected source of true single photon events in $\mathcal{O}$(GeV) energy neutrino-nucleus interactions. In this process, the neutrino interacts with a nucleon via a neutral $Z$ boson, exciting the nucleon to a $\Delta$ resonance. This $\Delta$ resonance decays to a nucleon and a pion 99.4\% of the time, but rarely, it can decay instead to a nucleon and a single photon. This process can happen by exciting a proton to a $\Delta^+$, or by exciting a neutron to a $\Delta^0$, as shown in Fig. \ref{fig:nc_delta_feynman}.

According to the Particle Data Group, the branching ratio of $\Delta \rightarrow N \gamma$ is 0.55-0.65\%, so the uncertainty in this rate is only 8.3\% \cite{ParticleDataGroup}. The branching ratio for $\Delta^+ \rightarrow p \gamma$ and $\Delta^0 \rightarrow n \gamma$ are known to be identical due to isospin symmetry. This rare branching ratio is primarily determined via a fitting to many different observation channels from high energy pion and photon scattering experiments \cite{resonances_global_fit}. In particular, measurements of $p \gamma \rightarrow \Delta^+ \rightarrow \pi^+ n$ allow us to access this physics, by directly measuring the inverse of this decay process. As an example, see Ref. \cite{p_gamma_to_n_pi_example}, which uses a fixed target inside a photon beam produced by Bremsstrahlung from an accelerated electron beam.

\begin{figure}[H]
    \centering
    \includegraphics[trim=0 2 0 0, clip, width=0.6\textwidth]{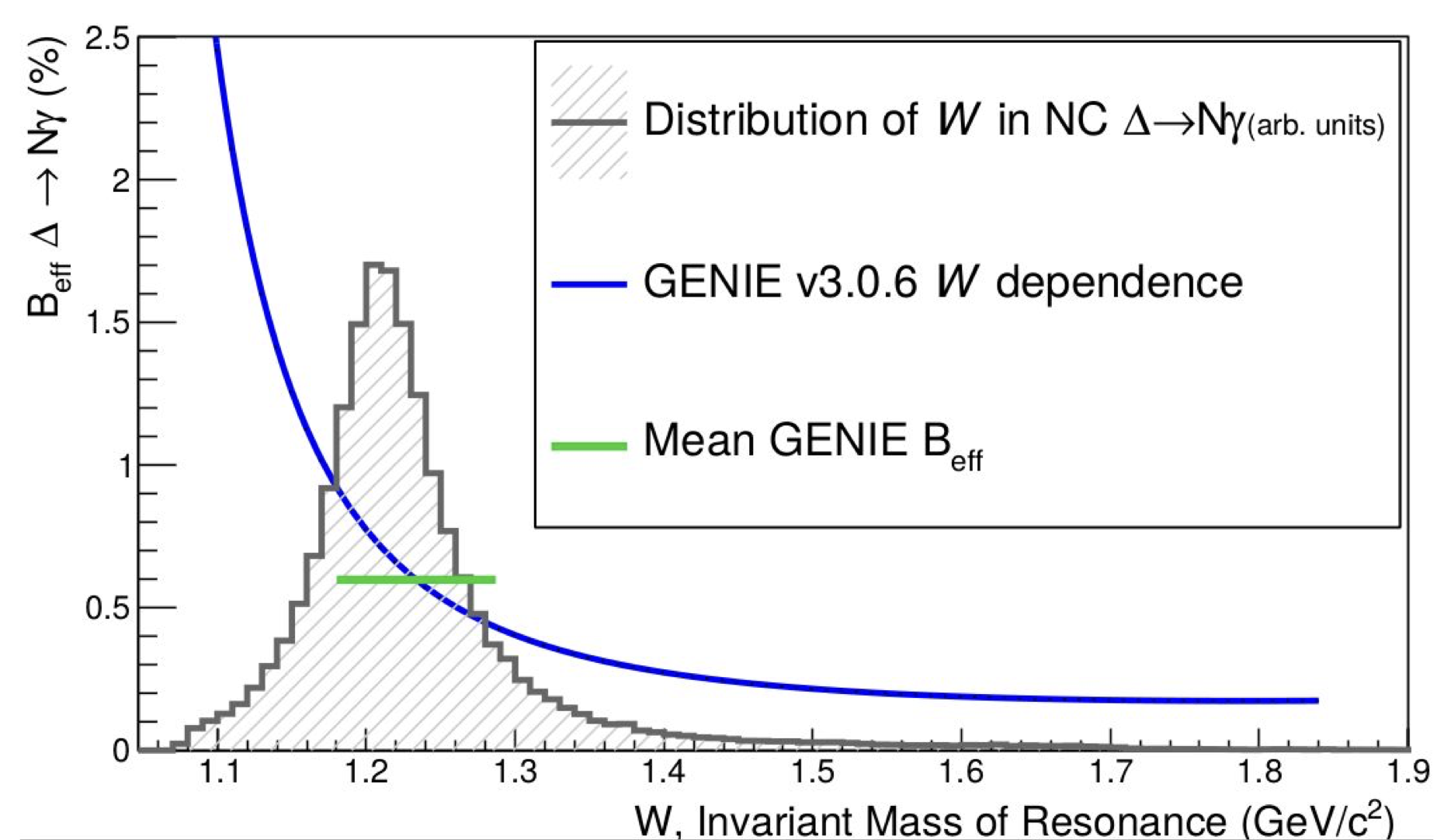}
    \caption[NC $\Delta\rightarrow N \gamma$ effective branching fraction]{NC $\Delta\rightarrow N \gamma$ effective branching fraction in MicroBooNE. Figure from Ref. \cite{mark_2021_nc_delta_wine_cheese}.}
    \label{fig:effective_branching_fraction}
\end{figure}

Note that in our experiments, we see a resonance consisting of virtual $\Delta$ particles which decay very quickly, and therefore this branching fraction is modified relative to an on-shell $\Delta$. We use GENIE to model the branching ratio as a function of $W$, the invariant mass of the $\Delta$ resonance, and to estimate the distribution of $W$ we expect to see in our interactions, as shown in Fig. \ref{fig:effective_branching_fraction}. Therefore, in neutrino-induced NC $\Delta \rightarrow N \gamma$ measurements, we will consider the effective branching fraction $\mathrm{B}_\mathrm{eff}\ \Delta \rightarrow N \gamma$. 

\subsection{NC \texorpdfstring{$\Delta$}{Delta} Radiative Decay in MiniBooNE}\label{sec:miniboone_318}

\begin{figure}[H]
    \centering
    \begin{subfigure}[b]{0.59\textwidth}
        \includegraphics[width=\textwidth]{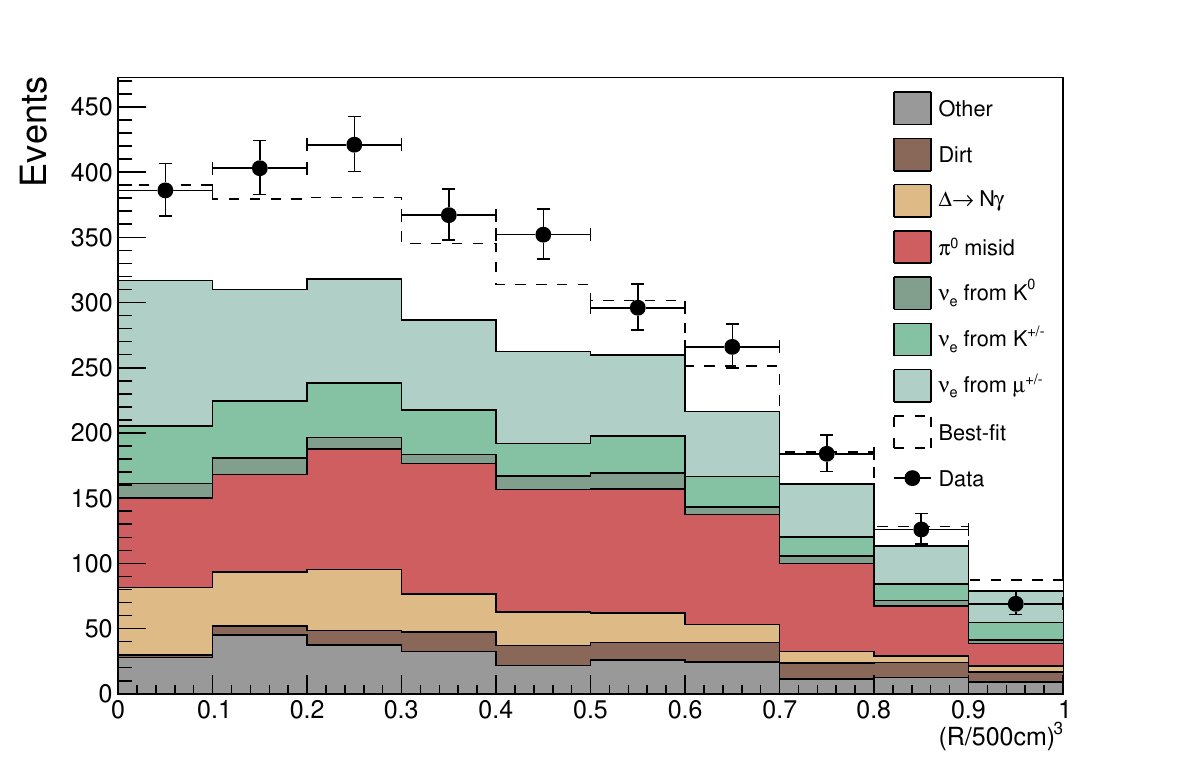}
        \caption{}
        \label{fig:miniboone_radius_distribution}
    \end{subfigure}
    \begin{subfigure}[b]{0.4\textwidth}
        \includegraphics[trim=15 600 1450 0, clip, width=\textwidth]{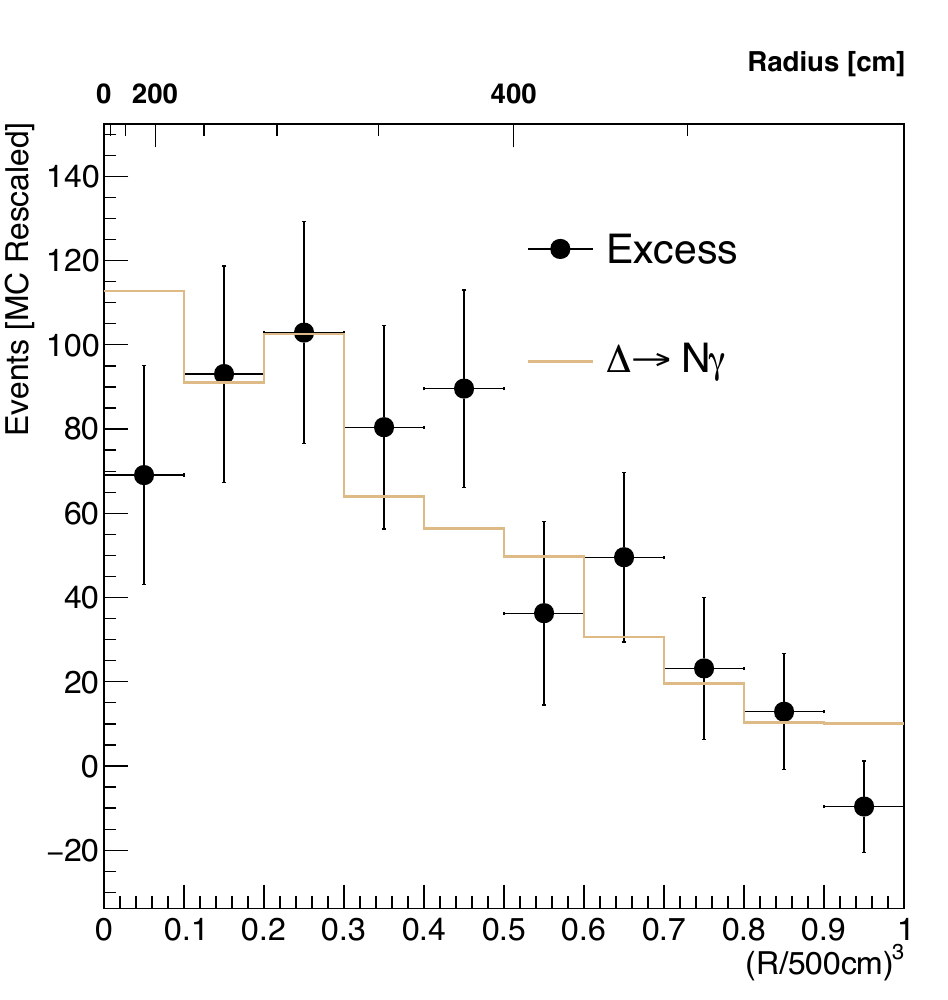}
        \caption{}
        \label{fig:miniboone_radial_fit_nc_delta}
    \end{subfigure}
    \caption[MiniBooNE radial fit to NC $\Delta$ scaling]{MiniBooNE radial fit to NC $\Delta$ scaling. Panel (a) shows the LEE radial distribution, and panel (b) shows the radial distribution after scaling up NC $\Delta$ radiative events by a factor of 3.18. Figures from Ref. \cite{miniboone_lee}.}
    \label{fig:miniboone_radial}
\end{figure}

The majority of the photon background in MiniBooNE's $\nu_e$CC search is mis-reconstructed NC $\pi^0$ events where one photon is not properly identified, but true single photons from the NC $\Delta\rightarrow N \gamma$ process also make up a sizable portion of the events. In addition to the shower energy and angle discussed previously, MiniBooNE uses the radial distribution of events. $\pi^0$ events make up a bigger fraction of the prediction near the surface of the detector at large radius values, while true single photons make up a bigger fraction of the prediction near the center of the detector at small radius values, as shown in Fig. \ref{fig:miniboone_radius_distribution}. The excess exists to a greater extent near the center than near the edges, which is one factor that disfavors a mismodeled $\pi^0$ explanation of the anomaly. To quantify this radial distribution more, different scalings of a variety of different event types were fitted to this radial distribution. The best fit was a factor of 3.18 multiplicative enhancement of the NC $\Delta\rightarrow N \gamma$ rate, as shown in Fig. \ref{fig:miniboone_radial_fit_nc_delta}.

\subsection{Previous Searches For NC \texorpdfstring{$\Delta$}{Delta} Radiative Decay}

Inspired by the MiniBooNE LEE, two neutrino experiments have previously set limits on the NC $\Delta \rightarrow N \gamma$ process.

\begin{figure}[H]
    \centering
    \begin{subfigure}[b]{0.45\textwidth}
        \includegraphics[angle=90, width=\textwidth]{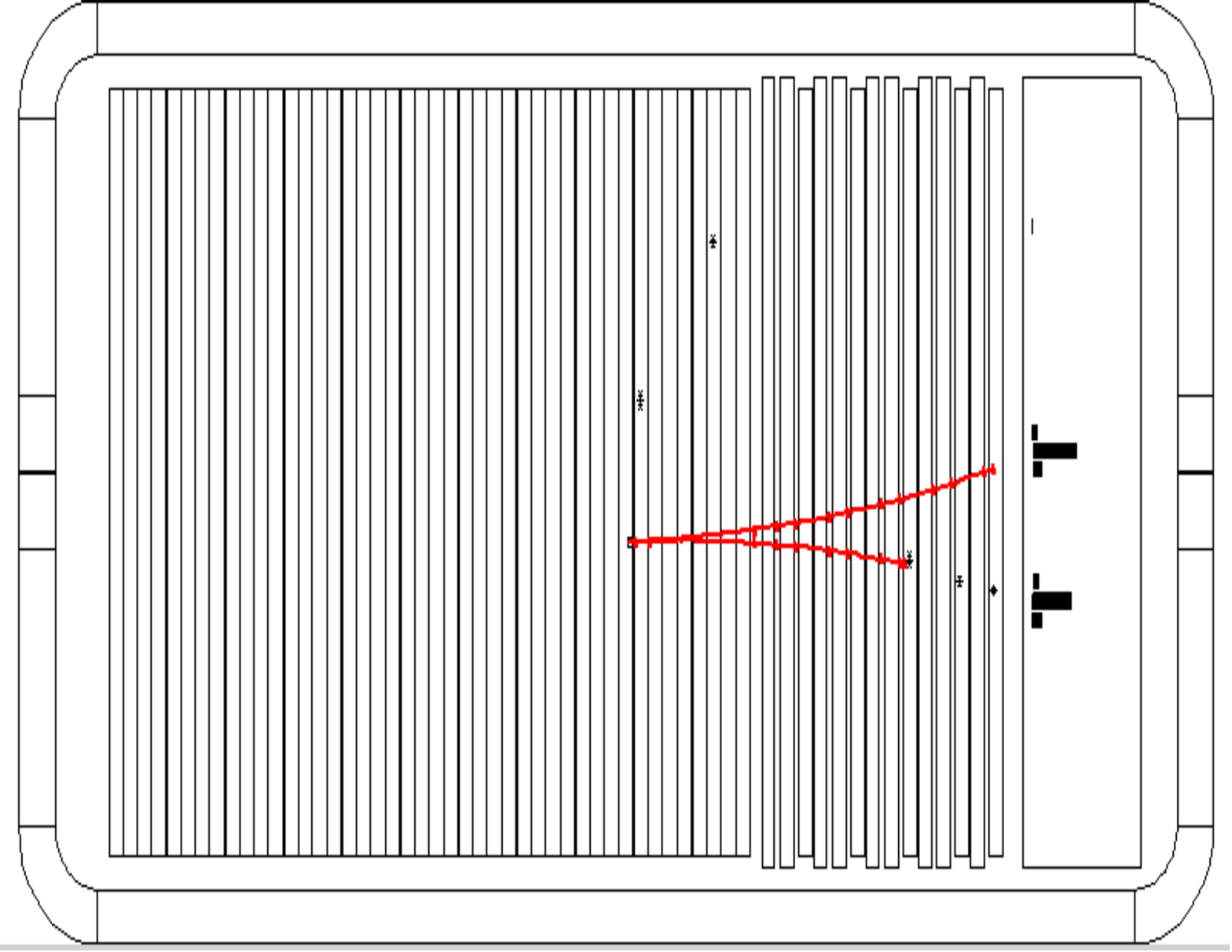}
        \caption{}
        \label{fig:nomad_event_display}
    \end{subfigure}
    \begin{subfigure}[b]{0.26\textwidth}
        \includegraphics[trim=0 0 30 0, clip, width=\textwidth]{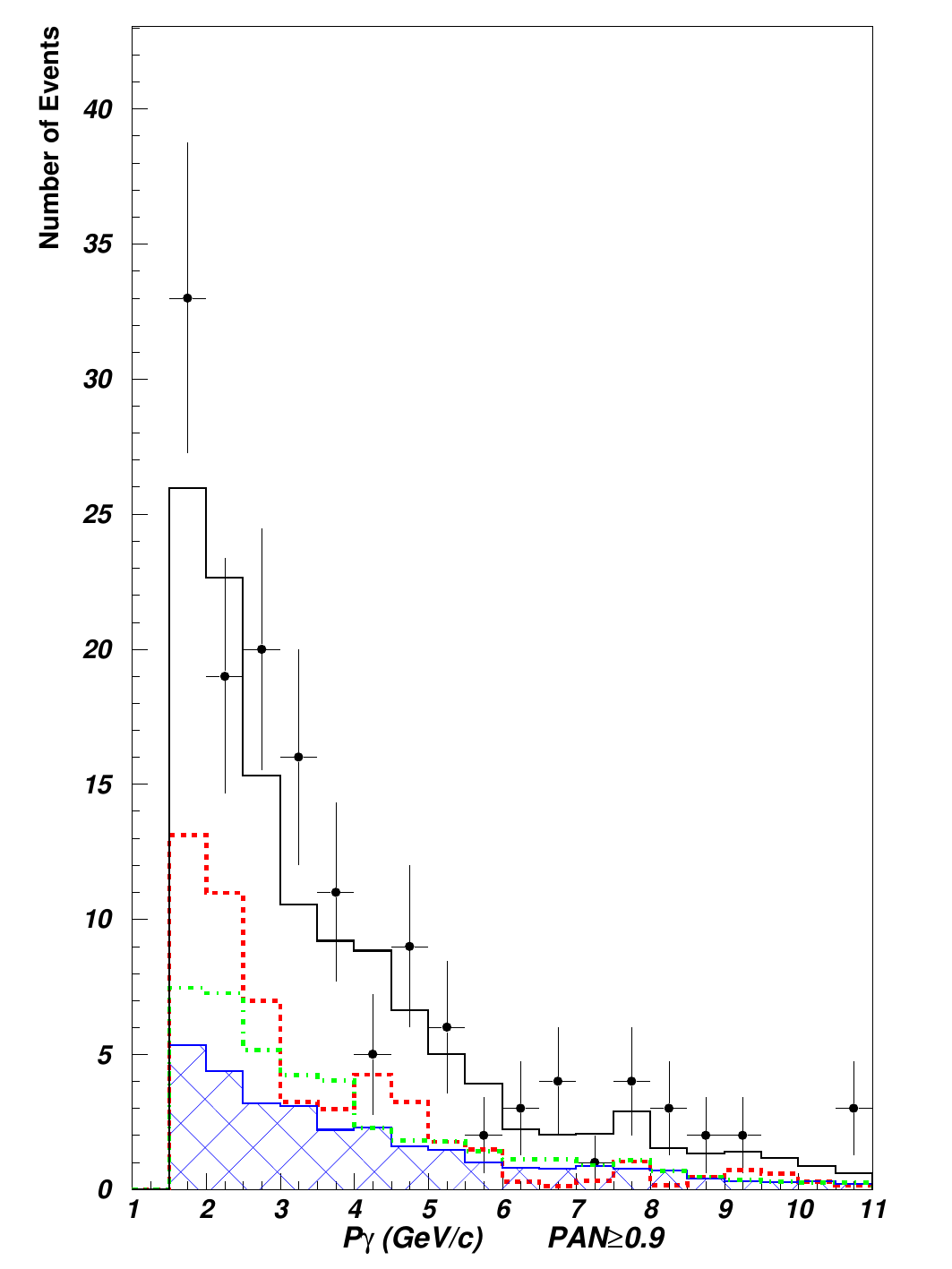}
        \caption{}
        \label{fig:nomad_energy}
    \end{subfigure}
    \begin{subfigure}[b]{0.26\textwidth}
        \includegraphics[trim=0 0 30 0, clip, width=\textwidth]{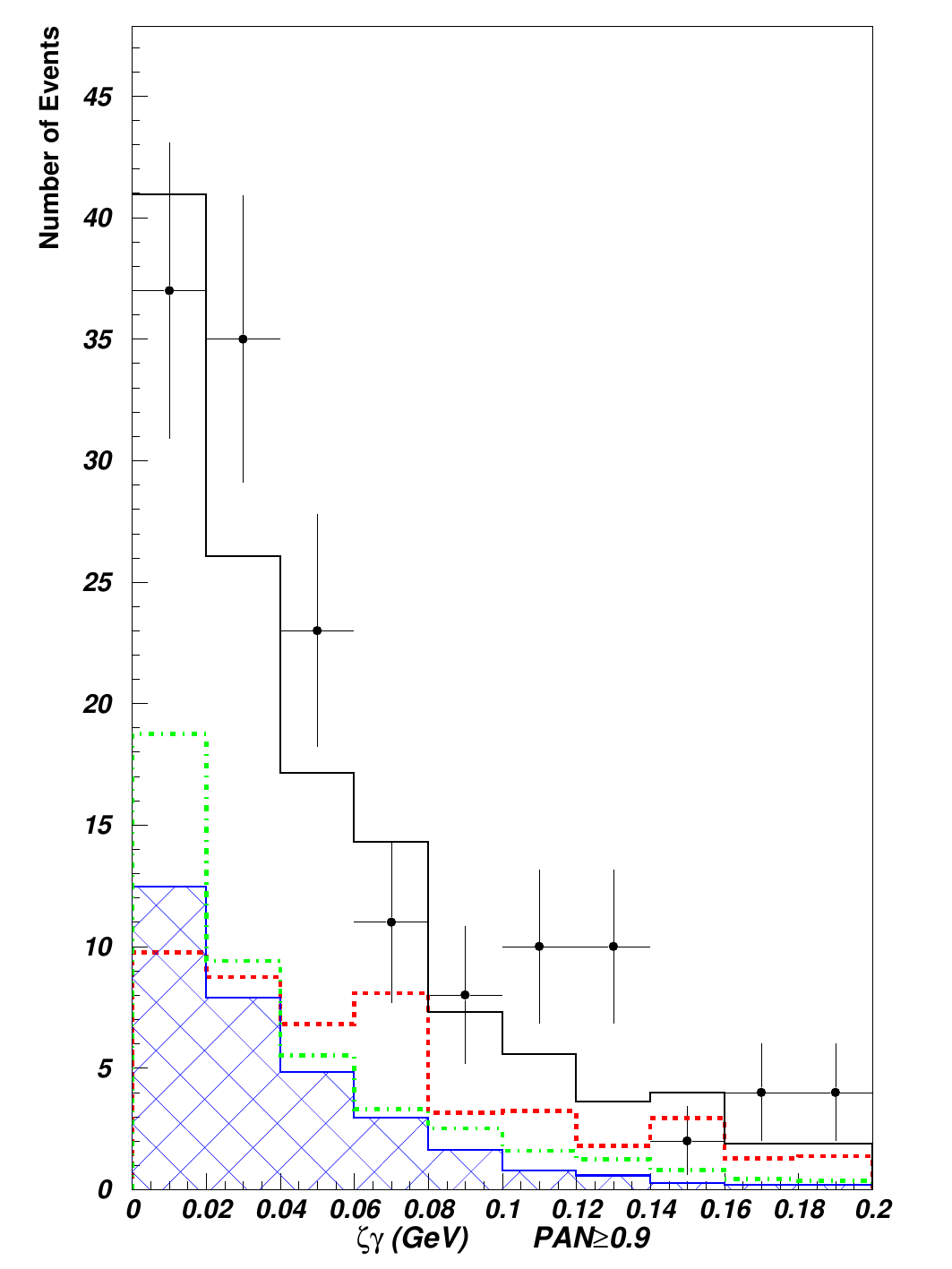}
        \caption{}
        \label{fig:nomad_angle}
    \end{subfigure}
    \caption[NOMAD single photon search]{NOMAD single photon search. Panel (a) shows a single photon event, with electron and positron tracks diverging due to the magnetic field. Panel (b) shows the single photon momentum distribution, and panel (c) shows a variable related to the angular distribution. The coherent $\pi^0$ prediction is in blue, the deep inelastic scattering prediction is in red, the out of fiducial volume prediction is in green, and the total prediction is in black. Figures from Ref. \cite{NOMAD_nc_delta}.}
    \label{fig:nomad_single_photon}
\end{figure}

The NOMAD experiment saw a significantly higher energy neutrino flux from the CERN SPS, with an average neutrino energy of around 25 GeV. Single photons are analyzed in the drift chambers, which have a low density of 0.1 $\mathrm{g}/\mathrm{cm}^3$, causing minimal scattering and allowing electron and positrons to appear as tracks rather than showers. The electron and positron tracks are split apart by a 0.4 T magnetic field so each track can be reconstructed as shown in Fig. \ref{fig:nomad_event_display}. The resulting data events are shown in Figs. \ref{fig:nomad_energy}-\ref{fig:nomad_angle}. No significant excess is observed, and an upper limit is placed on general NC $1\gamma$ processes, which includes NC $\Delta\rightarrow N \gamma$.

\begin{figure}[H]
    \centering
    \begin{subfigure}[b]{0.51\textwidth}
        \includegraphics[width=\textwidth]{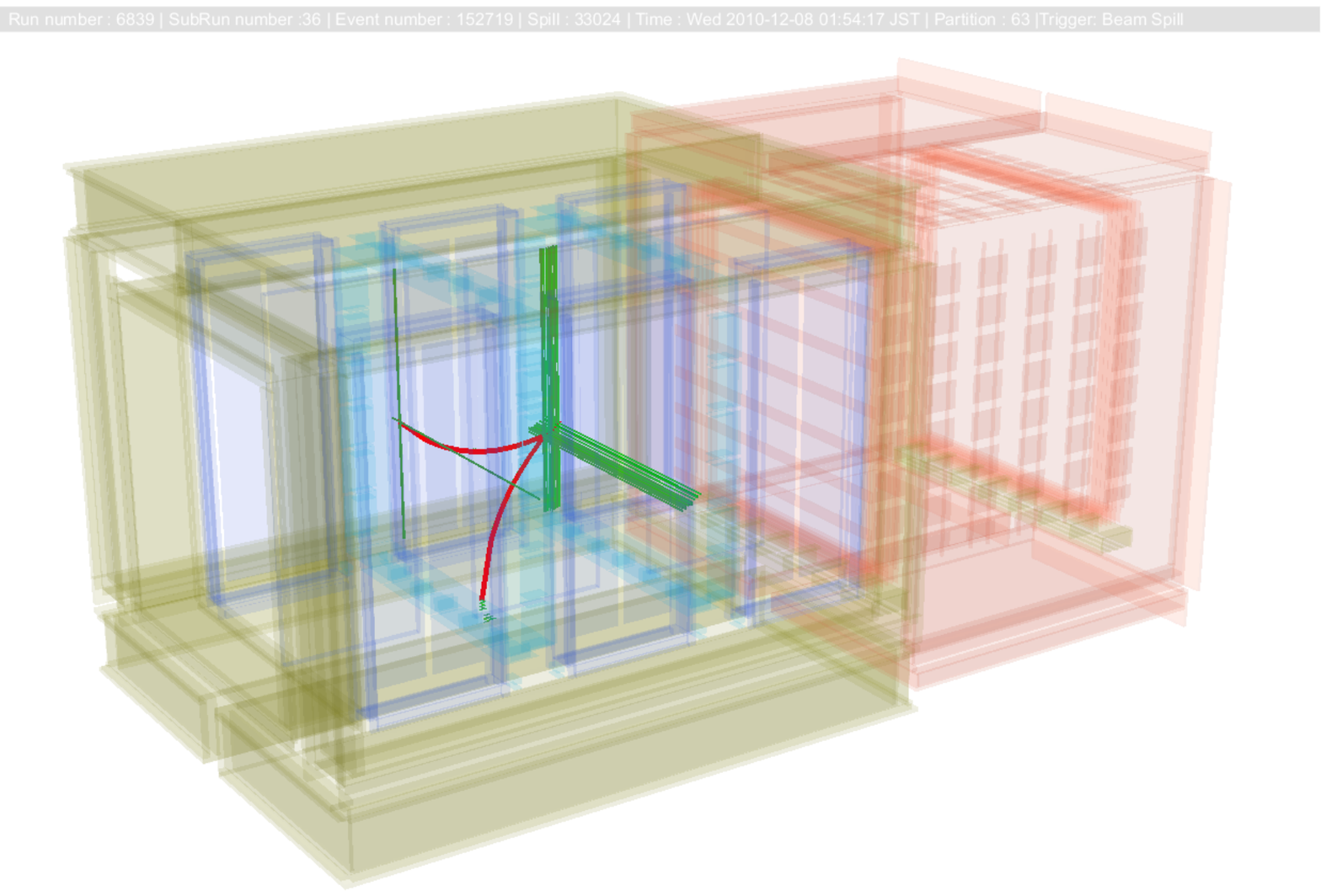}
        \caption{}
        \label{fig:t2k_event_display}
    \end{subfigure}
    \begin{subfigure}[b]{0.49\textwidth}
        \includegraphics[trim=10 0 50 20, clip, width=\textwidth]{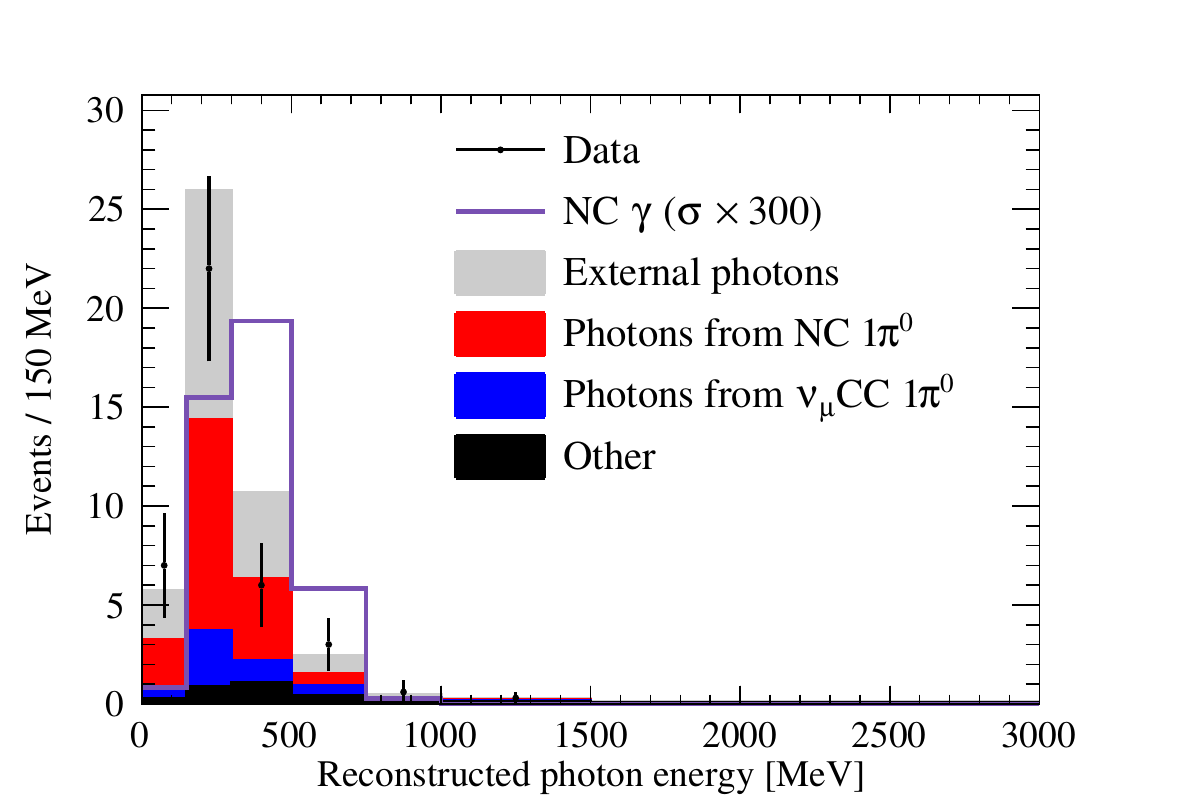}
        \caption{}
        \label{fig:t2k_energy}
    \end{subfigure}
    \begin{subfigure}[b]{0.49\textwidth}
        \includegraphics[trim=10 0 50 20, clip, width=\textwidth]{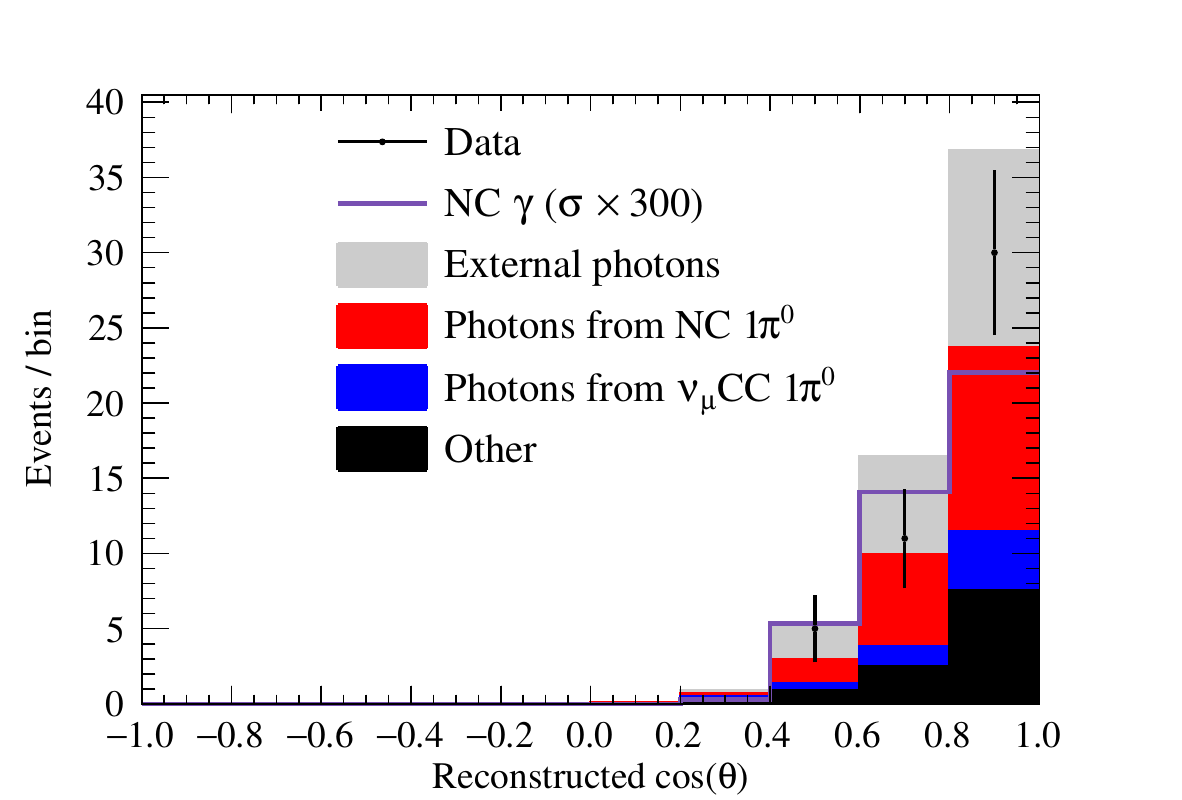}
        \caption{}
        \label{fig:t2k_angle}
    \end{subfigure}
    \caption[T2K single photon search]{T2K NC $\Delta\rightarrow N \gamma$ search results. Panel (a) shows the energy distribution, and panel (b) shows the angle distribution. Figures from Ref. \cite{t2k_nc_delta}.}
    \label{fig:t2k_nc_delta_energy_angle}
\end{figure}

The T2K experiment's ND280 near detector performed a similar search, using a neutrino beam with average energy of about 0.6 GeV, much closer to the relevant energy range to study the MiniBooNE LEE. It identifies photon events by plastic scintillators as well as a magnetized gas TPC, as shown in Fig. \ref{fig:t2k_event_display}. The resulting data events are shown in Figs. \ref{fig:t2k_energy}-\ref{fig:t2k_angle}, and no significant excess is observed.

\begin{figure}[H]
    \centering
    \includegraphics[width=0.7\textwidth]{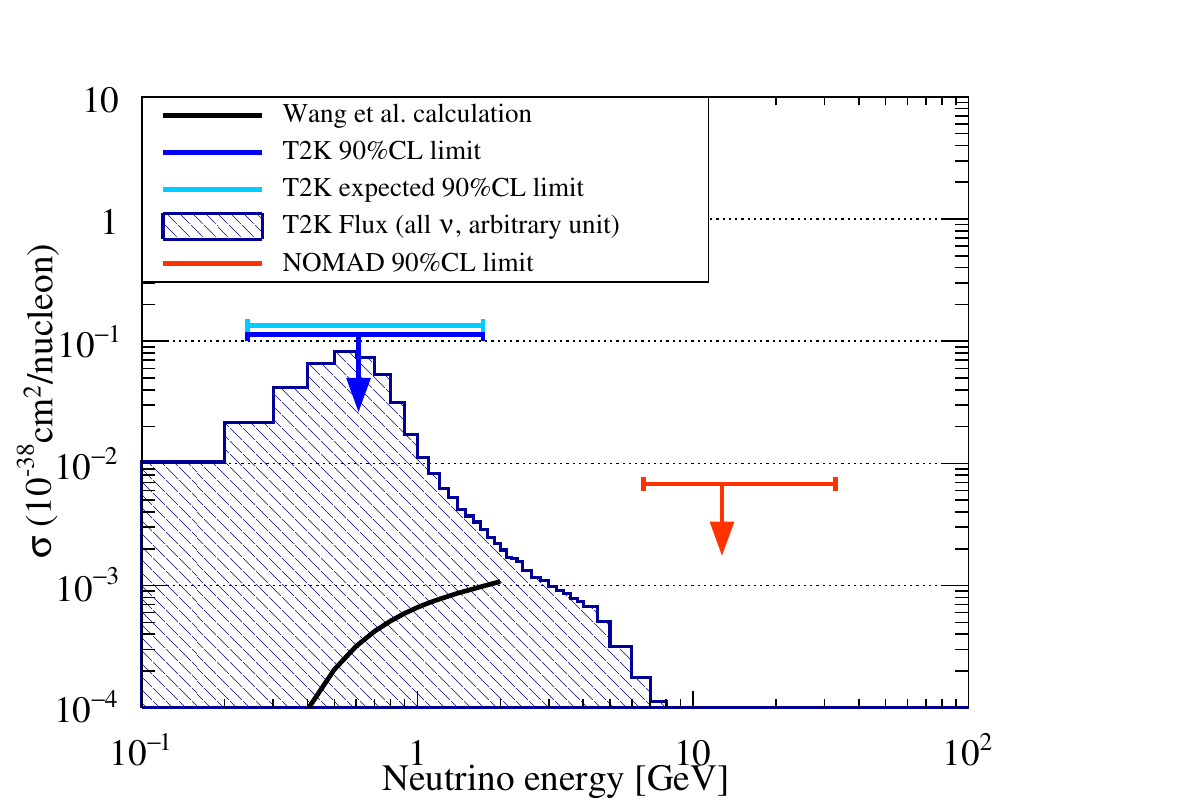}
    \caption[T2K and NOMAD NC $\Delta$ exclusions]{T2K and NOMAD NC $\Delta$ exclusions, considering their neutrino energy ranges. Figures from Ref. \cite{t2k_nc_delta}.}
    \label{fig:t2k_nomad_nc_delta_exclusions}
\end{figure}

Both the NOMAD and T2K results can be interpreted as limits on the total NC $1\gamma$ cross section, as shown in Fig. \ref{fig:t2k_nomad_nc_delta_exclusions}. The T2K results can also be compared to a theoretical prediction which is dominated by the NC $\Delta\rightarrow N \gamma$ prediction \cite{wang_nc1g_prediction} in the same general energy region as the MiniBooNE LEE, but the exclusion is more than a factor of 100 higher from the prediction, leaving much room for improvement.

\section{Pandora NC \texorpdfstring{$\Delta$}{Delta} Radiative Decay Search in MicroBooNE}\label{sec:pandora_nc_delta_search}

Now, I will describe the first MicroBooNE search for single photons, specifically for the NC $\Delta\rightarrow N \gamma$ process using Pandora reconstruction.

MiniBooNE's favored scaling consisting of a 318\% increase in the NC $\Delta\rightarrow N \gamma$ rate is far outside of the 8.3\% uncertainty on the $\Delta\rightarrow N \gamma$ branching fraction. However, it remains an interesting possibility to test. This process has never been measured in neutrino interactions, and it is the only significant predicted source of single photons in MiniBooNE and MicroBooNE. In addition to a literal increase in the NC $\Delta\rightarrow N \gamma$ rate, this factor can also be interpreted as a more general quantitative prediction for how many photons we should expect at these energy and angle ranges in order to explain the MiniBooNE LEE. In particular, this factor gives us a concrete prediction for how many photon events at these energy and angle ranges we should expect to appear in MicroBooNE. The NC $\Delta\rightarrow N \gamma$ process gives us a model for how event rates scale with the different baselines, beam exposures, volumes, and target nuclei between the MiniBooNE and MicroBooNE experiments. Of course, this is not the only possible event rate scaling between the two detectors; for example, coherent explanations would give significantly larger predictions in MicroBooNE due to the larger argon nucleus relative to carbon, and beyond-standard-model decays in flight would give significantly smaller predictions in MicroBooNE due to smaller volume relative to MiniBooNE. When we do interpret this in the context of a broader search for single photons, we must carefully consider hadronic system kinematics, as we will describe in Sec. \ref{sec:combined_pandora_wc_nc_delta}.

\begin{figure}[H]
    \centering
    \includegraphics[width=0.5\textwidth]{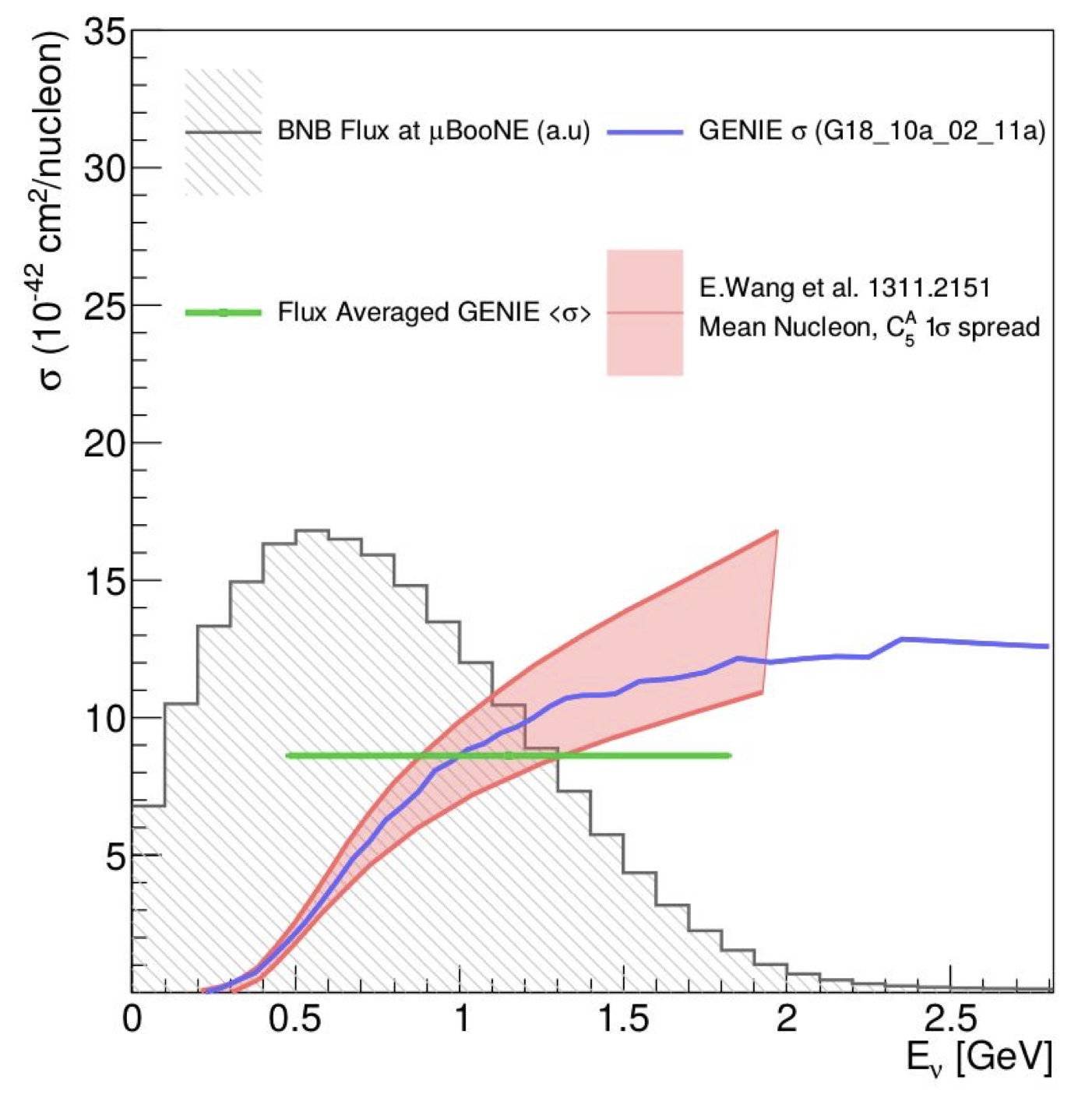}
    \caption[NC $\Delta\rightarrow N \gamma$ cross section predictions]{NC $\Delta\rightarrow N \gamma$ cross section predictions. Figure from Ref. \cite{mark_2021_nc_delta_wine_cheese}.}
    \label{fig:nc_delta_genie_xs}
\end{figure}

To predict the rate of NC $\Delta\rightarrow N \gamma$ in MicroBooNE, we use GENIEv3 \cite{genie_v3, genie-tune-paper}, which agrees with predictions from Ref. \cite{wang_nc1g_prediction}, as shown in Fig. \ref{fig:nc_delta_genie_xs}. The NC $\Delta\rightarrow N \gamma$ process is predicted to be very rare, with only 124.1 events expected in MicroBooNE's first three years of running consisting of $6.80\cdot 10^{20}$ POT, which makes this analysis consist of relatively low statistics, and makes $\pi^0$ backgrounds very challenging.

This first single photon search in MicroBooNE used Pandora reconstruction \cite{microboone_pandora}. Unlike the Wire-Cell reconstruction method described in Sec. \ref{sec:wire_cell}, Pandora performs particle clustering in each 2D view before attempting to reconstruct the full 3D particle activity in an event. This fundamental change in the basic reconstruction affects many downstream decisions about how the selection is performed.

\begin{figure}[H]
    \centering
    \begin{subfigure}[b]{0.48\textwidth}
        \includegraphics[width=\textwidth]{figs/event_displays/data_1g1p_9524_127_6375.pdf}
        \caption{}
    \end{subfigure}
    \begin{subfigure}[b]{0.50\textwidth}
        \includegraphics[width=\textwidth]{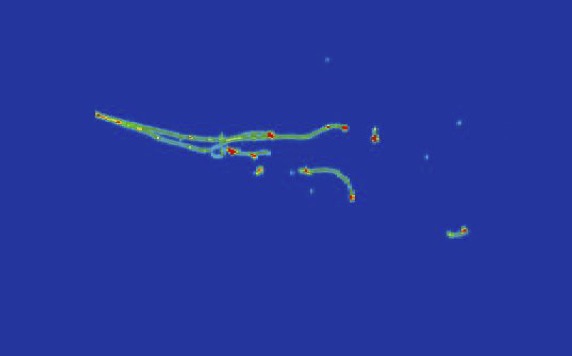}
        \caption{}
    \end{subfigure}
    \caption[NC $\Delta\rightarrow N \gamma$ event displays]{NC $\Delta\rightarrow N \gamma$ event displays. Panel (a) shows a candidate $1\gamma1p$ event, where one photon points back to the neutrino vertex which also produced a proton track, in run 9524 subrun 127 event 6375. Panel (b) shows a simulated $1\gamma0p$ event, where the neutrino interaction happened upstream of this photon shower with no visible activity.}
    \label{fig:nc_delta_event_displays}
\end{figure}

The Pandora NC $\Delta\rightarrow N \gamma$ search targets two specific topologies: events only containing a single photon and a single proton $1\gamma 1p$, and events containing only a single photon $1\gamma 0p$, as shown in Fig. \ref{fig:nc_delta_event_displays}. These two topologies match the decay products of the $\Delta^+$ and $\Delta^0$ resonances, although the specific observable topologies are affected by final state interactions, when particles produced by the neutrino scattering interact with the nuclear medium before escaping.

After particles are clustered on each projected view and matched between views, they are classified as tracks or showers and clustered into a candidate neutrino interaction. For this analysis, we then perform a topological selection, keeping only events with exactly one reconstructed shower and zero or one reconstructed tracks. Next, we form a pre-selection by requiring all particles to be contained within the fiducial volume, requiring a sufficiently high energy shower, requiring the track to be sufficiently short and highly-ionizing in order to be consistent with a proton, and requiring that the track and shower are not collinear in order to minimize badly reconstructed events where the shower stem could be confused with the track.

Next, we pass these events through several BDTs, each attempting to reject a specific category of background. This includes a cosmic BDT, an NC $\pi^0$ BDT, a $\nu_e$CC BDT, a $\nu_\mu$CC BDT, and a second-shower-veto BDT, which attempts to reject events in which a $\pi^0$ causes a second photon shower nearby which is visible on one or more projected views but was not successfully reconstructed in 3D. For $1\gamma 0p$, the $\nu_e$CC, $\nu_\mu$CC, and second-shower-veto BDTs were merged, since there is no track present. The training of these BDTs and the choosing of optimal cut values was performed separately for $1\gamma 1p$ and $1\gamma 0p$.

\begin{figure}[H]
    \centering
    \begin{subfigure}[b]{0.49\textwidth}
        \includegraphics[width=\textwidth]{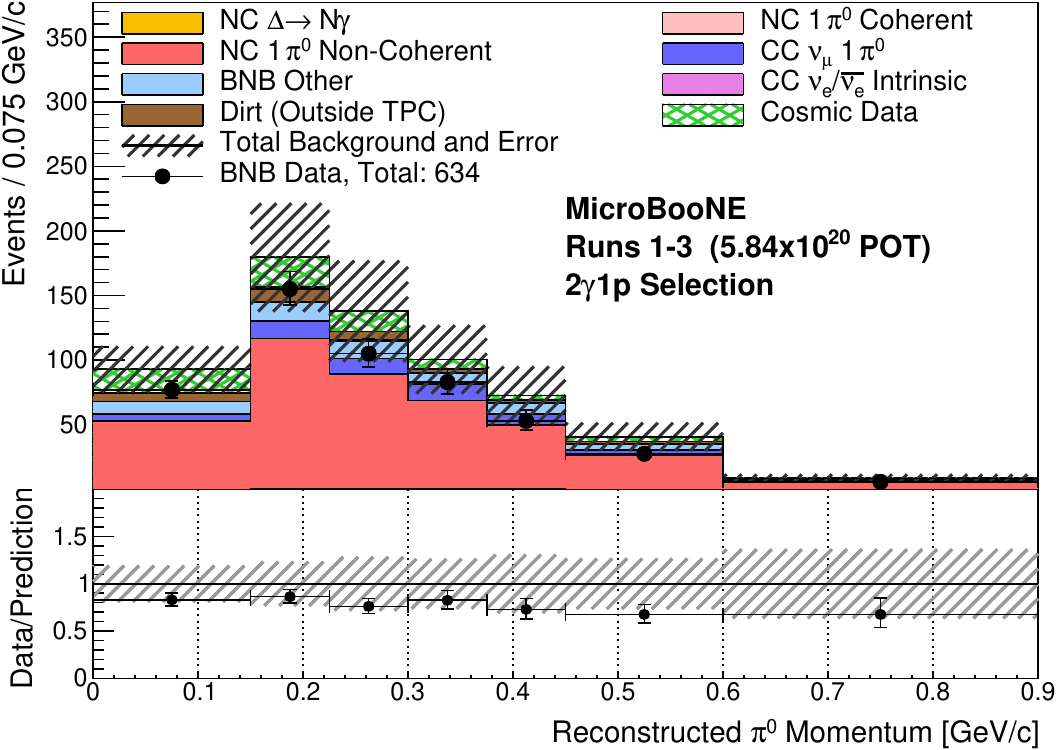}
        \caption{}
    \end{subfigure}
    \begin{subfigure}[b]{0.49\textwidth}
        \includegraphics[width=\textwidth]{figs/nc_delta/pigLEE_combined_datamc_Data2g1pFiltered_Reconstructedpi0MomentumGeVc_stage_2FinalFullSys_crop.pdf}
        \caption{}
    \end{subfigure}
    \caption[Pandora NC $\Delta$ constraining NC $\pi^0$ observations]{Constraining NC $\pi^0$ distributions used to constrain the Pandora NC $\Delta$ radiative decay selections. Panel (a) shows events with reconstructed protons, and panel (b) shows events without reconstructed protons. Figures from Ref. \cite{glee_prl}.}
    \label{fig:pandora_nc_delta_nc_pi0}
\end{figure}

\begin{figure}[H]
    \centering
    \includegraphics[width=0.6\textwidth]{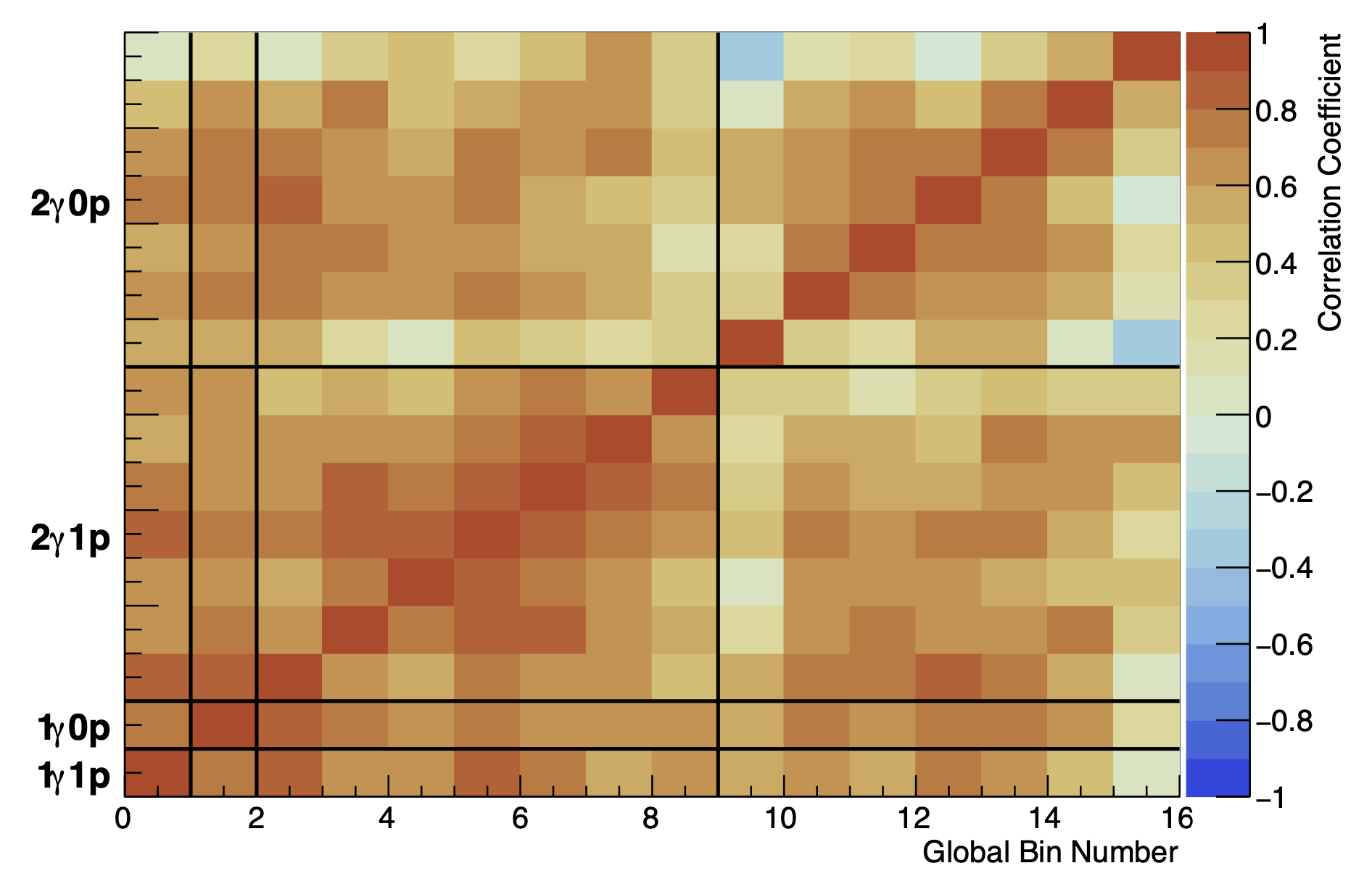}
    \caption[Pandora NC $\Delta$ correlations]{Correlation matrix between the Pandora NC $\Delta$ radiative decay selections ($1\gamma 1p$ and $1\gamma 0p$) and NC $\pi^0$ selections ($2\gamma 1p$ and $2\gamma 0p$). Figure from Ref. \cite{glee_prl}.}
    \label{fig:pandora_nc_delta_corr}
\end{figure}

Using the same reconstruction tools, constraining NC $\pi^0$ sidebands were developed, both with a proton ($2\gamma 1p$) and without a proton ($2\gamma 0p$), as shown in Fig. \ref{fig:pandora_nc_delta_nc_pi0}. Notably, we see a slight overprediction of the data in the entire energy range in both channels. Systematic uncertainties are highly correlated between the $1\gamma$ and $2\gamma$ samples as shown in Fig. \ref{fig:pandora_nc_delta_corr}. There are strong correlations between $\Delta\rightarrow N \pi$ in the sideband channels and $\Delta\rightarrow N \gamma$ in the signal channels, since the Delta resonance is the most common source of $\pi^0$ events in our detector. There are also strong correlations between $\pi^0$ events in the sideband channels and $\pi^0$ events in the signal channels, since this background makes up a majority of the selection even after all selection cuts have been performed. These sidebands play an important role in reducing systematic uncertainties through a conditional constraint, as described in Sec. \ref{sec:conditional_constraint}. 

\begin{figure}[H]
    \centering
    \begin{subfigure}[b]{0.49\textwidth}
        \includegraphics[width=\textwidth]{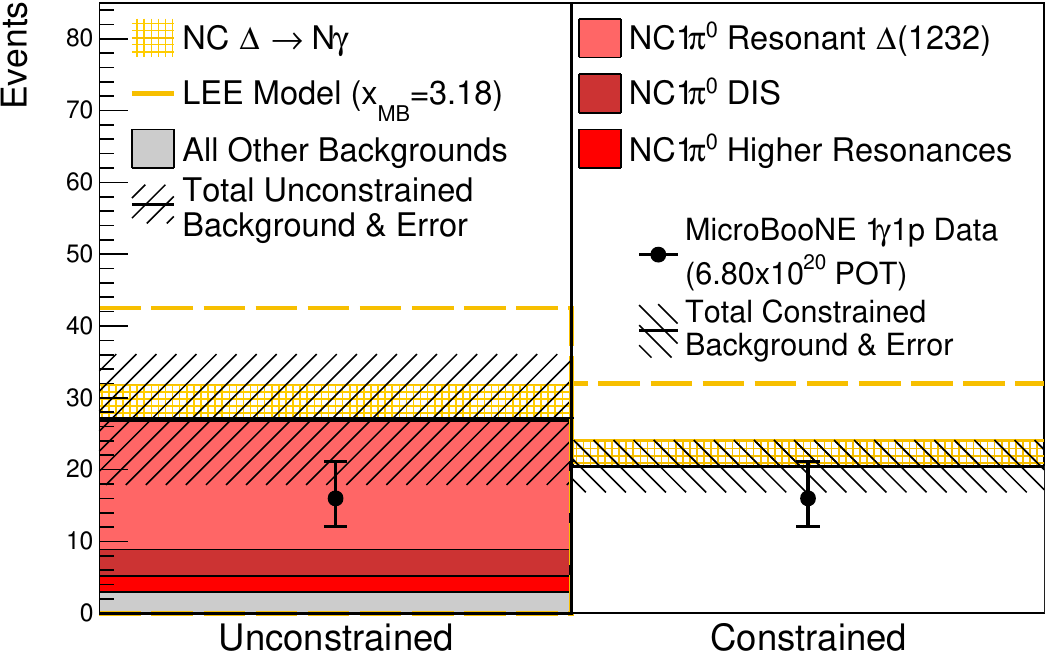}
        \caption{}
    \end{subfigure}
    \begin{subfigure}[b]{0.49\textwidth}
        \includegraphics[width=\textwidth]{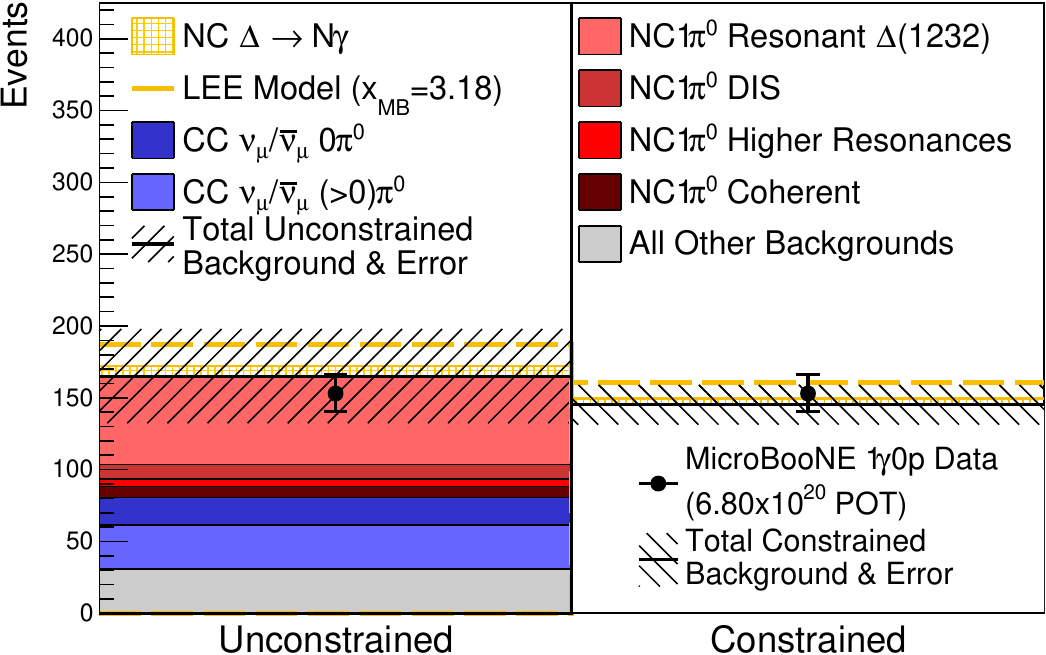}
        \caption{}
    \end{subfigure}
    \caption[Pandora NC $\Delta$ one bin]{Pandora NC $\Delta$ radiative decay results in one bin. The left of each panel shows the result before constraints, and the right of each panel shows the result after constraints. Panel (a) shows events with reconstructed protons, and panel (b) shows events without reconstructed protons. Figures from Ref. \cite{glee_prl}.}
    \label{fig:pandora_nc_delta_one_bin}
\end{figure}

Similarly to the $\nu_e$CC searches described earlier, this analysis was developed according to a blinding policy, where only a small fraction of the data was examined before the analysis was frozen. Several fake data studies were also performed in order to finalize all plots and procedures before looking at the real data. After all selection cuts, our results can be seen both with and without the application of the conditional constraint in Fig. \ref{fig:pandora_nc_delta_one_bin}. In red and blue colors, we see $\pi^0$ backgrounds which dominate the prediction in both the $1\gamma1p$ and $1\gamma0p$ channels. In black, we see the prediction with no NC $\Delta\rightarrow N \gamma$ events, with systematic uncertainties. In yellow, we see the predicted NC $\Delta\rightarrow N \gamma$ rate. In the yellow dashed line, we see the MiniBooNE LEE prediction of a 3.18 times enhancement of the NC $\Delta\rightarrow N \gamma$ rate. In each topology, we show the unconstrained prediction on the left, and the constrained prediction on the right, which shows how the observed NC $\pi^0$ deficit in our sidebands acts to reduce our prediction.

We see that our $1\gamma 1p$ channel has much more sensitivity to the LEE prediction, due to the much larger ratio of NC $\Delta\rightarrow N \gamma$ to NC $\pi^0$ in the prediction. This is because the presence of a proton gives us a much more distinctive topology and tells us the spatial location of the neutrino interaction, which gives us much more information that we use in order to reject $\pi^0$ backgrounds. In our $1\gamma 1p$ channel, we see a mild deficit, with data slightly below the no NC $\Delta\rightarrow N \gamma$ prediction, and significantly below the LEE prediction. In the $1\gamma 0p$ channel, we see data that is consistent with both the no NC $\Delta\rightarrow N \gamma$ prediction and the LEE prediction.

\begin{figure}[H]
    \centering
    \includegraphics[width=0.6\textwidth]{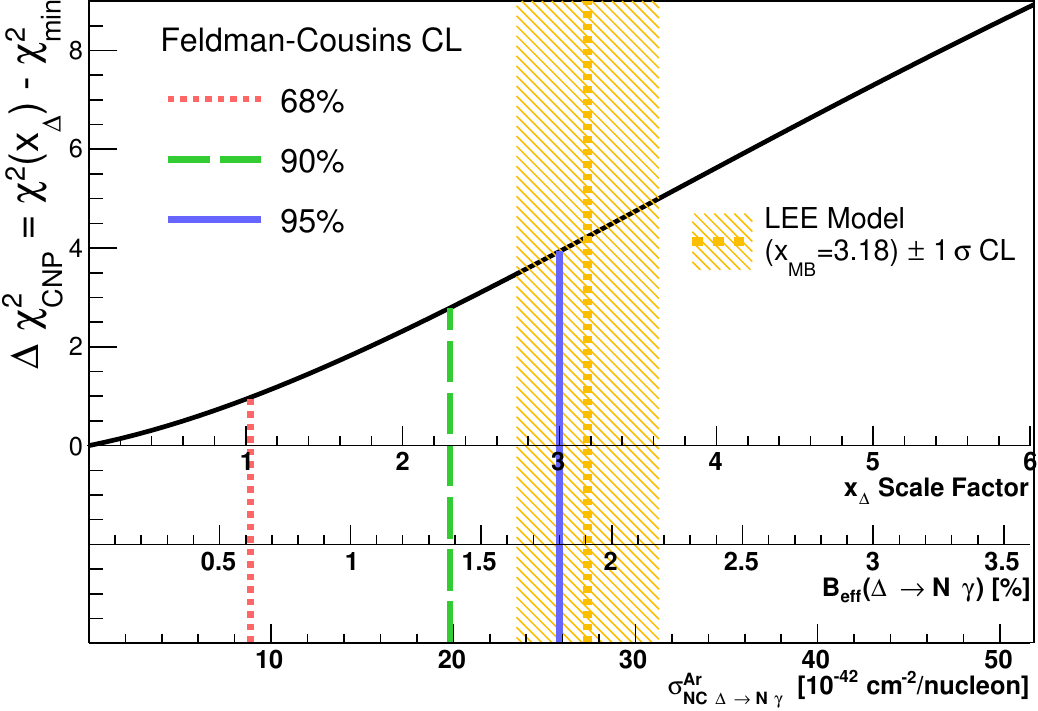}
    \caption[Pandora NC $\Delta$ scaling exclusion]{Pandora NC $\Delta$ radiative decay exclusion for $x_\Delta$ scaling. Figure from Ref. \cite{glee_prl}.}
    \label{fig:pandora_nc_delta_chi2}
\end{figure}

Similarly to our $\nu_e$CC search as shown in Fig. \ref{fig:LEEx_exclusion}, we use these results in order to analyze more general scalings of the NC $\Delta\rightarrow N \gamma$ rate. We call these multiplicative scalings $x_\Delta$, with $x_\Delta=0$ representing the no NC $\Delta\rightarrow N \gamma$ prediction, $x_\Delta=1$ representing the nominal prediction, and $x_\Delta=3.18$ representing the MiniBooNE LEE prediction. We can also estimate uncertainties on this 3.18 number by using the overall significance of the MiniBooNE LEE, creating an error band that corresponds to 4.8$\sigma$ from $x_\Delta=1$. This scale factor can also be interpreted in terms of the effective branching fraction $\mathrm{B}_\mathrm{eff}\ \Delta \rightarrow N \gamma$ as described in Fig. \ref{fig:effective_branching_fraction}, and in terms of the NC $\Delta \rightarrow N \gamma$ cross section. Note that both of these additional interpretations are performed as simple scalings according to GENIE, and do not take into account the conditional constraint which could, for example, update the flux prediction to slightly modify these values. As shown in Fig. \ref{fig:pandora_nc_delta_chi2}, our best-fit is at $x_\Delta=0$, due to the mild deficit observed in the $1\gamma 1p$ channel. In this test, we exclude $x_\Delta=3.18$ at greater than 95\% confidence level. We also perform a two-hypothesis test, similar to that shown in Fig. \ref{fig:simple_simple}, and rule out the MiniBooNE LEE 3.18 times enhanced NC $\Delta \rightarrow N \gamma$ model in favor of the nominal prediction at 94.8\% CL (1.9$\sigma$).

\begin{figure}[H]
    \centering
    \begin{subfigure}[b]{0.49\textwidth}
        \includegraphics[width=\textwidth]{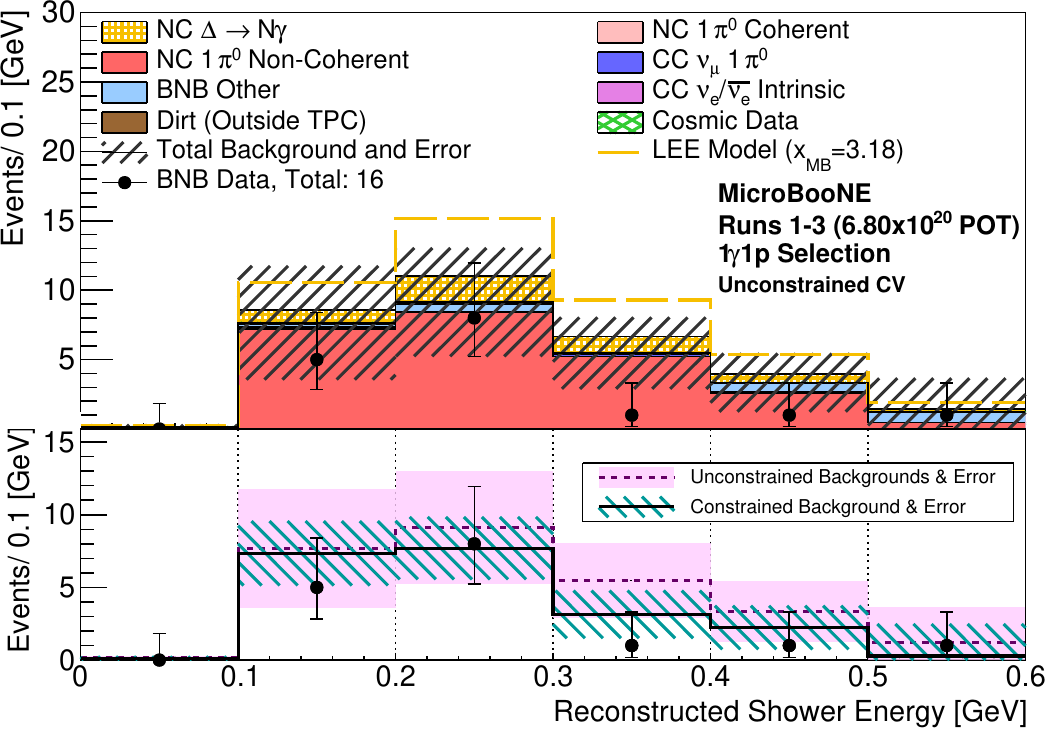}
        \caption{}
    \end{subfigure}
    \begin{subfigure}[b]{0.49\textwidth}
        \includegraphics[width=\textwidth]{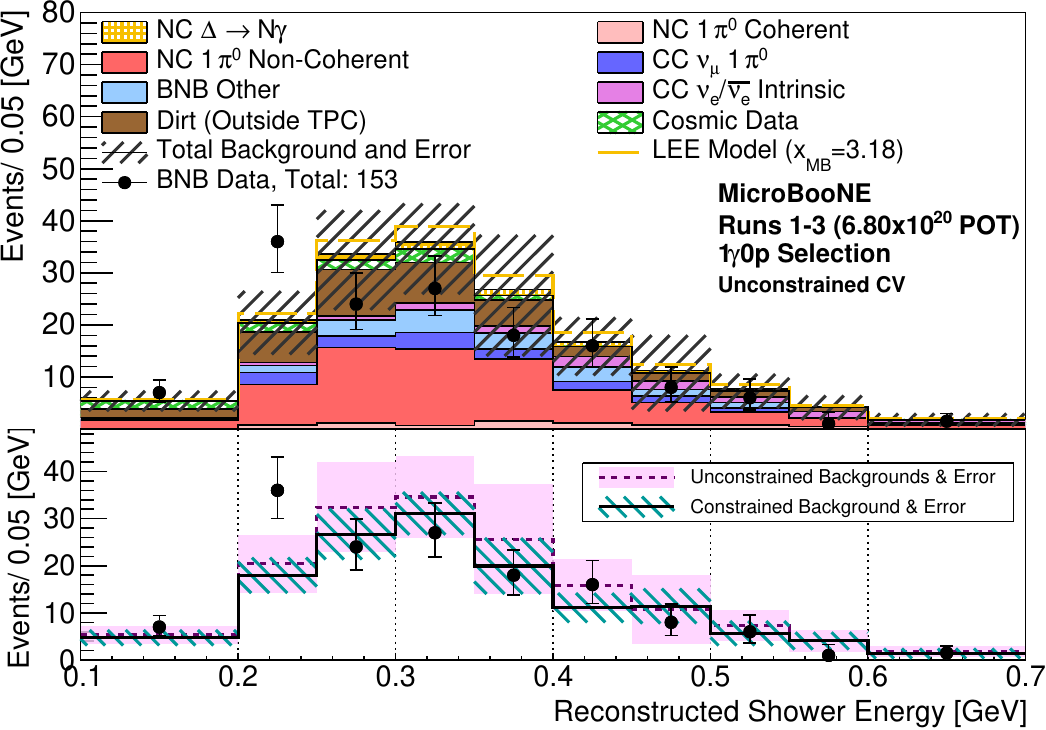}
        \caption{}
    \end{subfigure}
    \caption[Pandora NC $\Delta$ radiative decay energies]{Pandora NC $\Delta$ radiative decay shower energy distributions. Each top panel shows the result before constraints, and the bottom panel shows the result both before and after constraints. Panel (a) shows events with reconstructed protons, and panel (b) shows events without reconstructed protons. Figures from Ref. \cite{glee_prl}.}
    \label{fig:pandora_nc_delta_energy_distributions}
\end{figure}

In addition to the one-bin distributions, we also consider kinematic distributions, as shown in Fig. \ref{fig:pandora_nc_delta_energy_distributions}. We see general good agreement, however we do observe a local excess from 200-250 MeV in the $1\gamma0p$ channel. This corresponds to a local significance of 2.7$\sigma$ after constraint. Using an MC study, the probability of seeing a one bin excess this large due to statistical and systematic fluctuations is 4.74\%.

\section{Wire-Cell NC \texorpdfstring{$\Delta$}{Delta} Radiative Decay Selection in MicroBooNE}\label{sec:wc_nc_delta_selection}

Now, I will describe the selection I developed for my thesis, using Wire-Cell reconstruction to target NC $\Delta\rightarrow N \gamma$ events. This was initially intended as a cross-check of the Pandora $\Delta\rightarrow N \gamma$ selection described above, since we had three reconstruction chains being applied for $\nu_e$CC events, but only one being used for photons. Our goals for this analysis eventually evolved into an independent study of NC $\Delta\rightarrow N \gamma$ events, and then into an analysis which combines results of both the Pandora and Wire-Cell NC $\Delta\rightarrow N \gamma$ selections.

This Wire-Cell NC $\Delta\rightarrow N \gamma$ selection was developed according to our blinding policy, using only $5.3\cdot 10^{20}$ POT of BNB data while developing the selection, with $6.37 \cdot 10^{20}$ POT of BNB data from our first three years of data taking ultimately used for the final results. This selection was actually frozen a long time ago, in May 2021; this was just before the Pandora NC $\Delta\rightarrow N \gamma$ selections were unblinded in June 2021, and was also before any of our $\nu_e$CC selections were unblinded.

\begin{figure}[H]
    \centering
    \begin{subfigure}[b]{0.265\textwidth}
        \includegraphics[trim=200 0 190 0, clip, width=\textwidth]{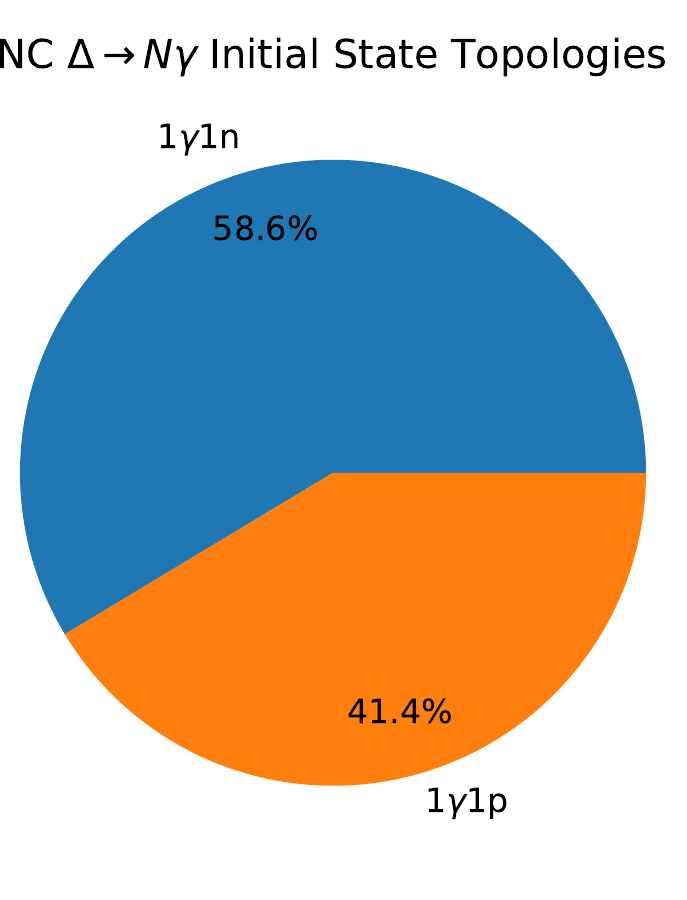}
        \caption{}
        \label{fig:nc_delta_initial_state}
    \end{subfigure}
    \begin{subfigure}[b]{0.375\textwidth}
        \includegraphics[trim=140 0 110 0, clip, width=\textwidth]{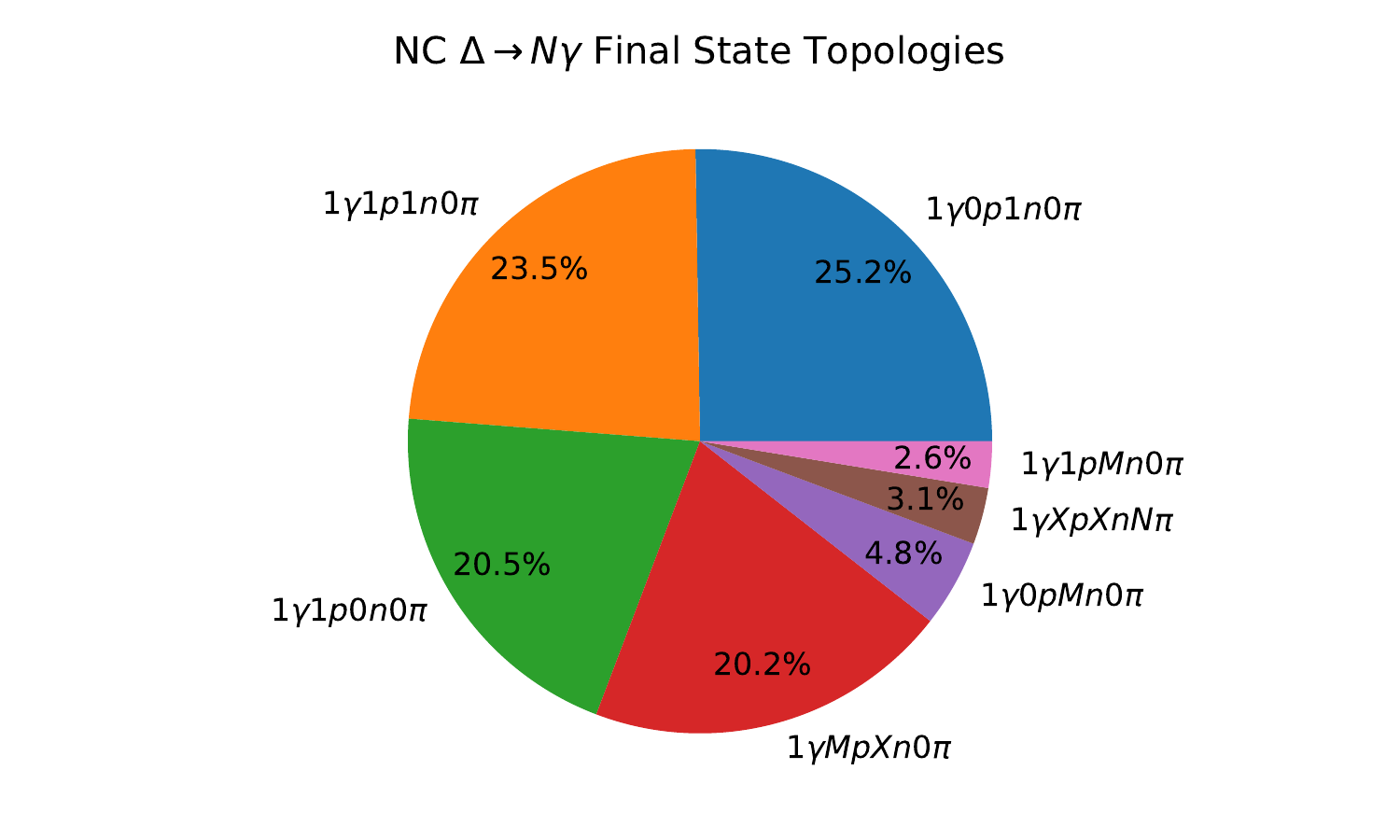}
        \caption{}
        \label{fig:nc_delta_final_state}
    \end{subfigure}
    \begin{subfigure}[b]{0.34\textwidth}
        \includegraphics[trim=170 0 140 0, clip, width=\textwidth]{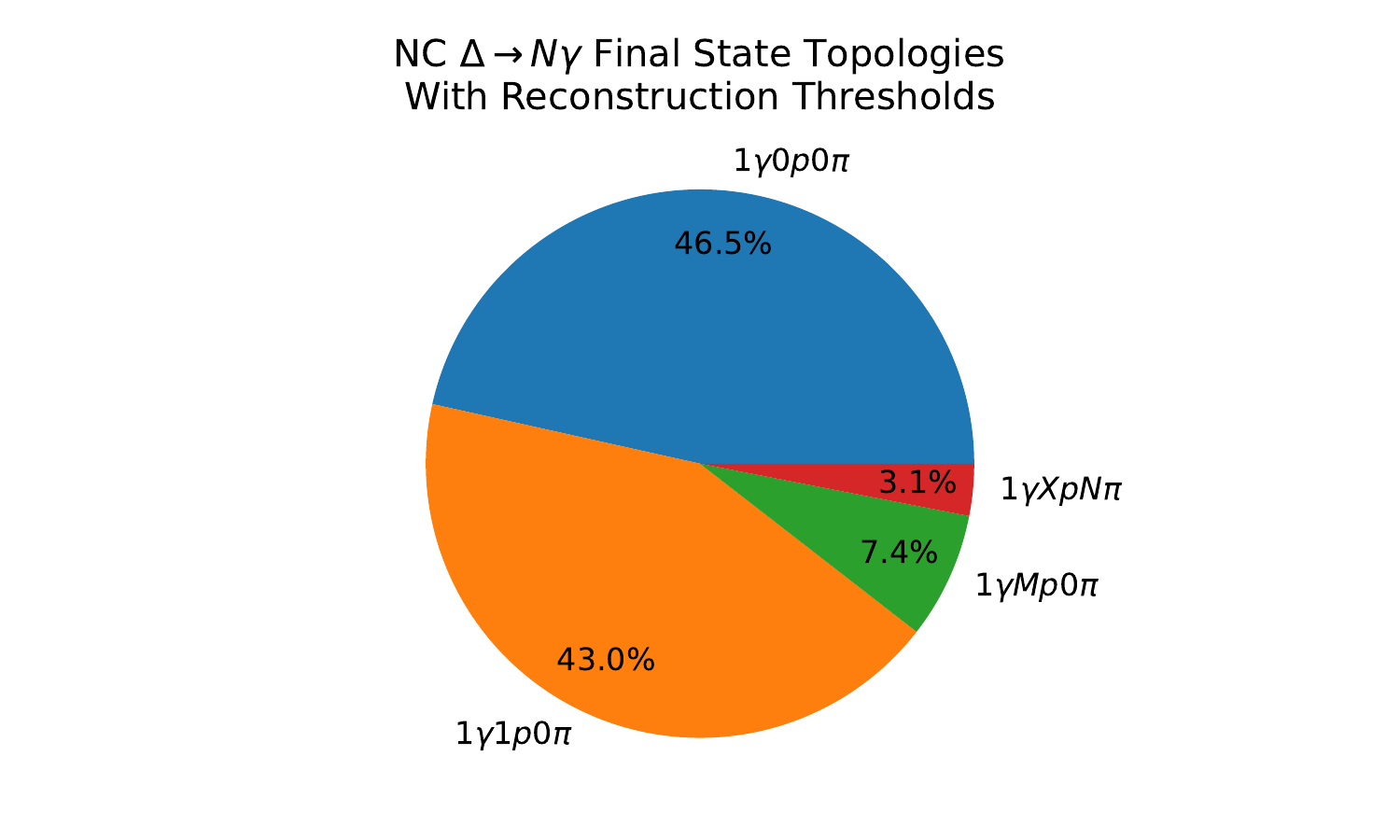}
        \caption{}
        \label{fig:nc_delta_final_state_threshold}
    \end{subfigure}
    \caption[NC $\Delta\rightarrow N \gamma$ topologies]{NC $\Delta\rightarrow N \gamma$ topologies predicted by GENIE. We break multiplicities down to $0$, $1$, $M$ for multiple (greater than or equal to two), and $X$ for any number (zero or greater than zero). Panel (a) shows the breakdown of initial state particles, just after the $\Delta\rightarrow N \gamma$ decay. Panel (b) shows the breakdown for all particle after final state interactions within the nucleus. Panel (c) shows a breakdown including reconstruction thresholds, where we ignore neutrons and very low energy particles. We assume a 35 MeV kinetic energy threshold for protons, a 10 MeV kinetic energy threshold for charged pions, no kinetic energy threshold for neutral pions, and we assume that neutrons will not be reconstructed.} 
    \label{fig:nc_delta_topologies}
\end{figure}

Our goal was to make this analysis as inclusive as possible while targeting the NC $\Delta\rightarrow N \gamma$ topology. Our Wire-Cell reconstructed variables are well suited for this, since they were primarily developed to target inclusive $\nu_e$CC selections. Although the dominant topologies are $1\gamma1p$ and $1\gamma0p$ as described above, there are rarer cases where final state interactions cause simulated NC $\Delta\rightarrow N \gamma$ events to have two or more exiting protons, or one or more exiting charged pions. In Fig. \ref{fig:nc_delta_initial_state}, we show the topologies for NC $\Delta\rightarrow N \gamma$ after the initial $\Delta$ decay, where we have $1\gamma 1n$ from $\Delta^0$ radiative decays and $1\gamma 1p$ from $\Delta^+$ radiative decays. The fraction of these is primarily driven by the fractions of neutrons and protons within the nucleus, with 18 protons and 22 neutrons. In Fig. \ref{fig:nc_delta_final_state}, we show the topologies after final state interactions. Here, we see that there are many events with the simple $1\gamma 1p$ and $1\gamma 1n$ topologies, but there are also about as many $1\gamma 1p 1n$ events that have both a proton and a neutron in the final state. There are also many events with multiple protons in the final state, some events with zero or one proton and multiple neutrons, and some events with pions in the final state. In Fig. \ref{fig:nc_delta_final_state_threshold}, we apply simple energy thresholds to different particles by assuming we do not detect neutrons and that we only detect protons and charged pions with enough kinetic energy. With these thresholds applied, we see that we primarily have events with just a single photon and one or zero protons, but there are some events with multiple protons and some events with charged and neutral pions.

We expect these topologies with multiple visible protons or with visible charged pions to be much rarer, but we choose to give our BDT the freedom to choose to select these topologies as well. This choice to include all reconstructed $1\gamma Xp X\pi^{\pm}$ topologies (where $X$ represents any number of particles, zero or greater than zero) rather than only the simpler $1\gamma 0p 0\pi$ and $1\gamma 1p 0\pi$ topology let us improve our relative efficiency by 9\%. This choice also lets us search for any potential anomalies in these even rarer channels; for example, Ref. \cite{miniboone_1g2p} describes a rare 2p2h$\gamma$ process which could potentially explain a fraction of the MiniBooNE LEE and would appear in MicroBooNE as $1\gamma 2p$ events. For most purposes in this analysis we split these topologies into those with ($1\gamma N p$) and without ($1\gamma 0 p$) reconstructed protons, using a 35 MeV reconstructed proton kinetic energy threshold.

\begin{figure}[H]
    \centering
    \begin{tikzpicture}[
        node distance=0.8cm and -1.5cm,
        every node/.style={draw, thick, align=center, font=\bfseries, minimum height=1.2cm, minimum width=4cm},
        arrow/.style={-{Latex[length=3mm, width=2mm]}, thick},
    ]
    
        \node (start) {Wire-Cell Generic\\Neutrino Selection};
        \node (cut1) [below=of start] {Fully Contained Cut};
        \node (cut2) [below=of cut1] {Reconstructed Shower Cut};
        \node (cut3) [below=of cut2] {NC $\Delta\rightarrow N \gamma$\\BDT Cut};
        \node (sel1) [below left=of cut3] {$1\gamma Np$\\Selection};
        \node (sel2) [below right=of cut3] {$1\gamma 0p$\\Selection};
        
        \draw[arrow] (start) -- (cut1);
        \draw[arrow] (cut1) -- (cut2);
        \draw[arrow] (cut2) -- (cut3);
        \draw[arrow] (cut3) -- (sel1);
        \draw[arrow] (cut3) -- (sel2);
    
    \end{tikzpicture}
    \caption{Flowchart of the Wire-Cell NC $\Delta\rightarrow N \gamma$ selection.}
    \label{fig:nc_delta_flow_chart}
\end{figure}
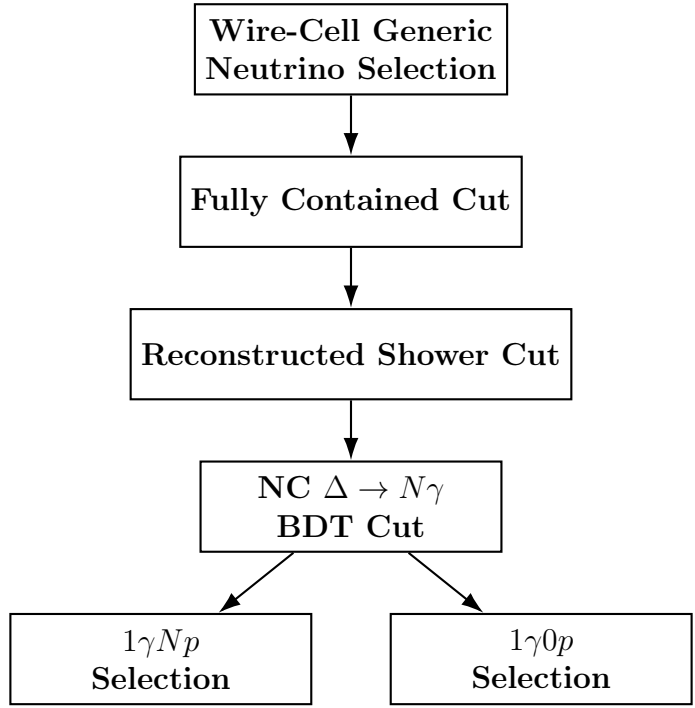

As we did for the Wire-Cell $\nu_e$CC selection, we first apply Wire-Cell generic neutrino selection, require that all particles are reconstructed as fully contained within the TPC, require the existence of a reconstructed shower, then apply unified BDT for all topologies, and only separate by final state topology (particularly proton multiplicity) in later stages of the analysis, as outlined in Fig. \ref{fig:nc_delta_flow_chart}. 

We train the BDT to select simulated NC $\Delta\rightarrow N \gamma$ events from simulated backgrounds, including a high-statistics sample of NC $\pi^0$ events, and measured beam-off cosmic ray backgrounds. This training consisted of 157,967 background events and 23,525 signal events. These background events were the same simulated events used for our $\nu_e$CC and $\nu_\mu$CC BDT trainings, so we can maximize statistics for our final predictions which exclude events used for training. These signal events correspond to half of our $4.438\cdot 10^{23}$ POT NC $\Delta\rightarrow N \gamma$ simulation file, in order to preserve half for high-statistics predictions and efficiency studies. We further preserve 1\% of the remaining events in order to have a test sample to evaluate during BDT training. We require all the training events to pass Wire-Cell generic neutrino selection as described in Sec. \ref{sec:generic_neutrino_selection}, and additionally require that each event is reconstructed as fully contained in the detector. We used all of the training variables and vector variable BDT outputs used for our $\nu_e$CC and $\nu_\mu$CC BDTs as described in Sec. \ref{sec:bdt_selections}, as well as variables used for our cut-based $\pi^0$ selections, which totals to 341 total variables, which are described in detail in Sec. \ref{sec:wc_bdt_vars}. We trained using XGBoost \cite{xgboost}. XGBoost can take input event weights, but it was empirically observed that performance was degraded when including POT weighting, which caused the signal events to have very low weights. So for the XGBoost training, we only used cross section weights which tend to be close to one. We used the same hyperparameters as were used for the $\nu_e$CC BDT:

\begin{spacing}{0.9}
\begin{lstlisting}

XGBClassifier(max_depth=30, 
              n_estimators=600, 
              learning_rate=0.1, 
              objective='binary:logitraw',
              colsample_bytree=1, 
              colsample_bylevel = 0.3, 
              colsample_bynode = 0.3
              subsample = 0.8, 
              gamma = 0.1, 
              seed = 5)

\end{lstlisting}
\end{spacing}

\begin{figure}[H]
    \centering
    \begin{subfigure}[b]{0.49\textwidth}
        \includegraphics[trim=360 0 0 0, clip, width=\textwidth]{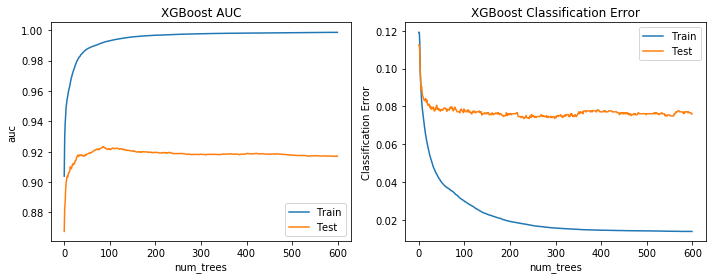}
        \caption{}
        \label{fig:loss}
    \end{subfigure}
    \begin{subfigure}[b]{0.49\textwidth}
        \includegraphics[trim=0 0 360 0, clip, width=\textwidth]{figs/nc_delta/nc_delta_train_curves.png}
        \caption{}
        \label{fig:auc}
    \end{subfigure}
    \caption[Wire-Cell NC $\Delta\rightarrow N \gamma$ BDT training curves]{Wire-Cell NC $\Delta\rightarrow N \gamma$ BDT training curves. Panel (a) shows the classification error, which functions as the loss being minimized. Panel (b) shows the area under the ROC curve.}
    \label{fig:training_curves}
\end{figure}

We show the classification error which functions as the loss being minimized during training as a function of the training time in Fig. \ref{fig:loss}. The training loss continuously decreases, and the test loss flattens out over time. We show the area under the ROC (Receiver Operating Characteristic) curve, which is a metric related to the maximum efficiency and purity achievable at different cut values, in Fig. \ref{fig:auc}. The number of trees along each x-axis is analogous to the number of epochs of training. We did not use any early stopping during the training.

\begin{figure}[H]
    \centering
    \begin{subfigure}[b]{0.49\textwidth}
        \includegraphics[width=\textwidth]{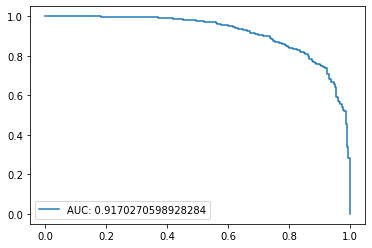}
        \caption{}
    \end{subfigure}
    \begin{subfigure}[b]{0.49\textwidth}
        \includegraphics[width=\textwidth]{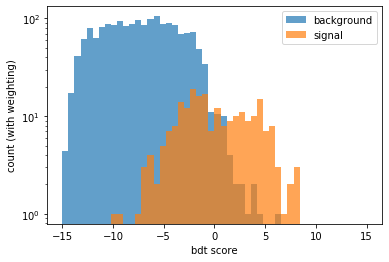}
        \caption{}
    \end{subfigure}
    \caption[Wire-Cell NC $\Delta\rightarrow N \gamma$ BDT training results]{Wire-Cell NC $\Delta$ training results. Panel (a) shows the ROC curve, showing all possible efficiency vs purity values. Panel (b) shows the distribution of background and signal events as a function of BDT scores. Note that this only displays the small amount of training events used for validation.}
    \label{fig:training_results}
\end{figure}

\begin{figure}[H]
    \centering
    \includegraphics[width=0.7\textwidth]{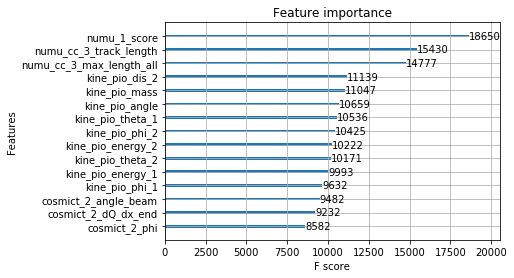}
    \caption[Wire-Cell NC $\Delta\rightarrow N \gamma$ BDT important variables]{Wire-Cell NC $\Delta\rightarrow N \gamma$ BDT important variables.}
    \label{fig:important_variables}
\end{figure}

The performance on the validation set after the BDT training is shown in Fig. \ref{fig:training_results}. The most important variables according to the XGBoost F score metric are shown in Fig. \ref{fig:important_variables}. Out of the 341 total variables in the BDT, the most influential are related to $\nu_\mu$CC identification, since these make up a large fraction of all events passing Wire-Cell generic neutrino selection. The next most influential are a series of variables related to $\pi^0$ identification, since this is the background which is most difficult to distinguish from the single photon signal.

\begin{figure}[H]
    \centering
    \begin{subfigure}[b]{0.49\textwidth}
        \includegraphics[trim=40 0 20 0, clip, width=\textwidth]{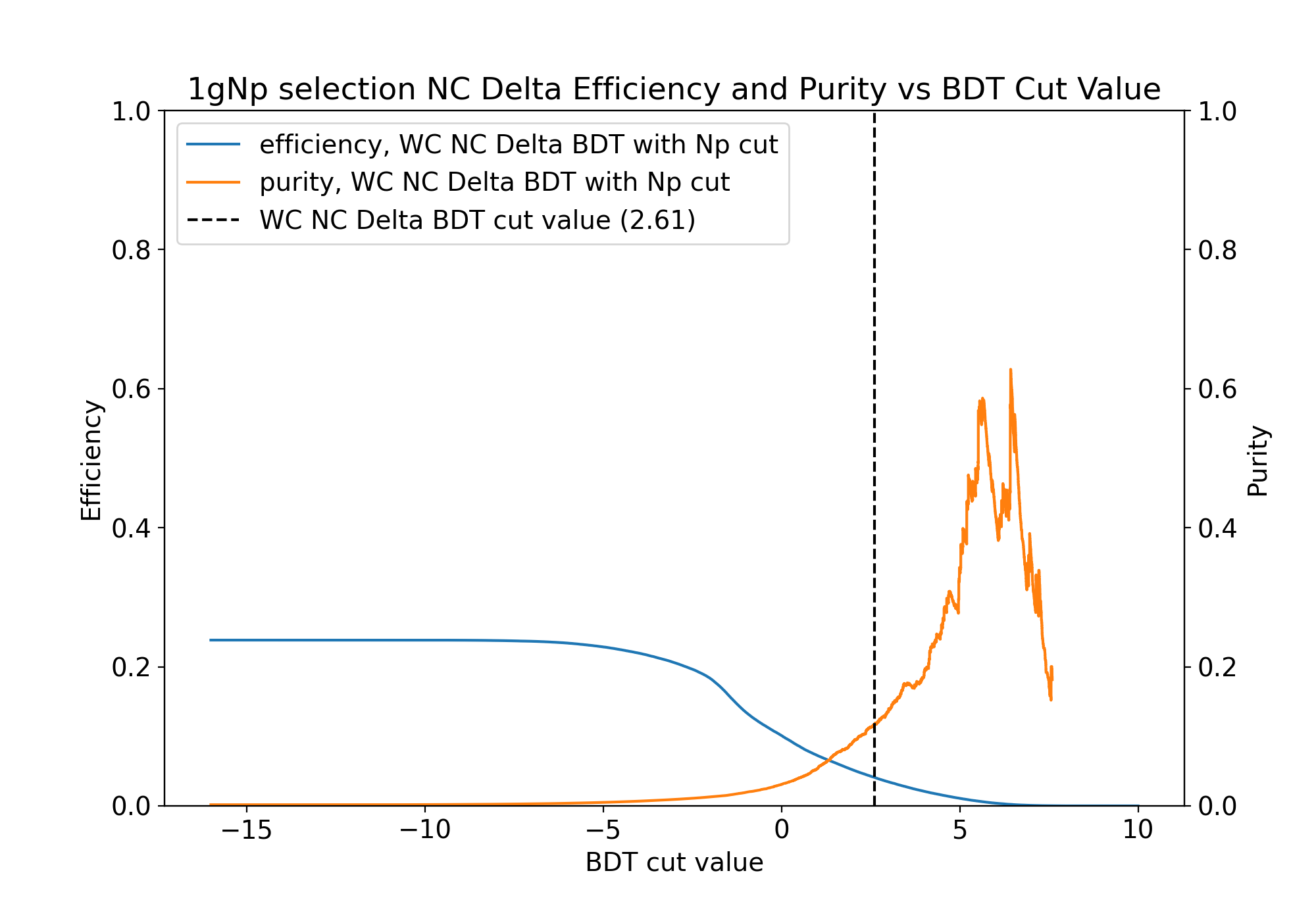}
        \caption{}
    \end{subfigure}
    \begin{subfigure}[b]{0.49\textwidth}
        \includegraphics[trim=40 0 20 0, clip, width=\textwidth]{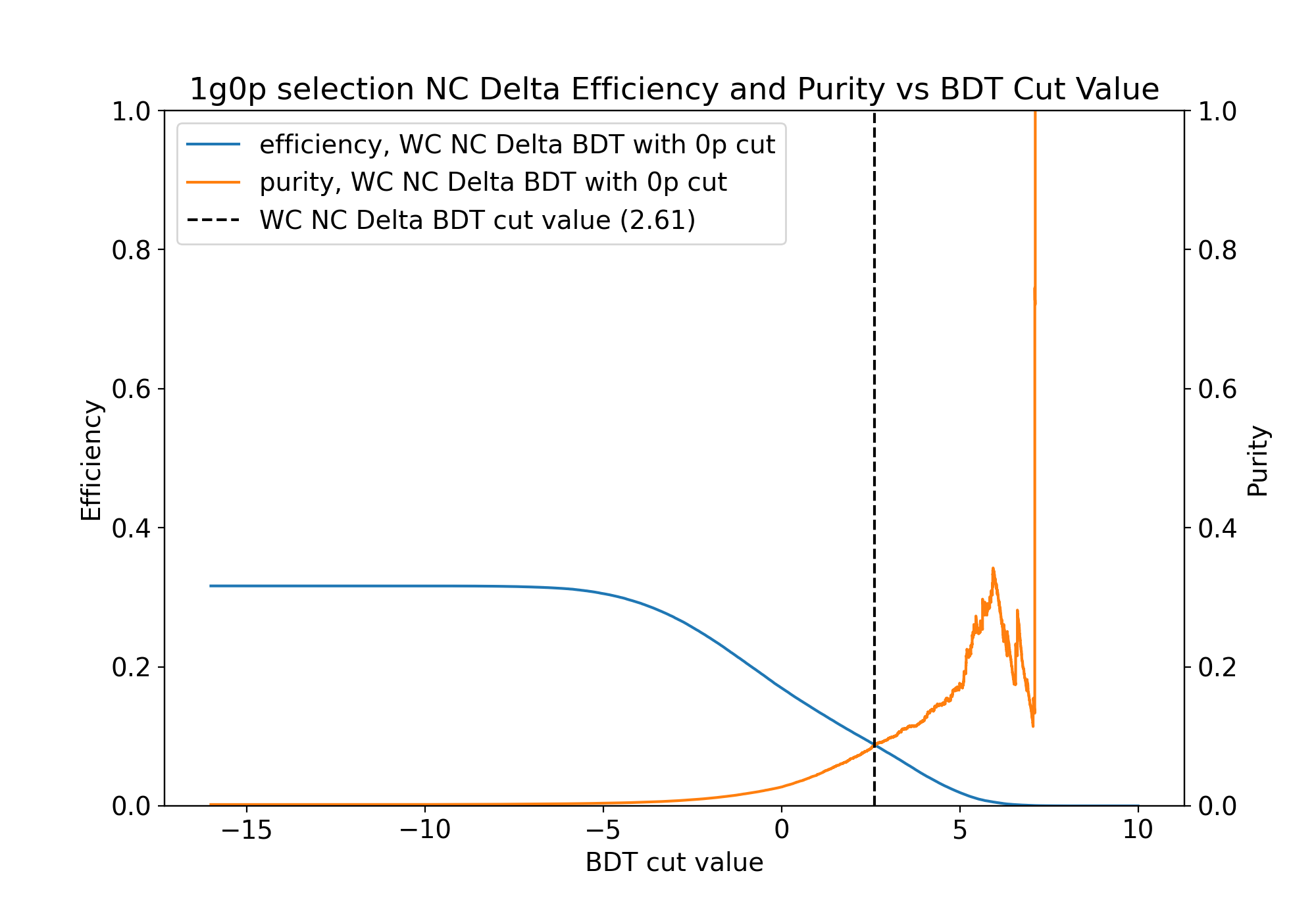}
        \caption{}
    \end{subfigure}
    \begin{subfigure}[b]{0.49\textwidth}
        \includegraphics[trim=25 0 70 0, clip, width=\textwidth]{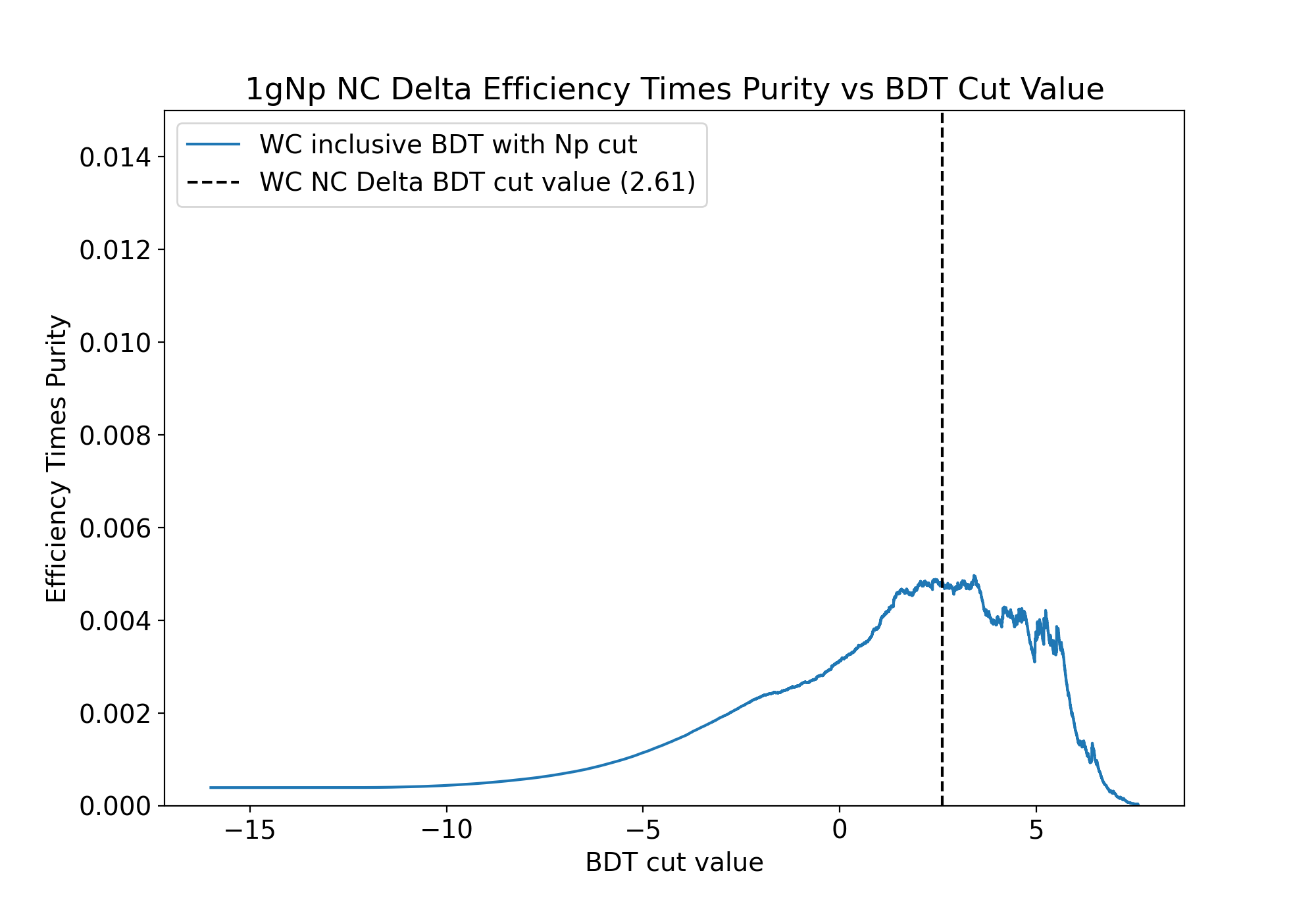}
        \caption{}
    \end{subfigure}
    \begin{subfigure}[b]{0.49\textwidth}
        \includegraphics[trim=25 0 70 0, clip, width=\textwidth]{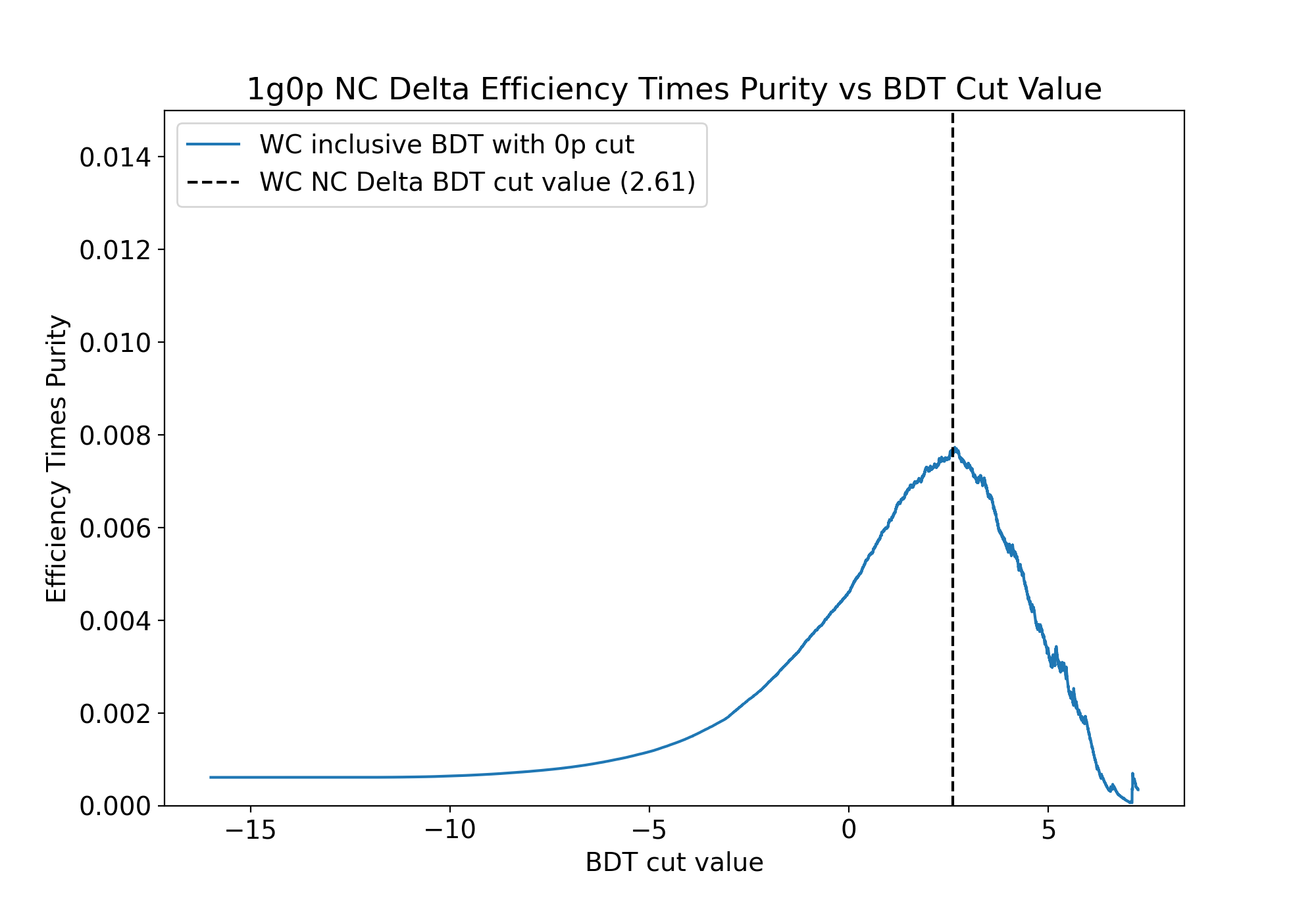}
        \caption{}
    \end{subfigure}
    \caption[Wire-Cell NC $\Delta\rightarrow N \gamma$ BDT efficiency vs purity]{Wire-Cell NC $\Delta\rightarrow N \gamma$ BDT efficiency, purity, and efficiency times purity vs BDT score. Panels (a) and (c) show the performance for events with one or more reconstructed protons, and panels (b) and (d) show the performance for events without reconstructed protons. Note that the efficiency is less than 100\% even with a very loose BDT cut, due to the application of Wire-Cell generic neutrino selection.}
    \label{fig:eff_vs_pur}
\end{figure}

A final cut value was chosen at a BDT score of 2.61. As shown in Fig. \ref{fig:eff_vs_pur}, this value simultaneously maximizes the efficiency times purity for events both with and without reconstructed protons.

\begin{figure}[H]
    \centering
    \begin{subfigure}[b]{0.49\textwidth}
        \includegraphics[trim=25 0 100 0, clip, width=\textwidth]{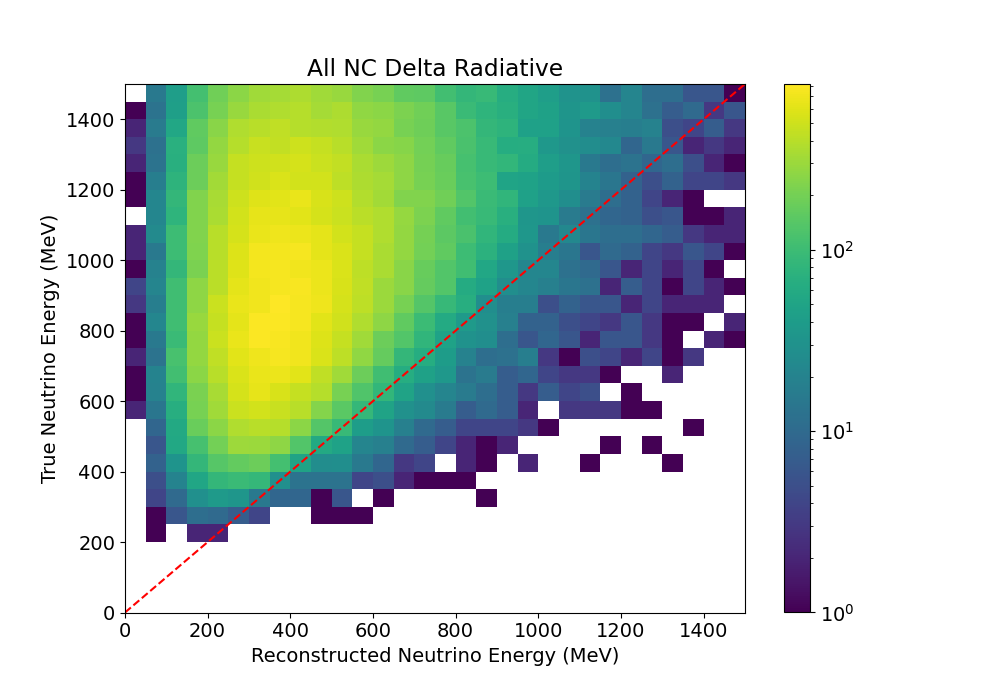}
        \caption{}
    \end{subfigure}
    \begin{subfigure}[b]{0.49\textwidth}
        \includegraphics[trim=25 0 100 0, clip, width=\textwidth]{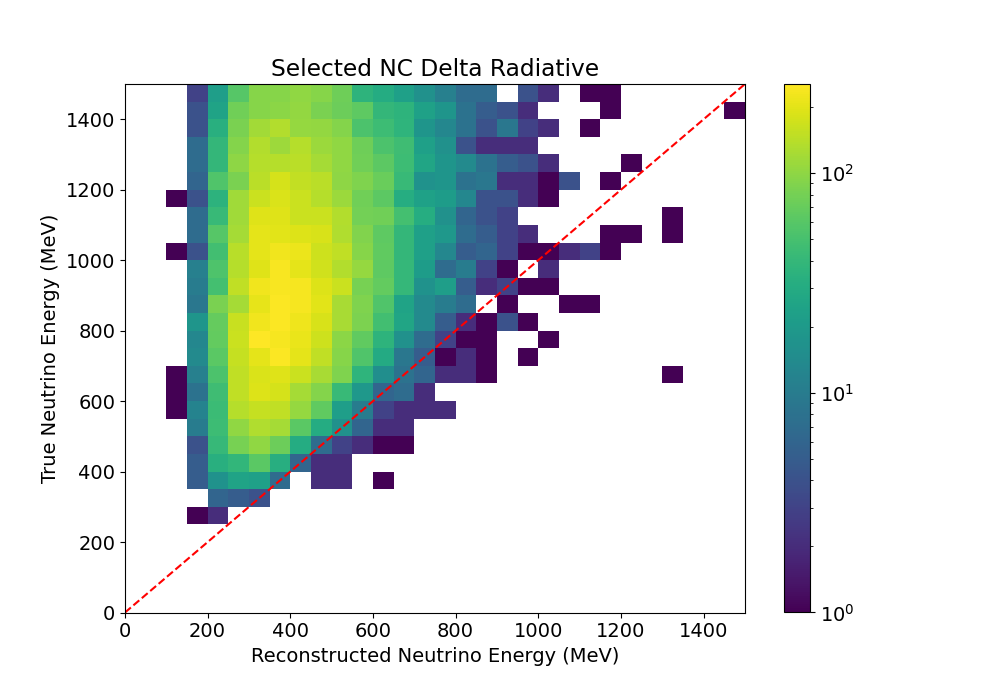}
        \caption{}
    \end{subfigure}
    \caption[Wire-Cell NC $\Delta\rightarrow N \gamma$ neutrino energy resolution]{Wire-Cell NC $\Delta\rightarrow N \gamma$ neutrino energy resolution. Panel (a) shows the resolution for all NC $\Delta\rightarrow N \gamma$ events, and panel (b) shows the resolution for only Wire-Cell NC $\Delta\rightarrow N \gamma$ selected events.}
    \label{fig:Enu_resolution}
\end{figure}

\begin{figure}[H]
    \centering
    \begin{subfigure}[b]{0.49\textwidth}
        \includegraphics[trim=25 0 100 0, clip, width=\textwidth]{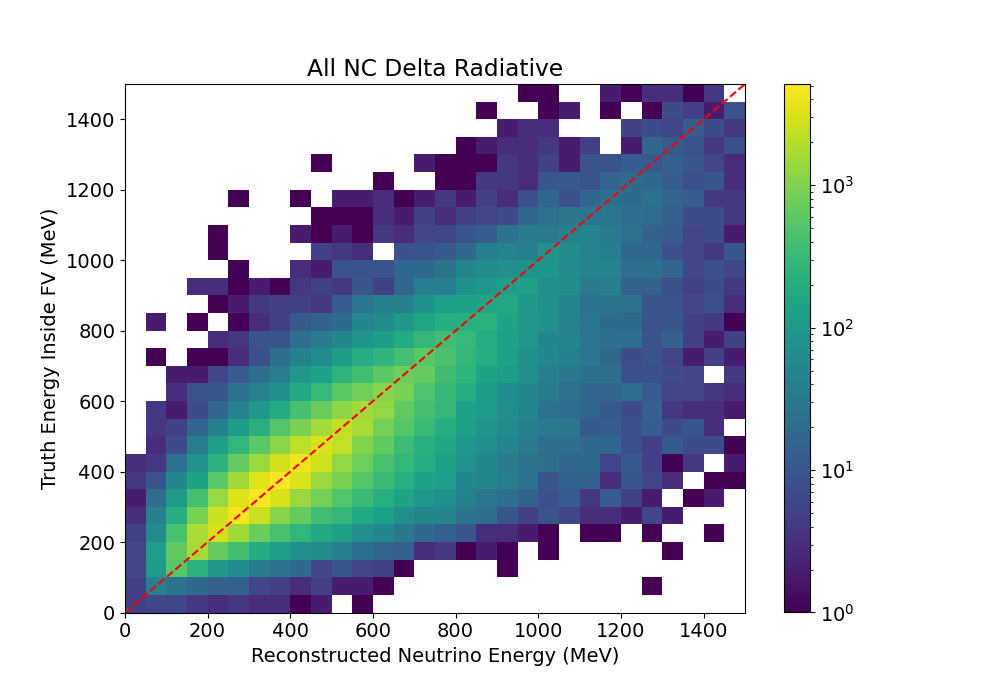}
        \caption{}
    \end{subfigure}
    \begin{subfigure}[b]{0.49\textwidth}
        \includegraphics[trim=25 0 100 0, clip, width=\textwidth]{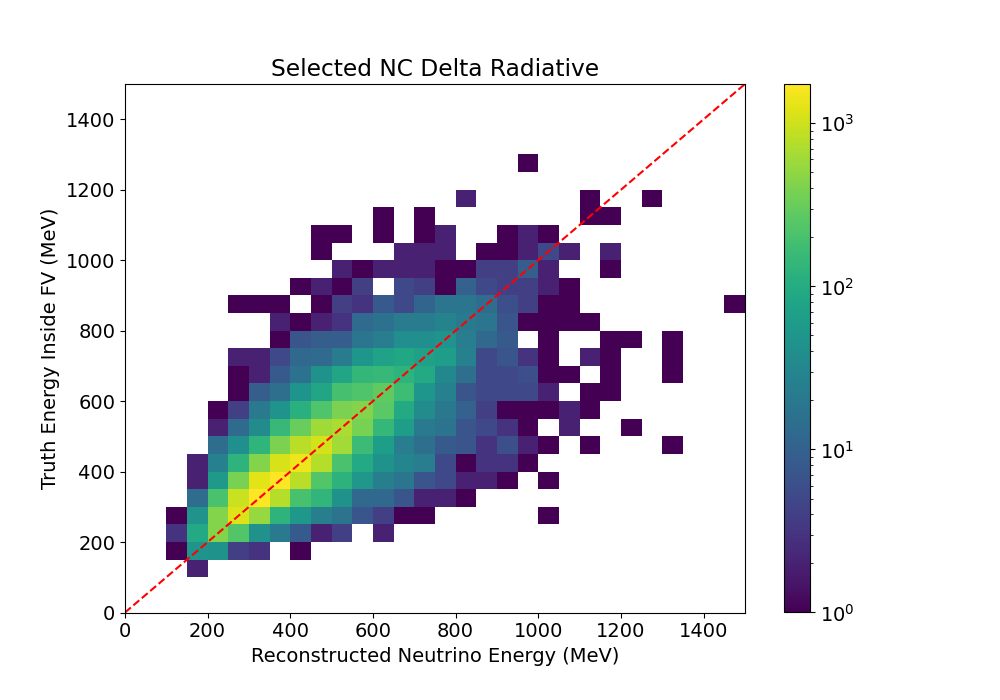}
        \caption{}
    \end{subfigure}
    \caption[Wire-Cell NC $\Delta\rightarrow N \gamma$ resolution in-FV deposited energy]{Wire-Cell NC $\Delta\rightarrow N \gamma$ resolution for deposited energy inside the TPC fiducial volume. Panel (a) shows the resolution for all NC $\Delta\rightarrow N \gamma$ events, and panel (b) shows the resolution for only Wire-Cell NC $\Delta\rightarrow N \gamma$ selected events.}
    \label{fig:E_inside_resolution}
\end{figure}

\begin{figure}[H]
    \centering
    \begin{subfigure}[b]{0.49\textwidth}
        \includegraphics[trim=25 0 100 0, clip, width=\textwidth]{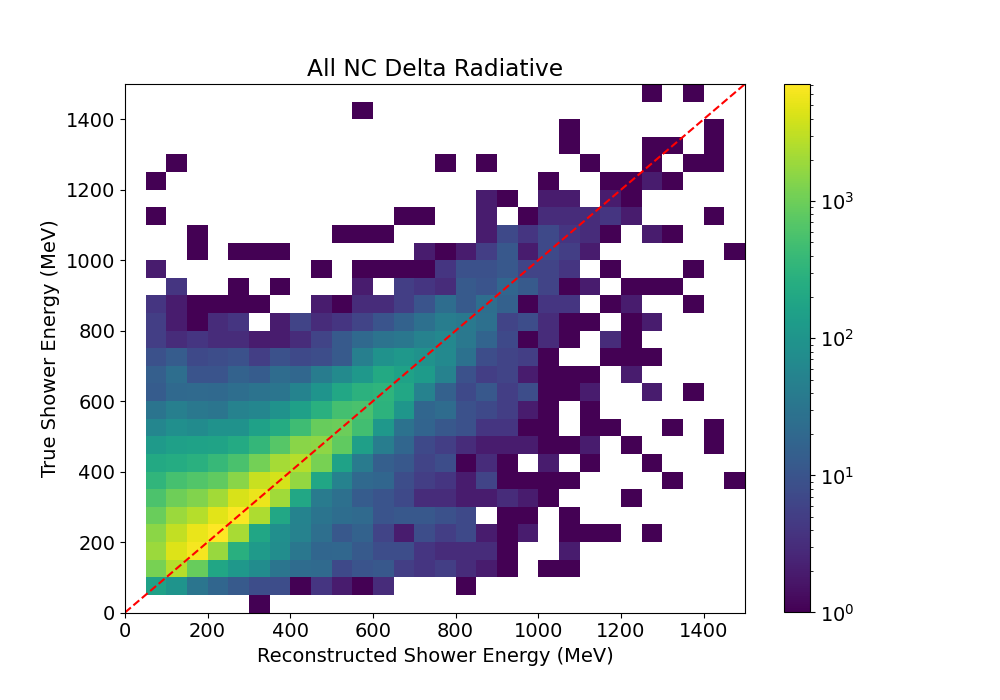}
        \caption{}
    \end{subfigure}
    \begin{subfigure}[b]{0.49\textwidth}
        \includegraphics[trim=25 0 100 0, clip, width=\textwidth]{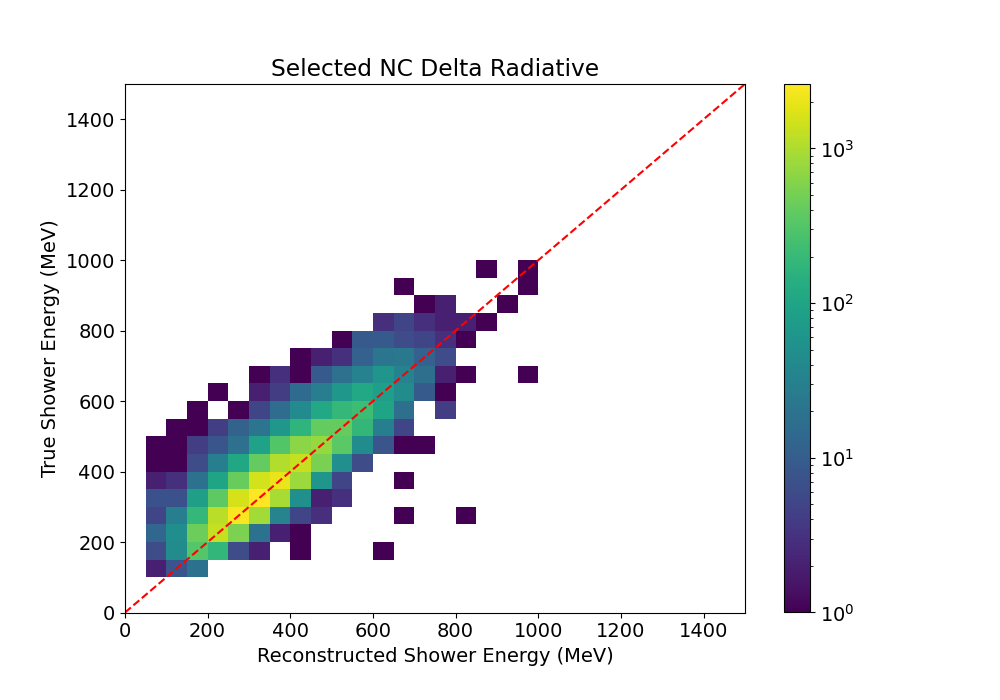}
        \caption{}
    \end{subfigure}
    \caption[Wire-Cell NC $\Delta\rightarrow N \gamma$ shower energy resolution]{Wire-Cell NC $\Delta\rightarrow N \gamma$ shower resolution. Panel (a) shows the resolution for all NC $\Delta\rightarrow N \gamma$ events, and panel (b) shows the resolution for only Wire-Cell NC $\Delta\rightarrow N \gamma$ selected events.}
    \label{fig:E_shower_resolution}
\end{figure}

\begin{figure}[H]
    \centering
    \begin{subfigure}[b]{0.49\textwidth}
        \includegraphics[trim=25 0 100 0, clip, width=\textwidth]{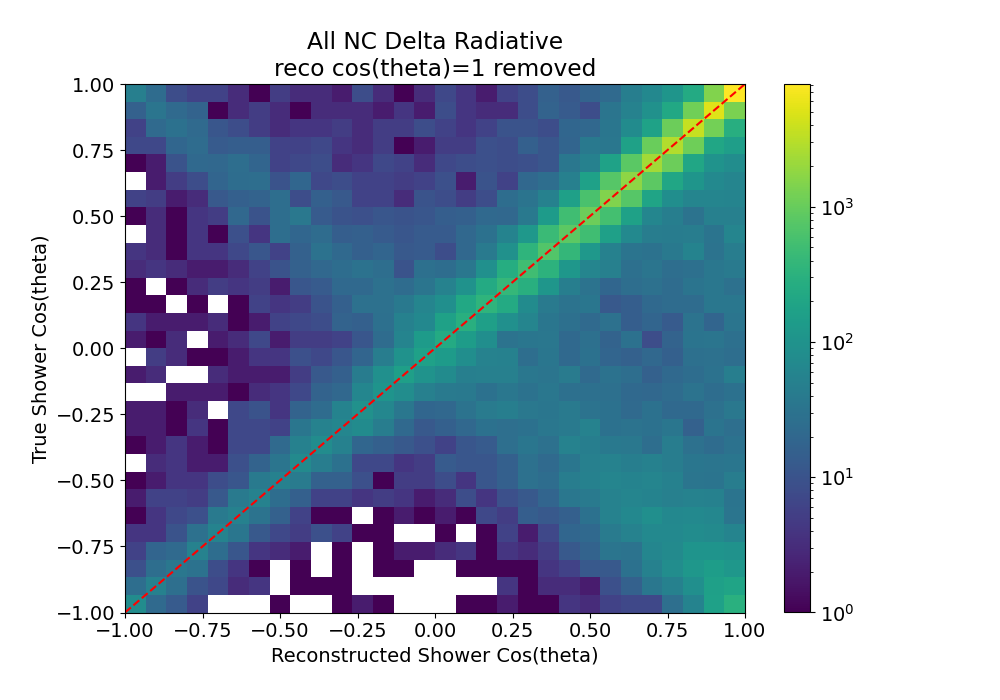}
        \caption{}
    \end{subfigure}
    \begin{subfigure}[b]{0.49\textwidth}
        \includegraphics[trim=25 0 100 0, clip, width=\textwidth]{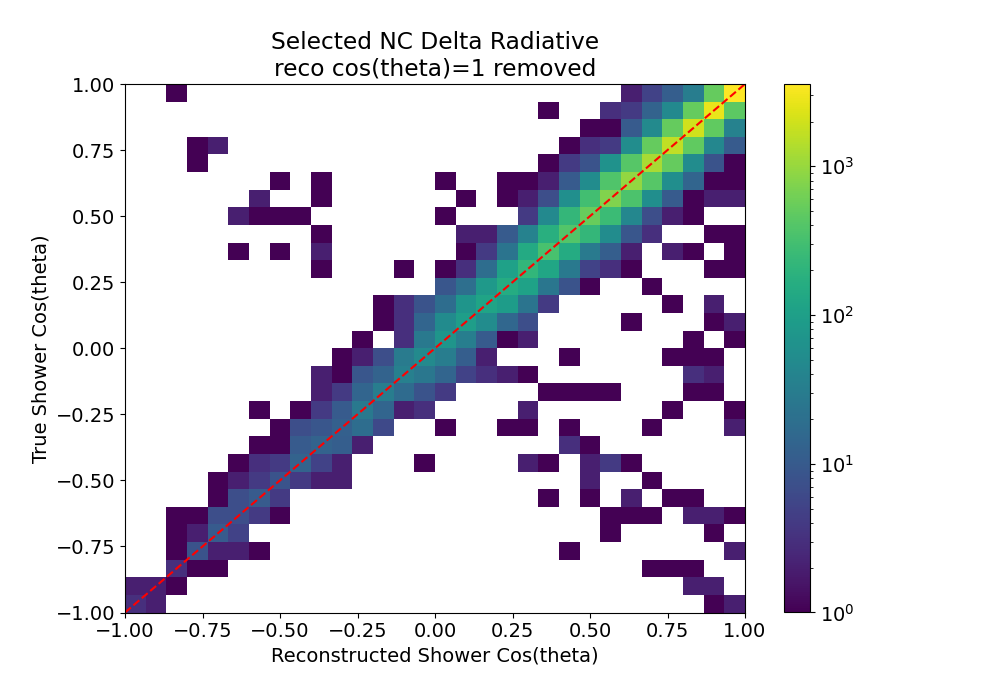}
        \caption{}
    \end{subfigure}
    \caption[Wire-Cell NC $\Delta\rightarrow N \gamma$ shower angle resolution]{Wire-Cell NC $\Delta\rightarrow N \gamma$ shower angle resolution. Panel (a) shows the resolution for all NC $\Delta\rightarrow N \gamma$ events, and panel (b) shows the resolution for only Wire-Cell NC $\Delta\rightarrow N \gamma$ selected events.}
    \label{fig:theta_shower_resolution}
\end{figure}

With this Wire-Cell reconstruction and selection, we study the resolution for measuring various kinematic quantities, both for all NC $\Delta\rightarrow N \gamma$ events as well as for selected NC $\Delta\rightarrow N \gamma$ events. In Fig. \ref{fig:Enu_resolution}, we see a large bias in the neutrino energy resolution; this will generally be the casy for any neutral current interactions, since we cannot measure the energy in the exiting neutrino. We account for this effect in Fig. \ref{fig:E_inside_resolution}, where we study the energy deposited inside the TPC fiducial volume, which is much more closely related to the observable energy that we reconstruct. Figure \ref{fig:E_shower_resolution} shows the our resolution for reconstructed shower energies, showing generally diagonal behavior. Note that this shower energy resolution is also calibrated on real data events using the 135 MeV peak in the reconstructed $\pi^0$ invariant mass. Figure \ref{fig:theta_shower_resolution} shows the resolution for reconstructed shower angles, showing generally diagonal behavior, but also including a small component along the anti-diagonal corresponding to events where the shower is mis-reconstructed along the backward direction.

Before unblinding, a number of validations were performed to make sure that the selection performs as we expect. First, we examined the performance on a small sample of $5.3\cdot 10^{20}$ POT BNB data, and confirmed that in all distributions, the prediction and data agree within statistical and systematic uncertainties. We also investigated the performance using several fake data sets, where a sample was simulated with changes which were only revealed to analyzers after carefully studying the results of selections. One of these simulated fake data sets used GENIE v2.12.10 with an empirical MEC model, one had increased NC coherent and non-coherent rates, and one used GENIE v3 without the MicroBooNE tune \cite{genie-tune-paper} applied. In all cases, our selection performed as we would expect.

We also validated the performance using a small sample of NuMI data, since this data set was not expected to have significant sensitivity to our initial models of the MiniBooNE LEE. This only used $2.099\cdot 10^{20}$ POT of forward horn current (FHC) data from the first year of MicroBooNE's beam operations. This only included a fairly small simulation sample with fairly large Monte-Carlo statistical uncertainties, and did not include any detector response systematic uncertainties. This used an older simulation of the NuMI beam flux, before the improvements described in Ref. \cite{microboone_numi_flux_public_note}. The NuMI beam at this far off-axis location contains a much larger fraction of $\nu_e$ compared to the BNB, which causes a much larger $\nu_e$CC background contribution to this NC $\Delta\rightarrow N\gamma$ selection. This $\nu_e$CC background would likely be reducible with stronger cuts on the deposited energy per unit length if we used an updated BDT training specifically targeting NC $\Delta\rightarrow N\gamma$ events in the NuMI beam. For this analysis, the NuMI beam data was used only in order to validate the selection performance on a larger sample of real data before unblinding the full statistics BNB data sample used for our final results.

\begin{figure}[H]
    \centering
    \begin{overpic}[trim=30 0 50 16.5, clip, page=9, width=0.48\textwidth]{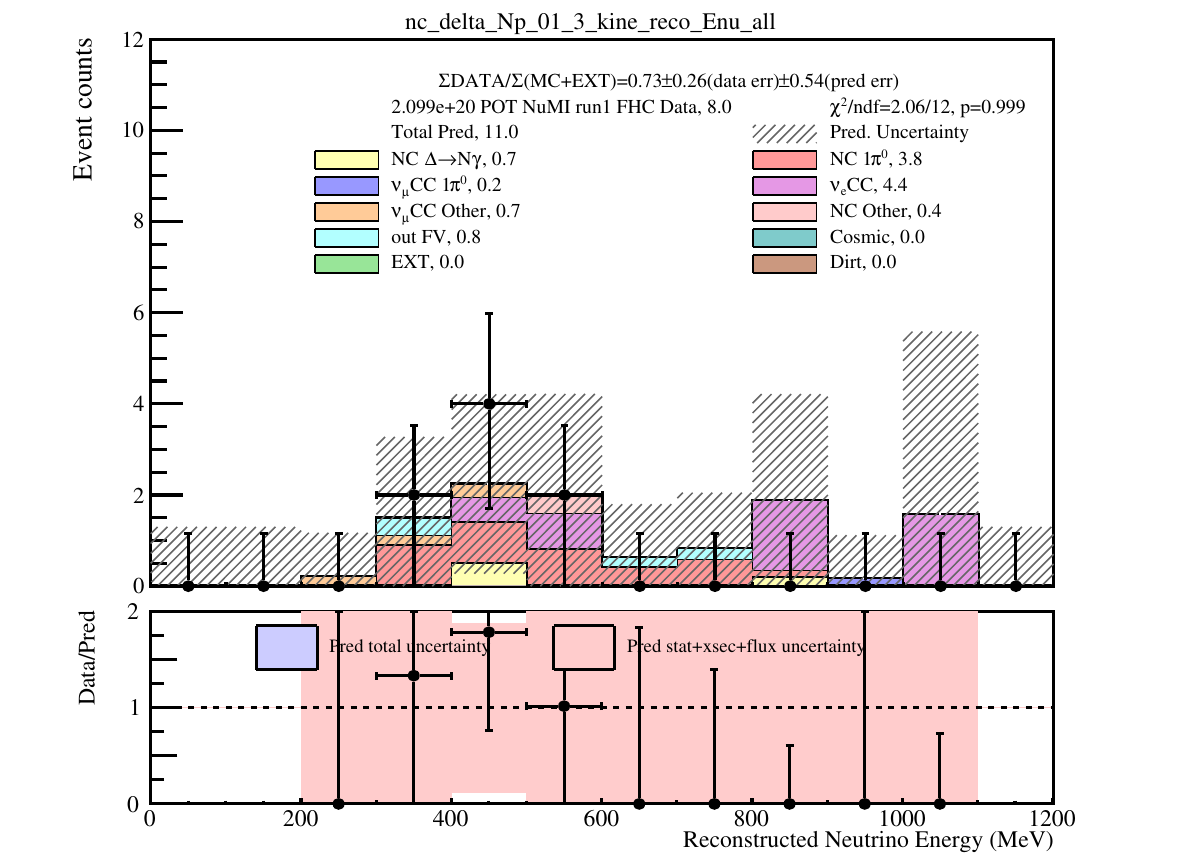}
        \put(60,30){\small \textbf{NuMI $1\gamma Np$}} 
    \end{overpic}
    \begin{overpic}[trim=30 0 50 16.5, clip, page=10, width=0.48\textwidth]{figs/nc_delta/numi_nc_delta_plots.pdf}
        \put(60,29){\small \textbf{NuMI $1\gamma 0p$}} 
    \end{overpic}
    \caption[One-bin Wire-Cell NC $\Delta\rightarrow N \gamma$ selection with NuMI data]{One-bin Wire-Cell NC $\Delta\rightarrow N \gamma$ selection with NuMI data. Panel (a) shows $1\gamma Np$ events, and panel (b) shows $1\gamma 0p$ events. Detector systematics have not been included.}
    \label{fig:one_bin_numi}
\end{figure}

\begin{figure}[H]
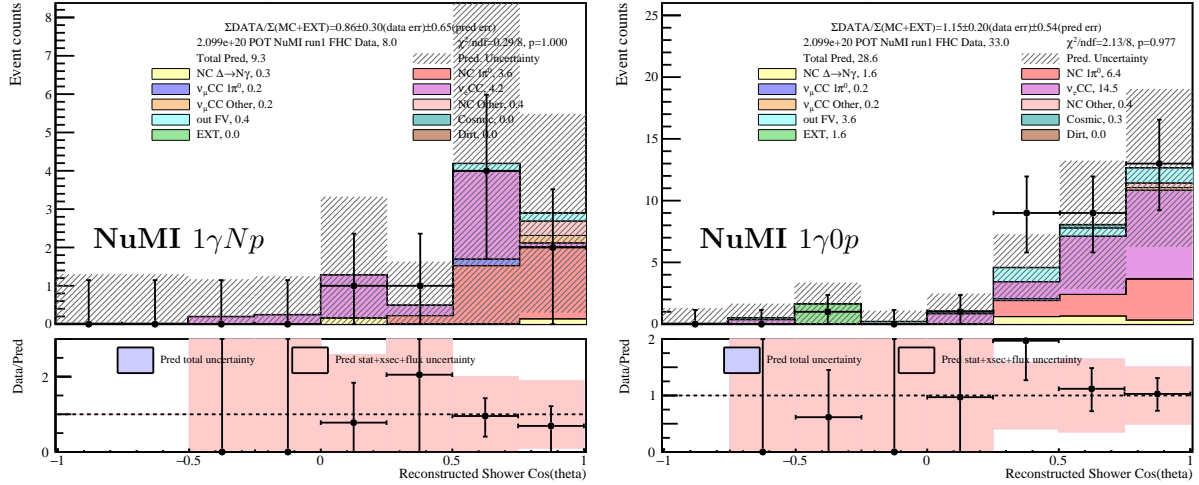

    \centering
    \begin{overpic}[trim=30 0 50 16.5, clip, page=5, width=0.48\textwidth]{figs/nc_delta/numi_nc_delta_plots.pdf}
        \put(15,40){\small \textbf{NuMI $1\gamma Np$}}
    \end{overpic}
    \begin{overpic}[trim=30 0 50 16.5, clip, page=6, width=0.48\textwidth]{figs/nc_delta/numi_nc_delta_plots.pdf}
        \put(15,40){\small \textbf{NuMI $1\gamma 0p$}}
    \end{overpic}
    \caption[Reconstructed shower $\cos(\theta)$ Wire-Cell NC $\Delta\rightarrow N \gamma$ selection with NuMI data]{Reconstructed shower $\cos(\theta)$ for the Wire-Cell NC $\Delta \rightarrow N \gamma$ selection with NuMI data. Panel (a) shows $1\gamma Np$ events, and panel (b) shows $1\gamma 0p$ events. Events reconstructed with default values have been removed. Detector systematics have not been included.}
    \label{fig:shower_angle_numi}
\end{figure}

\begin{figure}[H]
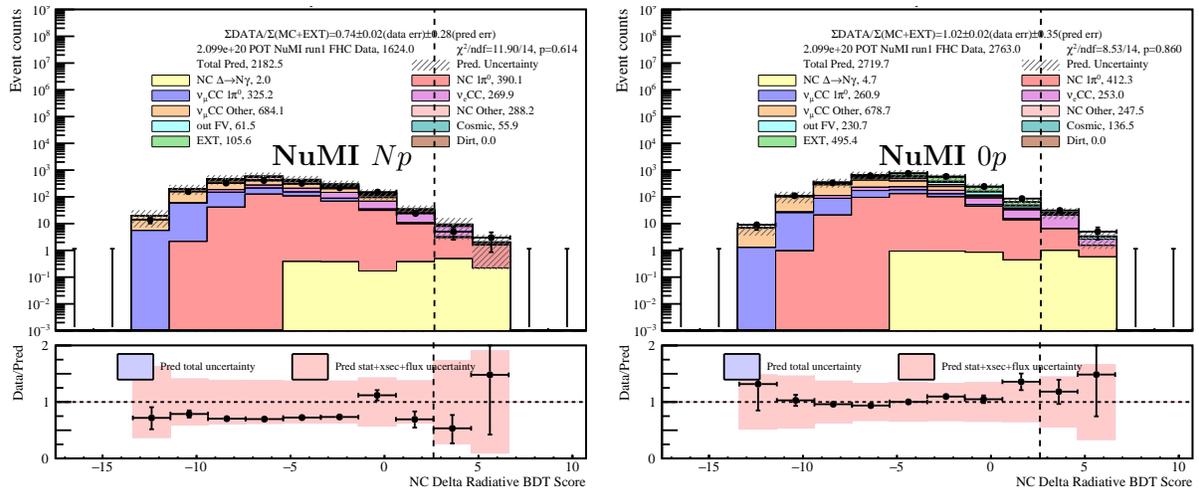

    \centering
    \begin{overpic}[trim=30 0 50 16.5, clip, page=7, width=0.48\textwidth]{figs/nc_delta/numi_nc_delta_plots.pdf}
        \put(45,55){\small \textbf{NuMI $Np$}}
    \end{overpic}
    \begin{overpic}[trim=30 0 50 16.5, clip, page=8, width=0.48\textwidth]{figs/nc_delta/numi_nc_delta_plots.pdf}
        \put(45,55){\small \textbf{NuMI $0p$}}
    \end{overpic}
    \caption[Reconstructed Wire-Cell NC $\Delta\rightarrow N \gamma$ BDT scores with NuMI data]{Reconstructed Wire-Cell NC $\Delta \rightarrow N \gamma$ BDT scores for generic neutrino selected events with a reconstructed shower in NuMI data. Panel (a) shows all $Np$ events, and panel (b) shows all $1\gamma 0p$ events. Events with a score above 2.61 (dashed line) are included in the selection. Detector systematics have not been included.}
    \label{fig:scores_numi}
\end{figure}

Figs. \ref{fig:one_bin_numi}, \ref{fig:neutrino_energy_numi}, \ref{fig:shower_energy_numi}, and \ref{fig:shower_angle_numi} show Wire-Cell NC $\Delta\rightarrow N\gamma$ selected events in this small sample of NuMI data. The systematic uncertainties are generally very large, even without detector uncertainties included. Within systematic uncertainties, the data agrees with our prediction in all of these cases. Fig. \ref{fig:scores_numi} shows the distribution of Wire-Cell NC $\Delta\rightarrow N\gamma$ BDT scores for $Np$ and $0p$ events passing our generic neutrino selection and reconstructed shower cuts, again showing good agreement within uncertainties even without detector uncertainties included.

\begin{figure}[H]
    \centering
    \begin{overpic}[trim=30 0 50 16.5, clip, page=1, width=0.48\textwidth]{figs/nc_delta/numi_nc_delta_plots.pdf}
        \put(50,50){\small \textbf{NuMI $1\gamma Np$}}
    \end{overpic}
    \begin{overpic}[trim=30 0 50 16.5, clip, page=2, width=0.48\textwidth]{figs/nc_delta/numi_nc_delta_plots.pdf}
        \put(65,40){\small \textbf{NuMI $1\gamma 0p$}}
    \end{overpic}
    \caption[Reconstructed $E_\nu$ Wire-Cell NC $\Delta\rightarrow N \gamma$ selection with NuMI data]{Reconstructed neutrino energy for the Wire-Cell NC $\Delta \rightarrow N \gamma$ selection with NuMI data. Panel (a) shows $1\gamma Np$ events, and panel (b) shows $1\gamma 0p$ events. Detector systematics have not been included.}
    \label{fig:neutrino_energy_numi}
\end{figure}

\begin{figure}[H]
    \centering
    \begin{overpic}[trim=30 0 50 16.5, clip, page=3, width=0.48\textwidth]{figs/nc_delta/numi_nc_delta_plots.pdf}
        \put(65,40){\small \textbf{NuMI $1\gamma Np$}}
    \end{overpic}
    \begin{overpic}[trim=30 0 50 16.5, clip, page=4, width=0.48\textwidth]{figs/nc_delta/numi_nc_delta_plots.pdf}
        \put(65,40){\small \textbf{NuMI $1\gamma 0p$}}
    \end{overpic}
    \caption[Reconstructed shower energy Wire-Cell NC $\Delta\rightarrow N \gamma$ selection with NuMI data]{Reconstructed primary shower energy for the Wire-Cell NC $\Delta \rightarrow N \gamma$ selection with NuMI data. Panel (a) shows $1\gamma Np$ events, and panel (b) shows $1\gamma 0p$ events. Detector systematics have not been included.}
    \label{fig:shower_energy_numi}
\end{figure}

\section{Joint Wire-Cell+Pandora NC \texorpdfstring{$\Delta$}{Delta} Radiative Decay Selections}

After unblinding the Wire-Cell NC $\Delta\rightarrow N \gamma$ selections on our full $6.369\cdot 10^{20}$ POT sample of BNB data, we decided to combine our results with the prior Pandora NC $\Delta\rightarrow N \gamma$ results. We chose to do this because each analysis has independent strengths, with Pandora generally offering better performance for events with a proton ($1\gamma 1p$) and Wire-Cell generally offering better performance for events without protons ($1\gamma 0p$).

\subsection{Efficiencies}\label{sec:nc_delta_efficiencies}

\begin{figure}[H]
    \centering
    \begin{subfigure}[b]{0.49\textwidth}
        \includegraphics[trim=15 0 50 0, clip, width=\textwidth]{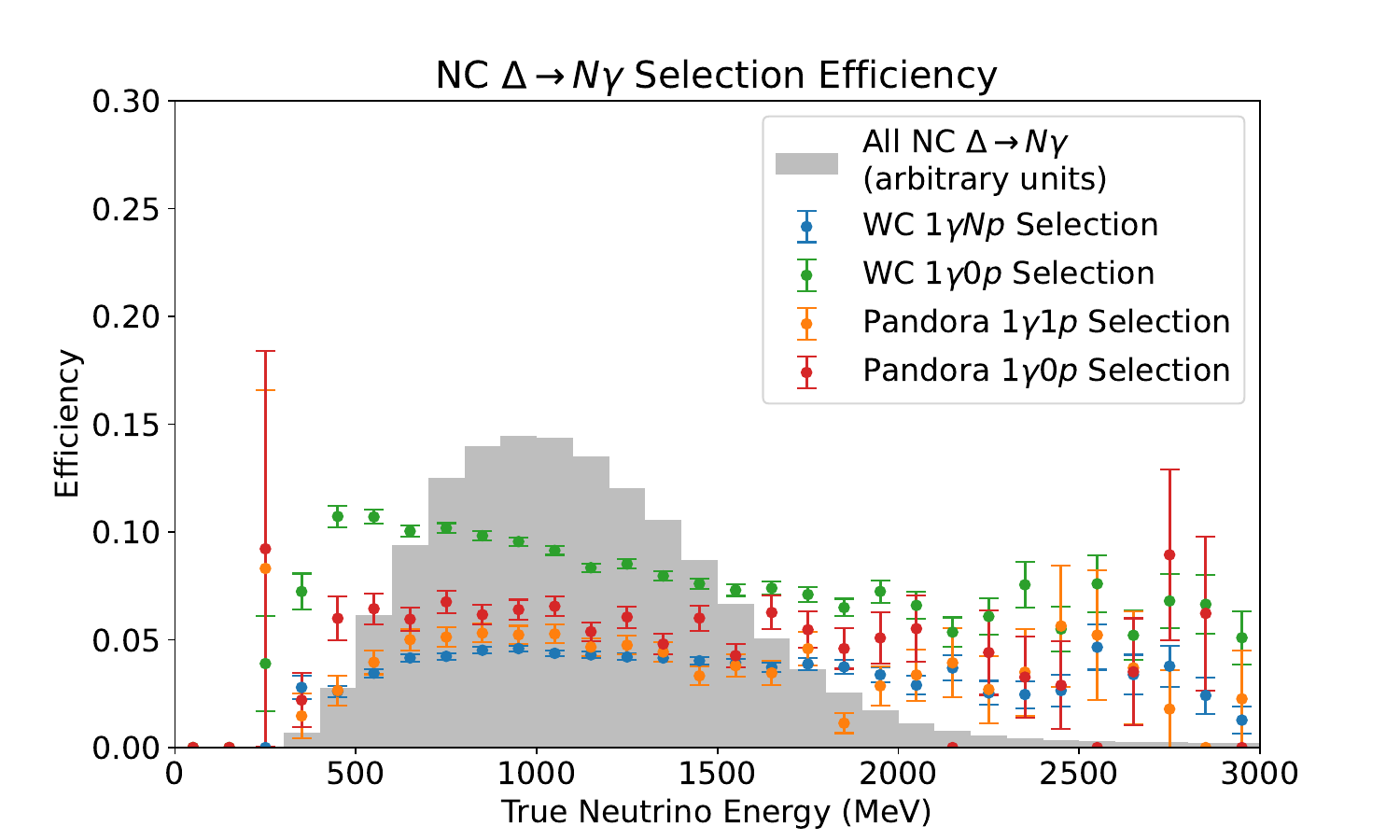}
        \caption{}
        \label{fig:Enu_eff}
    \end{subfigure}
    \begin{subfigure}[b]{0.49\textwidth}
        \includegraphics[trim=15 0 50 0, clip, width=\textwidth]{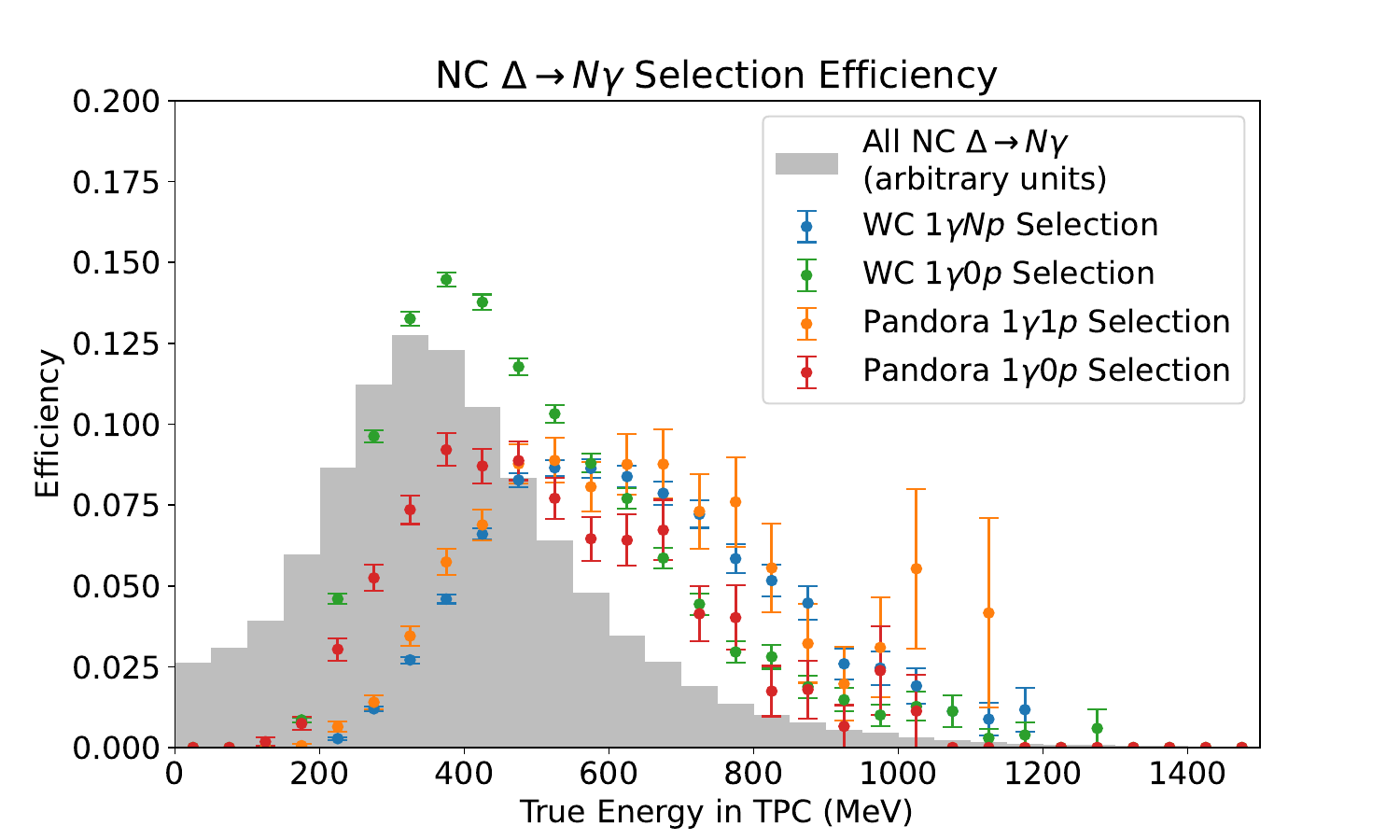}
        \caption{}
        \label{fig:Einside_eff}
    \end{subfigure}
    \caption[NC $\Delta\rightarrow N \gamma$ neutrino energy efficiencies]{Panel (a) shows efficiencies as a function of true neutrino energy. Panel (b) shows efficiencies as a function of true deposited energy in the TPC (not including the energy of the exiting neutrino). Error bars show binomial statistical uncertainties on each efficiency calculation. The gray histogram shows the shape of all true NC $\Delta\rightarrow N \gamma$ events.}
    \label{fig:E_efficiencies}
\end{figure}

\begin{figure}[H]
    \centering
    \includegraphics[width=0.7\textwidth]{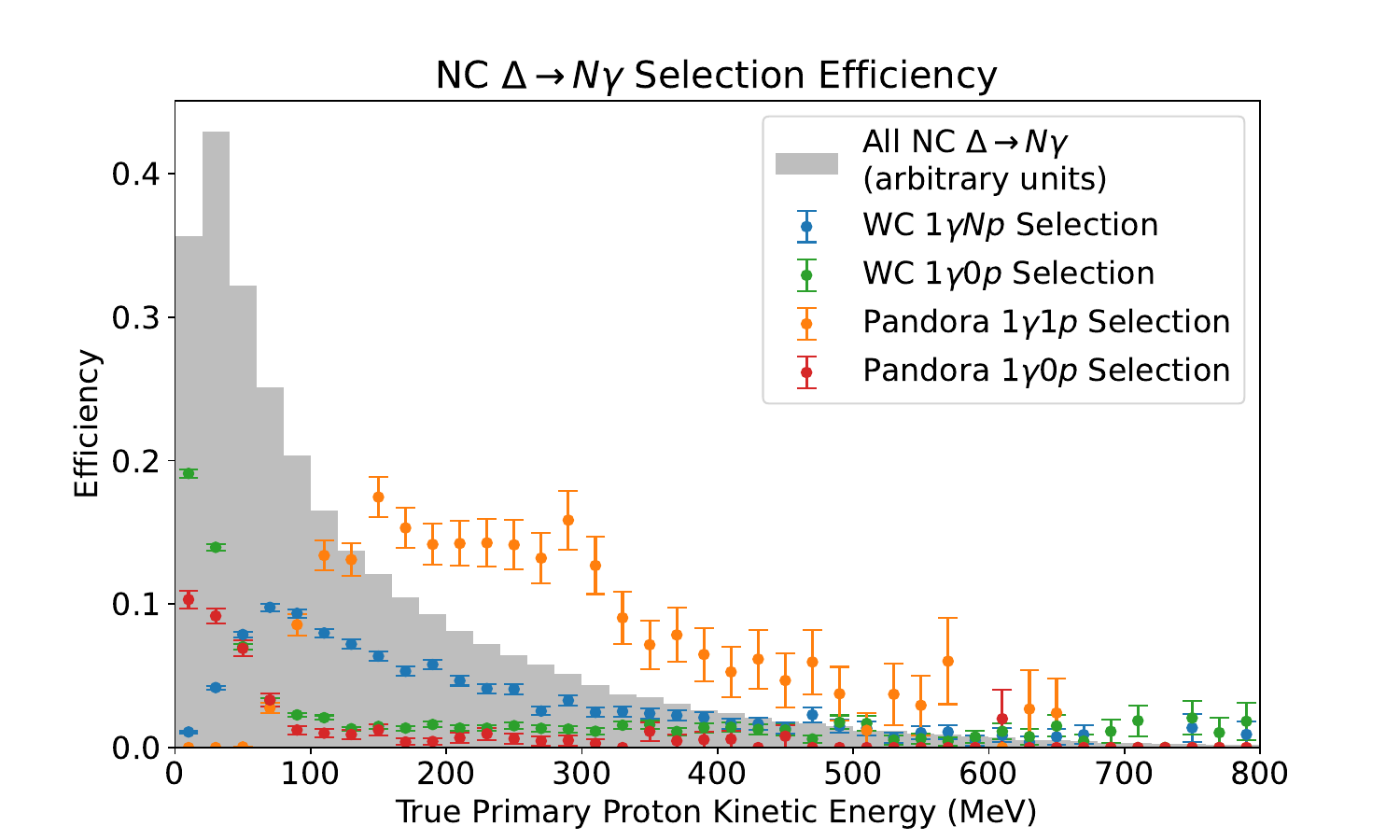}
    \caption[NC $\Delta\rightarrow N \gamma$ proton energy efficiencies]{NC $\Delta\rightarrow N \gamma$ proton energy efficiencies. Error bars show binomial statistical uncertainties on each efficiency calculation. The gray histogram shows the shape of all true NC $\Delta\rightarrow N \gamma$ events.}
    \label{fig:Ep_efficiencies}
\end{figure}

In this section, we show efficiencies for both the Wire-Cell and Pandora NC $\Delta\rightarrow N \gamma$ selections. Here, a ``primary'' particle refers to the particle of that type with the highest true kinetic energy. Figure \ref{fig:E_efficiencies} shows selection efficiencies which indicate that each selection accepts of a wide range of neutrino energies, and an intermediate range of true deposited energy in the TPC (not including the energy of the exiting neutrino). Figure \ref{fig:Ep_efficiencies} shows the efficiency as a function of the true primary proton kinetic energy, indicating that the Wire-Cell $1\gamma0p$ selection has higher efficiency at low proton energies, and the Pandora $1\gamma1p$ selection has higher efficiency at high proton energies.

\begin{figure}[H]
    \centering
    \begin{subfigure}[b]{0.49\textwidth}
        \includegraphics[trim=15 0 50 0, clip, width=\textwidth]{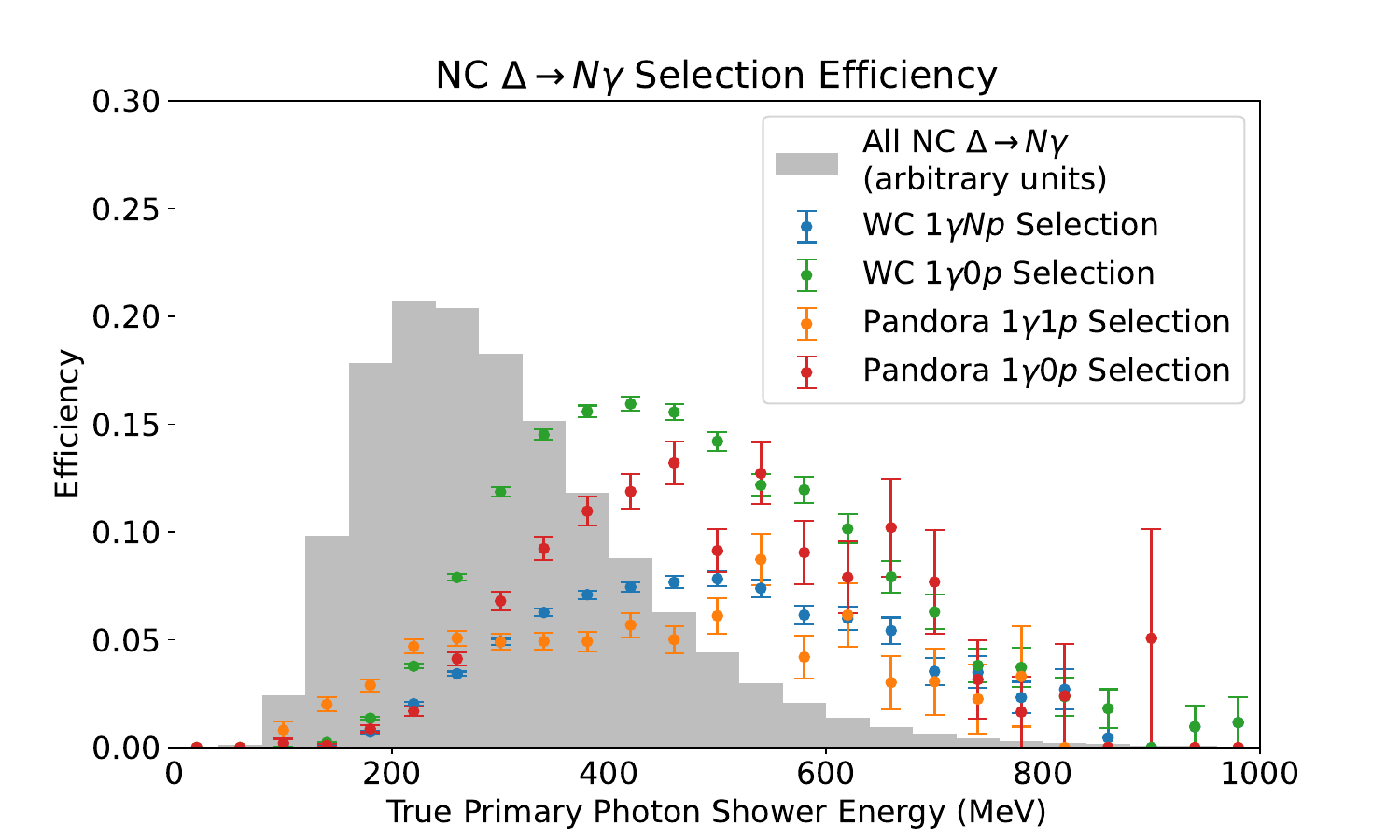}
        \caption{}
    \end{subfigure}
    \begin{subfigure}[b]{0.49\textwidth}
        \includegraphics[trim=15 0 50 0, clip, width=\textwidth]{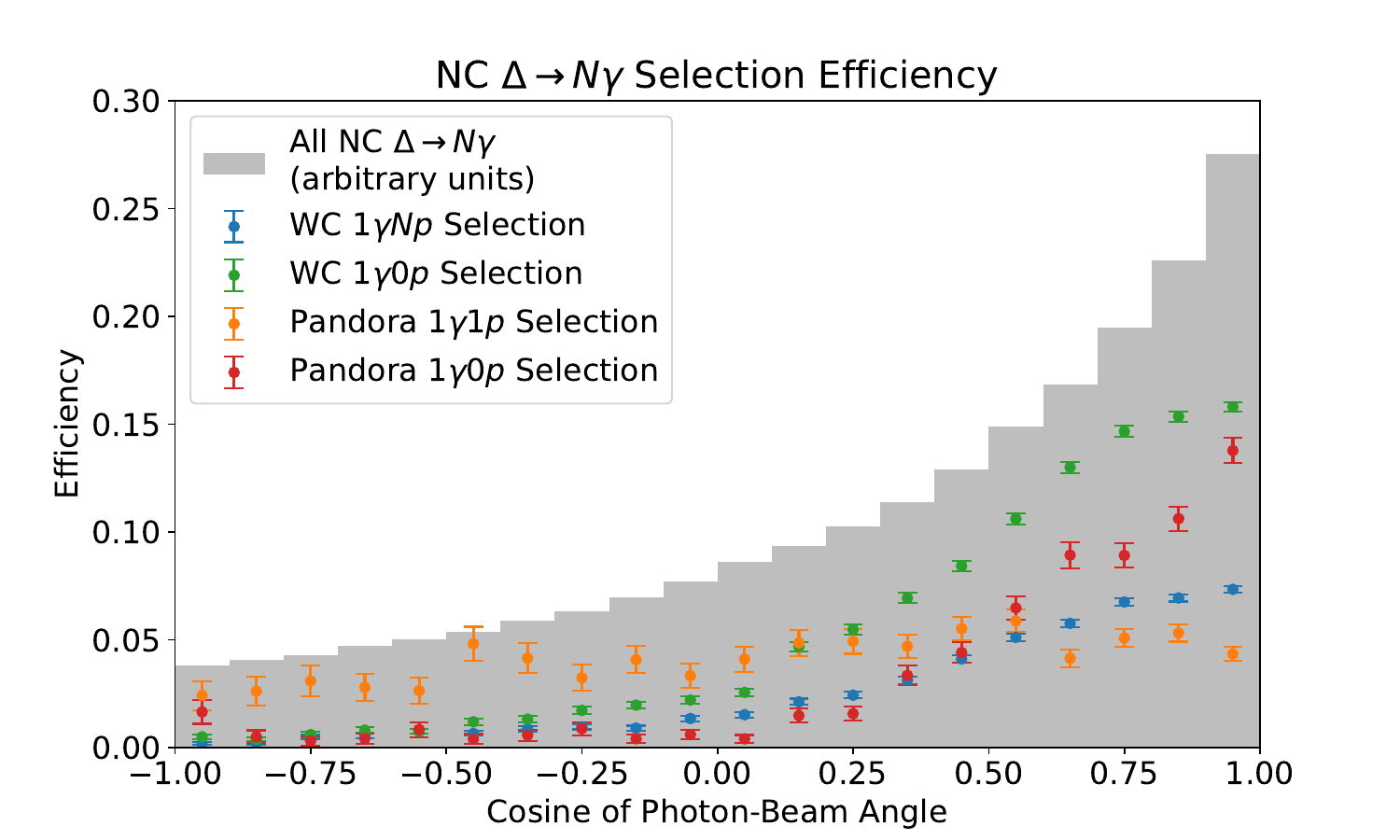}
        \caption{}
    \end{subfigure}
    \caption[NC $\Delta\rightarrow N \gamma$ 1D shower kinematic efficiencies]{NC $\Delta\rightarrow N \gamma$ 1D shower kinematic efficiencies. Panel (a) shows efficiencies as a function of true primary photon shower energy. Panel (b) shows efficiencies as a function of true primary photon angle with respect to the beam. Error bars show binomial statistical uncertainties on each efficiency calculation. The gray histogram shows the shape of all true NC $\Delta\rightarrow N \gamma$ events.}
    \label{fig:shower_1d_efficiencies}
\end{figure}

\begin{figure}[H]
    \centering
    \begin{subfigure}[b]{0.49\textwidth}
        \includegraphics[trim=0 0 80 0, clip, width=\textwidth]{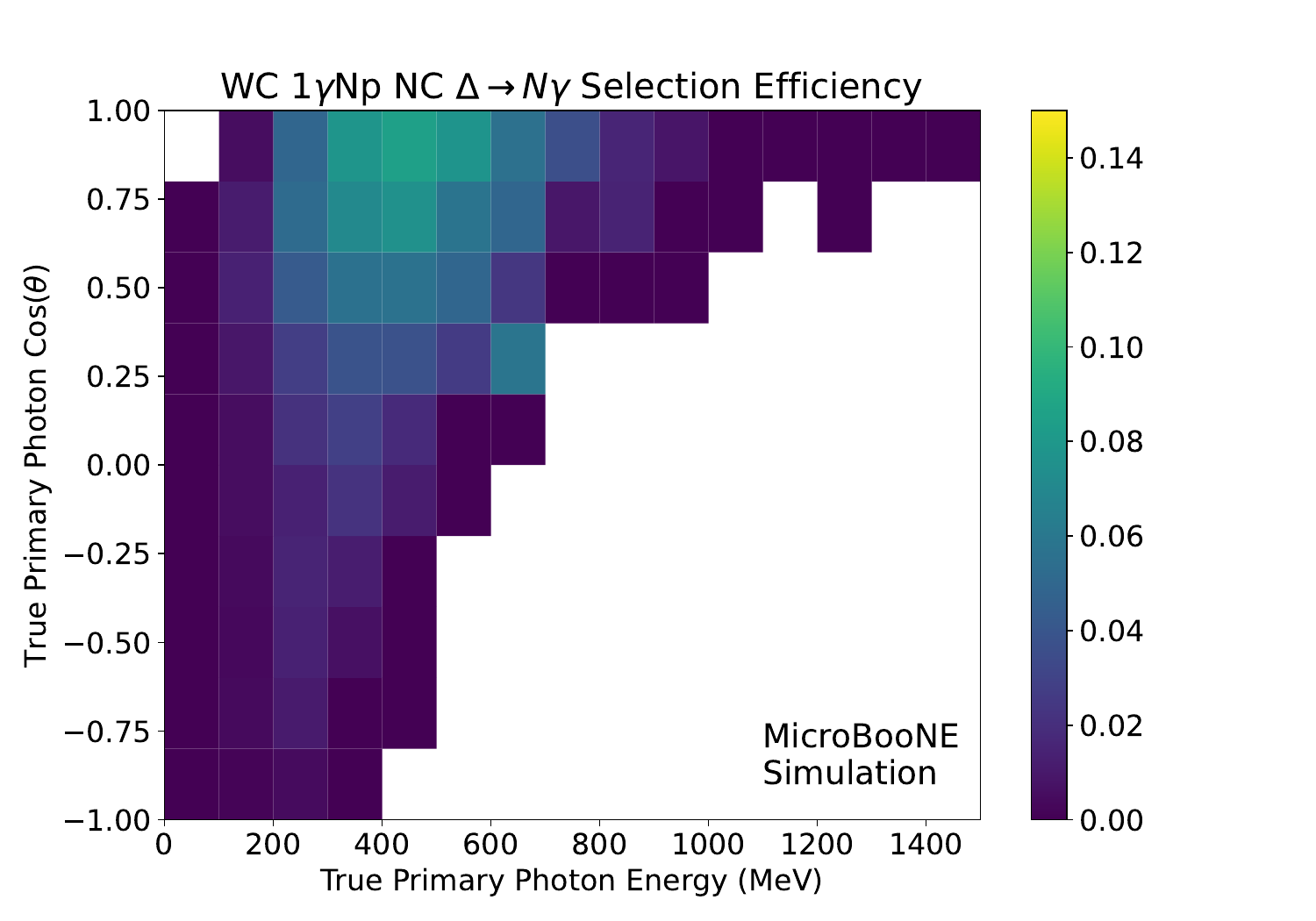}
        \caption{}
    \end{subfigure}
    \begin{subfigure}[b]{0.49\textwidth}
        \includegraphics[trim=0 0 80 0, clip, width=\textwidth]{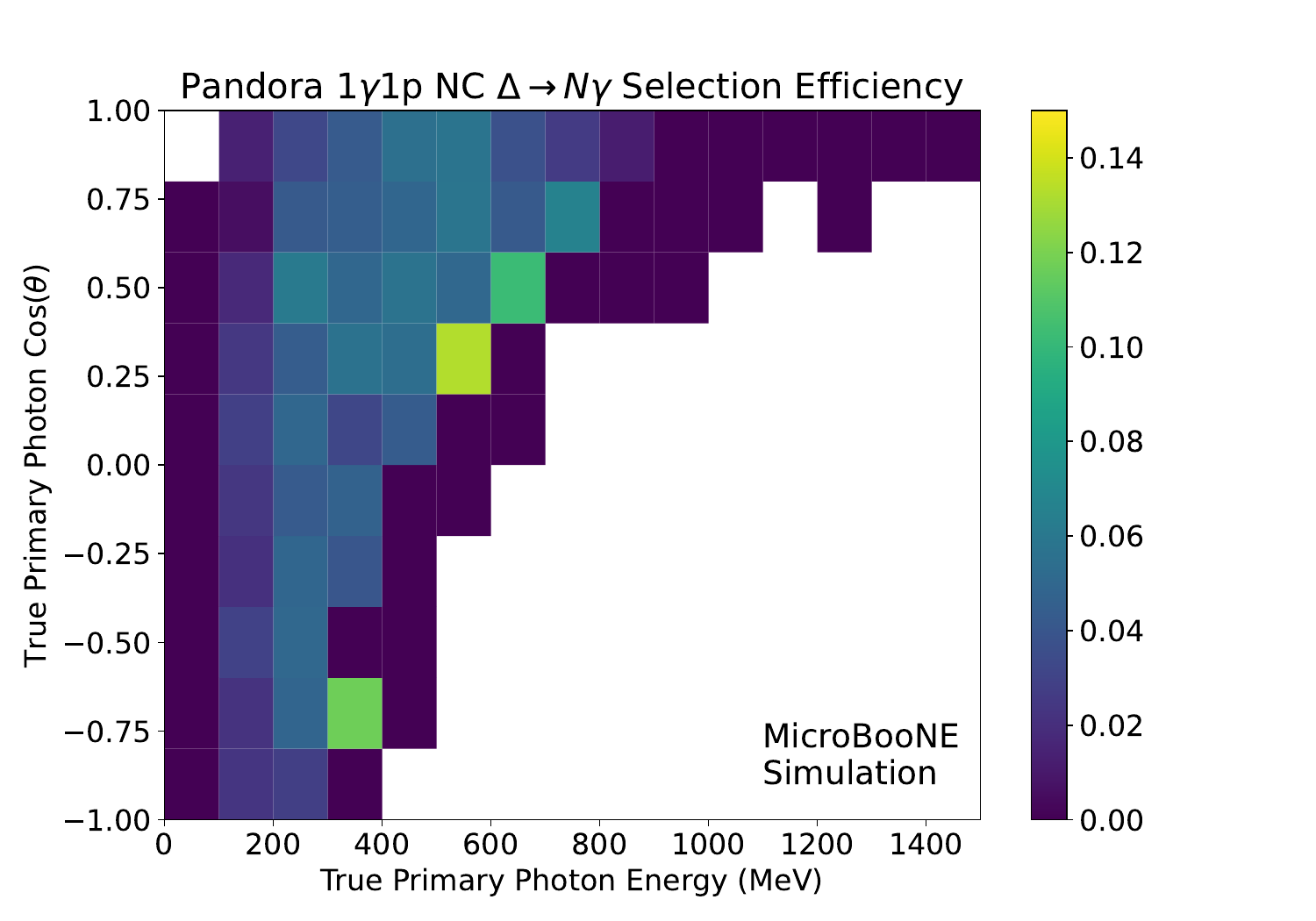}
        \caption{}
    \end{subfigure}
    \begin{subfigure}[b]{0.49\textwidth}
        \includegraphics[trim=0 0 80 0, clip, width=\textwidth]{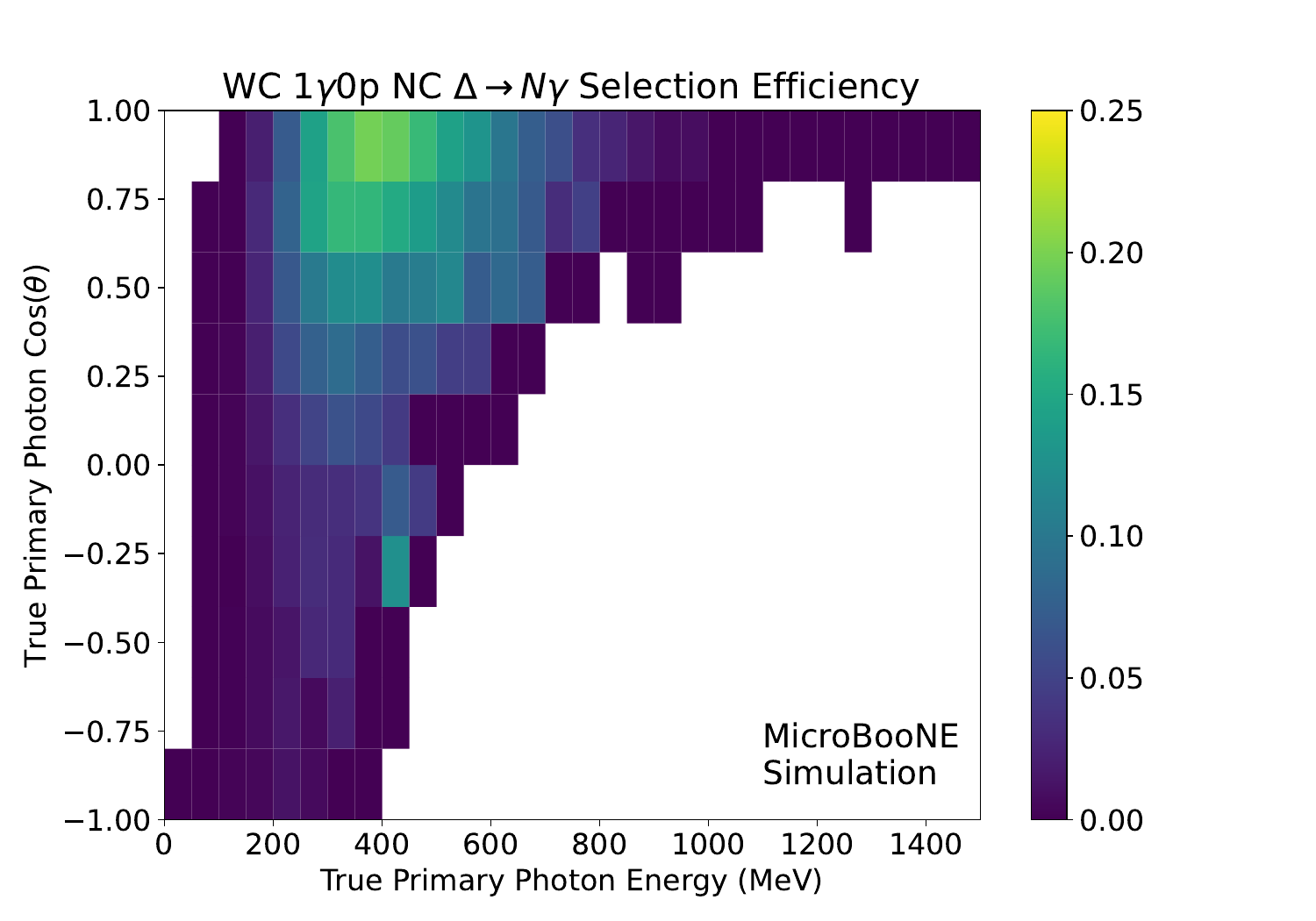}
        \caption{}
    \end{subfigure}
    \begin{subfigure}[b]{0.49\textwidth}
        \includegraphics[trim=0 0 80 0, clip, width=\textwidth]{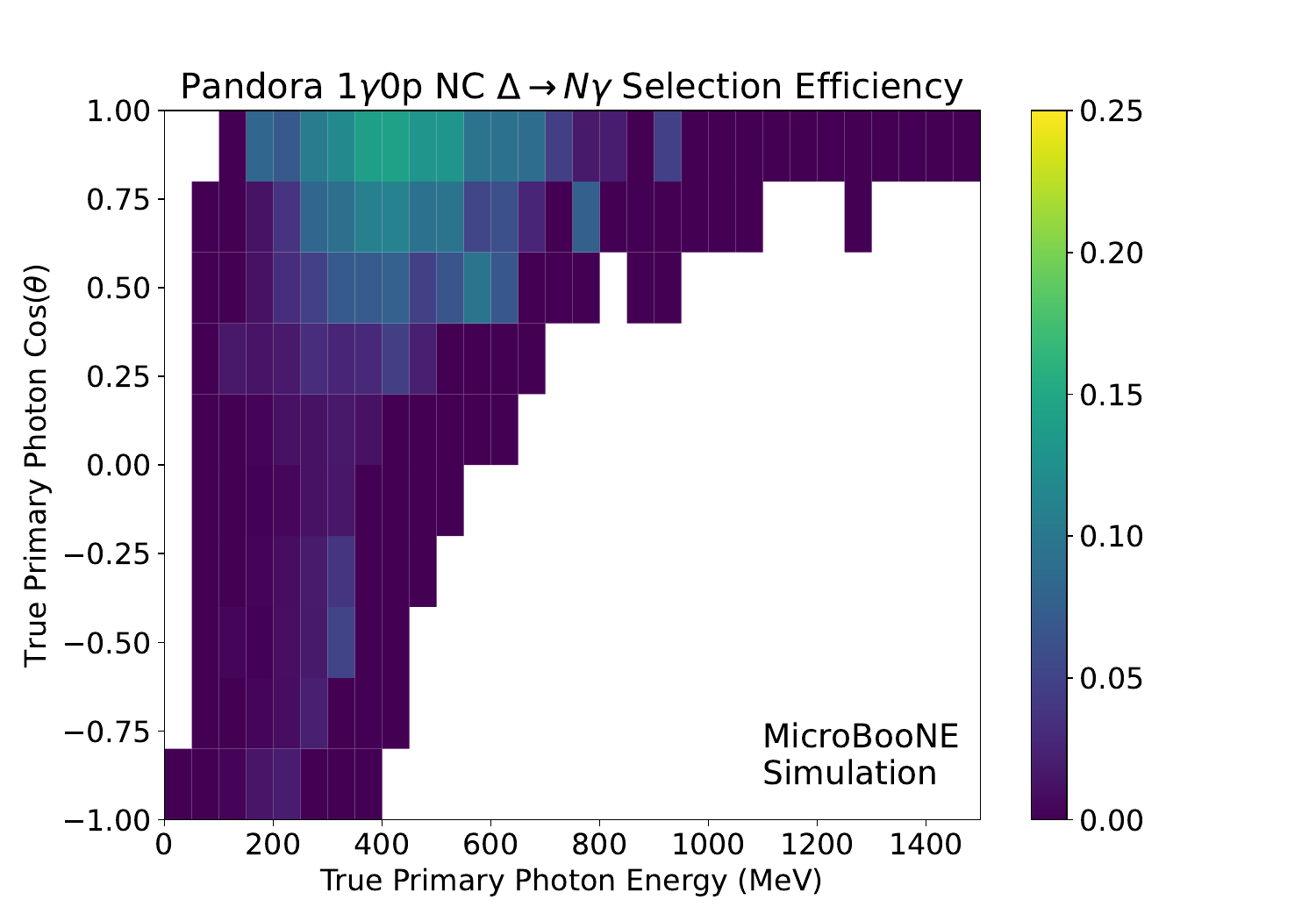}
        \caption{}
    \end{subfigure}
    \caption[NC $\Delta\rightarrow N \gamma$ 2D shower kinematic efficiencies]{2D shower kinematic efficiencies for all NC $\Delta\rightarrow N \gamma$ as functions of true primary shower energy and angle with respect to the beam. Panel (a) shows the Wire-Cell $1\gamma Np$ selection, panel (b) shows the Pandora $1\gamma Np$ selection, panel (c) shows the Wire-Cell $1\gamma 0p$ selection, and panel (d) shows the Pandora $1\gamma 0p$ selection.}
    \label{fig:shower_2d_efficiencies}
\end{figure}

Figure \ref{fig:shower_1d_efficiencies} shows the efficiency of each selection as a function of the primary shower energy and angle, and Fig. \ref{fig:shower_2d_efficiencies} shows these efficiencies as functions of the primary shower energy and angle simultaneously. We see that the Pandora $1\gamma 1p$ selection has the best efficiency for lower energy photons, and for photons traveling upstream back towards the neutrino beam source. The Wire-Cell selections are generally higher efficiency for events with higher photon energies and events with the photons traveling downstream with respect to the neutrino beam.

\begin{figure}[H]
    \centering
    \begin{subfigure}[b]{0.49\textwidth}
        \includegraphics[trim=0 0 80 0, clip, width=\textwidth]{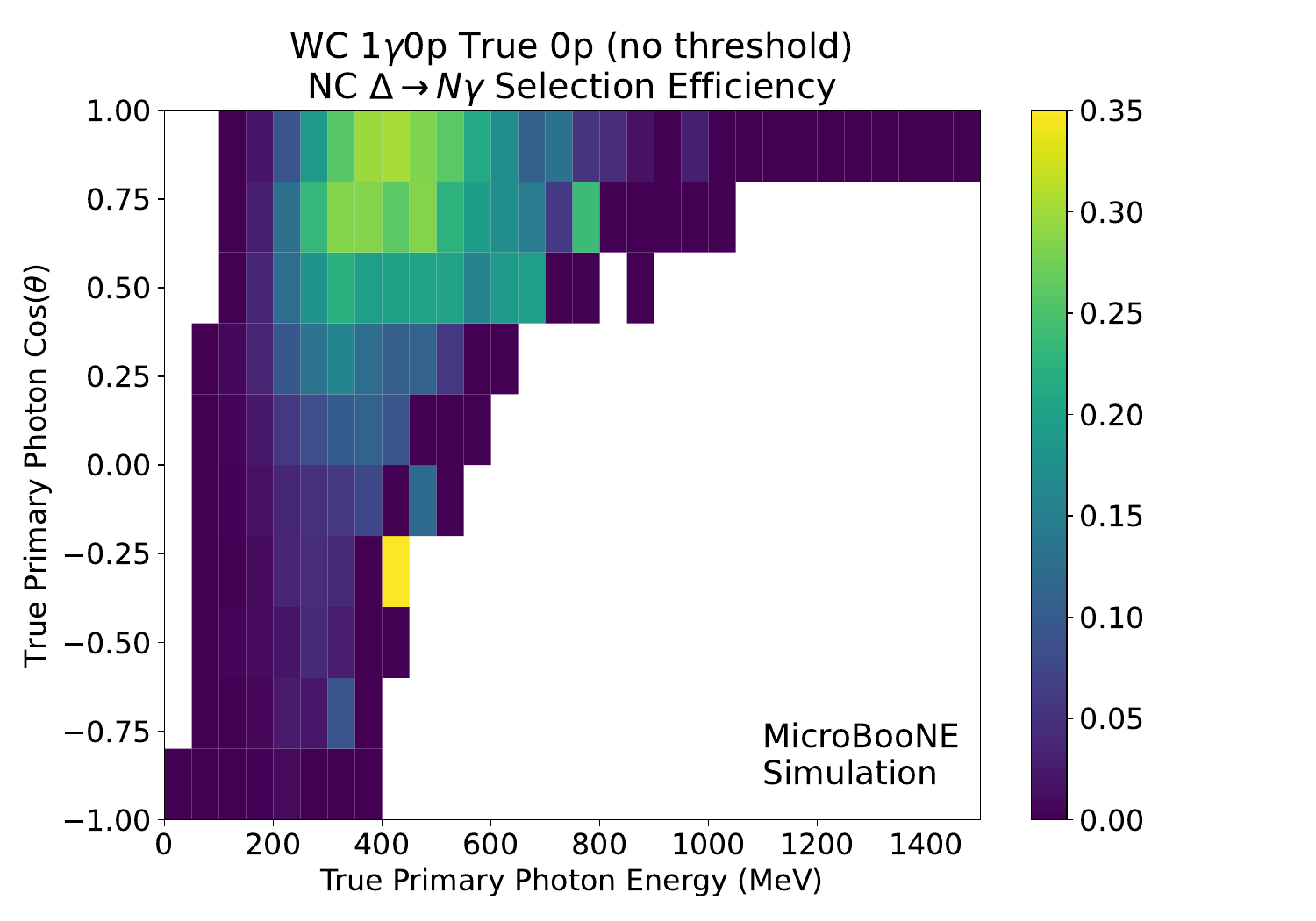}
        \caption{}
    \end{subfigure}
    \begin{subfigure}[b]{0.49\textwidth}
        \includegraphics[trim=0 0 80 0, clip, width=\textwidth]{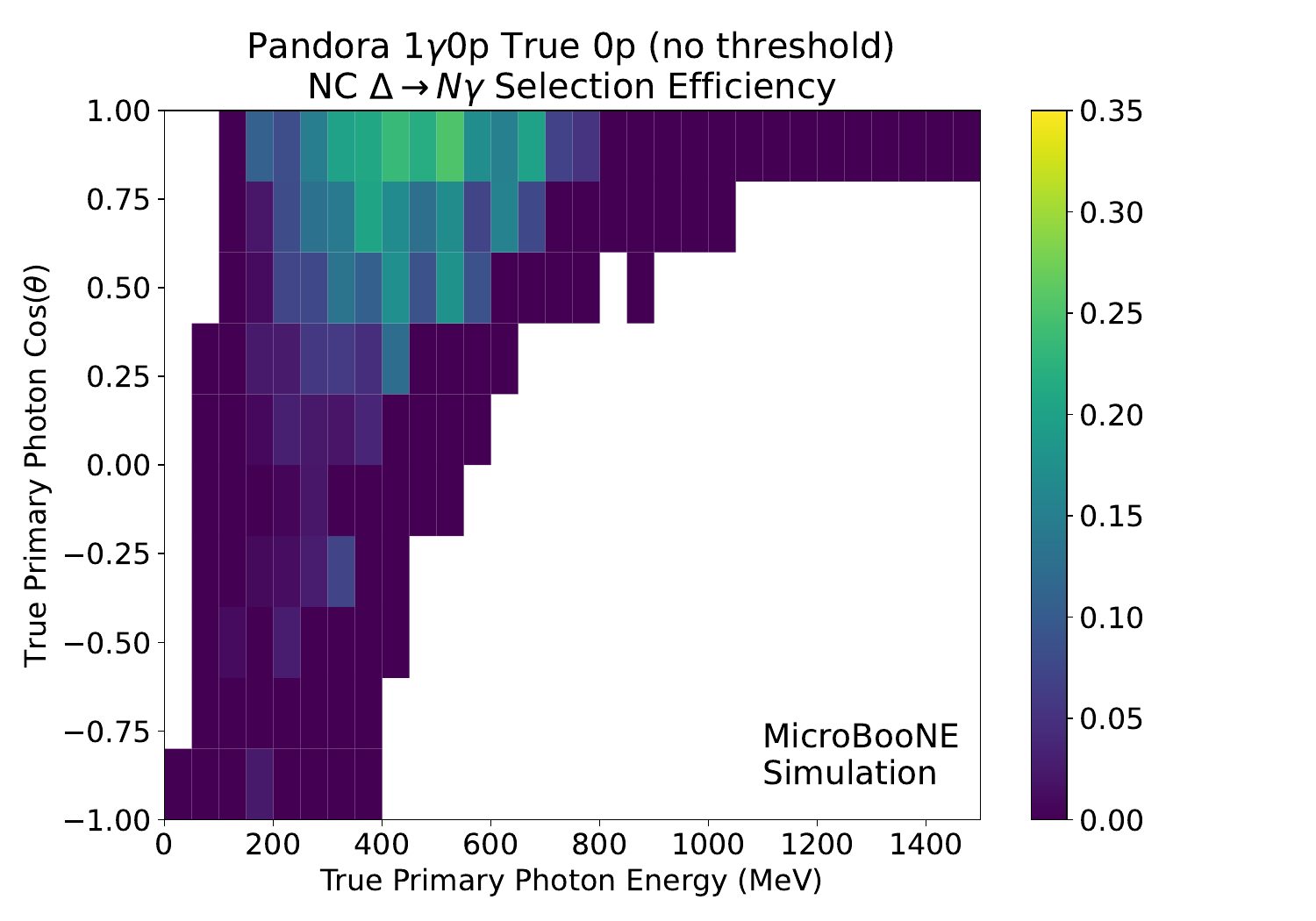}
        \caption{}
    \end{subfigure}
    \caption[NC $\Delta\rightarrow N \gamma$ true $0p$ 2D shower kinematic efficiencies]{2D shower kinematic efficiencies for NC $\Delta\rightarrow N \gamma$ with zero true primary protons of any energy as functions of true primary shower energy and angle with respect to the beam. Panel (a) shows the Wire-Cell $1\gamma 0p$ selection, and panel (b) shows the Pandora $1\gamma 0p$ selection.}
    \label{fig:shower_2d_true_0p_efficiencies}
\end{figure}

\begin{figure}[H]
    \centering
    \begin{subfigure}[b]{0.49\textwidth}
        \includegraphics[trim=15 0 50 0, clip, width=\textwidth]{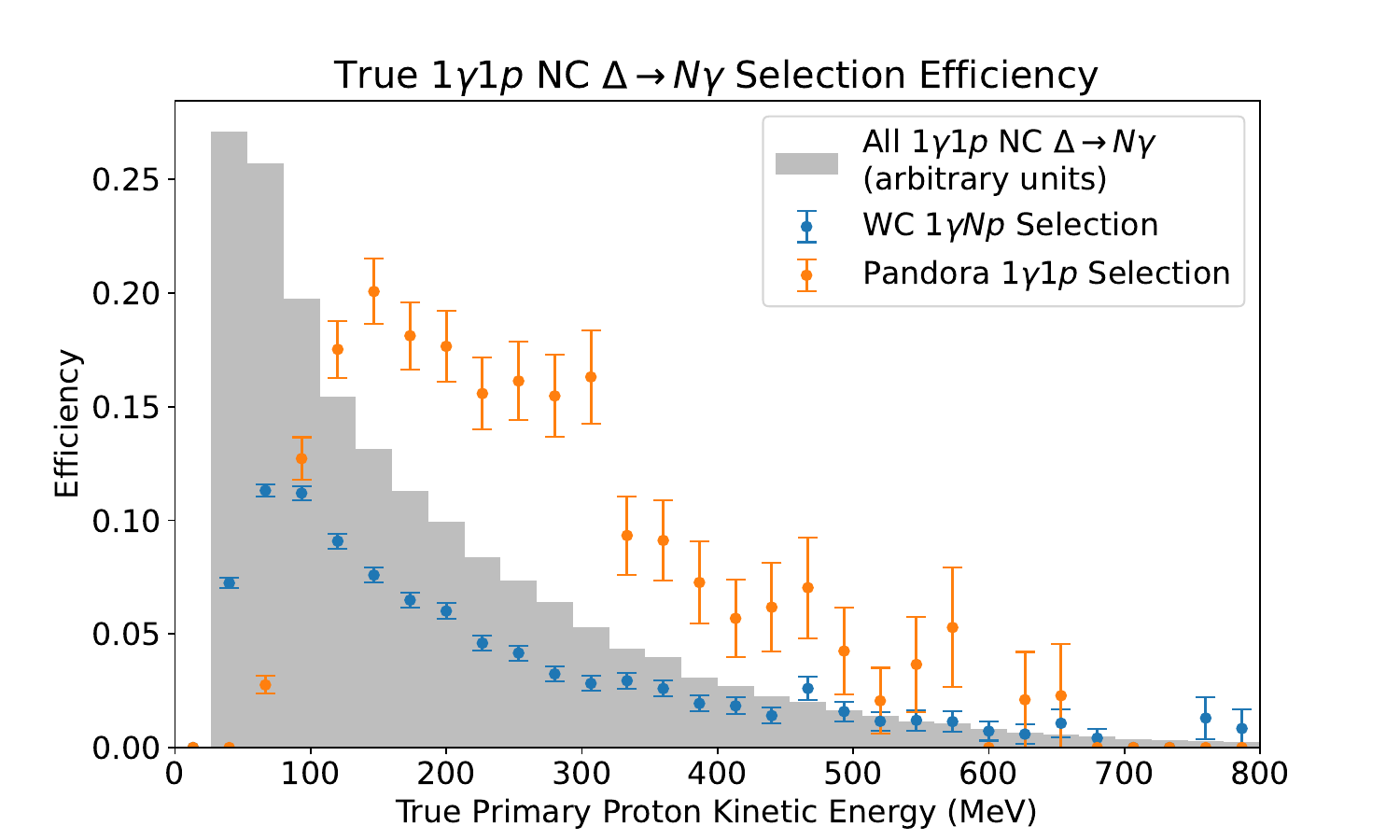}
        \caption{}
    \end{subfigure}
    \begin{subfigure}[b]{0.49\textwidth}
        \includegraphics[trim=15 0 50 0, clip, width=\textwidth]{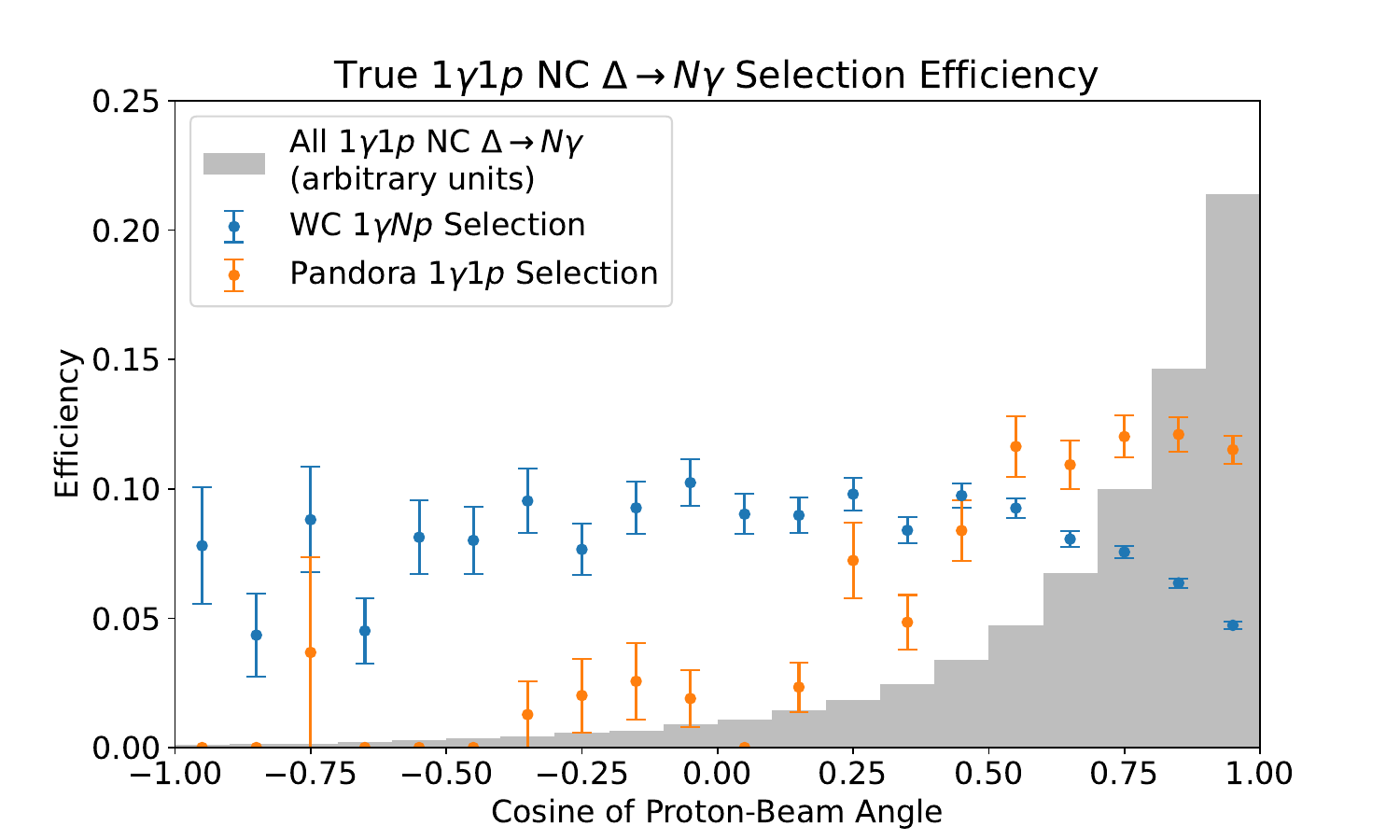}
        \caption{}
    \end{subfigure}
    \caption[$1\gamma 1p$ NC $\Delta\rightarrow N \gamma$ proton kinematic efficiencies]{$1\gamma 1p$ NC $\Delta\rightarrow N \gamma$ proton kinematic efficiencies. Panel (a) shows efficiencies as a function of the true primary proton kinetic energy. Panel (b) shows efficiencies as a function of the true primary proton angle with respect to the beam. Error bars show binomial statistical uncertainties on each efficiency calculation. The gray histogram shows the shape of all true $1\gamma 1p$ NC $\Delta\rightarrow N \gamma$ events.}
    \label{fig:1g1p_proton_kinematic_efficiencies}
\end{figure}

\begin{figure}[H]
    \centering
    \begin{subfigure}[b]{0.47\textwidth}
        \includegraphics[trim=50 0 50 0, width=\textwidth]{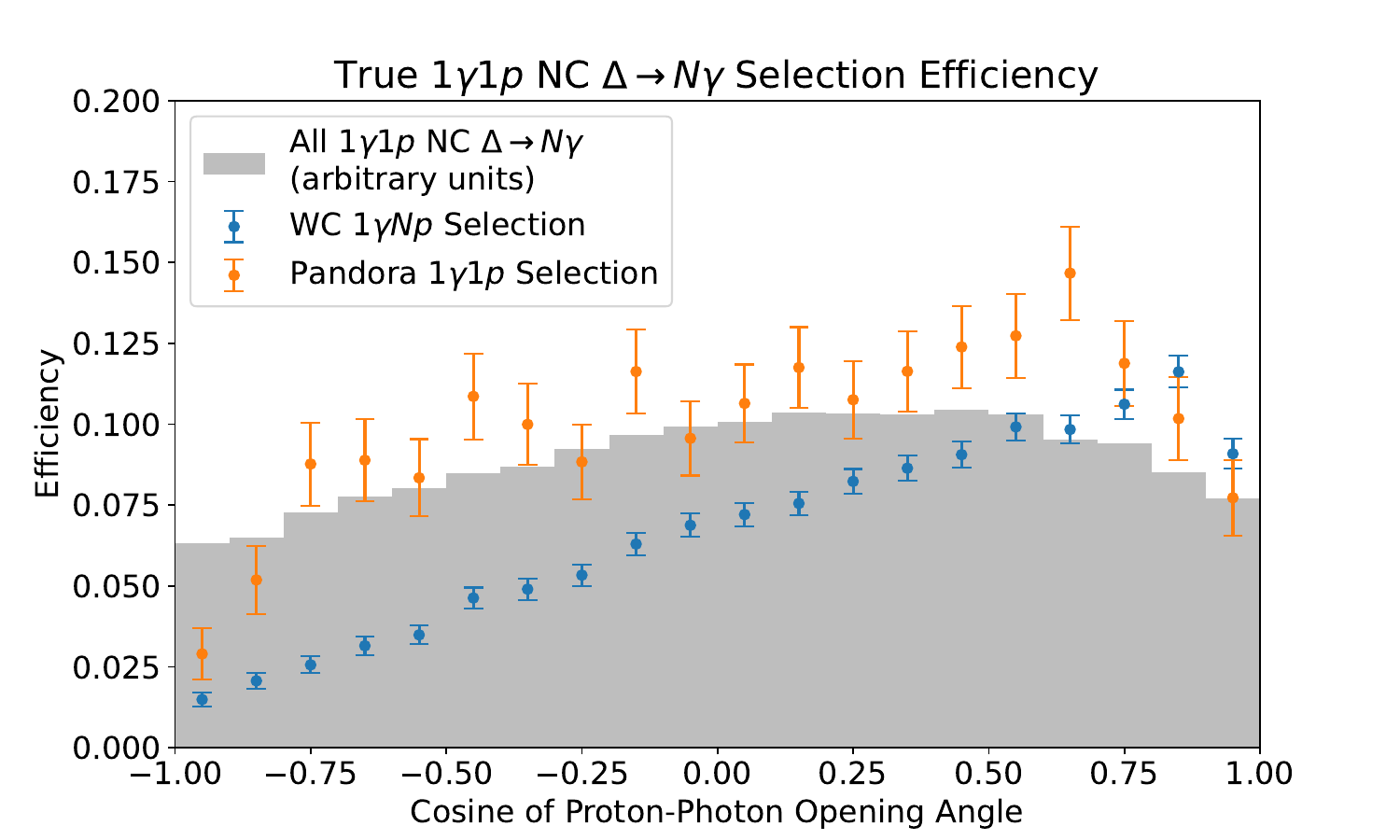}
        \caption{}
        \label{fig:proton_photon_angle_eff}
    \end{subfigure}
    \begin{subfigure}[b]{0.52\textwidth}
        \includegraphics[trim=30 0 30 0, clip, width=\textwidth]{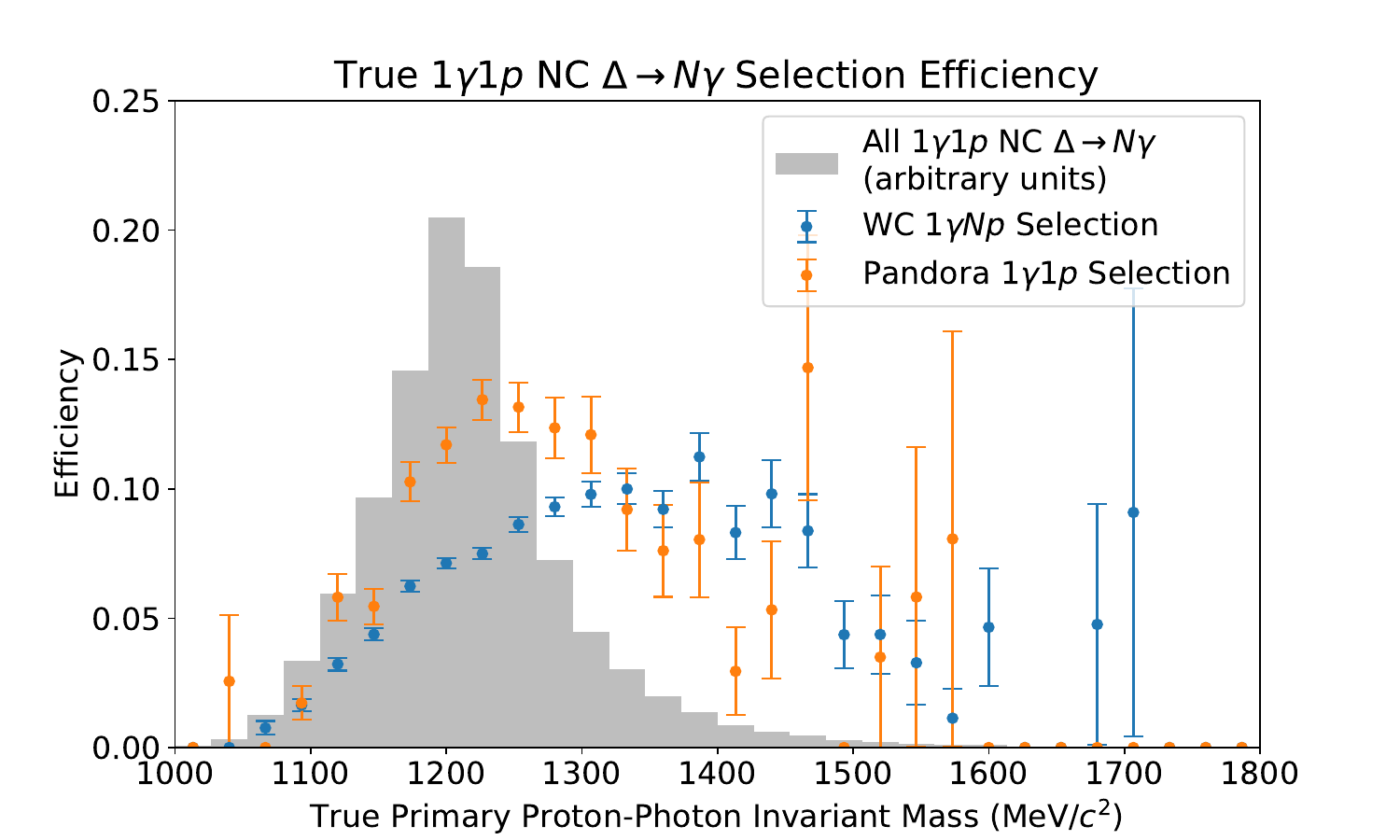}
        \caption{}
        \label{fig:invariant_mass_eff}
    \end{subfigure}
    \caption[$1\gamma 1p$ NC $\Delta\rightarrow N \gamma$ invariant mass and opening angle efficiencies]{True $1\gamma 1p$ NC $\Delta\rightarrow N \gamma$ efficiencies involving both the proton and photon. Panel (a) shows efficiencies as a function of the proton-photon opening angle. Panel (b) shows efficiencies as a function of the proton-photon invariant mass. The gray histogram shows the shape of all true $1\gamma 1p$ NC $\Delta\rightarrow N \gamma$ events.}
    \label{fig:proton_photon_effs}
\end{figure}

For events without protons, Fig. \ref{fig:shower_2d_efficiencies} essentially contains all the information about the selection performance towards any potential MiniBooNE LEE model. However, for events with protons, the efficiency phase space is more complicated, depending on the kinematics of both the shower and the hadronic activity. To illustrate this hadronic kinematic model dependence in the $1\gamma Np$ topology, we investigate our selection efficiencies for true $1\gamma1p$ events, defined by exactly one proton with true kinetic energy greater than 35 MeV, no charged pions with true kinetic energy greater than 10 MeV, and no true neutral pions. Figure \ref{fig:1g1p_proton_kinematic_efficiencies} shows efficiencies as functions of the primary proton energy and angle for these events. Figure \ref{fig:proton_photon_angle_eff} shows the efficiency as a function of the proton-photon opening angle. Figure \ref{fig:proton_photon_effs} shows efficiencies as functions of variables that involve both the proton and the shower. In Fig. \ref{fig:invariant_mass_eff}, we show the efficiency as a function of the proton-photon invariant mass. The true distribution in this variable peaks at around 1232 MeV, corresponding to the mass of the $\Delta^+$ resonance producing these particles. The selection efficiencies are highest at masses greater than this peak, which makes sense since higher invariant masses will tend to have higher energy photon showers which are more easily reconstructed. Despite limited Monte-Carlo statistics at these rare high invariant masses, we can see that the efficiency declines as these kinematics get further away from the common values for the NC $\Delta\rightarrow N \gamma$ process.

\subsection{Combining Wire-Cell and Pandora}\label{sec:combined_pandora_wc_nc_delta}

As shown in the previous section, the Wire-Cell and Pandora NC $\Delta\rightarrow N \gamma$ selections have different strengths in different parts of the phase space when analyzing efficiencies. Because of this fact, we decided to combine both reconstruction frameworks and all four selections into one enhanced analysis of NC $\Delta\rightarrow N \gamma$ events.

In addition to the efficiency which was carefully analyzed in the previous section, we must also consider the purity for each selection, and this is summarized in Table \ref{tab:effs_and_purs}. For events with reconstructed protons ($1\gamma Np$ and $1\gamma 1p$), we see than Wire-Cell has slightly lower efficiency and purity relative to Pandora. For events without reconstructed protons (1$\gamma 0p$), we see than Wire-Cell has significantly higher efficiency and purity relative to Pandora. We also can see the results after combining all four Wire-Cell and Pandora selections; in this case, our total efficiency is 19.64\%, more than double the 9.76\% efficiency for the combination of both Pandora selections. So, the addition of these Wire-Cell selections effectively doubles our expected number of NC $\Delta\rightarrow N \gamma$ events, equivalent to adding three additional years of MicroBooNE beam data. 

\begin{table}[H]
    \centering
    \small
    \begin{tabular}{c c c c c c c c} 
        \toprule
        & &\makecell{WC \\ $1\gamma Np$} & \makecell{Pandora \\ $1\gamma 1p$} & \makecell{WC \\ $1\gamma 0p$} & \makecell{Pandora \\ $1\gamma 0p$} & & \makecell{Combined} \\ 
    \hline
    \makecell{NC $\Delta\rightarrow N \gamma$ efficiency} & & 4.09\% & 4.24\% & 8.79\% & 5.52\% & &19.64\%\\
    \makecell{NC $\Delta\rightarrow N \gamma$ purity} & & 9.60\% & 14.84\% & 7.50\% & 3.98\% & &6.37\%\\
        \bottomrule
    \end{tabular}
    \caption[Wire-Cell and Pandora NC $\Delta\rightarrow N \gamma$ efficiency and purity summary]{Wire-Cell and Pandora NC $\Delta\rightarrow N \gamma$ efficiency and purity summary. The rightmost column shows the efficiency and purity for a union of all four selections; note that the combined efficiency is less than the sum of the four efficiencies, because some events can be selected by both reconstructions. Efficiency is calculated as the fraction of simulated true NC $\Delta\rightarrow N \gamma$ events in the fiducial volume which enter the final selection. Purity is calculated as the fraction of the predicted selected events which are from the NC $\Delta\rightarrow N \gamma$ process.}
    \label{tab:effs_and_purs}
\end{table}

\begin{figure}[H]
    \centering
    \includegraphics[width=0.7\textwidth]{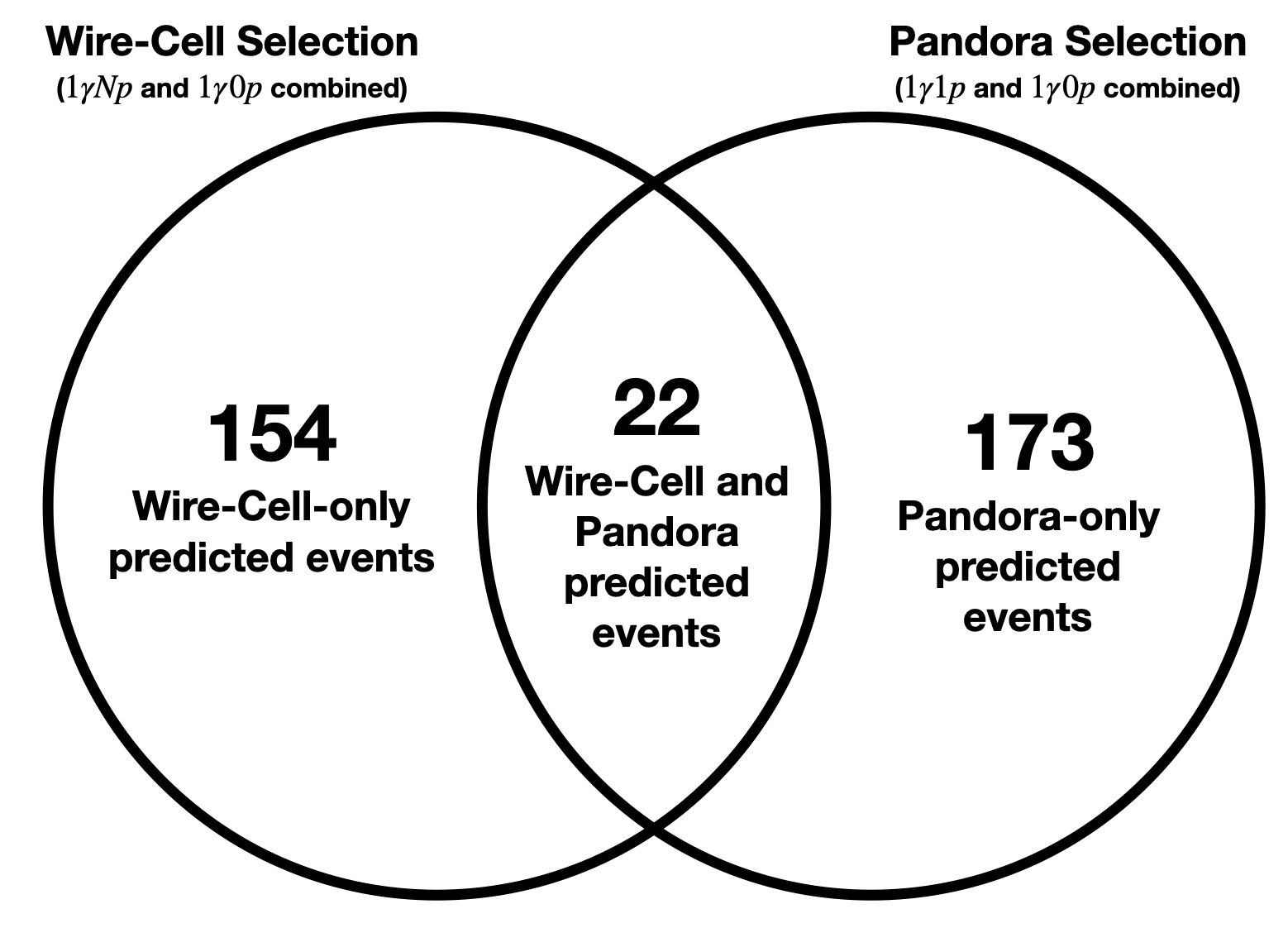}
    \caption[Wire-Cell and Pandora overlap Venn diagram]{Wire-Cell and Pandora overlap Venn diagram.}
    \label{fig:overlap_venn_diagram}
\end{figure}

This increased combined efficiency is a reflection of the fact that Wire-Cell and Pandora actually select mostly different events in the detector, as illustrated in Fig. \ref{fig:overlap_venn_diagram}. This is somewhat surprising, given that both selections in the same detector are trying to select the same topologies. The efficiency of each selection is fairly low, 4-9\%, so there is room for the different selections to choose different events. This orthogonality is due to efficiency differences between the two selections. Some of these differences are shown in Sec. \ref{sec:nc_delta_efficiencies}, but there are many more variables in the high dimensional phase space describing all reconstructed events which can also have differences. Note that the efficiencies may not vary only as functions of simple interaction topologies and kinematics, but also as functions of more detector-specific factors, such as the branching structure of the photon shower, or the angular deflections of the proton track, or the proximity of nearby cosmic rays in different projected wire planes. So there is not necessarily one simple explanation for this small overlap between selections. There is one conclusion we can draw from this: there is certainly room for improvement in both of these reconstruction and selection workflows. An ideal reconstruction and selection would be able to select all events in both the Wire-Cell and Pandora selections, and likely more events beyond those two as well. This helps motivate some of the ideas for reconstruction improvements I will describe in Sec. \ref{sec:future_photon_reconstruction}.

There were several technical challenges when combining Wire-Cell and Pandora analysis files. Both were developed separately and never intended to be combined. Different choices were made with respect to which filtered high-statistics simulation samples were processed and analyzed. In order to estimate correlated statistical and systematic uncertainties, we need to analyze a common set of events in both selections. This required some reprocessing, in particular since the Wire-Cell analysis never used a dedicated $\nu_\mu$CC $\pi^0$ simulation, instead using a more general neutrino simulation to analyze this topology, while the Pandora analysis only used this dedicated simulation and specifically removed this topology from the more general simulation. This required re-processing some Wire-Cell files to ensure overlapping processed events for all topologies.

Once we had common events in all topologies, we had to account for the rate of overlap between each set of files. For example, for certain filtered simulation file types, Wire-Cell used simulations of three different MicroBooNE beam data collection periods, known as runs 1-3, while Pandora only used simulations for run 1 and run 3, and weighted these events up to cover the missing run 2 simulation. As another example, Wire-Cell and Pandora independently chose certain portions of simulation files to use for BDT training, and removed these portions from all final predictions. This overlapping fraction and implied event weighting had to be carefully calculated for each run period and each simulation file in order to ensure that we could recreate the Pandora predictions using only simulated events in our processed Wire-Cell files. Another consideration is that Wire-Cell only processed a smaller sample of BNB data, $6.37\cdot 10^{20}$ POT, relative to the Pandora selections' $6.80\cdot 10^{20}$ POT; this is primarily due to rare failures in processing, and these failures are being addressed as we continue to process our full data set for future analyses. Here, we modified Wire-Cell files in order to add the rare data events which caused failures in Wire-Cell processing and were selected by the Pandora selections, in order to avoid any reduction in real data statistics when restricting to Wire-Cell processed events. After these steps, we have a set of files that we can naturally use in order to analyze both Pandora and Wire-Cell NC $\Delta\rightarrow N \gamma$ selected events with all statistical and systematic correlations properly considered. Note that this process resulted in some statistical fluctuations in the Pandora NC $\Delta\rightarrow N \gamma$ predictions, so we do not expect precisely identical predictions relative to the older Pandora NC $\Delta\rightarrow N \gamma$ results \cite{glee_prl}.

\subsection{Systematic Uncertainties}

The systematic uncertainties in this analysis are largely the same as those described in Sec. \ref{sec:systematic_uncertainties}, with considerations for variations in the detector response modeling, neutrino-argon interaction modeling, hadron-argon re-interaction modeling, and Monte-Carlo statistical uncertainties. However, there are some unique considerations to consider for this single photon analysis.

In our GENIEv3 with MicroBooNE tune \cite{genie_v3,genie-tune-paper}, we include a 50\% fractional uncertainty on the NC $\Delta\rightarrow N \gamma$ branching ratio. This uncertainty is conservative, given the Particle Data Group's assigned uncertainty of 8.3\% \cite{ParticleDataGroup}. For this analysis, where we attempt to measure compatibility with different branching ratios which scale this process, we do not want to use any uncertainty on this parameter. For the prior MicroBooNE NC $\Delta\rightarrow N \gamma$ search using just Pandora reconstruction \cite{glee_prl}, this uncertainty was explicitly turned off in the GENIE reweighting uncertainty calculations. In this analysis, we use the same files which were processed for the Wire-Cell $\nu_e$CC search, and therefore include all GENIE uncertainties. So, to remove this uncertainty, we can re-weight our events in different systematic variation universes in order to cancel out the effect of the NC $\Delta\rightarrow N \gamma$ branching ratio uncertainty which was applied during the initial processing.

\begin{figure}[H]
    \centering
    \begin{subfigure}[b]{0.49\textwidth}
        \includegraphics[width=\textwidth]{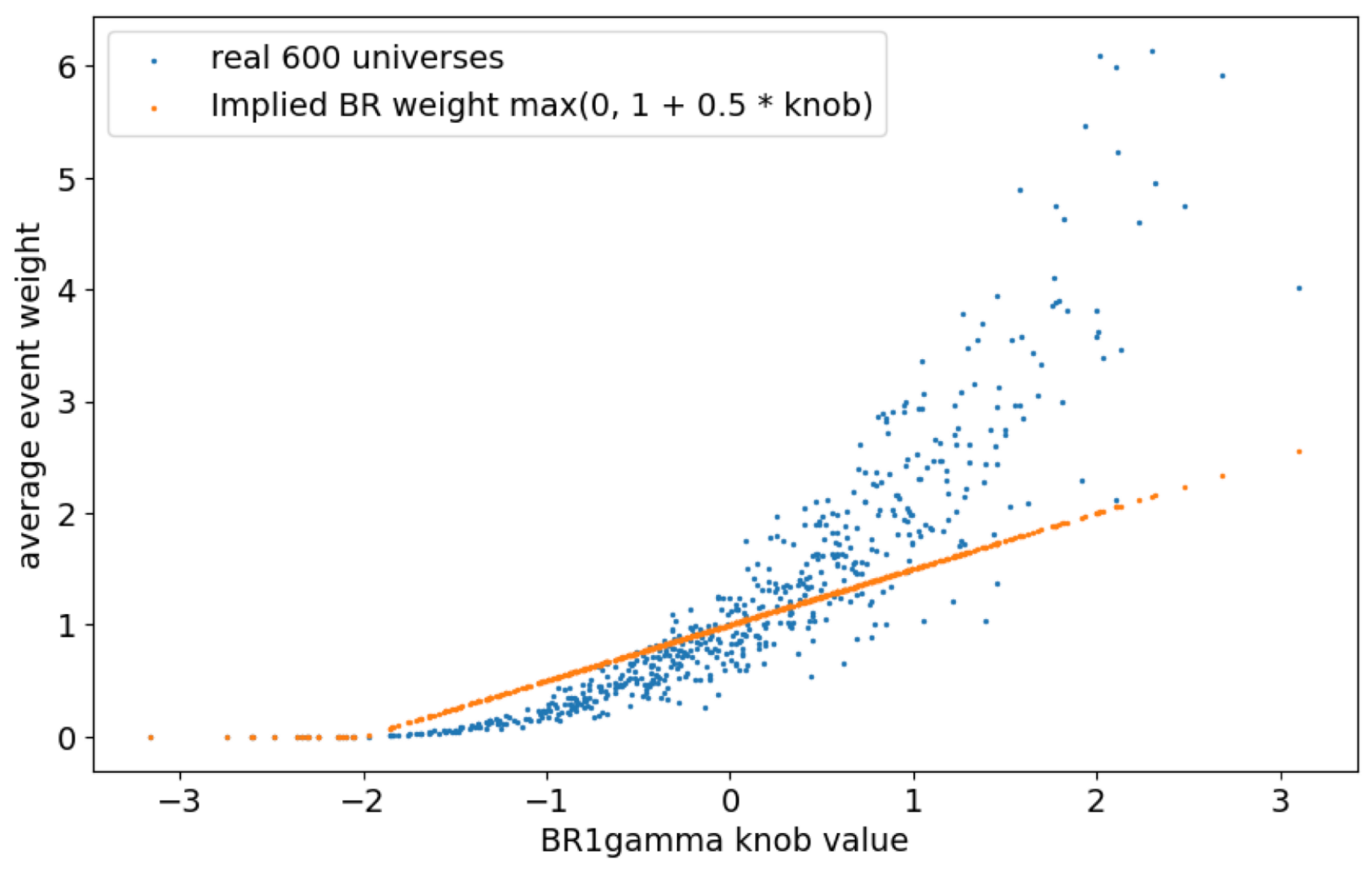}
        \caption{}
        \label{fig:bugged_genie_BR_corr}
    \end{subfigure}
    \begin{subfigure}[b]{0.49\textwidth}
        \includegraphics[width=\textwidth]{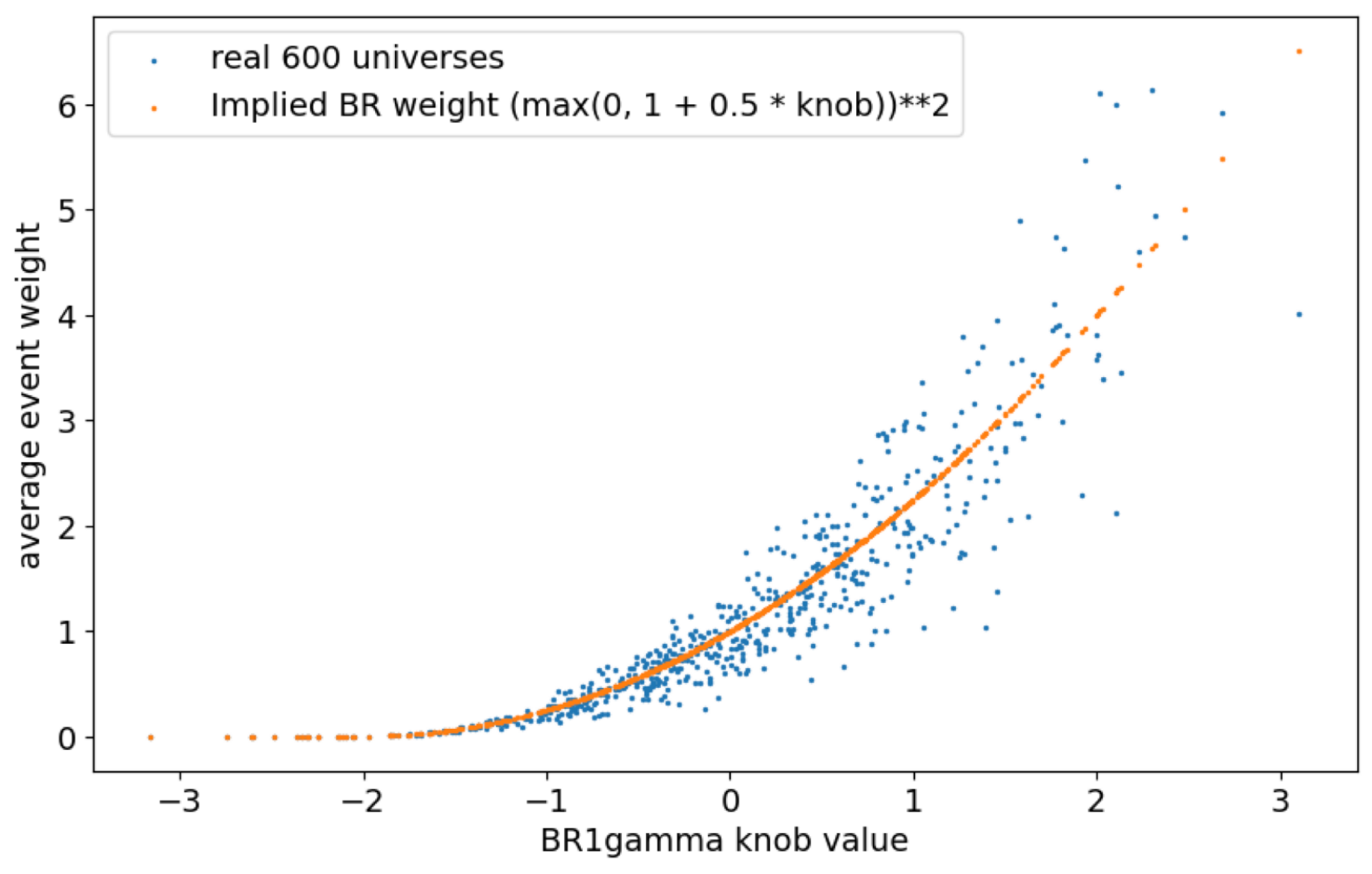}
        \caption{}
        \label{fig:fixed_genie_BR_corr}
    \end{subfigure}
    \begin{subfigure}[b]{0.49\textwidth}
        \includegraphics[width=\textwidth]{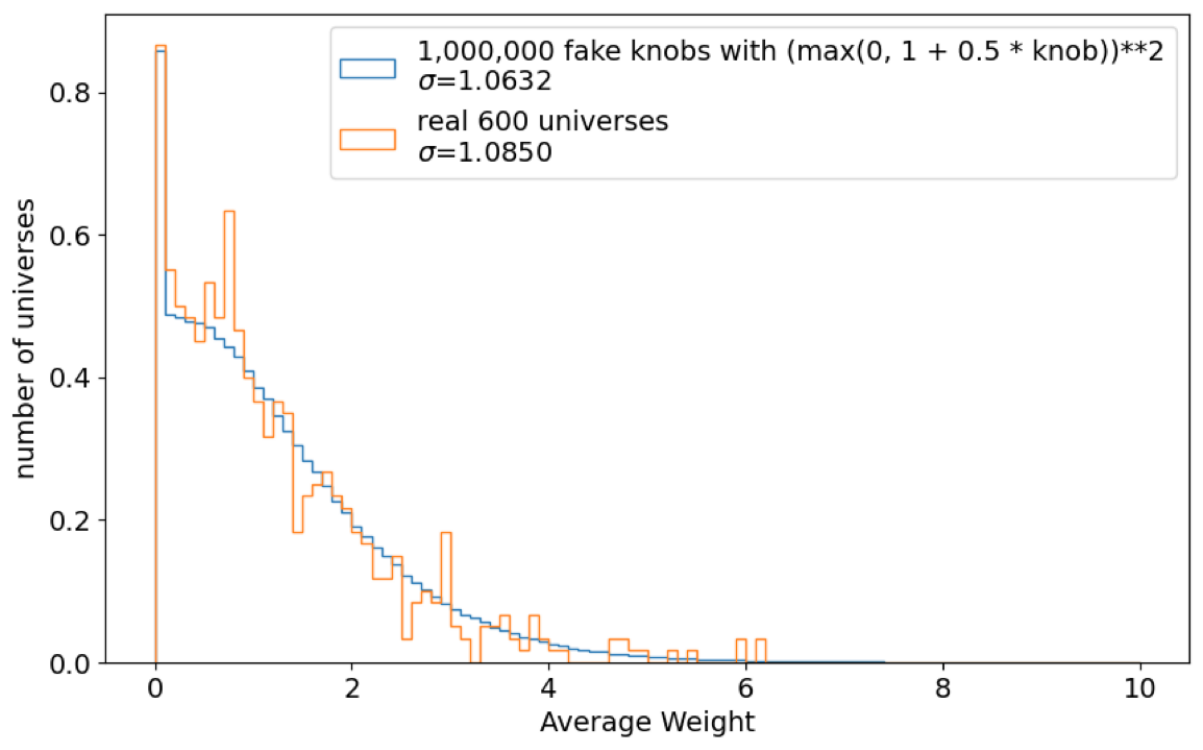}
        \caption{}
        \label{fig:average_weight_BR}
    \end{subfigure}
    \caption[GENIE branching ratio bugfix]{GENIE branching ratio bugfix.}
    \label{fig:genie_br_bug}
\end{figure}

However, in this process, we discovered an issue with the initial uncertainties. For our NC $\Delta\rightarrow N \gamma$ simulation file, we investigate each of our 600 variation universes which simultaneously fluctuate many GENIE cross section parameters. For each universe, there is a specific knob value called ``BR1gamma'' which is randomly assigned according to a Gaussian distribution with a mean of zero and a standard deviation of one. There is a specific relationship expected between the knob value and the average event weight for all NC $\Delta\rightarrow N \gamma$ events in the file. We show this correlation in Fig. \ref{fig:genie_br_bug}. The correlation is not exact, since there are also other GENIE parameters which affect NC $\Delta\rightarrow N \gamma$ events. We see that the real correlation in blue looks distinctly different from the expected correlation in orange, which comes from the equation $\mathrm{max}(0, 1+0.5\cdot\mathrm{knob})$ which describes how we assign a 50\% fractional uncertainty while forbidding negative event weights. We then noticed that if we square this weight, we get a relationship that matches the real correlation much better, as shown in Fig. \ref{fig:fixed_genie_BR_corr}. This formula can also be used to explain the distribution of average weight values in each universe as shown in Fig. \ref{fig:average_weight_BR}, which shows us that our real uncertainty on NC $\Delta\rightarrow N \gamma$ branching ratio is 106\% rather than the expected 50\%.

We then carefully tested our GENIE uncertainty code to try to understand why this weight which is supposed to assign a 50\% uncertainty to the $\Delta\rightarrow N \gamma$ branching ratio is being squared in our resulting files. This was eventually identified as an issue with the application of the GReWeightResonanceDecay calculator inside code which uses GENIE reweight software. This weight calculator was applied twice, once with the name ``res\_dk'' and once with the name ``hadro\_res\_decay'', leading to a squaring of the weight for these variations. In addition to the BR1gamma knob, this also affects BR1eta which affects branching ratios in $\eta$ events, and Theta\_Delta2Npi, which affects angular distributions of $\Delta\rightarrow N \pi$ decays. Now that this has been understood, we can calculate the bugged weights in order to remove the effect of the $\Delta\rightarrow N \gamma$ branching ratio knob as we wanted. This issue caused an expansion in uncertainties and therefore could be considered conservative, and only caused a serious issue in this analysis since we are manually trying to reverse these weights in order to remove the uncertainty. This code existed in GENIE reweight software, and was then copied to MicroBooNE, SBN, ANNIE, and DUNE software packages. The issue has since been fixed for all these experiments.

One consideration in this analysis is the potential for higher mass resonances than the $\Delta$ to similarly decay to a single photon. These processes are not included in our GENIE neutrino interaction simulation. In a simulation of NC resonant events, 82.6\% excite a $\Delta$ resonance, and 17.4\% of events excite a higher mass resonance. Branching ratios to single photons are taken from Particle Data Group \cite{ParticleDataGroup} in order to estimate that 92\% of NC resonant single photon production comes from $\Delta$ resonances, with 8\% coming from higher mass resonances. A more detailed description of this study can be found in Ref. \cite{kathryn_thesis}. These higher mass resonance decays to photons were studied in the context of the MiniBooNE LEE in Ref. \cite{altarelli_cocktail}. In particular, the next biggest contributing resonances are estimated to be $N^0(1520)\rightarrow n+\gamma$, $N^+(1520)\rightarrow p+\gamma$, $N^0(1535)\rightarrow n+\gamma$, and $N^+(1535)\rightarrow p+\gamma$. These higher mass resonances will have different decay kinematics, for example proton-photon invariant mass distributions, so we expect them to be selected at similar or lower efficiency relative to NC $\Delta\rightarrow N \gamma$. If they are selected with the same efficiency, considering our 4-14\% purities, this 8\% increase in signal prediction corresponds to a maximum of a 1.1\% increase in our total prediction, and therefore is deemed negligible in this analysis.

Another consideration in this analysis is coherent single photon production, which is predicted but not included in our GENIE neutrino interaction simulation. We have recently performed a dedicated search for this process in MicroBooNE, observing events consistent with our predictions \cite{microboone_coherent_photon}. This is expected to be a very rare process, with only around 11 events in our first three years of MicroBooNE data taking. This makes the process approximately ten times rarer than the NC $\Delta\rightarrow N \gamma$ process, and would therefore be expected to contribute only a 1-2\% change in our total prediction, so we consider coherent single photon production negligible in this analysis. 

\begin{figure}[H]
    \centering
    \begin{subfigure}[b]{0.49\textwidth}
        \includegraphics[width=\textwidth]{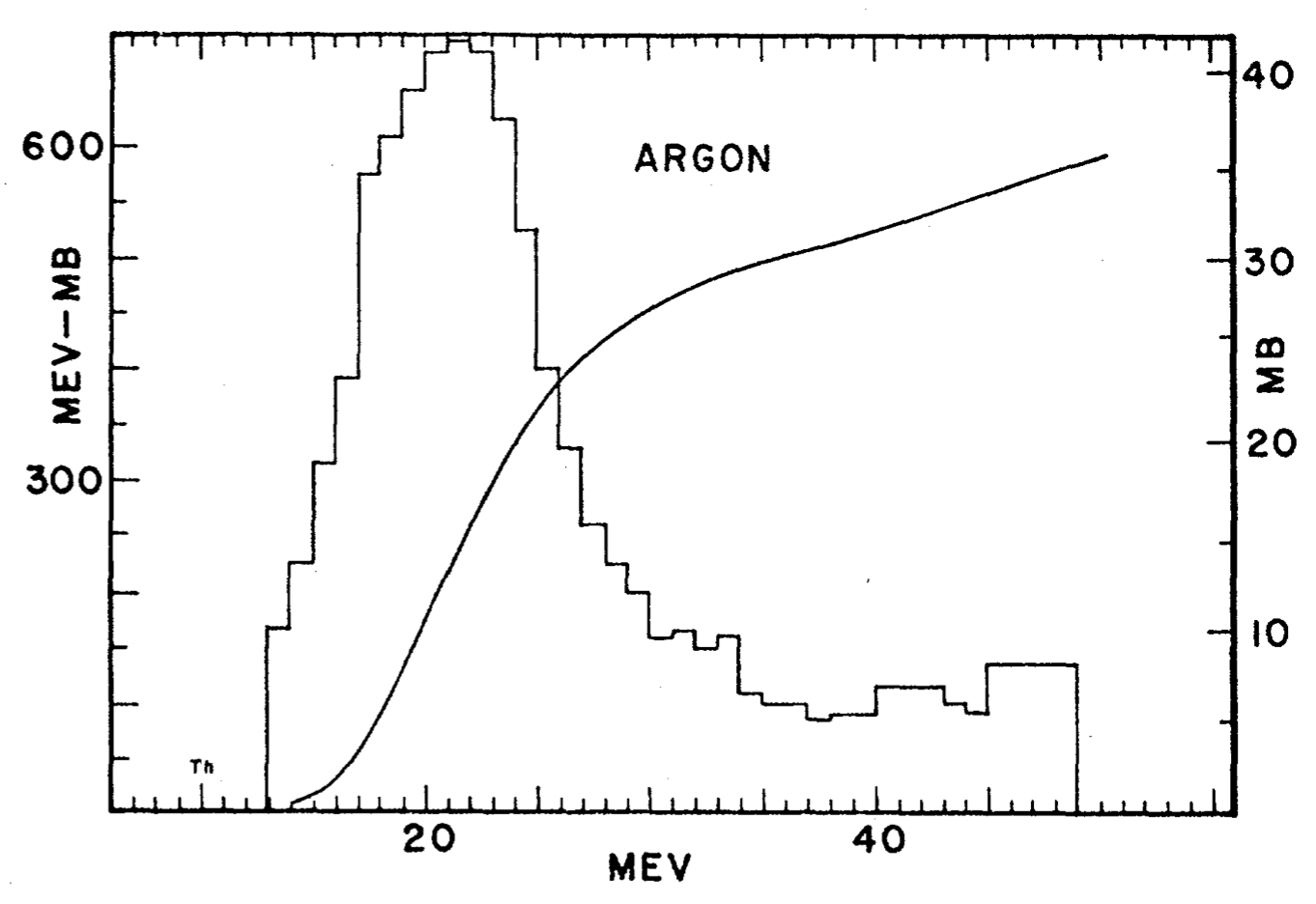}
        \caption{}
        \label{fig:photonuclear_argon}
    \end{subfigure}
    \begin{subfigure}[b]{0.49\textwidth}
        \includegraphics[width=\textwidth]{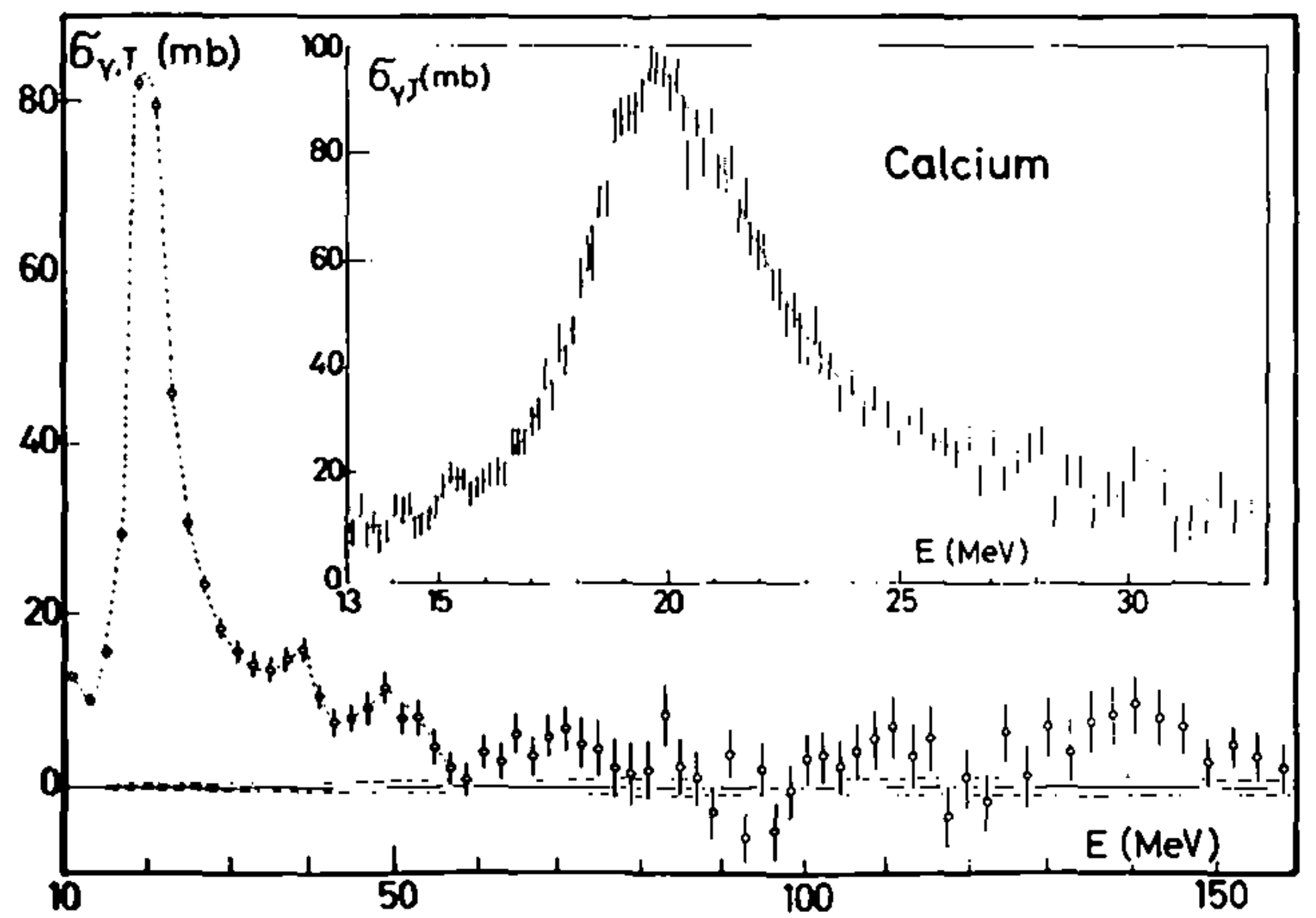}
        \caption{}
        \label{fig:photonuclear_calcium}
    \end{subfigure}
    \begin{subfigure}[b]{0.6\textwidth}
        \includegraphics[width=\textwidth]{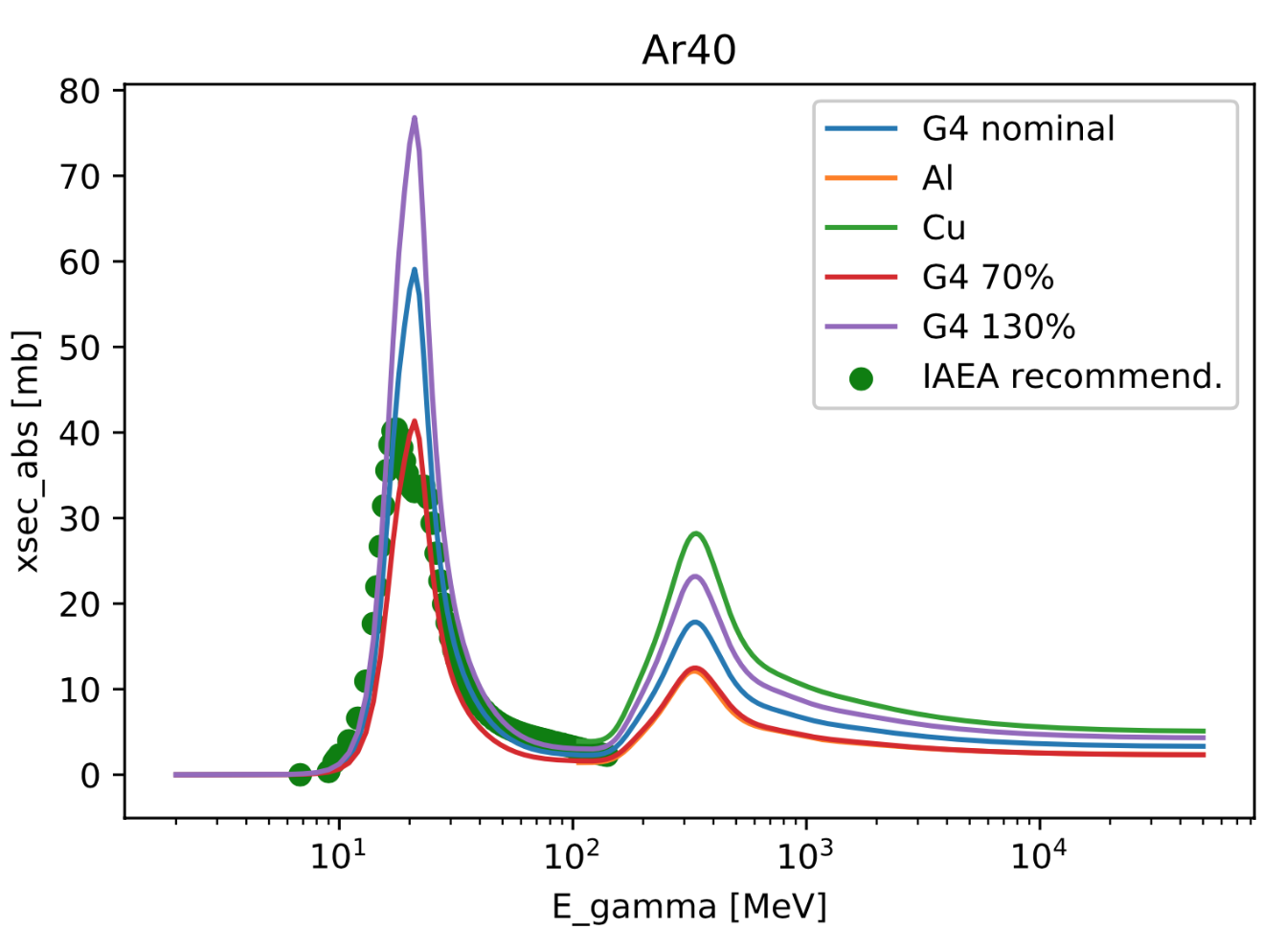}
        \caption{}
        \label{fig:photonuclear_sim}
    \end{subfigure}
    \caption[Photonuclear absorption in argon]{Photonuclear absorption in argon. Panel (a) shows a measurement of the photonuclear absorption cross section on argon, measured by photon-induced neutron emission, from Ref. \cite{photonuclear_argon}. Panel (b) shows a measurement of the photonuclear absorption cross section on calcium measured via attenuation of photons, from Ref. \cite{photonuclear_argon}. Panel (c) shows our modeling of the photonuclear absorption cross section on argon, from Ref. \cite{kathryn_thesis}.}
    \label{fig:photonuclear_absorption}
\end{figure}

There are several ways that NC $\pi^0$ photon pairs can resemble one photon in our detector, but there is one that we have not discussed in previous sections. Photonuclear absorption can occur, where a photon interacts with an argon nucleus and is absorbed before pair producing and creating a visible electromagnetic shower. As shown in Figs. \ref{fig:photonuclear_argon}-\ref{fig:photonuclear_calcium}, this process has been measured on argon via neutron emission \cite{photonuclear_argon} and on calcium via photon absorption \cite{photonuclear_calcium}, which is a nucleus with the same number of nucleons (40) as argon. There are two notable peaks in the photonuclear absorption cross section as a function of photon energy, one at lower energies corresponding to the giant dipole resonance, and one at higher energies corresponding to the Delta resonance, as shown in Fig. \ref{fig:photonuclear_sim}. This process is simulated in our detector, but there is no systematic uncertainty assigned to it. Our simulated cross section is higher than that recommended by the International Atomic Energy Agency \cite{iaea_photonuclear_data}, and from this we estimate that we could assign a 30\% systematic uncertainty to this process. In a simulation, it was found that 0.7\% of NC $\pi^0$ events contain a photon which undergoes photonuclear absorption. In our $6.37\cdot10^{20}$ POT of BNB data, we expect about 24,410 NC $\pi^0$ events before any selection, and therefore expect about 171 photonuclear absorption events. If we assume that these single photon events are selected with the same efficiency as NC $\Delta\rightarrow N \gamma$ events, this would correspond to about 36 expected photonuclear events across all four NC $\Delta\rightarrow N \gamma$ selections. If we then assume a 30\% systematic uncertainty to this process, this corresponds to an additional uncertainty of about 10 events spread across all four selections. Given the existing statistical and systematic uncertainties on these predictions, this was deemed negligible for this analysis. However, more detailed investigations of this process and the potentially visible low energy nuclear activity produced by an absorbed photon would be interesting in the future.

\begin{figure}[H]
    \centering
    \includegraphics[trim=0 0 0 10, clip, width=0.45\textwidth]{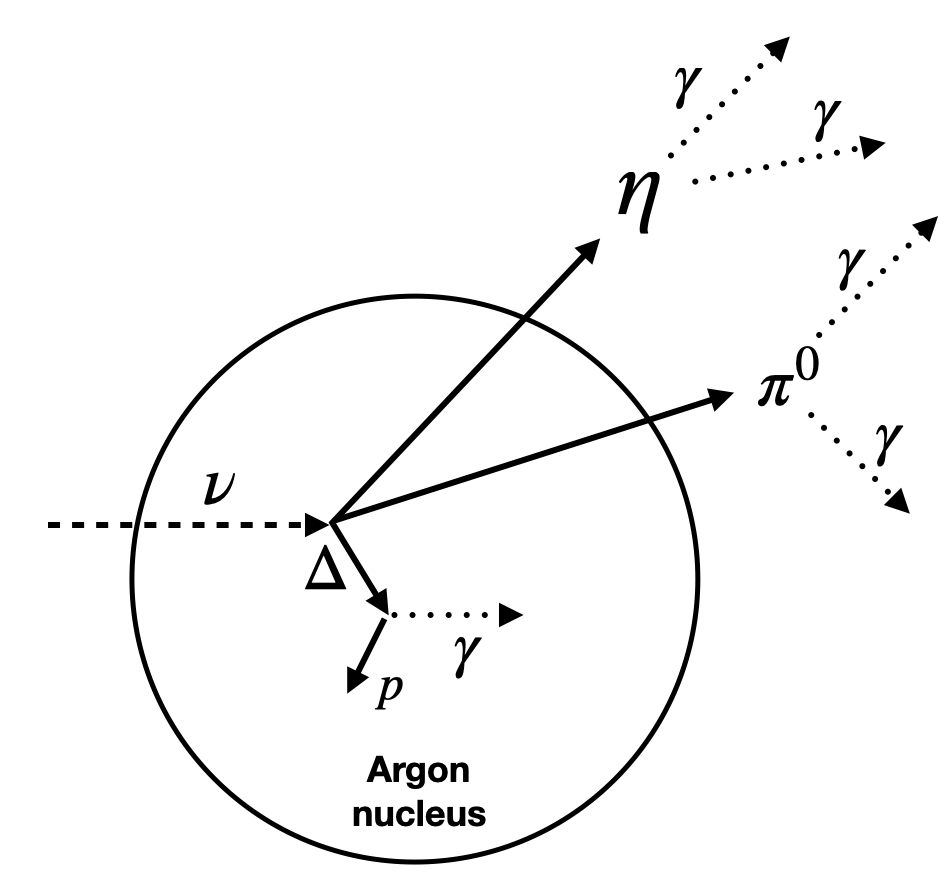}
    \caption[Photon FSI diagram]{A diagram illustrating the fact that $\pi^0$ and $\eta$ particles will produce photons outside of the nuclear medium, while $\Delta$ particles will produce photons inside of the nucleus, and therefore those photons could undergo final state interactions with the nuclear medium.}
    \label{fig:photon_fsi_diagram}
\end{figure}

The last type of systematic effect that is especially unique to this analysis is photon final state interactions (FSI). FSI refers to particles which are produced inside the nucleus, and can interact with the nuclear medium before escaping. Almost unique to this NC $\Delta\rightarrow N \gamma$ process is the potential for photonuclear absorption of photons. This is because the $\Delta$ has such a short lifetime that it will often decay and produce a photon inside the nucleus, unlike $\pi^0$ particles. We can calculate the mean propagation length $L$ as a function of the lifetime $\tau$ and the kinetic energy $E$: $L= v t = \beta c \gamma \tau = c \tau \sqrt{\gamma^2-1} = c\tau \sqrt{\left(\frac{E}{m c^2} + 1\right)^2-1}$. For illustration, this is shown for a 200 MeV particle in Table \ref{tab:photon_producing_prop_lengths}. We see that for $\pi^0$ and $\eta$ particles which produce the vast majority of two-photon events, the expected propagation length is far greater than the size of the argon nucleus, while the $\Delta$ particle is expected to decay before leaving the argon nucleus, as diagrammed in Fig. \ref{fig:photon_fsi_diagram}. This means that final state interactions between $\mathcal{O}$(100 MeV) photons and the nuclear medium will uniquely affect $\Delta\rightarrow N \gamma$ single photon events, and will not affect $\pi^0$ two-photon events.

\begin{table}[H]
    \centering
    \small
    \begin{tabular}{ccccc}
        \toprule
        \makecell{Particle} & \makecell{Mass} & \makecell{Lifetime} & \multicolumn{2}{c}{Propagation Length}\\
        \midrule
        $\pi^0$ & 135 MeV & $8.5\cdot 10^{-17}$ s & $5.8\cdot10^{-8}$ m & $1.7\cdot10^{7}$ Ar radii\\
        $\eta$ & 548 MeV & $5.0\cdot 10^{-19}$ s & $1.4\cdot10^{-10}$ & $4.1\cdot10^{4}$ Ar radii\\
        $\Delta$ & 1232 MeV & $5.6\cdot 10^{-24}$ s & $9.9\cdot10^{-16}$ & 0.29 Ar radii\\
        \bottomrule
    \end{tabular}
    \caption[Photon-producing particle expected propagation length before decay]{Table of expected propagation lengths before decay for 200 MeV kinetic energy photon-producing particles. For the argon nuclear radius, we use $3.4274\cdot 10^{-15}$ m from Ref. \cite{nuclear_radii}.}
    \label{tab:photon_producing_prop_lengths}
\end{table}

\begin{figure}[H]
    \centering
    \begin{subfigure}[b]{0.49\textwidth}
        \includegraphics[width=\textwidth]{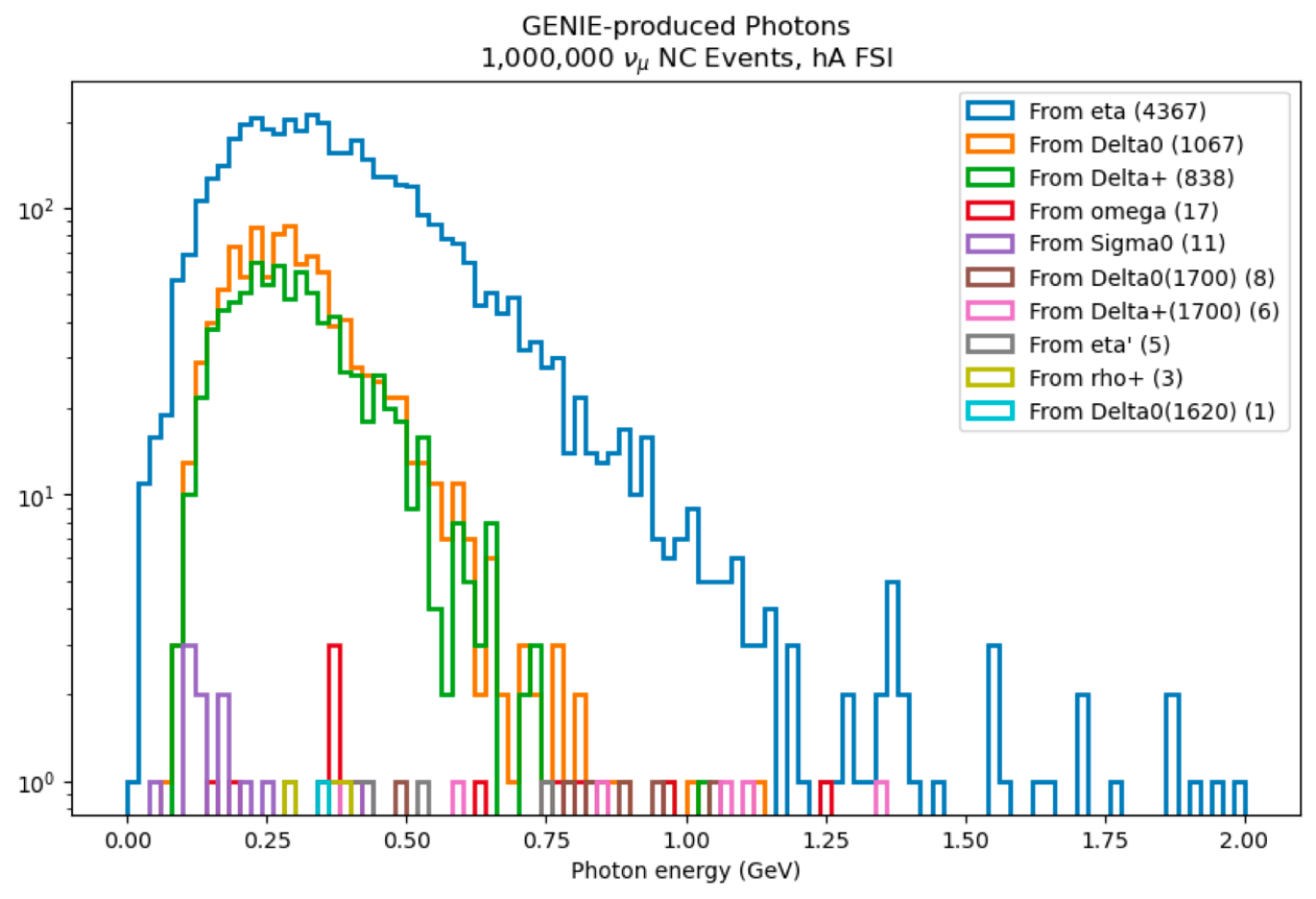}
        \caption{}
    \end{subfigure}
    \begin{subfigure}[b]{0.49\textwidth}
        \includegraphics[width=\textwidth]{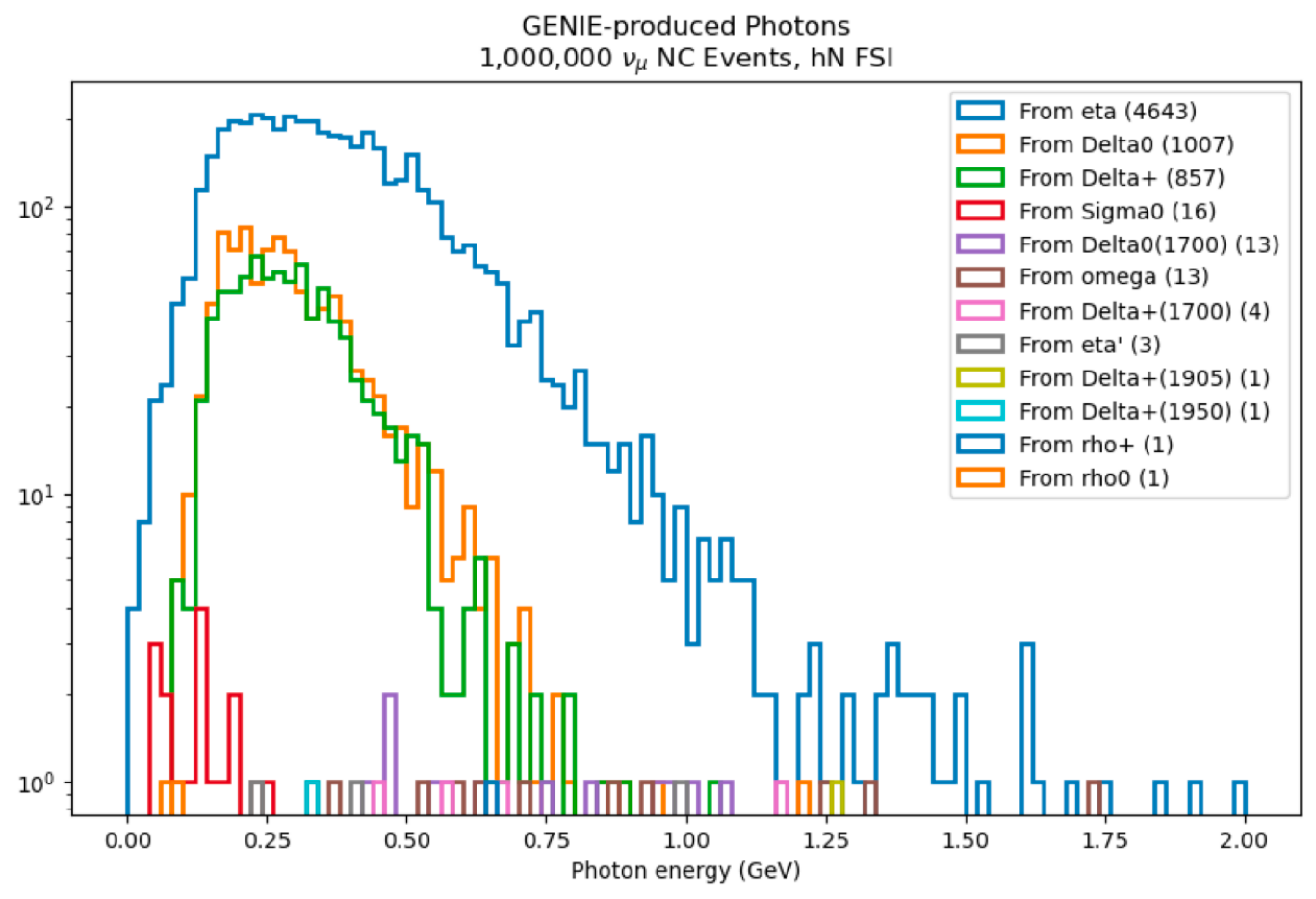}
        \caption{}
    \end{subfigure}
    \begin{subfigure}[b]{0.49\textwidth}
        \includegraphics[width=\textwidth]{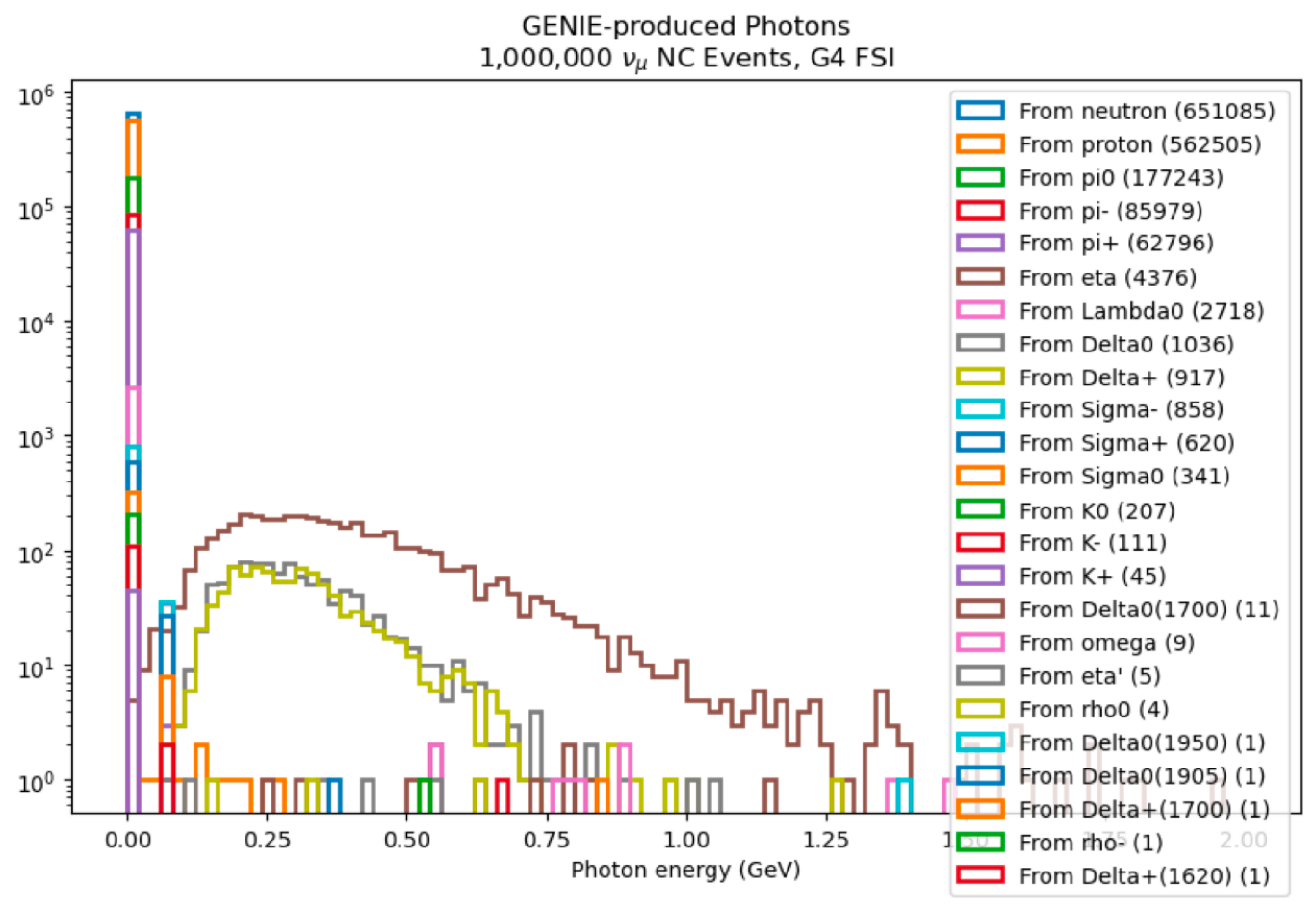}
        \caption{}
    \end{subfigure}
    \begin{subfigure}[b]{0.49\textwidth}
        \includegraphics[width=\textwidth]{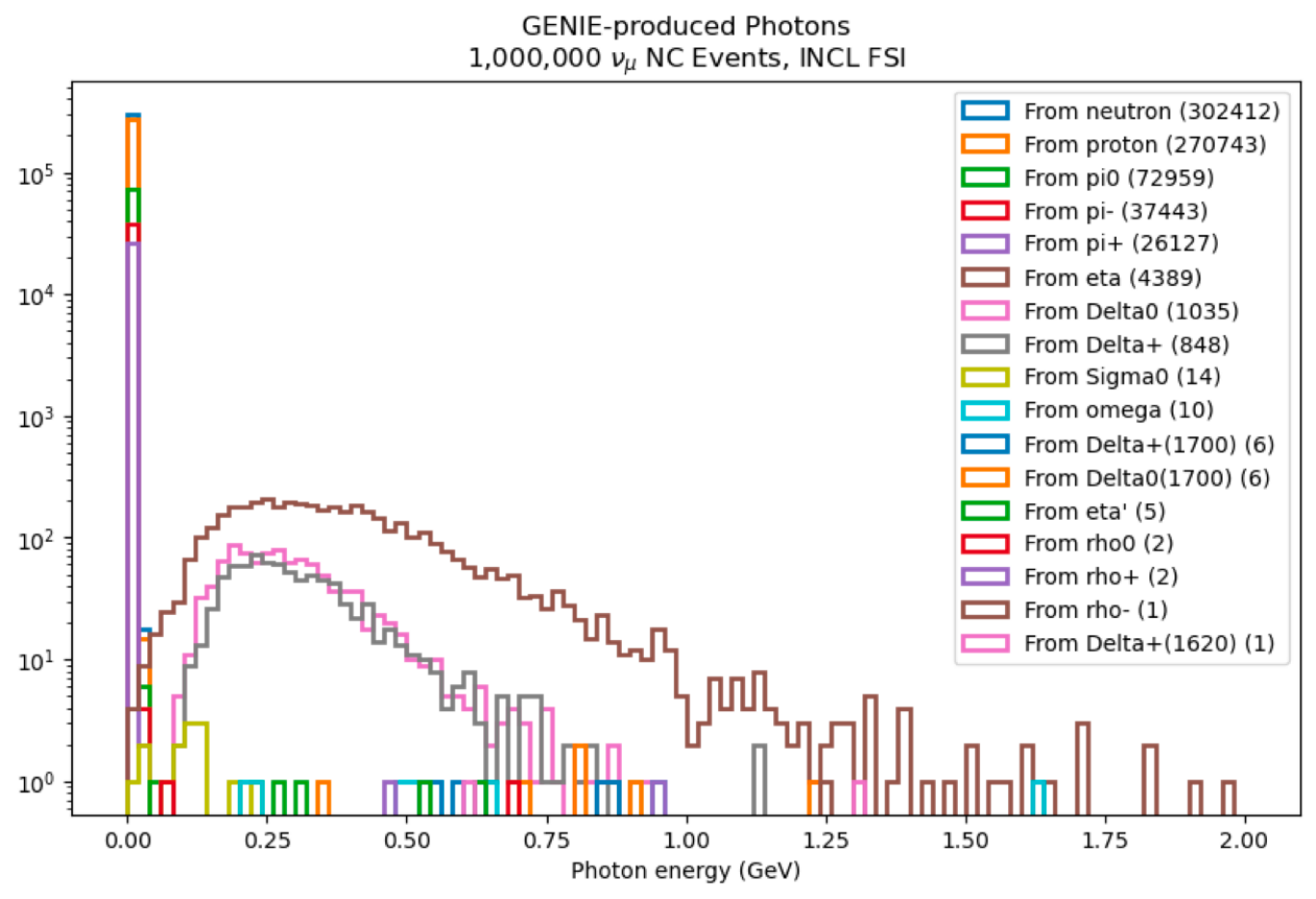}
        \caption{}
    \end{subfigure}
    \caption[Photons from different GENIE FSI models]{Photon energies and parent particles from 1,000,000 $\nu_\mu$ NC events simulated using different GENIE final state interaction models. Panel (a) shows the hA model, panel (b) shows the hN model, panel (c) shows the G4 model, and panel (d) shows the INCL model.}
    \label{fig:genie_FSI}
\end{figure}

Our nominal FSI model is hA2018, which only considers pion and nucleon final state interactions. There are three alternative FSI models available in GENIE. The hN model includes interactions of pions, nucleons, kaons, and photons with nuclei \cite{Dytman_hA_hN_FSI}. The INCL model uses the Li\`ege intranuclear cascade \cite{INCL_FSI}. The G4 model uses a \textsc{Geant4} Bertini cascade, which includes photonuclear interactions \cite{G4_FSI}. A detailed study of the exact physics models implemented in all of these models is beyond the scope of this work, but we check the distributions of resulting final state photons simulated from NC events and each of these physics models, as shown in Fig. \ref{fig:genie_FSI}. We see that in each case, the majority of the photons comes from $\eta$ particles. This is because GENIE simulates $\eta\rightarrow \gamma\gamma$ but not $\pi^0\rightarrow \gamma\gamma$, which is left to be decayed by \textsc{Geant4} in a later stage of our detector simulation. We can also notice that the G4 and INCL models each produce a lot of low energy photons from nuclear de-excitations, which are not considered in the hA and hN models.


\begin{figure}[H]
    \centering
    \begin{subfigure}[b]{0.52\textwidth}
        \includegraphics[width=\textwidth]{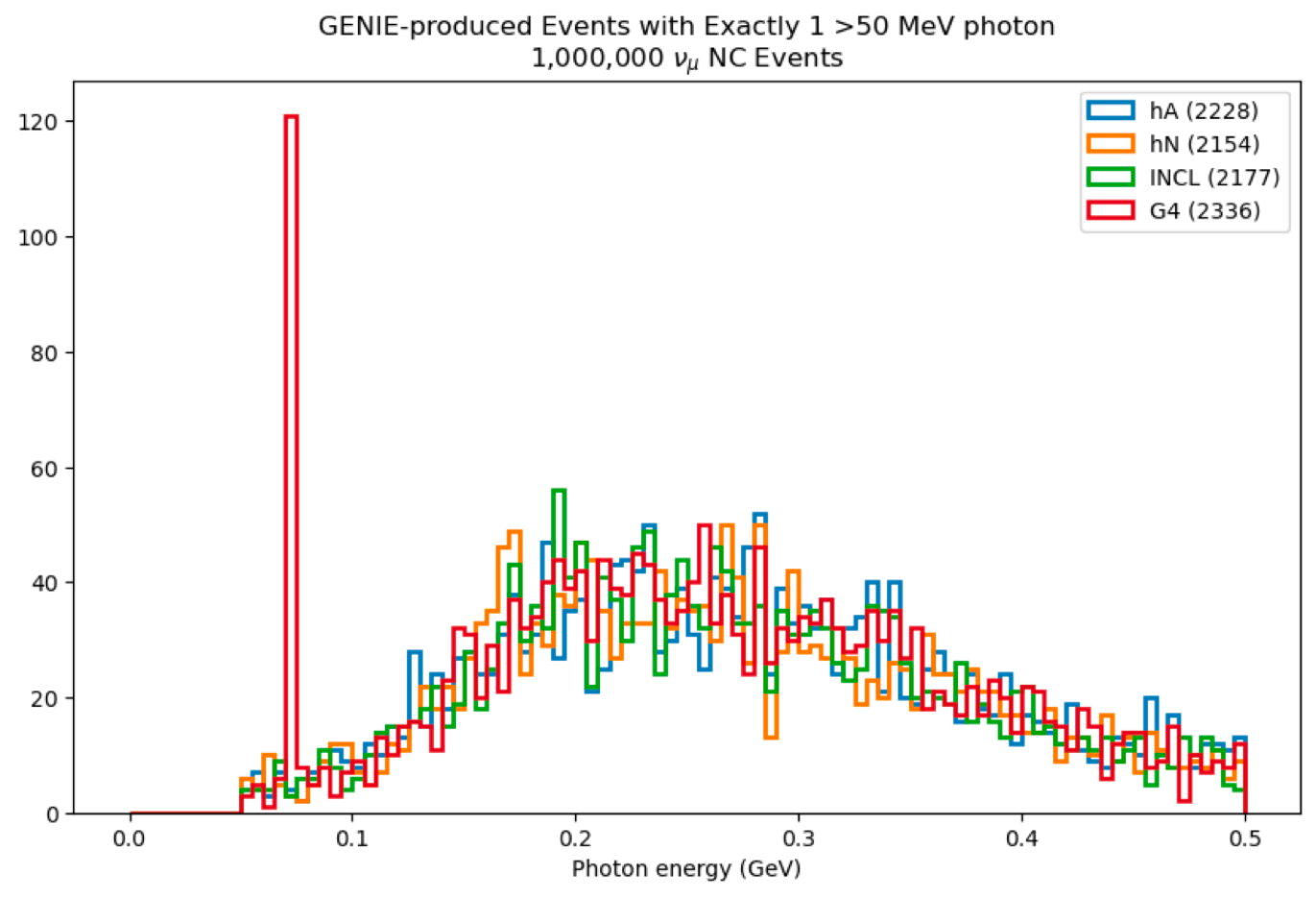}
        \caption{}
        \label{fig:weird_g4_photon_FSI}
    \end{subfigure}
    \begin{subfigure}[b]{0.49\textwidth}
        \includegraphics[width=\textwidth]{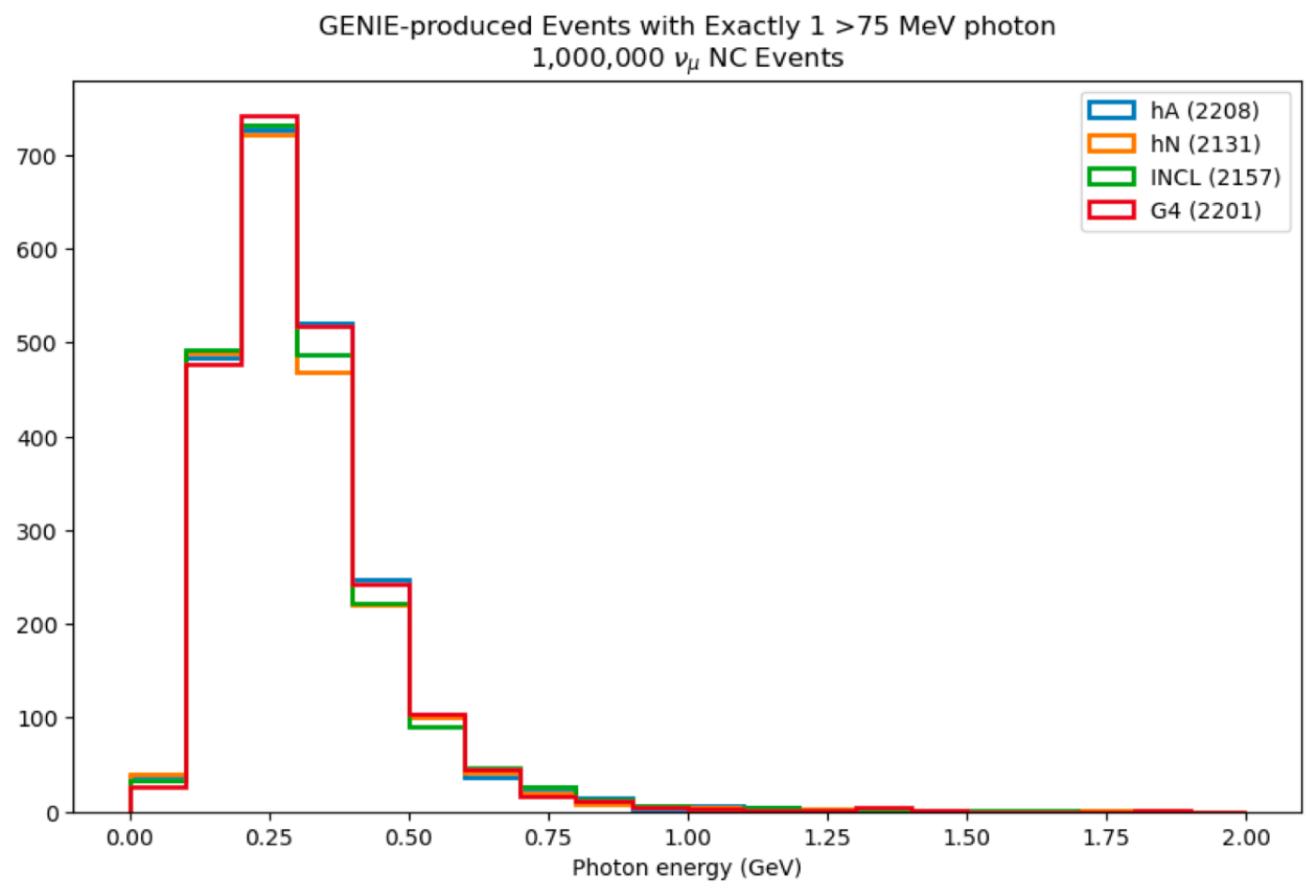}
        \caption{}
        \label{fig:good_photon_fsi_spectrum}
    \end{subfigure}
    \begin{subfigure}[b]{0.49\textwidth}
        \includegraphics[width=\textwidth]{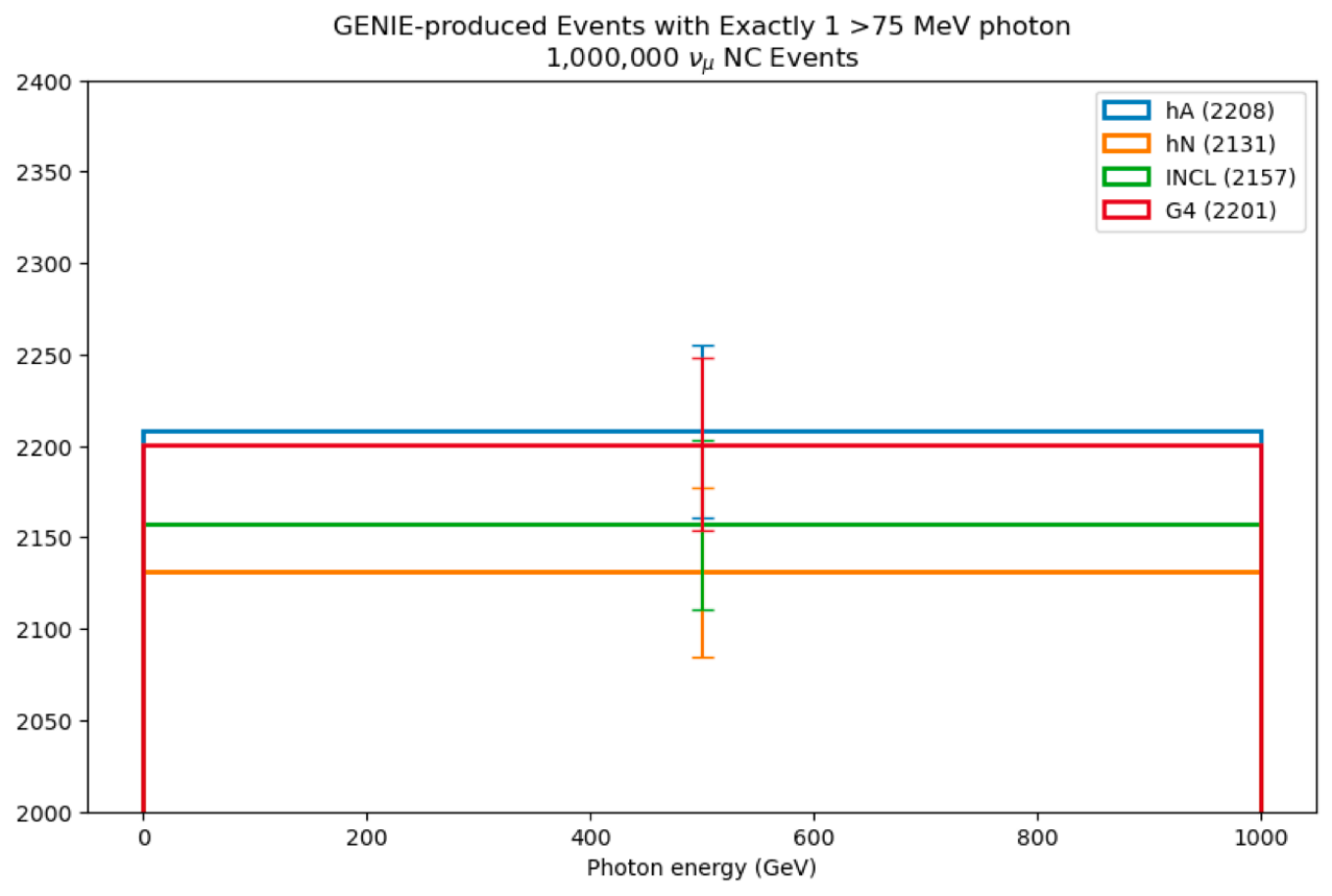}
        \caption{}
        \label{fig:good_photon_fsi_one_bin}
    \end{subfigure}
    \caption[Single photons from different GENIE FSI models]{Single photons from different GENIE FSI models. Panel (a) shows the energy spectrum for events with one photon greater than 50 MeV, showing a notable feature in the G4 model at 74.0 MeV. Panel (b) shows the energy spectrum for events with one photon greater than 75 MeV. Panel (c) shows one bin for events with one photon greater than 75 MeV, zoomed in on the peak to show that the different generators agree within statistical uncertainties.}
    \label{fig:single_photon_FSI}
\end{figure}

Most of these photons are produced in pairs from $\eta$ decays or at very low energies, so next we specifically investigate events with exactly one photon above a 50 MeV energy threshold, as shown in Fig. \ref{fig:weird_g4_photon_FSI}. We see one notable feature, a large spike from the G4 model for single photons at exactly 74.0 MeV. Further investigation showed that this comes specifically from strange baryon decays. This is an unphysical feature that should be ignored, so next we raise our energy threshold and investigate events with exactly one photon above a 75 MeV energy threshold, as shown in Figs. \ref{fig:good_photon_fsi_spectrum}-\ref{fig:good_photon_fsi_one_bin}. Here, we see very good agreement between all four models, with are each agreeing within statistical uncertainties after simulating 1,000,000 NC events. If we assume this difference is not statistical in order to estimate an upper bound on the uncertainty related to final state interaction modeling for single photon events, we get an upper bound of 3.5\% on our signal NC $\Delta\rightarrow N \gamma$ events. Considering our maximum purity of about 15\%, this corresponds to an upper bound of 0.5\% uncertainty on our final prediction, and we deem this negligible for this analysis. Further investigation of the precise photon interaction physics implemented in these FSI models is left for future studies.

\begin{figure}[H]
    \centering
    \includegraphics[width=0.6\textwidth]{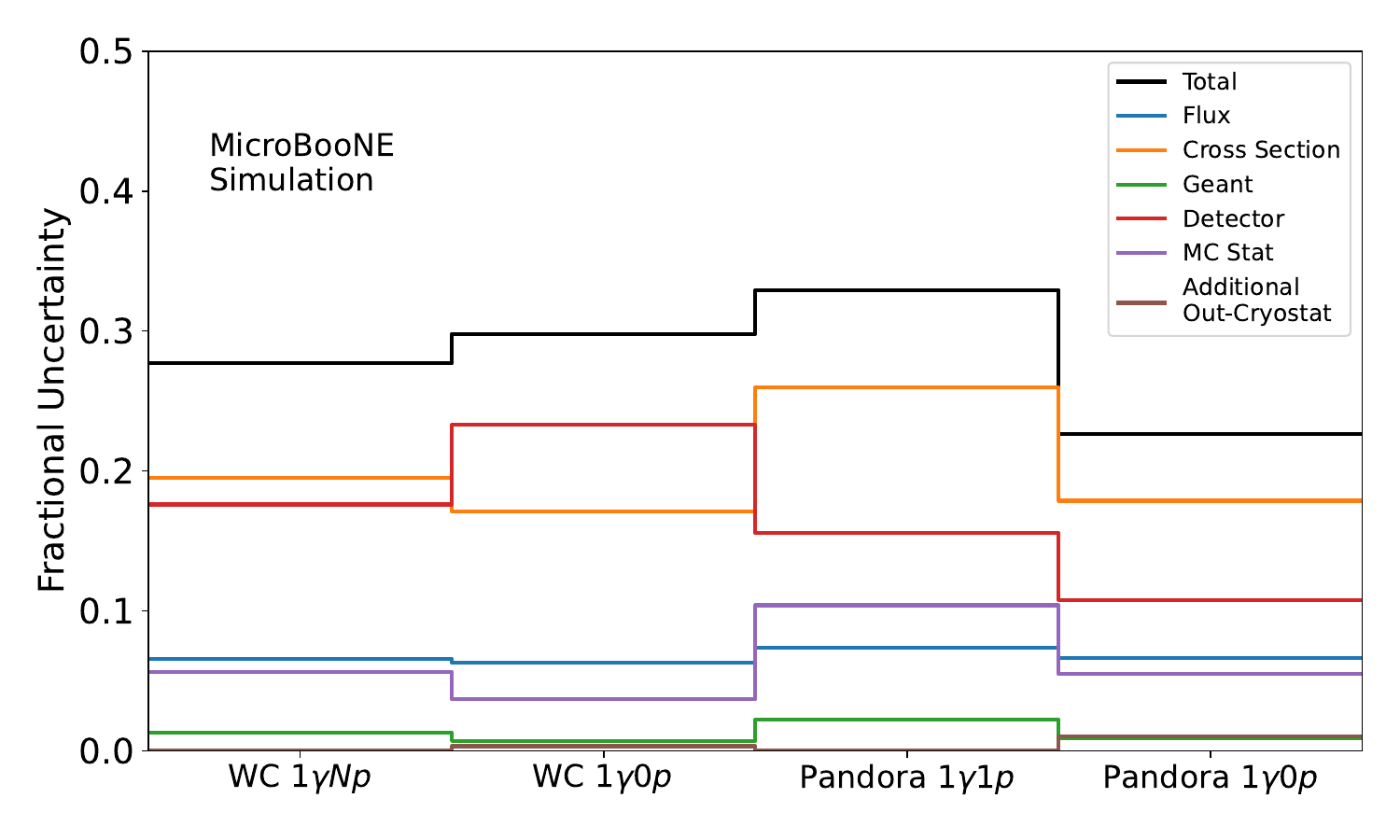}
    \caption[NC $\Delta\rightarrow N \gamma$ systematic uncertainty breakdown]{NC $\Delta\rightarrow N \gamma$ systematic uncertainty breakdown.}
    \label{fig:nc_delta_err_breakdown}
\end{figure}

The total systematic uncertainty from all sources is shown in Table \ref{tab:sys_breakdown} and Fig. \ref{fig:nc_delta_err_breakdown}. The biggest sources of systematic uncertainty are from GENIE neutrino-argon interaction cross section modeling, and detector response modeling.  

\begin{table}[H]
    \centering
    \small
    \begin{tabular}{c c c c c} 
        \toprule
        Uncertainty Type & \makecell{WC $1\gamma N p$} & \makecell{WC $1\gamma0p$} & \makecell{Pandora $1\gamma 1 p$} & \makecell{Pandora $1\gamma0p$} \\
        \midrule
        Flux model & 6.58\% & 6.29\% & 7.39\% & 6.66\% \\
        GENIE cross section & 19.49\% & 17.09\% & 25.96\% & 17.87\% \\
        Hadron re-interaction & 1.27\% & 0.70\% & 2.22\% & 0.89\% \\
        Detector modeling & 17.58\% & 23.35\% & 15.69\% & 10.96\% \\
        \makecell{Monte Carlo statistics} & 5.64\% & 3.67\% & 10.40\% & 5.47\% \\
        \makecell{Out-of-cryostat interactions} & 0.00\% & 0.33\% & 0.00\% & 1.02\% \\
        \midrule
        \makecell{Total uncertainty} & 27.65\% & 29.85\% & 32.94\% & 22.61\% \\
        \bottomrule
    \end{tabular}
    \caption[NC $\Delta\rightarrow N \gamma$ signal channel systematic uncertainty breakdown]{NC $\Delta\rightarrow N \gamma$ Signal channel systematic uncertainty breakdown. No conditional constraint has been applied.}
    \label{tab:sys_breakdown}
\end{table}

\begin{figure}[H]
    \centering
    \includegraphics[width=0.6\textwidth]{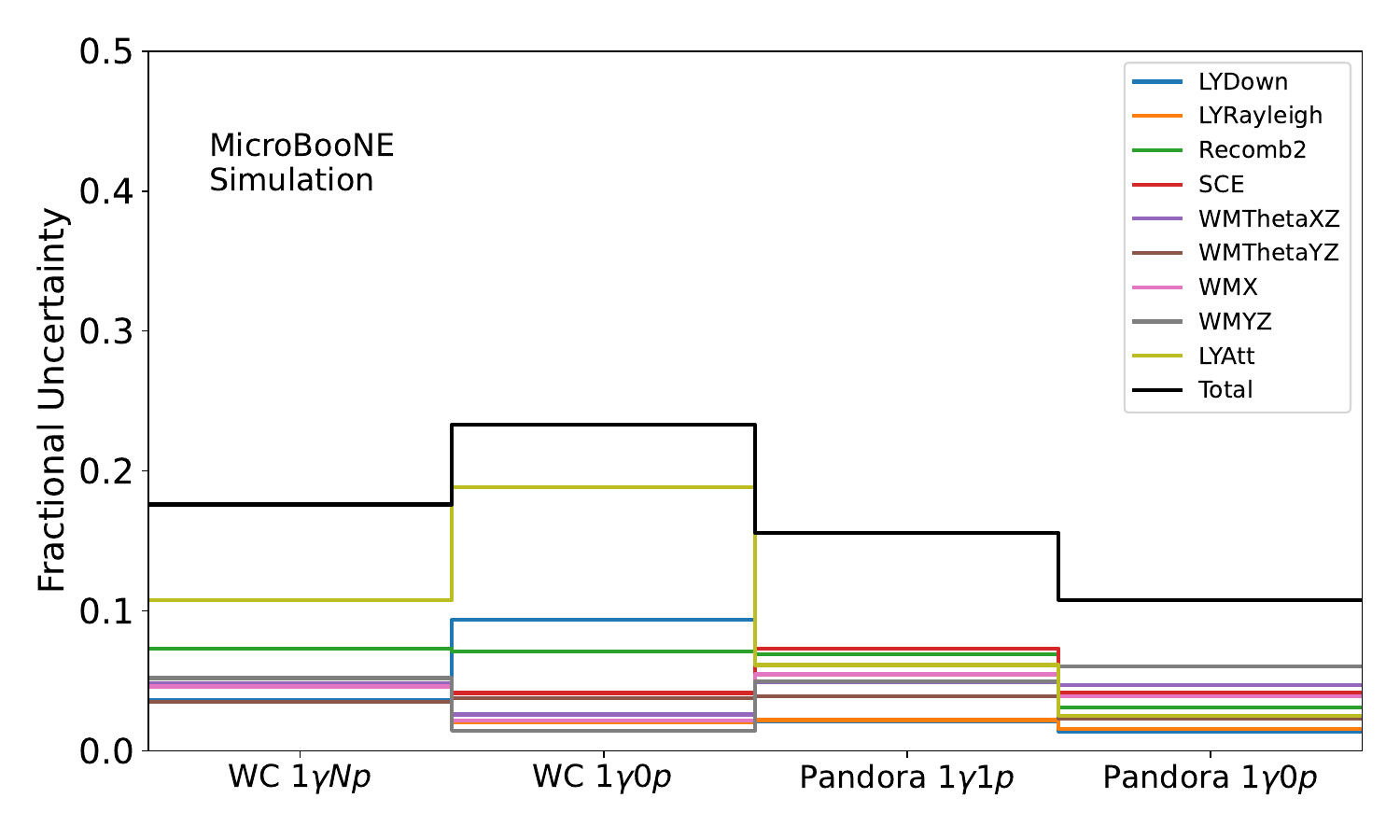}
    \caption[NC $\Delta\rightarrow N \gamma$ detector systematic uncertainty breakdown]{NC $\Delta\rightarrow N \gamma$ detector response systematic uncertainty breakdown.}
    \label{fig:nc_delta_detvar_breakdown}
\end{figure}

Within the detector response uncertainty, the biggest contributor is the light yield attenuation uncertainty, particularly in the Wire-Cell selections, as shown in Fig. \ref{fig:nc_delta_detvar_breakdown}. Evidently, this uncertainty in our reconstructed PMT light patterns affects the performance of the Wire-Cell charge-light matching step, which is particularly important since mis-clustering an event with two disconnected photons and only reconstructing one photon is an important type of background.

In order to constrain our systematic uncertainties, we use NC $\pi^0$ and $\nu_\mu$CC observations, both with and without reconstructed protons. Our NC $\pi^0$ constraining sidebands are updated relative to those described in Sec. \ref{sec:conditional_constraint}. A new selection was developed using a BDT trained on Wire-Cell variables, similarly to the NC $\Delta\rightarrow N \gamma$ selection described in Sec. \ref{sec:wc_nc_delta_selection}. This new NC $\pi^0$ selection was also developed in order to perform cross section measurements, and is described in more detail in Ref. \cite{wc_ncpi0_xs}. The $\nu_\mu$CC selection is the same described in Sec. \ref{sec:bdt_selections}. Note that these constraining sidebands are not the same as the Pandora NC $\pi^0$ selection to those used in the older Pandora NC $\Delta\rightarrow N \gamma$ search \cite{glee_prl}, and therefore we do expect some small differences in the effect of the conditional constraint on the Pandora signal channels relative to that result.

\begin{figure}[H]
    \centering
    \includegraphics[width=0.7\textwidth]{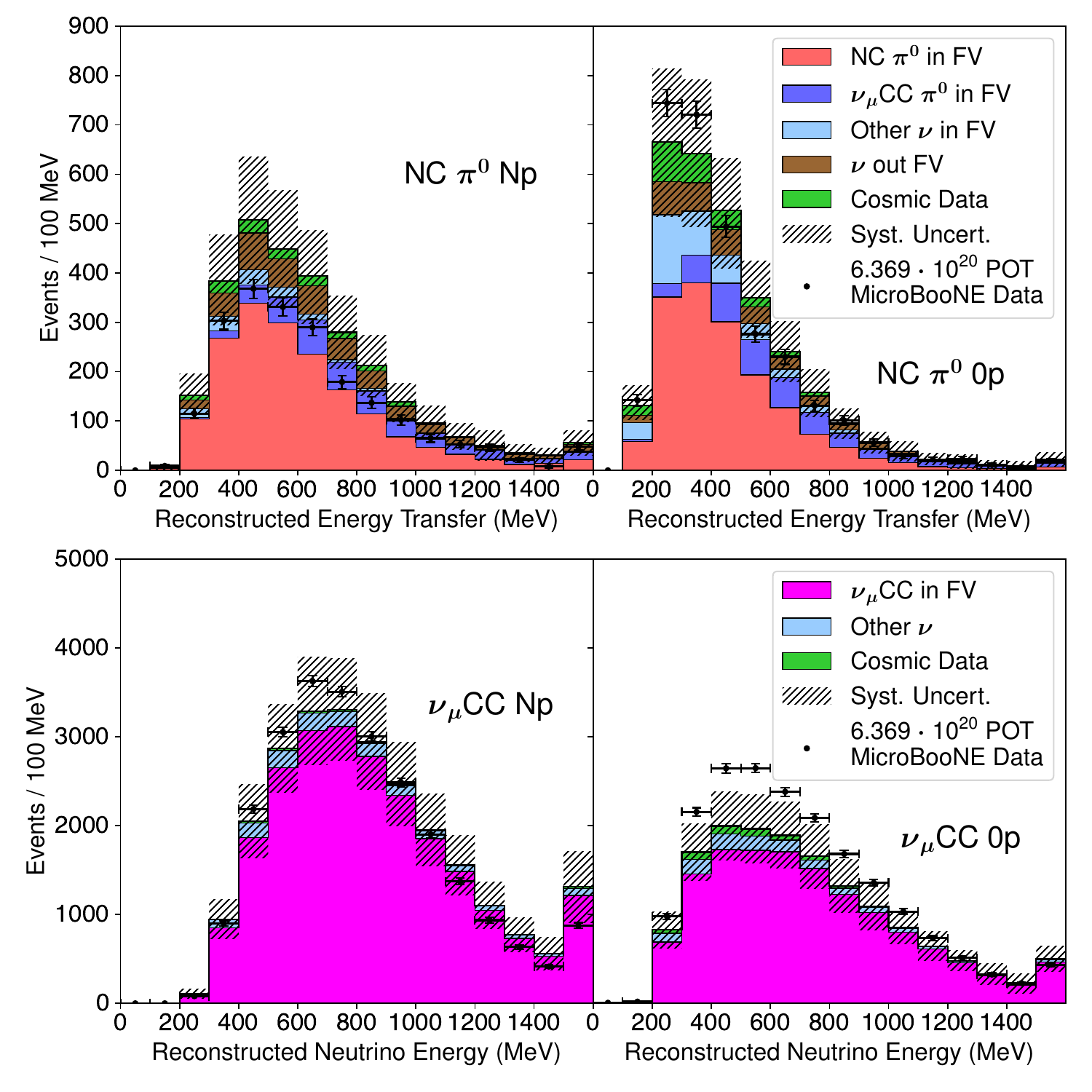}
    \caption[NC $\Delta\rightarrow N \gamma$ constraining channels]{NC $\Delta\rightarrow N \gamma$ constraining channels.}
    \label{fig:nc_delta_constraining_channels}
\end{figure}

We bin each NC $\pi^0$ selection in reconstructed energy transfer, which is constructed the same way as our reconstructed neutrino energy, but acknowledging that we do not reconstruct any energy from the exiting neutrino. These distributions are shown in Fig. \ref{fig:nc_delta_constraining_channels}, where we see a slight overprediction of $Np$ NC $\pi^0$ events, a slight underprediction of low energy $0p$ NC $\pi^0$ events, and a slight underprediction of $\nu_\mu$CC $0p$ events. Overall, we see good agreement within systematic uncertainties, with a joint $\chi^2$/ndf value for all four channels of $46.9/64$, corresponding to $0.07 \sigma$.

\begin{figure}[H]
    \centering
    \includegraphics[width=0.75\textwidth]{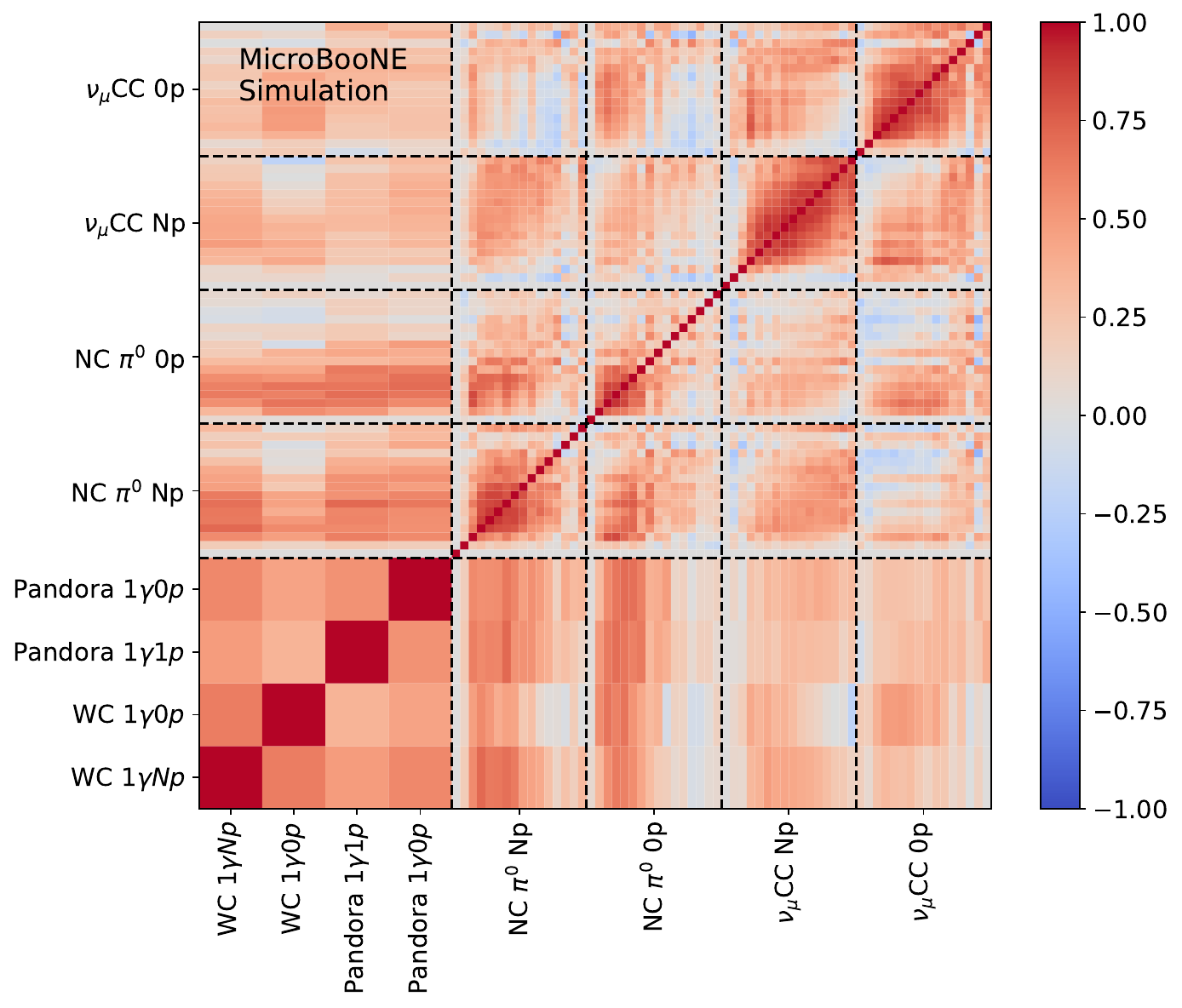}
    \caption[NC $\Delta\rightarrow N \gamma$ correlation matrix]{Correlation matrix between NC $\Delta\rightarrow N \gamma$ signal channels and constraining sideband channels. This correlation matrix $R$ is constructed from the covariance matrix $\Sigma$ by $\Sigma_{ij}/\sqrt{\Sigma_{ii}\Sigma_{jj}}$.}
    \label{fig:correlation_matrix}
\end{figure}

These uncertainties are highly correlated with our predictions in the signal channels, as shown in Fig. \ref{fig:correlation_matrix}. The strongest correlation is with the NC $\pi^0$ sidebands, since this is the same process making up the majority of the background prediction of our signal channels, and therefore shares common flux, neutrino-argon interaction cross section, and detector response uncertainties. The NC $\pi^0$ sidebands are also closely related to the signal NC $\Delta\rightarrow N \gamma$ events, since most NC $\pi^0$ events come from an NC interaction exciting a $\Delta$ resonance which decays to a pion and a nucleon. The $\nu_\mu$CC channels are also correlated due to common flux uncertainties as well as cross section uncertainties.

\begin{figure}[H]
    \centering
    \includegraphics[width=0.6\textwidth]{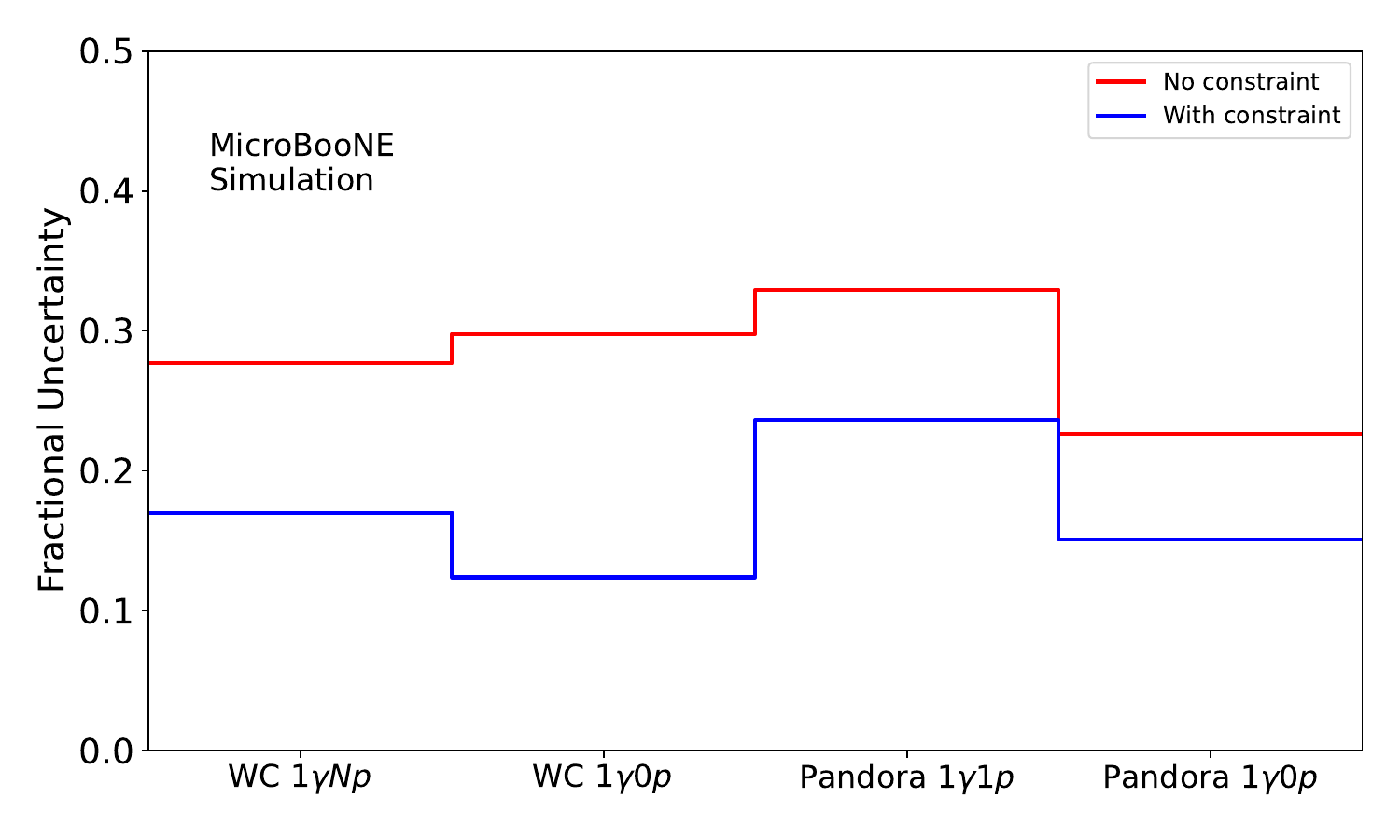}
    \caption[NC $\Delta\rightarrow N \gamma$ constrained systematic uncertainties]{NC $\Delta\rightarrow N \gamma$ constrained systematic uncertainties.}
    \label{fig:constr_errs}
\end{figure}

The resulting reduction in systematic uncertainties is shown in Table \ref{tab:sys_constraint} and Fig. \ref{fig:constr_errs}. Note that the effect of the constraint is more significant in the Wire-Cell channels than in the Pandora channels; this could be due to the fact that our constraining channels use Wire-Cell reconstruction, and therefore detector response uncertainties are more highly correlated with Wire-Cell signal channels.

\begin{table}[H]
    \centering
    \small
    \begin{tabular}{c c c c c} 
        \toprule
        Uncertainty Type & \makecell{WC $1\gamma N p$} & \makecell{WC $1\gamma0p$} & \makecell{Pandora $1\gamma 1 p$} & \makecell{Pandora $1\gamma0p$} \\
        \midrule
        \makecell{Total uncertainty (unconstrained)} & 27.65\% & 29.85\% & 32.94\% & 22.61\% \\
        \makecell{Total uncertainty (constrained)} & 16.80\% & 12.39\% & 23.96\% & 15.02\% \\
        \bottomrule
    \end{tabular}
    \caption[NC $\Delta\rightarrow N \gamma$ signal channel systematic uncertainty before and after constraint]{NC $\Delta\rightarrow N \gamma$ Signal channel systematic uncertainty before and after the application of the conditional constraint.}
    \label{tab:sys_constraint}
\end{table}

\section{Joint Wire-Cell+Pandora NC \texorpdfstring{$\Delta$}{Delta} Radiative Decay Results}

\begin{figure}[H]
    \centering
    \begin{subfigure}[b]{0.99\textwidth}
        \includegraphics[width=\textwidth]{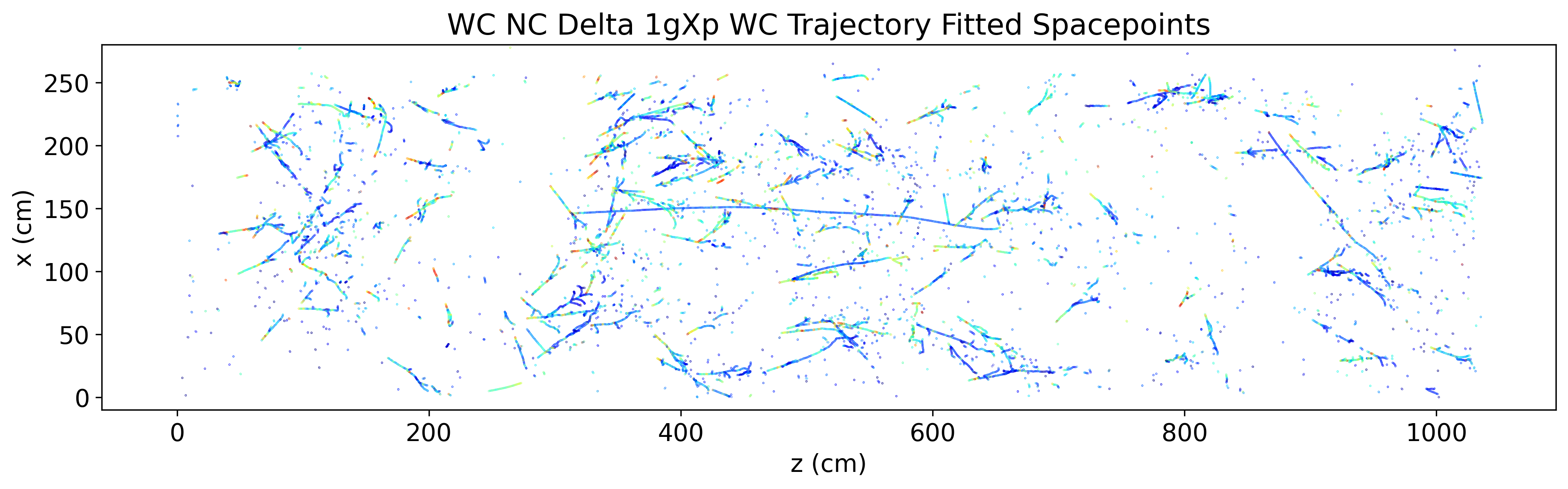}
        \caption{}
    \end{subfigure}
    \begin{subfigure}[b]{0.99\textwidth}
        \includegraphics[width=\textwidth]{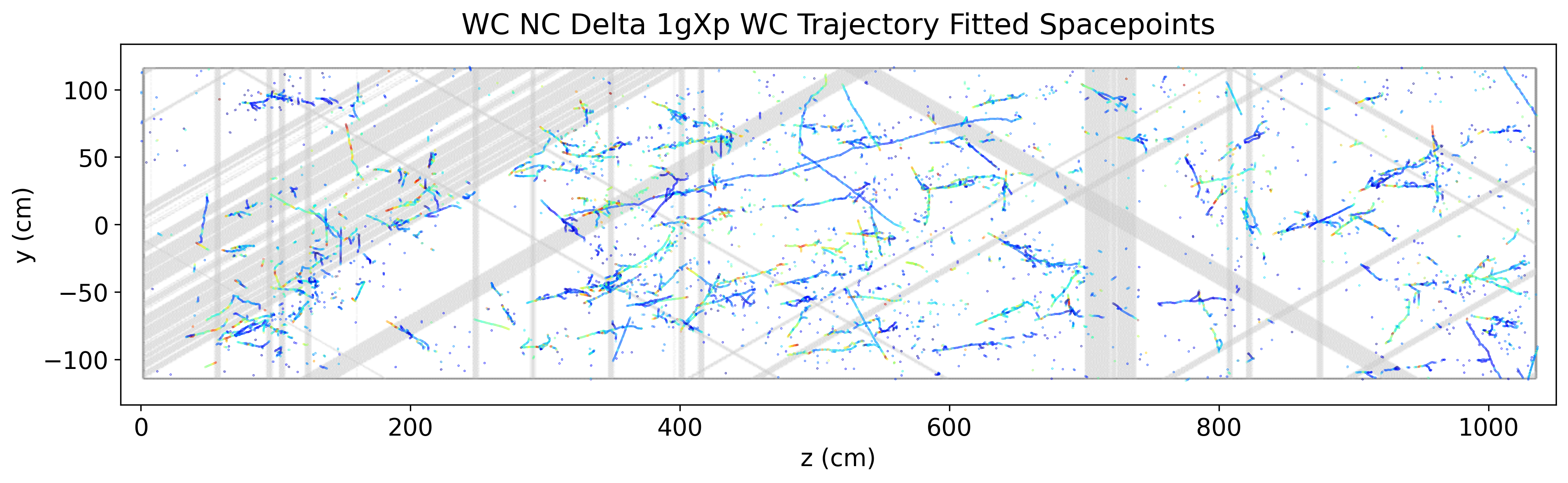}
        \caption{}
    \end{subfigure}
    \caption[Wire-Cell NC $\Delta\rightarrow N \gamma$ overlaid event displays]{Wire-Cell NC $\Delta\rightarrow N \gamma$ overlaid event displays. The color corresponds to the amount of ionization energy per unit length. Panel (a) shows the x-z view, and Panel (b) shows the y-z view, with dead wires indicated by gray shading.}
    \label{fig:wc_nc_delta_overlaid_event_displays}
\end{figure}

\begin{figure}[H]
    \centering
    \begin{subfigure}[b]{0.99\textwidth}
        \includegraphics[width=\textwidth]{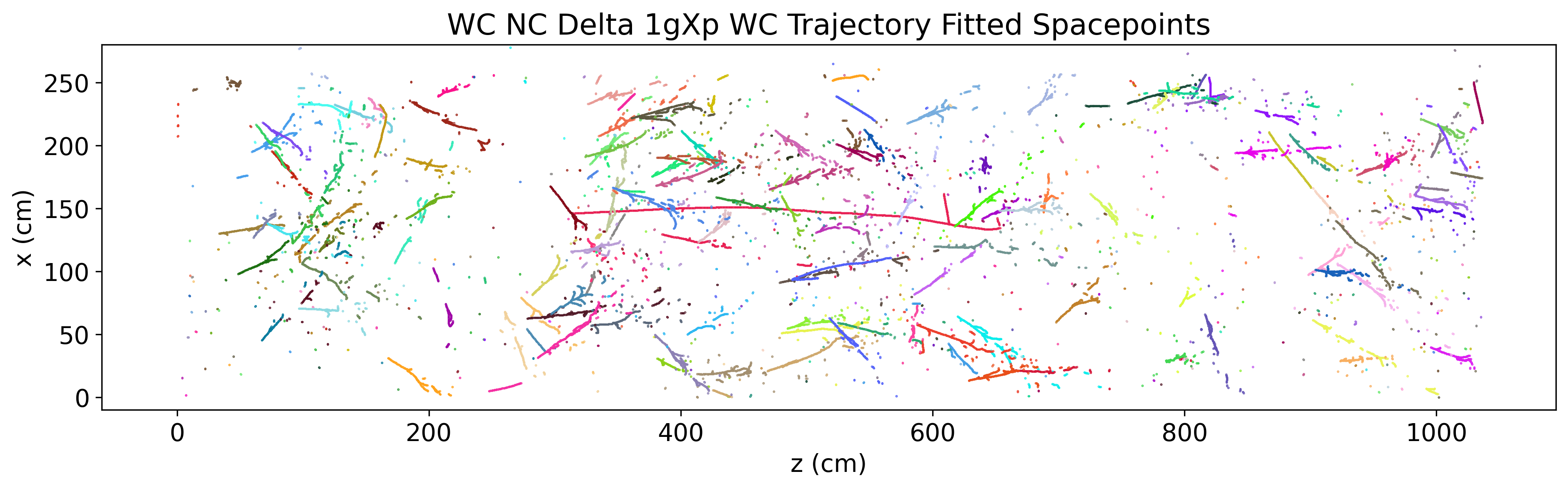}
        \caption{}
    \end{subfigure}
    \begin{subfigure}[b]{0.99\textwidth}
        \includegraphics[width=\textwidth]{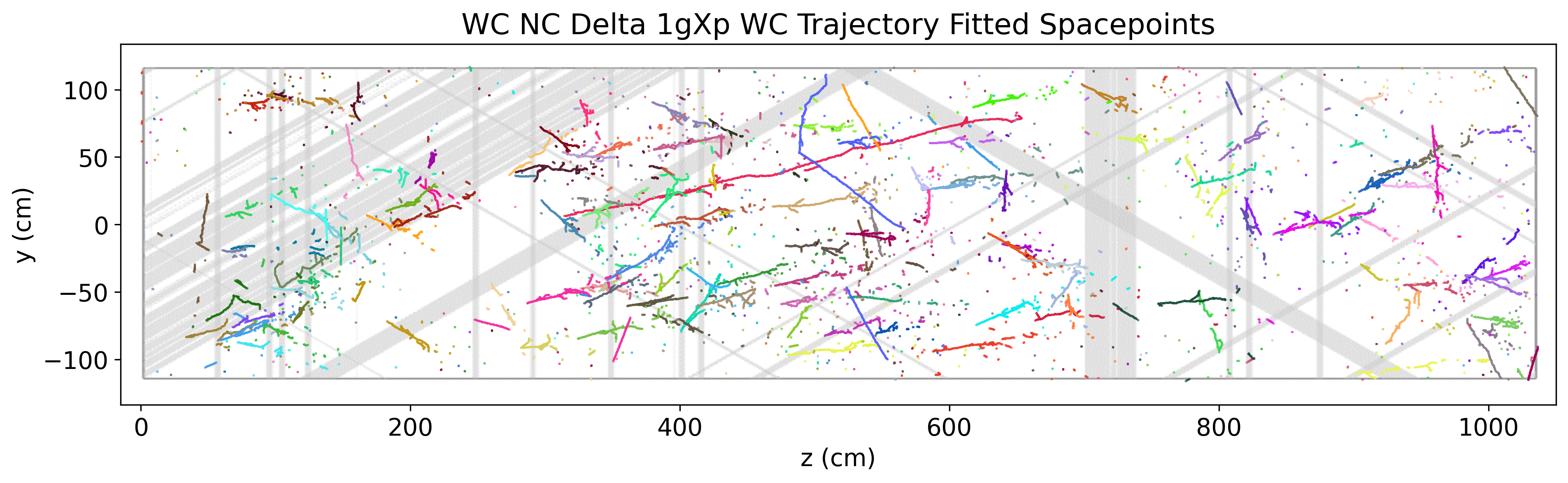}
        \caption{}
    \end{subfigure}
    \caption[Wire-Cell NC $\Delta\rightarrow N \gamma$ overlaid event display clusters]{Wire-Cell NC $\Delta\rightarrow N \gamma$ overlaid event displays. The color corresponds to the reconstructed event for each 3D space point. Panel (a) shows the x-z view, and Panel (b) shows the y-z view, with dead wires indicated by gray shading.}
    \label{fig:wc_nc_delta_overlaid_event_display_clusters}
\end{figure}

With our selection finalized and careful checks of small BNB data samples, NuMI data samples, and fake data studies, and careful consideration of all systematic uncertainties, we are ready to unblind our full runs 1-3 data set for the new Wire-Cell NC $\Delta\rightarrow N \gamma$ selection. We show a sampling of 116 out of the total 204 selected data events in Figs. \ref{fig:wc_nc_delta_overlaid_event_displays} and \ref{fig:wc_nc_delta_overlaid_event_display_clusters}. We primarily see small showers as we would expect. A few long tracks corresponding to reconstructed charged pions are visible, but the majority of selected data events have no such tracks.

\begin{figure}[H]
    \centering
    \begin{subfigure}[b]{0.49\textwidth}
        \includegraphics[trim=20 0 60 0, clip, width=\textwidth]{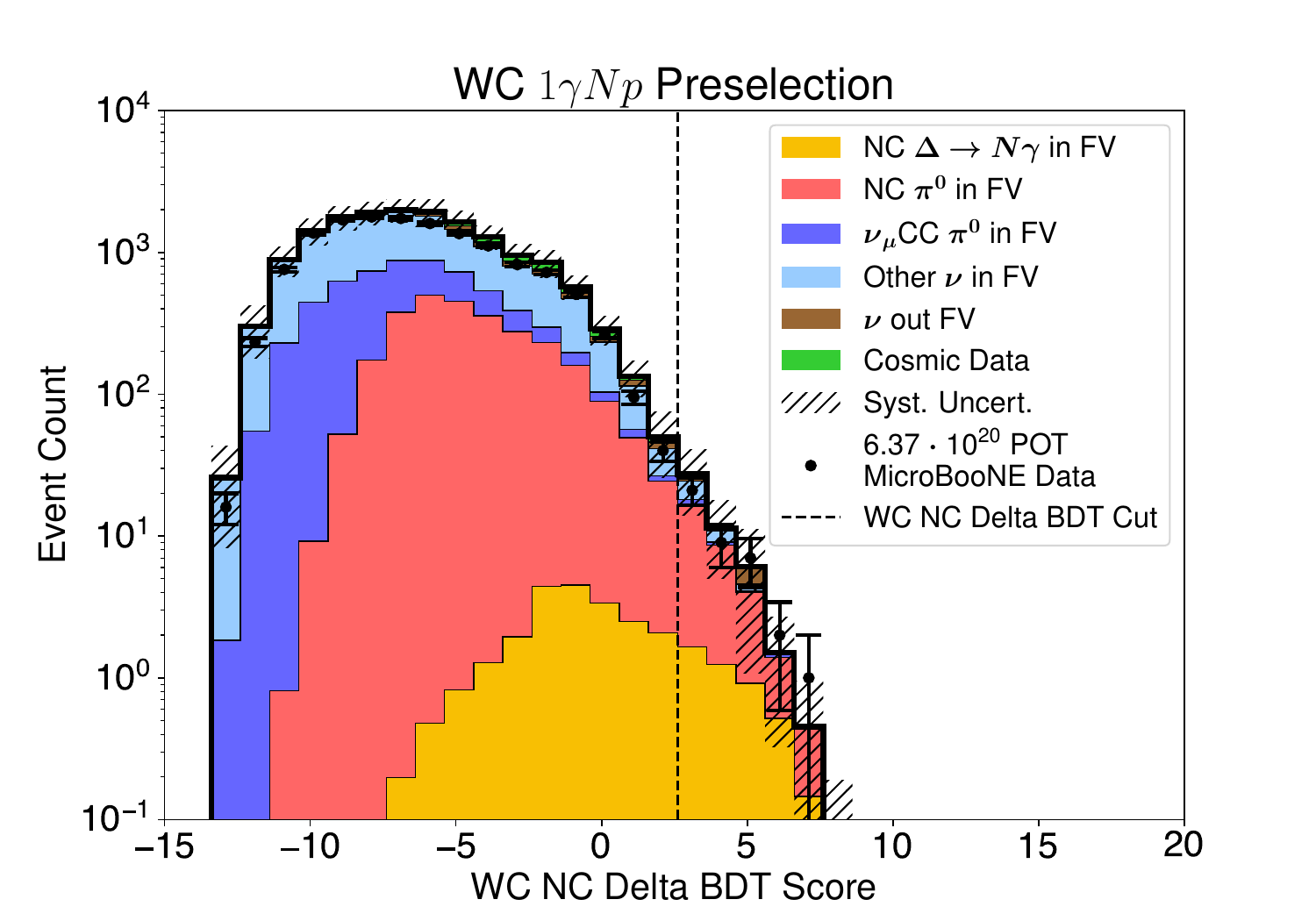}
        \caption{}
    \end{subfigure}
    \begin{subfigure}[b]{0.49\textwidth}
        \includegraphics[trim=20 0 60 0, clip, width=\textwidth]{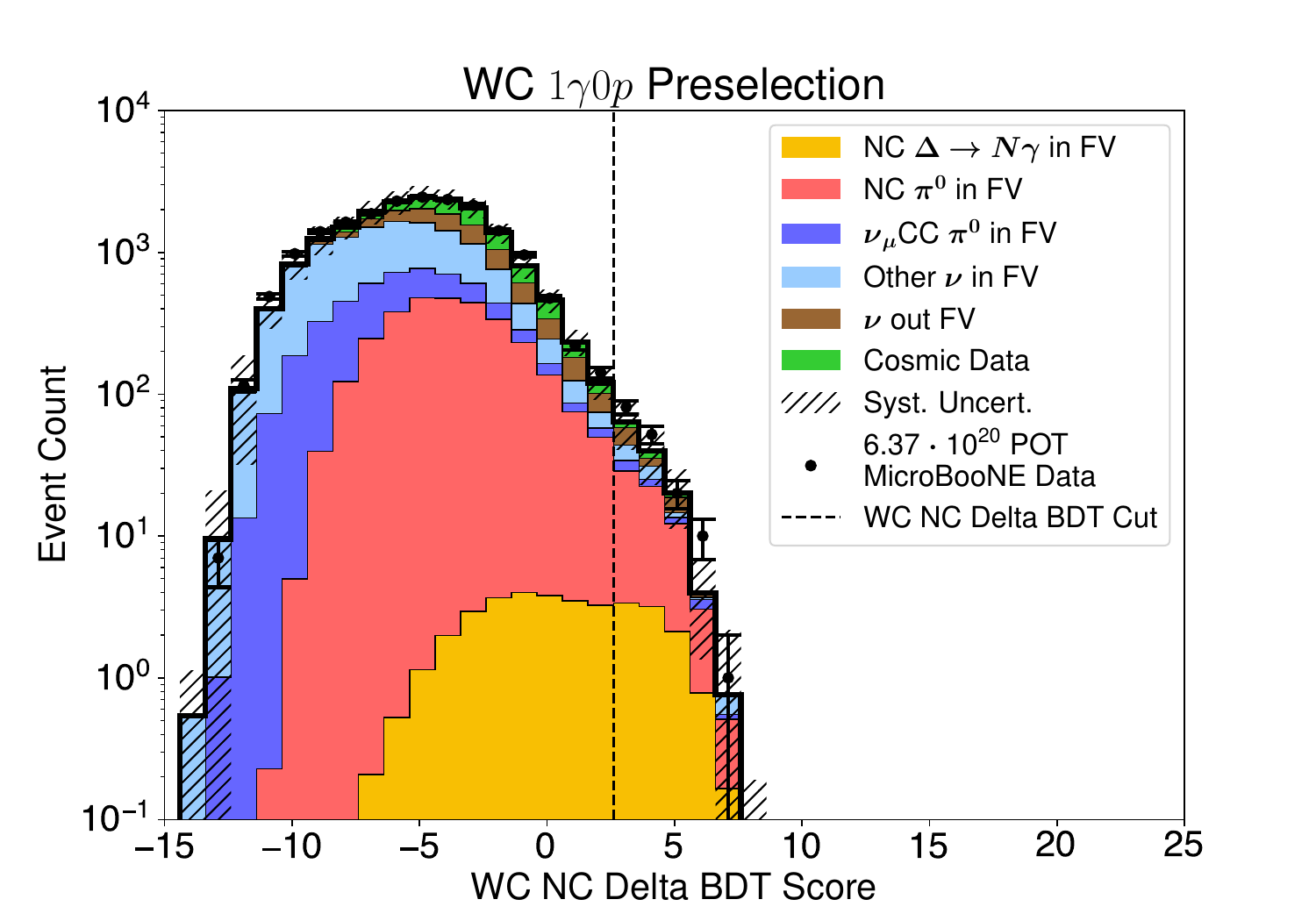}
        \caption{}
    \end{subfigure}
    \begin{subfigure}[b]{0.49\textwidth}
        \includegraphics[trim=20 0 60 0, clip, width=\textwidth]{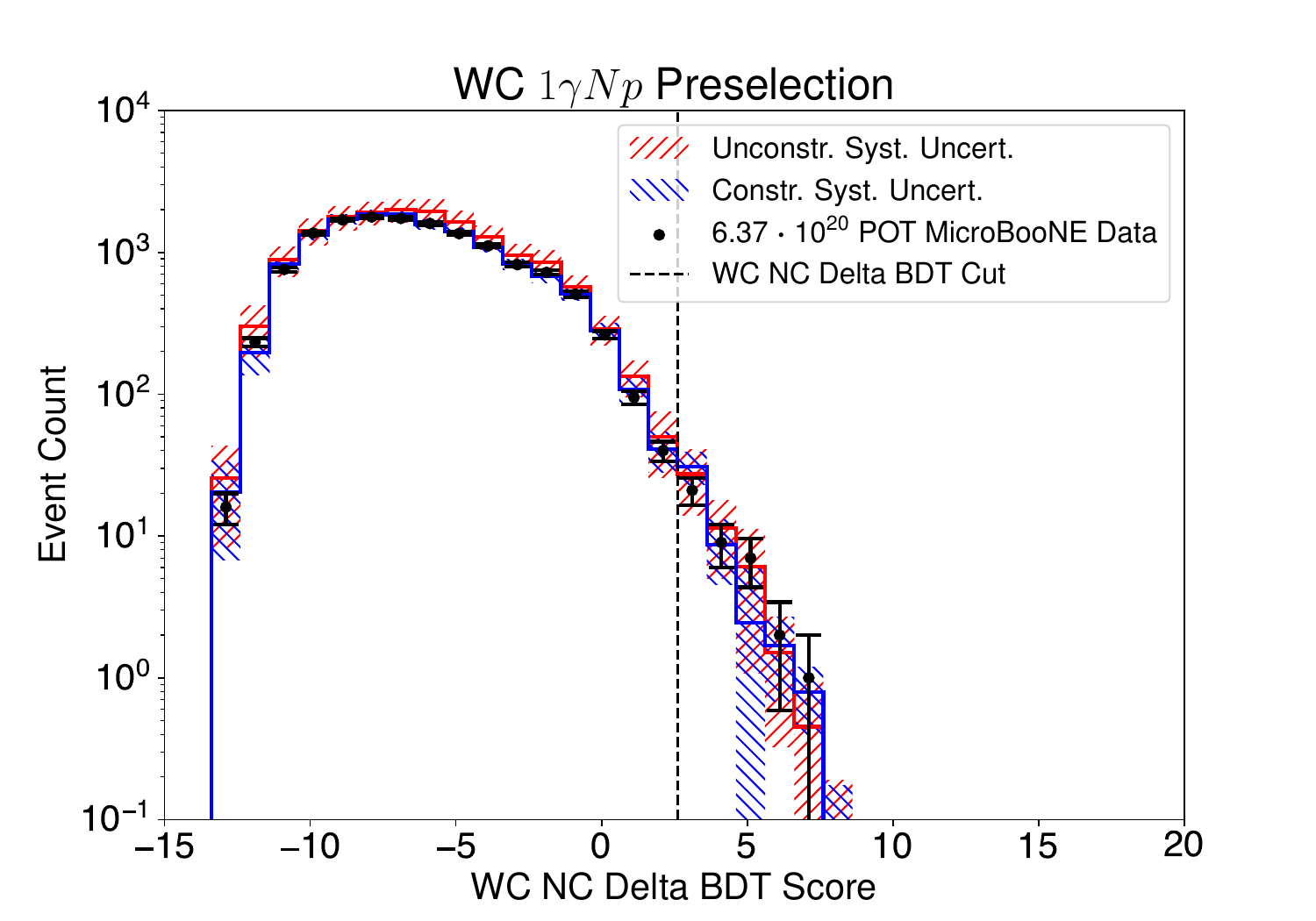}
        \caption{}
    \end{subfigure}
    \begin{subfigure}[b]{0.49\textwidth}
        \includegraphics[trim=20 0 60 0, clip, width=\textwidth]{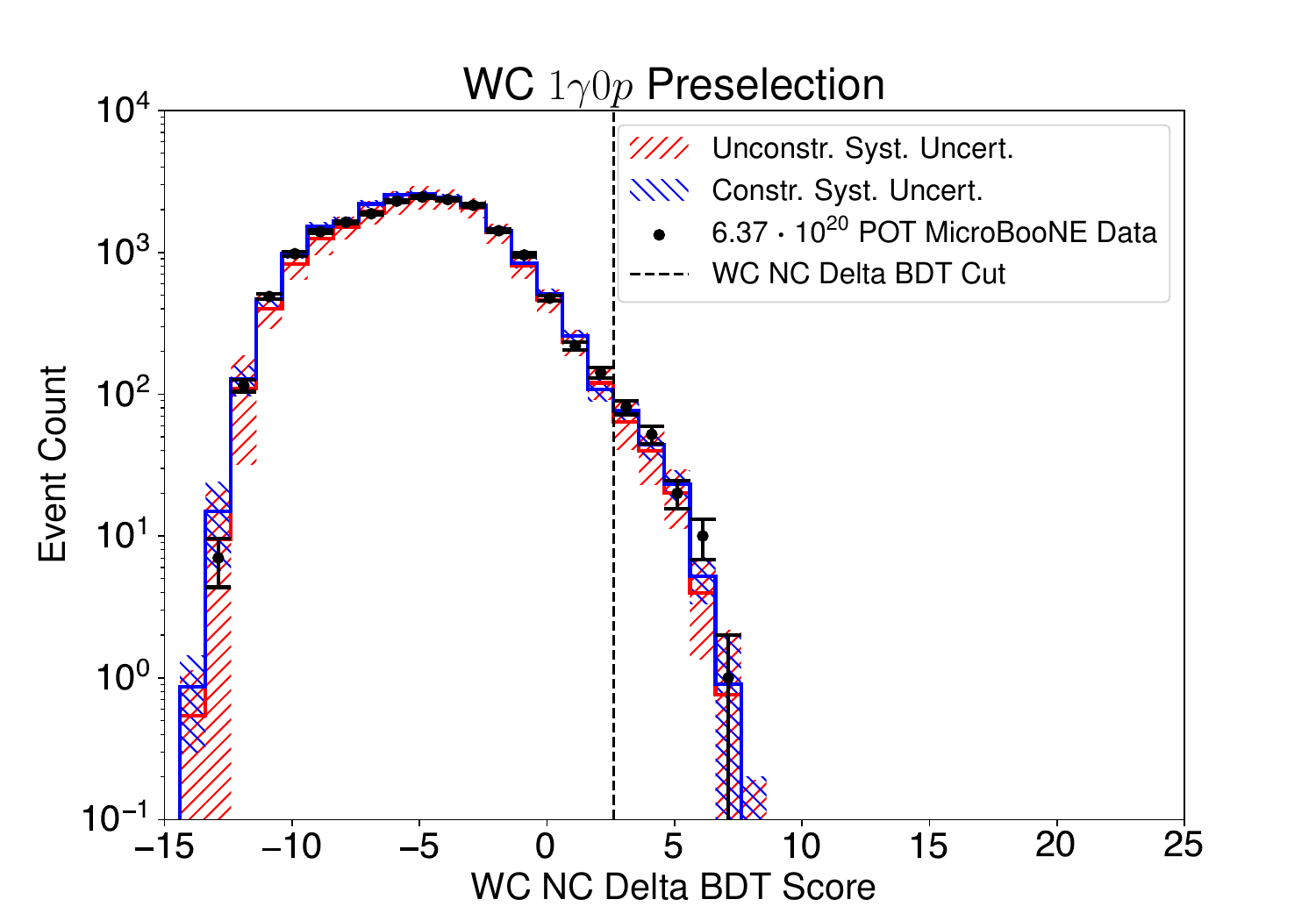}
        \caption{}
    \end{subfigure}
    \caption[NC $\Delta\rightarrow N \gamma$ BDT score distributions]{NC $\Delta\rightarrow N \gamma$ BDT score distributions. We show only events passing the preselection, which consists of Wire-Cell generic neutrino selection plus the existence of a reconstructed shower. Panels (a) and (c) shows events with one or more reconstructed protons, and panels (b) and (d) show events with zero reconstructed protons. Panels (a) and (b) show the distributions with a breakdown of different prediction types, and with no conditional constraint. Panels (c) and (d) show distributions both before and after the application of the conditional constraint.}
    \label{fig:nc_delta_bdt_score_distributions}
\end{figure}

Figure. \ref{fig:nc_delta_bdt_score_distributions} shows the Wire-Cell BDT score distributions for $Np$ and $0p$ events, and we see good agreement across the full range of BDT scores.

\begin{figure}[H]
    \centering
    \includegraphics[trim=30 70 60 70, clip, width=0.6\textwidth]{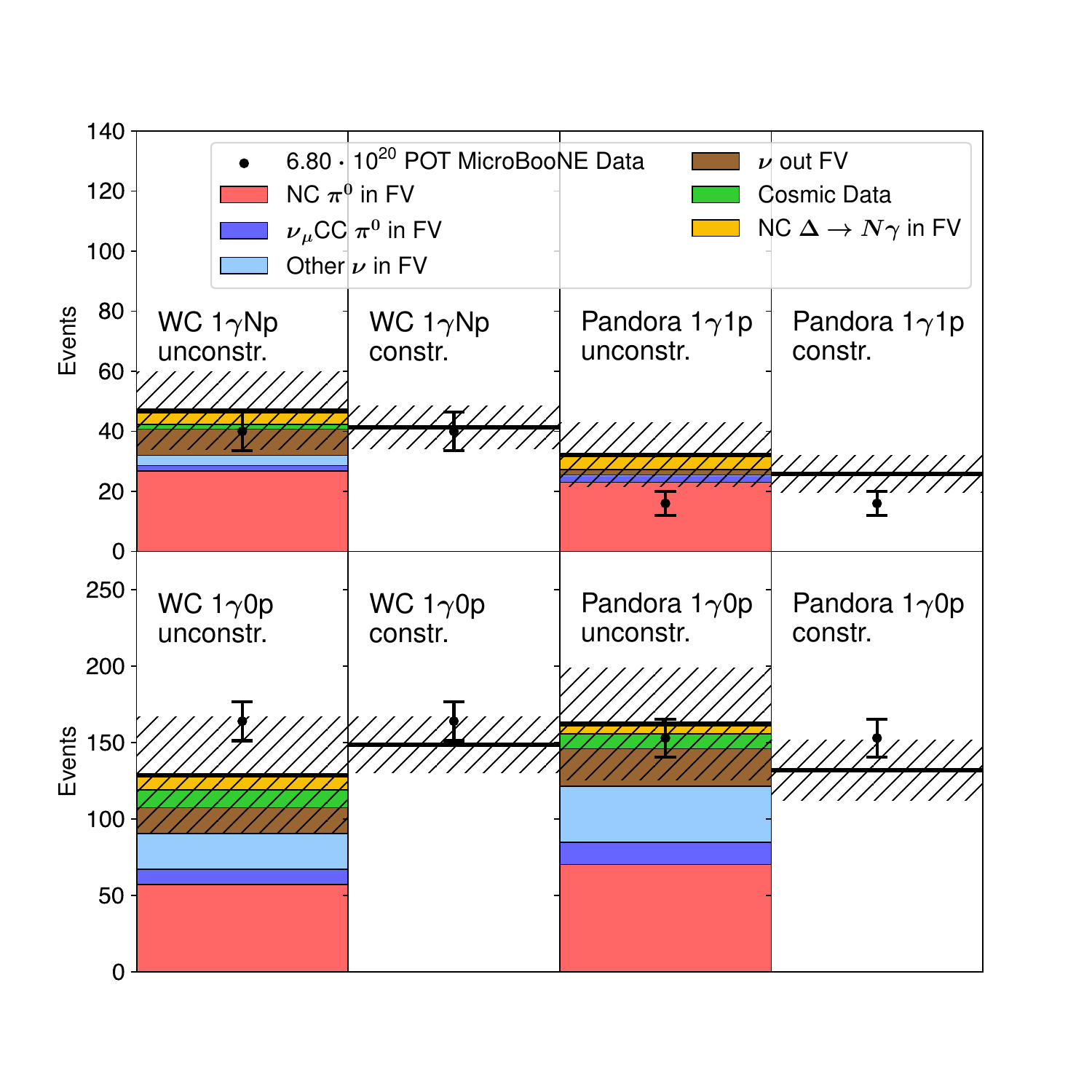}
    \caption[NC $\Delta\rightarrow N \gamma$ one bin observations]{NC $\Delta\rightarrow N \gamma$ one bin observations.}
    \label{fig:one_bin_no_lee}
\end{figure}

Table \ref{tab:nc_delta_results_components} and Fig. \ref{fig:one_bin_no_lee} show the data and prediction comparison for each of the four signal channel bins. We see generally good agreement between prediction and data, both before and after the conditional constraint. 

\begin{table}[H]
    \centering
    \small
    \begin{tabular}{c c c c c} 
        \toprule
        Process & \makecell{WC\\$1\gamma N p$} & \makecell{WC\\$1\gamma0p$} & \makecell{Pandora\\$1\gamma 1 p$} & \makecell{Pandora\\$1\gamma0p$} \\
        \midrule
        NC $1\pi^0$ in FV & 26.8 & 57.2 & 23.0 & 70.1 \\
        CC $1\pi^0$ in FV & 1.9 & 10.0 & 2.4 & 14.7 \\
        Other $\nu$ in FV & 8.7 & 16.9 & 1.9 & 24.6 \\
        Out FV & 3.4 & 23.3 & 0.0 & 36.6 \\
        Cosmic Beam-off Data & 1.6 & 11.7 & 0.0 & 9.8 \\
        \midrule
        NC $\Delta\rightarrow N \gamma$ in FV & 4.5 & 9.7 & 4.9 & 6.5 \\
        \midrule
        \makecell{Unconstrained total prediction} & 46.8 & 128.7 & 32.2 & 162.2 \\
        \makecell{Constrained total prediction} & 41.3 & 148.5 & 25.8 & 131.9 \\
        \midrule
        \makecell{Observed data} & 40 & 164 & 16 & 153 \\
        \bottomrule
    \end{tabular}
    \caption[NC $\Delta\rightarrow N \gamma$ one-bin results]{NC $\Delta\rightarrow N \gamma$ one-bin results. Categories are broken into those with true neutrino interaction vertices inside and outside the fiducial volume (FV).}
    \label{tab:nc_delta_results_components}
\end{table}

We performed a series of goodness of fit $\chi^2$/ndf tests, both before and after the conditional constraint, as shown in Table \ref{tab:gof_tests_no_LEE}. We perform separate tests on each signal channel bin, for each reconstruction framework's pair of bins, and for all four bins together, and see good agreement between data and prediction within uncertainties in each case.

\begin{table}[H]
    \centering
    \begin{tabular}{c c c} 
        \toprule
        Selection & \makecell{Unconstrained \\ $\chi^2$/ndf,  p-value, $\sigma$} & \makecell{Constrained \\ $\chi^2$/ndf,  p-value, $\sigma$} \\
        \midrule
        WC $1\gamma Np$ & 
            0.216/1, 0.642, 0.465 $\sigma$ & 
            0.018/1, 0.891, 0.137$\sigma$ \\
        WC $1\gamma 0p$ & 
            0.777/1, 0.378, 0.881$\sigma$ & 
            0.495/1, 0.482, 0.704$\sigma$\\
        Pandora $1\gamma 1p$ & 
            1.812/1, 0.178, 1.346$\sigma$ & 
            1.476/1, 0.224, 1.215$\sigma$ \\
        Pandora $1\gamma 0p$ & 
            0.056/1, 0.814, 0.236$\sigma$ & 
            0.838/1, 0.360, 0.916$\sigma$\\
        \midrule
        WC 1gNp+$1\gamma 0p$ & 
            1.990/2, 0.370, 0.897$\sigma$ & 
            0.565/2, 0.754, 0.313$\sigma$\\
        Pandora $1\gamma 1p$+$1\gamma 0p$ & 
            1.977/2, 0.372, 0.892$\sigma$ & 
            2.309/2, 0.315, 1.004$\sigma$ \\
        \midrule
        \makecell{WC $1\gamma Np$+$1\gamma 0p$\\+ Pandora $1\gamma 1p$+$1\gamma 0p$} & 
            4.096/4, 0.393, 0.854$\sigma$ & 
            2.652/4, 0.618, 0.499$\sigma$\\
        \bottomrule
    \end{tabular}
    \caption[Goodness of fit tests]{Goodness of fit tests.}
    \label{tab:gof_tests_no_LEE}
\end{table}

\subsection{Particle Multiplicity Distributions}

Next, we investigate particle multiplicity distributions. For these plots, we have not calculated systematic uncertainties, and just show the comparison between our observed data points and our central value prediction without any conditional constraint. 

\begin{figure}[H]
    \centering
    \begin{subfigure}[b]{0.49\textwidth}
        \includegraphics[trim=30 0 70 0, clip, width=\textwidth]{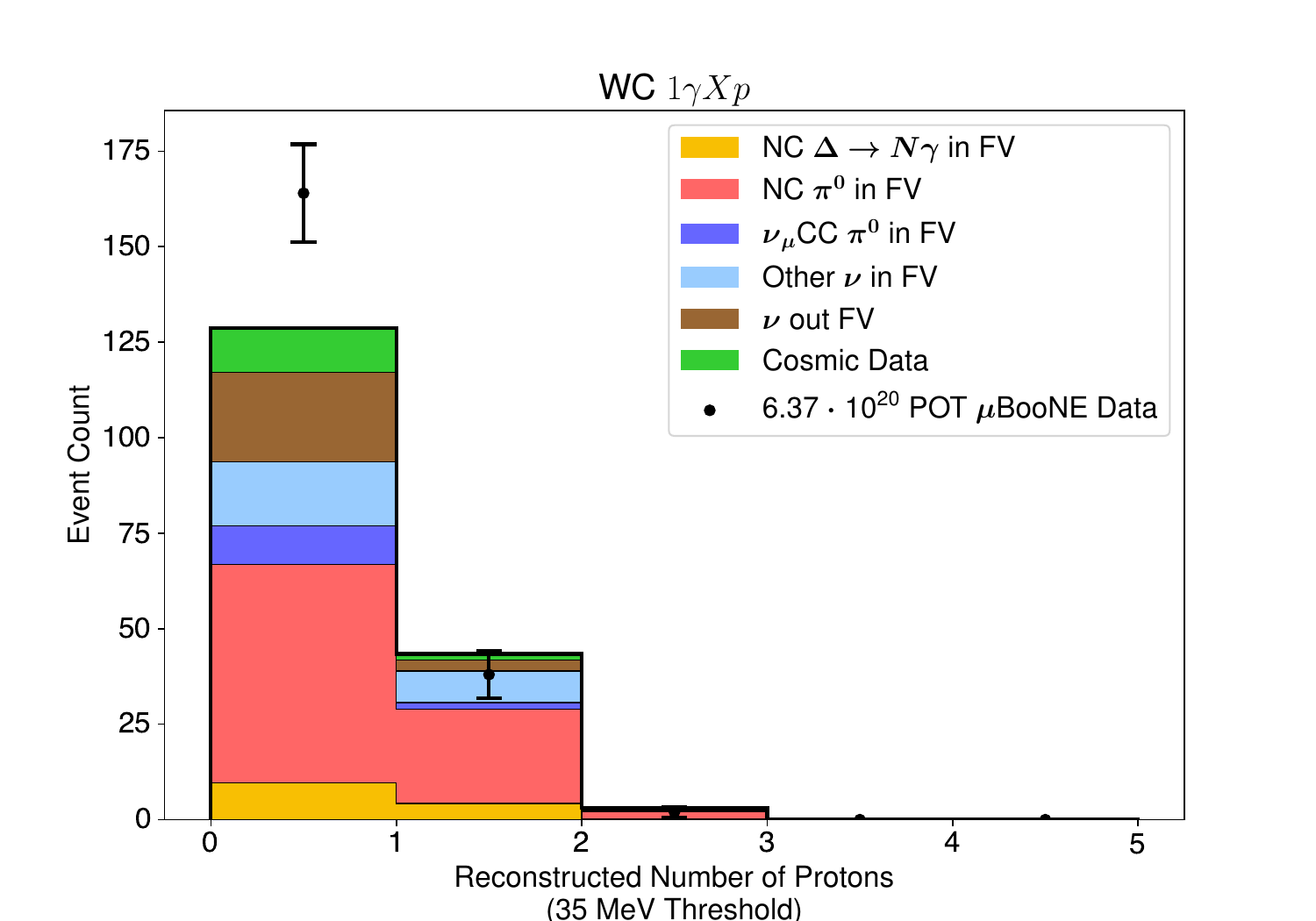}
        \caption{}
    \end{subfigure}
    \begin{subfigure}[b]{0.49\textwidth}
        \includegraphics[trim=30 0 70 0, clip, width=\textwidth]{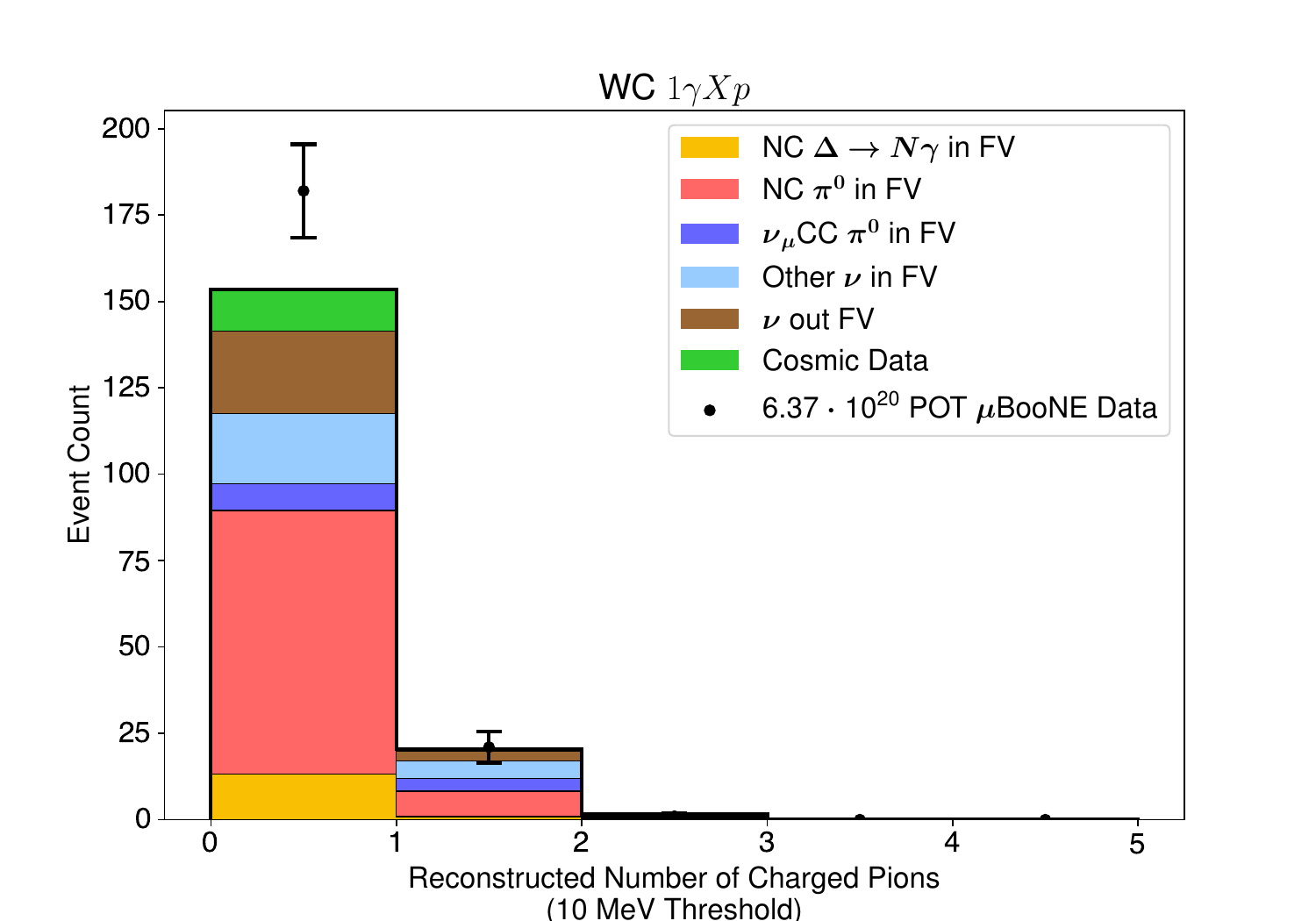}
        \caption{}
    \end{subfigure}
    \caption[Wire-Cell NC $\Delta\rightarrow N \gamma$ proton and charged pion multiplicities]{Wire-Cell NC $\Delta\rightarrow N \gamma$ reconstructed particle multiplicities. Panel (a) shows the reconstructed number of protons, and panel (b) shows the reconstructed number of charged pions. No systematic uncertainties are considered, and no conditional constraint has been applied.}
    \label{fig:proton_pion_multiplicities}
\end{figure}

Figure \ref{fig:proton_pion_multiplicities} shows distributions of reconstructed protons and charged pions. We see that as expected, most events have zero or one reconstructed proton, and most events have zero reconstructed charged pions. For the very rare events with multiple reconstructed protons or reconstructed charged pions, we do not see any notable data/prediction differences.

\begin{figure}[H]
    \centering
    \begin{subfigure}[b]{0.49\textwidth}
        \includegraphics[trim=30 0 70 0, clip, width=\textwidth]{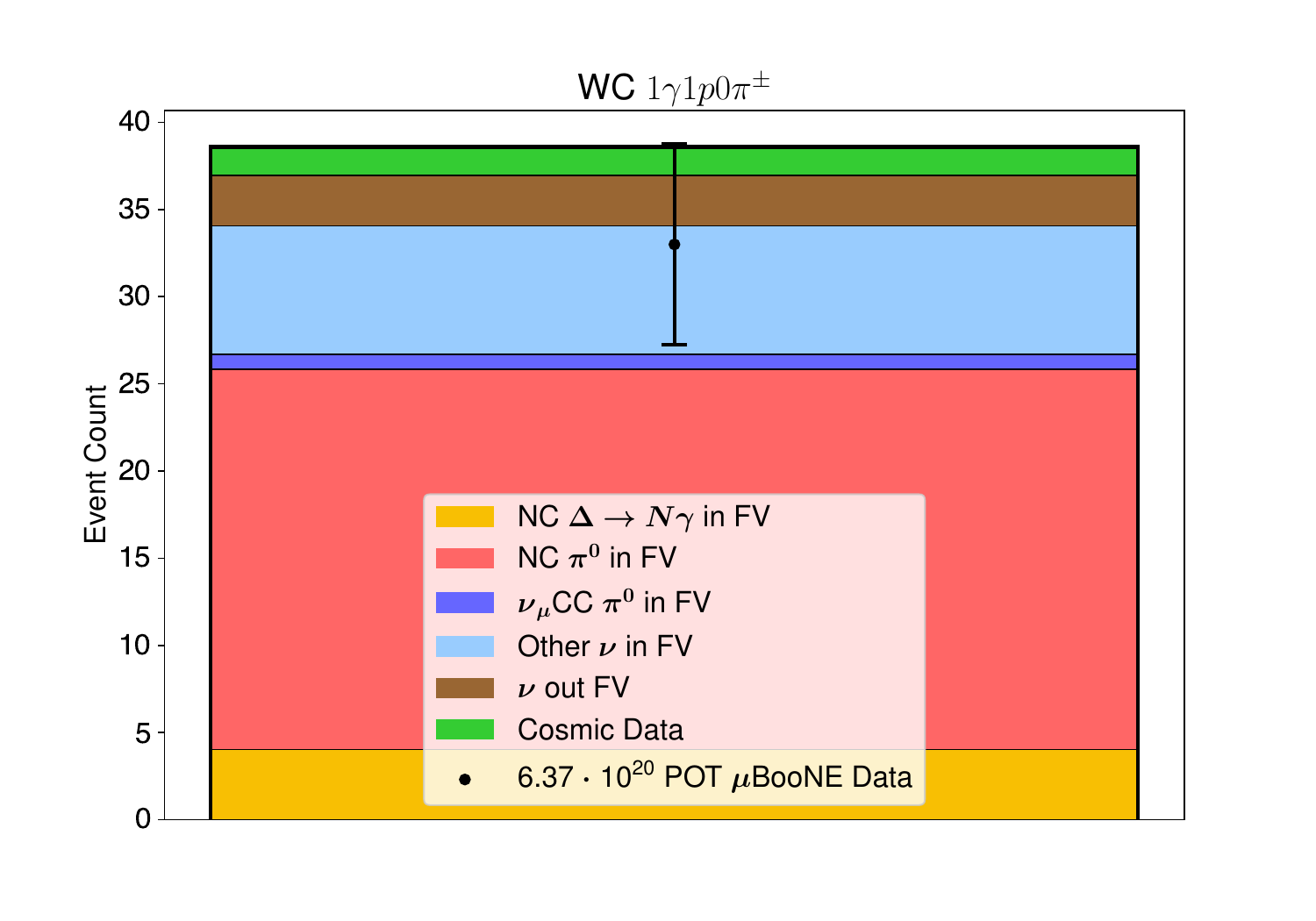}
        \caption{}
    \end{subfigure}
    \begin{subfigure}[b]{0.49\textwidth}
        \includegraphics[trim=30 0 70 0, clip, width=\textwidth]{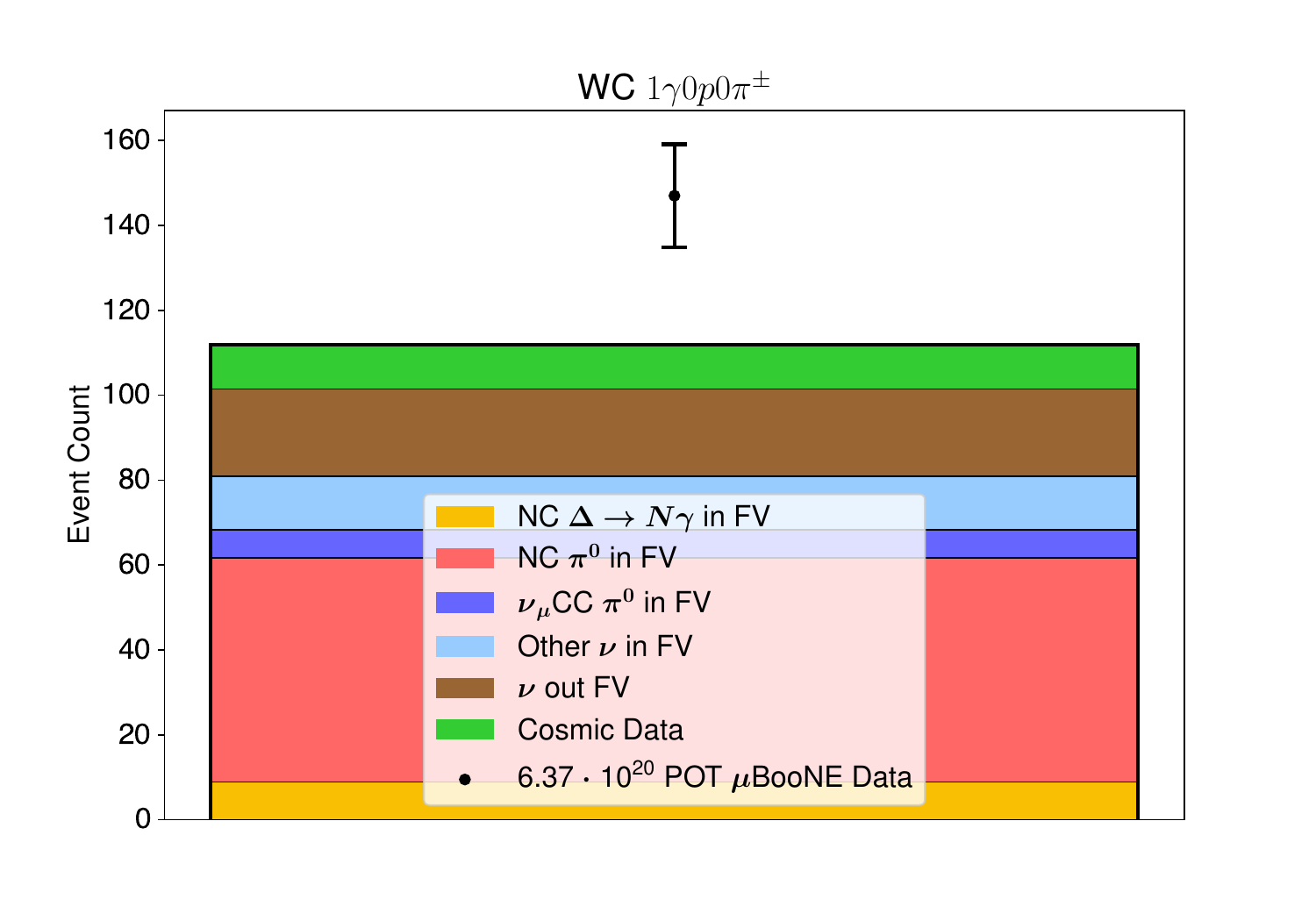}
        \caption{}
    \end{subfigure}
    \caption[Wire-Cell NC $\Delta\rightarrow N \gamma$ $1\gamma 1p 0\pi^\pm$ and $1\gamma 0p 0\pi^\pm$ selections]{Wire-Cell NC $\Delta\rightarrow N \gamma$ selections with simpler required reconstructed topologies. Panel (a) shows the $1\gamma 1p 0\pi^\pm$ selection and panel (b) shows the $1\gamma 0p 0\pi^\pm$ selection. No systematic uncertainties are considered, and no conditional constraint has been applied.}
    \label{fig:wc_1g0_1p0pi}
\end{figure}

We can also examine our selections when restricted to the simpler topologies that Pandora uses, specifically $1\gamma 1p 0\pi^\pm$ and $1\gamma 0p 0\pi^\pm$, as shown in Fig. \ref{fig:wc_1g0_1p0pi}. This shows similar conclusions as our full selection with the broader set of allowed topologies.

\subsection{Wire-Cell and Pandora Overlapping Distributions}

\begin{figure}[H]
    \centering
    \includegraphics[trim=30 30 70 30, clip, width=0.6\textwidth]{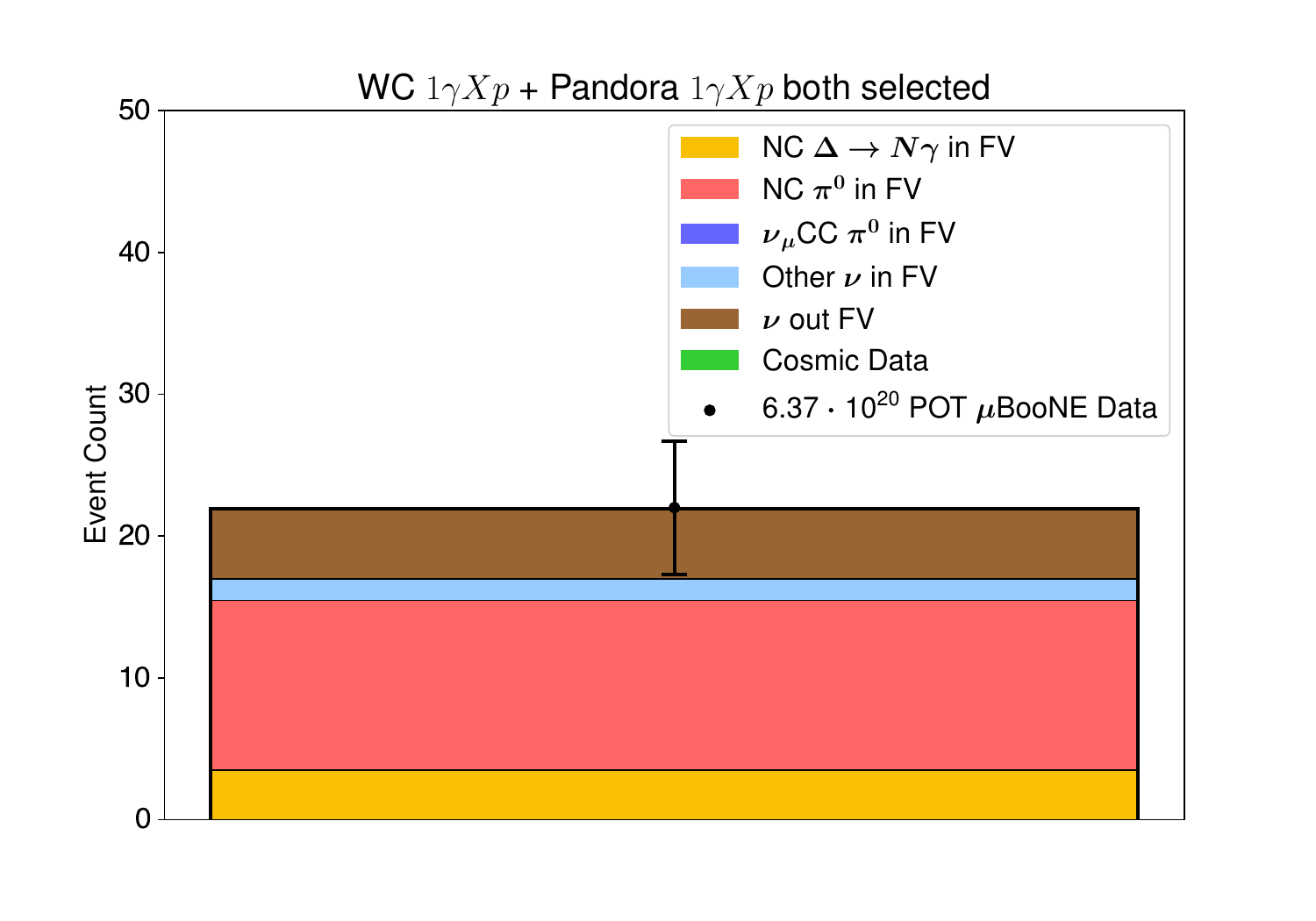}
    \caption[Wire-Cell and Pandora NC $\Delta\rightarrow N \gamma$ overlapping selection]{Wire-Cell and Pandora NC $\Delta\rightarrow N \gamma$ overlapping selection. No systematic uncertainties are considered, and no conditional constraint has been applied.}
    \label{fig:wc_pandora_overlaps}
\end{figure}

\begin{figure}[H]
    \centering
    \begin{subfigure}[b]{0.49\textwidth}
        \includegraphics[trim=30 30 70 30, clip, width=\textwidth]{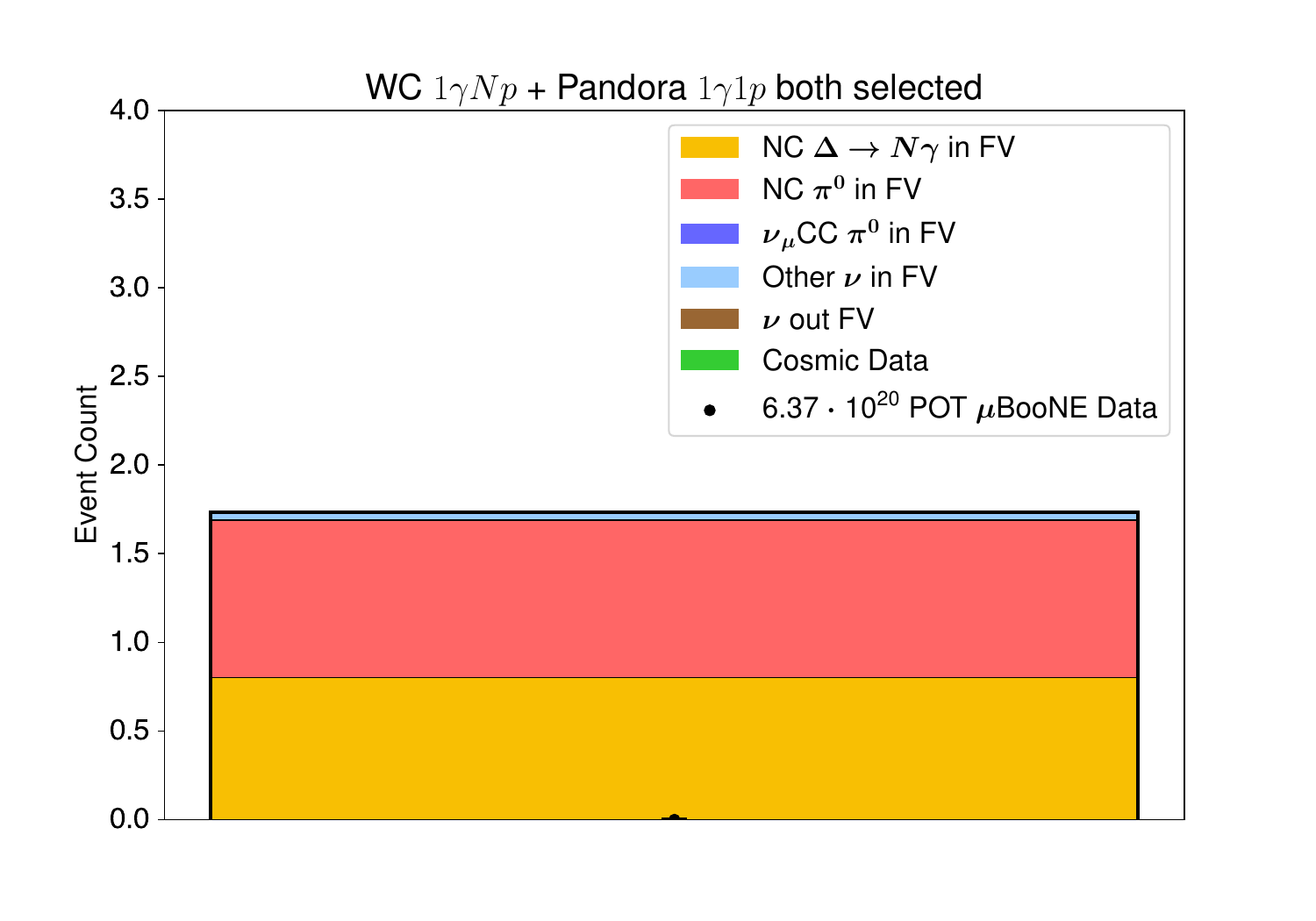}
        \caption{}
    \end{subfigure}
    \begin{subfigure}[b]{0.49\textwidth}
        \includegraphics[trim=30 30 70 30, clip, width=\textwidth]{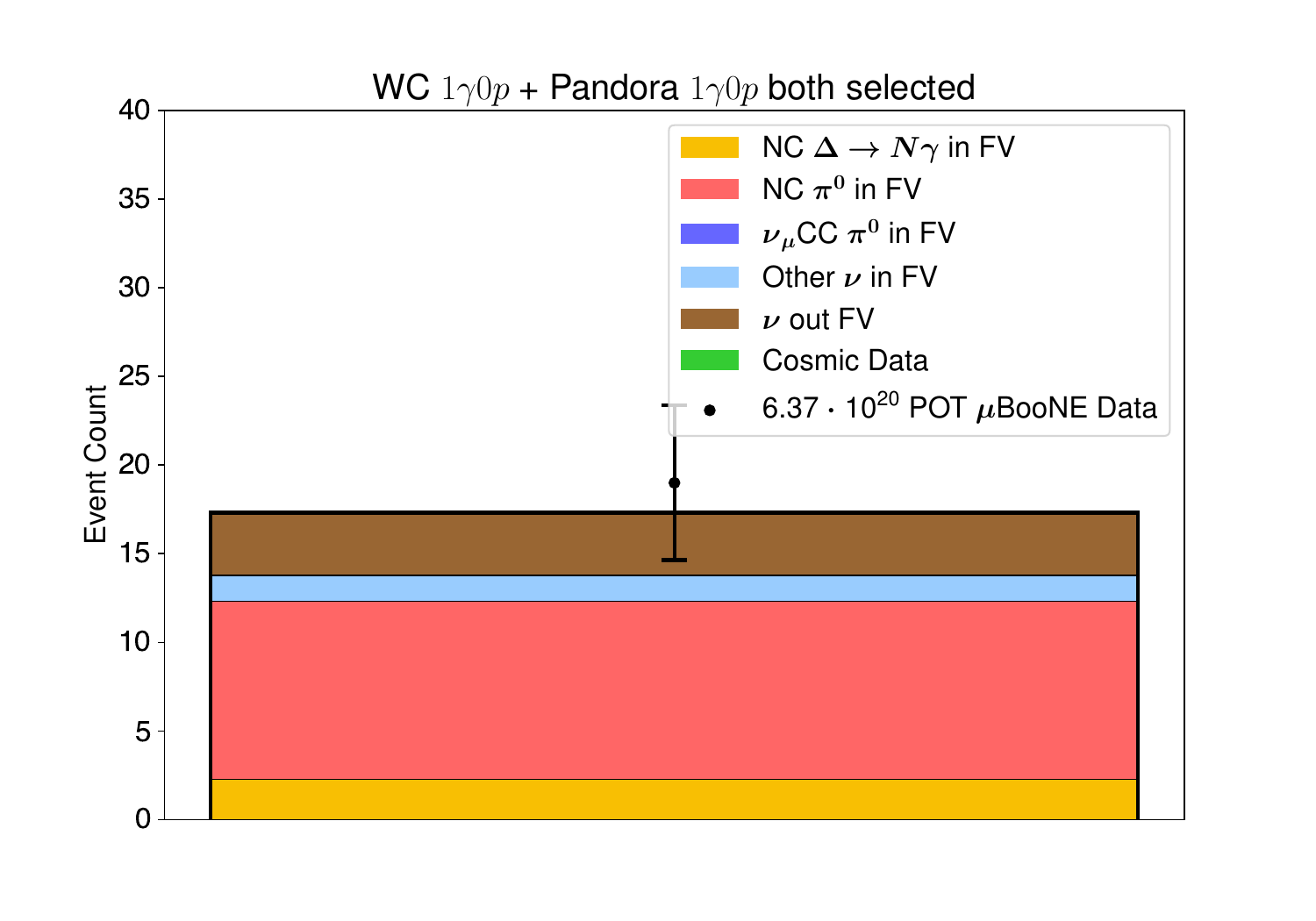}
        \caption{}
    \end{subfigure}
    \begin{subfigure}[b]{0.49\textwidth}
        \includegraphics[trim=30 30 70 30, clip, width=\textwidth]{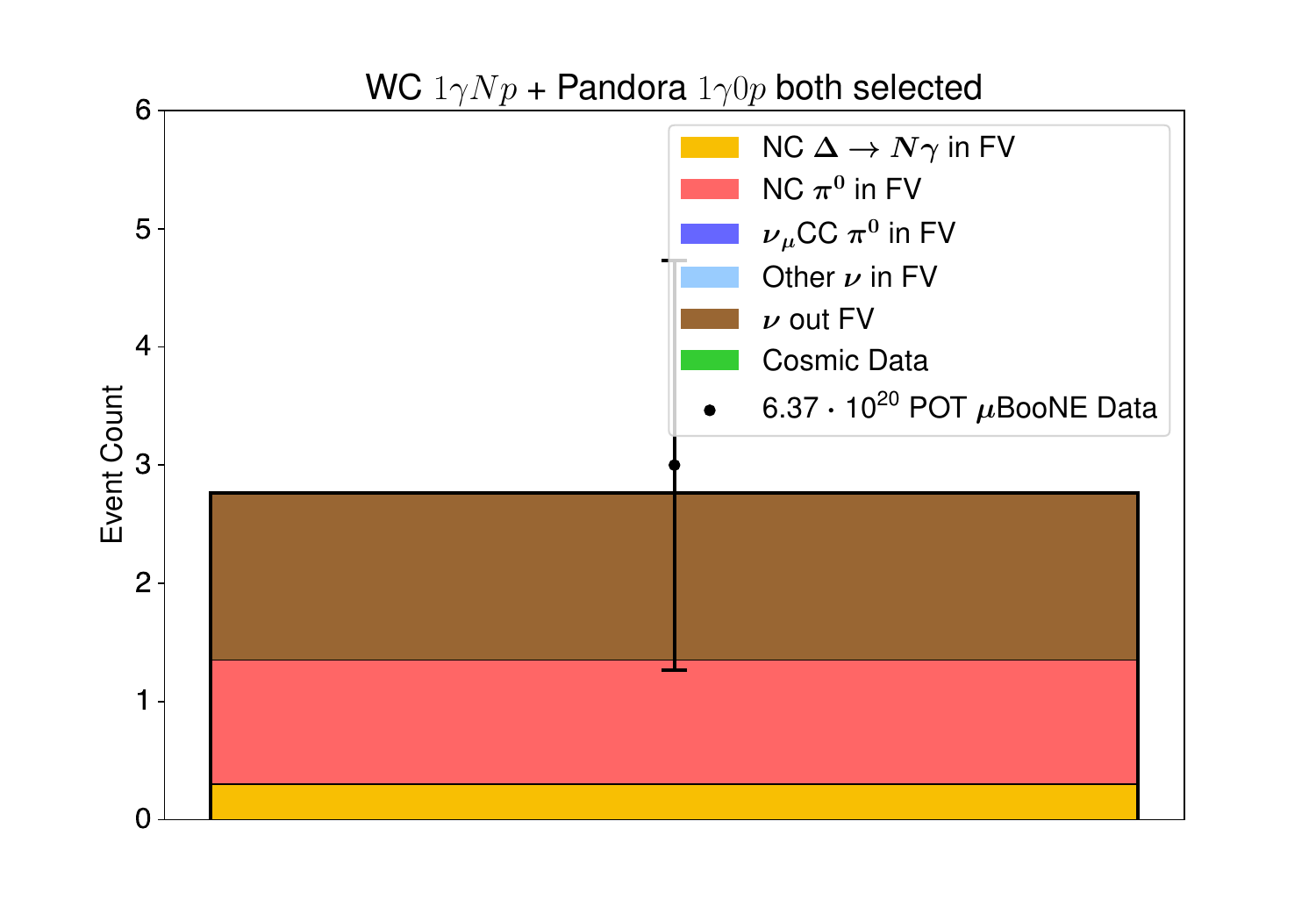}
        \caption{}
    \end{subfigure}
    \begin{subfigure}[b]{0.49\textwidth}
        \includegraphics[trim=30 30 70 30, clip, width=\textwidth]{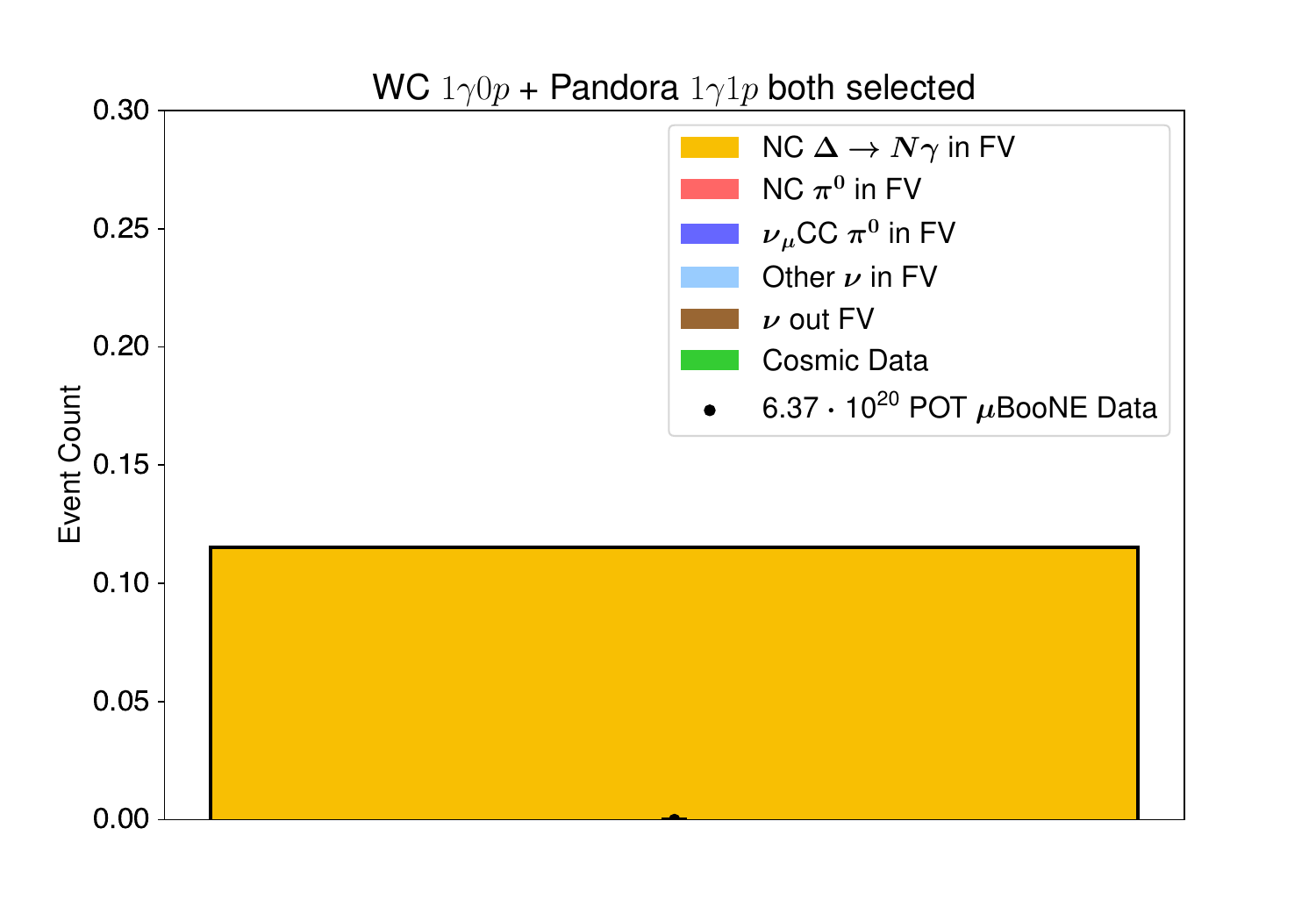}
        \caption{}
    \end{subfigure}
    \caption[Wire-Cell and Pandora NC $\Delta\rightarrow N \gamma$ partially overlapping selections]{Wire-Cell and Pandora NC $\Delta\rightarrow N \gamma$ partially overlapping selections. Panel (a) shows events which have a reconstructed proton in both reconstructions, panel (b) shows events which have no reconstructed protons in either reconstructions, panel (c) shows events which have a reconstructed proton only in the Wire-Cell reconstruction, and panel (d) shows events which have a reconstructed proton only in Pandora reconstruction (this is especially rare and therefore has especially low statistics for the prediction). No systematic uncertainties are considered, and no conditional constraint has been applied.}
    \label{fig:wc_pandora_partial_overlaps}
\end{figure}

Next, we investigate the overlaps between Wire-Cell and Pandora selected events. Figure \ref{fig:wc_pandora_overlaps} shows the overlapping selection consisting of events selected by both Wire-Cell and Pandora. This has lower efficiency, and good data/prediction agreement. Figure \ref{fig:wc_pandora_partial_overlaps} shows a more detailed comparison of the overlaps between different Wire-Cell and Pandora reconstructed topologies. In each case, we see good agreement within statistical uncertaintes.

\subsection{Shower Kinematic Distributions}

Next, we investigate shower kinematic distributions for these selected events. When splitting these selections into kinematic distributions, the systematic uncertainties can grow fairly large, particularly due to limited Monte-Carlo statistics in certain bins. For our quantitative conclusions about MiniBooNE LEE hypotheses, we will use only integrated one-bin selections.

\begin{figure}[H]
    \centering
    \begin{subfigure}[b]{0.49\textwidth}
        \includegraphics[trim=20 0 60 0, clip, width=\textwidth]{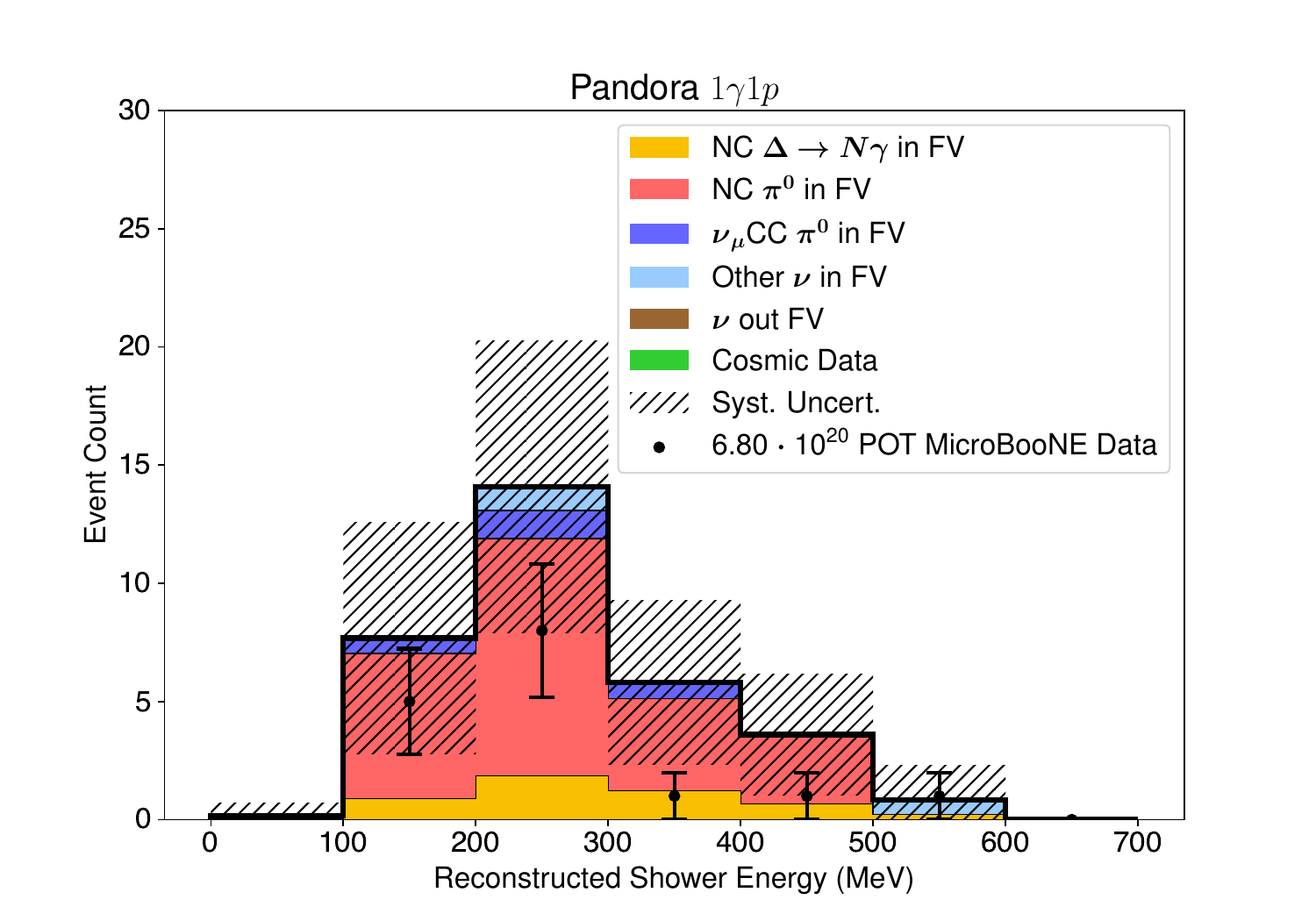}
        \caption{}
    \end{subfigure}
    \begin{subfigure}[b]{0.49\textwidth}
        \includegraphics[trim=20 0 60 0, clip, width=\textwidth]{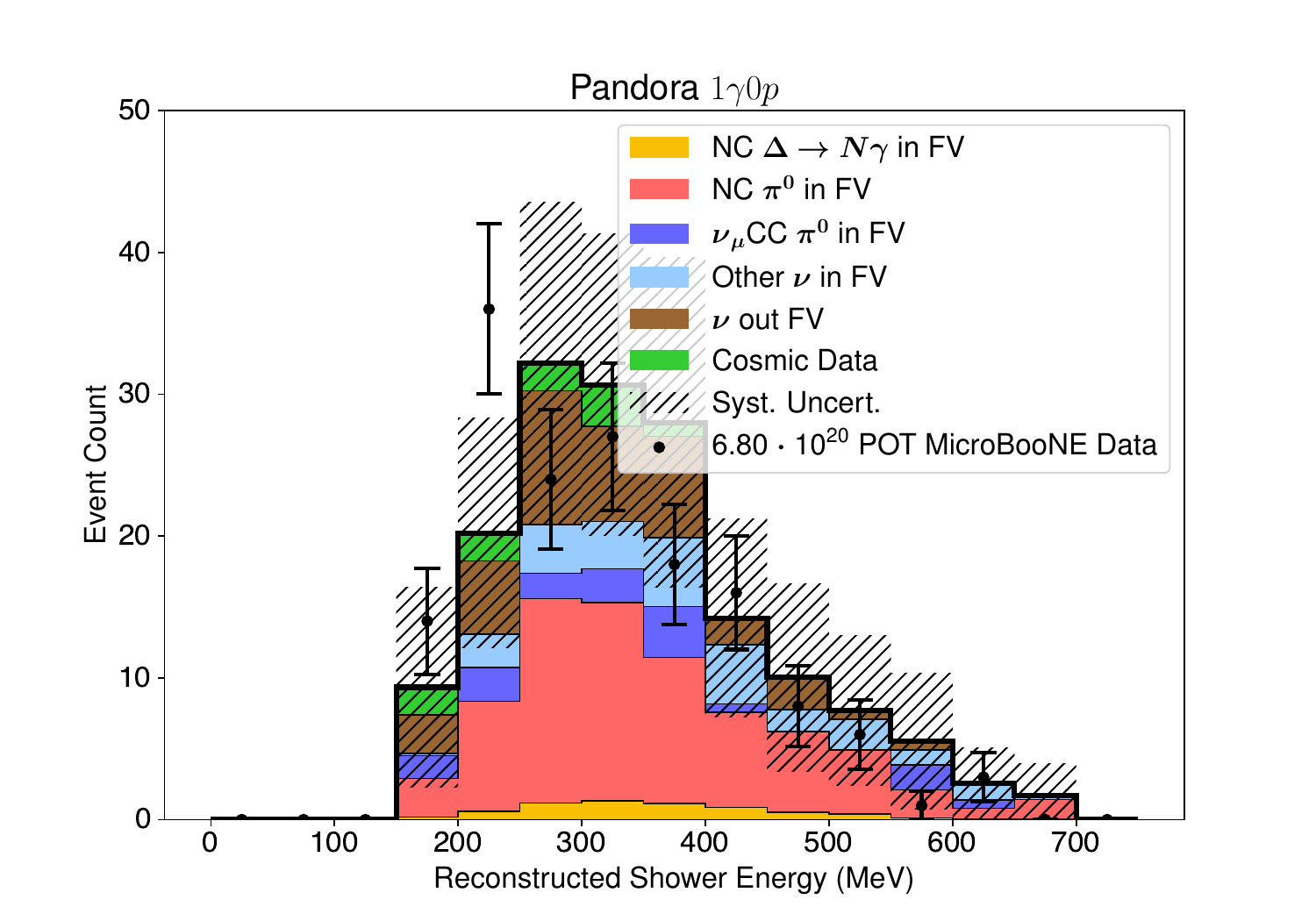}
        \caption{}
    \end{subfigure}
    \begin{subfigure}[b]{0.49\textwidth}
        \includegraphics[trim=20 0 60 0, clip, width=\textwidth]{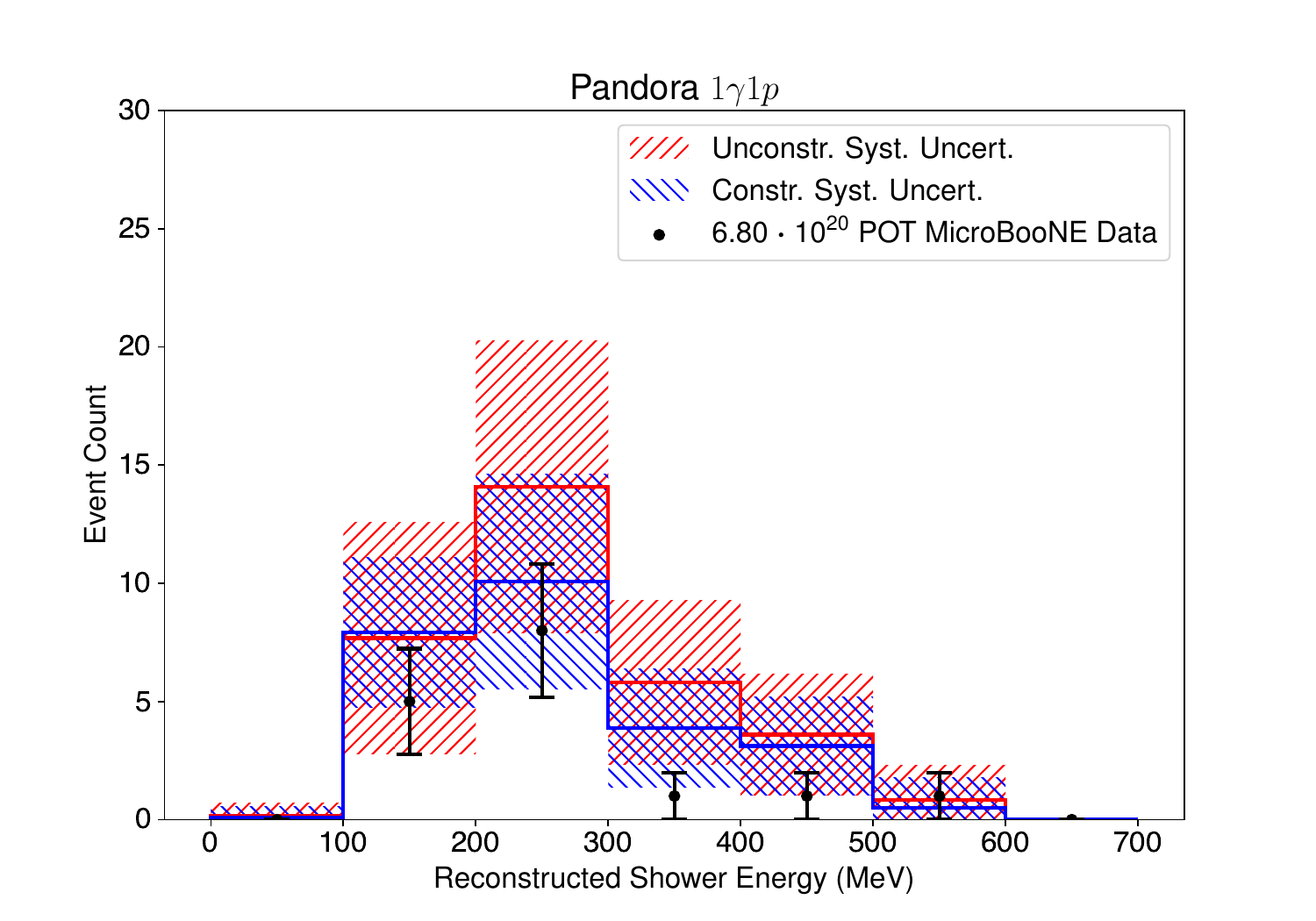}
        \caption{}
    \end{subfigure}
    \begin{subfigure}[b]{0.49\textwidth}
        \includegraphics[trim=20 0 60 0, clip, width=\textwidth]{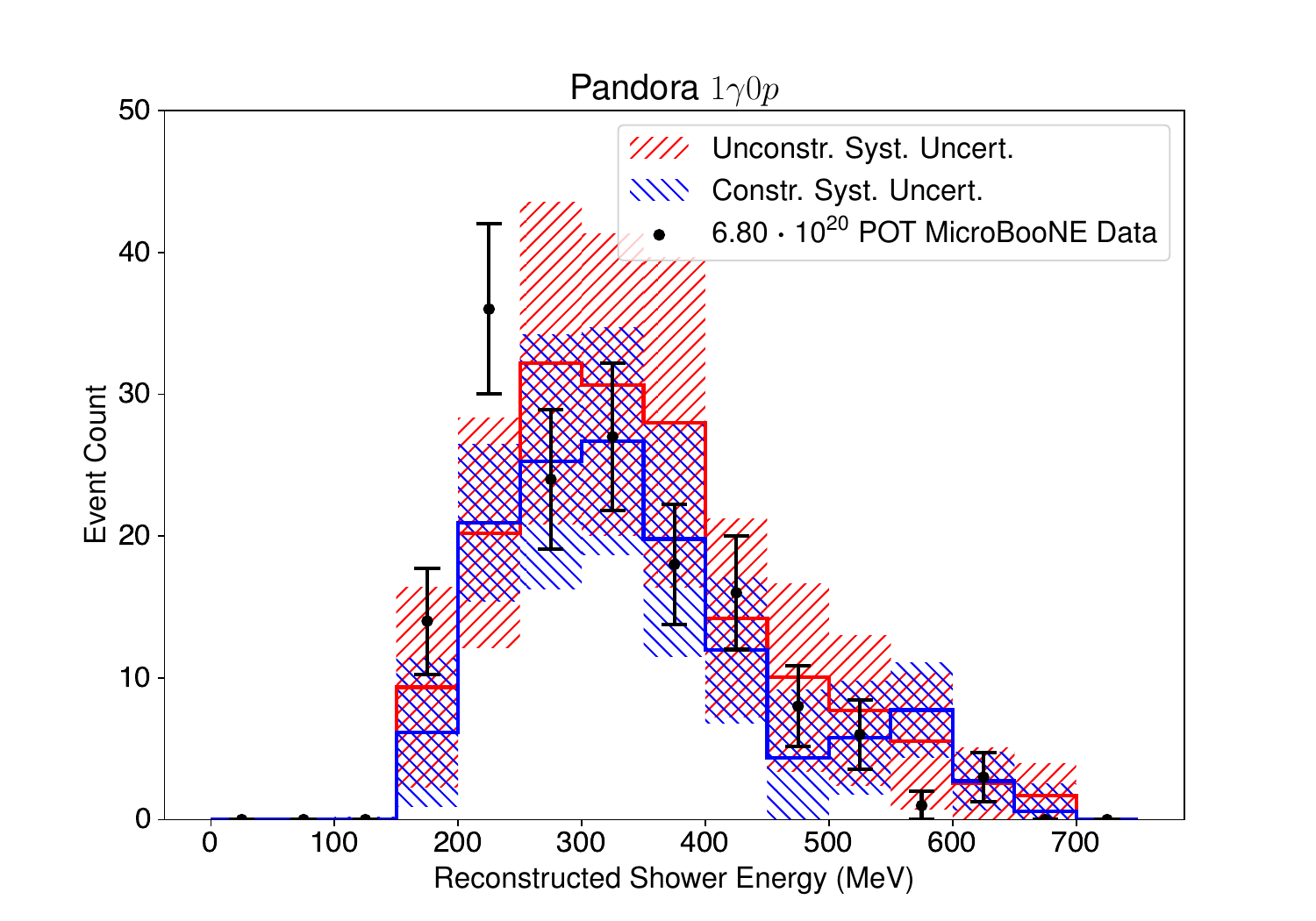}
        \caption{}
    \end{subfigure}
    \caption[Pandora NC $\Delta\rightarrow N \gamma$ shower energy distributions]{Pandora NC $\Delta\rightarrow N \gamma$ shower energy distributions. Panels (a) and (c) shows events with one reconstructed proton, and panels (b) and (d) show events with zero reconstructed protons. Panels (a) and (b) show the distributions with a breakdown of different prediction types, and with no conditional constraint. Panels (c) and (d) show distributions both before and after the application of the conditional constraint.}
    \label{fig:pandora_shower_energy_distributions}
\end{figure}

\begin{figure}[H]
    \centering
    \begin{subfigure}[b]{0.49\textwidth}
        \includegraphics[trim=20 0 60 0, clip, width=\textwidth]{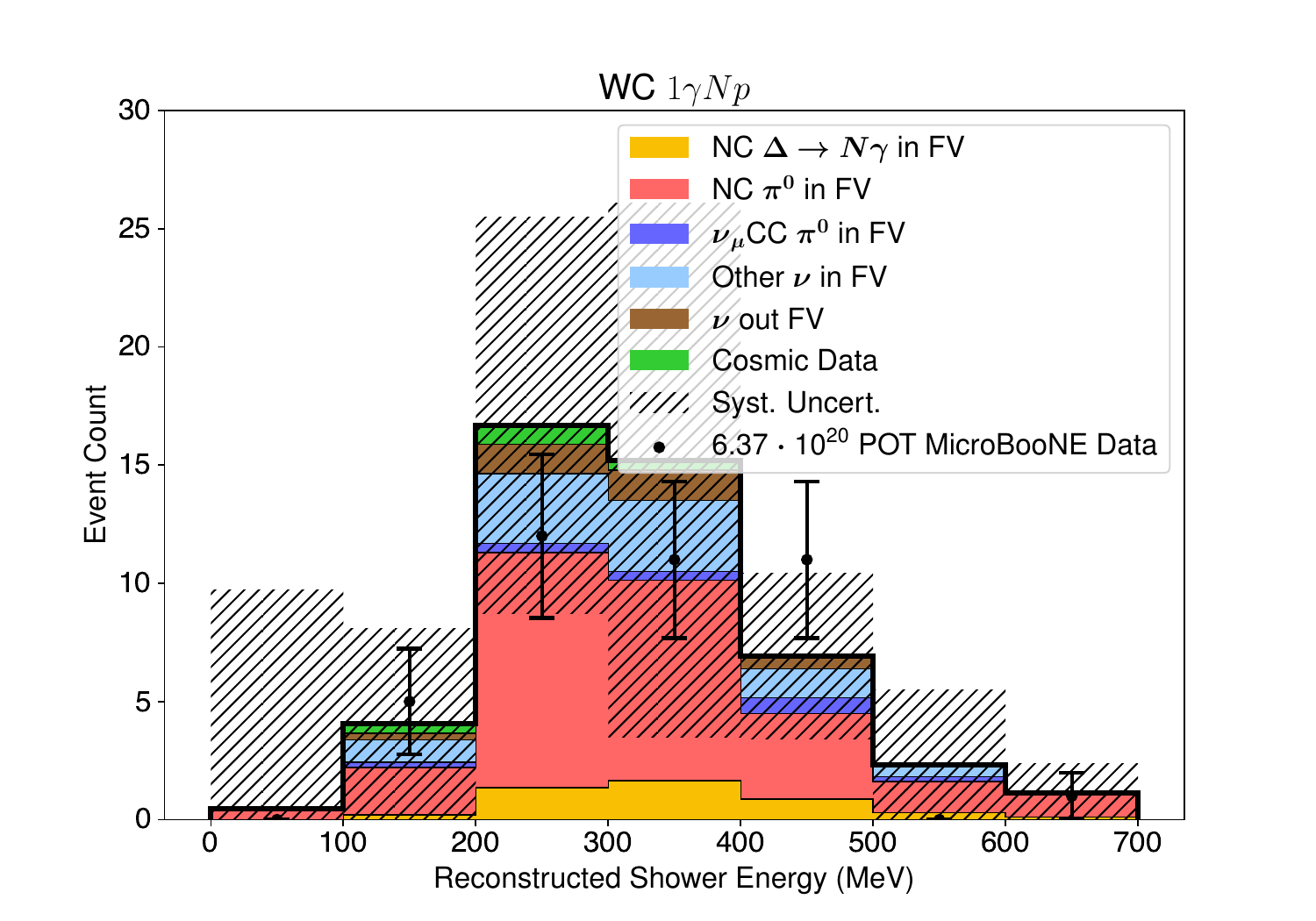}
        \caption{}
    \end{subfigure}
    \begin{subfigure}[b]{0.49\textwidth}
        \includegraphics[trim=20 0 60 0, clip, width=\textwidth]{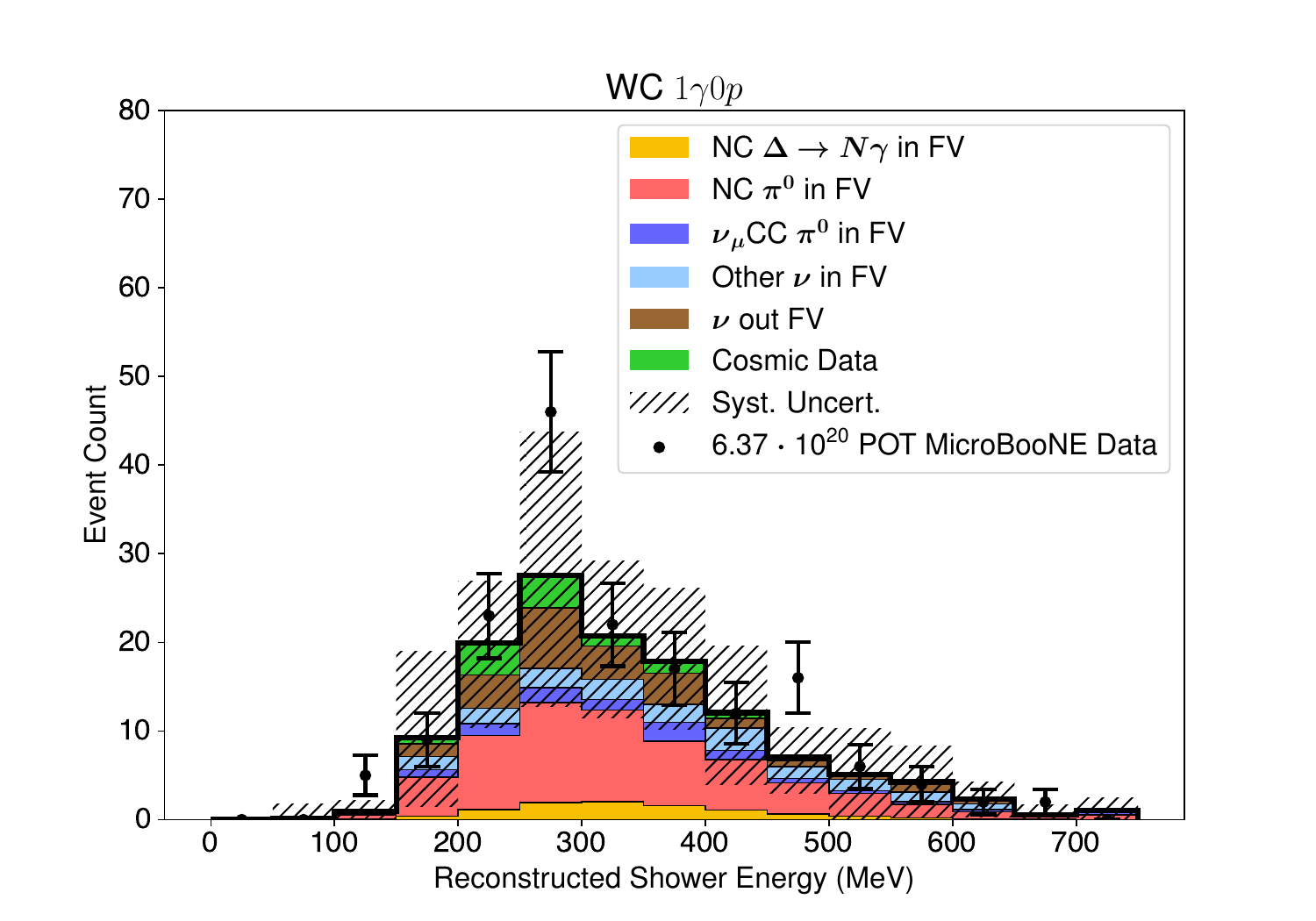}
        \caption{}
    \end{subfigure}
    \begin{subfigure}[b]{0.49\textwidth}
        \includegraphics[trim=20 0 60 0, clip, width=\textwidth]{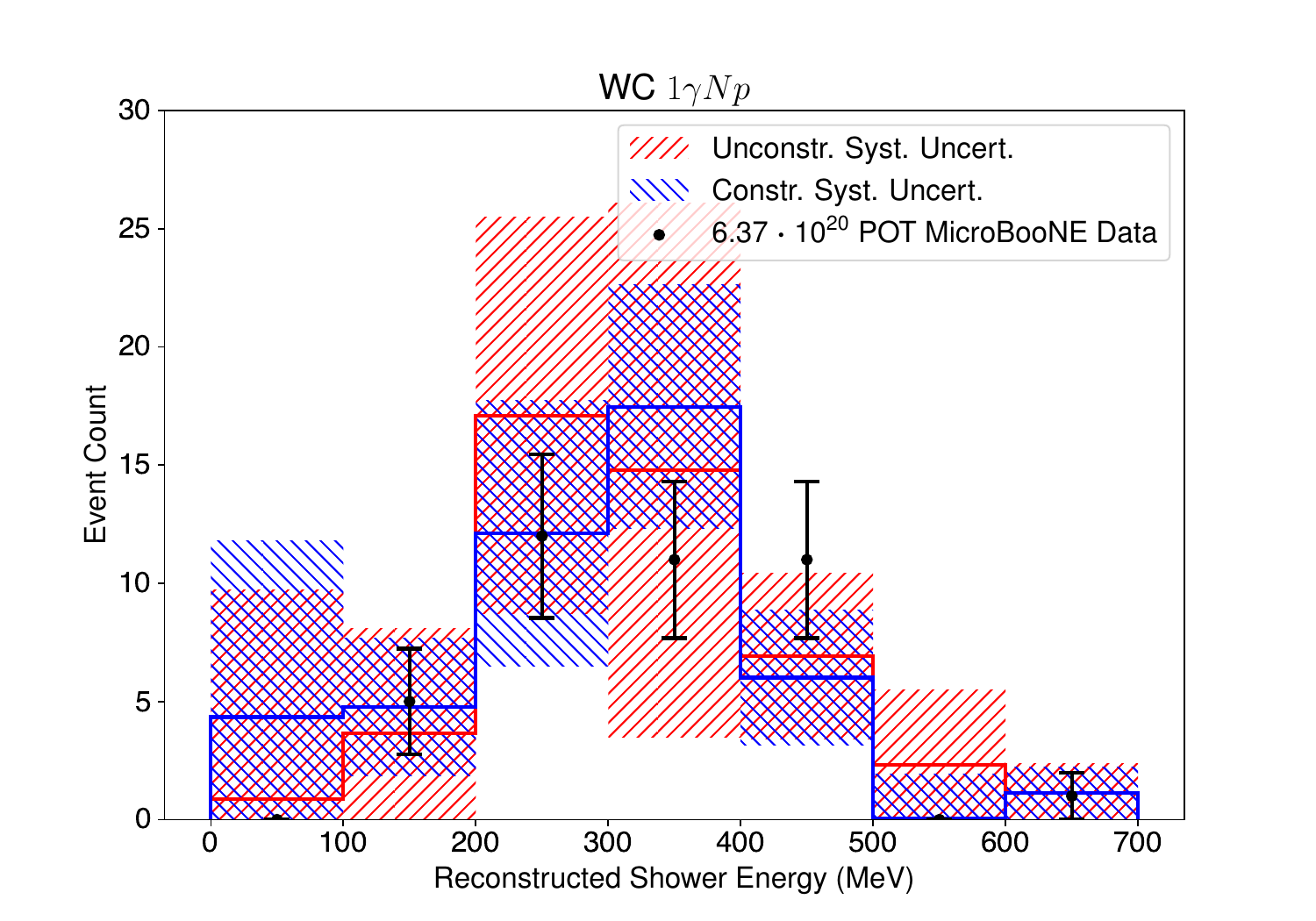}
        \caption{}
    \end{subfigure}
    \begin{subfigure}[b]{0.49\textwidth}
        \includegraphics[trim=20 0 60 0, clip, width=\textwidth]{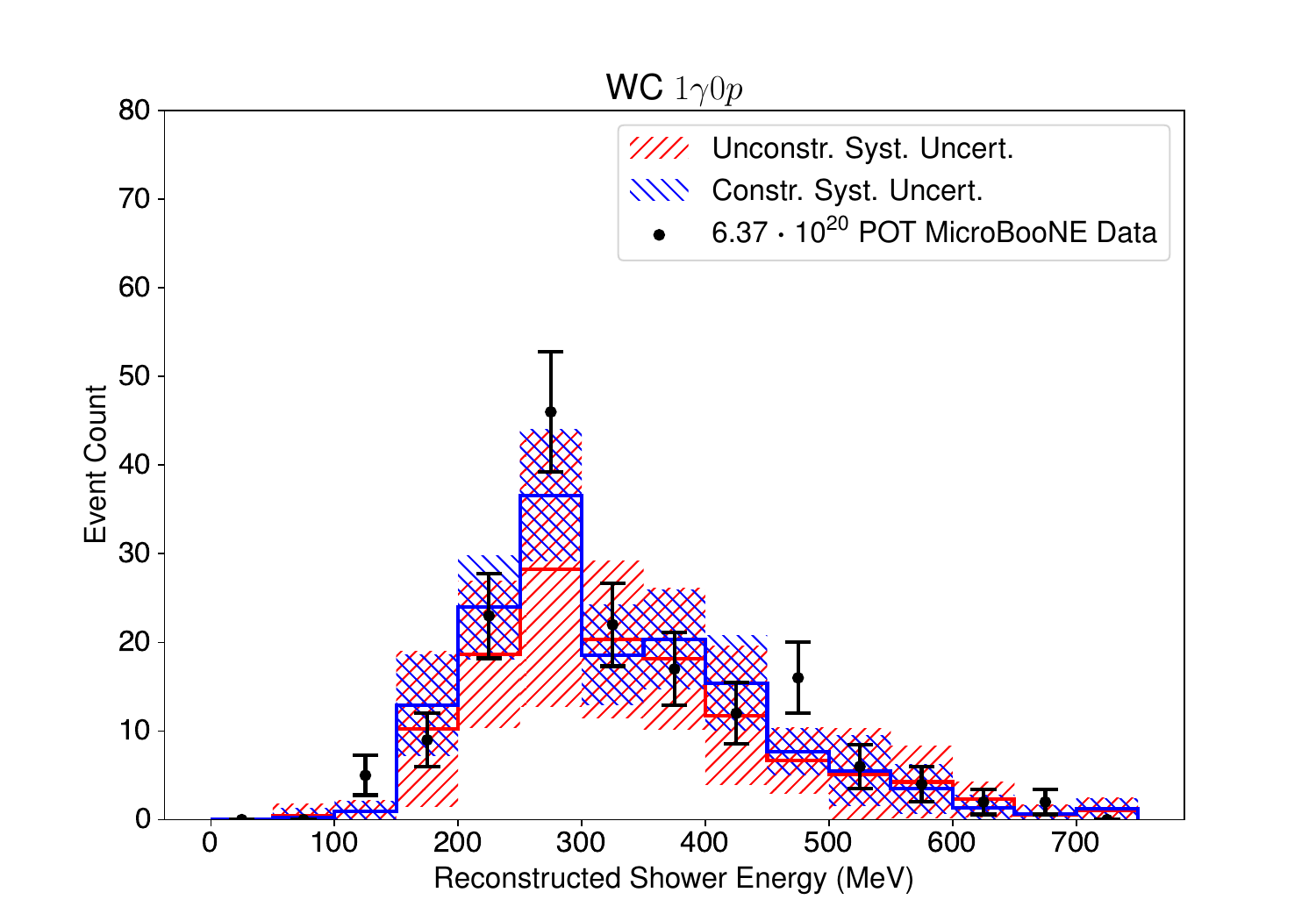}
        \caption{}
    \end{subfigure}
    \caption[Wire-Cell NC $\Delta\rightarrow N \gamma$ shower energy distributions]{Wire-Cell NC $\Delta\rightarrow N \gamma$ shower energy distributions. Panels (a) and (c) shows events with one or more reconstructed protons, and panels (b) and (d) show events with zero reconstructed protons. Panels (a) and (b) show the distributions with a breakdown of different prediction types, and with no conditional constraint. Panels (c) and (d) show distributions both before and after the application of the conditional constraint.}
    \label{fig:wc_shower_energy_distributions}
\end{figure}

Figure \ref{fig:pandora_shower_energy_distributions} shows reconstructed shower energy distributions for Pandora selected events. Figure \ref{fig:wc_shower_energy_distributions} shows reconstructed shower energy distributions for Wire-Cell selected events. We note that the local excess from 200-250 MeV in the Pandora $1\gamma 0p$ selection described in Sec. \ref{sec:pandora_nc_delta_search} does not appear in the corresponding Wire-Cell selection after the constraint is applied. We see generally good agreement within uncertainties in all distributions.

\begin{figure}[H]
    \centering
    \includegraphics[width=0.7\textwidth]{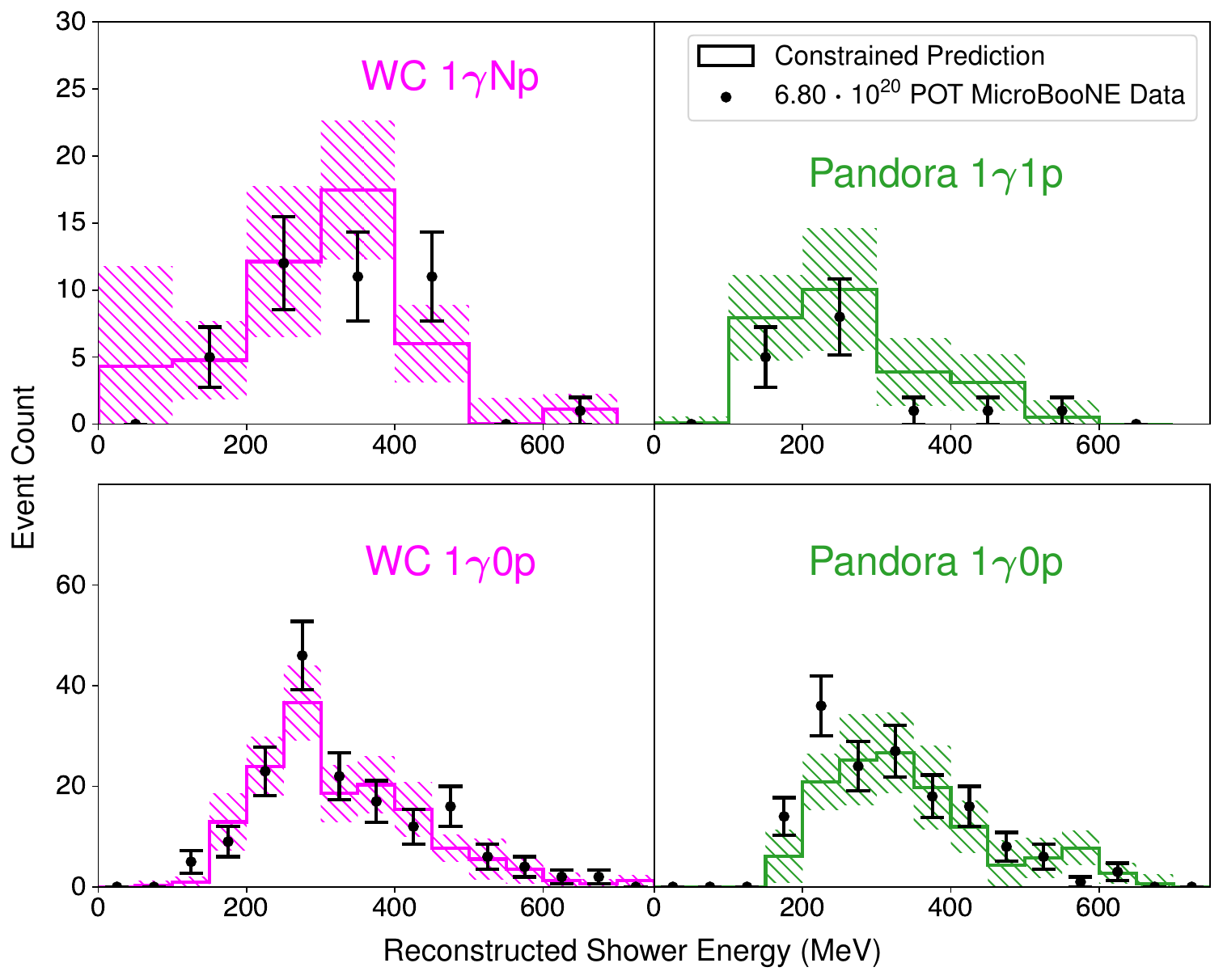}
    \caption[NC $\Delta\rightarrow N \gamma$ shower energy summary]{NC $\Delta\rightarrow N \gamma$ shower energy summary. The Pandora and Wire-Cell data samples correspond to $6.80\times 10^{20}$ and $6.37\times 10^{20}$ POT, respectively.}
    \label{fig:nc_delta_shower_energy_summary}
\end{figure}

Figure \ref{fig:nc_delta_shower_energy_summary} summarizes all four reconstructed shower energy distributions after constraints.

\begin{figure}[H]
    \centering
    \begin{subfigure}[b]{0.49\textwidth}
        \includegraphics[trim=20 0 55 0, clip, width=\textwidth]{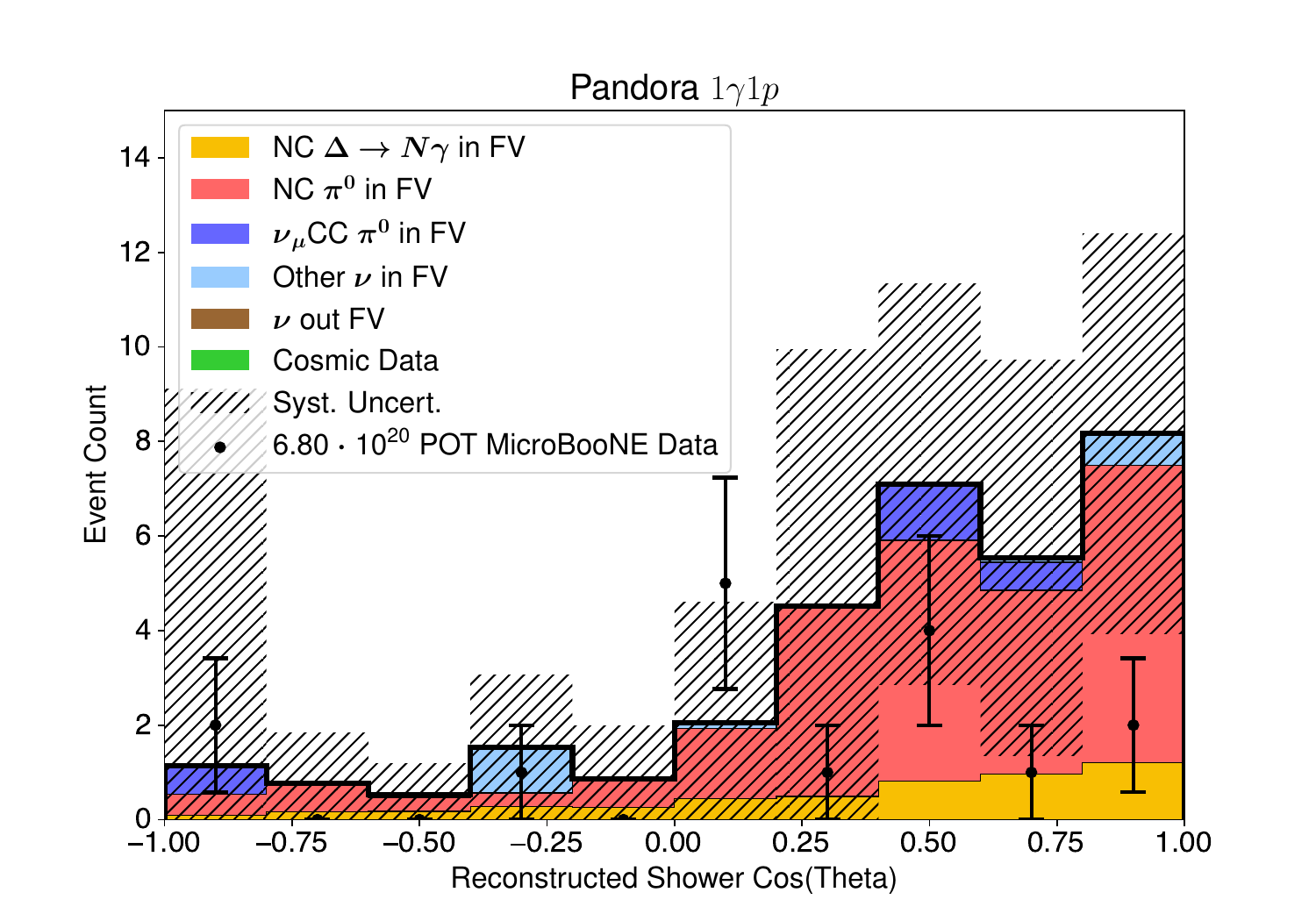}
        \caption{}
    \end{subfigure}
    \begin{subfigure}[b]{0.49\textwidth}
        \includegraphics[trim=20 0 55 0, clip, width=\textwidth]{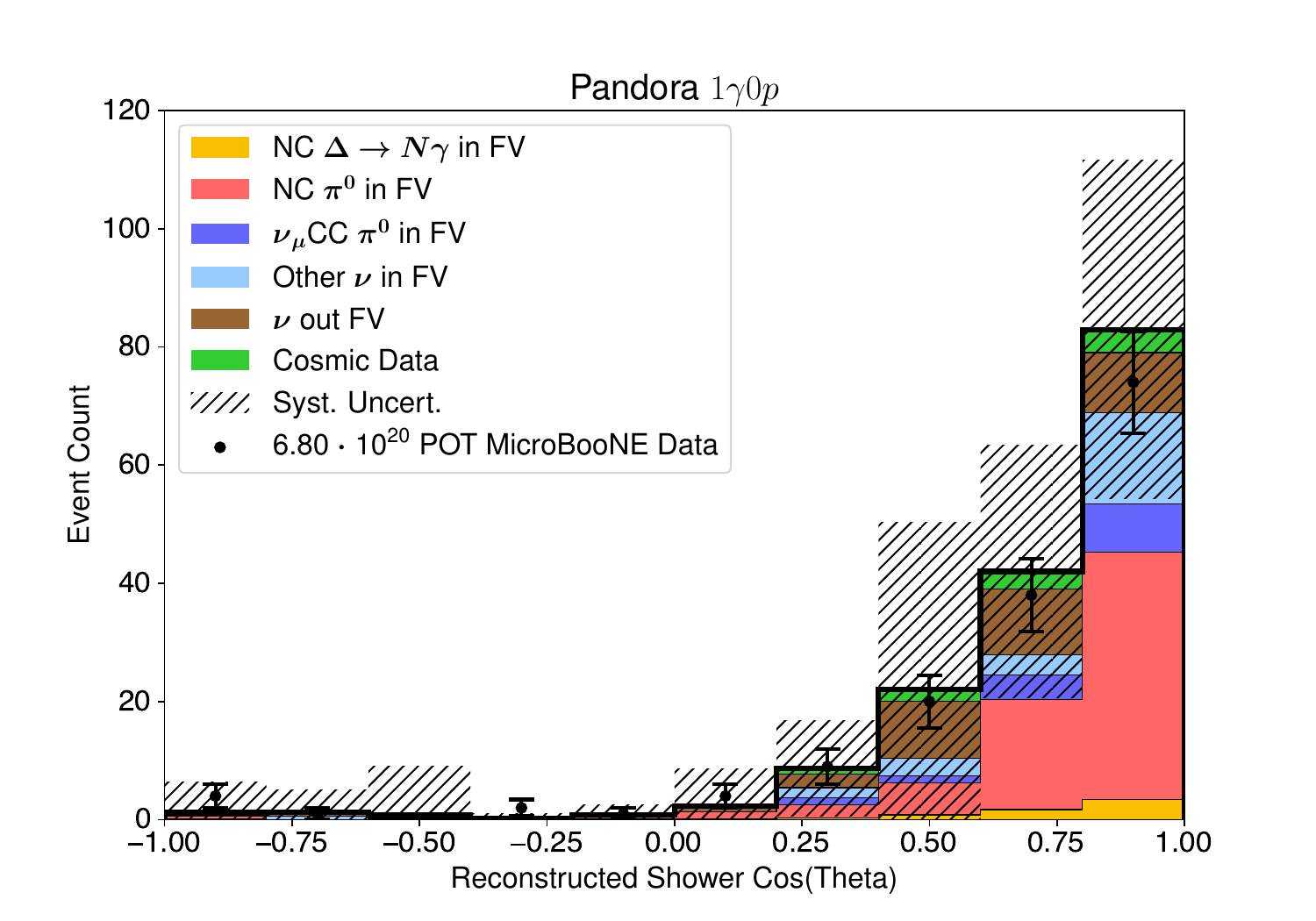}
        \caption{}
    \end{subfigure}
    \begin{subfigure}[b]{0.49\textwidth}
        \includegraphics[trim=20 0 55 0, clip, width=\textwidth]{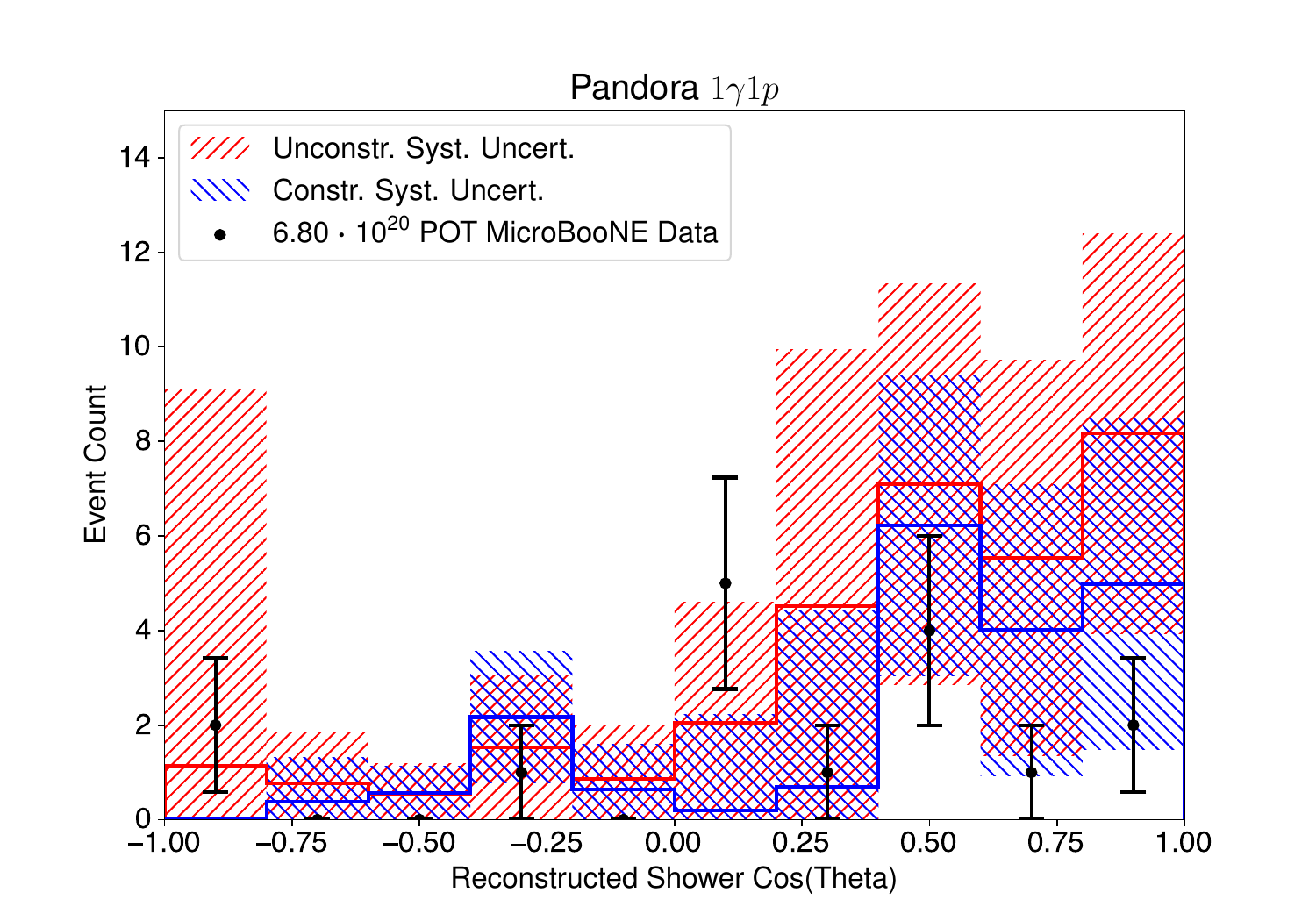}
        \caption{}
    \end{subfigure}
    \begin{subfigure}[b]{0.49\textwidth}
        \includegraphics[trim=20 0 55 0, clip, width=\textwidth]{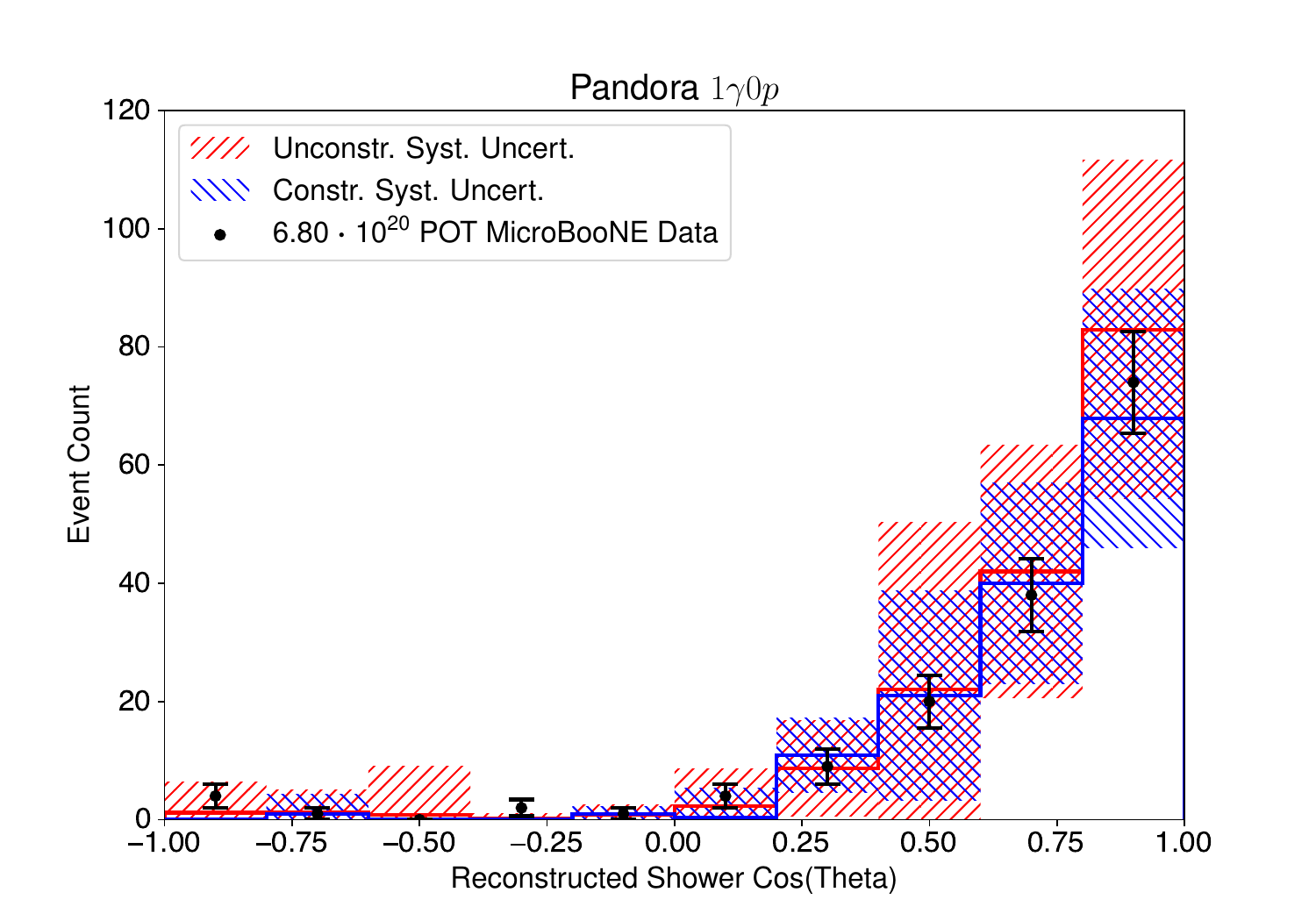}
        \caption{}
    \end{subfigure}
    \caption[Pandora NC $\Delta\rightarrow N \gamma$ shower angle distributions]{Pandora NC $\Delta\rightarrow N \gamma$ shower angle distributions. The x-axis shows the cosine of the angle between the reconstructed shower direction and the neutrino beam direction. Panels (a) and (c) shows events with one reconstructed proton, and panels (b) and (d) show events with zero reconstructed protons. Panels (a) and (b) show the distributions with a breakdown of different prediction types, and with no conditional constraint. Panels (c) and (d) show distributions both before and after the application of the conditional constraint.}
    \label{fig:pandora_shower_angle_distributions}
\end{figure}

\begin{figure}[H]
    \centering
    \begin{subfigure}[b]{0.49\textwidth}
        \includegraphics[trim=20 0 55 0, clip, width=\textwidth]{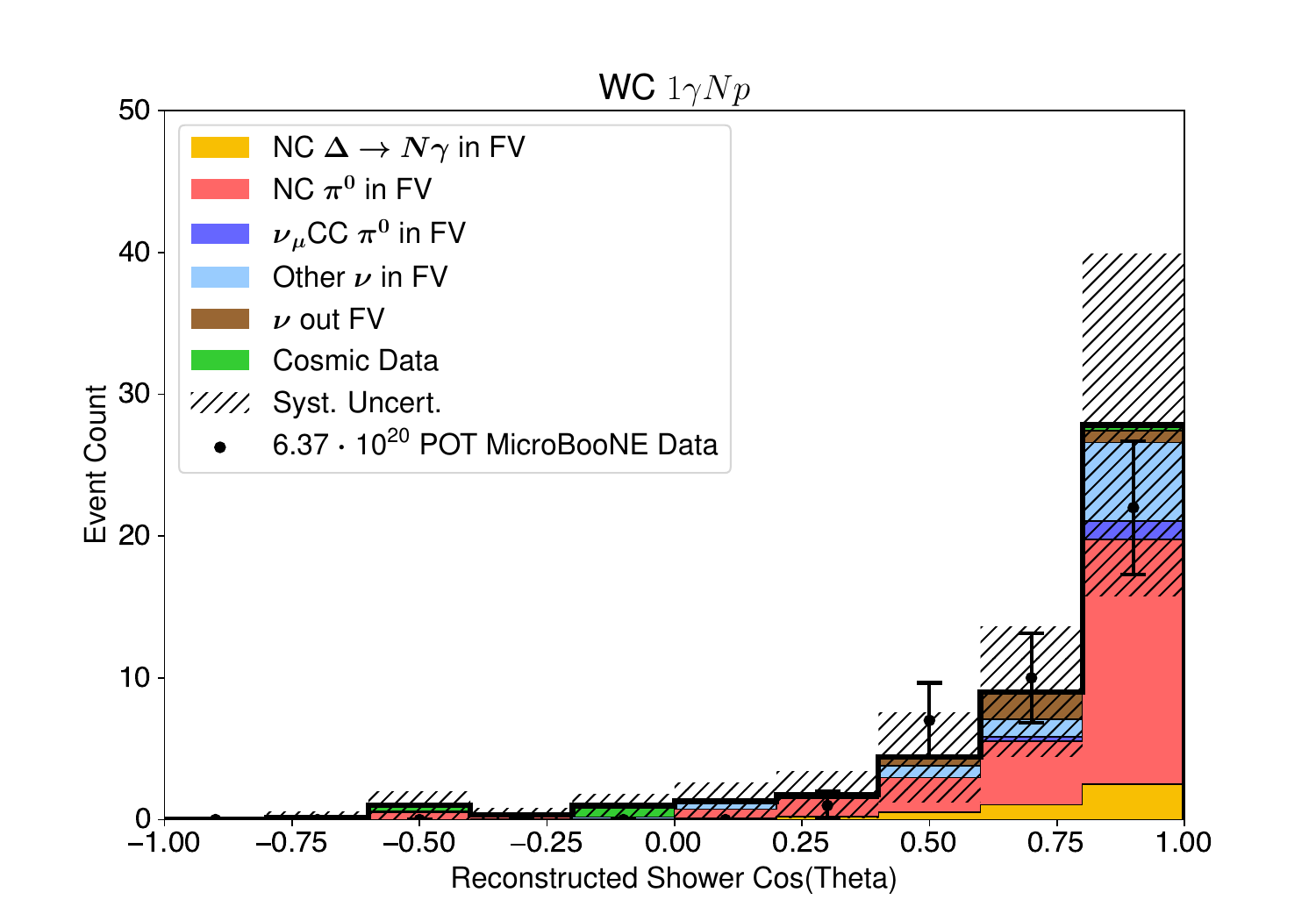}
        \caption{}
    \end{subfigure}
    \begin{subfigure}[b]{0.49\textwidth}
        \includegraphics[trim=20 0 55 0, clip, width=\textwidth]{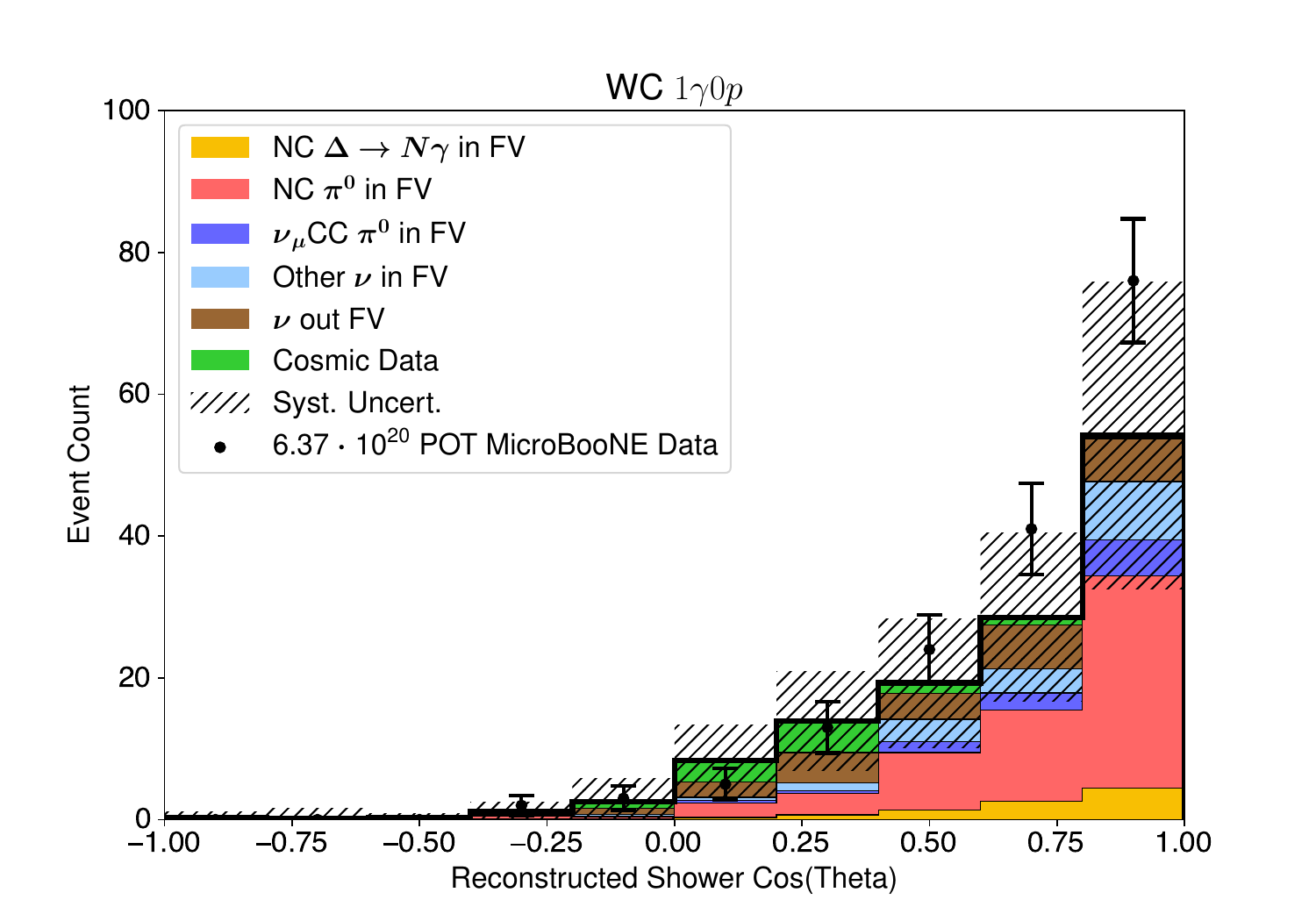}
        \caption{}
    \end{subfigure}
    \begin{subfigure}[b]{0.49\textwidth}
        \includegraphics[trim=20 0 55 0, clip, width=\textwidth]{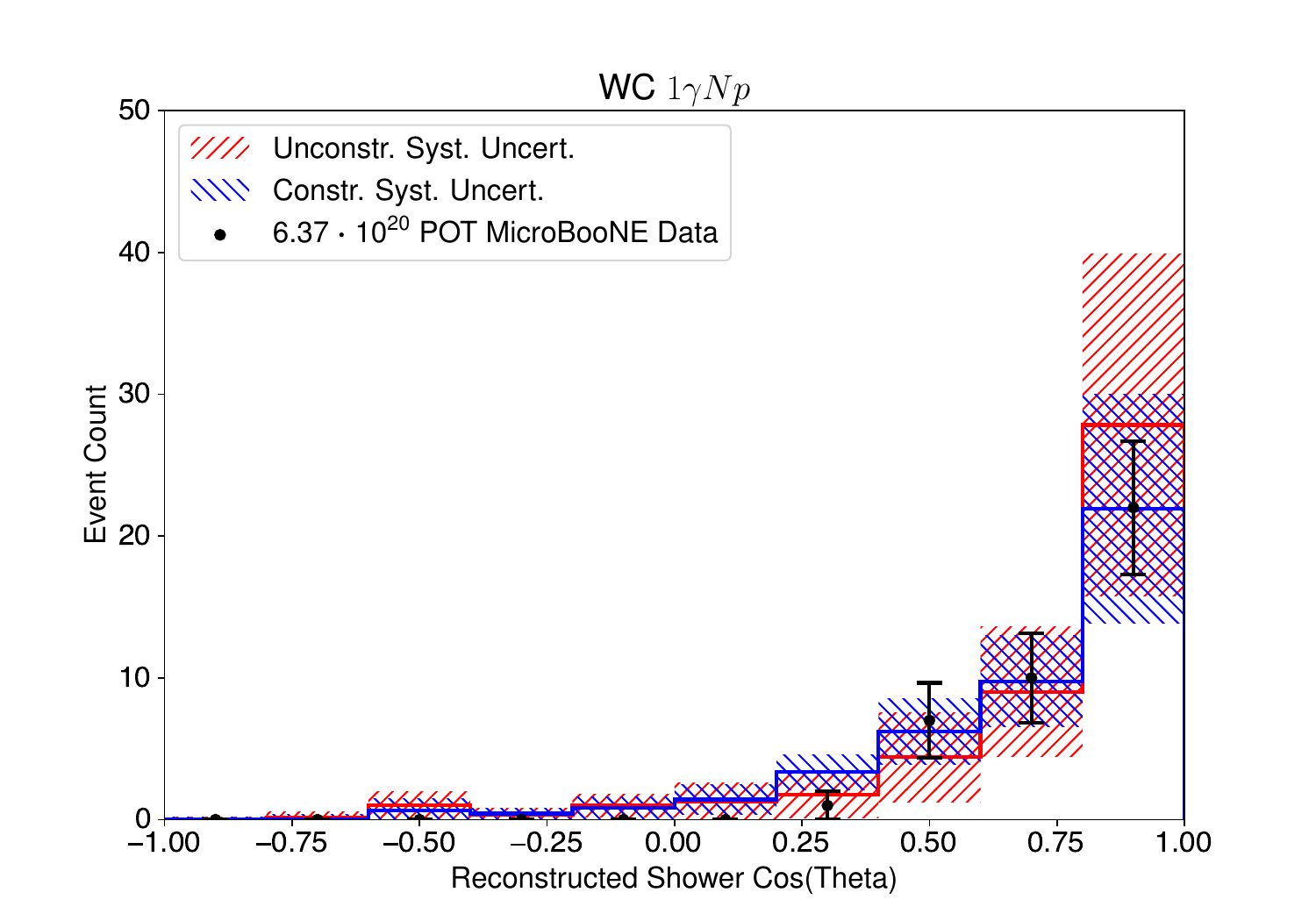}
        \caption{}
    \end{subfigure}
    \begin{subfigure}[b]{0.49\textwidth}
        \includegraphics[trim=20 0 55 0, clip, width=\textwidth]{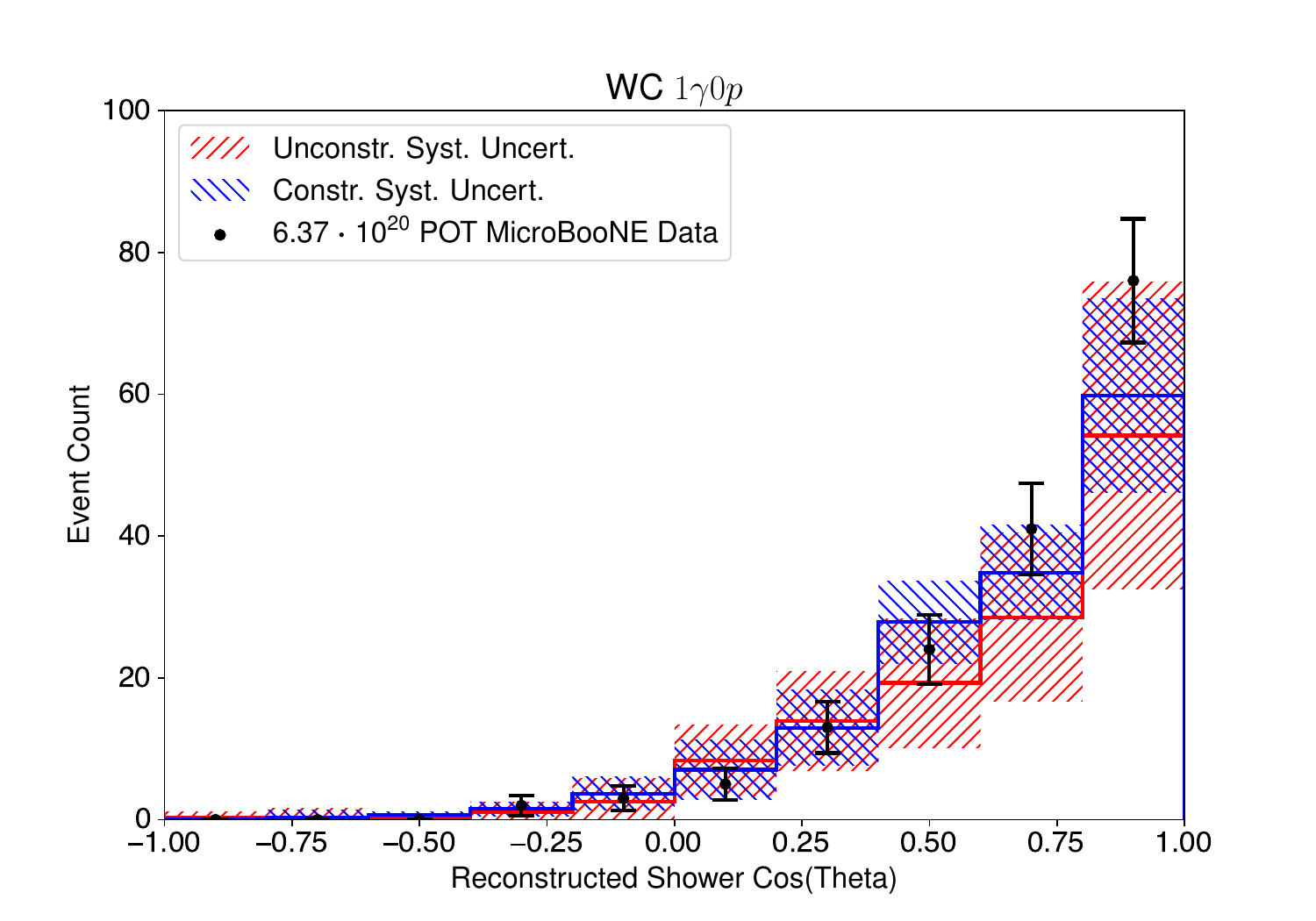}
        \caption{}
    \end{subfigure}
    \caption[Wire-Cell NC $\Delta\rightarrow N \gamma$ shower angle distributions]{Wire-Cell NC $\Delta\rightarrow N \gamma$ shower angle distributions. The x-axis shows the cosine of the angle between the reconstructed shower direction and the neutrino beam direction. Panels (a) and (c) shows events with one or more reconstructed protons, and panels (b) and (d) show events with zero reconstructed protons. Panels (a) and (b) show the distributions with a breakdown of different prediction types, and with no conditional constraint. Panels (c) and (d) show distributions both before and after the application of the conditional constraint.}
    \label{fig:wc_shower_angle_distributions}
\end{figure}

Figure \ref{fig:pandora_shower_angle_distributions} shows reconstructed shower angle distributions for Pandora selected events. Figure \ref{fig:wc_shower_angle_distributions} shows reconstructed shower angle distributions for Wire-Cell selected events. We see generally good agreement within uncertainties in all distributions.

\begin{figure}[H]
    \centering
    \begin{subfigure}[b]{0.32\textwidth}
        \includegraphics[trim=20 0 100 0, clip, width=\textwidth]{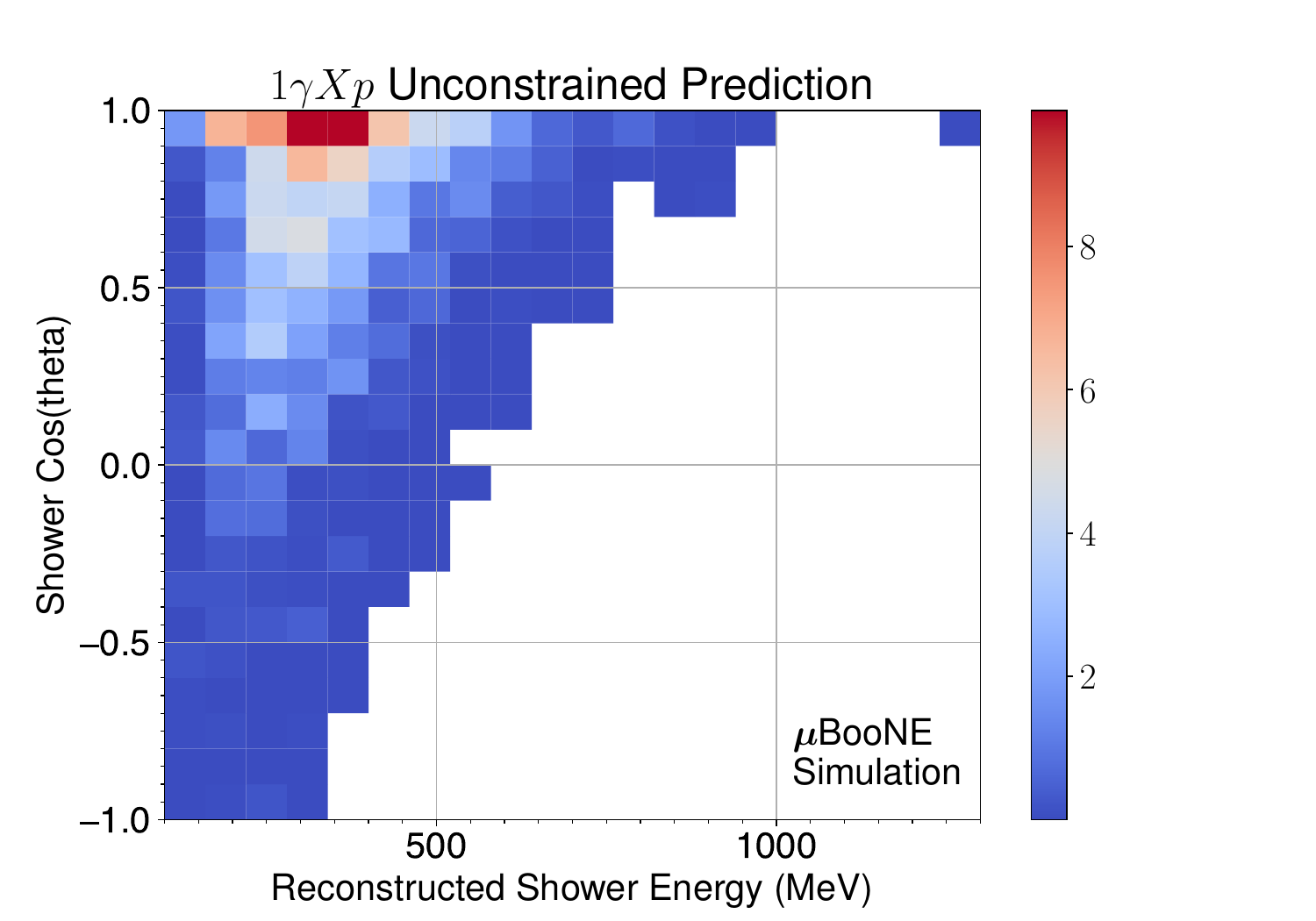}
        \caption{}
    \end{subfigure}
    \begin{subfigure}[b]{0.32\textwidth}
        \includegraphics[trim=20 0 100 0, clip, width=\textwidth]{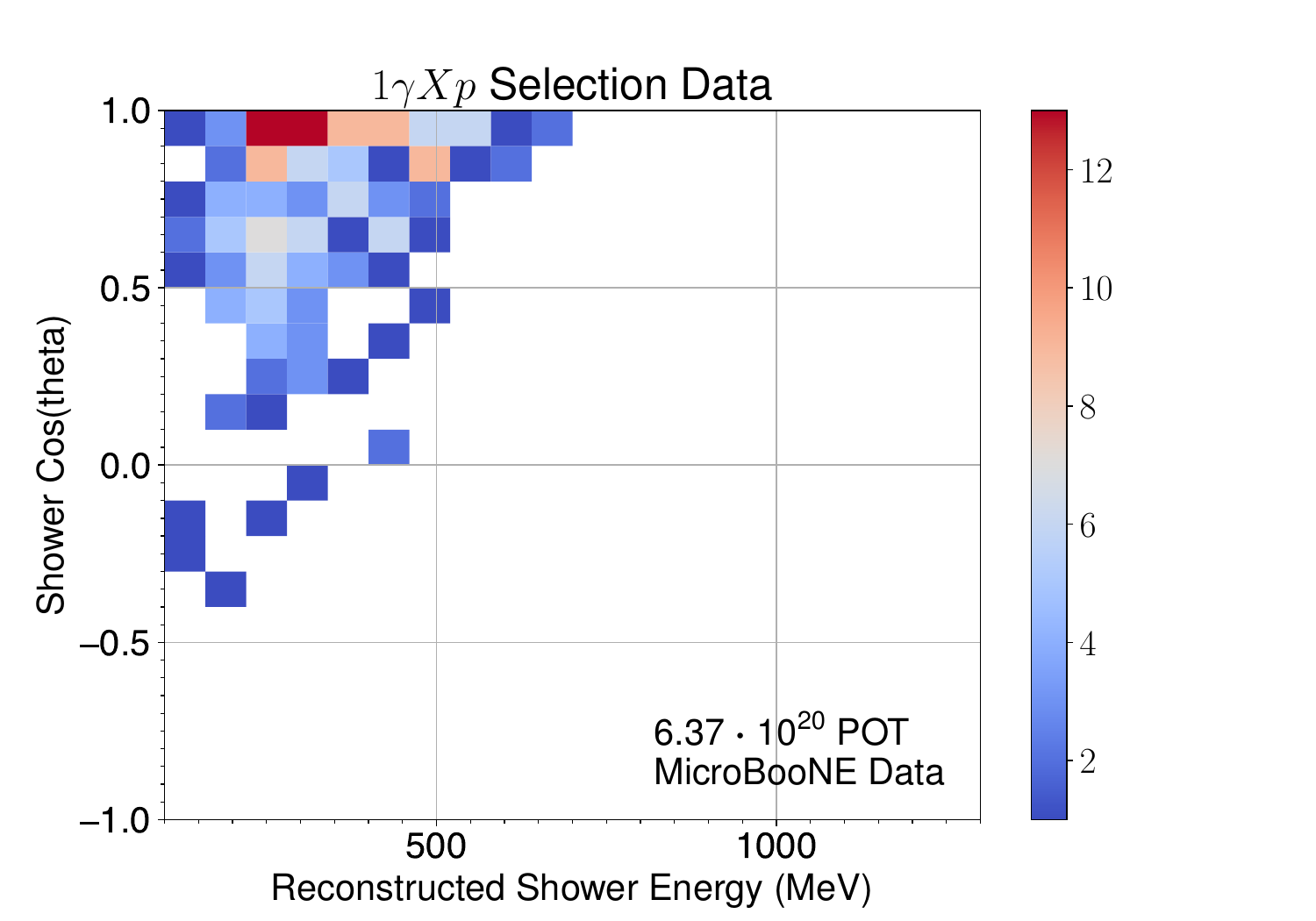}
        \caption{}
    \end{subfigure}
    \begin{subfigure}[b]{0.32\textwidth}
        \includegraphics[trim=20 0 100 0, clip, width=\textwidth]{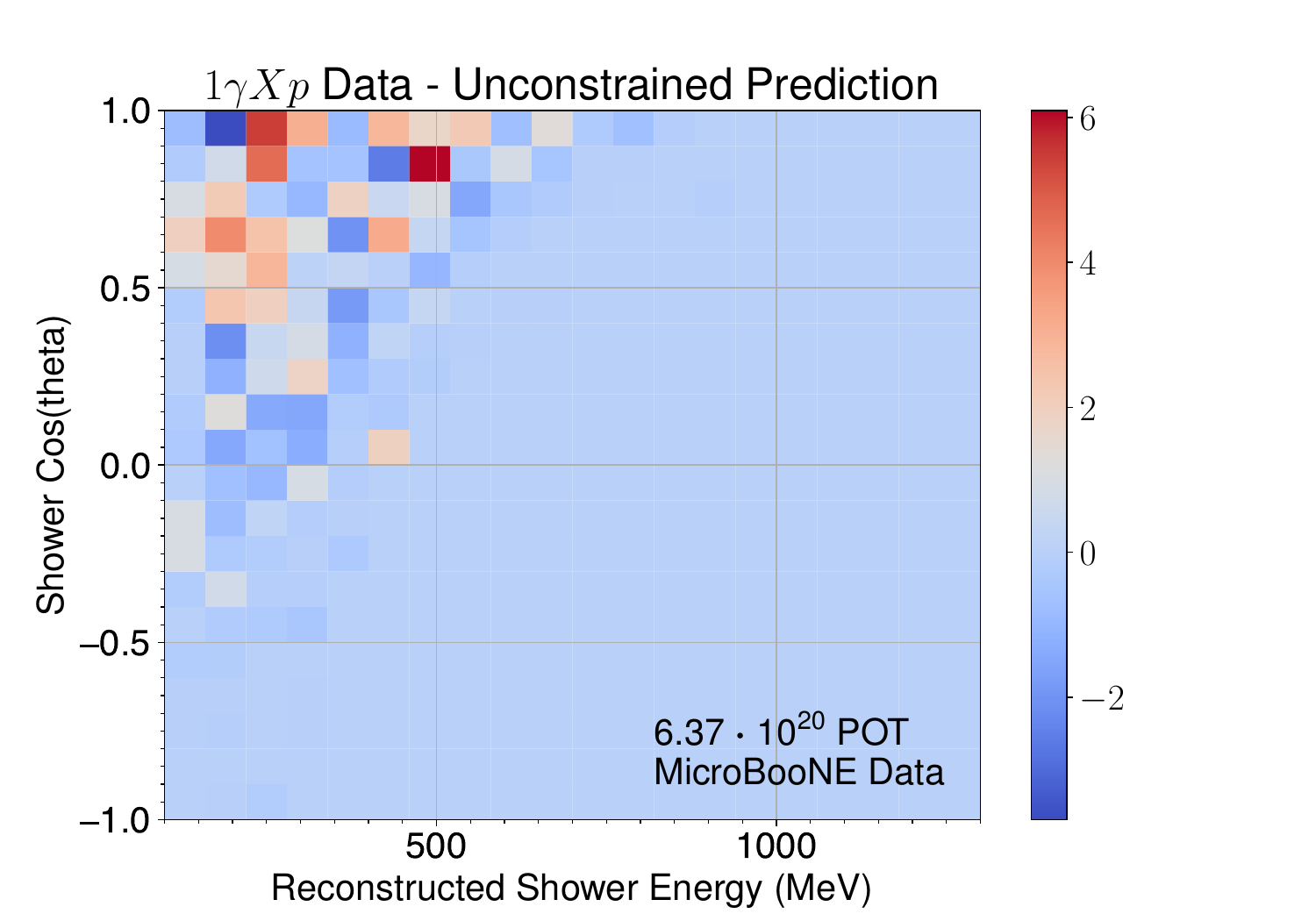}
        \caption{}
    \end{subfigure}
    \caption[Wire-Cell NC $\Delta\rightarrow N \gamma$ 2D shower kinematics]{Wire-Cell NC $\Delta\rightarrow N \gamma$ 2D shower kinematics, with no conditional constrained applied. Panel (a) shows the predicted events, panel (b) shows the observed data events, and panel (c) shows data minus prediction. Events with and without reconstructed final state protons have both been included. }
    \label{fig:nc_delta_2d_shower_kinematics_unconstr}
\end{figure}

\begin{figure}[H]
    \centering
    \begin{subfigure}[b]{0.32\textwidth}
        \includegraphics[trim=20 0 100 0, clip, width=\textwidth]{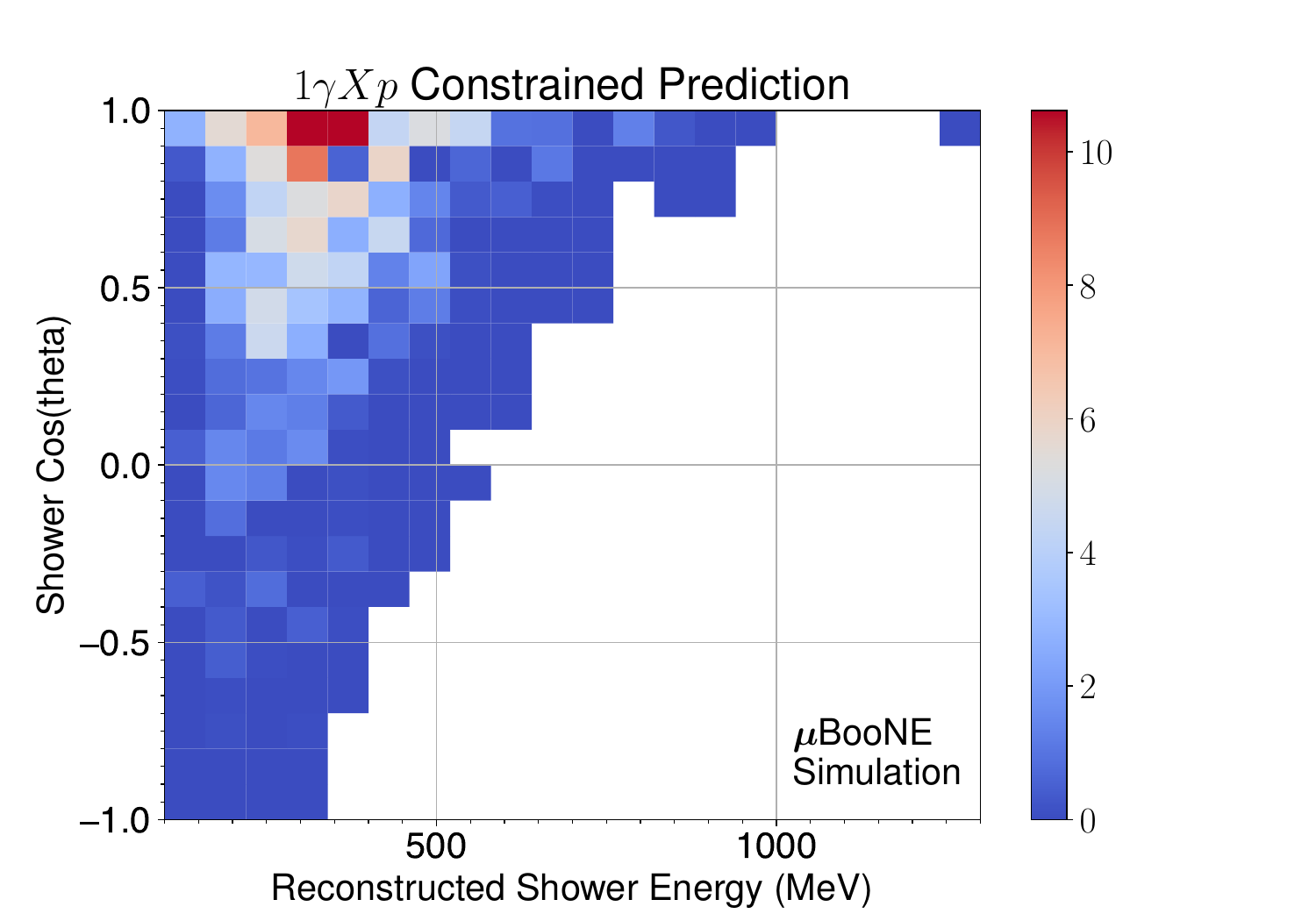}
        \caption{}
    \end{subfigure}
    \begin{subfigure}[b]{0.32\textwidth}
        \includegraphics[trim=20 0 100 0, clip, width=\textwidth]{figs/nc_delta/histogram_plots/1gXp_2d_energy_angle_data.pdf}
        \caption{}
    \end{subfigure}
    \begin{subfigure}[b]{0.32\textwidth}
        \includegraphics[trim=20 0 100 0, clip, width=\textwidth]{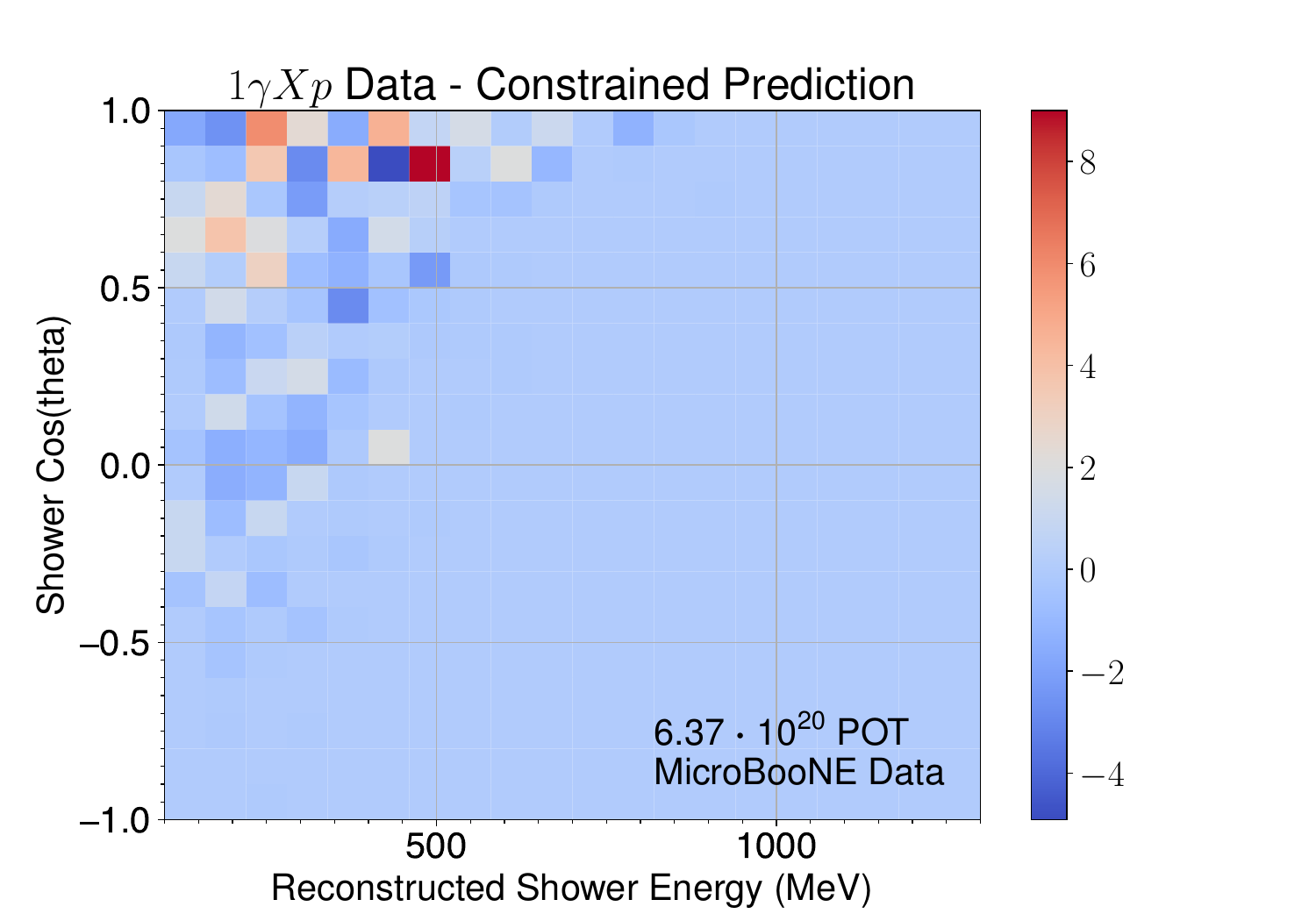}
        \caption{}
    \end{subfigure}
    \caption[Wire-Cell NC $\Delta\rightarrow N \gamma$ 2D shower kinematics with constraint]{Wire-Cell NC $\Delta\rightarrow N \gamma$ 2D shower kinematics, with the conditional constrained applied. Panel (a) shows the predicted events, panel (b) shows the observed data events, and panel (c) shows data minus prediction. Events with and without reconstructed final state protons have both been included. }
    \label{fig:nc_delta_2d_shower_kinematics_constr}
\end{figure}

\begin{figure}[H]
    \centering
    \begin{subfigure}[b]{0.32\textwidth}
        \begin{overpic}[trim=20 0 80 0, clip, width=\textwidth]{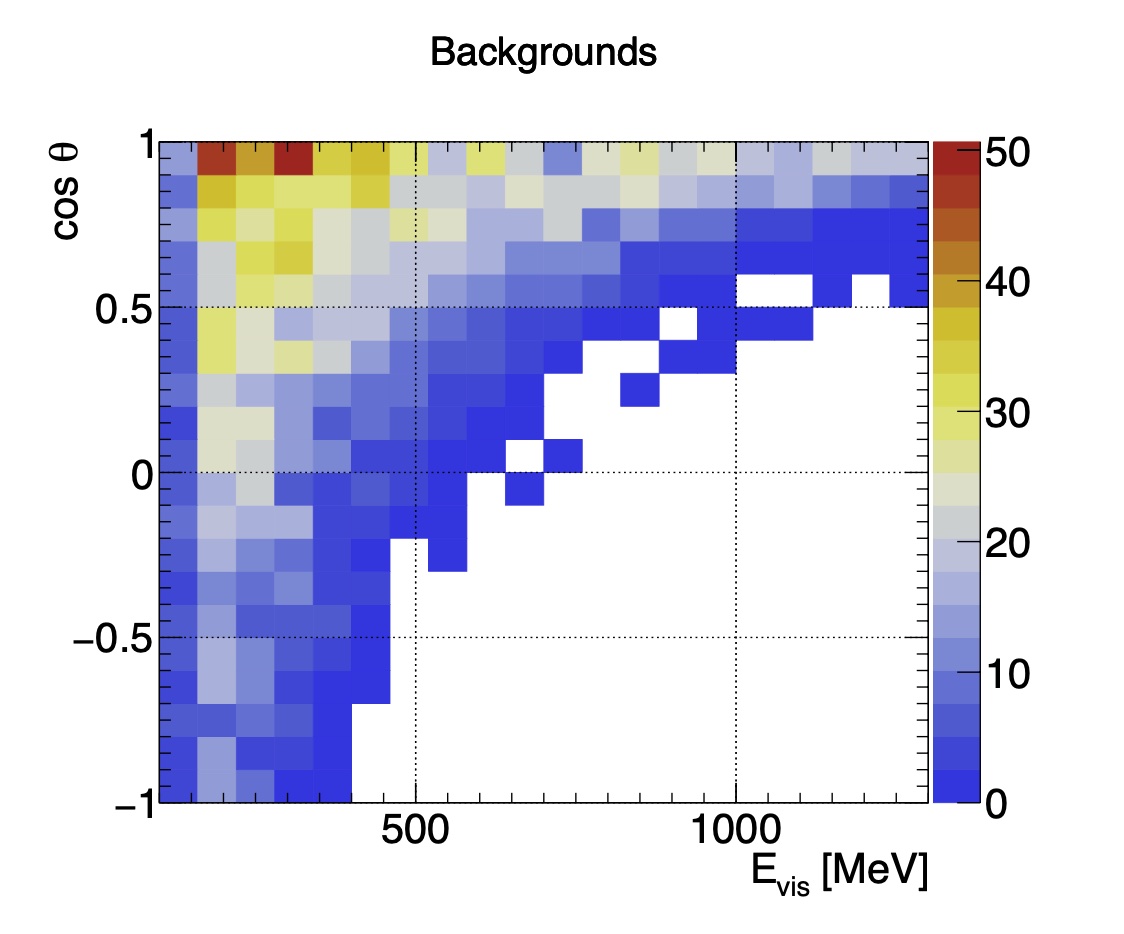}
            \put(45,20){\small MiniBooNE}
        \end{overpic}
        \caption{}
    \end{subfigure}
    \begin{subfigure}[b]{0.32\textwidth}
        \begin{overpic}[trim=20 0 80 0, clip, width=\textwidth]{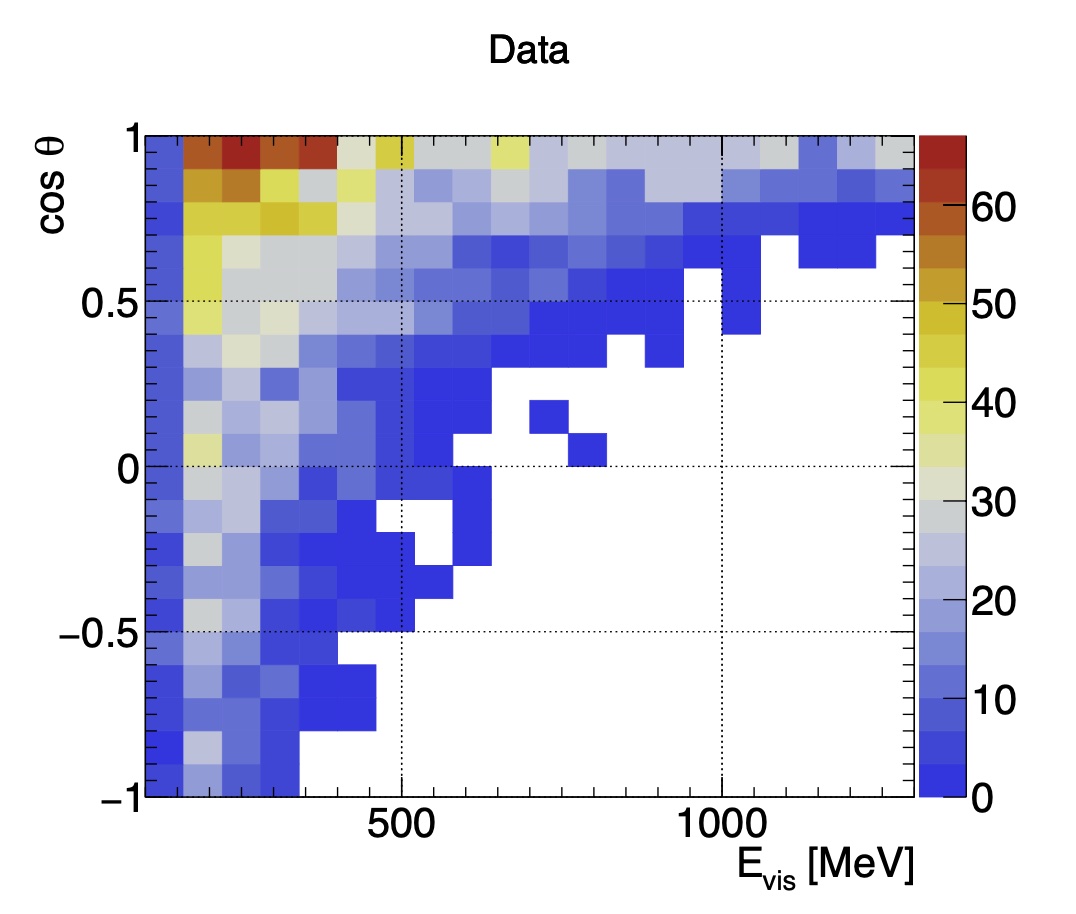}
            \put(45,20){\small MiniBooNE}
        \end{overpic}
        \caption{}
    \end{subfigure}
    \begin{subfigure}[b]{0.32\textwidth}
        \begin{overpic}[trim=20 0 80 0, clip, width=\textwidth]{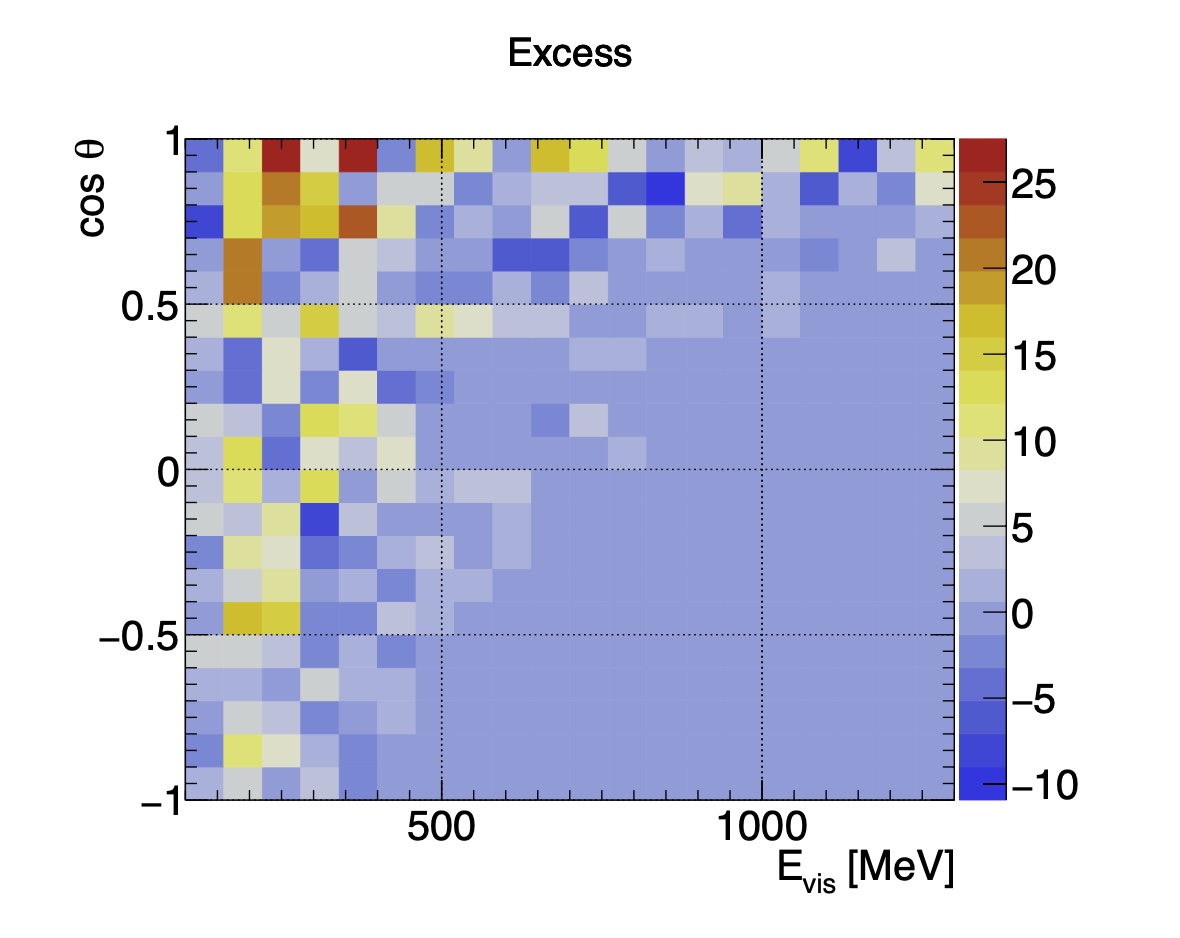}
            \put(45,20){\small MiniBooNE}
        \end{overpic}
        \caption{}
    \end{subfigure}
    \caption[MiniBooNE LEE 2D shower kinematics]{MiniBooNE LEE 2D shower kinematics. Panel (a) shows the predicted events, panel (b) shows the observed data events, and panel (c) shows data minus prediction. From Ref. \cite{miniboone_lee}.}
    \label{fig:miniboone_2d_shower_kinematics}
\end{figure}

Next, we consider 2D reconstructed shower energy and angle distributions for the Wire-Cell selection. Figure \ref{fig:nc_delta_2d_shower_kinematics_unconstr} shows the prediction, data, and excess distributions with no conditional constraint applied. Figure \ref{fig:nc_delta_2d_shower_kinematics_constr} shows the same result after the conditional constraints have been applied. Figure \ref{fig:miniboone_2d_shower_kinematics} shows the same visualization for the MiniBooNE LEE, from Ref. \cite{miniboone_lee}. Note that this 2D format does not directly allow for the visualization of any uncertainties, but statistical and other uncertainties remain large in these plots.

\begin{figure}[H]
    \centering
    \begin{subfigure}[b]{0.49\textwidth}
        \includegraphics[trim=20 0 60 0, clip, width=\textwidth]{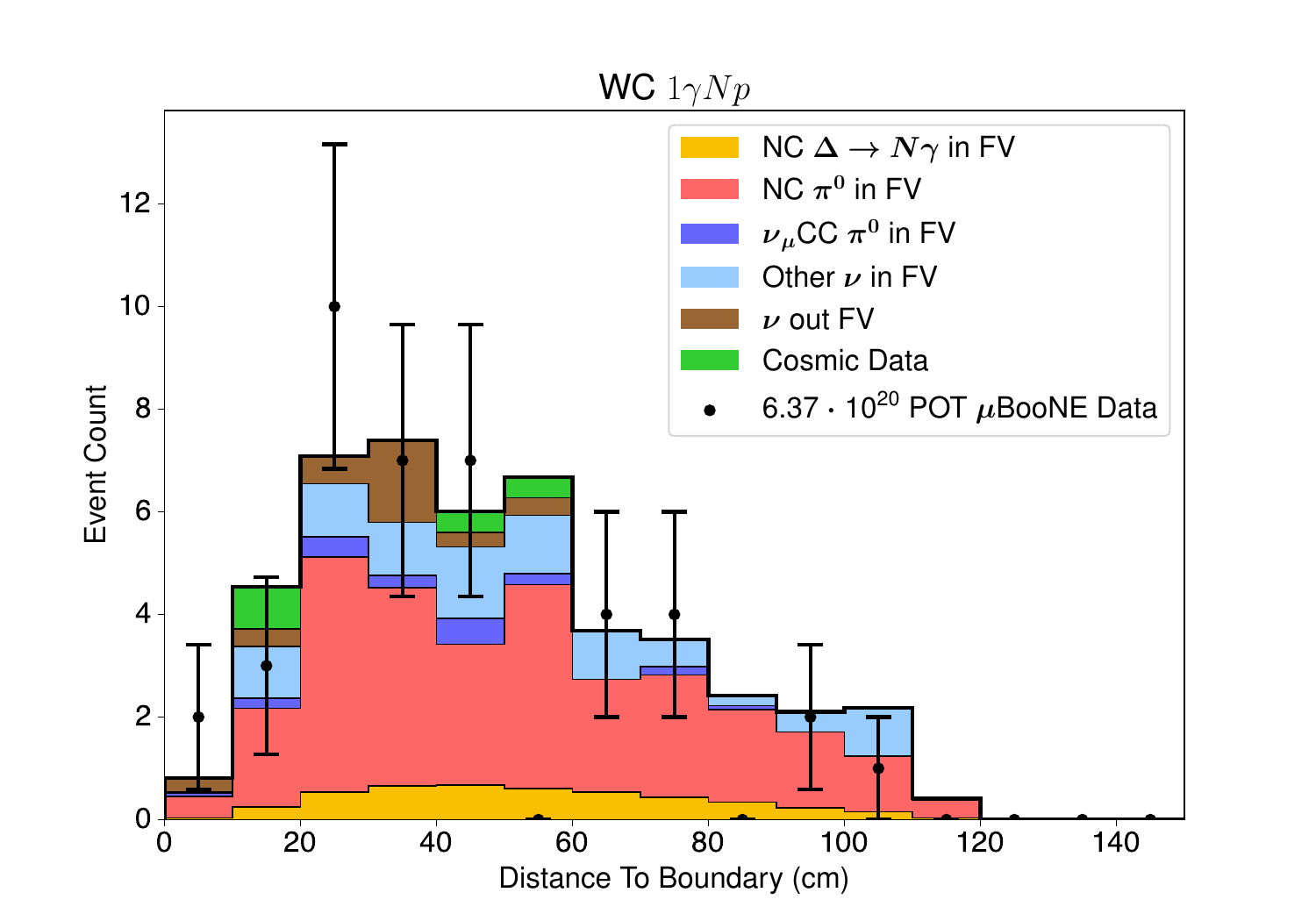}
        \caption{}
    \end{subfigure}
    \begin{subfigure}[b]{0.49\textwidth}
        \includegraphics[trim=20 0 60 0, clip, width=\textwidth]{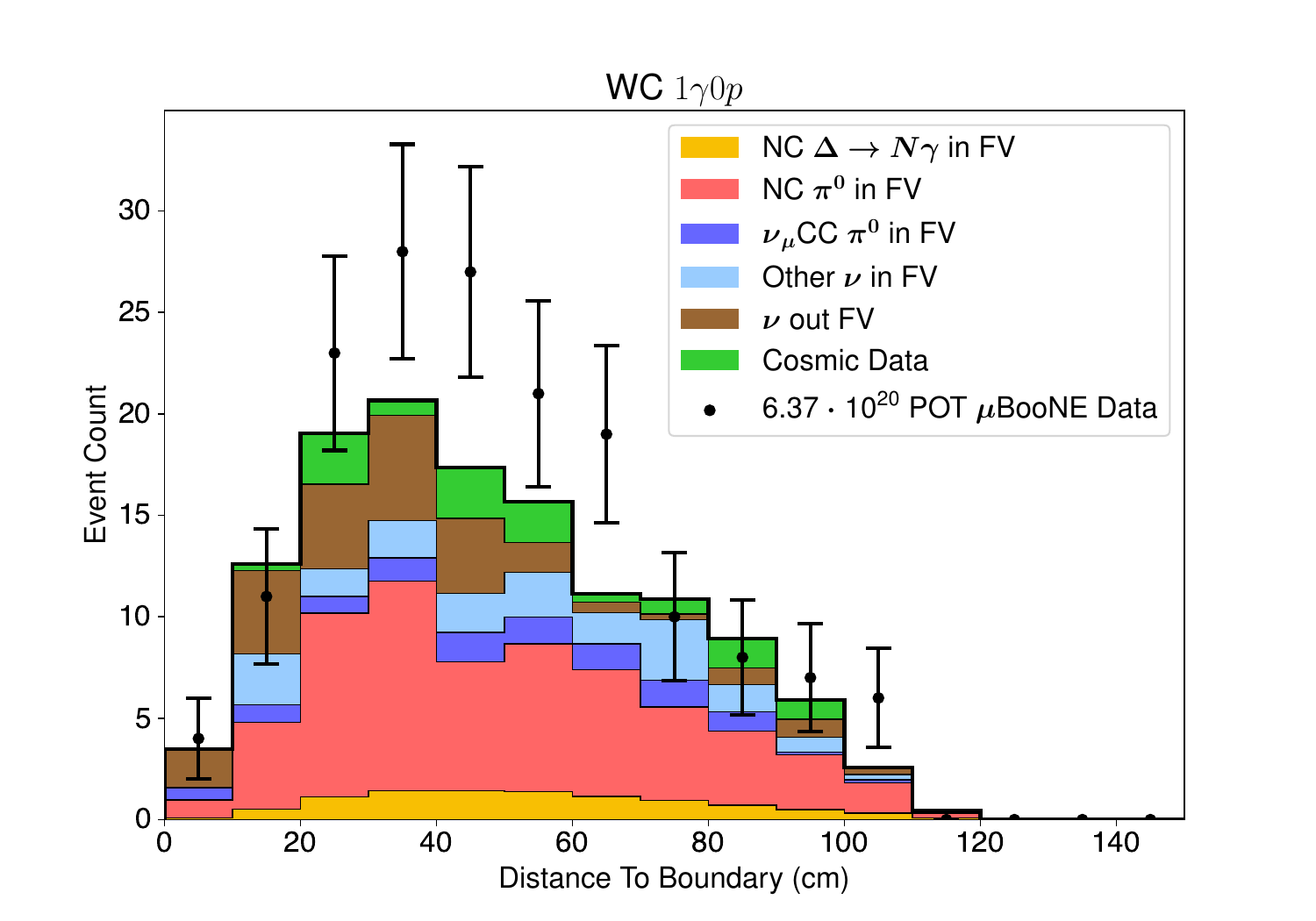}
        \caption{}
    \end{subfigure}
    \caption[Wire-Cell NC $\Delta\rightarrow N \gamma$ distance to TPC boundary]{Wire-Cell NC $\Delta\rightarrow N \gamma$ distance to the nearest TPC boundary. Panel (a) shows events with reconstructed protons, and panel (b) shows events without reconstructed protons. No systematic uncertainties are considered, and no conditional constraint has been applied.}
    \label{fig:distance_to_boundary}
\end{figure}

\begin{figure}[H]
    \centering
    \begin{subfigure}[b]{0.49\textwidth}
        \includegraphics[trim=20 0 50 0, clip, width=\textwidth]{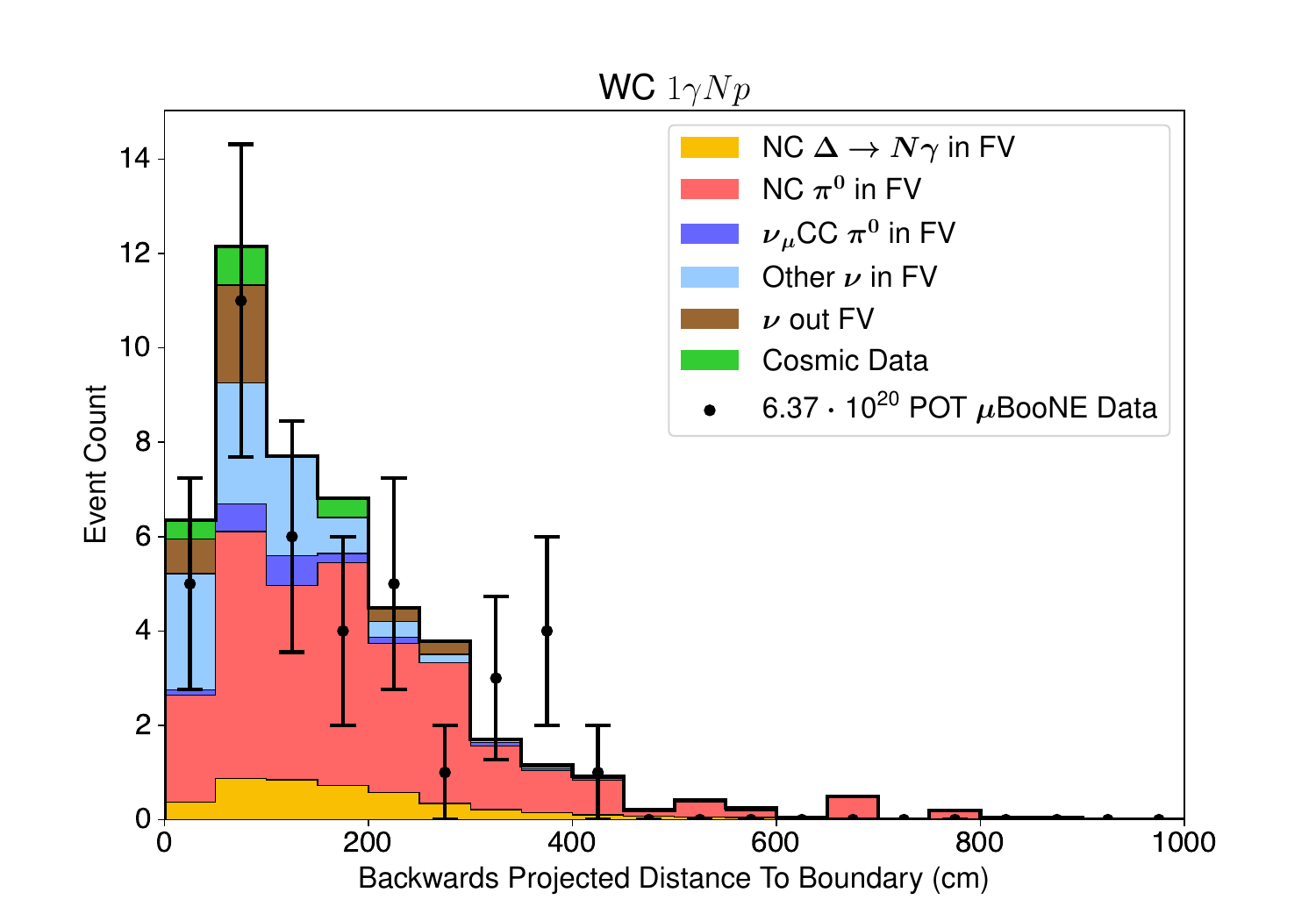}
        \caption{}
    \end{subfigure}
    \begin{subfigure}[b]{0.49\textwidth}
        \includegraphics[trim=20 0 50 0, clip, width=\textwidth]{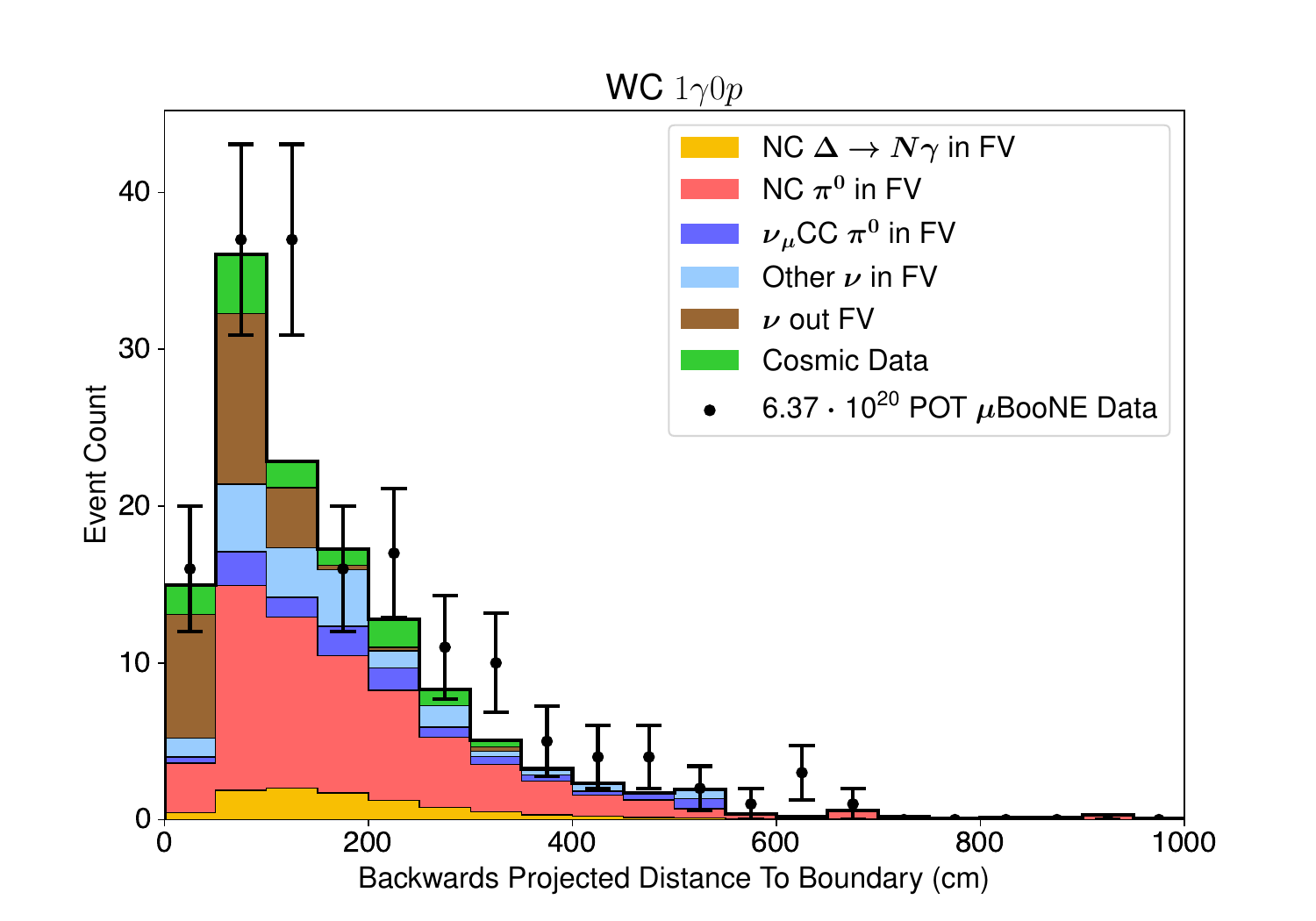}
        \caption{}
    \end{subfigure}
    \caption[Wire-Cell NC $\Delta\rightarrow N \gamma$ backwards projected distance to TPC boundary]{Wire-Cell NC $\Delta\rightarrow N \gamma$ backwards projected distance to the nearest TPC boundary. Panel (a) shows events with reconstructed protons, and panel (b) shows events without reconstructed protons. No systematic uncertainties are considered, and no conditional constraint has been applied.}
    \label{fig:backward_projected_distance}
\end{figure}

We also investigate the position dependence of the showers in Wire-Cell NC $\Delta\rightarrow N \gamma$ selected events. Figure \ref{fig:distance_to_boundary} shows the distance between the reconstructed shower vertex and the nearest TPC boundary, again highlighting how there are no points in the detector further than about 116 cm from a TPC boundary. We also investigate the distance to the detector boundary along the projected direction upstream of the reconstructed shower, as shown in Fig. \ref{fig:backward_projected_distance}. We see no notable data/prediction differences in any of these distributions.

\subsection{Tests of the Enhanced NC \texorpdfstring{$\Delta$}{Delta} MiniBooNE LEE Hypothesis}

Qualitatively, from these plots and goodness of fit tests, we do not see signs of a significant NC $\Delta\rightarrow N \gamma$ excess in our data. To quantify what exactly these results can tell us about the MiniBooNE LEE, we test the hypothesis that the MiniBooNE LEE is due to a $3.18$ times enhanced rate of NC $\Delta\rightarrow N \gamma$ as described in Sec. \ref{sec:miniboone_318}. We refer to the scaling of NC $\Delta\rightarrow N \gamma$ events as $x_\Delta$. We assume that this same multiplicative increase would apply in MicroBooNE, and search for an excess due to this $x_\Delta=3.18$ scaling.

Table \ref{tab:gof_tests} shows similar goodness of fit tests as shown earlier, but now also comparing to this $x_\Delta=3.18$ LEE hypothesis. In general, we see worse agreement with this LEE prediction relative to the nominal $x_\Delta=1$ hypothesis.

\begin{table}[H]
    \centering
    \resizebox{\textwidth}{!}{
    \begin{tabular}{c c c c c} 
        \toprule
        Selection & \makecell{$x_\Delta=1$ Unconstrained \\ $\chi^2$/ndf,  p-value, $\sigma$} & \makecell{$x_\Delta=1$ Constrained \\ $\chi^2$/ndf,  p-value, $\sigma$} & \makecell{$x_\Delta=3.18$ Unconstrained \\ $\chi^2$/ndf,  p-value, $\sigma$} & \makecell{$x_\Delta=3.18$ Constrained \\ $\chi^2$/ndf,  p-value, $\sigma$} \\
        \midrule
        WC $1\gamma Np$ & 
            0.216/1, 0.642, 0.465 $\sigma$ & 
            0.018/1, 0.891, 0.137$\sigma$ &
            0.973/1, 0.324, 0.986$\sigma$ & 
            0.551/1, 0.458, 0.742$\sigma$ \\
        WC $1\gamma 0p$ & 
            0.777/1, 0.378, 0.881$\sigma$ & 
            0.495/1, 0.482, 0.704$\sigma$ &
            0.106/1, 0.745, 0.325$\sigma$ & 
            0.013/1, 0.910, 0.113$\sigma$ \\
        Pandora $1\gamma 1p$ & 
            1.812/1, 0.178, 1.346$\sigma$ & 
            1.476/1, 0.224, 1.215$\sigma$ &
            3.333/1, 0.068, 1.826$\sigma$ & 
            3.373/1, 0.066, 1.837$\sigma$ \\
        Pandora $1\gamma 0p$ & 
            0.056/1, 0.814, 0.236$\sigma$ & 
            0.838/1, 0.360, 0.916$\sigma$ &
            0.318/1, 0.573, 0.564$\sigma$ & 
            0.226/1, 0.634, 0.476$\sigma$ \\
        \midrule
        WC 1gNp+$1\gamma 0p$ & 
            1.990/2, 0.370, 0.897$\sigma$ & 
            0.565/2, 0.754, 0.313$\sigma$ &
            2.098/2, 0.350, 0.934$\sigma$ & 
            0.552/2, 0.759, 0.307$\sigma$ \\
        Pandora $1\gamma 1p$+$1\gamma 0p$ & 
            1.977/2, 0.372, 0.892$\sigma$ & 
            2.309/2, 0.315, 1.004$\sigma$ &
            3.486/2, 0.175, 1.356$\sigma$ & 
            3.682/2, 0.159, 1.410$\sigma$ \\
        \midrule
        \makecell{WC $1\gamma Np$+$1\gamma 0p$\\+ Pandora $1\gamma 1p$+$1\gamma 0p$} & 
            4.096/4, 0.393, 0.854$\sigma$ & 
            2.652/4, 0.618, 0.499$\sigma$ &
            5.270/4, 0.261, 1.125$\sigma$ & 
            4.222/4, 0.377, 0.884$\sigma$ \\
        \bottomrule
    \end{tabular}
    }
    \caption[Goodness of fit tests with and without the LEE]{Goodness of fit tests with and without the LEE. Expanded from Table \ref{tab:gof_tests_no_LEE}.}
    \label{tab:gof_tests}
\end{table}

\begin{figure}[H]
    \centering
    \includegraphics[width=0.75\textwidth]{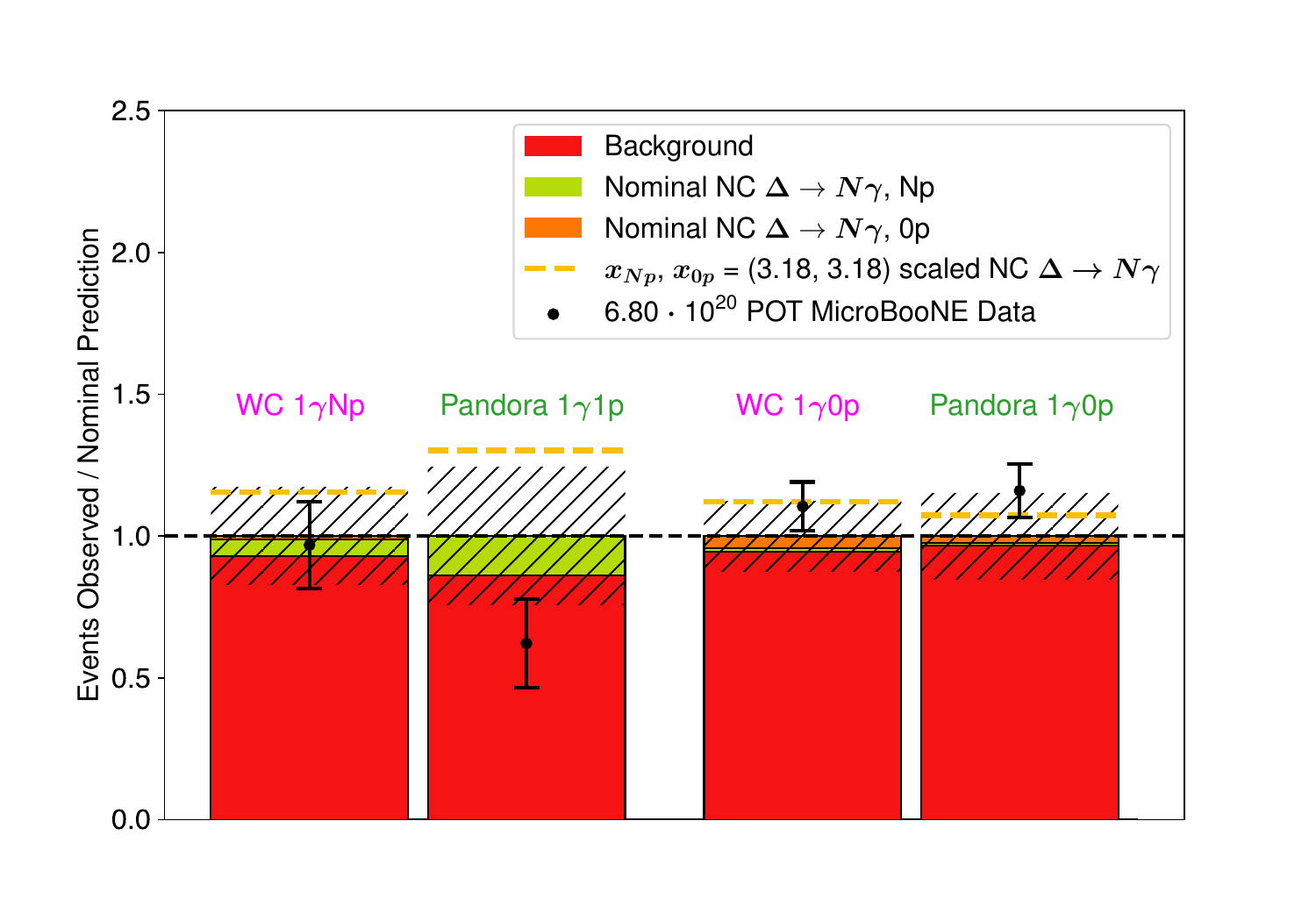}
    \caption[NC $\Delta\rightarrow N \gamma$ one bin ratio results]{NC $\Delta\rightarrow N \gamma$ one bin ratio results. The Pandora and Wire-Cell data samples correspond to $6.80\times 10^{20}$ and $6.37\times 10^{20}$ POT, respectively.}
    \label{fig:one_bin_ratio_results}
\end{figure}

Figure \ref{fig:one_bin_ratio_results} shows our data compared to this $x_\Delta=3.18$ LEE prediction, both before and after the application of the conditional constraint. We show systematic uncertainties on the no-NC $\Delta\rightarrow N \gamma$ prediction and stack the enhanced NC $\Delta\rightarrow N \gamma$ excess prediction on top. This plot is most directly comparable with the result of the Pandora-only NC $\Delta\rightarrow N \gamma$ analysis shown in Fig. \ref{fig:pandora_nc_delta_one_bin} \cite{glee_prl}. Figure \ref{fig:one_bin_ratio_results} shows the same results as ratios, with the $x_\Delta=3.18$ prediction present as a dashed line. In these plots, we show true NC $\Delta\rightarrow N \gamma$ events separated into those with and without true protons, using a true 35 MeV kinetic energy threshold.

\begin{figure}[H]
    \centering
    \includegraphics[trim=30 70 60 70, clip, width=0.6\textwidth]{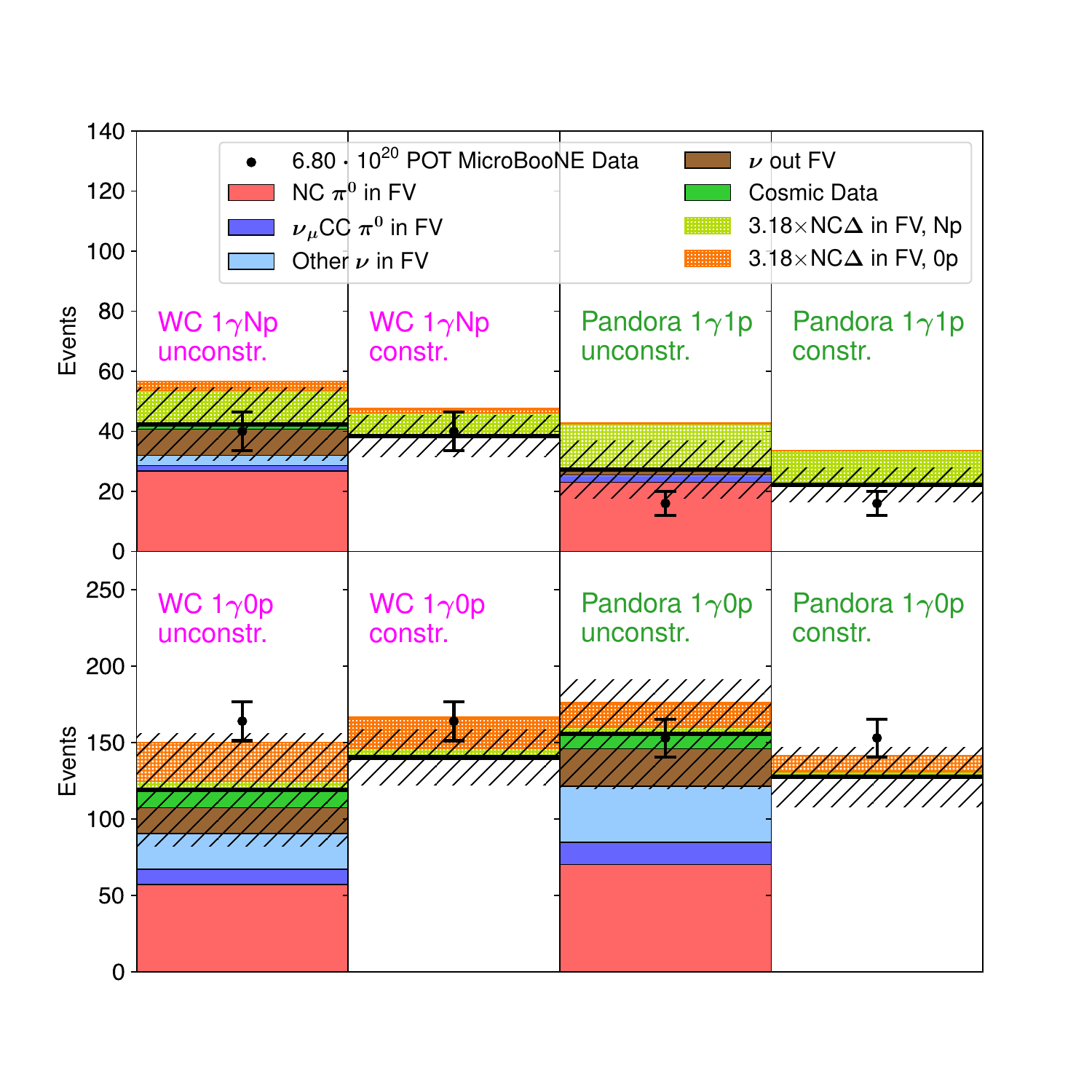}
    \caption[NC $\Delta\rightarrow N \gamma$ one bin results with LEE prediction]{NC $\Delta\rightarrow N \gamma$ one bin results with the $x_\Delta=3.18$ LEE prediction. The Pandora and Wire-Cell data samples correspond to $6.80\times 10^{20}$ and $6.37\times 10^{20}$ POT, respectively.}
    \label{fig:one_bin_results}
\end{figure}

\begin{figure}[H]
    \centering
    \begin{subfigure}[b]{0.49\textwidth}
        \includegraphics[trim=20 0 60 0, clip, width=\textwidth]{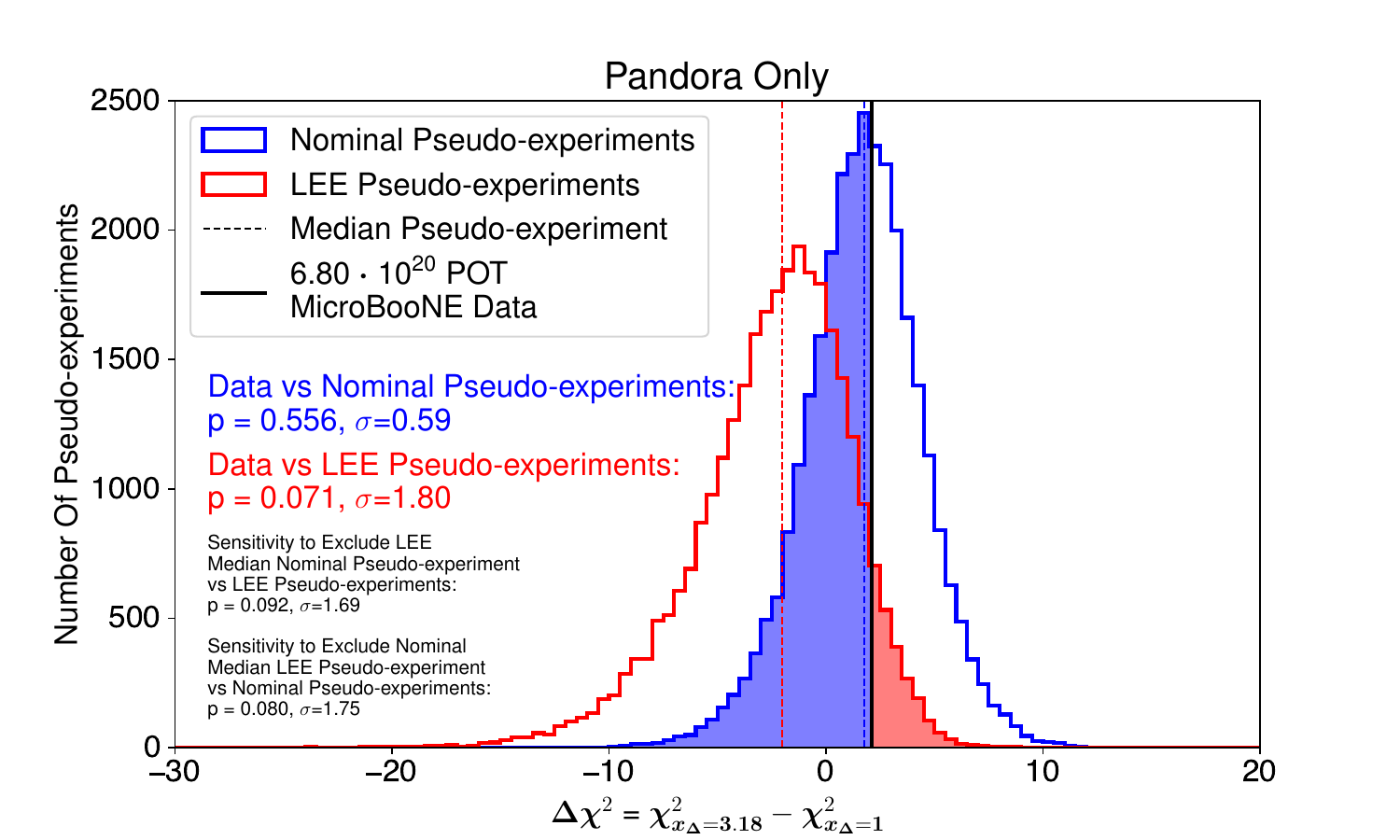}
        \caption{}
    \end{subfigure}
    \begin{subfigure}[b]{0.49\textwidth}
        \includegraphics[trim=20 0 60 0, clip, width=\textwidth]{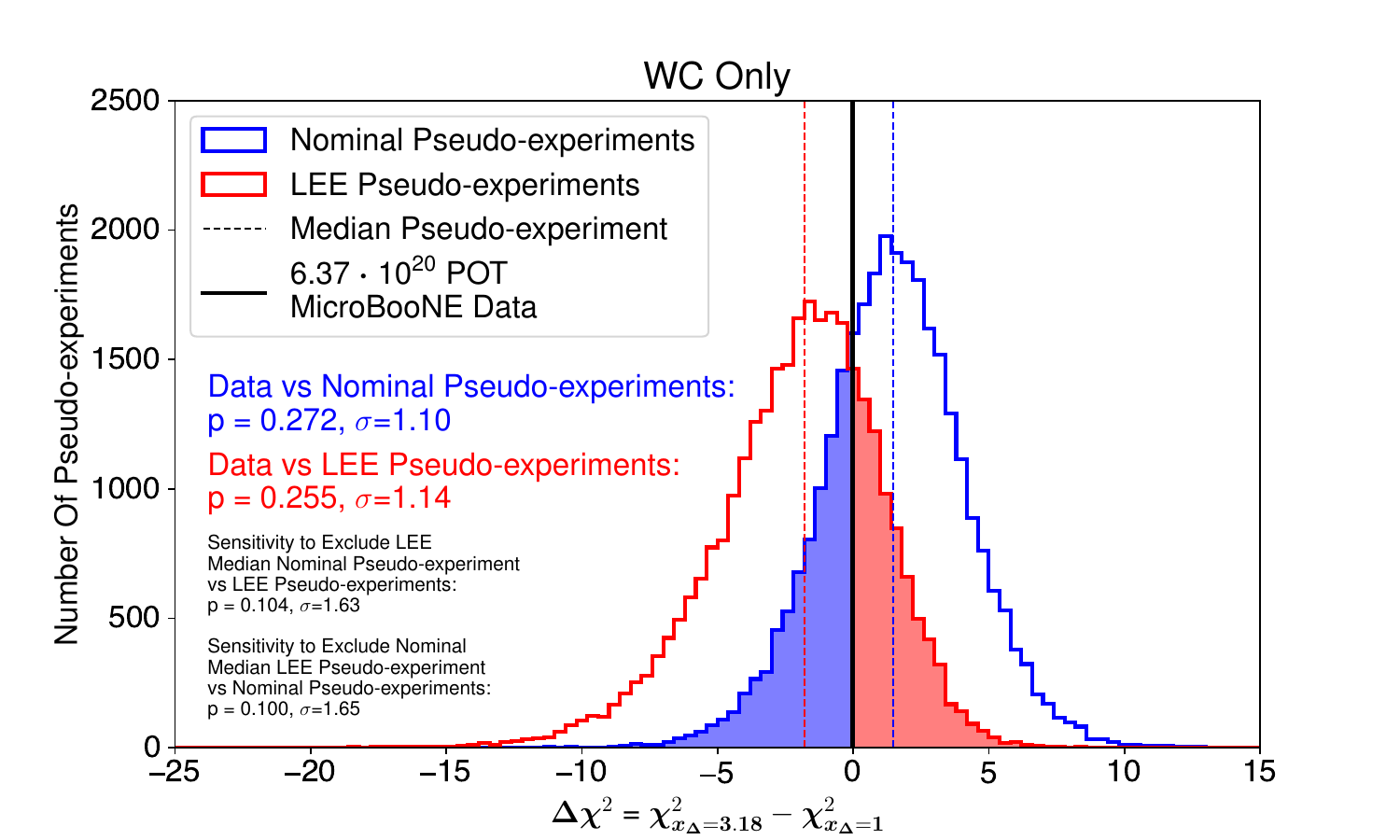}
        \caption{}
    \end{subfigure}
    \begin{subfigure}[b]{0.7\textwidth}
        \includegraphics[trim=20 0 60 0, clip, width=\textwidth]{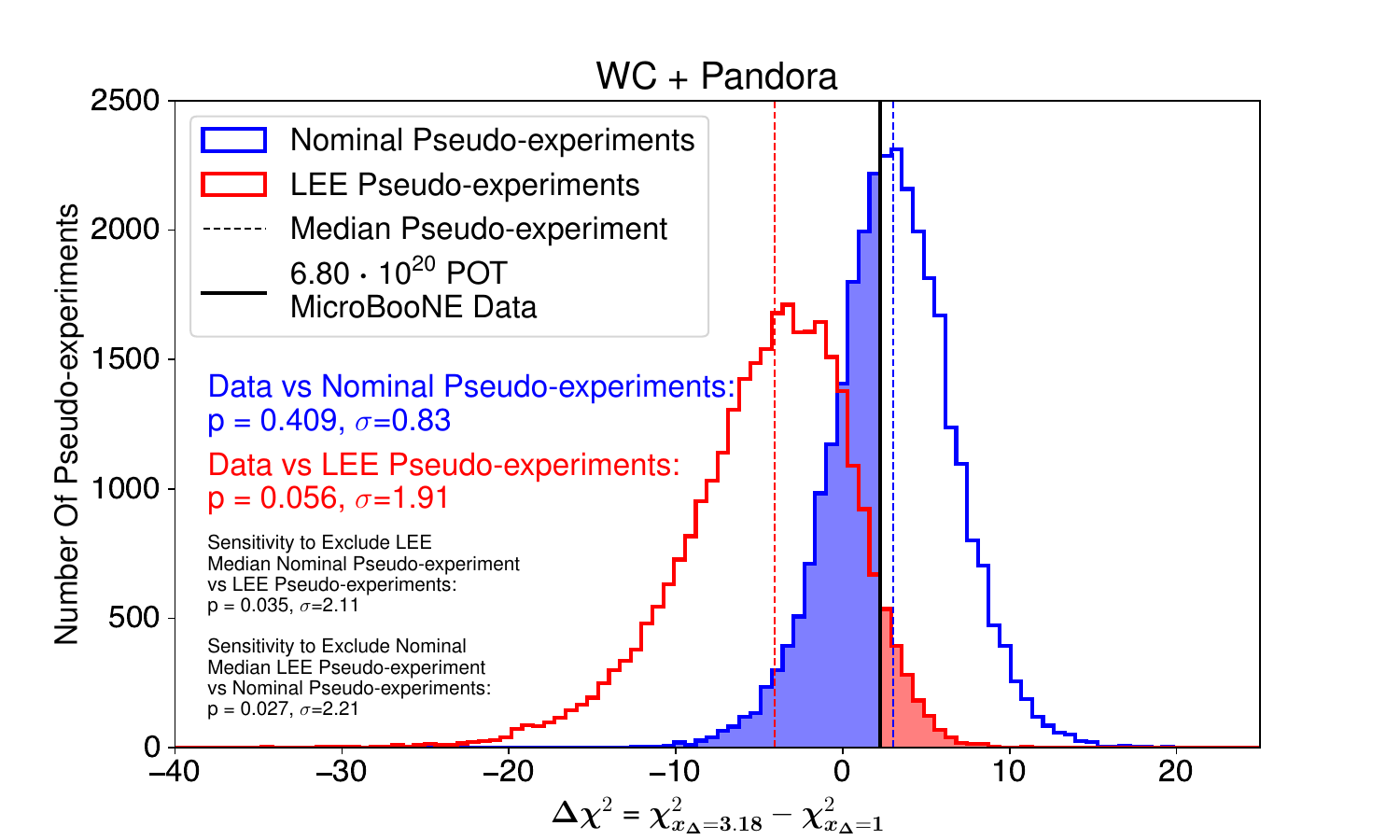}
        \caption{}
    \end{subfigure}
    \caption[NC $\Delta\rightarrow N \gamma$ two-hypothesis tests]{NC $\Delta\rightarrow N \gamma$ two-hypothesis tests. The Pandora and Wire-Cell data samples correspond to $6.80\times 10^{20}$ and $6.37\times 10^{20}$ POT, respectively.}
    \label{fig:two_hypothesis_tests}
\end{figure}

We perform several two-hypothesis tests, as shown in Fig. \ref{fig:two_hypothesis_tests}. These are performed using $\Delta\chi^2$ distributions which characterize whether our data prefer the nominal $x_\Delta=1$ prediction or the LEE $x_\Delta=3.18$ prediction, in the same way as described in Sec. \ref{sec:wc_nueCC_results}. In our Pandora selections, our data lies near the median expectation of our nominal hypothesis, in our Wire-Cell selections, our data lies in between the median nominal and LEE hypothesis expectations, and when we combine all four channels, our data lies near the median expectation of our nominal hypothesis. We exclude the $x_\Delta=3.18$ LEE hypothesis at 94.4\% CL, or 1.91 $\sigma$, which is very close to the 94.8\% CL value from our previous Pandora-only results as described in Sec. \ref{sec:pandora_nc_delta_search} \cite{glee_prl}.

\begin{figure}[H]
    \centering
    \begin{subfigure}[b]{0.495\textwidth}
        \includegraphics[trim=20 0 15 0, clip, width=\textwidth]{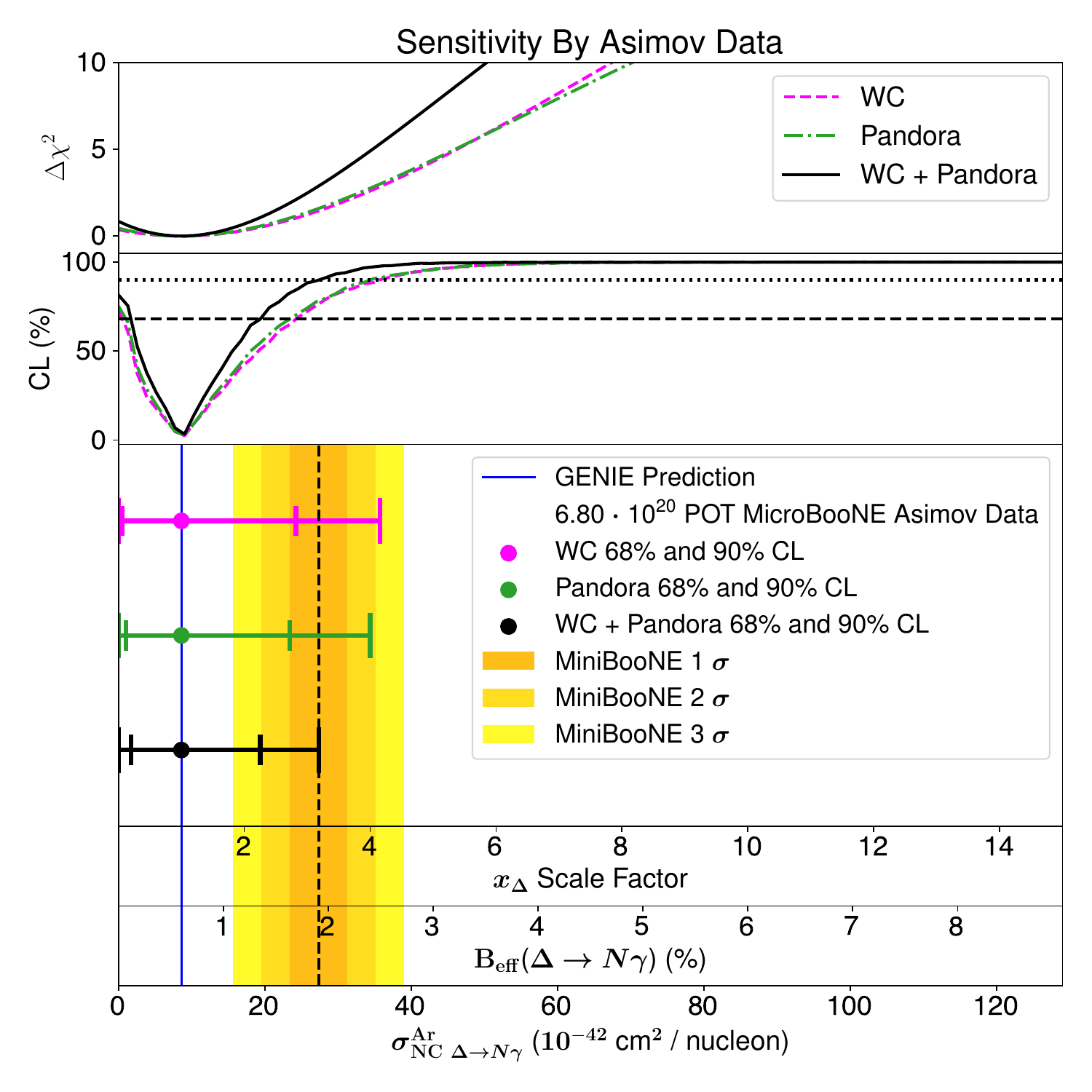}
        \caption{}
    \end{subfigure}
    \begin{subfigure}[b]{0.485\textwidth}
        \includegraphics[trim=20 0 15 0, clip, width=\textwidth]{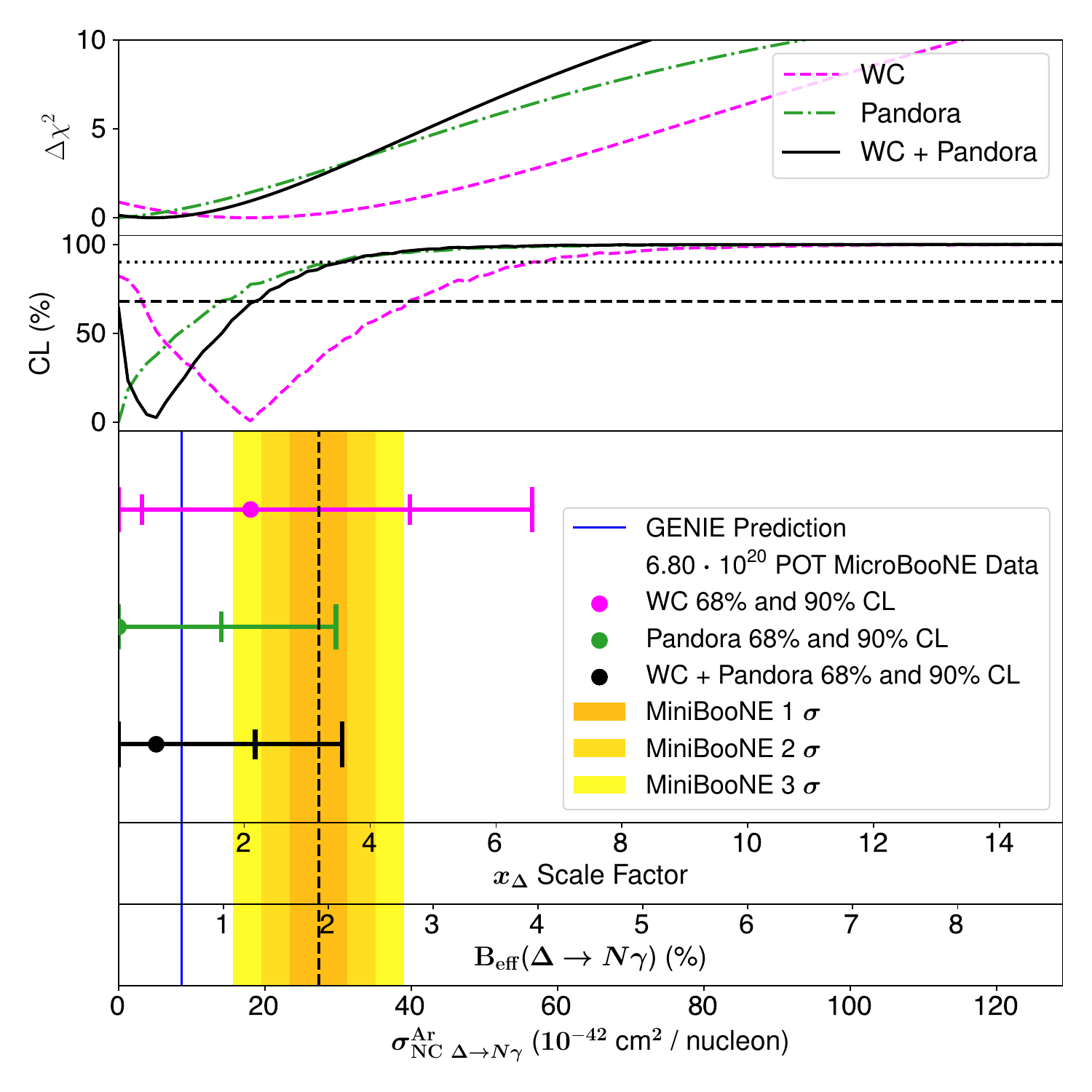}
        \caption{}
    \end{subfigure}
    \caption[NC $\Delta\rightarrow N \gamma$ 1D $x_\Delta$ LEE sensitivity and data exclusion]{NC $\Delta\rightarrow N \gamma$ 1D $x_\Delta$ LEE sensitivity and data exclusion. We show the $\Delta\chi^2$ values and CL values as functions of $x_\Delta$, for Wire-Cell, Pandora, and Wire-Cell+Pandora. We then show resulting error bars based on the 68\% and 90\% CL values. Panel (a) shows the sensitivity, calculated with an Asimov data set which exactly matches the prediction. Panel (b) shows the real data result. The Pandora and Wire-Cell data samples correspond to $6.80\times 10^{20}$ and $6.37\times 10^{20}$ POT, respectively.}
    \label{fig:1d_LEE_exclusion_data_sensitivity}
\end{figure}

Next, we consider expanded LEE hypotheses consisting of a wide range of possible $x_\Delta$ values beyond just $x_\Delta=1$ and $x_\Delta=3.18$. This is the same test as described in Sec. \ref{sec:pandora_nc_delta_search} and Fig. \ref{fig:pandora_nc_delta_chi2}. For each $x_\Delta$ value, we generate many pseudo-experiments and $\Delta\chi^2=\chi^2_{x_\Delta}-\chi^2_{x_{\Delta\mathrm{min}}}$, and compare the $\Delta\chi^2$ value from data with a distribution of pseudo-experimennts in order to get a CL value for each $x_\Delta$ value. We also interpret this $x_\Delta$ parameter in terms of the effective branching fraction and in terms of the flux-integrated cross section for the NC $\Delta\rightarrow N \gamma$ process. As explained in Sec. \ref{sec:pandora_nc_delta_search}, note that both of these additional interpretations are performed as simple scalings according to GENIE, and do not take into account the conditional constraint which could, for example, update the flux prediction to slightly modify these values. Also as explained in Sec. \ref{sec:pandora_nc_delta_search}, we can estimate an uncertainty on the MiniBooNE band by using the overall significance of the MiniBooNE LEE, creating an error band that corresponds to 4.8$\sigma$ from $x_\Delta=1$.  Figure \ref{fig:1d_LEE_exclusion_data_sensitivity} shows both the sensitivity and real data results in this test. We see that the Wire-Cell and Pandora channels have similar sensitivities, and the combination of the two has a higher-sensitivity. The Wire-Cell data is consistent with a fairly wide range of scalings, while the Pandora analysis prefers scalings at or below the LEE prediction. The combined result is fairly similar to the Pandora-only result, which is driven by the overprediction in the Pandora $1\gamma 1p$ channel which has the most influence on rejecting high $x_\Delta$ hypotheses.

\begin{figure}[H]
    \centering
    \begin{subfigure}[b]{0.495\textwidth}
        \includegraphics[trim=20 0 15 0, clip, width=\textwidth]{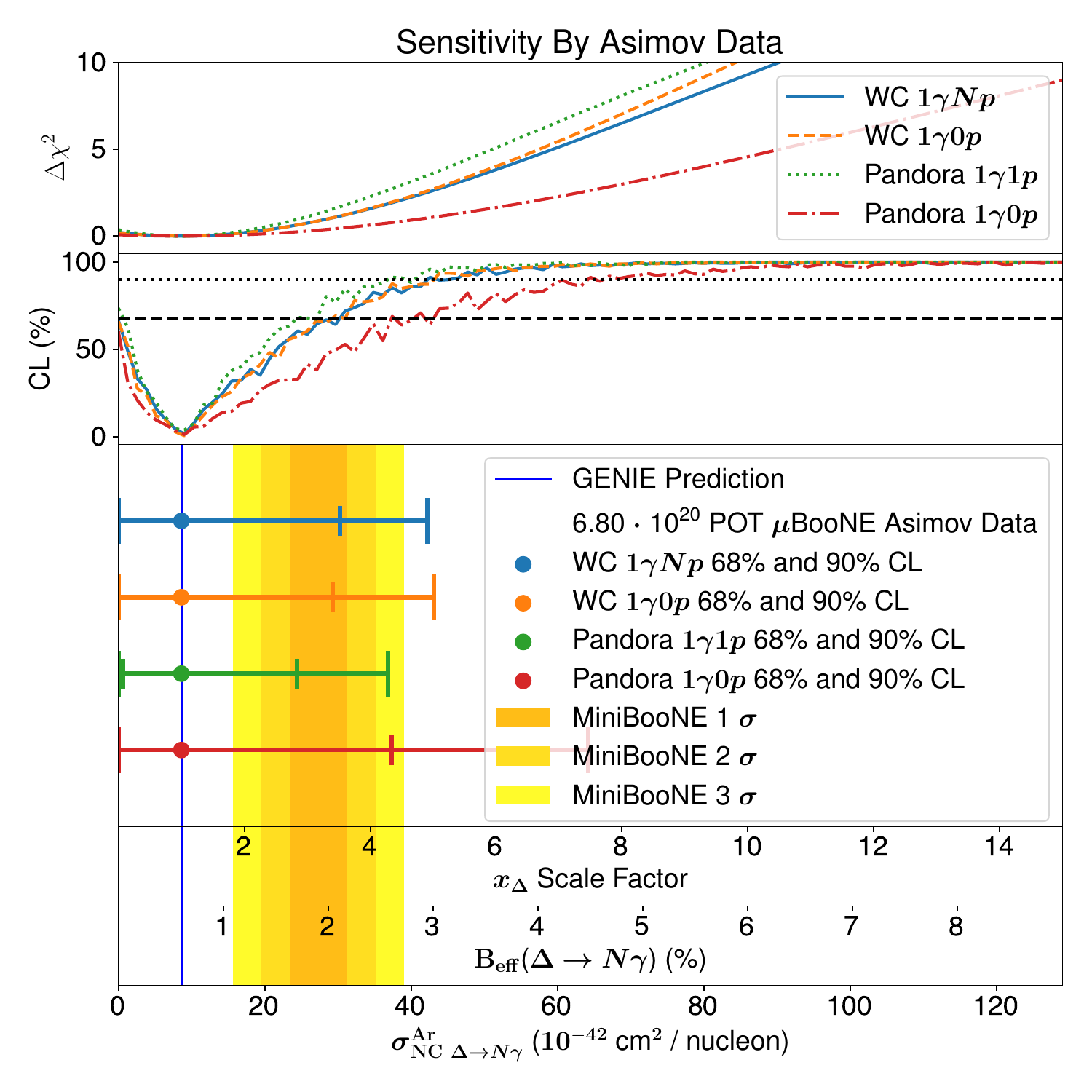}
        \caption{}
    \end{subfigure}
    \begin{subfigure}[b]{0.485\textwidth}
        \includegraphics[trim=20 0 15 0, clip, width=\textwidth]{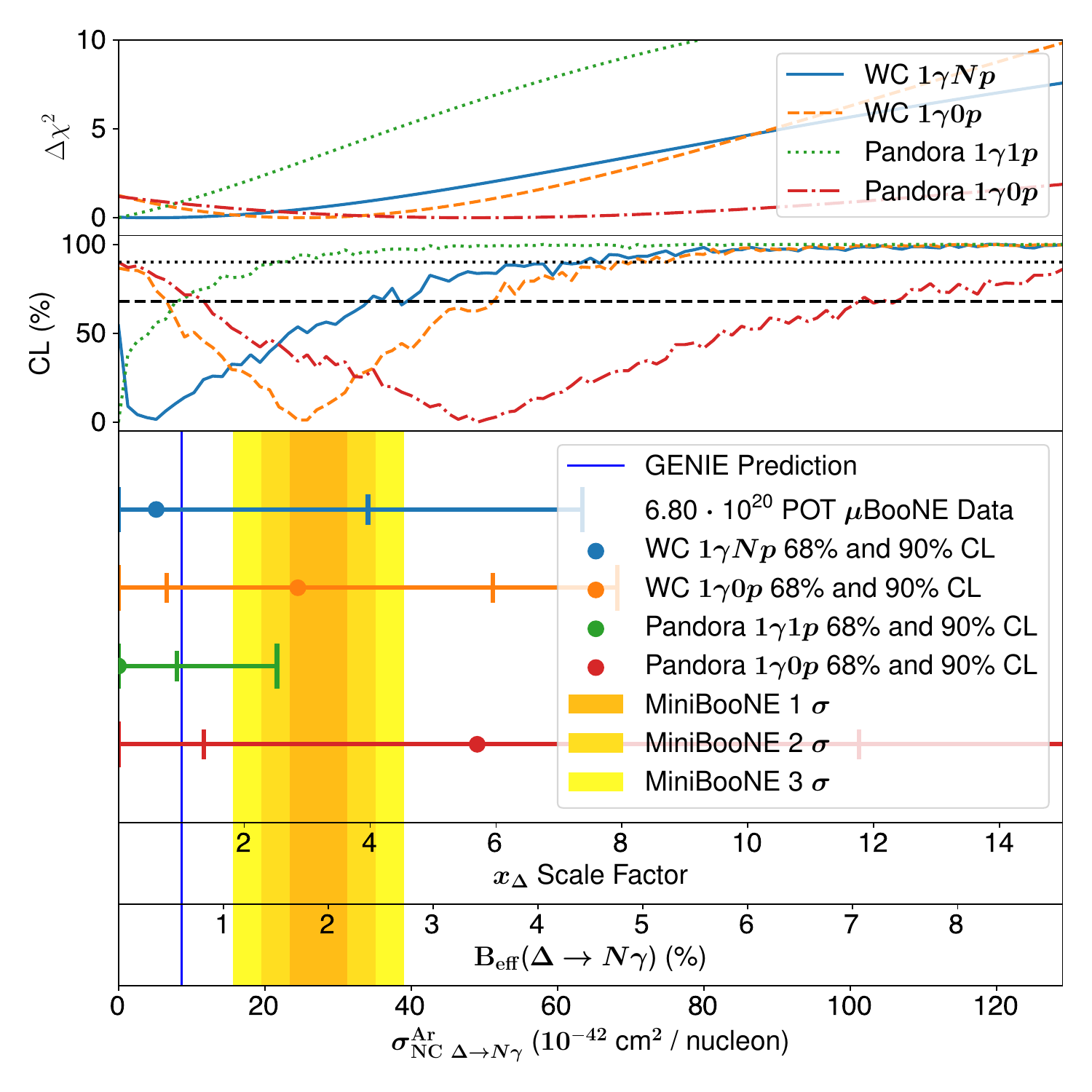}
        \caption{}
    \end{subfigure}
    \caption[NC $\Delta\rightarrow N \gamma$ 1D $x_\Delta$ LEE sensitivity and data exclusion by topology]{NC $\Delta\rightarrow N \gamma$ 1D $x_\Delta$ LEE sensitivity and data exclusion by topology. We show the $\Delta\chi^2$ values and CL values as functions of $x_\Delta$ for each of our four signal channels. We then show resulting error bars based on the 68\% and 90\% CL values. Panel (a) shows the sensitivity, calculated with an Asimov data set which exactly matches the prediction. Panel (b) shows the real data result. The Pandora and Wire-Cell data samples correspond to $6.80\times 10^{20}$ and $6.37\times 10^{20}$ POT, respectively.}
    \label{fig:topology_1d_LEE_exclusion_data_sensitivity}
\end{figure}

We can also perform this test on each of the four signal channels individually, as we show in Fig. \ref{fig:topology_1d_LEE_exclusion_data_sensitivity}. This again illustrates how the Pandora $1\gamma1p$ channel plays the most significant role in the final exclusion, even though the sensitivity of this bin is comparable to the Wire-Cell $1\gamma Np$ and $1\gamma 0p$ bins.

\subsection{Tests of More General NC \texorpdfstring{$\Delta$}{Delta}-like MiniBooNE LEE Hypotheses}

In the previous section, we described tests of the MiniBooNE LEE via an overall scaling of NC $\Delta\rightarrow N \gamma$ events. Now, we consider a broader set of hypotheses, which include different scalings for events with ($x_{Np}$) and without ($x_{0p}$) the existence of true protons using a 35 MeV kinetic energy threshold.

This hypothesis has less of a clear physical meaning, and must be interpreted as more of a benchmark for other types of non-$\Delta\rightarrow N \gamma$ single photons which could be produced in MiniBooNE and MicroBooNE. Note that this was also effectively the case in the 1D NC $\Delta\rightarrow N \gamma$ scaling model, since the Particle Data Group places an uncertainty on the $\Delta\rightarrow N \gamma$ of 8.3\%, while we considered a 318\% shift in this rate \cite{ParticleDataGroup}, and therefore a more realistic interpretation is that we were using this model as a benchmark for potential mis-modeled or beyond-standard-model single photon sources.

Considering $1\gamma Np$ and $1\gamma0p$ signal events separately is interesting, since each gives us different information about a potential signal source. The $1\gamma Np$ process is more specific to the NC $\Delta\rightarrow N \gamma$ process, and our selections depend on the kinematics of the proton as well as the shower, as shown in Fig. \ref{fig:proton_photon_effs} for example. In contrast, $1\gamma 0p$ is less specific to a specific model, and the performance of our selections will depend primarily on just the photon shower energy and angle as shown in Fig. \ref{fig:shower_2d_true_0p_efficiencies}. Therefore, a search for an excess of $1\gamma 0p$ specifically is potentially more easily applicable to a broader class of single photon excess hypotheses.

\begin{figure}[H]
    \centering
    \begin{subfigure}[b]{0.32\textwidth}
        \includegraphics[trim=15 0 25 0, clip, width=\textwidth]{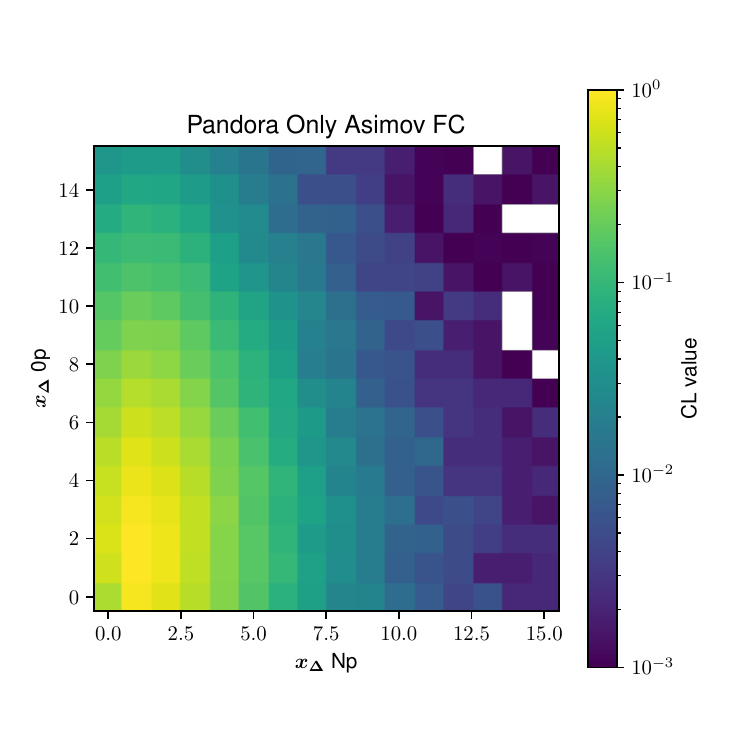}
        \caption{}
    \end{subfigure}
    \begin{subfigure}[b]{0.32\textwidth}
        \includegraphics[trim=15 0 25 0, clip, width=\textwidth]{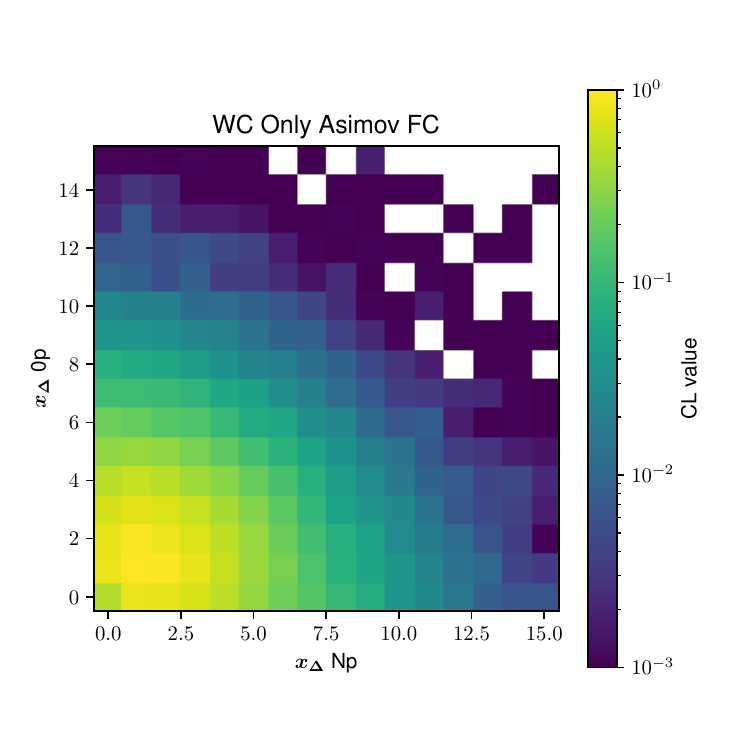}
        \caption{}
    \end{subfigure}
    \begin{subfigure}[b]{0.32\textwidth}
        \includegraphics[trim=15 0 25 0, clip, width=\textwidth]{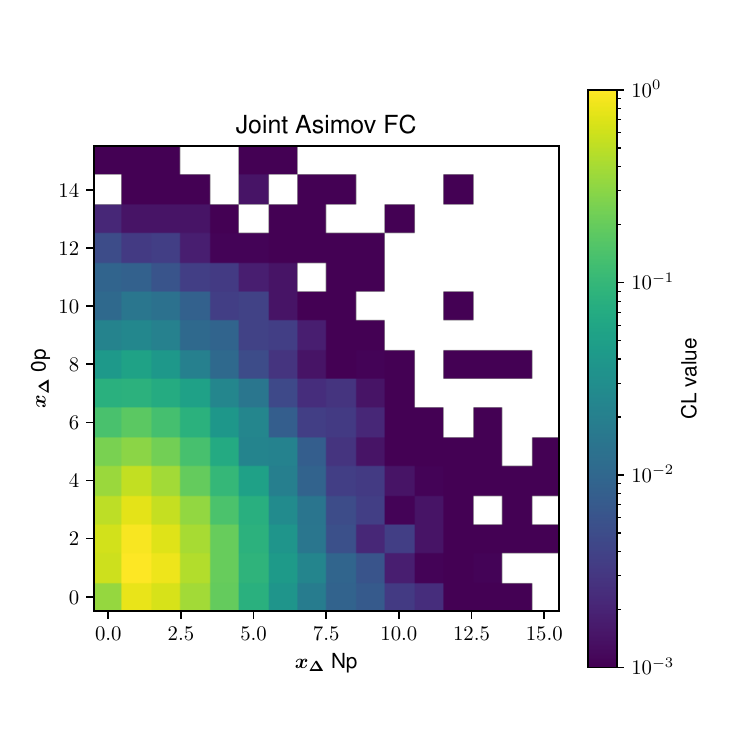}
        \caption{}
    \end{subfigure}
    \begin{subfigure}[b]{0.32\textwidth}
        \includegraphics[trim=15 0 25 0, clip, width=\textwidth]{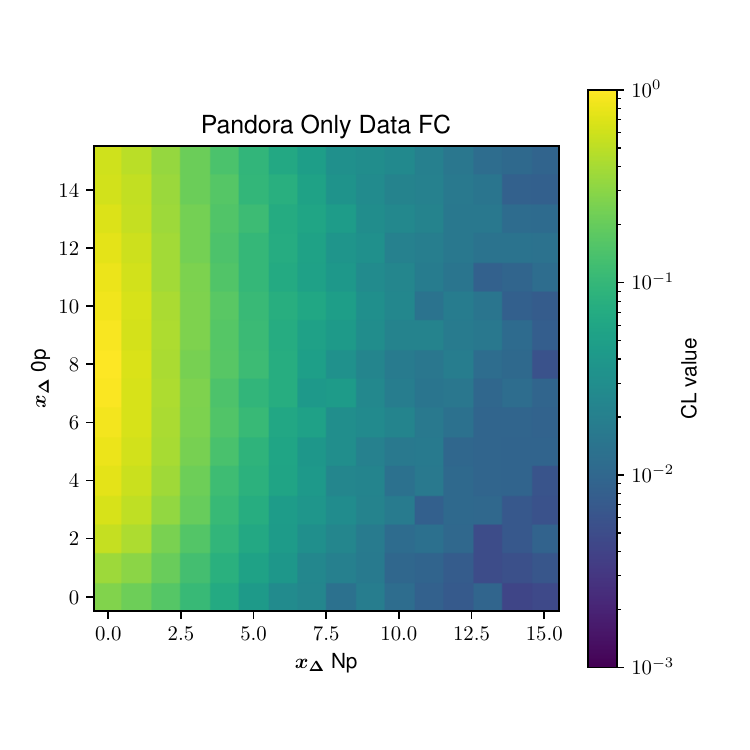}
        \caption{}
    \end{subfigure}
    \begin{subfigure}[b]{0.32\textwidth}
        \includegraphics[trim=15 0 25 0, clip, width=\textwidth]{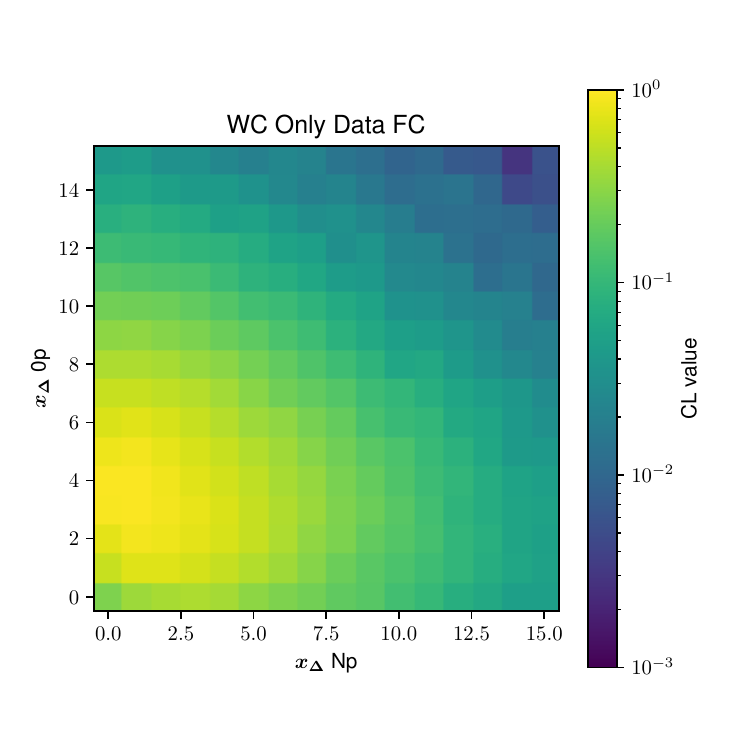}
        \caption{}
    \end{subfigure}
    \begin{subfigure}[b]{0.32\textwidth}
        \includegraphics[trim=15 0 25 0, clip, width=\textwidth]{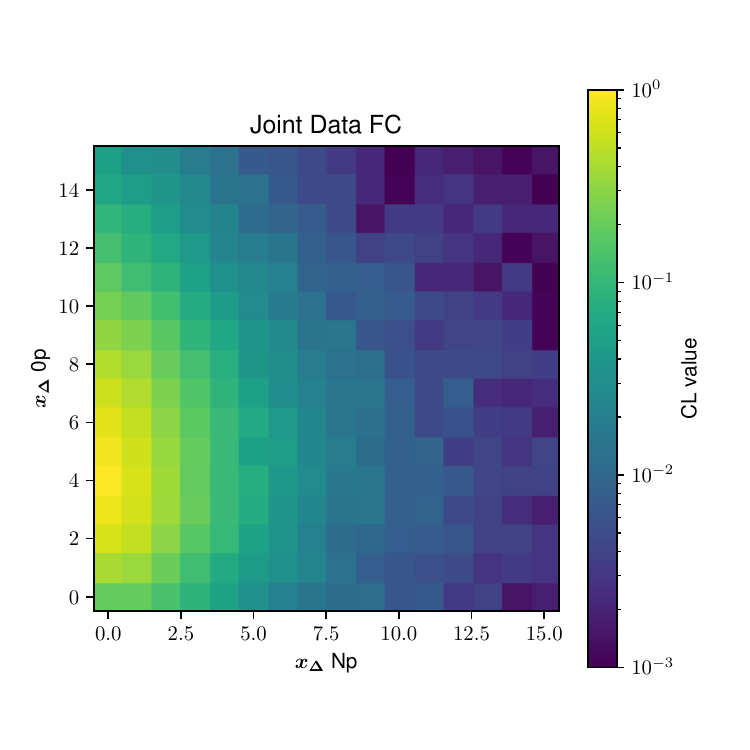}
        \caption{}
    \end{subfigure}
    \caption[NC $\Delta\rightarrow N \gamma$ 2D LEE exclusion CL values]{NC $\Delta\rightarrow N \gamma$ 2D LEE exclusion CL values. Panels (a), (b), and (c) show Wire-Cell, Pandora, and Wire-Cell+Pandora sensitivities via an Asimov data set. Panels (d), (e), and (f) show Wire-Cell, Pandora, and Wire-Cell+Pandora real data results.}
    \label{fig:2d_LEE_CL_values}
\end{figure}

We follow the same procedure as in the 1D scaling tests, performing a full Feldman-Cousins procedure in order to extract an exclusion in this phase space of LEE hypotheses. We compare data to a distribution of pseudo-experiments at each phase space point in order to extract a CL value for each point. Note that in this 2D phase space, the minimization required for each $\Delta\chi^2$ calculation is done in 2D. The resulting confidence levels for both an Asimov dataset used for sensitivity calculations and the real data is shown in Fig. \ref{fig:2d_LEE_CL_values}. We can see low CL values for especially large $(x_{Np}, x_{0p})$ scalings, as we would expect since we do not see a large excess.

To visualize these results more clearly, we form contour plots indicating 90\% CL exclusions. In this more general phase space, we can still interpret certain regions as consistent with the MiniBooNE LEE. To do this, we translate these scalings of true NC $\Delta\rightarrow N \gamma$ events with and without true protons into a total scaling of all NC $\Delta\rightarrow N \gamma$ in MicroBooNE: $x_\Delta = 0.53\cdot x_{Np} + 0.47\cdot x_{0p}$. These coefficients differ from 0.5 due to the unequal proportions of $1\gamma Np$ and $1\gamma 0p$ NC $\Delta\rightarrow N \gamma$ events after final state interactions and a 35 MeV true proton kinetic energy threshold. We then form preferred MiniBooNE regions with uncertainties using this $x_\Delta$ value in the same way as in the 1D case. Note that this procedure does not consider anything about MiniBooNE's relative sensitivities to events with or without protons, and just uses the total scaling of all NC $\Delta\rightarrow N \gamma$ events reported by MiniBooNE.

\begin{figure}[H]
    \centering
    \begin{subfigure}[b]{0.49\textwidth}
        \includegraphics[trim=15 0 25 0, clip, width=\textwidth]{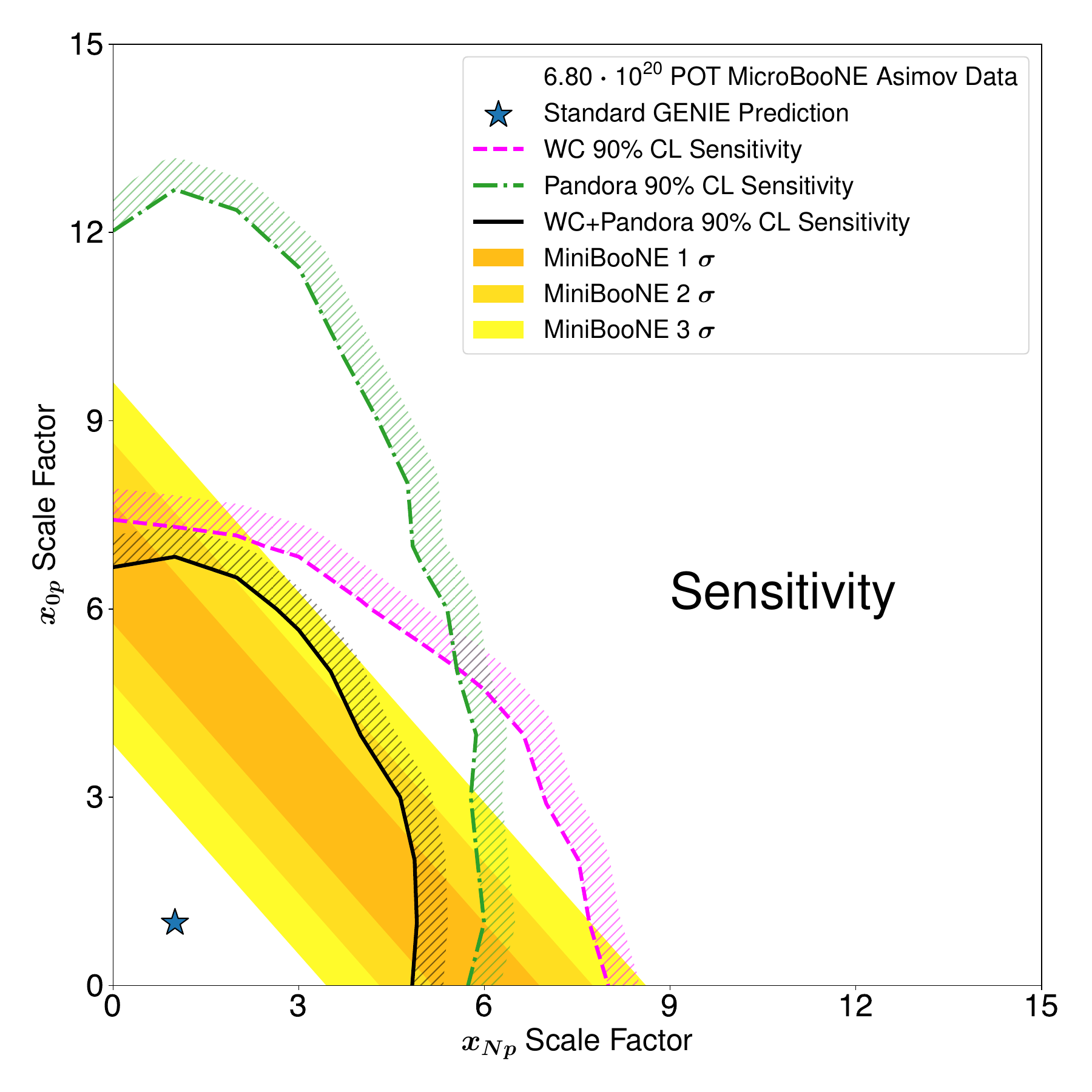}
        \caption{}
    \end{subfigure}
    \begin{subfigure}[b]{0.49\textwidth}
        \includegraphics[trim=15 0 25 0, clip, width=\textwidth]{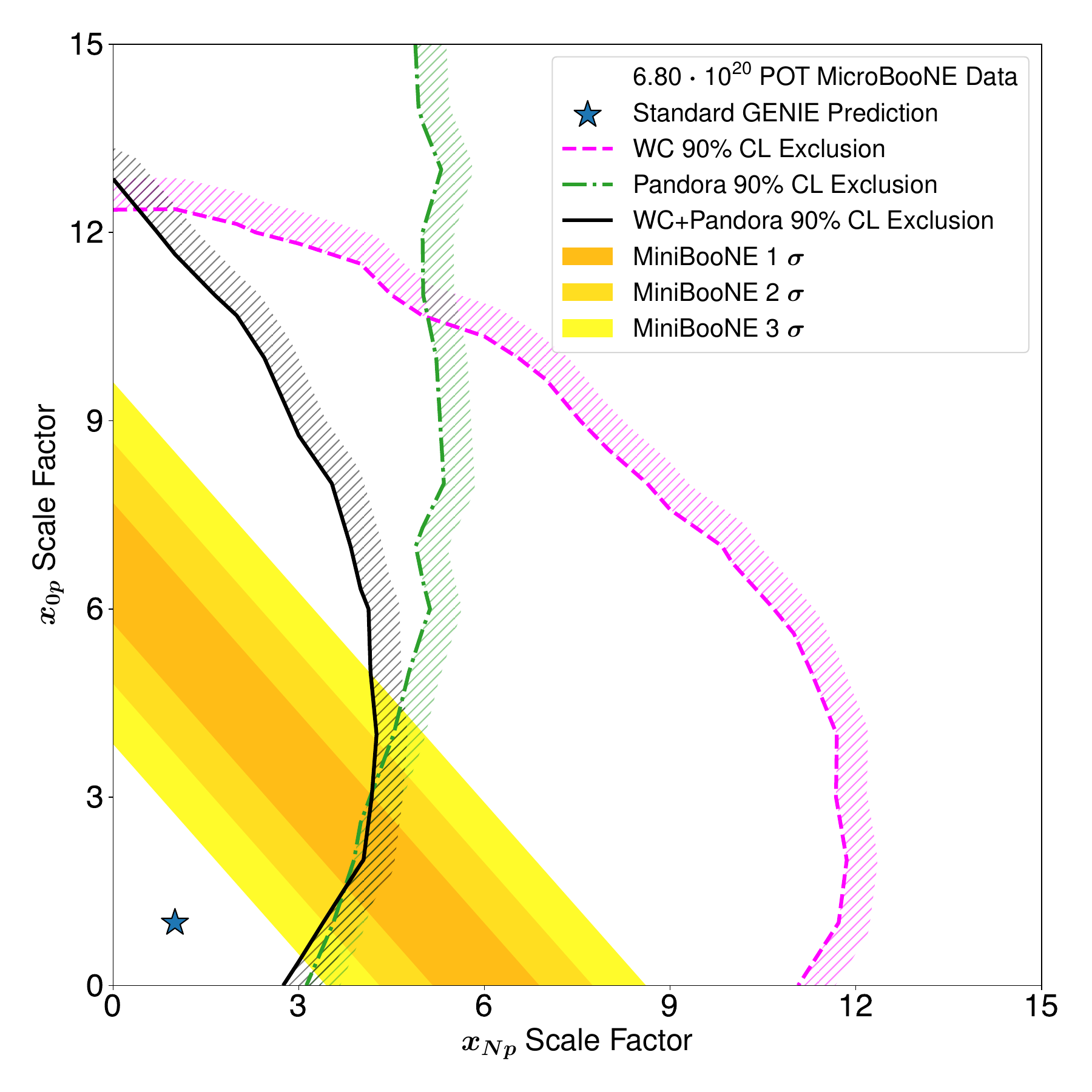}
        \caption{}
    \end{subfigure}
    \caption[NC $\Delta\rightarrow N \gamma$ 2D $(x_{Np}, x_{0p})$LEE exclusion sensitivity and data result]{NC $\Delta\rightarrow N \gamma$ 2D $(x_{Np}, x_{0p})$ LEE sensitivity and data exclusion. We show 90\% CL contours for Wire-Cell, Pandora, and Wire-Cell+Pandora. Panel (a) shows the sensitivity, calculated with an Asimov data set which exactly matches the prediction. Panel (b) shows the real data result. The Pandora and Wire-Cell data samples correspond to $6.80\times 10^{20}$ and $6.37\times 10^{20}$ POT, respectively.}
    \label{fig:2D_LEE_exclusion_sensitivity_and_data}
\end{figure}

These results are shown in Fig. \ref{fig:2D_LEE_exclusion_sensitivity_and_data}. In the sensitivity, we see that the combined Wire-Cell and Wire-Cell+Pandora contours are expected to be able to exclude significantly more phase than the Pandora curve alone, in particular for large $0p$ scalings. This is expected due to the Wire-Cell $1\gamma 0p$ selection's higher efficiency and purity relative to the Pandora $1\gamma 0p$ selection, as shown in Table \ref{tab:effs_and_purs}. In the real result, as was the case for the 1D test, most of the exclusion is driven by the overprediction in the Pandora $1\gamma1p$ channel, but the addition of the Wire-Cell channels is able to let us exclude significantly more phase space with large $0p$ scalings. The slight underprediction in both the Wire-Cell and Pandora $1\gamma 0p$ channels does cause our exclusion to be weaker than our sensitivity for $0p$ scalings. We also notice that the Pandora contour does not exclude any $0p$ scalings; this is possible since the fractional systematic uncertainty on the signal prediction becomes the majority of the total uncertainty at high scalings. For example, if the fractional systematic uncertainty on your signal is greater than 60.8\%, then a 100\% downfluctuation which gets rid of all of the excess events is possible within $1.64\sigma$ or 90\% CL, and therefore no amount of excess, no matter how high, will be excluded by an observation of data at the nominal rate.

\begin{figure}[H]
    \centering
    \includegraphics[width=0.75\textwidth]{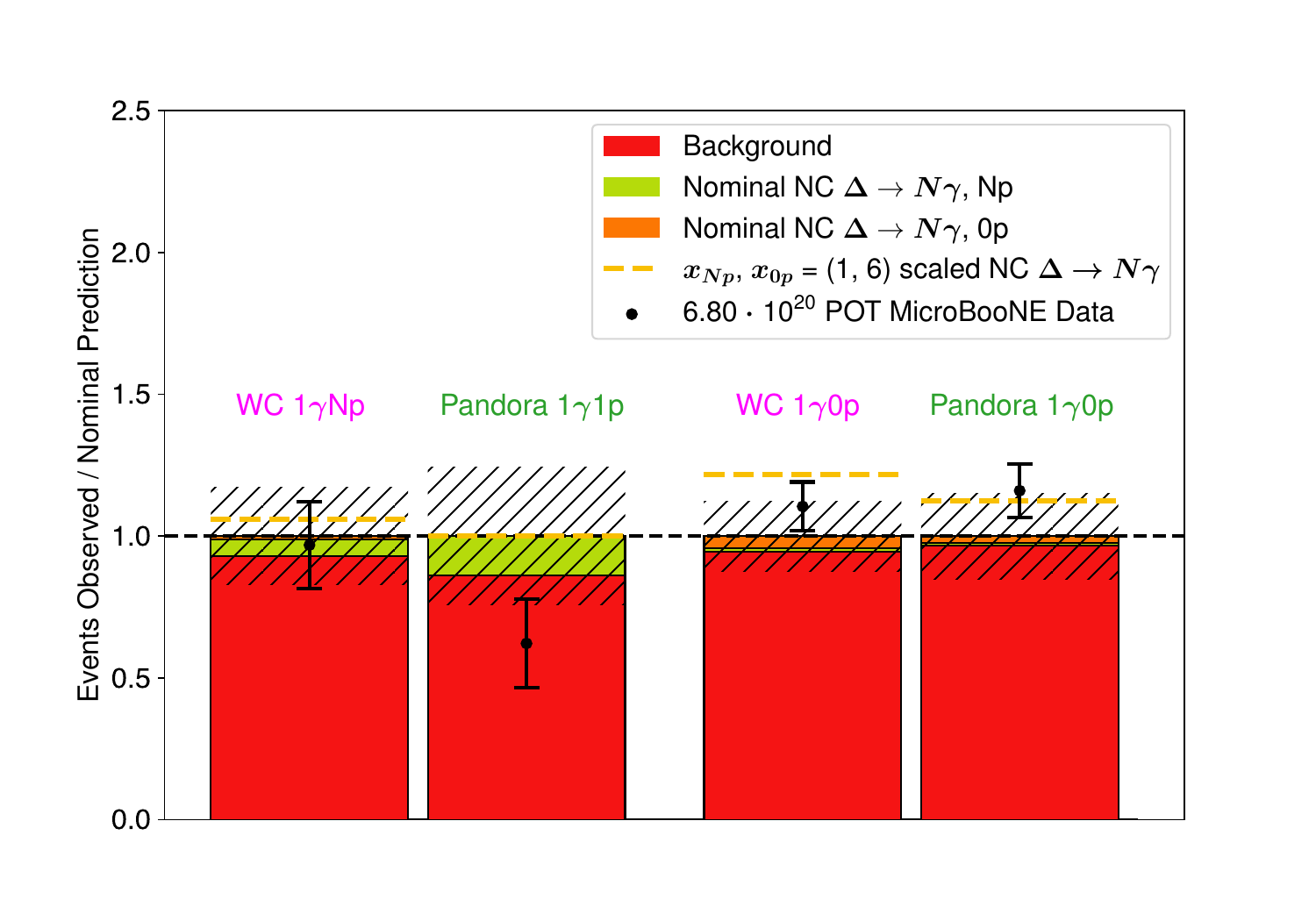}
    \caption[NC $\Delta\rightarrow N \gamma$ comparison with 0p-only LEE model]{NC $\Delta\rightarrow N \gamma$ comparison with 0p-only LEE model, corresponding to the $(x_{Np}, x_{0p}) = (1, 6)$ point in our 2D LEE model phase space, which is consistent with a total 3.18 times enhancement matching the MiniBooNE prediction for an NC $\Delta\rightarrow N \gamma$ enhancement.}
    \label{fig:one_bin_0p_LEE}
\end{figure}

To illustrate a point in this phase space which is allowed by our observations and also aligns with the MiniBooNE LEE prediction for the total NC $\Delta\rightarrow N \gamma$ rate, we consider the point $(x_{Np}, x_{0p}) = (1, 6)$, where we do not alter the scale of true $Np$ NC $\Delta\rightarrow N \gamma$ events and we scale up true $0p$ NC $\Delta\rightarrow N \gamma$ events by a factor of six. A comparison of this phase space point with our data is shown in Fig. \ref{fig:one_bin_0p_LEE}.

Naively, we would expect these 2D results in Fig. \ref{fig:one_bin_0p_LEE} to be consistent with the 1D results in Fig. \ref{fig:1d_LEE_exclusion_data_sensitivity} along the diagonal, where $x_{Np} = x_{0p}$. This is not exactly the case, and we see that the Wire-Cell+Pandora 90\% exclusion is at close to $x_\Delta=3.5$ in the 1D case, and close to $x_{Np} = x_{0p} = 4.2$ in the 2D case. This difference is not unexpected, due to the nature different statistical tests being performed. In the 1D case, we model perfectly correlated excesses across the true $Np$ and $0p$ channels, and this gives the four-bin fit significantly more constraining power. More technically, when considering a phase space point along this diagonal where $x_{Np} = x_{0p}$, the minimization that is performed when calculating $\Delta\chi^2=\chi^2_{(x_{Np}, x_{0p})}-\chi^2_{\mathrm{min}}$ values is free to choose points off of this diagonal.

\begin{figure}[H]
    \centering
    \begin{subfigure}[b]{0.49\textwidth}
        \includegraphics[trim=15 0 25 0, clip, width=\textwidth]{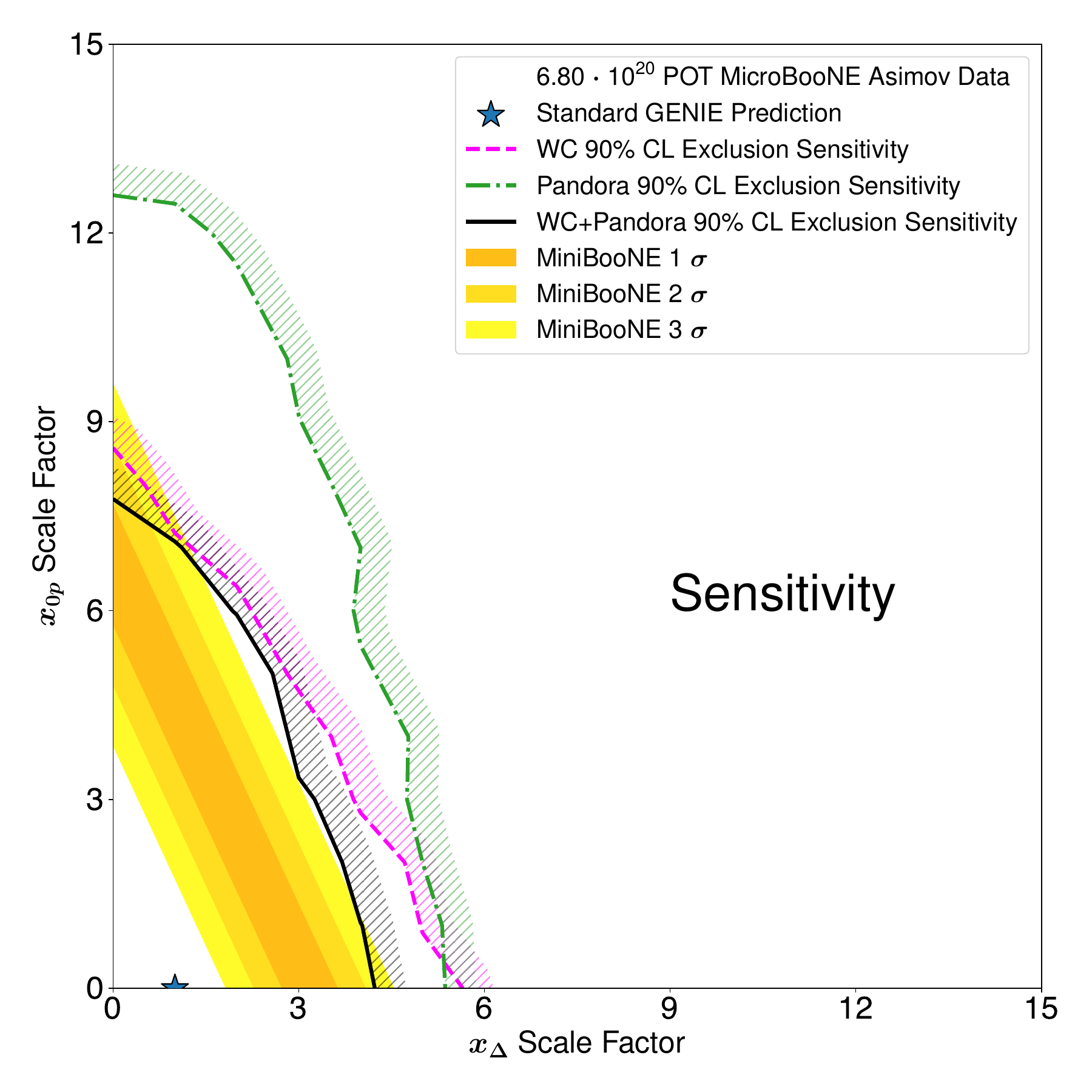}
        \caption{}
    \end{subfigure}
    \begin{subfigure}[b]{0.49\textwidth}
        \includegraphics[trim=15 0 25 0, clip, width=\textwidth]{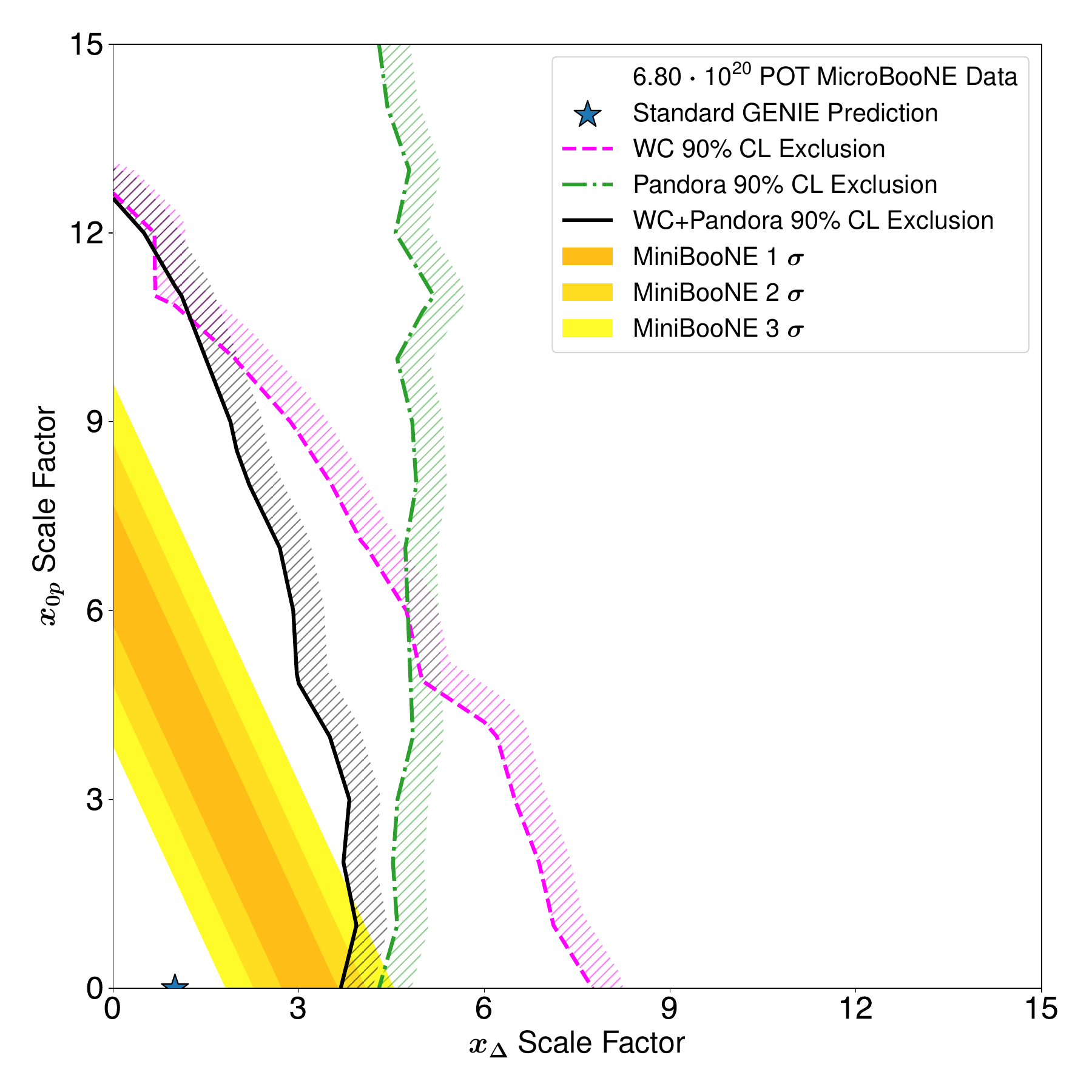}
        \caption{}
    \end{subfigure}
    \caption[NC $\Delta\rightarrow N \gamma$ 2D $(x_\Delta, x_{0p})$LEE exclusion sensitivity and data result]{NC $\Delta\rightarrow N \gamma$ 2D $(x_\Delta, x_{0p})$LEE exclusion sensitivity and data result}
    \label{fig:alt_2D_LEE_exclusion_sensitivity_and_data}
\end{figure}

We can choose an alternative 2D phase space that we expect to more closely match the behavior in the 1D case. Here, we parameterize our phase space by $(x_\Delta, x_{0p})$, which translates to the prior phase space $(x_{0p}', x_{Np}')$ by $(x_\Delta, x_{0p}) \rightarrow (x_{0p}', x_{Np}') = (x_\Delta + x_{0p}, x_\Delta)$. This means that $x_\Delta$ reflects an overall scaling of both the $Np$ and $0p$ NC $\Delta\rightarrow N \gamma$ components, and $x_{0p}$ represents an additional scaling of only the $0p$ component. This means that the $x_{0p}=0$ bottom edge corresponds to the simple scaling of both $Np$ and $0p$ events, playing the role of the diagonal in the previous 2D exclusion. The sensitivity and data result in this alternate phase space is shown in Fig. \ref{fig:alt_2D_LEE_exclusion_sensitivity_and_data}. Along this $x_{0p}=0$ bottom edge, the 90\% Wire-Cell+Pandora exclusion is at about 3.7, which is closer to the 1D result at 3.5.

A data release from the combined Wire-Cell+Pandora NC $\Delta\rightarrow N \gamma$ analysis is available at \url{https://www.hepdata.net/record/158531}.

%% file: chapters/05_more_photon.tex
\chapter{More MicroBooNE photon-like Analyses and Studies}

At the same time as I was developing my combined Wire-Cell+Pandora NC $\Delta\rightarrow N \gamma$ analysis, there were several other photon-like analyses being developed in MicroBooNE and releasing results on a similar timeline. Note that I call an $e^+e^-$ analysis ``photon-like'' because it produces an $e^+e^-$ shower like a photon does, although the position and direction relative to vertex activity, the opening angle, and the energy asymmetry can all have different properties. I made significant contributions in studying the consistency between these different results, as well as some other studies relevant to multiple photon-like analyses, and I briefly summarize some of that work in this chapter. I do not discuss a Wire-Cell $e^+e^-$ analysis which is still being developed, but many of the comparisons discussed in this chapter were also performed for that selection. 

\section{Other MicroBooNE Single-photon-like Analyses}

Here, I briefly summarize recent searches for NC coherent single photon production, inclusive single photon production, and dark neutrino $e^+e^-$ pair production.

\subsection{Coherent Single Photon Analysis}

\begin{figure}[H]
    \centering
    \includegraphics[width=0.6\textwidth]{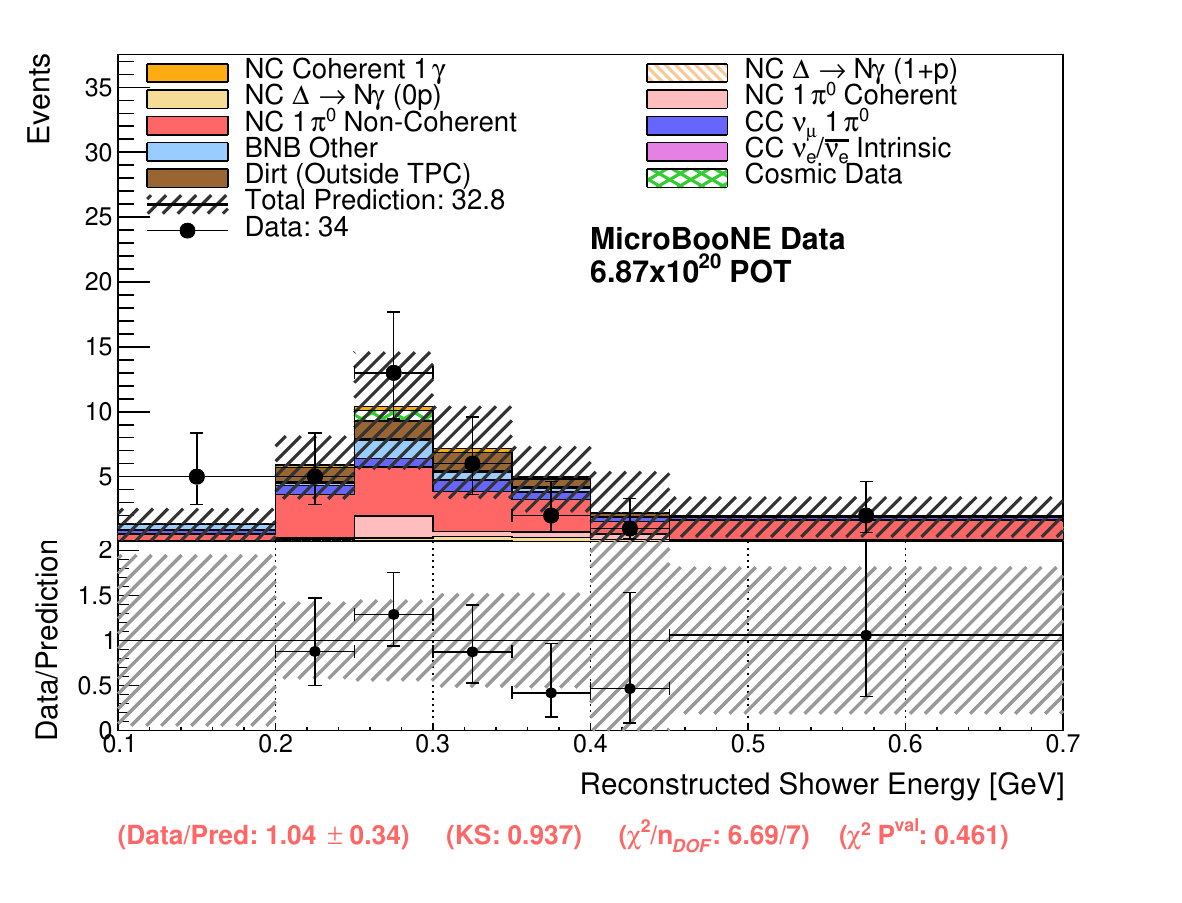}
    \caption[NC coherent single photon results]{NC coherent single photon results. From Ref. \cite{microboone_coherent_photon}.}
    \label{fig:microboone_coherent_single_photon_results}
\end{figure}

A search for coherent single photon production was developed. This topology has specifically a single photon and no other particles, and is predicted to be very rare, around ten times rarer than NC $\Delta\rightarrow N \gamma$ events. This selection built using the same tools as the Pandora NC $\Delta\rightarrow N \gamma$ selection, with additional tools developed in order to reject events with low energy proton activity upstream of the shower at the neutrino vertex. Overall, we see consistency between data and prediction, as shown in Fig. \ref{fig:microboone_coherent_single_photon_results}.

\subsection{Inclusive Single Photon Analysis}

\begin{figure}[H]
    \centering
    \begin{subfigure}[b]{0.49\textwidth}
        \includegraphics[trim=200 0 300 0, clip, width=\textwidth]{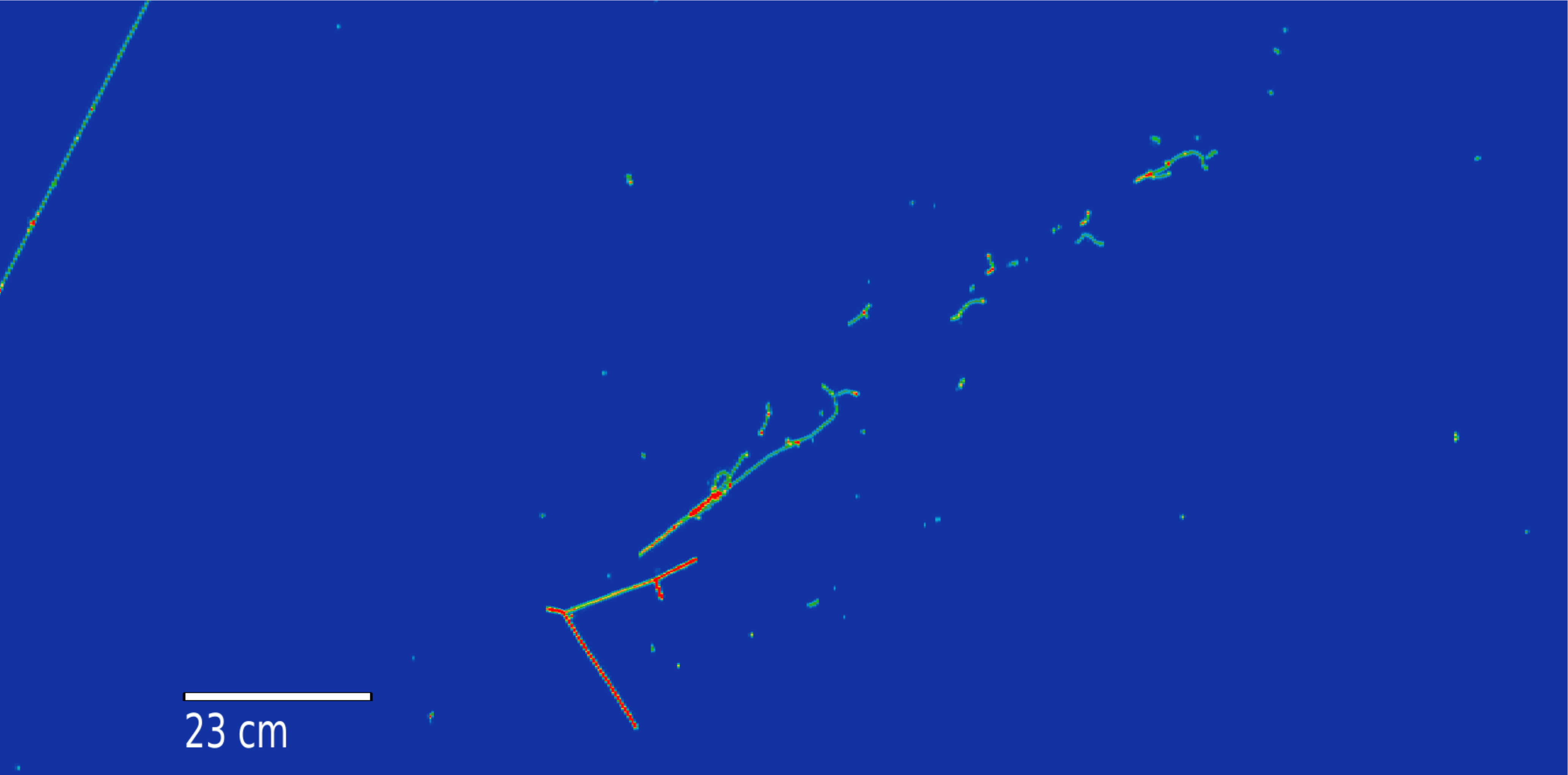}
        \caption{}
    \end{subfigure}
    \begin{subfigure}[b]{0.49\textwidth}
        \includegraphics[trim=15 15 15 15, clip, width=\textwidth]{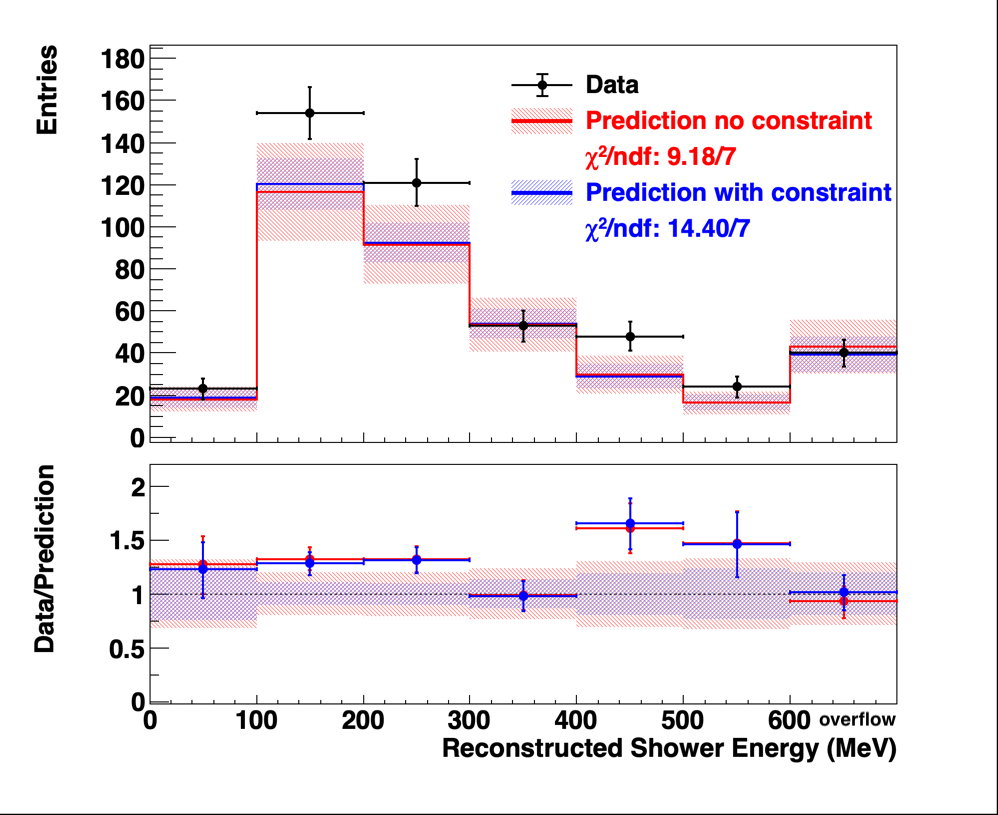}
        \caption{}
    \end{subfigure}
    \caption[Inclusive single photon candidate and $0p$ low energy results]{Panel (a) shows a candidate inclusive single photon event from BNB data, Run 9539, Subrun 72, Event 3634. Panel (b) shows the inclusive single photon $0p$ low shower energy spectrum from Ref. \cite{wc_inclusive_single_photon}.}
    \label{fig:inclusive_0p_low_energy}
\end{figure}

Using Wire-Cell reconstruction, an inclusive search for single photons was developed. The goal of this analysis was to target all photon events that could have been identified as single showers in MiniBooNE. Therefore, it targets true single photons as well as NC $\pi^0$ backgrounds where one photon is a low energy or exits the TPC, and where the two photons have a small opening angle. It also includes events with a low energy muon which would have been below MiniBooNE's Cherenkov threshold. This results in a selection which accepts a significantly wider range of topologies and a broader kinematic phase space than other analyses. This analysis saw overall good consistency with data in the fully inclusive channel, but when restricted to $0p$ events at low energy, a 2.2 $\sigma$ local excess is observed, as shown in Fig. \ref{fig:inclusive_0p_low_energy}.

\subsection{Dark Neutrino \texorpdfstring{$e^+e^-$}{e+e-} Analysis}

\begin{figure}[H]
    \centering
    \begin{subfigure}[b]{0.4\textwidth}
        \includegraphics[width=\textwidth]{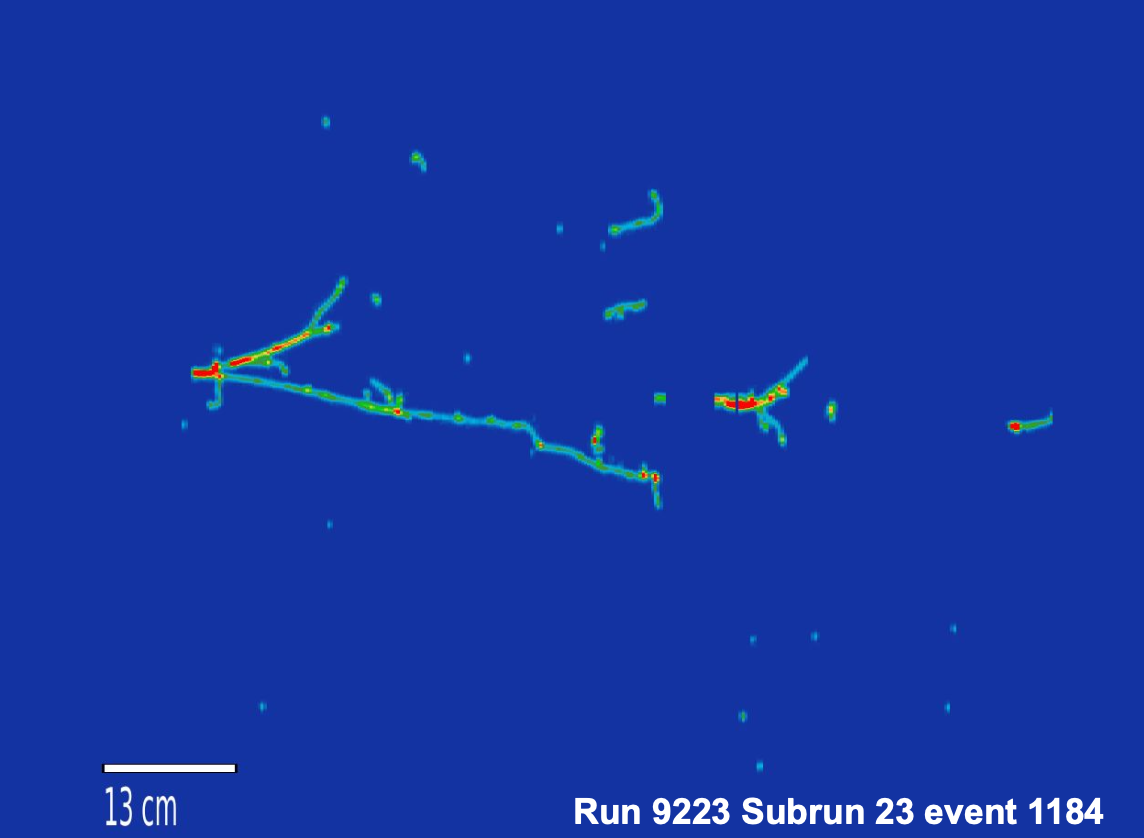}
        \caption{}
    \end{subfigure}
    \begin{subfigure}[b]{0.59\textwidth}
        \includegraphics[trim=15 630 15 15, clip, width=\textwidth]{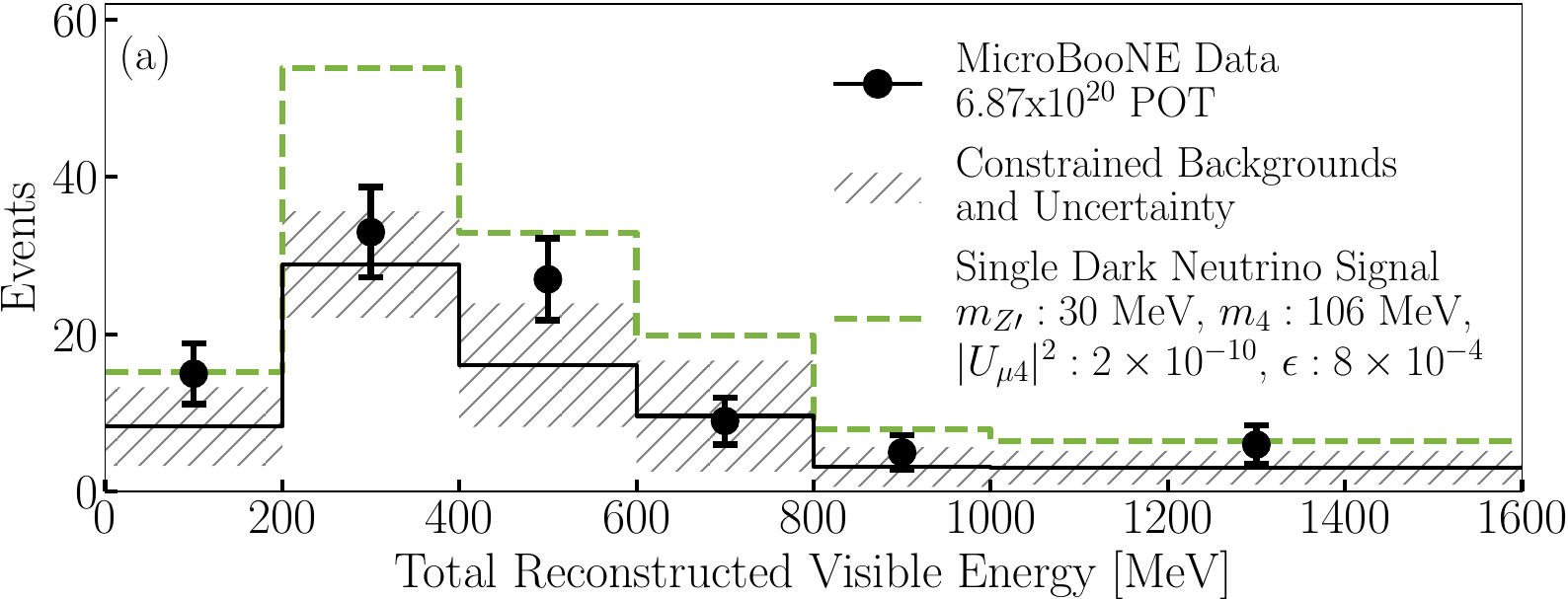}
        \caption{}
    \end{subfigure}
    \caption[$e^+e^-$ candidate and results]{Panel (a) shows a candidate $e^+e^-$ event from BNB data, Run 9223 Subrun 23 Event 1184. Panel (b) shows the resulting visible energy spectrum from Ref. \cite{uboone_epem}.}
    \label{fig:epem_data}
\end{figure}

Again using the same tools as the Pandora NC $\Delta\rightarrow N \gamma$ selection, a search for $e^+e^-$ events was developed. This targets dark sector explanations of the MiniBooNE LEE, and is able to exclude the majority of the allowed phase space in certain models, largely due to the coherent nature of the signal which enhances the prediction for the larger argon nucleus compared to carbon in MiniBooNE. Overall, we see good agreement with our prediction as shown in Fig. \ref{fig:epem_data}.

\section{Studies Comparing Single-photon-like Selections}

In this section, we compare the different selections in several ways.

\subsection{Overlapping Data Events Between Different Selections}

\begin{figure}[H]
    \centering
    \includegraphics[width=0.8\textwidth]{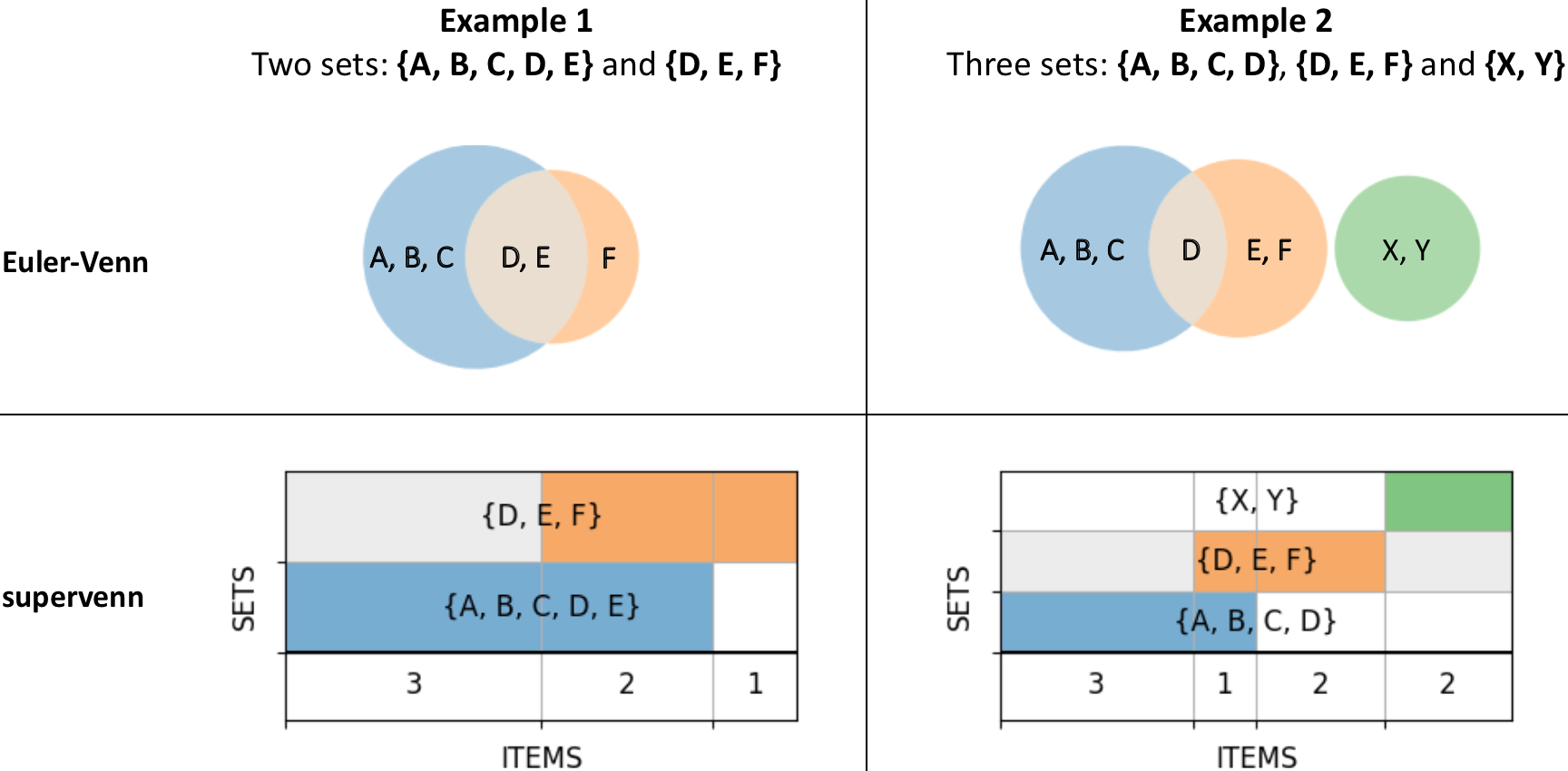}
    \caption[Super-Venn diagram example]{Super-Venn diagram example from Ref. \cite{supervenn}.}
    \label{fig:super_venn_example}
\end{figure}

In order to quantify the amount of overlapping data events between each selection, we use Ref. \cite{supervenn} in order to visualize a Venn diagram with a higher number of possible overlaps, as illustrated in Fig. \ref{fig:super_venn_example}.

\begin{figure}[H]
    \centering
    \begin{subfigure}[b]{\textwidth}
        \includegraphics[trim=120 20 100 40, clip, width=\textwidth]{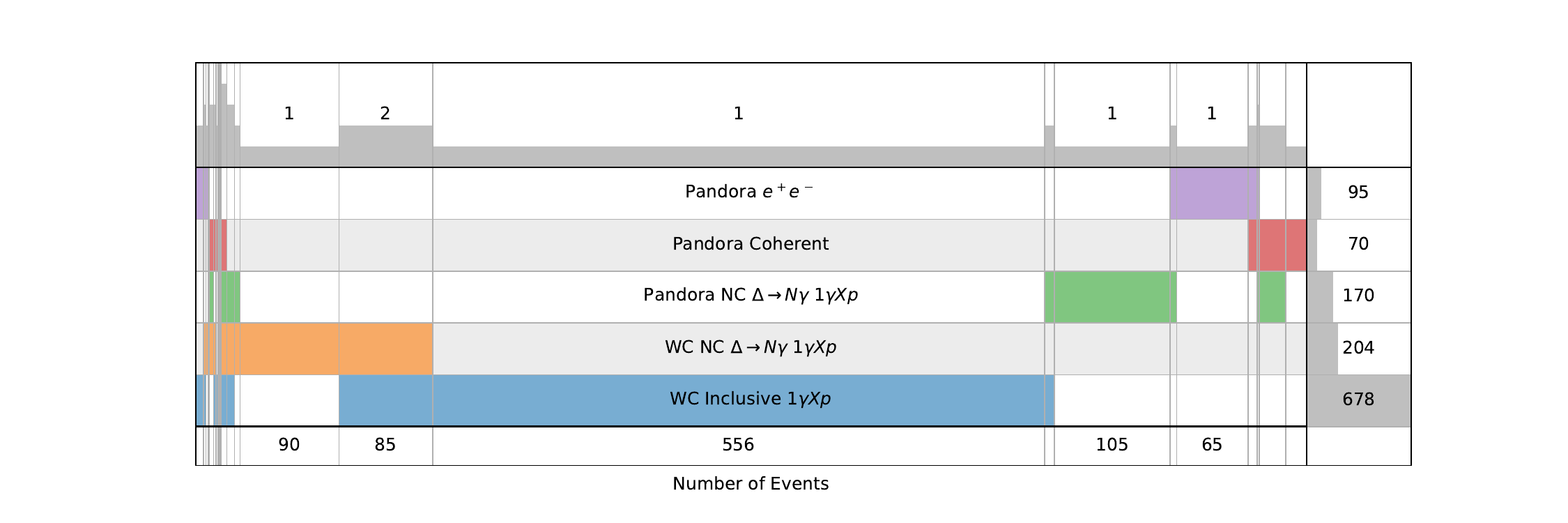}
        \caption{}
    \end{subfigure}
    \begin{subfigure}[b]{\textwidth}
        \includegraphics[trim=120 20 100 40, clip, width=\textwidth]{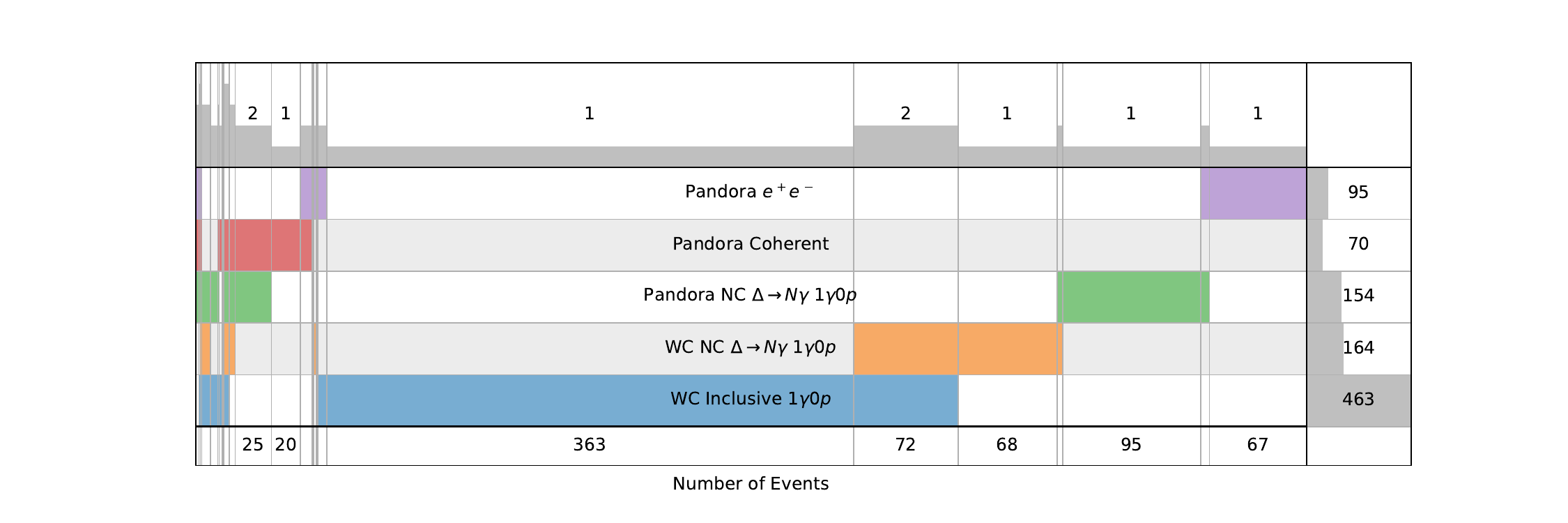}
        \caption{}
    \end{subfigure}
    \caption[Overlapping data events]{Overlapping data events. Panel (a) shows overlaps between each of the five selections with no particle multiplicity cuts. Panel (b) shows overlaps between each of the five selections with a zero reconstructed proton cut applied when applicable.}
    \label{fig:overlap_data}
\end{figure}

The resulting overlaps are shown for $1\gamma Xp$ selections and $1\gamma 0p$ selections in Fig. \ref{fig:overlap_data}. We see that in general, there is very little overlap between different selections. In particular, overlaps between Wire-Cell selected events and Pandora selected events are rare. Around half of the Pandora coherent selected events are selected by the Pandora NC $\Delta\rightarrow N \gamma$ selection, and around half of the Wire-Cell NC $\Delta\rightarrow N \gamma$ selected events are selected by the Wire-Cell inclusive selection.

\subsection{\texorpdfstring{$0p$}{0p} Shower Kinematic Efficiencies}

\begin{figure}[H]
    \centering
    \begin{subfigure}[b]{0.49\textwidth}
        \includegraphics[trim=15 15 30 20, clip, width=\textwidth]{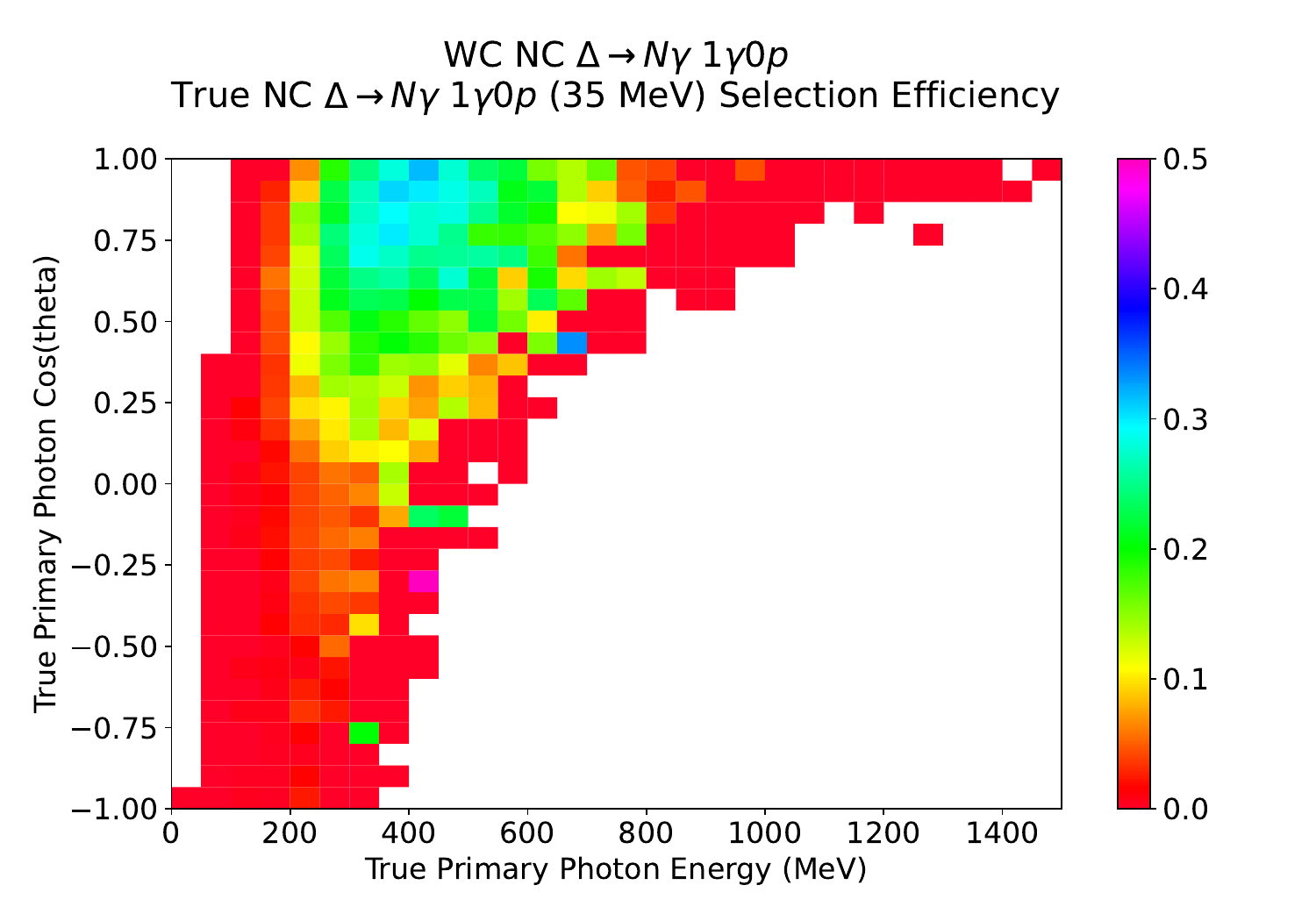}
        \caption{}
    \end{subfigure}
    \begin{subfigure}[b]{0.49\textwidth}
        \includegraphics[trim=15 15 30 20, clip, width=\textwidth]{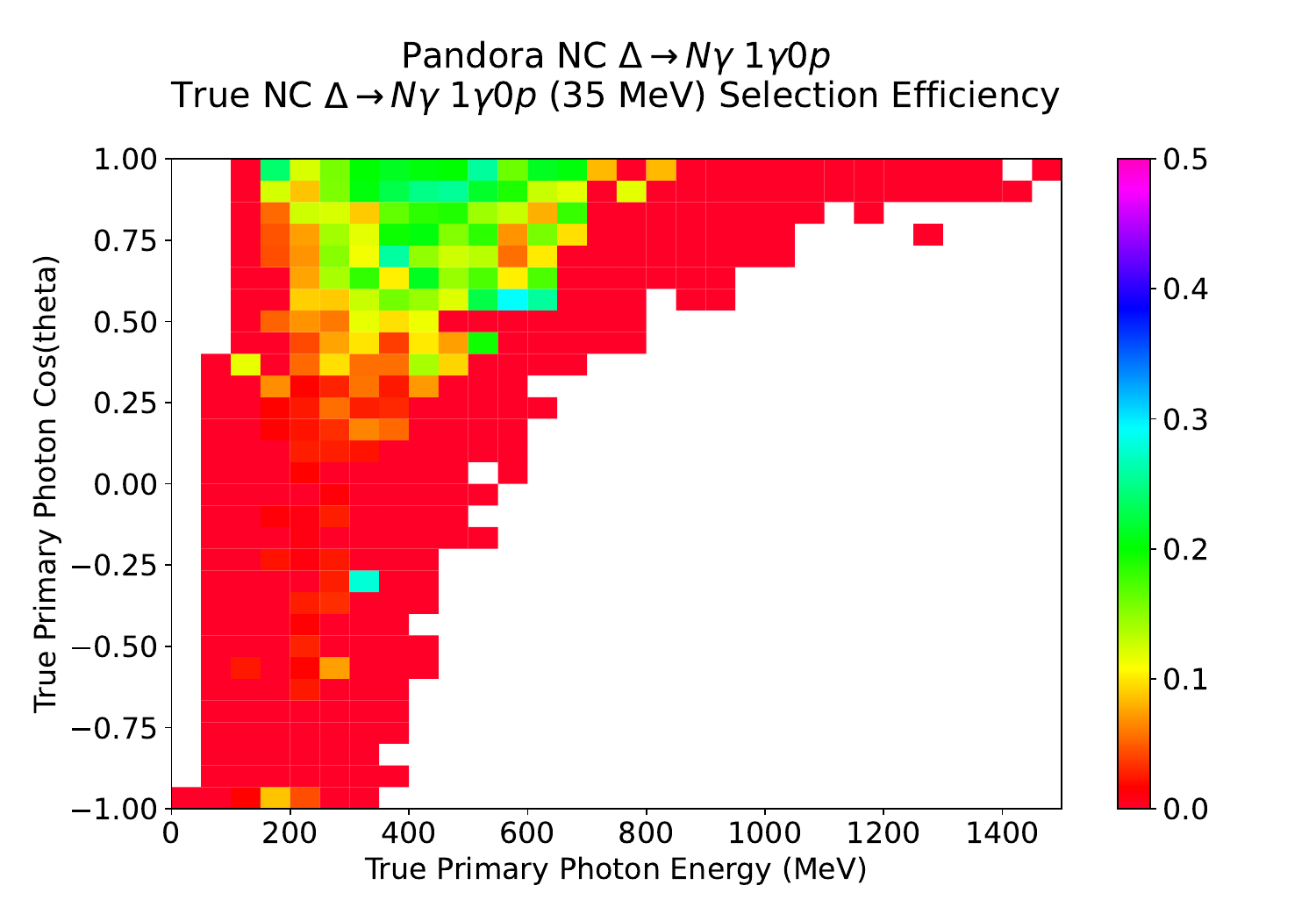}
        \caption{}
    \end{subfigure}
    \begin{subfigure}[b]{0.49\textwidth}
        \includegraphics[trim=15 15 30 20, clip, width=\textwidth]{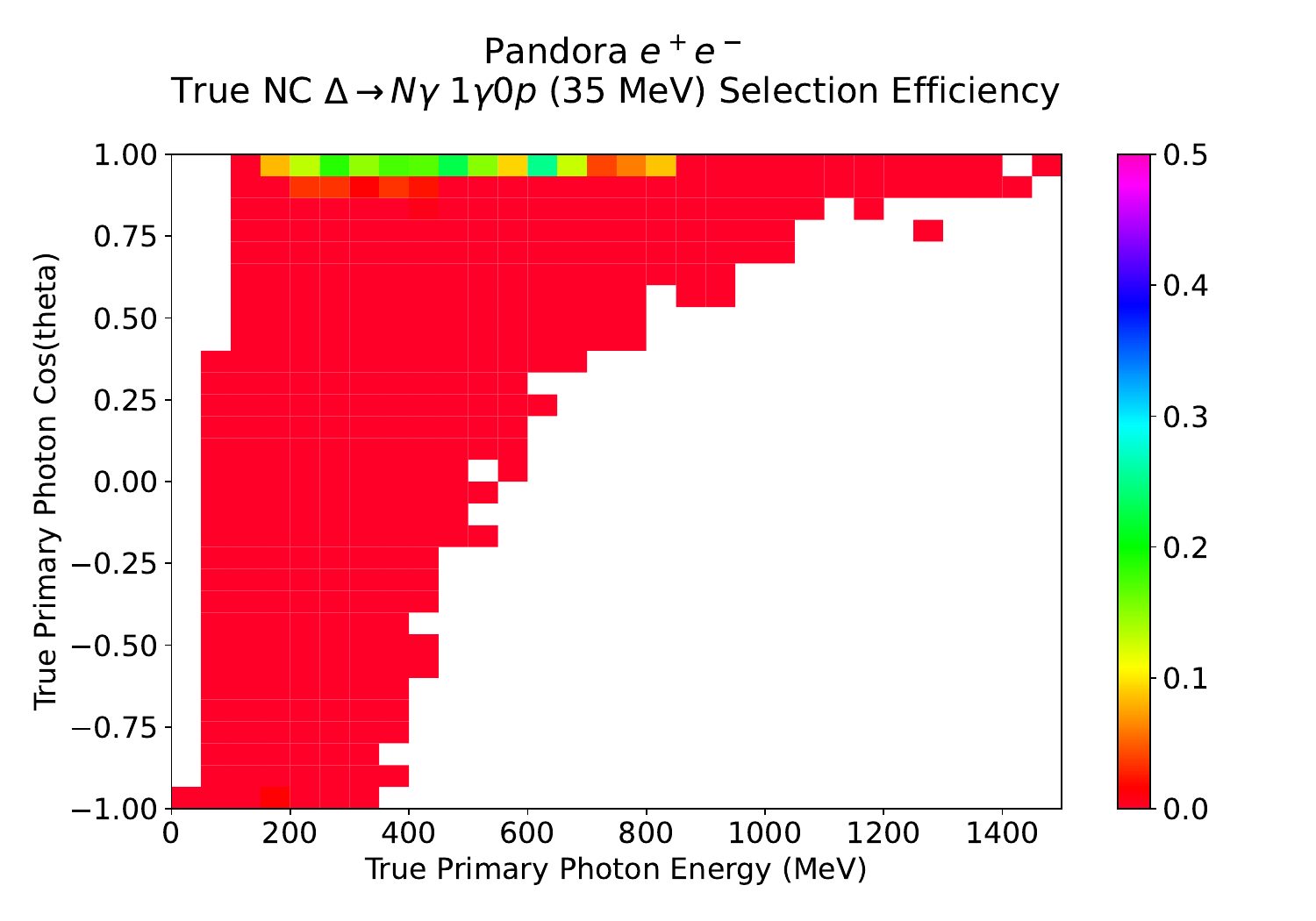}
        \caption{}
    \end{subfigure}
    \begin{subfigure}[b]{0.49\textwidth}
        \includegraphics[trim=15 15 30 20, clip, width=\textwidth]{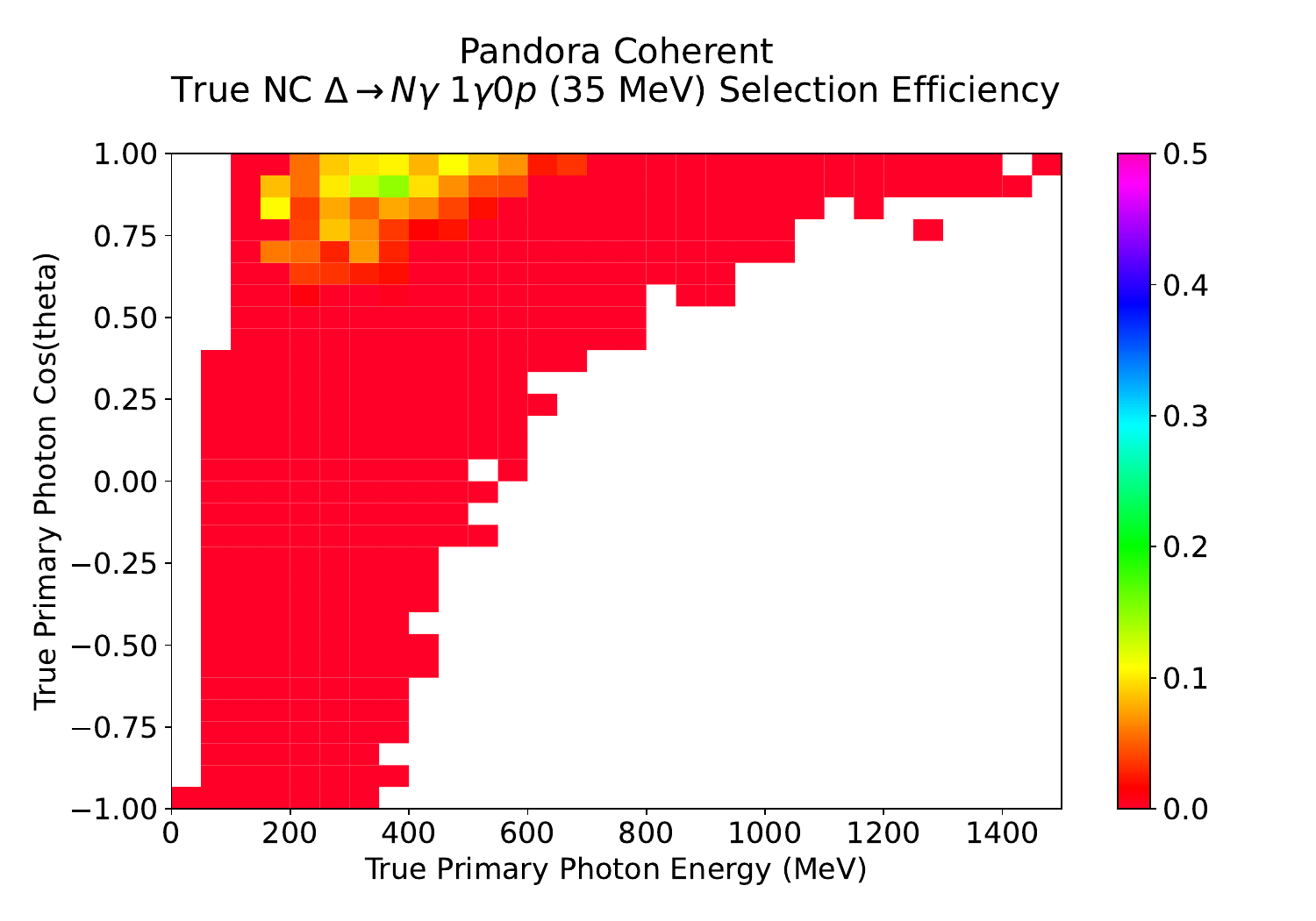}
        \caption{}
    \end{subfigure}
    \begin{subfigure}[b]{0.49\textwidth}
        \includegraphics[trim=15 15 30 20, clip, width=\textwidth]{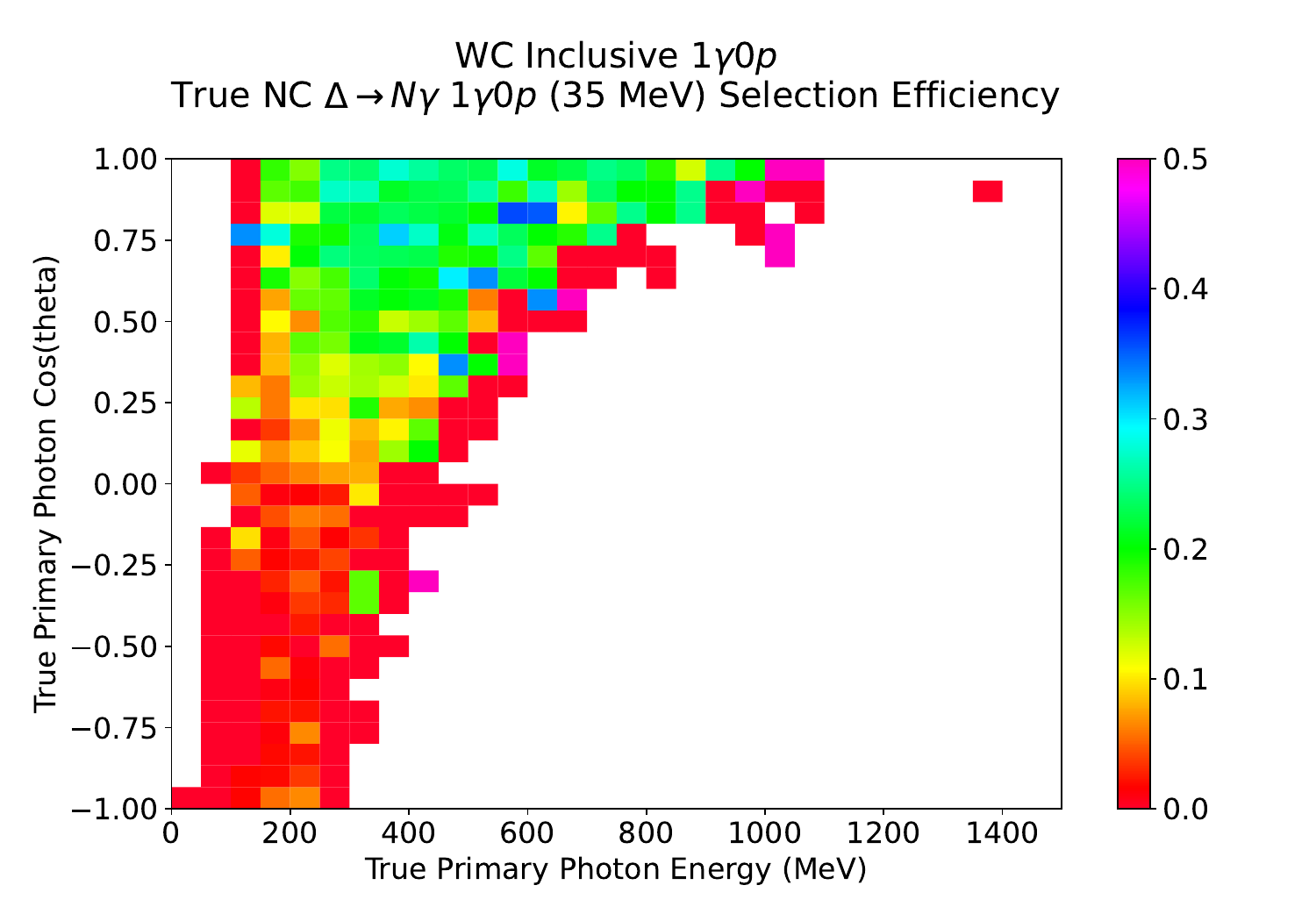}
        \caption{}
    \end{subfigure}
    \caption[True $1\gamma 0p$ (35 MeV threshold) efficiencies for different photon-like selections]{2D shower kinematic efficiencies for NC $\Delta\rightarrow N \gamma$ events with no true primary proton with kinetic energy greater than 35 MeV. Panel (a) shows the Wire-Cell NC $\Delta\rightarrow N \gamma$ $1\gamma0p$ selection, panel (b) shows the Pandora NC $\Delta\rightarrow N \gamma$ $1\gamma0p$ selection, panel (c) shows the Pandora $e^+e^-$ selection, panel (d) shows the Pandora coherent selection, and panel (e) shows the Wire-Cell inclusive selection.}
    \label{fig:true_1g0p_35_eff_comparisons}
\end{figure}

\begin{figure}[H]
    \centering
    \begin{subfigure}[b]{0.49\textwidth}
        \includegraphics[trim=15 15 30 20, clip, width=\textwidth]{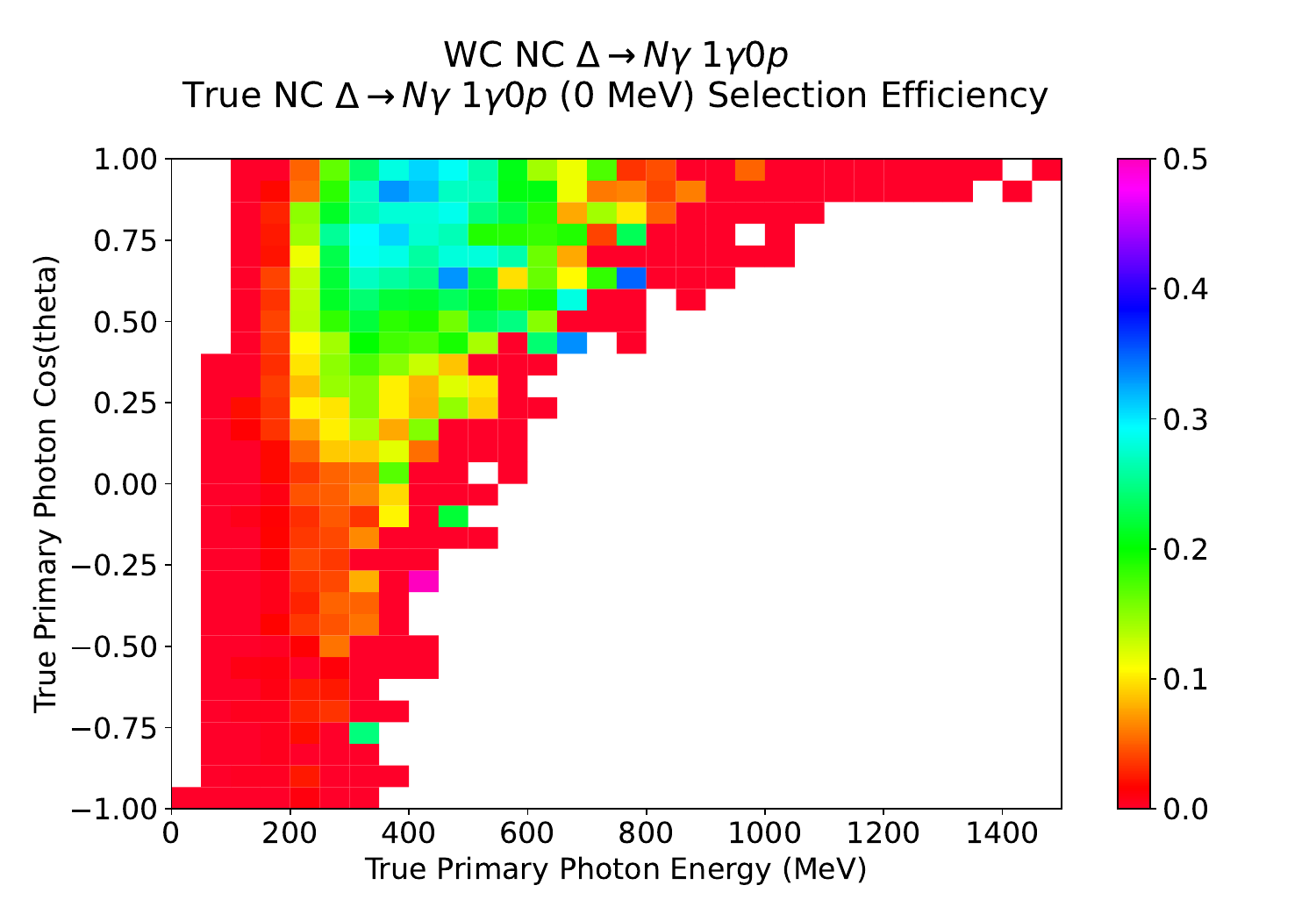}
        \caption{}
    \end{subfigure}
    \begin{subfigure}[b]{0.49\textwidth}
        \includegraphics[trim=15 15 30 20, clip, width=\textwidth]{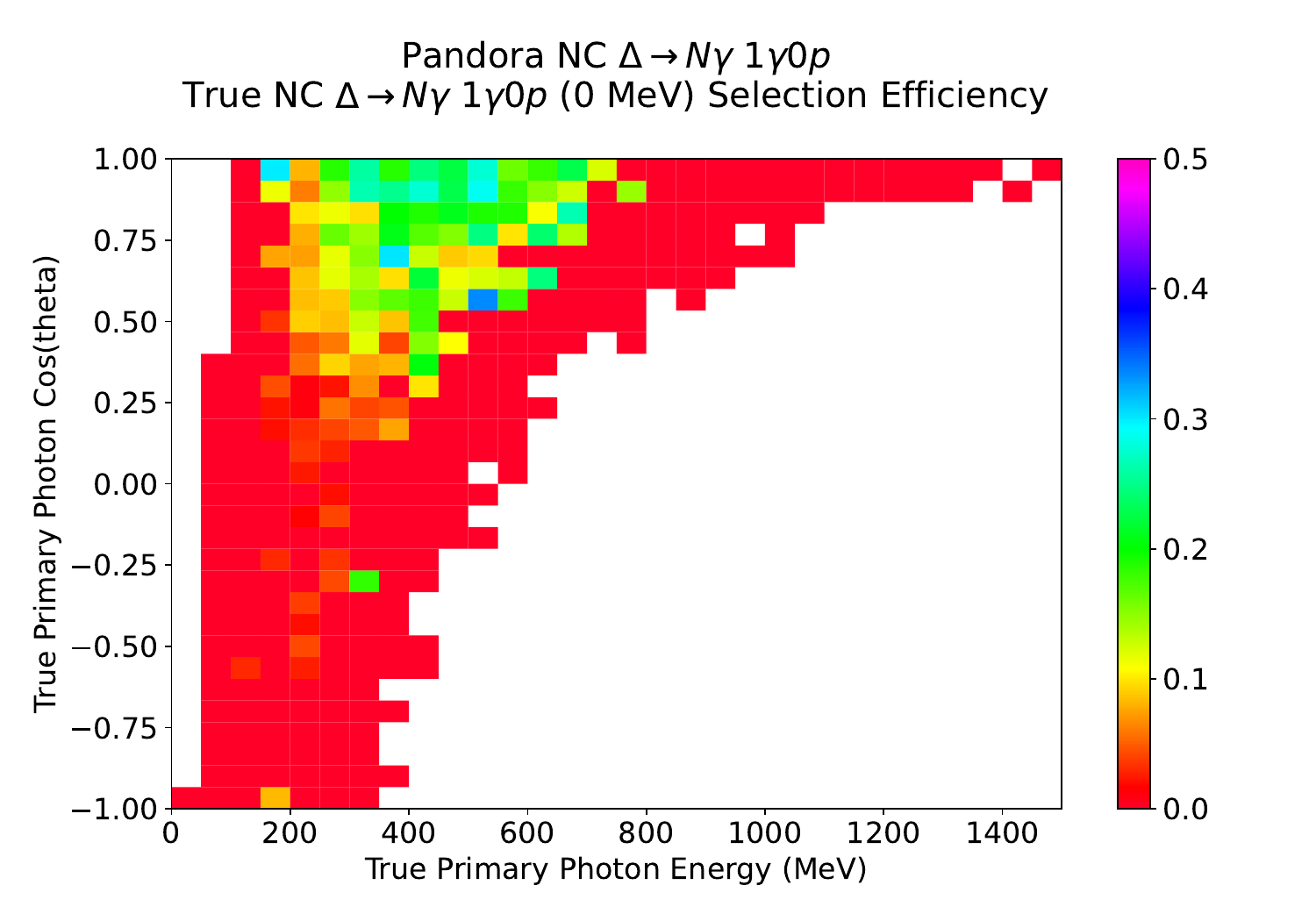}
        \caption{}
    \end{subfigure}
    \begin{subfigure}[b]{0.49\textwidth}
        \includegraphics[trim=15 15 30 20, clip, width=\textwidth]{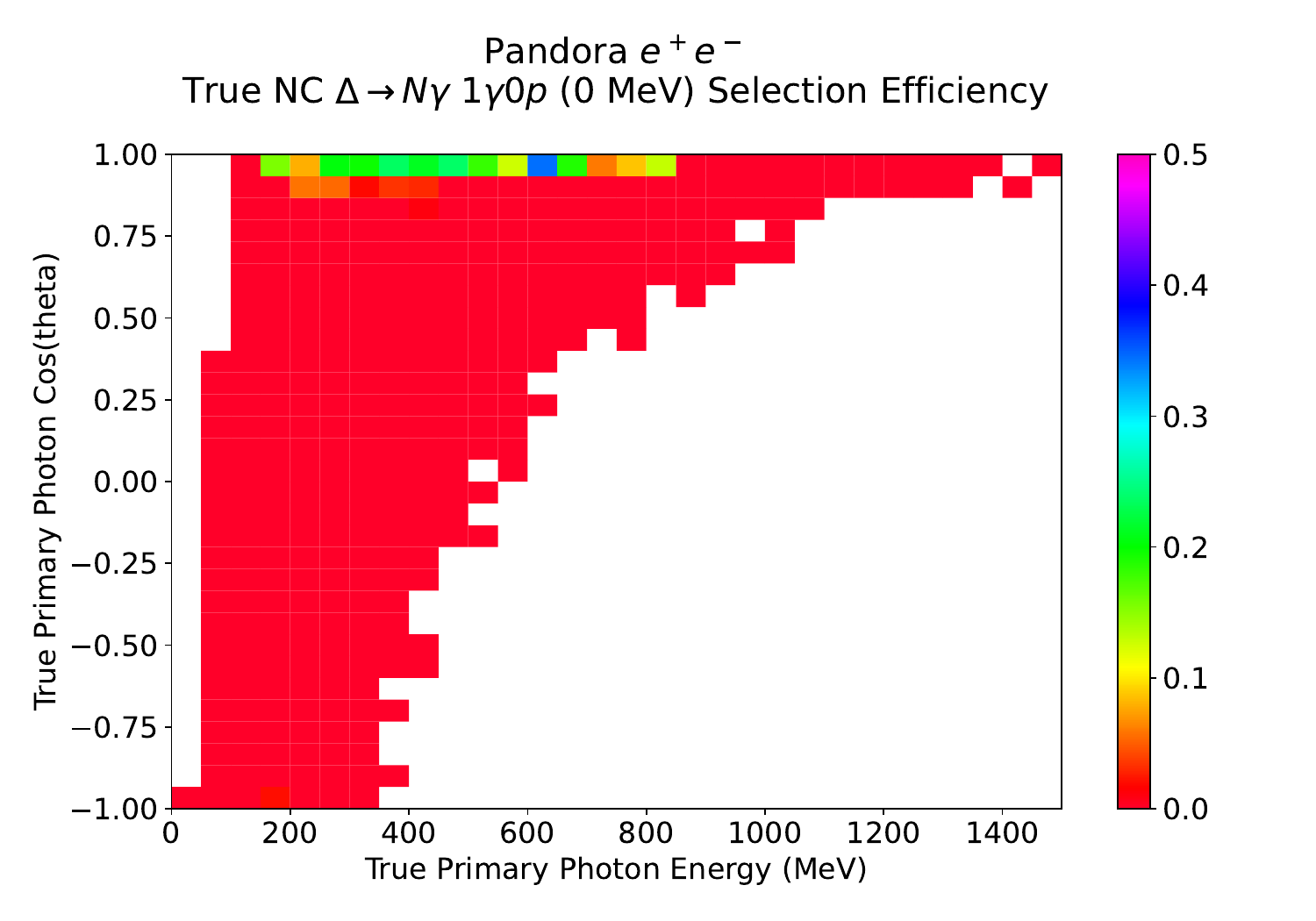}
        \caption{}
    \end{subfigure}
    \begin{subfigure}[b]{0.49\textwidth}
        \includegraphics[trim=15 15 30 20, clip, width=\textwidth]{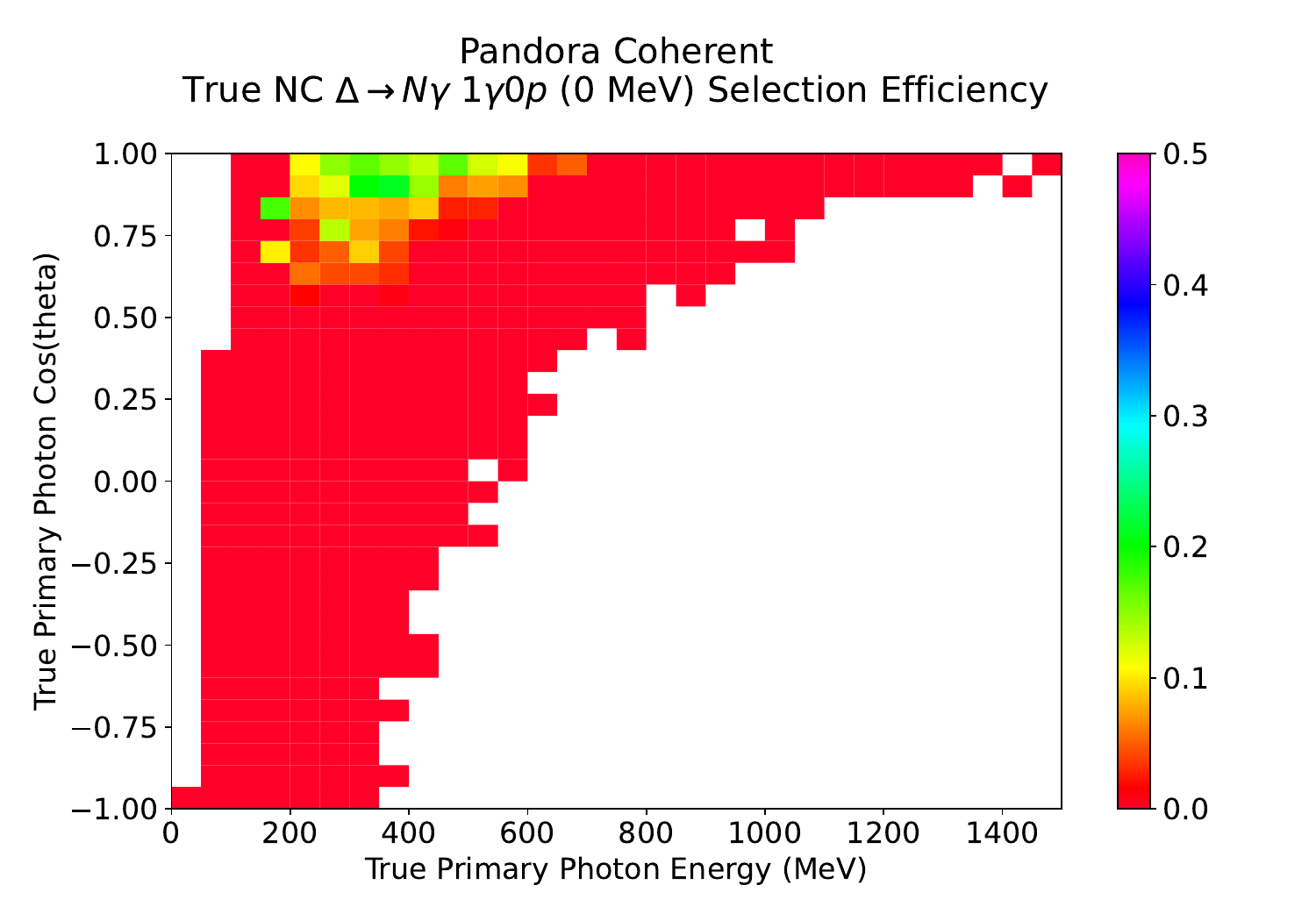}
        \caption{}
    \end{subfigure}
    \begin{subfigure}[b]{0.49\textwidth}
        \includegraphics[trim=15 15 30 20, clip, width=\textwidth]{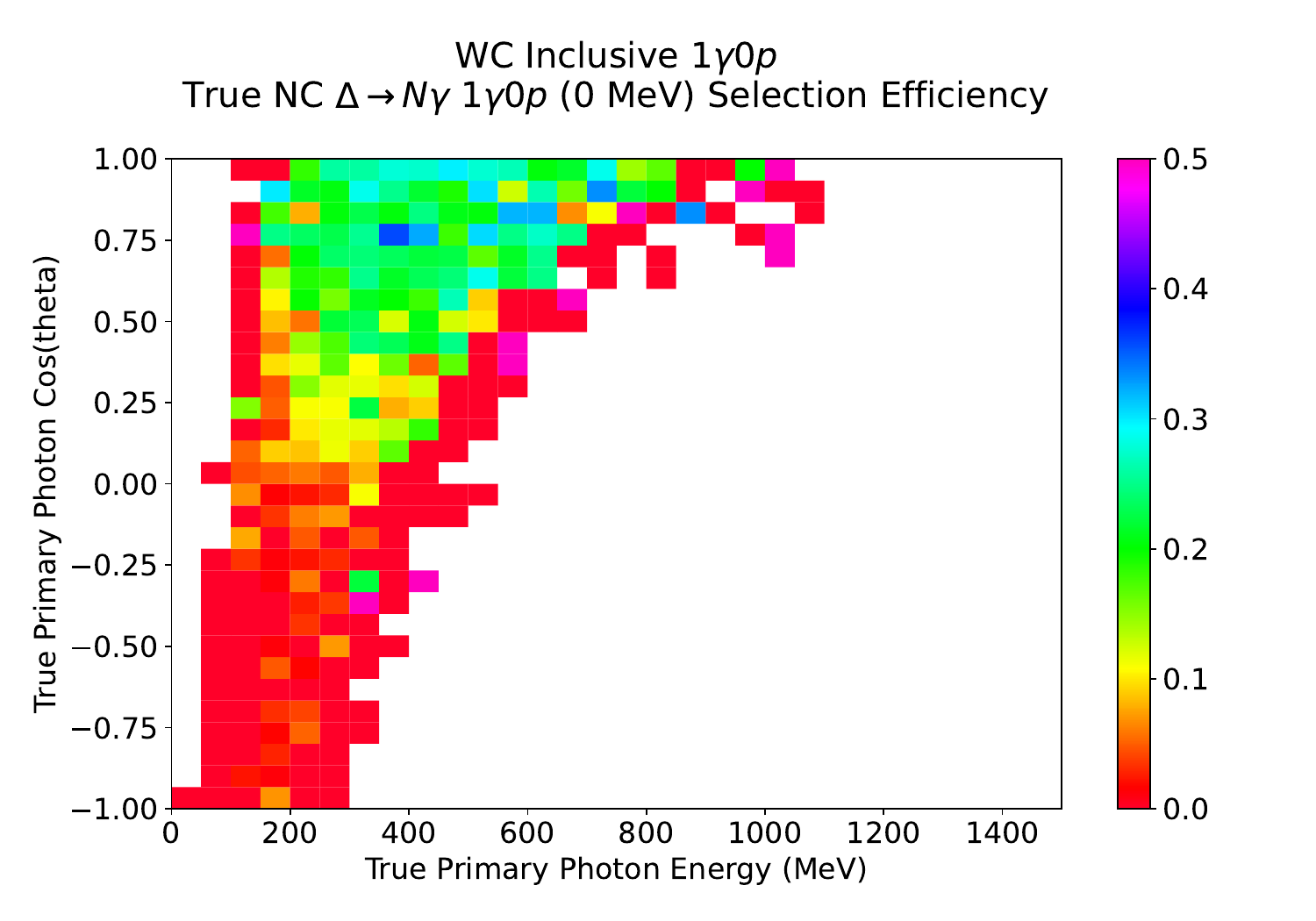}
        \caption{}
    \end{subfigure}
    \caption[True $1\gamma 0p$ (no threshold) efficiencies for different photon-like selections]{2D shower kinematic efficiencies for NC $\Delta\rightarrow N \gamma$ events with no true primary proton with any kinetic energy. Panel (a) shows the Wire-Cell NC $\Delta\rightarrow N \gamma$ $1\gamma0p$ selection, panel (b) shows the Pandora NC $\Delta\rightarrow N \gamma$ $1\gamma0p$ selection, panel (c) shows the Pandora $e^+e^-$ selection, panel (d) shows the Pandora coherent selection, and panel (e) shows the Wire-Cell inclusive selection.}
    \label{fig:true_1g0p_0_eff_comparisons}
\end{figure}

Something that all of these different analyses have in common is the ability to focus on events with a photon shower and zero protons. In this case, we can compare the efficiencies as functions of photon shower energy and angle. In Fig. \ref{fig:true_1g0p_35_eff_comparisons}, we show the efficiencies for NC $\Delta\rightarrow N \gamma$ events with no primary protons with greater than 35 MeV of kinetic energy, or no protons which will make a track long enough for our particle identification algorithms to consistently identify. In Fig. \ref{fig:true_1g0p_0_eff_comparisons}, we show the efficiencies for NC $\Delta\rightarrow N \gamma$ events with no primary protons of any kinetic energy, leaving no low-energy vertex activity at all. In these plots, we can note that the Wire-Cell NC $\Delta\rightarrow N \gamma$ has the highest peak efficiency at forward shower angles and intermediate energies. The Wire-Cell inclusive selection has the broadest efficiency with the most acceptance at low and high shower energies and in the backward direction. The Pandora $e^+e^-$ and coherent selections are the most focused on the forward direction, which is in line with the angular distribution of their expected signals, and also helps to reject $\pi^0$ backgrounds in these specific rare searches. For true events truly containing only a single photon shower, this represents the complete set of efficiencies as functions of underlying physics parameters. However, there are still more complex efficiencies related to the detector to consider, for example the position relative to the TPC boundaries, or the distance to the nearest cosmic ray cluster; these types of factors can cause additional efficiency differences between these selections which are not illustrated here.

\subsection{Proton Energy Efficiencies}

\begin{figure}[H]
    \centering
    \includegraphics[width=0.8\textwidth]{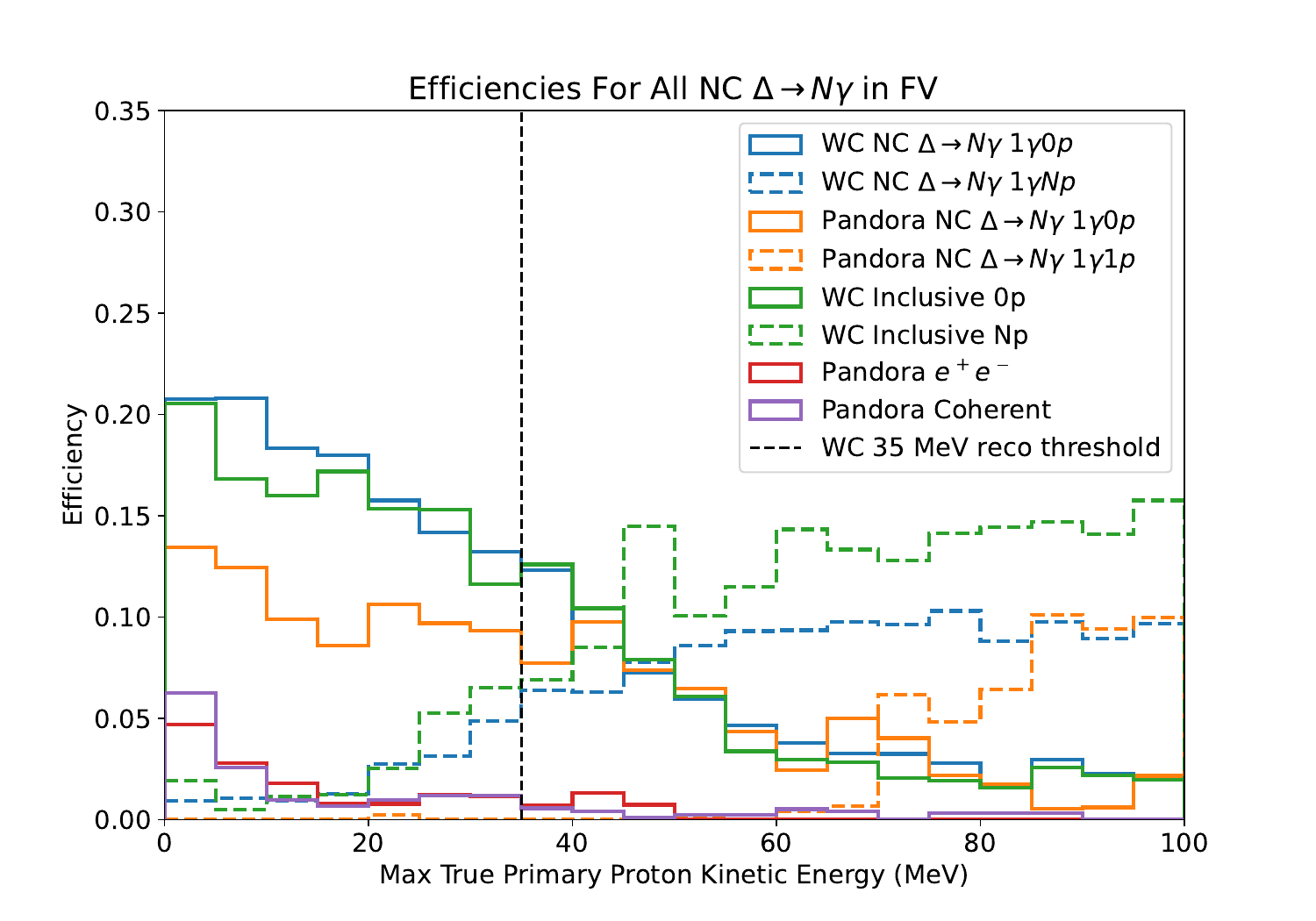}
    \caption[Proton energy efficiencies]{Proton energy efficiencies.}
    \label{fig:proton_energy_efficiencies}
\end{figure}

We can also compare each of these selections' performances as functions of the maximum true primary proton kinetic energy, as shown in Fig. \ref{fig:proton_energy_efficiencies}. We use NC $\Delta\rightarrow N \gamma$ simulated events in order to evaluate these efficiencies. For the selections with no reconstructed protons in solid lines, the efficiency rises at lower proton kinetic energies, as the protons get harder to reconstruct. For the Wire-Cell NC $\Delta\rightarrow N \gamma$ and inclusive selections, the efficiency rises around 35 MeV, which corresponds to our reconstructed proton kinetic energy threshold, and the Pandora NC $\Delta\rightarrow N \gamma$ selection rises similarly. For the Pandore $e^+e^-$ and coherent selections, the efficiency rises only at much lower energies, around 5-10 MeV. This is due to the fact that these analyses specifically target signals with no visible hadronic activity at the vertex, and the fact that these selections use a specific ``proton stub veto'' tool which was developed in order to reject events with low energy protons. For the events with reconstructed protons shown in dashed lines, the Wire-Cell NC $\Delta\rightarrow N \gamma$ and inclusive selections increase in efficiency around 35 MeV as expected, and the Pandora NC $\Delta\rightarrow N \gamma$ selection rises later, at around 70-80 MeV.

\subsection{Photon Conversion Distance Efficiencies}

\begin{figure}[H]
    \centering
    \includegraphics[width=0.8\textwidth]{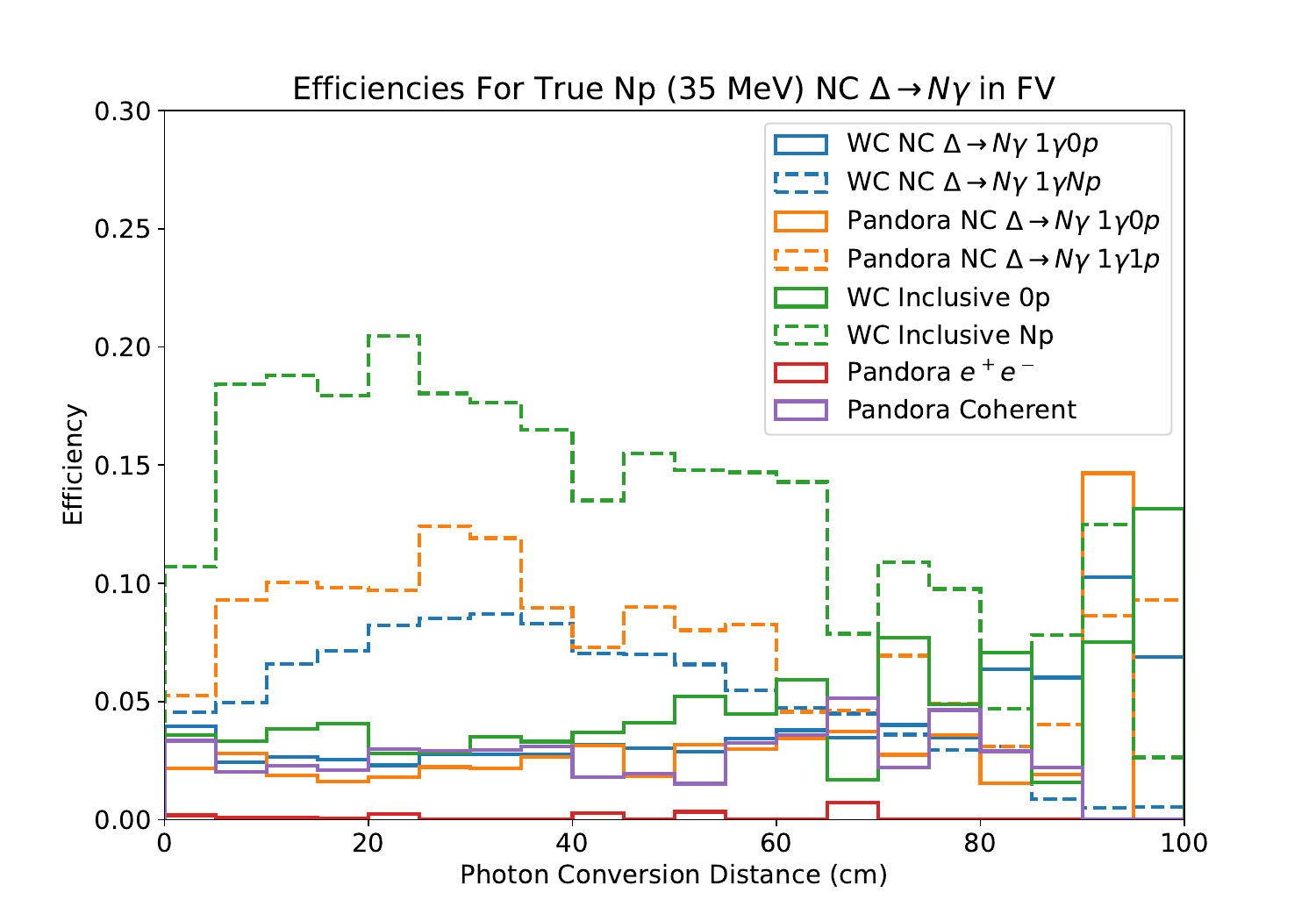}
    \caption[Photon conversion distance efficiencies]{Photon conversion distance efficiencies.}
    \label{fig:conversion_distance_efficiencies}
\end{figure}

Lastly, we consider the efficiency as functions of the photon conversion distance, as shown in Fig. \ref{fig:conversion_distance_efficiencies}. This is the distance between the neutrino interaction vertex and the photon pair production which starts the electromagnetic shower. To calculate these efficiencies, we use NC $\Delta\rightarrow N \gamma$ simulated events with reconstructed protons with greater than 35 MeV of kinetic energy, so that the neutrino vertex activity will have visible activity and influence the efficiency for different sizes of gaps between the neutrino vertex and the shower. For events with reconstructed protons in dashed lines, the efficiency decreases at short conversion distances, since for small gaps, the proton can blend in with a shower and fail to be reconstructed. At large conversion distances, the proton can fail to be clustered with the photon shower, which can cause a reduction in efficiency as we see. For events without reconstructed protons in solid lines, the distributions look fairly flat, but we can see subtle increases in efficiency at low and high conversion distances, which can be caused by the same factors affecting efficiencies for events with reconstructed protons as explained above.

\section{Detector Model Studies Relevant for Single Photon-like Analyses}

In this section, I describe two studies we performed in order to better understand how sensitive some of our selections are to different types of detector modeling.

\subsection{Geant4 Electromagnetic Model Modifications}

\begin{figure}[H]
    \centering
    \begin{subfigure}[b]{0.49\textwidth}
        \includegraphics[width=\textwidth]{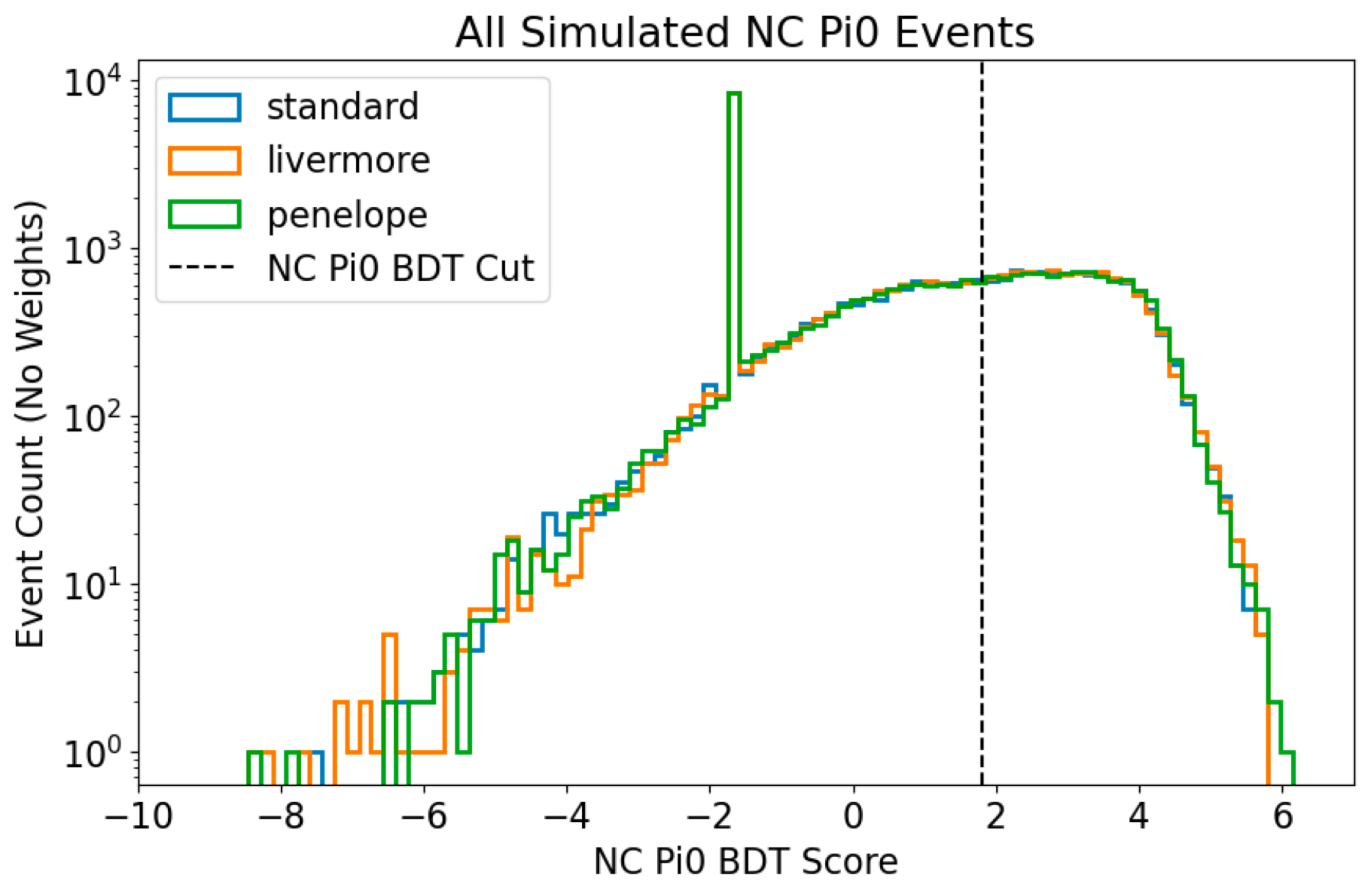}
        \caption{}
    \end{subfigure}
    \begin{subfigure}[b]{0.49\textwidth}
        \includegraphics[width=\textwidth]{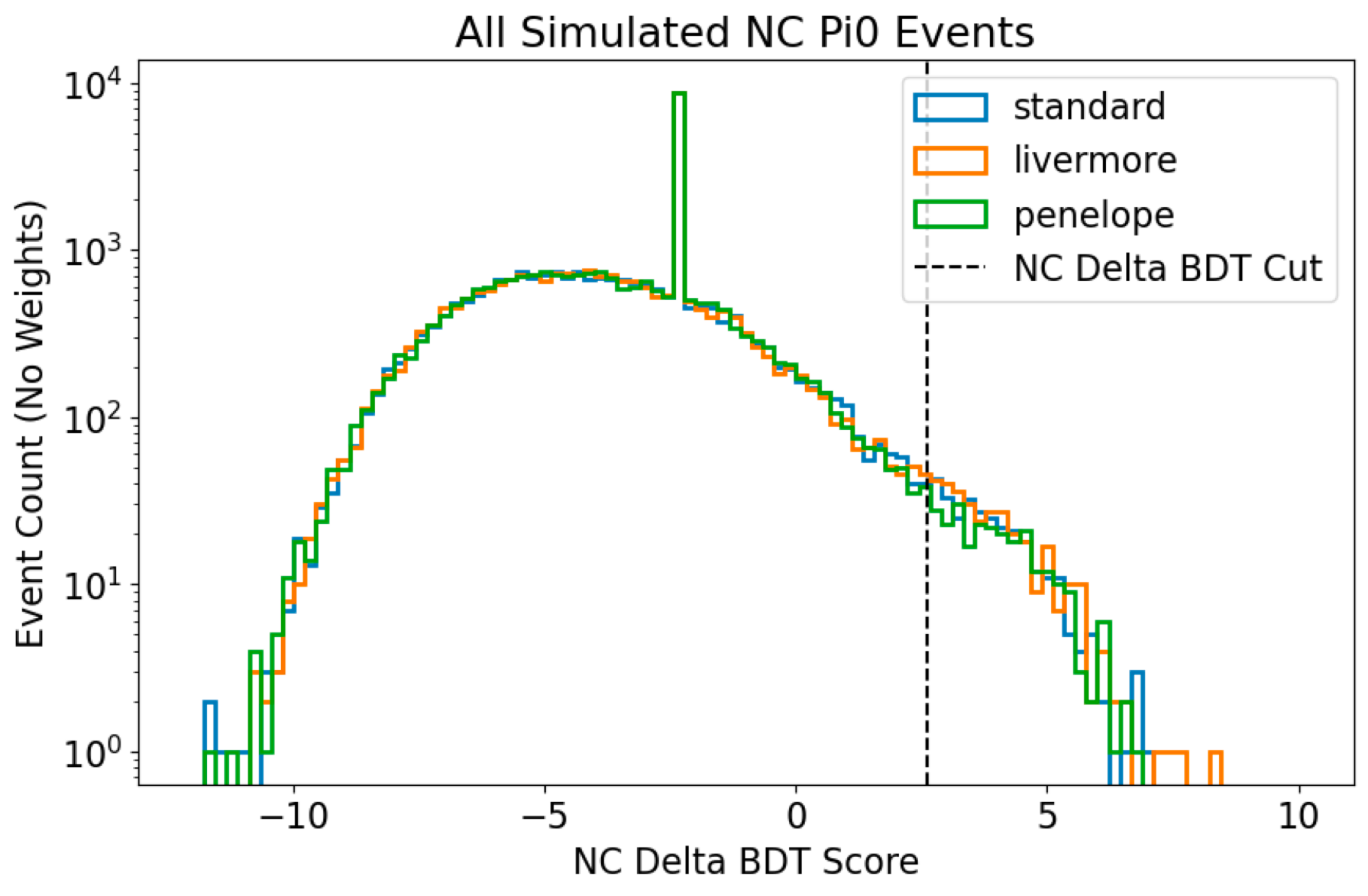}
        \caption{}
    \end{subfigure}
    \begin{subfigure}[b]{0.49\textwidth}
        \includegraphics[width=\textwidth]{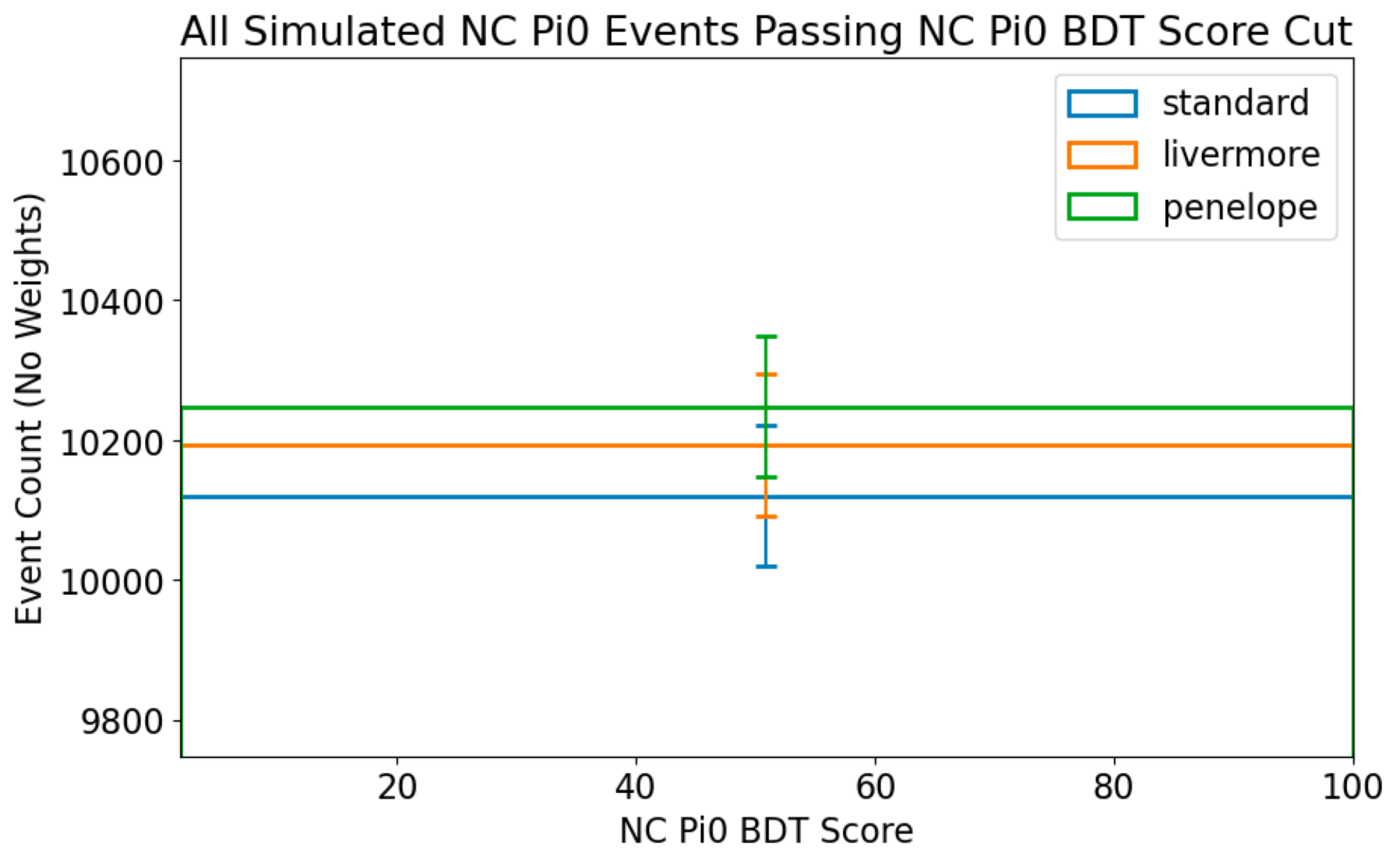}
        \caption{}
    \end{subfigure}
    \begin{subfigure}[b]{0.49\textwidth}
        \includegraphics[width=\textwidth]{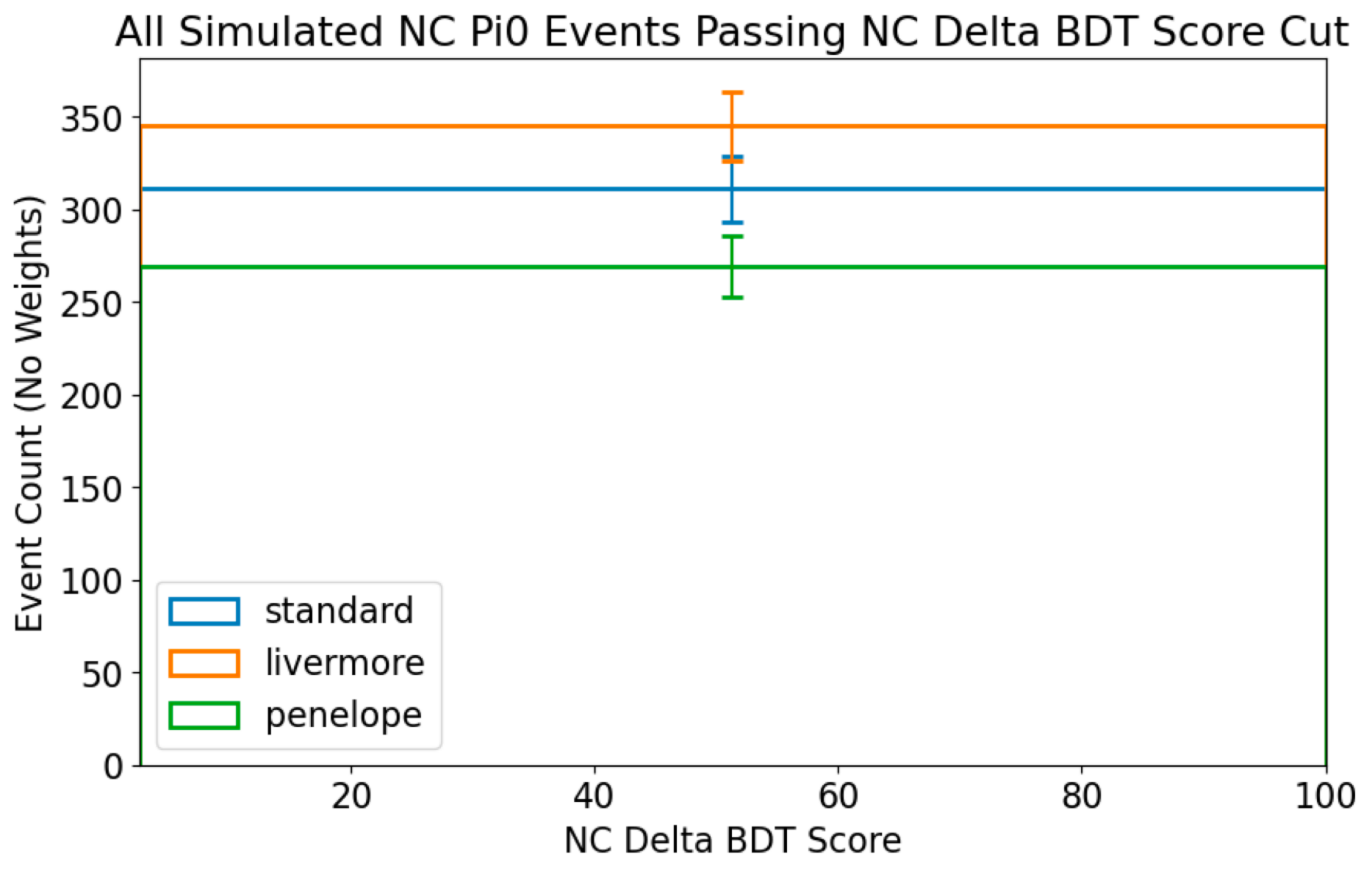}
        \caption{}
    \end{subfigure}
    \caption[Geant4 EM modeling modifications]{Geant4 EM modeling modifications. Panel (a) shows the Wire-Cell NC $\pi^0$ BDT score, and panel (b) shows the Wire-Cell NC $\Delta\rightarrow N \gamma$ BDT score. The cut values for each selection are indicated as a black dashed vertical line. Spikes in the BDT score distributions consist of events which fail Wire-Cell generic neutrino selection. Panels (c) and (d) show the resulting one-bin distributions for events passing the BDT cuts.}
    \label{fig:em_mod}
\end{figure}

In our \textsc{Geant4} detector simulation, we use reweighting to apply uncertainties to certain hadronic interactions, but we do not have any uncertainty on our electromagnetic modeling. Since the detailed behavior of the electromagnetic model is potentially very important to the behavior of electromagnetic showers, we study the effects of changing the underlying electromagnetic physics model to the Penelope and Livermore models, described in more detail in Ref. \cite{Geant4_PhysRefManual}. We re-simulated NC $\pi^0$ events using this alternate physics, and study the resulting Wire-Cell NC $\pi^0$ and NC $\Delta\rightarrow N \gamma$ selections in Fig. \ref{fig:em_mod}. We see consistency within statistical uncertainties for all three underlying electromagnetic models, which helps increase our confidence in the underlying simulation and the robustness of our selections to small changes in the electromagnetic shower physics.

\subsection{Out-TPC Light Yield Modifications}

The light propagation model is particularly important to understand and validate, for a few reasons. First, the light modeling is the largest detector uncertainty for some of our single photon searches, as shown in Fig. \ref{fig:nc_delta_detvar_breakdown}. Secondly, the detector response to scintillation light changes over time as the light yield has decreased over the years of MicroBooNE running, as described in Sec. \ref{sec:light_detvar}. Lastly, light modeling has a specific potential application for single photon searches, since scintillation light is the only possible way to identify a $\pi^0$ photon which showers outside of the TPC. This is not an application that we have explicitly made use of in our existing analyses, but the behavior of light being produced outside the TPC is still especially important for $\pi^0$ events which often have only part of the total activity identified in a cluster.

\begin{figure}[H]
    \centering
    \begin{subfigure}[b]{0.3\textwidth}
        \includegraphics[width=\textwidth]{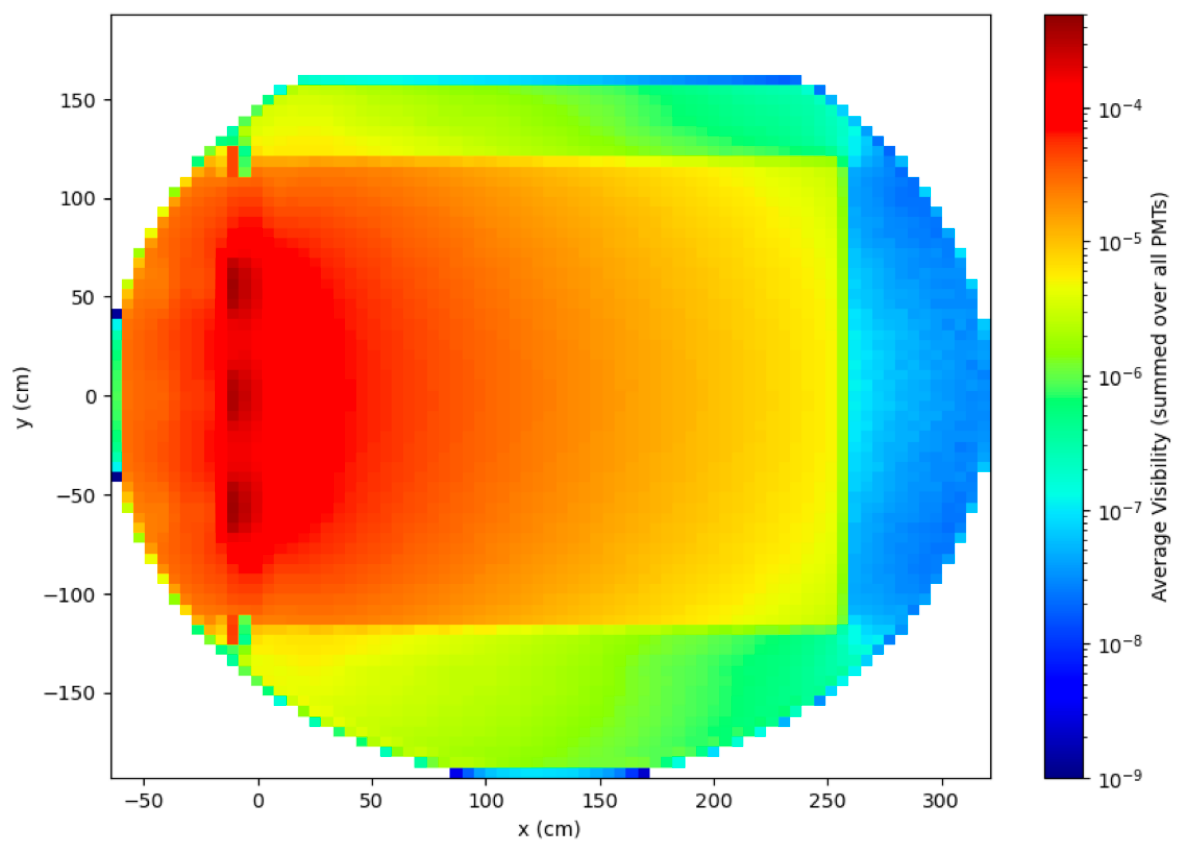}
        \caption{}
    \end{subfigure}
    \begin{subfigure}[b]{0.34\textwidth}
        \includegraphics[width=\textwidth]{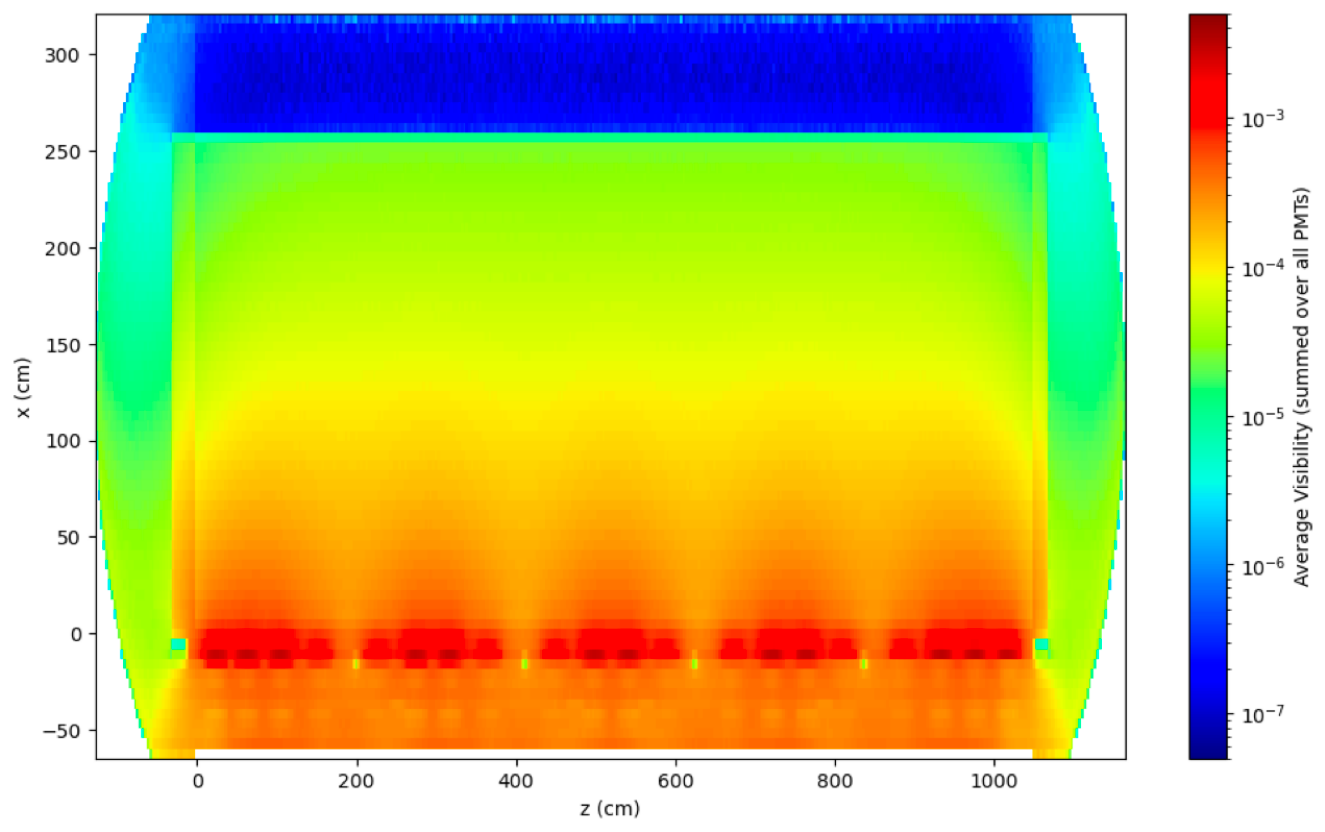}
        \caption{}
    \end{subfigure}
    \begin{subfigure}[b]{0.34\textwidth}
        \includegraphics[width=\textwidth]{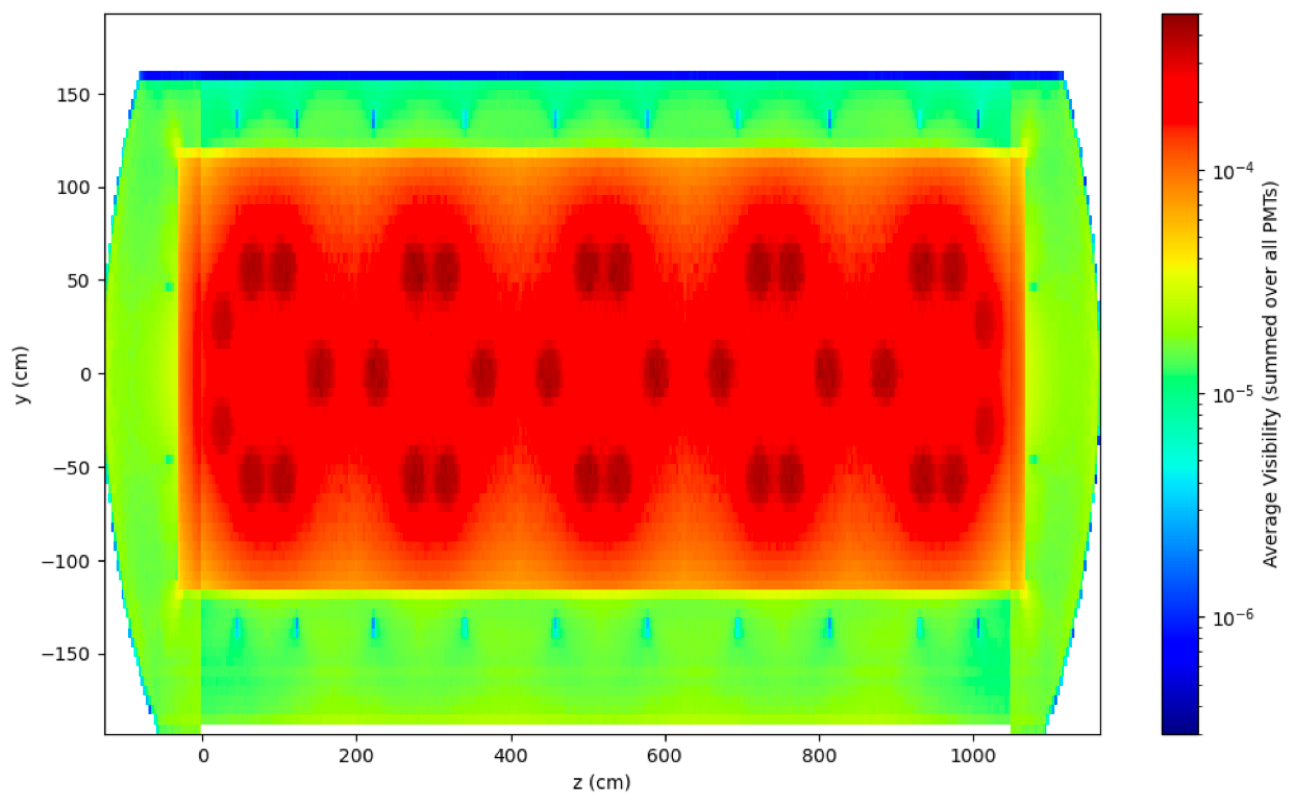}
        \caption{}
    \end{subfigure}

    \begin{subfigure}[b]{0.3\textwidth}
        \includegraphics[ width=\textwidth]{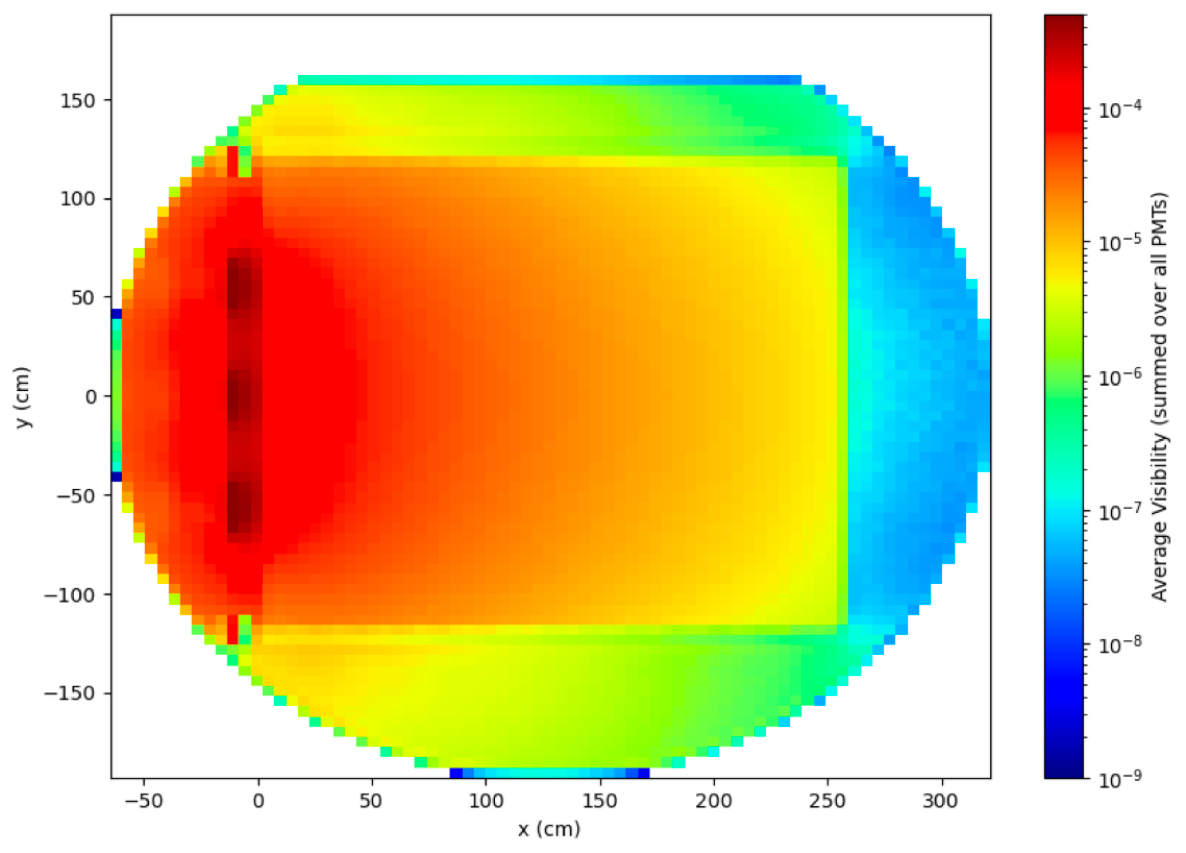}
        \caption{}
    \end{subfigure}
    \begin{subfigure}[b]{0.34\textwidth}
        \includegraphics[width=\textwidth]{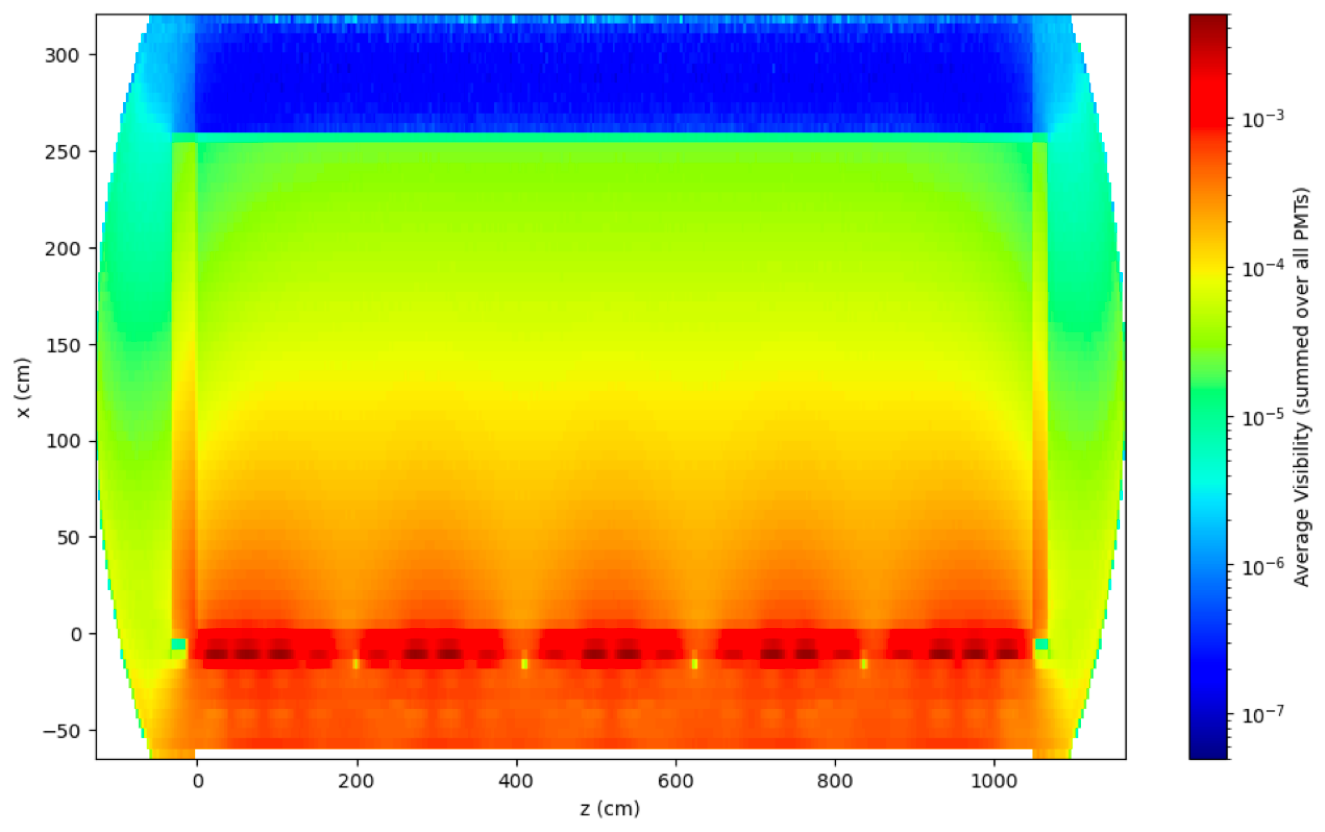}
        \caption{}
    \end{subfigure}
    \begin{subfigure}[b]{0.34\textwidth}
        \includegraphics[width=\textwidth]{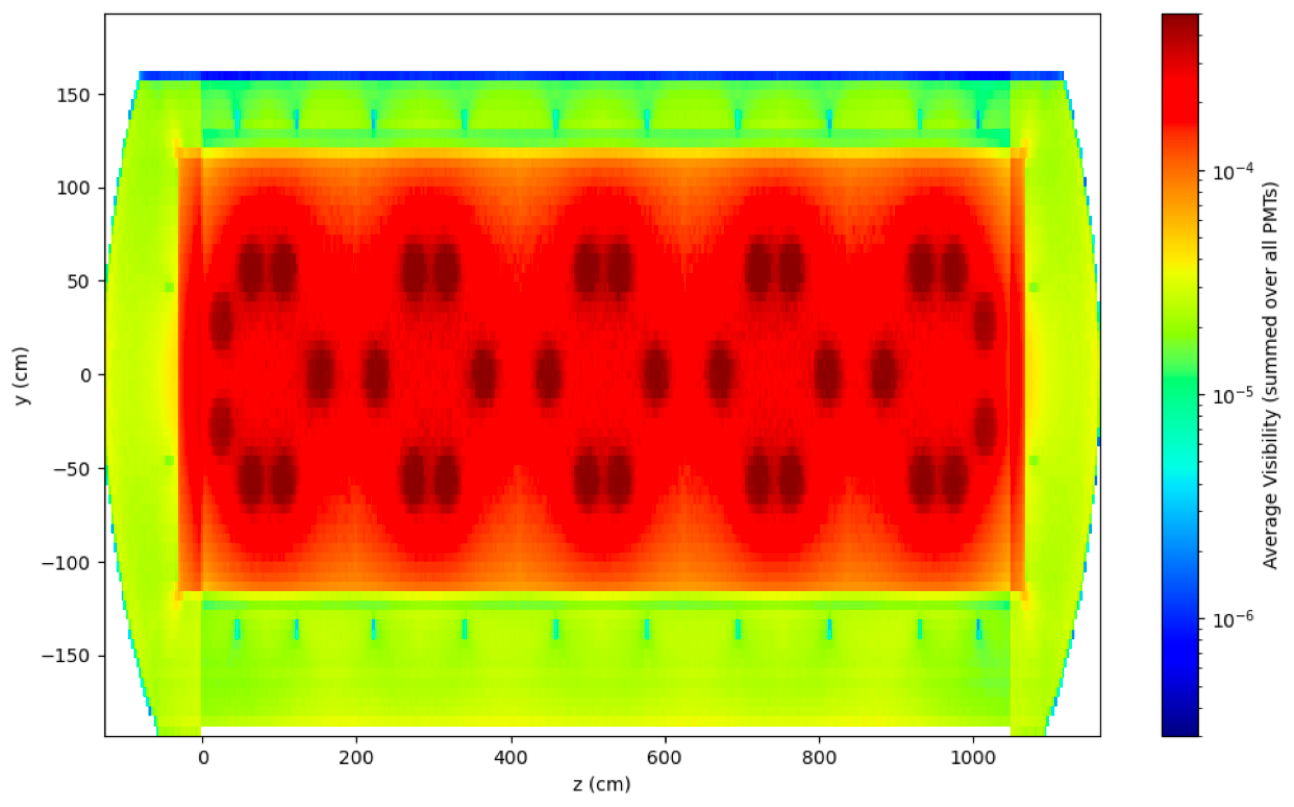}
        \caption{}
    \end{subfigure}
    \caption[PMT photon libraries with and without modification]{PMT photon libraries. Panels (a), (b), and (c) show the nominal light maps, integrated across all PMTs. Panels (e), (e), and (f) show the modified light maps, with 50\% increased light yield outside of the TPC. In these maps, we can clearly see the positions of the PMTs and the transparency of the field cage. Small differences are visible between these two light maps in certain locations.}
    \label{fig:light_maps}
\end{figure}

In order to better study the sensitivity of our single photon selections to our light modeling, we simulate NC $\pi^0$ events using an alternative photon library, which is used to simulate the amount of light reaching our PMTs from each position in the MicroBooNE detector. Specifically, we use a modified photon library which increases light production outside the TPC by 50\%. Each photon library, integrated over all PMTs, can be seen in Fig. \ref{fig:light_maps}.

\begin{figure}[H]
    \centering
    \begin{subfigure}[b]{0.49\textwidth}
        \includegraphics[width=\textwidth]{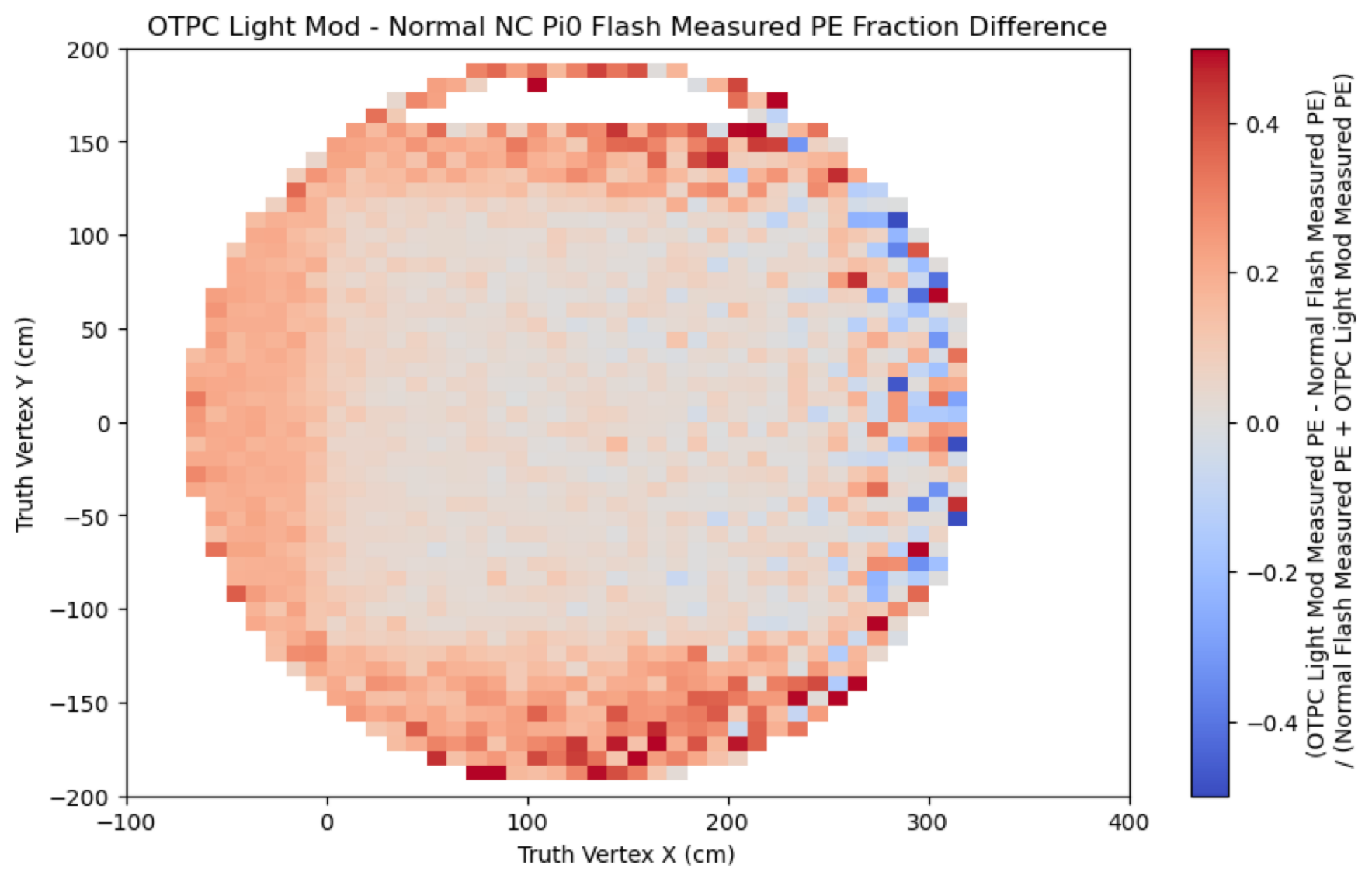}
        \caption{}
    \end{subfigure}
    \begin{subfigure}[b]{0.49\textwidth}
        \includegraphics[width=\textwidth]{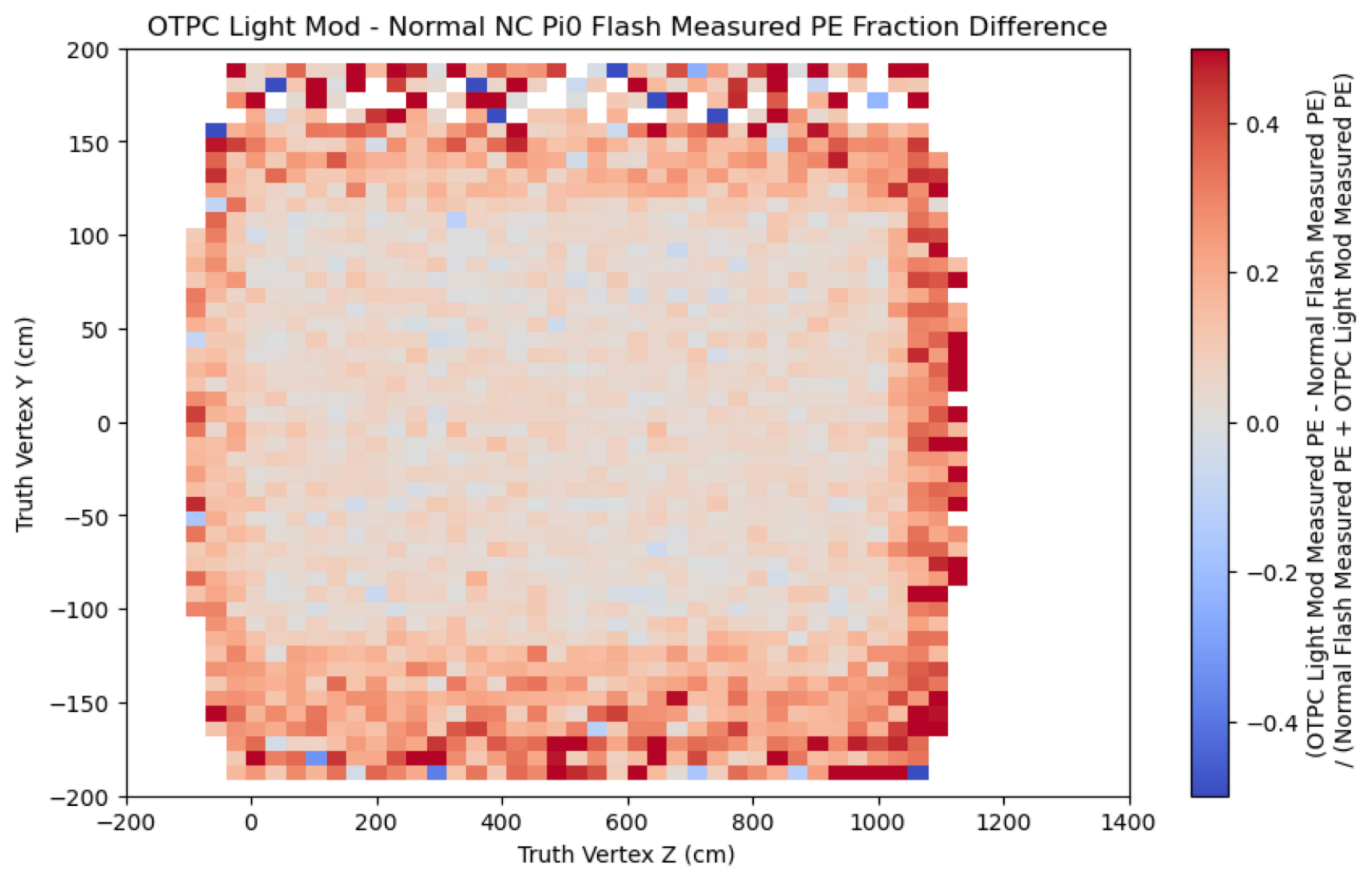}
        \caption{}
    \end{subfigure}
    \caption[Reconstructed light modification by position]{Reconstructed light modification by position. The exact same events are simulated with both photon libraries, and in each bin the color indicates the average fractional increase in light yield for events with a true neutrino vertex falling within that bin. Panel (a) shows the x-y plane, and panel (b) shows the y-z plane.}
    \label{fig:reco_light_mod_map}
\end{figure}

With this new NC $\pi^0$ simulation, we can compare how the measured light differs from the nominal photon library as a function of the true neutrino interaction location, as shown in Fig. \ref{fig:reco_light_mod_map}. We see increased measured light for events occurring outside the TPC as we expect.

\begin{figure}[H]
    \centering
    \begin{subfigure}[b]{0.49\textwidth}
        \includegraphics[width=\textwidth]{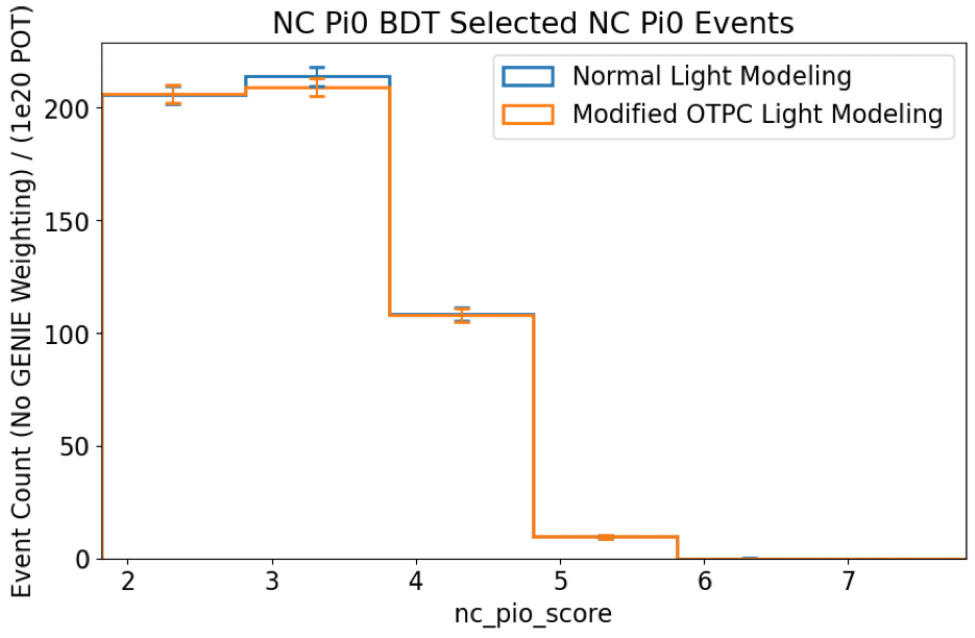}
        \caption{}
    \end{subfigure}
    \begin{subfigure}[b]{0.49\textwidth}
        \includegraphics[width=\textwidth]{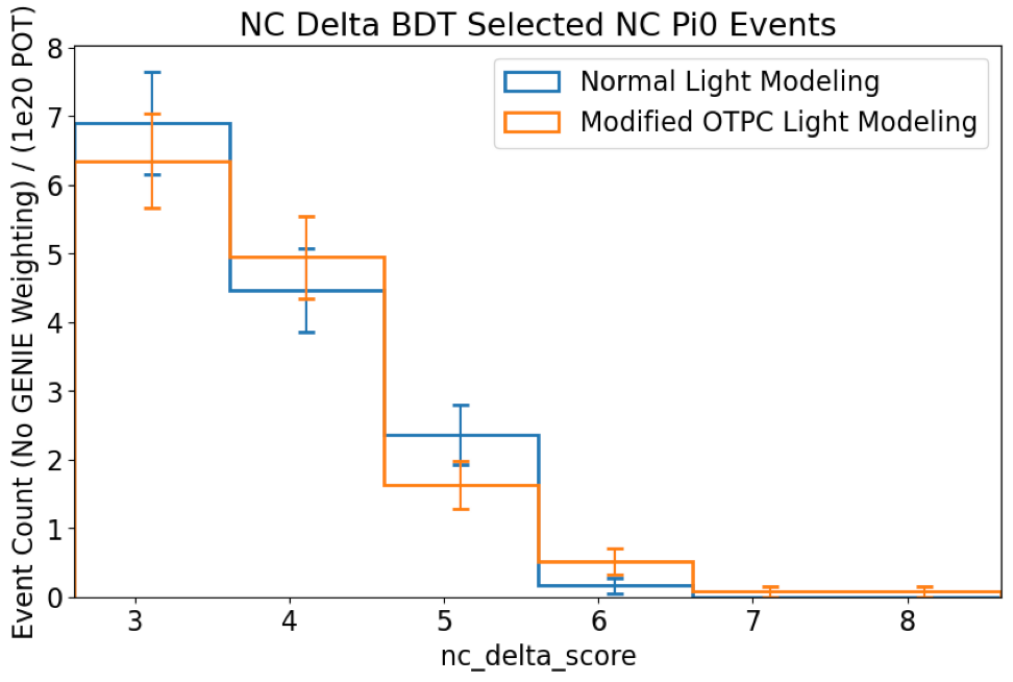}
        \caption{}
    \end{subfigure}
    \caption[Reconstructed light modification BDT scores]{Reconstructed light modification BDT scores. Panel (a) shows NC $\pi^0$ BDT scores, and panel (b) shows NC $\Delta\rightarrow N \gamma$ BDT scores.}
    \label{fig:light_mod_bdt_scores}
\end{figure}

Figure \ref{fig:light_mod_bdt_scores} shows the resulting Wire-Cell NC $\pi^0$ and NC $\Delta\rightarrow N \gamma$ BDT score distributions for both light models. We see good agreement within statistical uncertainties, indicating that our selections are not especially sensitive to this type of light modeling change.

\section{Future Reconstruction Opportunities}\label{sec:future_photon_reconstruction}

In this section, I briefly describe some directions we have been exploring for increasing the performance of future photon-like reconstructions and selections in MicroBooNE. As shown in Figs. \ref{fig:overlap_venn_diagram} and \ref{fig:overlap_data}, there is a significant number of events that are only selected by one selection or another, even if both selections are targeting the same events. This indicates a large potential for improved performance.

\subsection{In-FV vs Out-FV Studies}

Because of MicroBooNE's long rectangular prism shape, fiducializing our detector in order to reduce backgrounds from entering and exiting $\pi^0$ photons without losing significant volume is very difficult. This is why we instead investigate the distance to the detector boundary as well as the distance along the backward projected shower distance, as shown in Fig. \ref{fig:backward_projected_distance}. These are just a few of many potential variables that could use the geometry of the shower angle and direction in order to reject out of fiducial volume (out-FV) $\pi^0$ events. In order to generalize this and get a higher performance rejection of out-FV events, I trained a BDT on all Wire-Cell events with one reconstructed shower in order to separate in-FV from out-FV events. I use only geometric variables derived from the shower position and direction, described in Table \ref{tab:geometric_bdt_vars}.

\begin{table}[H]
    \centering
    \begin{tabular}{l p{10cm}}
        \toprule
        \textbf{Variable} & \textbf{Description} \\
        \midrule
        \texttt{shower\_theta} & Reconstructed shower angle with respect to the neutrino beam. \\
        \hline
        \texttt{shower\_phi} & Azimuthal reconstructed shower angle with respect to the neutrino beam. \\
        \hline
        \texttt{shower\_vertex\_x} & $x$-coordinate of the reconstructed shower vertex in cm. \\
        \hline
        \texttt{shower\_vertex\_y} & $y$-coordinate of the reconstructed shower vertex in cm. \\
        \hline
        \texttt{shower\_vertex\_z} & $z$-coordinate of the reconstructed shower vertex in cm. \\
        \hline
        \texttt{forward\_dist\_to\_boundary} & Distance from the shower vertex to the detector boundary projected forward along the shower direction. \\
        \hline
        \texttt{backward\_dist\_to\_boundary} & Distance from the shower vertex to the detector boundary projected opposite the shower direction. \\
        \hline
        \texttt{dist\_to\_boundary} & Minimum distance from the shower vertex to the detector boundary. \\
        \hline
        \texttt{inwardness\_2d} & Dot product between the shower direction unit vector and a unit vector pointing toward the central $z$ axis of the detector. \\
        \hline
        \texttt{inwardness\_3d} & Dot product between the shower direction unit vector and a unit vector pointing to the geometric center of the detector. \\
        \bottomrule
    \end{tabular}
    \caption[Geometric In-FV BDT variables]{Geometric In-FV BDT variables.}
    \label{tab:geometric_bdt_vars}
\end{table}

\begin{figure}[H]
    \centering
    \begin{subfigure}[b]{0.24\textwidth}
        \includegraphics[trim=7 5 8 5, clip, width=\textwidth]{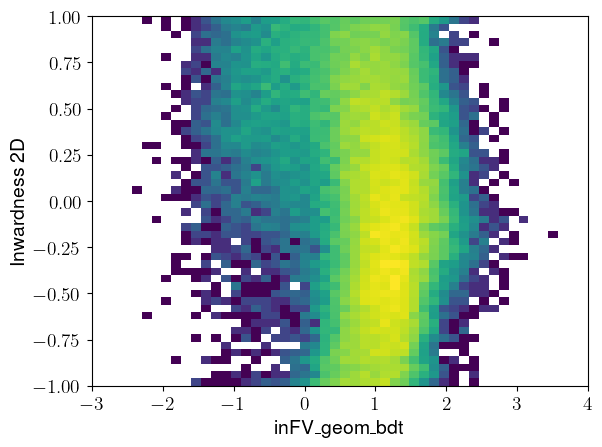}
        \caption{}
    \end{subfigure}
    \begin{subfigure}[b]{0.24\textwidth}
        \includegraphics[trim=7 5 8 5, clip, width=\textwidth]{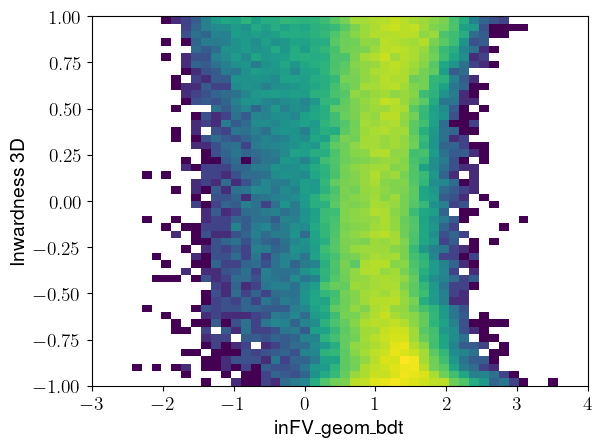}
        \caption{}
    \end{subfigure}
    \begin{subfigure}[b]{0.24\textwidth}
        \includegraphics[trim=7 5 8 5, clip, width=\textwidth]{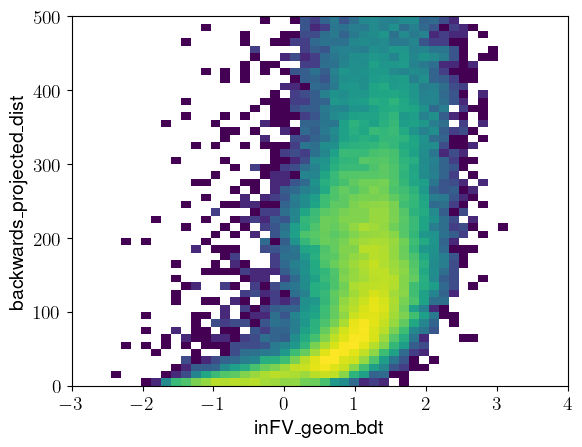}
        \caption{}
    \end{subfigure}
    \begin{subfigure}[b]{0.24\textwidth}
        \includegraphics[trim=7 5 8 5, clip, width=\textwidth]{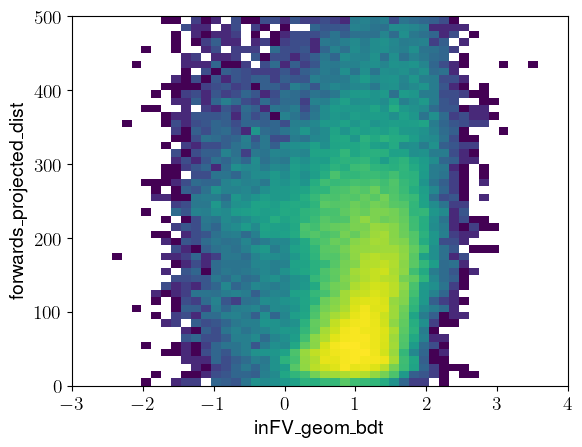}
        \caption{}
    \end{subfigure}

    \begin{subfigure}[b]{0.24\textwidth}
        \includegraphics[trim=7 5 8 5, clip, width=\textwidth]{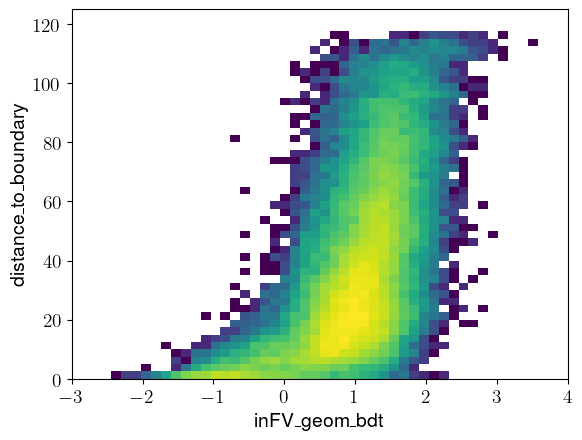}
        \caption{}
    \end{subfigure}
    \begin{subfigure}[b]{0.24\textwidth}
        \includegraphics[trim=7 5 8 5, clip, width=\textwidth]{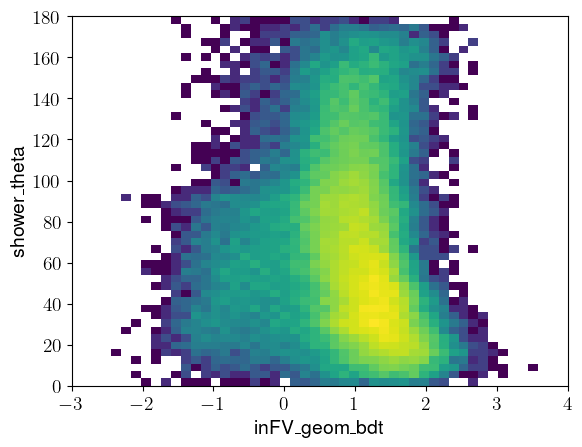}
        \caption{}
    \end{subfigure}
    \begin{subfigure}[b]{0.24\textwidth}
        \includegraphics[trim=7 5 8 5, clip, width=\textwidth]{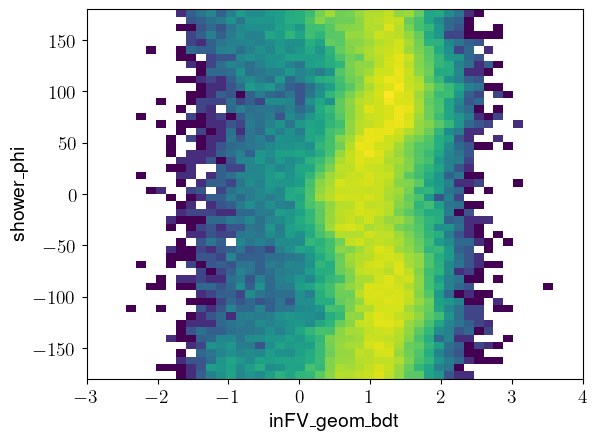}
        \caption{}
    \end{subfigure}

    \begin{subfigure}[b]{0.24\textwidth}
        \includegraphics[trim=7 5 8 5, clip, width=\textwidth]{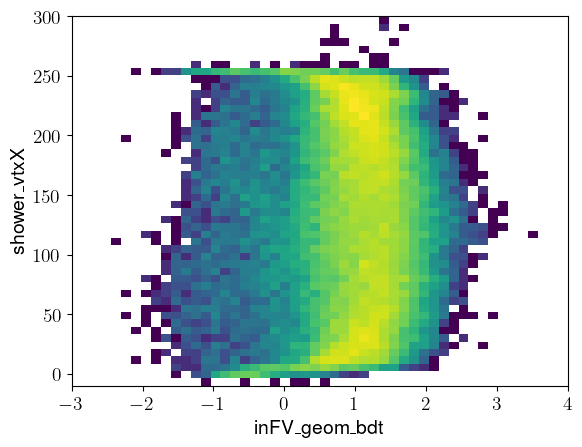}
        \caption{}
    \end{subfigure}
    \begin{subfigure}[b]{0.24\textwidth}
        \includegraphics[trim=7 5 8 5, clip, width=\textwidth]{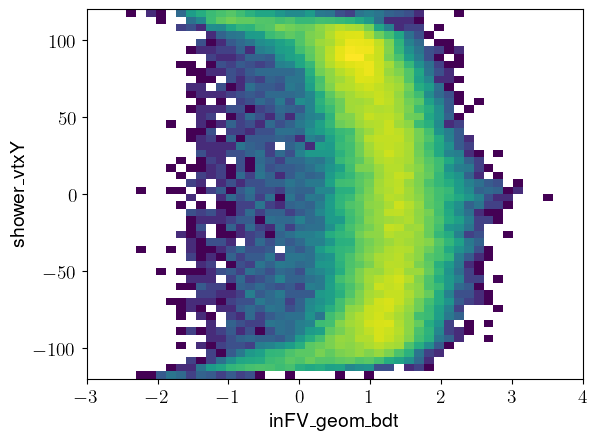}
        \caption{}
    \end{subfigure}
    \begin{subfigure}[b]{0.24\textwidth}
        \includegraphics[trim=7 5 8 5, clip, width=\textwidth]{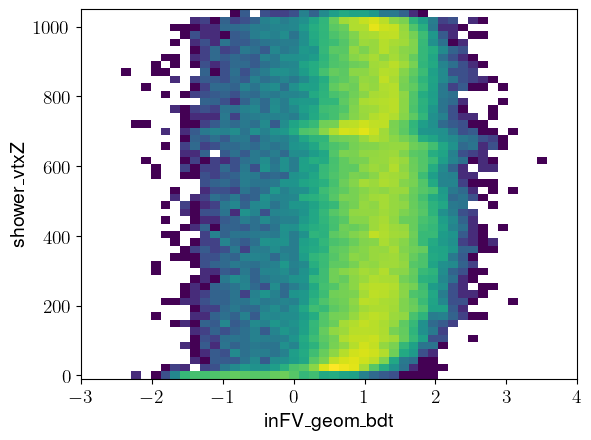}
        \caption{}
    \end{subfigure}
    \caption[Geometric in-fiducial-volume BDT]{Geometric in-fiducial-volume BDT. We show correlations between each input variable and the output in-FV BDT score.}
    \label{fig:geometric_bdt}
\end{figure}

The behavior of this BDT in a high dimensional phase space is hard to visualize and understand, so in Fig. \ref{fig:geometric_bdt} we show several slices illustrating the correlations between the input geometric variables and the output geometric in-FV BDT score. We can see that certain variable values, for example low \texttt{backward\_dist\_to\_boundary}, are associated with low BDT scores, corresponding to more likely out-FV origins, as we would expect. There is remaining work to be done in order to understand validate the use of this tool on real selections.

\subsection{Deep Learning For \texorpdfstring{$e^+e^-$}{e+e-} Opening Angle Reconstruction}

A ``smoking gun'' signature of a beyond the standard model $e^+e^-$ excess would be the reconstruction of two electromagnetic showers, each with a minimum ionizing shower stem, sharing a shower vertex, and the existence of a nonzero opening angle between the two. Photon showers always have very small opening angles between the two showers, but if the particle producing the $e^+e^-$ is massive, we expect the opening angle to be larger.

\begin{figure}[H]
    \centering
    \begin{subfigure}[b]{0.49\textwidth}
        \includegraphics[width=\textwidth]{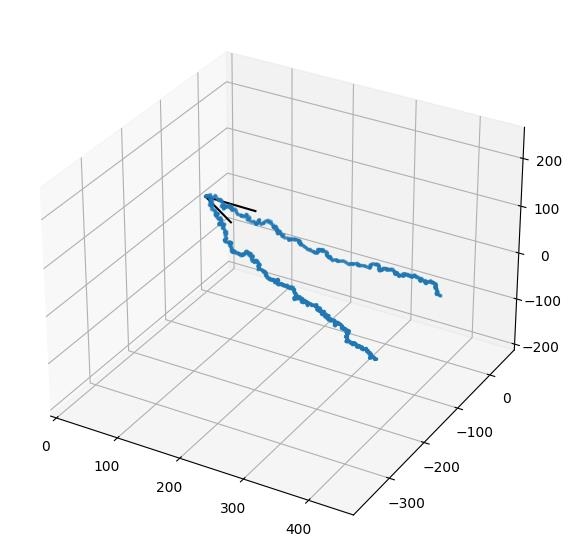}
        \caption{}
    \end{subfigure}
    \begin{subfigure}[b]{0.49\textwidth}
        \includegraphics[width=\textwidth]{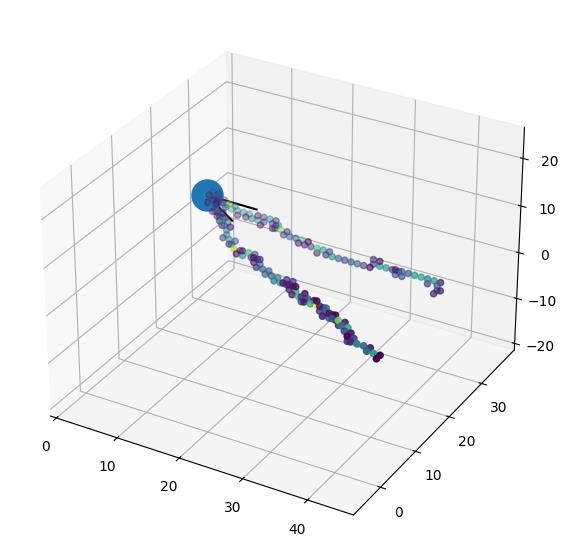}
        \caption{}
    \end{subfigure}
    \begin{subfigure}[b]{0.49\textwidth}
        \includegraphics[width=\textwidth]{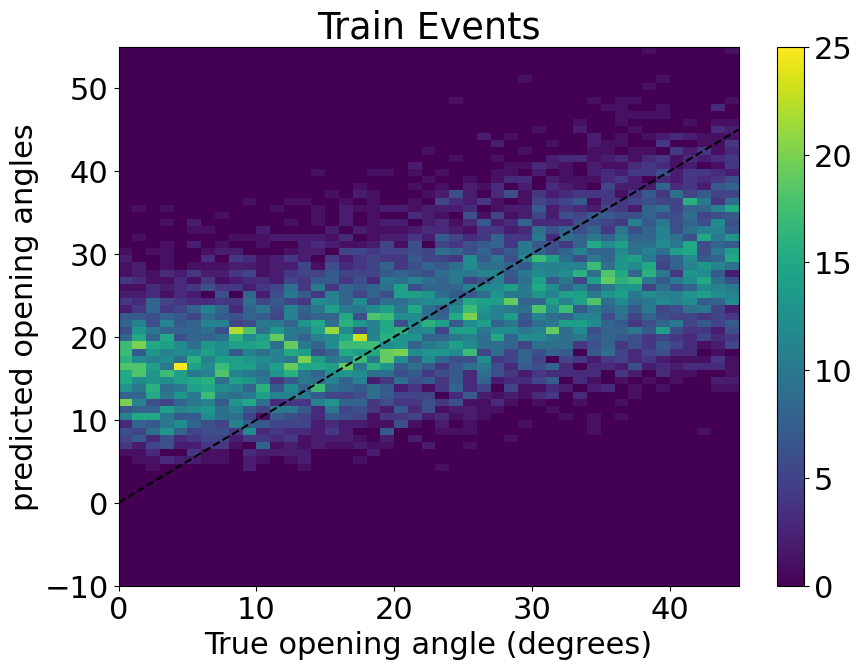}
        \caption{}
    \end{subfigure}
    \begin{subfigure}[b]{0.49\textwidth}
        \includegraphics[width=\textwidth]{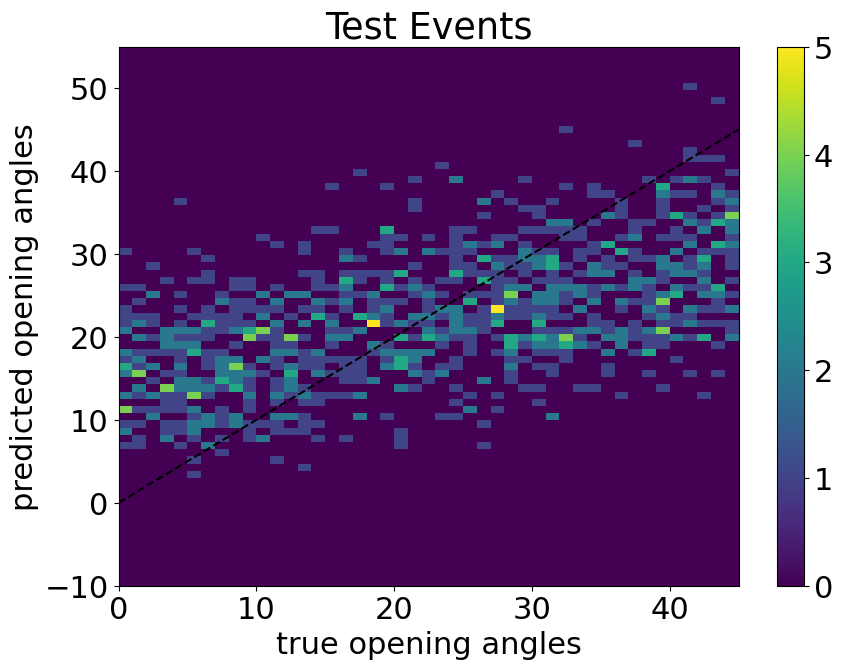}
        \caption{}
    \end{subfigure}
    \caption[Deep learning $e^+e^-$ opening angle identification]{Deep learning $e^+e^-$ opening angle identification.}
    \label{fig:dl_opening_angle}
\end{figure}

Reconstructing this opening angle is difficult, particularly for low energy showers which are most relevant for studies of the MiniBooNE LEE. This topology is very rare in the standard model (for example Dalitz decay, $\pi^0\rightarrow \gamma e^+e^-$), and our existing tools do not do a particularly good job at this reconstruction. Working with an University of Chicago undergraduate student, Lisa Solomey, we tried applying deep learning to this problem. We started with a similar approach as was used for Wire-Cell deep learning vertexing, as described in Sec. \ref{sec:pattern_rec}, where we use Wire-Cell reconstructed 3D spacepoints, and then voxelize them before passing the result to a sparse convolutional neural network \cite{sparseconvnet}. We simulate $e^+e^-$ pairs isotropically, then use only those with an opening angle of $45^\circ$ or less as a region of interest, and train the network to identify the opening angle as accurately as possible. The voxelization and resulting performance is shown in Fig. \ref{fig:dl_opening_angle}. Our current network certainly has room for improvement, but performs notably better than guessing. Others in the MicroBooNE collaboration have been developing a different network based on the PointNet architecture \cite{pointnet} for the same purpose, and have begun to see some promising results.

\subsection{3D Second Shower Veto For Misclustered \texorpdfstring{$\pi^0$}{pi0}s}

\begin{figure}[H]
    \centering
    \includegraphics[width=0.8\textwidth]{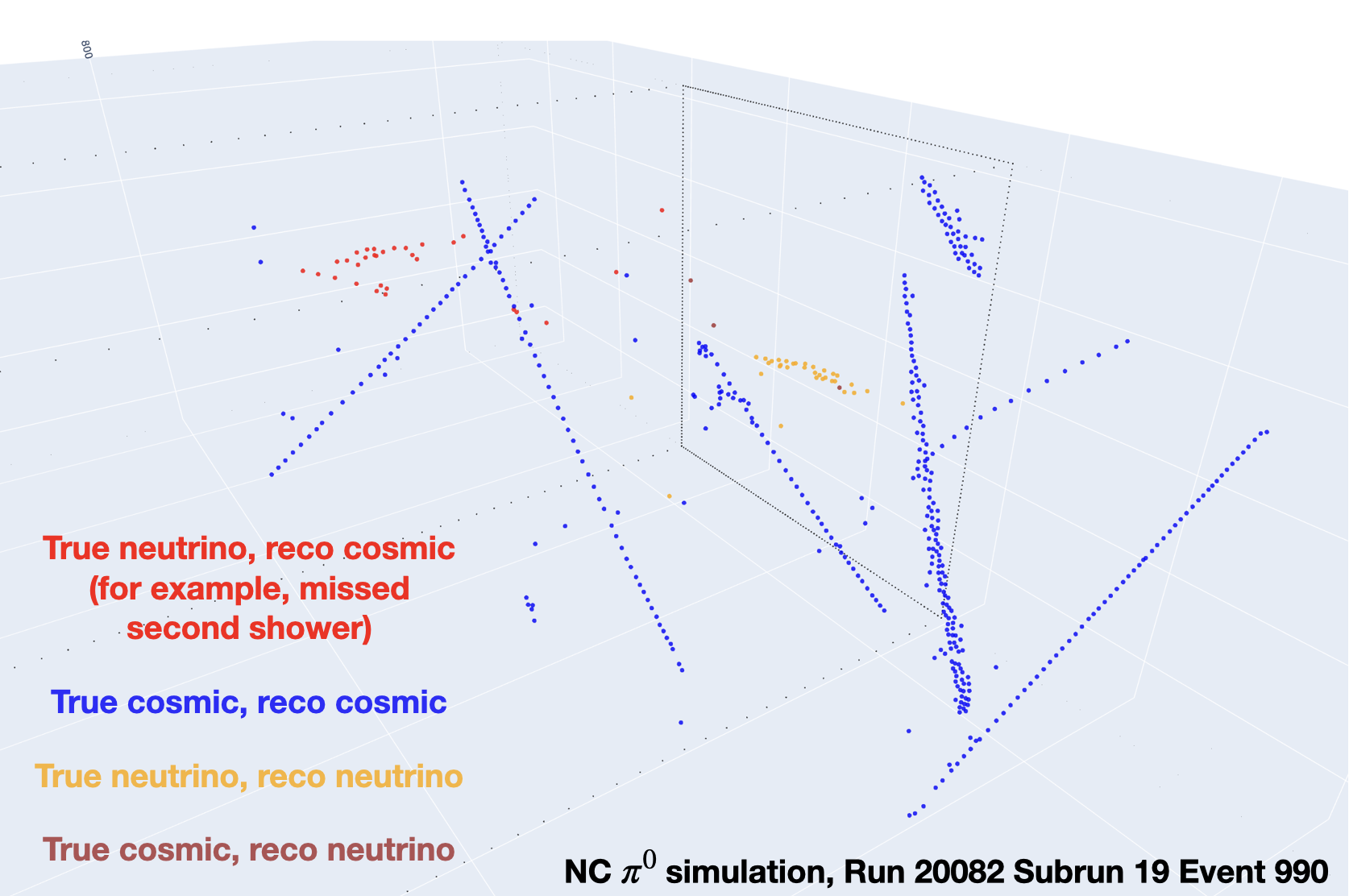}
    \caption[3D spacepoints in a mis-clustered $\pi^0$ event]{3D spacepoints in a mis-clustered $\pi^0$ event. Reconstructed cosmic and neutrino points have been downsampled using furthest-point-sampling, and then have been identified as true neutrino or true cosmic using proximity to simulated \textsc{Geant4} energy depositions.}
    \label{fig:3d_ssv}
\end{figure}

Another important background category for all of these single-photon-like selections is mis-clustered $\pi^0$ events. In these events, both photon showers are visible in the TPC, but clustering algorithms have failed, and therefore the reconstruction incorrectly thinks that one of the showers is cosmic-induced and the other is neutrino-induced. In Pandora reconstruction, there is a dedicated tool to reject events with improperly reconstructed $\pi^0$ showers, known as the ``second shower veto''. A potential future improvement would be to develop a similar tool using Wire-Cell 3D reconstruction. Potentially, this could be done by first downsampling reconstructed 3D points for both neutrinos and cosmic rays, and then training a deep neural network in order to identify cases in which true neutrino activity has been clustered with a cosmic ray. An example of this behavior is shown in Fig. \ref{fig:3d_ssv}. Thusfar, I have done some preliminary work in order to ensure that the necessary information with neutrino and cosmic reconstructed spacepoints will be present in future rounds of MicroBooNE data processing.

\subsection{Other New Reconstruction Methods}

There are also several more potential reconstruction improvements coming to MicroBooNE in the near future. There is a new deep learning based reconstruction paradigm in MicroBooNE \cite{LANTERN_public_note} which has shown very promising results in $\nu_\mu$CC and $\nu_e$CC selections, and is currently being developed to target single photon selections. Similarly, graph neural networks are also currently being developed to improve the Pandora reconstruction toolchain \cite{nugraph2}. This reconstruction could also be potentially integrated with Wire-Cell, as we have seen benefits when simultaneously using multiple reconstructions for single photons. Additionally, making use of our reconstruction of very low energy deposits, also known as ``blips'', has the potential to significantly decrease backgrounds related to low energy protons at the neutrino vertex \cite{microboone_ambient_blips}. The integration of nanosecond timing resolution has the potential to significantly increase signal and background discrimination for certain physics models \cite{microboone_nanosecond_timing}. Lastly, analyzing previously unstudied reconstructed topologies could significantly advance our understanding of backgrounds in single photon searches. For example, a $1\mu1\gamma$ selection would let us study high-statistics $\pi^0$ backgrounds both with and without reconstructed protons, while identifying the neutrino vertex in each case. This type of selection could let us study different types of cross section and detector effects for single photon events in new ways.

%% file: chapters/06_conclusions.tex
\chapter{Conclusions}

In this thesis, I presented measurements of neutrino interactions in the MicroBooNE experiment, in particular focusing on investigations of the MiniBooNE anomalous observation of excess electromagnetic showers at low energies in the Booster Neutrino Beam at Fermi National Accelerator Laboratory. 

First, we searched for an anomalous excess of charged current electron neutrino interactions, and saw data consistent with our expectations. This allowed us to place constraints on models predicting an enhanced electron neutrino flux, in particular short baseline oscillations caused by the existence of a sterile neutrino. 

Then, we focused on photon-like explanations of the MiniBooNE anomaly, consisting of a photon which which creates an electromagnetic shower from an $e^+e^-$ pair. In particular, our main result searches for events from neutral current Delta radiative decays, the largest source of predicted single photons in MicroBooNE. Combining two different event reconstruction frameworks, we significantly enhance previous results by adding data statistics and reducing backgrounds for events without reconstructed protons. We exclude an enhanced rate of Delta radiative decay events as an explanation of the MiniBooNE anomaly at 94.4\% CL. Considering a broader set of hypotheses, we cannot exclude the existence of a potential enhancement of events with no reconstructed protons as an explanation of the MiniBooNE anomaly, highlighting the difficulties of selecting the $1\gamma 0p$ topology among $2\gamma$ backgrounds from NC $\pi^0$ interactions. Other photon-like MicroBooNE analyses similarly constrain some hypotheses and leave the door open to others depending on the kinematics and topology being considered.

Looking forward, there is a lot of potential to further improve neutrino-induced single photon measurements. Adding nanosecond timing, MeV-scale activity, and more deep learning techniques to our reconstructions, we expect to be able to develop significantly improved selections over a variety of photon-like topologies in the near future in MicroBooNE. Additionally, the SBND and DUNE LAr-ND detectors will have even more advanced LArTPCs, larger and more compact volumes, and higher neutrino fluxes, allowing even more detailed investigations of the single photon topology in GeV-scale neutrino interactions. This thesis presents important milestones toward a more complete understanding of the origin of the MiniBooNE anomaly, and will serve as a stepping stone as we reach additional definitive conclusions about the origins of neutrino anomalies in the future.

%% file: chapters/07_cross_sections.tex
\chapter{Cross Sections}

In addition to my work on MicroBooNE single photon searches, I have also been involved in cross section studies, which I briefly summarize in this appendix. Neutrino-argon interaction cross section measurements are critically important for DUNE, which will need a high precision understanding of this cross section modeling in order to analyze observed events and extract neutrino oscillation parameters. This modeling will have to be understood at a much more precise level than the current understanding used for the NOvA and T2K experiments in order to meet DUNE's physics goals, in particular for a measurement of $\delta_{CP}$.

\section{\texorpdfstring{$\nu_\mu$}{numu}CC Inclusive Cross Section Measurements Using Wire-Cell}

\begin{figure}[H]
    \centering
    \includegraphics[width=0.7\textwidth]{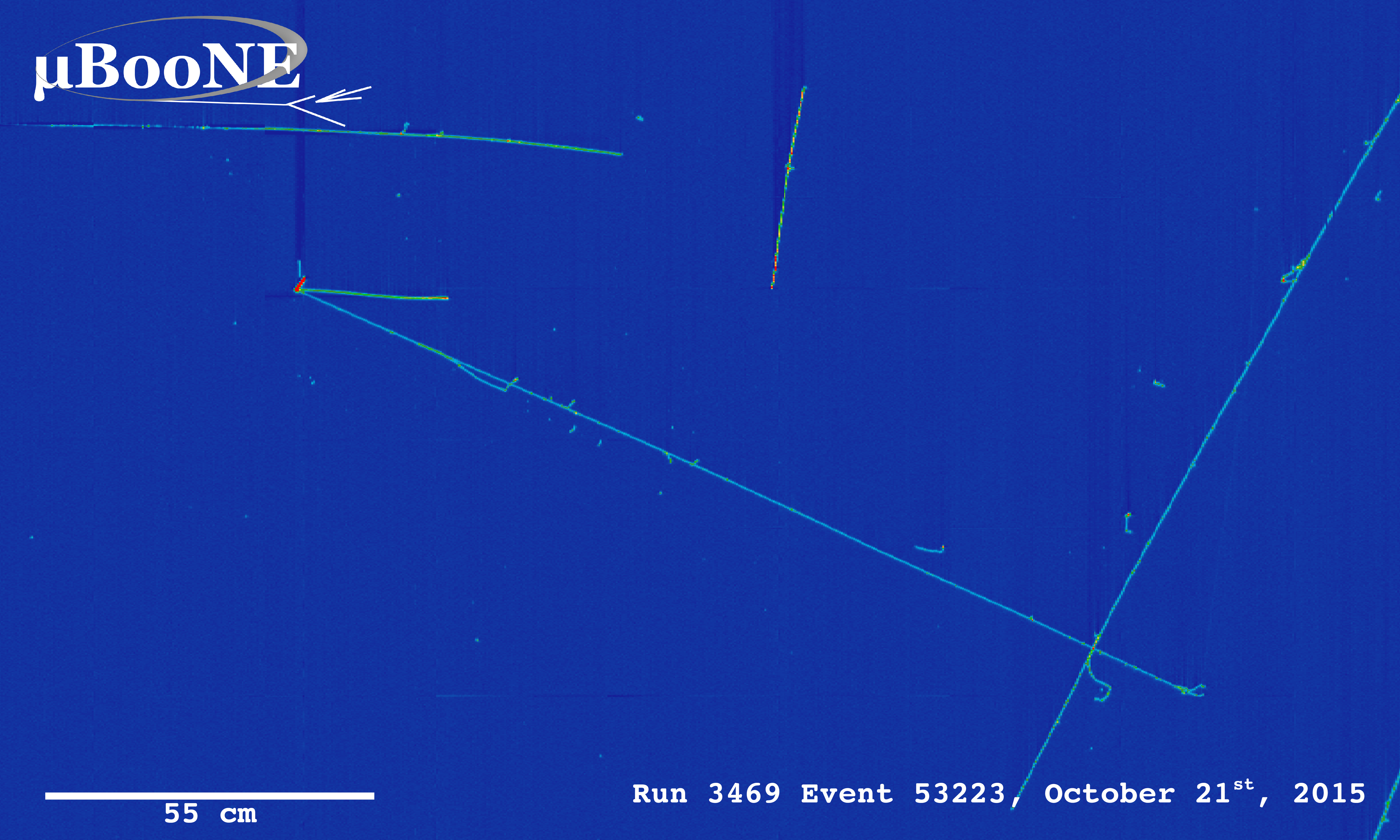}
    \caption[$\nu_\mu$CC event display]{$\nu_\mu$CC event display, from Run 3469 Subrun 1064 Event 53223. We can see a muon extending from the top left to the bottom right, with some other particle activity at the neutrino vertex, as well as some unrelated cosmic ray tracks surrounding this event.}
    \label{fig:numuCC_event_display}
\end{figure}

MicroBooNE has now published several analyses using the Wire-Cell inclusive $\nu_\mu$CC selection which I trained as described in Sec. \ref{sec:bdt_selections}. An example candidate $\nu_\mu$CC events is shown in Fig. \ref{fig:numuCC_event_display}. In this section, I briefly summarize these analyses.

\subsection{1D Cross Sections}

Our first cross section result using Wire-Cell reconstruction extracted inclusive $\nu_\mu$CC cross sections in one dimension, as functions of neutrino energy $E_\nu$, muon energy $E_\mu$, and energy transfer to the nuclear system $E_\nu=E_\nu-E_\mu$ \cite{wc_numu_1d}. Notably, the neutrino energy and energy transfer are not directly observable, since there is missing energy that can be carried away by neutrons for example. This means that we have to use our cross section model in order to perform this unfolding and translate our measurement from the reconstructed neutrino energy to the true neutrino energy. The cross section model we use for this could potentially be incorrect and have insufficient systematic uncertaintainties, and therefore could result in a biased extracted cross section result. We address this concern using detailed model validation tests, as we will describe in Sec. \ref{sec:model_validation}.

\begin{figure}[H]
    \centering
    \begin{subfigure}[b]{0.32\textwidth}
        \includegraphics[width=\textwidth]{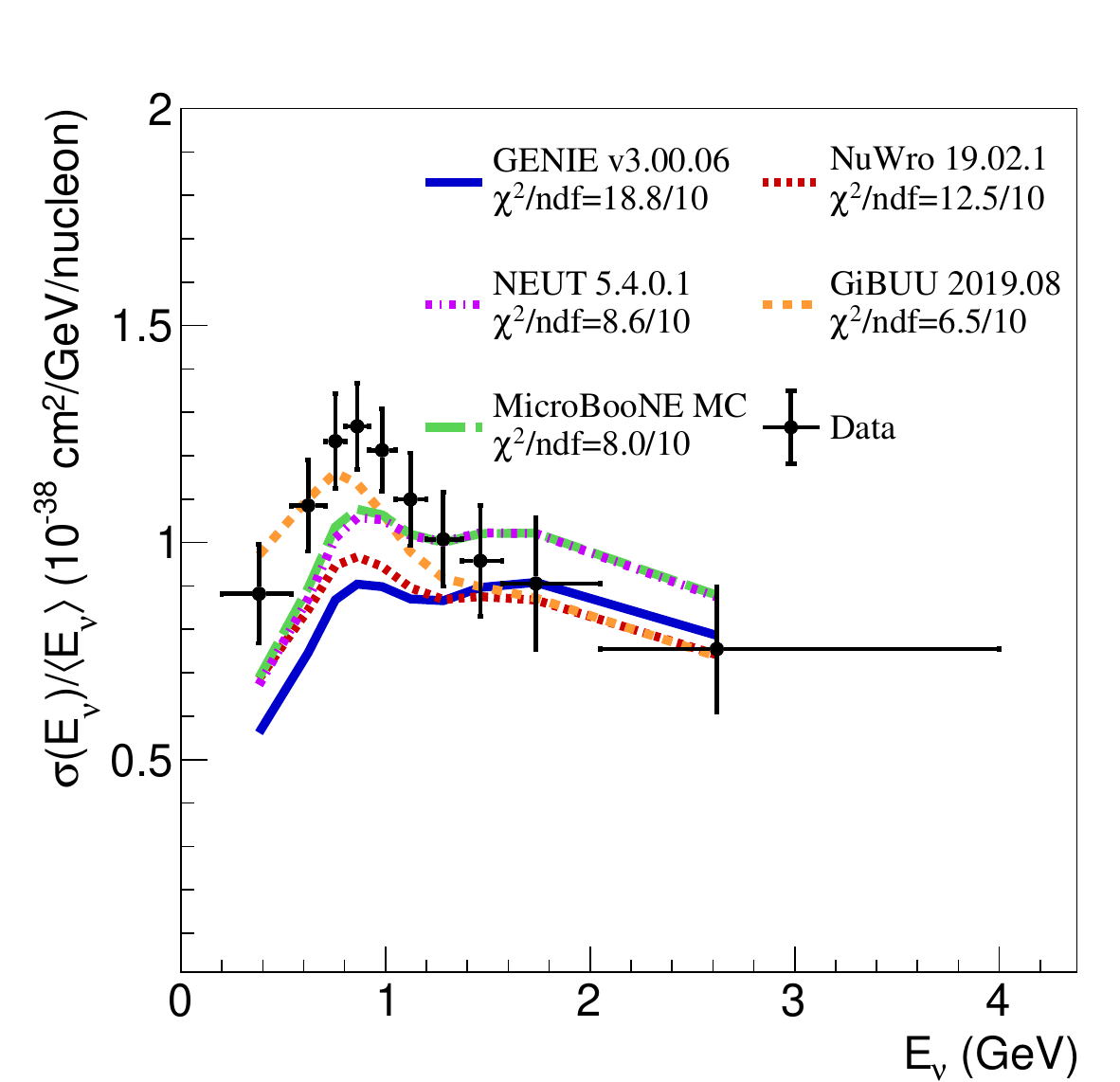}
        \caption{}
    \end{subfigure}
    \begin{subfigure}[b]{0.32\textwidth}
        \includegraphics[width=\textwidth]{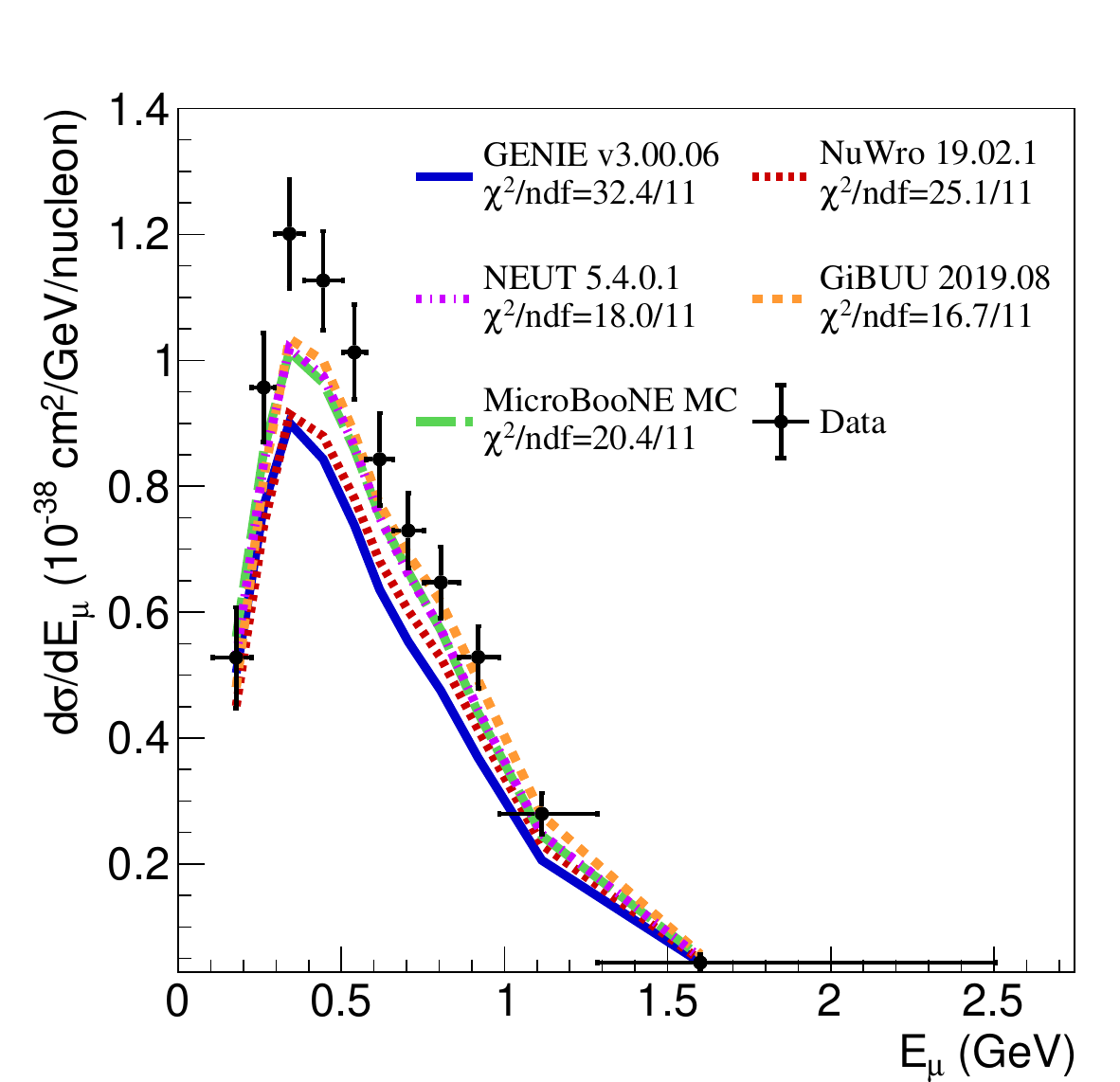}
        \caption{}
    \end{subfigure}
    \begin{subfigure}[b]{0.32\textwidth}
        \includegraphics[width=\textwidth]{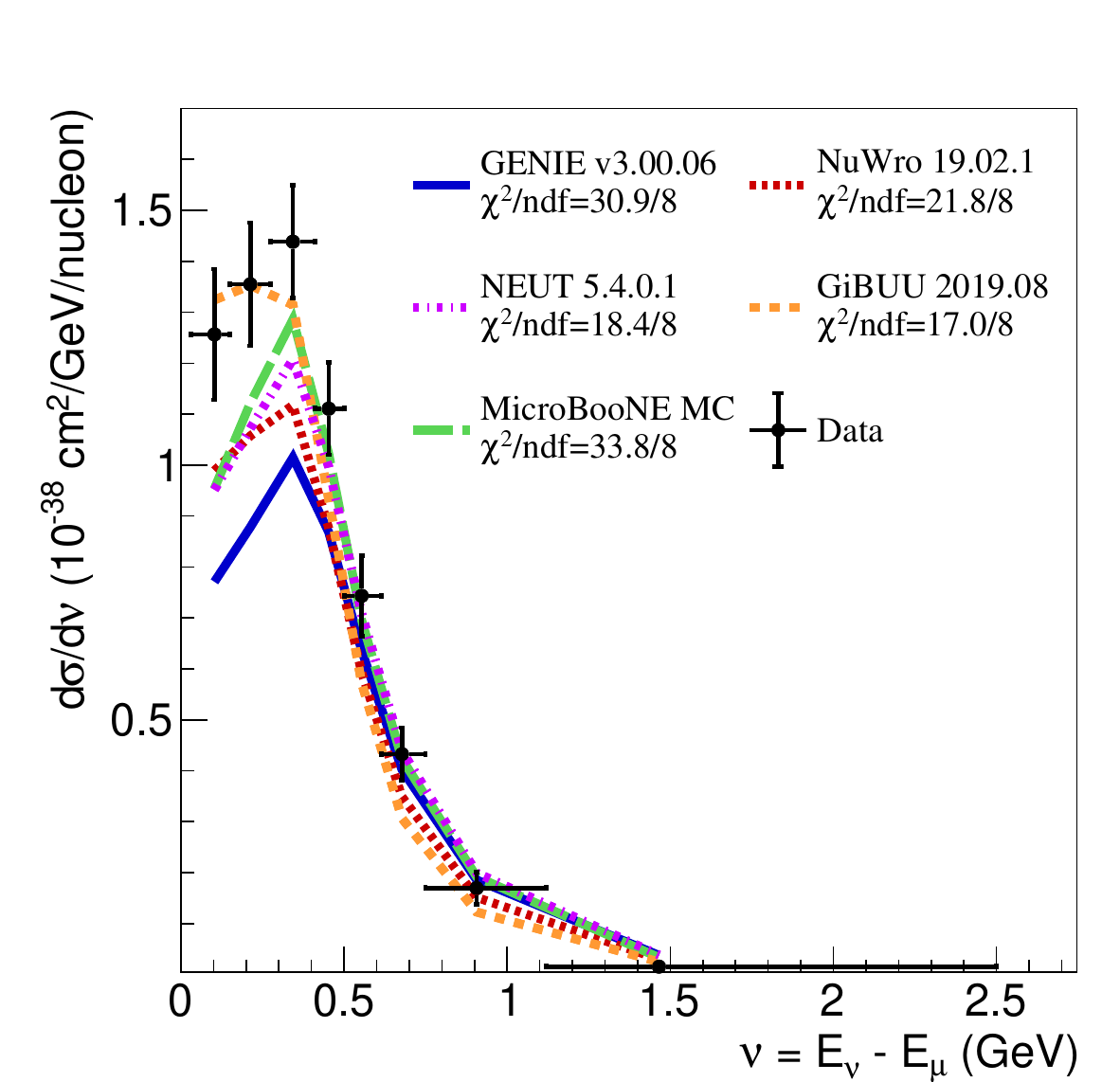}
        \caption{}
    \end{subfigure}
    \caption[1D $\nu_\mu$CC Inclusive Cross Section Results]{1D $\nu_\mu$CC inclusive cross section results. Panel (a) shows the unfolded cross section as a function of true neutrino energy, panel (b) shows the flux averaged differential cross section as a function of muon energy, and panel (c) shows the flux averaged differential cross section as a function of energy transfer to the nuclear system. From Ref. \cite{wc_numu_1d}.}
    \label{fig:1d_numuCC_XS}
\end{figure}

\subsection{Proton-related Cross Sections}

\begin{figure}[H]
    \centering
    \begin{subfigure}[b]{0.32\textwidth}
        \includegraphics[trim=8 0 7 0, clip, width=\textwidth]{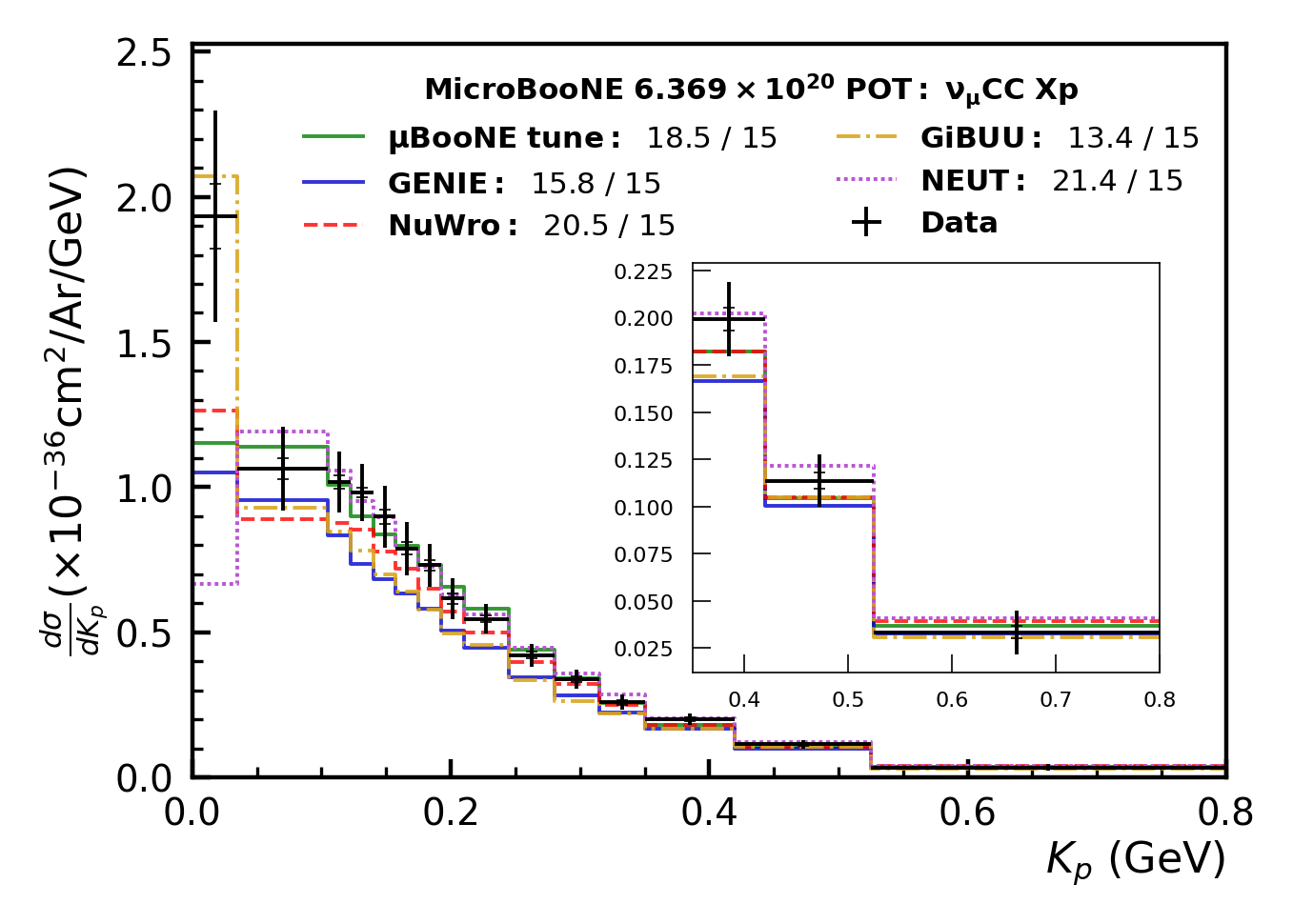}
        \caption{}
    \end{subfigure}
    \begin{subfigure}[b]{0.32\textwidth}
        \includegraphics[trim=8 0 6 0, clip, width=\textwidth]{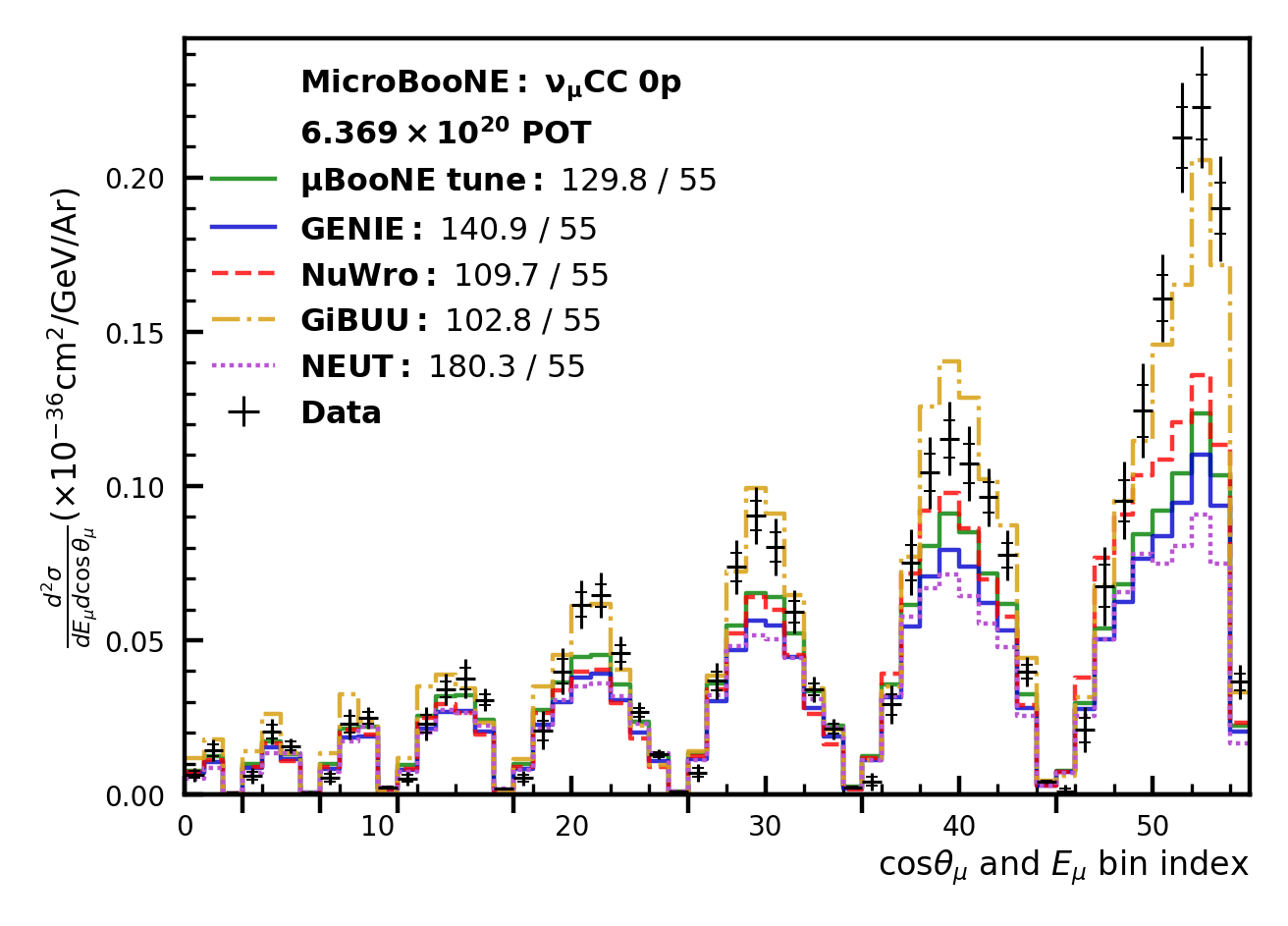}
        \caption{}
    \end{subfigure}
    \begin{subfigure}[b]{0.32\textwidth}
        \includegraphics[trim=8 0 7 0, clip, width=\textwidth]{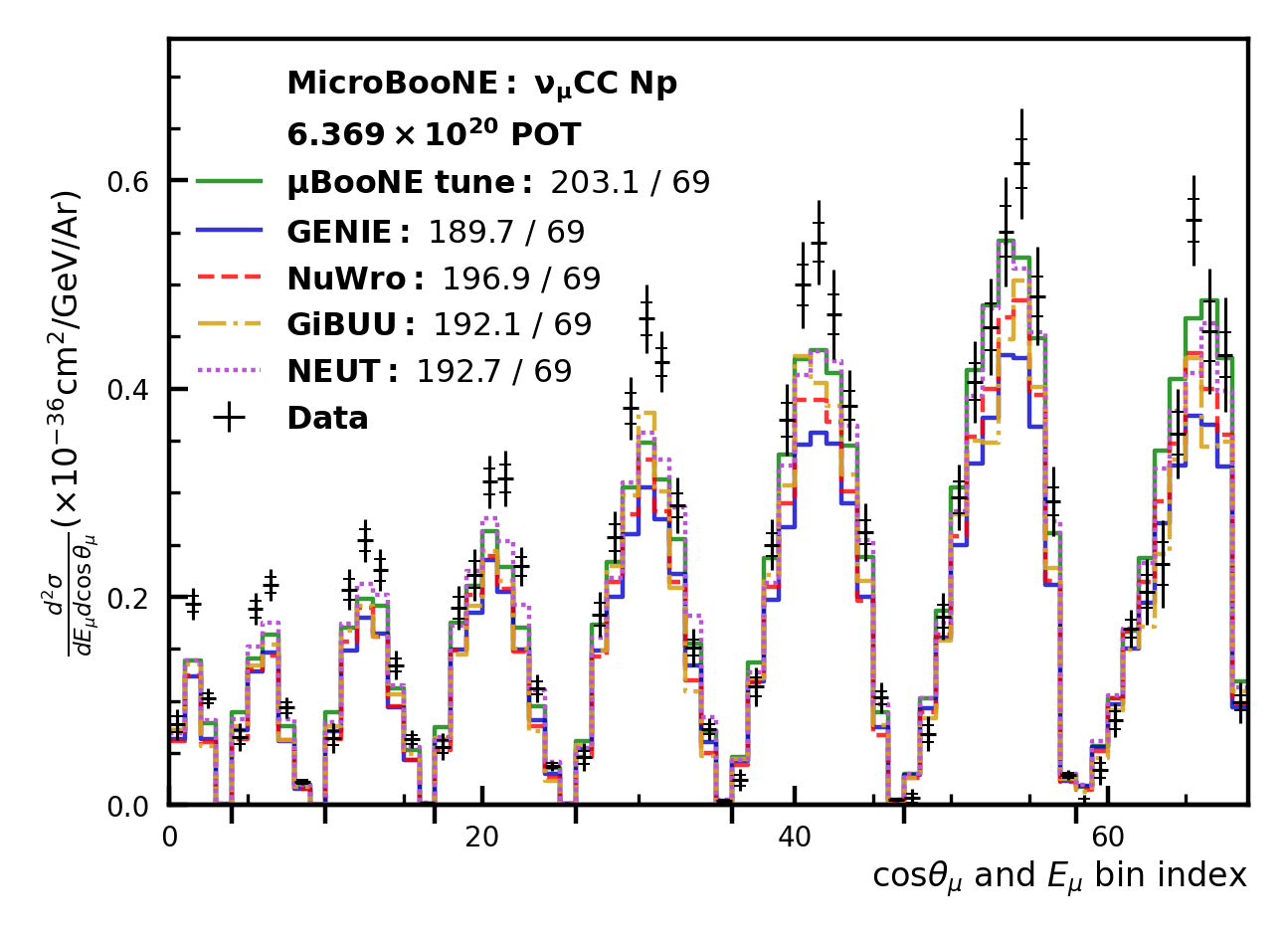}
        \caption{}
    \end{subfigure}
    \caption[Proton-related $\nu_\mu$CC inclusive cross section results]{Proton-related $\nu_\mu$CC inclusive flux averaged differential cross section results. Panel (a) shows the leading proton kinetic energy. Panels (b) and (c) show 2D results as a function of the muon energy and angle, for $0p$ and $Np$ events respectively, using a 35 MeV true proton kinetic energy threshold. From Ref. \cite{uboone_Np0p_PRL}.}
    \label{fig:0pNp_numuCC_XS}
\end{figure}

We expanded the previous inclusive $\nu_\mu$CC cross section results to include several more measurements, including separating by final state topology into events with and without final state protons \cite{uboone_Np0p_PRD,uboone_Np0p_PRL}, as shown in Fig. \ref{fig:0pNp_numuCC_XS}. This analysis revealed insufficient cross section model uncertainties related to low energy protons, a region of kinematic phase space which LArTPCs are especially capable in measuring, and used a data-driven method to expand these uncertainties before unfolding. This analysis similarly relied on detailed model validation tests.

\subsection{3D Cross Sections}

\begin{figure}[H]
    \centering
    \includegraphics[width=0.7\textwidth]{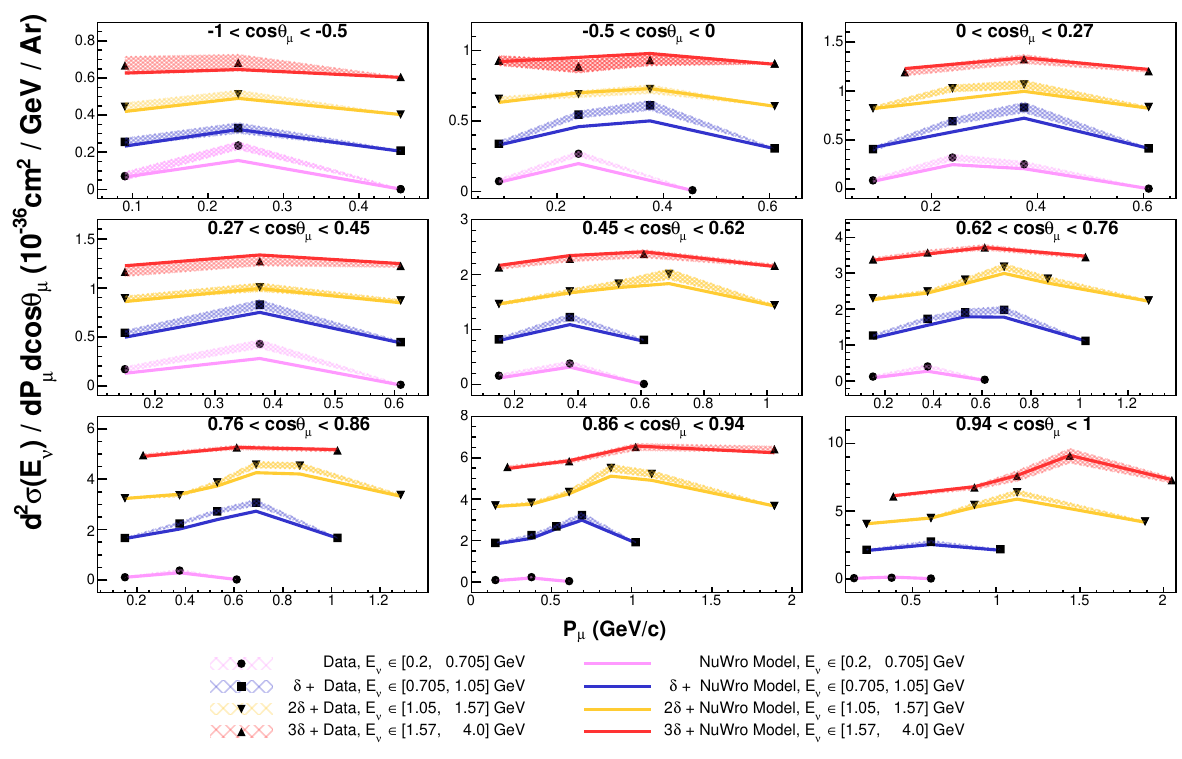}
    \caption[3D $\nu_\mu$CC inclusive cross section results]{3D $\nu_\mu$CC inclusive cross section results, in terms of neutrino energy, muon momentum, and muon angle, compared to the prediction from the NuWro event generator \cite{nuwro}. From Ref. \cite{uboone_Np0p_PRL}.}
    \label{fig:3D_numuCC_XS}
\end{figure}

We have also expanded our inclusive $\nu_\mu$CC cross section results to a 3D unfolding in terms of neutrino energy, muon momentum, and muon angle \cite{wc_numu_3D}, as shown in Fig. \ref{fig:3D_numuCC_XS}. Again, these results rely on detailed model validation tests.

\section{Data-driven Model Validation}\label{sec:model_validation}

In this section, I describe the discussion and results in Ref. \cite{uboone_model_validation}. We wrote this paper partly in response to feedback from the neutrino cross section community about the potential for bias when extracting cross sections as a function of neutrino energy in the results discussed above.

The cross section $S$ is related to the measurement in the detector $M$ by the following equation, where $POT$ is the number of protons on target in the neutrino beam during the data collection period, $T$ is the number of nuclear targets in the fiducial detector medium, $\phi(E_\nu)$ is the modeled neutrino beam flux as a function of the neutrino energy $E_\nu$, $D\left(T_\mathrm{true}\rightarrow T_\mathrm{rec}\right)$ is the detector response, describing the modeled mapping from a true observable $T_\mathrm{true}$ to a reconstructed observable $T_\mathrm{rec}$, $\varepsilon\left(E_\nu, T_\mathrm{true}\right)$ is the modeled efficiency of the selection as a function of the $E_\nu$ and $T_\mathrm{true}$, $B\left(T_\mathrm{rec}\right)$ is the modeled background, and $\sigma\left(E_\nu, T_\mathrm{true}\right)$ is the cross section to be extracted:

{\small
\begin{equation}
    M(T_\mathrm{rec})=
    POT \cdot T 
    \cdot \int
    \phi\left(E_\nu\right)
    \cdot\sigma\left(E_\nu, T_\mathrm{true}\right)
    \cdot D\left(T_\mathrm{true}\rightarrow T_\mathrm{rec}\right)
    \cdot \varepsilon\left(E_\nu, T_\mathrm{true}\right)
    \cdot dE_\nu
    + B\left(T_\mathrm{rec}\right)
\end{equation}
}

This integral equation can be written as a matrix equation:

\begin{equation}
    M_i = R_{i, j} \cdot S_j + B_i
\end{equation}

To solve for the cross section $S$, we invert this matrix equation. We normally do this with a regularization, in order to approximate this matrix inversion in a way which is less prone to large statistical fluctuations. In the Wire-Cell analyses discussed this chapter, this is done via Wiener-SVD unfolding \cite{wiener_svd}.

\begin{equation}
    S_j = R_{i, j}^{-1}\cdot \left(M_i - B_i\right)
\end{equation}

In this unfolding process, we leave the cross section $S$ as an unknown that we are solving for. However, we must also model the cross section in order to correctly estimate the detector response $D\left(T_\mathrm{true}\rightarrow T_\mathrm{rec}\right)$, the efficiency $\varepsilon\left(E_\nu, T_\mathrm{true}\right)$, and the background $B\left(T_\mathrm{rec}\right)$. To illustrate a few specific examples in the case of a muon energy cross section extraction: The detector response $D\left(E_{\mu, \mathrm{true}}\rightarrow E_{\mu, \mathrm{rec}}\right)$ could depend on the cross section model as a function of muon angle, since forward muons will be contained in the detector more often, which could lead to better energy resolution. The selection efficiency $\varepsilon(E_\nu, E_{\mu, \mathrm{true}})$ could depend on the cross section model as a function of proton multiplicity, since $\nu_\mu$CC events with observed protons could be easier to distinguish from cosmic muons, leading to higher efficiencies. The background $B(E_{\mu, \mathrm{rec}})$ could depend on the cross section for NC $1\pi^+$ production, since this topology can mimic the $\nu_\mu$CC topology in our detectors.
 
Ensuring that the cross section model used to estimate these quantities has sufficient uncertainties is an important part of any cross section extraction. Modeling these quantities is also important for understanding flux uncertainties when extracting to a nominal flux spectrum which differs from the real flux spectrum, as we discuss and illustrate in more detail in Ref. \cite{uboone_model_validation}. It is particularly important for extractions of quantities like the neutrino energy which are not directly observable, since the detector response $D\left(E_{\nu, \mathrm{true}}\rightarrow E_{\nu, \mathrm{rec}}\right)$ is harder to estimate, and therefore our model validation procedure is particularly important for the results described in the previous section.

In general, the idea behind model validation is to make many detailed data-prediction comparisons, in order to catch any potential mis-modeling that could affect modeling of the detector response, efficiency, or background, and if found, expand the relevant systematic uncertainty and avoid the extraction of any biased cross section results.

A different analysis method which similarly aims to ensure the robustness of extracted cross section results is fake data closure testing. In this method, a different neutrino event generator with a different cross section prediction is used in order to generate a fake data set, and then an extracted cross section measurement is performed, and the results are compared with the input truth. If a significant difference is found between the extracted cross section and the input true cross section, we determine that the systematic uncertainties on the cross section model used for the unfolding are insufficient to cover the input cross section. In this case, we can expand the cross section uncertainty in order to pass the fake data test, and this is something MicroBooNE has done in one of our $1\mu1p$ cross section results \cite{uboone_1mu1p_xs}.

\begin{figure}[H]
    \centering
    \includegraphics[width=\textwidth]{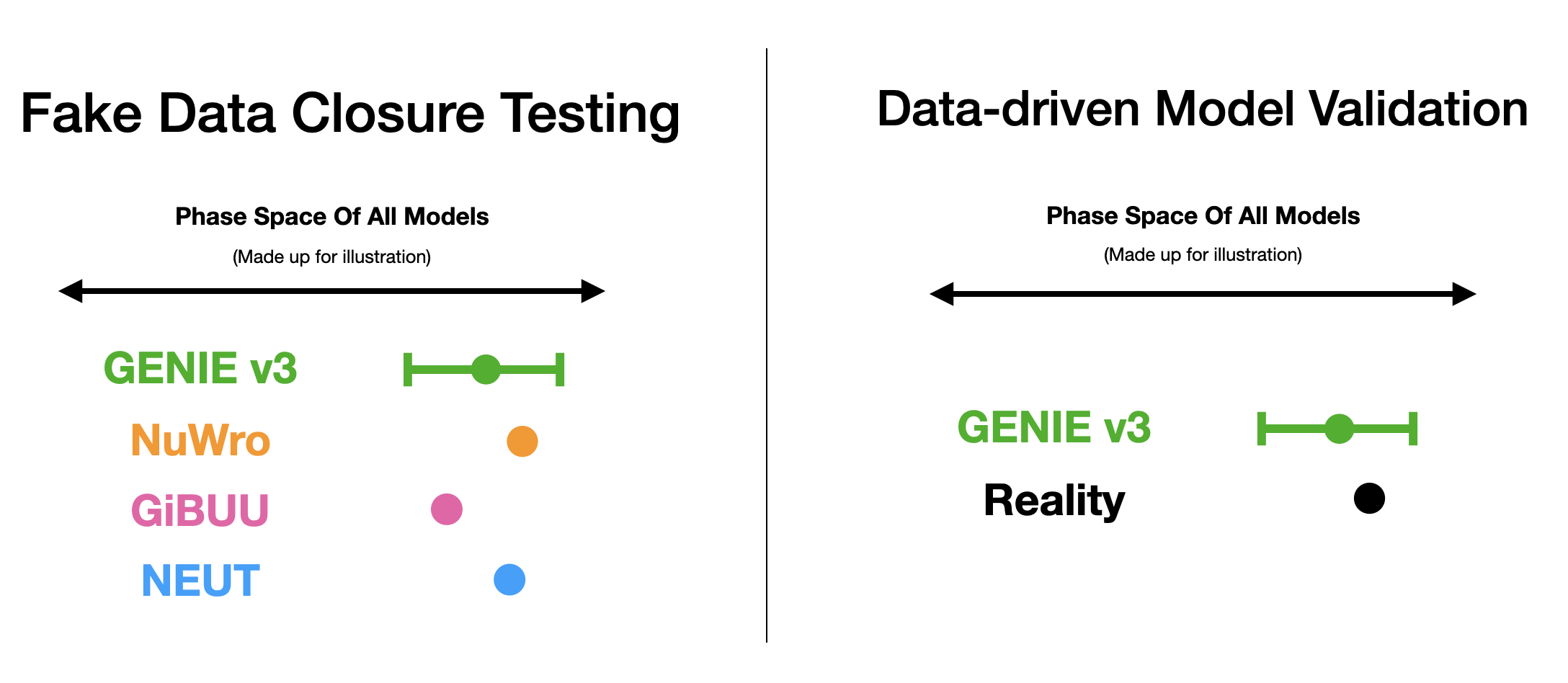}
    \caption[Illustration of fake data closure testing vs. data-driven model validation]{Illustration of fake data closure testing vs. data-driven model validation. This specific example is made up for illustration. In reality, the phase space of models is high dimensional, and the relevant comparisons between different generators will depend on the specific cross section analysis.}
    \label{fig:error_bar_illustration}
\end{figure}

Fake data closure testing and data-driven model validation both have the same goal, of testing the cross section model used for unfolding in order to prevent the extraction of unbiased cross sections. But they do this in two distinct ways, as illustrated in Fig. \ref{fig:error_bar_illustration}. Fake data closure testing tries to ensure that the cross section model used for unfolding has large enough uncertainties to cover other popular cross section models, while data-driven model validation tries to ensure that the cross section model used for unfolding has large enough uncertainties to cover the true cross section in reality. Additionally, unlike fake data closure testing, data-driven model validation tests not only the cross section modeling, but also the flux and detector response modeling. Both techniques have their merits, and both are used in MicroBooNE. 

\subsection{\texorpdfstring{$\chi^2$}{chi2} Tests}

\begin{figure}[H]
    \centering
    \begin{subfigure}[b]{0.32\textwidth}
        \includegraphics[width=\textwidth]{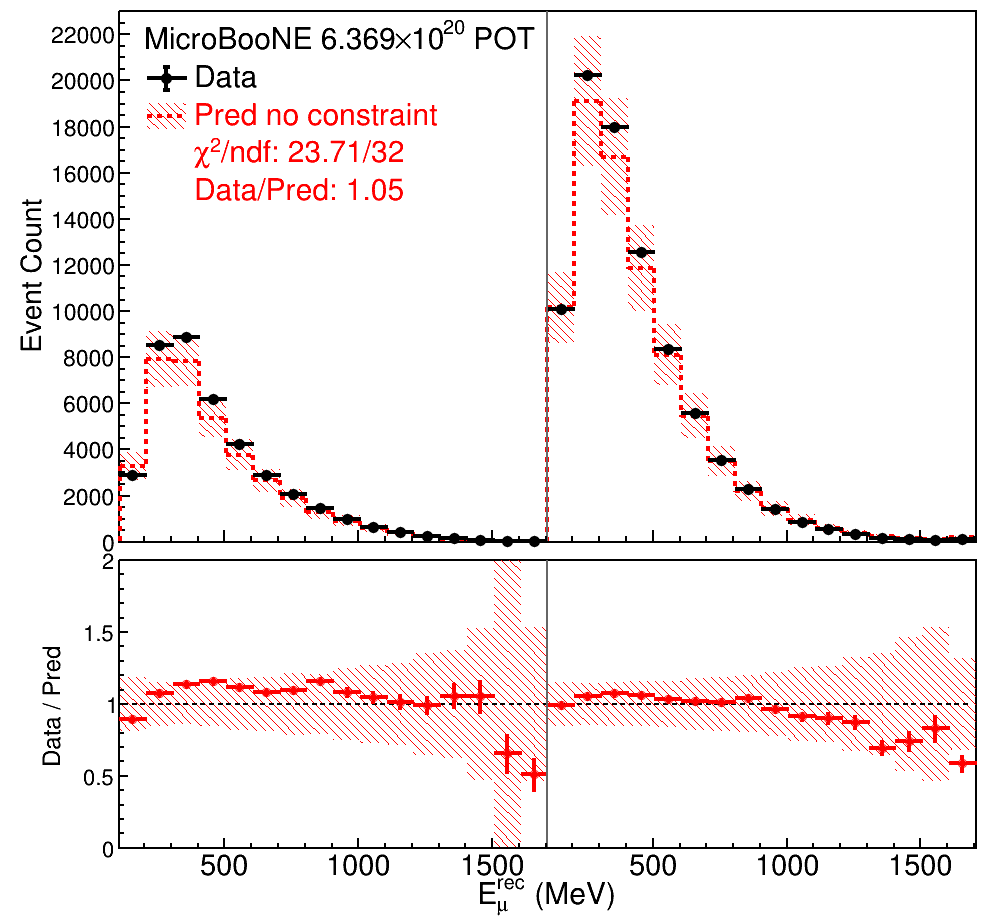}
        \put(-100,80){FC}
        \put(-30,80){PC}
        \caption{}
    \end{subfigure}
    \begin{subfigure}[b]{0.32\textwidth}
        \includegraphics[width=\textwidth]{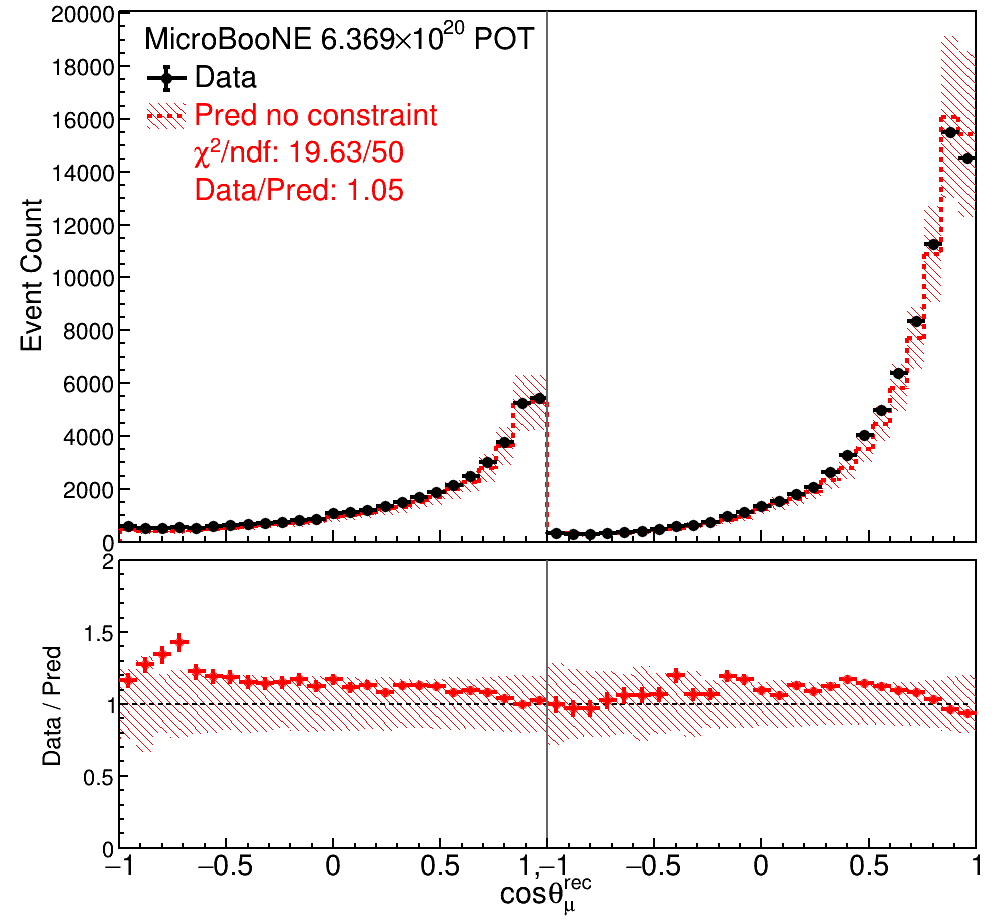}
        \put(-120,80){FC}
        \put(-50,80){PC}
        \caption{}
    \end{subfigure}
    \begin{subfigure}[b]{0.32\textwidth}
        \includegraphics[width=\textwidth]{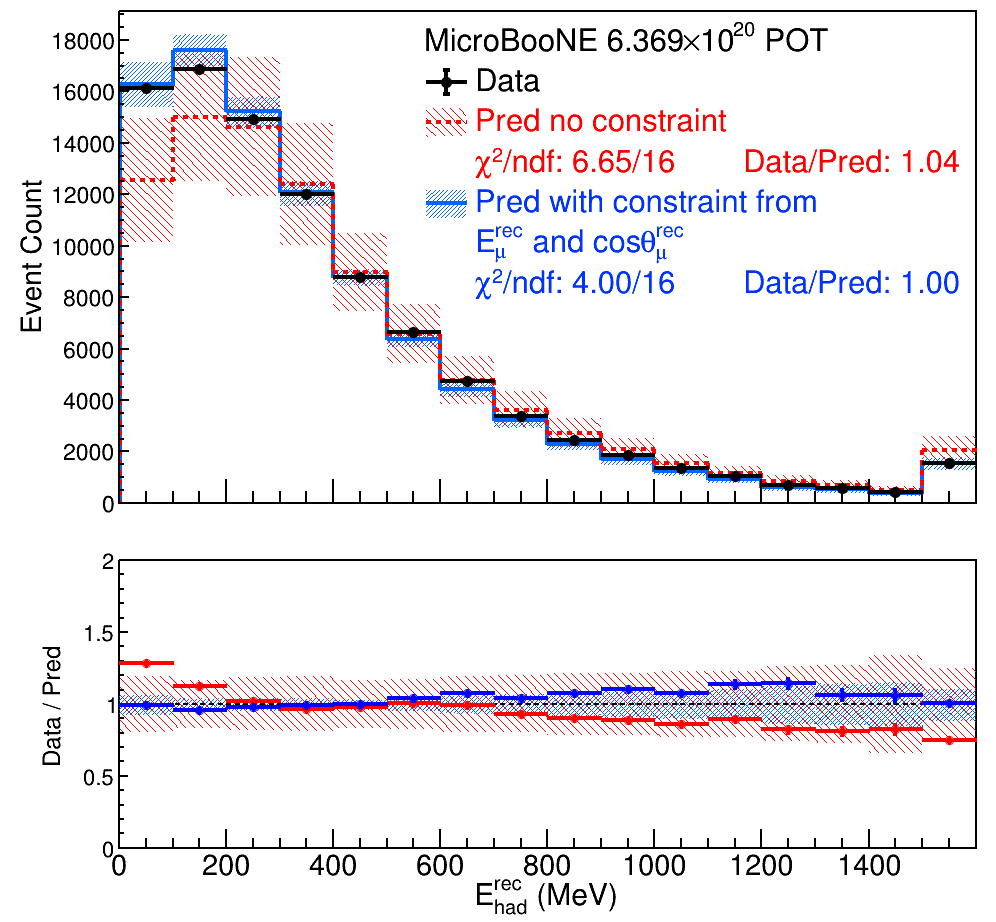}
        \put(-50,80){PC}
        \caption{}
    \end{subfigure}
    \caption[Hadronic energy conditional constraint model validation test]{Hadronic energy conditional constraint model validation test. Panel (a) shows our measurement of reconstructed muon energy, and panel (b) shows our measurement of reconstructed muon angles with respect to the neutrino beam, each for fully contained and partially contained events separately. Panel (c) shows the resulting reconstructed hadronic energy measurement for partially contained events, both unconstrained in red and constrained in blue. From Ref. \cite{uboone_model_validation}.}
    \label{fig:Ehad_model_validation}
\end{figure}

Perhaps the simplest way to test agreement between data and prediction is a $\chi^2$/ndf test, which uses the difference between data and prediction as well as a covariance matrix describing correlated uncertainties in order to evaluate agreement. As discussed in Sec. \ref{sec:conditional_constraint}, we can form a more sensitive test of data-prediction agreement when using a conditional constraint, reducing uncertainties by using observations of other distributions. We can apply a conditional constraint using the same events in both the constraining and target distributions as long as we properly account for correlated statistical uncertainties between the two distributions. As an example, we measure distributions of the reconstructed muon energy and angle for fully contained and partially contained events, and use that to constrain the distribution of reconstructed hadronic energy, as shown in Fig. \ref{fig:Ehad_model_validation}. In this case, we can see a dramatic reduction of the systematic uncertainty, as well as a significant shift in the prediction. After constraint, the prediction matches the data well, both visually and according to a constrained $\chi^2$/ndf test.

However, even after constraints, it is possible for a $\chi^2$/ndf to fail to detect mismodeling. For example, if a distribution has perfect shape agreement but a normalization which is very far from the prediction, a $\chi^2$/ndf test could fail to detect this mismodeling. This is an example of a more general class of possibile failure modes where an underestimated uncertainty in one feature of the distribution can be masked by conservative uncertainties in other features of the distribution. We developed a method of decomposing this $\chi^2$ according to different eigenvalues in order to mitigate this possibility. 

Specifically, we start with the normal $\chi^2$ formula in terms of the measurement $M$, the prediction $P$, and the covariance matrix describing uncertainties on the predictions $V$:

\begin{equation}
    \chi^2 = (M - P)^T \cdot V^{-1} \cdot (M - P)
\end{equation}

Then, we apply a this transformation, where $\Lambda$ is a diagonal matrix containing the eigenvalues of $V$ and $Q$ containing the corresponding eigenvectors as rows:

\begin{align*}
    V &= \tilde{Q} \cdot \Lambda \cdot \tilde{Q}^T \\
    Q &= \tilde{Q}^{-1} \\
\end{align*}

After also assigning $\Delta = (M - P)$, we get this equation:

\begin{equation}
    \chi^2 = (Q \cdot \Delta)^T \cdot (Q \cdot V \cdot Q^T)^{-1} \cdot (Q \cdot \Delta)
\end{equation}

By defining $\epsilon_i = \frac{\Delta'_i}{\sqrt{\Lambda_{ii}}}$ and $\Delta' = Q \cdot \Delta$, we can write this as 

\begin{equation}
    \chi^2 = \Delta'^T \cdot \Lambda^{-1} \cdot \Delta' = \sum_i \epsilon_i^2
\end{equation}

These $\epsilon_i^2$ values sum up to the full $\chi^2$, but each correspond to an independent eigenvector of the covariance matrix. Essentially, we have transformed our distribution into a space where all bins are completely uncorrelated. For example, for some covariance matrix, we might expect one of these transformed bins to be the normalization, in which case we can investigate that agreement independently and reject a model which fails to describe this data-prediction normalization difference, resolving the failure mode discussed above. By comparing each bin independently, we get a series of $p_\text{local}$ values, and can use the largest discrepancy to construct $p_\text{global} = 1 - (1 - p_\text{local})^n$, which will serve as a more sensitive data-prediction comparison in many cases. This $\chi^2$ decomposition is a very general procedure, and can be applied to any scenario with a covariance matrix, before or after a conditional constraint, and before or after an unfolding from a reconstructed variable to a truth variable.

\begin{figure}[H]
    \centering
    \begin{subfigure}[b]{0.3\textwidth}
        \includegraphics[width=\textwidth]{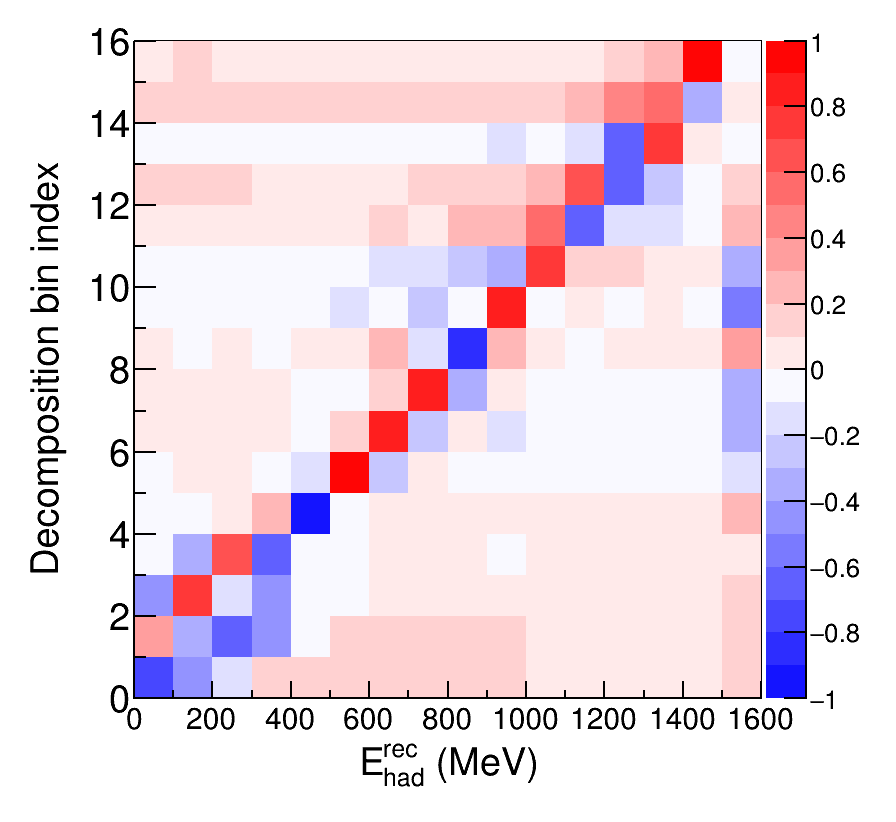}
        \caption{}
    \end{subfigure}
    \begin{subfigure}[b]{0.55\textwidth}
        \includegraphics[width=\textwidth]{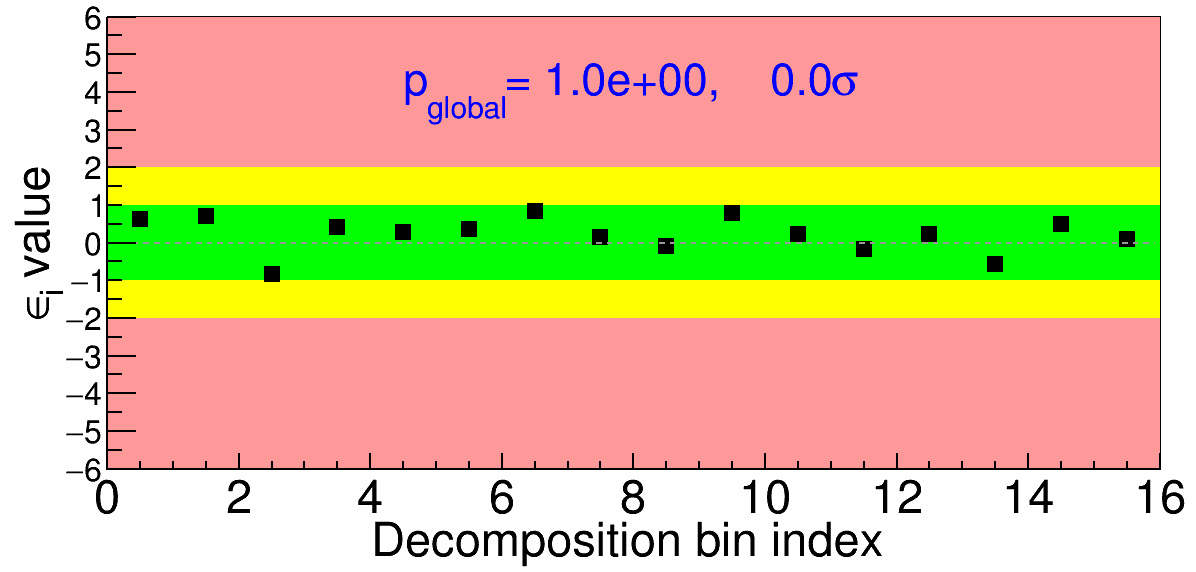}
        \caption{}
    \end{subfigure}
    \caption[Hadronic energy conditional constraint model validation test]{Hadronic energy conditional constraint model validation test, after the conditional constraint. Panel (a) shows the $Q$ matrix which translates between the original reconstructed hadronic energy bins and the decomposition bins. Panel (b) shows the decomposition data-prediction agreement. In this case, no data point exists more than one sigma away from the prediction in any bin. From Ref. \cite{uboone_model_validation}.}
    \label{fig:Ehad_decomposition}
\end{figure}

This $\chi^2$ decomposition is illustrated for the above example in Fig. \ref{fig:Ehad_decomposition}. We have transformed the hadronic energy distribution into a space where all bins are uncorrelated, and we see that each bin independently agrees with the prediction. We also can investigate the $Q$ matrix which translates between the original reconstructed hadronic energy bins and the decomposition bins. In this case, we see that the first decomposition bin corresponds to a decrease in the lowest hadronic energies and an increase in higher hadronic energies, indicating a type of energy shift which is possible according to our systematic uncertainty covariance matrix.

We see that for this hadronic energy model validation test, we see good agreement even after our conditional constraint and our $\chi^2$ decomposition. This specific fake data test is particularly relevant for our unfolding to the true neutrino energy. This is because we can use energy conservation in order to understand the relationship between the true neutrino energy $E_\nu$, the muon energy $E_\mu$, the visible hadronic energy (for example, protons) $E^\mathrm{vis}_\mathrm{had}$, and the invisible hadronic energy $E^\mathrm{invis}_\mathrm{had}$ (for example, neutrons):
\begin{equation}
    E_\nu = E_\mu + E^\mathrm{vis}_\mathrm{had} + E^\mathrm{invis}_\mathrm{had}
\end{equation}
The distribution of $E_\nu$ is known with uncertainties from our BNB flux model, the distribution of $E_\mu$ is measured as part of our conditional constraint, and $E^\mathrm{vis}_\mathrm{had}$ is measured as the target distribution of the constraint. Our model validation tells us that the distributions of these three variables all agree with our modeling within uncertainties, and since our model obeys energy conservation, this effectively tells us that our model also agrees with the distribution of $E^\mathrm{invis}_\mathrm{had}$ within uncertainties. This is especially valuable, since the modeling of $E^\mathrm{invis}_\mathrm{had}$ is something that could bias neutrino energy cross section measurements and which cannot be directly tested in simpler ways. Of course, this does not test $E^\mathrm{invis}_\mathrm{had}$ event-by-event, but only for the entire distribution, but this model validation test does serve as a powerful handle on potential $E^\mathrm{invis}_\mathrm{had}$ mis-modeling, as we will see in the next section. 

In this section, we have illustrated just one fake data test, but many different tests should be performed with different variables depending on the type of cross section being extracted.

\subsection{Proton Energy Shift Fake Data Studies}

In this section, I describe a series of fake data studies which effectively modify the amount of invisible hadronic energy $E^\mathrm{invis}_\mathrm{had}$ in each event, which corresponds to the type of mis-modeling which could potentially bias our neutrino energy cross sections.

Directly changing $E^\mathrm{invis}_\mathrm{had}$ in our simulation would do nothing, since each event would look identical when changing the amount of invisible activity. We have to consider energy conservation in the hadronic system; if the amount of invisible hadronic energy increases, the amount of visible hadronic energy decreases. Therefore, we investigate a fake data test where we decrease the amount of proton energy, and this effectively models the scenario where a larger fraction of proton energy goes instead to invisible neutrons.

\begin{figure}[H]
    \centering
    \includegraphics[width=0.5\textwidth]{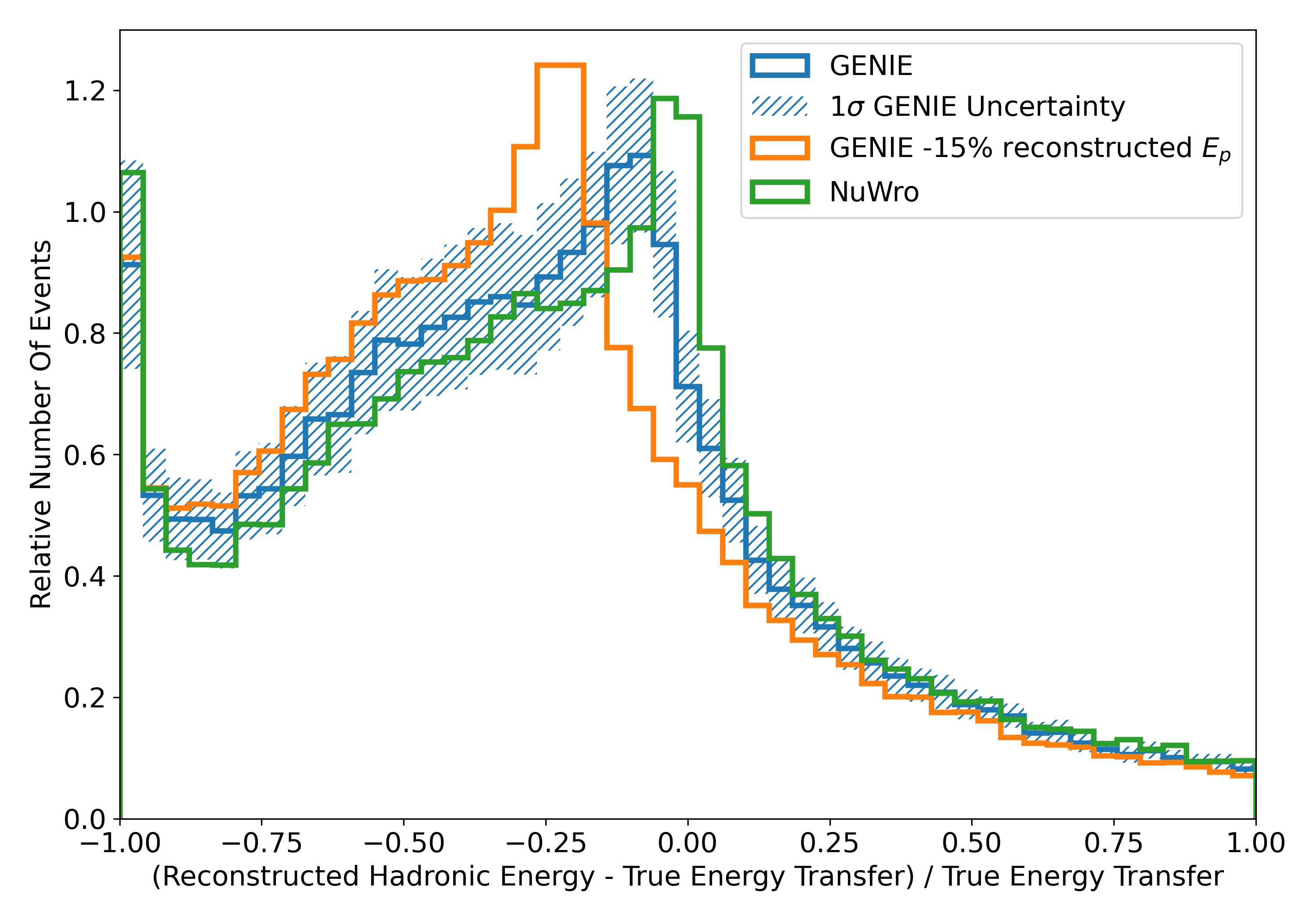}
    \caption[Energy transfer resolution with proton energy shift]{Energy transfer resolution with proton energy shift. From Ref. \cite{uboone_model_validation}.}
    \label{fig:energy_transfer_resolution}
\end{figure}

Figure \ref{fig:energy_transfer_resolution} illustrates this shift in terms of the reconstructed hadronic energy resolution. This biases the distribution significantly outside the GENIE 1 $\sigma$ band, showing that this fake data set is far beyond our currently modeled uncertainties. We also illustrate the distribution from a different cross section prediction, NuWro \cite{nuwro}, and we see that this generator similarly has a distribution outside of GENIE uncertainties.

\begin{figure}[H]
    \centering
    \begin{subfigure}[b]{0.49\textwidth}
        \includegraphics[width=\textwidth]{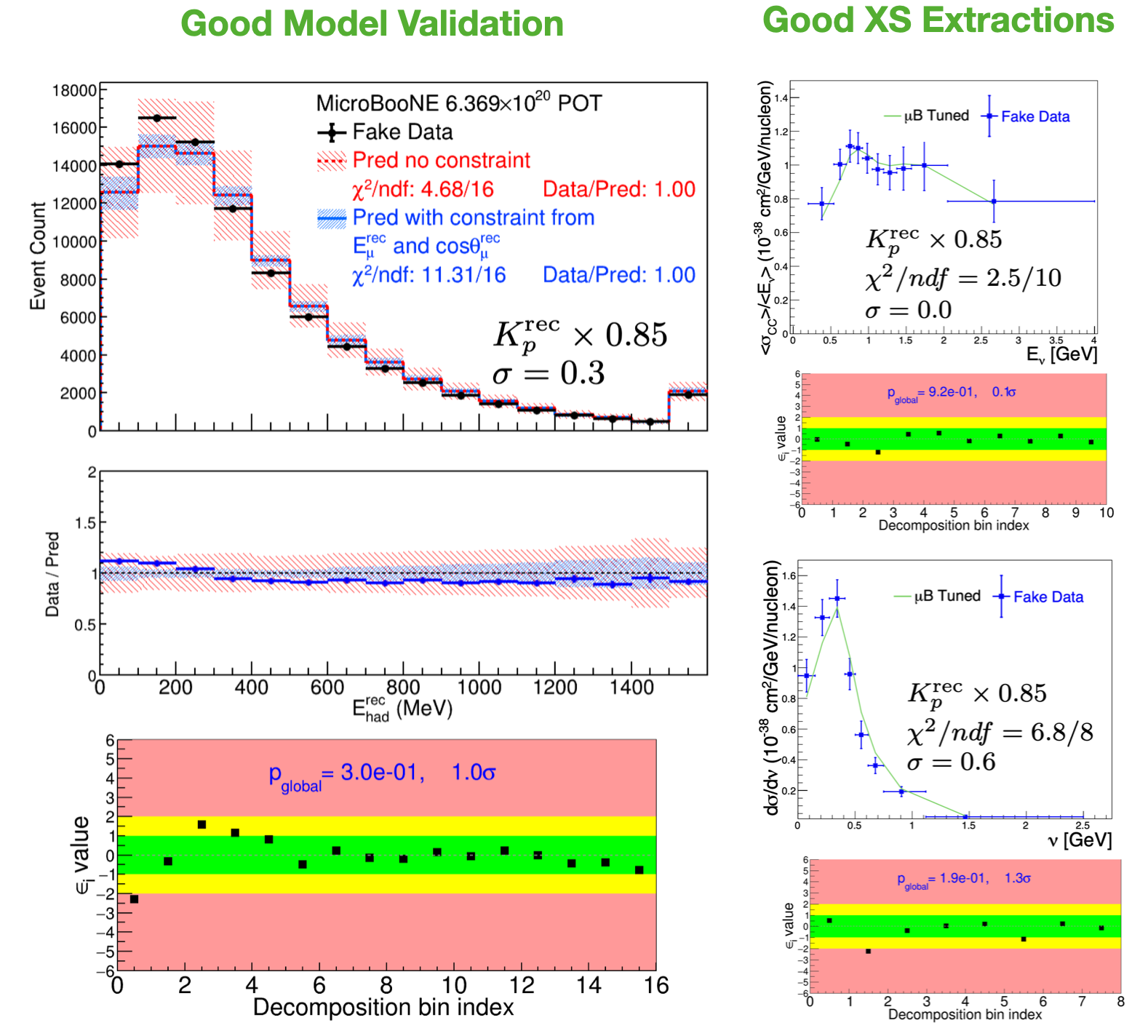}
        \caption{}
        \label{fig:good_proton_scaling}
    \end{subfigure}
    \begin{subfigure}[b]{0.49\textwidth}
        \includegraphics[width=\textwidth]{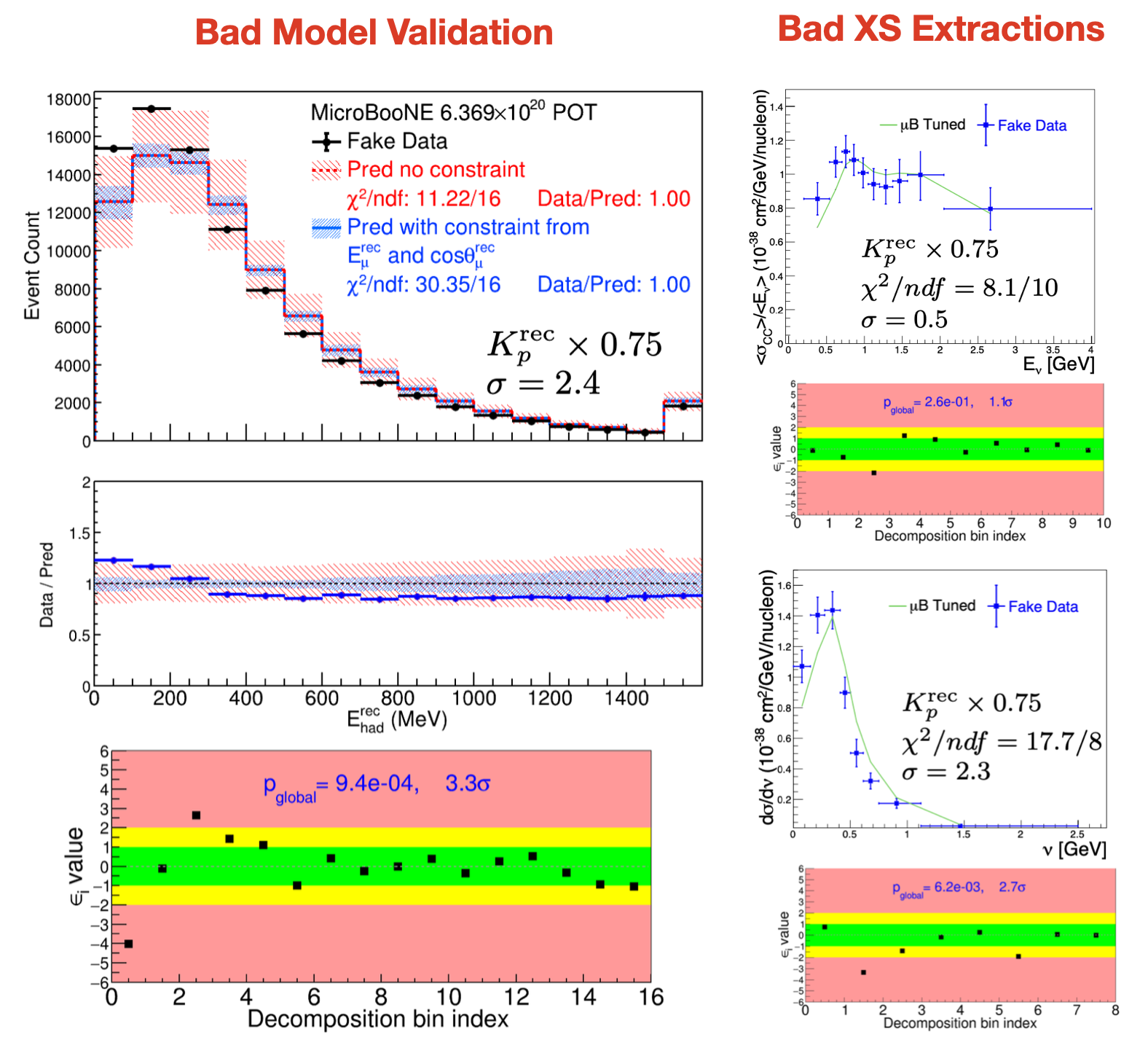}
        \caption{}
        \label{fig:bad_proton_scaling}
    \end{subfigure}
    \begin{subfigure}[b]{0.49\textwidth}
        \includegraphics[width=\textwidth]{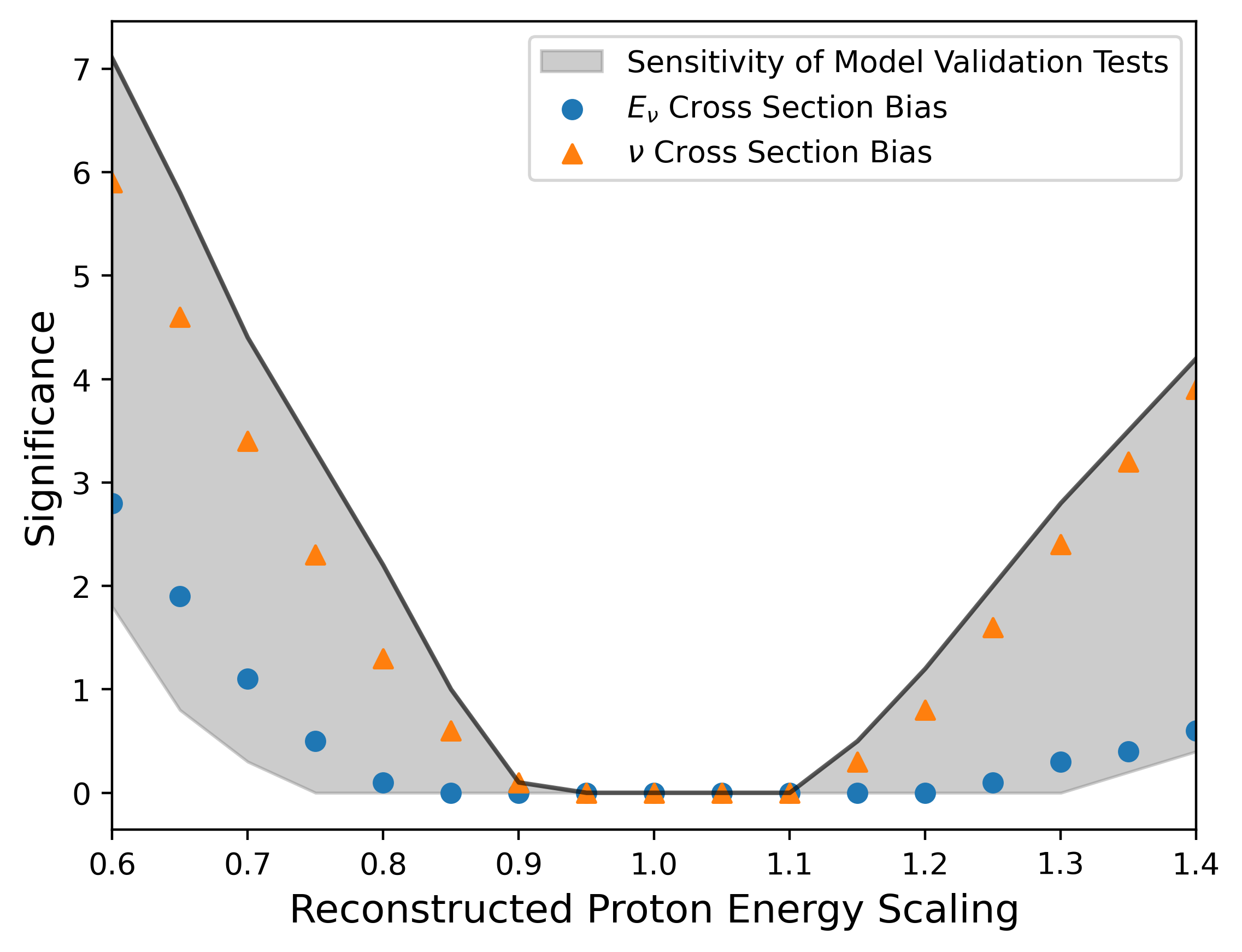}
        \caption{}
        \label{fig:distribution_proton_scaling}
    \end{subfigure}
    \caption[Proton energy scaling fake data model validation tests]{Proton energy scaling fake data model validation tests. We show the reconstructed hadronic energy model validation test both before and after conditional constraint from muon kinematic measurements, and after a $\chi^2$ decomposition. We also show extracted $E_\nu$ and $\nu$ cross section measurements compared with the input truth.
    In panel (a), we show this for a 15\% downward shift in proton energies, and see good agreement in the model validation and cross section extraction. In panel (b), we show this for a larger 25\% downward shift in proton energies, and see significant discrepancies in the model validation and significant bias in the extracted cross sections.
    In panel (c), we generalize the above results for a variety of downward and upward proton energy scalings, and compare with 50 different model validation tests, indicated by a gray band, with a black curve indicating the most sensitive test in each case.
    From Ref. \cite{uboone_model_validation}.}
    \label{fig:proton_scaling_model_validations}
\end{figure}

In Fig. \ref{fig:good_proton_scaling}, we apply a 15\% downward shift to proton energies and show the results of the hadronic energy model validation test as well as the neutrino energy and energy transfer cross section extractions. In this case, we see no disagreement outside of our uncertainties in the model validation test, and in the cross section extractions we see no bias with respect to the true input. In Fig. \ref{fig:bad_proton_scaling}, we increase the scale of this invisible hadronic energy mismodeling, applying a 25\% downward shift to proton energies. Now, we see significant disagreement in our model validation test, as well as significant bias in our extracted cross sections. These two cases are generalized with a wide range of both downward and upward scalings, and with 50 different model validation tests described in Table \ref{tab:validation_tests_list}, forming a gray band in Fig. \ref{fig:distribution_proton_scaling}. The important feature to note is that in this test, the most sensitive model validation test is always more significant than the extracted cross section bias. This means that if we encountered a scenario like this in real data, we would always be able to identify the modeling deficiency before reporting any biased cross section results to the community.

\begin{table}[H]
    \centering
    \footnotesize
    \begin{tabular}{p{0.9\linewidth}}
        \toprule
        Test Description \\
        \midrule
        Evaluation of the $E_\mu^{\text{rec}}$ FC, PC and FC\&PC distributions through overall $\chi^2$ GoF tests and $\chi^2$ decompositions (6 total tests). \\
        \midrule
        Evaluation of the $E_\mu^{\text{rec}}$ PC distribution after constraint from the analogous FC distribution. The overall $\chi^2$ GoF test and $\chi^2$ decomposition are examined (2 total tests). \\
        \midrule
        Evaluation of the $\cos\theta_\mu^{\text{rec}}$ FC, PC and FC\&PC distributions through overall $\chi^2$ GoF tests and $\chi^2$ decompositions (6 total tests). \\
        \midrule
        Evaluation of the $\cos\theta_\mu^{\text{rec}}$ PC distribution after constraint from the analogous FC distribution. The overall $\chi^2$ GoF test and $\chi^2$ decomposition are examined (2 total tests). \\
        \midrule
        Evaluation of the $\cos\theta_\mu^{\text{rec}}$ FC, PC and FC\&PC distributions after constraint from the FC\&PC $E_\mu^{\text{rec}}$ distribution. The overall $\chi^2$ GoF test and $\chi^2$ decomposition is examined for each distribution (6 total tests). \\
        \midrule
        Evaluation of the $E_\nu^{\text{rec}}$ FC, PC and FC\&PC distributions through overall $\chi^2$ GoF tests and $\chi^2$ decompositions (6 total tests). \\
        \midrule
        Evaluation of the $E_\nu^{\text{rec}}$ PC distribution after constraint from the analogous FC distribution. The overall $\chi^2$ GoF test and $\chi^2$ decomposition are examined (2 total tests). \\
        \midrule
        Evaluation of the $E_\nu^{\text{rec}}$ FC, PC and FC\&PC distributions after constraint from the FC\&PC $E_\mu^{\text{rec}}$ and $\cos\theta_\mu^{\text{rec}}$ distributions. The overall $\chi^2$ GoF test and $\chi^2$ decomposition is examined for each distribution (6 total tests). \\
        \midrule
        Evaluation of the $E_{\text{had}}^{\text{rec}}$ FC, PC and FC\&PC distributions through overall $\chi^2$ GoF tests and $\chi^2$ decompositions (6 total tests). \\
        \midrule
        Evaluation of the $E_{\text{had}}^{\text{rec}}$ PC distribution after constraint from the analogous FC distribution. The overall $\chi^2$ GoF test and $\chi^2$ decomposition are examined (2 total tests). \\
        \midrule
        Evaluation of the $E_{\text{had}}^{\text{rec}}$ FC, PC and FC\&PC distributions after constraint from the FC\&PC $E_\mu^{\text{rec}}$ and $\cos\theta_\mu^{\text{rec}}$ distributions. The overall $\chi^2$ GoF test and $\chi^2$ decomposition is examined for each distribution (6 total tests). \\
        \bottomrule
    \end{tabular}
    \caption[List of data-driven model validation tests]{List of data-driven model validation tests performed in the proton energy scaling fake data study.}
    \label{tab:validation_tests_list}
\end{table}

The process I have described is not completely infallible; it is theoretically possible to have two different mismodelings cancel each other out and result in good agreement in all reconstructed distributions, despite the existence of underlying mismodeling. For example, this possibility was considered in a DUNE-PRISM study in the DUNE near detector CDR \cite{DUNE_ND_CDR}. In this study, they considered a 20\% downward shift in proton energies, modeling an increase in $E^\mathrm{invis}_\mathrm{had}$ as described above, but then assumed that there was also a cross section mis-modeling, so this discrepancy would be removed by fitting the measurements to the near detector data. In this case, they would extract biased oscillation parameters, and this serves as a good motivation for the construction of DUNE-PRISM, which would move the near detector to experience different off-axis neutrino fluxes, making this type of mis-modeling much easier to identify. To illustrate and understand this type of effect in our context, I construct a similar study using MicroBooNE simulation and reconstruction.

\begin{figure}[H]
    \centering
    \begin{subfigure}[b]{0.49\textwidth}
        \includegraphics[width=\textwidth]{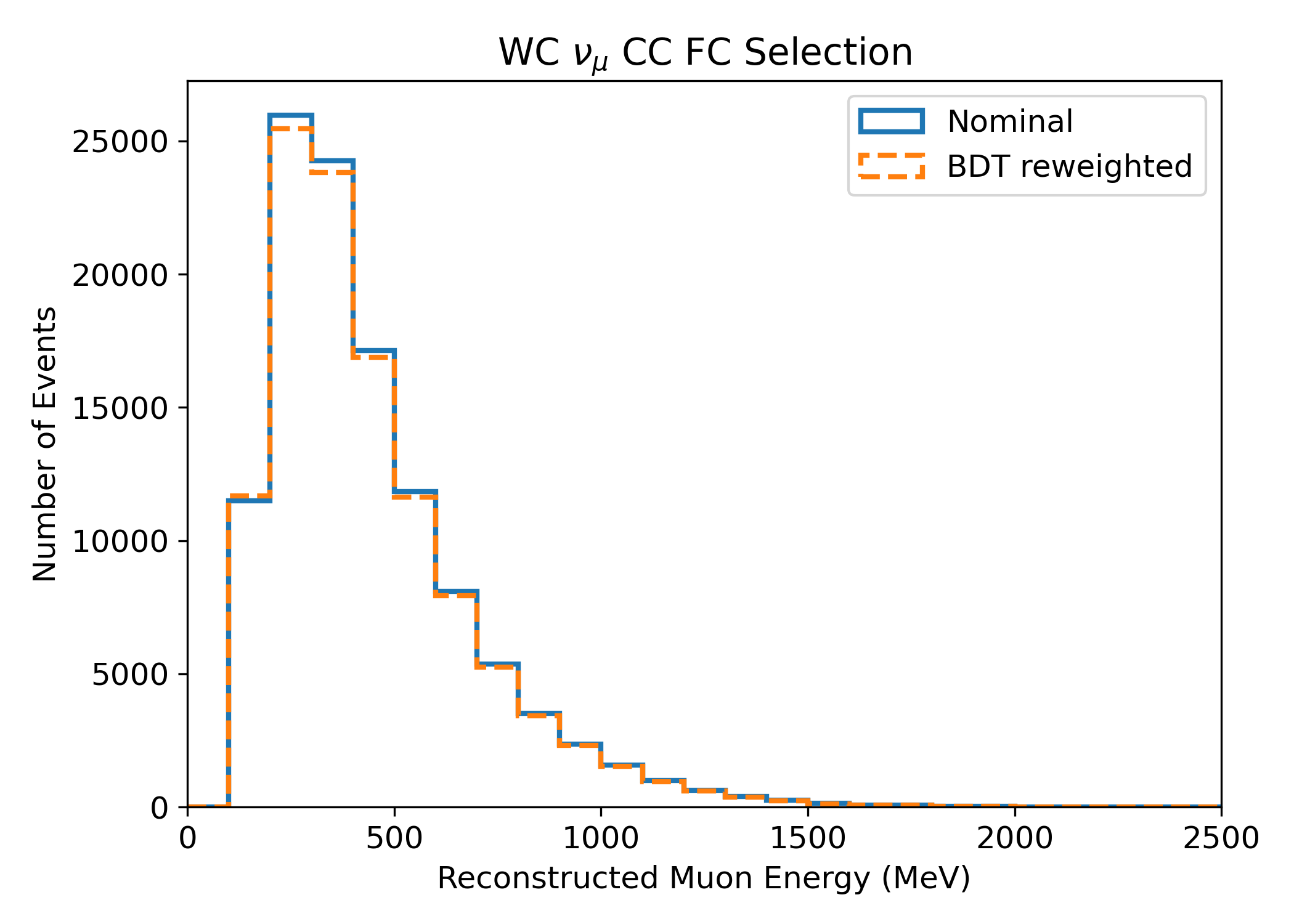}
        \caption{}
    \end{subfigure}
    \begin{subfigure}[b]{0.49\textwidth}
        \includegraphics[width=\textwidth]{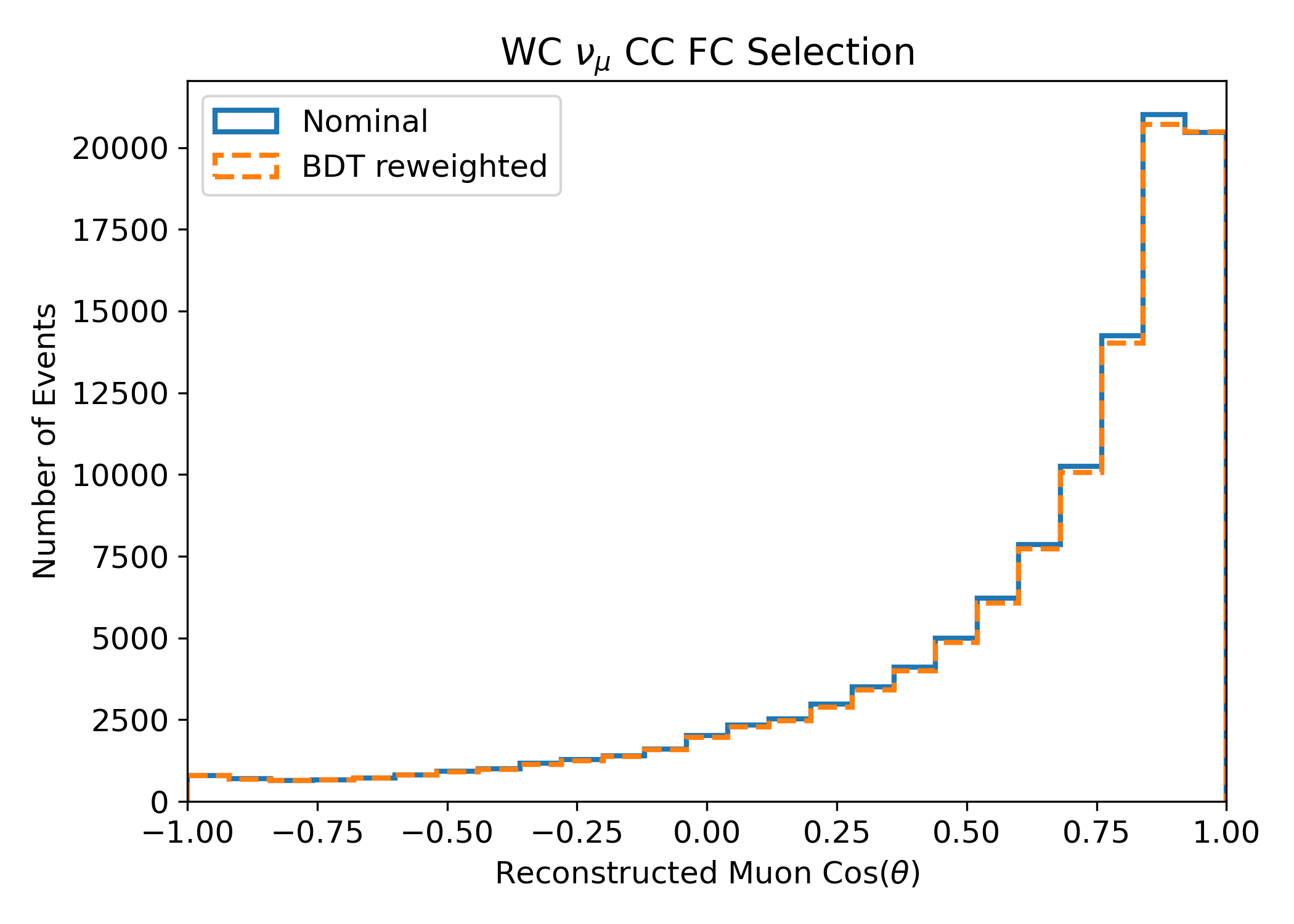}
        \caption{}
    \end{subfigure}
    \begin{subfigure}[b]{0.49\textwidth}
        \includegraphics[width=\textwidth]{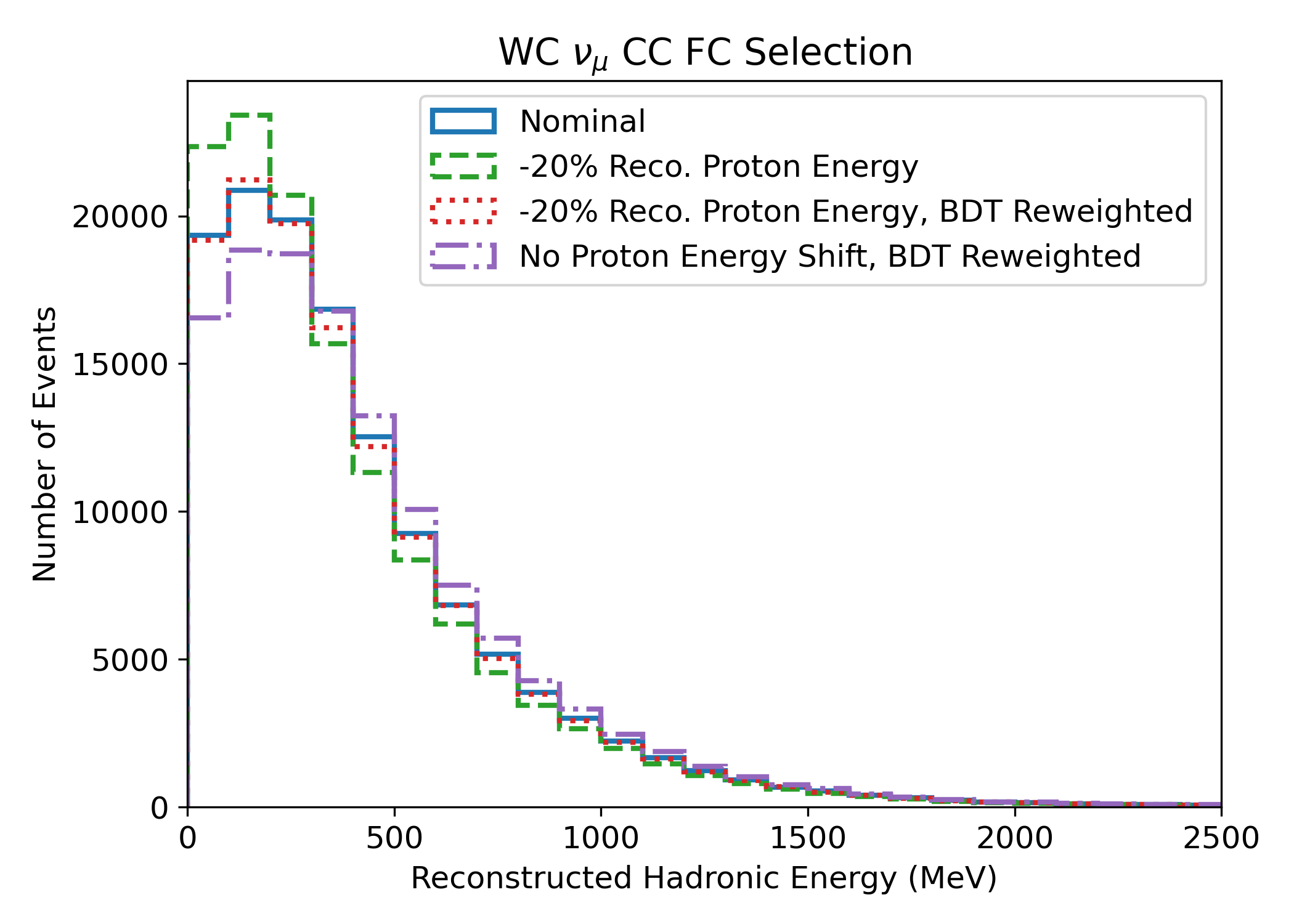}
        \caption{}
    \end{subfigure}
    \caption[BDT reweighted distributions]{BDT reweighted distributions. Panel (a) shows the reconstructed muon energy, and panel (b) shows the reconstructed muon angle with respect to the neutrino beam, showing good agreement between the two distributions in both cases. Panel (c) shows the reconstructed hadronic energy distribution, where the 20\% downward shift in reconstructed proton energy also affects the distribution. After the proton shift and the BDT reweighting, the distribution closely matches the original input distribution. From Ref. \cite{uboone_model_validation}.}
    \label{fig:bdt_reweight_1d}
\end{figure}

\begin{figure}[H]
    \centering
    \begin{subfigure}[b]{0.52\textwidth}
        \includegraphics[width=\textwidth]{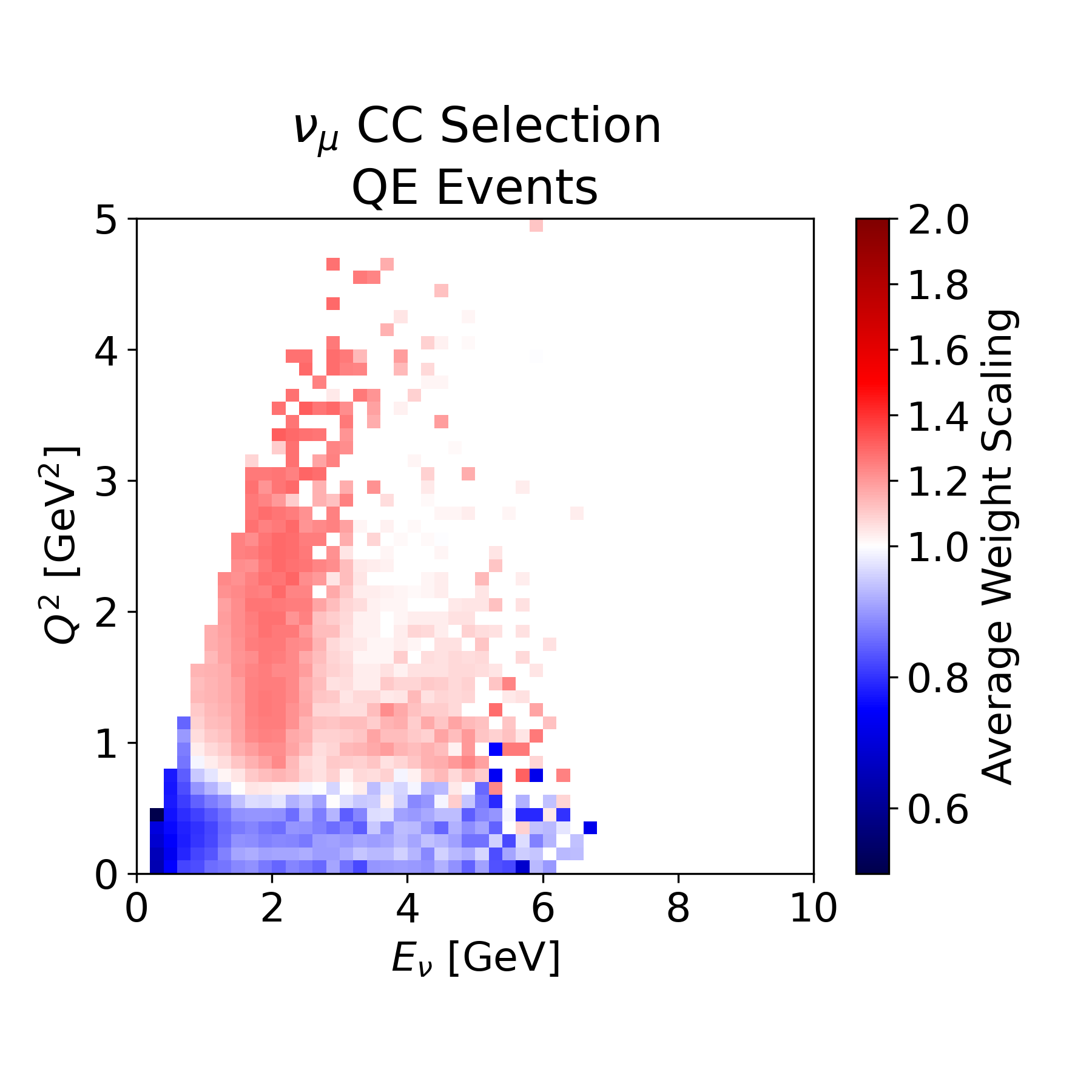}
        \caption{}
    \end{subfigure}
    \begin{subfigure}[b]{0.46\textwidth}
        \includegraphics[width=\textwidth]{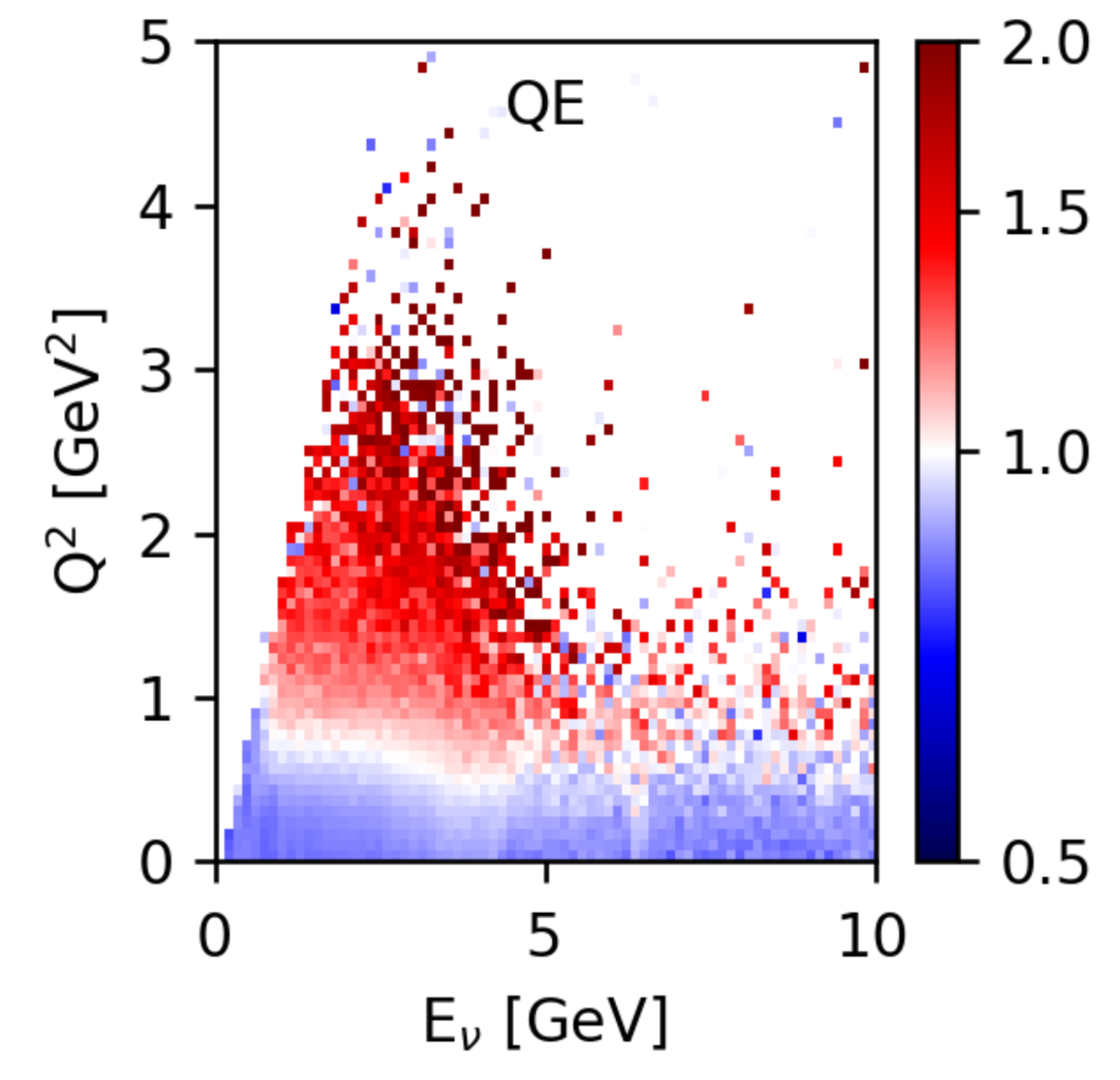}
        \caption{}
    \end{subfigure}
    \caption[Quasi-elastic BDT reweighting compared with DUNE-PRISM study]{Average BDT weight scaling for quasi-elastic events as a function of $Q^2$ and $E_\nu$. Panel (a) shows the BDT reweighting in this study for MicroBooNE, from Ref. \cite{uboone_model_validation}. Panel (b) shows the BDT reweighting from the DUNE study in Ref. \cite{DUNE_ND_CDR}.}
    \label{fig:bdt_reweight_2d}
\end{figure}

Like in the DUNE study, I shift proton energies down by 20\%, and then apply a multivariate BDT in order to alter the cross section as a function of true muon energy, true muon angle, and true energy transfer to the nuclear system in order to recover the same reconstructed distributions of muon energy, muon angle, and reconstructed hadronic energy. I used a GBReweighter with all default hyperparameters from the hep\_ml python package \cite{hep_ml}. This process was able to correct the distributions and effectively undo the proton energy shift in all distributions, as shown in Fig. \ref{fig:bdt_reweight_1d}. We also verify that the behavior of this BDT reweighting for quasi-elastic events as a function of neutrino energy $E_\nu$ and momentum transfer $Q^2$ closely matches the behavior in the DUNE study, as shown in Fig. \ref{fig:bdt_reweight_2d}.

So, the fact that we have recovered identical distributions for all relevant reconstructed variables means that this mis-modeling would be impossible to detect in any model validation test. In this test, we measured 0 $\sigma$ disagreement in our model validation tests, and 0.7 $\sigma$ bias in our extracted energy transfer cross section. This conspiracy between mismodeling of the visible/invisible split of hadronic energy and mismodeling of the cross section completely cancel each other out, and leave us in a worst-case scenario, where the extraction of biased cross section results is possible. This is an inherent limitation of the data-driven model validation procedure; it is not able to identify every possible scenario of mis-modeling. 

\begin{figure}[H]
    \centering
    \begin{subfigure}[b]{0.50\textwidth}
        \includegraphics[width=\textwidth]{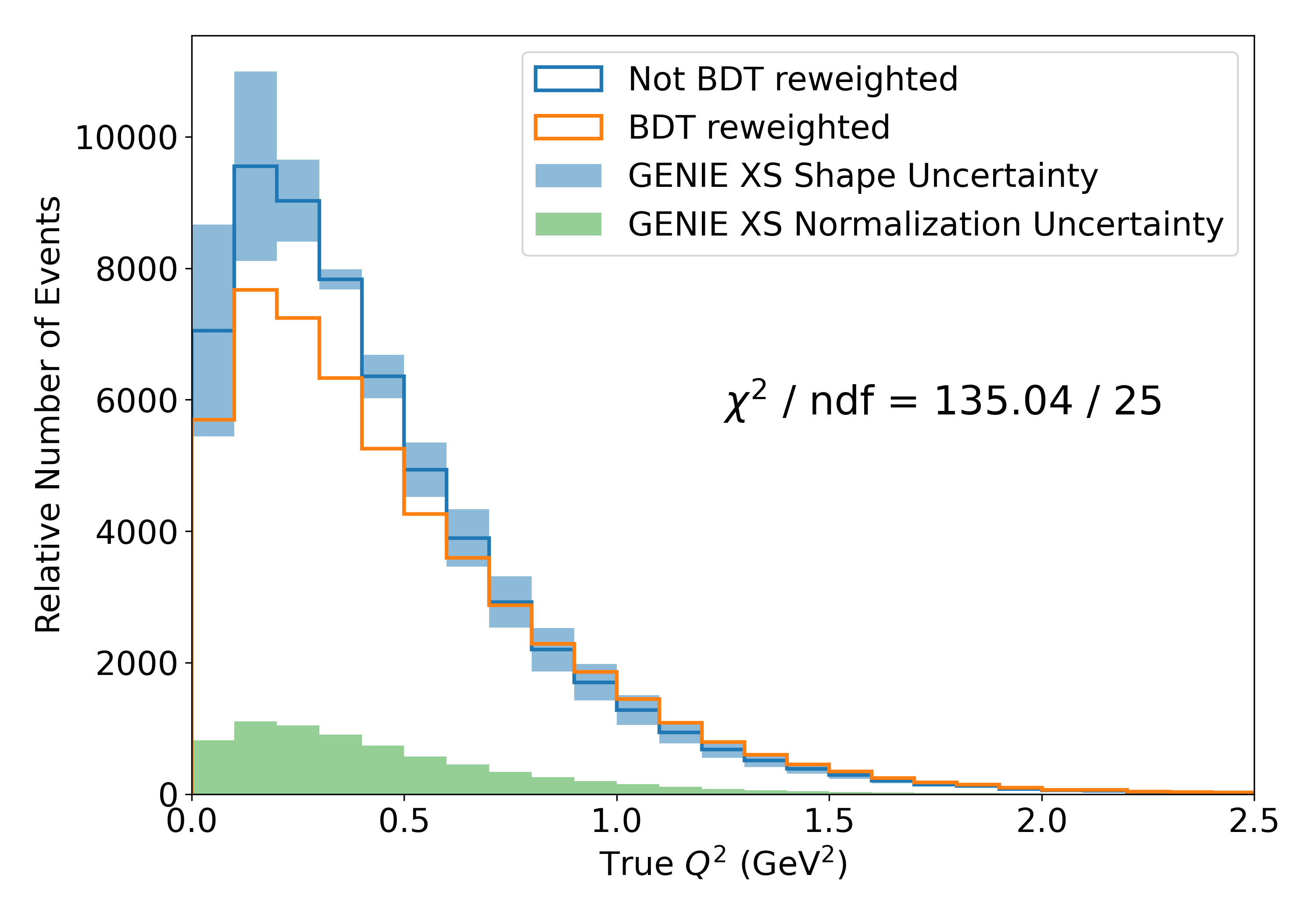}
        \caption{}
        \label{fig:bdt_reweighting_qe_q2}
    \end{subfigure}
    \begin{subfigure}[b]{0.48\textwidth}
        \includegraphics[width=\textwidth]{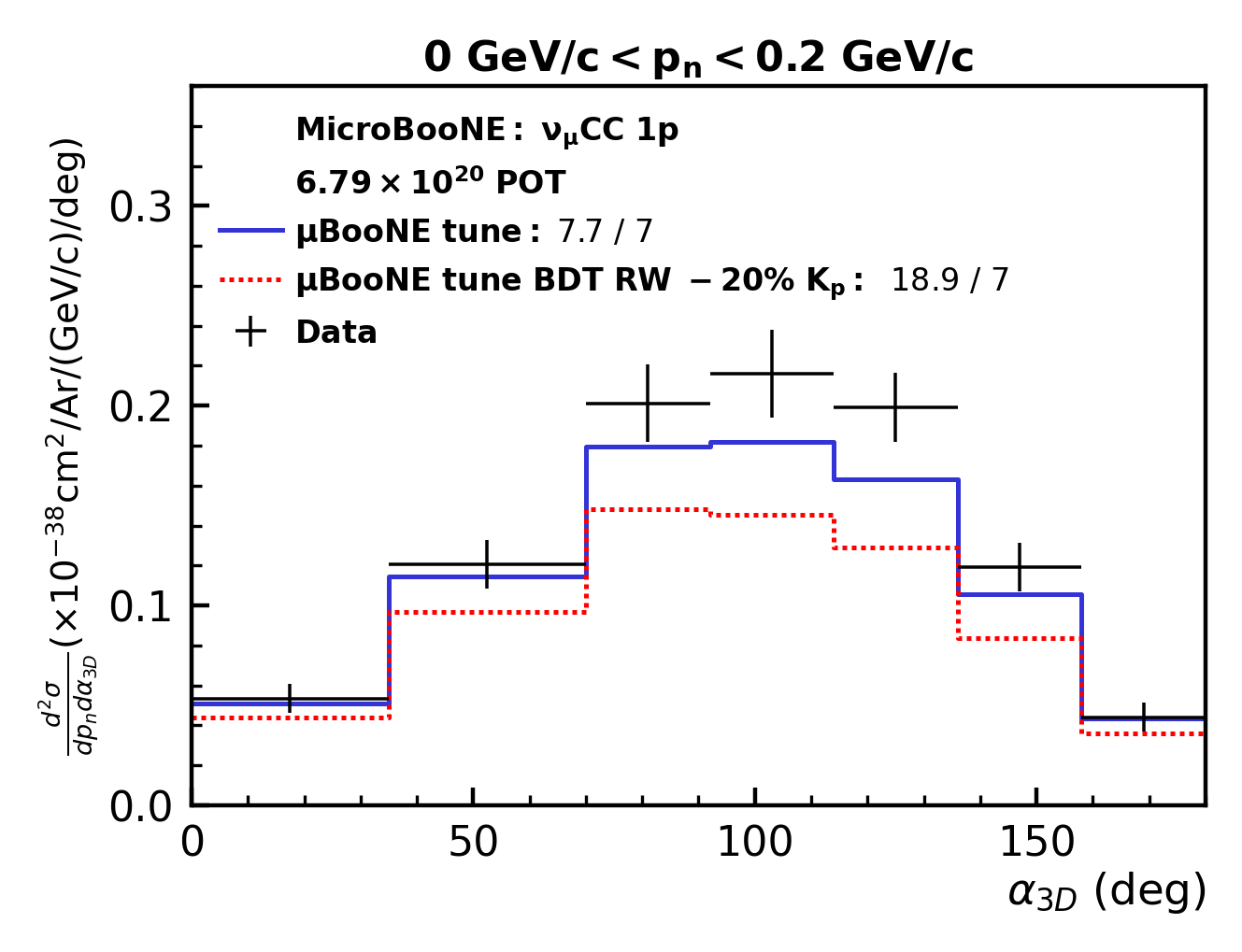}
        \caption{}
        \label{fig:bdt_reweighting_kinematic_imbalance}
    \end{subfigure}
    \caption[BDT reweighting comparisons]{BDT reweighting comparisons. Panel (a) shows a comparison of the BDT reweighted model to our GENIE model, with the GENIE uncertainties split into shape and normalization components, showing a large disagreement far outside of uncertainties. Panel (b) shows a comparison of the proton shifted BDT reweighted model to our nominal model and to our data result for a quasi-elastic-pure slice of our generalized kinematic imbalance $\nu_\mu$CC $1p$ cross section \cite{uboone_gki}. From Ref. \cite{uboone_model_validation}.}
    \label{fig:bdt_reweighting_comparisons}
\end{figure}

Now we ask, how likely are we to end up in this situation, where multiple mis-modelings cancel each other out to give good agreement to data? This is very hard to quantify, but for illustration, we can analyze this proton-shifted BDT-reweighted fake data from a few other perspectives. In Fig. \ref{fig:bdt_reweighting_qe_q2}, we compare the BDT reweighted model with the GENIE $Q^2$ distribution specifically for quasi-elastic events. This is a part of our cross section model with better theoretically understood uncertainties, avoiding dependence on nuclear final state interactions, and supported by many neutrino and electron scattering experiments. We see a very significant discrepancy, indicating that this specific cross section change would likely be in tension with a lot of prior experimental results. As another example, in Fig. \ref{fig:bdt_reweighting_kinematic_imbalance}, we show this proton energy shifted and BDT reweighted cross section prediction for $\nu_\mu$CC $1p$ events in a slice of phase space with small kinematic imbalance, which is able to isolate quasi-elastic events with very high purity \cite{uboone_gki}. In this measurement, we can see that this prediction moves further away from our data points, again indicating that this scenario with these two cross section mis-modelings canceling each other out is unlikely to be consistent with the broader set of exclusive neutrino measurements.

This represents one example of a case where different effects could conspire to make data-driven model validation fail. In this specific case, there are signs that such a scenario would be inconsistent with other types of measurements in MicroBooNE and other experiments. We cannot prove this in general for all possible model validation failures, but this example serves as an illustration of the fact that there are many constraints on our models, so it is not easy for them to be badly wrong and cancel out in such specific ways.

\subsection{NuWro Fake Data Model Validation}

\begin{figure}[H]
    \centering
    \begin{subfigure}[b]{0.49\textwidth}
        \includegraphics[width=\textwidth]{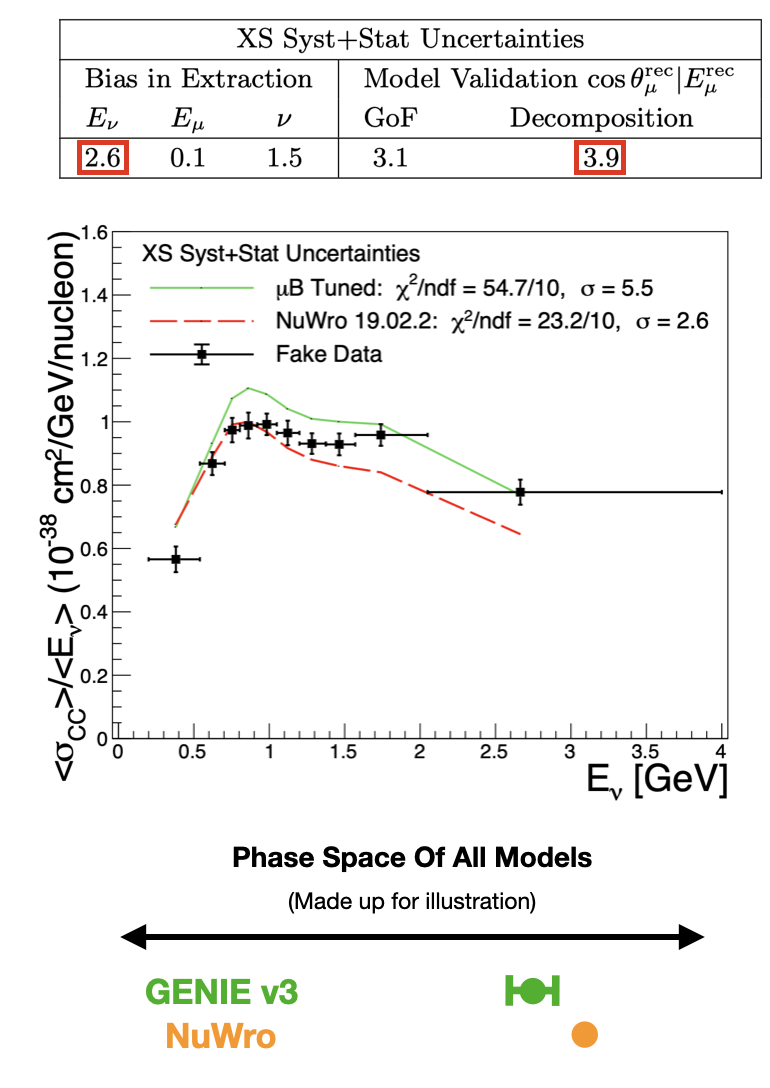}
        \caption{}
        \label{fig:nuwro_fake_data_set_just_xs}
    \end{subfigure}
    \begin{subfigure}[b]{0.49\textwidth}
        \includegraphics[width=\textwidth]{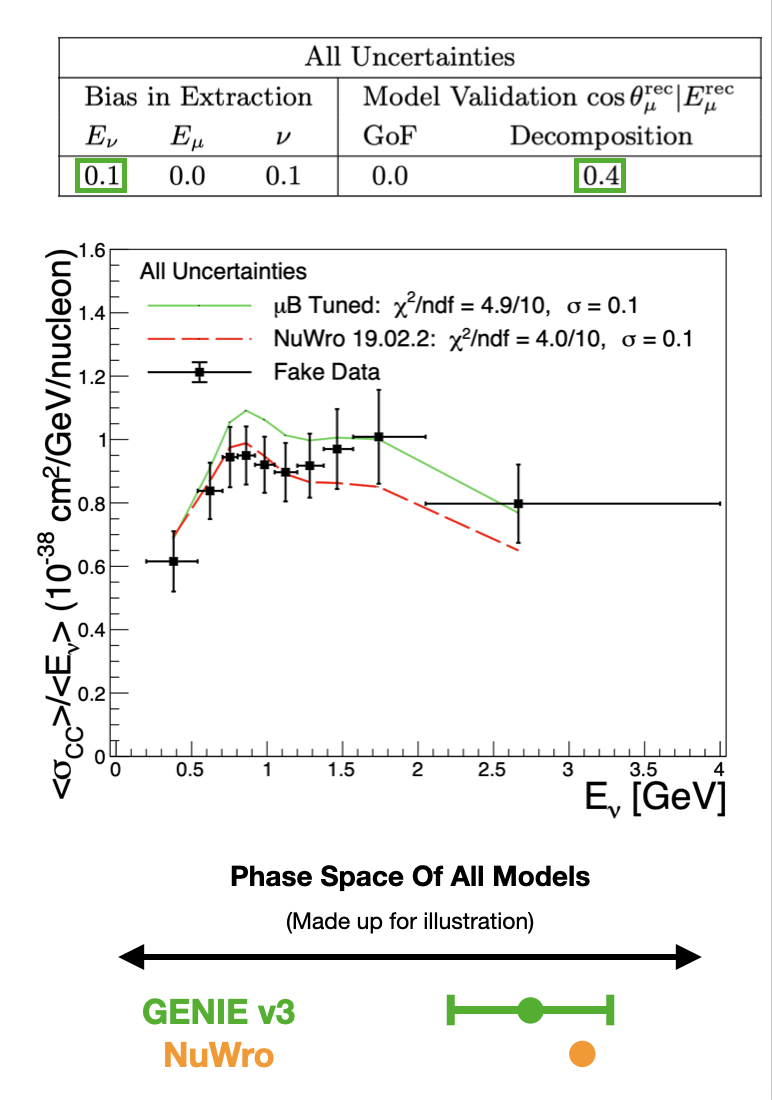}
        \caption{}
        \label{fig:nuwro_fake_data_set_full}
    \end{subfigure}
    \caption[NuWro fake data tests with cross section only and full uncertainties]{
    NuWro fake data tests, measuring the disagreement in model validation tests as well as the bias in extracted cross sections. Panel (a) shows the fake data test with only cross section and statistical uncertainties, and panel (b) shows the fake data test with all flux, detector response, cross section, and statistical uncertainties. From Ref. \cite{uboone_model_validation}.}
    \label{fig:nuwro_fake_data_set}
\end{figure}

We can also use alternate cross section models in order to test our model validation procedure. This is not exactly the same as a fake data closure test described earlier; in this case, we are not using alternative cross section models to test our cross section model used for unfolding, but are instead using them to test the validity of our model validation procedures. We demonstrate using the NuWro event generator \cite{nuwro}. There are two ways we can perform this fake data test. Figure \ref{fig:nuwro_fake_data_set_just_xs} shows a fake data test with just cross section and statistical uncertainties. In this case, we fail our model validation test, and also would extract a biased cross section. This indicates that the NuWro modeling of the cross section is outside of our GENIE modeled cross section uncertainties in this phase space. Figure \ref{fig:nuwro_fake_data_set_full} shows a fake data test with the full set of cross section, statistical, flux, and detector resonse uncertainties used for real data results. In this case, we pass our model validation test, and extract an unbiased cross section. This indicates that this fake data set is well described by our prediction with our full set of uncertainties. Most importantly, in both cases, the model validation disagreement basically matches or exceeds any bias in the extracted cross section, so in either case we would be able to identify mis-modeling before reporting any incorrect cross section results to the community.

\section{Nuclear Model Dependent CCQE Axial Form Factor Study}

In this section, I briefly discuss an in-progress analysis of nuclear-model-dependent CC quasi-elastic (CCQE) axial form factors. I will describe the concept and show some features of the reconstructed inputs. I will not show results or sensitivities, since at this stage we are still actively studying fake data tests.

In a CCQE interaction, we approximate a neutrino interacting with a single nucleon inside the nucleus. This process is described by a set of form factors which describe the structure of the nucleon. Elastic and magnetic form factors have been precisely measured in electron-nucleon scattering experiments \cite{electron_scattering_form_factors}, but the axial form factor which describes the distribution of weak charge inside the nucleon is more unique to neutrino interactions, and thus much harder to measure. One simple way to parameterize the axial form factor ($F_A$) as a function of the negative four-momentum transfer to the nucleus squared ($Q^2$) is via the axial mass $M_A$:
\begin{equation}
    F_A(Q^2) = g_A \cdot \frac{1}{\left(1 + Q^2/M_A^2\right)^2},
\end{equation}
with $g_A\approx 1.27$ \cite{ParticleDataGroup}. This is a simple parametrization, and some models instead use a $z$-expansion method with more parameters.

Many experiments and calculations have aimed to measure this $M_A$ parameter. Based on the partially conserved axial vector current (PCAC) hypothesis, $M_A = 1.077\pm 0.039$ GeV$/c^2$ was extracted from measurements of charged pion electro-production \cite{electroproduction_M_A}. Adding neutrino data from deuterium bubble chambers \cite{deuterium_M_A} to the charged pion data, the extracted value becomes $M_A = 1.014 \pm 0.014$ GeV$/c^2$ \cite{deuterium_and_electroproduction_M_A}. More recently, there have been a variety of measurements using neutrino scattering on more nuclear targets. NOMAD saw a value fairly consistent with those described above \cite{NOMAD_M_A}, but several other experiments have preferred larger values, as shown in Table \ref{tab:experimental_MAs}. These discrepancies in extracted $M_A$ values have triggered many theoretical studies, as detailed in reviews such as Refs. \cite{M_A_review_1,M_A_review_2}, among others. The apparent neutrino energy dependence has been noted and studied in Refs. \cite{MArun, MArun_test}.

\renewcommand{\baselinestretch}{1.5}
\begin{table}[H]
    \centering
    \begin{tabular}{c c c}
        \toprule
        Experiment & Nuclear Target & Measured $M_A$  \\ 
        \midrule
        NOMAD & Carbon & $1.07\pm0.07$ \\
        \midrule
        K2K & Oxygen & $1.2\pm0.12$ \\
        K2K & Carbon & $1.14\pm 0.11$ \\
        \midrule
        T2K & Carbon & $1.26^{+0.21}_{-0.18}$ \\
        T2K shape-only & Carbon & $1.43^{+0.28}_{-0.22}$ \\
        \midrule
        MINOS & Iron & $1.23\pm0.18$ \\ 
        \midrule
        MiniBooNE & Carbon & $1.35\pm0.17$ \\
        \bottomrule
    \end{tabular}
    \caption[Prior $M_A$ measurements]{Summary of prior measurements of $M_A$. These values come from Refs. \cite{NOMAD_M_A, K2K_MA, K2K_MA_2, t2k_ccqe_2015, MINOS_M_A, miniboone_ccqe2d}.}
    \label{tab:experimental_MAs}
\end{table}

The leading explanation of this difference is that these measured large $M_A$ values should be interpreted as more complex nuclear effects. In particular, the MiniBooNE measurement has been interpreted as a two-particle-two-hole (2p2h) effect \cite{miniboone_2p2h}, which describes a neutrino interacting with a correlated pair of nucleons, knocking both out of the nucleus. However, there have been two recent pieces of evidence that are free from nuclear effects: Lattice quantum chromodynamics (QCD) calculations of axial form factor \cite{LQCD_axial_form_factor} and MINERvA's measurement of the axial form factor with antineutrino-hydrogen scattering \cite{minerva_axial_form_factor} each see a preference for higher form factors than would be indicated by previous measurements with deuterium experiments. These two recent developments motivated this analysis which seeks to examine what $M_A$ values are most consistent with MicroBooNE data observations in argon. We cannot hope to be free of nuclear effects, but we can use our most modern understanding of nuclear models, and we can make high resolution measurements of muon kinematics as well as hadronic energies in order to constrain this modeling as much as possible. 

Technically, we plan to do this measurement using a detailed four-dimensional phase space of our Wire-Cell inclusive $\nu_\mu$CC selection. This is the same $\nu_\mu$CC BDT selection which I trained as described in Sec. \ref{sec:bdt_selections}, and the same binning which was used for our Wire-Cell 3D inclusive $\nu_\mu$CC cross section results \cite{wc_numu_3D}. This data is binned in containment within the detector, reconstructed neutrino energy, reconstructed muon angle, and reconstructed muon momentum, resulting in 1,152 total bins.

\begin{figure}[H]
    \centering
    \begin{subfigure}[b]{0.75\textwidth}
        \includegraphics[trim=0 5 0 50, clip, width=\textwidth]{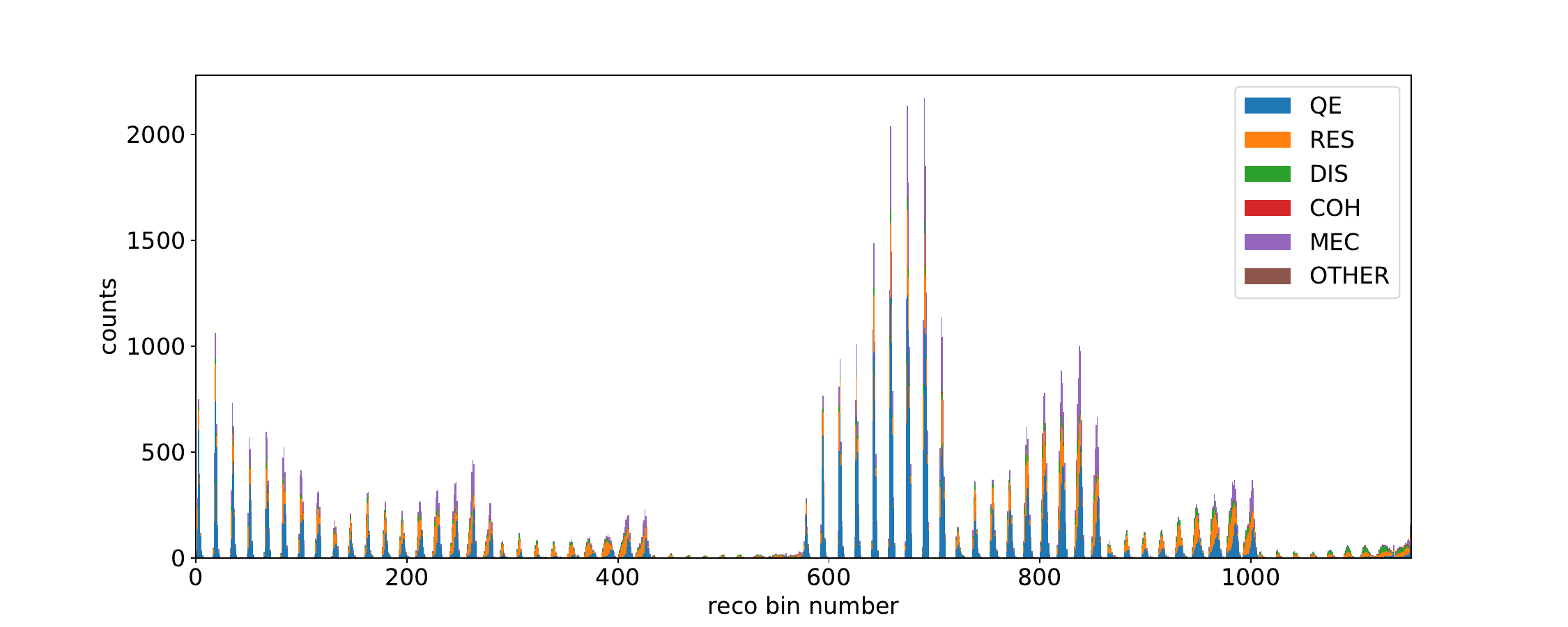}
        \caption{}
    \end{subfigure}
    \begin{subfigure}[b]{0.75\textwidth}
        \includegraphics[trim=0 5 0 50, clip, width=\textwidth]{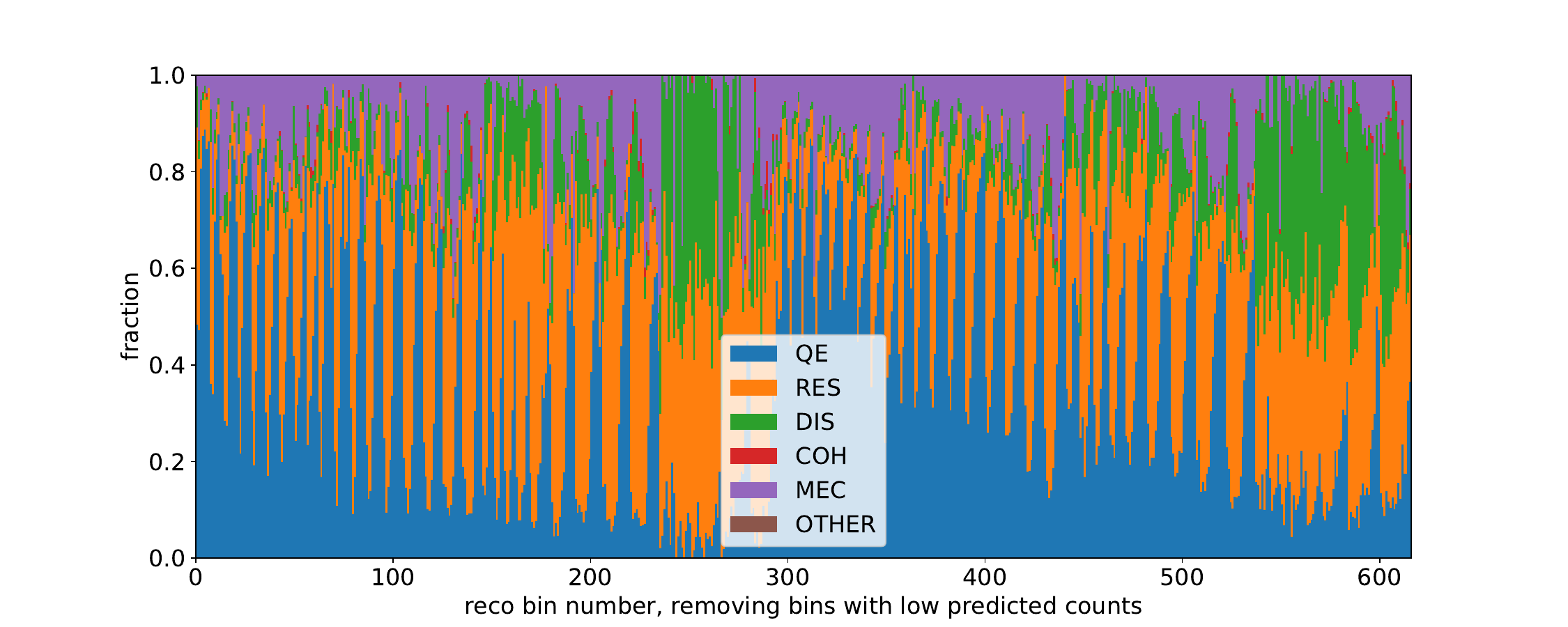}
        \caption{}
    \end{subfigure}
    \begin{subfigure}[b]{0.75\textwidth}
        \includegraphics[trim=0 5 0 50, clip, width=\textwidth]{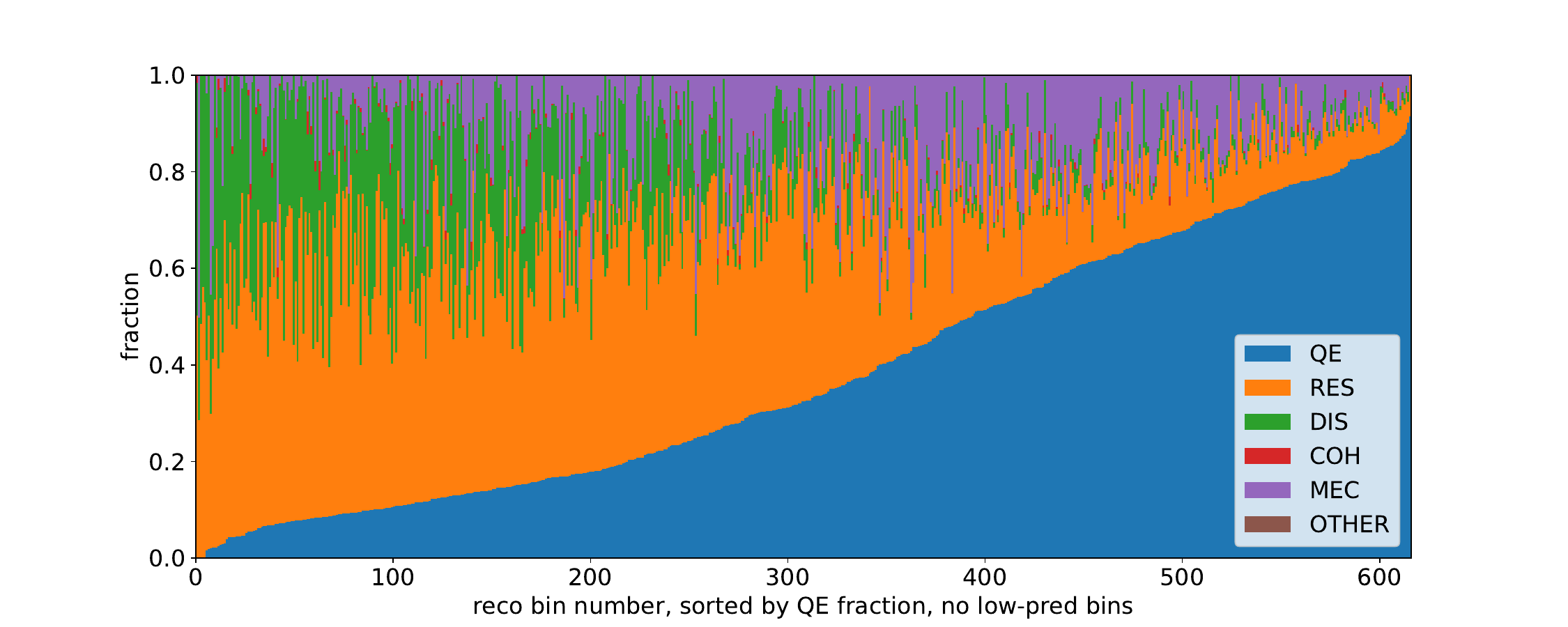}
        \caption{}
    \end{subfigure}
    \caption[CCQE and non-CCQE components of 4D $\nu_\mu$CC bins]{Panel (a) shows simulated event counts in each of the 1152 reconstructed bins, broken down by event type quasi-elastic (QE), resonant (RES), deep inelastic scattering (DIS), coherent (COH), and other. Panel (b) removes events with fewer than ten predicted events to lower statistical uncertainties, and shows the fractional contribution to each bin. Panel (c) sorts these bins by QE fraction, illustrating the existence of many bins which are fairly pure in QE events.}
    \label{fig:ccqe_nonccqe_components}
\end{figure}

This 4D reconstructed phase space is very detailed. In Fig. \ref{fig:ccqe_nonccqe_components}, we illustrate how different bins in this space have different purities of QE events. This is why we can use an inclusive analysis to focus on QE events while simultaneously measuring and constraining backgrounds from non-QE events. We also choose an inclusive selection because the modeling of specific hadronic states is more difficult and has the potential to introduce additional model dependence; for example, we have seen deficiencies in GENIE modeling of final state proton energies as described in Refs. \cite{uboone_Np0p_PRD, uboone_Np0p_PRL}.

\begin{figure}[H]
    \centering
    \includegraphics[width=0.7\textwidth]{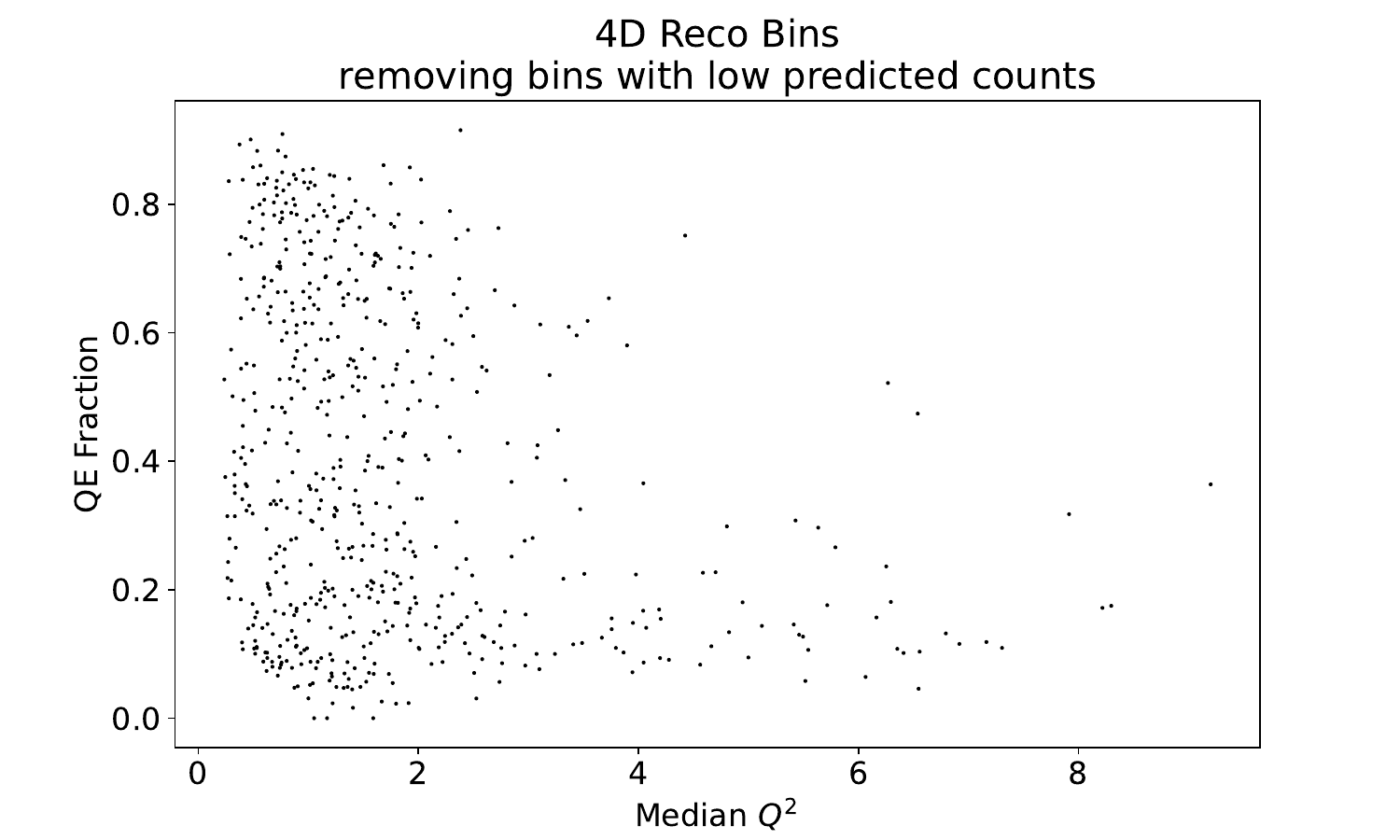}
    \caption[Median $Q^2$ and QE fraction for 4D $\nu_\mu$CC bins]{Median $Q^2$ vs QE fraction for 4D $\nu_\mu$CC bins with more than ten predicted events.}
    \label{fig:q2_vs_qe_frac}
\end{figure}

The $M_A$ value affects both the normalization of QE events as well as their shape as a function of $Q^2$. In Fig. \ref{fig:q2_vs_qe_frac}, we illustrate the fact that while our bins separate out QE fractions, they also have distinct $Q^2$ values, which allows us to use shape information in our $M_A$ extraction as well.

In order to extract $M_A$ from these 1,152 bins in reconstructed space, we expand the systematic covariance matrix with an extra row and column, built from fluctuations in the GENIE model's $M_A$ value in different systematic universes. The initial uncertainty in our MicroBooNE tune GENIEv3 model \cite{genie-tune-paper} for $M_A$ is $1.1\pm0.1$ GeV/$c^2$. This uncertainty has correlations with the prediction for these 1,152 bins, and we exploit that correlation via a conditional constraint acting on just this extra $M_A$ bin, as described in Sec. \ref{sec:conditional_constraint}. This effectively uses our observation in this 4D phase space in order to update our prior $M_A$ expectation of $1.1\pm0.1$ GeV/$c^2$ to a posterior with uncertainty. The fact that this process relies on a prior will slightly bias the measurement towards $1.1$ GeV/$c^2$, but the effect of this prior can be removed by a mathematical process; since these are all Gaussian uncertainties and the prior and posterior distributions are known, it is possible to extract a Gaussian uncertainty on the update alone.

We are currently evaluating the extent to which our inevitable nuclear model dependence will vary between different neutrino event generators using different fake data studies.

%% file: chapters/08_sbnd_dune.tex
\chapter{SBND and DUNE}

In addition to my primary work on the MicroBooNE experiment, I have made some contributions to the related LArTPC experiments SBND and DUNE, which I briefly describe in this appendix.

\section{SBND Drift Velocity}

The Short Baseline Near Detector (SBND) is a LArTPC neutrino experiment being constructed at Fermilab, upstream of MicroBooNE. SBND, MicroBooNE, and ICARUS form the Short-Baseline Neutrino (SBN) program, which will use observations at several baselines to conclusively answer the question of short baseline sterile neutrino oscillations in the range that might explain the MiniBooNE LEE.

I participated in early efforts preparing for the commissioning of SBND at Fermilab. In particular, I developed machinery to do an early test of the drift velocity using anode/cathode piercing tracks, which can function as a first test of the high voltage system, directly measuring the strength of the electric field in the detector.

\begin{figure}[H]
    \centering
    \includegraphics[width=\textwidth]{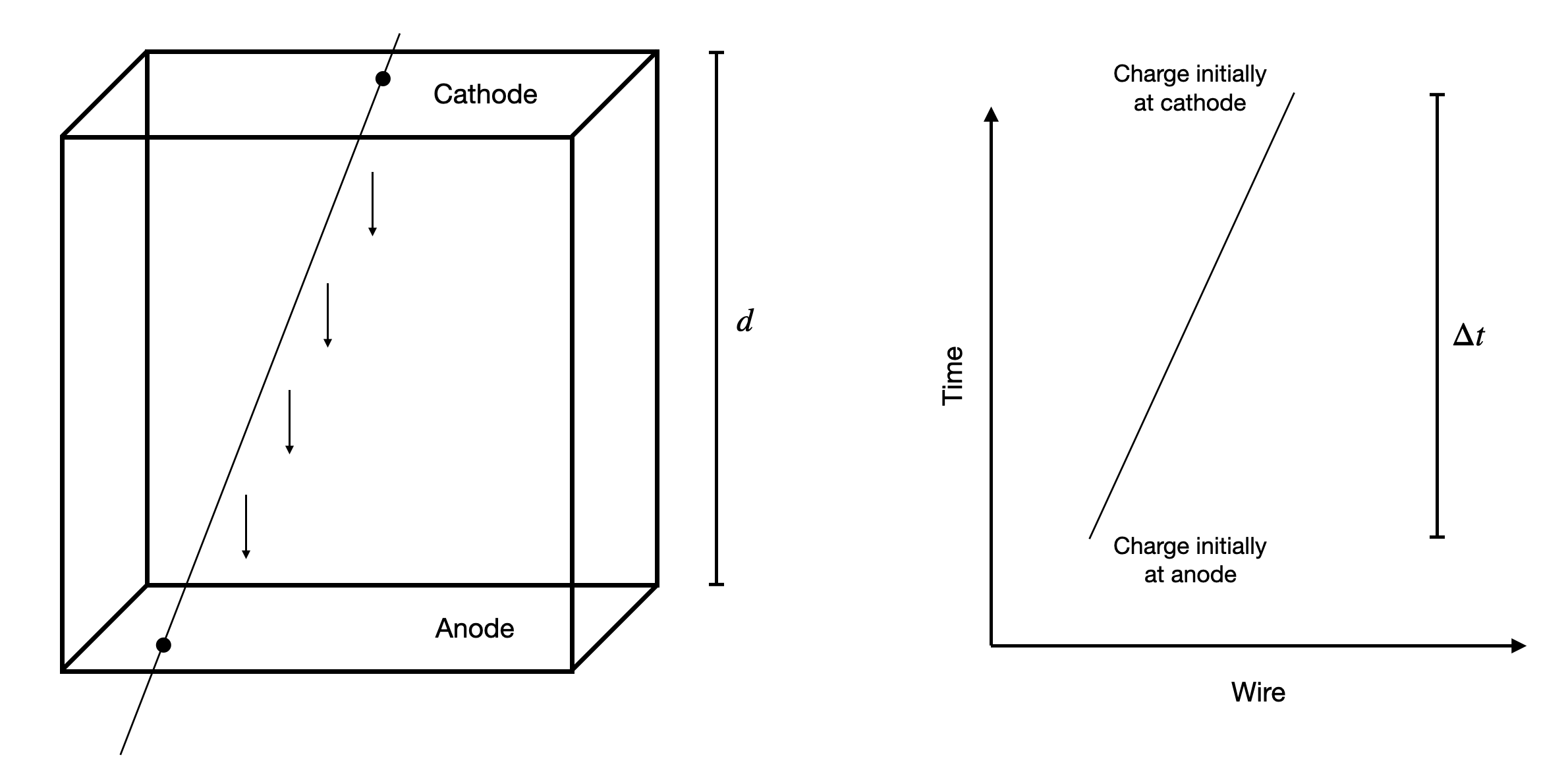}
    \caption{Diagram illustrating the SBND drift velocity measurement}
    \label{fig:drift_velocity_diagram}
\end{figure}

The drift velocity is extracted by measuring the beginning and end times of a cosmic muon track with known length. In this case, we plan to use anode/cathode crossing muons, so the known length is the length from the anode to the cathode. This process is illustrated in Fig. \ref{fig:drift_velocity_diagram}, and uses the following equation:

\begin{equation}
    v_\mathrm{drift}=\frac{d}{\Delta t}
\end{equation}

\begin{figure}[H]
    \centering
    \includegraphics[width=0.7\textwidth]{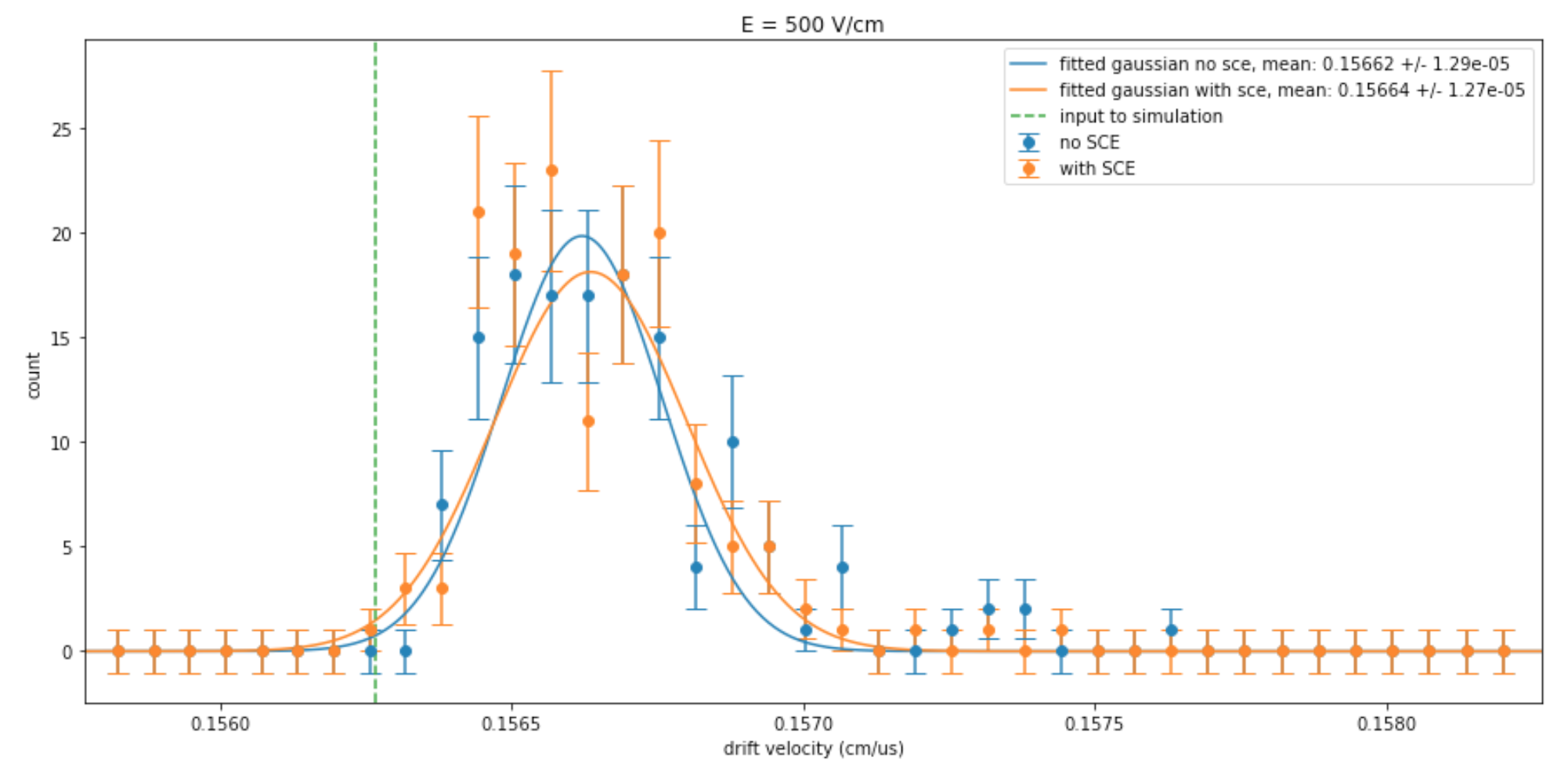}
    \caption[SBND simulated drift velocity measurement at 500 V/cm]{SBND simulated drift velocity measurement at 500 V/cm.}
    \label{fig:500_Vcm}
\end{figure}

I simulate anode-cathode-crossing muons in order to study this measurement using only reconstructed information. We simulate events both with and without the presence the space charge effect (SCE), which results in a position dependent modification to the electric field caused by significant amounts of ionization leading to positive ions building up in the detector. Initially, we saw a significant difference between measured and predicted values, and this was eventually tracked down as a geometric feature of the simulation, where ionization of argon atoms was not being simulated near the cathode and anode planes, causing shorter tracks relative to the actual detector geometry. After correcting for that, we show the distribution of measured drift velocities for different events in our nominal 500 V/cm simulation in Fig. \ref{fig:500_Vcm}. We see that the results are consistent both with and without the SCE, and the measured drift velocities are around 0.2\% larger than the input in our simulation. This is likely a result of diffusion effects, and this small offset is much smaller than the expected uncertainty in the anode to cathode distance, which will most likely dominate this measurement for real data.

\begin{figure}[H]
    \centering
    \includegraphics[width=0.8\textwidth]{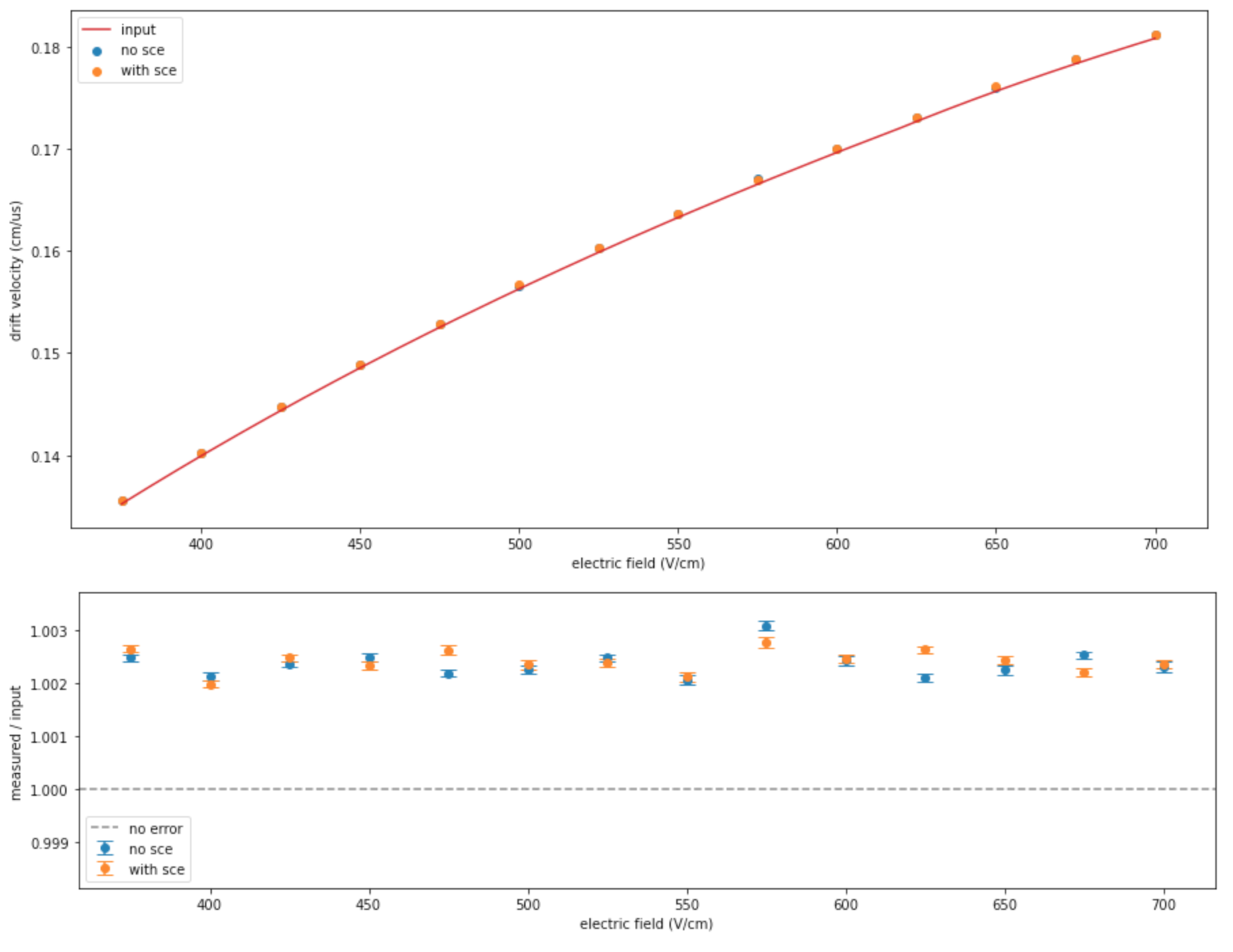}
    \caption[SBND simulated drift velocity measurement at different field strengths]{SBND simulated drift velocity measurement at different field strengths.}
    \label{fig:sbnd_drift_velocity_different_Efield}
\end{figure}

We repeat this process for a wide range of drift electric field strengths, as shown in Fig. \ref{fig:sbnd_drift_velocity_different_Efield}. We see that this 2\% offset is approximately constant both with and without the SCE across all electric fields simulated.

\begin{figure}[H]
    \centering
    \includegraphics[width=\textwidth]{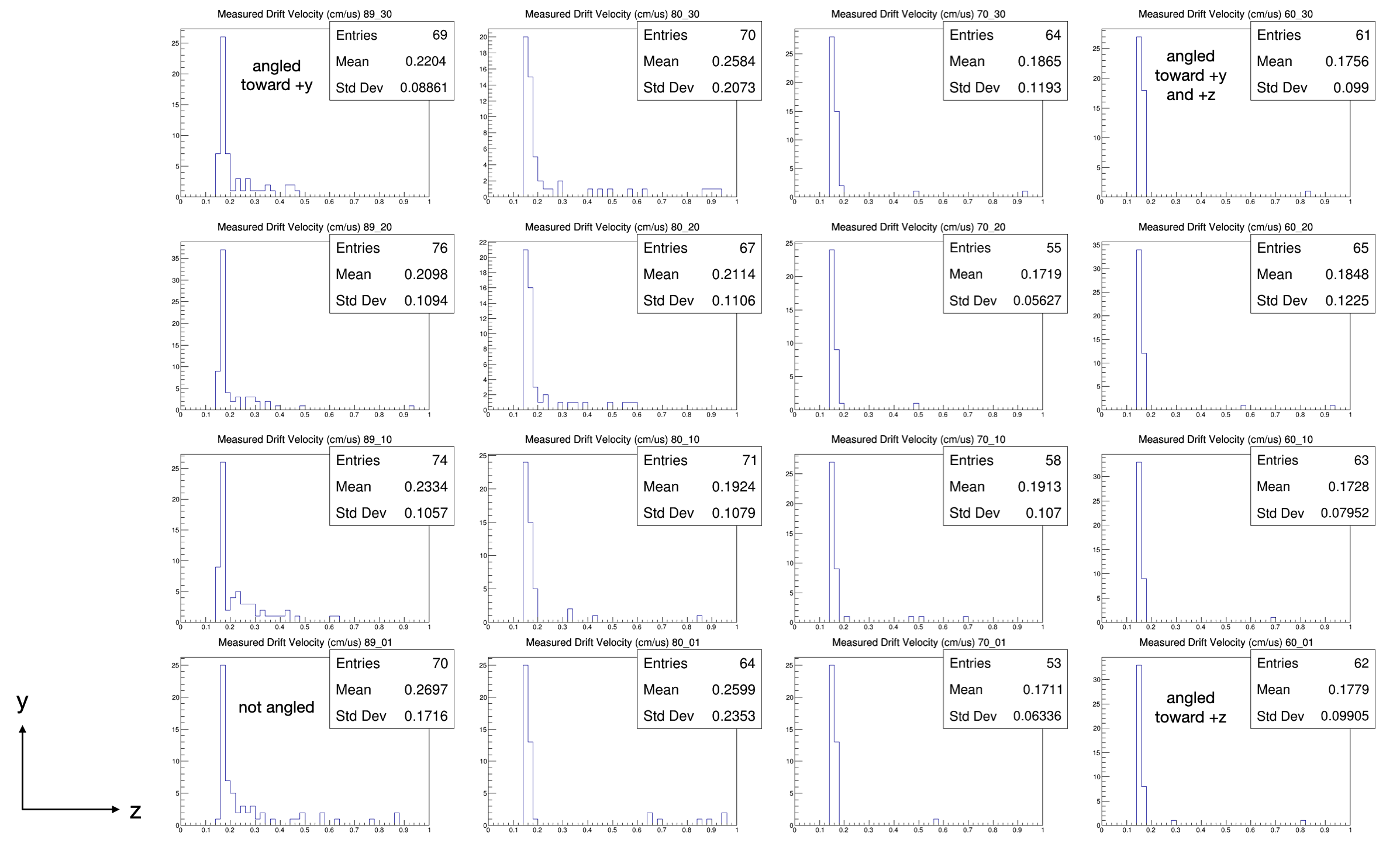}
    \caption[SBND simulated drift velocity measurement at different muon angles]{SBND simulated drift velocity measurement at different muon angles.}
    \label{fig:sbnd_drift_velocity_different_angles}
\end{figure}

\begin{figure}[H]
    \centering
    \includegraphics[width=\textwidth]{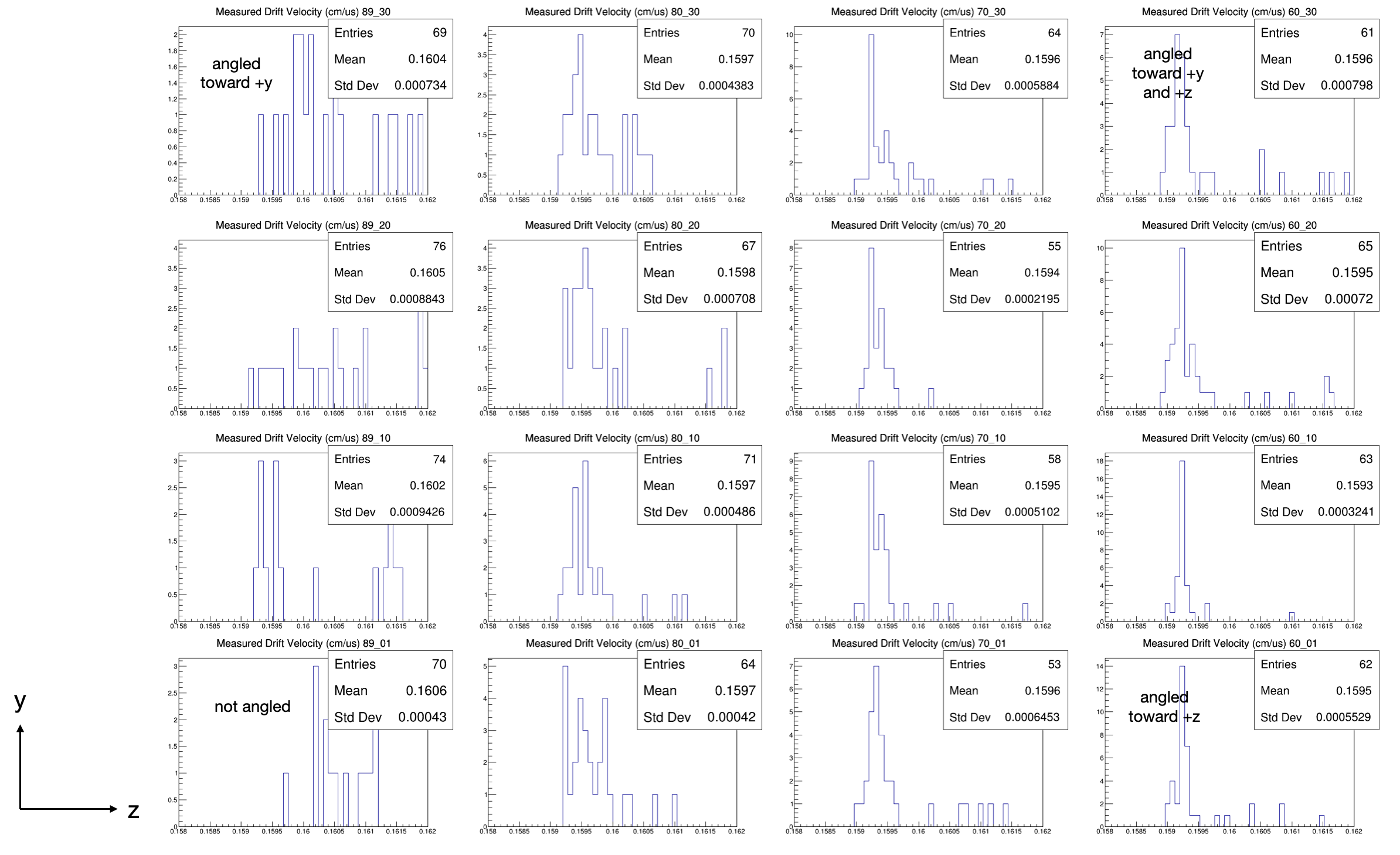}
    \caption[SBND simulated drift velocity measurement at different muon angles zoomed]{SBND simulated drift velocity measurement at different muon angles, zoomed in on each peak.}
    \label{fig:sbnd_drift_velocity_different_angles_zoomed}
\end{figure}

I also investigated how sensitive this measurement is to different muon angles within the detector. We consider only anode-cathode crossing muons, directed primarily along the x axis, but then simulate a variety of different angles toward the +y and +z axes and combinations of the two. The results are shown in Fig. \ref{fig:sbnd_drift_velocity_different_angles}. We see that when we consider angles with no z-extent, along the left edge of the figure, we have more variation in measured drift velocities. This is likely due to the fact that this study only uses the collection plane, and therefore tracks with a smaller extent along the z axis will cross a smaller number of wires; this results in prolonged pulses on each wire, which are more difficult to reconstruct cleanly. In Fig. \ref{fig:sbnd_drift_velocity_different_angles_zoomed}, we ignore these outlier points and zoom in on the peaks, seeing fairly consistent measurements for all of these muon angles.

\section{ArgonCube and DUNE LAr-ND}

The Deep Underground Neutrino Experiment (DUNE) is a long baseline neutrino experiment being developed at Fermilab and at Sanford Underground Research Facility (SURF). It is a massive experiment which will lead neutrino physics in the United States for the coming decades.

The DUNE near detector will include a LArTPC (LAr-ND) based on the ArgonCube design concept: highly segmented, with many optically isolated drift regions; high coverage photon detection; and pixelated charge readout.

\begin{figure}[H]
    \centering
    \includegraphics[trim=100 50 500 500, clip, width=0.6\textwidth]{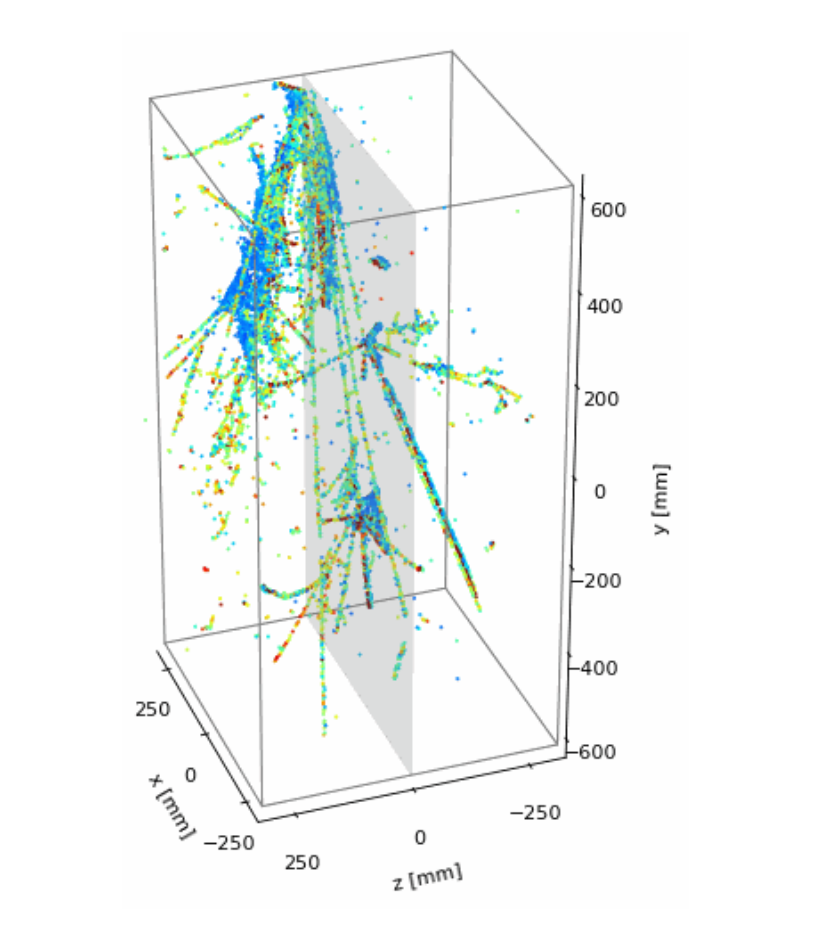}
    \caption[DUNE-ND Module-0 prototype event display]{DUNE-ND Module-0 prototype event display showing a multi-prong shower topology. From Ref. \cite{2x2_performance}.}
    \label{fig:module_0_event_display}
\end{figure}

I traveled to the University of Bern in Switzerland in March and April 2021 in order to help turn on and collect the first cosmic ray data from an early prototype of a DUNE LAr-ND module, referred to as ``Module-0'', consisting of two drift regions, a resistive field cage, silicon photomultipliers, and two pixellated charge readout planes. This resulted in high resolution 3D images of charge, as shown in Fig. \ref{fig:module_0_event_display}. The performance of the detector during this period was reported in Ref. \cite{2x2_performance}. In December 2021, I traveled to Fermilab to help prepare for the integration of this module with three others, forming the 2x2 DUNE near detector prototype, which has since been installed underground and taken NuMI beam neutrino data.

\begin{figure}[H]
    \centering
    \includegraphics[width=0.7\textwidth]{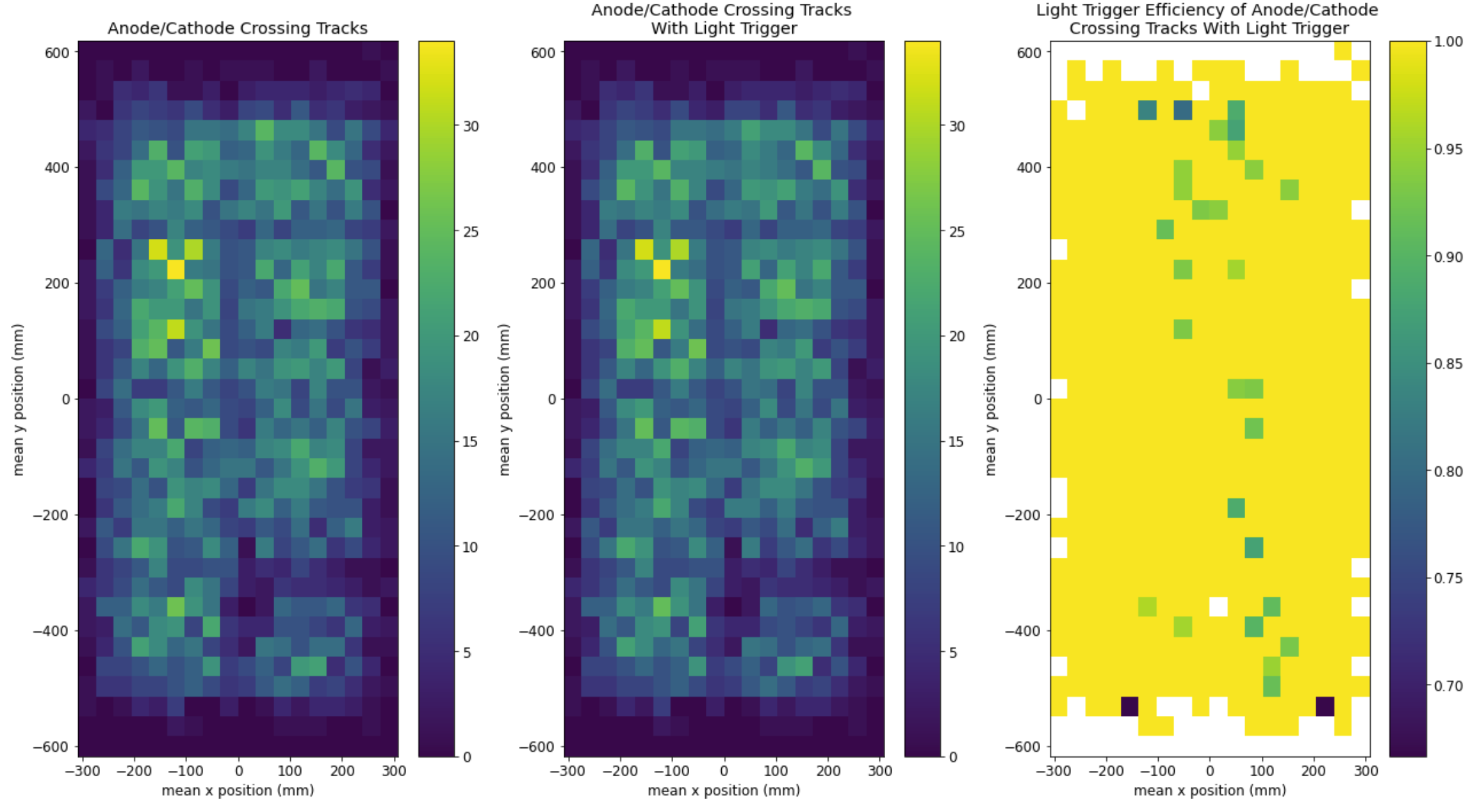}
    \caption[Module-0 light trigger spatial efficiency]{Left: Spatial position of anode-cathode crossing tracks in Module-0. Middle: Spatial position of anode-cathode crossing tracks with an associated light flash in Module-0. Right: Light trigger efficiency as a function of spatial position in Module-0 (obtained by dividing the left and middle histograms). Note that the axes here are orthogonal to the drift direction.}
    \label{fig:argoncube_anode_cathode_efficiency}
\end{figure}

\begin{figure}[H]
    \centering
    \includegraphics[width=0.7\textwidth]{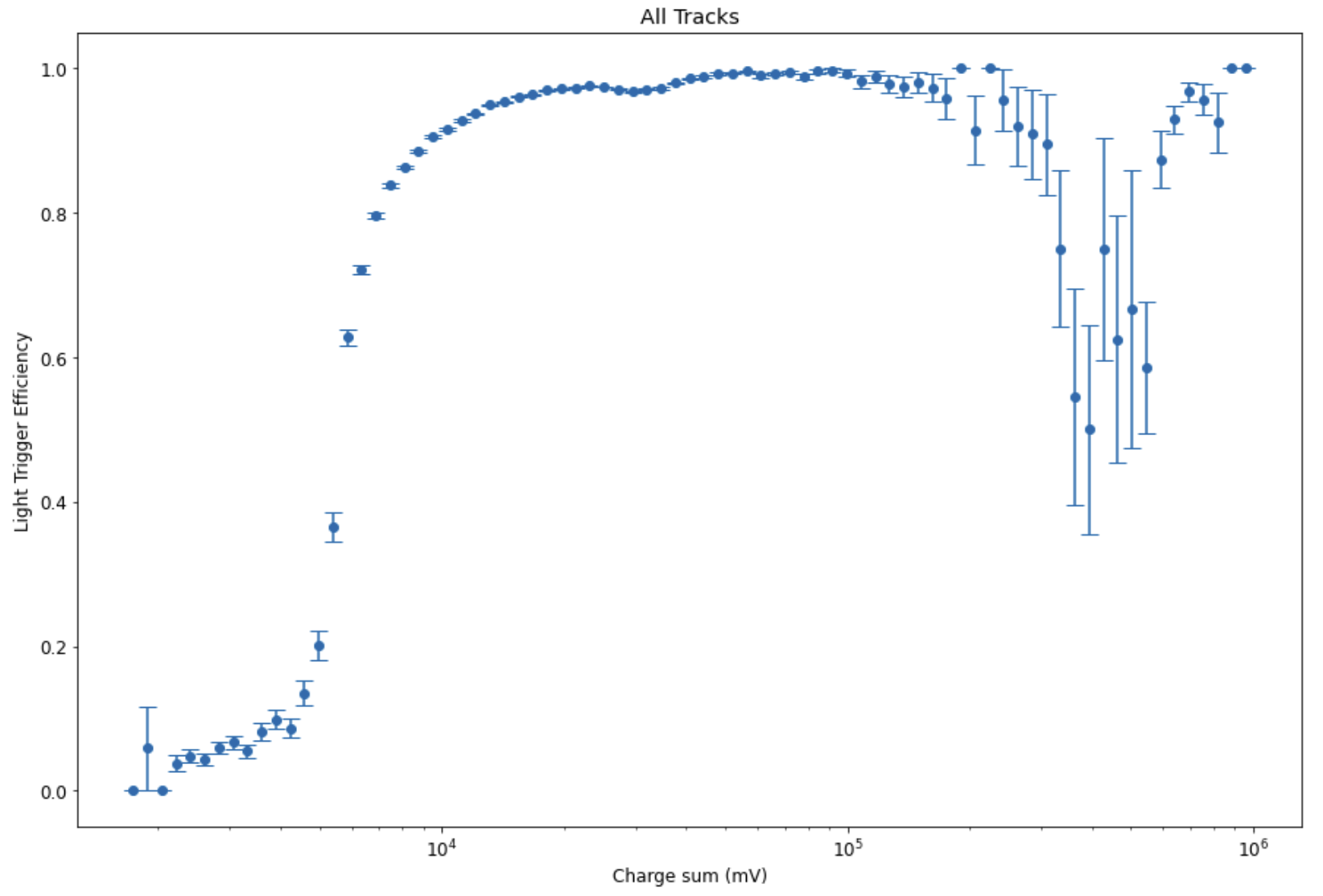}
    \caption[Module-0 light trigger charge efficiency]{Light trigger efficiency as a function of total charge for cosmic ray events in Module-0.}
    \label{fig:argoncube_charge_efficiency}
\end{figure}

I have also contributed to analysis of this early Module-0 data. Fig. \ref{fig:argoncube_anode_cathode_efficiency} shows an early study of the light system trigger efficiency as a function of spatial position for anode/cathode crossing cosmic muon tracks. Despite the fact that some pixel subsystems and some light detection subsystems are known to be inactive, the light trigger efficiency is fairly high and uniform across the detector, as expected given the large fractional coverage of the light detection systems. I also found that the light trigger efficiency rises sharply as a function of total charge, as expected due to the fact that larger tracks/showers will produce both more charge and more scintillation light, as shown in Fig. \ref{fig:argoncube_charge_efficiency}.

\section{DUNE Charge Readout Plane Construction And Testing}

I have also contributed to the construction of DUNE far detector charge readout planes (CRPs). These are a newer approach to charge readout, functioning like wire planes, but assembled like printed circuit boards, leading to a design which is much less fragile and easier to assemble. Currently, these CRPs are planned to be used for the first DUNE far detector module, with wire planes being used for the second module. The third and fourth module designs are still to be determined.

\begin{figure}[H]
    \centering
    \includegraphics[width=0.7\textwidth]{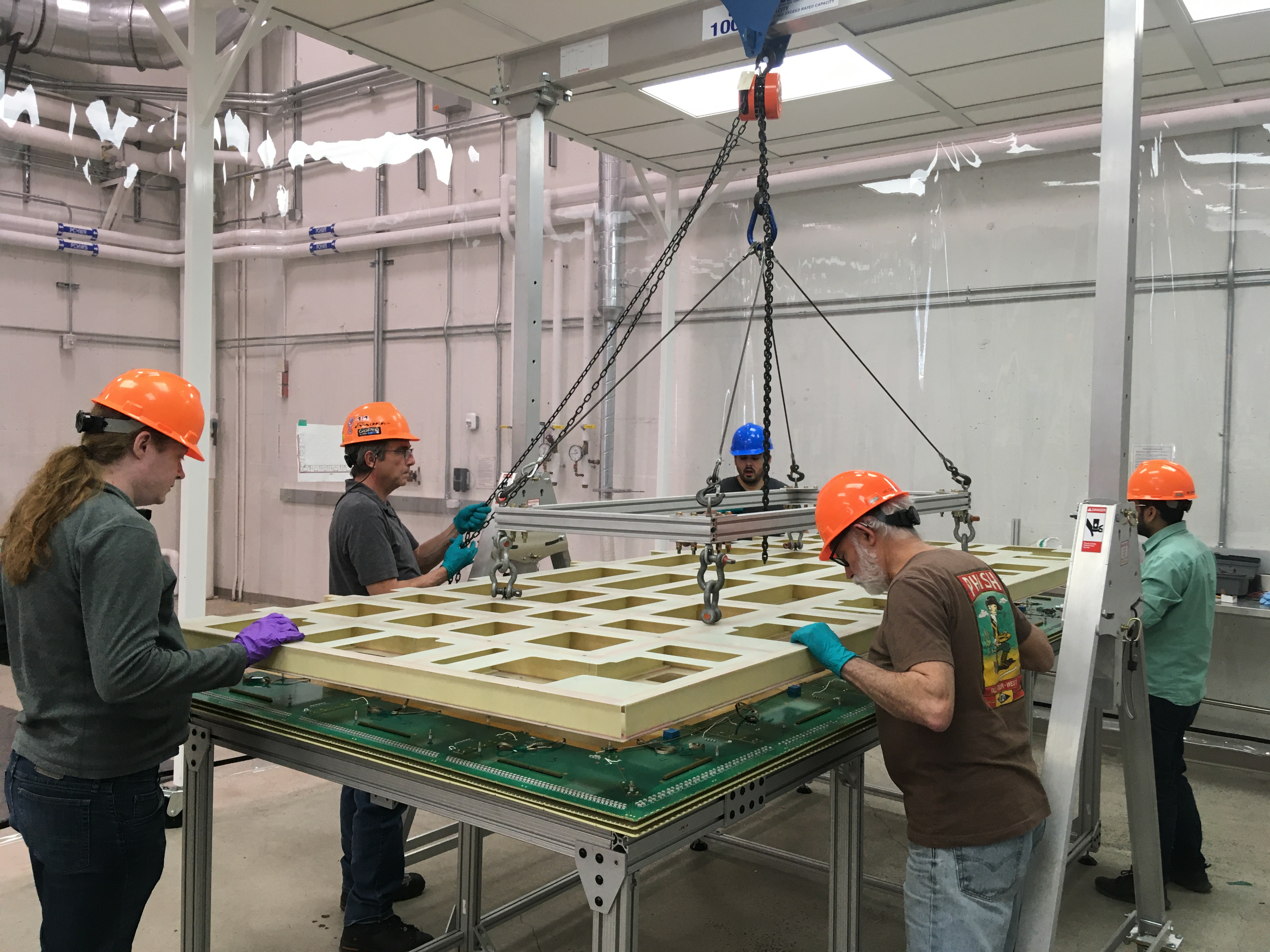}
    \includegraphics[width=0.7\textwidth]{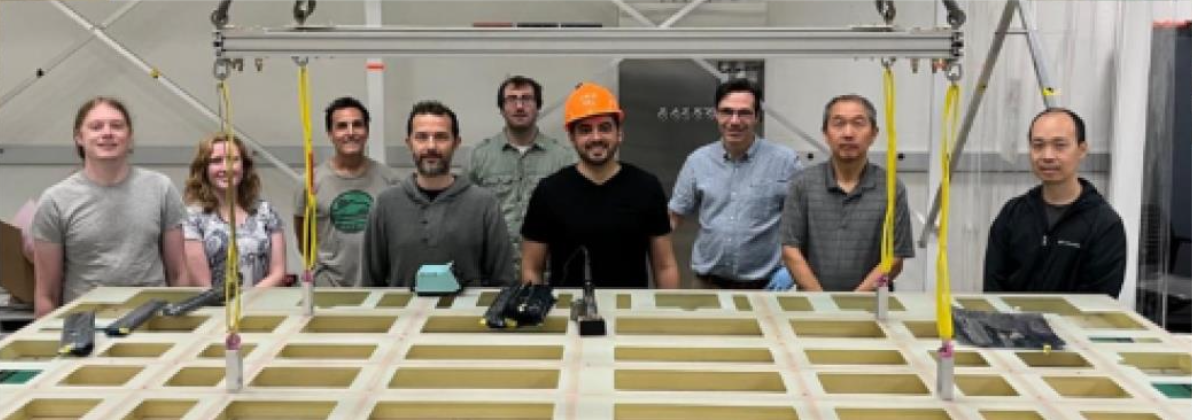}
    \caption[Charge Readout Plane construction]{Charge Readout Plane construction at Yale. More photons are available at \url{https://www.flickr.com/photos/yalewlab/albums/72177720306018664/}.}
    \label{fig:crp_construction}
\end{figure}

In May 2022, I traveled to CERN to help with CRP construction. These printed circuit board panels are constructed in smaller sections, before eventually being integrated into large 3x3 meter panels. I helped clean and prepare printed circuit board panels, and prepared for them to be glued and soldered together. In January 2023, I traveled from Brookhaven National Laboratory to Yale, where I helped assemble and test a CRP in liquid nitrogen at Yale, as shown in Fig. \ref{fig:crp_construction}.

\begin{figure}[H]
    \centering
    \begin{subfigure}[b]{0.49\textwidth}
        \centering
        \includegraphics[width=\textwidth]{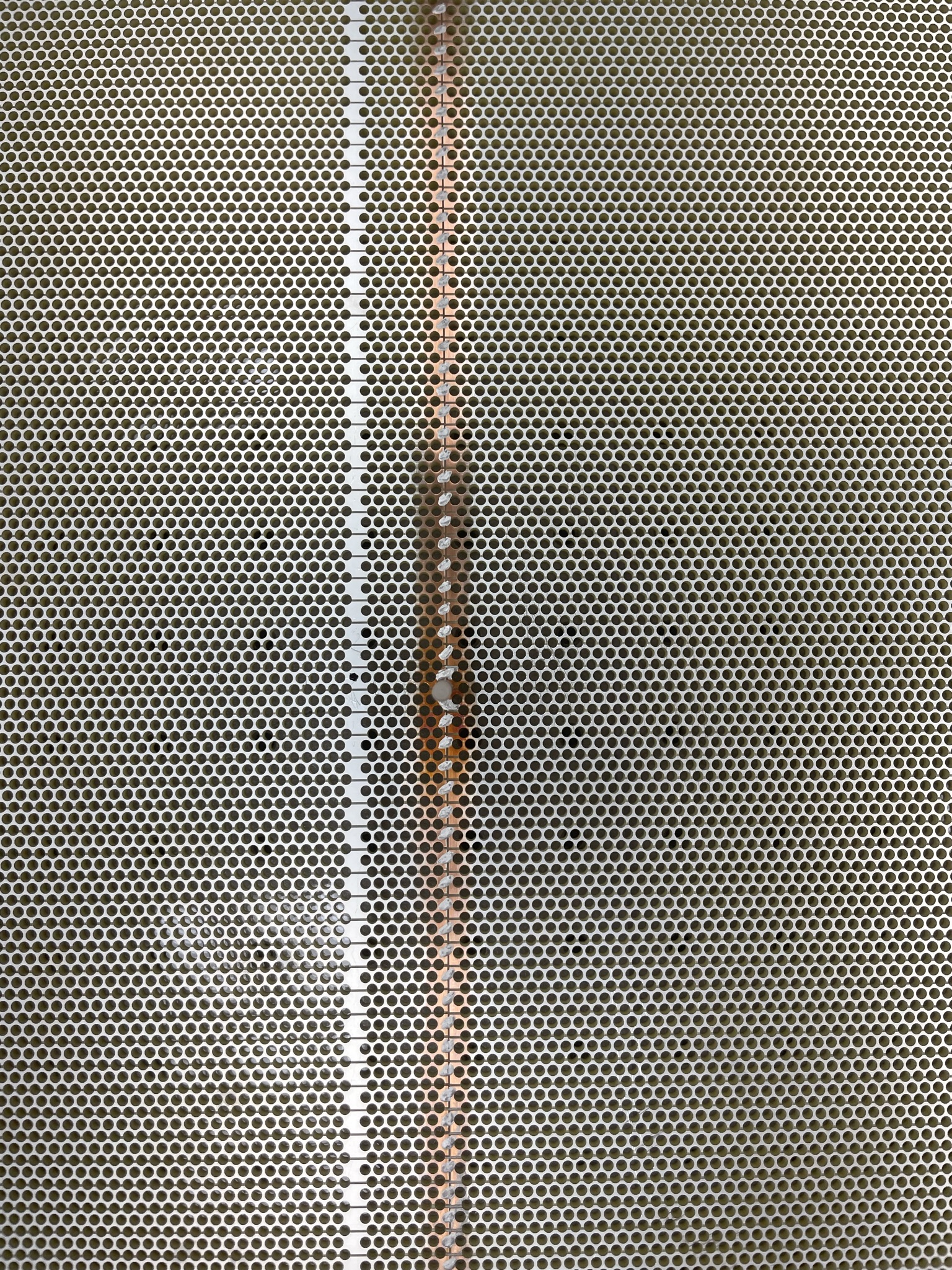}
        \caption{}
    \end{subfigure}
    \hfill
    \begin{subfigure}[b]{0.49\textwidth}
        \centering
        \includegraphics[width=\textwidth]{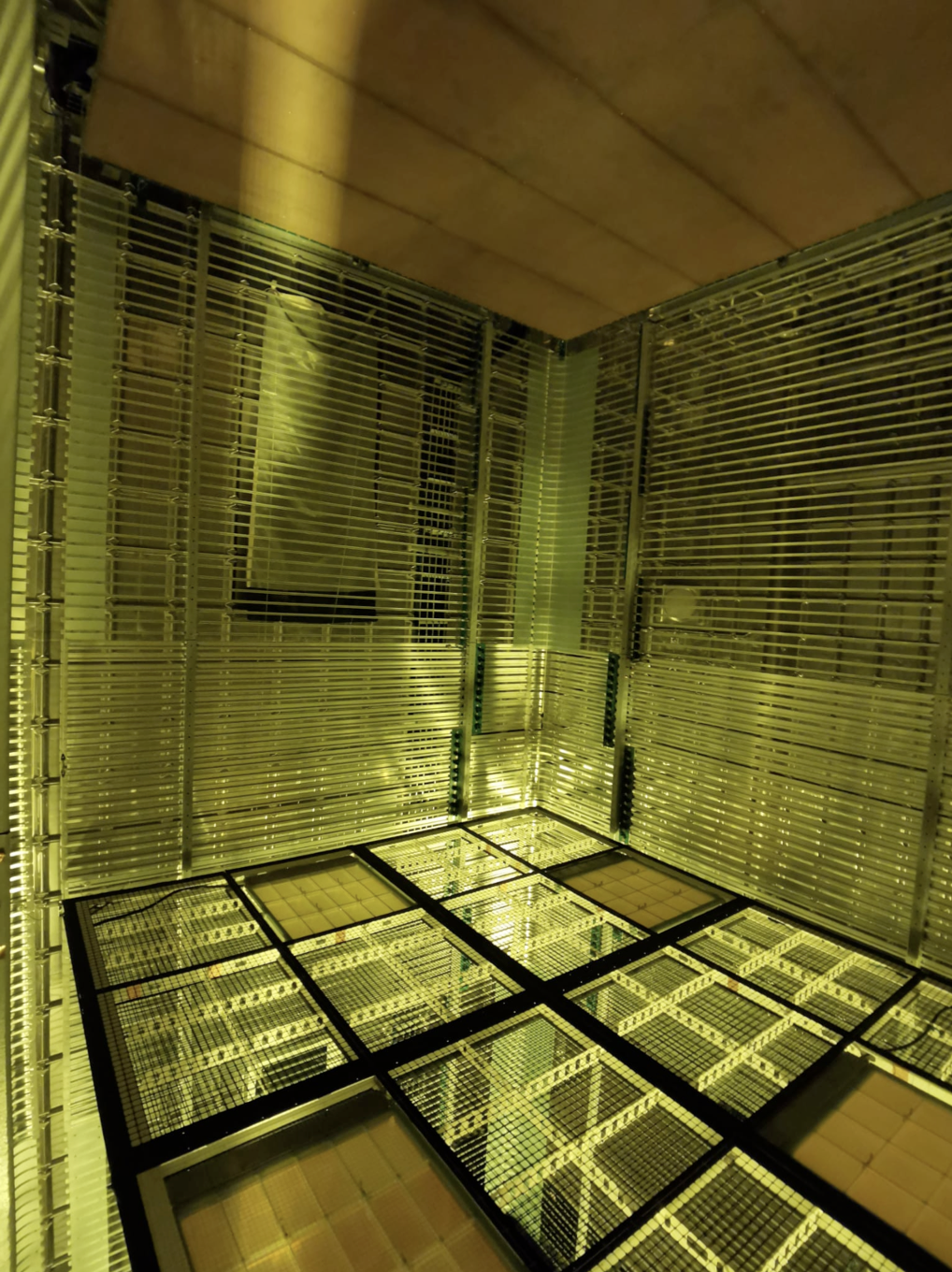}
        \caption{}
    \end{subfigure}
    \caption[Charge Readout Plane construction]{Panel (a) shows a close up view of a CRP panel, showing horizontal strips with solder joints connecting two sections of the panel. Panel (b) shows an installed CRP panel above a field cage and cathode at the bottom. Out of sight, there is another CRP panel installed below the cathode. From Ref. \cite{vertical_drift_design}.}
    \label{fig:crp_finished}
\end{figure}

A close-up view of a CRP as well as a view of a CRP installed in the ProtoDUNE vertical drift detector is shown in Fig. \ref{fig:crp_finished}.

\section{Local Lab Activities}

I have also contributed to hardware activities for the Fleming group local lab. This started at Wright Laboratory at Yale, and local lab activities are also planned University of Chicago in the near future.

\begin{figure}[H]
    \centering
    \begin{subfigure}[b]{0.49\textwidth}
        \centering
        \includegraphics[width=\textwidth]{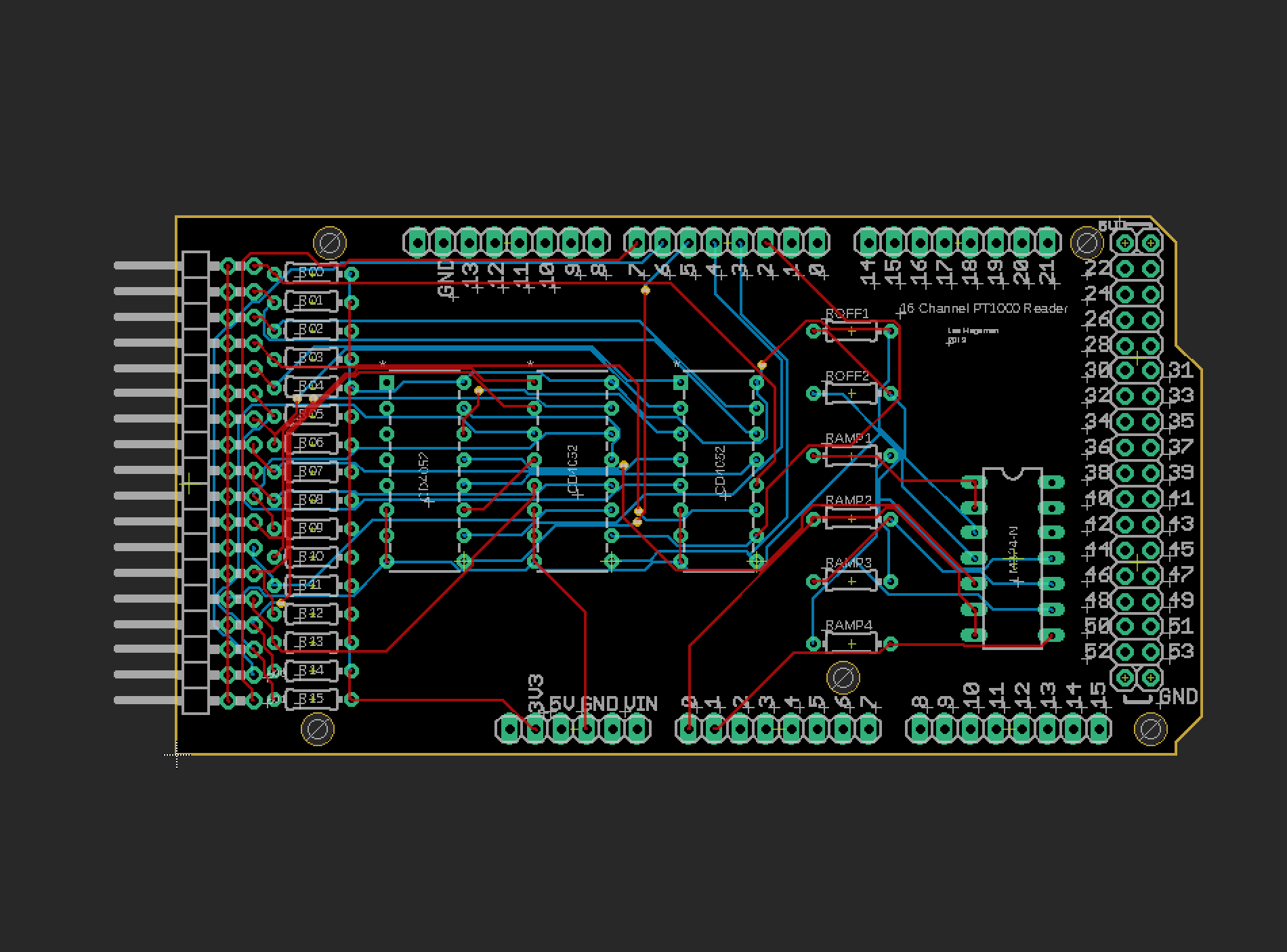}
        \caption{}
    \end{subfigure}
    \begin{subfigure}[b]{0.49\textwidth}
        \centering
        \includegraphics[width=\textwidth]{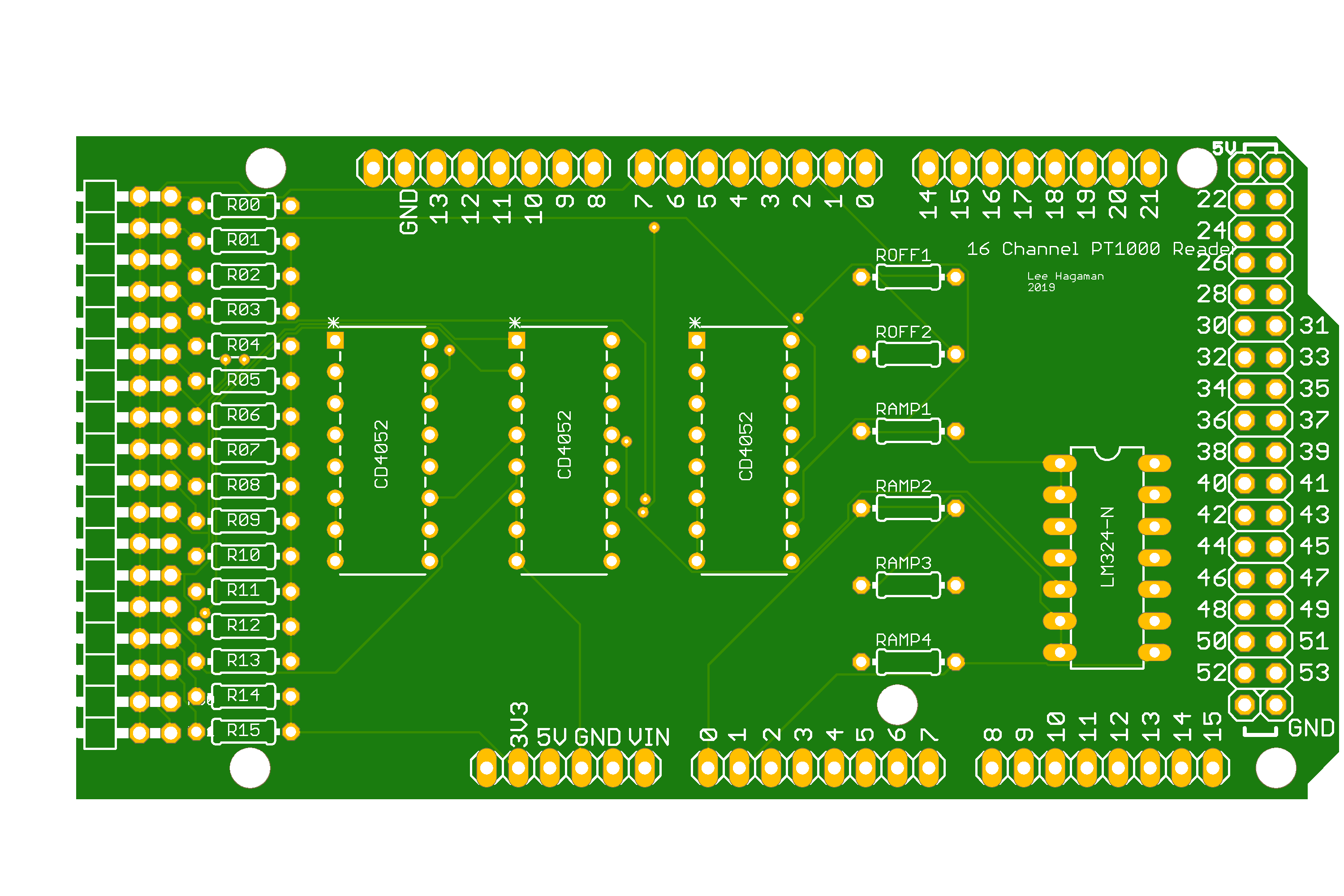}
        \caption{}
    \end{subfigure}
    \caption[RTD reader design]{16 Channel RTD reader. Panel (a) shows the design showing each two layer connections. Panel (b) shows a model of the PCB as it appears after printing.}
    \label{fig:rtd_reader}
\end{figure}

Any LArTPC requires careful monitoring of temperature, since there is a small $3.5^\circ$ C range where argon remains liquid at atmospheric pressure. Additionally, it is desirable to have an independent system which can control heaters in order to maintain these temperatures, even in the event of a computer failure. For this reason, I developed a 16 channel Resistive Temperature Device (RTD) reader as an Arduino shield, as shown in Fig.~\ref{fig:rtd_reader}. I make use of multiplexing and analog amplification in order to accurately and precisely measure temperatures from 16 separate PT-1000 resistive temperature sensors. In addition to communicating with a computer, the Arduino can also run its own simple logic in order to control a heater attached to the condenser of a pulse tube refrigerator. This printed circuit board which I designed, constructed, and tested is very cheap, and is potentially able to replace many expensive specialty commercial PT-1000 readers, especially when many readout channels are required. This design can potentially be applicable to a wide variety of applications, and it is available at \url{https://github.com/leehagaman/rtd-reader}.

%% file: chapters/09_WC_BDT_vars.tex
\chapter{Wire-Cell BDT Variables} \label{sec:wc_bdt_vars}


This appendix describes Wire-Cell reconstruction variables used for $\nu_e$, $\nu_\mu$, and $\pi^0$ selections described in Chapter \ref{sec:nueCC}, as well as for the Wire-Cell $\pi^0$ and NC Delta radiative selections described in Chapter \ref{sec:nc_delta}.

These variables are grouped into sets, originally for specific ``taggers'', which were each designed to reject a certain type of event using human-designed cuts. This causes some information to be duplicated, with the same variable existing with different names corresponding to different taggers. Some of these variables are ``vector variables'', and contain more than one entry per event, for example variables related to each shower in an event. Some of these vector variables are vectors for historical reasons, and actually contain only global information about the neutrino interaction, or always contain only one entry. Some of the variables developed for these taggers were not used for the final BDT input, and those variables are not included here.

Many of these variables are related to the ``primary shower'', meaning the highest energy shower connected to the reconstructed neutrino vertex, which will be the electron shower in correctly reconstructed $\nu_e$CC events. Some of these variables refer to ``attached'' segments, which directly touch other activity, or ``connected'' segments, which are connected in the same cluster but perhaps with other segments in between.

All of these variables refer only to reconstructed quantities, and are applied in the same way to data and simulation. Note that these are simplified descriptions, and there are many edge cases, kinematic thresholds, and candidate reconstructions with different confidence levels involved in these variable definitions; these details can only be fully described by directly analyzing the code that we reference in each table below.

\renewcommand{\baselinestretch}{1.1}

\begin{table}[H]
    \centering
    \begin{tabular}{l p{10.5cm}}
        \toprule
        \textbf{Variable} & \textbf{Description} \\
        \midrule
        \texttt{gap\_flag} & Indicates if a significant gap is present in the shower. \\
        \hline
        \texttt{gap\_flag\_prolong\_u} & Indicates that the shower has significant extent in the U wire plane. \\
        \hline
        \texttt{gap\_flag\_prolong\_v} & Indicates that the shower has significant extent in the V wire plane. \\
        \hline
        \texttt{gap\_flag\_prolong\_w} & Indicates that the shower has significant extent in the W (also called Z) wire plane. \\
        \hline
        \texttt{gap\_flag\_parallel} & Indicates that the shower is nearly parallel to the wire planes (isochronous activity). \\
        \hline
        \texttt{gap\_n\_points} & Total number of evaluated 3D shower points. \\
        \hline
        \texttt{gap\_n\_bad} & Total number of evaluated 3D shower points with missing information. \\
        \hline
        \texttt{gap\_energy} & Shower energy in MeV. \\
        \hline
        \texttt{gap\_num\_valid\_tracks} & Number of tracks connected to the neutrino vertex. \\
        \hline
        \texttt{gap\_flag\_single\_shower} & Indicates if the shower is the only activity at the reconstruced neutrino vertex. \\
        \hline
        \texttt{gap\_filled} & Flag indicating if the information in the above variables has been filled for this event. \\
        \bottomrule
    \end{tabular}
    \caption[Gap identification $\nu_e$CC Variables]{Description of $\nu_e$CC inclusive selection variables related to shower gap detection. The full code defining these is available at \url{https://github.com/BNLIF/wire-cell-pid/blob/master/src/NeutrinoID_nue_tagger.h}.}
    \label{tab:gap_variables}
\end{table}

\begin{table}[H]
    \centering
    \begin{tabular}{l p{10.5cm}}
        \toprule
        \textbf{Variable} & \textbf{Description} \\
        \midrule
        \texttt{pio\_flag} & Indicates if a $\pi^0$ candidate is identified. \\
        \hline
        \texttt{pio\_mip\_id} & Indicates if the primary shower $dQ/dx$ is consistent with a minimum ionizing particle. \\
        \hline
        \texttt{pio\_filled} & Indicates if $\pi^0$ variables have been filled. \\
        \hline
        \texttt{pio\_flag\_pio} & Indicates if the primary shower is one of the two showers originating from a $\pi^0$ candidate. \\
        \hline
        \texttt{pio\_1\_flag} & Indicates whether a $\pi^0$ candidate is identified by examining the two shower energies and the invariant mass. \\
        \hline
        \texttt{pio\_1\_mass} & Invariant mass of the identified $\pi^0$ candidate in MeV. \\
        \hline
        \texttt{pio\_1\_pio\_type} & Classifies the $\pi^0$ candidate based on the implied invariant mass difference from the expected $135\ \mathrm{MeV}/c^2$. \\
        \hline
        \texttt{pio\_1\_energy\_1} & Energy of the first shower in MeV. \\
        \hline
        \texttt{pio\_1\_energy\_2} & Energy of the second shower in MeV. \\
        \hline
        \texttt{pio\_1\_dis\_1} & Distance of the first shower from the vertex in cm. \\
        \hline
        \texttt{pio\_1\_dis\_2} & Distance of the second shower from the vertex in cm. \\
        \hline
        \texttt{pio\_2\_v\_flag} & Vector variable indicating if these showers correspond to a $\pi^0$ candidate is identified by examining the shower vertex locations. \\
        \hline
        \texttt{pio\_2\_v\_dis2} & Vector variable of the distance of the secondary vertex from the interaction vertex in cm. \\
        \hline
        \texttt{pio\_2\_v\_angle2} & Vector variable of the angular separation between shower directions in degrees. \\
        \hline
        \texttt{pio\_2\_v\_acc\_length} & Vector variable of the accumulated length of all track segments associated with a potential second shower. \\
        \bottomrule
    \end{tabular}
    \caption[$\pi^0$ variables]{Description of $\nu_e$CC inclusive selection variables related to $\pi^0$ identification. The full code defining these is available at \url{https://github.com/BNLIF/wire-cell-pid/blob/master/src/NeutrinoID_nue_tagger.h}.}
\end{table}

\begin{table}[H]
    \centering
    \begin{tabular}{l p{9.5cm}}
        \toprule
        \textbf{Variable} & \textbf{Description} \\
        \midrule
        \texttt{mip\_quality\_flag} & Indicates if the primary shower has a minimum ionizing particle (MIP) track included, either overlapping the shower or making up part of a split pair of showers that have been merged. \\
        \hline
        \texttt{mip\_quality\_energy} & Primary shower energy in MeV. \\
        \hline
        \texttt{mip\_quality\_overlap} & Indicates if the primary shower contains an overlapping MIP track. \\
        \hline
        \texttt{mip\_quality\_n\_showers} & Number of showers connected to the neutrino vertex. \\
        \hline
        \texttt{mip\_quality\_n\_tracks} & Number of tracks connected to the neutrino vertex. \\
        \hline
        \texttt{mip\_quality\_flag\_inside\_pi0} & Indicates if the primary shower is part of a $\pi^0$ candidate. \\
        \hline
        \texttt{mip\_quality\_n\_pi0\_showers} & Number of candidate $\pi^0$ showers. \\
        \hline
        \texttt{mip\_quality\_shortest\_length} & Length of the shortest start segment of all neutrino vertex-attached showers. \\
        \hline
        \texttt{mip\_quality\_acc\_length} & Accumulated length of all segments in the shower with the shortest start segment of all neutrino vertex-attached showers. \\
        \hline
        \texttt{mip\_quality\_flag\_proton} & Indicates if any connected shower start segments have a $dQ/dx$ consistent with a proton track. \\
        \hline
        \texttt{mip\_quality\_filled} & Indicates if MIP variables have been filled. \\
        \bottomrule
    \end{tabular}
    \caption[MIP quality $\nu_e$CC variables]{MIP quality $\nu_e$CC variables. Description of $\nu_e$CC inclusive selection variables related to minimum ionizing particle quality reconstruction. The full code defining these is available at \url{https://github.com/BNLIF/wire-cell-pid/blob/master/src/NeutrinoID_nue_tagger.h}.}
\end{table}

\begin{table}[H]
    \centering
    \scriptsize
    \begin{tabular}{l p{11.5cm}}
        \toprule
        \textbf{Variable} & \textbf{Description} \\
        \midrule
        \texttt{mip\_flag} & Indicates whether the primary shower is consistent with a minimum ionizing particle (MIP). \\
        \hline
        \texttt{mip\_energy} & Energy of the shower in MeV. \\
        \hline
        \texttt{mip\_n\_end\_reduction} & Number of shower trajectory segments continuously reducing in $dQ/dx$ above a certain cut value from the start. \\
        \hline
        \texttt{mip\_n\_first\_mip} & Index of the first shower trajectory segment dropping below a MIP $dQ/dx$ threshold. \\
        \hline
        \texttt{mip\_n\_first\_non\_mip} & Index of the first shower trajectory segment raising above a MIP $dQ/dx$ threshold after  \texttt{mip\_n\_first\_mip}. \\
        \hline
        \texttt{mip\_n\_first\_non\_mip\_1} & Index of the first pair of two connected shower trajectory segments both above a MIP $dQ/dx$ threshold after \texttt{mip\_n\_first\_mip}. \\
        \hline
        \texttt{mip\_n\_first\_non\_mip\_2} & Index of the first trio of three connected shower trajectory segments all above a MIP $dQ/dx$ threshold after \texttt{mip\_n\_first\_mip}. \\
        \hline
        \texttt{mip\_max\_dQ\_dx\_sample} & Maximum $dQ/dx$ in the first trio of three connected shower stem segments all above a MIP $dQ/dx$ threshold after \texttt{mip\_n\_first\_mip}. \\
        \hline
        \texttt{mip\_vec\_dQ\_dx\_[0-19]} & Vector of $dQ/dx$ values sampled along the shower trajectory segments. \\
        \texttt{mip\_n\_below\_threshold} & Number of shower trajectory segments with $dQ/dx$ below the expected threshold. \\
        \hline
        \texttt{mip\_n\_below\_zero} & Number of shower trajectory segments with negative $dQ/dx$ values. \\
        \hline
        \texttt{mip\_n\_lowest} & Number of shower trajectory segments which update the lowest $dQ/dx$ value from the start. \\
        \hline
        \texttt{mip\_n\_highest} & Number of shower trajectory segments which update the highest $dQ/dx$ value from the start. \\
        \hline
        \texttt{mip\_lowest\_dQ\_dx} & Lowest $dQ/dx$ value recorded in the shower trajectory. \\
        \hline
        \texttt{mip\_highest\_dQ\_dx} & Highest $dQ/dx$ value recorded in the shower trajectory. \\
        \hline
        \texttt{mip\_medium\_dQ\_dx} & Median $dQ/dx$ value along the shower trajectory. \\
        \hline
        \texttt{mip\_min\_dQ\_dx\_5} & Minimum $dQ/dx$ value among the first five shower trajectory segments. \\
        \hline
        \texttt{mip\_flag\_all\_above} & Indicates whether all shower trajectory segmentns $dQ/dx$ values are above a certain threshold. \\
        \hline
        \texttt{mip\_stem\_length} & Length of the initial shower stem, in cm. \\
        \hline
        \texttt{mip\_length\_main} & Total length of all segments in the primary cluster of the shower, in cm. \\
        \hline
        \texttt{mip\_length\_total} & Total length of all segments in the entire shower, in cm. \\
        \hline
        \texttt{mip\_angle\_beam} & Angle between the shower direction and the beam direction, in degrees. \\
        \hline
        \texttt{mip\_iso\_angle} & Angle between the shower direction and the isochronous plane (angle from the plane containing the wires), in degrees. \\
        \hline
        \texttt{mip\_n\_good\_tracks} & Number of shower-vertex-connected segments > 10 cm with confident direction reconstruction. \\
        \hline
        \texttt{mip\_n\_vertex} & Number of shower-vertex-connected segments. \\
        \hline
        \texttt{mip\_n\_other\_vertex} & Number of segments connected to the end of the first shower trajectory segment. \\
        \hline
        \texttt{mip\_E\_indirect\_max\_energy} & Maximum shower energy out of showers not connected to the neutrino vertex, in MeV. \\
        \hline
        \texttt{mip\_flag\_stem\_trajectory} & Indicates whether the first shower trajectory segment is tagged as shower-like. \\
        \hline
        \texttt{mip\_min\_dis} & Minimum distance between a primary shower segment and another particle, in cm. \\
        \hline
        \texttt{mip\_n\_stem\_size} & Number of segments in the primary shower. \\
        \hline
        \texttt{mip\_filled} & Indicates whether the MIP-related variables have been filled. \\
        \bottomrule
    \end{tabular}
    \caption[MIP $\nu_e$CC variables]{MIP $\nu_e$CC variables. Description of $\nu_e$CC inclusive selection variables related to minimum ionizing particle identification. The full code defining these is available at \url{https://github.com/BNLIF/wire-cell-pid/blob/master/src/NeutrinoID_nue_tagger.h}.}
\end{table}

\begin{table}[H]
    \centering
    \begin{tabular}{l p{9.5cm}}
        \toprule
        \textbf{Variable} & \textbf{Description} \\
        \midrule
        \texttt{sig\_1\_v\_angle} & Vector variable indicating the angle of each second shower. \\
        \hline
        \texttt{sig\_1\_v\_flag\_single\_shower} & Vector variable indicating if the primary shower is the only activity at the reconstruced neutrino vertex. \\
        \hline
        \texttt{sig\_1\_v\_energy} & Vector variable of the primary shower energy in MeV. \\
        \hline
        \texttt{sig\_1\_v\_energy\_1} & Vector variable of the second shower energy in MeV. \\
        \hline
        \texttt{sig\_1\_v\_flag} & Vector variable which is false if the following conditions are met: the second shower points back to the neutrino vertex, 
        the primary shower is the only activity at the neutrino vertex, the primary shower energy is less than 250 MeV, and the secondary shower energy is greater than 60 MeV. \\
        \hline
        \texttt{sig\_1\_score} & Minimum score of the \texttt{sig\_1} BDT across all entries in the vector. \\
        \hline
        \texttt{sig\_2\_v\_energy} & Vector variable of the primary shower energy in MeV. \\
        \hline
        \texttt{sig\_2\_v\_shower\_angle} & Vector variable of the primary shower angle with respect to the beam direction in degrees. \\
        \hline
        \texttt{sig\_2\_v\_flag\_single\_shower} & Vector variable indicating if the primary shower is the only activity at the reconstruced neutrino vertex. \\
        \hline
        \texttt{sig\_2\_v\_medium\_dQ\_dx} & Vector variable of median $dQ/dx$ value in the furthest cluster from the neutrino vertex.  \\
        \hline
        \texttt{sig\_2\_v\_start\_dQ\_dx} & Vector variable of starting $dQ/dx$ value in the furthest cluster from the neutrino vertex.  \\
        \hline
        \texttt{sig\_2\_v\_flag} & Vector variable which is false if certain cuts on the above variables indicate a likely $\pi^0$ origin. \\
        \hline
        \texttt{sig\_2\_score} & Minimum score of the \texttt{sig\_2} BDT across all entries in the vector. \\
        \bottomrule
    \end{tabular}
    \caption[SIG $\nu_e$CC variables]{SIG $\nu_e$CC variables. Description of $\nu_e$CC inclusive selection variables related to $\pi^0$ identification via the examination of second showers, meaning showers besides the primary shower candidate. The full code defining these is available at \url{https://github.com/BNLIF/wire-cell-pid/blob/master/src/NeutrinoID_nue_tagger.h}.}
\end{table}

\begin{table}[H]
    \centering
    \begin{tabular}{l p{9.5cm}}
        \toprule
        \textbf{Variable} & \textbf{Description} \\
        \midrule
        \texttt{mgo\_energy} & Maximum energy of the primary shower connected to the neutrino vertex in MeV. \\
        \hline
        \texttt{mgo\_max\_energy} & Maximum energy of all showers in MeV. \\
        \hline
        \texttt{mgo\_total\_energy} & Summed energy of all >50 MeV showers in MeV. \\
        \hline
        \texttt{mgo\_n\_showers} & Number of > 50 MeV showers. \\
        \hline
        \texttt{mgo\_max\_energy\_1} & Maximum energy of all $\pi^0$-tagged showers in MeV. \\
        \hline
        \texttt{mgo\_max\_energy\_2} & Maximum energy of all non-$\pi^0$-tagged showers in MeV.  \\
        \hline
        \texttt{mgo\_total\_other\_energy} & Summed energy of all non-$\pi^0$-tagged showers in MeV. \\
        \hline
        \texttt{mgo\_total\_other\_energy\_1} & Summed energy of all non-neutrino-vertex-connected non-$\pi^0$-tagged showers in MeV. \\
        \hline
        \texttt{mgo\_n\_total\_showers} & Number of non-$\pi^0$-tagged showers.\\
        \hline
        \texttt{mgt\_flag\_single\_shower} & Indicates if the primary shower is the only activity at the reconstruced neutrino vertex. \\
        \hline
        \texttt{mgt\_max\_energy} & Maximum energy of all non-neutrino-vertex-connected showers in MeV. \\
        \hline
        \texttt{mgt\_total\_other\_energy} & Summed energy of all non-neutrino-vertex-connected showers in MeV. \\
        \hline
        \texttt{mgt\_max\_energy\_1} & Maximum energy of all non-neutrino-vertex-connected $\pi^0$-tagged showers in MeV.  \\
        \hline
        \texttt{mgt\_e\_indirect\_max\_energy} & Maximum energy of all showers with no other activity connected to the shower vertex in MeV. \\
        \hline
        \texttt{mgt\_e\_direct\_max\_energy} & Maximum energy of all showers with other activity connected to the shower vertex in MeV. \\
        \hline
        \texttt{mgt\_e\_indirect\_total\_energy} & Summed energy of showers with no other activity connected to the shower vertex in MeV. \\
        \hline
        \texttt{mgt\_e\_direct\_total\_energy} & Summed energy of showers with other activity connected to the shower vertex in MeV. \\
        \hline
        \texttt{mgt\_n\_direct\_showers} & Number of showers with other activity connected to the shower vertex. \\
        \hline
        \texttt{mgt\_flag\_indirect\_max\_pio} & Indicates if all relatively high-energy showers with no other activity connected to the shower vertex are part of a $\pi^0$ candidate. \\
        \bottomrule
    \end{tabular}
    \caption[MGO $\nu_e$CC variables]{MGO $\nu_e$CC variables. Description of $\nu_e$CC inclusive selection variables related to $\pi^0$ identification from the ``multiple gamma one'' and ``multiple gamma two'' taggers. The full code defining these is available at \url{https://github.com/BNLIF/wire-cell-pid/blob/master/src/NeutrinoID_nue_functions.h}.}
\end{table}

\begin{table}[H]
    \centering
    \footnotesize
    \begin{tabular}{l p{10.5cm}}
        \toprule
        \textbf{Variable} & \textbf{Description} \\
        \midrule
        \texttt{stw\_1\_energy} & Primary shower energy in MeV. \\
        \hline
        \texttt{stw\_1\_dis} & Projected distance from the shower vertex opposite the shower direction to the fiducial volume boundary. \\
        \hline
        \texttt{stw\_1\_dQ\_dx} & Maximum $dQ/dx$ in the primary shower stem. \\
        \hline
        \texttt{stw\_1\_flag\_single\_shower} & Indicates if the primary shower is the only activity at the reconstruced neutrino vertex. \\
        \hline
        \texttt{stw\_1\_n\_pi0} & Number of >25 MeV showers tagged as $\pi^0$. \\
        \hline
        \texttt{stw\_1\_num\_valid\_tracks} & Number of tracks connected to the neutrino vertex. \\
        \hline
        \texttt{stw\_2\_v\_medium\_dQ\_dx} & Vector variable of the median $dQ/dx$ for the first segment of the primary shower.\\
        \hline
        \texttt{stw\_2\_v\_energy} & Vector variable of the primary shower energy in MeV. \\
        \hline
        \texttt{stw\_2\_v\_angle} & Vector variable of the minimum angle in degrees between two directions: the direction between the reconstructed neutrino vertex and the current shower vertex, and either the forward or reverse primary shower direction. \\
        \hline
        \texttt{stw\_2\_v\_dir\_length} & Vector variable of the distance between the reconstructed neutrino vertex and the current shower vertex in cm. \\
        \hline
        \texttt{stw\_2\_v\_max\_dQ\_dx} & Vector variable of the maximum $dQ/dx$ in the primary shower stem. \\
        \hline
        \texttt{stw\_2\_score} & Minimum score of the \texttt{stw\_2} BDT across all entries in the vector. \\
        \hline
        \texttt{stw\_3\_v\_angle} & Vector variable of the minimum angle in degrees between two directions: the direction between the reconstructed neutrino vertex and the current cluster vertex, and either the forward or reverse primary shower direction. \\
        \hline
        \texttt{stw\_3\_v\_dir\_length} & Vector variable of the distance between the neutrino vertex and the current cluster vertex in cm. \\
        \hline
        \texttt{stw\_3\_v\_energy} & Vector variable of the primary shower energy in MeV. \\
        \hline
        \texttt{stw\_3\_v\_medium\_dQ\_dx} & Vector variable of the maximum $dQ/dx$ in the primary shower stem. \\
        \hline
        \texttt{stw\_3\_score} & Minimum score of the \texttt{stw\_3} BDT across all entries in the vector. \\
        \hline
        \texttt{stw\_4\_v\_angle} & Vector variable of the angle between the current shower direction and the primary shower direction in degrees. \\
        \hline
        \texttt{stw\_4\_v\_dis} & Vector variable of the distance projected from the end of the current shower along the shower direction until it intersects the fiducial volume boundary. \\
        \hline
        \texttt{stw\_4\_v\_energy} &  Vector variable of the primary shower energy in MeV. \\
        \hline
        \texttt{stw\_4\_score} &  Minimum score of the \texttt{stw\_4} BDT across all entries in the vector. \\
        \bottomrule
    \end{tabular}
    \caption[STW $\nu_e$CC variables]{STW $\nu_e$CC variables. Description of $\nu_e$CC inclusive selection variables related to $\pi^0$ identification from the ``shower to wall'' taggers. The full code defining these is available at \url{https://github.com/BNLIF/wire-cell-pid/blob/master/src/NeutrinoID_nue_tagger.h}.}
\end{table}

\begin{table}[H]
    \centering
    \begin{tabular}{l p{8.5cm}}
        \toprule
        \textbf{Variable} & \textbf{Description} \\
        \midrule
        \texttt{stem\_dir\_energy} & Primary shower energy in MeV. \\
        \hline
        \texttt{stem\_dir\_angle} & Angle between the principle component analysis (PCA) direction of the primary shower and the direction of the first 5 cm of the primary shower in degrees.\\
        \hline
        \texttt{stem\_dir\_angle1} & Angle between the direction of the first 30 cm of the primary shower and the drift direction in degrees, minus 90 degrees.\\
        \hline
        \texttt{stem\_dir\_angle2} & Angle between the principle component analysis (PCA) direction of the primary shower and the drift direction in degrees.\\
        \hline
        \texttt{stem\_dir\_angle3} & Angle between the first 5 cm of the primary shower and the first 100 cm of the primary shower in degrees.\\
        \hline
        \texttt{stem\_dir\_ratio} & The ratio of two measures of the distance between the first and 10th shower trajectory points: the straight line distance over the path length distance. \\
        \hline
        \texttt{stem\_dir\_flag\_single\_shower} & Indicates if the primary shower is the only activity at the reconstruced neutrino vertex. \\
        \hline
        \texttt{stem\_len\_energy} & Primary shower energy in MeV. \\
        \hline
        \texttt{stem\_len\_length} & Length of the first segment of the primary shower in cm.\\
        \hline
        \texttt{stem\_len\_flag\_avoid\_muon\_check} & Indicates if this shower has properties such that a check for a contained muon track is not performed.\\
        \hline
        \texttt{stem\_len\_num\_daughters} & Number of daughter track segments from the first primary shower segment. \\
        \hline
        \texttt{stem\_len\_daughter\_length} & Total summed length of daughter track segments from the first primary shower segment in cm.\\
        \bottomrule
    \end{tabular}
    \caption[STEM $\nu_e$CC variables]{STEM $\nu_e$CC variables. Description of $\nu_e$CC inclusive selection variables related to shower stem taggers. The full code defining these is available at \url{https://github.com/BNLIF/wire-cell-pid/blob/master/src/NeutrinoID_nue_tagger.h} and \url{https://github.com/BNLIF/wire-cell-pid/blob/master/src/NeutrinoID_nue_functions.h}.}
\end{table}

\begin{table}[H]
    \centering
    \begin{tabular}{l p{8.5cm}}
        \toprule
        \textbf{Variable} & \textbf{Description} \\
        \midrule
        \texttt{lem\_shower\_main\_length} & Total summed length of segments in the first cluster of the primary shower in cm.\\
        \hline
        \texttt{lem\_n\_3seg} & Number of clusters in the primary shower with at least three segments, not including the first cluster. \\
        \hline
        \texttt{lem\_e\_charge} & Total energy of the primary shower as calculated by summing the total charge.\\
        \hline
        \texttt{lem\_e\_dQdx} & $dQ/dx$ of the primary shower start segment. \\
        \hline
        \texttt{lem\_shower\_num\_main\_segs} & The number of primary shower segments in the same cluster as the start segment.\\
        \hline
        \texttt{lem\_shower\_main\_length} & Total summed length of all primary shower segments in the same cluster as the start segment, in cm.\\
        \bottomrule
    \end{tabular}
    \caption[LEM $\nu_e$CC variables]{LEM $\nu_e$CC variables. Description of $\nu_e$CC inclusive selection variables related to the low energy Michel electron tagger. The full code defining these is available at \url{https://github.com/BNLIF/wire-cell-pid/blob/master/src/NeutrinoID_nue_tagger.h}.}
\end{table}

\begin{table}[H]
    \centering
    \begin{tabular}{l p{9.5cm}}
        \toprule
        \textbf{Variable} & \textbf{Description} \\
        \midrule
        \texttt{brm\_n\_mu\_segs} & Number of segments in the muon candidate contained within the primary shower.\\
        \hline
        \texttt{brm\_acc\_length} & Summed path length of all segments in the muon candidate contained within the primary shower in cm. \\
        \hline
        \texttt{brm\_acc\_direct\_length} & Summed distance between the start and end points for all segments in the muon candidate contained within the primary shower in cm. \\
        \hline
        \texttt{brm\_Ep} & Energy of the muon candidate contained within the primary shower based on the length in MeV. \\
        \hline
        \texttt{brm\_shower\_total\_length} & Total summed length of all primary shower segments in cm. \\
        \hline
        \texttt{brm\_connected\_length} & Total summed length of all segments connected to the primary shower vertex. \\
        \hline
        \texttt{brm\_n\_size} & Number of separate clusters in the muon candidate contained within the primary shower, which can be broken apart in the reconstruction. \\
        \hline
        \texttt{brm\_n\_shower\_main\_segs} & Number of shower segments connected to the primary shower vertex.\\
        \hline
        \texttt{brm\_n\_mu\_main} & Number of segments connected to the primary shower vertex that are part of the muon candidate contained within the primary shower.\\
        \bottomrule
    \end{tabular}
    \caption[BRM $\nu_e$CC variables]{BRM $\nu_e$CC variables. Description of $\nu_e$CC inclusive selection variables related to the broken muon tagger. The full code defining these is available at \url{https://github.com/BNLIF/wire-cell-pid/blob/master/src/NeutrinoID_nue_tagger.h}.}
\end{table}

\begin{table}[H]
    \centering
    \begin{tabular}{l p{11.5cm}}
        \toprule
        \textbf{Variable} & \textbf{Description} \\
        \midrule
        \texttt{cme\_mu\_energy} & Muon candidate energy estimated by the length in MeV. \\
        \hline
        \texttt{cme\_energy} & Primary shower energy.\\
        \hline
        \texttt{cme\_mu\_length} & Muon candidate length in cm. \\
        \hline
        \texttt{cme\_length} & Summed total length of all segments in the primary shower in cm. \\
        \hline
        \texttt{cme\_angle\_beam} & Angle between the primary shower direction and the beam direction in degrees.\\
        \bottomrule
    \end{tabular}
    \caption[CME $\nu_e$CC variables]{CME $\nu_e$CC variables. Description of $\nu_e$CC inclusive selection variables related to the ``compare muon energy'' tagger, which tries identify cases where the length and energy of a shower is consistent with a mis-reconstructed muon. The full code defining these is available at \url{https://github.com/BNLIF/wire-cell-pid/blob/master/src/NeutrinoID_nue_functions.h}.}
\end{table}

\begin{table}[H]
    \centering
    \begin{tabular}{l p{10.5cm}}
        \toprule
        \textbf{Variable} & \textbf{Description} \\
        \midrule
        \texttt{anc\_angle} & Angle between the primary shower direction and the beam direction in degrees. \\
        \hline
        \texttt{anc\_max\_angle} & Maximum angle between the direction of a segment in the primary shower and the overall primary shower direction in degrees. \\
        \hline
        \texttt{anc\_max\_length} & The length of the segment with the maximum angle between the direction of a segment in the primary shower and the overall primary shower direction in cm. \\
        \hline
        \texttt{anc\_acc\_forward\_length} & Total summed length of segments of the primary shower which point in the forward direction relative to the beam in cm. \\
        \hline
        \texttt{anc\_acc\_backward\_length} & Total summed length of segments of the primary shower which point in the backward direction relative to the beam in cm.  \\
        \hline
        \texttt{anc\_acc\_forward\_length1} & Total summed length of segments of the primary shower which point in the forward direction with an angle less than $85^\circ$ relative to the beam in cm. \\
        \hline
        \texttt{anc\_shower\_main\_length} & Total summed length of all segments connected to the primary shower vertex in cm. \\
        \hline
        \texttt{anc\_shower\_total\_length} & Total summed length of all segments in the primary shower in cm. \\
        \hline
        \texttt{anc\_flag\_main\_outside} & Boolean variable indicating if any part of the shower extends to edge of the TPC, outside the fiducial volume. \\
        \bottomrule
    \end{tabular}
    \caption[ANC $\nu_e$CC variables]{ANC $\nu_e$CC variables. Description of $\nu_e$CC inclusive selection variables related to the ``angular cut'' tagger, which tries identify cases where the shower angle is more consistent with backgrounds rather than $\nu_e$CC. The full code defining these is available at \url{https://github.com/BNLIF/wire-cell-pid/blob/master/src/NeutrinoID_nue_tagger.h}.}
\end{table}

\begin{table}[H]
    \centering
    \small
    \begin{tabular}{l p{10.5cm}}
        \toprule
        \textbf{Variable} & \textbf{Description} \\
        \midrule
        \texttt{vis\_1\_n\_vtx\_segs} & Number of segments connected to the neutrino vertex. \\
        \hline
        \texttt{vis\_1\_energy} & Primary shower energy in MeV. \\
        \hline
        \texttt{vis\_1\_num\_good\_tracks} & Number of tracks connected to the neutrino vertex. \\
        \hline
        \texttt{vis\_1\_max\_angle} & Maximum angle between a neutrino vertex attached segment direction and the primary shower direction in degrees. \\
        \hline
        \texttt{vis\_1\_max\_shower\_angle} & Maximum angle between the primary shower direction and a non-primary shower direction in degrees. \\
        \hline
        \texttt{vis\_1\_tmp\_length1} & Length of the neutrino vertex attached segment with the maximum angle between its direction and the primary shower direction in cm. \\
        \hline
        \texttt{vis\_1\_tmp\_length2} & Length of the first segment of the primary shower. \\
        \hline
        \texttt{vis\_2\_n\_vtx\_segs} & Number of segments connected to the neutrino vertex. \\
        \hline
        \texttt{vis\_2\_min\_angle} & Minimum supplementary angle of the angles between the first segment of the primary shower direction or the overall shower direction and any other shower vertex-attached segments. \\
        \hline
        \texttt{vis\_2\_max\_angle} & Maximum supplementary angle of the angles between the first segment of the primary shower direction or the overall shower direction and any other shower vertex-attached segments. \\
        \hline
        \texttt{vis\_2\_min\_weak\_track} & Indicates of the segment associated with \texttt{vis\_2\_min\_angle} has uncertain direction information. \\
        \hline
        \texttt{vis\_2\_max\_weak\_track} & Indicates of the segment associated with \texttt{vis\_2\_max\_angle} has uncertain direction information. \\
        \hline
        \texttt{vis\_2\_angle\_beam} & Angle between the primary shower direction and the neutrino beam direction in degrees. \\
        \hline
        \texttt{vis\_2\_min\_angle1} & Complement of the angle between the segment associated with \texttt{vis\_2\_min\_angle} and the drift direction in degrees. \\
        \hline
        \texttt{vis\_2\_iso\_angle1} & Complement of the angle between the primary shower direction and the drift direction in degrees. \\
        \hline
        \texttt{vis\_2\_min\_medium\_dQ\_dx} & $dQ/dx$ of the segment associated with \texttt{vis\_2\_min\_angle}. \\
        \hline
        \texttt{vis\_2\_min\_length} & Length of the segment associated with \texttt{vis\_2\_min\_angle} in degrees. \\
        \hline
        \texttt{vis\_2\_sg\_length} & Length of the first segment of the primary shower in cm. \\
        \bottomrule
    \end{tabular}
    \caption[VIS $\nu_e$CC variables]{VIS $\nu_e$CC variables. Description of $\nu_e$CC inclusive selection variables related to the ``vertex inside shower'' tagger, which tries identify cases where the neutrino vertex is incorrectly reconstructed inside a shower. The full code defining these is available at \url{https://github.com/BNLIF/wire-cell-pid/blob/master/src/NeutrinoID_nue_functions.h}.}
\end{table}

\begin{table}[H]
    \centering
    \small
    \begin{tabular}{l p{10.5cm}}
        \toprule
        \textbf{Variable} & \textbf{Description} \\
        \midrule
        \texttt{br1\_1\_shower\_type} & This value is 1 if the primary shower has a direct connection, 2 if it has an indirect connection or gap, and 3 if it has an ambiguously reconstructed connection type. \\
        \hline
        \texttt{br1\_1\_vtx\_n\_segs} & Number of segments attached to the neutrino vertex. \\
        \hline
        \texttt{br1\_1\_energy} & Primary shower energy in MeV. \\
        \hline
        \texttt{br1\_1\_n\_segs} & Number of segments in the primary shower. \\
        \hline
        \texttt{br1\_1\_flag\_sg\_topology} & Flag indicating that the first primary shower segment is identified as part of a shower by examining topology information. \\
        \hline
        \texttt{br1\_1\_flag\_sg\_trajectory} & Flag indicating that the first primary shower segment is identified as part of a shower by examining trajectory information. \\
        \hline
        \texttt{br1\_1\_sg\_length} & Length of the first shower segment in cm. \\
        \hline
        \texttt{br1\_2\_n\_connected} & The number of segments connected to the endpoints of the longest track-like trajectory in the primary shower. \\
        \hline
        \texttt{br1\_2\_n\_connected\_1} &  The number of segments connected to the longest track-like trajectory in the primary shower.  \\
        \hline
        \texttt{br1\_2\_max\_length} & Length of the longest track-like trajectory in the primary shower in cm. \\
        \hline
        \texttt{br1\_2\_n\_shower\_segs} & Total number of segments in the primary shower. \\
        \hline
        \texttt{br1\_2\_max\_length\_ratio} & The ratio of two measures of the distance for the longest track-like trajectory: the start-to-end direct distance over the path length. \\
        \hline
        \texttt{br1\_2\_shower\_length} & Summed length of all segments in the primary shower in cm. \\
        \hline
        \texttt{br1\_3\_n\_connected\_p} & Number of primary shower segments nearly parallel to the first primary shower segment. \\
        \hline
        \texttt{br1\_3\_max\_length\_p} & Summed length of primary shower segments nearly parallel to the first primary shower segment in cm. \\
        \hline
        \texttt{br1\_3\_n\_shower\_main\_segs} & Number of shower segments connected to the primary shower vertex.\\
        \bottomrule
    \end{tabular}
    \caption[BR1 $\nu_e$CC variables]{BR1 $\nu_e$CC variables. Description of $\nu_e$CC inclusive selection variables related to the ``bad reconstruction one'' tagger. The full code defining these is available at \url{https://github.com/BNLIF/wire-cell-pid/blob/master/src/NeutrinoID_nue_tagger.h}.}
\end{table}

\begin{table}[H]
    \centering
    \small
    \begin{tabular}{l p{10.5cm}}
        \toprule
        \textbf{Variable} & \textbf{Description} \\
        \midrule
        \texttt{br2\_num\_valid\_tracks} & Number of tracks tracks connected to the neutrino vertex. \\
        \hline
        \texttt{br2\_n\_shower\_main\_segs} & Number of shower segments connected to the primary shower vertex.\\
        \hline
        \texttt{br2\_max\_angle} & Maximum angle between the first primary shower segment and a connected shower segment. \\
        \hline
        \texttt{br2\_sg\_length} & Length of the first primary shower segment in cm. \\
        \hline
        \texttt{br2\_flag\_sg\_trajectory} & Flag indicating that the first primary shower segment is identified as part of a shower by examining trajectory information. \\
        \bottomrule
    \end{tabular}
    \caption[BR2 $\nu_e$CC variables]{BR2 $\nu_e$CC variables. Description of $\nu_e$CC inclusive selection variables related to the ``bad reconstruction two'' tagger. The full code defining these is available at \url{https://github.com/BNLIF/wire-cell-pid/blob/master/src/NeutrinoID_nue_tagger.h}.}
\end{table}

\begin{table}[H]
    \centering
    \small
    \begin{tabular}{l p{10.5cm}}
        \toprule
        \textbf{Variable} & \textbf{Description} \\
        \midrule
        \texttt{br3\_1\_energy} & Primary shower energy in MeV. \\
        \hline
        \texttt{br3\_1\_n\_shower\_segments} & Total number of segments in the primary shower. \\
        \hline
        \texttt{br3\_1\_sg\_flag\_trajectory} & Flag indicating that the first primary shower segment is identified as part of a shower by examining trajectory information. \\
        \hline
        \texttt{br3\_1\_sg\_direct\_length} & Direct distance between the first primary shower segment start and endpoint in cm.  \\
        \hline
        \texttt{br3\_1\_sg\_length} & Distance along the trajectory between the first primary shower segment start and endpoint in cm. \\
        \hline
        \texttt{br3\_1\_total\_main\_length} & Total summed length of all segments connected to the primary shower vertex in cm. \\
        \hline
        \texttt{br3\_1\_total\_length} & Total summed length of all primary shower segments in cm. \\
        \hline
        \texttt{br3\_1\_iso\_angle} & Angle between the first primary shower segment and the drift direction in degrees. \\
        \hline
        \texttt{br3\_1\_sg\_flag\_topology} & Flag indicating that the first primary shower segment is identified as part of a shower by examining topology information. \\
        \hline
        \texttt{br3\_2\_n\_ele} & Number of electron-like segments in the primary shower, determined by topology and $dQ/dx$ and wiggliness. \\
        \hline
        \texttt{br3\_2\_n\_other} & Number of non-electron-like segments in the primary shower, determined by topology and $dQ/dx$ and wiggliness. \\
        \hline
        \texttt{br3\_2\_other\_fid} & Indicates if the end of the first primary shower segment ends at the TPC boundary outside the fiducial volume. \\
        \hline
        \texttt{br3\_4\_acc\_length} & Summed length of primary shower segments connected to the neutrino vertex which are aligned backwards with the overall primary shower direction in cm. \\
        \hline
        \texttt{br3\_4\_total\_length} & Summed length of primary shower segments connected to the neutrino vertex in cm. \\
        \hline
        \texttt{br3\_7\_min\_angle} & Minimum angle between the first primary shower segment and a segment connected to the endpoint in degrees. \\
        \hline
        \texttt{br3\_8\_max\_dQ\_dx} & Maximum median $dQ/dx$ from a rolling window of five points along each primary shower segment connected to the neutrino vertex. \\
        \hline
        \texttt{br3\_8\_n\_main\_segs} & Number of segments connected to the neutrino vertex in the primary shower. \\
        \bottomrule
    \end{tabular}
    \caption[Scalar BR3 $\nu_e$CC variables]{Scalar BR3 $\nu_e$CC variables. Description of $\nu_e$CC inclusive selection scalar variables related to the ``bad reconstruction three'' tagger. The full code defining these is available at \url{https://github.com/BNLIF/wire-cell-pid/blob/master/src/NeutrinoID_nue_tagger.h}.}
\end{table}

\begin{table}[H]
    \centering
    \scriptsize
    \begin{tabular}{l p{11cm}}
        \toprule
        \textbf{Variable} & \textbf{Description} \\
        \midrule
        \texttt{br3\_3\_v\_energy} & Vector variable of the primary shower energy in MeV. \\
        \hline
        \texttt{br3\_3\_v\_angle} & Vector variable of the angle between the first primary shower segment and each other primary shower segment. \\
        \hline
        \texttt{br3\_3\_v\_dir\_length} & Vector variable of the distance between the start and end points of each non-first primary shower segment in cm. \\
        \hline
        \texttt{br3\_3\_v\_length} & Vector variable of the path length of each non-first primary shower segment in cm. \\
        \hline
        \texttt{br3\_3\_score} & Minimum score of the \texttt{br3\_3} BDT across all entries in the vector. \\
        \hline
        \texttt{br3\_5\_v\_dir\_length} & Vector variable of the distance between the centroid of the vertex-attached segments in the primary shower and the endpoint of the first primary shower segment. \\
        \hline
        \texttt{br3\_5\_v\_total\_length} & Vector variable of the total distance of all vertex-attached segments of the primary shower. \\
        \hline
        \texttt{br3\_5\_v\_flag\_avoid\_muon\_check} & Vector variable that indicates if this shower has properties such that a check for a contained muon track is not performed. \\
        \hline
        \texttt{br3\_5\_v\_n\_seg} & Vector variable of the number of vertex-attached segments in the primary shower. \\
        \hline
        \texttt{br3\_5\_v\_angle} & Vector variable of the angle between the primary shower direction and the direction from the centroid of the vertex-attached segments in the primary shower to the endpoint of the first primary shower segment. \\
        \hline
        \texttt{br3\_5\_v\_sg\_length} & Vector variable of the length of the first segment of the primary shower in cm. \\
        \hline
        \texttt{br3\_5\_v\_energy} & Vector variable of the primary shower energy in MeV. \\
        \hline
        \texttt{br3\_5\_v\_n\_segs} & Vector variable of the number of segments in the primary shower. \\
        \hline
        \texttt{br3\_5\_v\_shower\_main\_length} & Vector variable of the summed length of all vertex-connected segments in the primary shower. \\
        \hline
        \texttt{br3\_5\_v\_shower\_total\_length} & Vector variable of the summed length of all segments in the primary shower. \\
        \hline
        \texttt{br3\_5\_score} & Minimum score of the \texttt{br3\_5} BDT across all entries in the vector. \\
        \hline
        \texttt{br3\_6\_v\_angle} & Vector variable of the angle between the primary shower direction and the direction of each segment which is attached to the end of the first primary shower segment. \\
        \hline
        \texttt{br3\_6\_v\_angle1} & Vector variable of the maximum angle between either the primary shower direction or the direction of each segment which is attached to the end of the first primary shower segment to the drift direction. \\
        \hline
        \texttt{br3\_6\_v\_flag\_shower\_trajectory} & Vector variable of indicating whether each segment attached to the end of the first primary shower segment is identified as part of a shower by examining trajectory information. \\
        \hline
        \texttt{br3\_6\_v\_direct\_length} & Vector variable of the distance between the start and end points of each segment attached to the end of the first primary shower segment in cm. \\
        \hline
        \texttt{br3\_6\_v\_length} & Vector variable of the path length distance of each segment attached to the end of the first primary shower segment in cm. \\
        \hline
        \texttt{br3\_6\_v\_n\_other\_vtx\_segs} & Vector variable of the number of segments attached to the end of the first primary shower segment. \\
        \hline
        \texttt{br3\_6\_v\_energy} & Vector variable of primary shower energy in MeV. \\
        \hline
        \texttt{br3\_6\_score} & Minimum score of the \texttt{br3\_6} BDT across all entries in the vector. \\
        \bottomrule
    \end{tabular}
    \caption[Vector BR3 $\nu_e$CC variables]{Vector BR3 $\nu_e$CC variables. Description of $\nu_e$CC inclusive selection vector variables related to the ``bad reconstruction three'' tagger. The full code defining these is available at \url{https://github.com/BNLIF/wire-cell-pid/blob/master/src/NeutrinoID_nue_tagger.h}.}
\end{table}

\begin{table}[H]
    \centering
    \scriptsize
    \begin{tabular}{l p{11cm}}
        \toprule
        \textbf{Variable} & \textbf{Description} \\
        \midrule
        \texttt{br4\_1\_shower\_main\_length} & Summed distance of all vertex-connected primary shower segments in cm. \\
        \hline
        \texttt{br4\_1\_shower\_total\_length} & Summed distance of all primary shower segments in cm. \\
        \hline
        \texttt{br4\_1\_min\_dis} & Minimum distance between the primary shower vertex and a shower segment vertex which is not connected to the first primary shower segment in cm. \\
        \hline
        \texttt{br4\_1\_energy} & Primary shower energy in MeV. \\
        \hline
        \texttt{br4\_1\_flag\_avoid\_muon\_check} & Indicates if the primary shower has properties such that a check for a contained muon track is not performed.\\
        \hline
        \texttt{br4\_1\_n\_vtx\_segs} & Number of segments connected to the neutrino vertex. \\
        \hline
        \texttt{br4\_1\_n\_main\_segs} & Number of primary shower segments connected to the neutrino vertex. \\
        \hline
        \texttt{br4\_2\_ratio\_45} & Fraction of the neutrino vertex-connected primary shower trajectory points and segment endpoints that are contained within a $45^\circ$ cone from the primary shower vertex and direction. \\
        \hline
        \texttt{br4\_2\_ratio\_35} & Fraction of the neutrino vertex-connected primary shower trajectory points and segment endpoints that are contained within a $35^\circ$ cone from the primary shower vertex and direction. \\
        \hline
        \texttt{br4\_2\_ratio\_25} & Fraction of the neutrino vertex-connected primary shower trajectory points and segment endpoints that are contained within a $25^\circ$ cone from the primary shower vertex and direction. \\
        \hline
        \texttt{br4\_2\_ratio\_15} & Fraction of the neutrino vertex-connected primary shower trajectory points and segment endpoints that are contained within a $15^\circ$ cone from the primary shower vertex and direction. \\
        \hline
        \texttt{br4\_2\_ratio1\_45} & Fraction of the primary shower trajectory points and segment endpoints that are contained within a $45^\circ$ cone from the primary shower vertex and direction. \\
        \hline
        \texttt{br4\_2\_ratio1\_35} & Fraction of the primary shower trajectory points and segment endpoints that are contained within a $35^\circ$ cone from the primary shower vertex and direction. \\
        \hline
        \texttt{br4\_2\_ratio1\_25} & Fraction of the primary shower trajectory points and segment endpoints that are contained within a $25^\circ$ cone from the primary shower vertex and direction. \\
        \hline
        \texttt{br4\_2\_ratio1\_15} & Fraction of the primary shower trajectory points and segment endpoints that are contained within a $15^\circ$ cone from the primary shower vertex and direction. \\
        \hline
        \texttt{br4\_2\_iso\_angle} & Angle between the primary shower direction and the drift direction in degrees. \\
        \hline
        \texttt{br4\_2\_iso\_angle1} & Angle between the primary shower first segment direction and the drift direction in degrees. \\
        \hline
        \texttt{br4\_2\_angle} & Angle between the primary shower first segment direction and the primary shower direction in degrees. \\
        \hline
        \bottomrule
    \end{tabular}
    \caption[BR4 $\nu_e$CC variables]{BR4 $\nu_e$CC variables. Description of $\nu_e$CC inclusive selection variables related to the ``bad reconstruction four'' tagger. The full code defining these is available at \url{https://github.com/BNLIF/wire-cell-pid/blob/master/src/NeutrinoID_nue_tagger.h}.}
\end{table}

\begin{table}[H]
    \centering
    \small
    \begin{tabular}{l p{10cm}}
        \toprule
        \textbf{Variable} & \textbf{Description} \\
        \midrule
        \texttt{tro\_1\_v\_particle\_type} & Vector variable of the PDG code for each segment connected to the neutrino vertex in the primary shower. \\
        \hline
        \texttt{tro\_1\_v\_flag\_dir\_weak} & Vector variable indicating if there is large uncertainty on the direction of each segment connected to the neutrino vertex in the primary shower. \\
        \hline
        \texttt{tro\_1\_v\_min\_dis} & Vector variable of the closest distance between either the start point or the end point for each segment connected to the neutrino vertex in the primary shower in cm. \\
        \hline
        \texttt{tro\_1\_v\_sg1\_length} & Vector variable of the length of each segment connected to the neutrino vertex in the primary shower in cm. \\
        \hline
        \texttt{tro\_1\_v\_shower\_main\_length} & Vector variable of summed length of all neutrino vertex connected primary shower segments in cm. \\
        \hline
        \texttt{tro\_1\_v\_max\_n\_vtx\_segs} & Vector variable of the maximum of the number of segments attached to either the start point or end point for each segment connected to the neutrino vertex in the primary shower in cm. \\
        \hline
        \texttt{tro\_1\_v\_tmp\_length} & Vector variable of the length of each segment connected to the neutrino vertex in the primary shower in cm. \\
        \hline
        \texttt{tro\_1\_v\_medium\_dQ\_dx} & Vector variable of the median $dQ/dx$ for each segment connected to the neutrino vertex in the primary shower in cm. \\
        \hline
        \texttt{tro\_1\_v\_dQ\_dx\_cut} & Vector variable of $0.8866$ + $0.9533 \cdot (18\ \mathrm{cm}$ / $\mathrm{\texttt{segment\_length}})^{0.4234}$ for each segment connected to the neutrino vertex in the primary shower. \\
        \hline
        \texttt{tro\_1\_v\_flag\_shower\_topology} & Vector variable indicating if each segment connected to the neutrino vertex in the primary shower is identified as part of a shower by examining topology information. \\
        \hline
        \texttt{tro\_1\_score} & Minimum score of the \texttt{tro\_1} BDT across all entries in the vector. \\
        \bottomrule
    \end{tabular}
    \caption[TRO1 $\nu_e$CC variables]{TRO1 $\nu_e$CC variables. Description of $\nu_e$CC inclusive selection variables related to the ``track overclustering one'' tagger. The full code defining these is available at \url{https://github.com/BNLIF/wire-cell-pid/blob/master/src/NeutrinoID_nue_tagger.h}.}
\end{table}

\begin{table}[H]
    \centering
    \begin{tabular}{l p{11.5cm}}
        \toprule
        \textbf{Variable} & \textbf{Description} \\
        \midrule
        \texttt{tro\_2\_v\_energy} & Vector variable of the primary shower energy in MeV. \\
        \hline
        \texttt{tro\_2\_v\_stem\_length} & Vector variable of the length of the primary shower stem in cm. \\
        \hline
        \texttt{tro\_2\_v\_iso\_angle} & Vector variable of the maximum of the angle minus $90^\circ$ for each opening angle between the drift direction and the direction of either the last segment of shower stem or the current shower stem segment in degrees. \\
        \hline
        \texttt{tro\_2\_v\_max\_length} & Vector variable of the maximum distance between a primary shower segment and the each shower stem segment in cm. \\
        \hline
        \texttt{tro\_2\_v\_angle} & Vector variable of the supplementary angle of the angle between the last shower stem segment direction and the current shower stem segment in degrees. \\
        \hline
        \texttt{tro\_2\_score} & Minimum score of the \texttt{tro\_2} BDT across all entries in the vector. \\
        \hline
        \texttt{tro\_3\_stem\_length} & The primary shower stem length when allowing gaps between segments in cm. \\
        \hline
        \texttt{tro\_3\_n\_muon\_segs} & The number of segments in the primary shower stem when allowing gaps between segments. \\
        \bottomrule
    \end{tabular}
    \caption[TRO2 and TRO3 $\nu_e$CC variables]{TRO2 and TRO3 $\nu_e$CC variables. Description of $\nu_e$CC inclusive selection variables related to the ``track overclustering two'' and ``track overclustering three'' taggers. The full code defining these is available at \url{https://github.com/BNLIF/wire-cell-pid/blob/master/src/NeutrinoID_nue_tagger.h}.}
\end{table}

\begin{table}[H]
    \centering
    \small
    \begin{tabular}{l p{9.5cm}}
        \toprule
        \textbf{Variable} & \textbf{Description} \\
        \midrule
        \texttt{tro\_4\_v\_dir2\_mag} & Vector variable of the length between the neutrino vertex and each ``dangling vertex'' (an endpoint with no continuing segments) in the shower in cm. \\
        \hline
        \texttt{tro\_4\_v\_angle} & Vector variable of the angle between the shower direction and the direction of the vector from the neutrino vertex to each dangling vertex in degrees. \\
        \hline
        \texttt{tro\_4\_v\_angle1} & Vector variable of the maximum of the absolute value of the complementary angle of the angle between the drift direction and either the shower direction or the direction from the neutrino vertex to each dangling vertex in degrees. \\
        \hline
        \texttt{tro\_4\_v\_angle2} & Vector variable of the minimum of the absolute value of the complementary angle of the angle between the drift direction and either the shower direction or the direction from the neutrino vertex to each dangling vertex in degrees. \\
        \hline
        \texttt{tro\_4\_v\_length} & Vector variable of length of the first primary shower segment in cm. \\
        \hline
        \texttt{tro\_4\_v\_length1} & Vector variable of the length of each primary shower segment segment containing a dangling vertex in cm. \\
        \hline
        \texttt{tro\_4\_v\_medium\_dQ\_dx} & Vector variable of the median $dQ/dx$ of each primary shower segment segment containing a dangling vertex in cm. \\
        \hline
        \texttt{tro\_4\_v\_end\_dQ\_dx} & Vector variable of the end $dQ/dx$ of each primary shower segment segment containing a dangling vertex in cm. \\
        \hline
        \texttt{tro\_4\_v\_energy} & Vector variable of the primary shower energy in MeV. \\
        \hline
        \texttt{tro\_4\_v\_shower\_main\_length} & Vector variable of summed length of all neutrino vertex connected primary shower segments in cm. \\
        \hline
        \texttt{tro\_4\_v\_flag\_shower\_trajectory} & Vector variable indicating if each primary shower segment containing a dangling vertex is identified as part of a shower by examining trajectory information. \\
        \hline
        \texttt{tro\_4\_score} & Minimum score of the \texttt{tro\_4} BDT across all entries in the vector. \\
        \bottomrule
    \end{tabular}
    \caption[TRO4 $\nu_e$CC variables]{TRO4 $\nu_e$CC variables. Description of $\nu_e$CC inclusive selection variables related to the ``track overclustering four'' tagger. The full code defining these is available at \url{https://github.com/BNLIF/wire-cell-pid/blob/master/src/NeutrinoID_nue_tagger.h}.}
\end{table}

\begin{table}[H]
    \centering
    \begin{tabular}{l p{11cm}}
        \toprule
        \textbf{Variable} & \textbf{Description} \\
        \midrule
        \texttt{tro\_5\_v\_max\_angle} & Vector variable of the supplementary angle of the angle between the direction of the first primary shower segment and the direction of the shower segment with the maximum total length of daughter segments for each segment downstream of the first segment of the primary shower in degrees. \\
        \hline
        \texttt{tro\_5\_v\_min\_angle} & Vector variable of the minimum supplementary angle of the angle between the direction of the first primary shower segment and the direction of each segment downstream of the first segment of the primary shower in degrees. \\
        \hline
        \texttt{tro\_5\_v\_max\_length} & Vector variable of the maximum total length of daughter segments for each segment downstream of the first segment of the primary shower in cm. \\
        \hline
        \texttt{tro\_5\_v\_iso\_angle} & Vector variable of the complementary angle of the angle between the direction of the first primary shower segment and the drift direction. \\
        \hline
        \texttt{tro\_5\_v\_n\_vtx\_segs} & Vector variable of the number of segments attached to the endpoint of the first primary shower segment. \\
        \hline
        \texttt{tro\_5\_v\_min\_count} & Vector variable of number of daughter segments for the segment downstream of the first segment of the primary shower corresponding to the minimum supplementary angle of the angle between the direction of the first primary shower segment and the direction of each segment downstream of the first segment of the primary shower. \\
        \hline
        \texttt{tro\_5\_v\_max\_count} & Vector variable of number of daughter segments for the segment downstream of the first segment of the primary shower corresponding to the maximum total length of daughter segments. \\
        \hline
        \texttt{tro\_5\_v\_energy} & Vector variable of the primary shower energy in MeV. \\
        \hline
        \texttt{tro\_5\_score} & Minimum score of the \texttt{tro\_5} BDT across all entries in the vector. \\
        \bottomrule
    \end{tabular}
    \caption[TRO5 $\nu_e$CC variables]{TRO5 $\nu_e$CC variables. Description of $\nu_e$CC inclusive selection variables related to the ``track overclustering five'' tagger. The full code defining these is available at \url{https://github.com/BNLIF/wire-cell-pid/blob/master/src/NeutrinoID_nue_tagger.h}.}
\end{table}

\begin{table}[H]
    \centering
    \begin{tabular}{l p{11cm}}
        \toprule
        \textbf{Variable} & \textbf{Description} \\
        \midrule
        \texttt{hol\_1\_n\_valid\_tracks} & Number of tracks connected to the primary shower vertex. \\
        \hline
        \texttt{hol\_1\_min\_angle} & Minimum angle between the first primary shower segment and another primary shower segment connected to the primary shower vertex in degrees. \\
        \hline
        \texttt{hol\_1\_energy} & Primary shower energy in MeV. \\
        \hline
        \texttt{hol\_1\_flag\_all\_shower} & Indicates the lack of existence of a track-like segment inside the primary shower. \\
        \hline
        \texttt{hol\_1\_min\_length} & Length of the track-like primary shower segment with the minimum angle with respect to the first primary shower segment in cm. \\
        \hline
        \texttt{hol\_2\_min\_angle} & Minimum angle between the first primary shower segment and another primary shower segment attached to the primary shower vertex in degrees. \\
        \hline
        \texttt{hol\_2\_medium\_dQ\_dx} & Median $dQ/dx$ for the primary shower segment which is most closely aligned with the first primary shower segment. \\
        \hline
        \texttt{hol\_2\_ncount} & Number of primary shower points within 0.6 cm of a vertex-attached primary shower segment. \\
        \bottomrule
    \end{tabular}
    \caption[HOL $\nu_e$CC variables]{HOL $\nu_e$CC variables. Description of $\nu_e$CC inclusive selection variables related to the ``high energy overlap'' tagger. The full code defining these is available at \url{https://github.com/BNLIF/wire-cell-pid/blob/master/src/NeutrinoID_nue_tagger.h}.}
\end{table}

\begin{table}[H]
    \centering
    \begin{tabular}{l p{11cm}}
        \toprule
        \textbf{Variable} & \textbf{Description} \\
        \midrule
        \texttt{lol\_1\_v\_energy} & Vector variable of the primary shower energy in MeV. \\
        \hline
        \texttt{lol\_1\_v\_vtx\_n\_segs} & Vector variable of the number of primary shower vertices attached to the neutrino vertex. \\
        \hline
        \texttt{lol\_1\_v\_nseg} & Vector variable of the number of primary shower segments connected to the neutrino vertex. \\
        \hline
        \texttt{lol\_1\_v\_angle} & Vector variable of the angle between the directions of two segments for each vertex in the neutrino vertex connected cluster of the primary shower in degrees. \\
        \hline
        \texttt{lol\_1\_score} & Minimum score of the \texttt{lol\_1} BDT across all entries in the vector. \\
        \bottomrule
    \end{tabular}
    \caption[LOL1 $\nu_e$CC variables]{LOL1 $\nu_e$CC variables. Description of $\nu_e$CC inclusive selection variables related to the ``low energy overlap one'' tagger. The full code defining these is available at \url{https://github.com/BNLIF/wire-cell-pid/blob/master/src/NeutrinoID_nue_tagger.h}.}
\end{table}

\begin{table}[H]
    \centering
    \begin{tabular}{l p{10cm}}
        \toprule
        \textbf{Variable} & \textbf{Description} \\
        \midrule
        \texttt{lol\_2\_v\_length} & Vector variable of the length of each primary shower segment attached to the end of the first primary shower segment in cm. \\
        \hline
        \texttt{lol\_2\_v\_angle} & Vector variable of the angle between the direction of the first primary shower segment and  the direction of each primary shower segment attached to the end of the first primary shower segment in degrees. \\
        \hline
        \texttt{lol\_2\_v\_type} & Vector variable of the PDG code for each primary shower segment attached to the end of the first primary shower segment. \\
        \hline
        \texttt{lol\_2\_v\_vtx\_n\_segs} & Vector variable of the number of segments attached to the end of the first primary shower segment. \\
        \hline
        \texttt{lol\_2\_v\_energy} & Vector variable of the primary shower energy in MeV. \\
        \hline
        \texttt{lol\_2\_v\_shower\_main\_length} & Vector variable of the summed length of all primary shower segments connected to the neutrino vertex in cm. \\
        \hline
        \texttt{lol\_2\_v\_flag\_dir\_weak} & Vector variable indicating if each primary shower segment attached to the end of the first primary shower segment has uncertain direction reconstruction. \\
        \hline
        \texttt{lol\_2\_score} & Minimum score of the \texttt{lol\_2} BDT across all entries in the vector. \\
        \bottomrule
    \end{tabular}
    \caption[LOL2 $\nu_e$CC variables]{LOL2 $\nu_e$CC variables. Description of $\nu_e$CC inclusive selection variables related to the ``low energy overlap two'' tagger. The full code defining these is available at \url{https://github.com/BNLIF/wire-cell-pid/blob/master/src/NeutrinoID_nue_tagger.h}.}
\end{table}

\begin{table}[H]
    \centering
    \begin{tabular}{l p{10cm}}
        \toprule
        \textbf{Variable} & \textbf{Description} \\
        \midrule
        \texttt{lol\_3\_angle\_beam} & Angle between the direction of the first primary shower segment and the beam direction in degrees. \\
        \hline
        \texttt{lol\_3\_n\_valid\_tracks} & Number of tracks connected to the primary shower vertex. \\
        \hline
        \texttt{lol\_3\_min\_angle} & Minimum angle between the directions of two segments for each vertex in the neutrino vertex connected cluster of the primary shower in degrees. \\
        \hline
        \texttt{lol\_3\_vtx\_n\_segs} & Number of segments attached to the end of the first primary shower segment. \\
        \hline
        \texttt{lol\_3\_shower\_main\_length} & Summed length of all primary shower segments connected to the neutrino vertex in cm. \\
        \hline
        \texttt{lol\_3\_n\_sum} & Number of segments and vertices in the primary shower. \\
        \hline
        \texttt{lol\_3\_n\_out} & Number of segments and vertices in the primary shower which lie within a $15^\circ$ opening angle cone from the first primary shower segment. \\
        \bottomrule
    \end{tabular}
    \caption[LOL3 $\nu_e$CC variables]{LOL3 $\nu_e$CC variables. Description of $\nu_e$CC inclusive selection variables related to the ``low energy overlap three'' tagger. The full code defining these is available at \url{https://github.com/BNLIF/wire-cell-pid/blob/master/src/NeutrinoID_nue_tagger.h}.}
\end{table}

\begin{table}[H]
    \centering
    \begin{tabular}{l p{10cm}}
        \toprule
        \textbf{Variable} & \textbf{Description} \\
        \midrule
        \texttt{spt\_shower\_main\_length} & Summed length of all primary shower segments connected to the neutrino vertex in cm. \\
        \hline
        \texttt{spt\_shower\_total\_length} &  Summed length of all primary shower segments in cm. \\
        \hline
        \texttt{spt\_angle\_beam} & Angle between the direction of the primary shower and the beam direction in degrees. \\
        \hline
        \texttt{spt\_angle\_vertical} & Angle between the direction of the primary shower and the vertical direction in degrees. \\
        \hline
        \texttt{spt\_max\_dQ\_dx} & Maximum $dQ/dx$ in the shower stem. \\
        \hline
        \texttt{spt\_angle\_beam\_1} & Angle between the direction of the primary shower calculated using the first 15 cm of the shower stem and the beam direction in degrees. \\
        \hline
        \texttt{spt\_angle\_drift} & Complementary angle of the angle between the direction of the primary shower and the drift direction. \\
        \hline
        \texttt{spt\_angle\_drift\_1} & Complementary angle of the angle between the direction of the primary shower calculated using the first 15 cm of the shower stem and the drift direction. \\
        \hline
        \texttt{spt\_num\_valid\_tracks} & Number of track-like segments attached to the neutrino vertex.  \\
        \hline
        \texttt{spt\_n\_vtx\_segs} & Number of segments attached to the neutrino vertex.  \\
        \hline
        \texttt{spt\_max\_length} & Maximum length of track-like segments attached to the neutrino vertex. \\
        \bottomrule
    \end{tabular}
    \caption[SPT $\nu_e$CC variables]{SPT $\nu_e$CC variables. Description of $\nu_e$CC inclusive selection variables related to the ``shower pattern'' tagger. The full code defining these is available at \url{https://github.com/BNLIF/wire-cell-pid/blob/master/src/NeutrinoID_nue_tagger.h}.}
\end{table}

\begin{table}[H]
    \centering
    \begin{tabular}{l p{10cm}}
        \toprule
        \textbf{Variable} & \textbf{Description} \\
        \midrule
        \texttt{kine\_reco\_add\_energy} & Additional energy added to the total reconstructed neutrino energy, including 8.6 MeV per proton and the particle mass for each pion, muon, and electron. \\
        \hline
        \texttt{kine\_pio\_mass} & Reconstructed invariant mass of two showers from a $\pi^0$ using the shower energies $E_1$, $E_2$ and the opening angle between the two $\theta$ using the formula $m=\sqrt{4 E_1 E_2 \sin^2\theta/2}$. \\
        \hline
        \texttt{kine\_pio\_flag} & This value is 0 if a $\pi^0$ is not reconstructed, 1 if a $\pi^0$ is identified with visible activity at the neutrino vertex, and 2 if a $\pi^0$ is identified with no visible activity at the neutrino vertex. \\
        \hline
        \texttt{kine\_pio\_vtx\_dis} & Distance between the $\pi^0$ vertex and the neutrino vertex. \\
        \hline
        \texttt{kine\_pio\_energy\_1} & Energy of the first $\pi^0$ photon shower in MeV. \\
        \hline
        \texttt{kine\_pio\_theta\_1} & Angle with respect to the beam for the first $\pi^0$ photon shower in degrees. \\
        \hline
        \texttt{kine\_pio\_phi\_1} & Azimuthal angle with respect to the beam for the first $\pi^0$ photon shower in degrees. \\
        \hline
        \texttt{kine\_pio\_dis\_1} & Distance between the $\pi^0$ vertex and the first photon shower vertex in MeV. \\
        \hline
        \texttt{kine\_pio\_energy\_2} & Energy of the second $\pi^0$ photon shower in MeV. \\
        \hline
        \texttt{kine\_pio\_theta\_2} & Angle with respect to the beam for the second $\pi^0$ photon shower in degrees. \\
        \hline
        \texttt{kine\_pio\_phi\_2} & Azimuthal angle with respect to the beam for the second $\pi^0$ photon shower in degrees. \\
        \hline
        \texttt{kine\_pio\_dis\_2} & Distance between the $\pi^0$ vertex and the second photon shower vertex in MeV. \\
        \hline
        \texttt{kine\_pio\_angle} & Opening angle between the two $\pi^0$ photon showers. \\
        \bottomrule
    \end{tabular}
    \caption[KINE $\nu_e$CC variables]{KINE $\nu_e$CC variables. Description of $\nu_e$CC inclusive selection variables related to kinematic reconstructions. The full code defining these is available at \url{https://github.com/BNLIF/wire-cell-pid/blob/master/src/NeutrinoID_kine.h} and \url{https://github.com/BNLIF/wire-cell-pid/blob/master/src/NeutrinoID_shower_clustering.h}.}
\end{table}

\begin{table}[H]
    \centering
    \begin{tabular}{l p{9.5cm}}
        \toprule
        \textbf{Variable} & \textbf{Description} \\
        \midrule
        \texttt{numu\_cc\_1\_particle\_type} & Vector variable of the PDG code for each segment attached to the neutrino vertex. \\
        \hline
        \texttt{numu\_cc\_1\_length} & Vector variable of the path length of each segment attached to the neutrino vertex in cm. \\
        \hline
        \texttt{numu\_cc\_1\_direct\_length} & Vector variable of the start to endpoint length of each segment attached to the neutrino vertex in cm. \\
        \hline
        \texttt{numu\_cc\_1\_medium\_dQ\_dx} & Vector variable of the median $dQ/dx$ of each segment attached to the neutrino vertex. \\
        \hline
        \texttt{numu\_cc\_1\_dQ\_dx\_cut} & Vector variable of $0.8866$ + $0.9533 \cdot (18\ \mathrm{cm}$ / $\mathrm{\texttt{segment\_length}})^{0.4234}$ for each segment attached to the neutrino vertex. \\
        \hline
        \texttt{numu\_cc\_1\_n\_daughter\_tracks} & Vector variable of the number of daughter tracks for each segment attached to the neutrino vertex. \\
        \hline
        \texttt{numu\_cc\_1\_n\_daughter\_all} & Vector variable of the number of daughter segments for each segment attached to the neutrino vertex. \\
        \hline
        \texttt{numu\_1\_score} & Maximum score of the \texttt{numu\_1} BDT across all entries in the vector. \\
        \bottomrule
    \end{tabular}
    \caption[NUMUCC1 $\nu_\mu$CC variables]{NUMUCC1 $\nu_\mu$CC variables. Description of $\nu_\mu$CC inclusive selection variables related to the ``$\nu_\mu$CC one'' tagger. The full code defining these is available at \url{https://github.com/BNLIF/wire-cell-pid/blob/master/src/NeutrinoID_numu_tagger.h}.}
\end{table}

\begin{table}[H]
    \centering
    \begin{tabular}{l p{9cm}}
        \toprule
        \textbf{Variable} & \textbf{Description} \\
        \midrule
        \texttt{numu\_cc\_2\_length} & Vector variable of the summed length of all track-like segments in each shower in cm. \\
        \hline
        \texttt{numu\_cc\_2\_total\_length} & Vector variable of the summed length of all segments in each shower in cm. \\
        \hline
        \texttt{numu\_cc\_2\_n\_daughter\_tracks} & Vector variable of the number of daughter tracks for each shower. \\
        \hline
        \texttt{numu\_cc\_2\_n\_daughter\_all} & Vector variable of the number of daughter segmentns for each shower. \\
        \hline
        \texttt{numu\_2\_score} & Maximum score of the \texttt{numu\_2} BDT across all entries in the vector. \\
        \bottomrule
    \end{tabular}
    \caption[NUMUCC2 $\nu_\mu$CC variables]{NUMUCC2 $\nu_\mu$CC variables. Description of $\nu_\mu$CC inclusive selection variables related to the ``$\nu_\mu$CC two'' tagger. The full code defining these is available at \url{https://github.com/BNLIF/wire-cell-pid/blob/master/src/NeutrinoID_numu_tagger.h}.}
\end{table}

\begin{table}[H]
    \centering
    \begin{tabular}{l p{9cm}}
        \toprule
        \textbf{Variable} & \textbf{Description} \\
        \midrule
        \texttt{numu\_cc\_3\_particle\_type} & PDG code for the longest muon candidate track. \\
        \hline
        \texttt{numu\_cc\_3\_max\_length} & Length of the longest muon candidate track in cm. \\
        \hline
        \texttt{numu\_cc\_3\_track\_length} & Summed length of all tracks connected to the neutrino vertex in cm. \\
        \hline
        \texttt{numu\_cc\_3\_max\_length\_all} & Length of the longest track in cm. \\
        \hline
        \texttt{numu\_cc\_3\_max\_muon\_length} & Length of the longest muon candidate when considering possible broken muon tracks in cm. \\
        \hline
        \texttt{numu\_cc\_3\_n\_daughter\_tracks} & Number of daughter tracks for the muon candidate. \\
        \hline
        \texttt{numu\_cc\_3\_n\_daughter\_all} & Number of daughter segments for the muon candidate. \\
        \hline
        \texttt{numu\_cc\_flag\_3} & Indicates if the above variables indicate the likely presence of a real muon track. \\
        \bottomrule
    \end{tabular}
    \caption[NUMUCC3 $\nu_\mu$CC variables]{NUMUCC3 $\nu_\mu$CC variables. Description of $\nu_\mu$CC inclusive selection variables related to the ``$\nu_\mu$CC three'' tagger. The full code defining these is available at \url{https://github.com/BNLIF/wire-cell-pid/blob/master/src/NeutrinoID_numu_tagger.h}.}
\end{table}

\begin{table}[H]
    \centering
    \begin{tabular}{l p{9cm}}
        \toprule
        \textbf{Variable} & \textbf{Description} \\
        \midrule
        \texttt{cosmict\_2\_filled} & Indicates if COSMICT2 variables have been filled. \\
        \hline
        \texttt{cosmict\_2\_particle\_type} & The PDG code of the candidate muon track. \\
        \hline
        \texttt{cosmict\_2\_n\_muon\_tracks} & The number of daughter tracks coming from the candidate muon track. \\
        \hline
        \texttt{cosmict\_2\_total\_shower\_length} & Summed length of candidate muon daughter particles. \\
        \hline
        \texttt{cosmict\_2\_flag\_inside} & Indicates if the candidate muon is contained within the fiducial volume. \\
        \hline
        \texttt{cosmict\_2\_angle\_beam} & The angle between the candidate muon and the beam direction in degrees. \\
        \hline
        \texttt{cosmict\_2\_flag\_dir\_weak} & Indicates if the candidate muon has uncertain direction reconstruction. \\
        \hline
        \texttt{cosmict\_2\_dQ\_dx\_end} & $dQ/dx$ at the end of the candidate muon track. \\
        \hline
        \texttt{cosmict\_2\_dQ\_dx\_front} & $dQ/dx$ at the front of the candidate muon track. \\
        \hline
        \texttt{cosmict\_2\_theta} & Polar angle of the candidate muon with respect to the beam direction in degrees. \\
        \hline
        \texttt{cosmict\_2\_phi} & Azimuthal angle of the candidate muon with respect to the beam direction in degrees. \\
        \hline
        \texttt{cosmict\_2\_valid\_tracks} & Number of tracks attached to the neutrino vertex. \\
        \hline
        \texttt{cosmict\_flag\_2} & Indicates if the above variables indicate the likely presence of a cosmic ray muon track. \\
        \bottomrule
    \end{tabular}
    \caption[COSMICT2 $\nu_\mu$CC variables]{COSMICT2 $\nu_\mu$CC variables. Description of $\nu_\mu$CC inclusive selection variables related to the ``cosmic two'' tagger. The full code defining these is available at \url{https://github.com/BNLIF/wire-cell-pid/blob/master/src/NeutrinoID_cosmic_tagger.h}.}
\end{table}

\begin{table}[H]
    \centering
    \begin{tabular}{l p{9cm}}
        \toprule
        \textbf{Variable} & \textbf{Description} \\
        \midrule
        \texttt{cosmict\_3\_filled} & Indicates if COSMICT3 variables have been filled. \\
        \hline
        \texttt{cosmict\_3\_flag\_inside} & Indicates if the endpoint of the last segment in a split-segment muon candidate is contained within the fiducial volume. \\
        \hline
        \texttt{cosmict\_3\_angle\_beam} & Angle between the candidate muon direction and the beam direction in degrees. \\
        \hline
        \texttt{cosmict\_3\_flag\_dir\_weak} & Indicates if the last segment in a split-segment muon candidate has uncertain direction reconstruction. \\
        \hline
        \texttt{cosmict\_3\_dQ\_dx\_end} & $dQ/dx$ at the end of the candidate muon track. \\
        \hline
        \texttt{cosmict\_3\_dQ\_dx\_front} & $dQ/dx$ at the start of the candidate muon track. \\
        \hline
        \texttt{cosmict\_3\_theta} & Polar angle of the candidate muon with respect to the beam direction in degrees. \\
        \hline
        \texttt{cosmict\_3\_phi} & Azimuthal angle of the candidate muon with respect to the beam direction in degrees. \\
        \hline
        \texttt{cosmict\_3\_valid\_tracks} & Number of tracks attached to the neutrino vertex. \\
        \hline
        \texttt{cosmict\_flag\_3} & Indicates if the above variables indicate the likely presence of a cosmic ray muon track. \\
        \bottomrule
    \end{tabular}
    \caption[COSMICT3 $\nu_\mu$CC variables]{COSMICT3 $\nu_\mu$CC variables. Description of $\nu_\mu$CC inclusive selection variables related to the ``cosmic three'' tagger. The full code defining these is available at \url{https://github.com/BNLIF/wire-cell-pid/blob/master/src/NeutrinoID_cosmic_tagger.h}.}
\end{table}

\begin{table}[H]
    \centering
    \begin{tabular}{l p{9cm}}
        \toprule
        \textbf{Variable} & \textbf{Description} \\
        \midrule
        \texttt{cosmict\_4\_filled} & Indicates if COSMICT4 variables have been filled. \\
        \hline
        \texttt{cosmict\_4\_flag\_inside} & Indicates if the endpoint of the candidate muon track is contained within the fiducial volume. \\
        \hline
        \texttt{cosmict\_4\_angle\_beam} & Angle between the candidate muon track direction and the beam direction in degrees. \\
        \hline
        \texttt{cosmict\_4\_connected\_showers} & Number of showers associated with the neutrino vertex. \\
        \hline
        \texttt{cosmict\_flag\_4} & Indicates if the above variables indicate the likely presence of a cosmic ray muon track. \\
        \bottomrule
    \end{tabular}
    \caption[COSMICT4 $\nu_\mu$CC variables]{COSMICT4 $\nu_\mu$CC variables. Description of $\nu_\mu$CC inclusive selection variables related to the ``cosmic four'' tagger. The full code defining these is available at \url{https://github.com/BNLIF/wire-cell-pid/blob/master/src/NeutrinoID_cosmic_tagger.h}.}
\end{table}

\begin{table}[H]
    \centering
    \begin{tabular}{l p{9cm}}
        \toprule
        \textbf{Variable} & \textbf{Description} \\
        \midrule
        \texttt{cosmict\_5\_filled} & Indicates if COSMICT5 variables have been filled. \\
        \hline
        \texttt{cosmict\_5\_flag\_inside} & Indicates if the endpoint of the last segment in a split-segment muon candidate is contained within the fiducial volume. \\
        \hline
        \texttt{cosmict\_5\_angle\_beam} & Angle between the candidate muon track direction and the beam direction in degrees. \\
        \hline
        \texttt{cosmict\_5\_connected\_showers} & Number of showers associated with the neutrino vertex. \\
        \hline
        \texttt{cosmict\_flag\_5} & Indicates if the above variables indicate the likely presence of a cosmic ray muon track. \\
        \bottomrule
    \end{tabular}
    \caption[COSMICT5 $\nu_\mu$CC variables]{COSMICT5 $\nu_\mu$CC variables. Description of $\nu_\mu$CC inclusive selection variables related to the ``cosmic five'' tagger. The full code defining these is available at \url{https://github.com/BNLIF/wire-cell-pid/blob/master/src/NeutrinoID_cosmic_tagger.h}.}
\end{table}

\begin{table}[H]
    \centering
    \begin{tabular}{l p{9cm}}
        \toprule
        \textbf{Variable} & \textbf{Description} \\
        \midrule
        \texttt{cosmict\_6\_filled} & Indicates if COSMICT6 variables have been filled. \\
        \hline
        \texttt{cosmict\_6\_flag\_dir\_weak} & Indicates if the 2nd longest muon candidate track has uncertain direction reconstruction. \\
        \hline
        \texttt{cosmict\_6\_flag\_inside} & Indicates if the 2nd longest muon candidate track endpoint is contained within the fiducial volume. \\
        \hline
        \texttt{cosmict\_6\_angle} & Angle between the 2nd longest candidate muon track direction and the longest candidate muon track direction in degrees. \\
        \hline
        \texttt{cosmict\_flag\_6} & Indicates if the above variables indicate the likely presence of the two candidate muon tracks coming from a single cosmic ray muon track. \\
        \bottomrule
    \end{tabular}
    \caption[COSMICT6 $\nu_\mu$CC variables]{COSMICT6 $\nu_\mu$CC variables. Description of $\nu_\mu$CC inclusive selection variables related to the ``cosmic six'' tagger. The full code defining these is available at \url{https://github.com/BNLIF/wire-cell-pid/blob/master/src/NeutrinoID_cosmic_tagger.h}.}
\end{table}

\begin{table}[H]
    \centering
    \begin{tabular}{l p{9cm}}
        \toprule
        \textbf{Variable} & \textbf{Description} \\
        \midrule
        \texttt{cosmict\_7\_filled} & Indicates if COSMICT7 variables have been filled. \\
        \hline
        \texttt{cosmict\_7\_flag\_sec} & Indicates if a 2nd longest candidate muon track is likely to be back-to-back with the longest candidate muon track. \\
        \hline
        \texttt{cosmict\_7\_n\_muon\_tracks} & Number of daughter tracks from the candidate muon. \\
        \hline
        \texttt{cosmict\_7\_total\_shower\_length} & Summed length of the daughter tracks from the candidate muon in cm. \\
        \hline
        \texttt{cosmict\_7\_flag\_inside} & Indicates if the candidate muon track is contained within the fiducial volume. \\
        \hline
        \texttt{cosmict\_7\_angle\_beam} & Angle between the candidate muon track direction and the beam direction in degrees. \\
        \hline
        \texttt{cosmict\_7\_flag\_dir\_weak} & Indicates if the candidate muon track has uncertain direction reconstruction. \\
        \hline
        \texttt{cosmict\_7\_dQ\_dx\_end} & $dQ/dx$ at the end of the candidate muon track. \\
        \hline
        \texttt{cosmict\_7\_dQ\_dx\_front} & $dQ/dx$ at the start of the candidate muon track. \\
        \hline
        \texttt{cosmict\_7\_theta} & Polar angle of the candidate muon with respect to the beam direction in degrees. \\
        \hline
        \texttt{cosmict\_7\_phi} & Azimuthal angle of the candidate muon with respect to the beam direction in degrees. \\
        \hline
        \texttt{cosmict\_flag\_7} & Indicates if the above variables indicate the likely presence of a cosmic ray muon track. \\
        \bottomrule
    \end{tabular}
    \caption[COSMICT7 $\nu_\mu$CC variables]{COSMICT7 $\nu_\mu$CC variables. Description of $\nu_\mu$CC inclusive selection variables related to the ``cosmic seven'' tagger. The full code defining these is available at \url{https://github.com/BNLIF/wire-cell-pid/blob/master/src/NeutrinoID_cosmic_tagger.h}.}
\end{table}

\begin{table}[H]
    \centering
    \begin{tabular}{l p{10cm}}
        \toprule
        \textbf{Variable} & \textbf{Description} \\
        \midrule
        \texttt{cosmict\_8\_filled} & Indicates if COSMICT8 variables have been filled. \\
        \hline
        \texttt{cosmict\_8\_flag\_out} & Indicates if the candidate muon track is not contained within the fiducial volume. \\
        \hline
        \texttt{cosmict\_8\_muon\_length} & Length of the candidate muon track in cm. \\
        \hline
        \texttt{cosmict\_8\_acc\_length} & Summed length of all tracks attached to the neutrino vertex in cm. \\
        \hline
        \texttt{cosmict\_flag\_8} & Indicates if the above variables indicate the likely presence of a cosmic ray muon track. \\
        \bottomrule
    \end{tabular}
    \caption[COSMICT8 $\nu_\mu$CC variables]{COSMICT8 $\nu_\mu$CC variables. Description of $\nu_\mu$CC inclusive selection variables related to the ``cosmic eight'' tagger. The full code defining these is available at \url{https://github.com/BNLIF/wire-cell-pid/blob/master/src/NeutrinoID_cosmic_tagger.h}.}
\end{table}

\begin{table}[H]
    \centering
    \begin{tabular}{l p{10cm}}
        \toprule
        \textbf{Variable} & \textbf{Description} \\
        \midrule
        \texttt{cosmict\_10\_vtx\_z} & Vector variable of the z coordinate of the endpoint of each segment attached to the neutrino vertex in cm. \\
        \hline
        \texttt{cosmict\_10\_flag\_shower} & Vector variable indicating if each segment attached to the neutrino vertex is identified as part of a shower. \\
        \hline
        \texttt{cosmict\_10\_flag\_dir\_weak} & Vector variable indicating if each segment attached to the neutrino vertex has uncertain direction information. \\
        \hline
        \texttt{cosmict\_10\_angle\_beam} & Vector variable of the angle between the beam direction and the direction of each segment attached to the neutrino vertex. \\
        \hline
        \texttt{cosmict\_10\_length} &  Vector variable of the length of each segment attached to the neutrino vertex in cm. \\
        \hline
        \texttt{cosmict\_10\_score} & Minimum score of the \texttt{numu\_2} BDT across all entries in the vector. \\
        \bottomrule
    \end{tabular}
    \caption[COSMICT10 $\nu_\mu$CC variables]{COSMICT10 $\nu_\mu$CC variables. Description of $\nu_\mu$CC inclusive selection variables related to the ``cosmic ten'' tagger. The full code defining these is available at \url{https://github.com/BNLIF/wire-cell-pid/blob/master/src/NeutrinoID_cosmic_tagger.h}.}
\end{table}

\begin{table}[H]
    \centering
    \begin{tabular}{l p{12cm}}
        \toprule
        \textbf{Variable} & \textbf{Description} \\
        \midrule
        \texttt{cosmict\_flag\_1} & Indicates if the neutrino vertex is outside the fiducial volume. \\
        \hline
        \texttt{cosmict\_flag\_9} & Indicates if an event is likely to have a cosmic ray origin using topological, geometric, and energy-based information. \\
        \hline
        \texttt{cosmic\_filled} & Indicates if cosmic ray tagging variables have been filled. \\
        \hline
        \texttt{cosmic\_flag} & Indicates if an event was initially suspected as a cosmic ray, but then this decision was reversed after certain features indicated that it is potentially a $\nu_e$CC candidate. \\
        \hline
        \texttt{cosmict\_flag} & Indicates if any of \texttt{cosmict\_flag\_1}, \texttt{cosmict\_flag\_2}, \texttt{cosmict\_flag\_3}, \texttt{cosmict\_flag\_4}, \texttt{cosmict\_flag\_5}, \texttt{cosmict\_flag\_6}, \texttt{cosmict\_flag\_7}, \texttt{cosmict\_flag\_8}, or \texttt{cosmict\_flag\_9} are true, indicating a likely cosmic origin for the event. \\
        \hline
        \texttt{numu\_cc\_flag} & A value of -1 indicates that an event has failed Wire-Cell generic neutrino selection, a value of 0 indicates that an event has failed the cut-based Wire-Cell $\nu_\mu$CC selection, and a value of 1 indicates that an event has passed the cut-based Wire-Cell $\nu_\mu$CC selection. \\
        \bottomrule
    \end{tabular}
    \caption[Other NUMUCC and COSMIC $\nu_\mu$CC variables]{Other NUMUCC and COSMIC $\nu_\mu$CC variables. Description of other $\nu_\mu$CC inclusive selection variables related to the other cosmic and $\nu_\mu$CC taggers. The full code defining these is available at \url{https://github.com/BNLIF/wire-cell-pid/blob/master/src/NeutrinoID_cosmic_tagger.h} and \url{https://github.com/BNLIF/wire-cell-pid/blob/master/src/NeutrinoID_numu_tagger.h}.}
\end{table}